**Future Circular Collider**

**Feasibility Study Report**

# Volume 2

# Accelerators, Technical Infrastructure and Safety

**March 31, 2025**

Submitted to the European Physics Journal ST, a joint publication of EDP Sciences, Springer Science+Business Media, and the Società Italiana di Fisica.

## Note from the Editors

One of the recommendations of the 2020 update of the European Strategy for Particle Physics was that "Europe, together with its international partners, should investigate the technical and financial feasibility of a future hadron collider at CERN with a centre-of-mass energy of at least 100 TeV and with an electron-positron Higgs and electroweak factory as a possible first stage.

In June 2021, the CERN Council launched the FCC Feasibility Study to be completed by 2025, in time for the next update of the European Strategy for Particle Physics. The study results are made publicly available through this FCC Feasibility Study Report, as input to the European Particle Physics Strategy update process, initiated by the CERN Council in March 2024. The studies presented in this FCC Feasibility Study Report do not imply any commitment by the CERN Member or Associate Member States to build the Future Circular Collider.

This report and the assumptions contained in it do not prejudge further territorial feasibility analysis by the Host States, France and Switzerland, as well as the outcome of their respective public debate and concertation processes, and future decisions of their relevant authorities.

## Acknowledgements

We would like to thank the **International Steering Committee members:**

| |
|---|
| F. Gianotti (Chair), CERN |
| R. Bello, CERN |
| P. Chomaz, CEA, France |
| M. Cobal, INFN and University of Udine, Italy |
| B. Heinemann, DESY, Germany |
| T. Koseki, KEK, Japan |
| M. Lamont, CERN |
| L. Merminga, FNAL, United States |
| J. Mnich, CERN |
| M. Seidel, PSI and EPFL, Switzerland |
| C. Warakaulle, CERN |

and the **Scientific Advisory Committee members:**

| |
|---|
| A. Parker (Chair), Cambridge University, UK |
| R. Bartolini, DESY, Germany |
| A. Chabert, SFTRF, France |
| H. Ehrbar, Heinz Ehrbar Partners LLC, Switzerland |
| B. Gavela Legazpi, UAM Madrid, Spain |
| G. Hiller, TU Dortmund, Germany |
| S. Krishnagopal, FNAL, U.S. |
| P. Križan, University of Ljubljana, Slovenia |
| P. Lebrun, ESI, France |
| P. McIntosh, STFC, ASTeC, UKRI, UK |
| M. Minty, BNL, U.S. |
| R. Tenchini, INFN Sezione di Pisa, Italy |

for their continued guidance and careful reviewing that helped to complete this report successfully.

The research carried out by the international FCC collaboration hosted by CERN, which led to this publication, has received funding from the European Union's Horizon 2020 research and innovation programme under the grant numbers 951754 (FCCIS), 654305 (EuroCirCol), 764879 (EASITrain), 730871 (ARIES), 777563 (RI-Paths), 101086276 (EAJADE), 101004730 (iFAST), 101131435 (iSAS), 101131850 (RF2.0) and from FP7 under grant number 312453 (EuCARD-2).

This work has also benefited from the support of CHART (Swiss Accelerator Research and Technology, founded in 2016 as an umbrella collaboration for accelerator research and technology activities. Present partners in CHART are CERN, PSI, EPFL, ETH-Zurich and the University of Geneva.

## List of Editors at 31 March 2025

M. Benedikt[1] (Study Leader), F. Zimmermann[1] (Deputy Study Leader), B. Auchmann[1,2], W. Bartmann[1], J.P. Burnet[1], C. Carli[1], A. Chancé[3], P. Craievich[2], M. Giovannozzi[1], C. Grojean[4,5], J. Gutleber[1], K. Hanke[1], A. Henriques[1], P. Janot[1], C. Lourenço[1], M. Mangano[1], T. Otto[1], J. Poole[1], S. Rajagopalan[6], T. Raubenheimer[7], E. Todesco[1], L. Ulrici[1], T. Watson[1], G. Wilkinson[1,8].

## List of Contributors at 31 March 2025

A. Abada[9,10,11], M. Abbrescia[12,13], H. Abdolmaleki[14,15], S.H. Abidi[6], A. Abramov[1], C. Adam[9,16,17], M. Ady[1], P.R. Adžić[18], I. Agapov[4], D. Aguglia[1], I. Ahmed[19], M. Aiba[2], G. Aielli[20,21], T. Akan[22], N. Akchurin[23], D. Akturk[24], M. Al-Thakeel[1,25,26], G.L. Alberghi[25], J. Alcaraz Maestre[27], M. Aleksa[1], R. Aleksan[3], F. Alharthi[9,10,28], J. Alimena[4], A. Alimenti[29], S. Alioli[30,31], L. Alix[1,9,16], B.C. Allanach[32], L. Allwicher[4], A.A. Altintas[33], M. Altınlı[33,34], M. Alviggi[35,36], G. Ambrosio[37], Y. Amhis[9,10,11], A. Amiri[38,39], G. Ammirabile[40], T. Andeen[41], K.D.J. André[1], J. Andrea[9,42,43], A. Andreazza[44,45], M. Andreini[1], T. Andriollo[46], L. Angel[47], M. Angelucci[48], S. Antusch[49], M.N. Anwar[12,50], L. Apolinário[51], G. Apollinari[37], R.B. Appleby[52,53], A. Apresyan[37], Aram Apyan[54], Armen Apyan[55], A. Arbey[9,56,57], B. Argiento[35,36], V. Ari[58], S. Arias[59], B. Arias Alonso[1], O. Arnaez[9,16,17], R. Arnaldi[60], F. Arneodo[61], H. Arnold[62], P. Arrutia Sota[1], M.E. Ascioti[63,64], K.A. Assamagan[6], S. Aumiller[65], G. Aydın[66], K. Azizi[38,67], P. Azzi[68], N. Bacchetta[68], A. Bacci[44], B. Bai[69], Y. Bai[70], L. Balconi[44,45], G. Baldinelli[63,64], B. Balhan[1], A.H. Ball[1,71], A. Ballarino[1], S. Banerjee[72], S. Banik[2,73], D.P. Barber[4,74], M.B. Barbero[9,75,76], D. Barducci[40,77], D. Barna[78], G.G. Barnaföldi[78], M.J. Barnes[1], A.J. Barr[8], R. Bartek[79], H. Bartosik[1], S.A. Bass[80], U. Bassler[9,81,82], M.J. Basso[83,84], A. Bastianin[45,85], P. Bataillard[86], M. Battistin[1], J. Bauche[1], L. Baudin[1], J. Baudot[9,42,43], B. Baudouy[3], L. Bauerdick[37], C. Bayındır[87,88], H.P. Beck[89], F. Bedeschi[40], C. Bee[62], M. Begel[6], M. Behtouei[48], L. Bellagamba[25], N. Bellegarde[1], E. Belli[1,90], E. Bellingeri[91], S. Belomestnykh[37], A.D. Benaglia[30], G. Bencivenni[48], J. Bendavid[1], M. Benmergui[92], M. Benoit[93], D. Benvenuti[1,40], T. Bergauer[94], N. Bernachot[95], G. Bernardi[9,96,97], J. Bernardi[98], Q. Berthet[99,100,101], S. Bertoni[102], C. Bertulani[103], M.I. Besana[2], A. Besson[9,42,43], M. Bettelini[104], S. Bettoni[2], S. Beuvier[†105], P.C. Bhat[37], S. Bhattacharya[106], J. Bhom[107], M.E. Biagini[48], A. Bibet-Chevalier[108], M. Bicrel[109], M. Biglietti[110], G.M. Bilei[63], B. Bilki[111,112], K. Bisgaard Christensen[1], T. Biswas[113], F. Blanc[114], F. Blekman[4,115,116], A. Blondel[9,101,117], J. Blümlein[4], D. Boccanfuso[35,118], A. Bogomyagkov[119], P. Boillon[108], P. Boivin[100], M.J. Boland[120], S. Bologna[121], O. Bolukbasi[33], R. Bonnet[102], J. Borburgh[1], F. Bordry[1], P. Borges de Sousa[1], G. Borghello[1], L. Borriello[35], D. Bortoletto[8], M. Boscolo[48], L. Bottura[1], V. Boudry[9,81,82], R. Boughezal[122], D. Bourilkov[123], M. Boyd[83,124], D. Boye[6], G. Bozzi[125,126], V. Braccini[91], C. Bracco[1], B. Bradu[1], A. Braghieri[127], S. Braibant[25,26], J. Bramante[128], G.C. Branco[129], R. Brenner[130], N. Brisa[102], D. Britzger[131], G. Broggi[1,90], L. Bromiley[1], E. Brost[6], Q. Bruant[3], R. Bruce[1], E. Bründermann[132], L. Brunetti[9,16,17], O. Brüning[1], O. Brunner[1], X. Buffat[1], E. Bulyak[133], A. Burdyko[44,134], H. Burkhardt[1,135], P.N. Burrows[136], S. Busatto[44,90], S. Buschaert[86], D. Buttazzo[40], A. Butterworth[1], D. Butti[1], G. Cacciapaglia[137,138,139], Y. Cai[7], B. Caiffi[140], V. Cairo[1], O. Cakir[58], P. Calafiura[141], R. Calaga[1], S. Calatroni[1], D.G. Caldwell[142], A. Çalışkan[143], C. Calpini[144], M. Calviani[1], E. Camacho-Pérez[145], P. Camarri[20,21], L. Caminada[2,73], M. Campajola[35,36], A.C. Canbay[58], K. Canderan[1], S. Candido[1], F. Canelli[73], A. Canepa[37], S. Cantarella[48], K.B. Cantún-Avila[145], L. Capriotti[146,147], A. Caram[148], A. Carbone[44], J.M. Carceller[1], G. Carini[6], F. Carlier[1], C.M. Carloni Calame[127], F. Carra[1], C. Cartannaz[86], S. Casenove[1], G. Catalano[149], V. Cavaliere[6], C. Cazzaniga[150], C. Cecchi[63,64], F.G. Celiberto[151], M. Cepeda[27], F. Cerutti[1], F. Cetorelli[30,31], G. Chachamis[51], Y. Chae[4], F. Chagnet[152], I. Chaikovska[9,10,11], M. Chalhoub[86], M. Chamizo-Llatas[6], M. Champagne[153], H. Chanal[9,154,155], G. Chapelier[108], P. Charitos[1], C. Charles[105], T.K. Charles[156], C. Charlot[9,81,82], S. Chatterjee[4], A. Chaudhuri[157], R. Chehab[9,10,11], S.V. Chekanov[158], H. Chen[6], T. Chesne[105], F. Chiapponi[25,26], G. Chiarello[159,160], M. Chiesa[127], P. Chiggiato[1], Ph. Chomaz[3], M. Chorowski[161], J.P. Chou[162],

M. Chrzaszcz[107], W. Chung[163], S. Ciarlantini[68,164], A. Ciarma[48], D. Cieri[131], A.K. Ciftci[165], R. Ciftci[166], R. Cimino[48], F. Cirotto[35,36], M. Ciuchini[110], M. Cobal[167,168], A. Coccaro[140], R. Coelho Lopes De Sa[169], J.A. Coleman-Smith[1], F. Collamati[170], C. Colldelram[171], P. Collier[1], P. Collins[1], J. Collot[9,172,173], M. Colmenero[1], L. Colnot[149], G. Coloretti[73], E. Conte[9,42,43], F.A. Conventi[35,174], A. Cook[1], L. Cooley[175,176], A.S. Cornell[177], C. Cornella[1], G. Cornette[105], I. Corredoira[178], P. Costa Pinto[1], F. Couderc[3], J. Coupard[1], S. Coussy[86], R. Crescenzi[179], I. Crespo Garrido[1,180], T. Critchley[1,101], A. Crivellin[73], T. Croci[63], C. Cudré[105], G. Cummings[37], F. Cuna[12], R. Cunningham[1], B. Curé[1], E. Curtis[181], M. D'Alfonso[182], L. D'Aloia Schwartzentruber[183], G. D'Amen[6], B. D'Anzi[12,13], A. D'Avanzo[35,36], D. d'Enterria[1], A. D'Onofrio[35], M. D'Onofrio[184], M. Da Col[149], M. Da Rocha Rolo[60], C. Dachauer[185], B. Dağli[24], A. Dainese[68], B. Dalena[3], W. Dallapiazza[186], M. Dam[187], H. Damerau[1], V. Dao[62], A. Das[188], M.S. Daugaard[1], S. Dauphin[108], A. David[1], T. Davídek[189], G.J. Davies[181], S. Dawson[6], J. de Blas[190], A. de Cosa[150], S. De Curtis[191], N. De Filippis[12,50], E. De Lucia[48], R. De Maria[1], E. De Matteis[44], A. De Roeck[1], A. De Santis[48], A. De Vita[1,68,164], A. Deandrea[9,56,57], C.J. Debono[192], M. Deeb[100], M.M. Defranchis[1], J. Degens[184], S. Deghaye[1], V. Del Duca[48], C.L. Del Pio[6], A. Del Vecchio[90], D. Delikaris[1], A. Dell'Acqua[1], M. Della Pietra[35,36], M. Delmastro[9,16,17], L. Delprat[1], E. Delugas[149], Z. Demiragli[193], L. Deniau[1], D. Denisov[6], H. Denizli[194], A. Denner[195], A. Denot[108], G. Deptuch[6], A. Desai[196], H. Deveci[1], A. Di Canto[6], A. Di Ciaccio[20,21], L. Di Ciaccio[9,16,17], D. Di Croce[1,114], C. Di Fraia[35,36], B. Di Micco[29,110], R. Di Nardo[29,110], T.B. Dingley[8], F. Djama[9,75,76], F. Djurabekova[197], D. Dockery[37], S. Doebert[1], D. Domange[1,198], M. Donegà[150], U. Dosselli[68], H.A. Dostmann[1,199], J.A. Dragovich[37], I. Drebot[44], M. Drewes[200], T.A. du Pree[201], Z. Duan[202], C. Duarte-Galvan[203], O. Duboc[204], M. Duda[2], P. Duda[161], H. Duran Yildiz[58], H. Durand[105], P. Durand[105], G. Durieux[200], Y. Dutheil[1], I. Dutta[37], J.S. Dutta[205], S. Dutta[206], F. Duval[1], F. Eder[1], M. Eisterer[98], Z. El Bitar[9,42,43], A. El Saied[207], M. Elisei[44], J. Ellis[1,208], W. Elmetenawee[12], J. Elmsheuser[6], V. Daniel Elvira[37], S.C. Eno[209], Y. Enomoto[210], B.A. Erdelyi[68,164], O.E. Eruteya[101,211], M. Escobar[212], O. Etisken[213], I. Eymard[144], J. Eysermans[182], D. Falchieri[25], C. Falkenberg[204], F. Fallavollita[1,131], A. Afalou[1,9,10], J. Faltova[189], J. Fanini[1], L. Fanò[63,64], K. Fanti[105], R. Farinelli[25], M. Farino[163], S. Farinon[140], H. Fatehi[38], J. Fatterbert[105], A. Faure[214], A. Faus-Golfe[9,10,11], G. Favia[1], L. Favilla[35,118], W.J. Fawcett[32], A. Federowicz[37], L. Feligioni[9,75,76], L. Felsberger[1], Y. Feng[23], A. Fernández Téllez[215], R. Ferrari[127], L. Ferreira[1], F. Ferro[140], M. Fiascaris[1], C. Fiorio[45], S.A. Fleury[1], L. Florez[186], M. Florio[45,149], A. Fondacci[63], B. Fontimpe[212], K. Foraz[1], R. Fortunati[2], M. Fouaidy[9,10,11], A. Foussat[1], A. Fowler[1], J.D. Fox[216], M. Francesconi[35], B. Francois[1], R. Franqueira Ximenes[1], F. Fransesini[48], A. Frasca[1,184], A. Freitas[217], J.A. Frost[8], K. Furukawa[210], A. Gabrielli[25,26], A. Gaddi[1], F. Gaede[4], A. Gallén[130], R. Galler[218,219], E. Gallice[105], E. Gallo[4,115], H. Gamper[1], G. Ganis[1], S. Ganjour[3], S. Gao[6], A. Garand[148], C. Garaus[204], D. Garcia[1], R. García Alía[1], R. García Gil[220], C.M. Garcia Jaimes[1,114], H. Garcia Rodrigues[2,221], C. Garion[1], M. Garlaschè[1], D. Garnier[152], M.V. Garzelli[115], S. Gascon-Shotkin[9,56,57], M. Gasior[1], G. Gaudino[35,118], G. Gaudio[127], V. Gaur[222], K. Gautam[73,116], V. Gawas[1], T. Gehrmann[73], A. Gehrmann-De Ridder[73,150], K. Geiger[1], M. Genco[149], F. Gerigk[1], H. Gerwig[1], A. Ghribi[1,9,223], P. Giacomelli[25], S. Giagu[90,170], E. Gianfelice[37], S. Giappichini[132], D. Gibellieri[1,224], F. Giffoni[149], G. Gil da Silveira[225], S.S. Gilardoni[1], M. Giovannetti[48], T. Girardet[105], S. Girod[1,105], P. Giubellino[60], P. Giubilato[68,164], F. Giuli[20,21], M. Giuliani[102], E.L. Gkougkousis[1,73], S. Glukhov[226], J. Gluza[227], B. Goddard[1], C. Goffing[1,132], D. Goldsworthy[1], T. Golling[101], R. Gonçalo[51,228], V.P. Gonçalves[47,229], T. Gonçalves Da Silva[212], J. Gonski[7], R. Gonzalez Suarez[130], S. Gorgi Zadeh[1], S. Gori[230], E. Gorini[159,231], L. Gouskos[232], M. Gouzevitch[9,56,57], E. Granados[1], F. Grancagnolo[159], S. Grancagnolo[159,231], A. Grassellino[37], A. Grau[132], E. Graverini[40,77,114], F.G. Gravili[159,231], H.M. Gray[141,233], M. Grazzini[73], Mario Greco[29,110], Michela Greco[60,234], A. Greljo[49], J-L. Grenard[1], A.V. Gritsan[235], R. Gröber[68,164], A. Grudiev[1], E. Gschwendtner[1], J. Gu[236], D. Guadagnoli[17,137,237], G. Guerrieri[1], A. Guiavarch[207], G. Guillermo Canton[1,238], M. Guinchard[1], Y.O. Günaydin[239], K. Gurcel[92], L.X. Gutierrez Guerrero[240,241], D. Gutiérrez Rueda[1], A. Gutiérrez-Rodríguez[242], V. Guzey[197,243], C. Haber[141], T. Hacheney[244], B. Hacışahinoğlu[33],

K. Hahn[122], J. Hajer[129], T. Hakulinen[1], J.C. Hammersley[245], M. Hance[230], J.B. Hansen[187], B. Härer[132], E. Hauzinger[218], M. Haviernik[189], B. Hegner[1], C. Helsens[114], Ana Henriques[1], C. Hernalsteens[1], H. Hernández-Arellano[215], R.J. Hernández-Pinto[203], M.A. Hernández-Ruíz[242], J. Hernández-Sánchez[215], J.W. Heron[1], L.M. Herrmann[1], R. Hirosky[246], J.F. Hirschauer[37], J.D. Hobbs[62], K. Hock[6], S. Höche[37], M. Hofer[1], G. Hoffstaetter[6,247], W. Höfle[1], M. Hohlmann[248], F. Holdener[249], B. Holzer[1], C.G. Honorato[215], H. Hoorani[250], A. Houver[105], E. Howling[1,8,136], X. Huang[7], F. Hug[251], B. Humann[1], P. Hunchak[120], Y. Husein[1], A. Hussain[1,252], G. Iadarola[1], G. Iakovidis[6], G. Iaselli[12,50], P. Iengo[35], A. Ilg[73], M. Iodice[110], A.O.M. Iorio[35,36], V. Ippolito[170], U. Iriso[171], J. Isaacson[37], G. Isidori[73], R. Islam[253], A. Istepanyan[105], S. Izquierdo Bermudez[1], V. Izzo[35], P.D. Jackson[196], R. Jafari[1,38], S.S. Jagabathuni[1,101], S. Jana[254,255], C. Järmyr Eriksson[1], P. Jausserand[152], M. Jensen[256], J.M. Jimenez[1], F.R. Joaquim[129], O.R. Jones[1], J. Joos[108], E. Jourd'huy[9,257], E. Jourdan[212], J.M. Jowett[1,258], A. Jueid[259], A.W. Jung[205], M. Kagan[7], I. Kahraman[58], V. Kain[1], J. Kalinowski[260], J.F. Kamenik[261,262], A. Kanso[263], T. Kar[264], S.O. Kara[265], H. Karadeniz[266], S.R. Karmarkar[205], V. Karpati[267], I. Karpov[1], M. Karppinen[1], P. Karst[9,75,76], S. Kartal[33], V.V. Kashikhin[37], U. Kaya[58], A. Kehagias[1,268], J. Keintzel[1], M. Kennouche[1], M. Kenzie[32], M. Kerréveur-Lavaud[46], R. Kersevan[1,269], V. Keus[197,270], H. Khanpour[14,271,272], V.V. Khoze[273], V.A. Khoze[273], P. Kicsiny[1], R. Kieffer[1], C. Kiel[114], J. Kieseler[132], A. Kilic[274], B. Kilminster[73], S. Kim[275], Z. Kırca[274], M. Klein[†184], A. Klimentov[6], M. Klute[132], V. Klyukhin[119,276], M. Knecht[137,277,278], B. Kniehl[115], P. Ko[279], S. Ko[1], F. Kocak[274], T. Koffas[280], C. Kokkinos[281,282], K. Kołodziej[227], K. Kong[283], P. Kontaxakis[101], I.A. Koop[119], P. Kopciewicz[1], P. Koppenburg[201], M. Koratzinos[1,2], K. Kordas[284], A Korsun[9,10,11], O. Kortner[131], S. Kortner[131], B. Korzh[101], T. Koseki[210], J. Kosse[2], P. Kostka[1,184], S. Kostoglou[1], A.V. Kotwal[80], G. Kozlov[1,276], I. Kozsar[1], T. Kramer[1], P. Krkotić[1], H. Kroha[131], K. Kröninger[244], S. Kuday[1,58], G. Kuhlmann[285], O. Kuhlmann[1,286], M. Kuhn[287], A. Kulesza[288], M. Kumar[289], F. Kurian[6], A. Kurtulus[1,150], T.H. Kwok[73], S. La Mendola[1], M. Lackner[98,290], T. Ładziński[1], D. Lafarge[1], P. Laïdouni[1], G. Lamanna[9,16,17], N. Lamas[19], G. Landsberg[232], C. Lange[2], D.J. Lange[163], A. Langner[1], A.J. Lankford[291], L. Lari[6], M.S. Larson[292], K. Lasocha[1], A. Latina[1], S. Lauciani[48], M. Laufenberg[105], G. Lavezzari[1], L. Lavezzi[60], L. Lavezzo[1], M. Le Garrec[1,9,16], A. Le Jeune[102], Ph. Lebrun[1,293], Y. Léchevin[1], A. Lechner[1], E. Lecointe[105], J.S.H. Lee[294], S.W. Lee[295], S.J. Lee[279,296], T. Lefevre[1], C. Leggett[141], T. Lehtinen[297], S. Leone[40], C. Leonidopoulos[298], S. Leontsinis[73], G. Leprince-Maillère[299], G. Lerner[1], O. Leroy[9,75,76], T. Lesiak[107], P. Levai[78], A. Leveratto[91], R. Levi[152], A. Li[6], S. Li[300,301], D. Liberati[302], G.L. Lichtenstein[47], M. Liepe[247], Z. Ligeti[141], H. Lin[303], S. Linda[144], E. Lipeles[304], Z. Liu[305], S.M. Liuzzo[306], T. Loeliger[287], A. Loeschcke Centeno[307], A. Lorenzetti[73], C. Lorin[3], R. Losito[1], M. Louka[12,308], M.L. Loureiro García[180], I. Low[122,158], K. Lubonis[152], M.T. Lucchini[30,31], V. Lukashenko[73], G. Luminati[48], A.J.G. Lunt[1,309], A. Lusiani[40,310], M. Luzum[311], H. Ma[6], A. Maas[312], E. Macchia[1,90,170], A. Macchiolo[73], G.E. Machinet[263], R. Madar[9,154,155], T. Madlener[4], C. Madrid[23], A. Magalotti[29], M. Maggiora[60,234], A.-M. Magnan[181], M.A. Mahmoud[313], Y. Mahmoud[314,315], F. Mahmoudi[1,9,56], H. Mainaud Durand[1], J. Maitre[108], Y. Makhloufi[101], B. Malaescu[9,117,316], A. Malagoli[91], C.H. Malan[108], M. Malekhosseini[38], A. Maloizel[1,96,97], S. Malvezzi[30], A. Malzac[148], G. Manco[127], L.S. Mandacarú Guerra[163], P. Manfrinetti[91,317], E. Manoni[63], J. Mans[305], L. Mantani[318], S. Manzoni[1], L. Marafatto[167], C. Marcel[1], T. Marcel[109], R. Marchevski[114], G. Marchiori[9,96,97], F. Mariani[44,90], V. Mariani[63,64], S. Marin[1], C. Marinas[318], V. Marinozzi[37], S. Mariotto[44,45], C. Marquis[105], J. Martelain[319], G. Martelli[63,64], A. Martens[9,10,11], I. Martin-Melero[1], V.I. Martinez Outschoorn[169], F. Martinez[215], C.M. Jardim[27], L. Marzola[320,321], S. Masciocchi[258,264], A. Mashal[14], A. Masi[1], I. Masina[146,147], P. Mastrapasqua[200], V. Mateu[322], S. Mattiazzo[68,164], M. Maugis[102], D. Mauree[144], G.H.I. Maury-Cuna[323], A. Mayoux[1], E. Mazzeo[1], S. Mazzoni[1], M. McCullough[1], M. Meena[9,42,43], E. Meftah[101], Andrew Mehta[184], Ankita Mehta[1], B. Mele[170], R. Mena-Andrade[1], M. Mentink[1], D. Mergelkuhl[1], V. Mertinger[267], L. Mether[1], S. Meylan[105], T. Michel[102], T. Michlmayr[2], M. Migliorati[90,170], A. Milanese[1], C. Milardi[48], G. Milhano[51],

M. Minty[6], C. Mirabelli[324], T. Miralles[9,154,155], L. Miralles Verge[1], D. Mirarchi[1], K. Mirbaghestan[73], N. Mirian[4,325], V.A. Mitsou[318], D.S. Mitzel[244], M. Mlynarikova[1], S. Möbius[89], M. Mohammadi Najafabadi[1,14], G.B. Mohanty[326], R. N. Mohapatra[209], S. Moneta[63], P.F. Monni[1], E. Monnier[9,75,76], S. Monteil[9,154,155], I. León Monzón[203], F. Moortgat[1,327], N. Morange[9,10,11], M. Moretti[146,147], S. Moretti[71], T. Mori[1,210], I. Morozov[119], A. Morozzi[63], M. Morrone[1], A. Moscariello[101], F. Moscatelli[63,328], I. Moulin[214], N. Mounet[1], A. Mueller[329], A.-S. Müller[132], B.O. Müller[285], J. Mundet[220], E. Musa[1,4], V. Musat[1,8], R. Musenich[140], E. Musumeci[318], M. Mylona[1], V.V. Mytrochenko[9,10,133], B. Nachman[141], S. Nagaitsev[6], T. Nakamoto[210], M. Napsuciale[323], M. Nardecchia[90,170], G. Nardini[330], G. Narváez-Arango[331], S. Naseem[61], A. Natochii[6], A. Navascues Cornago[1], B. Naydenov[1], G. Nergiz[1], A.V. Nesterenko[276], C. Neubüser[332], H.B. Newman[333], F. Niccoli[1,334], O. Nicrosini[127], U. Niedermayer[226], G. Niehues[132], J. Nielsen[1], G. Nigrelli[1,90,170], S. Nikitin[119], I.B. Nikolaev[119], A. Nisati[170], N. Nitika[167,168], J.M. No[335], M. Nonis[1], Y. Nosochkov[7], A. Novokhatski[1,7], J.M. O'Callaghan[336], S.A. Ochoa-Oregon[203], K. Ohmi[202,210], K. Oide[1,101,210], V.A. Okorokov[119], C. Oleari[30,31], D. Oliveira Damazio[1,6], Y. Onel[112], A. Onofre[337,338,339], P. Osland[340], Y.M. Oviedo-Torres[341,342,343], A. Ozansoy[58], F. Ozaydin[87,344], K. Ozdemir[345], A. Ozturk[1], M.A. Pérez de León[203], S. Pacetti[63,64], H. Pacey[8], J. Paciello[108], C.E. Pagliarone[346,347], A. Paillex[105], H.F. Pais da Silva[1], F. Palla[40], A. Pampaloni[140], C. Pancotti[149], M. Pandurović[348], O. Panella[63], G. Panizzo[167,168], C. Pantouvakis[68,164], L. Panwar[9,117,316], P. Paolucci[35], Y. Papa[105], A. Papaefstathiou[349], Y. Papaphilippou[1], A. Paramonov[158], A. Pareti[127,350], B. Parker[6], V. Parma[1], F. Parodi[140,317], M. Parodi[1], B. Paroli[44,45], J.A. Parsons[351], D. Passarelli[37], D. Passeri[63,64], B. Pattnaik[318], A. Patwa[352], C. Paus[182], F. Pauss[150], F. Peauger[1], I. Pedraza[215], R. Pedro[51], J. Pekkanen[1], G. Peon[1], A. Perez[109], E. Perez[1], F. Pérez[171], J.C. Perez[1], J.M. Pérez[27], R. Perez-Ramos[137,138,353], G. Pérez Segurana[1], A. Perillo Marcone[1], S. Perna[35,36], K. Peters[4], S. Petracca[35,354], A.R. Petri[44], F. Petriello[122], A. Petrovic[1], L. Pezzotti[25], G. Piacquadio[62], G. Piazza[179], A. Piccini[1], F. Piccinini[127], A. Pich[318], T. Pieloni[114], J. Pierlot[1], A.D. Pilkington[52], M. Pillet[324], M. Pinamonti[167,168], N. Pinto[235], L. Pintucci[167,168], F. Pinzauti[1], K. Piotrzkowski[271], C. Pira[48], M. Pitt[1], R. Pittau[190], S. Pittet[1], P. Placidi[63,64], W. Płaczek[355], S. Plätzer[312,356], M.-A. Pleier[6], E. Ploerer[73,116], H. Podlech[357,358], F. Poirier[9,16,17], G. Polesello[127], M. Poli Lener[48], J. Polinski[161], Z. Polonsky[73], N. Pompeo[29], M. Pont[171], G. Alexandru-Popeneciu[359], W. Porod[195], L. Porta[1], L. Portales[3], T. Portaluri[307], M.A.C. Potenza[45], C. Prasse[285], E. Premat[183], M. Presilla[132], S. Prestemon[141], A. Price[355], M. Primavera[159], R. Principe[1], M. Prioli[44], F.M. Procacci[12], E. Proserpio[44,134], A. Provino[91,317], C. Pueyo[1], T. Puig[19], N. Pukhaeva[276], S. Pulawski[227], G. Punzi[40,77], A. Pyarelal[360], J. Qian[303], H. Quack[361], F. S. Queiroz[47], G. Quintas-Neves[299], H. Rafique[71], J.-Y. Raguin[2], J. Raidal[320], M. Raidal[320], P. Raimondi[37], A. Rajabi[4], S. Ramírez-Uribe[203], S. Randles[184], T. Rao[6], C.Ø. Rasmussen[6], A. Ratkus[362], P.N. Ratoff[53,363], P. Razis[364,365], P. Rebello Teles[1,366], M.N. Rebelo[129], M. Reboud[9,10,11], S. Redaelli[1], C. Regazzoni[105], L. Reichenbach[1,367], M. Reissig[132], E. Renou[105], A. Rentería-Olivo[318], J. Reuter[4], S. Rey[105], A. Ribon[1], D. Ricci[1], W. Riegler[1], M. Rignanese[68,164], S. Rimjaem[368], R.A. Rimmer[369], R. Rinaldesi[1], L. Rinolfi[1,293], O. Rios[1], G. Ripellino[130], B. Rivas[370], A. Rivetti[60], T. Robens[371], F. Robert[183], E. Robutti[140], C. Roderick[1], G. Rodrigo[318], M. Rodríguez-Cahuantzi[215], L. Röhrig[154,155,244], M. Roig[372], F. Rojat[108], J. Rojo[201,373], J. Roloff[232], P. Roloff[1], A. Romanenko[37], A. Romero Francia[1], H. Romeyer[374], N. Rompotis[184], N. Rongieras[102], G. Rosaz[1], K. Roslon[375], M. Rossetti Conti[44], A. Rossi[63,64], E. Rossi[35,36], L. Rossi[44,45], A.N. Rossia[68,164], S. Rostami[38], G. Roy[1], B. Rubik[37], I. Ruehl[1], A. Ruiz-Jimeno[376], R. Ruprecht[132], J.P. Rutherfoord[360], L. Rygaard[4], M.S. Ryu[295], L. Sabato[1,114], G. Sadowski[9,42,43], D. Saez de Jauregui[132,377], M. Sahin[378], A. Sailer[1], M. Saito[379], P. Saiz[1], G.P. Salam[380,381], R. Salerno[9,81,82], T. Salmi[297], B. Salvachua[1], J.P.T. Salvesen[1,8,136], B. Salvi[299], D. Sampsonidis[284], Y. Villamizar[137,138,139], C. Sandoval[331], S. Sanfilippo[2], E. Santopinto[140], R. Santoro[44,134], X. Sarasola[114], L. Sarperi[287], I.H. Sarpün[382], S. Sasikumar[1], M. Sauvain[383], A. Savoy-Navarro[3,9], R. Sawada[379], G. Sborlini[384], J. Scamardella[35,36], M. Schaer[2], M. Schaumann[1,4], M. Schenk[1], C. Scheuerlein[1], C. Schiavi[140,317], A. Schloegelhofer[1], D. Schoerling[1],

A. Schöning[264], S. Schramm[101], D. Schulte[1], P. Schwaller[251,385], A. Schwartzman[7], Ph. Schwemling[3], R. Schwienhorst[386], A. Sciandra[6], L. Scibile[1], I. Scimemi[387], E. Scomparin[60], C. Sebastiani[1], B. Seeber[388], J.T. Seeman[7], F. Sefkow[4], M. Seidel[2,114], S. Seidel[74], J. Seixas[339,389,390], N. Selimović[68], M. Selvaggi[1], C. Senatore[101], A. Senol[194], N. Serra[73], A. Seryi[369], A. Sfyrla[101], Pramond Sharma[391], Punit Sharma[6], C.J. Sharp[1], L. Shchutska[114], V. Shiltsev[392], M. Siano[44,45], R. Sierra[1], E. Silva[29], R.C. Silva[47,343], L. Silvestrini[170], F. Simon[132], G. Simonetti[1], R. Simoniello[1], B.K. Singh[393], S. Singh[6], B. Singhal[79], A. Siodmok[1,355], Y. Sirois[9,81,82], E. Sirtori[149], B. Sitar[394], D. Sittard[1], E. Sitti[150], T. Sjöstrand[59], P. Skands[395], L. Skinnari[292], K. Skoufaris[1], K. Skovpen[327], M. Skrzypek[107], P. Slavich[137,138,139], V. Slokenbergs[23], V. Smaluk[6], J. Smiesko[1,396], S.S. Snyder[6], E. Solano[171], P. Sollander[1], O.V. Solovyanov[1,9,154], M. Son[397], F. Sonnemann[1], R. Soos[1,9,10], F. Sopkova[189], T. Sorais[398], M. Sorbi[44,45], S. Sorti[44,45], R. Soualah[399], M. Souayah[1], L. Spallino[48], S. Spanier[400], P. Spiller[258], M. Spira[2], D. Stagnara[102], M. Stallmann[186], D. Standen[1], J.L. Stanyard[1], B. Stapf[1], G.H. Stark[230], M. Statera[44], C. Staudinger[1,204], G. Streicher[401], N.P. Strohmaier[2], R. Stroynowski[106], S. Stucci[6], G. Stupakov[7], S. Su[360], A. Sublet[1], K. Sugita[258], M.K. Sullivan[7], S. Sultansoy[24], I. Syratchev[1], R. Szafron[6], A. Sznajder[402], W. Tachon[403], N.D. Tagdulang[37,171,336], N.A. Tahir[258], Y. Takahashi[123], J. Tamazirt[9,10,11], S. Tang[6], Y. Tanimoto[210], I. Tapan[274], G.F. Tassielli[12,404], A.M. Teixeira[9,154,155], V.I. Telnov[119], H.H.J. Ten Kate[1,405], V. Teotia[6], J. ter Hoeve[298], A. Thabuis[1], G.T. Telles[19], A. Tishelman-Charny[6], S. Tissandier[108], S. Tizchang[14,406], J.-P. Tock[1], B. Todd[1], L. Toffolin[1,167,407], A. Tolosa-Delgado[1], R. Tomás García[1], T. Tomasini[408], G. Tonelli[40,77], T. Tong[409], F. Toral[27], T. Torims[1,362], L. Torino[171], K. Torokhtii[29], R. Torre[140], E. Torrence[410], R. Torres[53,184], T. Mitsuhashi[210], A. Tracogna[149], O. Traver[171], D. Treille[1], A. Tricoli[6], P. Trubacova[1], E. Tsesmelis[1], G. Tsipolitis[268], V. Tsulaia[141], B. Tuchming[3], C.G. Tully[163], I. Turk Cakir[58], C. Turrioni[63], J. Tynan[105], F.P. Ucci[127,350], S. Udongwo[411], C.S. Ün[274], A. Unnervik[1], A. Upegui[99,100], J.P. Uribe-Ramírez[203], J. Uythoven[1], R. Vaglio[36,91], F. Valchkova-Georgieva[412], P. Valente[170], R.U. Valente[170], A.-M. Valente-Feliciano[369], G. Valentino[1,192], C.A. Valerio-Lizarraga[203,323], S. Valette[1], J.W.F. Valle[318], L. Valle[1], N. Valle[127], N. Vallis[1,2,114], G. Vallone[141], P. van Gemmeren[158], W. Van Goethem[1], P. van Hees[59], U. van Rienen[411], L. van Riesen-Haupt[1,114], P. Van Trappen[1], M. Vande Voorde[413,414], A.L. Vanel[1], E.W. Varnes[360], J.-L. Vay[141], F. Veit[285], I. Veliscek[6], R. Veness[1], A. Ventura[159,231], M. Verducci[40,77], C.B. Verhaaren[415], C. Vernieri[7], A.P. Verweij[1], J.-F. Vian[416], A. Vicini[44,45], N. Vignaroli[159,231], S. Vignetti[149], M.C. Villeneuve[218], I. Vivarelli[25,26], E. Voevodina[1,131], D.M. Vogt[417], B. Voirin[418], S. Voiriot[105], J. Voiron[144], P. Vojtyla[1], V. Völkl[1], L. von Freeden[1], Z. Vostrel[1,419], N. Voumard[1], E. Vryonidou[52], V. Vysotsky[119], R. Wallny[150], L.-T. Wang[420], Y. Wang[9,10,11], R. Wanzenberg[4], B.F.L. Ward[421], N. Wardle[181], Z. Wąs[107], L. Watrelot[1], A.T. Watson[422], M.F. Watson[422], M.S. Weber[89], C.P. Welsch[53,184], M. Wendt[1,6], J. Wenninger[1], B. Weyer[1], G. White[423], S. White[306], B. Wicki[1], M. Widorski[1], U.A. Wiedemann[1], A.R. Wiederhold[52], A . Wiedl[132], H.-U. Wienands[158], A. Wieser[150], C. Wiesner[1], H. Wilkens[1], D. Willi[424], P.H. Williams[53,425], S.L. Williams[32], A. Winter[422], R.B. Wittwer[73], D. Wollmann[1], Y. Wu[114], Z. Wu[9,16,17], J. Xiao[9,56,57], K. Xie[386], S. Xie[37,333], M. Yalvac[22], F. Yaman[425,426], W.-M. Yao[141], M. Yeresko[9,154,155], A. Yilmaz[194], H.D. Yoo[275], T. You[208], F. Yu[251,385], S.S. Yu[79], T.-T. Yu[410], S. Yue[1], A. Zaborowska[1], M. Zahnd[105], C. Zamantzas[1], G. Zanderighi[65,131], C. Zannini[1], R. Zanzottera[44,45], P. Zaro[102], R. Zennaro[2], M. Zerlauth[1], H. Zhang[202], J. Zhang[158], Y. Zhang[202], Z. Zhang[9,10,202], Y. Zhao[1], Y.-M. Zhong[427], B. Zhou[303], D. Zhou[210], J. Zhu[303], G. Zick[372], M.A. Zielinski[1], E. Zimmermann[105], A. Zingaretti[68,164], J. Zinn-Justin[3], A.V. Zlobin[37], M. Zobov[48], F. Zomer[9,10,11], S. Zorzetti[37], X. Zuo[132], J. Zurita[318], V.V. Zutshi[392], M. Zykova[2].

[†] deceased

[1] Switzerland - CERN, European Organization for Nuclear Research

[2] Switzerland - PSI, Paul Scherrer Institute

[3] France - CEA/Irfu, Commissariat à l'Energie Atomique et aux Energies Alternatives, Institut de recherche sur les lois fondamentales de l'Univers

[4] Germany - DESY, Deutsches Elektronen-Synchrotron
[5] Germany - Humboldt-Universität zu Berlin
[6] United States - BNL, Brookhaven National Laboratory
[7] United States - SLAC National Accelerator Laboratory
[8] United Kingdom - University of Oxford
[9] France - CNRS/IN2P3, Centre National de la Recherche Scientifique, Institut National de Physique Nucléaire et de Physique des Particules
[10] France - IJCLab, Laboratoire de Physique des 2 Infinis Irène Joliot Curie
[11] France - Université Paris-Saclay et Université Paris-Cité
[12] Italy - INFN, Istituto Nazionale di Fisica Nucleare, Sezione di Bari
[13] Italy - Università di Bari
[14] Iran - IPM, Institute for Research in Fundamental Science
[15] Iran - Malayer University
[16] France - LAPP, Laboratoire d'Annecy de Physique des Particules
[17] France - Université Savoie Mont Blanc
[18] Serbia - University of Belgrade
[19] Spain - ICMAB/CISC, Institut de Ciència de Materials de Barcelona, Consejo Superior de Investigaciones Científicas
[20] Italy - INFN, Istituto Nazionale di Fisica Nucleare, Sezione di Roma Tor Vergata
[21] Italy - Università Roma Tor Vergata
[22] Türkiye - Yozgat Bozok Üniversitesi
[23] United States - Texas Tech University
[24] Türkiye - TOBB ETU, TOBB Ekonomi ve Teknoloji Üniversitesi
[25] Italy - INFN, Istituto Nazionale di Fisica Nucleare, Sezione di Bologna
[26] Italy - Università di Bologna
[27] Spain - CIEMAT, Centro de Investigaciones Energéticas, Medioambientales y Tecnológicas
[28] Saudi Arabia - KACST, King Abdulaziz City for Science and Technology
[29] Italy - Università Roma Tre
[30] Italy - INFN, Istituto Nazionale di Fisica Nucleare, Sezione di Milano-Bicocca
[31] Italy - Università di Milano-Bicocca
[32] United Kingdom - University of Cambridge
[33] Türkiye - İstanbul Üniversitesi
[34] Türkiye - Eskişehir Teknik Üniversitesi
[35] Italy - INFN, Istituto Nazionale di Fisica Nucleare, Sezione di Napoli
[36] Italy - Università di Napoli Federico II
[37] United States - FNAL, Fermi National Accelerator Laboratory
[38] Iran - University of Tehran
[39] Iran- FUM, Ferdowsi University of Mashhad
[40] Italy - INFN, Istituto Nazionale di Fisica Nucleare, Sezione di Pisa
[41] United States - University of Texas Austin
[42] France - IPHC, Institut Pluridisciplinaire Hubert Curien
[43] France - Université de Strasbourg
[44] Italy - INFN, Istituto Nazionale di Fisica Nucleare, Sezione di Milano
[45] Italy - Università di Milano

[46] Switzerland - PIBG, Pôle Invertébrés du Basin Genevois
[47] Brazil - UFRN, Universidade Federal do Rio Grande do Norte
[48] Italy - INFN, Istituto Nazionale di Fisica Nucleare, Laboratori Nazionali di Frascati
[49] Switzerland - UNIBAS, University of Basel
[50] Italy - Politecnico di Bari
[51] Portugal - LIP, Laboratório de Instrumentação e Física Experimental de Partículas
[52] United Kingdom - University of Manchester
[53] United Kingdom - CI, Cockcroft Institute
[54] United States - Brandeis University
[55] Armenia - A. Alikhanyan National Laboratory
[56] France - IP2I, Institut de Physique des 2 Infinis de Lyon
[57] France - Université Claude Bernard Lyon 1
[58] Türkiye - Ankara Üniversitesi
[59] Sweden - Lund University
[60] Italy - INFN, Istituto Nazionale di Fisica Nucleare, Sezione di Torino
[61] United Arab Emirates - New York University Abu Dhabi
[62] United States - Stony Brook University
[63] Italy - INFN, Istituto Nazionale di Fisica Nucleare, Sezione di Perugia
[64] Italy - Università di Perugia
[65] Germany - Technische Universität München
[66] Türkiye - Hatay Mustafa Kemal Üniversitesi
[67] Türkiye - Doğuş Üniversitesi
[68] Italy - INFN, Istituto Nazionale di Fisica Nucleare, Sezione di Padova
[69] People's Republic of China - Harbin Institute of Technology
[70] United States - University of Wisconsin-Madison
[71] United Kingdom - RAL, Rutherford Appleton Laboratory, Science and Technology Facilities Council
[72] India - IMSc, Institute of Mathematical Sciences, Chennai
[73] Switzerland - Universität Zürich
[74] United States - University of New Mexico
[75] France - CPPM, Centre de Physique des Particules de Marseille
[76] France - Aix-Marseille Université
[77] Italy - Università di Pisa
[78] Hungary - HUN-REN Wigner Research Centre for Physics
[79] United States - Catholic University of America
[80] United States - Duke University
[81] France - LLR, Laboratoire Leprince-Ringuet
[82] France - École Polytechnique, Institut Polytechnique de Paris
[83] Canada - TRIUMF, Canada's National Laboratory for Particle and Nuclear Physics
[84] Canada - Simon Fraser University
[85] Italy - FEEM, Fondazione Ente Nazionale Idrocarburi (ENI) Enrico Mattei
[86] France - BRGM, Bureau de Recherches Géologiques et Minières
[87] Türkiye - Işık Üniversitesi
[88] Türkiye - İstanbul Teknik Üniversitesi

[89] Switzerland - UNIBE, University of Bern
[90] Italy - Università di Roma la Sapienza
[91] Italy - CNR-SPIN, Consiglio Nazionale delle Ricerche
[92] France - Expert naturaliste et entomologiste
[93] United States - ORNL, Oak Ridge National Laboratory
[94] Austria - HEPHY, Institut für Hochenergiephysik
[95] Switzerland - Geos, Bureau d'ingénieurs conseils en géotechnique, génie civil, hydraulique et environnement
[96] France - APC, Laboratoire AstroParticule et Cosmologie
[97] France - Université Paris Cité
[98] Austria - TUWIEN, Technische Universität Wien
[99] Switzerland - HEPIA, Haute École du Paysage, d'Ingénierie et d'Architecture de Genève
[100] Switzerland - HES-SO University of Applied Sciences and Arts Western Switzerland
[101] Switzerland - UNIGE, Université de Genève
[102] France - SETEC ALS, Société d'ingénierie conseil en infrastructures de transport, génie civil et environnement
[103] United States - East Texas A&M University
[104] Switzerland - Amberg Engineering Ltd
[105] Switzerland - ECOTEC Environnement SA, Bureau d'études et de conseil en environnement
[106] United States - Southern Methodist University
[107] Poland - IFJ PAN, Institute of Nuclear Physics, Polish Academy of Sciences
[108] France - Cerema, établissement public pour l'élaboration, le déploiement et l'évaluation de politiques publiques d'aménagement et de transport
[109] United Kingdom - Rendel Ltd, Engineering design consultancy firm
[110] Italy - INFN, Istituto Nazionale di Fisica Nucleare, Sezione di Roma Tre
[111] Türkiye - İstanbul Beykent Üniversitesi
[112] United States - University of Iowa
[113] India - Indian Institute of Technology Kanpur
[114] Switzerland - EPFL, École Polytechnique Fédérale de Lausanne
[115] Germany - Universität Hamburg, Fakultät für Mathematik, Informatik und Naturwissenschaften
[116] Belgium - VUB, Vrije Universiteit Brussel
[117] France - LPNHE, Laboratoire de Physique Nucléaire et de Hautes Énergies
[118] Italy - Scuola Superiore Meridionale
[119] Affiliated with an institute formerly covered by a cooperation agreement with CERN
[120] Canada - University of Saskatchewan and the Canadian Light Source
[121] United Kingdom - University of Bristol
[122] United States - Northwestern University
[123] United States - University of Florida
[124] Canada - York University
[125] Italy - INFN, Istituto Nazionale di Fisica Nucleare, Sezione di Cagliari
[126] Italy - Università di Cagliari
[127] Italy - INFN, Istituto Nazionale di Fisica Nucleare, Sezione di Pavia
[128] Canada - Queen's University
[129] Portugal - CFTP-IST, Centro de Física Téorica de Partículas, Instituto Superior Tecnico,

Universidade de Lisboa
[130] Sweden - Uppsala University
[131] Germany - MPP, Max-Planck-Institut für Physik Garching
[132] Germany - KIT, Karlsruher Institut für Technologie
[133] Ukraine - NSC KIPT, National Science Center Kharkiv Institute of Physics and Technology
[134] Italy - Università degli Studi dell'Insubria
[135] Germany - Albert-Ludwigs-Universität Freiburg
[136] United Kingdom - JAI, John Adams Institute for Accelerator Science, University of Oxford
[137] France - CNRS/INP, Centre National de la Recherche Scientifique, Institut de Physique
[138] France - LPTHE, Laboratoire de Physique Théorique et Hautes Energies
[139] France - Sorbonne Université
[140] Italy - INFN, Istituto Nazionale di Fisica Nucleare, Sezione di Genova
[141] United States - LBNL, Lawrence Berkeley National Laboratory
[142] Italy - IIT, Instituto Italiano di Tecnologia
[143] Türkiye - Gümüşhane Üniversitesi
[144] Switzerland - WSP Ingénieurs Conseils SA
[145] Mexico - UADY, Autonomous University of Yucatan
[146] Italy - INFN, Istituto Nazionale di Fisica Nucleare, Sezione di Ferrara
[147] Italy - Università di Ferrara
[148] France - MARCELEON, Cabinet d'ingénierie juridique et foncière
[149] Italy - CSIL (Economic Research Institute)
[150] Switzerland - ETHZ, Swiss Federal Institute of Technology Zurich
[151] Spain - UAH, Universidad de Alcalá Madrid
[152] France - CIA, Conseil Ingénierie Acoustique
[153] France - Evinerude, Bureau d'études environnementales
[154] France - LPCA, Laboratoire de Physique de Clermont Auvergne
[155] France - Université Clermont Auvergne
[156] Australia - ANSTO, Australian Synchrotron
[157] India - Brahmananda Keshab Chandra College
[158] United States - ANL, Argonne National Laboratory
[159] Italy - INFN, Istituto Nazionale di Fisica Nucleare, Sezione di Lecce
[160] Italy - Università di Palermo
[161] Poland - Wrocław University of Science and Technology
[162] United States - Rutgers University
[163] United States - Princeton University
[164] Italy - Università di Padova
[165] Türkiye - IUE, İzmir Ekonomi Üniversitesi
[166] Türkiye - Ege Üniversitesi
[167] Italy - INFN, Istituto Nazionale di Fisica Nucleare, Gruppo Collegato di Udine
[168] Italy - Università di Udine
[169] United States - University of Massachusetts Amherst
[170] Italy - INFN, Istituto Nazionale di Fisica Nucleare, Sezione di Roma
[171] Spain - CELLS/ALBA, Consortium for the Construction, Equipment and Exploitation of the Synchrotron Light Laboratory

[172] France - LPSC, Laboratoire de Physique Subatomique et de Cosmologie
[173] France - Université Grenoble Alpes
[174] Italy - Università degli Studi di Napoli Parthenope
[175] United States - National High Magnetic Field Laboratory
[176] United States - Florida State University
[177] South Africa - University of Johannesburg
[178] Spain - IGFAE, Instituto Galego de Fisica de Altas Enerxías, Universidade de Santiago de Compostela
[179] United Kingdom - LSE, London School of Economics
[180] Spain - Universidade de Santiago de Compostela
[181] United Kingdom - Imperial College London
[182] United States - MIT, Massachusetts Institute of Technology
[183] France - CETU, Centre d'Etude des Tunnels
[184] United Kingdom - University of Liverpool
[185] Switzerland - Linde Kryotechnik AG
[186] Switzerland - ILF Consulting Engineers
[187] Denmark - NBI, Niels Bohr Institute
[188] Japan - Hokkaido University
[189] Czech Republic - CUNI, Charles University
[190] Spain - Universidad de Granada
[191] Italy - INFN, Istituto Nazionale di Fisica Nucleare, Sezione di Firenze
[192] Malta - University of Malta
[193] United States - BU, Boston University
[194] Türkiye - IBU, Bolu Abant İzzet Baysal Üniversitesi
[195] Germany - Julius-Maximilians-Universität Würzburg
[196] Australia - University of Adelaide
[197] Finland - HIP, Helsinki Institute of Physics, University of Helsinki
[198] Belgium - ULB, Université Libre de Bruxelles
[199] Germany - IMA, Institut für Maschinenelemente, Universität Stuttgart
[200] Belgium - CP3, Centre de Cosmologie, de Physique des Particules et de Phénoménologie, Université Catholique de Louvain
[201] Netherlands - NIKHEF, Nationaal instituut voor subatomaire fysica
[202] People's Republic of China - IHEP, Chinese Academy of Sciences
[203] Mexico - UAS, Universidad Autónoma de Sinaloa
[204] Austria - BOKU, Universität für Bodenkultur Wien
[205] United States - Purdue University
[206] India - University of Delhi
[207] France - Ginger BURGEAP, bureau d'études en environnement
[208] United Kingdom - King's College London
[209] United States - University of Maryland
[210] Japan - KEK, High Energy Accelerator Research Organization
[211] Switzerland - Geoenergy, Reservoir Geology and Basin Analysis Group
[212] France - SETEC International, Société d'ingénierie en charge des transports et des infrastructures
[213] Türkiye - KKU, Kırıkkale Üniversitesi

[214] France - SETEC LERM, Société d'ingénierie conseil en matériaux de construction
[215] Mexico - BUAP, Benemérita Universidad Autónoma de Puebla
[216] United States - Stanford University
[217] United States - University of Pittsburgh
[218] Austria - MUL, Montanuniversität Leoben, Lehrstuhl für Subsurface Engineering, Geotechnik und unterirdisches Bauen
[219] Austria - MUL-ZaB, Underground Research Center, Zentrum am Berg
[220] Spain - IFAE, Institut de Física d'Altes Energies
[221] Switzerland - FHNW, University of Applied Sciences Northwestern Switzerland
[222] India - UPES, University of Petroleum and Energy Studies
[223] France - GANIL, Grand Accélérateur National d'Ions Lourds
[224] France - Université Caen Normandie
[225] Brazil - UFRGS, Universidade Federal do Rio Grande do Sul
[226] Germany - Technische Universität Darmstadt
[227] Poland - University of Silesia in Katowice
[228] Portugal - Universidade de Coimbra
[229] Brazil - UFPel, Universidade Federal de Pelotas
[230] United States - University of California Santa Cruz
[231] Italy - Università del Salento
[232] United States - Brown University
[233] United States - University of California Berkeley
[234] Italy - Università di Torino
[235] United States - Johns Hopkins University
[236] People's Republic of China - Fudan University
[237] France - LAPTh, Laboratoire d'Annecy-le-Vieux de Physique Théorique
[238] People's Republic of China - Dongguan University of Technology
[239] Türkiye - Kahramanmaraş Sütçü İmam Üniversitesi
[240] Mexico - UNACH, Universidad Autónoma de Chiapas
[241] Mexico - MCTP, Mesoamerican Centre for Theoretical Physics
[242] Mexico - UAZ, Universidad Autónoma de Zacatecas
[243] Finland - University of Jyväskylä
[244] Germany - Technische Universität Dortmund
[245] United Kingdom - Overleaf
[246] United States - University of Virginia
[247] United States - Cornell University
[248] United States - FIT, Florida Institute of Technology
[249] Switzerland - Shirokuma GmbH
[250] Pakistan - National Centre for Physics
[251] Germany - Johannes Gutenberg Universität Mainz
[252] Pakistan - PAEC, Pakistan Atomic Energy Commission
[253] India - Mathabhanga College
[254] India - Harish-Chandra Research Institute
[255] Germany - MPIK, Max-Planck-Institut für Kernphysik Heidelberg
[256] Sweden - European Spallation Source ERIC

[257] France - Centre de calcul de l'IN2P3
[258] Germany - GSI, Helmholtzzentrum für Schwerionenforschung GmbH
[259] Republic of Korea - IBS, Institute for Basic Science, Center for Theoretical Physics of the Universe
[260] Poland - University of Warsaw
[261] Slovenia - University of Ljubljana
[262] Slovenia - Jozef Stefan Institute
[263] France - Microhumus, Bureau d'étude et d'ingénierie spécialisé dans la gestion des sols dégradés
[264] Germany - Fakultät für Physik und Astronomie, Universität Heidelberg
[265] Türkiye - Niğde Ömer Halisdemir Üniversitesi
[266] Türkiye - Giresun Üniversitesi
[267] Hungary - University of Miskolc
[268] Greece - NTUA, National Technical University of Athens
[269] Switzerland - Transmutex SA
[270] Ireland - DIAS, Dublin Institute for Advanced Studies, School of Theoretical Physics
[271] Poland - AGH, University of Science and Technology
[272] Iran - University of Science and Technology of Mazandaran
[273] United Kingdom - IPPP, Institute for Particle Physics Phenomenology, Durham University
[274] Türkiye - Bursa Uludağ Üniversitesi
[275] Republic of Korea - YU, Yonsei University
[276] Affiliated with an international laboratory covered by a cooperation agreement with CERN
[277] France - CPT, Centre de Physique Théorique
[278] France - Aix-Marseille Université et Université du Sud Toulon Var
[279] Republic of Korea - KIAS, Korea Institute for Advanced Study
[280] Canada - Carleton University
[281] Greece - FEAC Engineering P.C.
[282] Greece - UPATRAS, University of Patras
[283] United States - University of Kansas
[284] Greece - AUTH, Aristotle University of Thessaloniki
[285] Germany - IML, Fraunhofer-Institut für Materialfluss und Logistik
[286] Germany - RWTH Aachen, Rheinisch-Westfälische Technische Hochschule Aachen
[287] Switzerland - ZHAW, Zurich University of Applied Sciences
[288] Gernany - Universität Münster
[289] South Africa - University of the Witwatersrand
[290] Austria - Fachhochschule Technikum Wien
[291] United States - University of California Irvine
[292] United States - Northeastern University
[293] France - ESI, European Scientific Institute
[294] Republic of Korea - UOS, University of Seoul
[295] Republic of Korea - KNU Kyungpook National University
[296] Republic of Korea - KU, Korea University
[297] Finland - Tampere University
[298] United Kingdom - University of Edinburgh
[299] Switzerland - BG Ingénieurs Conseils

[300] People’s Republic of China - T.-D. Lee Institute
[301] People’s Republic of China - Shanghai Jiao Tong University
[302] Italy - CNR, Consiglio Nazionale delle Ricerche
[303] United States - University of Michigan
[304] United States - University of Pennsylvania
[305] United States - University of Minnesota
[306] France - ESRF, European Synchrotron Radiation Facility
[307] United Kingdom - SUSSEX, University of Sussex
[308] Italy - Università di Bari Aldo Moro
[309] United Kingdom - University of Bath
[310] Italy - Scuola Normale Superiore di Pisa
[311] Brazil - Universidade de São Paulo
[312] Austria - Universität Graz
[313] Egypt - Center for High Energy Physics, Fayoum University
[314] Egypt - Center of theoretical physics, British University in Egypt
[315] Egypt - Cairo University
[316] France - Sorbonne Université et Université Paris Cité
[317] Italy - Università di Genova
[318] Spain - IFIC-CSIC/UV, Instituto de Física Corpuscular, Consejo Superior de Investigaciones Científicas/Universidad de Valencia
[319] Switzerland - Service de géologie, sols et déchets du canton de Genève
[320] Estonia - NICPB, National Institute for Chemical Physics and Biophysics
[321] Estonia - UT, University of Tartu
[322] Spain - Universidad de Salamanca
[323] Mexico - UGTO, Universidad de Guanajuato
[324] Switzerland - Edaphos engineering
[325] Germany - Helmholtz-Zentrum Dresden-Rossendorf
[326] India - Tata Institute of Fundamental Research Mumbai
[327] Belgium - Universiteit Gent
[328] Italy - CNR-IOM, Consiglio Nazionale delle Ricerche
[329] Austria - JKU, Johannes Kepler Universität Linz
[330] Norway - University of Stavanger
[331] Colombia - Universidad Nacional de Colombia
[332] Italy - Trento Institute for Fundamental Physics and Applications
[333] United States - Caltech, California Institute of Technology
[334] Italy - Università dalla Calabria
[335] Spain - IFT, Instituto de Física Teórica, Universidad Autónoma de Madrid
[336] Spain - UPC, Universitat Politècnica de Catalunya
[337] Portugal - Departamento de Física, Universidade do Minho
[338] Portugal - Centro de Física das Universidades do Minho e do Porto
[339] Portugal - LaPMET, Laboratory of Physics for Materials and Emergent Technologies
[340] Norway - University of Bergen
[341] Chile - SAPHIR, Instituto Milenio de Física Subatómica en la Frontera de Altas Energías
[342] Chile - Universidad Andres Bello

[343] Brazil - IIP, International Institute of Physics
[344] Japan - Tokyo International University
[345] Türkiye - İzmir Bakırçay Üniversitesi
[346] Italy - INFN, Istituto Nazionale di Fisica Nucleare, Laboratori Nazionali del Gran Sasso
[347] Italy - Universitá degli Studi di Cassino e del Lazio Meridionale
[348] Serbia - Vinča Institute of Nuclear Sciences
[349] United States - Kennesaw State University
[350] Italy - Università di Pavia
[351] United States - Columbia University
[352] United States - DOE, Department of Energy of the United States of America
[353] France - IPSA, Institut Polytechnique des Sciences Avancées
[354] Italy - Università degli Studi del Sannio
[355] Poland - UJ, Jagiellonian University
[356] Austria - Universität Wien
[357] Germany - Goethe-Universität Frankfurt, Institut für Angewandte Physik
[358] Germany - HFFH, Helmholtz Forschungsakademie Hessen für FAIR
[359] Romania - INCDTIM, National Institute for Research and Development of Isotopic and Molecular Technologies
[360] United States - University of Arizona
[361] Germany - Technische Universität Dresden
[362] Latvia - RTU, Riga Technical University
[363] United Kingdom - Lancaster University
[364] Cyprus - University of Cyprus
[365] Cyprus - Cosmos Open University
[366] Brazil - CBPF, Centro Brasileiro de Pesquisas Físicas
[367] Germany - Universität Bonn
[368] Thailand - CMU, Chiang Mai University
[369] United States - JLAB, Thomas Jefferson National Accelerator Facility
[370] Ecuador - ESPOL, Escuela Superior Politécnica del Litoral
[371] Croatia - IRB, Rudjer Boskovic Institute
[372] France - Air Liquide Advanced Technologies
[373] Netherlands - VU Amsterdam
[374] France - INGÉROP ,Groupe d'ingénierie et de conseil en mobilité durable, transition énergétique et cadre de vie
[375] Poland - Warsaw University of Technology
[376] Spain - IFCA, Instituto de Física de Cantabria
[377] Germany - Institut für Beschleunigerphysik und Technologie
[378] Türkiye - Uşak Üniversitesi
[379] Japan - ICEPP, International Center for Elementary Particle Physics, University of Tokyo
[380] United Kingdom - Rudolf Peierls Centre for Theoretical Physics, University of Oxford
[381] United Kingdom - All Souls College, University of Oxford
[382] Türkiye - Akdeniz Üniversitesi
[383] Switzerland - Latitude Durable SARL
[384] Spain - USAL, Universidad de Salamanca

[385] Germany - PRISMA+ Cluster of Excellence
[386] United States - Michigan State University
[387] Spain - Universidad Complutense Madrid
[388] Switzerland - scMetrology SARL
[389] Portugal - IST, Instituto Superior Tecnico, Universidade de Lisboa
[390] Portugal - CeFEMA, Center of Physics and Engineering of Advanced Materials
[391] India - Indian Institute of Science Education and Research Mohali
[392] United States - NIU, Northern Illinois University
[393] India - Banaras Hindu University
[394] Slovakia - Comenius University
[395] Australia - Monash University
[396] Slovakia - Slovak Academy of Sciences
[397] Republic of Korea - KAIST, Korea Advanced Institute of Science and Technology
[398] France - Amberg Engineering Chambéry
[399] United Arab Emirates - Khalifa University of Science and Technology
[400] United States - University of Tennessee
[401] Austria - WIFO, Österreichisches Institut für Wirtschaftsforschung
[402] Brazil - Universidade do Estado do Rio de Janeiro
[403] France - Mélica, NATURA SCOP, Études et expertises environnementales
[404] Italy - Università LUM, Casamassima
[405] Netherlands - University of Twente
[406] Iran - Arak University
[407] Italy - Università di Trieste
[408] France - ForestAllia, Cabinet de gestion et d'expertise forestières
[409] Germany - Universität Siegen
[410] United States - University of Oregon
[411] Germany - Universität Rostock
[412] Switzerland - CEGELEC SA
[413] Sweden - KTH, Royal Institute of Technology, Stockholm
[414] Sweden - OKC, Oskar Klein Centre for Cosmoparticle Physics
[415] United States - Brigham Young University
[416] France - Expert foncier et agricole
[417] Germany - ITSM, Institut für Thermische Strömungsmaschinen und Maschinenlaboratorium, Universität Stuttgart
[418] France - École Normale Supérieure de Lyon
[419] Czech Republic - CTU, Czech Technical University
[420] United States - University of Chicago
[421] United States - Baylor University
[422] United Kingdom - University of Birmingham
[423] United Kingdom - University of Southampton
[424] Switzerland - Swisstopo, Federal Office of Topography
[425] United Kingdom - Daresbury Laboratory, Science and Technology Facilities Council
[426] Türkiye - IZTECH, İzmir Yüksek Teknoloji Enstitüsü
[427] Hong Kong - City University of Hong Kong

## Abstract

In response to the *2020 Update of the European Strategy for Particle Physics*, the Future Circular Collider (FCC) Feasibility Study was launched as an international collaboration hosted by CERN. This report describes the FCC *integrated programme*, which consists of two stages: an electron-positron collider (FCC-ee) in the first phase, serving as a high-luminosity Higgs, top, and electroweak factory; followed by a proton-proton collider (FCC-hh) at the energy frontier in the second phase.

The FCC-ee is designed to operate at four key centre-of-mass energies: the Z pole, the WW pair production threshold, the ZH production peak, and the top/anti-top production threshold—each delivering the highest possible luminosities to four experiments. Over 15 years of operation, FCC-ee will produce more than 6 trillion Z bosons, 200 million WW pairs, nearly 3 million Higgs bosons, and 2 million top anti-top pairs. Precise energy calibration at the Z pole and WW threshold will be achieved through frequent resonant depolarisation of pilot bunches. The sequence of operation modes between the Z, WW, and ZH substages remains flexible.

The FCC-hh will operate at a centre-of-mass energy of approximately 85 TeV—nearly an order of magnitude higher than the LHC—and is designed to deliver 5 to 10 times the integrated luminosity of the upcoming High-Luminosity LHC. Its mass reach for direct discovery extends to several tens of TeV. In addition to proton-proton collisions, the FCC-hh is capable of supporting ion-ion, ion-proton, and lepton-hadron collision modes.

This second volume of the Feasibility Study Report presents the complete design of the FCC-ee collider, its operation and staging strategy, the full-energy booster and injector complex, required accelerator technologies, safety concepts, and technical infrastructure. It also includes the design of the FCC-hh hadron collider, development of high-field magnets, hadron injector options, and key technical systems for FCC-hh.

## Preface from CERN's Director-General

In 2021, in response to the 2020 update of the European Strategy for Particle Physics, the CERN Council initiated the Future Circular Collider (FCC) Feasibility Study.

This report summarises an immense amount of work carried out by the international FCC collaboration over several years. It covers, inter alia, physics objectives and potential, geology, civil engineering, technical infrastructure, territorial implementation, environmental aspects, R&D needs for the accelerators and detectors, socio-economic benefits and cost. It constitutes important input for the ongoing update of the European Strategy for Particle Physics.

The Feasibility Study required engagement with a broad range of stakeholders. In particular, throughout the Study, CERN has been accompanied by its two Host States, France and Switzerland, and has been working with entities at local, regional and national level. I am very grateful to the Host State authorities and teams for their invaluable help. Furthermore, significant sections of the Study were supported by the European Union under the Horizon 2020 and Horizon Europe framework programmes. The Study also greatly benefited from contributions from accelerator laboratories and universities from across Europe, such as the Swiss Accelerator Research and Technology (CHART) initiative, and from the Americas, Asia, Africa and Australia.

The proposed FCC integrated programme consists of two possible stages: an electron–positron collider serving as a Higgs-boson, electroweak and top-quark factory running at different centre-of-mass energies, followed at a later stage by a proton–proton collider operating at an unprecedented collision energy of around 100 TeV. The complementary physics programmes of each stage match the physics priorities expressed in the 2020 update of the European Strategy for Particle Physics.

A major achievement of the Feasibility Study is the choice of placement of the collider ring and the entire infrastructure, including the surface sites and the access shafts, which was developed and optimised over several years following the principle 'avoid, reduce, compensate'. Sustainability studies have assessed energy efficiency, land use, water and resource management, and socio-economic impact, ensuring that the FCC is designed in accordance with the latest environmental and societal standards.

I would like to thank all contributors to this report for their hard work and commitment, which allowed the outstanding results presented here to be achieved.

**Fabiola Gianotti**
CERN, Director-General

## Preface from the FCC Collaboration Board Chair

Building on the earlier Future Circular Collider (FCC) Conceptual Design Study conducted between 2014 and 2018, the FCC Feasibility Study (2021–2025) has been undertaken by a robust international collaboration, now comprising over 160 institutes worldwide. The FCC 'integrated programme', developed in the framework of the Feasibility Study, consists of an initial electron-positron collider, the FCC-ee, which could be followed by a proton-proton collider, the FCC-hh. This staging takes into account the physics priorities as formulated in the updates of the European Strategy for Particle Physics of 2012 and 2020, as well as the relative technology readiness and costs of the FCC-ee and FCC-hh.

Over the years, I have closely followed the steady progress of the study, representing the FCC collaboration at the international steering committee and participating in annual FCC Week meetings, which include sessions of the International Collaboration Board. The commitment and enthusiasm of the members of the collaboration has always been impressive. The collective effort is clearly visible. Participation by students and early-career researchers is increasing. There is a shared determination and momentum to move forward.

The strong international collaboration around the FCC and its global network provide a solid foundation for the future of this project. The FCC community continues to grow, with increasing engagement from new institutes and partners worldwide. This broad support will be essential as the project enters its next phase.

The FCC Feasibility Study demonstrates not only the technical viability of the project, but also the strength of the international community that supports it. As we move towards the next step in the decision-making phase, this collective effort is key to showing a possible path forward. The FCC promises far-reaching scientific opportunities and long-term benefits for innovation, training, and global collaboration in science and technology.

**Philippe Chomaz**
CEA, Chair of the FCC International Collaboration Board

# Contents

# Introduction to the FCC integrated project

## FCC design and layout considerations

The Future Circular Collider (FCC) ‘integrated programme’ consists of an initial electron-positron collider FCC-ee, which is later followed by a proton-proton collider, FCC-hh. This comprehensive programme is well matched to the current scientific landscape after 15 years of LHC operation. The proposed staging takes into account: (1) the physics priorities as developed and stated by the Updates of the European Strategy for Particle Physics in 2013 and 2020; and (2) the relative technology readiness and costs of FCC-ee and FCC-hh.

Both FCC-ee and FCC-hh are installed in the same 91 km circumference tunnel close to CERN, which allows reusing all of the FCC-ee civil engineering and much of the technical infrastructure for the subsequent FCC-hh, thereby maximising the return on investment and ensuring guaranteed physics deliverables along with the broadest and most versatile exploration potential of the intensity and energy frontiers. Taking advantage of m a perfect four-fold superperiodicity, FCC-ee and FCC-hh each accommodate four detectors. The two FCC stages, FCC-ee and FCC-hh, are optimised so as to enable the widest possible physics programme, with ample complementarity and synergies between stage 1 and stage 2.

The FCC-ee does not only serve as a Higgs and top factory, but it also produces several $10^{12}$ Z bosons, opening another access to new physics. The hadron collider, FCC-hh, operates at a centre-of-mass energy of about 85 TeV, extending the energy frontier by almost an order of magnitude compared with the LHC, and providing a 5–10 times higher integrated luminosity than the upcoming High-Luminosity LHC. The mass reach for direct discovery at FCC-hh amounts to several tens of TeV, and it allows, for example, the direct production of new particles, whose existence could already be indirectly exposed by precision measurements at FCC-ee. The FCC-hh hadron collider can also accommodate ion and lepton-hadron collision options, allowing for complementary physics explorations.

The layouts of the FCC-ee electron-positron collider and its injector design are fully compatible with the demands of, and do not compromise the performance of, the future hadron collider FCC-hh. The main ring optics and RF configurations for both colliders were refined both to simplify operation across the energy range of FCC-ee and to enable a smooth transition to FCC-hh after the completion of the FCC-ee research programme.

The FCC-hh baseline assumes 14 Tesla $Nb_3Sn$ magnets which provide for a collision centre-of-mass energy of 85 TeV, with an R&D path towards HTS magnets, that would enable higher collision energies and/or reduced energy consumption. Designing a detector for a $\sim$100 TeV hadron collider is a challenging enterprise. Recent detailed studies prove that it should be possible to build a detector that can fully exploit the physics potential of such a machine, provided there is investment in the necessary detector R&D. The experience gained from the Phase-II upgrades of the LHC detectors for the HL-LHC, developments for further exploitation of the LHC, and ongoing detector R&D for future Higgs factories will be important stepping stones in this endeavour.

As noted, the FCC layout is designed to accommodate, first, an $e^+/e^-$ and, then, a hadron-hadron collider in the same tunnel. The arc tunnel diameter of 5.5 metres is larger than the LEP/LHC tunnel and satisfies the integration requirements of both colliders while also being fully compatible with the safety concept. Tunnel length and location are chosen to be consistent with the placement constraints described in Volume 3 and to allow hadron injection from either the SPS tunnel or the LHC tunnel at CERN, and lepton injection from an injector situated on the CERN Prévessin site. The FCC tunnel has also been developed so that the lepton and hadron colliders can house four experiments each.

A four-fold super-periodicity, with a particle physics experiment located at each quarter of the circumference, makes the ring appear like a four times smaller machine with a single interaction point. The super-periodicity, therefore, reduces the density of strong resonances in the betatron tune diagram, allowing a maximum beam-beam tune shift and optimum performance, for both lepton and hadron machines. Technical straight sections are located half-way between experiment insertion and these are used to accommodate important accelerator equipment and systems. The FCC layout accommodates 8 arcs of equal length, 4 technical straights (around the points PB, PF, PH and PL) and 4 experimental straights (around points PA, PD, PG and PJ), with lengths listed in Table 1. The final layout is illustrated in Figs. 1 and 2 for FCC-ee and FCC-hh, respectively. The collision points of FCC-hh lie on top of the FCC-ee interaction points, facilitating the sharing of some experimental infrastructure between FCC-ee and FCC-hh.

Table 1: Parameters of the tunnel layout for the two FCC colliders.

| circumference [m] | arc ($\times 8$) [m] | technical straight ($\times 4$) [m] | experimental straight ($\times 4$) [m] |
|---|---|---|---|
| 90 657.400 | 9616.175 | 2032.000 | 1400.000 |

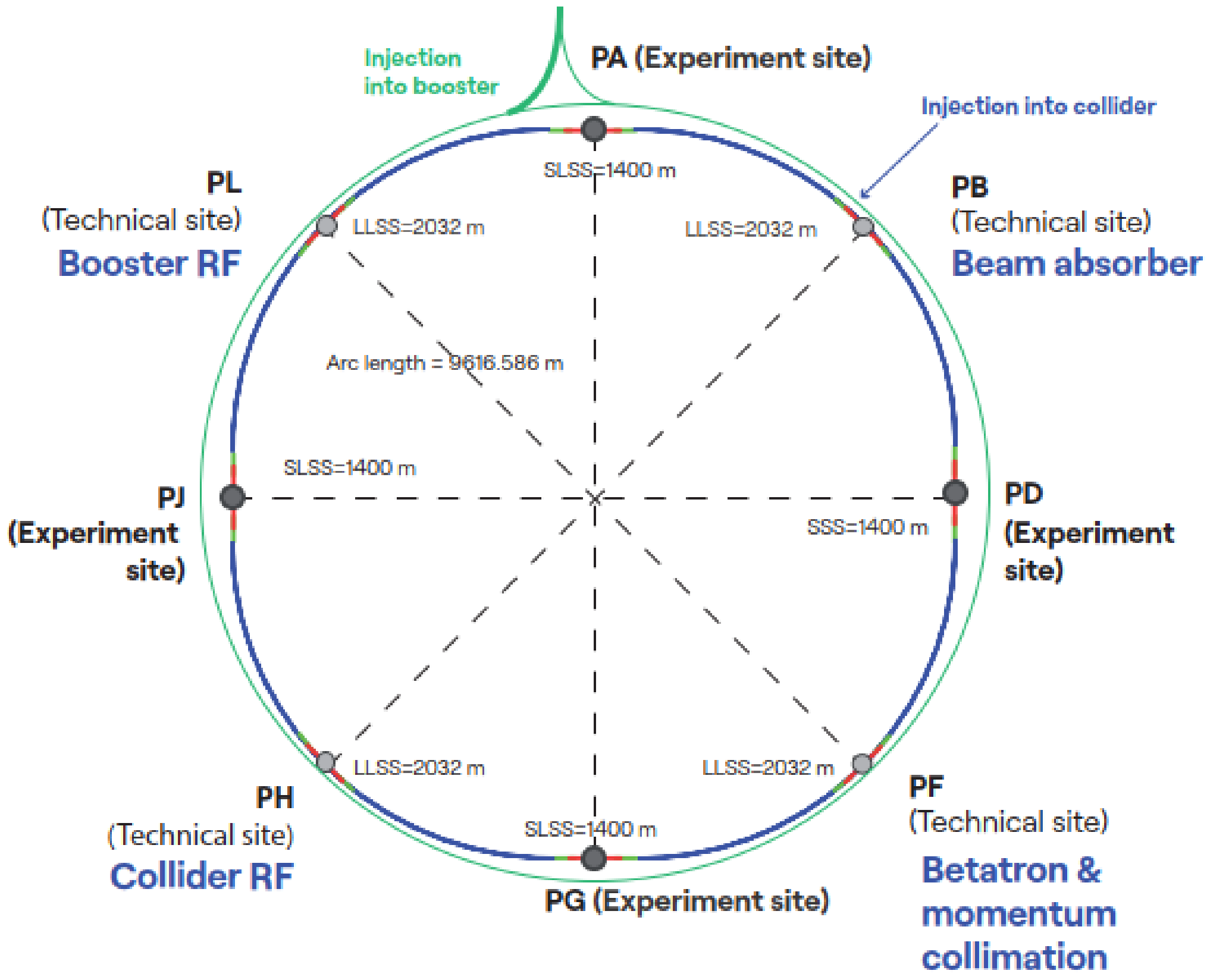

Fig. 1: The layout of the FCC-ee illustrating the four collision points and the four technical insertions.

The FCC-ee is conceived as a double-ring collider with separate beam pipes for electrons and positrons. Its luminosity is maximised by regular top-up injection from a full-energy booster synchrotron, which must also be located in the collider tunnel (sketched schematically by the green circle in Fig. 1). Transfer lines connecting from the FCC-ee injector on the CERN Prévessin site to the booster are indicated schematically around PA. For the FCC-ee, the straight around PL houses the booster radiofrequency (RF) system, and the one at PH accommodates the main-ring RF systems, as indicated in Fig. 1. Injection into the collider and extraction are integrated in a single technical straight section at PB. Both betatron cleaning and momentum collimation systems are located in the remaining straight section around PF, as shown in Fig. 1.

For the FCC-hh, two high-luminosity experiments can be installed in the diametrically opposed locations PA and PG, and two special-purpose experiments in PD and PJ. The betatron collimation is accommodated in the technical straight at PH, while the momentum collimation is housed in the technical straight around PB. In case of an abort, the beam is extracted in the straight around PF. The FCC-hh RF system is installed at PL, which for FCC-ee houses the booster RF. Note that the FCC-hh injection systems are installed in PB and PL, which are, therefore, shared between two accelerator systems. The hadron transfer line tunnels also connect to the main tunnel in the vicinity of PA. The transfer lines themselves continue inside the ring tunnel, on top of the arcs connecting PA with PL and PB where the counter-clockwise and clockwise beams are injected into the collider rings, respectively.

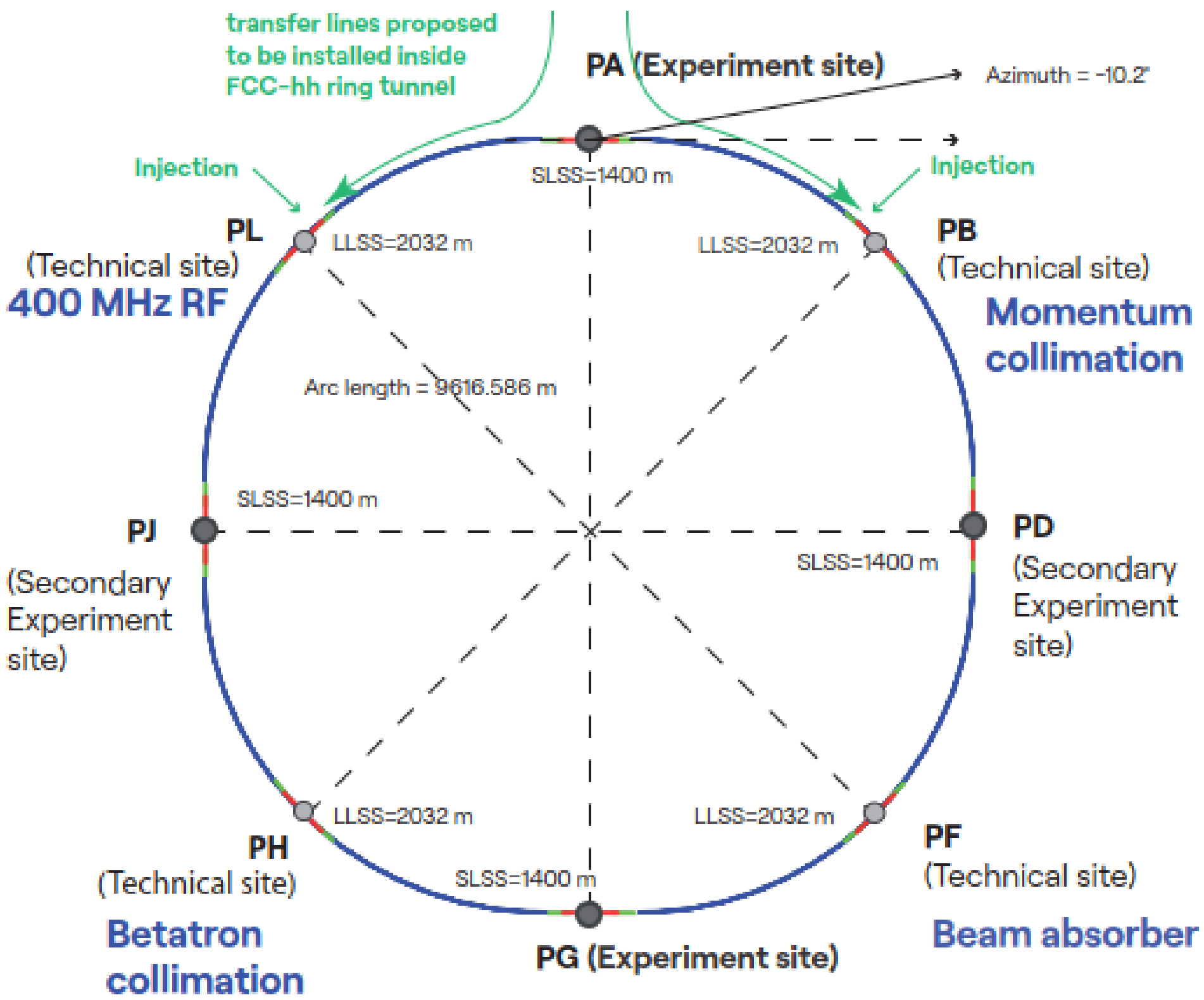

Fig. 2: The layout of the FCC-hh illustrating the four collision points and the four technical insertions.

## FCC-ee goals and parameters

The FCC-ee achieves maximum peak and integrated luminosities at four main working points, corresponding to the Z pole, the WW threshold, the (Z)H production peak and the $t\bar{t}$ threshold. Integrated luminosity targets amount to the production of $6 \times 10^{12}$ Z bosons (as required by the search for sterile right-handed neutrinos), more than $2 \times 10^{8}$ WW pairs, in excess of $2 \times 10^{6}$ Higgs bosons, and $2 \times 10^{6}$ $t\bar{t}$ pairs. All these objectives can be achieved during a 15-year period, which includes a one-year shutdown necessary to reconfigure the machine and install additional RF systems for the highest-energy $t\bar{t}$ running. With these integrated luminosities, the FCC-ee physics programme improves the precision of all electro-weak observables bytwo orders of magnitude, allows measurements of Higgs couplings (in a model-independent way) by up to an order of magnitude more precise than the HL-LHC, provides ten times the Belle-2 design statistics for bottom quarks, charm quarks and tau leptons, boosts the indirect discovery potential up to approximately $100\,\mathrm{TeV}$, and unlocks direct discovery potential for feebly-interacting particles (e.g., heavy sterile right-handed neutrinos) over the $5-100\,\mathrm{GeV}$ mass range.

The lepton-collider beam parameters are limited by various constraints and effects [1], such as beamstrahlung [2], a coherent beam-beam instability in collision with a large crossing angle [3], synchro-betatron resonances, polarisation requirements, and finally impedance effects, as discussed in Section 1.1. The parameters were optimised under these constraints [1]. Further refined simulations combining the effect of the full nonlinear lattice and either weak-strong or quasi-strong-strong beam-beam simulations, and the reverse-phase operation scheme for the RF cavities (with the implied transient beam loading effects) have led to parameter adjustments. The latest set of parameters is presented in Table 2. The parameters have been largely stable since 2021.

The bunch population is held approximately constant for all modes of operation, while the number of bunches is varied to adjust the beam current. The beam current is limited by the synchrotron radiation power, and strongly decreases at higher beam energies. For the Z operating point, a large number of $11\,200$ bunches are stored in each of the two collider rings. In this case, the bunches can be separated into 40 bunch trains of 280 bunches, with a bunch-to-bunch separation of $25\,\mathrm{ns}$ and a train-to-train separation of roughly $0.6\,\mu\mathrm{s}$. Similarly, at the WW threshold with 1780 bunches, bunches might be separated into 20 trains of 89 bunches having a bunch-to-bunch separation of roughly $150\,\mathrm{ns}$ and a train-to-train separation of roughly $2\,\mu\mathrm{s}$. At the ZH and $\mathrm{t\bar{t}}$ it is likely that the bunches are uniformly distributed around the ring. Two contributions to the beam lifetime are indicated separately: (1) the effect of lattice dynamic aperture and beamstrahlung plus quantum fluctuation, and (2) the unavoidable luminosity-related radiative Bhabha scattering. The total beam lifetime is the inverse of the sum of the individual inverse lifetimes. Finally, the luminosity and integrated luminosity at each operating point are discussed further in Section 1.1.

## FCC-hh goals and parameters

The hadron collider FCC-hh should provide proton–proton collisions with a centre-of-mass energy of the order of $85\,\mathrm{TeV}$ and an integrated proton-proton luminosity of about $20\,\mathrm{ab}^{-1}$ in each of the two multi-purpose experiments during 25 years of operation. Two specialised detectors are located in the remaining two experiment straights. In addition to colliding protons with protons, also proton-ion and ion-ion collisions, as with the LHC [5–7] but at much higher energy, are envisaged. Furthermore, an interaction point could be upgraded to electron–proton and electron–ion collisions, in which case an additional recirculating energy-recovery linac would provide the electron beam. The FCC–hh would use (modified) parts of the existing CERN accelerator complex for its injector chain.

The design of the FCC-hh hadron collider is based on LHC experience. The key challenges are the magnet technology and power consumption in the presence of strong synchrotron radiation. To limit the latter and also because the radiation damping during the store is significant, the beam current and the bunch population are relaxed compared to those of the HL-LHC [8, 9].

The key parameters for FCC-hh are compiled in Table 3. The bunch population (close to $1 \times 10^{11}$ protons per bunch) and the beam current ($0.5\,\mathrm{A}$) are kept the same as in the 2018 CDR [10]. Table 3 presents a baseline design with $14\,\mathrm{T}$ dipole magnets, which could be based on $Nb_3Sn$ technology. The table illustrates how the synchrotron radiation strongly increases with higher magnetic field.

The synchrotron-radiation heat, which must be extracted from inside the cold magnets, is a major contribution to cryogenic power. For the CDR, with $16\,\mathrm{T}$ $Nb_3Sn$ magnets, the FCC-hh cryogenics required around $250\,\mathrm{MW}$ of electrical power. With the lower field of $14\,\mathrm{T}$, this power can be significantly reduced. However, the cryogenic power might also be lowered for the higher-field magnets based on HTS technology, as these could conceivably be operated at higher temperature, together with an elevated temperature of the beamscreen intercepting the synchrotron radiation.

Table 2: Parameters of FCC-ee. Peak luminosity values are given per interaction point (IP), for a total of 4 IPs, integrated luminosities refer to the sum over four IPs. Both natural bunch lengths due to synchrotron radiation (SR) and collision values including beamstrahlung (BS) are shown. The FCC-ee collider rings feature a combination of 400 MHz RF systems (at the first three energies) and 800 MHz (additional cavities for $t\bar{t}$ operation), with voltage strengths respectively indicated. For the integrated luminosity, 185 days of operation per year, and luminosity production at 75% efficiency with respect to the ideal top-up running is assumed, as in the report [4].

| Running mode | Z | WW | ZH | $t\bar{t}$ |
|---|---|---|---|---|
| Number of IPs | 4 | 4 | 4 | 4 |
| Beam energy (GeV) | 45.6 | 80 | 120 | 182.5 |
| Bunches/beam | 11200 | 1856 | 300 | 60 |
| Beam current [mA] | 1292 | 135 | 26.8 | 5.1 |
| Luminosity/IP [$10^{34}$ cm$^{-2}$ s$^{-1}$] | 144 | 20 | 7.5 | 1.45 |
| Energy loss / turn [GeV] | 0.039 | 0.369 | 1.86 | 9.94 |
| Synchrotron Radiation Power [MW] | 100 | 100 | 100 | 100 |
| RF Voltage 400/800 MHz [GV] | 0.09/0 | 1.0/0 | 2.1/0 | 2.1/9.2 |
| Rms bunch length (SR) [mm] | 5.15 | 3.46 | 3.26 | 1.91 |
| Rms bunch length (+BS) [mm] | 15.2 | 5.28 | 5.59 | 2.33 |
| Rms relative momentum spread (SR) [%] | 0.039 | 0.069 | 0.102 | 0.152 |
| Rms relative momentum spread (+BS) [%] | 0.115 | 0.105 | 0.176 | 0.186 |
| Rms horizontal emittance $\varepsilon_x$ [nm] | 0.71 | 2.16 | 0.66 | 1.65 |
| Rms vertical emittance $\varepsilon_y$ [pm] | 2.1 | 2.0 | 1.0 | 1.32 |
| Longitudinal damping time [turns] | 1171 | 218 | 65.4 | 19.6 |
| Horizontal IP beta $\beta_x^*$ [mm] | 110 | 220 | 240 | 900 |
| Vertical IP beta $\beta_y^*$ [mm] | 0.7 | 1.0 | 1.0 | 1.4 |
| Hor. IP beam size $\sigma_x^*$ [µm] | 9 | 22 | 13 | 37 |
| Vert. IP beam size $\sigma_y^*$ [nm] | 40 | 45 | 32 | 44 |
| Beam lifetime (q+BS+lattice) [min.] | 87 | 75 | 100 | 105 |
| Beam lifetime (lum.) [min.] | 22 | 16 | 10 | 11 |
| Total beam lifetime [min.] | 18 | 13 | 9 | 10 |
| Total int. annual luminosity [ab$^{-1}$/yr] | 68$^\dagger$ | 9.6 | 3.6 | 0.67$^\ddagger$ |

$^\dagger$ The integrated luminosity in the first two years of Z running is assumed to be half this value to account for the machine commissioning and beam tuning; for WW and ZH running no additional commissioning time is allocated since the machine configuration and hardware are unchanged from the Z operation.

$^\ddagger$ The integrated luminosity in the first year of $t\bar{t}$ running, at the slightly lower beam energy of 170 – 175 GeV, is assumed to be about 65% of this value to account for the machine commissioning and beam tuning. The shorter time for commissioning compared with the lower energy running reflects the LEP/LEP-2 experience.

Table 3: Parameters of FCC-hh compared with the HL-LHC and LHC. For the integrated luminosity, 160 days of operation per year, and luminosity production at 75% efficiency with respect to the ideal running is assumed, as in the report [4]. The regular bunch spacing is 25 ns for all three colliders.

| | FCC-hh | HL-LHC | LHC |
|---|---|---|---|
| Centre-of-mass energy [TeV] | 85 | 14 | |
| Circumference [km] | 90.7 | 26.7 | |
| Dipole field [T] | 14 | 8.33 | |
| Beam current [A] | 0.5 | 1.1 | 0.58 |
| Bunch Intensity [$10^{11}$] | 1.0 | 2.2 | 1.15 |
| Number of bunches per beam | 9500 | 2760 | 2808 |
| Total synchrotron radiation power [kW] | 2400 | 15 | 7 |
| Synchrotron radiation power power per unit length [W/m/aperture] | 6.5 | 0.33 | 0.17 |
| Longitudinal emittance damping time [h] | 0.75 | 12.9 | |
| Interaction Point (IP) beta function $\beta^*_{x,y}$ [m] | 0.3 | 0.15 (min.) | 0.55 |
| Normalised rms emittance [µm] | 2.2 | 2.5 | 3.75 |
| Peak luminosity [$10^{34}$ cm$^{-2}$s$^{-1}$] | 30 | 5 (lev.) | 1 |
| Peak number of events per bunch crossing | 1000 | 132 | 27 |
| Stored energy per beam [GJ] | 6.5 | 0.7 | 0.36 |
| Integrated annual luminosity per IP [ab$^{-1}$/yr] | 0.9 | 0.25 | 0.05 |

## Sustainability goals

The general FCC implementation is driven by the principles of cost optimisation and sustainability, as described in Volume 3 of this Feasibility Study Report. Examples include the minimisation of the number of surface sites and the length of access roads by the placement optimisation, the planned processing of the excavated spoil for multiple purposes of reuse, and the local use of waste heat and cooling water.

The existence of a single such tunnel serving the global particle physics community until the end of the century exemplifies the discipline's commitment to sustainability. Moreover, the choice to propose the project in France and Switzerland – where electricity is already largely decarbonised – demonstrates a serious and proactive approach to environmental considerations. Finally, the fact that the FCC-ee is about hundred thousand times more efficient than its predecessor LEP, in terms of luminosity per unit electrical power, underscores the significant efforts of the community to advance responsibly.

As for the accelerators, storage rings are intrinsically sustainable machines, as they collide the same beams again and again, at multiple interaction points and over millions of turns. Their efficiency is limited only by the energy loss from synchrotron radiation, which, in the case of the FCC, is minimised by its large circumference.

Concerning the technical systems of the FCC-ee accelerators, their energy consumption is restricted by a variety of measures and design choices, such as novel ultra-high efficient continuous-wave (cw) RF power sources like tristrons, with a projected efficiency exceeding 90%, or twin dipole and quadrupole magnets for the FCC-ee arcs, and an increased $Q_0$ value of the superconducting RF cavities, by deploying exactly the same RF system for the first three modes of operation, and choosing a moderate accelerating gradient around $20-22\,\mathrm{MV/m}$ for the injector linacs. With regard to materials, by fabricating the main $400\,\mathrm{MHz}$ RF cavities using thin-film (niobium on copper) coating technology, the total amount of niobium is greatly reduced and the cavities also become more robust.

For the FCC-hh, the R&D aims at developing magnets with a cold-bore temperature higher than $1.9\,\mathrm{K}$, along with an elevated beamscreen temperature, which will relax the cryogenic power required, while the optimisation of the FCC-hh injector will further reduce the overall energy consumption.

# Chapter 1

# FCC-ee collider design and performance

## 1.1 Beam-beam effects, parameter choices, and luminosity

The FCC-ee will run at different beam energies, spanning the range from 45.6 to 182.5 GeV, so as to cover the Z pole, the WW threshold, the ZH production peak, and the $\mathrm{t\bar{t}}$ threshold. The latest parameter sets for the various operating energies are listed in Table 1.2. The parameter optimisation process for the FCC-ee was discussed in Ref. [1].

The FCC-ee design for all energies is based on the 'crab-waist' collision scheme, following its successful implementation at both DAΦNE and SuperKEKB [11, 12]. This scheme features flat beams ($\sigma_x^* \gg \sigma_y^*$), a large crossing angle between the two colliding beams, and (either 'virtual' or real) crab sextupoles at suitable betatron phase advance on either side of each of the interaction points (IPs). The crab waist scheme not only allows a small vertical beta function, but it also avoids coupling the vertical and horizontal betatron motion through the crossing-angle collision. It, thereby, ensures good stability of particle trajectories even when the beam-beam interaction is strong, thus enabling a high luminosity.

The luminosity at one IP may be expressed as

$$\mathcal{L} = \frac{\gamma}{2er_e}\frac{I_{tot}\xi_y}{\beta_y^*}R_G \ , \tag{1.1}$$

with the relativistic factor $\gamma$, the elementary charge $e$, the classical electron radius $r_e$, the total beam current $I_{tot}$, the vertical beam-beam parameter per IP $\xi_y$, the vertical optical $\beta$ function at the IP $\beta_y^*$ and the geometric reduction factor $R_G$, which represents both the 'hourglass' effect and the reduced overlap due to the crossing angle. The total current is constrained by the design limit for the synchrotron radiation power of 50 MW per beam. Therefore, the highest vertical beam-beam parameter is desired for maximum luminosity. However, in order to keep the tune footprint away from low-order resonances, all FCC-ee configurations feature $\xi_y \approx 0.1$ (Fig. 1.1).

The smallest $\beta_y^*$ is also desired, yet it is limited by the reduction of the luminosity due to the hourglass effect. The hourglass reduction factor is historically determined by the bunch length. However, in the crab waist collision scheme, the length of the overlap of the colliding beams is minimised through the crossing angle, leading to the condition $\beta_y^* \approx L_i$ with the interaction length defined by

$$L_i = \frac{\sigma_z}{\sqrt{1+\phi^2}}, \ \text{with} \quad \phi = \frac{\sigma_z}{\sigma_x^*}\tan\left(\frac{\theta_c}{2}\right) \tag{1.2}$$

the so-called Piwinski angle, related to the rms bunch length $\sigma_z$ the horizontal rms beam size at the collision point $\sigma_x^*$, and the full crossing angle between the beams at the IP $\theta_c$. In the small angle approximation ($\theta_c \ll 1$) for the large Piwinski angle regime ($\phi \gg 1$), the interaction length reduces to $L_i \approx 2\sigma_x^*/\theta_c$. The crossing angle $\theta_c$ is imposed by choice of the layout, in particular, due to the short common chamber around the IP specifically designed to avoid parasitic beam-beam encounters, and to allow a small $\beta_y^*$ with the lowest vertical chromaticity possible. Maximum luminosity, therefore, is achieved by minimising the horizontal beam size at the collision point, to allow for a reduction of $\beta_y^*$. The beam-beam tune shifts are given by

$$\xi_x = \frac{N_b r_e}{2\pi\gamma}\frac{\beta_x^*}{\sigma_x^2(1+\phi^2)}, \ \xi_y = \frac{N_b r_e}{2\pi\gamma}\frac{\beta_y^*}{\sigma_x\sigma_y\sqrt{1+\phi^2}} \tag{1.3}$$

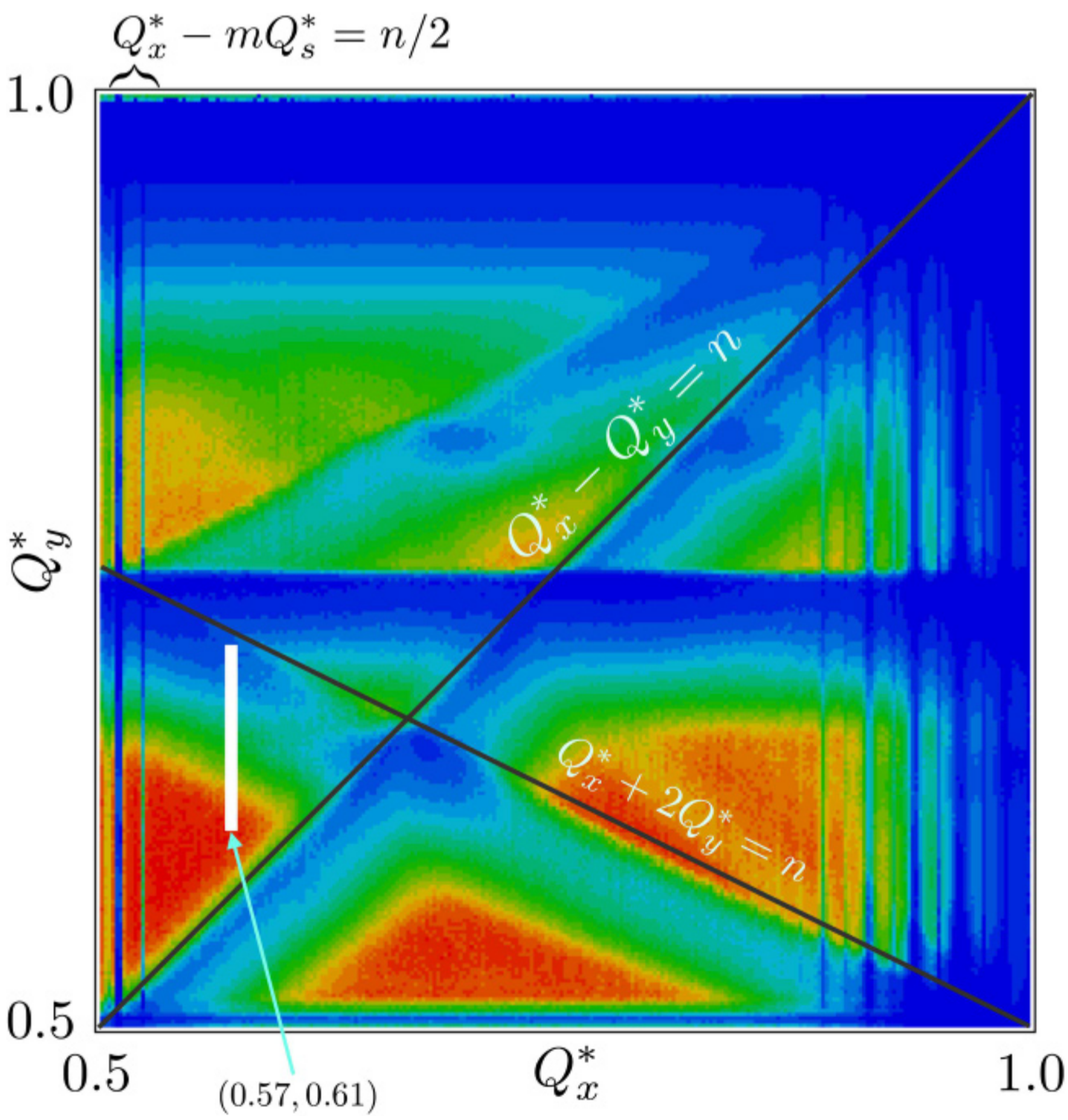

Fig. 1.1: Luminosity at the Z energy as a function of betatron tunes for the CDR configuration [13], represented by a single arc and a single IP. The colour scale extends from zero (blue) to $2.3{\cdot}10^{36}$ cm$^{-2}$s$^{-1}$ (red) [14]. The white narrow rectangle above (0.57, 0.61) indicates the footprint due to the beam-beam interaction

.

with $N_b$ the number of electron or positron per bunch, given by

$$N_b = \frac{I_{tot}}{e f_{rev} n_b} \tag{1.4}$$

with the revolution frequency $f_{rev}$ and the maximum number of bunches $n_b$. The number of bunches cannot be arbitrarily large due to the need for a minimal spacing between bunches (about 25 ns) in order to avoid adverse effects such as electron clouds (Section 1.4) as well as the need for gaps for injection and extraction (Section 1.8). This constraint is mostly relevant for the Z, yielding a minimum bunch charge of about $2 \cdot 10^{11}$ e$^{\pm}$.

In the small crossing angle and large Piwinski angle approximation, the beam-beam parameters reduce to

$$\xi_x \approx \frac{N_b r_e}{\pi\gamma} \frac{2\beta_x^*}{(\sigma_z\theta_c)^2}, \; \xi_y \approx \frac{N_b r_e}{\pi\gamma} \frac{1}{\sigma_z\theta_c} \sqrt{\frac{\beta_y^*}{\epsilon_y}}. \tag{1.5}$$

The bunch length, therefore, is another key parameter. It is determined by the effect of beamstrahlung, the synchrotron radiation in the arcs, the RF voltage and RF frequency, and by the choice of lattice mainly through the momentum compaction factor $\alpha_C$. It can be expressed as:

$$\sigma_z = \sigma_\delta \frac{\alpha_C C}{2\pi Q_s}, \;\; \text{with} \;\; Q_s = \frac{1}{2\pi}\left(\frac{eV_{RF}\omega_{RF}C_0}{p_0 c^2}\alpha_C \cos\phi_s\right)^{1/2} \;\; \text{and} \;\; \sin\phi_s = \frac{U_0}{eV_{RF}} \tag{1.6}$$

where $V_{RF}$ and $\omega_{RF}$ designate the RF voltage and frequency, $C$ the ring circumference, $p_0$ the reference momentum, and $U_0$ the total energy loss per turn. The synchrotron tune and synchronous phase are $Q_s$ and $\phi_s$, respectively. The momentum spread is obtained from

$$\sigma_\delta^2 = \sigma_{\delta,SR}^2 + \sigma_{\delta,BS}^2, \quad \text{with} \quad \sigma_{\delta,BS}^2 = \frac{n_{IP}\,\tau_{E,SR}\,c}{4C_0} \frac{55}{24\sqrt{3}} \frac{r_e^2\gamma^5}{\alpha_e} \int ds \left\langle \frac{1}{\rho^3} \right\rangle_{x,y,z} \tag{1.7}$$

where $\sigma_{\delta,SR}$ denotes the rms relative momentum spread caused by synchrotron radiation in the arcs and $\tau_{E,SR}$ the corresponding longitudinal radiation damping time, $n_{IP}$ the number of IPs, $\alpha_e$ the fine structure constant and $\rho$ the local bending radius of the particles' trajectories caused by the beam-beam interaction. Under certain assumptions, the integral over the bending radius can be approximated as follows [2]

$$\int ds \left\langle \frac{1}{\rho^3} \right\rangle \approx \frac{0.77562}{\sqrt{2\pi}^3 \sigma_z^2 \phi} \left( \frac{2N_b r_e}{\gamma \sigma_x} \sqrt{\frac{2}{\pi}} \right)^3. \tag{1.8}$$

Using the equations above, and introducing the parameter $\alpha_{BS} \geq 0$ to represent the effect of beamstrahlung [1], the equation for the momentum spread can be written as

$$\sigma_\delta^5 - \sigma_{\delta,SR}^2 \sigma_\delta^3 - \alpha_{BS} = 0, \quad \text{with} \; \alpha_{BS} \propto \frac{n_{IP}\tau_{E,SR}\gamma^{\frac{1}{2}} N_b^3}{\sigma_x^2 \theta_c} \left( \frac{V_{RF}\omega_{RF}}{\alpha_C} \right)^{\frac{3}{2}}. \tag{1.10}$$

The corresponding bunch length follows from Eq. (1.6). Due to its approximate nature, the equation is hardly used in the design process, and the impact of beamstrahlung is often obtained via tracking simulations. Yet this equation reveals that, in the regime of strong beamstrahlung ($\sigma_\delta \gg \sigma_{\delta,SR}$), the sensitivity to RF parameters and momentum compaction factor is reduced from the usual square root dependence due to fact that the beam-beam force, and consequently the strength of the beamstrahlung, depends on the bunch length. In the regime of strong beamstrahlung, the main drivers for the bunch length and momentum spread are the bunch intensity and the horizontal beam size ($\sigma_\delta \propto N_b^{3/5} \sigma_x^{*-2/5}$). These quantities must be adjusted to minimise the magnitude of the beamstrahlung.

As discussed above, the bunch intensity is constrained on the low-energy side, namely at the Z, by the electron-cloud instability. At higher energies, it needs to be kept large enough to maintain the beam-beam parameter (and consequently the luminosity) at the specified level. Also, a low horizontal emittance is required to achieve a low vertical emittance. Indeed, the two quantities are bound by the quality of the optics correction: currently, it is assumed that $\epsilon_y \approx \epsilon_x/10^3$ can be achieved (Section 1.4). The low vertical emittance enters directly into the luminosity, but it is also important for maintaining a good beam lifetime in the presence of a limited vertical dynamic aperture that comes along with the low $\beta_y^*$. Thus, maintaining a high $\beta_x^*$ is key to reducing beamstrahlung. Nevertheless, the horizontal $\beta^*$ is limited on the high side by the corresponding increase of the horizontal beam-beam parameter as well as of the strength of horizontal synchrobetatron resonances. These aspects are critical as the transverse tunes are set just above the half-integer in the horizontal plane and above the coupling resonance but below the third-order resonance in the vertical plane (Fig. 1.1), thus minimising the impact of low-order resonances on the beam quality. In this area, synchro betatron sidebands of the half-integer resonance in the horizontal plane are strongly excited due to the beam-beam interaction with a large Piwinski angle, leading to coherent instabilities, so-called $x$-$z$ instabilities [3], as well as by incoherent blow up [15]. Consequently, the horizontal $\beta^*$ must be chosen to maintain the strength of synchrobetatron resonances and the horizontal beam-beam parameter ($\xi_x \ll Q_s$) at an acceptable level. This optimisation, coupled to the relevant longitudinal aspects treated in the next paragraph, is done based on tracking simulation

[1]
$$\alpha_{BS} \cong 0.77562 \cdot \frac{220}{3\sqrt{3}} \frac{r_e^5 \gamma^2}{\alpha_e} \frac{n_{IP}\tau_{E,SR}}{T_{rev}} \left( \frac{Q_s}{\alpha_C C_0} \right)^3 \frac{N_b^3}{\sigma_x^2 \theta_c} \; \text{with } Q_s \text{ from Eq. (1.6)} \tag{1.9}$$

(Section 1.4). Thanks to the increase in radiation damping and the shorter bunch length at higher energies, these effects become less severe, and higher $\beta_x^*$ are allowed.

The tune space depicted in Fig. 1.1 corresponds to the CDR configuration, yet the main features have not fundamentally changed. The main difference is the lowering of the tune per quarter of the machine towards the half-integer (layout with 4 IPs), with respect to the half of the machine in the CDR (layout with 2 IPS) in order to maintain the total tune in the same area, thereby avoiding important resonances when considering the impact of the real lattice (Fig. 1.3). The total tune spread is large and a tight control of the resonances driven by the lattice including imperfections is important.

The longitudinal parameters are set to ensure a sufficiently large RF bucket height. At the same time, the spin tune spread needs to remain smaller than the synchrotron tune to allow for energy calibration via resonant depolarisation (Section 1.7). These two conditions are

$$\left(\frac{\Delta p}{p}\right)_{\max} = \left(\frac{eV_{RF}\omega_{RF}C_0}{2\pi^2 c\alpha_C E_0}(2\cos(\phi) + (2\phi_s - \pi)\sin(\phi_s))\right)^{1/2} > 0.01, \text{ and } Q_s > a\gamma\sigma_{\delta,SR} \tag{1.11}$$

with $a$ the anomalous magnetic moment of the electron. Aiming at a high voltage and a low momentum compaction factor, a high synchrotron tune is favoured. The RF voltage available is an important cost driver, it is kept at a level required to compensate for the energy lost by synchrotron radiation, and to maintain a sufficiently large RF momentum acceptance.

At the Z, an additional constraint on the RF voltage arises from the choice of operating 2-cell cavities in reverse polarity mode in order to keep the same RF system for the Z, WW and ZH energies (Section 3.4). The voltage may need to be kept higher than otherwise necessary to minimise the impact of transient beam loading.

Two different lattices are considered, with the one for the two highest energies (ZH and $\mathrm{t\bar{t}}$) featuring a lower momentum compaction factor than the optics for the two lower energies (Z and WW). The optics change is required to maintain a low transverse emittance at the higher energies.

While at the Z energy the maximum number of bunches is constrained by the electron cloud instability, at other energies it can be fully optimised to obtain the highest luminosity. This is achieved by choosing the highest bunch charge that does not lead to a significant lifetime degradation or emittance growth (corresponding to a vertical beam-beam tune shift of about $\xi_y \approx 0.1$) [16].

A key difference from existing colliders is the fact that beam parameters are defined by an equilibrium condition that is mostly driven by the beam-beam interaction itself, and in particular by the beamstrahlung. Transverse and longitudinal beam sizes are, thus, strongly coupled, introducing additional constraints on the design and operation of the collider.

An important aspect is the need for a reasonably adiabatic ramp up of the beam-beam force such that the energy spread and the bunch length may increase from the lattice equilibrium to the new equilibrium with beamstrahlung avoiding uncontrolled losses in the process. Indeed, considering, for example, the situation where one nominal beam would be circulating, with the lattice equilibrium emittances, and a second lower intensity beam would be injected from the booster, a 3D flip-flop mechanism could occur where the injected beam blows up significantly, while the other one remains smaller than nominal. As a mitigation measure, the so-called bootstrap injection scheme was devised, where the intensity difference between colliding bunches does not exceed a few percent [16] (Section 1.8).

The 3D flip-flop mechanism is a direct consequence of the strong coupling between the equilibria in the different planes. In the case of an asymmetry in the strength of beamstrahlung between the two colliding bunches, the bunch experiencing lesser beamstrahlung sees its length decrease, thus increasing the strength of beamstrahlung for the other beam. As a result, beamstrahlung decreases further on the shorter bunch, thus again enhancing the effect. These dynamics may reach a stable equilibrium where one beam is large and the other small in the longitudinal direction, a so-called 'flip-flop' effect. Similar 'flip-flop' phenomena, albeit without beamstrahlung and in the transverse direction, have been seen in $e^+e^-$ colliders for the past half a century, e.g., Ref. [17]. At FCC-ee, due to the finite momentum acceptance, as well as the strong beam-beam force generated by the short bunch, the long, 'weak' bunch may also experience transverse blow-up and possibly significant beam losses, making the mechanism three dimensional. The '3D flip-flop' situation is considered irreversible so the affected bunches would need to be dumped. Avoiding this mechanism imposes tight tolerances on the symmetry of the two beams, regarding bunch intensity and optics control [16, 18].

## 1.2 Optics design

This section describes the main electron and positron collider rings. Subsections describe the arc optics, the experimental insertions, and the design of the various technical straight sections. The baseline optics is called the Global Hybrid Correction (GHC) scheme, since the vertical chromaticity for the low-beta insertion is corrected locally, with two sextupole magnets in the final focus, while the horizontal chromaticity is globally corrected globally using the arc sextupoles.

The baseline collider optics corresponds to a layout that can accommodate four interaction points (IPs), with a super-periodicity of four and fourfold symmetry, comprising eight arcs, and eigt long straight sections (LSS), which are further divided into four technical "long long straight sections" (LLSS, at points PB, PF, PH, and PL), and four experiment "short long straight sections" (SLSS, at points PA, PD, PG, and PJ), as are defined in Table 1.1 and illustrated in Fig. 1.2.

Table 1.1: Parameters of the layout in metres.

| circumference | arc | technical LSS (LLSS) | experiment LSS (SLSS) |
|---|---|---|---|
| 90 657.400 | 9616.175 | 2032.000 | 1400.000 |

The experiment long straight sections accommodate the optics leading to the interaction points (IPs) and the detectors. The beam crossing angle and the distance between the IP and the face of the first quadrupole ($\ell^*$) are maintained at 30 mrad and 2.2 m, respectively. Each beam must arrive towards the IP from the inside to minimise the synchrotron radiation going into the detector; therefore the beams must cross in each technical long straight section. The technical LSSs are used for RF, injection, extraction, and collimation as illustrated in Fig. 1.2. The technical LSS at PH is used for the RF for the collider at all energies, while the LSS at PL is used for the booster RF. The beam optics of the collider RF section will change at the transition from WW to ZH, so as to accommodate RF cryomodules common to both $e^+$ and $e^-$ for ZH, and separate for WW.

The PL long straight section is used for the booster RF. The RF section for the collider is concentrated in the PH long straight section for all energies. This seems to induce an additional non-structural synchro-betatron resonance, limiting the choice of the transverse tune space at some energies.

The design goal of the LLSS is to use identical optics for the RF, injection/extraction, and collimation at Z and WW, in order to maintain the superperiodicity of the ring in terms of path length, phase advance and chromaticity. Thus, no special tuning of the optics and sextupoles is required. At higher energies, for the ZH and $t\bar{t}$ modes of operation, the optics in the RF section changes, so as to accommodate the RF cavities common to both $e^+$ and $e^-$ beams.

The optics in the other technical LSSs remain the same for all energies. At the experiment LSSs, the IP localisation and beamline layouts deviate from the reference layout line defined by Table 1.1 to generate the crossing angle for the collision with reduced synchrotron radiation (SR) toward the IP.

The ring vertical chromaticities are set to $+5$ and $+2$ at Z and WW, respectively, and to 0 at other energies. This is necessary to suppress the transverse mode coupling beam instability. The dynamic aperture (DA) and beam lifetime are optimised under these chromaticities.

The choice of the betatron tunes and parameters related to beam-beam such as bunch intensity,

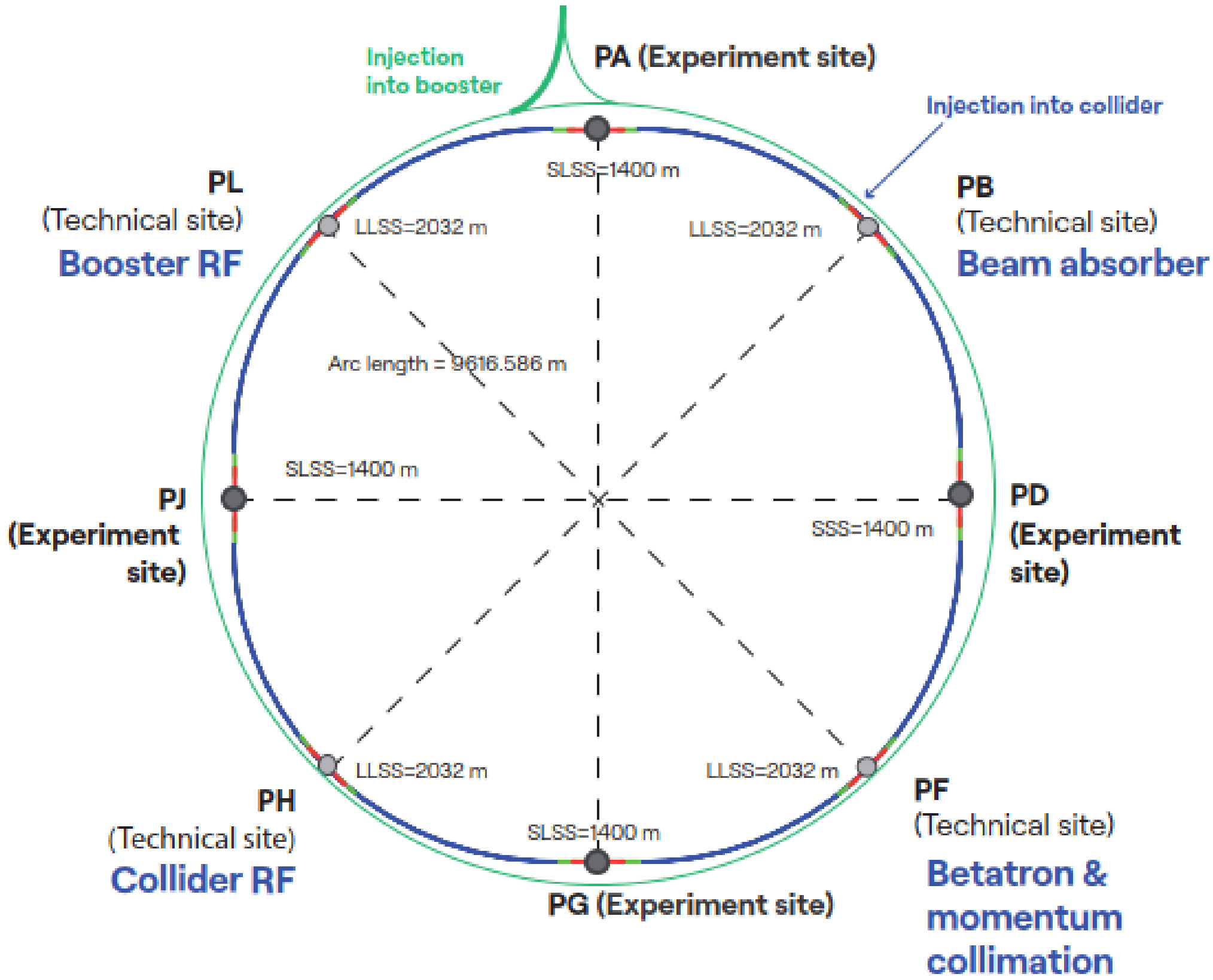

Fig. 1.2: The layout of the FCC-ee illustrating the 4 collision points in the SLSS and the four technical insertions at the LLSS.

$\beta^*$, bunch length, chromaticities, etc., assumes the reverse-phase operation of the RF cavities. There are limitations on the combination of parameters as discussed in Section 1.1.

Table 1.2 lists machine parameters associated with the baseline design for four beam energies. The effect of the full non-linear lattice and the beam-beam collisions with beamstrahlung is now simulated simultaneously. These simulations include the full lattice, synchrotron radiation from all components with realistic photon spectra, tapering, beam-beam collisions, and beamstrahlung, using the simulation codes SAD and BBWS. As a result, the beam lifetime is considered realistic, and the blow-up of the vertical emittance due to beam-beam effects and beamstrahlung is quantitatively estimated, as reported in the table.

It is important to distinguish the vertical emittance as ‘emittance after collision’ and ‘emittance by lattice’, as shown in Table 1.2. Roughly a factor of 2 blowup of the vertical emittance is expected at each energy according to the simulations above. The required lattice emittance will be the smallest at Z, 0.75 pm, including the vertical emittance generated by the interaction point (IP) solenoid, $\sim 0.43$ pm. Therefore, if such a small lattice vertical emittance must be achieved at Z, the lattice emittance at higher energies should be reduced to a comparable level. Note that the solenoid emittance scales as $B_z/E^5$, so it will be easier to achieve the same level of lattice emittance at higher energies. The vertical emittances in Table 1.2 are smaller than those in the CDR, which they had been chosen as 2% of the horizontal emittance in collision at each energy. At ZH, the lattice vertical emittance is required to be smaller than 0.65 pm, which is even smaller than at the Z. However, at the Z there is the large, inevitable vertical emittance due to the solenoid, $\sim 0.43$ pm, which is reduced by $\sim 1/E^5$, so that the emittance from the rest of the ring can still be larger at ZH than at Z.

Table 1.2: FCC-ee collider parameters for the GHC lattice. SR: synchrotron radiation, BS: +beamstrahlung.

| Beam energy | [GeV] | 45.6 | 80 | 120 | 182.5 |
|---|---|---|---|---|---|
| Layout | | PA31-3.0 | | | |
| # of IPs | | 4 | | | |
| Circumference | [km] | 90.658509 | | | |
| Bend. radius of arc dipole | [km] | 10.021 | | | |
| Energy loss / turn | [GeV] | 0.0387 | 0.369 | 1.86 | 9.93 |
| SR power / beam | [MW] | 50 | | | |
| Beam current | [mA] | 1292 | 135 | 26.8 | 5.0 |
| Colliding bunches / beam | | 11200 | 1856 | 300 | 60 |
| Colliding bunch population | $[10^{11}]$ | 2.18 | 1.38 | 1.69 | 1.58 |
| Hor. emittance at collision $\varepsilon_x$ | [nm] | 0.71 | 2.16 | 0.66 | 1.65 |
| Ver. emittance at collision $\varepsilon_y$ | [pm] | 2.1 | 2.0 | 1.0 | 1.32 |
| Lattice v. emittance $\varepsilon_{y,\mathrm{lattice}}$ | [pm] | 0.87 | 1.20 | 0.57 | 0.82 |
| Arc cell | | Long 90/90 | | 90/90 | |
| Momentum compaction $\alpha_p$ | $[10^{-6}]$ | 28.52 | 28.67 | 7.52 | 7.57 |
| Arc sext families | | 73 | | 144 | |
| $\beta^*_{x/y}$ | [mm] | 110 / 0.7 | 220 / 1 | 240 / 1 | 900 / 1.4 |
| Transverse tunes $Q_{x/y}$ | | 218.168 / 222.200 | 218.185 / 222.220 | 398.150 / 398.220 | 394.148 / 390.218 |
| Chromaticities $Q'_{x/y}$ | | +5 / +5 | 0 / +5 | 0 / 0 | 0 / 0 |
| Energy spread (SR/BS) $\sigma_\delta$ | [%] | 0.039 / 0.115 | 0.069 / 0.105 | 0.102 / 0.176 | 0.152 / 0.186 |
| Bunch length (SR/BS) $\sigma_z$ | [mm] | 5.15 / 15.2 | 3.46 / 5.28 | 3.26 / 5.59 | 1.91 / 2.34 |
| RF voltage 400/800 MHz | [GV] | 0.0885 / 0 | 1.00 / 0 | 2.09 / 0 | 2.10 / 9.20 |
| Harm. number for 400 MHz | | 121200 | | | |
| RF frequency (400 MHz) | MHz | 400.788026 | | | |
| Synchrotron tune $Q_s$ | | 0.0310 | 0.0809 | 0.0334 | 0.0892 |
| Long. damping time | [turns] | 1179 | 218 | 65.4 | 19.4 |
| RF acceptance | [%] | 1.21 | 3.32 | 2.06 | 3.07 |
| Energy acceptance (DA) | [%] | $\pm1.0$ | $\pm1.0$ | $\pm1.9$ | -2.8/+2.5 |
| Beam crossing angle at IP $\theta_x$ | [mrad] | $\pm15$ | | | |
| Crab waist ratio | [%] | 60 | 55 | 50 | 40 |
| Beam-beam $\xi_x/\xi_y$ [2] | | 0.0023 / 0.098 | 0.013 / 0.129 | 0.0108 / 0.130 | 0.066 / 0.144 |
| Piwinski ang. $(\theta_x\sigma_{z,\mathrm{BS}})/\sigma^*_x$ | | 25.8 | 3.6 | 6.6 | 0.91 |
| Lifetime (q + BS + lattice) | [sec] | 5200 | 4500 | 6000 | 6300 |
| Lifetime (lum)[3] | [sec] | 1330 | 960 | 600 | 640 |
| Luminosity / IP $/10^{34}$ | $[/\mathrm{cm}^2\mathrm{s}]$ | 144 | 20 | 7.5 | 1.45 |

Tune scans for the lattices at the different working points are plotted in Fig. 1.3. It is interesting to note that the resonance $Q_x + 2Q_y - Q_s = \mathrm{int.}$ disappears at ZH and $\mathrm{t\bar{t}}$. One may speculate that it is due to the faster damping rate at higher energies, but it is not. It has been found that even if the energy of the ZH lattice is reduced to the WW energy, this resonance does not appear at all. So, this resonance seems to be related to the lattice itself.

Figure 1.4 shows the vertical emittance in the collision and the beam lifetime as functions of the lattice emittance at each energy. The parameters in Table 1.2 are optimised within the range over which the lifetime due to nonlinear lattice and beamstrahlung is much longer than the luminosity lifetime. However, it is important to note that neither machine errors nor corrections are included in this calculation. The resulting lifetime for the lattice and beamstrahlung may seem long at some energies, such as Z. More luminosity may be obtained by increasing the bunch current, but it was decided not to push the parameters and leave some room for future improvements. Considering that the luminosity simulations do not yet include machine errors and corrections, leaving some margin appears reasonable for this design study phase.

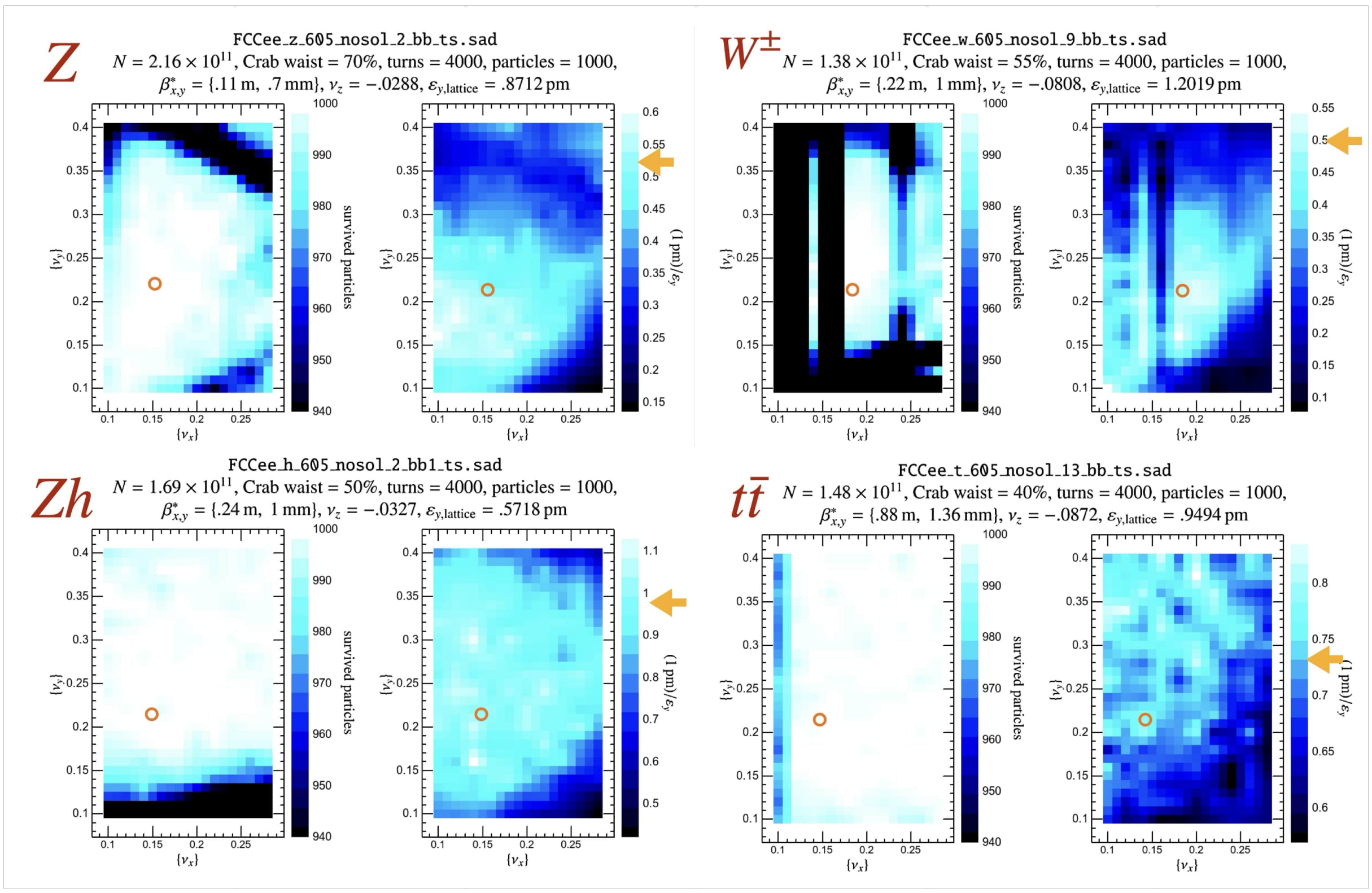

Fig. 1.3: The tune scan of the beam-beam effect with the full lattice, upper-left: Z, upper-right: WW, lower-left: ZH, lower-right: $\mathrm{t\bar{t}}$. At each energy, the left/right plots show the beam loss/vertical emittance blowup, respectively. The whiter areas correspond to longer lifetimes and higher luminosity. At the Z and WW, a strong resonance line $Q_x + 2Q_y - Q_s = \mathrm{int.}$ is seen. Also, at WW, strong vertical lines appear at $Q_x + nQ_s = \mathrm{int.}$, $(n = 1, 2)$. The red circles show the design tunes. Each orange arrow on the colour scale of the blowup indicates the level at the design working point.

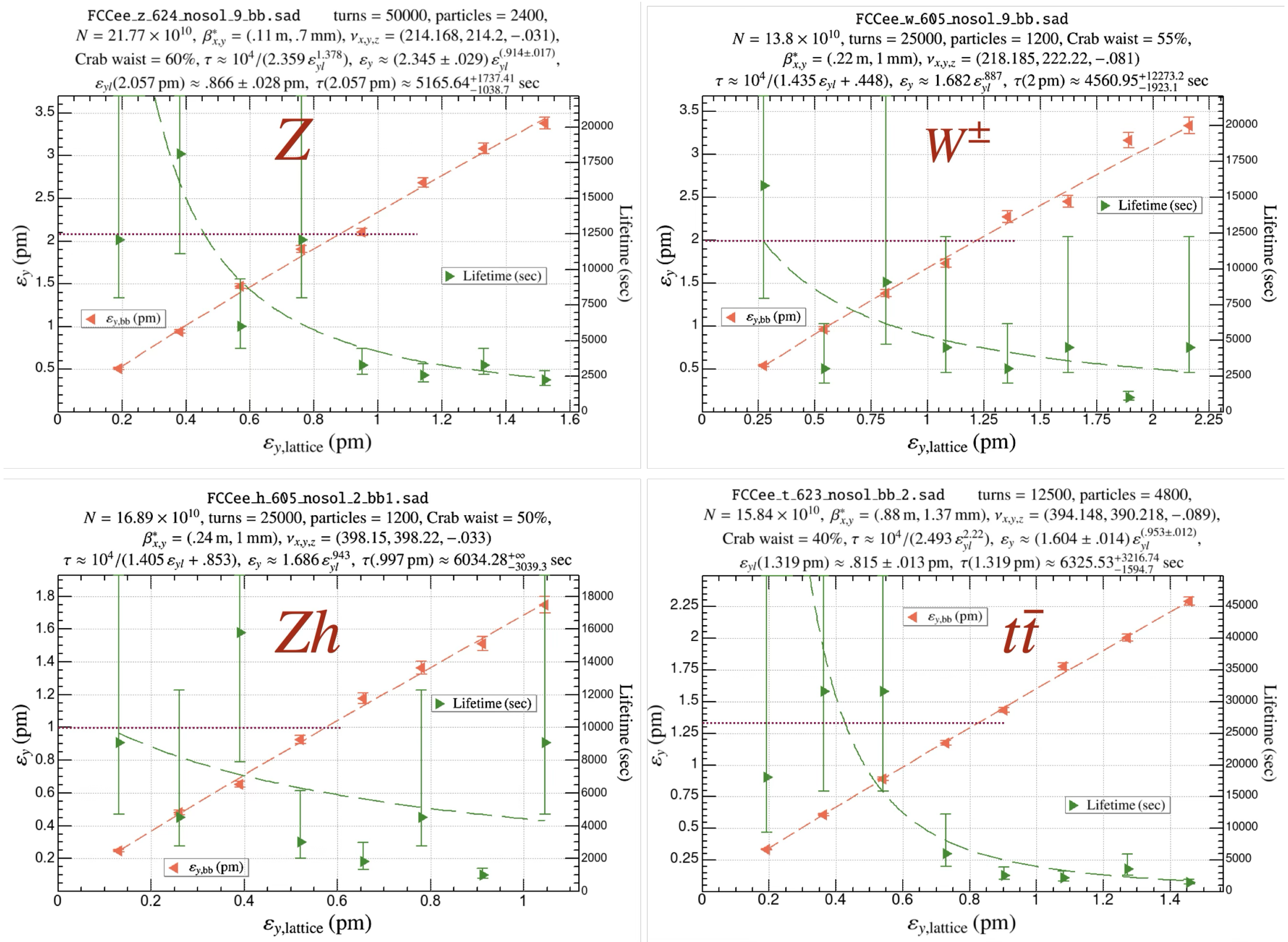

Fig. 1.4: Results of beam-beam tracking with lattice and beamstrahlung for each energy of FCC-ee. Each plot shows the vertical emittance after collision (red) and the lifetime (green) against the lattice vertical emittance at each collision energy. The purple horizontal dashed line shows the target vertical emittance at collision, where the vertical emittance of the strong beam is set at.These results with SAD as well as the DA have been reproduced by independent simulations with XSUITE.

### 1.2.1 Arc optics

The arc FODO cell phase advance at lower energies (Z, WW) is 90°/90° with twice the cell length (long 90/90) of the ZH and $\mathrm{t\bar{t}}$ machine, as shown in Fig. 1.5. Twin aperture dipoles and quadrupoles are used in the arcs. The separation between the two beams is 35 cm. Multi-family $-I$-paired sextupoles are deployed in the arc for global chromatic correction, respecting the fourfold superperiodicity. This multifamily scheme provides great flexibility for controlling additional parameters such as the chromatic behaviour at the IP or RF, and the dynamic aperture.

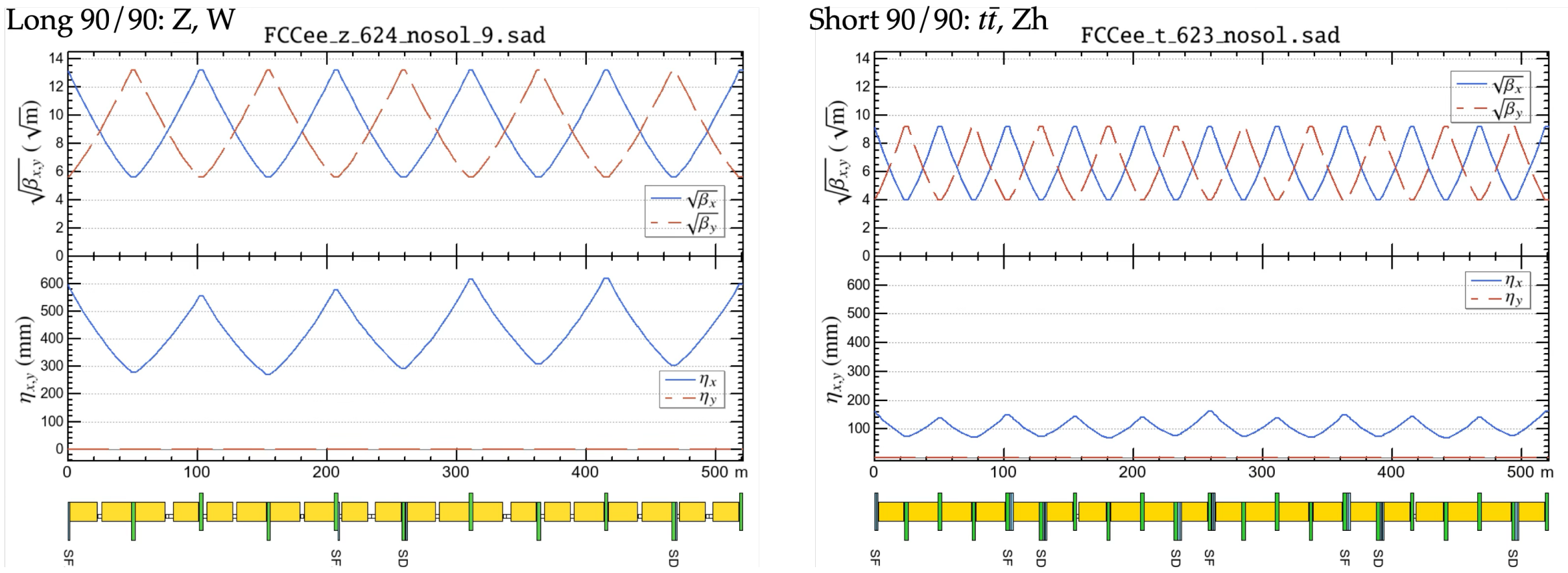

Fig. 1.5: Optics for the arc cell: Z/W operation modes (left) and Zh/$\mathrm{t\bar{t}}$ operation modes (right). Labels show the $-I$ paired sextupoles.

### 1.2.2 Experiment insertion optics

The critical photon energy of incoming synchrotron radiation from the dipoles in the interaction region is kept below 100 keV at $\mathrm{t\bar{t}}$ up to $\sim 450$ m upstream of the IP, as in Ref. [19]. There are 32 sextupoles in the interaction regions: 2 per side and per IP $\times$ 2 sides $\times$ 4 IPs $\times$ 2 beams. The optics uses the 'virtual' crab sextupoles scheme, using the vertical local chromaticity correction sextupoles as described in [20].

The lattice assumes a perfect solenoid compensation with counter-solenoids between the face of the last quadrupole (QC1L1/R1) and the IP ('local scheme'). This scheme guarantees a perfect achromatic coupling correction with no leak of vertical orbit and dispersion to the outside. It is known that this scheme also guarantees the perfect removal of harmful beam-beam effects coupled to the chromatic coupling, from which SuperKEKB has been suffering so far. This compensation scheme, due to dispersion and synchrotron radiation in the solenoid fields with crossing angle, generates a vertical emittance of 0.5 pm at the Z.

The optics of the experiment LSS incorporates the polarisation wigglers and a space for a Compton polarimeter. The latter is essential to detect the spin precession angle at each IP to provide additional constraints for the beam energy calibration.

The optics for the experimental straight sections (short long straight section — SLSS) is shown in Fig. 1.6.

### 1.2.3 Collimation insertion optics

A global beam halo collimation system will be required in the FCC-ee to protect the machine hardware from unavoidable beam losses and for detector background control. Two global collimation systems are foreseen: one betatron collimation system to remove large amplitude particles and one momentum cleaning section to intercept particles with a significant momentum deviation. Both systems are foreseen to be located in the LSS at PF, as shown in Fig. 1.2. Additional collimation will be located in each interaction region (IR) and is described in Section 1.6.4.

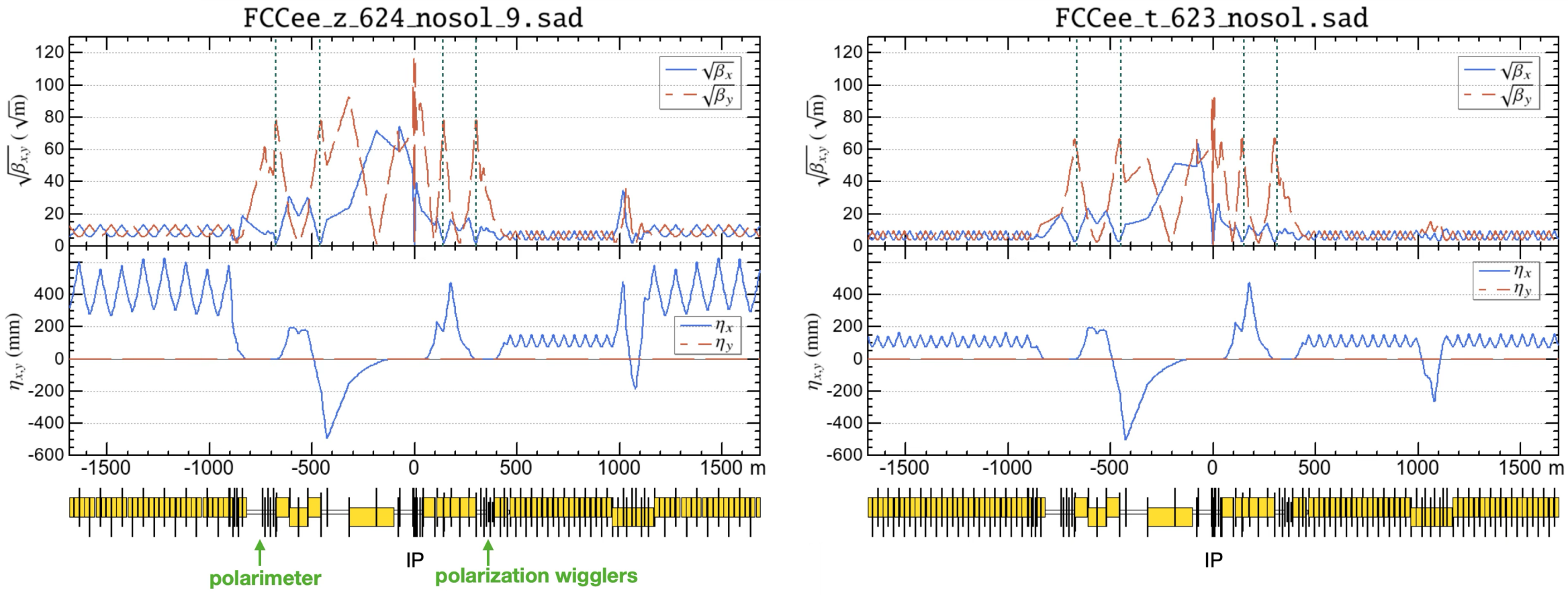

Fig. 1.6: Layout and optics for the experiment straight section: Z/W operation modes (left) and ZH/$t\bar{t}$ operation modes (right). This section includes sections for an inverse-Compton polarimeter and polarisation wigglers. The vertical lines show the location of the sextupoles for vertical local chromaticity correction and crab waist.

The layout and optics for a betatron and momentum collimation section are presented in Fig. 1.7 for both the Z and $t\bar{t}$ operation mode. A beam crossing section with a length of $200\,\mathrm{m}$ is located at the centre of the insertion, where the incoming beam will cross from the outside to the inside of the ring. The betatron collimation is installed upstream of the crossing, whereas the momentum collimation is located downstream. The optics is matched to obtain collimator half-gaps of more than $2\,\mathrm{mm}$ for the required betatron cuts in units of beam sigma. Moreover, the horizontal dispersion is kept low in the betatron collimation section to avoid the collimators there becoming the momentum bottleneck. In the momentum collimation section, the ratio of the horizontal dispersion and $\beta$-function is matched such that the collimators there can provide a sufficiently small momentum cut without becoming the aperture bottleneck.

The collimation settings and performance of the collimation system is presented in Section 1.5 and R&D challenges for the collimators are described in Section 3.6.3.

### 1.2.4 RF insertion optics

Two long-long straight sections (LLSS) of roughly 2000 metres in length will be used for SRF systems. In particular, the LLSS at PL will house the booster RF systems and LLSS PH will house the main ring SRF systems as indicated in Fig. 1.2. The layouts of these insertions are based on initial models for the 400 and 800 MHz cryomodules, which are based on existing cryomodules at CERN. These cryomodule models have not yet been optimised for the FCC-ee.

The requirements on the SRF insertions include:

– accommodate the required number of SRF cavities into the 2032 metre long straight section; the details of the SRF systems can be found in Section 3.4.
– provide a beam crossing, e.g., the beam entering on the outside of the tunnel leaves the LLSS on the inside of the tunnel.
– minimise synchrotron radiation reaching the SRF cavities.
– provide the appropriate focusing and diagnostics to control the beam through the SRF cavities.

In addition, effort was made to ensure that the SRF systems along with the accelerators, the booster and

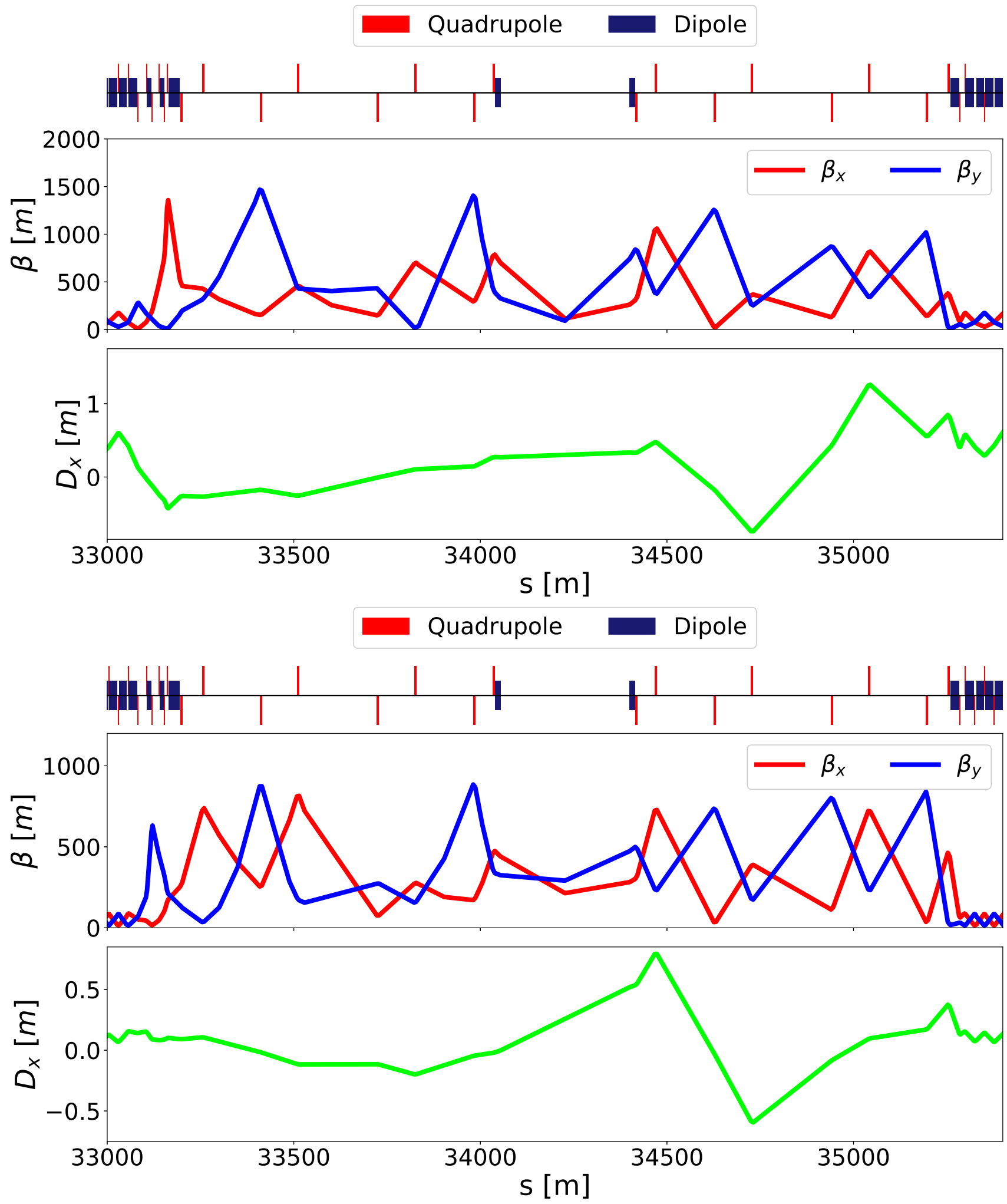

Fig. 1.7: Layout and optics for a collimation insertion: Z operation mode (top) and $t\bar{t}$ operation mode (bottom).

e+/e- main rings, and the technical infrastructure required could be accommodated into the 5.5 metre diameter tunnel that is used elsewhere in the collider arcs and other LLSS.

Example 400 MHz and 800 MHz cryomodule designs are shown in Section 3.4.11. Figures 1.9 and 1.10 present tunnel cross sections including SRF cryomodules at PH and PL, respectively.

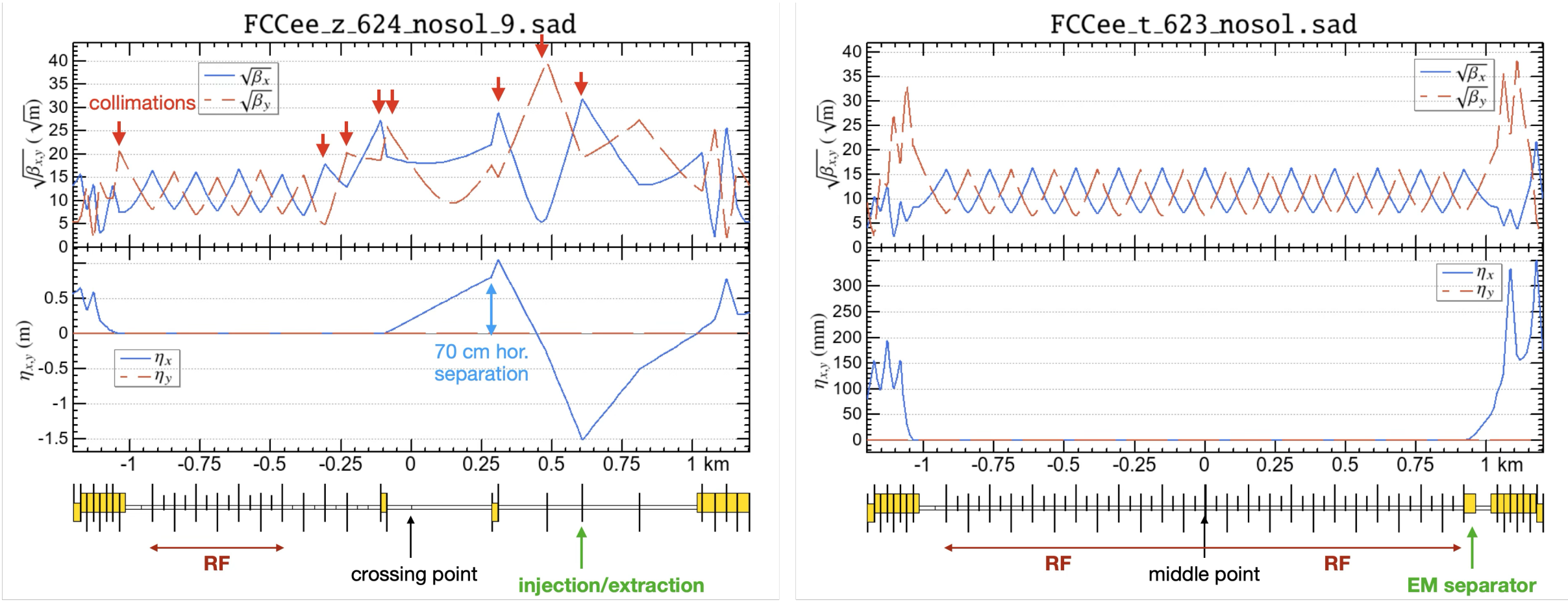

Fig. 1.8: Layout and optics for LLSS: Z/W operation modes for RF, injection/extraction, collimation (left) and ZH/$t\bar{t}$ operation modes for RF (right). The left optics is identical for all four LLSSs at Z/W. The left RF section is replaced by the right optics at ZH/$t\bar{t}$ for the common-RF scheme with an electromagnetic separator. The space for the RF components is 1890 m except the quadrupoles.

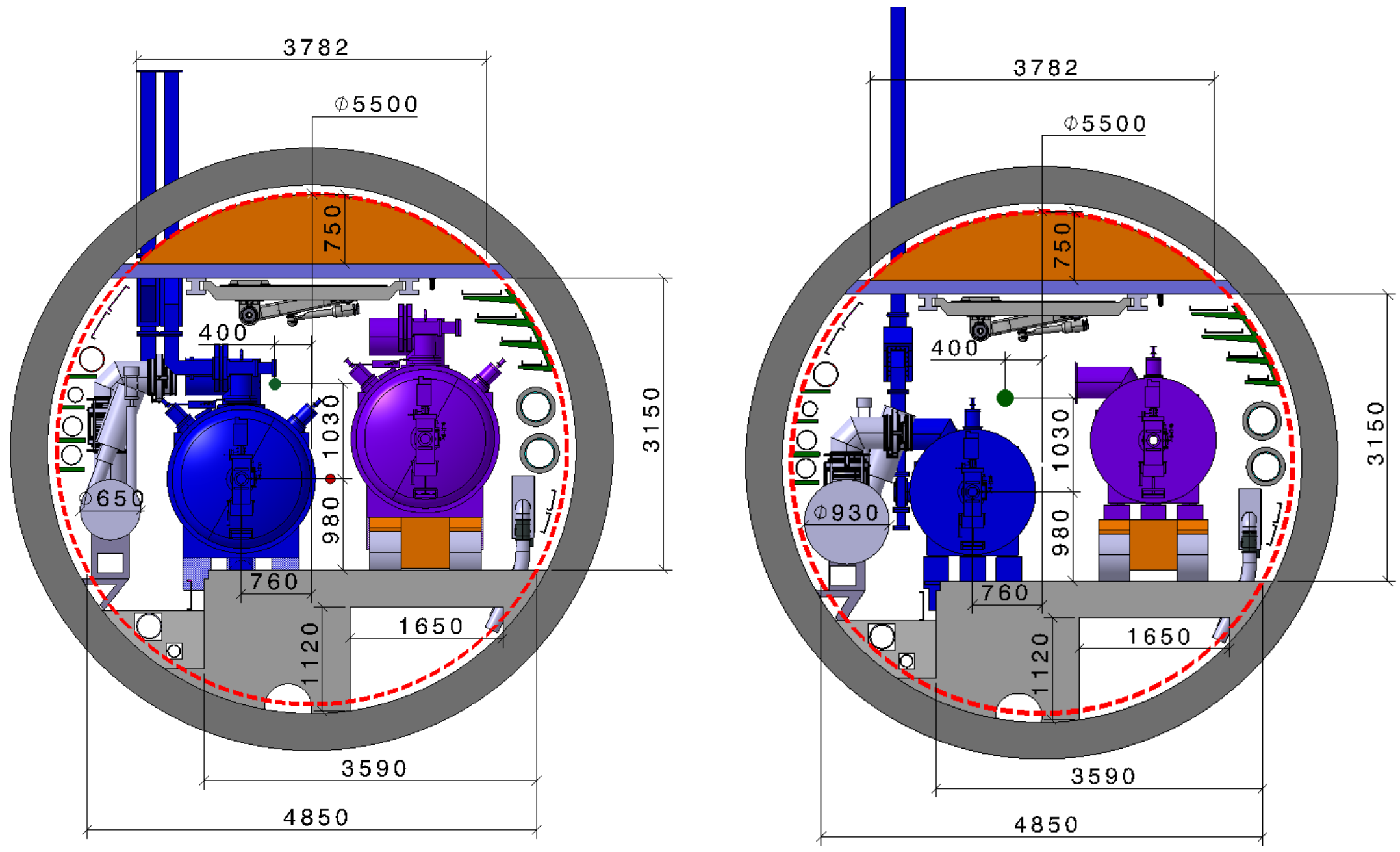

(a) 400 MHz SRF cryomodules

(b) 800 MHz SRF cryomodules

Fig. 1.9: Tunnel cross sections with 400 MHz (left) and 800 MHz SRF cryomodules (right) for the collider rings at point PH. The cryomodule on the right side of the tunnel indicatest he transport path.

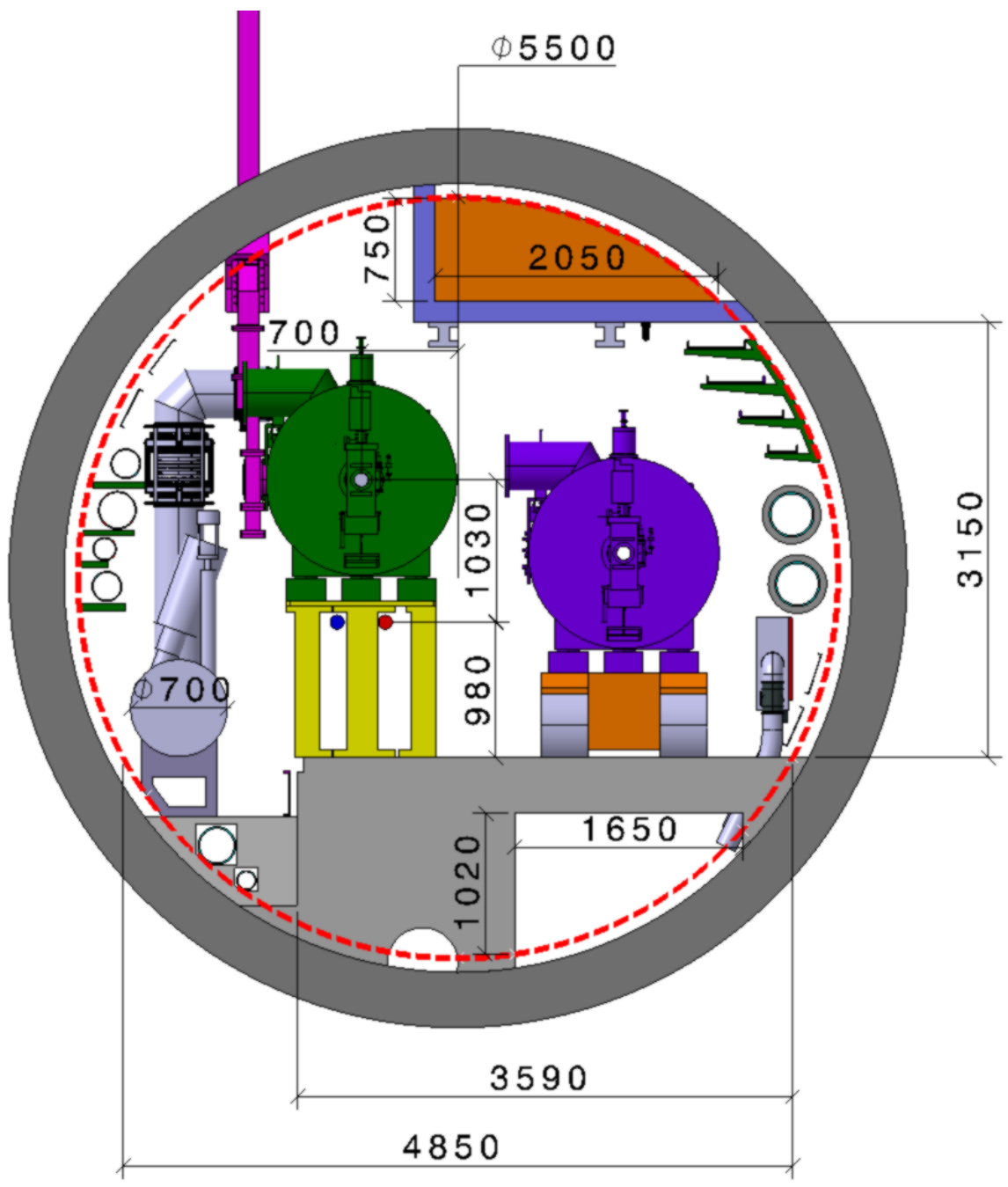

Fig. 1.10: Tunnel cross section with 800 MHz SRF cryomodule for the booster ring at PL. The cryomodule on the right side of the tunnel indicatest he transport path.

## 1.3 Impact of misalignments and field errors

Lattice imperfections may restrict the performance of the FCC-ee collider by affecting momentum acceptance (MA), dynamic aperture (DA), beam lifetime, luminosity at the interaction points (IPs), emittance blow-up, injection efficiency, polarisation, energy calibration, and machine protection. The feasibility of the FCC-ee operation in the presence of realistic imperfections is studied via computer simulations.

Early studies [21, 22] already indicated that linear optics corrections alone, as are applied, e.g., in the LHC [23], do not ensure a good DA. The tuning simulations need to be extended by beam-based alignment (BBA) techniques and dispersion-free steering (DFS), as was the case for LEP [24, 25] and at the SLC [26, 27]. In addition, starting with a relaxed, or even ballistic, optics in the interaction regions (IRs), allows establishing a circulating beam in the uncorrected machine, as an intermediate step towards the collision optics [28]. A ballistic optics was also used for the triplet alignment at the SLC [29, 30].

The currently considered alignment and magnetic tolerances for the FCC-ee magnets and relevant beam instrumentation are discussed in Section 1.3.1. The preliminary tolerances on the magnetic field quality are described in Section 1.3.4. The field quality tolerances likely require dedicated correction circuits, and further studies are needed to finalise these. A first look at alignment tolerances for the Interaction Regions (IRs) is reported below and in Section 2.3.4.

Simulations should, as much as possible, replicate the actual steps to be followed during FCC-ee commissioning, as described in Section 2.3.4. Detailed descriptions of the different commissioning steps are presented in Sections 2.3.4 and 1.3.3. The simulations are mostly carried out for the baseline Global Hybrid Correction (GHC) lattice [31] of Section 1.2. In general, the alternative Local Chromatic Correction (LCC) lattice [32], described in Section 1.10.1, is less sensitive to lattice imperfections in the arcs.

### 1.3.1 Main tolerances for magnets and beam instrumentation

The target alignment and magnetic strength tolerances for the arc elements are compiled in Table 1.3.

It is assumed that the group of quadrupole and sextupole magnets located between main arc dipoles are supported by a common girder. Therefore, the total misalignment of these elements needs to take into account the independent alignment errors of the girder and of the elements on the girder.

The tolerance quoted for the BPM-to-quadrupole alignment refers to the offset between the electric and magnetic centres of the BPM and the quadrupole, respectively. Simulations have shown that by applying Beam Based Alignment (BBA), the initial BPM-to-quadrupole alignment tolerance could be relaxed to 150 µm, if required.

The final-doublet quadrupoles, most of the other IR quadrupoles and the IR sextupoles will require tighter alignment tolerances than the arc components. Table 1.4 presents a preliminary set of tolerances for the IR magnets. Details are discussed in Section 2.3.4.

Spin dynamics studies demonstrate that, for most error seeds and after orbit correction only, with rms alignment errors of 100 µm in the arcs and 25 µm in the IR, the stringent energy calibration demands can be met. Section 1.7.1 provides further information.

Table 1.5 summarizes the arc BPM performance requirements. DA studies indicate that the phase advance should be measured with a resolution better than $10^{-3}\ 2\pi$ [33]. Simulated optics measurements based on a turn-by-turn technique demonstrate that this goal is achieved in the arcs uusing 50 000 turns and a BPM resolution of 1 um [34]. To sustain betatron oscillations for 50 000 turns it is mandatory to use an AC dipole as in RHIC and at the LHC [35–41]. Location and design specifications of the AC dipoles still need to be defined.

In the IR, the sextupoles used for the local chromatic correction represent one of the most critical places where the phase advance must be determined to within $10^{-4}\ 2\pi$. By using only BPMs next to the defocusing quadrupoles in the region of interest and under the same measurement assumptions, this target is achieved, too. The BPM performance requirements so far still need to be validated given transverse coupling and non-linear measurements.

Table 1.3: Arc alignment and strength tolerances. These values correspond to the $1\sigma$ standard deviation for a Gaussian distribution truncated at $2.5\sigma$. Quadrupoles and sextupoles are placed on top of common girders.

| **Element** | $\sigma_{x/y}$ [µm] | $\sigma_{\theta/\psi/\phi}$ [µrad] | $\Delta k/k$ [$10^{-4}$] |
|---|---|---|---|
| Arc quads & sext. | 50 | 50 | 2 |
| Dipoles | 1000 | 1000 | 2 |
| Girders | 150 | 150 | - |
| BPMs-to-quad | 100 | - | - |

Table 1.4: Preliminary IR alignment tolerances, determined with reduced arc misalignments of 100 µm and neglecting BPM misalignments.

| **Element** | $\sigma_{x/y}$ [µm] | $\sigma_{\theta}$ [µrad] |
|---|---|---|
| Final Doublet (FD) quads | 10 | 10 |
| IR quads (excluding FD) | 100 | 100 |
| Sextupoles | 30 | 30 |

The need for orbit stability determines the closed orbit measurement resolution. For BBA performance, a closed-orbit BPM resolution of 1 µm suffices for optimal results as is described in Section 2.3.4.

Table 1.5: Arc BPM performance specifications.

| | |
|---|---|
| Closed orbit resolution | 0.1 µm |
| Turn-by-turn (TbT) position resolution | 1 µm |
| Number of turns in TbT mode | 50 000 |

### 1.3.2 Location of arc corrector magnets and BPMs

Table 1.6 reports the baseline locations of arc magnets and BPMs. The skew quadrupole embedded in the sextupole magnet generates $65\times10^{-4}$ units of skew octupole at 10 mm when powered at its maximum strength. This presents a concern for the DA and requires further studies with realistic distributions of the skew quadrupole strengths.

Tuning studies performed using the GHC lattice have demonstrated that placing the BPMs attached to quadrupoles yields better results than when attached to sextupoles.

Table 1.6: Location of arc magnet correctors and BPMs.

| **Device** | **Location** |
|---|---|
| Horizontal orbit corrector | Embedded at the edge of the main dipole next to main quadrupole |
| Vertical orbit corrector | Embedded at the sextupole or stand-alone |
| Quadrupolar corrector | Trim coil in all main quadrupoles |
| Skew quadrupole | Embedded at the sextupole |
| BPM (H & V) | Attached to the main quadrupole |

Table 1.7 summarises the closed-orbit deviations, beta beating, and spurious dispersion errors before and after the linear optics correction. The simulated dynamic aperture and momentum acceptance are presented in Section 1.3.3.

### 1.3.3 Performance after global optics tuning

Figure 2.5 shows the general sequence of the commissioning steps to be followed in FCC-ee. First orbit threading steps require the sextupole magnets to be switched off. However, a reduced dynamic aperture for this configuration [42,43] renders this approach impractical for the baseline optics. A new optics has been designed with both quadrupoles and sextupoles switched off near the IP [44]. This optics features a lower natural chromaticity, lower beta functions in the IR, no synchrotron radiation in the final doublet (FD), and it allows establishing a straight reference trajectory across the IP in the absence of focusing elements. This is why this optics is named ballistic. Refined optics measurements are not possible immediately after threading but dispersion-free steering (DFS) is effective at allowing the commissioning to then proceed with beam-based alignment (BBA). Global optics corrections are applied iteratively as the optics transitions to the collision optics, following a $\beta^*$ squeeze. Various relaxed optics have been developed, e.g., Ref. [45]. The optics commissioning is further detailed in Section 2.3.4 and in Ref. [28].

For the optics at Z energy (GHC v22), the simulated dynamic aperture and momentum acceptance after global corrections, including only arc errors, are presented in Fig. 1.11. While alignment errors do not significantly affect the vertical DA, some seeds experience a clear degradation in the horizontal DA and all seeds suffer from a reduced momentum acceptance (MA). These degradations could significantly affect lifetime and injection efficiency. Dedicated nonlinear corrections will be required at this stage, despite the fact that non-linear aberrations have not yet been introduced in the simulations. Tables 1.7 and 1.8 summarise the residual orbit and optics errors after global corrections.

In the absence of IR errors, the IP optics parameters are under control thanks to the global corrections, yet they are at the edge of the imposed tolerances. For example, a vertical IP dispersion $D_y^*$ of 1 µm was determined as the tolerance for a 5% increase of the vertical beam size [46], and a 3% $\beta$-beating causes a measurable luminosity loss.

Transverse coupling is characterised via the sum and difference resonance driving terms, $f_{1010}$ and $f_{1001}$, see e.g., Ref. [47]. Early tolerance estimates on the average values of $f_{1010}$ and $f_{1001}$ pointed at a few $10^{-3}$ [48]. However, a specific tolerance for errors in the IP itself is still to be quantified.

These observations point to the need for robust IP tuning and luminosity optimisation techniques. Pertinent studies are described in Section 2.3.4.

### 1.3.4 Tolerances on magnet field quality

The field quality tolerances for the collision optics and also during top-up injection are studied using 6D tracking simulations with XSUITE. These studies incorporate various random and systematic relative field errors in dipole, quadrupole, and sextupole magnets within the arcs and in the interaction regions (IRs), along with beam-beam kicks and synchrotron radiation. This ongoing work complements previous studies [43, 49, 50]. Field error tolerances are defined as the value at which a noticeable change is observed in dynamic aperture (DA) or momentum acceptance (MA). Results for ‘nominal’ colliding beams with the GHC lattice are summarised in Table 1.9. The deterioration of the simulated injection efficiency by 10% or more is used as an additional criterion to define field quality tolerances. Findings for top-up injected beams are presented in Table 1.10. The top-up injection studies consider both the baseline on-axis injection ($\delta p/p$ offset: 0.95%) and also a hybrid injection mode ($\delta p/p$ offset: 0.85%, x offset: 1.5 mm). The injection efficiency and DA/MA criteria yield similar tolerances for the systematic dipolar multipoles ($b_3$, $b_4$ and $b_5$) on the order of $10^{-5}$, without any specific corrections.

These tolerances concerning magnetic design and manufacturing [51,52] are considered challenging. Future work will explore mitigation strategies, such as utilising dedicated corrector coils in the arcs and the IRs as implemented for the LHC [53, 54]. The additional nonlinear correctors, which might require dedicated space in the lattice, would significantly relax the field quality tolerances.

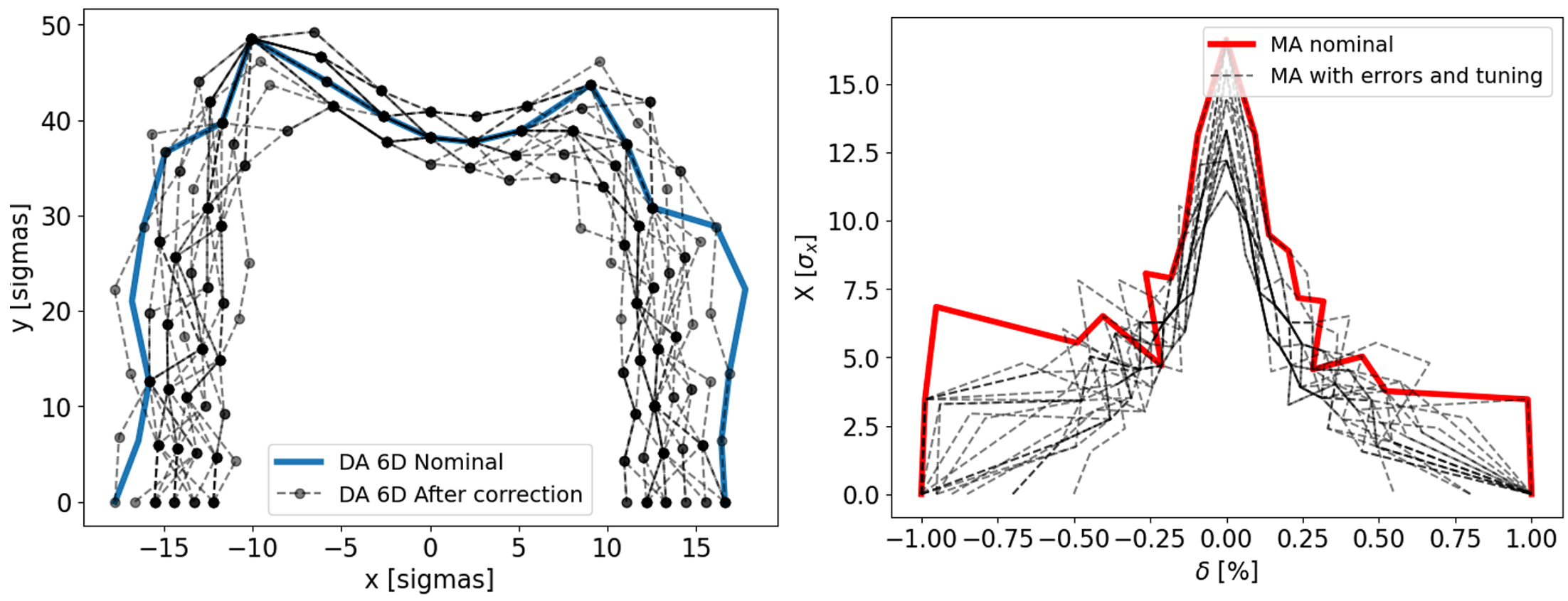

Fig. 1.11: DA (left) and MA (right) after linear optics corrections having used the errors described in Table 1.3, which neglect IR errors and magnetic multipolar errors, for betatron tunes of 217.77 and 220.369.

Table 1.7: Median rms values of several optics parameters before (after sextupole ramping) and after linear optics correction (nominal lattice). $\Delta\psi$ stands for phase advance deviations between nearby BPMs.

| **Parameter** | **Before correction (rms)** | **After correction (rms)** |
|---|---|---|
| horizontal orbit (µm) | 120.2 | 120.5 |
| vertical orbit (µm) | 217.5 | 217.6 |
| $\Delta\beta_x/\beta_x$ (%) | 7.41 | 0.29 |
| $\Delta\beta_y/\beta_y$ (%) | 15.79 | 2.81 |
| $\Delta D_x$ (mm) | 57.79 | 0.28 |
| $\Delta D_y$ (mm) | 62.24 | 2.80 |
| $\varepsilon_h$ (nm) | 0.72 | 0.71 |
| $\varepsilon_v$ (pm) | 26.01 | 0.57 |
| horiz. $\Delta\psi$ $[2\pi]$ | $1.1 \times 10^{-2}$ | $2.9 \times 10^{-4}$ |
| vert. $\Delta\psi$ $[2\pi]$ | $1.9 \times 10^{-2}$ | $2.3 \times 10^{-3}$ |
| Re $f_{1001}$ | $4.9 \times 10^{-2}$ | $1.7 \times 10^{-4}$ |
| Im $f_{1001}$ | $4.4 \times 10^{-2}$ | $5.2 \times 10^{-5}$ |
| Re $f_{1010}$ | $3.7 \times 10^{-2}$ | $1.3 \times 10^{-4}$ |
| Im $f_{1010}$ | $3.7 \times 10^{-2}$ | $1.3 \times 10^{-4}$ |

### 1.3.5 Impact of errors on spin dynamics and polarisation

The precise beam energy measurement in FCC-ee relies on depolarising previously polarised low-intensity pilot bunches using resonant depolarisation (RDP). This method is suitable for beam energies up to about 80 GeV beam energy, above which the beam energy spread becomes too large and polarisation is expected to be lost. Section 1.7 presents more information related to the energy-calibration studies.

Table 1.8: Values of IP optics parameters after a global correction for the nominal lattice from simulations that did not include any IR errors, nor dedicated IP corrections.

| Parameter at the IP | After correction (rms) |
|---|---|
| $\Delta\beta_x/\beta_x$ (%) | 0.34 |
| $\Delta\beta_y/\beta_y$ (%) | 3.08 |
| $\Delta D_x$ (mm) | 0.003 |
| $\Delta D_y$ (mm) | 0.001 |
| Re $f_{1001}$ | $6.5 \times 10^{-6}$ |
| Im $f_{1001}$ | $1.7 \times 10^{-5}$ |
| Re $f_{1010}$ | $1.6 \times 10^{-5}$ |
| Im $f_{1010}$ | $2.2 \times 10^{-5}$ |

Table 1.9: Preliminary bare field quality tolerances, without correction, in units of $10^{-4}$ at a reference radius of 1 cm, inferred from 6D tracking studies for the GHC Z lattice.

| Error | Arc Dipoles | | Arc Quadrupoles | | Arc Sextupoles | |
|---|---|---|---|---|---|---|
| | Random | Systematic | Random | Systematic | Random | Systematic |
| $b_3$ | 0.25 | 0.1 | 1.5 | 1.5 | — | — |
| $a_3$ | — | — | 1 | 2 | — | — |
| $b_4$ | 0.5 | 0.25 | — | — | — | — |
| $a_4$ | — | — | — | — | 30 | 25 |
| $b_5$ | 0.3 | 0.1 | — | — | 36 | 25 |
| $a_5$ | — | — | — | — | 30 | 25 |
| $b_6$ | — | — | 1 | 0.5 | — | — |
| | IR Dipoles | | IR Quadrupoles | | | |
| | Random | Systematic | Random | Systematic | | |
| $b_3$ | 1 | 1 | — | — | | |
| $b_4$ | — | — | 0.1 | 0.4 | | |
| $a_4$ | — | — | — | — | | |
| $b_5$ | 1.5 | 0.6 | — | — | | |
| $a_5$ | — | — | — | — | | |

Resonant depolarisation (RDP) requires a minimum polarisation level of approximately 10%. While the theoretical equilibrium polarisation exceeds 90%, machine errors significantly reduce the achievable polarisation. Additionally, these errors can induce a spin-tune shift that does not correspond to a genuine beam energy shift, but instead leads to an error in the beam energy inferred from the spin-tune measurement.

On the Z pole, the expected uncertainty in the measured collision energy is 100 keV. To assess the impact of misalignments on the spin dynamics, simulation studies [55] are performed using the code BMAD for an IR optics with a local solenoid compensation scheme. Arc and IR misalignments are included with varying magnitude of misalignment errors. Sextupole strengths are ramped up in a stepwise manner with orbit and tune corrections applied on each step. In simulation studies, up to 100 µm and 25 µm rms misalignments are applied to arc and IR magnets, respectively, along with other errors. For the vast majority of seeds at various beam energies around the Z-pole, the resulting bias in the beam energy extracted from the spin tune stays well below 100 keV, and an equilibrium polarisation above 10% is reached. Further details are reported in Section 1.7.1.

Table 1.10: Preliminary bare field quality tolerances for arc dipoles, considering top-up injection without correction, in $10^{-4}$ units at a reference radius of 1 cm, deduced from 6D tracking studies, for the GHC Z lattice.

| Error | On-axis Injection Arc Dipoles | | Hybrid Injection Arc Dipoles | |
|---|---|---|---|---|
| | Random | Systematic | Random | Systematic |
| $b_3$ | — | 0.12 | — | 0.14 |
| $b_4$ | — | 0.07 | — | 0.15 |
| $b_5$ | — | 0.07 | — | 0.1 |

Future spin dynamics studies should include even larger misalignment errors along with magnetic field imperfections, as in Table 1.3. The polarisation simulations do not yet include specific correction techniques, which could reduce the spurious spin tune shifts or increase equilibrium polarisation, such as harmonic spin matching. Furthermore, alternative optics featuring a non-local solenoid compensation scheme and the latter's impact on polarisation remain to be investigated in greater detail.

## 1.4 Collective effects

This section describes the impact of collective effects on the main electron and positron rings. The first subsection describes the impedance, while the second and third subsections show its impact on the beam dynamics for both non-colliding and colliding beams, respectively.

The fourth subsection discusses electron-cloud effects, which limit the bunch spacing in the collider. The last two subsections review the minor impact of intra-beam scattering and the increase in emittance due to interaction with residual gas.

### 1.4.1 Impedance and wakefield model

The low-energy machine, operating at 45.6 GeV, is the most affected by collective effects because of the lowest beam energy combined with the highest beam current, the lowest emittances and the longest damping times. The design of this machine is still in progress and also the coupling impedance budget is continuously evolving in parallel with the updates of the vacuum chamber components. Correspondingly, the collective effects and instability thresholds need constant revision. The latest impedance model includes vacuum chambers, collimators, bellows, taper transitions and initial models of BPMs and RF cavities.

The resistive wall (RW) beam coupling impedance of the arc vacuum chamber is one of the key contributions to be assessed. The beam pipe, shown in Fig. 1.12, with a radius $b$ of 30 mm, is made of copper coated with a 150 nm thin layer of NEG (recently, it has been proposed to increase the thickness to 200 nm) used for pumping purposes and electron cloud suppression. There are two lateral winglets for placing synchrotron radiation absorbers. The impedance model was initially obtained with the electromagnetic code IW2D [56] in the circular vacuum chamber approximation. As a following step, the calculation has been extended to include the effect of the winglets by using appropriate form factors, which have been numerically estimated with CST PARTICLE STUDIO. The results were also compared with those of the 2D electromagnetic solver VACI [57] that can simulate the geometry, including the winglets.

The collimation system is another important impedance source. The impedance model that is being evaluated accounts for the geometric dimensions, materials, and beta functions at collimator locations. Since a mechanical design of a collimator still does not exist at the moment, the geometric impedance has been evaluated assuming a linear taper with an angle of 15° to go from the collimator

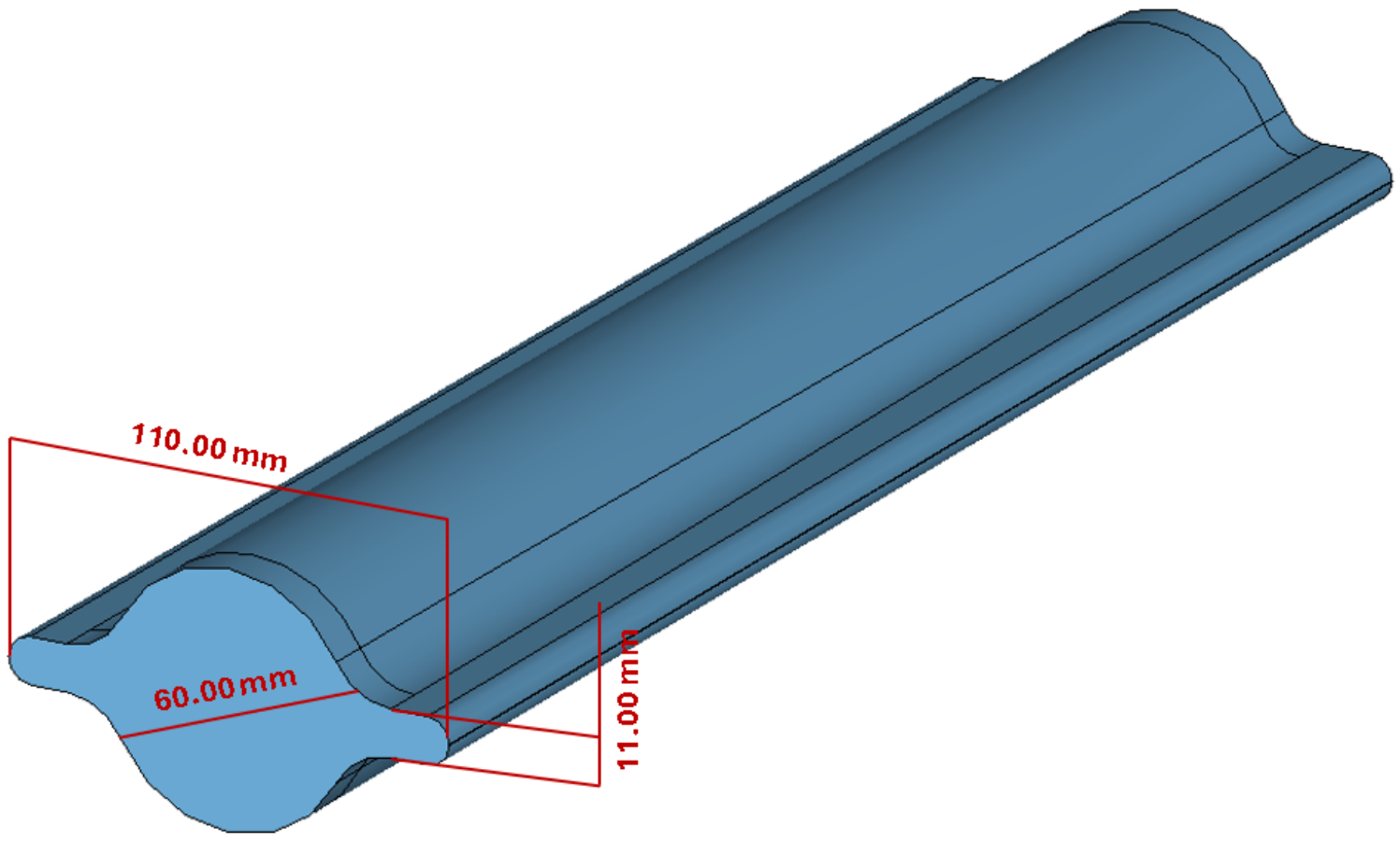

Fig. 1.12: Beam pipe shape.

aperture to the chamber aperture of 30 mm.

The overall impedance of collimators, including material losses and tapers, has been obtained using CST simulations. Additionally, the resistive part of the collimators' impedance has been evaluated with IW2D using a flat chamber model. Therefore, by combining CST and IW2D results, resistive and geometric contributions can be disentangled. The resistive wall impedance of the collimators is between 5% and 10% of the vacuum chamber contribution, depending on the choice of material. The primary vertical collimator (tcp.v.b1) makes the highest contribution to this impedance source due to the very small gap.

The preliminary collimator model with linear tapers gives the largest impedance contribution in the transverse plane. An optimised design of the collimator tapers, for example, by minimising the taper angle (increasing the taper length) and considering optimised nonlinear taper modelling, is crucial.

Bellows are another important source of impedance. A crucial component of the device is the RF shielding with comb-type fingers and small electric fingers to ensure electric contact between the two sides of the shielding. These fingers are shown on the left-hand side of Fig. 1.13 for a model similar to that of SuperKEKB, designed by the vacuum group and shown on the right-hand side of the same figure. The contribution of the shielding is fundamental to suppress the low-frequency resonances due to the bellows, which otherwise would lead to a high impedance contribution.

An important aspect of the bellows' contribution to the impedance model is related to their number. For dipole arcs, quadrupoles/sextupoles sections, and including additional bellows for the RF system, injection system, collimation, etc., a total number of 10 000 devices is assumed. Other geometries are under study by the vacuum group.

Additionally, the impedance contribution due to the 400 MHz RF system has been evaluated. This comprises 132 two-cell cavities per beam, arranged in groups of 4 for each cryomodule, with 50 cm long tapers on both ends, guaranteeing a transition from 50 mm to 150 mm circular pipe inside the cryomodule. Finally, 4000 BPMs have also been taken into account.

The total impedance is shown in Fig. 1.14 for the longitudinal and transverse cases. The contribution of each device to the longitudinal and vertical dipolar impedance is displayed in Fig. 1.15.

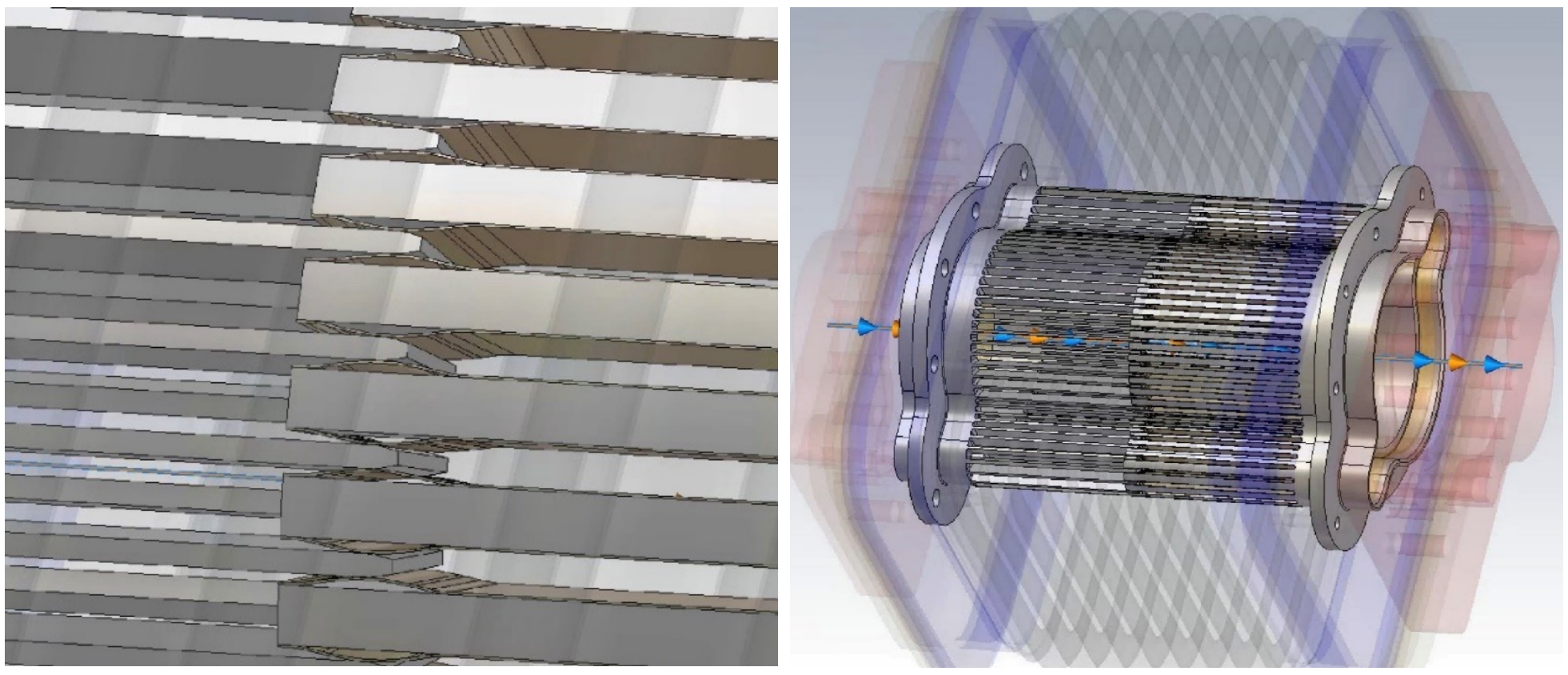

Fig. 1.13: Simulated models of FCC-ee beam vacuum chamber including bellows.

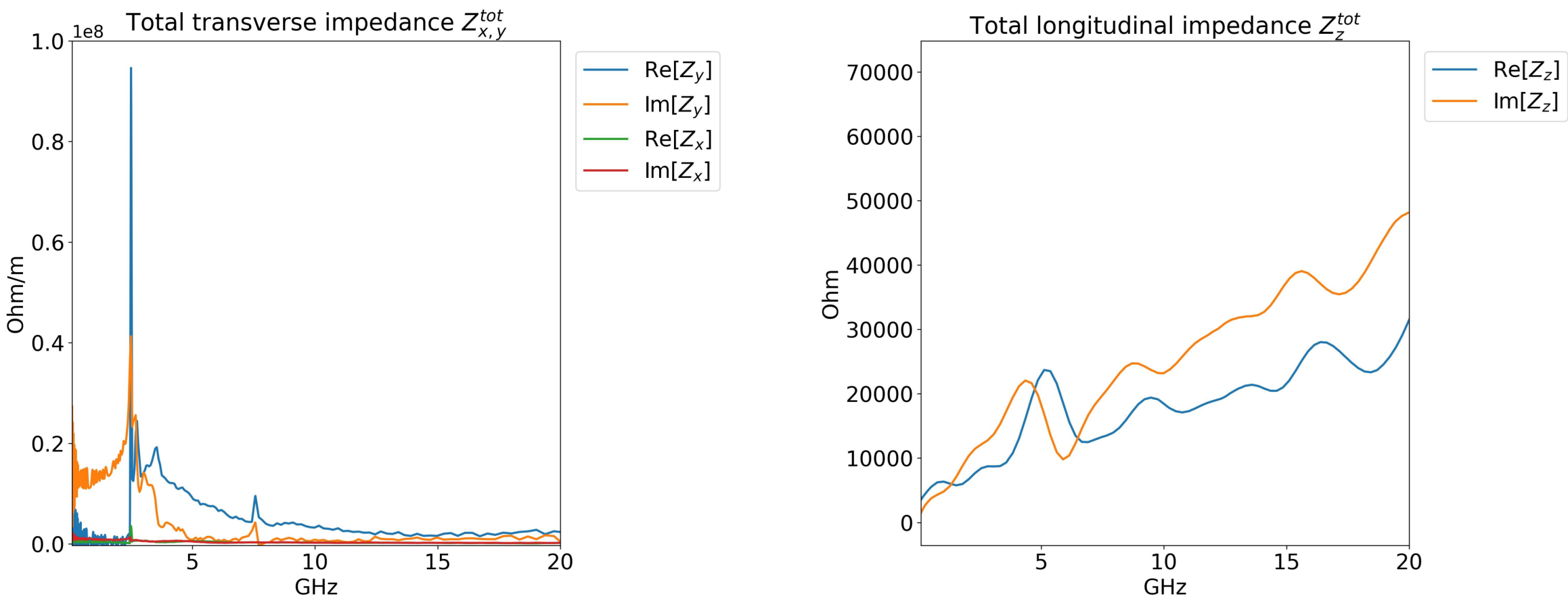

Fig. 1.14: Total longitudinal (on the right) and transverse (on the left) impedance model.

### 1.4.2 Impedance induced collective effects

***Longitudinal effect***

Below the microwave instability threshold, longitudinal wakefields result in bunch lengthening and bunch shape distortion. Above the instability threshold, the energy spread also starts growing, and the internal bunch motion becomes more turbulent. The internal bunch oscillations can be harmful to the process of reaching the nominal luminosity. However, in the FCC-ee the longitudinal dynamics is strongly affected by the beam-beam interaction with a large Piwinski angle and beamstrahlung. Both the bunch length and the energy spread increase in collision due to beamstrahlung help mitigate the longitudinal collective effects.

Longitudinal beam dynamics simulations have been performed with the PyHEADTAIL code [58] which was compared with other tracking codes [59,60], for the FCC-ee case, giving excellent agreement.

Two regimes are being considered for beam dynamics studies: the single-beam mode and the colliding beams mode. Both regimes are important for collider operations. While the single beam mode is important for the machine commissioning and tuning, the collision mode must be considered for the luminosity production runs. In the first case, the bunch length at the nominal intensity is strongly affected by the potential well distortion, but the energy spread is essentially constant. In the second case, thanks to the beamstrahlung, the potential well distortion due to the wakefield is small and the collision predominantly defines the bunch lengthening and the energy spread growth. In this condition, however, a self-consistent study considering the beam-beam effects is necessary. This aspect is discussed

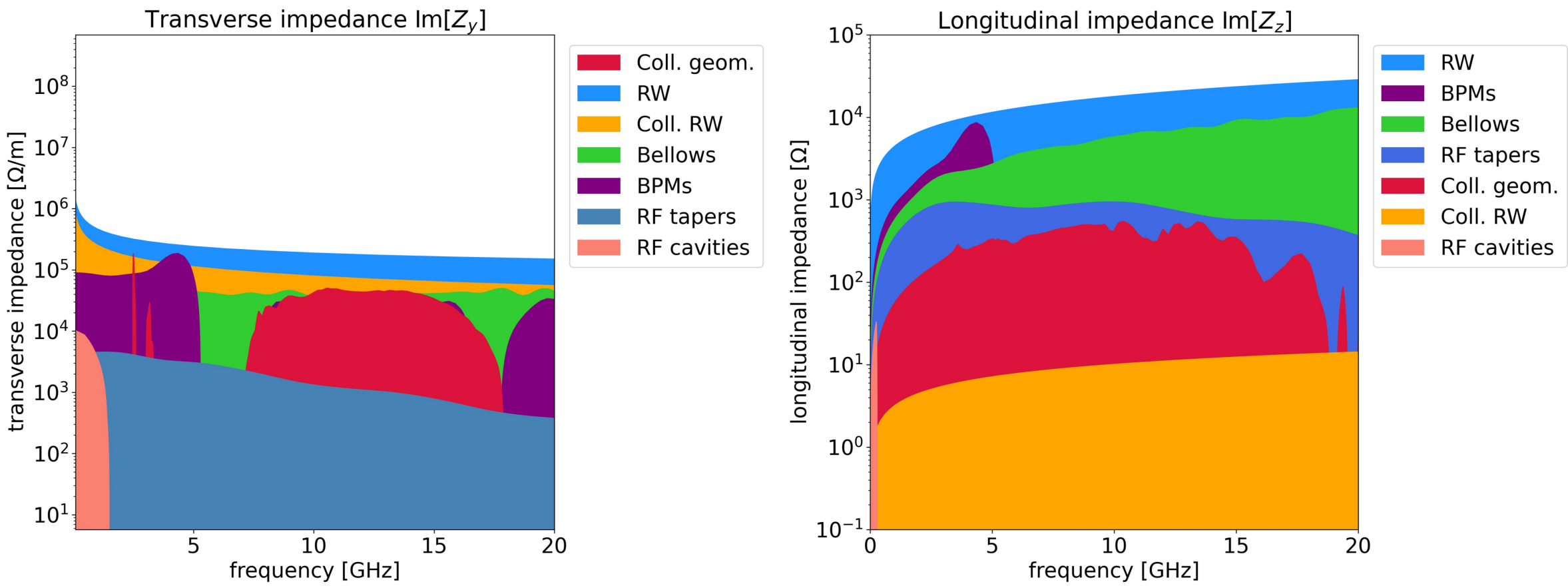

Fig. 1.15: Contribution of each device included in the model. Longitudinal impedance on the right, vertical dipolar impedance on the left.

in Section 1.4.3.

#### *Transverse effects*

The main effect of the short-range transverse wakefield on the single bunch dynamics is the excitation of the so-called transverse mode coupling instability (TMCI). Under certain conditions, the frequencies of some coherent transverse oscillation modes of a bunch can shift and couple together. In particular, for FCC-ee, the '0' mode shifts towards the '-1' mode. When they couple together, the instability occurs with a consequent loss of the beam (or a part thereof).

The coherent frequencies of the lowest order coherent oscillation modes can be obtained from the results of XSUITE/PYHEADTAIL with a proper Fourier analysis [61]. Additionally, it has been found that for FCC-ee the TMCI threshold depends on the longitudinal wakefield. In Fig. 1.16, left-hand side, the real part of the tune shift of the first azimuthal transverse oscillation modes normalised by the synchrotron tune $Q_{s0}$ is shown as a function of bunch population. As can be seen, the instability threshold is about $3 \times 10^{11}$, which is well above the nominal bunch intensity. This instability is not of the 'mode coupling' type, but it is due to the single '-1' oscillation mode. These results were obtained without accounting for the beamstrahlung due to collisions, but a transverse damper for stabilising the coupled bunch instabilities and a value of the chromaticity equal to 5 were introduced.

The geometrical impedance of the collimators is not included since their design is still in progress. However, to take into account all the devices not evaluated so far, and to check for a possible upper limit on the machine stability, simulations under the same conditions as for the left-hand side of Fig. 1.16 were performed, but assuming an overall two times larger impedance. The results are shown in the right-hand picture.

In addition to the single bunch dynamics studies, a coupled bunch instability can be excited, driven essentially by the real part of the resistive wall impedance at low frequency. Its study can be performed by considering the motion of the entire beam (not of the single bunch) as a sum of coherently coupled bunch oscillation modes. Under some conditions, the growth rate of the $\mu^{th}$ mode ($\mu = 0, 1, ..., N_b - 1$) is

$$\alpha_{\mu,\perp} = -\frac{cI}{4\pi(E_0/e)Q_\beta} \sum_{q=-\infty}^{\infty} Re\left[Z_\perp\left(\omega_q\right)\right] \tag{1.12}$$

where $I$ is the total beam current, $Q_\beta$ the betatron tune, $\sigma_\tau$ the rms bunch length in time, and $\omega_q$ are frequencies spaced by the revolution period and depending on the coupled bunch mode excited and on chromaticity. When $\alpha_\mu$ is positive, the corresponding mode is unstable. This occurs when the transverse impedance is evaluated at negative frequencies. The most unstable mode has a rise time of about 1.3 ms, corresponding to a few turns. This instability depends on the fractional part of the betatron tune and on chromaticity, and it can be mitigated by a bunch-by-bunch feedback system, like that used in other circular accelerators (DAΦNE, SuperKEKB, . . . ). Such feedback, in combination with the longitudinal wakefield, also has a mitigating effect on the TMCI. Without feedback, the TMCI threshold is expected to be below a bunch population of $1.0 \times 10^{11}$.

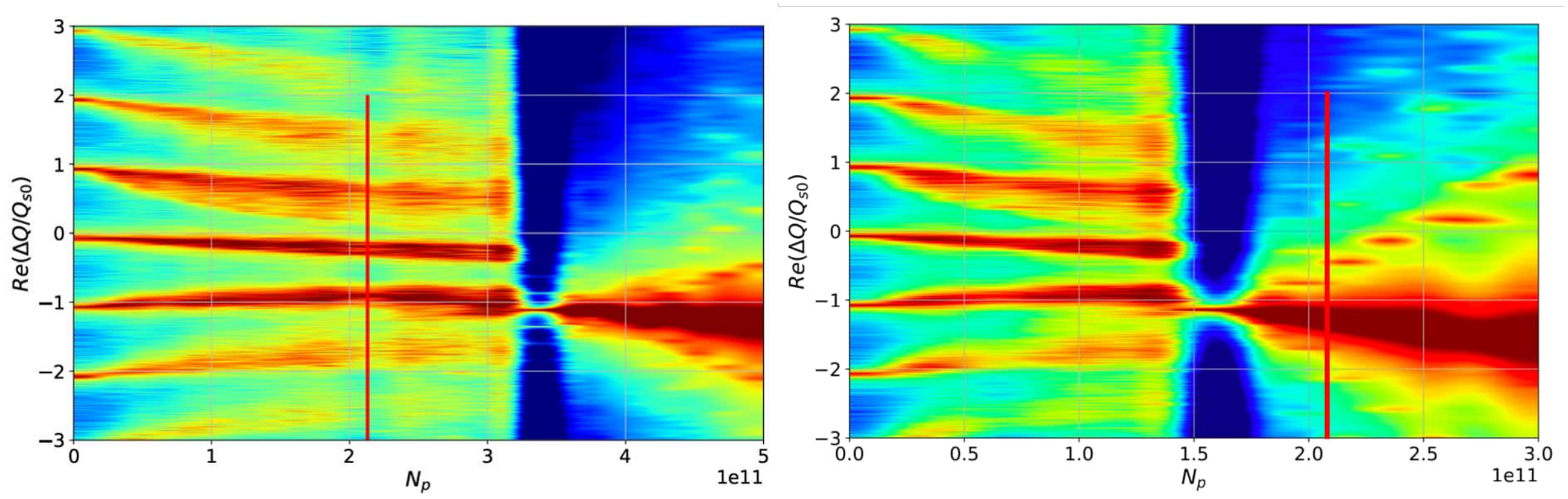

Fig. 1.16: Real part of the tune shift of the first azimuthal transverse coherent oscillation modes normalised by the synchrotron tune $Q_{s0}$ as a function of bunch population with a bunch-by-bunch feedback system and chromaticity = 5. On the right-hand side, the impedance is multiplied by a factor of 2 to explore the margins with the current impedance model.

Finally, it must be noted that the results related to the transverse wakefield are valid in the single-beam regime, without the beamstrahlung effect. For self-consistent results in a collision, the beam-beam effects must also be included.

### 1.4.3 Interplay between beam-beam and beam coupling impedance effects

***The $x$-$z$ instability***

Among the new effects caused by beam-beam collisions, the large Piwinski angle causes the coherent horizontal-longitudinal ($x$-$z$) beam-beam instability to become a critical limiting phenomenon for the collider design performance [3]. Unlike impedance-induced collective instabilities, the $x$-$z$ instability is driven by the beam–beam force itself. This new coherent instability also differs from the classic incoherent synchro-betatron resonances excited by the beam–beam interaction. The instability manifests itself as a horizontal beam position variation along the bunch length, similar to a head-tail instability. It limits the ranges of horizontal tunes where the design luminosity can be achieved.

The interplay between the beam–beam interaction, beamstrahlung and the longitudinal and transverse beam coupling impedance may affect both the $x$-$z$ instability and the beam parameters in the stable betatron tune areas. It has been observed in numerical simulations that the stable areas get narrower and the stable regions on the betatron tune diagram are shifted because of the impedance-related synchrotron tune reduction. On the other hand, the horizontal beam blow-up becomes somewhat weaker due to the synchrotron frequency spread and bunch lengthening induced by the longitudinal impedance [62].

Figure 1.17 shows the results of strong-strong beam-beam simulations, featuring the effect of both the longitudinal and transverse impedance for different horizontal tunes. The luminosity loss at each synchrotron sidebands of the horizontal half integer is clearly visible in the configuration without chromaticity. A chromaticity of 5 units is sufficient to stabilise the instability over a wide range of horizontal tune. Here, it is important to consider the impact of transient beam loading, which is stronger with the new reversed polarity operation of the double cell RF cavities at the Z energy. Different bunches will experience difference RF voltages ranging from 79 to 93 MV (Chap. 3.4). The stability has to be ensured for all bunches with a given horizontal tune. A slightly higher chromaticity (6 units) is required to stabilise the $x$-$z$ instability with higher voltages (Fig. 1.17, right). When considering the existing impedance model, the horizontal tune range is constrained above ~0.55 by bunches featuring high RF voltage. However, when considering a stronger impedance, the constraint on the horizontal tune is relaxed, thanks to the bunch lengthening by the impedance. This shows that bunch lengthening is an efficient way to mitigate this instability. An alternative mitigation consists in reducing the horizontal $\beta^*$.

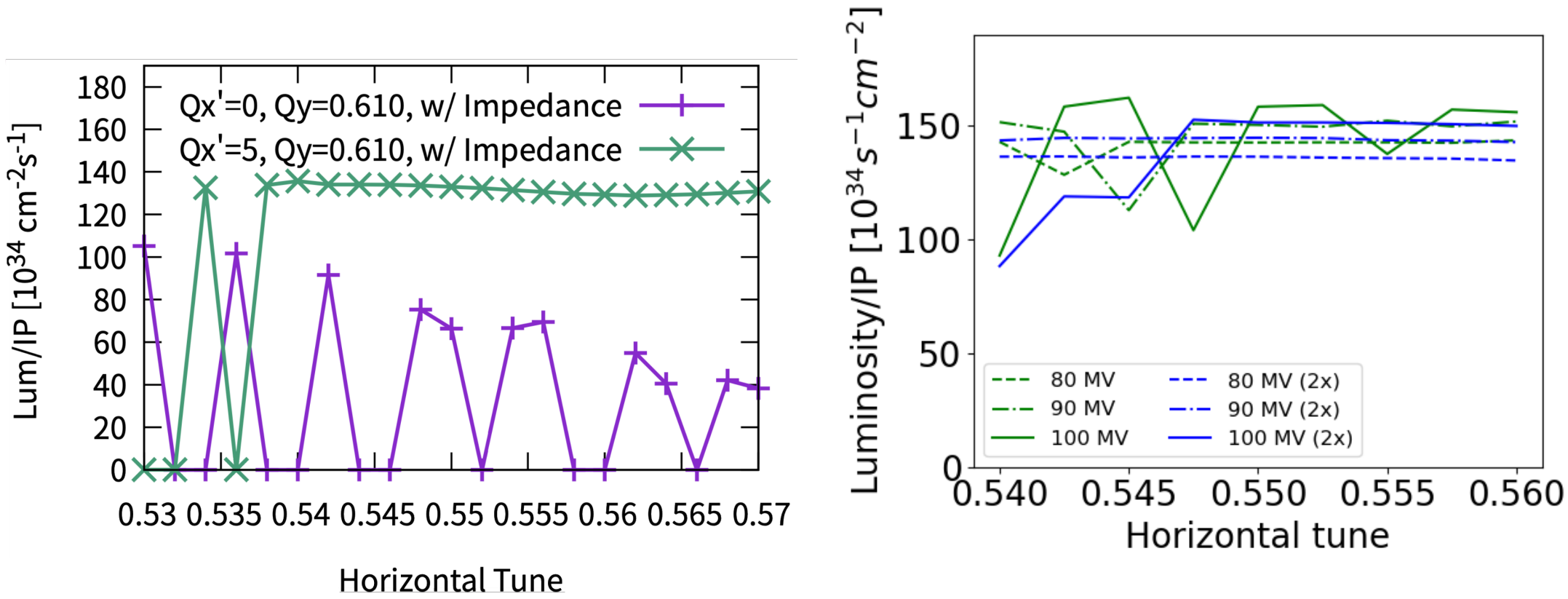

Fig. 1.17: Luminosity dependence on the horizontal tune due to the $x$-$z$ instability. The left plot corresponds to the nominal configuration at the Z energy without active damper simulated by IBB [63] taking into account the effect of the strong-strong beam-beam interaction as well as transverse and longitudinal wakefields [64]. The right plot corresponds to the same configuration but with a chromaticity of 6 units, including a transverse damper featuring a damping time of 10 turns and is obtained with XSUITE [65]. The results are shown for different RF voltages. Blue curves, marked (2x) feature an impedance twice as strong as the existing impedance model.

***The mode coupling instability of colliding beams***

With a vertical beam-beam tune shift larger than the synchrotron tune, it is expected that the mode coupling instability of colliding beams [66] may occur at the FCC-ee. This was confirmed with two semi-analytical models as well as with tracking simulations [67, 68]. The mechanism is illustrated in Fig. 1.18. The lower synchrotron sideband (head-tail mode -1) couples with the beam-beam $\pi$-mode leading to a strong instability. This instability can be partially mitigated by active feedback, yet a sizeable vertical chromaticity is required, as shown in Fig. 1.19. Based on these tracking simulations, it appears that a rather low vertical chromaticity (2 units) is sufficient to maintain the beam stability considering the present impedance model. Higher chromaticities are required if an impedance larger by a factor two is considered. It is clear that this instability constrains the vertical impedance budget.

### 1.4.4 Electron Cloud

As observed in several accelerators [69–72], electron clouds may cause several unwanted effects, in particular beam instabilities, emittance growth, tune shifts, additional heat load on the vacuum chambers and vacuum degradation. Since electron cloud formation depends strongly on the bunch spacing, it is a concern primarily at the Z operating point, where the high beam current requires a large number of closely spaced bunches.

Electron clouds can be created by secondary electron emission through a beam-induced multipacting process, through an accumulation of photoelectrons, or a combination thereof. For their mitigation, it is important to identify the required constraints on the corresponding material properties to avoid elec-tron cloud formation. This has been achieved by simulating the process of electron cloud formation in the beam chamber, using the PyECLOUD code [73].

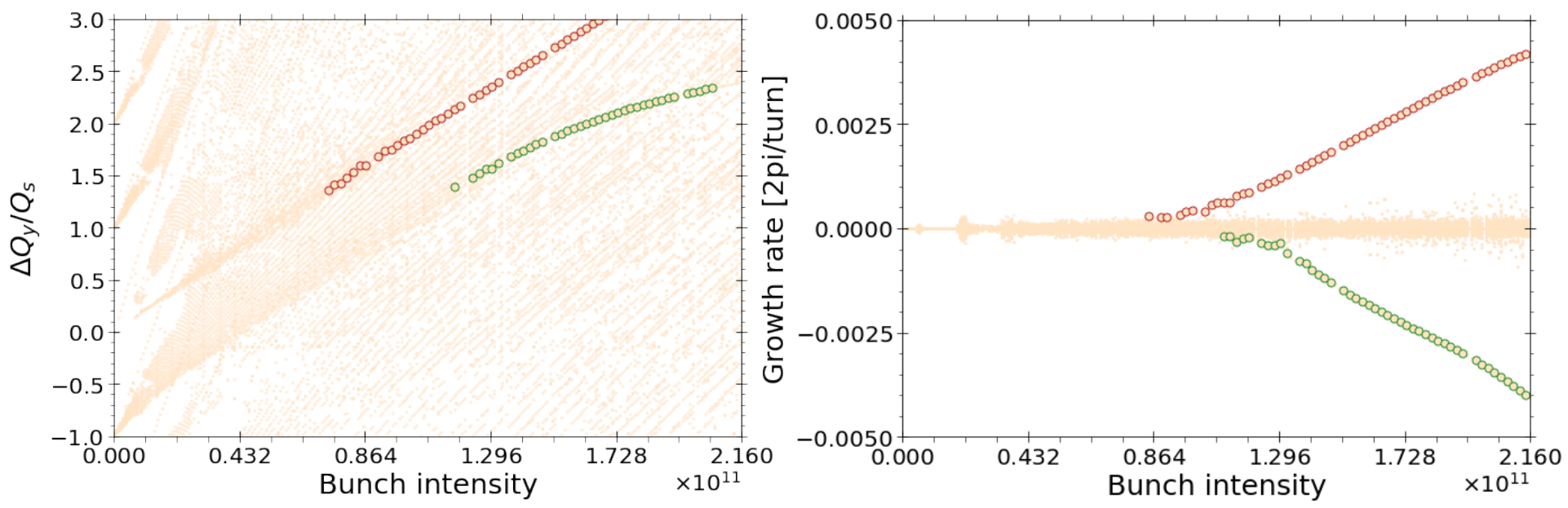

Fig. 1.18: Tune (left) and growth rate (right) for all coherent modes obtained with the circulant matrix model for the nominal configuration with varying bunch intensity without active damper or chromaticity. The impact of the longitudinal impedance is also neglected. The most unstable low order mode of oscillation resulting from the interplay of the head-tail mode -1 with the beam-beam $\pi$-mode is highlighted in light blue.

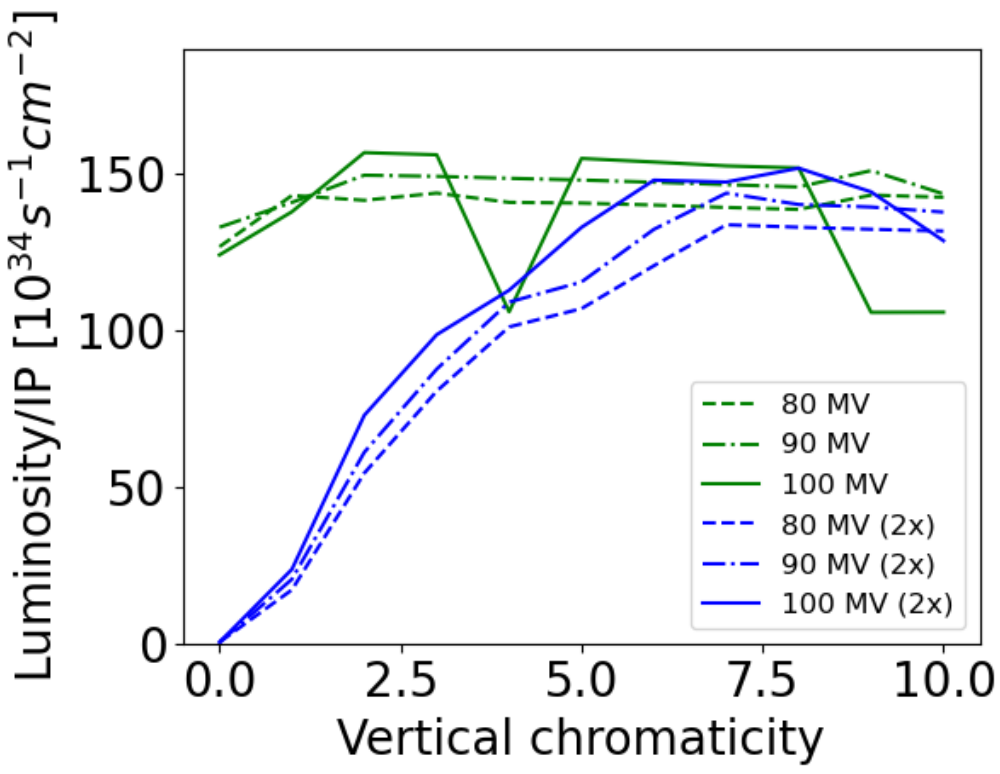

Fig. 1.19: Luminosity dependence on the vertical chromaticity used to mitigate the mode coupling instability of colliding beams for different RF voltages. The blue curves marked (2x) feature and impedance twice as strong as the existing impedance model.

#### *Secondary electron emission*

Electron cloud formation strongly depends on the secondary electron yield (SEY) of the beam chamber surface, defined as the ratio between the emitted and the impinging electron currents. The tendency for electron cloud build-up for different values of the SEY has been determined in drift spaces, as well as in the presence of the main arc dipolar, quadrupolar and sextupolar fields [74].

The build-up is found to depend strongly on the bunch intensity in addition to the SEY and the bunch spacing. The bunch intensity dependence is non-monotonic, so that electron cloud formation occurs more easily with bunch intensities that are lower than the nominal intensity. In particular, the most critical intensities fall within the range of 1.0 to $1.5 \times 10^{11}$ $e^+$ per bunch, as shown in Fig 1.20. Since the machine relies on a top-up injection scheme, with individual injections of one-tenth of the nominal bunch population, the most critical intensities will be encountered when filling the machine. The multipacting thresholds, i.e., the highest maximum SEY guaranteed to suppress build-up, with nominal intensity as well as any intensity between $0.2 \times 10^{11}$ $e^+$ per bunch and the nominal value of $2.14 \times 10^{11}$ $e^+$ per bunch are summarised in Table 1.11. These results are obtained using the so-called ECLOUD secondary emission model [75], parametrising measurements of LHC Cu co-laminated beam screen samples [76–78]. Simulations using the alternative Furman-Pivi model [79] reveal even tighter material constraints [80]. The SEY requirements, especially for intensities below the nominal value, are not guaranteed to be achieved with the planned copper surface with a thin NEG coating [81, 82]. Several mitigation measures have been considered to alleviate these constraints, as discussed below.

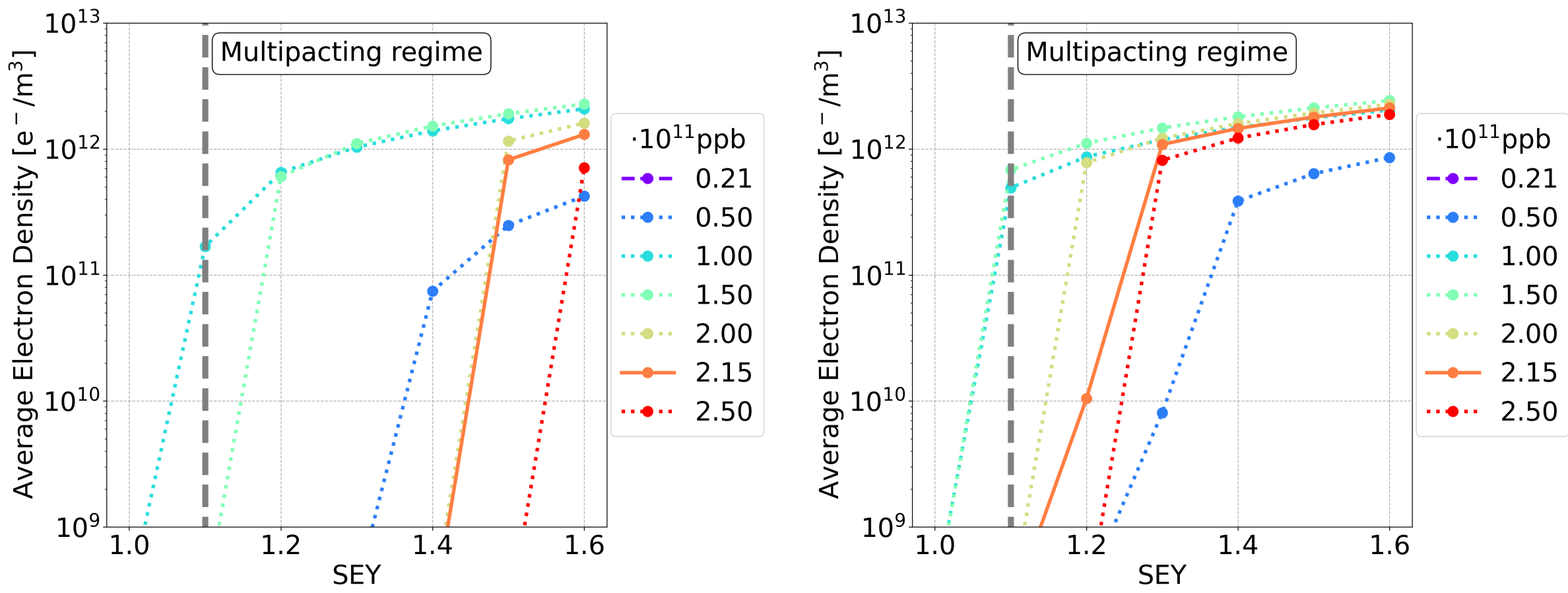

Fig. 1.20: Average electron density in the vacuum chamber versus SEY for different bunch intensities (the nominal value is shown as a solid line) in dipole (left) and quadrupole (right) magnets. The SEY multipacting threshold, considering all bunch intensities during the charge accumulation phase, is shown by the vertical dashed grey line.

Table 1.11: SEY multipacting thresholds for the main arc elements.

| **Element** | **Field** | **Bunch population** | **Thresholds** |
|---|---|---|---|
| Drift | - | nominal | 1.4 |
| | | below nominal | 1.2 |
| Dipole | $15.2\,\mathrm{mT}$ | nominal | 1.4 |
| | | below nominal | 1.0 |
| Quadrupole | $1.45\,\mathrm{T/m}$ | nominal | 1.1 |
| | | below nominal | 1.0 |
| Sextupole | $72.5\,\mathrm{T/m^2}$ | nominal | 1.1 |
| | | below nominal | 1.0 |

***Photoelectron emission***

The previous results do not take into account the photoelectrons. These are primary electrons produced through the photoemission from the chamber walls due to the synchrotron radiation emitted by the circulating beam. Photoelectrons enhance the electron cloud build-up process and, in large quantities, can induce electron cloud effects even in the absence of beam-induced multipacting. The quantity of photoelectrons emitted is determined by the photoelectron yield (PY) of the beam chamber, defined as the ratio between the emitted photoelectrons and the number of impinging photons, along with the quantity of photons scattered into the main beam chamber.

Results from electron cloud build-up simulations (see Fig. 1.21) indicate that an acceptable number of photoelectrons is around $n_{\mathrm{pe}} = 1.0 \times 10^{-4}$ $(\mathrm{e}^{+}\mathrm{m})^{-1}$. The PY and the number of photoelectrons generated inside the central chamber are related through the following equation:

$$\mathrm{PY} = \frac{I n_{\mathrm{pe}}}{\phi L_{\mathrm{pipe}} e}\,, \tag{1.13}$$

where $I$ is the beam current, $\phi$ is the photon flux and $L_{\mathrm{pipe}}$ is the perimeter of the vacuum chamber. Alternatively,

$$n_{\mathrm{pe}} = \frac{5\pi\alpha\gamma\; R\; PY}{\sqrt{3}\; L_{\mathrm{arc}}} \approx 0.08\; R\; PY \quad [e^{+}\mathrm{m}]^{-1} \tag{1.14}$$

with $\alpha_{\mathrm{e}}$ the fine-structure constant, $L_{\mathrm{arc}} \approx 77$ km the total length of the FCC-ee arcs, $\gamma$ the Lorentz factor ($\gamma \approx 90\,000$ for Z running) and $R$ the fraction of primary photons absorbed (possibly after multiple reflections) on the main circular part of the vacuum chamber. Preliminary ray-tracing simulations indicate that the photon flux on the central part of the chamber is in the order of $\phi = 10^{13}$ - $10^{14}$ photons/$(\mathrm{cm}^2{\cdot}\mathrm{s})$, except immediately around the photon absorbers where an even higher flux is expected [83]. Combining this information with the results from electron cloud build-up simulations using Eq. (1.13), a limit on PY in the range 3% - 3‰ is obtained. Photoelectron yields of order 2% have been measured for NEG coated chambers [84]. For FCC-ee, many primary and reflected photons are absorbed at the photon stops or inside the winglets of the vacuum chamber, where they do not contribute to electron-cloud buildup. A new design of the photon absorbers is predicted to reduce further the photon flux inside the main circular part of the vacuum chamber [85]; see Section 3.2.7).

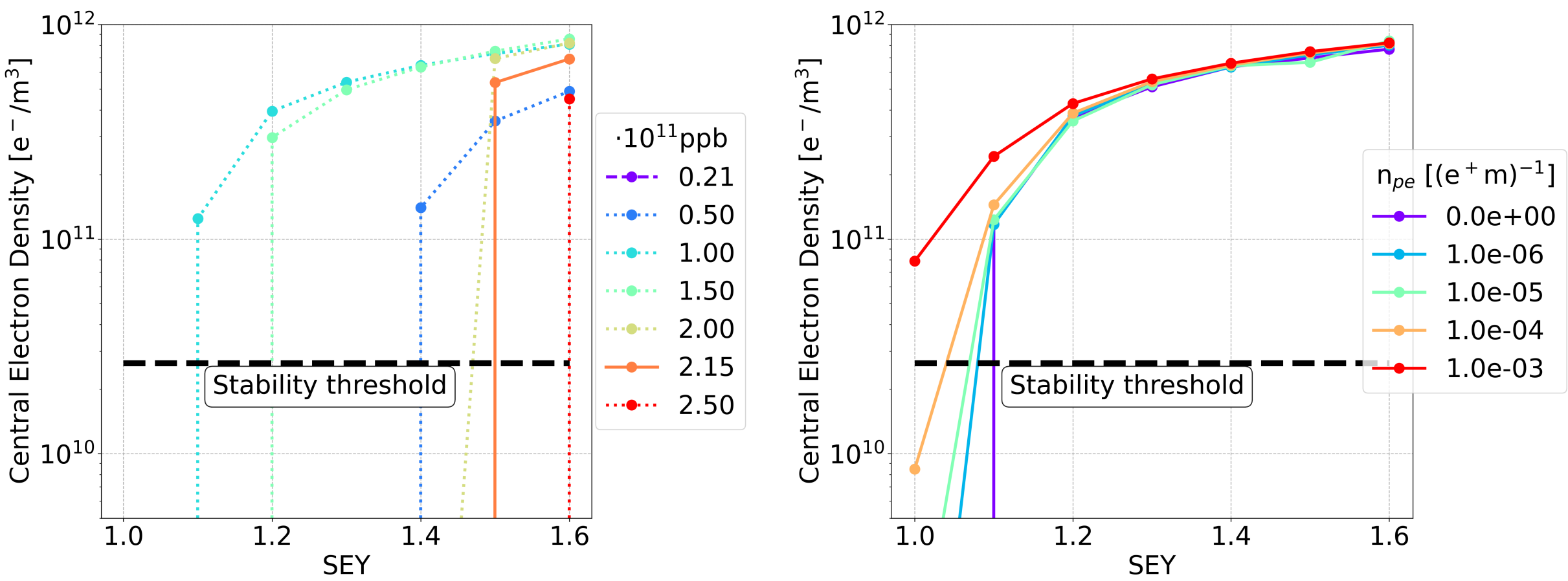

Fig. 1.21: Central electron density versus SEY in the arc dipoles for different bunch intensities (left) and for different photoelectron numbers $n_{\mathrm{pe}}$ (right), with the most critical bunch intensity of $1 \times 10^{11}$ $\mathrm{e}^{+}$. The theoretical stability threshold (Eq. 1.15) is indicated by the horizontal dashed black line.

### *Electron cloud effects*

Electron cloud effects can be fully avoided only by ensuring the suppression of primary and secondary electron emission according to the constraints above. In case the properties of the vacuum chamber cannot meet these constraints, the effects of the electron cloud on the beam and the machine environment must be assessed.

The electron cloud can trigger beam instabilities as the beams pass through the dense cloud [86]. The electron cloud density corresponding to the single-bunch instability threshold can be estimated as

[87–89]

$$\rho_{\text{thr}} = \frac{2\gamma Q_s \omega_e \sigma_z / c}{\sqrt{3} K Q r_e \beta_y L} \tag{1.15}$$

where

$$\omega_e = \left( \frac{N_b r_e c^2}{\sqrt{2\pi}\sigma_z \sigma_y (\sigma_x + \sigma_y)} \right)^{1/2} \tag{1.16}$$

is the electron angular oscillation frequency, $K = \omega_e \sigma_z / c$ characterises how many electrons contribute to the instability, $Q = \min(K, 7)$ is the quality factor of the effective wake field and $L$ is the length of the particle accelerator, or the considered element. The stability threshold has also been evaluated through simulations, using the PyECLOUD-PyHEADTAIL suite [90], for drift spaces and dipole fields. The results between the theoretical estimate and the simulation studies are consistent to the level of the order of magnitude. This stability threshold must be compared with the electron cloud density close to the vacuum chamber centre before a bunch passage, as shown for the dipole magnets in Fig 1.21. The build-up studies show that the electron density exceeds the stability threshold whenever the SEY is above the multipacting threshold in all the elements considered, except the sextupole magnets. In other words, if the material constraints are not met, beam instabilities are expected to occur. In addition, other effects caused by the interaction of the beam with a dense electron cloud, such as emittance growth, tune shift and tune spread, can also be expected.

The electron cloud impinging on the chamber surface can also cause environmental effects, such as outgassing and heat load. Build-up simulations estimate the total additional heat load in the arcs due to electron cloud to be in the order of a percentage of the synchrotron radiation power (50 MW per beam) when multipacting occurs.

***Further mitigation measures***

If the beam chamber material cannot be made to satisfy the multipacting thresholds, further mitigation measures are needed. Since the electron cloud build-up depends strongly on the bunch spacing, various modifications to the beam train pattern can raise the multipacting thresholds and ease the material constraints.

One approach to obtain larger SEY multipacting thresholds is choosing filling schemes with larger bunch spacing. A bunch spacing of 50 ns results in SEY multipacting thresholds that are larger than or equal to 1.3 for all the arc elements considered, see Fig. 1.22. However, increasing the bunch spacing would require increasing the bunch intensity to maintain a constant beam current. Larger bunch intensities, in turn, could lead to problems due to other collective effects, such as exceeding the beam-beam tune shift limit and instabilities driven by the beam-coupling impedance [91].

Because the strictest constraints on the SEY arise during the charge accumulation phase, a filling scheme with non-uniform bunch intensities during this stage only, as discussed in Chapter 2, is sufficient to significantly relax the constraints. This approach leads to SEY multipacting thresholds over the full charge accumulation phase that are equal or close to those for the nominal bunch intensity, as seen in Fig. 1.22, since the effective spacing between bunches of intermediate intensity is significantly increased. As with the nominal bunch intensity, the lowest SEY multipacting threshold is found in the quadrupoles at 1.1. A potential concern with this approach is having bunches of significantly different intensities in the collider at the same time, which may make it difficult to find a working point that ensures stability for all bunches due to the $x$-$z$ instability discussed in Section 1.4.3.

Another possibility is to use filling schemes with permanently non-uniform bunch spacing, as already successfully used for electron cloud mitigation in the LHC [92] and at the former PEP-II B factory [93]. Such filling schemes have an internal structure, with a few closely spaced bunches followed by a larger gap, which repeats itself over the duration of the train. A simulation study has been done to identify effective structures, keeping the duration and total number of bunches in the train equal to the filling pattern with uniform bunch spacing. The study shows that the largest electron cloud suppression can be achieved by reducing the bunch spacing of the closely spaced bunches as much as possible in order to maximise the length of the following gap. For example, the filling pattern shown in Fig. 1.23, using an internal structure with 5 ns bunch spacing, consisting of 4 consecutive bunches followed by 16 empty 5 ns bunch slots repeated over the full bunch train, gives good electron cloud suppression with SEY multipacting thresholds that are larger than or equal to 1.2, see Fig. 1.24. Such bunch train patterns could even allow decreasing the nominal bunch population, with a corresponding increase in the total number of bunches, while keeping the surface requirements achievable. This approach could also have the benefit of reducing the severity of other collective effects, such as beam-coupling impedance and beam-beam effects.

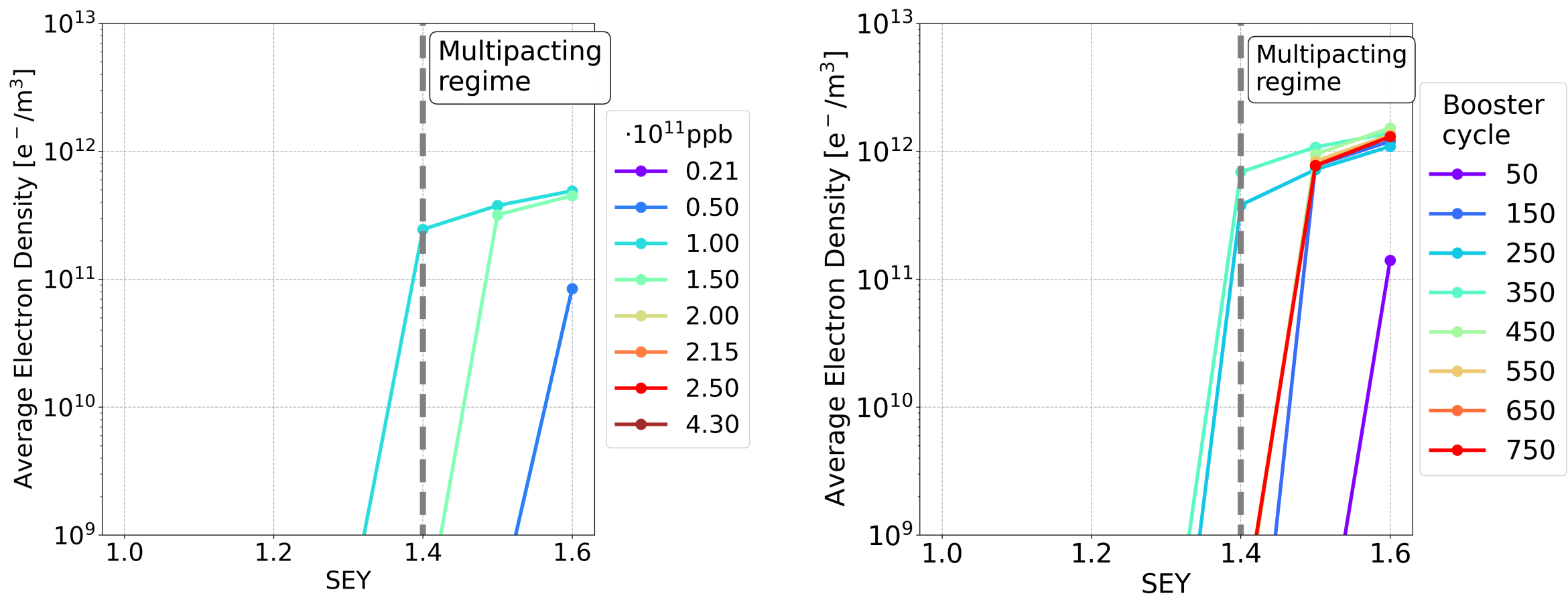

Fig. 1.22: Average electron density versus SEY in the arc dipole magnets at different stages of the charge accumulation phase with 50 ns uniform bunch spacing (left) and a filling scheme with non-uniform bunch intensity during charge accumulation (right). The SEY multipacting threshold, considering all the bunch intensities during the charge accumulation phase, is shown by the vertical dashed grey line.

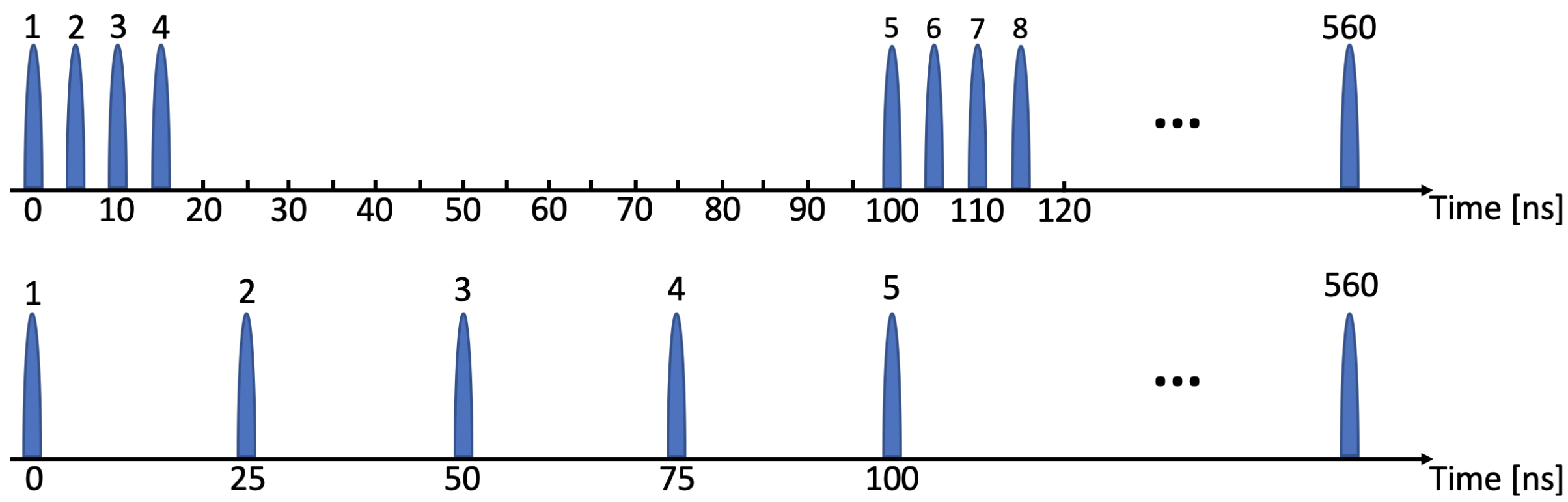

Fig. 1.23: Schematics of a non-uniform filling pattern (4e+16e) with a 5 ns bunch spacing and an internal structure consisting of 4 consecutive bunches followed by 16 empty bunch slots (top) and the nominal filling scheme with 25 ns uniform bunch spacing (bottom).

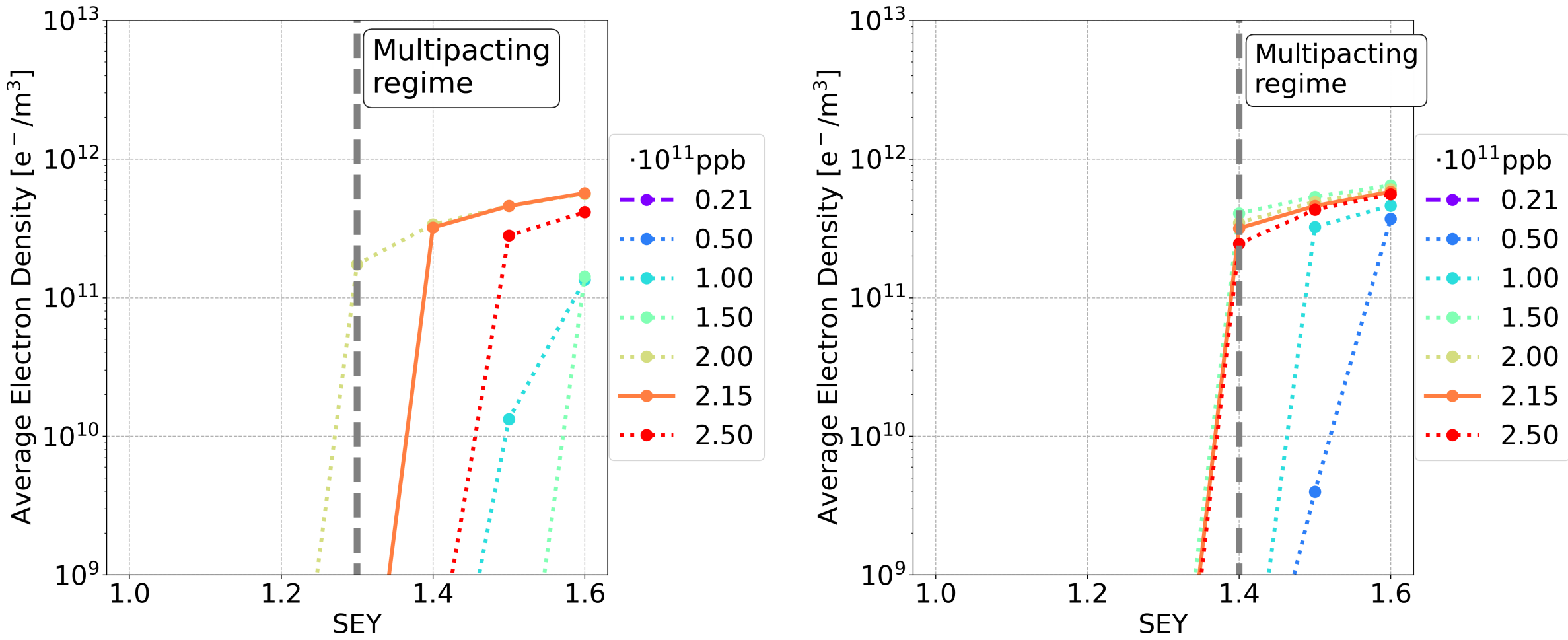

Fig. 1.24: Average electron density versus SEY at different stages of the charge accumulation phase with a non-uniform filling pattern (4b+16e) in the arc dipoles (left) and quadrupoles (right). The SEY multipacting threshold, considering all the bunch intensities during the charge accumulation phase, is shown by the vertical dashed grey line.

### 1.4.5 Space Charge

Given the large size of the FCC-ee ring and the small emittance, the space-charge tune shift is noticeable for the Z running. The vertical space-charge tune shift is [94]

$$\Delta Q_{\mathrm{SC},y} \approx \frac{N_b r_e C}{(2\pi)^{3/2}\gamma^3\sigma_z} \left\langle \frac{\beta_y}{\sigma_y\sigma_x} \right\rangle \approx \frac{N_b r_e C}{(2\pi)^{3/2}\gamma^3\sigma_z\varepsilon_x\kappa^{1/2}\left(1+\kappa^{1/2}\right)}\,, \tag{1.17}$$

where $\kappa = \varepsilon_y/\varepsilon_x$. Table 1.12 shows that the space charge tune shift approaches 0.01.

| | PETRA IV | SOLEIL II | FCC-ee collider |
|---|---|---|---|
| Beam energy [GeV] | 6.0 | 2.75 | 45.6 |
| Circumference [km] | 2.305 | 0.354 | 90.7 |
| Max. bunch charge [nC] | 8 | 7.4 | 35 |
| Rms bunch length rms [mm] | 20 | 15 | 15.2 |
| Rms vert. emittance $\varepsilon_x$ [pm] | 20 | 83 | 710 |
| Rms vert. emittance $\varepsilon_y$ [pm] | 2 | 8 | 2.1 |
| Emittance ratio $\kappa$ | 0.1 | 0.1 | 0.1003 |
| $\Delta Q_{\mathrm{SC},y}$ | 0.057 | 0.036 | 0.008 |

Table 1.12: Estimated SC tune shifts in PETRA IV, SOLEIL II, and the FCC-ee collider rings on the Z pole.

### 1.4.6 Intrabeam scattering

Intrabeam scattering (IBS) is a possible issue for the Z running, since, here, the beam energy is the lowest and the radiation damping the weakest. The largest effect is in the horizontal plane. For the nominal beam parameters the horizontal amplitude growth time due to IBS, $\tau_{\mathrm{IBS},x}$, amounts to 142 000 turns, compared with a horizontal radiation damping time, $\tau_{\mathrm{SR},x}$ of about 2400 turns. Consequently, at the Z, IBS increases the rms horizontal equilibrium emittance by $\tau_{\mathrm{SR},x}/\tau_{\mathrm{IBS},x} \approx 1.5\%$. IBS will also generate a horizontal non-Gaussian beam halo. The IBS effect is negligible in the other two planes or at other beam energies.

### 1.4.7 Vacuum and ion effects

The vacuum chamber of the two collider rings features winglets with regular photon stops to intercept and efficiently absorb synchrotron radiation. The chamber itself has a continuous coating of ultra-thin NEG material, in order to ensure an adequate vacuum pressure without long conditioning.

Baseline parameters for the collider rings relevant for vacuum and ion effects in the arcs are compiled in Table 1.13. As discussed in Section 2.3.7, at the FCC-ee in Z running mode, the beam consists of 40 trains, each containing 280 bunches spaced by $t_{\mathrm{sep}} = 25$ ns. After 1 hour of operation at nominal beam current the average vacuum pressure in the collider arcs is expected to be below $10^{-7}$ mbar (see Fig. 3.8).

Table 1.13: Collider ring parameters for Z running (45.6 GeV).

| Parameter | Symbol | Value | Unit |
|---|---|---|---|
| Bunch population | $N_b$ | 2.18 | $10^{11}$ |
| Bunch spacing | $t_{\mathrm{sep}}$ | 25 | ns |
| No. bunches/train | $n_b$ | 280 | — |
| Beam energy | $E_b$ | 45.6 | GeV |
| Hor. emittance | $\varepsilon_x$ | 0.70 | nm |
| Vert. emittance | $\varepsilon_y$ | 1.05 | pm |
| RMS rel. momentum spread with BS | | 0.121 | % |
| Av. hor. $\beta$ function | $\langle\beta_x\rangle$ | ∼100 | m |
| Av. hor. dispersion function | $\langle D_x\rangle$ | ∼0.45 | m |
| Av. vert. $\beta$ function | $\langle\beta_y\rangle$ | ∼100 | m |
| Av. hor. beam size | $\langle\sigma_x\rangle$ | ∼600 | µm |
| Av. vert. beam size | $\langle\sigma_y\rangle$ | ∼10 | µm |
| Transv. ampl. damping time | $\tau_{x,y}$ | 0.7 | s |
| Average vacuum pressure | $< P >$ | $< 10^{-9}$ | mbar |
| Emittance damping rate | $-d\varepsilon_x/dt$ | 2.5 | nm/s |
| IBS emittance growth rate | $d\varepsilon_x/dt$ | 0.1 | nm/s |

A representative residual gas component for the NEG-coated vacuum system of the FCC-ee is $H_2$. However, here we pessimistically consider CO, which has a shorter radiation length than $H_2$. In a Gaussian approximation [95], the emittance growth due to multiple gas scattering is

$$\left\langle \frac{d\varepsilon_{x,y}}{dt} \right\rangle \approx \frac{1}{2} \langle\beta_{x,y}\rangle \left(\frac{14.1\mathrm{MeV}/c}{p}\right)^2 \frac{m_{\mathrm{CO}} p_{\mathrm{CO}} c}{k_b T X_{0,\mathrm{CO}}} \, . \tag{1.18}$$

Assuming $m_{\mathrm{CO}} = 14$ g/mol, $T = 300$ K, $X_{0,\mathrm{CO}} \approx 40$ g cm$^{-2}$ (similar for $N_2$ and $CO_2$) gives

$$\langle d\varepsilon_{x,y}/dt\rangle \approx 2 \times 10^{-5} \ (\mathrm{m/s}) \ p_{\mathrm{CO}} \ [\mathrm{Pa}] \, . \tag{1.19}$$

At a pressure of $10^{-7}$ mbar, or $10^{-5}$ Pa, as reached after 1 hour of nominal operation, this emittance growth of $\sim 2$ nm/s, due to multiple gas scattering, is a few percent of the horizontal emittance damping or quantum excitation. However, in the vertical plane the growth is significant, and leads to a new rms equilibrium emittance of

$$\epsilon_{y,\mathrm{eq}} \approx \left\langle \frac{d\varepsilon_{x,y}}{dt} \right\rangle \frac{\tau_y}{2} \; . \tag{1.20}$$

Consequently, to achieve the target equilibrium emittance of 1 pm, the average vacuum pressure must not exceed $10^{-7}$ Pa or $10^{-9}$ mbar.

Another limit on the vacuum pressure is set by the beam lifetime due to bremsstrahlung [96]:

$$\frac{1}{\tau_{\mathrm{brems}}} = \sigma_{\mathrm{brems}}\, c\, n \; , \tag{1.21}$$

where the integrated Bethe-Heitler cross section for particle loss is [97]

$$\sigma_{\mathrm{brems}} \approx \frac{4}{3} \frac{\rho}{n X_0} \left( \ln \frac{1}{\epsilon_m} - \frac{5}{8} \right) \; , \tag{1.22}$$

with $n = p/(k_b T)$ the molecular density, $\rho$ the mass density, so that $\rho/n = 28$ g/$N_A$ for CO (with $N_A$ Avogadro's constant), and $\epsilon_m$ the fractional momentum acceptance, e.g. about 1% at the Z. For CO we obtain $\sigma_{\mathrm{brems}} \approx 6$ barn; for $H_2$ the cross section is $\sigma_{\mathrm{brems}} \approx 0.3$ barn. At a residual gas pressure of $10^{-9}$ mbar ($10^{-7}$ Pa), the beam lifetime would be 60 hours in case of CO and 1300 hours for $H_2$.

The ionisation cross section of carbon monoxide (CO) molecules impacted by high-energy charged particles is $\sigma_{\mathrm{iion}} \approx 2$ Mbarn [98], which translates to an ion generation rate of $\lambda'_{\mathrm{ion}} \approx 6$ m$^{-1}$ per electron at 1 Torr and 300 K. For hydrogen molecules ($H_2$) the ionization cross section is about 0.4 Mbarn [98], so that in this case $\lambda'_{\mathrm{ion}} \approx 1$ m$^{-1}$ per electron at 1 Torr and 300 K.

Ions are trapped between bunches if their atomic (or molecular) mass $A$ (in units of proton mass) exceeds a critical mass $A_c$ defined as [99]

$$A_c \equiv \frac{N_b r_p c\, \Delta t_{\mathrm{sep}}}{2\sigma_y(\sigma_x + \sigma_y)} \; . \tag{1.23}$$

For the FCC-ee parameters of Table 1.13, $A_c \approx 200$, which is significantly larger than the mass of any typical molecule of the residual gas. The Carli-Bartosik filling scheme described in Section 2.3.7 will also help prevent the trapping of ions between bunches when filling from zero after a beam abort.

## 1.5 Collimation

The FCC-ee has a target highest stored beam energy of 17.5 MJ for the most critical Z mode, and 0.3 MJ for the $\mathrm{t\bar{t}}$ mode, leading to a risk of experiment backgrounds, superconducting magnet quenches, equipment damage, radiation damage, and material activation, as a result of unavoidable beam losses. A robust collimation system is therefore needed in FCC-ee, not only for controlling the backgrounds of the physics experiments as in previous $e^+e^-$ colliders, but also to protect the machine. As a comparison, in SuperKEKB [100, 101], with a design stored beam energy of only 0.18 MJ, collimator damage and quenching of superconducting magnets have occurred as a result of sudden, unexpected beam losses, as well as high backgrounds during injections [102, 103].

Both beam losses and synchrotron radiation (SR) can cause detector backgrounds, with the latter expected to be the dominating source of machine-induced background. A distinction is hence made between the beam halo collimation system, designed to protect the machine against beam losses, and the SR collimation system, designed to protect the detectors against SR photons. The SR system consists of dedicated collimators and masks upstream of the interaction points (IPs), studied for the CDR [13] and further optimised for the present collider layout, as discussed in Section 1.6.

The halo collimation must be designed to protect the aperture bottlenecks from regular and anomalous beam losses and safely dissipate the loss power away from the superconducting final focus quadrupoles and other sensitive equipment. The beam halo collimation system must also protect the SR collimators. Excessive beam losses on these SR collimators might induce detector backgrounds or even damage them, as they are made of Inermet180 (a tungsten heavy alloy with high-Z chosen to optimise absorption) and hence less robust to beam losses than the beam-halo collimators (see below).

The beam halo collimation system was not studied for the CDR, and the following describes the first baseline design. It includes two-stage betatron and off-momentum collimation systems with specialised optics in PF [104, 105]. The betatron collimation system is located upstream of the auxiliary (non-collision) beam crossing in the middle of PF and consists of 1 primary collimator (TCP) to intercept the primary beam halo and 2 secondary collimators (TCS) to intercept particles out-scattered by the TCP, in each of the transverse planes, while the off-momentum collimation system is located downstream of the crossing and consists of 1 TCP and 2 TCSs in the horizontal plane. Secondary particle shower absorbers are placed in between the TCP and the TCS, as will be described more in detail in Section 1.5.3. The studies shown in the following are carried out for the Z mode, as this is the most critical for collimation. It should be noted that in this optics version, a vertical emittance blow-up resulting from the combination of the beam-beam interactions and the super-periodicity breaking due to the specialised collimation optics in PF has been discovered [106], as well as a reduction of the momentum acceptance. Work is ongoing to update the optics and hence also the layout, using a common LSS optics as discussed in Section 1.2. It is expected that a solution can be found, but it will imply some modifications to the presented layout.

The settings for the betatron TCPs, shown in Table 1.14, are selected to protect the aperture bottlenecks in the final focus doublets. Including alignment and beam tolerances [107] they are estimated to be 14.6 $\sigma$ in the horizontal plane ($\sigma$ is the rms betatron beam size) and 84.2 $\sigma$ in the vertical plane due to the asymmetric emittances ($\epsilon_x = 0.71\,\mathrm{nm}, \epsilon_y = 1.9\,\mathrm{pm}$). The minimum gap of the TCP is also constrained by requirements of the top-up injection scheme [108], as well as by impedance and beam lifetime considerations. Therefore, the betatron TCP cuts were chosen as 11 $\sigma$ in the horizontal plane and 65 $\sigma$ in the vertical plane, protecting the aperture bottleneck while ensuring a half-gap of at least 2 mm for impedance reasons. The physical opening is most challenging for the vertical TCP due to the optics and the flat beams. The horizontal TCP has an opening of 6.7 mm. The off-momentum TCP is set to a momentum cut $\delta_c$=1.3%, slightly outside of the RF bucket and momentum acceptance of about 1% (target value obtained without collimation optics). The betatron secondary collimators are set to provide a minimum retraction of 1$\sigma$ or 0.6 mm in the horizontal and of 10$\sigma$ or 0.3 mm in the vertical plane from the corresponding TCP setting as preliminary values to ensure the collimation hierarchy is maintained even in the case of orbit drifts, $\beta$-beating or other dynamic effects [109, 110]. The hierarchy margins are tight with the presently assumed settings, especially in the vertical plane. However, the vertical TCP is currently placed relatively far from the vertical DA of about 30 $\sigma$. Tightening their gaps is an option that is being considered, and that will relax the hierarchy margin constraints, although it is not sure that this still gives an acceptable impedance. The phase advance $\mu$ between the TCP and TCS is set by the optimal phase advance condition [111],

$$\mu = \arctan\left(\frac{\sqrt{n_{\mathrm{TCS}}^2 - n_{\mathrm{TCP}}^2}}{n_{\mathrm{TCP}}}\right), \tag{1.24}$$

where $n_{\mathrm{TCP}}$ and $n_{\mathrm{TCS}}$ are the gaps in units of $\sigma$ for the TCP and TCS. Two tertiary collimators (TCT), one for each transverse plane, are placed upstream of each IP to provide local protection of the SR collimators and the aperture bottlenecks. The settings of the TCTs have been selected to be 13$\sigma$ in the horizontal plane and 80 $\sigma$ in the vertical plane. The SR collimators have mechanical gaps in the range of 8–17 mm, corresponding to a minimum aperture of 14 $\sigma$ in the horizontal and 84.2$\sigma$ in the vertical plane, above the TCT apertures.

### 1.5.1 Collimator design parameters

The beam halo collimators must be robust enough to handle the loss of a significant fraction of the beam energy. Carbon-based materials like CFC, graphite, or molybdenum carbide-graphite (MoGr) [112] are hence considered for the TCPs and TCTs due to their potential exposure to direct large beam impacts. MoGr has about a factor 10 times better conductivity than graphite, while graphite is more robust to beam impacts. Higher-density, higher-Z materials, like Mo, or TZM (Ti-Mo-Zr), are considered for the TCSs, which intercept out-scattered particles from the TCPs. The design hence consists of 25 cm long carbon-based TCPs and 30 cm long Mo TCS. The lengths were selected based on the first studies, aiming to achieve a good balance between impedance and collimation efficiency [113–115] (impedance calculations are discussed in Section 1.4.1). In the tracking studies presented below, MoGr is assumed as TCP and TCT material, Mo as TCS material and Inermet180 as SR collimator material.

As for the mechanical design, described in detail in Section 3.6, is it assumed that the collimators have two movable jaws as in the LHC [116], with built-in BPMs and separate motors for the upstream and downstream edges to allow control of the jaw tilt angle. An LHC-like minimum step of 5 µm at each motor is tentatively assumed, although it will be refined in future studies. SR collimators are also movable but currently have no requirement for tilt angle adjustment, while the SR masks are fixed. The design of these devices is part of the MDI studies discussed in Section 1.6.

The collimator parameters and settings for all collimators, also including the SR collimators, are shown in Table 1.14. These parameters should be considered preliminary and might evolve in the future when detailed constraints from robustness and impedance are quantified.

Table 1.14: Summary table of collimator parameters and settings for the FCC-ee Z operation mode. The momentum cut $\delta_{cut}$ is not reported for collimators in nearly dispersion-free regions. All collimators are assumed to have two movable jaws and built-in BPMs.

| Type | Count | Plane | Material | Length [m] | Gap [$\sigma$] | Gap [mm] | $\delta_{cut}$ [%] | Angular adjustment |
|---|---|---|---|---|---|---|---|---|
| $\beta$ TCP | 1 | H | C-based | 0.25 | 11.0 | 6.7 | 8.9 | yes |
| $\beta$ TCS | 2 | H | Mo-based | 0.3 | 12.0 | 5.0, 7.0 | 6.0, 22.8 | yes |
| $\beta$ TCP | 1 | V | C-based | 0.25 | 65.0 | 2.4 | – | yes |
| $\beta$ TCS | 2 | V | Mo-based | 0.3 | 75.0 | 2.5, 2.9 | – | yes |
| $\delta$ TCP | 1 | H | C-based | 0.25 | 18.5 | 4.2 | 1.3 | yes |
| $\delta$ TCS | 2 | H | Mo-based | 0.3 | 21.5 | 4.6, 16.7 | 2.1, 1.6 | yes |
| TCSA | 1 | H | Mo-based | 0.3 | 15 | 8.2 | – | yes |
| TCSA | 1 | V | Mo-based | 0.3 | 91 | 3.2 | – | yes |
| TCT | 4 | V | C-based | 0.25 | 80.0 | 3.4 | – | yes |
| TCT | 4 | H | C-based | 0.25 | 13.0 | 6.1 | – | yes |
| SR BWL | 4 | H | W-based | 0.1 | 14.0 | 16.9 | – | no |
| SR QC3 | 4 | H | W-based | 0.1 | 14.0 | 17.1 | – | no |
| SR QC0 | 4 | V | W-based | 0.1 | 84.2 | 8.2 | – | no |
| SR QC0 | 4 | H | W-based | 0.1 | 14.0 | 17.4 | – | no |
| SR QC2 | 4 | V | W-based | 0.1 | 84.2 | 8.0 | – | no |
| SR QC2 | 4 | H | W-based | 0.1 | 14.0 | 17.0 | – | no |

### 1.5.2 Collimation performance studies

To judge if the cleaning performance of the collimation system is sufficient, a number of beam loss scenarios have been simulated, and further beam loss scenarios will be studied in the future to ensure that the machine is never at risk. For a full quantitative assessment of the adequacy of the collimation performance, tolerances to beam losses for different impacted elements are also needed, which are under study.

The collimation performance is evaluated using the XSUITE-BDSIM coupling simulation framework [117–122], integrating particle tracking in the magnetic lattice with particle-matter interactions in the collimators. The scenarios studied so far include betatron and off-momentum generic beam halo losses, losses from beam-residual-gas interactions, spent beam losses, and losses caused by fast instabilities for the most critical Z operation mode. The simulation outputs for these scenarios, presented in the following, provide the loss distribution along the longitudinal coordinate $s$. With the present state of knowledge, the performance is fully adequate in all scenarios studied, with the possible exception of the fast instability. In this case, interlocks or redundant damper design should be considered, as explained below. Further details on the machine-protection aspects are found in Section 2.6. A first iteration of studies has been done on the spent beam losses. However, further iterations need to be coupled to the collimation optics due to the issues discovered with vertical emittance blow-up and reduced MA.

### *Generic beam halo losses*

The scenario considered is that of a generic loss, impacting on the collimation system. The initial mechanism causing the loss is not simulated—as in LHC studies [123], instead, the beam distribution is sampled directly at the impacted TCP. The maximum impact parameter assumed is 1 μm, which may be further refined in the future. This allows to model the effect of collimator edge scattering and provides a sufficiently pessimistic estimate of the collimation performance [105]. $5 \times 10^6$ primary particles are tracked for 500 turns with SR, RF cavities and tapering of the magnets included. The loss maps in Fig. 1.25 show the power load distribution from generic horizontal betatron losses around the collider ring, normalised to a beam lifetime drop to 5 min, corresponding to 58.3 kW of loss power. This is assumed as a design specification that the system should be able to handle.

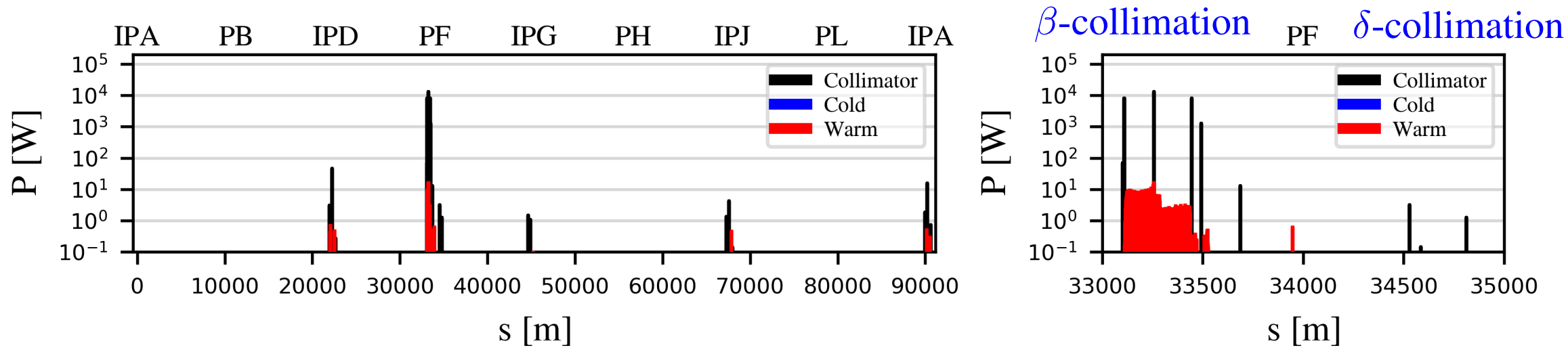

Fig. 1.25: Power load distribution loss map for generic beam halo losses of the FCC-ee positron beam, shown for horizontal betatron losses. The power loads are evaluated assuming a lifetime drop to 5 min. The beam circulates from left to right. On the right, a magnification of the collimation insertion PF is presented.

Figure 1.25 demonstrates an excellent cleaning performance, with the vast majority of losses (>99.5%) confined within the collimation insertion PF. The losses leaking out are safely intercepted by the TCTs upstream of the IPs. The collimation performance can be further enhanced by angularly aligning the collimator jaws to the beam divergence at the collimator locations [114, 115]. The more significant the beam divergence, the higher is the performance gain obtained.

### *Beam-gas losses*

The scenario considered is that of beam losses caused by bremsstrahlung interactions with the residual gas in the vacuum chamber. The assumed gas composition (85% $H_2$, 10% CO and 5% $CO_2$) and pressure distribution along the ring comes from dedicated vacuum studies see Section 3.2. To simulate the interaction with the residual gas, 10 000 beam-gas scattering centres are included in the tracking [124] and $10 \times 10^6$ primary particles are tracked for $17 \times 10^6$ equivalent turns with SR, RF cavities and magnet tapering enabled.

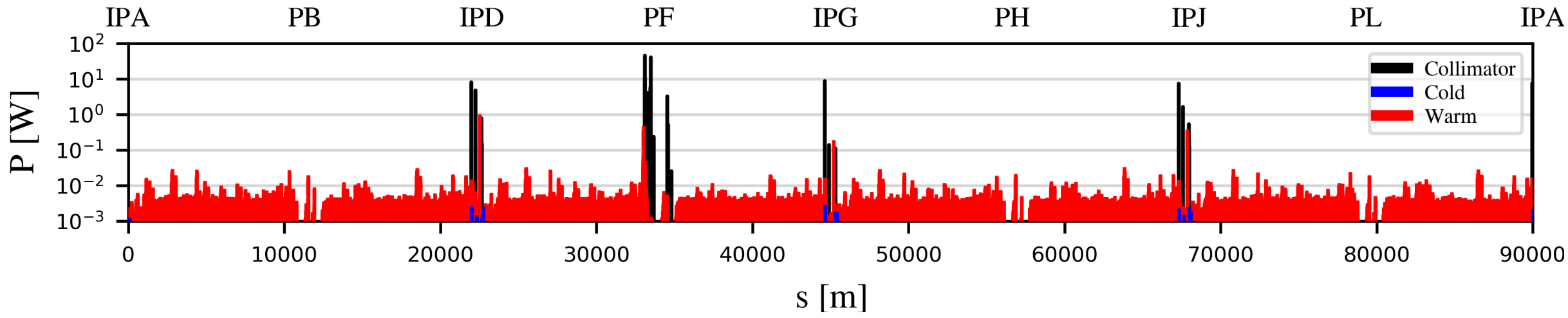

Fig. 1.26: Power load distribution loss map for beam-gas beam losses of the FCC-ee positron beam. The beam circulates from left to right. The power loads are evaluated considering a 5 h lifetime resulting from the expected pressure after 1 h of beam conditioning at full nominal current of 1.27 A.

The loss map in Fig. 1.26 shows the power load distribution from beam-gas losses, normalised to a beam-gas lifetime of 5 h that results from the expected pressure after 1 h of beam conditioning at full nominal current of 1.27 A. This pessimistic scenario represents the start of FCC-ee operation, and the pressure is expected to condition down by a factor of up to 100 over time (Section 3.2). Consequently, beam-gas interactions are unlikely to significantly affect the lifetime of the FCC-ee, which is primarily determined by Bhabha scattering at the IPs. Even in this pessimistic scenario, low power loads (<0.1 W) are expected on most components, with the highest loads recorded on the halo collimators (10-100 W) and SR collimators (1 W). Such power load levels are not a concern.

***Fast instability***

The fast instability scenario assumes the failure of the feedback system designed to mitigate the coupled bunch instability described in Section 1.4.2. This instability is not modelled using the beam interactions with the impedance but through eight synchronised dipole kickers instead, one per arc, to reproduce a smooth exponential growth of the betatron oscillation amplitude, either in the vertical or horizontal plane. The kicker strengths are defined as $k = (A_0/\sigma_{x,y}) \cos\big(2\pi Q_{x,y} t\big) \exp(t/\tau)$, where $A_0$ denotes an arbitrary amplitude, $\sigma_{x,y}$ the local beam size, $Q_{x,y}$ the betaron tunes, and $\tau$ the instability rise time. Two scenarios are studied: a rise time of either three turns (representing the worst case) or six turns. The dependence on the phase advance is also analysed. In each case, $5 \times 10^5$, 45.6 GeV primary electrons are simulated with SR and magnet tapering enabled. Several phase advances between the initial kick and the TCP have also been studied (0°, 30°, 60°, 90°).

The simulated beam oscillates coherently until the collimator apertures are reached, after which it is entirely lost within a few turns. At the Z mode, this results in the release of 17.5 MJ on collimators over a few turns. Due to the short rise time, losses of the order of MJ can be expected in the collimators as shown in Fig. 1.27. It is under study whether the collimators can sustain this impact, but the FCC-ee should be designed in such a way that the instability does not occur in this way, for example by interlocking the damper or the orbit, or through a redundant damper system that would increase the time constants in case of failure.

### 1.5.3 Studies of energy deposition and shower absorbers

In addition to the studies of the global cleaning performance, FLUKA [125–127] radiation transport simulations for the PF insertion were carried out in order to quantify the beam-induced power deposition in the machine and the environment. A specific geometry model of the betatron collimation system, including collimators, beam pipes, magnets and the machine tunnel, was implemented. The collimator jaws and tanks were represented by simplified models since no technical design existed at this early stage. The absorber blocks of the primary and secondary collimators were assumed to be made of graphite and TZM, respectively, with a similar cross section to the LHC collimators. The blocks were embedded in a metallic frame made of TZM. A generic beam loss scenario was simulated assuming that beam particles impact on the front face of the primary collimators (vertical and horizontal), at a distance of 1 μm from the collimator edge. The simulations were carried out for operation at the Z pole, since the stored beam energy and hence the expected power loss in the betatron collimation system is much higher than for the other beam modes. The interaction of the 45.6 GeV electrons and positrons in the blocks results in the production of electromagnetic showers, which are not contained in the primary collimator jaws because of the significant shower length and the small impact parameter; the simulations show that primary collimators absorb less than 0.5% of the impacting energy, while most of the energy is carried away by secondary photons, electrons and positrons. Most of these shower particles are lost on the downstream vacuum chambers in PF, or they are intercepted by the secondary collimators. The simulations show that the first secondary collimator absorbs about 40% of the power, while almost half of the power is deposited in the chamber walls or leaks into the tunnel environment.

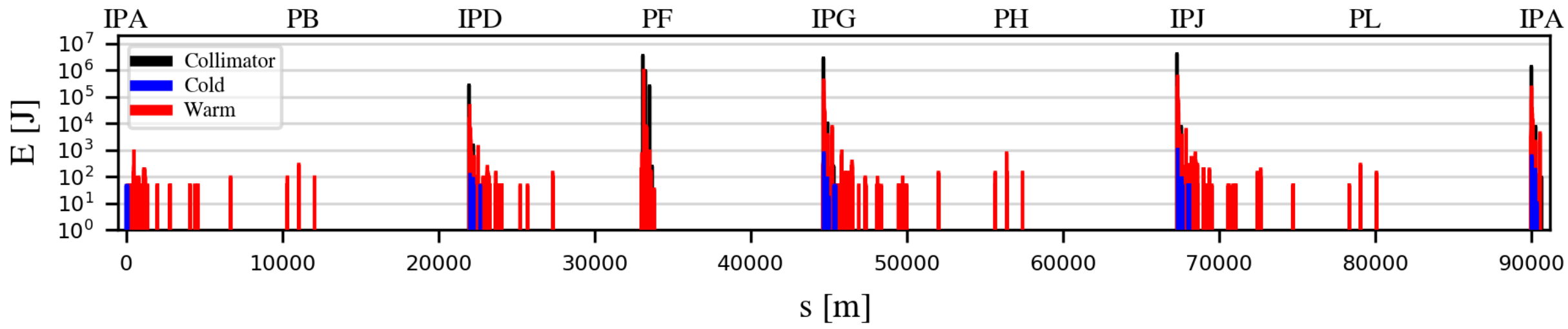

Fig. 1.27: Integrated loss map over all turns for a fast instability in the horizontal plane with a rise time of 3 turns.

Considering the significant particle leakage from the collimators, it is necessary to install additional shower absorbers in PF, which reduce the distributed production of radionuclides in the surroundings and mitigate radiation effects in other equipment. In addition, the shower absorbers dissipate the heat created in a more controlled way. The effects of placing two shower absorbers (one vertical and one horizontal) between the primary and secondary collimators have been investigated. Their position was optimised in such a way as to minimise the energy escaping to the vacuum chambers and the environment. Using the same design as the TCS but with a wider gap (15$\sigma$ and 91$\sigma$ in the x- and y-plane, respectively), the shower absorbers proved to be an effective strategy for mitigating the energy leakage; just two shower absorbers can reduce the power deposition in the vacuum chambers and tunnel to 15%. The optimal position and number of shower absorbers in PF are expected to evolve in the future, depending on the final collimation layout and distances between the collimators. In addition, shower absorbers or masks are also likely to be needed for the momentum collimation system and possibly also near the tertiary collimators in the experiment insertions.

## 1.6 Machine-detector interface (MDI)

The MDI of the FCC-ee has a compact and complex design [128–131] that fulfils constraints given both by the machine and the detector requirements.

A common IR layout is requested for all FCC-ee energies and is shown in Fig. 1.28. The flexibility of the IR optics is obtained by splitting the final focus quadrupoles (FFQs) QC1 and QC2 into three and two segments, respectively, and by modulating their sign and strength according to the beam energy.

Tens of nanometres in the vertical beam size and a few micrometres horizontally require small $\beta$-functions at the IP, as reported in Table 1.2. The distance of the face of QC1 from the IP ($\ell^*$) is 2.2 m, well inside the detector volume. The crab-waist collision scheme requires a small horizontal beam size and a relatively large crossing angle at the IP, set to 30 mrad, which results with the beams entering/exiting with separate beam pipes at about 1.32 m from the IP. In addition to the optical constraints on the interaction region (IR) layout, physics reconstruction benefits from a stay-clear cone of 100 mrad from the interaction point (IP) along the $z$-axis. [4]

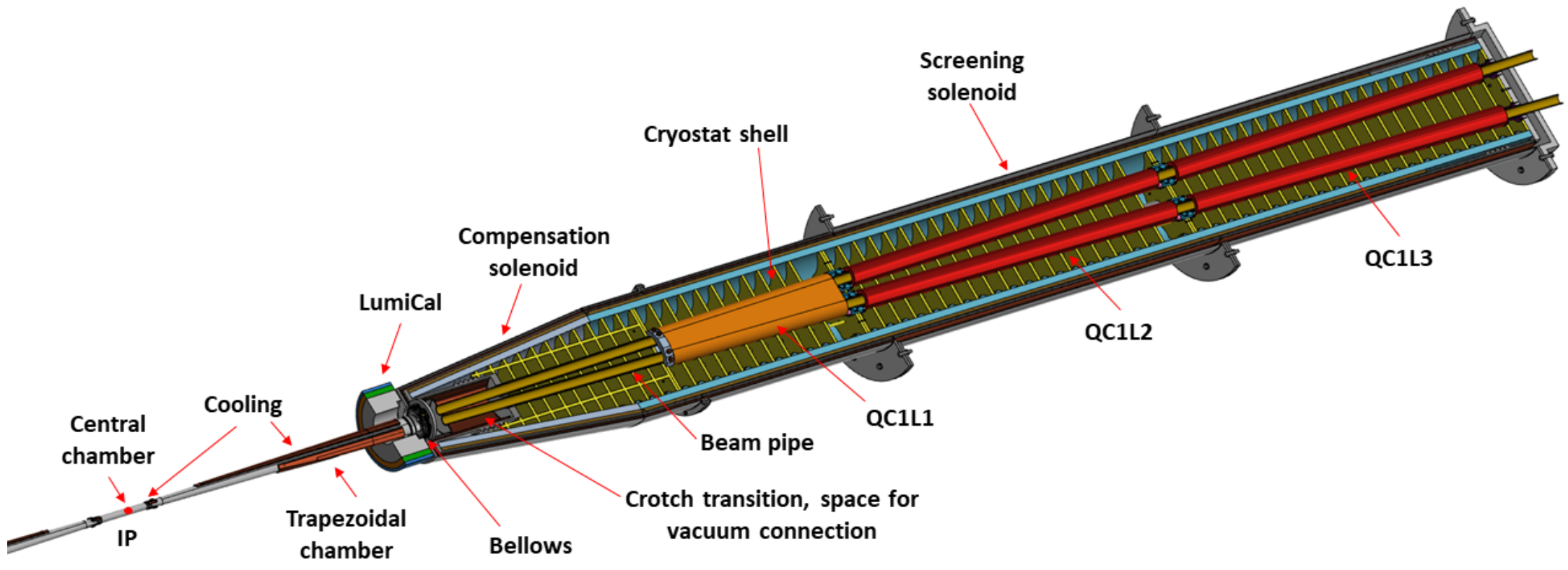

Fig. 1.28: Section view of the accelerator components from the IP to the end of the first final focus quadrupole (QC1), at about 5.6 m.

Two schemes are considered to compensate for the coupling induced by the detector solenoidal magnetic field and the crossing angle. The baseline one, called *local scheme*, uses a couple of strong compensating solenoids with opposite sign with respect to the one of the detector and placed either side of the IP, within $\pm\ell^*$ (shown in Fig. 1.28); this scheme requires a maximum detector field of 2 T, and 5 T for the compensating one, in order to limit the vertical emittance growth to about 30%. An alternative scheme, named *non-local* , envisages an equal and opposite strength magnetic field to that of the detector, positioned at either side of the IP several metres outside of the detector, possibly tolerating higher detector magnetic fields to about 2.5 T, inducing however, a high spin-depolarisation effect whose mitigation is currently under study. In the local scheme the field integral $\int \vec{B}ds$ is cancelled before the FFQs, while in the non-local one the compensating solenoids need additional very weak local dipole correctors and skew quadrupoles around the FFQs, to correct orbit and dispersion. In both schemes a screening solenoid is requested, around the QC1 portion inside the detector, to cancel the effects of the detector magnetic field on both beams. In the non-local scheme the emission of synchrotron radiation due to the compensating solenoid is much reduced with respect to the local scheme. These studies will be completed in the next design phase.

Two calorimeters, known as LumiCal, are positioned in front of the compensating solenoids, as illustrated in Fig. 1.29. They are designed to measure the integrated luminosity with an accuracy of $10^{-4}$.

Achieving this level of precision requires the relative positioning of the two calorimeters to be known within $\pm110$ µm, necessitating long-term mechanical stability. Additionally, to avoid compromising the luminosity measurement, the material budget within their angular acceptance range (50 to 110 mrad) must be minimised. To further reduce uncertainties in energy reconstruction, the LumiCal must be assembled and installed as a single, mechanically rigid unit.

The vertex detector, also shown in Fig. 1.29, is placed as close as possible to the interaction region (IR) beampipe, which has an internal radius of 1 cm. It covers an angular range of approximately $|\cos\theta| < 0.99$. The above topics are discussed in more detail in Volume 1 of this report.

The sections below describe the main features of the MDI. A more detailed discussion can be found in Ref. [132].

[4] The detector's coordinate system is defined with its origin at the nominal collision point. The $z$-axis is aligned along the bisector of the incoming and outgoing beam directions, with its positive direction corresponding to that of the outgoing positrons. The $y$-axis points vertically upward, while the $x$-axis extends radially outward from the centre of the FCC. The azimuthal angle $\varphi$ is measured from the $x$-axis in the $x$-$y$ plane, with the radial coordinate in this plane denoted as $r$. The polar angle $\vartheta$ is measured from the $z$-axis.

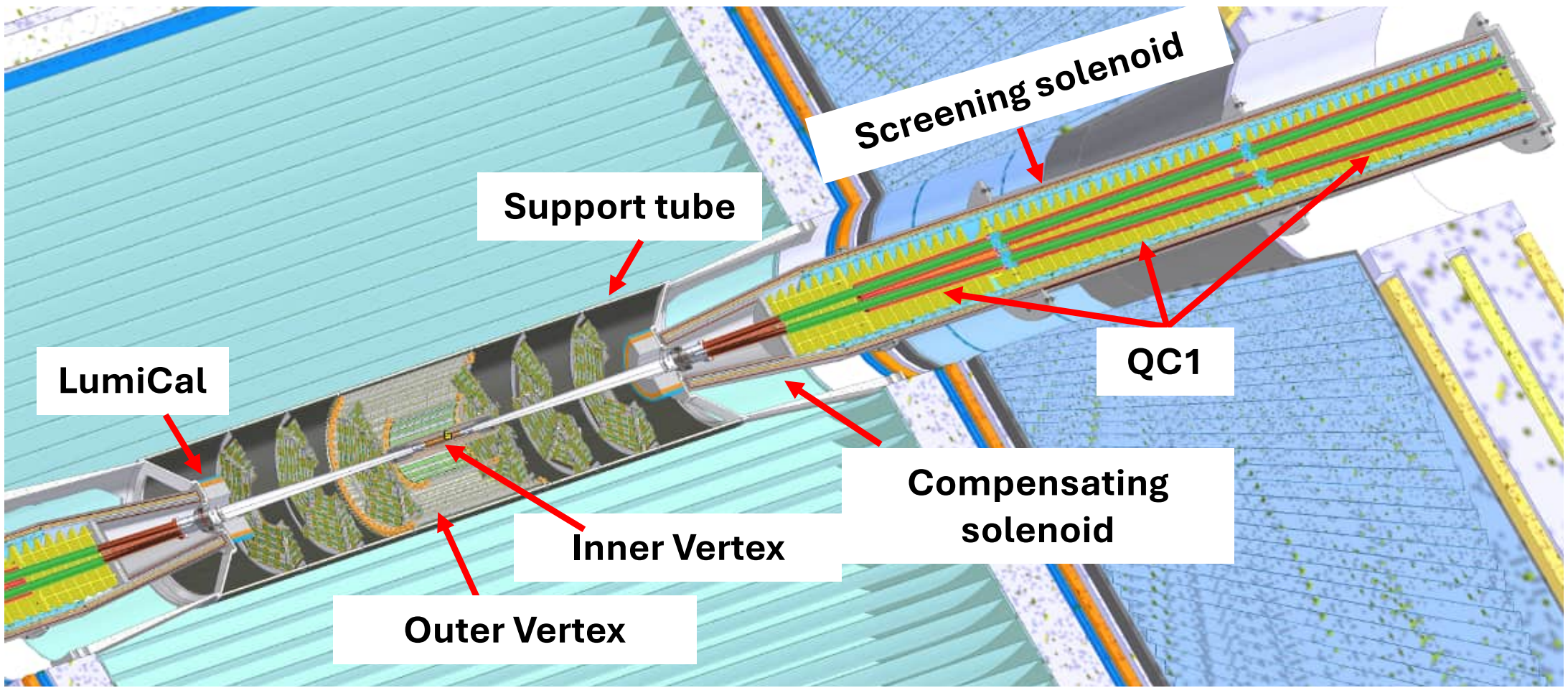

Fig. 1.29: Layout of the interaction region. The support tube allows the integration of the luminosity calorimeter (LumiCal) and the vertex detector. The three segments of the final focus quadrupoles (QC1) are shown with the screening and compensating solenoids.

### 1.6.1 Interaction Region layout

The central beam pipe is 18 cm long with 10 mm inner radius, followed by a pair of ellipto-conical beam pipes 1064 mm long on either side, as shown in Fig. 1.30. All of these are made in AlBeMet162, an alloy of 62% of beryllium and 38% aluminium, chosen for its high modulus and low-density characteristics. The mechanical model of the IR vacuum chamber is designed to provide low impedance, low material budget, and mechanical resistance while guaranteeing thermal stability by removing heat load with a suitable cooling system. The impedance was minimised by carefully designing the transverse section of the beam pipe, with a smooth transition from a circular to an elliptical transverse shape, as discussed in Ref. [133]. An internal coating layer of 5 µm gold, inside the central beam pipe, ensures a good thermal and electrical conductivity to minimise the beam heat load, with a maximum value of nearly 60 W expected at the Z pole [133], and shields the vertex detector from residual high energy synchrotron radiation photons. The central vacuum chamber envisages a double layer structure made of two concentric cylinders, each with a thickness of 0.35 mm and assembled with a 1 mm gap for the liquid paraffin cooling system, thus bringing its effective diameter to 23.4 mm.

The ellipto-conical vacuum chamber extends between 90 mm and 1154.5 mm from the IP, and its thickness is tapered from the central value until it reaches 2 mm. Water flowing into the AlbeMet cooling channels on top of the ellipto-conical beam pipe refrigerates it, extracting an expected heat load of about 130 W at the Z pole; an asymmetric design is needed to comply with the angular acceptance of the luminosity calorimeter, which is centred around the outgoing beam pipe axis.

The beam pipes will be supported by an external lightweight carbon-fibre structure (support tube) by means of two bellows [129], inspired by the DAΦNE and ESRF designs [134]. The support tube consists of an empty cylindrical multilayered wall rigid structure, that eases the integration of the MDI components, including the vertex and the LumiCal detector, providing a cantilevered support for the central beam pipe, as shown in Fig. 1.29.

A thermo-structural analysis has been performed to calculate the temperature distribution, stress, strain and displacement of the beam pipes. The temperature distribution for the two chambers is shown in Fig. 1.31 for operation at the Z pole which has the highest thermal load. The maximum temperature of the central chamber reaches 29 °C, cooled with paraffin entering at 18 °C, and reaches 50 °C in the conical chamber which is cooled with water entering at 16 °C. The maximum stress has been calculated considering the constraint from the configuration of a cantilevered support, which results in a maximum displacement of 0.5 mm, and a maximum stress ten times lower than the AlBeMet162 yield strength (193 MPa).

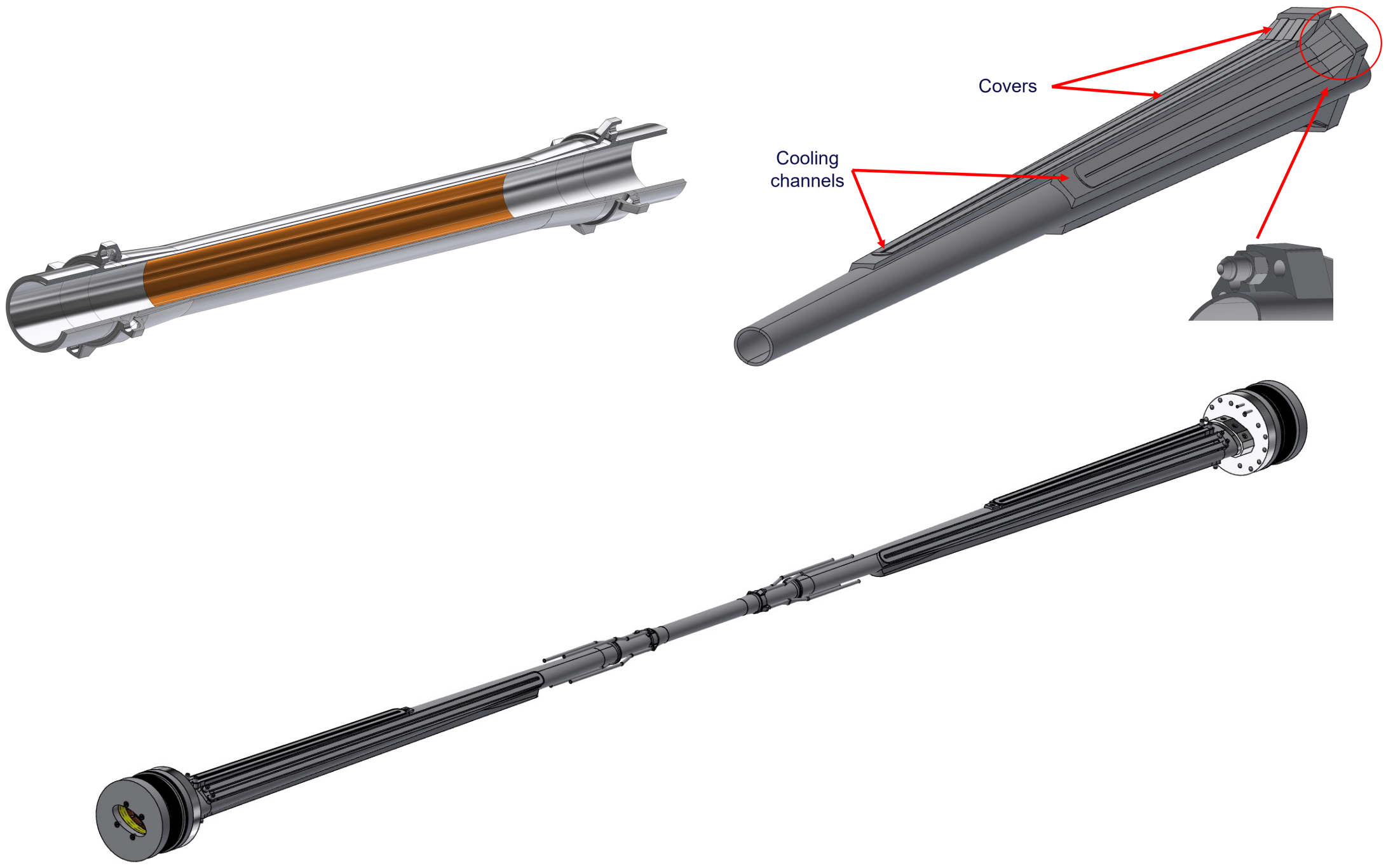

Fig. 1.30: Top Left: Central chamber including cooling inlets and outlets for the paraffin cooling circuit housed in a double layer and its internal gold coating layer; Top Right: ellipto-conical vacuum chamber with asymmetric cooling channels; Bottom: assembly of the IR chambers.

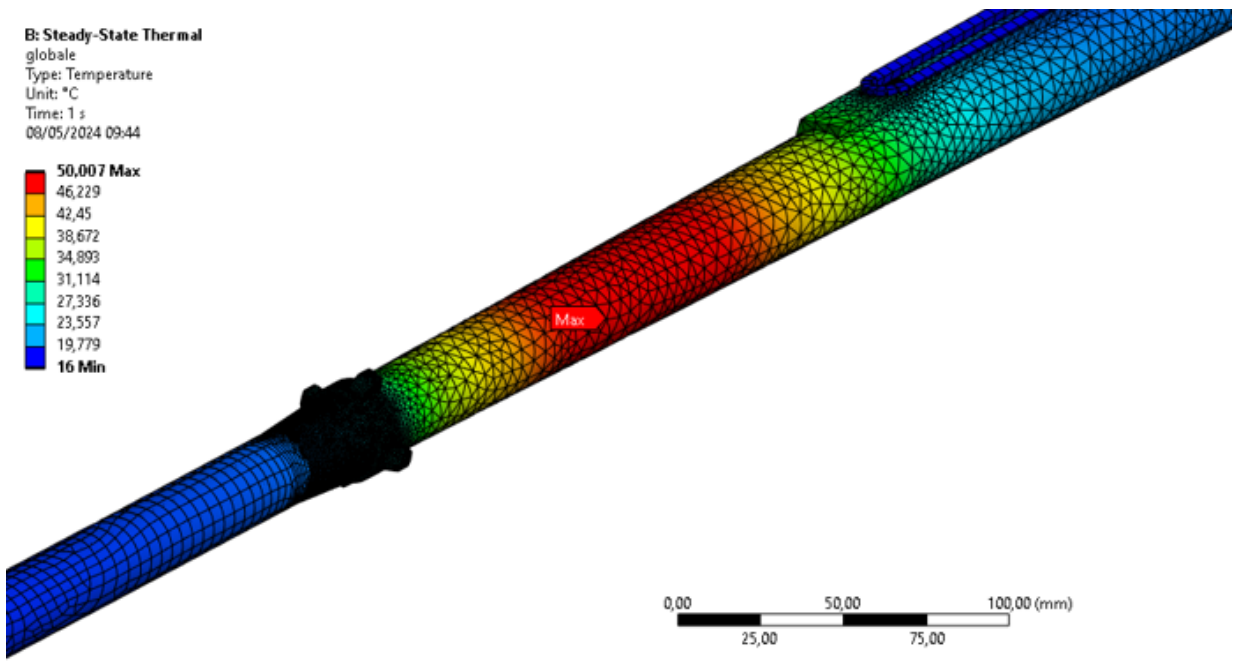

Fig. 1.31: ANSYS simulation of the temperature distribution along the ellipto-conical chamber for a deposited power of 54 W over the central chamber and 130 W over the conical chamber.

The integration of the vertex detector has been studied for the IDEA detector concept. The detector is placed on top of the vacuum chamber: two thin peek-based rings, anchored on either side at about 170 mm from the IP on the ellipto-conical chamber, hold a carbon fibre structure supporting three layers of silicon vertex detectors located at about 13.7, 23.7, and 35 mm radii, as shown in Fig. 1.32. Additional cylindrical layers and disks of silicon detectors complement the vertex detector in the reconstruction of charged particles. For a more comprehensive description, see Ref. [131].

A comprehensive calculation of beam induced HOM power deposition on the IR chambers including the effects on the bellows, and BPMs has to be finalised.

A remote vacuum connection device, located inside the niche of the front side of the QC1 cryostat (see Section 1.6.2), needs to be studied to provide a vacuum to the IR chambers once assembled inside the detector. The tight space and accessibility in that area pose considerable challenges and a solution is being studied.

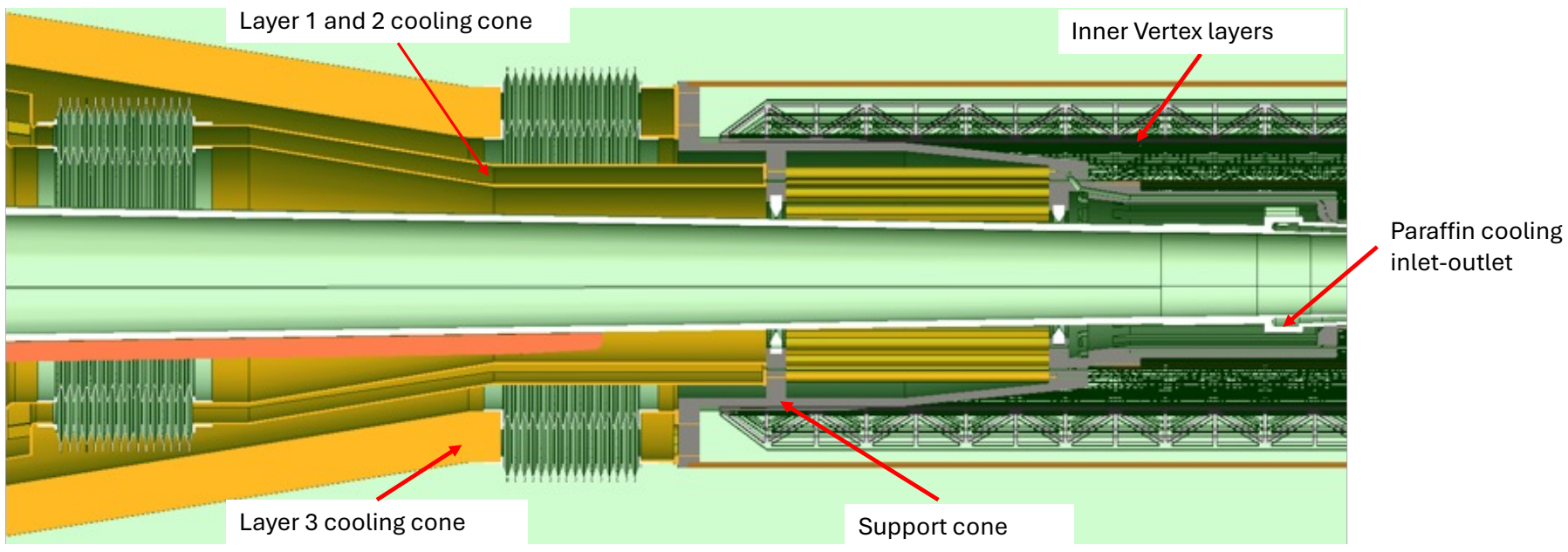

Fig. 1.32: Longitudinal section of the beam pipe and the inner vertex. The dark grey object is the conical support of the vertex detector, which is supported by the conical beam pipe. At the right edge of the support cone, the inlet/outlet paraffin cooling manifolds are visible.

### 1.6.2 IR magnet system

The IR magnet system consists of the superconducting (SC) FFQs, the solenoids, and the correctors, all housed in a cryostat. The optimal solution is to house QC1 and QC2 in separate cryostats, since the QC1 is entirely inside the detector, thus allowing it to be accessed separately from QC2, which is outside. A critical issue concerns the limited space available inside the QC1 cryostat, in which the two beam pipes come very close to each other due to the crossing angle. At one end, the challenge is to allow sufficient thermal shield between the warm beam pipes and the cold quadrupole coils. At the other end, the challenge is to allow sufficient space for the winding of the corrector coils, especially around the QC1 segment closest to the IP. As shown in Fig. 1.28, the shape of the QC1 cryostat closest to the IP features a niche to house warm elements, such as bellows, remote vacuum connection, and a beam position monitor (BPM). Other BPMs should be placed at the entrances of QC1 and QC2 for each beam. Widening the angular size of the cryostat, as seen from the IP, currently set to 100 mrad, could alleviate the problems mentioned above, but the impact on the detector calorimeter acceptance placed behind it needs to be evaluated.

The FFQs are based on the canted cosine theta design (CCT), with Nb-Ti conductors. Three different options to operate the FFQ cryostat at different temperatures are envisaged: pressurised He II at 1.9-2.1 K, supercritical He at 4.5 K, or He gas forced flow at 10-20 K. The first option allows a superfluid regime, minimising any possible induced vibrations. The third one minimises the power budget, while the second is intermediate. Yet another solution under study considers using high-temperature superconducting (HTS) magnets, which would allow even further power reduction.

The anti-solenoid and screening solenoid needed to compensate for the coupling induced by the detector field and the crossing angle, bring additional constraints to the IR magnet system. As said at the beginning of this chapter, two coupling compensation schemes are under study.

### 1.6.3 Alignment, detector integration and maintenance

An alignment strategy and monitoring system is under study, and envisages three main components [135]. The first one monitors the shape deformation of the screening solenoid's support [136], and uses in-line multiplexed and distributed frequency scanning interferometry (IMD-FSI) [137] to monitor sections of optical fibres firmly installed on the inner surface of the screening solenoid support.

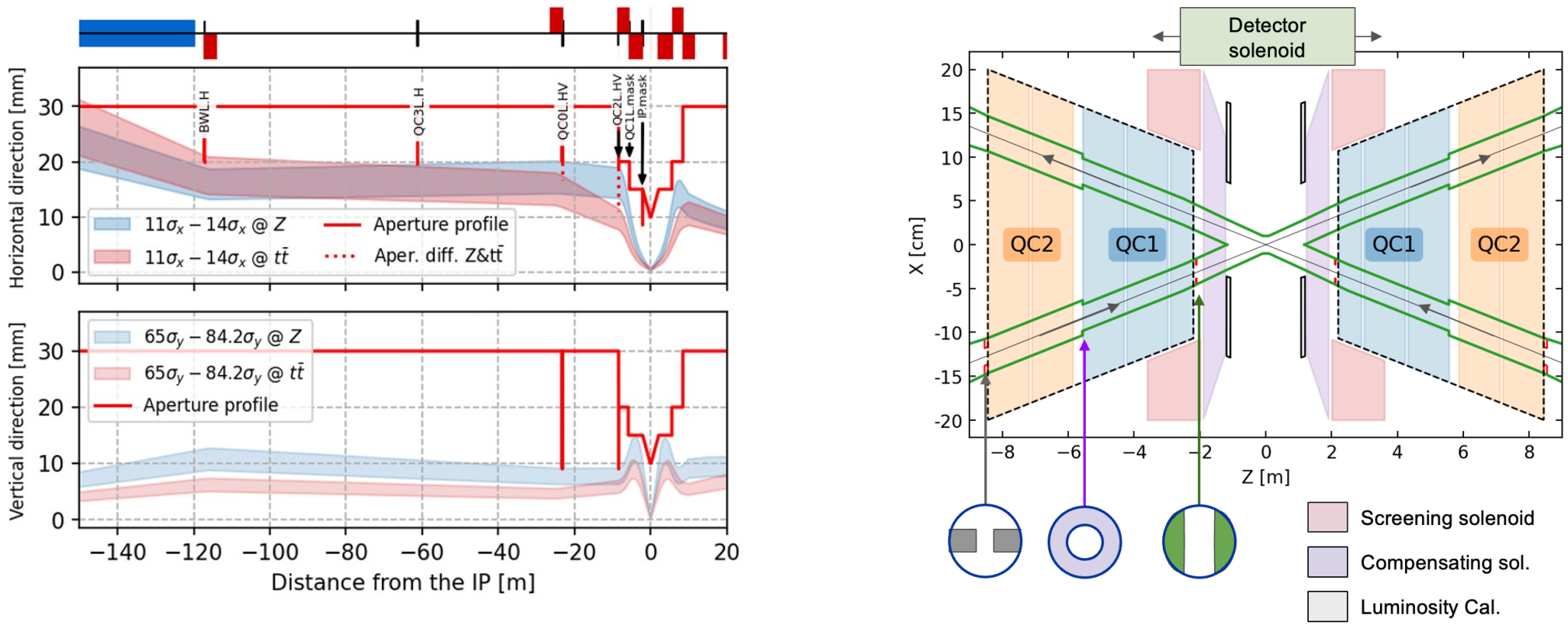

Fig. 1.33: SR collimators and masks upstream the IR (left); SR masks shapes and locations at the FFQs (right).

The second system utilises a mirror, installed at the end of each fibre, to redirect the laser beam towards the centre of the assembly, targeting the FFQs, the BPMs, LumiCal. This is very similar to the FSI heads installed on the low-$\beta$ quadrupoles in the HL-LHC MDI, as described in [138]. These distance measurements will monitor the position of the inner components relative to the cryostat. Finally, to ensure the alignment of both sides of the MDI, a long-range alignment system will be installed, also based on FSI but using a different optical setup to enable longer-distance measurements. A set-up at CERN is currently studying the experimental validation of the alignment system of the FFQs with the FSI system, using a 1:2 mock-up of the beam pipes and cryostat.

The accessibility in the FCC-ee MDI region may be limited by accelerator components, such as the FFQ cryostats and the booster ring (Section 4.1). Three opening scenarios for the detectors installed in either the large or small experiment caverns are described in Section 5.3 of Volume 1.

### 1.6.4 Beam induced backgrounds

Backgrounds in the IR arise either from processes where particles from one beam lose energy or deviate from their trajectory or interact with those of the opposite beam at the IP. Collimators and absorbers, as described in Section 1.5, remove the bulk of the particles (electrons/positrons and photons) which eventually would hit the detectors, but some effects may remain due to the scattering of the particles of these devices.

The effect of synchrotron radiation (SR) in the MDI region has been simulated with the latest version of BDSIM [139, 140], utilising the GEANT4 toolkit, which includes the X-ray reflection. The bulk of the SR is almost collinear with the beam, and thanks to the final focus optics design, does not enter the detector, however a fraction of it can still reach it as a result of magnet misalignments, imperfections, and beam tails [141]. SR masks are designed to stop such radiation close to the detector, located before and after the FFQs, as shown in the right plot of Fig. 1.33. The mask apertures are 15 mm for the circular ones between QC1 and QC2, and 7 mm horizontal aperture for the mask after QC1, closest to the IP. In realistic conditions, some scattering at their edges can still enter the detector region. Studies are ongoing to evaluate the effects in the various sub-detectors, and eventually optimise the masks and tertiary collimators; as an example Fig. 1.34 shows the power deposited in the IR, assuming 5 minutes beam lifetime, non-zero closed orbits with a transverse deviation of 100 µm, transverse divergence of 6 µrad, and beam tails.

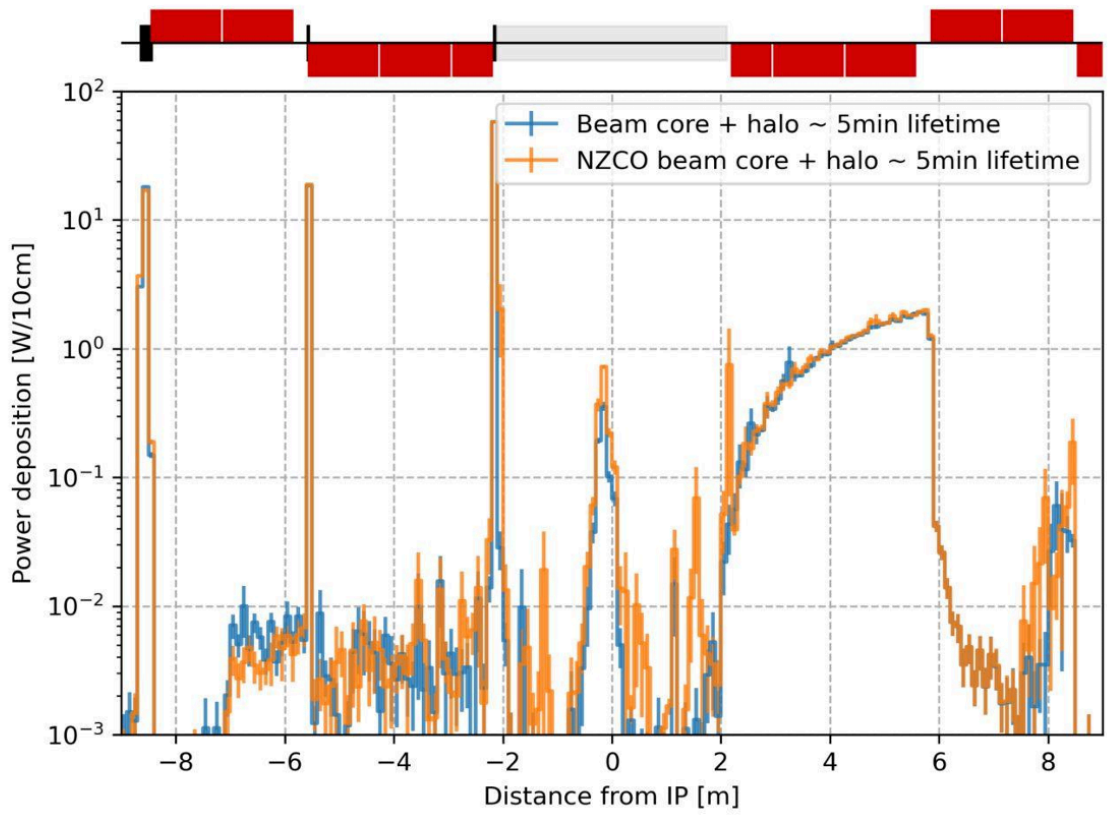

Fig. 1.34: SR power deposition in realistic conditions. The blue line represents the beam core with a halo corresponding to 5 minutes lifetime, the yellow line represents a beam core with a non zero closed orbit (NZCO) and halo.

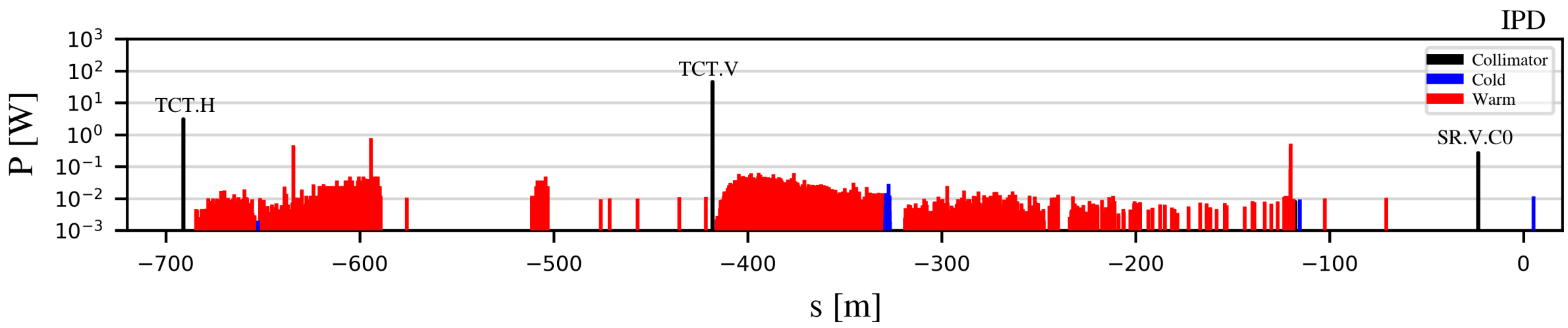

Fig. 1.35: Power load distribution loss map for generic beam halo losses of the FCC-ee positron beam, illustrating horizontal betatron losses in the region spanning 700 m upstream of IPD. Power loads are evaluated assuming a beam lifetime drop to 5 min. The beam circulates from left to right.

Electrons and positrons may lose energy or deviate from the central trajectory, thus populating the tails of the phase space, and are commonly known as halo beam losses [142]. These particles are mostly intercepted by the collimators. This effect has been simulated with XSUITE-BDSIM simulation tool [143]. Some of these particles may scatter off the edges of the collimators and reach the interaction region. The vast majority of the beam halo losses are intercepted by the tertiary collimators (TCTs) which have been added upstream of the SR collimators, and minimal losses ($\mathcal{O}(10\mathrm{mW})$ beyond the last SR collimators before the IPs are observed in all cases, as shown in Fig. 1.35. In this way, the background contribution from beam halo particles that leak from the beam halo collimation system (located in PF) is not expected to be an issue. Nevertheless, the background contribution from particle showers arising from the interaction of beam halo particles with the SR collimators might not be negligible and should be studied in the future.

Background and energy deposition due to beam-gas interactions, incoherent pair production, and radiative Bhabha scattering are discussed in Section 5.4 of Volume 1.

The bulk of the radiation emitted in the IR is collinear with the incoming and outgoing beams, and is dumped at the end of a 500 m long tunnel. This radiation is composed of two main contributions, beamstrahlung and synchrotron radiation [144]. In addition, radiative Bhabha events produced at the IP generate off-energy electrons that are lost in the magnetic elements within 150 m of the IP. All these sources have been studied at the Z pole and above the $t\bar{t}$ threshold. Radiative Bhabha are the main source of radiation in the first 200 m at the Z, while beamstrahlung dominates at around 500 m at the dump. Due to the harder spectrum of synchrotron radiation at $t\bar{t}$ energies, which has a critical energy of the order of 1 MeV, this source overwhelms the others. For more technical discussions, see Section 1.9. Studies are ongoing to evaluate the possibility of using the spatial profile of the beamstrahlung radiation for beam-beam fine tuning, as proposed for other colliders.

The effects of the thermal photons and injection backgrounds will be studied in the next phase of the project.

### 1.6.5 Experimental implementation of the interaction region

The finalisation of the system engineering of the IR is of paramount importance for both the machine and the detector layouts.

The construction of a full-scale mock-up of the IR beam pipes, including the results of the study of the integration with the vertex and LumiCal detectors and using the concept of the support tube, is ongoing at INFN-Frascati in collaboration with INFN-Pisa and CERN.

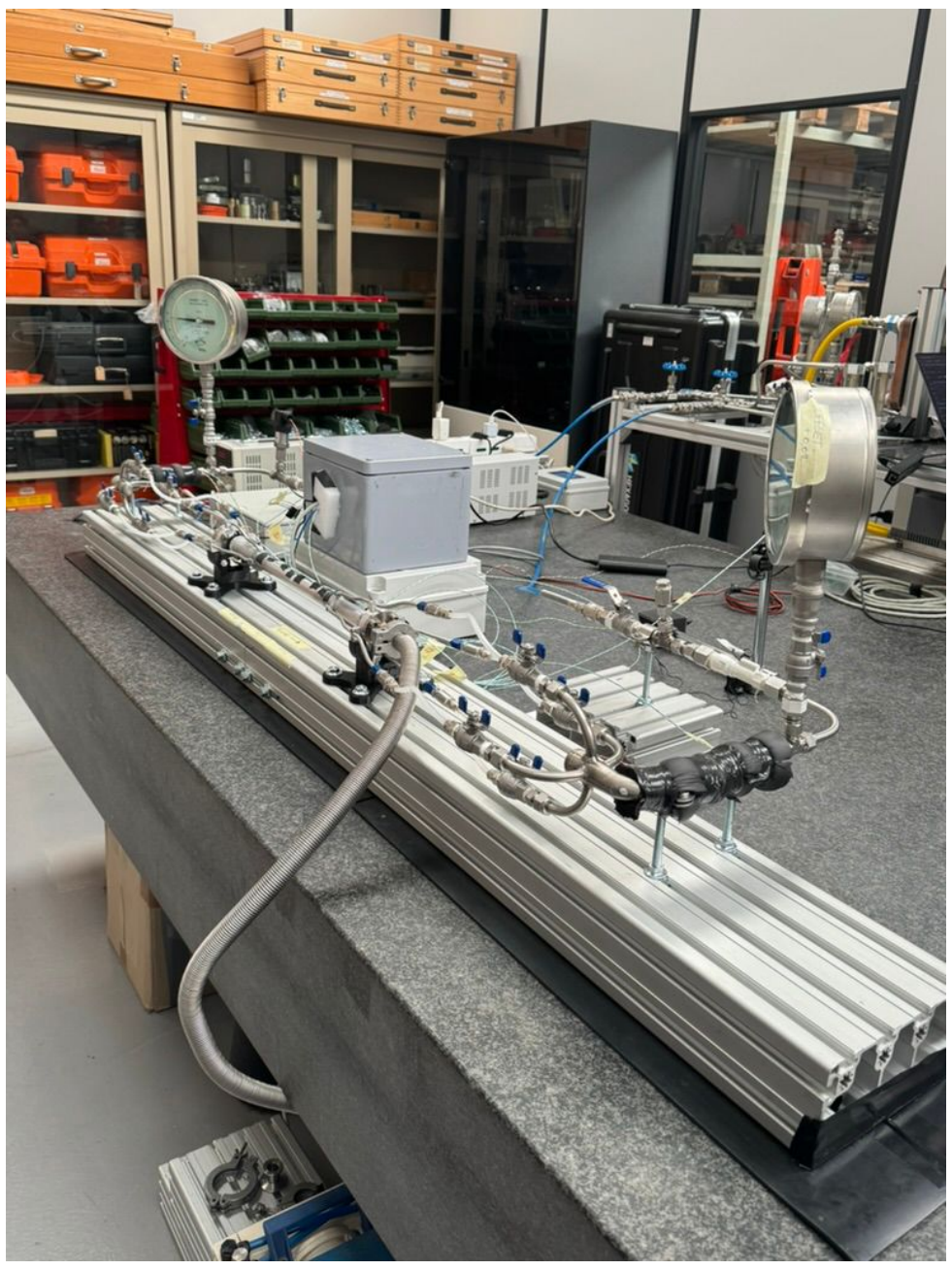

Fig. 1.36: Measurement setup for central beam pipe cooling system.

The first prototype of the central beam pipe was manufactured in aluminium and equipped with flanges to validate the cooling performance. The assembly of its components (see Fig. 1.30) has been performed using laser beam-welding in the ENEA-Casaccia laboratory. The test system comprises a cooling circuit with an operating range between 0.08 and 0.033 kg/s of liquid paraffin, temperature sensors, precision pressure gauges, and flow-meters. An ohmic internal heater provides the heat load expected from the wakefields, up to a total power of nearly 100 W allowing a factor of two safety margin. The setup is shown in Fig. 1.36. Initial measurements performed using water as a coolant confirm the expected behaviour, showing that for a nominal power of 54 W, flowing water at 0.017 kg/s with an inlet temperature of 18°C and 19°C at the exit, the beam pipe external surface temperature remains at 19°C, with a pressure drop of 0.13 kPa. Raising the power to 100 W and reducing the flow to 0.08 kg/s the beam pipe temperature increases to 24°C, still within the margins of the system.

The elliptic-conical beam pipe prototypes are being fabricated in aluminium and will eventually be welded to the central beam pipe, also made in aluminium. The cooling manifolds of the elliptic-conical beam pipes will be soldered using an electro-beam welding technique, and their performance will be validated using a similar system to that for the central beam pipe.

The bellows prototype will be fabricated in aluminium, and will allow the study of the assembly procedure, the welding of an elliptical geometry, and the effectiveness of the thermal/electrical contact.

At the same time, a mock-up of the inner vertex detector, along with the air-cooling cones, is being fabricated using carbon fibre. This aims to validate both the mechanical assembly procedure and the cooling performance in a dedicated experimental setup. The outer tracker and disks will be constructed from aluminium and integrated into the support tube, which will also house a LumiCal mock-up made from 3D-printed material.

The full-scale IR mock-up will ultimately be used to study the integration sequence of various components, helping to identify potential critical issues. Additionally, it may serve as a platform for investigating the alignment system.

## 1.7 Energy calibration and polarisation

A principal task of FCC-ee is to probe for physics beyond the Standard Model by making ultra-precise measurements of a wide range of electroweak observables, whose overall consistency can then be assessed. Knowledge of the collision energy $\sqrt{s}$ is a key input to many of these measurements, and this is obtained through measurements of the mean beam energy $E_b$. Corrections must then be applied to the naive relation $\sqrt{s} = 2E_b$ to obtain the centre-of-mass energy at each interaction point.

In electron or positron storage rings, transverse polarisation naturally builds up through the Sokolov-Ternov effect. The spin tune, defined as the ratio of the spin precession frequency to the revolution frequency, is proportional to the average beam energy $E_b$. The spin tune can be directly measured by the procedure of resonant depolarisation (RDP), in which the frequency of a depolariser kicker magnet is varied until the polarisation is found to vanish, when the depolariser frequency corresponds to the spin precession frequency. This technique has been exploited at many facilities, such as VEPP-2M [145], VEPP-4M [146], CESR [147], DORIS [148], and, most notably, at LEP in scans of the Z resonance [149]. Alternatively, in a free spin precession (FSP) measurement the depolariser may be used to rotate the spin vector into the horizontal plane, and the precession frequency can then be measured directly.

These polarisation-based precision measurements, however, will only be possible for Z-pole operation and at energies up to and including the $\mathrm{W^+W^-}$ threshold. At higher energies, the polarisation level will be too small for RDP and FSP measurements to be practical, and, here, the energy scale will have to be determined from physics processes at the experiments, such as $\mathrm{e^+e^-} \to f\bar{f}\gamma$ production, as it was done by the LEP experiments, e.g. Ref. [150].

For the ZW and $\mathrm{t\bar{t}}$ modes of operation, these energy-calibration data from the detectors could be complemented by dipole-magnet spectrometer techniques, as carried out at LEP [151] or based on the distribution of laser-Compton back-scattered electrons, utilising the polarimeter set up [152] (Subsection 1.7.5). Yet another possibility will be to regularly inject pilot bunches pre-polarised in the injector complex, and to measure their FSP after injection. Simulations predict that, up to the ZH energy, a polarisation level of $\sim 10\%$ can be preserved during the FCC-ee booster energy ramp [153].

When calculating $\sqrt{s}$ it is necessary to have good knowledge of the crossing angle of the two beams, to account for local energy variations from synchrotron radiation, the RF system and impedance, and to consider the effects of opposite sign vertical dispersion at the interaction points.

The knowledge of $E_b$ at LEP was dominated by the sampling rate of RDP measurements, which were performed outside physics operation with a periodicity of around a week. The energy was found to vary significantly between measurements due to several effects, for example, earth tides [149]. In order to enable the much greater degree of systematic control that the vastly larger sample sizes at FCC-ee warrants, the operational strategy will be very different to LEP. Measurements of $E_b$ will be performed several times an hour on non-colliding pilot bunches. Around 160 pilot bunches per beam will be injected at the start of the fill, and wiggler magnets will be activated to speed up the polarisation time. One to two hours will be required for the polarisation to build, after which the wigglers will be turned off and physics (colliding) bunches will be injected. The RF frequency will be continually adjusted to keep the beams centred in the quadrupoles, thus suppressing tide-driven energy changes, which would otherwise be $\mathcal{O}(100\,\mathrm{MeV})$. A model will be developed to track residual energy variations between measurements.

In Ref. [154], it was demonstrated that the systematic uncertainty from the knowledge of the collision energy on the key electroweak observables can be greatly reduced compared to what was possible at LEP. More recent studies have confirmed this conclusion and showed that further improvements are possible. For example, energy-related uncertainties of around 100 keV and 12 keV are envisaged for the Z mass and width, respectively, to be compared to the equivalent numbers of 1.7 MeV and 1.2 MeV at LEP [149]. These estimates should, however, be regarded as provisional, and efforts are underway to reduce them further. The uncertainty expected on the W mass from the knowledge of $\sqrt{s}$ is around 160 keV, which is sufficient for the statistical precision. Steps towards improvements and greater robustness in the energy calibration were explored at a workshop at CERN in autumn 2022 [155].

The following provides a brief status report of the key components required for the $\sqrt{s}$ calibration. The discussion is focused on the accelerator, but remarks are also included on important inputs that will come from the experiments, in particular, the measurement of the spread in the centre-of-mass energy $\delta_{\sqrt{s}}$, which must also be known to a high degree of precision. More details are given in Volume 1 of this Report. A fuller discussion on all these topics may be found in Ref. [156].

### 1.7.1 Beam polarisation and optimisation

The polarisation of electron and positron beams naturally builds up over time. The maximum theoretical polarisation is 92.4% and it is oriented anti-parallel and parallel to the magnetic field, for electrons and positrons, respectively. In an error-free flat machine, i.e. in the absence of vertical bending magnets, or solenoids, this means that the polarisation is fully vertical [157], and, hence, $\vec{n_0} \parallel \hat{y}$. The design-orbit spin tune is equal to $\nu_0 = a\gamma_{\mathrm{rel}}$, where $a$ is the gyro-magnetic anomaly and $\gamma_{\mathrm{rel}}$ the relativistic Lorentz-factor. The spin precesses around $\vec{n_0}$. Due to strong synchrotron radiation, the local beam energy varies significantly along the circumference. $\nu_0$ corresponds, therefore, to the average beam energy over one revolution with an error below 0.3 keV.

In practice, the level of polarization is lowered by several mechanisms, including magnetic and alignment errors, which can lead to resonance excitation between the spin-orbit and the betatron- and synchrotron motion. Since depolarising effects are stronger for larger vertical, closed orbits, well-optimised orbit correction and optics tuning techniques are required to achieve sufficient polarisation. In addition to optics and emittance tuning techniques (see Section 1.3), dedicated spin-matching bumps have been studied, and these show an improvement in polarisation [158]. Furthermore, errors not only reduce the achievable polarisation, but can also lead to a shift between $a\gamma$ and the measured spin tune $\nu_0$. To study the Z-line shape measurements at beam energies in the range of 43.85 to 47.37 GeV are foreseen. In recent studies, no systematic offset between $a\gamma$ and $\nu_0$ is found between the studied beam energies around the Z-pole. Additionally, simulation studies which assume up to 100 µm arc and 25 µm IR rms misalignments (plus 5% randomly missing BPMs, 1% BPM random scaling errors, and 1 µm random BPM resolution error) [159, 160], respectively, yield an absolute offset between $a\gamma$ and $\nu_0$ well below 100 keV for the vast majority of seeds [159]. These studies only included orbit corrections. Additional optics tuning techniques are likely to reduce the remaining offset further.

The alternative IR layout with a non-local solenoid compensation scheme leads to interleaved spin deflections around the longitudinal and the horizontal axis, resulting in a deviation of $\vec{n_0}$ estimated to roughly 10 µrad. Nevertheless, this deflection reduces the asymptotic polarisation, simulated in SAD, to about 1 %. Current studies aim at increasing the level of polarisation by introducing vertical pi-bumps, as was performed in LEP [161]. Although these bumps could increase the vertical emittance, this effect is presumed to be minor since they only need to correct $\vec{n_0}$ of 10 µrad, and thus, the required bump strength is expected to be rather small.

### 1.7.2 Wigglers

At 45.6 GeV, the natural polarisation time is 250 h. Thus, achieving a polarisation level in an error-free machine of 5-10% requires 15 to 30 hours, which is an unacceptably long period without calibration at the start of the fill. Reducing the polarisation rise time to about 12 h is feasible using wigglers, which also increase the rms energy spread to 64 MeV. In the currently planned operational scenario, low-intensity ($\approx 10^{10}$ particles) pilot bunches are injected at the start-of-fill and polarised using asymmetric polarisation wigglers [162, 163], similar to those at LEP [164]. When roughly 10% polarisation is achieved, after approximately 100 minutes at the Z-mode, the wigglers are switched off, and all the nominal-intensity colliding bunches are then injected and brought into collision. Around 160 pilot bunches are injected at the start of each fill, with an estimated lifetime of roughly 20 h. This corresponds to roughly one pilot bunch being available for an RDP scan every 7.5 min. By the time the last pilot bunch has been depolarised for the first time, the pilot bunches that were depolarised first will have naturally reacquired sufficient polarisation to be measured again.

The wiggler design for FCC-ee follows the three-pole design of the LEP damping wigglers. Wigglers will be grouped in packages of three units, and two packages will be installed in consecutive 16 m long drift spaces. Their current placement in the FCC lattice is in the straight section downstream of each IP. The required number of polarisation wigglers and their specifications are given in Table 1.15.

Table 1.15: Specification for the polarisation wigglers.

| | |
|---|---|
| Number of units per beam | 24 |
| Central field $B_+$ [T] | 0.7 |
| Central pole length $L_+$ [mm] | 430 |
| Asymmetry ration $r = B_+/B_- = L_-/L_+$ | 6 |
| Critical energy of SR photons $E_c$ [keV] | 968 |

### 1.7.3 Pre-polarised pilot bunches

In the current baseline approximately 100 min are required to polarise the pilot bunches in the main rings before commencing with injecting nominal physics bunches after every beam dump. Availability studies presented in Section 2.1.2 suggest that injecting pre-polarised pilot bunches could significantly enhance the time available for physics, especially when considering failure scenarios. Furthermore, this scheme would ease constraints on the maximum achievable polarisation and, hence, would limit the necessity of additional spin bumps, which could introduce additional vertical emittance. Nevertheless, injecting pre-polarised pilot bunches into the main rings demands a careful evaluation of the injector design, which must be suitable for generating polarised electrons and positrons and allow sufficient polarisation transport through the full injector chain and energy ramp. These studies have begun and will continue in the next phase of the project.

### 1.7.4 Depolariser

The pilot bunches are depolarised with an electromagnetic kicker using transverse fields (RF-kicker), with a TEM-wave travelling towards the beams and a varying excitation frequency. Once the driving frequency is equal to the spin-tune, the polarisation vector is rotated away from the vertical direction, leading to depolarisation or spin flip. The proposed tune-changing rate corresponds to 1 keV/s.

Radiative diffusion gives the spin resonance a natural width of about 200 keV at the Z-pole and 1.4 MeV at the W-pair-threshold. These values are significantly larger than the desired precision. Recent studies suggest that by alternating the scanning direction an uncertainty of a few keV at the Z-pole is achievable. Furthermore, RDP has recently been successfully simulated for the W-energy for the first time. RDP only yields a sufficiently large change in polarisation if the spin-modulation index $B =$ $\nu_0\sigma_E/Q_s < 1.5$, with the energy spread $\sigma_E$ and the synchrotron tune $Q_s$, which ensures a low number of synchrotron-tune sidebands inside the distribution of the spin tune [165, 166].

To achieve a sufficient spin rotation, a vertical kick of 10 µrad is required. A single kick would lead to a propagating orbit through the machine and thus this bump must be closed. In order to achieve a spin rotation, dipoles must be located within the closed orbit bump. It is found that the regular arc optics would allow sufficient rotation over four FODO cells [167]. Since one closed orbit bump of 10 µrad would lead to a vertical peak orbit above 1 mm, it is proposed to distribute it over four closed-orbit bumps per beam, each providing 2.5 µrad. At least two kickers per orbit bump are required, constraining the phase advance to 180°. A weaker third kicker would ease this constraint. Hence, a total of 16 to 24 kickers are required. The third correction kicker could be designed as a slightly shorter strip-line with less RF power installed, as it only has to provide corrections to the bump.

Each kicker providing 2.5 µrad features a strip-line design of 1 m length with four electrodes, operating around 40 MHz. Keeping the nominal vacuum chamber diameter of 70 mm requires an RF power per kicker port (electrode) of 35 kW, and it is therefore suggested to reduce the vacuum chamber at the kicker to 26 mm with electrodes at a distance of 9 mm from the beam. This reduces the RF power per kicker port to 2.26 kW (9.04 kW per kicker). Studies are underway to ensure that the overall impedance remains at an acceptable level [168]. The following Table 1.16 shows the configuration in point PA, where one-quarter of the necessary depolariser kickers are proposed to be installed. Similar configurations are proposed in the other three experiment points. All or part of the depolariser kickers can also be used as kickers for transverse feedback systems for instability mitigation.

Table 1.16: Possible configuration of depolariser kickers in point PA generating a local 2.5 µrad bump, assuming that the depolariser kicker system will be distributed over all four experiment points in a similar way to provide the total effect required for RDP. A shorter kicker is used for bump correction between the two main depolariser kickers.

| **Location** | **Beam** | **Function** | **Kicker length** | **Power per kicker** |
|---|---|---|---|---|
| Point PA left | electron | open bump | 1.0 m | 9.04 kW |
| Point PA left | electron | correction | 0.75 m | 4.5 kW |
| Point PA left | electron | close bump | 1.0 m | 9.04 kW |
| Point PA right | positron | open bump | 1.0 m | 9.04 kW |
| Point PA right | positron | correction | 0.75 m | 4.5 kW |
| Point PA right | positron | close bump | 1.0 m | 9.04 kW |

The technique of FSP is being investigated as a complementary approach to RDP. Here, the vertically orientated spin is flipped into the horizontal plane, and the coherent (free-spin) precession is then observed. The spin tune is then retrieved by a Fourier transform, which also yields the full spin spectrum of the spin motion. This technique would require a kicker pulse about ten times stronger than that planned for the RDP measurement [169].

Since residual longitudinal polarisation in colliding bunches would modify the cross-section and forward-backward asymmetries, it must be controlled to a level below $10^{-5}$. Hence, it is envisaged to regularly depolarise colliding bunches, requiring a selective RF-kicker.

### 1.7.5 Inverse Compton scattering polarimeter

Measurements of the polarisation of the FCC-ee beams will be performed through the process of inverse Compton scattering. The physics goals of FCC-ee set various requirements that the polarimeter system must fulfil.

- The requirement to calibrate and study the electron and positron beams separately means that each ring requires at least one polarimeter. In order to provide redundancy and meet the target availability for the polarimeter measurements, which is set at 95%, it is foreseen to have two instruments in each ring.
- The polarimeters will be deployed on the pilot bunches for both RDP and FSP measurements. During Z-pole operation they will also perform measurements on physics bunches in order to ensure that any longitudinal polarisation is kept sufficiently low.
- The statistical precision on the measurements of the transverse polarisation of the pilot bunches should be around 1% per second. A single bunch will be probed in each measurement.
- Many more ($\sim$ 100) physics bunches will be probed in a single measurement period, the laser temporal pattern being only indicative here and will be further optimised in the future. The physics-bunch studies place the most stringent demands on the systematic control of the absolute polarisation measurement: it is desirable to measure a polarisation level consistent with zero to a precision of $\sim 10^{-4}$.
- FSP and longitudinal-polarisation studies require that the complete spin vector be characterised. This can be achieved through measuring the spatial distribution of the scattered electrons (positrons), as well as the backscattered photons [170].
- Knowledge of the relative positions of the scattered electrons (positrons), back-scattered photons and electrons (positrons) allows real-time measurement of the beam energy, which can attain a statistical precision of $10^{-3}$ per second [170]. This capability will be valuable for many physics studies and will be optimised in the ongoing design of the polarimeter system.

A schematic drawing of the FCC-ee polarimeter is shown in Fig. 1.37.

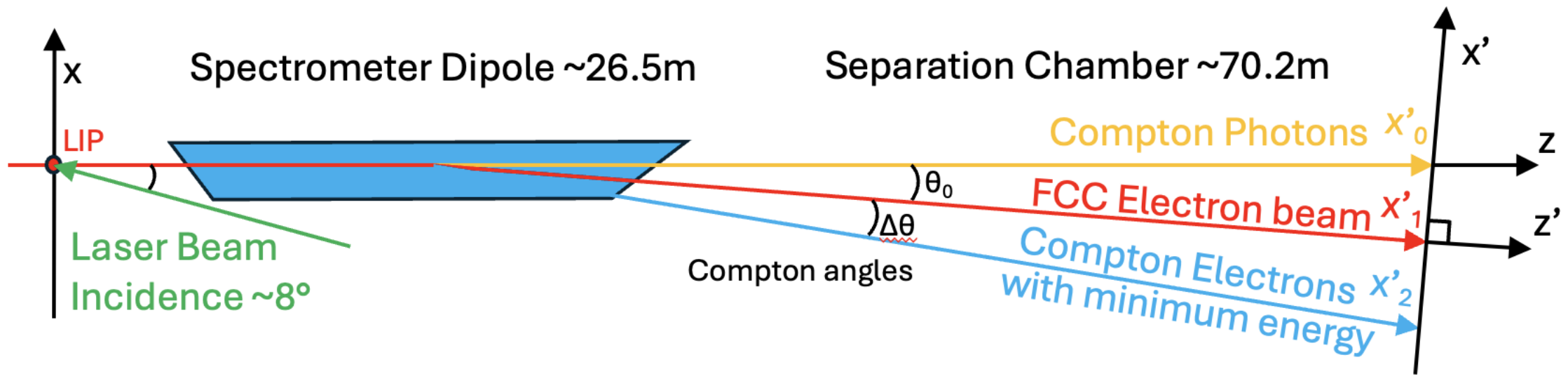

Fig. 1.37: Schematic drawing of the FCC-ee polarimeter. More details can be found in Ref. [170].

The most suitable location for the polarimeter using the most recent machine optics design is in the straight section 830 m upstream of the experiment IPs. The dispersion suppression dipole can then also be used as the system spectrometer magnet. This location is followed by 100 m of field-free propagation, allowing sufficient separation of the Compton products from the main beam. The room that houses the laser would be installed in a shielded region as close as possible to the laser interaction point (< 50 m). To meet the targeted availability for the energy-calibration measurements in the next phase of the project, it will be necessary to thoroughly investigate and validate the reliability of a system based on fully remote laser control. Studies of the homogeneity requirements on the dispersion suppression magnet are also planned.

The laser will operate at a green wavelength of around 515 nm, a choice which provides the optimum compromise between the field-free distance required and the reliability and versatility of operation. Both Q-switched Nd:YAG and Yb mode-lock technologies are under consideration. Currently the latter is favoured, as it seems best adapted to providing suitable pulses to both the pilot and physics bunches. Table 1.17 shows the key parameters for such a choice. The crossing angle is implemented in both the horizontal and vertical planes, with the vertical crossing angle necessary to take the scattered photons out of the plane of the synchrotron radiation.

Table 1.17: Preliminary laser parameters for pilot and colliding bunches. Note that single-bunch charges are different for pilot and colliding bunches.

| **Technology** | **Q-switch** | **Modelock Yb** | **Modelock Yb** |
|---|---|---|---|
| Bunch type | Pilot | Pilot | Colliding |
| Repetition frequency | 3 kHz | 3 kHz | 3 kHz |
| Number of targeted bunches | 1 | 1 | 10 |
| Pulse energy | 3 mJ | 3 mJ | 50 µJ |
| Average power | 9 W | 9 W | 1.5 W |
| Pulse duration | 3 ns | 30 ps | 30 ps |
| Beam width ($\sigma_{x/y,l}$) | 1 mm | 1 mm | 1 mm |
| Crossing angle | 2 mrad | 8 deg | 8 deg |
| Scatters per bunch crossing | 260 | 290 | 94 |
| Scatters per second | 8 $10^5$/s | 9 $10^5$/s | 28 $10^5$/s |

The polarimeter will contain two detector systems. The first will record the transverse ellipse of the Compton scattered electrons over a surface of about $5 \times 300\,\text{mm}^2$ (for Z-pole operation). The second will record the peak-shaped distribution from the Compton gammas on a detector of about $10 \times 10\,\text{mm}^2$ transverse area. The baseline design for the detectors are pixelated sensors of 18-50 µm pitch in the transverse direction.

Monte Carlo simulation models are under development to evaluate and optimise the expected polarimeter capabilities. These models include a description of the laser/beam interaction chamber, the separation chamber of almost 100 m, and the detector systems foreseen to record the Compton products. A preliminary study has been performed using the Toy Monte Carlo model described in Ref. [171]. Further investigations and optimisation of the system are now being pursued based on a complementary model developed using BDSIM [120], a GEANT4-based package that includes a description of particle-matter interactions.

### 1.7.6 IP-specific corrections to the collision energy

The resonant-depolarisation measurements determine the mean beam energy around the ring. A variety of mechanisms either induce local variations in the beam energy or lead to other corrections that need to be applied when calculating $\sqrt{s}$, the collision energy.

The beams experience continuous energy losses around the lattice from synchrotron radiation (SR) and resistive-wall impedance, which are compensated by the RF-cavities. It has been demonstrated in Ref. [154] that to minimise the $\sqrt{s}$ shifts at the IPs, both beams must have all RF-cavities located in one straight section, identical for the electron and positron beam. This is planned for both Z pole and $W^+W^-$ operation. While SR radiation losses are identical for pilot and colliding bunches, being, for example, 39 MeV at 45.6 GeV, impedance losses increase with increasing bunch intensity. Recent studies estimate a loss per revolution of about 0.8 MeV and 1.6 MeV at the Z pole, respectively, for $3 \times 10^{10}$ and $2.6 \times 10^{10}$ particles per bunch [172]. Furthermore, beamstrahlung energy losses, which range from 0.31 MeV per IP at $\sqrt{s} = 91.2$ GeV up to 14 MeV for $\sqrt{s} = 365$ GeV, must be monitored, for example by using a beamstrahlung monitor. Beamstrahlung losses increase with the beam energy and the bunch population, and lead to bunch lengthening, together with an increased energy spread. This beam-beam interaction accelerates the particles before the collision, decelerates them afterwards, and also modifies the crossing angle, but in a manner that leads to no net change in the collision energy.

The collision energy is also modified by opposite-sign vertical dispersion (OSVD) and collision offsets at the IP. At each IP, the shift in collision energy is [154]

$$\Delta\sqrt{s} = -2u_0 \frac{\sigma_E(D_{u,B1} - D_{u,B2})}{E_0(\sigma_{B1}^2 + \sigma_{B2}^2)} , \tag{1.25}$$

where $D_{u,B1,B2}$ is the dispersion at the IP for each beam, $\sigma_E$ is the beam-energy spread, which is assumed to be identical between electrons and positrons, $\sigma_{B1,B2}$ are the beam sizes, $E_0$ is the nominal beam energy and $u_0$ is the offset. Assuming 1 µm of spurious OSVD at the IP, as may occur according to optics tuning studies, the $\sqrt{s}$ is shifted by roughly 100 keV per nm of offset. The vertical offset can be determined and controlled via luminosity scans, where the beam positions at the IPs are scanned against each other until the optimum luminosity is achieved. At LEP such scans yielded a precision better than 0.1 µm at the Z-pole. However, at the FCC-ee an improvement by a factor 100 will be necessary. Complementarily to observing the luminosity increase with vertical beam positions, beam-beam deflection scans are envisaged; these rely on observing the change of crossing-angle from beam-beam with varying vertical beam position. In principle, the dispersion of the colliding bunches at the IP can be obtained by changing the momentum by applying a small RF-frequency shift and measuring the resulting orbit shift in the BPMs closest to the collision point. However, the maximum allowed RF-trim which does not alter the optics, as well as the interplay of orbit shifts due to beam-beam and dispersion, especially at these BPMs, remains to be studied.

An alternative approach to determining the dispersion at the interaction point (IP) involves using a transverse kicker to modify the path length of the pilot bunches and measuring the resulting orbit change. This measurement enables the inference of the dispersion for the colliding bunches, assuming they share the same OSVD as the pilot bunches. Since this method does not require altering the RF frequency, it prevents the potential induction of the flip-flop effect in the colliding bunches.

A crucial final input for the $\sqrt{s}$ calculation is the precise determination of the crossing angle at each interaction point (IP), which is provided by the experiments themselves.

### 1.7.7 Input from the experiments

Measurements conducted using collision data from the experiments provide essential input for determining key parameters related to the energy calibration.

The principal data set for these studies are $e^+e^- \to \mu^+\mu^-(\gamma)$ events. Reconstruction of the decay topology allows the crossing angle to be determined, as well as providing insight as to how the crossing angle varies with bunch intensity. In addition, it is possible to determine the longitudinal boost and $\delta_{\sqrt{s}}$, the spread in collision energy. Good knowledge of $\delta_{\sqrt{s}}$ is vital for many key objectives of the FCC-ee programme, such as the measurement of $\Gamma_Z$. Current studies indicate that all these quantities can be determined from collision data with the required precision. Finally, reconstruction of $e^+e^- \to f\bar{f}(\gamma)$ events and other processes provide a relative measurement of $\sqrt{s}$ that can be of great interest at high energies where RDP is not feasible. More details are given in Volume 1 of this report.

## 1.8 Injection and extraction

The technical straight section in point B (PB) of the FCC is dedicated to the beam transfer from the booster to the collider and the dump systems. In total, eight systems are installed in this area, and this section focuses on the collider system and the top-up injection and dump concepts.

### 1.8.1 Top-up injection

The integrated and peak luminosity targets of the collider necessitate a continuous injection scheme from a full-energy booster, as the beam lifetime during collisions is well below 1 h. Due to the high bunch charge in the collider and the limitations of the injector complex, a swap-out scheme is not feasible; thus, a top-up injection scheme is required.

Among the four operation modes, the Z mode is particularly challenging, with approximately 18 MJ of stored beam energy per collider ring at an energy of 45.6 GeV. As such, the Z mode is the current focus of the injection scheme design. In this mode, at each injection occurring every 3 s, up to 10 % of the collider's maximum bunch intensity and up to 1120 bunches are injected totalling up to 1 % of the collider nominal intensity.

This intensity is deemed reasonably safe for transfer from the booster to the collider, provided that a series of collimators is installed in the booster extraction region and along the transfer line to the collider to intercept any mis-kicked bunches.

Several potential schemes for implementing top-up injection at FCC-ee have been investigated [108]. Longitudinal injection schemes, which require an individual kick for each injected bunch trailing circulating bunches, have been ruled out due to the complexity of the necessary kicker system. For the conventional off-axis injection scheme, modelling of the synchrotron radiation (SR) cone produced by the beamlet, along with its impact on the SR mask and the aperture limitations around the experimental insertion, indicated an unacceptable level of interference [173].

### *On axis injection*

For on-axis injection, the beam is placed onto the chromatic closed orbit, where the energy offset together with the ring's optics dispersion provides the horizontal separation between the injected and circulating beams. The beamlet undergoes synchrotron oscillations, and around IPs it overlaps with the circulating beam due to the zero dispersion, thus preventing any increase in the SR cones or of the experiment background.

The beamlet benefits from faster damping compared to off-axis injection since synchrotron oscillations damp twice as fast as betatron oscillations. Another advantage of on-axis injection was observed at the LEP collider, where this scheme provided higher efficiency and lower sensitivity to errors at the injection point [174]. Therefore, the conventional on-axis injection has been selected for the present baseline concept.

The distance between injected and circulating beams establishes the requirements on the energy offset and dispersion at the injection septum:

$$|D_x \Delta| = 5\sigma_{cir} + S + 5\sigma_{inj} \tag{1.26}$$

where $D_x$ is the dispersion at the injection point, $\Delta$ is the relative energy offset of the injected beam, $S$ is the blade thickness of the septum, $\sigma_{cir}$ and $\sigma_{inj}$ are the beam sizes of the circulating and injected beams at the injection point.

As shown Eq. 1.26, the required energy offset increases with the septum blade thickness. Hence, an initial concept aimed at using an electrostatic septum to minimise its thickness but there are significant uncertainties on the reliability of such a system in the presence of synchrotron radiation. Therefore, the present concept focuses on thin magnetic septa with $S = 2.8\,\mathrm{mm}$.

During injection, the circulating beam closed orbit is placed at $5\sigma_{cir}$ from the septum blade (see Eq. 1.26). This condition must be fulfilled for the shortest possible duration because the septum, or a protection absorber placed immediately upstream, becomes the primary aperture of the ring. Therefore, the baseline concept features a fast bump to bring the circulating beam close to the injection septum for one single turn. Two sets of fast bumper magnets placed at a relative phase advance of $\pi$ are used to produce the orbit bump with a height of $10\sigma_{inj} + S$ at the injection point. The nominal position of the circulating beam is $15\sigma_{cir}$ from the edge of the injection septum. The failure of one of the bumpers would cause an open oscillation of all circulating bunches, potentially resulting in significant losses at downstream machine aperture bottlenecks (i.e., ideally at the collimators). The number of bumpers has to be defined in order to keep the oscillation amplitude small enough to avoid major damage to the machine in case of failure. Moreover, an absorber must be installed upstream of the septum blade to shield it from accidental and continuous losses. The absorber and the septum should never be the primary aperture bottleneck; this role should always remain with the collimators. The amplitude of the bump might need to be reviewed if this condition is not fulfilled.

The collider ring baseline optics for technical straight sections have been optimised so as to include the on-axis injection requirements. This optimsied collider optics in the PB straight section is shown in Fig. 1.38 with the injection region on the right side of the IP and the crossing dipoles in the centre. The injection point is located at $s = 800\,\mathrm{m}$ where it achieves a dispersion of $D_x = -1.5\,\mathrm{m}$ and $\beta_x = 1000\,\mathrm{m}$. This allows on-axis injection with an energy offset of $\sim 1\%$, providing sufficient space for the magnetic septum blade of approximately $3\,\mathrm{mm}$.

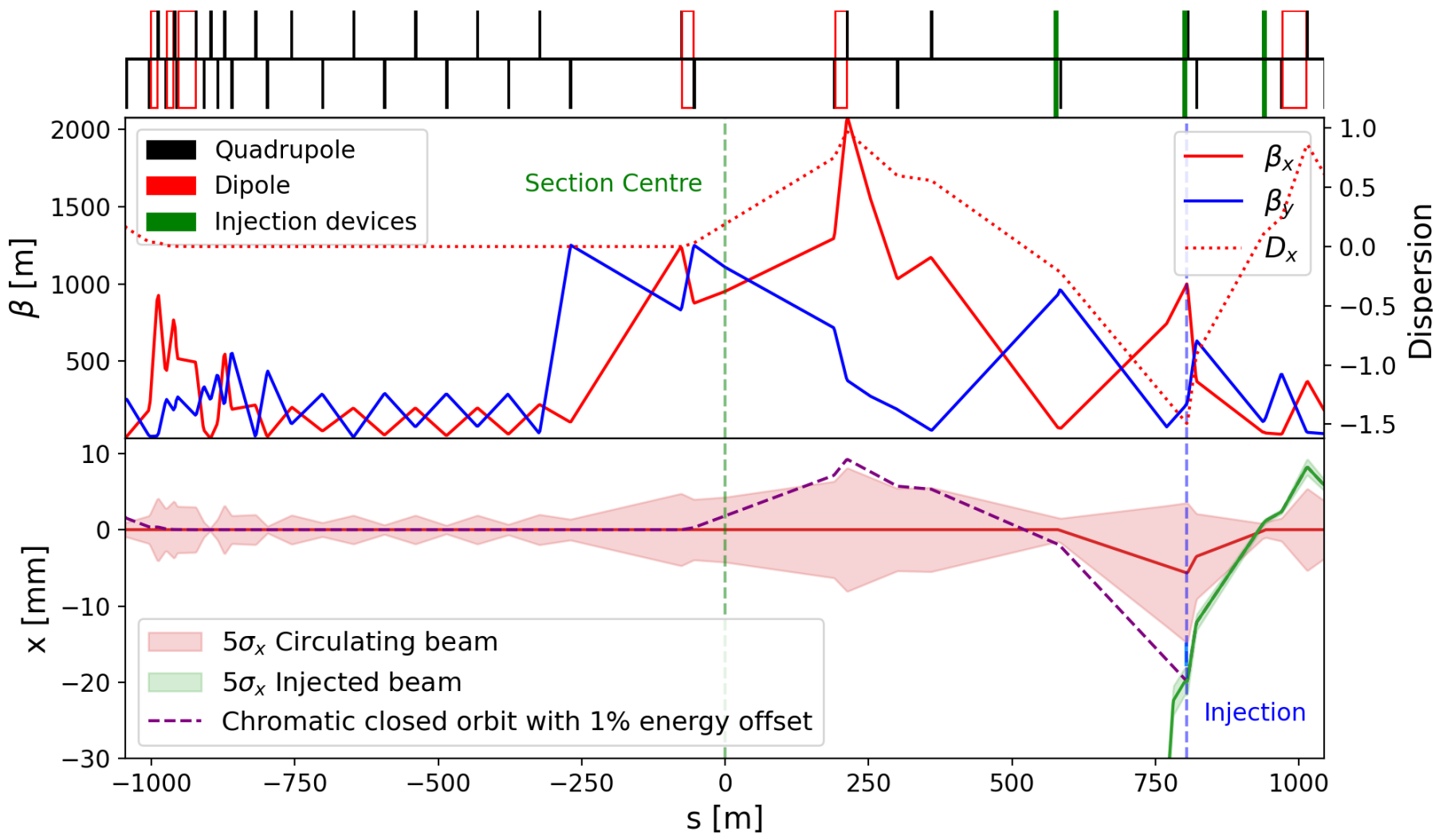

Fig. 1.38: The collider optics in the PB straight section with the longitudinal position relative to the IP.

The hardware requirements for the baseline injection scheme are summarised in Table 1.18.

Table 1.18: Collider injection hardware requirements.

| System | Value | unit |
|---|---|---|
| Beam energy | 45.6 – 182.5 | GeV |
| Thick septum apparent thickness | 10 | mm |
| Thick septum deflection | 0.1 | mrad |
| Thin septum apparent thickness | 2.8 | mm |
| Thin septum deflection | 100 | µrad |
| Fast bump kicker angles | 40 and 60 | µrad |
| Fast bump kicker max. rise/fall time | 600 | ns |
| Fast bump kicker flattop | 304 | µs |
| Fast bump kicker maximum ripple | 1.5 | % |

While the zero-dispersion condition was introduced to simplify the solution of Eq. 1.26 and to reduce the size of the injected beam at the injection septum, it also causes a mismatch between injected and circulating beams. The energy offset constrains the distance of the injected beam from the septum. In the Z and WW modes, in particular, the lattice momentum acceptance limits the energy offset of the injected beam to approximately 1%. On the other hand, the dispersion mismatch causes betatron oscillations in the injected particles away from the chromatic closed orbit. This effect remains moderate, as the momentum spread of the injected beam $\delta_{inj}$ is small. Additional studies will be needed to quantify this effect and investigate different matching for each operation mode.

A complete line design will need to be developed for the transfer from the booster to the collider. Presently, this line is approximately $500\,\mathrm{m}$ long from the booster extraction at the centre of the straight section to the collider injection point (see Fig. 1.38). Since the booster ring is positioned $1030\,\mathrm{mm}$ above the collider plane, special care must be taken to maintain the small vertical emittance and ensure precise matching of the vertical optics to the collider ring. In the horizontal plane, just upstream of the thin collider injection septum, a thicker magnetic septum is being considered for the beam trajectory, with specifications detailed in Table 1.18.

Both booster and collider beam parameters depend on the operation mode (see Tables 1.2 and 4.1). In the present concept, the lattice configuration of the injection straight section, as well as the injection devices' specifications, remain the same for all operation modes. With fixed optics and septum thickness, the energy offset is optimised for each energy mode to accommodate varying beam characteristics and the ring's momentum acceptance. The baseline injection settings for the four operation modes are shown in Fig. 1.39.

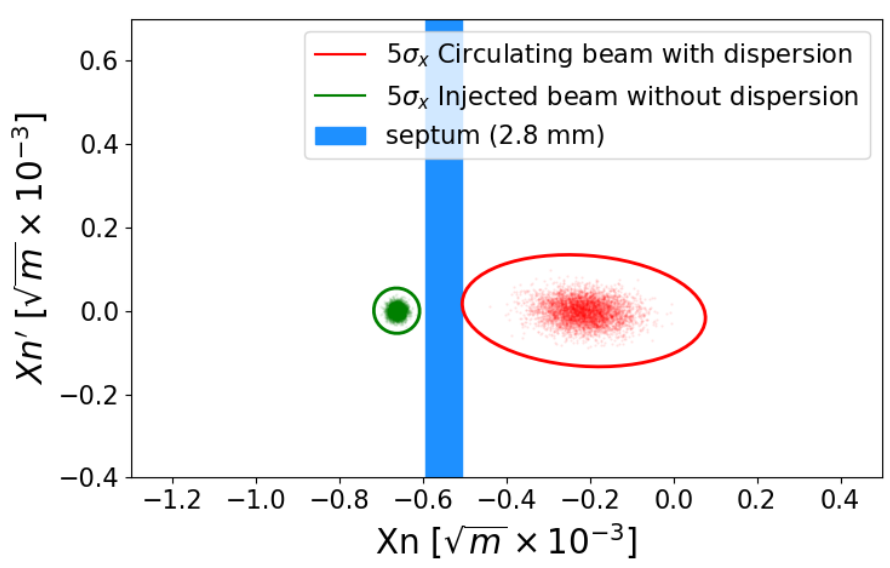

(a) Z mode, $\Delta = 0.95\,\%$.

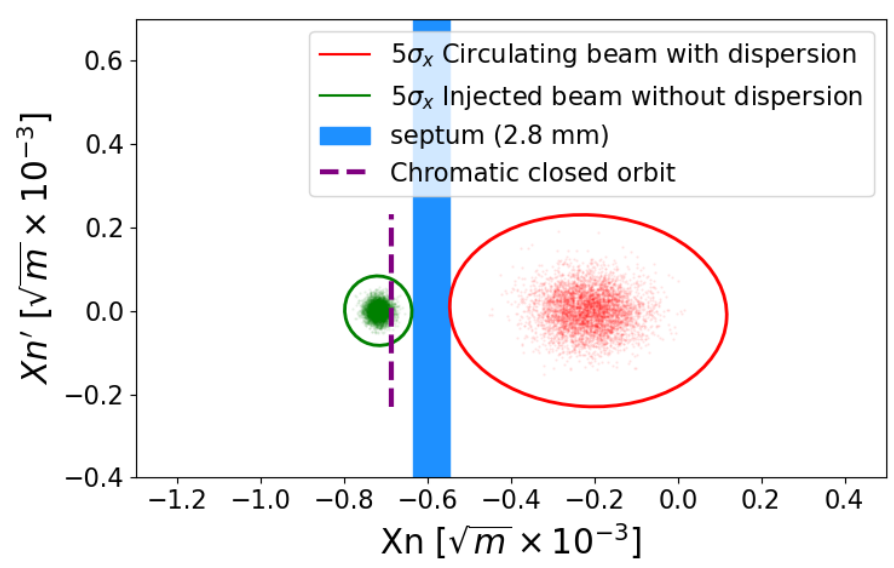

(b) WW mode, $\Delta = 1\,\%$.

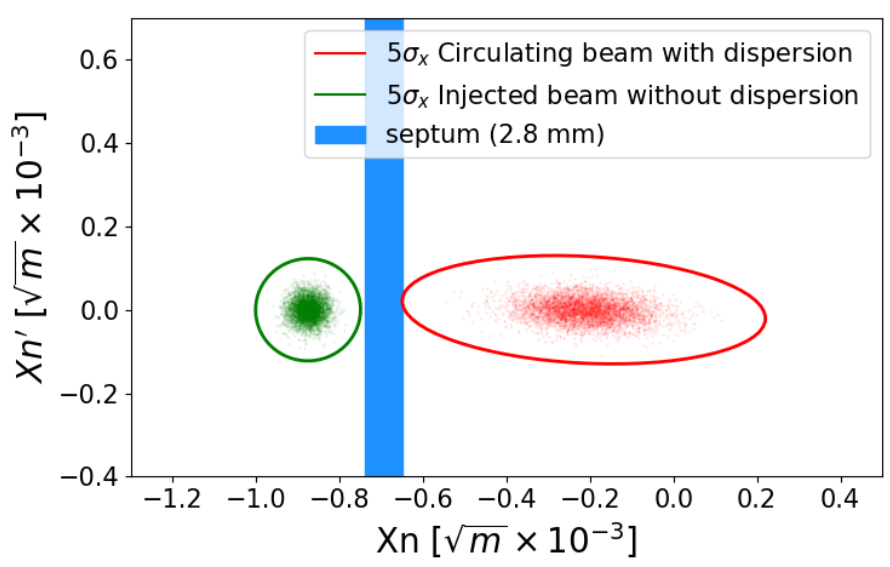

(c) H mode, $\Delta = 1.4\,\%$.

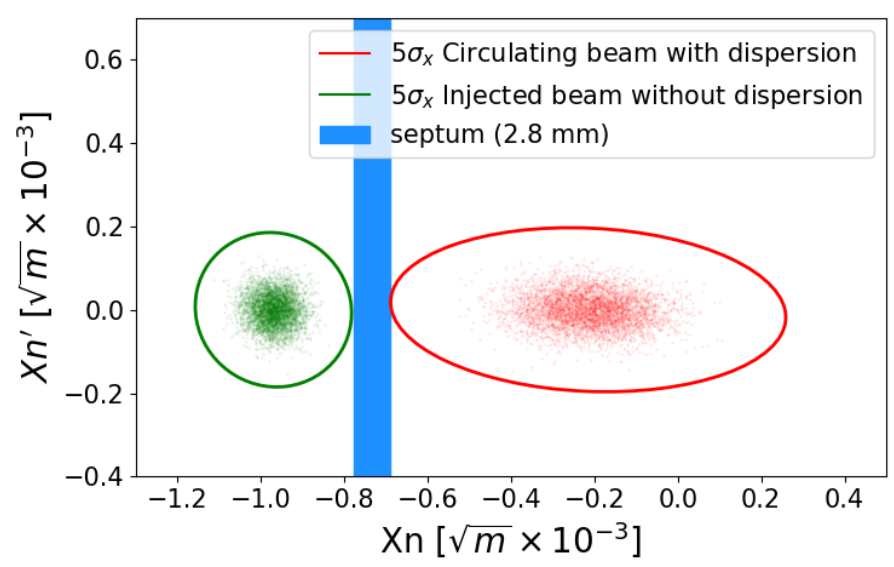

(d) $\mathrm{t\bar{t}}$ mode, $\Delta = 1.6\,\%$.

Fig. 1.39: Normalised horizontal phase space at the collider injection point for each operation mode. Beam distributions and associated $5\,\sigma$ envelopes around the injection septum are shown for each mode.

In order to maintain a clearance of $5\,\sigma$ for both the injected and circulating beam a larger energy offset is required for modes with higher beam emittance (see Eq. 1.26), and the energy offset in H and $\mathrm{t\bar{t}}$ mode is increased to $1.4\,\%$ and $1.6\,\%$, respectively. This remains compatible with the large momentum acceptance in Higgs and $\mathrm{t\bar{t}}$ modes, which are $\pm1.6\%$ and -2.8/+2.5%, respectively [31]. For Z mode, the RF acceptance is 1.06% and the momentum acceptance is approximately 1%, so the baseline injected beam energy offset is set to 0.95% to ensure the entire injected beam fits within the ring acceptance.

The collider's and booster's equilibrium emittances in the WW mode are significantly larger, but the energy acceptance of the ring is still limited to $1\,\%$. This prevents increasing the energy offset for the higher energy modes and makes the on-axis scheme unable to provide sufficient clearance for the $2.8\,\mathrm{mm}$ septum blade. The present concept introduces a small betatron offset to the scheme and moves towards a hybrid injection for the WW mode. The purple dashed line in Fig. 1.39b represents the chromatic orbit for the energy offset considered, and the injected beam is offset by $1\,\mathrm{mm}$ further away from the septum, which corresponds to $0.5\,\sigma_{\mathrm{cir}}$.

The hybrid injection scheme increases the separation between injected and circulating beams without increasing the energy offset, making it partially on-axis and off-axis. While the off-axis injection is not possible for the FCC collider, a small betatron oscillation is not incompatible with the SR absorption around the experiment IPs [173]. Other effects discussed earlier, such as experiment background sensitivity and errors, may still play a significant role and will need to be quantified.

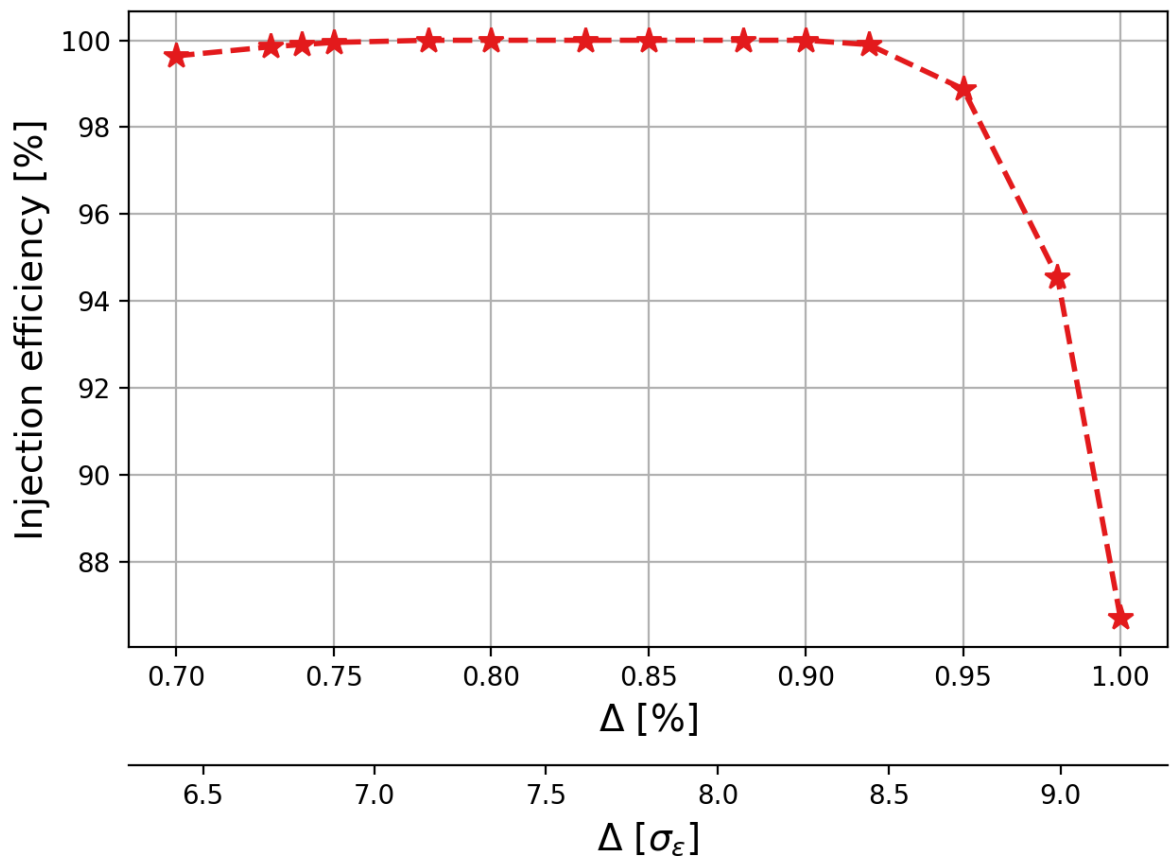

Fig. 1.40: Injection efficiency versus injected beam energy offset for the Z mode.

Presently, the injection efficiency is modelled with particle tracking of the injected beam in the collider ring lattice. The tracking consists of a Gaussian 6D distribution of 2500 injected particles, which are tracked for 3000 turns using the XSUITE code [175]. The synchrotron radiation model used accounts for quantum excitation to accurately predict the behaviour of particles injected near the edge of stability. Additional studies including strong-strong beam-beam effects are discussed Section 2.3.9. Further comprehensive studies will need to include beam-beam interaction, collective effects and lattice errors.

The evolution of the simulated injection efficiency for on-axis and hybrid injection schemes is shown in Fig. 1.40. The energy offset of the baseline on-axis injection is $0.95\,\%$, with an injection efficiency of $99\,\%$. This confirms that the baseline injection scheme is achievable with the present lattice, and the injected beam is within the acceptance of the ring.

At other injection offsets, only the energy of the injected beam is adjusted, but the optics is not optimised, and the physical beam position is unchanged. Below the baseline on-axis scheme energy offset, the injection efficiency remains high, which indicates that the DA and MA are sufficient to capture an injected beam with some betatron offset. At an energy offset below about $0.7\,\%$ the injection efficiency decreases, which shows that a maximum betatron offset is reached and part of the injected beam is outside the lattice DA. Due to the energy acceptance of the lattice, of $\pm 1\,\%$, the injection efficiency drops quickly for energy offset above the baseline scheme.

### ***Collider dump***

The stored beam energy can reach $18\,\mathrm{MJ}$ per beam during Z mode operation, which is the highest stored energy among the lepton machines worldwide. Due to the synchrotron radiation damping, the beam sizes in FCC-ee will be much smaller than in typical hadron machines, leading to a much higher energy density. The vertical beam size, in particular, is in the order of tens of micrometres, corresponding to energy densities around $5\,\mathrm{GJ/mm^2}$, which cannot be absorbed safely [176]. The design and operation of the beam abort system must accommodate this destructive potential, requiring multiple safety measures to prevent damage to accelerator components in the event of hardware failure. Protection elements strategically positioned at a precise phase advance from the extraction kickers must be installed to intercept any mis-kicked beam in case of an unintended kicker firing that is not synchronised with the abort gap. Additionally, the voltage of the kickers, as well as the current of the septa and ring dipoles, must be continuously monitored to ensure they remain within strict limits relative to the reference value for a given energy.

A retriggering system, which fires all the remaining kickers in case of the spurious firing of one kicker, has to be envisaged, and the reaction time must be defined based on the consequences of such an event. Redundancy in the powering scheme, the controls, and interlock logic have to be implemented to ensure that the required level of reliability and availability of the system is achieved.

The number of magnets has to be defined with the aim of reducing the maximum operational voltage and thus minimising the risk of spurious firing and the sensitivity to failures. Out-of-vacuum designs should be preferred to eliminate the risk of flashovers. The position of the beam at extraction has to be constantly monitored, and the beam dumped before orbit drifts translate into losses at extraction above the level that can be considered safe (to be defined). Similar measures to the present LHC beam dump system will need to be considered, and adapted to the specificities of the FCC but presently no major fundamental obstacles have been identified.

The collider dump design is implemented at the entrance of the PB straight section and extracted outwards with the dump placed on the other side of the IP. The beam dump is located $5\,\mathrm{m}$ away from the ring to allow sufficient space for shielding and to limit radiation to the ring equipment [177]. The present geometry uses a small deflection angle to achieve the required separation and a long transport line of $700\,\mathrm{m}$ from the ring extraction point to the beam absorber.

In order to reduce the energy density on the dump, the present design aims for a large beta function and dispersion in both horizontal and vertical planes to maximise the beam size. Despite the significant length of the line, the natural divergence of the beam at the extraction point is insufficient to produce a beam spot that is large enough on the dump. Therefore, a set of four dedicated quadrupole magnets is installed in the transfer line to increase the beam divergence further. In the horizontal plane, the dump kicker and septa provide the deflection to reach the dump but also a significant dispersion that is further amplified by the dump line quadrupoles to reach $20\,\mathrm{m}$ at the dump. Along with the beta function of $121\,\mathrm{km}$ and the beam parameters in the collision, the present design achieves a horizontal beam size on the dump of $10\,\mathrm{mm}$ for the Z mode [178].

For the vertical plane, a $1\,\mathrm{mrad}$ vertical dipole is placed at the start of the dump line to create a vertical dispersion that is further amplified by the dump line quadrupoles to reach $23\,\mathrm{m}$ at the dump. Between the betatron and dispersive contribution, this scheme also provides a large beam size in the vertical plane of $10\,\mathrm{mm}$ for the Z mode using the beam characteristics expected in a collision. The vertical deflection also places the dumped beam at the height of the booster.

The extraction design follows a traditional fast extraction scheme in the horizontal plane. The circulating beam is extracted in one turn so that the kicker flattop must be $304\,\mu\mathrm{s}$ and its rise time should be smaller than the filling scheme abort gap of $0.6\,\mu\mathrm{s}$. The dump design hardware requirements are summarised in Table 1.19.

## 1.9 Radiation environment

The emission of synchrotron radiation and the generation of secondary particles by other processes (beam-beam effects etc.) creates an intense radiation environment, which can have a significant impact on machine components and other equipment in the tunnel. A thorough study of the expected radiation load and its associated effects is essential for designing the FCC-ee machine. Radiation can lead to substantial heat deposition and, consequently, to thermal stress on the accelerator components. The power deposited, in particular from synchrotron radiation, needs to be extracted in a controlled way otherwise equipment can be damaged. Excessive heat deposition in the tunnel environment would also pose a significant challenge to the ventilation system. Another important aspect is the lifetime of equipment, which can be affected by cumulative radiation damage. One of the main concerns for FCC-ee is the total ionising dose in cables, cable connectors, optical fibres, insulation materials, seals, etc. As already observed in LEP, the ionising dose can severely compromise the organic insulation of magnet coils and cables and can rapidly degrade the functionality of optical fibres [179]; even covers of electrical junction boxes and hoses of fire extinguishers were found to be damaged [179], which demonstrates the challenge expected for FCC-ee. Other long-term radiation effects include radiation-induced corrosion due to the dissociation of molecules in the tunnel atmosphere. Another concern for FCC-ee are stochastic and cumulative effects in electronics, which can impede the machine performance (e.g., premature beam aborts due to single event effects) and can limit the lifetime of equipment electronics.

Table 1.19: Summary of the collider dump scheme hardware requirements.

| | Kicker | Septum |
|---|---|---|
| Beam energy (GeV) | 45.6 – 182.5 | |
| Deflection angle per system (mrad) | 0.3 | 5 |
| Maximum repetition frequency (Hz) | 0.3 | 0.3 |
| Kicker pulse flatness (%) | ±20 | N/A |
| Rise/fall time (µs) | 0.6 | N/A |
| flattop time (µs) | 304 | N/A |
| Blade thickness (mm) | N/A | 25 |
| Aperture (H×V mm) | N/A | 30×10 |
| Longitudinal available space (m) | 20 | 20 |

In order to mitigate radiation effects and manage their release, it is essential to sufficiently shield the power FCC-ee machine equipment and infrastructure from the radiation load. This requires dedicated synchrotron radiation absorbers (see Section 3.2), as well as additional shielding on the magnets and in the tunnel. Due to the sheer size of the FCC-ee machine, continuous shielding of the vacuum chamber, as it was implemented in LEP, may be challenging to implement. In order to develop an adequate shielding configuration, it is necessary to create a detailed inventory of radiation-sensitive components (e.g., cables, electronics, insulation) for all accelerator systems like magnets, power converters, beam instrumentation, and the beam interlock system. Different components can exhibit different levels of sensitivity to radiation, which needs to be accounted for in the shielding design. For example, organic insulation of busbars can typically sustain several tens of MGy, whereas general-purpose cables can only be used up to a few 100 kGy. Electronics systems based on commercial-off-the-shelf components can typically not withstand levels exceeding 1 kGy, even when designed to be radiation tolerant through commercial component selection and architectural mitigation solutions; for doses above 1 kGy it is necessary to rely on dedicated radiation-hard component designs that require investment in time and resources. Similar considerations apply to optical fibre cables. Finding a trade-off between shielding solutions and a radiation-hard component design is an important objective for the FCC-ee design phase to avoid radiation-induced equipment failures or degradation of the collider performance.

This section provides a first overview of the expected power deposition and the radiation environment in the arcs and experiment insertion regions, based on FLUKA Monte Carlo simulations [125–127]. A preliminary shielding design for the arc dipoles is presented, together with a possible electronics bunker near the arc quadrupoles. The present shielding and bunker design is only a first concept, with the purpose to quantify the achievable reduction of the radiation levels. A further evolution of the shielding and bunker design is expected in the engineering design phase (see also Section 3.3). No studies have been carried out so far for the technical insertion regions; the radiation levels will depend on different beam loss mechanisms (e.g., beam losses during top-up injection or beam collimation), which have to be assessed in detail. It is expected that some shielding installations will also be needed in these insertion regions.

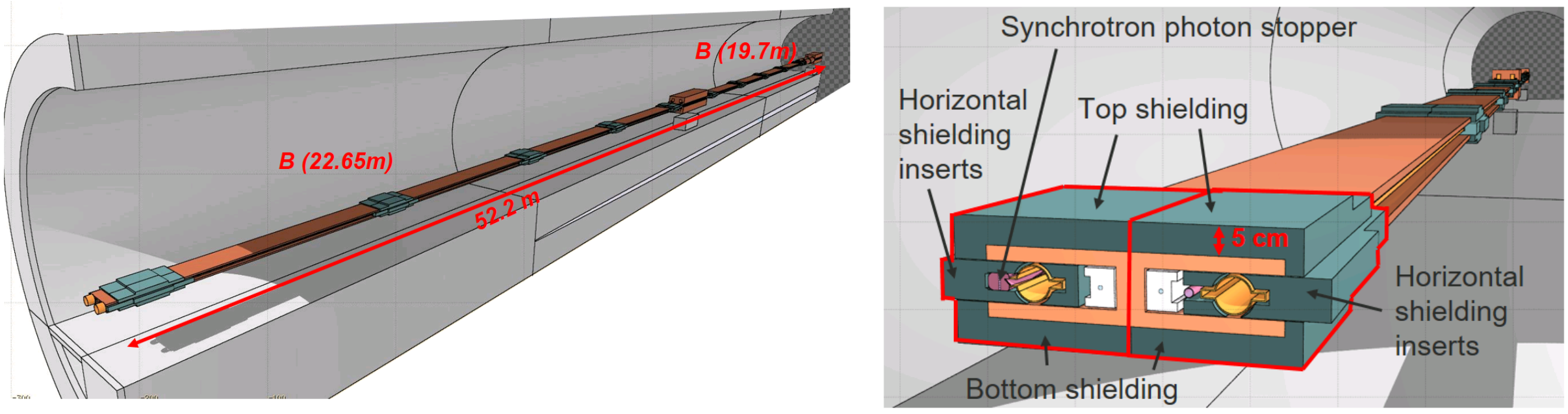

Fig. 1.41: Preliminary conceptual radiation shielding design for the collider dipoles in the FCC-ee arcs. The left figure shows a model of a representative FODO cell for ZH and $\mathrm{t\bar{t}}$ operation, whereas the right figure presents a more detailed view of the shielding elements on the dipoles. The shielding encloses tightly the synchrotron radiation absorbers. The shielding is assumed to be made of antimonial lead and weighs about $400\,\mathrm{kg}$ per synchrotron radiation absorber.

### 1.9.1 Radiation levels in the arcs and first shielding design

The main source of radiation in the FCC-ee arcs is the synchrotron radiation produced in the bending dipoles of the collider ring. Synchrotron radiation is emitted tangentially from the stored beams and the resulting energy loss increases steeply with the beam energy ($\propto E^4/(m^4\rho)$). The bending radius $\rho$ of the FCC-ee arc dipoles is about $10\,\mathrm{km}$, which is about three times larger than for LEP. Considering the important impact of synchrotron radiation, the synchrotron power emitted in FCC-ee is limited by design to $100\,\mathrm{MW}$ ($50\,\mathrm{MW/beam}$) for all operation modes (see Table 1.2). The resulting radiation levels in the tunnel are nevertheless more pronounced at higher beam energies since the emitted photons become more penetrating. A key figure is the critical energy, $E_c \propto E^3/\rho$, which divides the photon spectrum into two parts of equal power emission. For a bending radius of $10\,\mathrm{km}$, the critical energy is $0.02\,\mathrm{MeV}$ at the Z-pole ($45.6\,\mathrm{GeV}$), but it increases to $0.1\,\mathrm{MeV}$ in WW operation ($80\,\mathrm{GeV}$), to $0.4\,\mathrm{MeV}$ in ZH operation ($120\,\mathrm{GeV}$), and further to $1.3\,\mathrm{MeV}$ in $\mathrm{t\bar{t}}$ operation ($182.5\,\mathrm{GeV}$). The resulting radiation environment in the tunnel is composed of secondary photons and electrons. In addition, photo-neutron production becomes possible when the photon energies exceed the ($\gamma$,n)-threshold, which is around $10\,\mathrm{MeV}$ for copper. Neutron production by synchrotron photons is mostly relevant for $\mathrm{t\bar{t}}$ operation, where the high-energy tail of the synchrotron spectrum extends beyond the giant dipole resonance of the photo-nuclear cross section.

The vacuum system of the collider incorporates localised photon absorbers made of a copper alloy (CuCrZr), which intercept the synchrotron radiation fan (see Section 3.2). The absorbers have a length of about $35\,\mathrm{cm}$; they are placed every four to five metres in the winglets of the dipole chambers and shadow, as well as the short straight sections. At higher energies, the radiation leakage from these absorbers becomes significant, which requires the integration of additional shielding elements in the dipoles. A first conceptual design of the collider shielding configuration is shown in Fig. 1.41. The presently considered baseline material is antimonial lead with a density of about $10.88\,\mathrm{g/cm^3}$. The shielding tightly encloses the radiation absorbers and consists of horizontal inserts, as well as shielding plates above and below the dipole yoke. The integration of the shielding is a complex task and requires design iterations with the vacuum system and magnets. A description of technical aspects of the material design, including material selection and engineering considerations, is presented in Section 3.3.

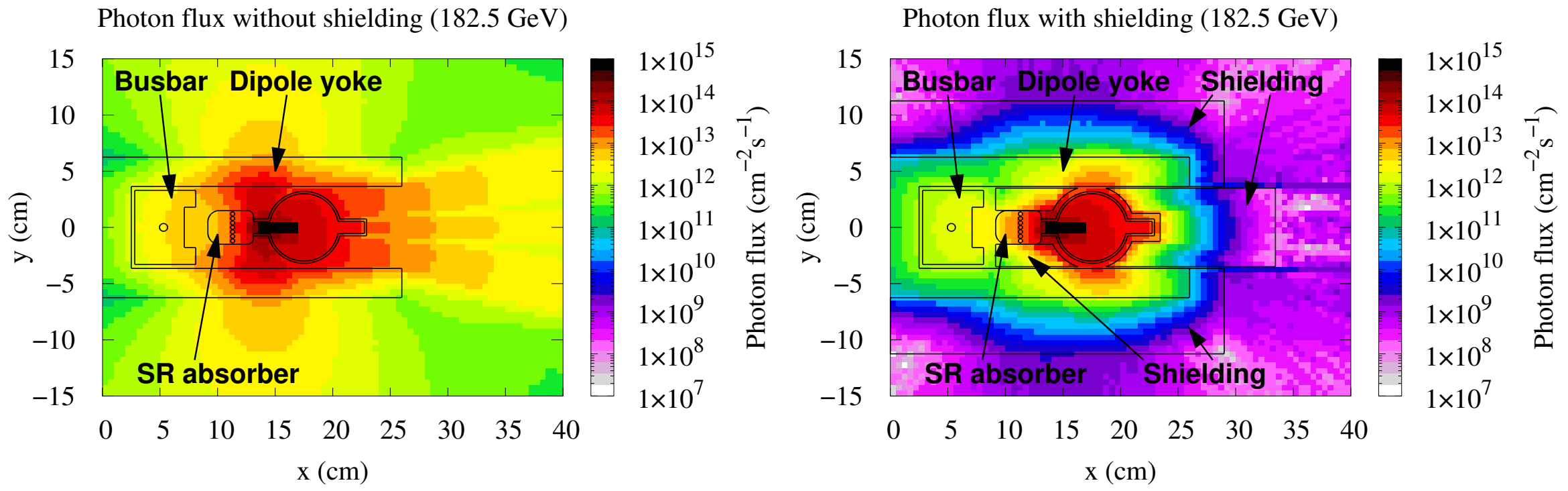

Fig. 1.42: Synchrotron radiation-induced photon flux around the dipole yoke in $\mathrm{t\bar{t}}$ operation. The two plots compare the photon leakage without radiation shielding (left) and with the conceptual shielding illustrated in Fig. 1.41 (right). The flux was calculated with the FLUKA code, assuming a photon transport cut of $10\,\mathrm{keV}$. In this picture the $x$-axis exceptionally points towards the centre of the collider ring, which is opposite to the convention adopted elsewhere in this document.

A first optimisation of the shielding topology has been carried out, to maximise the power absorption and to reduce the total ionising dose in the tunnel sufficiently. Given the complexity of the shielding integration and assembly, it is preferable to install the shielding before the start of FCC-ee operation. In this case, the shielding design must account for all beam modes, including $\mathrm{t\bar{t}}$. A suitable figure of merit for optimising the shielding geometry is the flux of secondary photons escaping vertically and horizontally from the dipoles. The attenuation length for $1\,\mathrm{MeV}$ photons in lead is about $1.3\,\mathrm{cm}$, but it decreases to less than $1\,\mathrm{mm}$ for photon energies below $100\,\mathrm{keV}$. With several centimetres of antimonial lead, the photon flux in the tunnel can be reduced by multiple orders of magnitude, even at the highest beam energy ($182.5\,\mathrm{GeV}$). Figure 1.42 shows the simulated secondary photon flux around the dipole yoke for $\mathrm{t\bar{t}}$ operation. The left plot is without shielding, whereas the right plot demonstrates the reduction achievable in the photon leakage with the preliminary shielding design shown in Fig. 1.41. The shielding must be sufficiently long ($130\,\mathrm{cm}$ in the present design) to reduce the backscattering of photons into the tunnel. The resulting shielding weight is about $400\,\mathrm{kg}$ per synchrotron radiation absorber.

Table 1.20 summarises the resulting power deposition in the machine and the tunnel for all beam modes, assuming the same shielding design for all beam energies. For comparison, the table also shows the power deposition without shielding. The shielding has only a limited function at the Z mode since most of the synchrotron radiation power (>98%) is dissipated by the synchrotron radiation absorbers themselves. With increasing energy, the power absorption by the absorbers decreases to about 83% in WW operation and to less than 70% in ZH and $\mathrm{t\bar{t}}$ operation. In these cases, the surrounding shielding is very efficient in reducing the power leakage to the dipoles and the environment by absorbing between 10% (WW) and 20% (ZH and $\mathrm{t\bar{t}}$) of the power. Since only the dipole busbars are actively cooled but not the yoke, a fraction of the heat deposited in the dipoles would also be dissipated in the air. Equipped with a dedicated cooling circuit, the shielding substantially reduces the heat to be evacuated by the ventilation system.

Figure 1.43 presents the corresponding annual dose in the arc tunnel. The plots illustrate the effect of the shielding for ZH and $\mathrm{t\bar{t}}$ operation, which yield the highest contribution to the annual dose. The shielding reduces the dose levels in tunnel by more than two orders of magnitude. At the location of the upper cable trays on the walls ($>2\,\mathrm{m}$ above floor level), the dose is $<1\,\mathrm{kGy/year}$ for the ZH operation and $<10\,\mathrm{kGy/year}$ for $\mathrm{t\bar{t}}$ operation, down from several hundreds of kGy/year without shielding. With the present shielding design, it seems feasible that most cables in the cable trays receive $<100\,\mathrm{kGy/year}$ in the full FCC-ee era, including $t\bar{t}$. This is compatible with the criteria for general-purpose cables presently adopted for HL-LHC and other projects at CERN. The radiation level specifications and required safety factors for FCC-ee cables and other equipment need to be scrutinised during the technical design phase and will depend on the chosen technologies. The results nevertheless demonstrate that radiation shielding can significantly reduce the need for expensive radiation-hard equipment in the tunnel. Near the machine, radiation-hard cables and cable connectors qualified for MGy dose levels can likely not be avoided. Depending on the evolution of equipment technology in the next years, the radiation resistance criteria are still expected to evolve. The shielding requirements need to be adapted accordingly.

Even with the dipole shielding, the annual dose in tunnel remains significant for electronics, i.e., the levels are too high for a radiation-tolerant system design based on commercial-off-the-shelf components. In addition, the expected neutron flux in $t\bar{t}$ operation increases the likelihood of single event effects and enhances the expected displacement damage. A conceptual design of a possible electronics bunker has been devised (see Section 3.3), in order to reduce the radiation levels for electronics. In this very first design, the bunker is assumed to be made of 10–20 cm-thick concrete walls, which are lined with borated polyethylene sheets. The latter is needed for moderating and capturing neutrons. The bunkers are assumed to be located below or near the arc quadrupoles. Table 1.21 compares the radiation levels inside and outside the bunker for one year of $t\bar{t}$ operation. The results demonstrate that similar levels as in the HL-LHC arcs can be achieved (see last column), with the exception of the total ionising dose which remains somewhat higher. With the bunker, using custom radiation tolerant electronics systems based on commercial-off-the-shelf semiconductor devices could be feasible for FCC-ee.

### 1.9.2 Radiation levels in the experiment insertion regions

The primary radiation sources in the tunnel of FCC-ee experimental insertions are beamstrahlung radiation, radiative Bhabha and synchrotron radiation, and their impact has been studied with FLUKA simulations. Other sources of radiation, such as beam-gas interactions or others, may cause additional radiation showers, but they are not considered in the present study. Concerning beamstrahlung, a dedicated absorber is needed to safely dispose of the photons because of the high power carried by the outgoing photon beam (several hundreds of kW at Z pole). The absorber will be placed about 500 m downstream of the IP and needs to be shielded to limit the radiation leakage to the tunnel. A second source of radiation is the off-momentum electrons from radiative Bhabha interactions; a fraction of these electrons is lost on the vacuum chamber when entering the first dipoles of the outgoing beam. Additionally, the stored beams emit a non-negligible amount of synchrotron radiation in the bending dipoles downstream of the IP, with more than $160\,\mathrm{kW}$ of power produced over the first $500\,\mathrm{m}$. On the other hand, the synchrotron radiation power emitted by the incoming beams is only $1\,\mathrm{kW}$, yielding a much smaller contribution to the radiation levels in the tunnel.

Table 1.20: Relative power deposition by synchrotron radiation in the machine and the tunnel. The results were obtained with FLUKA simulations for a representative arc cell, assuming 10 photon stoppers between successive quadrupole magnets, or roughly one photon stop every 5 metre for each beam. The table compares two configurations for each beam mode, with and without radiation shielding around the synchrotron radiation absorbers. The power deposition in the environment includes the power dissipated in the air, the tunnel walls and the surrounding earth or rock.

| | Z | | WW | | ZH | | $t\bar{t}$ | |
|---|---|---|---|---|---|---|---|---|
| | w/o | with | w/o | with | w/o | with | w/o | with |
| SR absorbers | 98.1% | 98.1% | 82.8% | 82.8% | 69.9% | 69.7% | 68.3% | 68.1% |
| Radiation shielding | - | | - | | - | 19.1% | - | 20.4% |
| Vacuum chambers | 1.8% | | 8.2% | | 9.0% | 8.9% | 8.1% | 8.0% |
| Dipoles | 0.1% | | 7.8% | | 16.7% | 2.3% | 17.4% | 3.5% |
| Quadrupoles | <0.001% | | <0.01% | | <0.02% | <0.01% | <0.1% | <0.02% |
| Sextupoles | <0.001% | | <0.01% | | <0.02% | <0.01% | <0.1% | <0.02% |
| Environment | 0.01% | | 1.2% | | 4.3% | <0.01% | 6.1% | <0.03% |

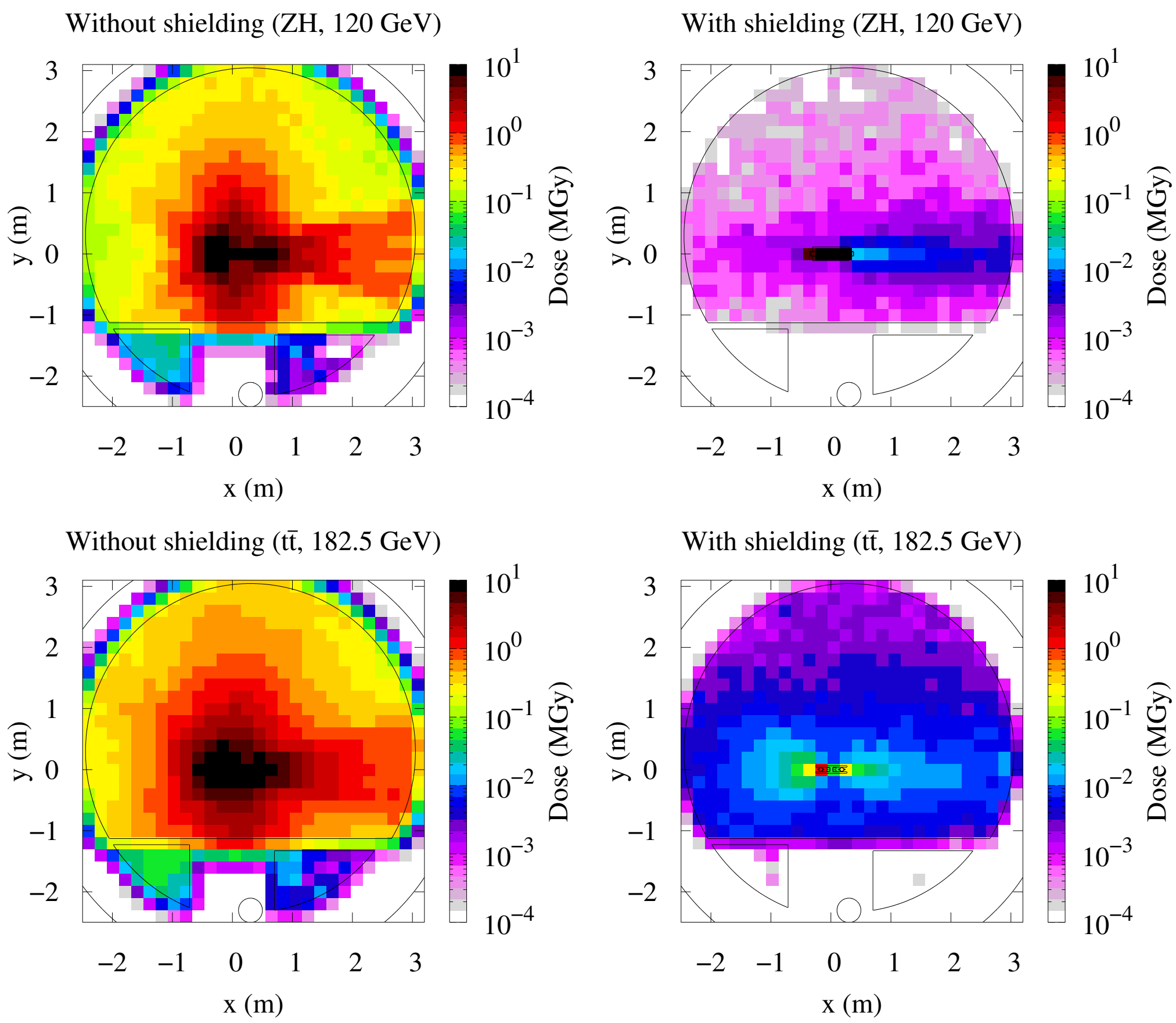

Fig. 1.43: Annual dose in the arc tunnel due to synchrotron radiation emission by the stored beams in the collider (top: ZH, bottom: $t\bar{t}$). The left figures are without radiation shielding, whereas the right figures were derived with the conceptual shielding illustrated in Fig. 1.41. The dose maps correspond to the position of a synchrotron radiation absorber, where the dose reaches its maximum value. The maps were calculated with FLUKA, assuming 185 days of operation with 75% operational efficiency. The collider is located in the origin. The x-axis points towards the centre of the collider ring.

Table 1.21: Radiation levels inside and outside a possible electronics bunker near lattice quadrupoles in the arcs. The first two quantities describe cumulative effects, whereas the two last quantities are relevant for single-event effects. The values correspond to one year of $t\bar{t}$ operation (185 days with 75% operational efficiency). The last column shows the radiation level specifications for the HL-LHC arcs (one year) [180].

| | FCC-ee $t\bar{t}$ (outside bunker) | FCC-ee $t\bar{t}$ (inside bunker) | HL-LHC arcs (below magnets) |
|---|---|---|---|
| Total ionising dose | few kGy | <10 Gy | 1.4 Gy |
| Si 1 MeV neutron-equiv. fluence | $6 \times 10^{11}$ cm$^{-2}$ | $\sim 1 \times 10^{10}$ cm$^{-2}$ | $1.6 \times 10^{10}$ cm$^{-2}$ |
| High-energy hadron-equiv. fluence | $8 \times 10^{8}$ cm$^{-2}$ | $\sim 1 \times 10^{7}$ cm$^{-2}$ | $2.4 \times 10^{9}$ cm$^{-2}$ |
| Thermal neutron-equiv. fluence | $5 \times 10^{11}$ cm$^{-2}$ | few $1 \times 10^{9}$ cm$^{-2}$ | $1.2 \times 10^{10}$ cm$^{-2}$ |

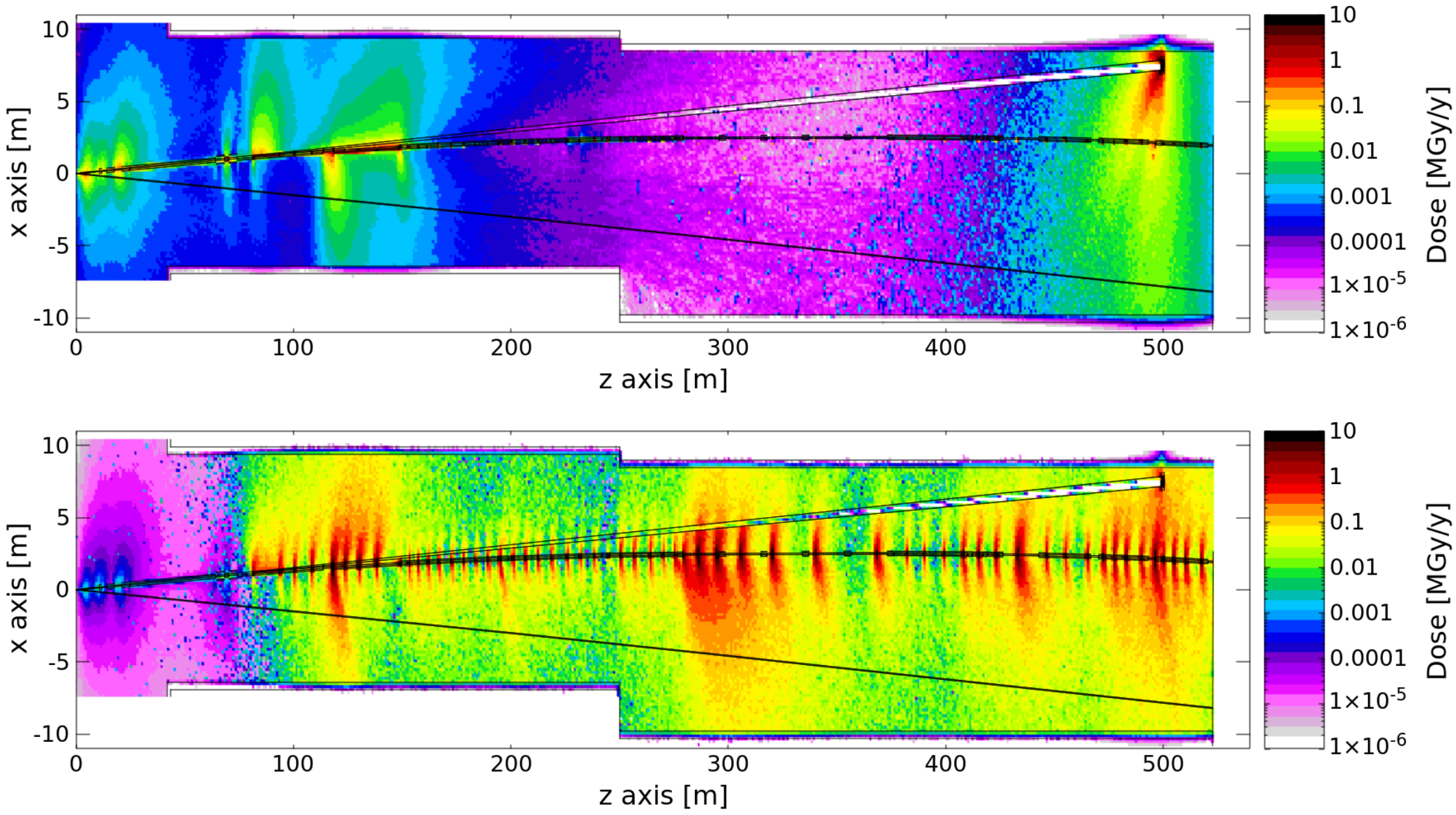

Fig. 1.44: Top view of the annual ionising dose at the beamline level (average for $y$ in $[-20, 20]$ cm) caused by beamstrahlung radiation, radiative Bhabha electrons and synchrotron radiation from the outgoing beam. The top figure corresponds to Z mode operation, whereas the bottom figure is for $\mathrm{t\bar{t}}$. The maps were simulated with FLUKA.

The secondary radiation fields by these three radiation sources have been simulated using FLUKA. The studies included similar CuCrZr synchrotron radiation absorbers as in the arcs but no additional shielding around the absorbers. The annual ionising dose from the IP (z=0) up to $\sim 520\,\mathrm{m}$ downstream is displayed in Fig. 1.44, comparing Z and $\mathrm{t\bar{t}}$ operation. The dose maps show that the radiation environments for the two operation modes are very different since different source terms dominate. At the Z pole, losses from radiative Bhabha give high dose levels up to $200\,\mathrm{m}$ downstream of the IP; another hot spot occurs around $500\,\mathrm{m}$ due to leakage of radiation from the beamstrahlung absorber. The conceptual shielding for the absorber assumed in these preliminary simulations ($20\,\mathrm{cm}$ of concrete) yields dose levels of several hundreds of kGy/y in the vicinity of the absorber and several tens of kGy/y at the beamline of the outgoing beam. Further optimisation of the shielding will be performed based on radiation protection studies. It is expected that with thicker shielding, the dose can be reduced by orders of magnitude. The synchrotron radiation emitted in the dipoles is less relevant for the radiation levels at the Z pole due to its soft spectrum; with critical energies ranging from $2\,\mathrm{keV}$ to $23\,\mathrm{keV}$, 95% of its power is absorbed in the synchrotron radiation absorbers.

The situation is the opposite at $\mathrm{t\bar{t}}$, where the dose from synchrotron radiation is dominant while the contributions from beamstrahlung and radiative Bhabha are smaller than for Z pole operation. The harder synchrotron spectra feature critical energies of the order of $1\,\mathrm{MeV}$, reducing the power fraction absorbed by the photon stoppers down to 68% (similar to the arcs). Dose hot spots of the order of 0.1–$1\,\mathrm{MGy/y}$ are observed around each synchrotron radiation absorber. This underlines that, like in the arcs, synchrotron radiation absorbers alone are not enough to suppress the radiation leakage into the tunnel; it is expected that additional shielding similar to the arc shielding is needed to reduce the dose. In the close proximity of the IP the dose levels are determined by radiative Bhabha, but they are two orders of magnitude lower than at Z pole because of the lower intensity.

For both operational modes, other sources of radiation (e.g., beam-gas interactions) are expected to give a negligible contribution to the radiation hotspots displayed in Fig. 1.44, but dedicated studies would be needed to assess a possible impact in the portions of the tunnel where the sources studied so far are yielding lower radiation levels.

## 1.10 Ongoing studies

This section will summarise five of the ongoing studies for the FCC-ee optics design. These include the local chromatic correction (LCC) optics which has the potential to improve the dynamic aperture and reduce the optics sensitivity to errors; a combined function optics design which improves the dipole packing fraction and reduces the amount of synchrotron radiation; a non-local detector solenoid compensation scheme which reduces the vertical emittance growth; a monochromatic optics configuration which would reduce the energy width of the collisions and could enable direct measurement of the Higgs width through s-channel production; polarised $e^+e^-$ sources which could eliminate the downtime needed to polarise the pilot bunches and improve the integrated luminosity.

### 1.10.1 LCC optics

Local Chromatic Correction (LCC) optics have been proposed for the FCC-ee collider and are detailed in [32]. The structure of the accelerator is the same as the baseline optics, including four final focus systems (FF) and four long straight sections (LSS) separated by eight arcs.

The LCC optics are designed to obtain *anharmonic* and *achromatic* beam dynamics for each of these components independently. The layout of the optics in the arcs, FF and LSS are shown in Figs. 1.45, 1.46 and 1.47 respectively.

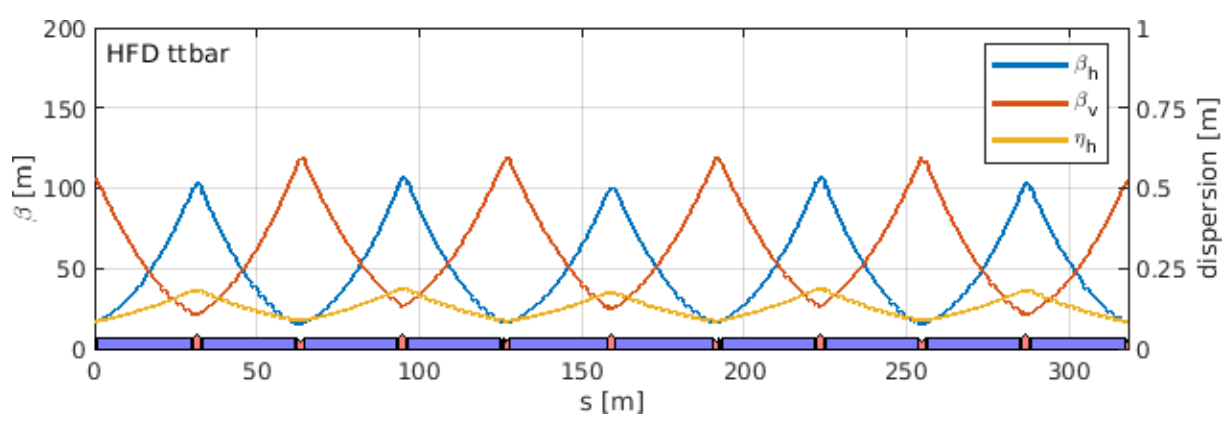

Fig. 1.45: LCC HFD arc cell optics at $\mathrm{t\bar{t}}$. The same layout is used at Z.

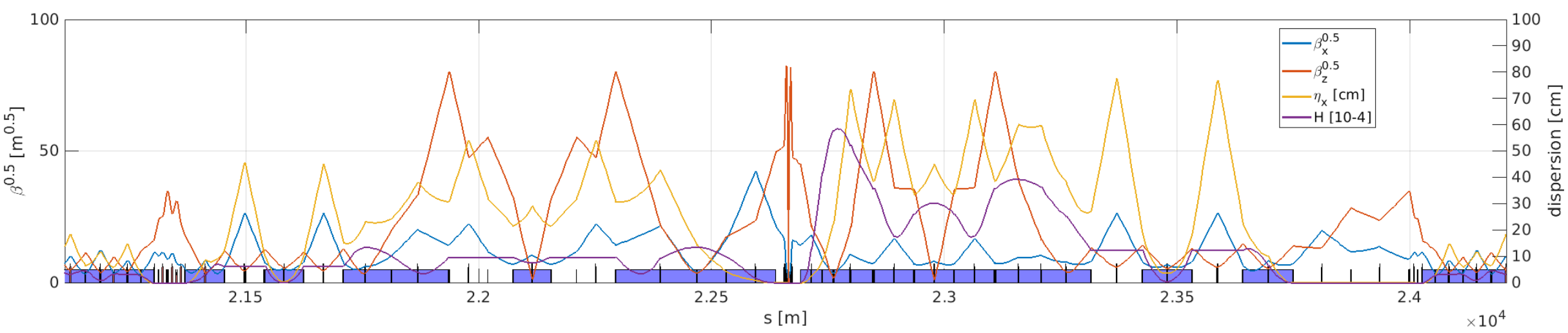

Fig. 1.46: LCC FF optics at $\mathrm{t\bar{t}}$. The same layout is used at Z.

The arcs optics are based on novel hybrid focusing de-focussing optics (HFD). Five standard FODO cells are matched as a single one (10 quadrupoles and 10 dipoles) in order to obtain the desired first and second order amplitude and momentum detuning coefficients by setting optics values at specific locations. Only two sextupoles families are needed. The HFD optics are extremely versatile and allow for large dynamic aperture and momentum acceptance. The same magnetic layout is used to operate at $(\mu_h, \mu_v) = (100, 74)$ deg for $\mathrm{t\bar{t}}$ and $(\mu_h, \mu_v) =(50,44)$ deg for Z.

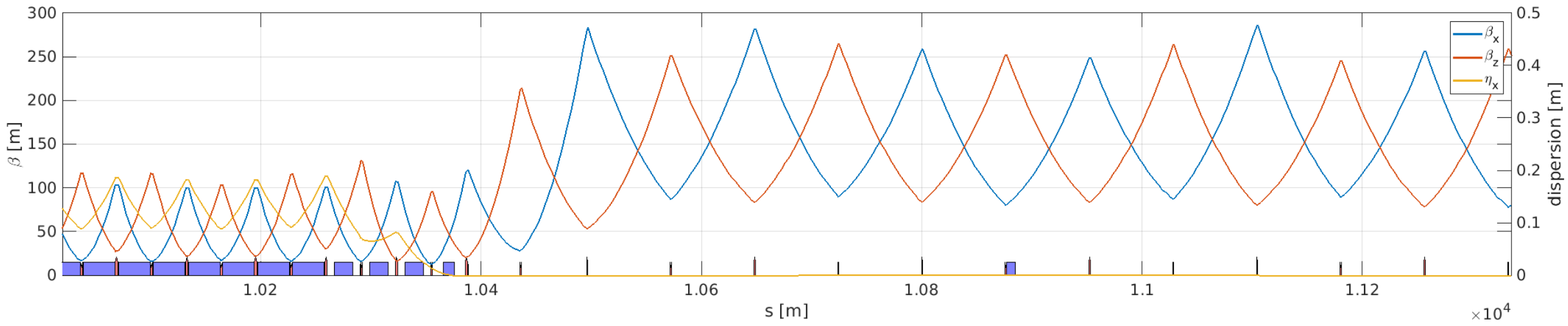

Fig. 1.47: LCC LSS optics at $t\bar{t}$. The same layout is used at Z.

The four LSS and the FF systems are designed following the *transparency conditions* detailed in [181]. These conditions mean that the beam dynamics on- and off-energy, and with or without sextupoles, are periodic and matched to the arcs. The lattice beam dynamics properties are then as close as possible to a fully periodic lattice (arcs only).

The FF optics meet all the baseline specifications in terms of crossing angle, synchrotron radiation handling and other geometrical constraints. Following the LCC principles, the chromaticity generated by the low-$\beta$ interaction points is fully corrected within the FF. Sextupoles at locations in phase with the IP are used to cancel first and second-order amplitude detuning coefficients. Decapoles placed in these exact locations minimise the detrimental effect of synchrotron radiation in the high gradient final doublet quadrupoles.

The same magnet specifications used for the baseline optics may be used. No reverse bends are needed for the LCC optics. No superconducting magnets are required except for the final doublets. The number of magnets, length and integrated strengths for the LCC optics are remarkably lower compared to the baseline optics.

The HFD cell dipoles are about 29 m long. QD and QF quadrupoles are 1.8 m and 2.4 m long respectively. 2240 quadrupoles per ring are needed for the arcs. SD and SF sextupoles are 50 cm and 35 cm long respectively, making use of the sextupole design already available for the baseline optics. Dipoles all have the same length, so the $e^+e^-$ arcs can be shifted longitudinally to align opposite polarity quadrupoles for the two rings. Twin quadrupoles and an additional short QF (60 cm long) can be used to replace the QF/QDs of the two arcs. If high-temperature superconducting magnets are used, quadrupoles do not need to be paired, and the sextupole coils can be wrapped around the quadrupole ones, thus improving the dipole filling ratio and reducing the horizontal natural equilibrium emittance. Trim coils are foreseen on the sextupoles for orbit (horizontal and vertical correctors) and optics correction (normal and skew quadrupole). No additional correctors are needed.

The LCC lattice optics at Z and $t\bar{t}$ energies have been analysed in terms of sensitivity to alignment errors in the arcs and in the final focus. As detailed in [32] the LCC lattice is for most parameters more tolerant or equivalent to the baseline optics and to other similar lattice designs such as CEPC [182].

The sensitivity to errors in the FF is dominant, and more studies on correction schemes and tuning techniques will be addressed in the future to fully evaluate the tolerated alignment errors.

Some relevant parameters for the Z and $t\bar{t}$ lattice options are reported in Table 1.22.

All non-linear magnets have been optimised using multi-objective minimisation tools in order to maximise the final lattice performance. The transverse dynamic aperture computed in the centre of the straight sections is well above $15\sigma$ in the horizontal plane and $70\sigma$ in the vertical plane for Z and $t\bar{t}$ energies. The 6D tracking for 2350 turns at Z and 40 turns at $t\bar{t}$ includes quantum diffusion, synchrotron radiation, tapering, and crab sextupoles at the optimal value for maximum luminosity. For the same tracking conditions, the momentum acceptance is close to 2% for Z and 4% for $t\bar{t}$.

The LCC optics are ready to be used for FCC-ee. Further work will be needed for the specification of optics at other energies. This work will benefit from the precise and deterministic strategies defined

Table 1.22: Full ring general parameters, for the two energies considered.

| | Z | $t\bar{t}$ |
|---|---|---|
| C [km] | 90.659 | 90.659 |
| Energy [GeV] | 45.6 | 182.5 |
| Num. IP per ring | 4 | 4 |
| Crossing angle [mrad] | 30 | 30 |
| $\beta$-tron tunes | (198.26,174.38) | (350.224,266.36) |
| Chromaticity | (0.20,0.21) | (0.23,1.66) |
| $\epsilon_h$ [pm rad] | 684.72 | 2100.9 |
| $J$ | (1,1,2) | (1,1,2) |
| $\alpha_c$ | 2.894e-05 | 0.946e-05 |
| $U_0$ [MeV/turn] | 34.3 | 8808.2 |
| $\sigma_E$ | 3.715e-4 | 14.9e-4 |
| $\beta_h^*$ [mm] | 0.7 | 1.6 |
| $\beta_v^*$ [mm] | 100 | 1000 |
| bunch length [mm] | 3.4 | 2.7 |
| RF Voltage [GV] | 0.17 | 10.4 |
| RF frequency [MHz] | 400 | 400 |
| Long. damp. time [ms] | 401.6 | 6.3 |
| Synchrotron tune | 0.045 | 0.074 |

during the LCC optics design as well as the matching and optimisation scripts available.

### 1.10.2 Optics with nested magnets

An alternative design considers nesting dipoles with the arc FODO quadrupoles and sextupoles to increase the dipole fill factor and reduce the overall synchrotron radiation [183]. This could be achieved by using dedicated superconducting arc magnets that nest dipoles with quadrupoles and sextupoles that are designed in close collaboration with the optics development [184]. These nested magnets would replace the regular quadrupoles and sextupoles in the GHC lattice, while otherwise keeping the average cell length and phase advance unchanged. The magnet design allows for individually tuning the overlapping fields, so that it can directly be used for energy tapering along with orbit and optics corrections.

The overlapping dipole and quadrupole fields directly impact the damping partitions, leading to significant changes in the horizontal equilibrium emittance and under certain conditions unstable beam sizes, for example when the dipole fields in the nested magnets match the arc dipole strength. The stability and the equilibrium emittance can be adjusted by carefully adjusting the dipole field nested with the focusing arc quadrupoles and compensating for this by redistributing the integrated strength among the arc dipoles and the dipoles overlapping with the defocusing quadrupoles. This was done to optimise the equilibrium emittance of the nested $t\bar{t}$ lattice design to be 1.39 nm, roughly equal to that of the baseline GHC design. This was achieved by having a dipole field in the focusing quadrupoles about 46% lower than in the rest of the arc. The design lends itself to easily further optimising the horizontal equilibrium emittance by tuning the dipole strength in the focusing quadrupoles [183].

The phase advance of the FODO cell can be preserved by performing a small adjustment on the quadrupole strengths, leading to a change in the $\beta$-function of about 1 % compared to that of the baseline GHC design. This small change in the optics can be easily absorbed by a slight re-matching of the first few magnets in the straight sections to allow a largely unchanged design of the experiment and RF insertions. The overall decrease in dipole field, due to the redistribution to the nested magnets results in a significant reduction of the synchrotron radiation by 16.5% from about 10.0 GeV/turn to 8.4 GeV/turn

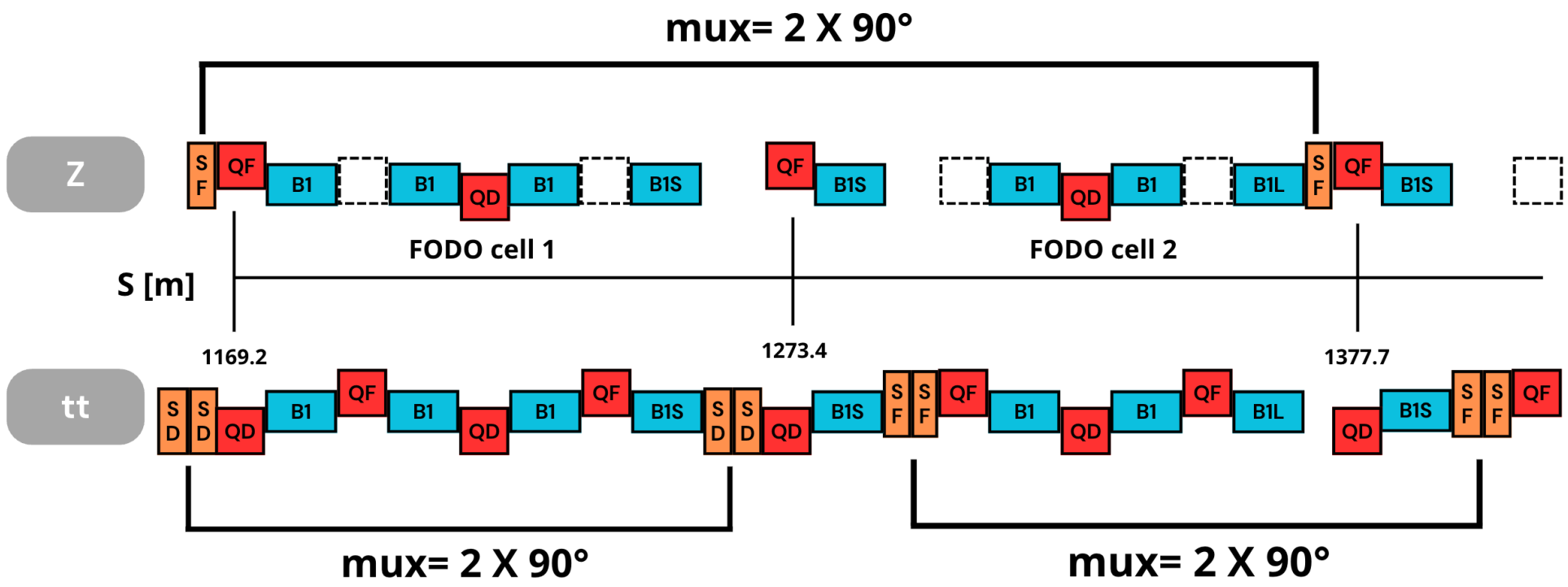

Fig. 1.48: Illustration of arc cell layouts for GHC lattices in Z and $\mathrm{t\bar{t}}$ operation.

at the $\mathrm{t\bar{t}}$ energy.

When basing the design entirely on the GHC design, a complication arises due to the unequal dipole field distribution when taking into account the second beam. The baseline design assumes a twin aperture design sharing the same yoke for the magnets of both beams, with a focusing quadrupole in one beam paired with a defocusing quadrupole in the other beam and vice-versa. The unequal dipole fields and bending angle between focusing and defocusing quadrupole would result in the design trajectories of the two beams diverging in the arcs. In the case of the z-lattice, this would result in a maximum deviation of about 1.5 mm, which would most likely be tolerable. However, an alternative arrangement with matching polarities between the beams is also feasible and could be implemented with very few changes to the design.

A second complication is due to the changing arc cell length difference between the Z and $\mathrm{t\bar{t}}$ operation. As is illustrated by Fig. 1.48, the shortening of the FODO cell by factor two between Z and $\mathrm{t\bar{t}}$ operation results in the change in the polarity of the focusing arc quadrupoles. This, along with the nested dipole field that depends on the polarity of the quadrupole for stability reasons, results in a change in geometry between Z and $\mathrm{t\bar{t}}$ operation. This effect can be compensated by a physical geometry that deviates from the magnetic geometry when operating at Z energy. This can be modelled by assigning a different magnetic $k_0$ than the geometric bending angle. The effect can be largely mitigated by utilising the dipoles in the gaps initially reserved for focusing quadrupoles, as orbit kickers in $\mathrm{t\bar{t}}$ operation. Such an approach would result in a maximum orbit deviation of approximately $2\,\mathrm{mm}$ while also reducing the horizontal equilibrium emittance.

An alternative, though more labour-intensive, solution would be to realign the magnets when transitioning between the two operational modes, which might preclude the use of dual-aperture magnets. While the first approach is more practical and cost-effective, realigning the magnets would be less challenging from a beam dynamics perspective, as it eliminates the need to manage sextupole feed-down effects. Both solutions should be explored to determine the optimal strategy.

For the sextupole placement, two options are considered, as shown in Fig. 1.49b. The first option envisions nesting the sextupoles together with the quadrupole and dipole coils, whilst the second option would nest the sextupoles with only dipoles and in series with the nested quadrupoles. Both options result in a similar dynamic aperture to the GHC lattice by linearly scaling the sextupole strength to recover the correct chromaticity. However, the momentum acceptance requires further sextupole optimisation [185]. Apart from requiring fewer superconducting magnets, a key advantage of Option 1 is that it could facilitate the alignment of the quadrupole and sextupole coils with each other, potentially reducing sources of coupling.

Overall, the nested magnet design based on the GHC offers a performance comparable to that of the GHC and is fully compatible with the insertion regions. A summary of the design parameters of

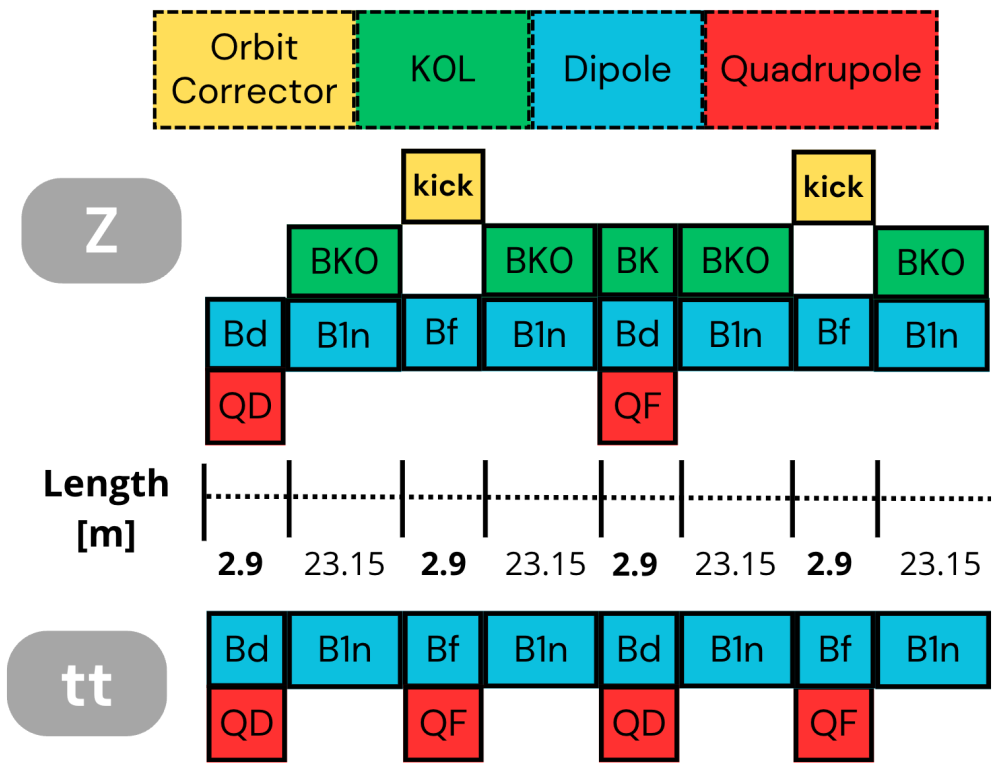

(a) Magnetic nested circuits proposed for Z and $t\bar{t}$ operation in nested magnet configuration.

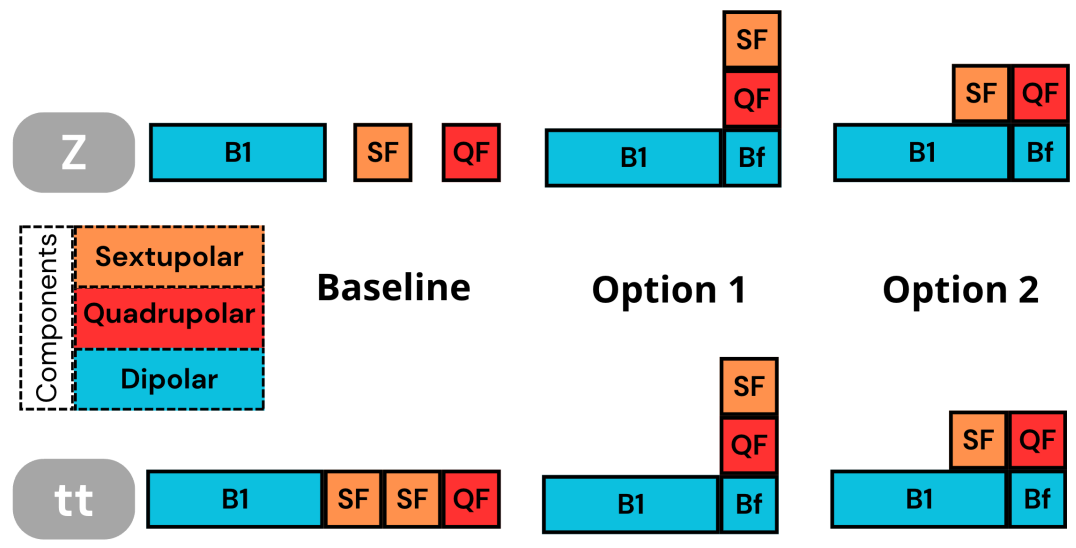

(b) Two options for nested sextupoles.

Fig. 1.49: Options for nested circuits.

the GHC and the nested magnet arc cell design is shown in Table 1.23. This was achieved by having a non-uniform dipole field in the arc cells, which brings additional challenges that have been identified and solved. Compared to the nominal GHC design, this solution offers reduced power loss due to ohmic heating by having superconducting magnets in place of the quadrupoles and reduced synchrotron radiation, leading to an estimated decrease in power consumption of up to 20%, even when taking into account the cooling for the cryogenic systems. Moreover, this solution reduces the number of dipole families required, and the individual circuits that are necessary by default could be utilised to taper the lattice efficiently using only the superconducting magnets. Similar methods should apply to the LCC design; however, the more compact quadrupoles would diminish the advantage of reduced synchrotron radiation.

Table 1.23: Overview of the different design parameters for the Z and $t\bar{t}$ lattices with nested magnets, including optical and synchrotron radiation properties.

| | **Z GHC** | **Z NMs Realigned** | **Z NMs K0** | **$t\bar{t}$ GHC** | **$t\bar{t}$ NMs** |
|---|---|---|---|---|---|
| $D_{x_{\max}}$ [m] | 0.634 | 0.638 | 0.722 | 0.559 | 0.559 |
| $I_2$ [$10^{-4}$] | 6.417 | 5.372 | 5.367 | 6.40 | 5.35 |
| $I_5$ [$10^{-10}$] | 1.484 | 1.027 | 1.101 | 0.194 | 0.138 |
| $U_0$ [MeV/turn] | 39.06 | 32.70 | 32.69 | 9994.85 | 8353.07 |
| $\epsilon_x$ [nm] | 0.705 | 0.605 | 0.500 | 1.478 | 1.388 |
| Damping | 0.709 | 0.880 | 0.675 | 0.0110 | 0.0146 |
| times[s] | 0.709 | 0.848 | 0.847 | 0.0110 | 0.0132 |
| | 0.354 | 0.416 | 0.486 | 0.0055 | 0.0063 |
| $J_x$ | 1.000 | 0.963 | 1.255 | 0.999 | 0.903 |
| $J_y$ | 1.000 | 1.000 | 1.001 | 1.000 | 1.000 |
| $J_z$ | 1.999 | 2.036 | 1.745 | 2.000 | 2.096 |

### 1.10.3 Non-local detector solenoid compensation

For cancelling the effect of the detector solenoid field, as an alternative to the baseline scheme, which concentrates the compensation solenoid within $\pm\ell^*$ around the IP (local scheme), the compensation solenoid can also be placed behind the final quadrupole unit (non-local scheme). Both local and non-local schemes can compensate for the $x$-$y$ coupling and the vertical dispersions without degrading the performance (including the beam-beam effects). In the non-local scheme, the required compensation solenoid field is weaker. While the non-local scheme, thereby, provides a better residual vertical emittance, it also seems to induce stronger spin depolarisation. Studies are needed to determine whether, for the non-local scheme, introducing targeted spin-orbit bumps can recover the desired level of equilibrium polarisation.

It is noted that the former LEP collider, without a large crossing angle, used a non-local scheme consisting of four pairs of anti-symmetrically powered tilted quadrupoles over one side of each insertion to compensate the betatron coupling induced by the detector solenoid [186, 187].

### 1.10.4 Monochromatic operation mode

One of the most fundamental outstanding measurements, since the Higgs boson discovery [188, 189], is determining its Yukawa couplings [13,190]. Measuring the coupling of first-generation fermions presents significant experimental challenges due to their low masses and, consequently, small Yukawa couplings to Higgs fields. The measurement of this coupling is virtually impossible at hadron colliders because the $\mathrm{H} \rightarrow \mathrm{e}^+\mathrm{e}^-$ decay has a tiny branching ratio, completely swamped by the Drell-Yan dielectron continuum with many orders of magnitude larger cross section. The FCC-ee, with unrivalled integrated luminosities of 10 $\mathrm{ab}^{-1}$ per year at 125 GeV could enable observing the resonant $s$-channel production of the scalar Higgs boson, namely the reaction $\mathrm{e}^+\mathrm{e}^- \rightarrow \mathrm{H}$ on the Higgs pole [191, 192]. This possibility motivated physics [193] and accelerator studies [194–202] towards implementing this new operation mode.

Such a measurement is more easily feasible if the centre-of-mass (CM) energy spread of $\mathrm{e}^+\mathrm{e}^-$ collisions, which is approximately 50 MeV due to energy spread from synchrotron radiation (SR) alone, and further enhanced by beamstrahlung, in a conventional collision scheme, can be reduced to a level comparable to the natural width of the Standard Model Higgs boson $\Gamma_\mathrm{H} = 4.1$ MeV. To reduce the collision-energy spread and enhance the CM energy resolution in colliding-beam experiments, the concept of *monochromatisation* has long been proposed [203]. The basic idea consists of creating opposite correlations between spatial position and energy deviation within the colliding beams, which can be accomplished in beam-optics terms by introducing a non-zero dispersion function with opposite signs for the two beams at the interaction point (IP), as sketched in Fig. 1.50 for a crossing-angle configuration.

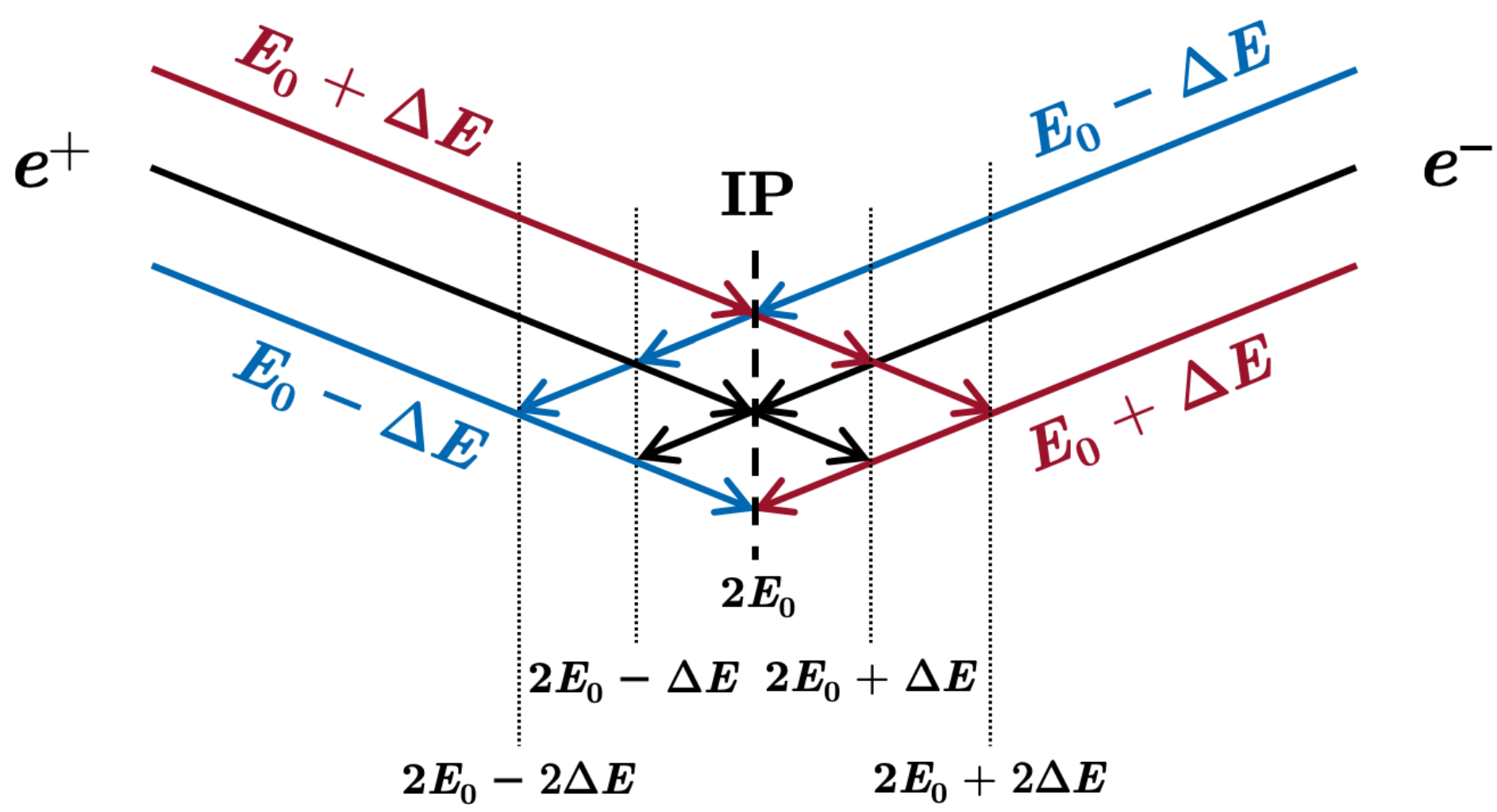

Fig. 1.50: Schematic of crossing-angle collision with monochromatisation based on nonzero horizontal IP dispersion, showing trajectories at the nominal energy $E_0$ and with an energy offset of $\pm\Delta E$.

Taking as a starting point the GHC optics [204–207], for the $\mathrm{t\bar{t}}$ mode, different monochromatisation schemes implying non-zero horizontal or vertical or both types of dispersion function at the IP ($D^*_{x,y}$) have been studied. All newly proposed IR optics has been designed to remain compatible with a standard operation mode without dispersion at the IP and also with the present tunnel configuration.

Given the presence of horizontal bending magnets in the vertical local chromaticity correction of the FCC-ee GHC Interaction Region (IR), the most natural way to implement monochromatisation in this FCC-ee lattice type is reconfiguring these IR dipoles so as to generate a non-zero $D^*_x$ while maintaining the same crossing angle $\theta_c$. Indeed, a wide $\sigma^*_x$ helps mitigate the impact of the beamstrahlung (BS) on the energy spread $\sigma_\delta$, while preserving a small $\sigma^*_y$ is crucial for attaining high $\mathcal{L}$. Taking into account the baseline parameters for the FCC-ee GHC lattice with horizontal betatron sizes ($\sigma^*_{x,\beta} = \sqrt{\varepsilon_x \beta^*_x}$) at the IP of the order of 10 µm and a $\sigma_{\delta,SR}$ of $\sim 0.05$ % at $s$-channel Higgs production energy ($\sim$125 GeV), a $D^*_x$ of around 10 cm is required to achieve a monochromatisation factor ($\lambda$) of $\sim$5-8.

Because $\sigma^*_{y,\beta}$ ($\sim$ nm) $\ll \sigma^*_x$ ($\sim$µm) for getting high luminosities, about 100 times smaller $D^*_y$ ($\sim$ mm) is needed to get a similar $\lambda$. A nonzero $D^*_y$ of this magnitude could be generated by simply using skew quadrupole correctors around the IP [208–210]. These quadrupoles could be located close to the sextupole pairs in the IR.

As an illustration, a monochromatisation IR optics with combined 0.105 m of horizontal and 1 mm vertical IP dispersion) based on the FCC-ee GHC $\mathrm{t\bar{t}}$ optics as starting point has been developed using MAD-X [211]. It is shown in Fig. 1.51. Different monochromatised beam-optics designs, including ones based on the lower-energy Z lattice, are detailed in Ref. [212].

After global implementation, the results of the analytical global performance evaluation for the monochromatisation IR optics based on the 'FCC-ee GHC $\mathrm{t\bar{t}}$' are summarised in Table 1.24, for the crossing-angle configuration. Parameters due only to synchrotron radiation are marked with 'SR', while those including the impact of beamstrahlung are marked with 'BS'. For comparison, the first column labelled Standard ZES, presents an energy-scaled (ES) optics configuration. Its performance parameters were calculated after increasing the FCC-ee V22 $\mathrm{t\bar{t}}$ optics from 45.6 GeV to 62.5 GeV, followed by completing all corrections and synchrotron radiation power loss compensation. The optics labelled MonochroM ZH4IP integrate the IP horizontal dispersion generation monochromatisation optics at all four IPs, while MonochroM ZH2IP does so at only two of the four IPs. The designation MonochroM ZHS refers to the re-matched standard optics design that is orbit-compatible with the MonochroM ZH4IP optics. The number of bunches per beam $n_b$ is constrained by the maximum beam-beam tune shift, taken to be 0.14, and by a minimum bunch spacing of 25 ns at FCC-ee. To select an appropriate $n_b$, studies optimising the luminosity per IP $\mathcal{L}$ and the CM energy spread $\sigma_W$ of the MonochroM ZH4IP optics, as a function of $n_b$ were conducted, including the beamstrahlung impact under the crossing-angle collision configuration.

To accurately assess the performance of the FCC-ee monochromatisation IR optics, which features non-zero dispersion at the IP, the $\sigma_W$ and luminosity per IP $\mathcal{L}$ were calculated for the different configurations using the simulation tool GUINEA-PIG [213] and taking into account the impact of beamstrahlung. In these calculations, the particle distribution at the IP was modelled as an ideal Gaussian distribution, characterised by the global optical parameters of each optics configuration.

It is noted that while the physics performance of a nonzero-$D^*_y$ scheme is less favourable, it would be easier to implement without altering the IR orbit, rendering it an attractive option for existing low-energy $\mathrm{e^+e^-}$ colliders. Without the $\epsilon_y$ blow-up due to BS, this scheme could potentially achieve better performance in such settings.

Looking ahead, the optical parameters for the monochromatisation mode will be further optimised for enhanced performance. Second, the dynamic aperture optimisation for these new types of monochromatisation optics will be carried out step-by-step by adjusting arc sextupole families according to particle tracking results. This will allow beam-beam simulations with non-zero IP dispersion in the code XSUITE, incorporating the hourglass and BS effects rather than relying solely on analytical evaluations. Implementations of monochromatisation in more symmetric IRs, such as those in the FCC-ee LCC optics, will also be explored. Finally, studies are underway to validate the monochromatisation concept experimentally in existing low-energy circular $e^+e^-$ colliders, such as BEPC II, DAΦNE, and SuperKEKB [100,214,215]. These efforts will be essential for achieving the full potential of monochromatisation in future collider projects.

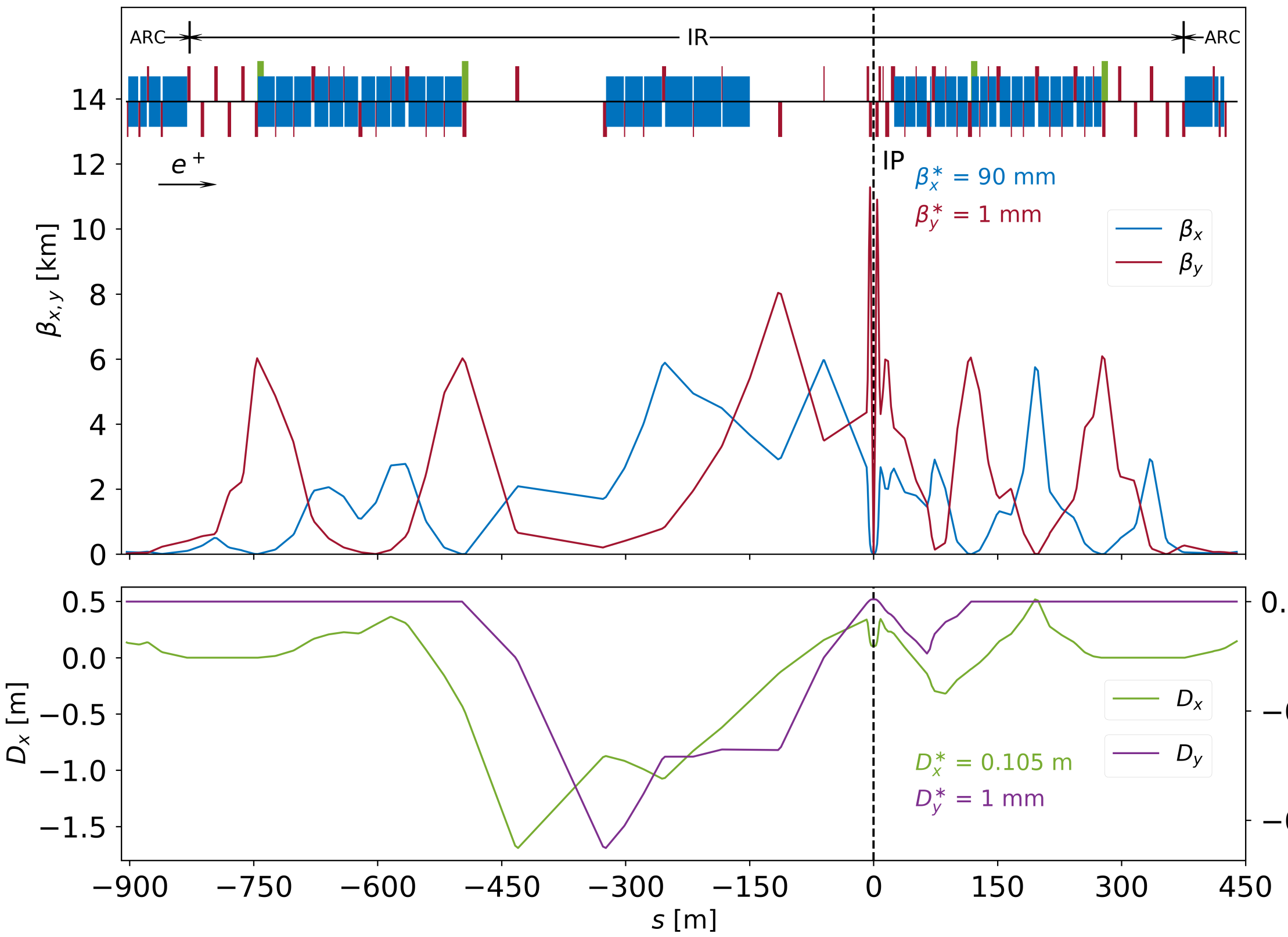

Fig. 1.51: Monochromatisation IR lattices and optics with combined 0.105 m horizontal and 1 mm vertical IP dispersion, based on the FCC-ee GHC $t\bar{t}$ optics, developed using MAD-X. The beam direction is from left to right and the dashed line $s = 0$ marks the location of IP. In the lattice, dipoles, quadrupoles and sextupoles are shown in blue, red and green respectively, while focusing and defocusing elements are positioned above and below the orbit. In the optics, horizontal and vertical betatron functions are displayed in blue and red respectively, while horizontal and vertical dispersion functions are shown in green and purple, respectively.

### 1.10.5 Alternate polarisation studies

As discussed in Sections 1.7.3 and 2.1.2, injecting pre-polarised pilot bunches could significantly enhance the available time for physics, especially in case of frequent beam aborts, and in addition, ease constraints on the maximum achievable polarisation. The injector would then need to generate polarised pilot bunches of electrons and positrons, with sufficient transport of polarisation through the full injector chain and booster energy ramp. Polarised electron guns providing the required bunch intensity exist, e.g., the dc polarised electron gun developed for the Electron Ion Collider [216]. For the positrons, a small polarising ring located in, or inside, the damping ring [217] could pre-polarise and store a set of pilot bunches until they are needed for energy calibration in the collider.

Table 1.24: Global performance parameters of monochromatisation IR optics with nonzero horizontal IP dispersion based on the 'FCC-ee GHC $\mathrm{t\bar{t}}$' optics under the crossing-angle configuration.

| Parameter | [Unit] | Standard ZES | MonochroM ZH4IP | MonochroM ZH2IP | MonochroM ZHS |
|---|---|---|---|---|---|
| # of IPs $n_{\mathrm{IP}}$ | | 4 | | | |
| Full crossing angle $\theta_c$ | [mrad] | 30 | | | |
| SR power / beam $P_{\mathrm{SR}}$ | [MW] | 50 | 50 | 49 | 50 |
| Beam Energy $E_0$ | [GeV] | 62.5 | | | |
| Energy loss / turn $U_0$ | [GeV] | 0.138 | 0.143 | 0.141 | 0.143 |
| Beam Current $I$ | [mA] | 360 | 350 | 350 | 350 |
| Bunches / beam $n_b$ | | 12000 | | | |
| Bunch population $N_b$ | [$10^{11}$] | 0.57 | 0.55 | 0.55 | 0.55 |
| Hor. emittance (SR/BS) $\varepsilon_x$ | [nm] | 0.17 / 0.17 | 1.48 / 7.27 | 0.84 / 4.23 | 0.35 / 0.35 |
| Vert. emittance (SR/BS) $\varepsilon_y$ | [pm] | 0.35 / 0.35 | 2.96 / 2.96 | 1.68 / 1.68 | 0.71 / 0.71 |
| Momentum compaction factor $\alpha_{\mathrm{C}}$ | [$10^{-6}$] | 7.31 | 6.92 | 7.12 | 7.06 |
| $\beta^*_{x/y}$ | [mm] | 1000 / 1.6 | 90 / 1 | 90 / 1 | 1000 / 1.6 |
| $D^*_{x/y}$ | [m] | 0 / 0 | 0.105 / 0 | 0.105 / 0 | 0 / 0 |
| Rel. energy spread (SR/BS) $\sigma_\delta$ | [%] | 0.054 / 0.076 | 0.055 / 0.057 | 0.054 / 0.057 | 0.055 / 0.068 |
| Bunch length (SR/BS) $\sigma_z$ | [mm] | 3.86 / 5.49 | 4.05 / 4.20 | 3.95 / 4.12 | 4.09 / 5.07 |
| RF voltage 400/800 MHz $V_{\mathrm{RF}}$ | [GV] | 0.170 / 0 | | | |
| RF frequency (400MHz) $f_{\mathrm{RF}}$ | [MHz] | 399.994581 | | | |
| Synchrotron tune $Q_s$ | | 0.015 | 0.014 | 0.014 | 0.014 |
| Long. damping time $\tau_E/T_{\mathrm{rev}}$ | [turns] | 454 | 436 | 445 | 436 |
| Horiz. beam-beam (SR/BS) $\xi_x$ | | 0.059 / 0.030 | 0.0025 / 0.0022 | 0.0027 / 0.0024 | 0.049 / 0.033 |
| Vert. beam-beam (SR/BS) $\xi_y$ | | 0.24 / 0.17 | 0.044 / 0.041 | 0.060 / 0.056 | 0.15 / 0.12 |
| CM energy spread (SR/BS) $\sigma_W$ | [MeV] | 47.45 / 67.58 | 13.41 / 25.75 | 10.25 / 20.95 | 48.80 / 60.47 |
| Luminosity / IP (SR/BS) $\mathcal{L}$ | [$10^{34}$ cm$^{-2}$s$^{-1}$] | 72.8 / 51.9 | 20.9 / 19.5 | 28.3 / 26.6 | 44.6 / 36.6 |

# Chapter 2

# FCC-ee collider operation concept

## 2.1 Operation requirements

### 2.1.1 Physics requirements

Table 2.1 shows the target integrated luminosities with four interaction points, allocated running times, and the total number of events, for the four baseline centre-of-mass energy stages at the Z pole, the WW production threshold, the ZH production cross-section maximum, and the top-pair production, in this chronological order.

The nominal integrated luminosity is computed by assuming 185 days of physics run time per year, a hardware availability of at least 80%, and a corresponding 'physics efficiency' $E$ of 75% (also see Section 2.1.2). In addition, for determining the total integrated luminosity and the number of events expected to be produced (in the last two rows of the table), the luminosity is assumed to be half the design value during the first two years at the Z pole and for the first year at the $\mathrm{t\bar{t}}$ threshold. At the Z pole, the integrated luminosity is distributed as follows: $40\,\mathrm{ab}^{-1}$ at 87.9 GeV, $125\,\mathrm{ab}^{-1}$ at 91.2 GeV, and $40\,\mathrm{ab}^{-1}$ at 94.3 GeV. The number of Z decays results from this setup. At the WW threshold, the run time is evenly distributed between 157.5 GeV and 162.5 GeV. The number of WW events include all $\sqrt{s}$ values from 157.5 GeV up.

Figure 2.1 displays the corresponding baseline sequence of events [218]. However, other mode sequences would be possible (see Section 2.2.1) and might be preferred. For example, scheduling a Z pole run after the ZH run or the WW threshold run can be considered, ideally with an initial Z pole run during the early phase of FCC-ee operation. Indeed, while the Z pole run offers an exceptionally rich set of physics opportunities, it is also the most ambitious and demanding part of the programme from all perspectives, including accelerator performance, energy calibration, detector systematic biases, and theoretical calculations. It will be extremely challenging to achieve all the goals of the Z pole run during the first four years of the collider operation. The new versatile RF system designed accordingly, enables a quasi-total flexibility to choose the running sequence. For example, it would allow short initial Z pole and WW threshold runs, to commission the collider and the detectors, to establish the resonant depolarisation procedures, etc. The ZH run could then proceed, before going back to the Z pole and the WW threshold, both now at full luminosity, with fully functional resonant depolarisation, and complete understanding of the collider.

### 2.1.2 Target availability and efficiency

Annual integrated luminosity estimates for FCC-ee at each mode of operation are derived from three or four parameters:

- Nominal Instantaneous Luminosity $L$: The first two years of Z and the first year of $\mathrm{t\bar{t}}$ are assumed to achieve, on average, 50% and 65% of this value, respectively, to account for machine commissioning and beam tuning. These reductions reflect LEP/LEP-2 experience. Nominal luminosity is then assumed from the third year onward in Z pole operation and from the second year onward at the $\mathrm{t\bar{t}}$ threshold.
- Annual Scheduled Physics Time $T$: It is assumed that 185 days per year are scheduled for physics. This is obtained by subtracting from one year (365 days): 17 weeks of extended shutdowns (120 days), 30 days of annual commissioning, 20 days for machine development and 10 days for technical stops.

Table 2.1: The baseline FCC-ee operation model with four interaction points, showing the centre-of-mass energies, design instantaneous luminosities for each IP, integrated luminosity per year summed over 4 IPs [218]. The integrated luminosity values correspond to 185 days of physics per year and 75% operational efficiency (i.e., 1.2 $10^7$ seconds per year) [4], in the Z, WW, ZH, $\mathrm{t\bar{t}}$ baseline sequence. The last two rows indicate the total integrated luminosity and the total number of events expected to be produced in the four detectors.

| Working point | Z pole | WW thresh. | ZH | $\mathrm{t\bar{t}}$ | |
|---|---|---|---|---|---|
| $\sqrt{s}$ (GeV) | 88, 91, 94 | 157, 163 | 240 | 340–350 | 365 |
| Lumi/IP ($10^{34}\,\mathrm{cm^{-2}s^{-1}}$) | 140 | 20 | 7.5 | 1.8 | 1.4 |
| Lumi/year ($\mathrm{ab^{-1}}$) | 68 | 9.6 | 3.6 | 0.83 | 0.67 |
| Run time (year) | 4 | 2 | 3 | 1 | 4 |
| Integrated Lumi ($\mathrm{ab^{-1}}$) | 205 | 19.2 | 10.8 | 0.42 | 2.70 |
| Number of events | $6\,10^{12}$ Z | $2.4\,10^{8}$ WW | $2.2\,10^{6}$ HZ + 65k WW $\rightarrow$ H | $2\,10^{6}\,\mathrm{t\bar{t}}$ +370k HZ +92k WW $\rightarrow$ H | |

– Availability $A$: Represents the percentage of scheduled physics time when the collider and injectors are able to deliver beam, as opposed to a hardware failure leading to downtime and need to repair. An overall machine availability of 80% is assumed.
– Efficiency $E$: The efficiency factor $E$ is an empirical factor, whose value can be extrapolated from other similar machines, or by simulations with average failure rate and average downtime. Thanks to the top-up mode of operation, it is expected that $E$ will be, within five percent, equal to the availability $A$ of the collider complex. The target availability is at least 80% and, thereby, a corresponding efficiency $E > 75\%$ is expected [4]. In the case of FCC-ee, no time is lost for acceleration and the efficiency only reflects the relative downtime due to technical problems and associated re-filling and recovery time. Therefore, the efficiency will be roughly equal to the hardware availability, taken to be at least 80%, minus 5% reduction for beam recovery after a failure. The assumed efficiency value of 75% with respect to the daily peak luminosity is lower than achieved with top-up injection at KEKB and PEP-II [4]. However, at FCC-ee the need for pilot bunches to be pre-polarised in the collider after each beam abort may pose a challenge for reaching the target collider efficiency in the Z and W modes of operation, which could be largely mitigated by generating polarised pilot bunches already in the injector (Fig. 2.24).

From the above factors, the annual achieved integrated luminosity at each interaction point (IP) is calculated, assuming top-up operation during stable beams, as

$$L_{\mathrm{int}} = E\,T\,L\,. \tag{2.1}$$

Using the above parameters, the integrated luminosity target in each energy mode is reached or exceeded. A detailed study to explore the implied challenges is summarised in Section 2.4, which also presents example simulations for the collider-ring systems. Similar studies for the booster, injector complex and technical infrastructure are presented in Sections 5.3, 7.9 and 8.10, respectively.

## 2.2 Changing operation modes

### 2.2.1 Switching between Z, WW, and ZH modes

The ability to easily switch between Z, WW and ZH modes of operation depends upon the following conditions and requirements.

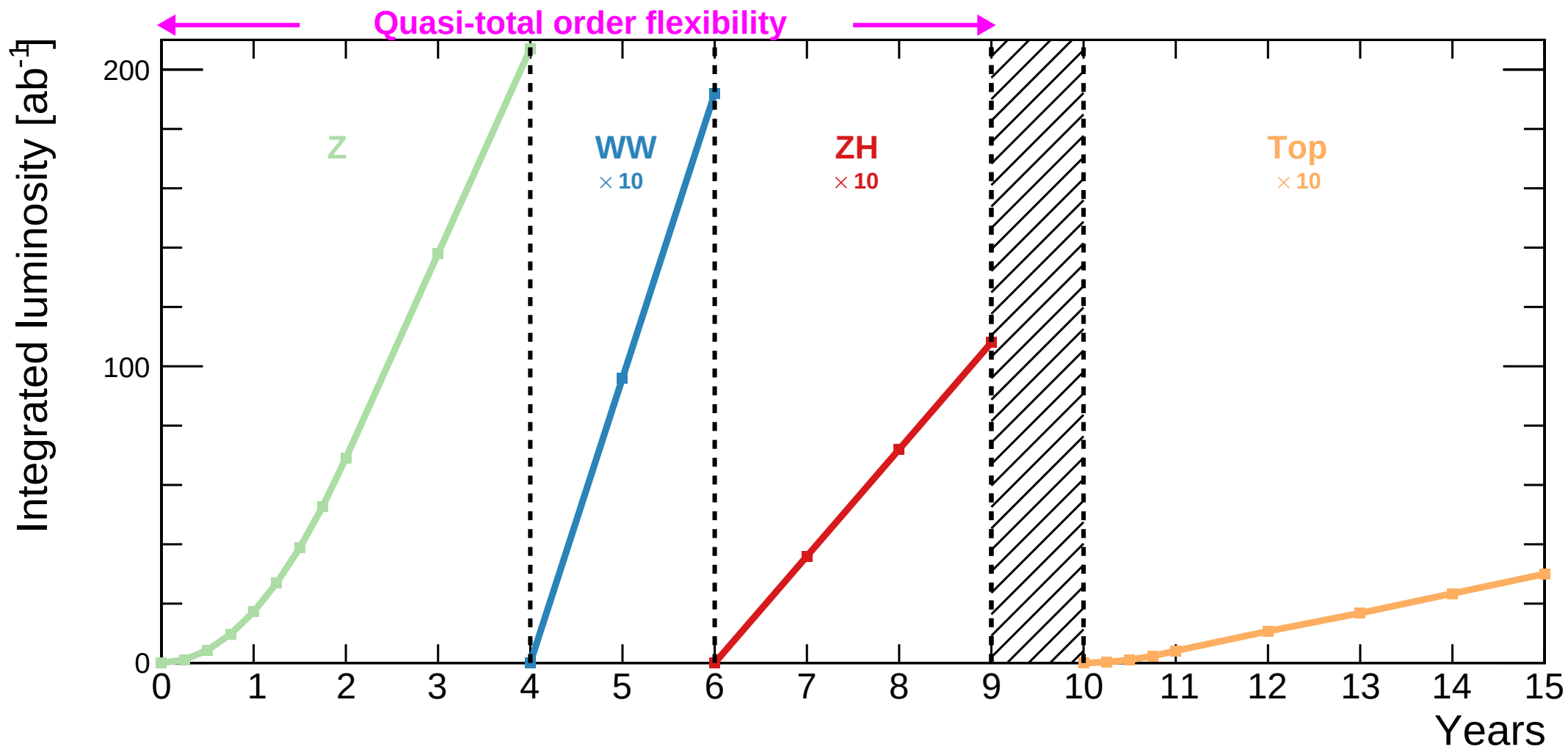

Fig. 2.1: Operation sequence for FCC-ee with four interaction points, showing the integrated luminosity at the Z pole (pink), the WW threshold (blue), the Higgs factory (red), and the top-pair threshold (green) as a function of time. In this baseline model, the sequence of events goes with increasing centre-of-mass energy, but there is quasi-total flexibility in the sequence all the way to 240 GeV. The integrated luminosity delivered during the first two years at the Z pole and the first year at the $t\bar{t}$ threshold is half the annual design value. The hatched area indicates the shutdown time needed to prepare the collider for the higher energy runs at the top-pair production threshold and above.

The 400 MHz cavities and cryomodules must all be installed in their final location from the start of operation; this represents 33 cryomodules on either side of point PH. This approach poses constraints for the early procurement and installation of all cryomodules, but it has the great merit of avoiding a staged installation of the cryogenics systems and later interventions in the tunnel for installing additional cryomodules.

At the Z and WW operating points, the incoming beam from the arc is arriving from the outer aperture and first passes through the 33 cryomodules on the incoming side of the insertion; it is then deviated towards a bypass line that goes around the other 33 cryomodules on the outgoing side of the insertion, before being brought back towards the inside aperture of the outgoing arc. Hence, the two counter-rotating beams are crossing at the middle of the RF insertion.

With the scheme of reverse phase operation for the 400 MHz RF, the switch between the Z and WW modes of operation involves only a reconfiguration of the RF system without any hardware intervention; the beamlines and beam paths stay the same.

The crossing of the beams in the middle of the insertion is done over a long length to avoid string dipole bends that could generate synchrotron radiation towards the cavities. The adverse effect of this shallow horizontal crossing is that the electron and positron beams share a common vacuum chamber over a certain length, which, in turn, generates long-range beam-beam interactions that would be very detrimental at the Z and WW modes of operation. For this reason, the horizontal crossing is supplemented by a vertical bump in opposite directions for both beams, such that they can be in separate vacuum chambers over the horizontal crossing, avoiding all long-range beam-beam interactions.

The separation scheme and recombination at the Z and WW operating points involve only magnetic dipoles (see Fig. 2.2, top picture), the only point of attention being that the dipoles should not generate synchrotron radiation towards the RF cavities and ancillary equipment and that any synchrotron radiation generated should be minimal, both in terms of power radiated and critical energy of the photons.

The switch to the ZH operating point requires that both beams go through the entire set of two times 33 cryomodules (Fig. 2.2, centre picture). The incoming beam from the arc coming from the outer aperture first goes through the 33 cryomodules on the incoming side of the insertion. The deviation dipoles towards the bypass line are now switched off, and the beam goes straight in a different beampipe towards the other side of the insertion, where it goes through the other 33 cryomodules on the outgoing side of the insertion. At the exit of the second set of cryomodules, the beam is deviated towards the inside aperture of the outgoing arc.

Because of the symmetry between the two beams, it is obvious that they share the same beam path across the whole insertion in opposite directions. The first consequence is that the timing of the bunches in the beams must ensure that no collision occurs in the RF insertion. This can be done by having only two trains of bunches for each beam, colliding in experimental insertions only.

The second consequence is that the deviation of the beam on the outgoing side of the insertion must be completely transparent for the incoming beam coming straight from the outer arc into the cryomodules. With opposite particle charges and opposite directions of the two beams, this cannot be achieved with simple magnetic elements. A single element can combine electrostatic and magnetic forces that add up for one charge and direction and cancel exactly for the opposite charge and opposite direction. The incoming beam from the outer aperture of the arc is not deviated and the outgoing beam is deviated to the inner aperture of the arc.

Finally, the ability to switch between Z, WW, and ZH operating points implies that both layouts coexist in a single insertion, as is sketched in Fig. 2.2 (bottom picture). Equal path lengths for each layout are required to maintain the same distance between interaction points. Further, the phase advances for each layout also need to be equal, unless the phase advances in the other technical insertions at points PB, PF, and PL can be re-tuned.

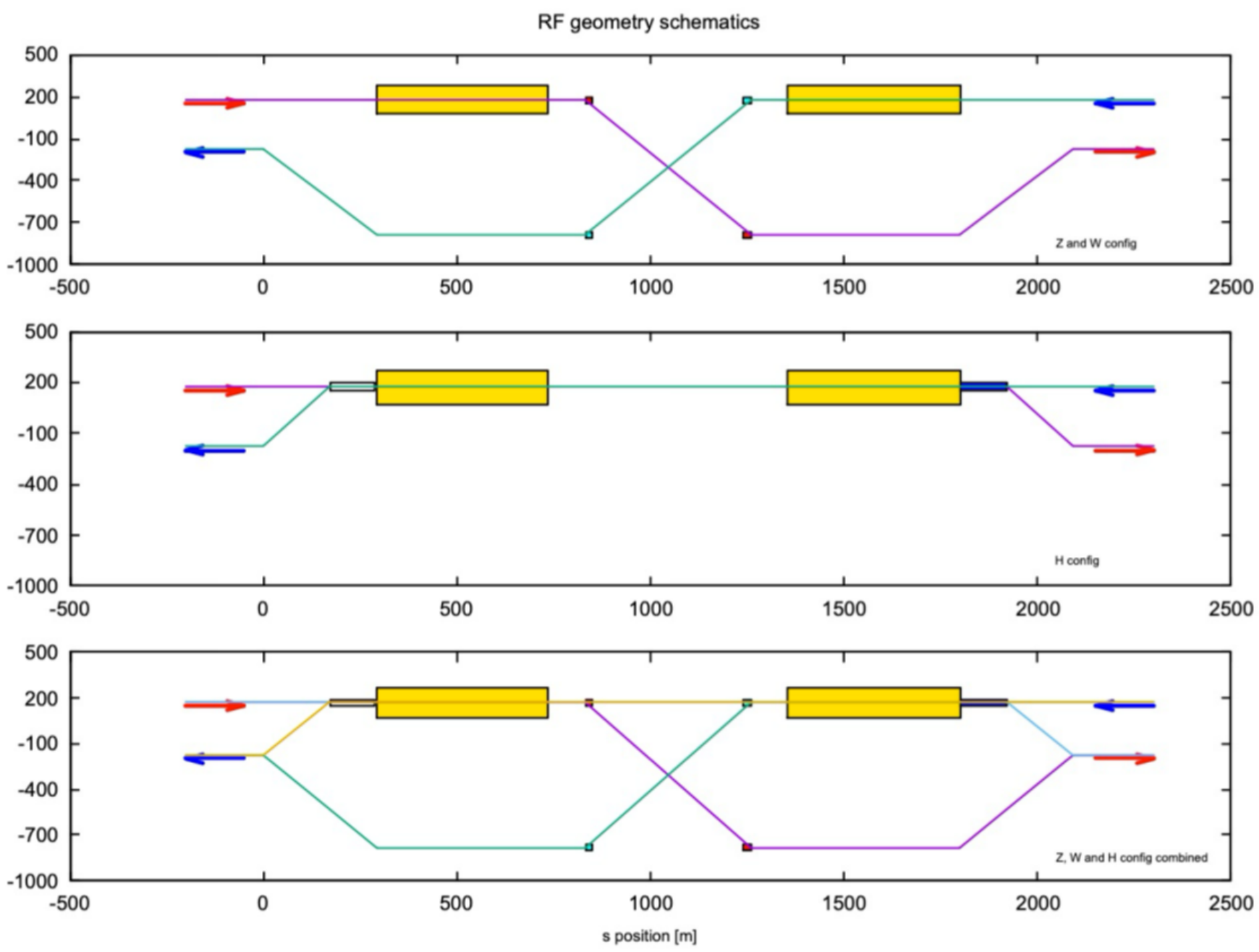

Fig. 2.2: Schematics of the beam path configurations in the RF section at FCC Point PH for the Z and WW operating points (top), the ZH operating point (centre) and the proposed combined layout for Z, WW and ZH operating points with seamless transition (bottom). The yellow rectangles represent the set of cryomodules on each side of the insertion. The smaller rectangles on the outside of cryomodules on the ZH and the combined schematics represent the combined electrostatic and magnetic separators.

### 2.2.2 Optimizing RF performance for $\mathrm{t\bar{t}}$ collisions

To complete the 2.1 GV delivered by the 400 MHz RF systems used at the ZH operating mode, additional cavities and RF sources are added during the year of shutdown to reach a total RF voltage of 11.3 GV required at the $\mathrm{t\bar{t}}$ mode.

For the collider, an additional 102 cryomodules and 204 microwave vacuum tube amplifiers at 800 MHz are installed in point PH. Each cryomodule hosting four 6-cell elliptical cavities will run at their maximum performances of 22.5 MV in order to produce 9.18 GV RF voltage. Each RF source will power two cavities at a level of 200 kW RF power each.

For the booster RF system in point PL, the same RF frequency of 800 MHz is used for all operational modes from Z to $\mathrm{t\bar{t}}$. The 28 cryomodules used at the Z-W-H modes will be completed by 84 additional cryomodules of the same type to reach a total RF voltage of 10.18 GV. The 112 cryomodules corresponding to 448 cavities will be powered individually by 10 kW solid state RF amplifiers after removing the 28 RF sources used at the previous modes.

Figure 2.3 shows the installation sequence of the RF cryomodules and high-power sources for the collider and booster.

It is important to note that once the full SRF system for $\mathrm{t\bar{t}}$ is installed, it will be impossible to run at the Z, W and H energy with the nominal beam current.

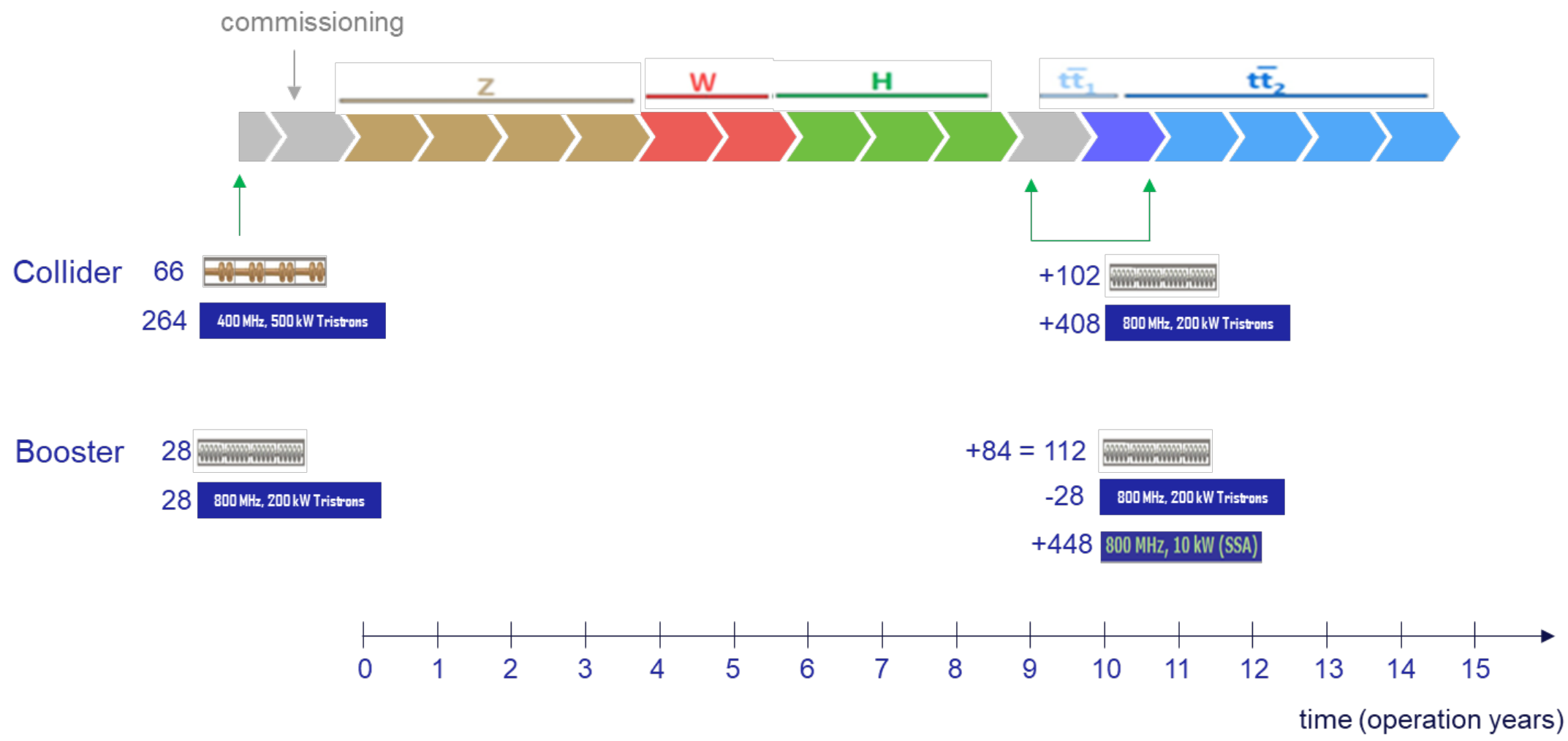

Fig. 2.3: RF system installation sequence for collider and high energy booster.

## 2.3 Operation and performance

### 2.3.1 Energy calibration

A principal goal of the FCC-ee is the ultra-precise measurement of electroweak (Z and W) observables, for which an accurately determined collision energy is key. This involves beam energy calibration every 10-15 minutes using non-colliding polarised pilot bunches, which circulate simultaneously with the main colliding bunches. The energy of these pilot bunches is measured by resonant depolarisation (RDP), where the frequency of a kicker magnet is adjusted until the pilot bunch's polarisation vanishes.

Pilot bunches are polarised in the main ring at the start of every fill using wiggler magnets, a process that takes roughly 90 minutes. The wigglers are then turned off before the injection of the main colliding bunches. During physics operation, the pilot bunches have a combined Touschek and gas scattering lifetime of less than 20 h [163]. Due to the long natural polarisation time (150 h in Z mode), it is presently unclear whether these pilot bunches will naturally achieve sufficient polarisation for RDP before expiring. The baseline simulation pessimistically expects that after 20 h the beam must be dumped to refill again.

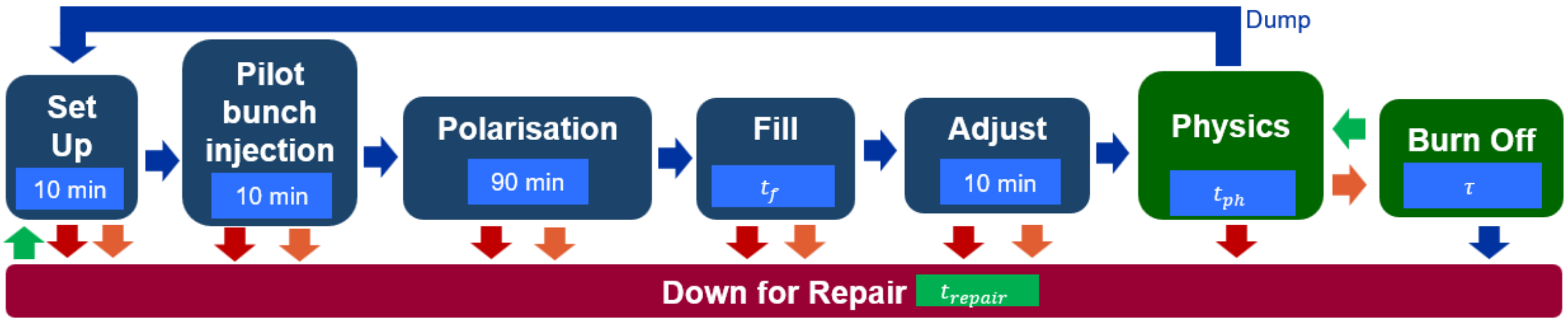

(a) Z and WW modes.

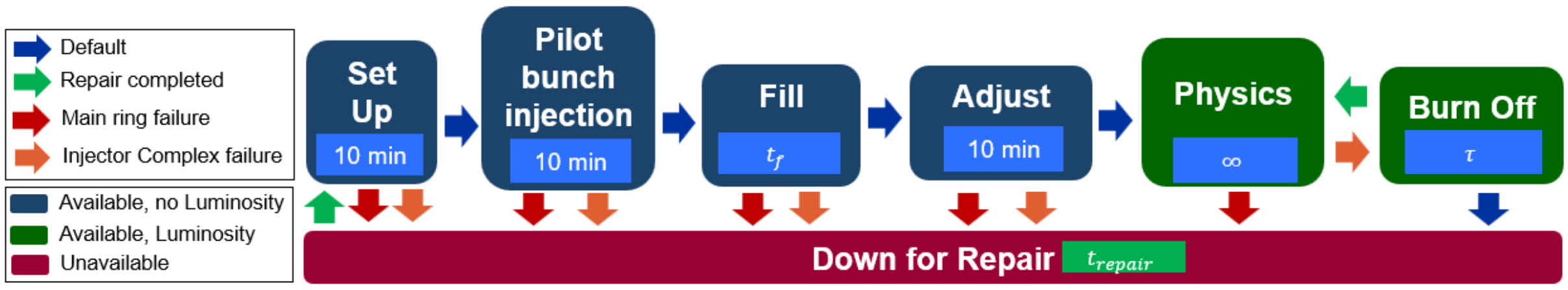

(b) ZH and $t\bar{t}$ modes

Fig. 2.4: Baseline operation cycles in the FCC-ee.

In ZH and $t\bar{t}$ modes, the energy spread makes RDP impossible but the required accuracy of the beam energy is also significantly less demanding. Energy measurement is instead achieved by observing collisions at the interaction point (IP). This removes the need for polarisation at the start of every fill. Further, with a top-up injection, physics can continue uninterrupted until a beam dump occurs due to a machine fault or schedule end. In these modes, pilot bunches are used only to verify optics before the injection of bunches with nominal intensities.

### 2.3.2 Operation cycle

Energy calibration imposes distinct operation cycles for the electroweak (Z, WW) and high energy (ZH, $t\bar{t}$) modes, shown inFig. 2.4. Phases in this cycle are as follows:

1. *Set Up:* Main magnets are cycled, RF system is set and other equipment is prepared for injection.
2. *Pilot Bunch Injection:* Pilot bunches are injected and equipment/optics are adjusted.
3. *Polarisation:* Z and WW modes only. Wigglers are turned on to begin polarising the pilot bunches, a process taking approximately 90 minutes.
4. *Fill:* Wigglers are turned off and the main colliding bunches are injected. Fill time $t_f$ in each energy mode is shown in Table 2.2. These times are indicative, and must be studied in detail according to constraints in the injector complex.
5. *Adjust:* Time to bring the beams into collision and make final adjustments to equipment before the detectors can begin to observe useful collisions.
6. *Physics:* Collisions begin at nominal luminosity. With top-up injection, flat luminosity can be maintained. In Z and W modes, an upper limit on the duration of physics $t_{ph} = 20$ h is applied corresponding to the Touschek and gas scattering lifetime of the pilot bunches.
7. *Burn Off:* If the injector complex fails, the beam can be preserved in the main colliding rings with decaying luminosity corresponding to lifetime $\tau$ (see Table 2.2). If the injector complex cannot be restored, after the time $\tau$ the beam is taken to be dumped.
8. *Down for Repair:* On equipment failure, the accelerator is stopped for repair.

Table 2.2: Operation cycle times.

| | Z | WW | ZH | $t\bar{t}$ |
|---|---|---|---|---|
| Fill time*, $t_f$ / minutes | 7.7 | 2.5 | 1.52 | 1.45 |
| Burn off lifetime, $\tau$ / minutes | 15 | 12 | 12 | 11 |
| Max. physics duration, $t_{ph}$ / hours | 20 | 20 | $\infty$ | $\infty$ |

* To be updated according to filling bunch intensity.

### 2.3.3 Luminosity

Luminosity is produced only in Physics or Burn Off phases in the operation cycle. By tracking the time $t$ spent in these operation phases, achieved integrated luminosity is calculated.

$$L_{int} = \begin{cases} N_{\mathrm{IP}} L t, & t \in \text{physics} \\ N_{\mathrm{IP}} L \tau (1 - e^{-\frac{t}{\tau}}), & t \in \text{burn off} \\ 0, & \text{otherwise} \end{cases} \tag{2.2}$$

where $N_{\mathrm{IP}} = 4$ is the number of interaction points and $L$ is the nominal instantaneous luminosity as per Table 2. In the first two years of Z operation, and first year at $t\bar{t}$, reduced luminosity is expected at 50 % and 65 %, respectively to account for machine commissioning and beam tuning.

### 2.3.4 Optics commissioning strategy

The optics commissioning strategy [21, 28, 50, 219] takes into account arc misalignment and strength errors. The various available optics and the associated tolerances are described in Section 1.3. IR optics imperfections and non-linear errors are being assessed separately and both require further developments.

#### *Commissioning sequence*

Figure 2.5 presents the general sequence of beam optics commissioning steps for the FCC-ee. Initial beam steering around the ring requires the sextupole magnets to be switched off. In this condition, storing the beam with a reasonable lifetime is not possible for the nominal optics, due to a big reduction in dynamic aperture without the sextupoles [42, 43]. Therefore, a commissioning optics was developed, in which both thequadrupole and sextupole magnets near the IP are switched off [44]. This optics features a lower natural chromaticity and lower peak beta functions across the IR. It also implies no synchrotron radiation from the final doublet (FD). Since this optics allows establishing a straight reference trajectory across the IP in the absence of focusing elements, it is called “ballistic” optics.

Right after threading the beam for the first time in the FCC-ee with the ballistic optics, and sextupoles switched off, large dispersion errors and too large a vertical emittance are observed (Table 2.3). At this point, refined optics measurements are not possible, but dispersion free steering (DFS), or simply dispersion correction using orbit correctors, is very effective at mitigating these imperfections and allowing the commissioning to continue. The resulting DA for the ballistic optics is shown in Fig. 2.6.

Turning to the nominal optics, Table 2.4 summarises the closed orbit deviations, beta beating, and spurious dispersion errors before and after the linear optics correction. The simulated dynamic aperture and momentum acceptance are presented in Section 1.3.3.

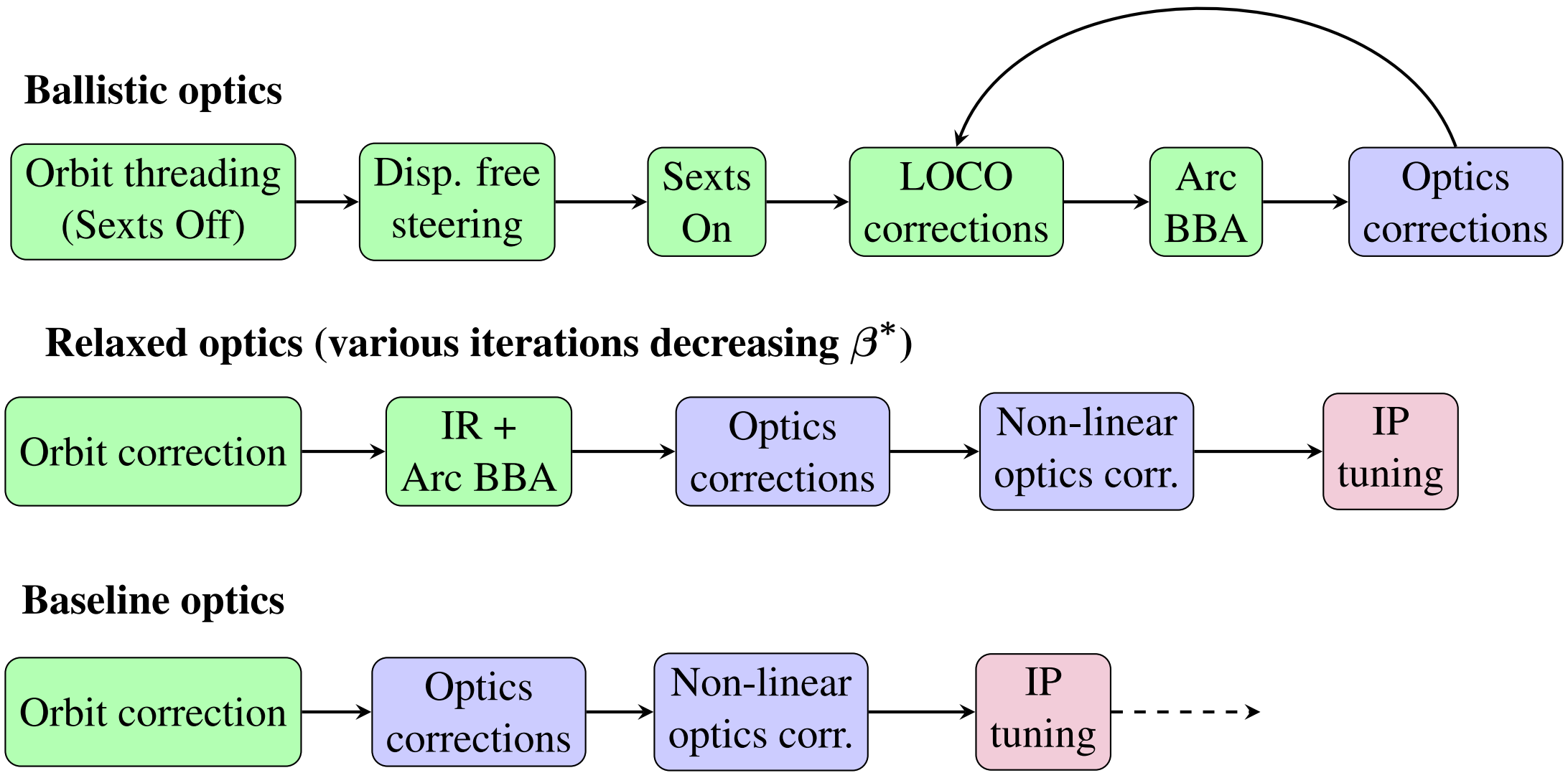

Fig. 2.5: Steps during the FCC-ee optics commissioning starting from the ballistic optics, followed by a sequence of relaxed optics [45] and, finally, the nominal collision optics.

Table 2.3: Median RMS values of several optics parameters right after the first beam threading with and without being followed by dispersion free steering (DFS) using the orbit correctors for the ballistic optics.

| Parameter | Without DFS (rms) | With DFS (rms) |
|---|---|---|
| ver. orbit (µm) | 222 | 255 |
| $\Delta D_x$ (mm) | 446 | 70 |
| $\Delta D_y$ (mm) | 416 | 39 |
| $\varepsilon_v$ (pm) | 659 | 4 |

***Beam based alignment***

In the commissioning sequence Beam Based Alignment (BBA) can be performed after sextupoles have been switched on.

Due to the sheer size of the FCC performing BBA to individual magnets would be a very time-consuming procedure and, hence, various promising parallel approaches are explored, aiming to achieve approximately 10 to 20 µm effective alignment after BBA.

Using Parallel Quadrupole Modulation System (PQMS), a technique which has already successfully been tested at SPEAR, is applied to the FCC-ee. Modulating 10 quadrupoles in parallel with a $\Delta K/K$ of 2 %, distributed equally over one arc, and using a calibrated lattice with 1 µm BPM resolution, an accuracy below 20 um for vertical and horizontal arc quadrupole BBA is achieved [220].

A different BBA technique is also studied for a relaxed $\beta^*$-optics with $\beta^*$ of 7 mm based on modulating 8 arc quadrupoles in parallel with a $\Delta K/K$ of 1 %, again with a BPM resolution of 1 µm, inducing orbit shifts with vertical orbit correctors nested to the quadrupoles. This yields a BBA rms accuracy achieving the target value. Other BBA studies assuming individual horizontal and vertical orbit correctors, located next to, respectively, focusing and defocusing quadruples, are also performed for the same optics modulating 20 quadrupoles in parallel. All studied seeds achieve an accuracy below 20 µm [221]. In addition, parameter scans confirm this performance even when processing a larger

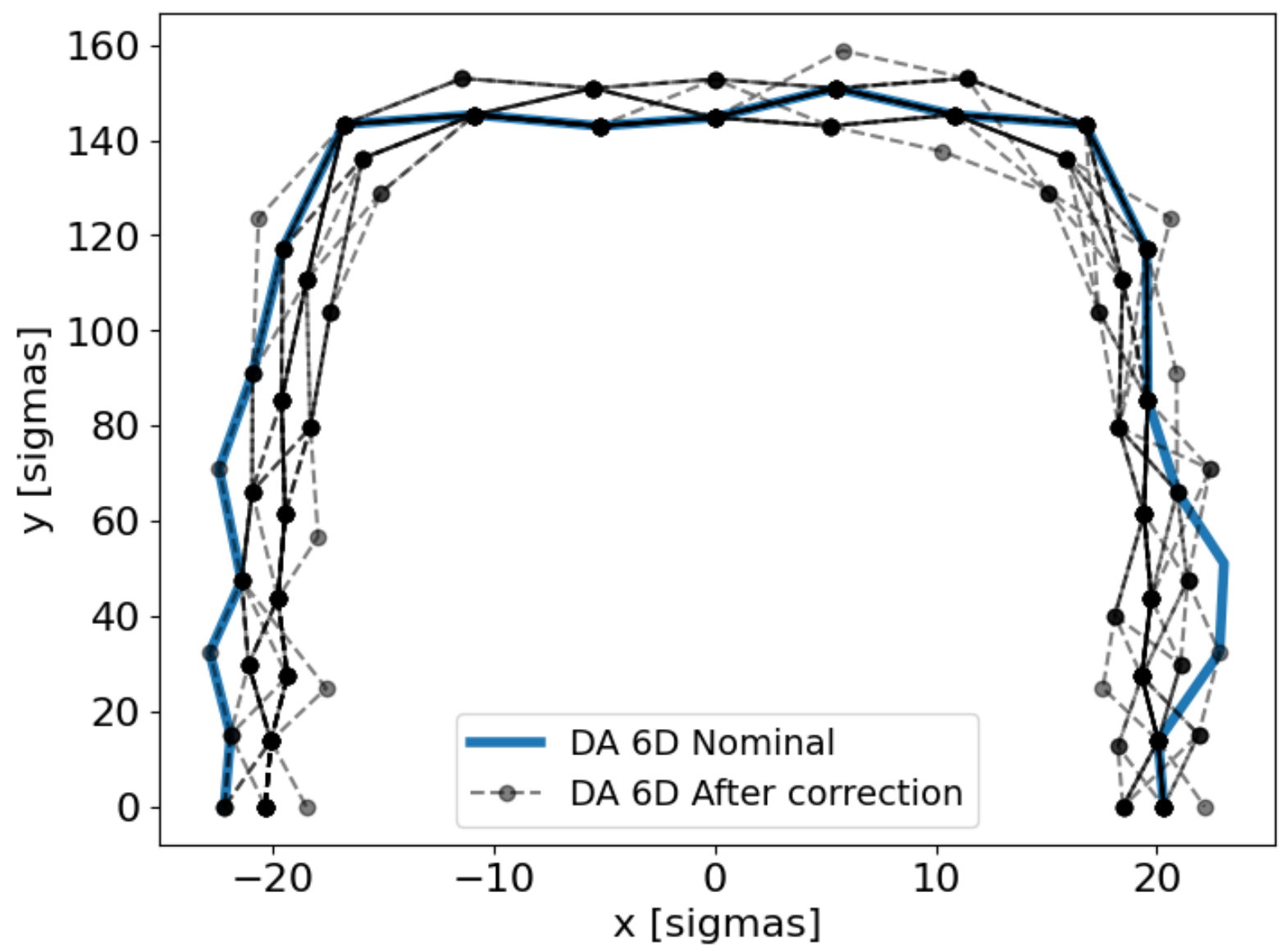

Fig. 2.6: DA after optics correction with the ballistic optics and having included IR alignment errors of 50 µm in addition to the arc errors in Section 1.3.1.

Table 2.4: Median RMS values of several optics parameters before (after sextupole ramping) and after linear optics correction (nominal lattice). $\Delta\psi$ stands for phase advance deviations between nearby BPMs.

| **Parameter** | **Before correction (rms)** | **After correction (rms)** |
|---|---|---|
| hor. orbit (µm) | 120.25 | 120.46 |
| ver. orbit (µm) | 217.53 | 217.56 |
| $\Delta\beta_x/\beta_x$ (%) | 7.41 | 0.29 |
| $\Delta\beta_y/\beta_y$ (%) | 15.79 | 2.81 |
| $\Delta D_x$ (mm) | 57.79 | 0.28 |
| $\Delta D_y$ (mm) | 62.24 | 2.80 |
| $\varepsilon_h$ (nm) | 0.72 | 0.71 |
| $\varepsilon_v$ (pm) | 26.01 | 0.57 |
| hor. $\Delta\psi$ [$2\pi$] | $1.13 \times 10^{-2}$ | $2.91 \times 10^{-4}$ |
| ver. $\Delta\psi$ [$2\pi$] | $1.93 \times 10^{-2}$ | $2.29 \times 10^{-3}$ |
| Re F1001 | $4.93 \times 10^{-2}$ | $1.66 \times 10^{-4}$ |
| Im F1001 | $4.43 \times 10^{-2}$ | $5.18 \times 10^{-5}$ |
| Re F1010 | $3.72 \times 10^{-2}$ | $1.35 \times 10^{-4}$ |
| Im F1010 | $3.68 \times 10^{-2}$ | $1.31 \times 10^{-4}$ |

number of magnets in parallel.

In addition to the quadrupoles, beam-based alignment (BBA) can also be performed for the sextupoles. In simulation studies, the center of the sextupole magnets is estimated using a response matrix method. Modulating six arc sextupoles in parallel yields a precision on the order of 20 to 50 µm [222]. Each arc sextupole shares a common girder with the adjacent quadrupole magnet, so that BBA for both may possibly yield redundant information.

***IP tuning***

Simulations are also performed including IR alignment errors for the FCC-ee GHC lattice using `pyAT`. In this IR study, the arc alignment tolerances are taken to be 100 µm and 100 µrad, which is slightly better than the numbers presented in Section 1.3.1. Final-focus doublet (FD) quadrupoles must meet strict alignment tolerances. The simulations in this section consider transverse shifts and rotation errors to 10 µm and 10 µrad, respectively. IR sextupoles, including crab-sextupoles required to correct the vertical chromaticity at the IP ($SY^*$), exhibit intermediate sensitivity. Their errors are set to 30 µm and 30 µrad. The few strong non-linear magnets, combined with the alignment errors, make proper tuning difficult for most seeds, resulting in a 75% success rate out of 100 seeds. The median vertical emittance for the successful seeds is 1.8 pm.

The tuning and correction of optics in the IP region of FCC-ee are essential for reaching the desired luminosity levels. Dedicated IP tuning knobs such as $\beta^*_{x,y}$, $W^*_{x,y}$ and $D^*_y$ are employed to correct lattice errors across multiple IPs, ensuring the restoration of the intended optics design and facilitating DA analysis on the fully corrected lattice. The tuning knobs have no effect on the dynamic aperture as shown in Fig. 2.7. The resulting dynamic aperture (DA) is comparable to that obtained in the previous study, which considered only arc errors (see Section 1.3). Relaxing alignment tolerances leads to a reduction in the DA and a decrease in the number of successful seeds in the simulations. To enhance both DA and momentum acceptance (MA), additional correction algorithms beyond linear corrections will be required.

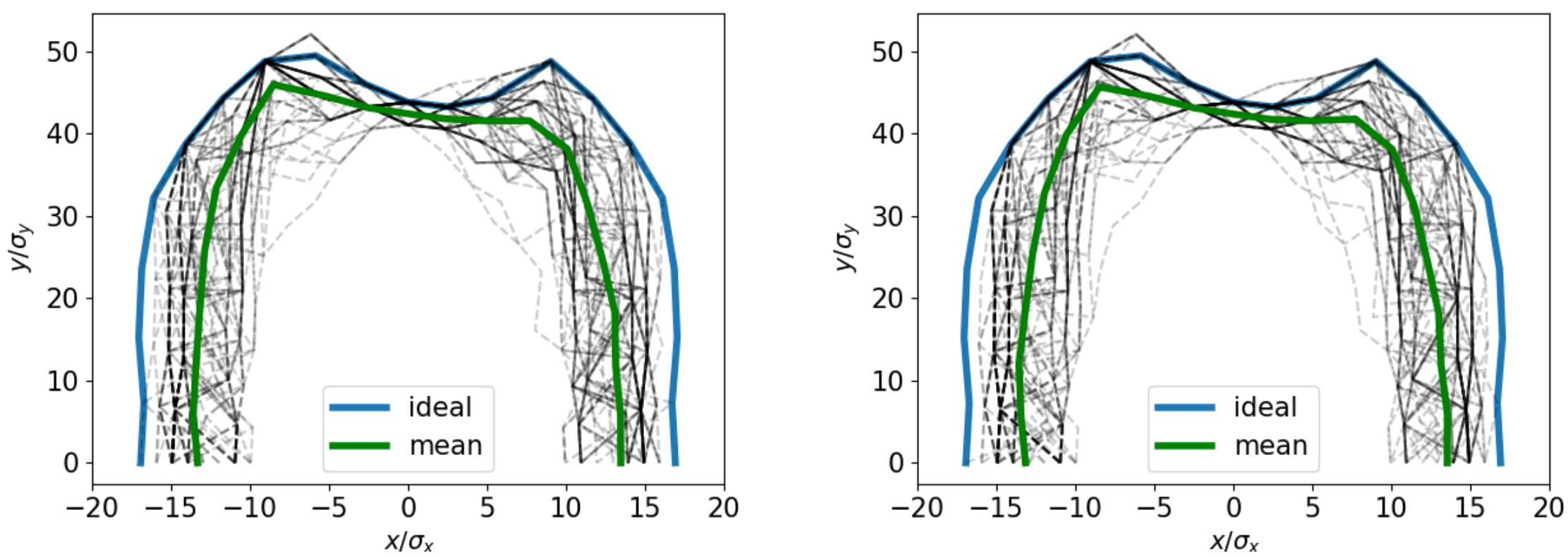

Fig. 2.7: Dynamic aperture after 512 turns for 75% of successful seeds with (right) and without (left) IP tuning knobs.

### 2.3.5 Luminosity optimisation and interaction-point tuning

FCC-ee luminosity optimisation relies on measuring realistic signals from Bhabha scattering, beamstrahlung and vertex detector hits. Initial assessments of these signals examine the variations in luminosity, beamstrahlung power and vertex detector hits in response to waist shifts, vertical dispersion and transverse coupling at the collision point. Waist shifts ($y^* \rightarrow y^* + l y'^*$, with $l$ referring to the waist-shift) and vertical-dispersion ($y^* \rightarrow y^* + D\delta_i$, where $D$ is the dispersion and $\delta_i$ is the relative energy deviation) are IP sport size aberrations, and should be corrected over the time of few minutes.

Figures 2.8 and 2.9 show the scans of the mentioned signals varying with waist-shifts produced in GUINEA-PIG for the Z and $t\bar{t}$ working points. At the $t\bar{t}$ working point, the luminosity drops to about 85% for the maximum waist shift of $\pm$ 1 mm and to 40% for a vertical dispersion of $\pm$ 0.1 mm. At the Z working point luminosity is slightly more sensitive to waist shifts and vertical dispersion. These studies do not take into account the degradation of the emittances due to the beam-beam interactions in presence of the IP aberrations which could be the dominant effect, as few micrometres of $D_y$ can increase the vertical beam size by 5% [46].

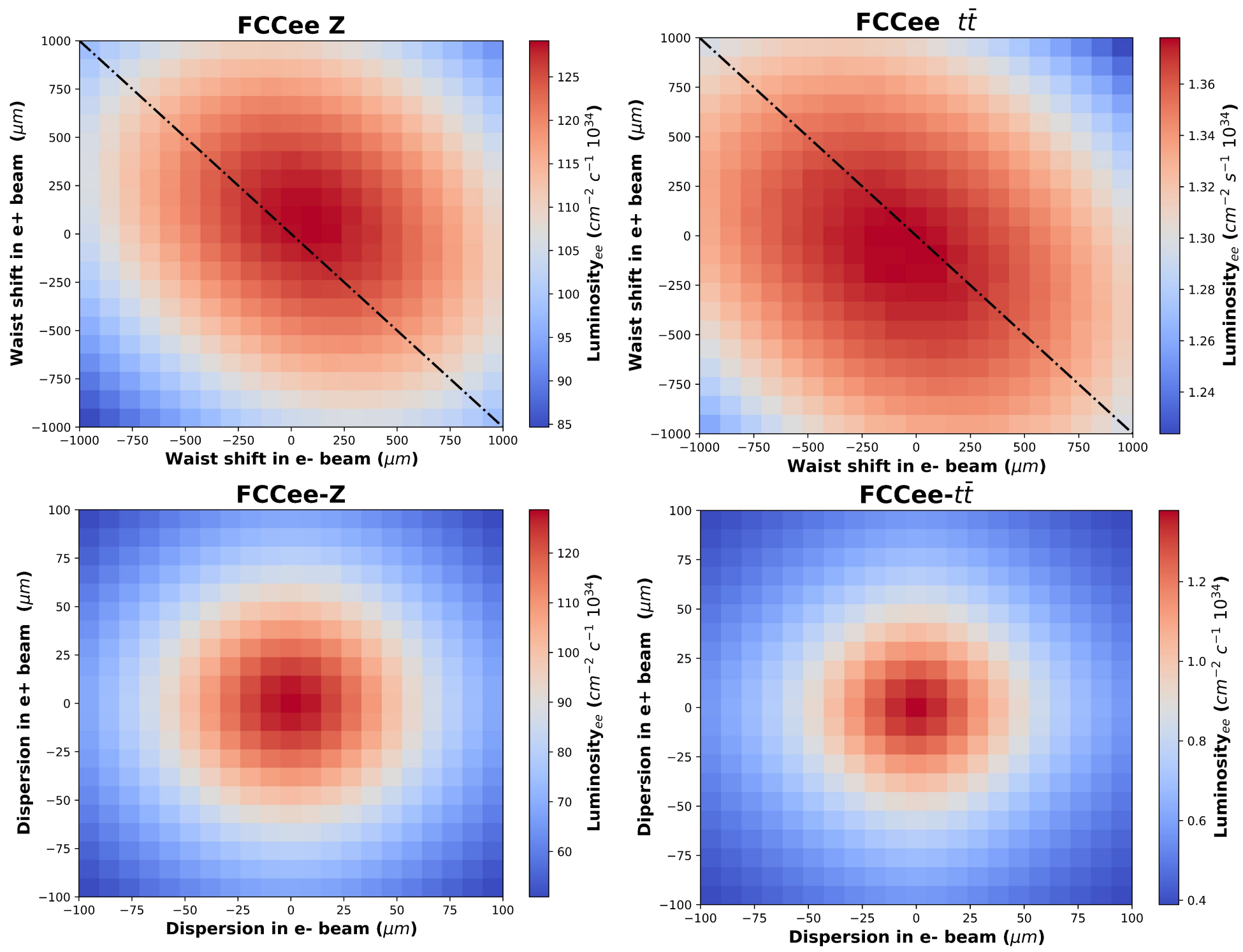

Fig. 2.8: The variation of luminosity with a residual waist shift (top) and residual dispersion (bottom) in the electron (horizontal axis) and positron beams (vertical axis) for the Z pole (left) and $t\bar{t}$ running, from simulations with the code GUINEA-PIG.

The beamstrahlung power emissions increase with the residual waist shift and vertical dispersion. The impact on total power due to both aberrations is within a similar range. However, residual dispersion in the $e^-(e^+)$ beam has a stronger effect, increasing the beamstrahlung from the $e^-(e^+)$ beam while decreasing emissions from the $e^+(e^-)$ beam, compared to waist shifts.

Furthermore, these studies aim to extract IP-aberration-related signals and integrate them into a machine-learning-based approach for luminosity tuning and optimisation. A Gaussian process-based model could be trained to predict the residual waist-shift, dispersion and coupling at the IP, based on the simulated or observed luminosity, beamstrahlung power and hits of secondaries in the vertex detector.

### 2.3.6 Operational feedbacks

***Global orbit feedback***

The beam orbit around the ring must be stabilised by a global orbit feedback system, based on averaged beam-position monitor (BPM) readings and orbit corrector magnets. The orbit feedback systems at the LHC and at the Swiss Light Source (SLC) serve as examples.

The LHC global orbit and energy feedback is of similar size and complexity as the future orbit feedback for FCC-ee. The LHC system bandwidth of only about 1 Hz is limited by the maximum excitation change permitted in the superconducting orbit-corrector magnets [223]. The rms orbit stability in the LHC arcs during long physics fills is of order 10 μm [224].

At the Swiss Light Source (SLS), a global fast orbit feedback (FOFB) system based on the digital beam position monitor (DBPM) system has been in use during user operation since 2003. The singular value decomposition (SVD)-based correction scheme operates at a sampling rate of 4 kHz, utilizing position data from all 72 DBPM stations and applying corrections through 72 horizontal and 72 vertical corrector magnets [225].

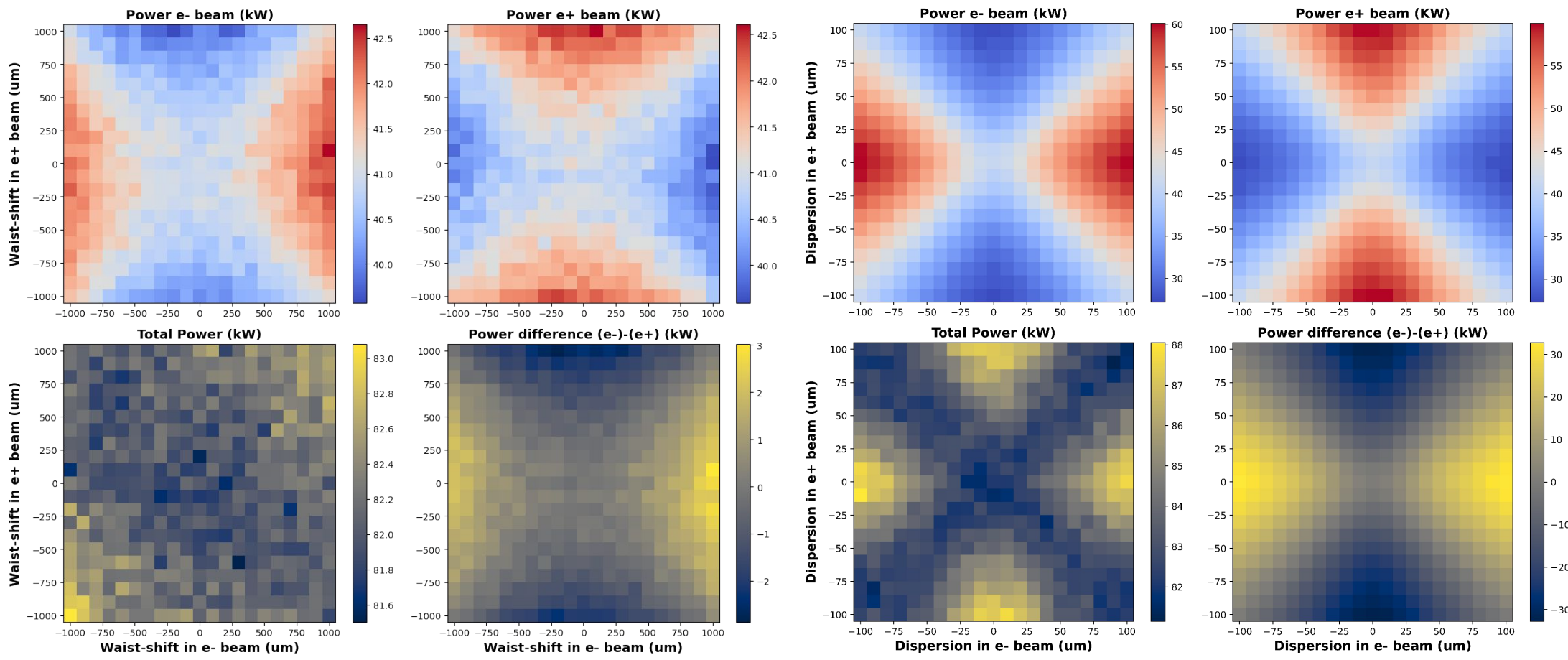

Fig. 2.9: The variation of beamstrahlung power with a residual waist shift (left) and residual dispersion (right) in the electron (horizontal axis) and positron beams (vertical axis) for $t\bar{t}$ running, from simulations with the code GUINEA-PIG.

This fast orbit feedback effectively mitigates orbit distortions, which are primarily induced by ground and girder vibrations, as well as a 3 Hz crosstalk from booster cycles. Additionally, it enables rapid and independent gap adjustments for the insertion devices, ensuring complete transparency for SLS users. With top-up as the standard operation mode, the system has achieved global beam stability at the µm level over timescales ranging from milliseconds to days.

For the FCC-ee, it is assumed that an orbit feedback efficiently suppresses all beam oscillations below a critical frequency of about 1 Hz [226]. This assumption appears fairly conservative, if compared with the orbit feedback performance at modern light sources like the SLS.

#### *Interaction point feedback*

Interaction point (IP) feedback systems are essential for maintaining luminosity in the presence of machine perturbations, such as magnet vibrations, ground motion, and fluctuations or drifts in magnet strengths. At each IP, beam position monitors detect offsets between the two colliding beams, and corrections are computed and applied by correctors in the interaction region. These corrections create a closed orbit bump, forming a dedicated local correction system.

For small IP offsets, the deflection is well described by the linear centre-of-mass beam-beam kick formula

$$\left\langle \Delta x'^{*} \right\rangle = \pm \frac{2\pi}{\beta_x^*} \xi_x \, \langle \Delta x^* \rangle \qquad \text{and} \qquad \left\langle \Delta y'^{*} \right\rangle = \pm \frac{2\pi}{\beta_y^*} \xi_y \, \langle \Delta y^* \rangle \tag{2.3}$$

where $\xi_{x,y}$ is the beam beam parameter in the horizontal or vertical plane, respectively, $\Delta x^*$ (or $\Delta y^*$) denotes the horizontal (vertical) offset from the centre of the opposing beam, and the angular brackets $< ... >$ represent an average over the bunch distribution. The nano-beam scheme and large horizontal crossing angle of FCC-ee results in sensitivity to errors far greater in the vertical than in the horizontal plane, due to the much larger vertical beam-beam parameter (see Table 1.2). The deflections can be measured by comparing orbits for colliding and non-colliding (pilot) bunches, taking into account any possible systematic errors due to different bunch intensity and bunch length.

*Performance requirements*

Beam-beam collisions with IP offsets reduce luminosity and lifetime. Previous studies have determined the resulting IP offset limit as $0.05\sigma_y$ [227]. Even tighter limits may be imposed by energy calibration and polarisation requirements (Section 1.7). The alignment of the IP to the detector is limited by the resolution of the luminosity calorimeter. Frequency response requirements are dependent on the performance of the global orbit feedback.

*Observables*

IP offset can be indirectly measured from a variety of signals:

From the interaction region beam position monitors and a knowledge of the transfer matrices to and from the IP, the IP offset can be calculated. The proposed IR BPMs are elliptical button BPMs situated on the common elliptical IR beampipe, described further in Section 3.9.1.

Luminosity is detected at the luminosity calorimeter and additional fast luminosity monitors 3.9.5. As a scalar signal, only the magnitude of the offset is identified.

Beamstrahlung radiation power and centroid position both depend upon IP offset, in a scalar and vector way respectively. Monitoring beamstrahlung is challenging due to the high radiation power. The proposed beamstrahlung monitors discussed in 3.9.5.

Initial studies are set out in [228] and Fig. 2.10 shows studies results at the Z working point for the GHC lattice.

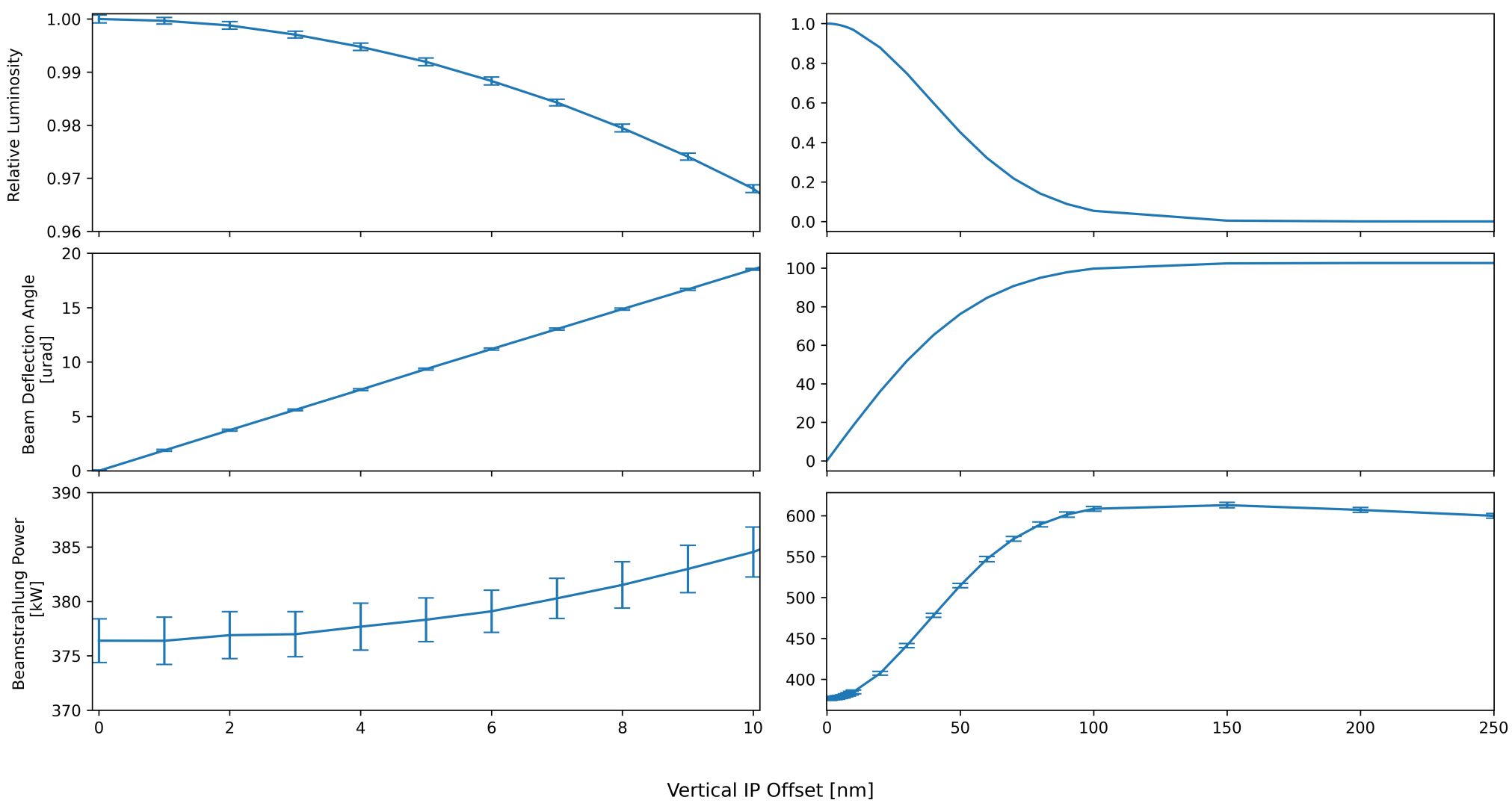

Fig. 2.10: The variation of relative luminosity, outgoing deflection angle and beamstrahlung radiation power with offset in the vertical plane. Simulations were performed with the Particle in cell solver GUINEA-PIG. Results for the Z working point of the 'GHC 24.3' lattice. The impact of the detector solenoid and its compensation is not currently taken into account.

*Beam-beam deflection feedback*

Due to the high vertical beam-beam parameter, vertical IP offsets result in strong beam deflections, changing the outgoing angle. This change to the outgoing angle results in detectable differences at the BPMs. Such a beam-beam deflection feedback system is most promising to address offsets in the vertical plane and has been successfully operated at SuperKEKB [229], KEKB [230], and other machines.

*Dither feedback*

The low horizontal beam-beam parameter implies that beam-beam inducted deflections will not be detectable at the IR BPMs. An alternative approach is 'dithering': driving one beam at a known frequency and minimising the IP offset by finding the phase of optimal luminosity. This approach relies solely on maximisation, and therefore, the absolute offset values are never required to be calculated. This type of dither feedback has been successfully operated at SuperKEKB (and others) [231].

### 2.3.7 Filling patterns

The number of bunch slots in FCC-ee for 25 ns bunch spacing is $h_{25\mathrm{ns}} = 12120 = 2^3 \cdot 3 \cdot 5 \cdot 101$. In these sections, so-called filling patterns indicating which slots are filled with bunches are described. As the required number of bunches depends on the operational mode, a filling pattern for each of the energies has been devised. Note that the harmonic number corresponding to the RF frequency is higher such that,many more filling patterns and shorter minimum spacing of, e.g., 5 ns (slightly reducing electron cloud threshold) are in principle possible. Such schemes are not proposed.

The main requirements and assumptions to design filling patterns are:

- The periodicity of the filling pattern for high-intensity bunches must contain a factor of two (filling the machine in half) to ensure that all physics bunches collide in all four IPs with bunches from the counter-rotating beam.
- Presence of gaps for the injection kicker rise and fall times and beam dump kicker rise time. This requirement is satisfied with a regular pattern consisting of bunch trains and gaps between them. The minimum length of the gaps has been reduced to 600 ns in order to mitigate beam loading transients being a potential limitation for the Z mode.
- Batches coming from the injectors and injected into the Booster contain four bunches spaced by 25 ns for the Z and WW mode and two bunches with the same spacing for the higher energies. For the Z mode, operation of the injectors with the maximum number of four bunches with the maximum repetition rate of 100/s is mandatory.
- Filling patterns for the H and $\mathrm{t\bar{t}}$ modes must contain appropriate long gaps such that no encounters between counter-rotating bunches occur in the common RF section.
- The filling patterns for operation at Z and WW energies must contain positions for low intensity bunches required for energy calibration. They are placed inside the gaps between trains of high intensity bunches. The minimum spacing between a polarised low intensity bunch and other bunches is 100 ns, as required, for the rise and fall time of the fast deflector used to excite a depolarising resonance for energy calibration. Calibration bunches of the two counter-rotating beams must not collide.

***Filling pattern for Z mode***

For Z mode operation, each of the two counter-rotating beams must contain $N_B = 11200 = 2^6 \cdot 5^2 \cdot 7$ colliding high-intensity bunches. The number of bunch trains must be a product of common prime factors of $N_B$ and the number of bunch slots $h_{25\mathrm{ns}}$. Factor four of the former is taken to allow the booster to be filled with injector trains consisting of four bunches. The number of trains chosen is $N_T = 2^3 \cdot 5 = 40$, each containing $2^3 \cdot 5 \cdot 7 = 280$ high intensity bunches, as sketched in Fig. 2.11. There are no gaps between trains from the injectors in order to maximise the total number of injected bunches. There are $h_{25\mathrm{ns}}/N_T - 280 = 23$ empty positions between trains corresponding to the minimum spacing of $24 \cdot 25\,\mathrm{ns} = 600\,\mathrm{ns}$ between trains required for kicker gaps.

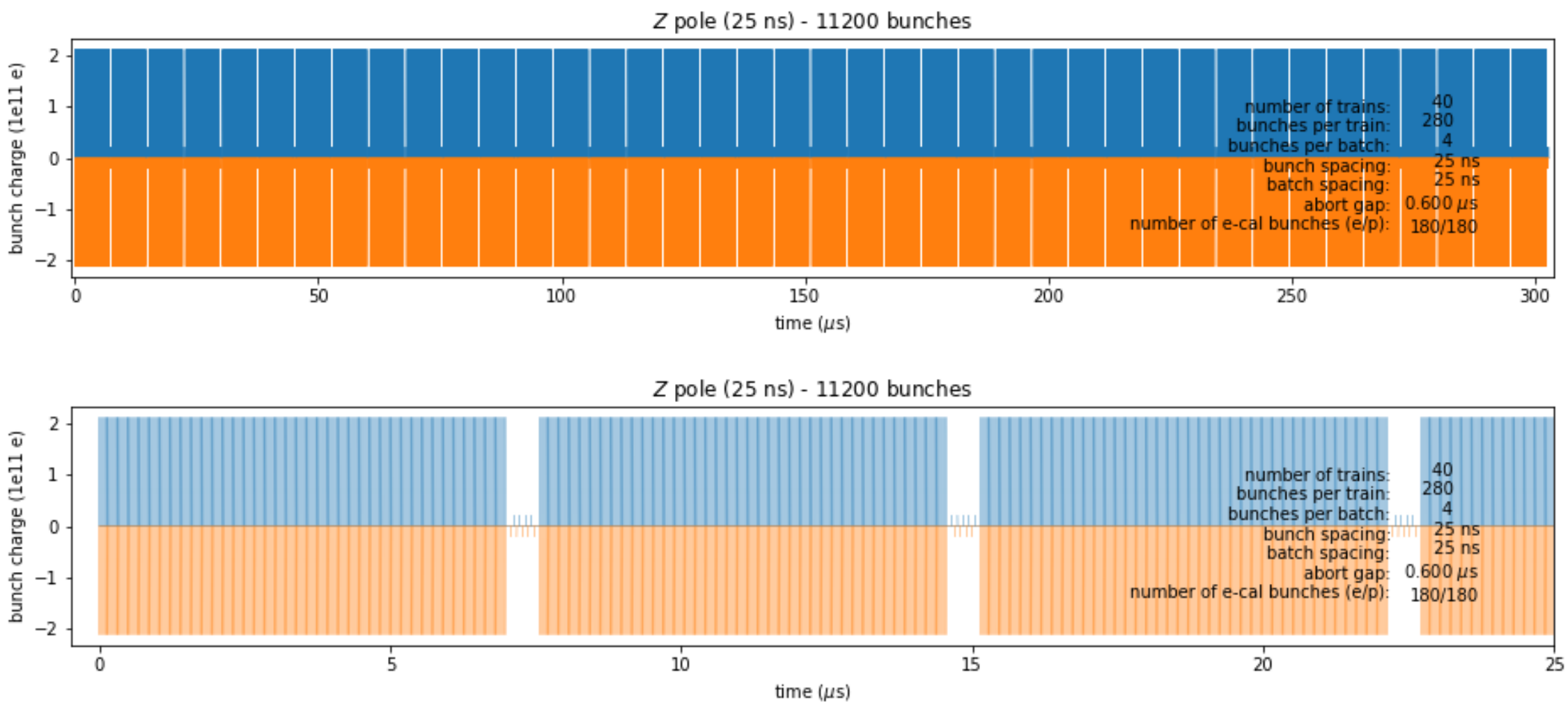

Fig. 2.11: Filling pattern for Z-mode (top) and a zoom into the first few trains (bottom), for the e$^+$ beam shown in blue and the e$^-$ beam in orange.

A scheme to fill the gaps between high-intensity physics bunches with low-intensity energy calibration bunches is sketched in Fig. 2.12 and refers to one of the four IPs. A maximum of five bunches (in the example for the $e^+$ beam) and four bunches (in the example for the $e^-$ beam) can be placed for the two beams such that the requirements are fulfilled. In case all gaps can be filled with calibration bunches and exchanging positions of e$^+$ and e$^-$ bunches for half of the gaps, 180 calibration bunches can be injected per beam. Note that strict gap-to-gap alternation of calibration bunch patterns leads to fillings avoiding collisions between calibration bunches in all four IPs. In case one gap has to be kept free of low-intensity calibration bunches for the rise and fall times of injection and beam dump kickers, 175 calibration bunches of one type and 174 of the other types can be injected. However, the additional beam loading transients may necessitate keeping additional gaps without beam (e.g. four equidistant gaps).

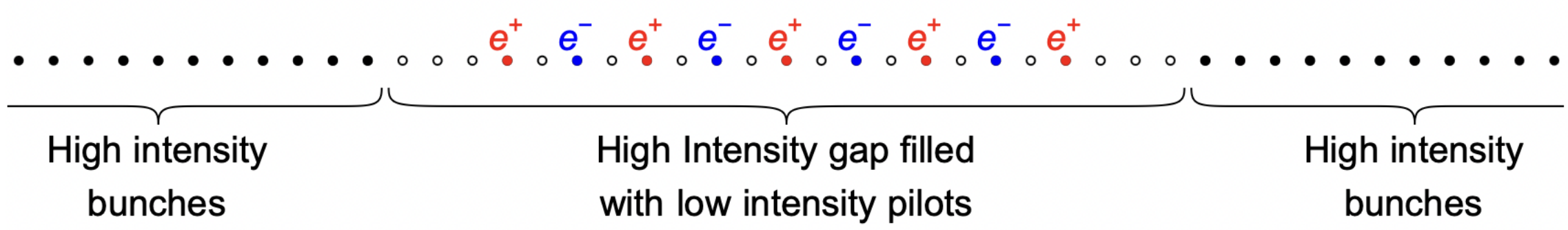

Fig. 2.12: Gap between high-intensity physics bunch trains filled with low-intensity energy calibration bunches for the Z-mode.

***Filling pattern for WW mode***

The filling pattern described and sketched in Fig. 2.13 is one out of many possible ones. The number of bunch trains is a product of common prime factors of the $N_B$ and the number of bunch slots $h_{25\mathrm{ns}}$.

Choosing 8 trains with 232 high-intensity bunches each, with 16 empty positions between injector batches with a tain, leaves 371 empty positions between trains as sketched in Fig. 2.13.

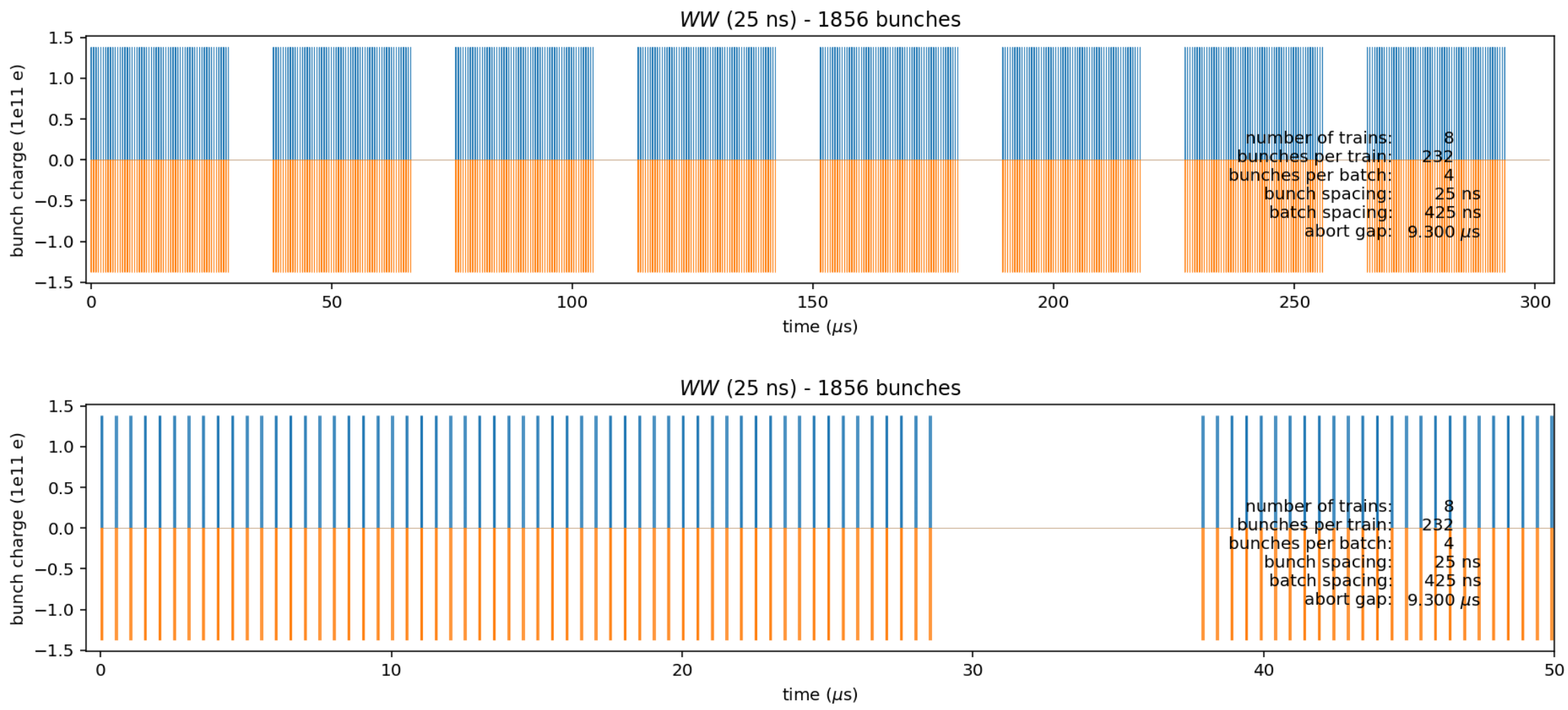

Fig. 2.13: Filling pattern for W-mode (top) and a zoom into the first few trains (bottom), for the $e^+$ beam shown in blue and the $e^-$ beam in orange.

A scheme to fill the gaps containing 371 available positions between high intensity physics bunches with low intensity energy calibration bunches is presented in Fig. 2.14, referring to one of the IPs. A maximum of 92 bunches (in the example for the $e^+$ beam) and 91 bunches (in the example for the $e^-$ beam) can be placed for the two counter-rotating beams. Filling two opposite gaps between trains allows placing of 182 and 184 bunches for the two beams. If needed, additional gaps can be filled.

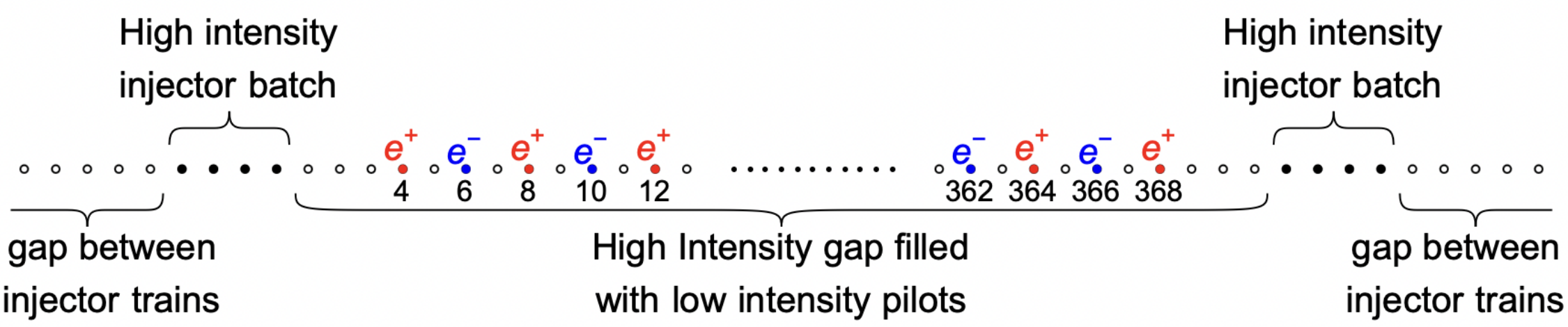

Fig. 2.14: Gap between intensity physics bunch trains filled with low intensity energy calibration bunches for Z-mode.

#### *Filling pattern for ZH mode*

For ZH mode, an additional requirement with respect to the lower energy modes to be taken into account is that the RF section is common to the two counter-rotating beams and that bunch crossings in this section must be avoided. The number of bunches per beam in the collider $N_B = 300 = 2^2 \cdot 3 \cdot 5^2$ is low and allows many different filling schemes. For the one proposed here, the number of bunches per injector cycle is reduced to two, and the injector repetition rate is reduced to 50 Hz. Filling the collider with two identical bunch trains opposite to each other ensures that all bunches collide at all four IPs. If the gaps between trains are sufficiently long, no bunch crossings take place in the RF section. The number of bunches per train is 150 and corresponds to 75 bunch pairs from the injector complex. Along a bunch train, injector bunch pairs are separated by 19 empty positions. This leaves long gaps with 3764 empty positions between trains and corresponding to a distance of $3765 \cdot 25\,\mathrm{ns} = 91.125\,\mu\mathrm{s}$ between trains. The resulting filling pattern is sketched in Fig. 2.15.

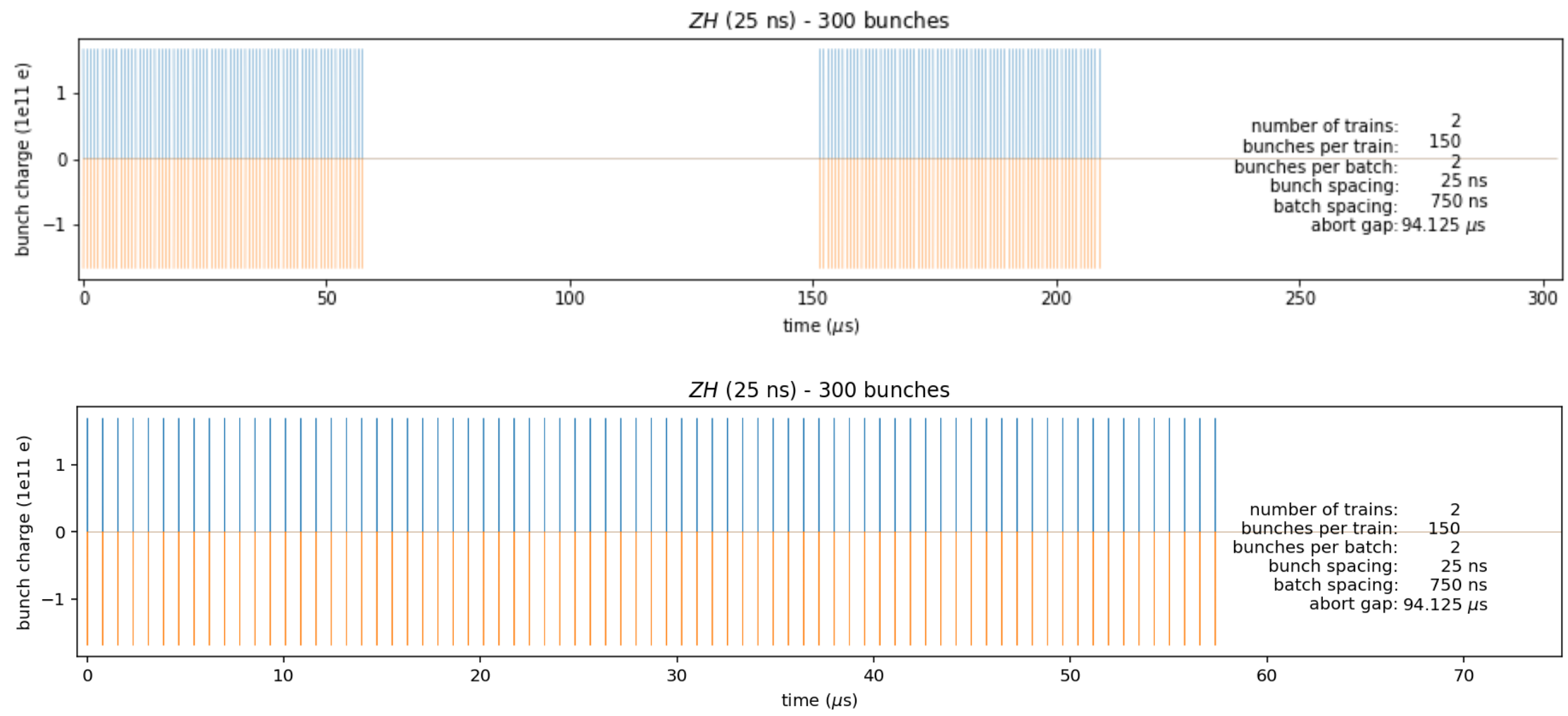

Fig. 2.15: Filling pattern for ZH mode (top) and a close-up view of the first train (bottom), with the $e^+$ beam shown in blue and the $e^-$ beam in orange.

#### *Filling pattern for* $\mathrm{t\bar{t}}$ *mode*

Since the definition of the $\mathrm{t\bar{t}}$ mode filling pattern described here, optimisations led to slight change of the number of bunches from 64 to 60 given in Table 2. The filling pattern will be adjusted accordingly in a future iteration. Many possible filling pattern exist for the $\mathrm{t\bar{t}}$ mode with a low number of bunches $N_B = 64 = 2^8$. The pattern proposed and sketched in Fig. 2.16 requires bunch pairs from the injectors assumed to be accumulated in the Booster with a repetition rate of 50 Hz. Two identical trains, consisting each of 32 bunches or 16 injector bunch pairs are foreseen per beam to ensure that no bunch crossings occur in the common RF section. Assuming 151 empty positions between bunch pairs from the injector in the same train gives the pattern shown in Fig. 2.16. The gaps between the trains comprise 3763 empty positions and have a length of $(3763 + 1) \cdot 25\,\mathrm{ns} = 94.1\,\mu\mathrm{s}$.

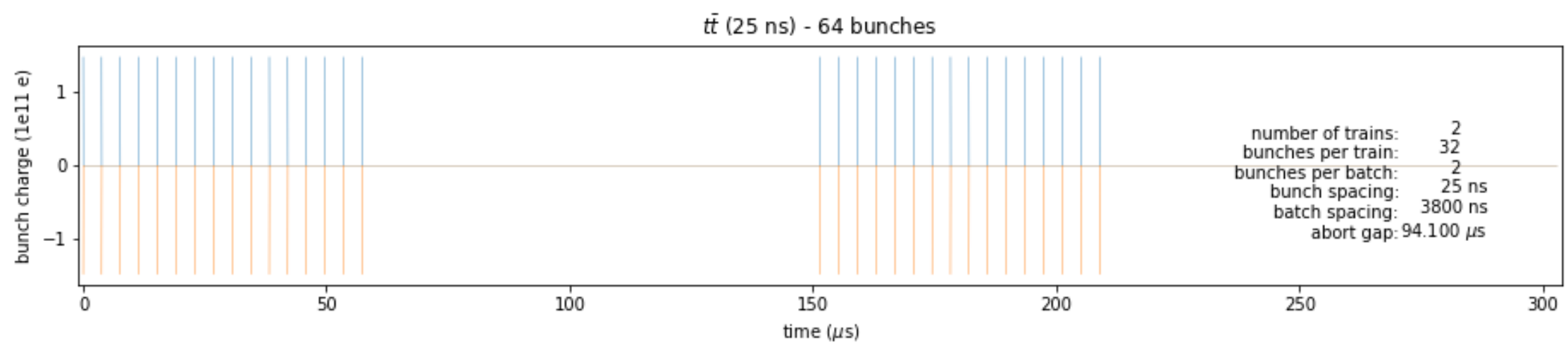

Fig. 2.16: Filling pattern for the $\mathrm{t\bar{t}}$ mode.

### 2.3.8 Bootstrapping injection scheme

This section describes the procedures to ramp up the intensity of colliding high intensity bunches from an empty machine to a steady state situation with periodic top-up injections in order to keep the intensities stable. For the two lower energy modes Z and WW, it is assumed that the low intensity bunches are already injected and have been circulated over a sufficiently long period to generate polarisation levels suitable for energy calibration. The asymmetric wigglers used to speed up the polarisation build-up for low intensity calibration bunches are switched off before starting the intensity build-up of bunches.

Constraints to be fulfilled by the intensity ramp-up procedure are

- The intensity of bunches from the counter-rotating beams interacting via beam-beam encounters have to be ramped up together. Large intensity imbalances between bunches interacting via the beam-beam effect result in the flip-flop phenomenon with imbalances of other parameters: small emittances and bunch length for the higher intensity bunches lead to large emittances and bunch length of the lower intensity bunch. The resulting lower life-time of the lower intensity bunch further enhances the intensity imbalance.
- Limitations of the intensity of injected bunches: to avoid excessive requirements to the injector complex, the maximum intensity increase of bunches circulating in the collider is limited to $\Delta n_{i,max} = 2.14 \cdot 10^{10}$ per injection per bunch, which is one tenth of the nominal bunch intensity in the collider ring for the Z mode (all other modes have lower bunch intensities). Maximum bunch intensities in the booster must be somewhat higher to accommodate for losses in the booster and at collider injection.
- Limitation of the total beam intensity circulating in the booster and being transferred to the collider to mitigate machine protection issues and effects related to the total intensity in the booster.
- For operation at the lower energy modes and, in particular, for the Z mode, electron cloud build-up has to be limited. Electron cloud build-up is more severe at intermediate intensities than at low and close to nominal intensities. Thus, situations with many or even all bunches at intermediate intensity during the ramp-up procedure must be avoided. A few bunches, say 10% of the total for Z operation, with intermediate intensity and not concentrated at a particular position of the ring is expected to mitigate electron cloud build-up.

***Intensity ramp up for the Z mode***

The total number of colliding bunches is divided into a sufficient number of groups, say 10 for the proposal described below.

The intensity of the 10 groups is not ramped up simultaneously to avoid having multiple bunches with intermediate intensities in the collider ring during the ramp-up process. The proposed procedure is one of several possible schemes and is best illustrated using Fig. 2.17, which displays the bunch pattern for two out of 40 identical trains, and Fig. 2.18, which depicts the evolution of beam intensities for each family along with the luminosity.

The topmost image in Fig. 2.17 shows the intensity of circulating bunches in the collider ring after the first injection, corresponding to the filling pattern of the booster for all high-intensity physics bunches. The evolution of bunch intensities is plotted in the upper image of Fig. 2.18. At the start of the process, the intensity of bunch family 1 is ramped up for both beams. Once its intensity exceeds a predefined threshold, the ramp-up for family 4 begins. Figure 2.17b illustrates the state after six injections for family 1 and three injections for family 4.

Each time all circulating bunches surpass the threshold intensity, the ramp-up of an additional family begins. The booster cycles are then distributed between injections for families with intensities above the threshold and injections for the most recently added family with intensity still below the threshold. As a result, the ramp-up duration is longer for families added later in the process. Once the transition to regular top-up injection occurs, the intensity of the injected bunches is reduced and adjusted such that the collider target bunch intensity is met after beam transfer.

A drawback of the described scheme is that, at the same time, low and high intensity bunches circulate while undergoing beam-beam interactions. Due to the potential well distortion, the low (high) intensity bunches have a higher (lower) synchrotron tune. This makes the choice of a suitable working point (mainly horizontal tune) more difficult, with the phenomena described in Section 1.4.3 [62]. Studies are ongoing to devise solutions; the first results using a small positive chromaticity are promising.

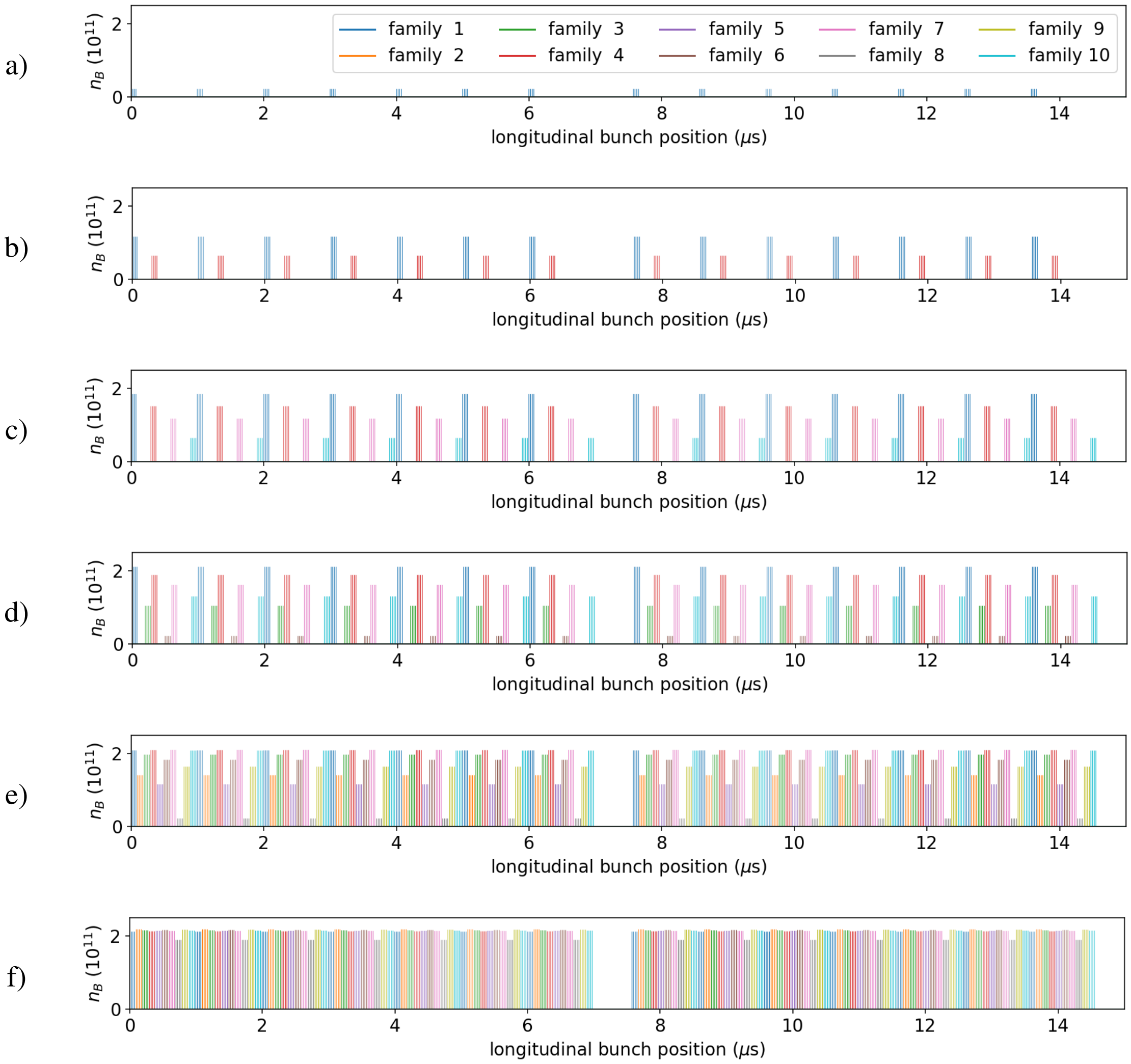

Fig. 2.17: Bunch intensity patterns in one ring at various stages during the intensity ramp procedure for Z mode operation. Patterns for only two out 40 identical trains are shown.

#### *Intensity ramp up for the W mode*

No issues are expected related to electron cloud in the collider for WW operation due to the larger gaps with the trains of the proposed filling pattern. In order to limit the maximum total intensity in the booster, separation of the collider bunches into two families is proposed for both intensity ramp-up and top-up injections. The principle is documented in Fig. 2.19 showing bunch patterns for one out of the eight (identical) trains in one of the collider rings and Fig. 2.20 showing the evolution of the beam intensities for the two families and the luminosity. Figure 2.19a shows the bunch pattern in the collider after one transfer and for one out of 8 identical trains. Every second injector batch of the complete filling pattern shown in Fig. 2.13 is filled. The next booster cycle injects into the positions belonging to the family two of the same ring leading to the pattern shown in Fig. 2.19b. The next two booster cycles inject into the counter-rotating beam. The same sequence is repeated periodically for intensity ramp-up and top-up injection.

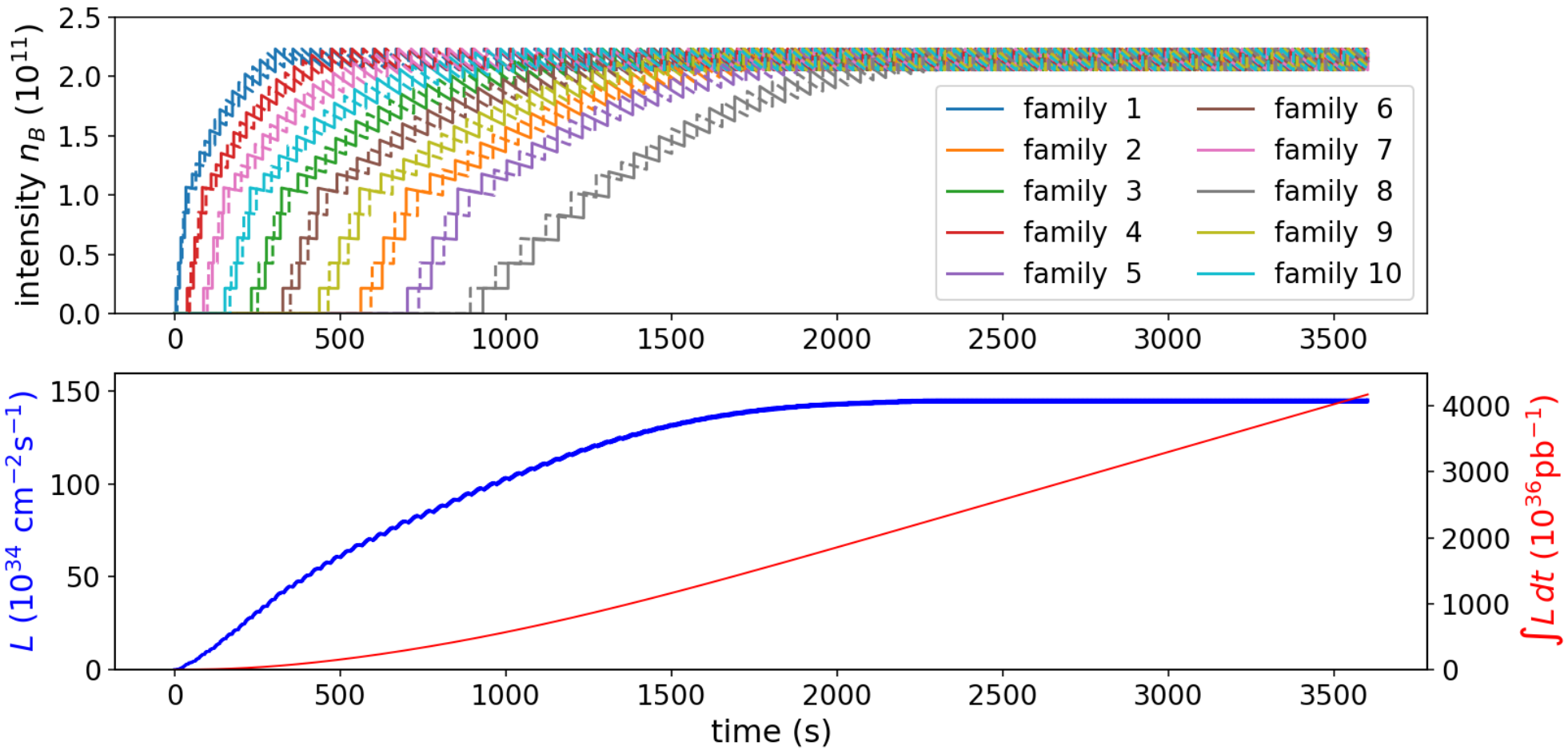

Fig. 2.18: Bunch intensity evolution for the various families (upper plot) and evolution of the luminosity for Z mode operation. Bunch intensities are plotted using solid and dashed lines for the two counter-rotating beams.

#### *Intensity ramp up for the ZH and $\mathrm{t\bar{t}}$ modes*

Due to the low number of bunches in the collider, no limitations related to electron cloud build-up or total intensity circulating in the booster exist for FCC-ee operation at the higher energies. Thus, the intensity of all collider bunches is ramped up simultaneously. The resulting filling procedure is obvious and leads to a booster filling pattern identical to that in the collider.

### 2.3.9 Top-up injection

The top-up injection scheme (described in Section 1.8.1) was simulated with a perfect lattice, without consideration of collective effects or machine imperfections. In practice, however, lower-than-expected injection efficiencies and significant penalties on injection performance at higher circulating currents have been major challenges at other lepton colliders [232, 233].

While the baseline injection scheme has a unique setting for on-axis injection, it is possible to use a hybrid injection scheme in the range of injected beam energy offset between $0.7\,\%$ and $0.95\,\%$ (Fig. 1.40). In later revisions of the collider lattice, different injected beam energy offsets may become feasible with the on-axis injection scheme by modifying the optics. In contrast, due to the septum thickness, the hybrid injection scheme must maintain a constant physical separation between the injected and circulating beams at the injection point. Since the optics remains unchanged, this constraint results in a geometric condition in the DA-MA plane, represented by red dots in Fig. 2.21b. Injection is physically possible but constrained by the dynamic aperture (DA) and momentum acceptance (MA) in the region above and to the right of this line. Conversely, injection is geometrically impossible in the region below and to the left of this line.

The injection efficiency for different hybrid schemes, different betatron offsets, is simulated and the most promising results, showing improved efficiency over the baseline (on-axis injection), are plotted in Fig. 2.21a. In these simulations, the weak-strong model of the beam-beam interactions is used. Figure 2.21b shows that a range of injection settings are possible between the geometrical constraints

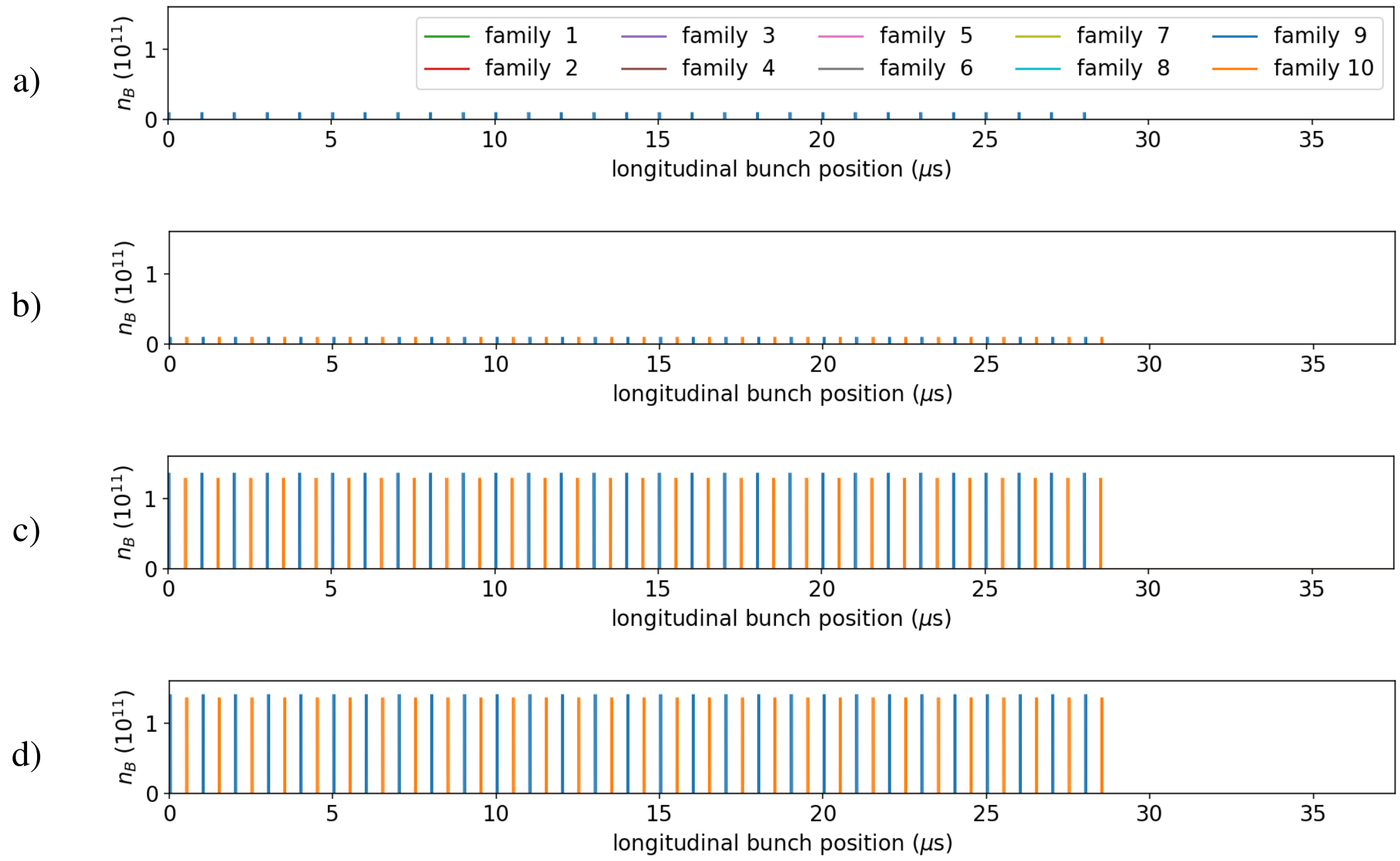

Fig. 2.19: Bunch intensity patterns in one ring at various stages during the intensity ramp procedure for WW mode operation. Patterns for only two out 8 identical trains are shown.

represented as the red dots and the DA and MA limits of the lattice shown as the blue lines [234].

At SuperKEKB, the injection efficiency reaches approximately 80 % with the aid of bunch-by-bunch (BbB) feedback. Without BbB feedback, the efficiency drops to around 50 %. The primary causes of this degradation are the larger-than-design emittance of the injected beam and the use of off-axis injection. In contrast, FCC-ee's baseline design features on-axis injection and large curvature radii in the beam transport from the booster to the collider ring, making the degradation of injection efficiency observed at SuperKEKB unlikely. SuperKEKB has also experienced beam chamber deformation due to heat, which may contribute to injection efficiency degradation. Therefore, effective temperature management is crucial.

This validates the feasibility of the present injection concept, but in the absence of comprehensive tracking that includes both errors and collective effects, it is not possible to be sure at this stage that high injection efficiency can be reached and maintained up to nominal intensity. Therefore, the present feasibility study establishes an injection efficiency goal of 80% based on the status of the concept and experience from other facilities [232, 233]. This injection efficiency target is used in the feasibility study for the sizing of the injector chain (Table 7.1).

### 2.3.10 Injection requirements

The top-up injection scheme strongly relies on the injected beam being considerably smaller horizontally than the circulating one (Fig. 1.39). By contrast, in the vertical plane, the injected beam could be several times larger than the stored beam. The horizontal beam parameters provided by the booster (see Table 4.1) must be maintained up to the nominal intensity, but also along all the bunches of the injected trains, and stably over time. The beam transfer systems for both the booster extraction and collider injection have to ensure a high level of stability and reproducibility to prevent any beam jitter that would increase emittances and reduce the injection efficiency.

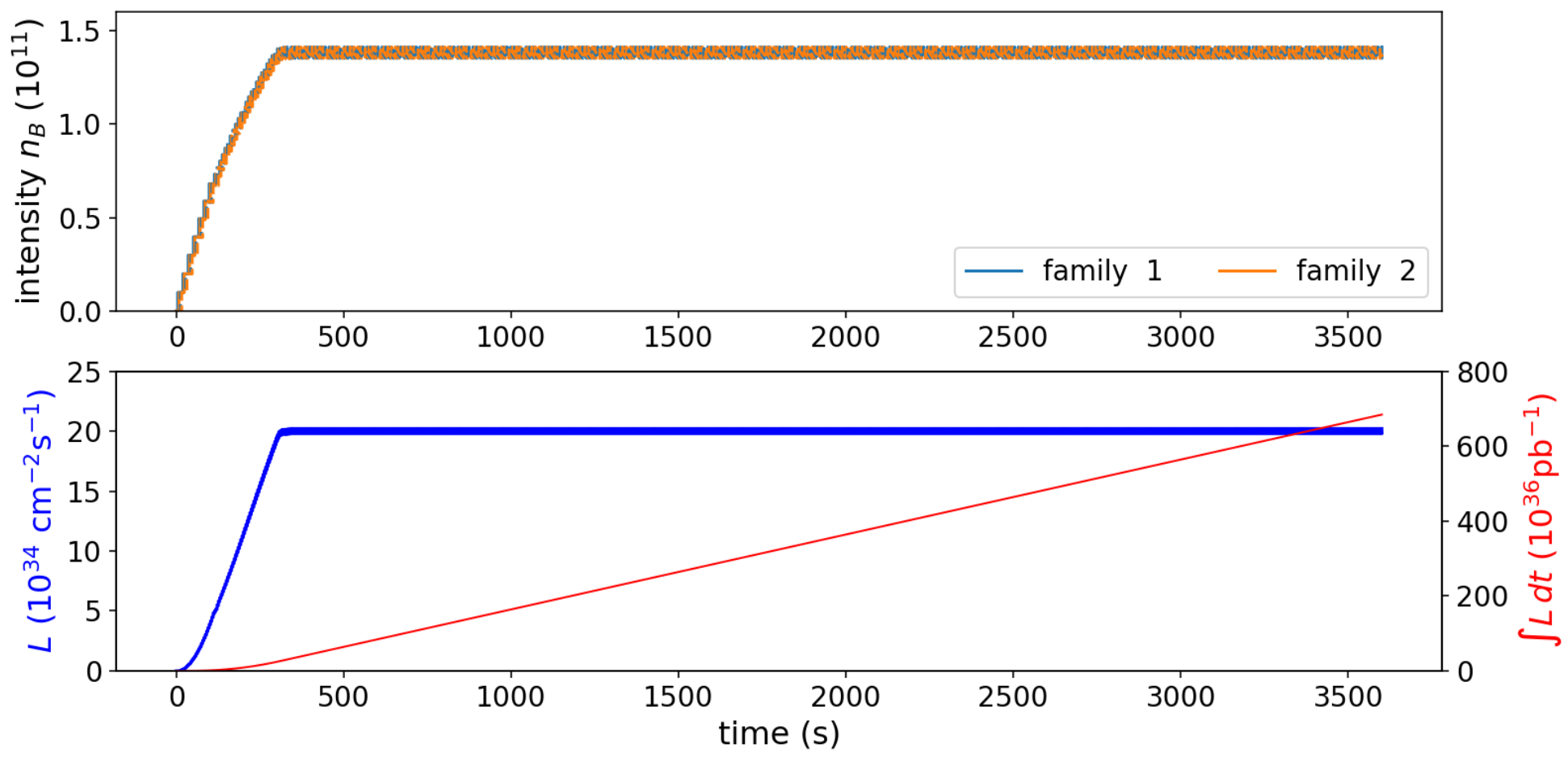

Fig. 2.20: Bunch intensity evolution for the various families (upper plot) and evolution of the luminosity. Bunch intensities are plotted using solid and dashed lines for the two counter-rotating beams.

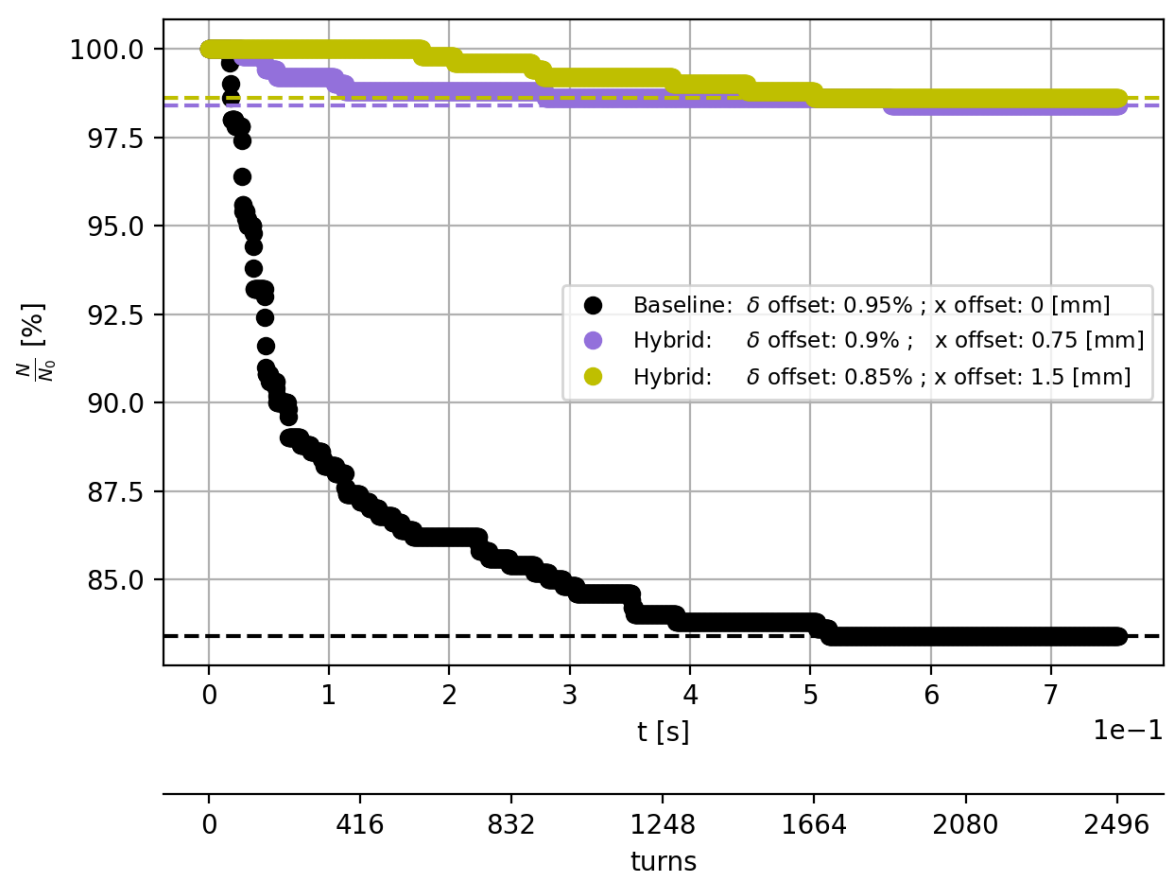

(a) Evolution of the injected beam intensity for three different injection betatron offset with beam-beam.

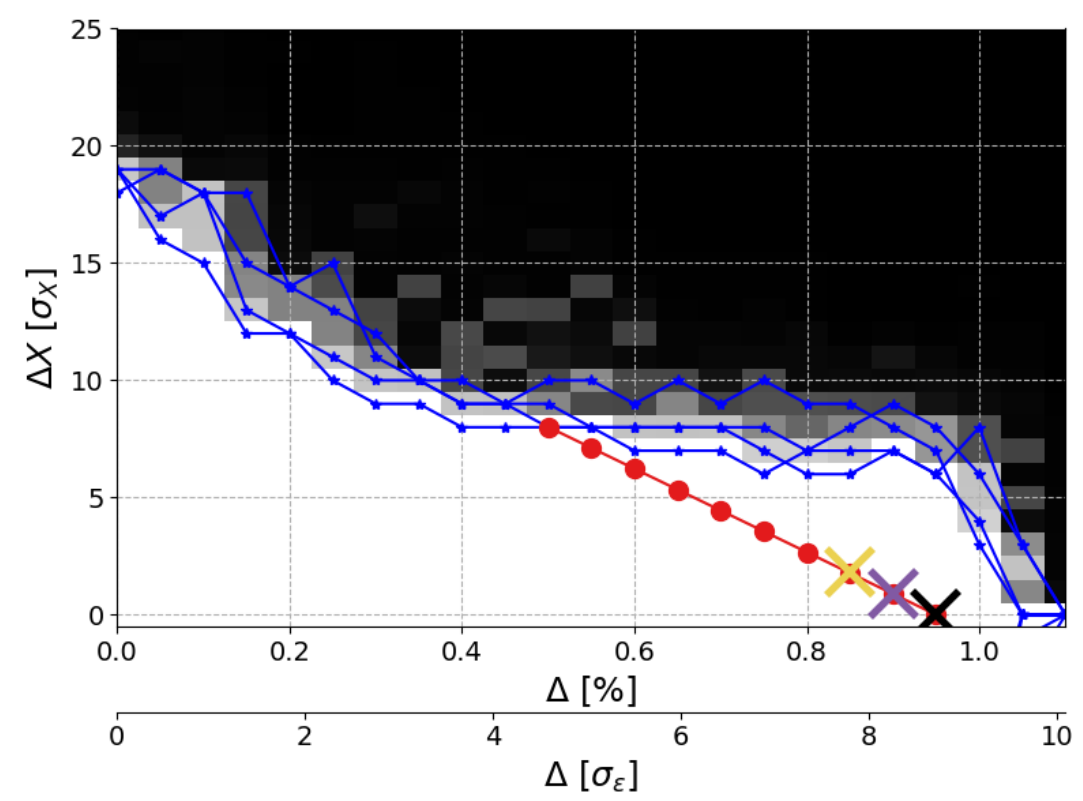

(b) Quarter quadrant representation of the DA/MA without beam-beam with crosses denoting the three injection settings simulated with beam-beam.

Fig. 2.21: Collider injection efficiency at various betatron offsets and their corresponding location in the DA/MA for the Z mode.

### 2.3.11 Refilling and pre-polarising pilot bunches after a beam abort

At the ZH and $\mathrm{t\bar{t}}$ energies, the FCc-ee can be filled with the booststrapping injection scheme detailed in subsection 2.3.8. At the Z and WW energies, prior to this, first pre-polarised must be produced, which are required for precise energy calibration from the start of each fill.

At 45.6 GeV the natural polarisation time is 250 h. Thus, achieving a polarisation level in an error-free machine of 5-10% requires 15 to 30 hours, which is an unacceptably long period without calibration at the start of fill. Reducing the polarisation rise time to about 12 h is feasible using special polarisation wigglers [162], which also increase the energy spread to 64 MeV. In the currently planned operational scenario, low-intensity ($\approx 10^{10}$ particles) pilot bunches are injected at the start-of-fill and polarised using wigglers. When roughly 5-10% polarisation is achieved, after 45 to 90 minutes, these wigglers are switched off, and all the nominal-intensity colliding bunches are then injected, according to the bootstrapping scheme, and brought into collision. Around 250 pilot bunches per beam are required in total, a number that follows from the assumption that five energy calibration measurements with two pilot bunches are performed every hour. By the time the last pilot bunch has been depolarised for the first time, the pilot bunches that were depolarised first will have naturally reacquired sufficient polarisation to be measured again.

At 80 GeV, the polarisation time is approximately 15 hours, and polarisation wigglers are not required [162]. In this case, the pilot bunches take about 1.5 hours to reach 5% polarisation. Significantly fewer pilot bunches - of the order of 25 - are needed compared to operation at the Z pole, as they re-polarise ten times faster.

## 2.4 Availability

Availability is defined as the percentage of scheduled physics time during which the machine successfully delivers beam, as opposed to being in downtime or undergoing repairs. The availability of FCC-ee is closely tied to its operational efficiency (Section 2.1.2) and the achieved integrated luminosity. Additionally, it is influenced by various performance factors, including operational costs, safety considerations, and system design constraints.

An enhanced Monte Carlo simulation framework is used to model availability by deconstructing the accelerator's main constituent systems. The simulation model is detailed in Sections 2.3.1-2.4.2, with the results presented in Section 2.4.3.

### 2.4.1 Contributing systems

In collaboration with relevant system experts, each main constituent system in FCC-ee was approximated for availability. If fault data from a comparable representative system could be found, this was scaled to an equivalent system in FCC-ee using a generalised framework. For some systems, no such fault data could be identified and a placeholder was used. The methodology in each case is as follows.

#### *Systems with representative fault data*

A generalised framework was designed for consistent and comparable representation of systems in the FCC-ee. Specific details relating to relevant peculiarities were then applied.

*Generalised framework for system availability estimation*

1. A similar or comparable state-of-the-art representative system is identified that exists currently in a working accelerator.
2. Where possible, representative system is deconstructed into subcomponents and fault mechanisms to achieve better granularity of reliability estimation.
3. Fault mechanisms are categorised by:
   - Repair type:
     - Remote Repairs: can be completed from the control room, e.g., by re-setting or by-passing the failed component.
     - Human Repairs: require on-site intervention by personnel. In addition to the time it takes to complete the repair, the approach time to get a person to the site of the fault must be considered. This is taken as the time to get from PA to the relevant access point by car, summarised in Table 2.5.
   - Equipment location:
     - Surface: Equipment located on the surface can be accessed for repair immediately when the technician arrives.

- Tunnel: Equipment located inside the accelerator tunnel cannot be accessed until a suitable cool-down time has passed at the relevant access point, corresponding to the time required to fully ventilate the tunnel. This is summarised Table 2.5 (Section 9.4.2).

4. Fault data is extracted from the representative system to gain Mean Time Between Failures (MTBF) and Mean Time to Repair (MTTR) information for the system and, where possible, subsystems for finer granularity.
5. MTBF and MTTR are scaled from the representative system to the study system in the FCC-ee:
   - MTBF scales inversely proportionate to the relative number of components. The more components in the FCC-ee system relative to the representative system, the lower the MTBF. If the number of components in FCC-ee is not yet precisely known, an approximation is made, assuming a design specification similar to that of the representative system.
   - MTTR for remote repair faults is left unchanged. An approach time and/or cool downtime is added to human repair times corresponding to the equipment location, Table 2.5.
6. If applicable, a formulation for redundancy in the FCC-ee is defined.

*System specifics for availability approximation*

Specifics used to approximate availability for each constituent system are also considered, in addition to the above generalised framework. Only faults leading to down time in the representative system are included, thereby assuming a similar degree of redundancy for each basic component family as currently exists in the working accelerator. Details of subsystems, scaling numbers, and any additionally implemented system-level redundancy are provided in Table 2.6. Fault data was taken from CERN's Accelerator Fault Tracking (AFT) database [235] and is specific to LHC physics operation 2015-2024, unless otherwise stated.

If no representative fault data could be found, but the system is still deemed a high risk for machine availability, it was given an availability placeholder pending a more rigorous reliability assessment. This applied to two systems, the beamstrahlung dump absorbers, and the polarimeter.

The remaining accelerator systems specific to the booster, injector complex and technical infrastructure are covered in Sections 5.3, 7.9 and 8.10, respectively.

### 2.4.2 Repair schedule

When redundant and non-redundant systems are combined, the schedule with which repairs are implemented has significant impact on availability. A schedule similar to the LHC was simulated: Following a dump, operators attempt for one hour to restore operation remotely from the control room. If operation cannot be restored during this time, only then people are called to begin human repairs. Once humans are in the tunnel they will finish all repairs, including all redundant components.

Table 2.5: Approach time (the drive time from PA) and cool down time (see Section 9.4.2) relevant to human repair faults at each access point.

| | PA | PB | PD | PF | PG | PH | PJ | PL | arc |
|---|---|---|---|---|---|---|---|---|---|
| Approach Time (h) | 0.00 | 0.75 | 0.55 | 0.8 | 0.63 | 0.600 | 0.416 | 0.283 | |
| Cool Down Time (h) | 1.54 | 1.70 | 1.54 | 24* | 1.54 | 0.717† | 1.540 | 0.717† | 1.7 |

*Collimation straight section without bypass tunnel.
†Klystron gallery only. A longer cool down time is required to access the beam tunnel.

Table 2.6: Parameters used to simulate availability of systems and subsystems in the FCC-ee Collider.

| **System** | **Subsystem** | **Representative System** | | **FCCee** | **Redundancy**[*] | **Location** | **Unit MTBF** | **Group MTBF** |
|---|---|---|---|---|---|---|---|---|
| | | Machine | # | # | # (%) | AP | days | days |
| Accelerator Controls | Software | LHC | 1 | 1 | 0 (0) | all | | 7.1 |
| | Hardware | | 607 | 2400 | 0 (0) | all | 15 840 | 6.6 |
| Access System | Access Points | LHC | 8 | 8 | 0 (0) | all | 178 | 22.4 |
| Beam Instrumentation | BPM | LHC | 1000 | 5906 | 0 (0) | all | 20 671 | 3.5 |
| | BLM | | 3600 | 1205 | 0 (0) | all | 36 632 | 30.4 |
| | Other | | 1 | 1 | 0 (0) | all | | 24.5 |
| Beam Losses | UFOs | LHC | 54 | 182 | 0 (0) | all | | 1.2 |
| | Beam Instability | | 1 | 1 | 0 (0) | all | | 2.1 |
| Beamstrahlung Dump | | SPS TIDVG | 1 | 8 | 0 (0) | PA, PD, PG, PJ | 68 | 8.5 |
| Collimation | Moveable Collimator | LHC | 108 | 32 | 0 (0) | PF | 1549 | 48.4 |
| Elevators & Handling Equip. | Elevator Shafts per AP | LHC | 1 | 2 | 1 (50) | all | 74 | 332.8 |
| Experiments | IPs | LHC | 4 | 4 | 0 (0) | PA, PD, PG, PJ | 17 | 4.3 |
| Extraction & Beam Dump | Dump Block | LHC | 2 | 2 | 0 (0) | PB | 224 | 112 |
| | Controls | | 2 | 2 | 0 (0) | PB | 36 | 18.2 |
| | Hardware | | 2 | 2 | 0 (0) | PB | 65 | 32.7 |
| | Other | | 2 | 2 | 0 (0) | PB | 71 | 35.6 |
| Injection Systems | MKI | LHC | 2 | 2 | 0 (0) | PB | 20 | 10.1 |
| | TDI | | 2 | 2 | 0 (0) | PB | 71 | 35.6 |
| Machine Protection | FMCM | LHC | 1 | 10 | 0 (0) | all | | 39.2 |
| | BIS | | 1 | 2 | 0 (0) | all | | 32.7 |
| | SMP | | 1 | 1 | 0 (0) | all | | 87.1 |
| | PIC | | 1 | 0.2 | 0 (0) | all | | 261.4 |
| | WIC | | 1 | 6 | 0 (0) | all | | 21.8 |
| Magnets | Normal-conducting | CERN | 3532 | 12 500 | 0 (0) | all | $71 \times 10^6$ | 5681 |
| Operation | | LHC | 1 | 1 | 0 (0) | all | | 5.8 |
| Polarimeter | | N/A | - | 1 | 0 (0) | PA, PG | | 3.2 |
| Power Converters | Dipole | LHC | 194 | 16 | LHC | all | 3672 | 229.5 |
| | Quadrupole | | 194 | 32 | LHC | all | 3672 | 114.8 |
| | Sextupole | | 194 | 1152 | LHC | all | 3672 | 3.2 |
| | Dipole Tapering | | 1032 | 710 | LHC | all | 15 485 | 21.8 |
| | Quadrupole Tapering | | 1032 | 709 | LHC | all | 15 485 | 21.8 |
| | Horizontal Corrector | | 1032 | 2824 | LHC | all | 15 485 | 5.5 |
| | Vertical Corrector | | 1032 | 2824 | LHC | all | 15 485 | 5.5 |
| | Skew Quadrupole | | 1032 | 2824 | LHC | all | 15 485 | 5.5 |
| Radio Frequency | Cavity Circuit $Z$ | LHC | 16 | 132[2] | 1 (0.75) | PH | 61 | 0.9 |
| | Cavity Circuit $W$ | | 16 | 132[2] | 6 (4.5) | PH | 61 | 169.1 |
| | Cavity Circuit $ZH$ | | 16 | 264[3] | 26 (10) | PH | 61 | 314.6 |
| | Cavity Circuit $t\bar{t}$ | | 16 | 752[3] | 75 (10) | PH | 61 | 465.2 |
| Smoke Detection | Smoke Alarms | SPS | 7 | 91 | 0 (0) | all | | 4.3 |
| Transverse Damper | | LHC | 2 | 2 | 0 (0) | PH | 65.4 | 32.7 |
| Vacuum | Pumps | LHC | 891 | 3344 | 0 (0) | all | 349 448 | 104.5 |
| | Gauges | | 1052 | 1808 | 0 (0) | all | 274 997 | 152.1 |
| | Valves | | 323 | 452 | 0 (0) | all | 84 434 | 186.8 |
| | Controllers | | 789 | 1584 | 0 (0) | all | 55 915 | 35.3 |

[*] System-level redundancy in addition to that already implemented in the component family of the representative system.

[†] Per beam.

[‡] Shared by both beams.

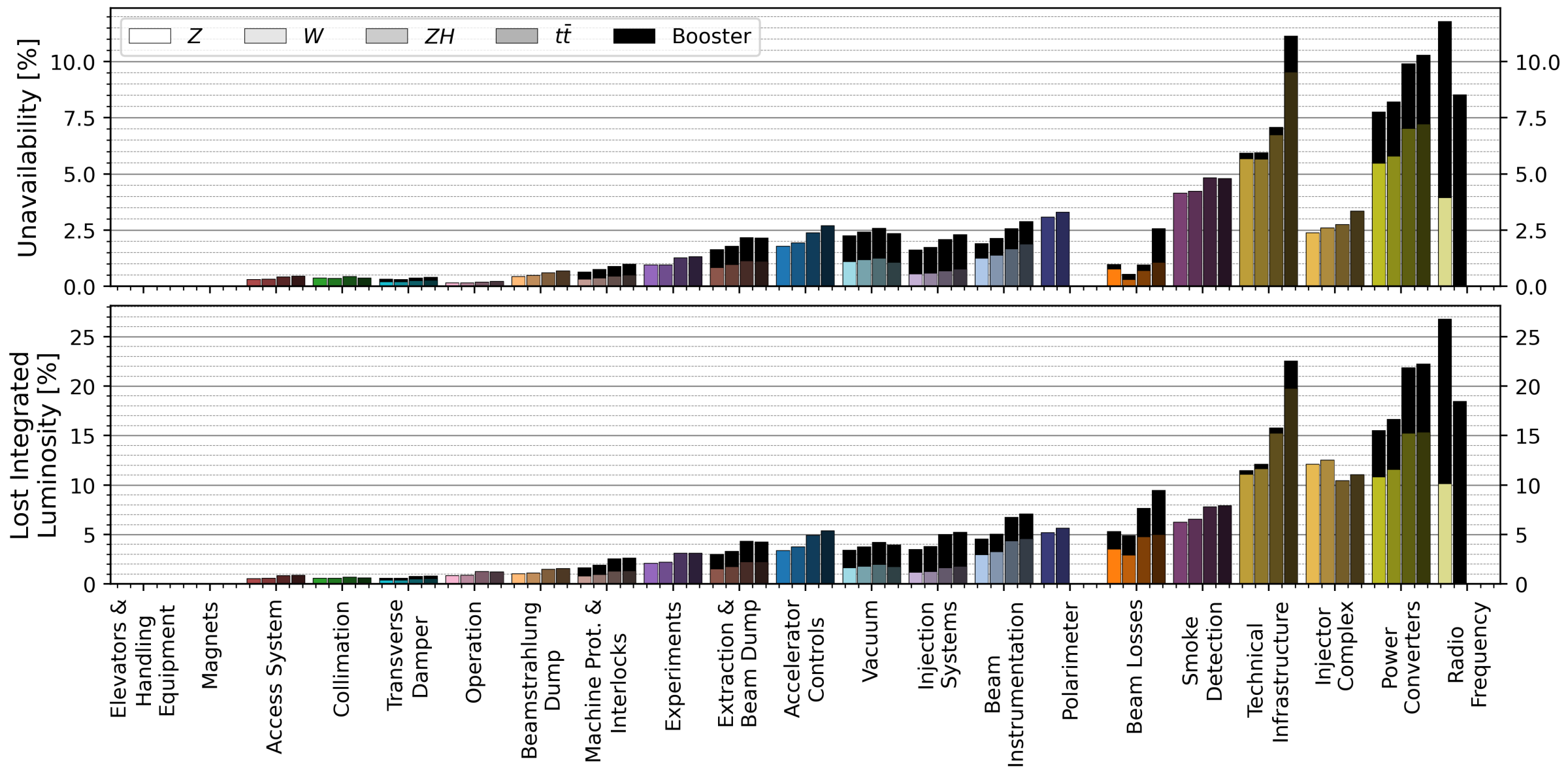

Fig. 2.22: Unavailability and lost luminosity contribution from each main constituent system in the FCC-ee. Systems are ordered according to Z mode lost luminosity contribution. contribution.

### 2.4.3 Availability simulations

FCC-ee availability is simulated using AVAILSIM4 [236], a discrete-event Monte Carlo tool developed in-house at CERN for analysing availability and reliability in complex systems. Each energy mode was simulated for its full operation term (four years in Z, two years in WW, etc.), with performance averaged over 100 iterations.

The breakdown for unavailability and lost luminosity in each system under certain assumptions and based on data from LHC sytems is shown in Fig. 2.22. The booster, injector complex and technical infrastructure are included for illustration; but treated in detail in Sections 5.3, 7.9 and 8.10, respectively. For the collider, the main contributors to lost luminosity are the RF, power converters, beam losses and the alarm system.

The RF system experiences the highest unavailability in Z mode due to the strict limit of single-cavity redundancy assumed. In WW mode, the ability to sustain beam operation despite the loss of up to six cavities eliminates the collider's unavailability contribution from the RF system. The booster significantly contributes to unavailability in both energy modes, as it lacks any redundancy. In ZH and $t\bar{t}$ modes, a 10% redundancy ensures zero downtime from the RF system, providing sufficient reserve to accommodate failed cavities until they can be repaired in the shadow of downtime scheduled for other systems.

Power converters show high contributions in all energy modes. This is predominantly due to the low MTBF seen from sextupole and corrector converters and the high number of different magnet circuits used in the assumed optics configuration.

Beam losses have lower contribution to unavailability as the recovery time to restore operation is short. But their contribution to lost luminosity is high as they cause frequent dumps, each of which incurs a turnaround penalty related to the time required to restore stable beams. Turnaround penalty is two hours in Z, W modes, and 30 minutes in ZH, $t\bar{t}$.

The alarm system features relatively high due to the scaling factor from SPS (7 km) to the FCC-ee tunnel (91 km). The number of alarms combined with mandatory on-site safety evaluation in the event of fire detection leads to large amounts of down time.

### 2.4.4 R&D opportunities

Significant improvement in reliability is required in order to meet luminosity targets, especially in the lower energy modes. Several R&D opportunities have been identified:

#### *Radiofrequency*

Increasing redundancy in the RF system for the Z, W modes presents challenges due to high beam loading effects but should be further investigated given the potential performance gains. Additionally, RF redundancy in the booster system should be explored further, as it remains a significant contributor to unavailability in both energy modes.

It is important to note that zero RF downtime is currently assumed in the simulation due to the idealised assumption of perfect redundancy between cavity circuits. In reality, common-mode failures would still occur, requiring reconditioning or replacement, which cannot always be immediately compensated by a redundant cavity. Further modelling is necessary to better understand these fault types. In the coming years, close collaboration with the RF design team will be essential to deconstruct the RF circuit into subsystems, identify problematic fault types, and analyse high-failure component families.

Analysis of the auxiliary systems around the superconducting cavities would unearth further reliability opportunities. For example, the klystron gallery is accessible to personnel while the main colliding rings are in operation. If more high-failing auxiliary systems can be located there and designed to be redundant, modular and hot-swappable, significant gains could be achieved.

#### *Power converters*

The current optics configuration powers sextupoles in groups of four. One failed converter would therefore lead to four magnets failing simultaneously. Collaboration with the optics working group is required to consider how to preserve the beam if combinations of converters and magnets fail. A system-level redundancy approach could significantly improve reliability in this case.

It is important to note that the current simulation assumes the same level of redundancy in power converters as is presently implemented in the LHC. Therefore, for an effective solution, the FCC-ee sextupole and corrector converters must be designed to be more robust than those used in the LHC. A detailed study on power converter reliability will be essential in the coming years.

Reducing the number of corrector families, especially for the sextupole magnets, can drastically reduce the number of power converters required and in turn reduce their overall fault rate.

#### *Operation cycle*

Opportunities also exist within the operation cycle:

1. **Indefinite physics:** If the lifetime of pilot bunches in the main ring can be made longer than the natural polarisation time, pilot bunches could be topped up immediately after being used for measurement and allowed to naturally re-polarise. This avoids the need to dump after 20 h in order to re-fill with polarised pilot bunches. Physics can then continue uninterrupted in Z, W modes, until equipment failure dumps the beam.
   The utility of this proposal is best illustrated by the distribution of stable beams durations at each energy mode, shown in Fig. 2.23. This shows that, under the assumptions made here, stable beams can rarely be sustained longer than 20 h without beam abort due to system failure.
2. **Polarised bunch injection:** If pilot bunches can be polarised prior to injection, and their polarisation preserved through the injector complex and booster, the 90-minute polarisation phase could be eliminated from the Z, W operation cycle entirely. This would effectively result in the same operation cycle as the higher energy modes ZH, $t\bar{t}$ while maintaining the capability for precise energy calibration.

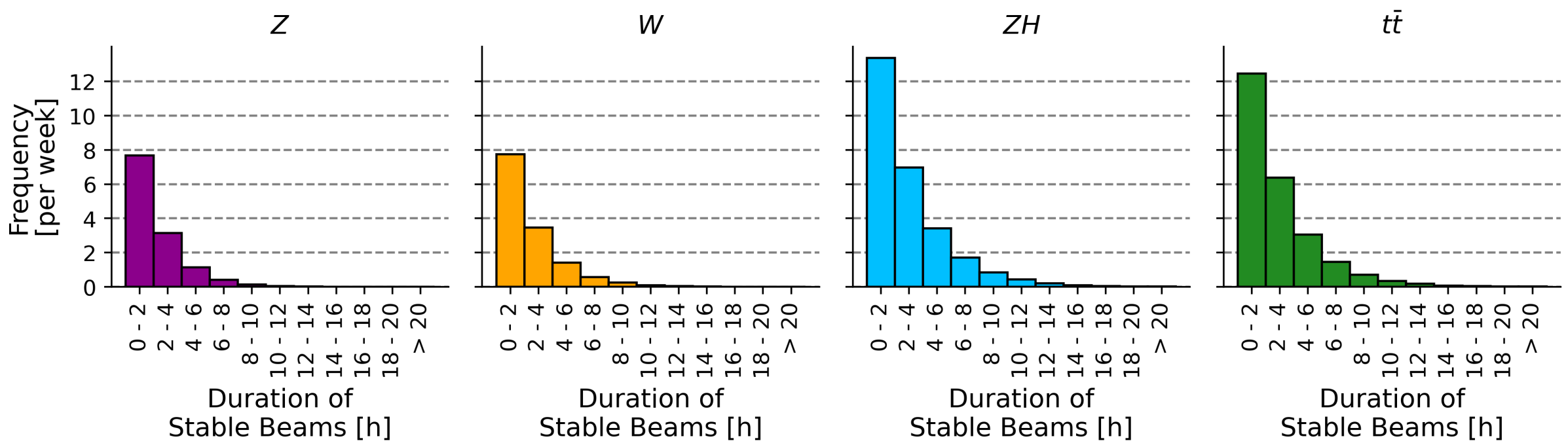

Fig. 2.23: Distributions for duration of stable beams in each energy mode.

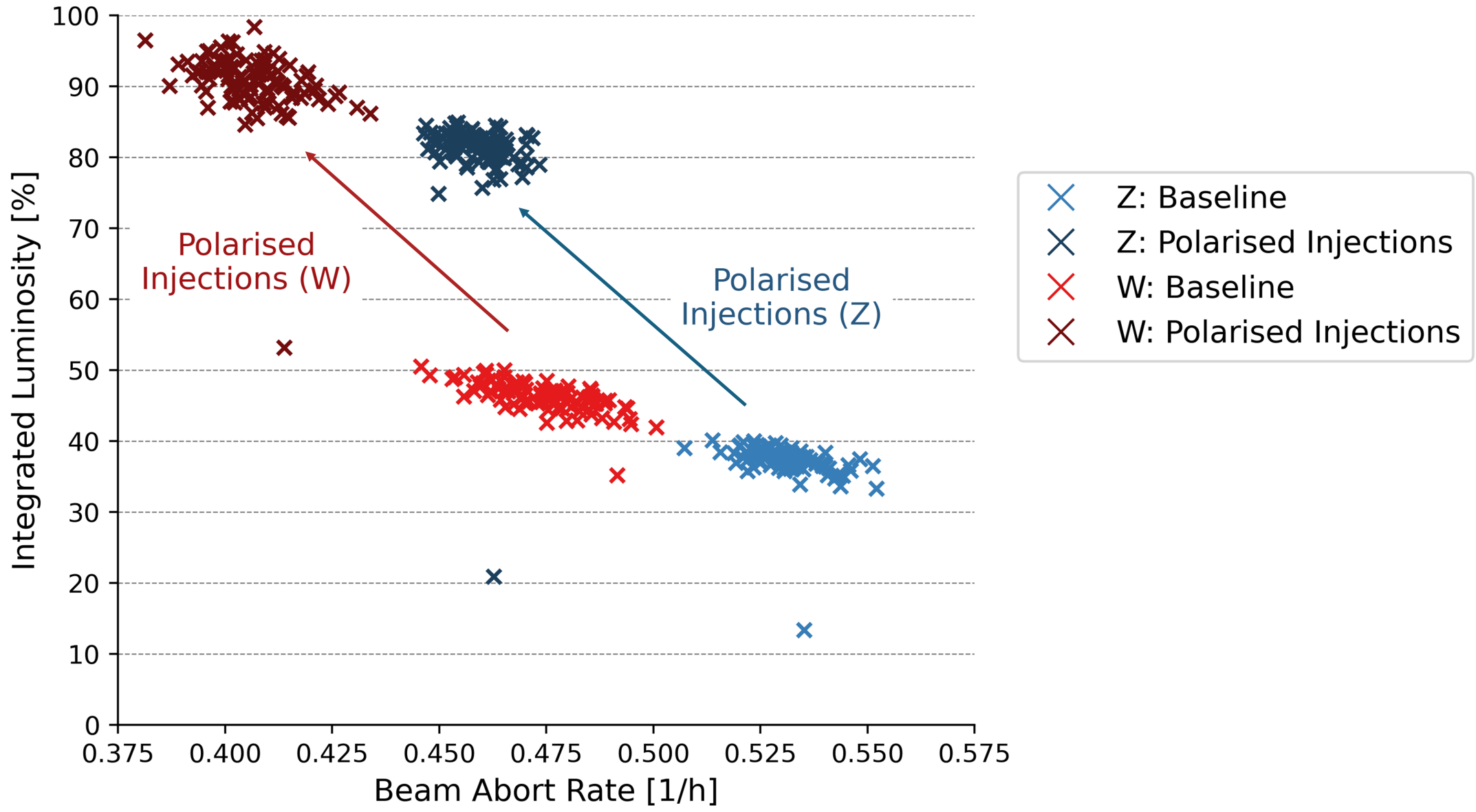

Fig. 2.24: Effect of polarised injections on achieved integrated luminosity and fault rate, under certain assumptions.

The potential of this scheme is illustrated in Fig. 2.24. In both Z, WW modes, this approach more than doubles the achieved integrated luminosity. This is because, by removing the 90-minute polarisation phase at the start of each fill, an equivalent duration of stable beams is effectively gained. This presents a potentially game-changing R&D opportunity to support physics objectives in this challenging reliability environment.

## 2.5 Operational model

CERN has a long-standing tradition of designing, building, and operating accelerators, with well-defined groups structured according to a clear division of responsibilities and tasks. The expectations regarding interfaces between equipment, controls, and human operators - whether in control rooms or via remote access - have been long established and are now considered immutable, ensuring maximum efficiency within the current CERN accelerator complex, which relies on a high degree of specialisation across various groups.

However, with the unprecedented scale of the FCC accelerator, new constraints such as energy provision from intermittent sources, and the assumption that CERN's human resources during the FCC era will remain at levels similar to today, necessitate a new approach to designing, building, and operating particle accelerators at CERN.

The necessary equipment paradigm can be summarised as:

– Full digitalisation: Equipment needs to be fully digital and remotely controllable as well as analysable with common interfaces and protocols.
– Automation towards fully autonomous systems: All equipment needs to be designed with automation in mind (across systems and within systems) to e.g., auto-configure, auto-stabilise, auto-analyse, auto-recover etc. An example would be designing cars without steering wheels and asking which control algorithms, additional instrumentation and software layers would have to be in place to safely and efficiently drive and maintain the car while maximising the mobility of the car user. The next design iteration will then address the new possibilities with the re-imagined system (with the car analogy this could be things like "Are close-by car parks still a necessity?", "Are personal cars still required or can they all be shared?"). A key additional aspect in this discussion are digital twins for training and constraining control algorithms and differentiable simulations.
– Full virtualisation: Space telescopes have the additional constraint that one cannot simply go there to fix them if things break. With the size of the FCC and the number of components, maintainability is similarly constrained as for space telescopes. Humans should not be considered for on-site repair at all or at least not during the run. This can be addressed with built-in margin, low failure rate, modularity, redundancy (possibility of degraded mode or auto-reconfiguration), robotics and other new technologies.

The next generation accelerator control system will have to foresee easy plug&play solutions and frameworks to implement the above and provide integration of artificial intelligence (AI). This concept is often referred to as *AI-ready* accelerators or control systems. Examples of the required capabilities include frameworks for sharing and storing ML/AI models (including continual learning functionalities), integrated digital twins, platforms for IoT, optimisation frameworks, time-aligned data across the accelerator, PYTHON (or equivalent) code infrastructure, and easily configurable virtual device services for continuous analysis and anomaly detection.

Clearly, the three elements of the proposed equipment paradigm outlined above are interlinked and mutually dependent. Many of today's systems in the LHC and other CERN accelerators already implement item 1 and, in some cases, partial implementations of items 2 and 3. For the FCC, however, the complete paradigm must become the standard for equipment integration, meaning it must be imposed at the design stage. This approach will significantly reduce the resources required for commissioning, operation, and maintenance, though it may increase construction costs due to factors such as redundancy, additional connectivity, and additional sensors.

While space telescopes have no choice but to adhere to such stringent design principles, for the earth-based FCC, trade-offs may be considered. These will depend on whether the increased construction costs are deemed unacceptable and/or whether the potential increase in operations and maintenance costs, along with documentation, training, long-term sustainability, and impact on availability, remains within acceptable limits. Forecast availability of the main constituent systems in the baseline FCC-ee design is provided in Section 2.4, based on similar equipment operating today.

The new equipment paradigm should allow:

– Preparing the FCC-ee injector complex and collider with a single person on shift; and
– keeping the global maintenance and exploitation effort level (in total FTE) for equipment teams for FCC-ee and injector complex at or below today's level (for LHC and the existing injector complex).

Even today, the most sophisticated and relatively recent particle accelerator at CERN, the LHC, is operated with a single person on shift. This supports the vision of automating the FCC to such an extent that the entire electron complex could be run single-handedly. In the current setup, this would correspond to 7–9 full-time shift personnel. However, further automation and the integration of AI assistants in the control room should enable an even more advanced approach, raising the question of whether full-time shift personnel will still be necessary in the 2040s.

Already today, LHC experiments operate with part-time, non-expert shifters drawn from a large pool of short-term shift workers. A similar approach could be explored for the FCC. Nevertheless, approximately five daytime experts dedicated to particle accelerator operation will still be required, ensuring sufficient expertise to support the rest of the accelerator complex as well.

### 2.5.1 Reducing the operation effort for equipment teams

Operating the FCC electron complex will require relatively few resources. The baseline availability model for FCC-ee, discussed in Section 2.4, provides an estimate of the number of on-site repairs needed per week, assuming the FCC-ee is built and operated following the same paradigm as the LHC today. Table 2.7 indicates that this number ranges between 12 and 20 interventions per week.

Table 2.7: Estimated number of on-site interventions for the FCC-ee complex, if it were built and operated as a large-scale LHC [237].

| Mode | Number of weekly on-site interventions |
|---|---|
| Z | 14.8 |
| WW | 15.0 |
| ZH | 15.5 |
| $t\bar{t}$ | 20.3 |

Equipment groups will greatly benefit from automating fault detection, analysis, repair and recovery. The goal should be to automate equipment to the extent that all 'typical' faults for equipment groups during the run do not require human intervention and hence will not require standby services. Only equipment which is designed, built and tested to this extent can be be considered suitable for operation. This requires acceptance testing in realistic mock-ups, well-planned and sufficiently long first commissioning phases and the requirement to be fully automated from the start of physics production. A comprehensive analysis on what availability is needed per system to reach the physics goals and how this can be reached in a robust and affordable manner is a key pre-requisite for equipment design. The right compromise between adequate operational margins through machine parameter choice and passive mitigation of e.g., radiation, automating fault detection and repair, and reliable design through redundancy, component margins, etc. has to be established on a system by system basis.

The target of a constant workforce will require a significant reduction in the maintenance effort per equipment compared with today. The suggested target number of at least a 50% reduction in maintenance effort comes from the analysis of the evolution of fault numbers for some key accelerator systems and from the assumption that remote interventions could be fully automated already today, given recent advances in technology. For example for the SY-ABT kicker systems, currently more than 50% of their interventions can be done remotely. Similarly, for the LHC QPS system the ratio of remote interventions remained roughly 50% in recent years, Fig. 2.25.

Setting-up/commissioning procedures (accTesting [238], re-conditioning procedures etc.) that need to be done regularly (even if only annually) should also all be automated. This has the additional benefit of increasing the flexibility for planning (as it allows for parallelisation without concern for sharing experts) and operational schedules.

The operation costs of the FCC-ee were assessed to amount to about 600 MCHF per annum. The approach consisted in assigning the nature of equipment with back-tested operation and maintenance percentages and related costs resulting in a bulk number of 200 MCHF, including a typical spending ratio of staff per materials budget giving about 300 MCHF. As a result the operation cost estimate for the FCC-ee consists of about 200 MCHF materials, 300 MCHF personnel and 100 MCHF on average for the electricity costs, an average total of 600 MCHF per annum.

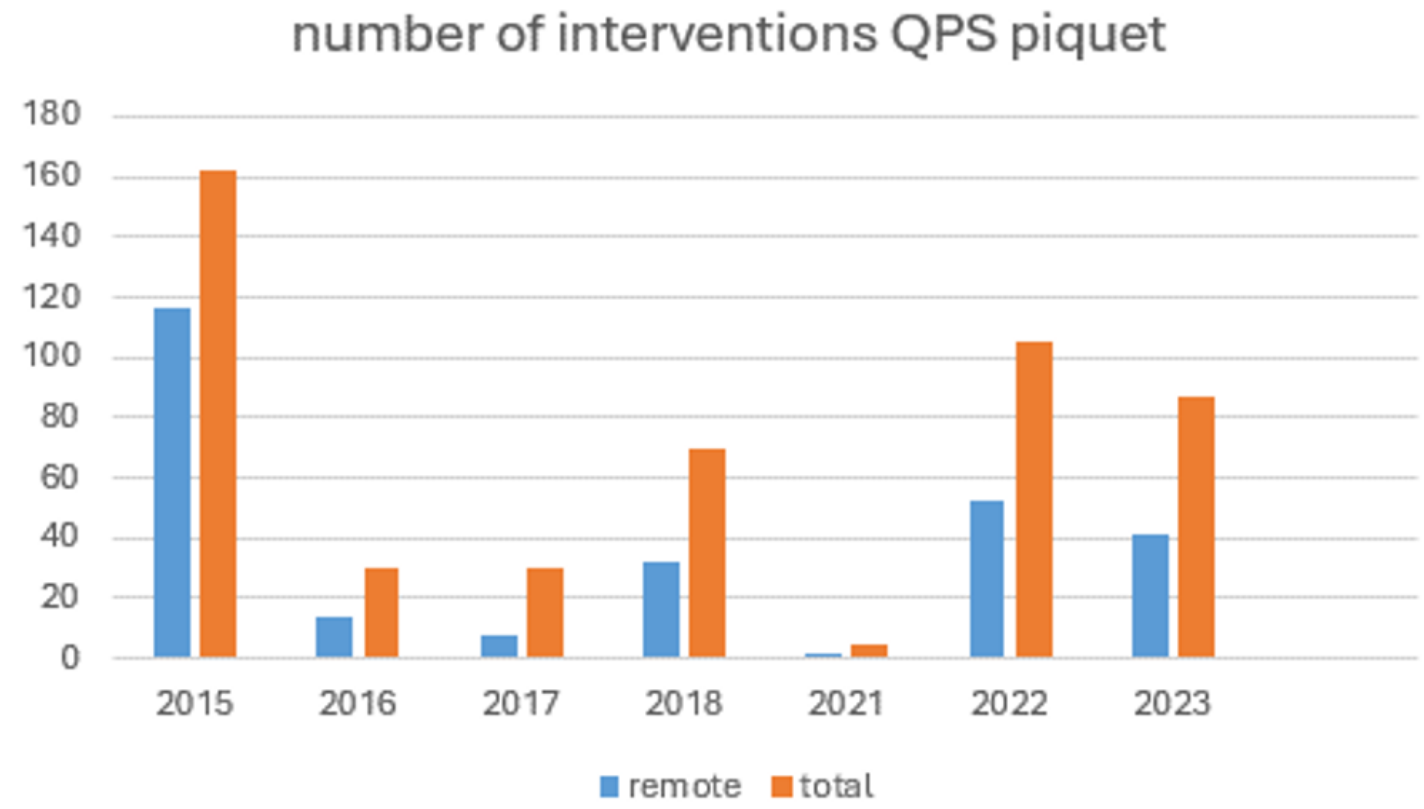

Fig. 2.25: Graph showing the number of remote and total standby interventions on the LHC QPS systems. Data extracted from the QPS standby logbook.

## 2.6 Machine Protection

### 2.6.1 Machine protection requirements

The machine protection aspects of FCC-ee are challenging in several ways.

The transverse beam sizes are small because of the small equilibrium emittance in a ring that is so large. The typical vertical beam size can be smaller than 10 µm. The horizontal beam size varies between 260 µm and 400 µm (assuming $\beta_{x,y} = 100\,\mathrm{m}$). Because of the small emittances, the energy density of FCC-ee beams at the Z mode reaches up to 17 % of the value for the HL-LHC beams (7 TeV), although the stored energy of the FCC-ee beams at Z is only 2.5 % of the HL-LHC's.

The risk of damage strongly depends on the actual loss scenario, where the shower development for a beam hitting any material needs to be taken into account. Compared to the HL-LHC proton beams, the energy is deposited over a much shorter distance, as illustrated in Fig. 2.26. The maximum energy deposition density induced by a single FCC-ee bunch in copper is about 4 to 10 times lower than for a HL-LHC bunch, but it is still sufficient to reach the melting point. This highlights the need for a sophisticated machine protection architecture that relies on a combination of active and passive protection systems. Machine protection aspects also have to be considered when designing accelerator hardware, in order to reduce the likelihood of beam losses and to mitigate the severity of any beam loss events.

The beams are stabilised by various means. Among them, the transverse feedback is particularly critical as in case of its failure, multi-bunch instabilities leading to significant beam losses can develop within a few turns only. The bunch lengthening generated by beamstrahlung is also key for maintaining beam stability and could be lost if the beams are not kept in collision. Due to the large circumference, with a single beam dump system, the signal transmission time of the beam abort request is significant.

The synchrotron radiation itself will be challenging as 100 MW of synchrotron radiation power will need to be absorbed and efficiently dissipated. At the higher beam energy modes, the effect of radiation on electronics will need to be taken into account in the design of the many electronic systems or of appropriate shielding.

#### *Machine protection system inputs*

The machine protection system interfaces with many hardware systems. Failure of these systems risks affecting the beams in such a way that damage can occur. There are also processes which can lead to damage of the machine without being related to a specific hardware item. Such processes are also listed below. The machine can generally be protected against these events by surveying the beam properties via beam diagnostics. Therefore, the beam diagnostics is part of the machine protection system and is listed in the next section.

- **Beam instabilities** are induced by the impedance seen by the beam. Both the transverse coupled bunch instability (TCBI) and collective effects in the transverse plane (TMCI) have been assessed. The most dangerous mode has a rise time of about $1.3\,\mathrm{ms}$ and a bunch-by-bunch feedback system is used to suppress the TCBI. The estimated required damping time of $1\,\mathrm{ms}$ corresponds to about three turns, which will be a challenge for the feedback system. The rise times for the TMCI are about $8\,\mathrm{ms}$ (i.e. without beamstrahlung). At the Z pole, such short bunch length could be reached within about ten thousand turns if the collisions between the beams is lost. This needs to be taken into account for example when dumping the two high intensity beams.
  More detailed simulations of the instabilities and the interplay with the feedback systems are outstanding. The values quoted here correspond to the Z-mode which is most critical since the beam currents here are the highest
- **The Transverse feedback system** is vital for stabilising the high intensity beams. In case of failure of the transverse feedback system the beam will need to be aborted within about 3 turns, which is challenging. Considering the criticality of the system, a redundant and distributed system would be preferred. However, multiple feedback systems risk to counteract each other, leading to an unstable situation.

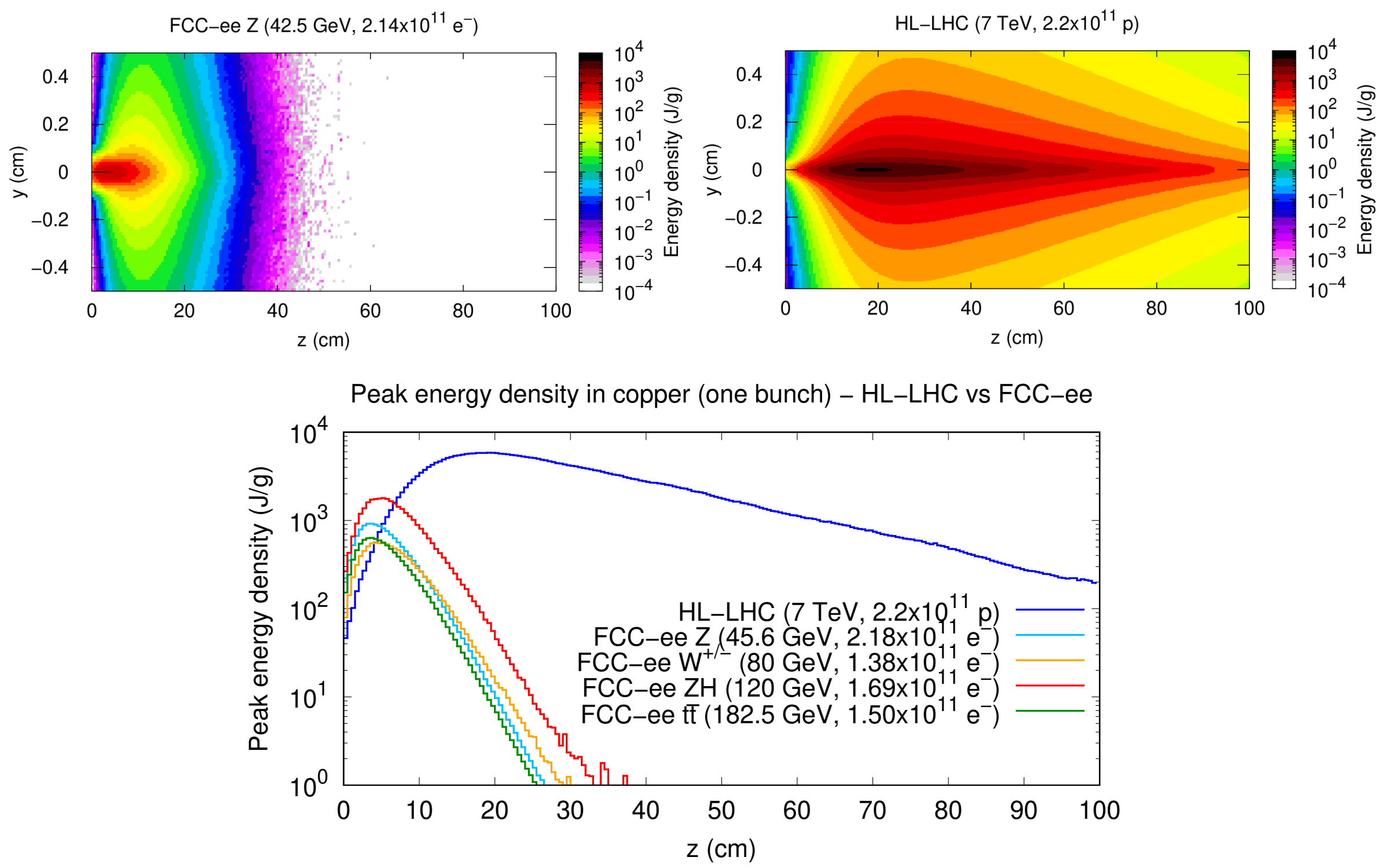

Fig. 2.26: Energy deposited in copper by a single FCC-ee (Z) electron bunch (top left) and HL-LHC proton bunch (top right). The maps show the energy density in the horizontal plane. The bottom figure compares the corresponding peak energy density profiles, for all FCC-ee beam modes, assuming a $\beta$-function of $100\,\mathrm{m}$ in both planes. The melting point of copper is reached if the energy density exceeds about $0.45\,\mathrm{kJ/g}$.

The same transverse feedback system is also foreseen to be used for depolarisation of the beam. The cohabitation of systems for excitation and feedback needs further study. Strong transverse momentum kicks of 10 μrad at 45 GeV are required for depolarisation and for this reason are critical for safe operation of the machine, especially in case of failures like non-synchronised excitations.

Protection against damper failures will require monitoring both the damper hardware and the beam itself. This will be achieved using beam loss monitors and beam position monitors, strategically placed at the appropriate phase advance relative to the kicker elements.

– **The RF system**

The collider hardware is dominated by the RF system. To avoid hardware modification between the Z, WW and ZH modes, it is foreseen to run in Reverse Phase Operation. First simulation results show no risk of too high induced voltages in case of a cavity trip, which means that operation can continue with a small number of cavities that have tripped.

If a cavity voltage or power is missing, the beam will need to be dumped. The time constant to be taken into account here is about 1/4 of a synchrotron motion, which is about 10 turns. The RF interlock — most likely based on phase detection — is backed up by the Beam Loss Monitors.

The RF system must be able to control the RF parameters during the duration of the abort gap. Preliminary studies indicate that a 600 ns abort gap duration is feasible.

– **The injection and extraction systems**

Injection and extraction of high intensity beams are machine-protection critical as the failures of the hardware involved is by definition extremely fast, and there is no time for the machine protection system to react to these failures. The injection and extraction systems are described in detail in Section 1.8.

For injection in the collider, two kicker magnets create a closed bump to bring the stored beam trajectory close to the injection septum. The bump is created by two kicker magnets, $180^\circ$ phase advance apart and originally planned to be constant over a single turn and having a rise and fall time of 1100 ns. In case of a kicker failure, all circulating bunches can be affected, leading to unacceptable losses. From a machine protection point of view, a system with rise and fall times over several turns is preferred, as it will allow the machine protection system to react in case of failures. The option that the injection bump is present over many turns will need to be counterbalanced against arguments related to machine and dynamic aperture and also the beam impedance.

The extraction to the dump must be highly reliable, as it must function without failure in the event of any issue arising in the machine. Fast extraction kickers are planned, with a rise time of 0.6 μs. The rise time of the kicker magnets must be precisely synchronised with the passage of the particle-free beam abort gap. The reaction time between a beam abort request and the full extraction of the beam is critical due to the rapid nature of potential failure scenarios. Therefore, it is anticipated that multiple beam abort gaps will be necessary.

The impact of failures in many injection and extraction related systems can be mitigated by segmenting them, for example, by dividing the injection kicker systems into multiple smaller, independent units. Additionally, the likelihood of failure can be minimised through redundancy and highly reliable designs.

To provide an additional layer of protection, multiple absorbers or passive diluters will be required as a final safety measure to shield septa and other machine components from fast failures of the injection and extraction pulsed magnets. These elements will need to be studied in further detail.

A preliminary list of typical fast system failures has been compiled, along with possible protection measures. However, a more comprehensive study is still required, incorporating the detailed design of the hardware systems involved.

- **Collider electromagnetic separators**
  Electromagnetic separators are required upstream and downstream of the RF section to separate the beam in $\mathrm{t\bar{t}}$ mode, as both beams will pass through the centre of the same cavities. Possible failures include spontaneous discharges of the electrodes, a mismatch between the magnetic and electric field in the longitudinal plane and synchrotron light shining on the HV electrodes, inducing discharges and resulting in very fast and large kicks to the beams. These failures can possibly be detected by using very fast interlocked beam position monitors and dedicated masks for the synchrotron light.
- **Powering failures**
  First analysis of a powering failure of one of the main dipole chains, shows that the beam can reach the horizontal limiting aperture in only a few turns. This will be very challenging for the machine protection system, even if the failure can be rapidly detected by the use of a fast magnet current change monitor (FMCM), described below.
- **Beam-dust interaction**
  Beam loss events due to the interaction of dust grains with high-energy particle beams have been observed in hadron and lepton accelerators over many decades. In the LHC, beam-dust interactions have an important detrimental impact on machine availability. They caused more than 70 premature beam dumps and more than ten quenches of superconducting magnets during Run 2 and 3. At the electron-positron collider SuperKEKB, a significant amount of beam aborts related to beam-dust interactions were observed and reproduced using a mechanical knocker. In recent years, very fast beam losses occurred at SuperKEKB, causing damage to machine components. Presently, the root cause of this phenomenon remains unknown. However, beam-dust interactions have been hypothesised as a possible explanation. While the dynamics of the dust movement has been studied in detail for the LHC, these models have to be extended to lepton colliders to evaluate the effect of beam-dust interactions on the protection, availability, and performance reach of FCC-ee.
- **Superconducting magnets**
  A quench protection system will be required for the superconducting final focus magnets. A detailed description of this system is provided in Section 3.11.
  The impact of beam-related failures on the superconducting magnet structures has not yet been studied. However, such failures are not expected to be critical due to the relatively long time constants of superconducting circuits and the limited number of superconducting magnets.
  The effects of the quench protection system on the beam, including potential erratic firing of quench heaters or CLIQ systems, must be modelled in detail.

***Machine protection system elements***

The machine elements listed below are part of the machine protection system and prevent the machine to be damaged by the beams.

- **Machine interlock systems** The dedicated hardware related to machine protection is described in Section 8.2. As in the current large CERN accelerators, the core of the machine protection system relies on the Beam Interlock System (BIS). This system connects user inputs that detect failures or beam losses in the machine to the beam extraction systems. The Beam Interlock System (BIS) must be highly reliable and, for that reason, is fully redundant. Due to the large machine circumference, a significant delay occurs between detecting a beam abort request and triggering the beam extraction.
  The cooling of the synchrotron radiation absorbers will need to be interlocked. This interlocking can be based on thermal switching and cooling water flow measurements. Given the high segmentation of the synchrotron radiation absorbers, a cost-effective method for identifying locations with excessive temperature or reduced cooling flow must be developed, as running individual cables to numerous controllers would be too expensive.

The main dipole circuits, along with other primary magnet circuits, will be water-cooled. While the impact of a loss of water cooling requires further analysis, initial assessments suggest that it is not highly critical. Water cooling is primarily chosen due to its significantly higher energy efficiency compared to air cooling. If the thermal time constants in the absence of cooling are of the order of days, it may be possible to operate without an interlock on magnet cooling, instead relying solely on tunnel temperature monitoring. In such a scenario, temperature measurements would need to be highly granular to accurately locate potential cooling issues.

The need of a fast magnet current change monitor (FMCM) can be determined through an analysis of powering failures across different systems. Strong, individual magnets with short time constants may require FMCM protection. Preliminary analyses indicate that failures in the main magnet circuits pose significant challenges, even when rapidly detected by an FMCM.

– **The beam extraction system** must operate with extreme reliability to ensure the beam can be aborted whenever required by the machine interlock system. Any degradation in the beam-dumping system, such as a reduction in redundancy or safety margins, must be detected early enough to allow safe beam disposal without losses. For this reason, the beam extraction system also functions as a machine protection system user.

  The number and length of abort gaps are critical system parameters that influence the RF system and the reaction time to a beam dump request. Since beam stability is maintained through collisions, the dumping of both beams must be synchronised. The specifics of this synchronisation require further study.

– **The collimation system** is described in detail in Section 1.5. In addition to cleaning the beam, it plays a crucial role in the machine protection system by safeguarding the machine aperture against irregular and accidental beam losses. The collimators must act as the aperture bottleneck, intercepting all primary losses. Depending on the analysis of failure scenarios, a distributed collimation system may need to be considered. This approach would represent a new regime compared to previous lepton colliders.

  Particular attention is required to prevent beam-induced damage to the collimators. Fast failures have been observed at SuperKEKB, where a beam with 100 times less stored energy caused damage to the collimators within just 2 to 3 turns. For this reason, fast and sensitive beam loss measurements, integrated with the beam interlock system, are essential at the collimator locations.

– **Beam Instrumentation**

  Ideally, any machine failure should be detected at the source, such as a power converter failure or an incorrect collimator position. A global safety net based on beam measurements is highly recommended to enhance reliability and, in the case of beam instabilities, may serve as the only protection. For this reason, highly reliable beam instrumentation systems, integrated with the beam interlock system, are essential for the safe operation of the accelerator.

  Historically this protection is based on a distributed system of fast beam loss monitors (BLMs), like in the LHC. A scaling of the number of BLMs as presently installed in the LHC with the FCC-ee circumference would lead to a total of 12’000 BLMs. Even larger numbers of BLMs have been proposed in initial studies. The BLM electronic systems will need to be extremely reliable to limit the number of false dumps initiated by the BLM system.

  Due to the extremely small vertical beam size, detecting the onset of instabilities in the vertical plane is unlikely to be feasible using beam loss monitors. Instead, the use of fast, interlocked beam position monitors must be studied in detail.

  A third layer of protection is the beam current change monitor, which aborts the beams upon detecting a reduction in measured beam current. Further investigation is required to determine whether this system is adequate for protecting against certain fast failure modes.

***Machine protection system strategy***

Several FCC-ee failure modes with time scales of only a few turns have been identified. The reaction time of the machine protection system can be slightly shortened by filling patterns with multiple abort gaps. The detection of the onset of failure modes will have to be optimised to detect any malfunctioning of equipment or the onset of instability as early as possible. As the vertical beam size is small, one can most likely not rely on interlocking on beam losses in the vertical plane, since the limiting apertures will be far away from the beam expressed in numbers of sigma. Fast interlocked beam position monitors will most likely need to be distributed along the accelerator.

Protection against fast failures is achieved through redundancy, segmentation and highly reliable designs. Also fixed absorbers are part of the protection against fast failure modes. These absorbers are challenging due to the high beam brightness and shower development. Where no suitable materials can be found, a possible remedy would be the use of disposable absorbers for rare failure cases.

A full and complete analysis of failure modes of all FCC-ee components requires a more advanced hardware design. This will allow a detailed definition of mitigation steps, including the detection systems for the various failures. At this stage in the machine protection study, not a single showstopper has been identified.

# Chapter 3

# FCC-ee collider technical systems

## 3.1 Main magnets

### 3.1.1 Introduction

The proposed collider magnet system meets the requirements of the FCC-ee V24.3 GHC optics. The magnets have been designed to operate below iron saturation at $\mathrm{t\bar{t}}$, so that they behave linearly at all intermediate energy levels, Z, WW and ZH.

The regular FODO lattice of the collider ring features 2840 arc half-cells composed of a short straight section (SSS) followed by a series of dipole magnets. Depending on the position along the lattice, the SSS hosts none, one or two sextupole magnets. Where there is either no sextupole or only one sextupole, the total dipole length is adapted to occupy the space, maximising the dipole filling factor in the machine.

The dipole and quadrupole magnets are designed as twin aperture units, as was already proposed in the CDR [13].

Compared to using separate magnetic elements, the twin-aperture solution significantly reduces power requirements by half, as the return ampere-turns in the coils supply the second aperture. Since the dipoles and quadrupoles are powered in long series, they include locally powered trim windings that enable independent adjustments of the magnetic field in each aperture. This allows fine-tuning, corrections, and compensation for local beam energy losses caused by synchrotron radiation (SR).

The requirements for the arc magnets are based on the FODO lattice described in Section 1.2 and summarised in Table 3.1. The total number of magnets in the lattice accounts for the arcs and long strait sections (LSS).

Table 3.1: Magnet requirements for the FODO lattice V24.3 GHC.

| | Dipole | Quadrupole | Sextupole |
|---|---|---|---|
| Total number in lattice... | 6128 | 3324 | 4672 |
| ... of which in the arcs | 5680 | 2836 | 4672 |
| Bore aperture | 74 mm | 74 mm | 66 mm |
| Magnetic length | 9.7 - 11.2 m | 2.9 m | 1.3 m |
| Max strength†, arc ($\mathrm{t\bar{t}}$, 182.5 GeV) | 61.0 mT | 11.9 T m$^{-1}$ | 880 T m$^{-2}$ |

† Sextupole strength given as $B''(B'' = 2S)$

### 3.1.2 Dipoles

The dipole yoke features an I-shape geometry corresponding to two back-to-back C-shape dipoles, with a common powering circuit around the central part. It allows a compact design which can be built in a cost-efficient way. The assembly is composed of three parts that can be machined out of solid iron (low-carbon steel) plates. As all three parts have a simple rectangular cross-section, the amount of machining can be minimised by selecting raw material plates in a thickness close to the final dimensions.

Currently, it is planned to split the dipole section of each arc half-cell into two dipole units to minimise the number of interconnections while keeping a feasible and practical unit length for the manufacture and transport of the magnet components and vacuum chambers.

The use of solid iron material is possible since the magnets will be DC-powered during operation. Compared to a laminated assembly, this also helps to maximise the magnet stiffness limiting its natural sag, in particular for a 12-m-long magnet with low second moment of area.

The dipole is powered by two busbars made of extruded aluminium, one in each aperture. The ground insulation around the busbars has to be radiation-hard due to the energy deposited by the synchrotron radiation. There are several options, such as inorganic coatings (e.g., hard anodisation), using the surrounding air only, or a combination of both. The separation between the busbars and the surrounding components - magnet yoke, vacuum chamber and SR shielding - can be ensured by ceramic spacers placed at regular intervals along the magnet. The layer of air insulation has to be adapted to the peak voltage of the circuit, which in the case of the collider dipoles is very low (only 210 volts), as a consequence of the low current density, the DC powering mode and the low resistance of the single busbars. Therefore, spacers between 3 and 5 mm thick are compatible with relaxed tolerances for manufacturing the busbars at low cost and limiting the number of spacers along the magnets. A similar insulation technology can also be used for the field tapering trim conductors wound around each pole.

The total field harmonics simulated in 2D are below $1 \times 10^{-4}$ relative field error at the reference radius of $10\,\mathrm{mm}$.

The general parameters of the dipole are summarised in Table 3.2.

The field map in the dipole cross-section at $\mathrm{t\bar{t}}$ operation (peak field) is shown in Fig. 3.1.

Table 3.2: General parameters of the arc dipole magnets.

| Parameter | Unit | Value |
|---|---|---|
| Strength, $B$, $45.6 - 182.5\,\mathrm{GeV}$ | mT | 15.2 - 61.0 |
| Bore aperture | mm | 74 |
| Magnetic length | m | 9.7 to 11.2 |
| Outer envelope | mm | $520 \times 133$ |
| Peak current | A | 3665 |
| Magnet resistance (at 32°C op. temp.) | mΩ | 0.27 |
| Peak voltage, magnet | V | 0.98 |
| Peak voltage, half-octant (incl. busbars) † | V | 420 |
| Conductor (Aluminium) | $\mathrm{mm}^2$, mm | $65 \times 35$, $\varnothing 7.7$ |
| Turns (busbar) | - | 1 |
| Turns per coil (trim) | - | 7 |
| Current density (busbar), $\mathrm{t\bar{t}}$ | $\mathrm{A/mm^2}$ | 1.61 |
| Temperature rise (5 bar) | °C | 14.5 |
| Yoke active mass | kg | 2621 |
| Busbar active mass | kg | 131 |
| Trim coil active mass | kg | 6.7 |
| Magnet active mass | kg | 2909 |

† One half-octant corresponds to one series circuit, which includes 355 magnets. The circuit voltage will be balanced around the middle point of the circuit so that the voltage to ground will not exceed half the circuit voltage.

### 3.1.3 Quadrupoles

The quadrupole magnet is designed as two joined figure-of-eight units, which allows powering the two apertures with only two simple racetrack coils. This approach minimises the production costs but imposes opposite polarities seen by the $e^+$ and $e^-$ beams, as they travel through the magnet, a constraint

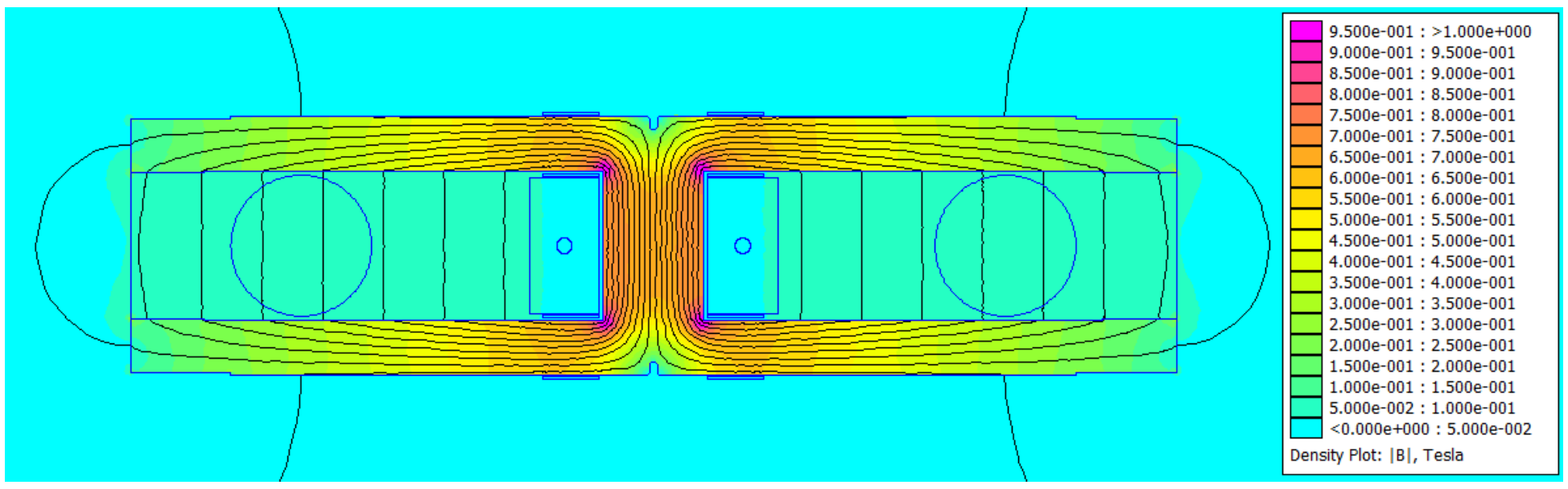

Fig. 3.1: Field map in the dipole cross-section at $t\bar{t}$ operation.

which has been included in the design of the beam line optics from the early stages of the study.

For cost reasons and due to the complex shape of the poles, a laminated construction of the yoke parts is planned. Each of the top and bottom yokes is split in two parts to allow the integration of the racetrack coils. The symmetry of the assembly is controlled by the tight tolerances on the mating faces and a system of V-grooves and pins for precise relative localisation of the assembled parts. Non-magnetic spacers are placed between the half-yoke assemblies to control the top-bottom pole symmetry.

The twin-aperture double figure-of-eight configuration generates coupling between the apertures due to the left-right asymmetry of the magnetic circuits of each aperture. This translates into a shift of the magnetic at different powering levels (i.e., between different operation phases of the machine). The geometry of the magnetic circuit has been designed with straight poles where the racetrack trim coils for the field tapering circuits have been placed. This allows streamlining of the magnetic flux, and minimising the coupling so that the axis shift can be compensated by using adjacent horizontal orbit correction circuits.

The total field harmonics as simulated in 2D are below $2 \times 10^{-4}$ relative field error at the reference radius of $10\,\mathrm{mm}$ when only the main coils are powered. When, in addition, the field tapering trim coils are activated to the peak value of + 3.5 percent of the main field on one aperture and - 3.5 percent of the main field on the other aperture, the b1 relative field harmonic is up to $12 \times 10^{-4}$, and the b3 relative field harmonic is up to $1.5 \times 10^{-4}$, due to the cross-talk between apertures and the asymmetry created by the opposite field differences. These errors can be compensated by the adjacent horizontal orbit correctors for the b1, and by the adjacent lattice sextupoles for the b3.

The parameters of the quadrupole magnet are summarised in Table 3.3. The field map in the quadrupole cross-section at $t\bar{t}$ operation (peak field) is given in Fig. 3.2.

### 3.1.4 Sextupoles

The sextupole magnet is designed with a conventional geometry of 6 symmetrical poles and hosts trim windings, which can be used for horizontal and vertical orbit correction, as well as to generate a skew quadrupole component like is done in synchrotron light sources. Although the sextupole pairs are not magnetically coupled between beam lines, they will be assembled as a single mechanical unit, either by fixing them on a common crib, or by merging the adjacent poles in a single piece lamination. Both solutions will be studied during the pre-TDR phase to determine the most cost-effective. The assembly of sextupole pairs will also minimise the fiducialisation and alignment operations during the construction and installation phases.

The magnet design is relatively compact since the sextupole pairs have to fit the 350 mm intra-beam distance. Consequently, the flux density in iron and current density in the conductors are high. The magnet operates below saturation at peak field, as the flux density does not exceed 1.6 T. This choice allows the trim circuits to be operated in a linear regime at all field levels.

The simulated 2D total field harmonics of the main circuit of the sextupole are below a $1 \times 10^{-4}$

Table 3.3: General parameters for the quadrupole.

| Parameter | Unit | Value |
|---|---|---|
| Strength, $B'$ | $\mathrm{T\,m^{-1}}$ | 11.8 |
| Bore aperture diameter | mm | 74 |
| Magnetic length | m | 2.9 |
| Overall width x height | mm | 590 x 610 |
| Peak current, $\mathrm{t\bar{t}}$ | A | 366 |
| Magnet resistance (at 35°C op. temp.) | mΩ | 51.3 |
| Peak voltage magnet | V | 18.8 |
| Peak voltage, half-octant (incl. cables) † | kV | 1.91 |
| Conductor (copper) | $\mathrm{mm^2}$, mm | $14.4 \times 14.4$, ⌀7.5 |
| Turns per coil (main) | - | 36 |
| Turns per coil (trim) | - | 26 |
| Current density, $\mathrm{t\bar{t}}$ | $\mathrm{A/mm^2}$ | 2.25 |
| Temperature rise (5 bar) | °C | 19.0 |
| Yoke active mass | kg | 5789 |
| Main coil active mass | kg | 334 |
| Trim coil active mass | kg | 9.7 |
| Magnet active mass | kg | 6535 |

† One half-octant corresponds to one series circuit (focusing or defocusing), which includes 89 magnets. The circuit voltage will be balanced around the middle point of the circuit so that the voltage to ground will not exceed half the circuit voltage.

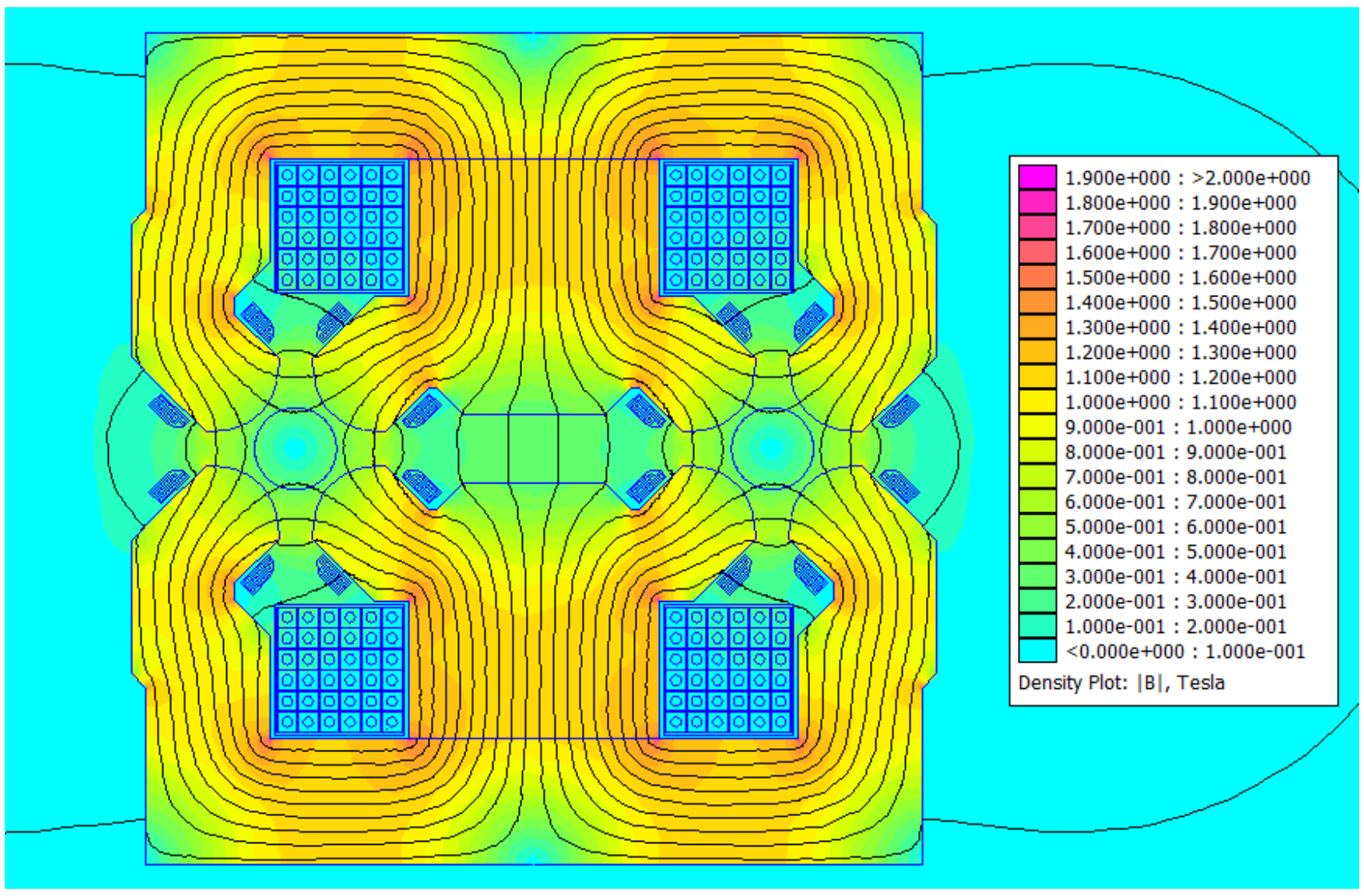

Fig. 3.2: Field map in the quadrupole cross-section at $\mathrm{t\bar{t}}$ operation.

relative field error at the reference radius of $10\,\mathrm{mm}$ when only the main coils are powered.

The field quality of the correction circuits is naturally low, as expected since the field is generated by the six poles of the sextupole, which are significantly different from the ideal lines of the constant scalar potential of each correction field. Both the b5 field harmonic (relative to B1) of the horizontal orbit correction circuits and the a5 field harmonic (relative to A1) of the vertical orbit correction circuits reach $60 \times 10^{-4}$. For the skew quadrupole circuit, the a4 field harmonic (relative to A2) reaches $775 \times 10^{-4}$. The effect of these field errors on the beam dynamics is being evaluated to understand if the solution to embed the correction circuits in the sextupole is viable. Otherwise, separate corrector magnets will have to be designed and integrated in the lattice. They would be relatively short since the field strengths of these correctors are relatively small.

The parameters of the sextupole magnet are summarised in Table 3.4. The field map in the sextupole cross-section at $\mathrm{t\bar{t}}$ operation (peak field) is given in Fig. 3.3.

Table 3.4: General parameters for the sextupole main circuit.

| Parameter | Unit | Value |
|---|---|---|
| Strength, $B''$ | $\mathrm{T\,m^{-2}}$ | 880 |
| Bore aperture diameter | mm | 66 |
| Length | m | 1.3 |
| Overall width x height (one beam) | mm | 350 x 350 |
| Peak current, $\mathrm{t\bar{t}}$ | A | 178 |
| Magnet resistance (at 35°C op. temp.) | mΩ | 274 |
| Peak voltage magnet | V | 49 |
| Peak voltage, circuit (incl. cables) † | V | 284 |
| Conductor dimensions (copper) | $\mathrm{mm^2}$, mm | $6.15 \times 6.15$, ⌀4.0 |
| Turns per coil | - | 24 |
| Current density, $\mathrm{t\bar{t}}$ | $\mathrm{A/mm^2}$ | 7.0 |
| Temperature rise (6 bar) | °C | 20 |
| Yoke active mass | kg | 498 |
| Main coil active mass | kg | 14.6 |
| Magnet active mass (incl. trim coils) | kg | 635 |

† One circuit corresponds to 8 magnets. The circuit voltage will be balanced around the middle point of the circuit so that the voltage to ground will not exceed half the circuit voltage.

The parameters of the sextupole correction circuits are summarised in Table 3.5. The field maps for the sextupole correction circuits at peak correction field are given in Fig. 3.4.

### 3.1.5 Coils and busbars

The conductor materials, operational current densities, and number of turns of the coils and busbars have been selected based on global optimisation of the lifetime cost of the magnet system, in combination with other systems in the machine, in particular the electrical distribution, power converters, and technical infrastructure. This optimisation has balanced the capital versus operational costs of the magnet system considering the powering needs and duration of each phase of the machine (from Z to $\mathrm{t\bar{t}}$), the integration constraints and dissipated power of the cabling in the tunnel, and the integration of the power converters in the alcoves, as to minimise the total cost of all these systems over the lifetime of the machine. As a result, the quadrupole and sextupole coils use copper conductors since they require a larger current density, whereas the dipole busbars use aluminium. Since aluminium and copper cannot be cooled by water in a common cooling circuit due to galvanic corrosion, it is considered to cool the busbar with a 1 mm thick copper tube embedded in the aluminium bulk. This solution avoids the need of a dedicated demineralised water-cooling network for aluminium. The global optimisation tool will be further developed during the pre-TDR phase also to include the cooling network costs. More details on this global optimisation can be found in Section 3.8.1.

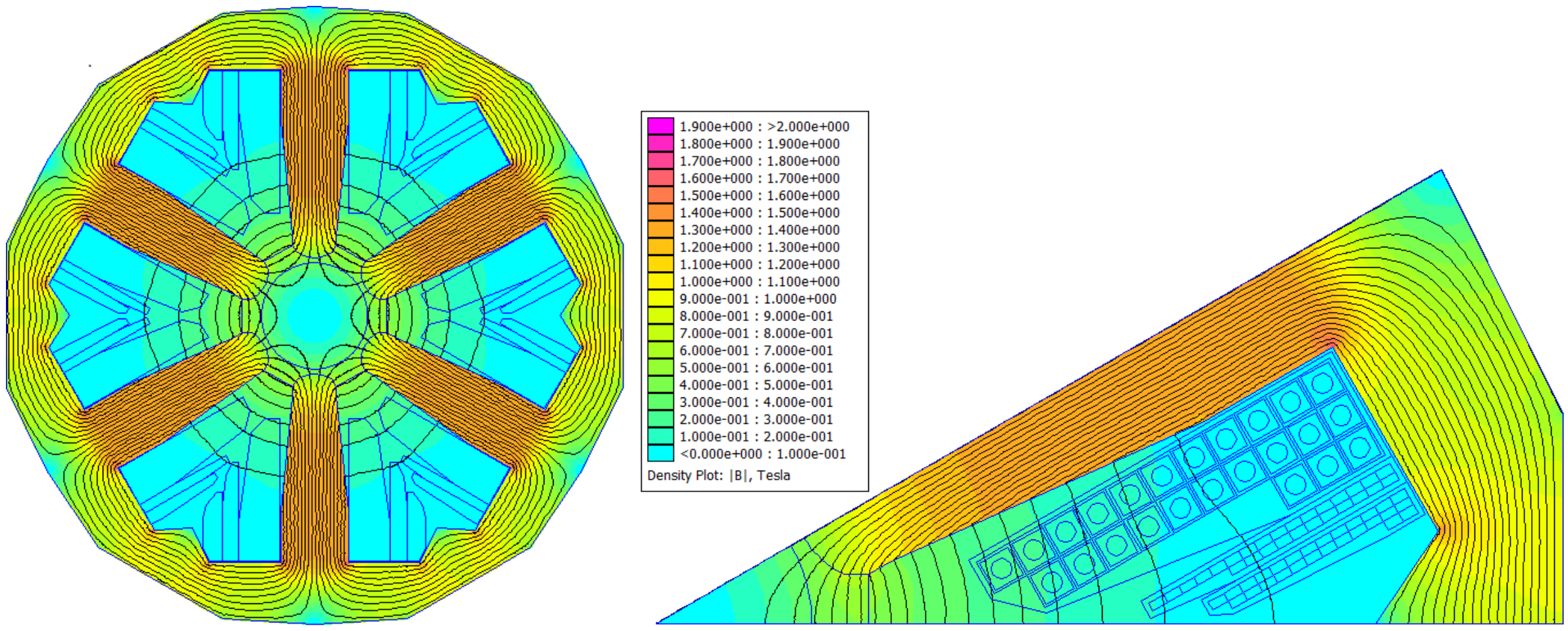

Fig. 3.3: Left: Field map in the sextupole cross-section at $t\bar{t}$ operation (peak field). Right: Detailed view of a half-sextant. The conceptual positioning of the conductor has been generated from parametric modelling for checking integration feasibility. It will be optimised for industrial production during the pre-TDR phase.

Table 3.5: General parameters for the sextupole correction circuits.

| Parameter | Unit | H orbit | V orbit | Skew quad |
|---|---|---|---|---|
| Strength, $B.l$ | mT m | 20 | 20 | – |
| Strength, $G.l$ | mT | – | – | 600 |
| Turns per pole | - | 54-27 | 27 | 43 |
| Peak current | A | 8.3 | 14.3 | 9.9 |
| Resistance per magnet | Ω | 2.15 | 1.06 | 0.84 |
| Peak voltage per magnet | V | 17.8 | 15.2 | 8.4 |
| Conductor dimensions (copper) | $mm^2$ | 3.2 x 1.6 | 3.2 x 1.6 | 3.2 x 1.6 |
| Copper mass per magnet | kg | 25.9 | 12.9 | 10.4 |

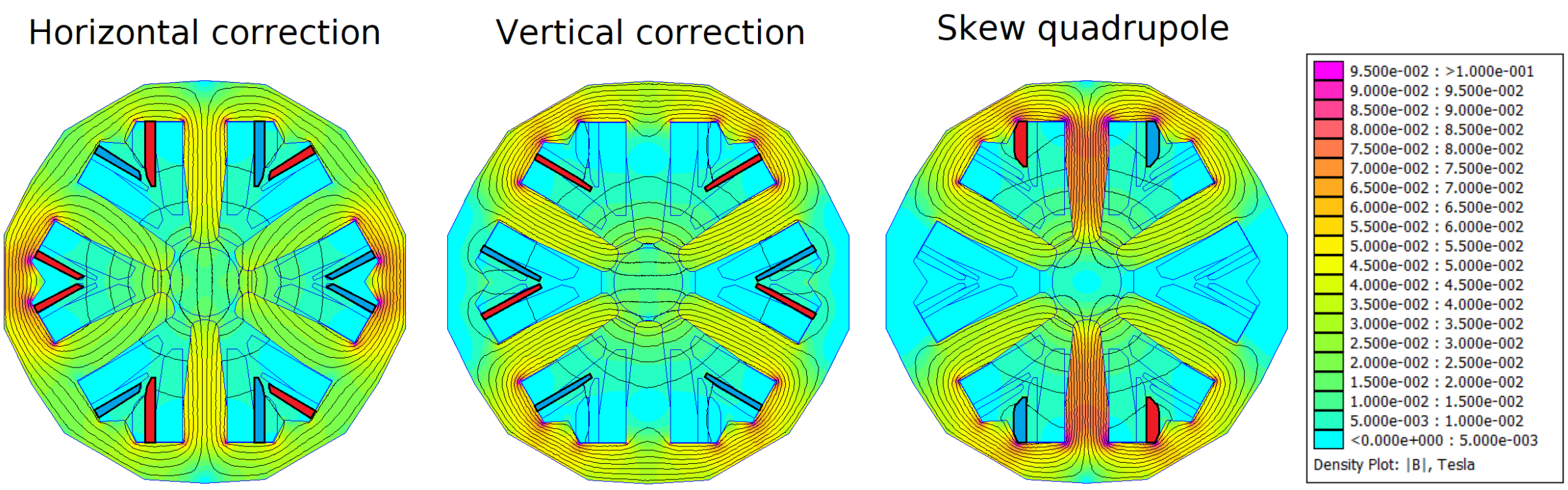

Fig. 3.4: Field maps of the correction circuits at peak correction fields. Left: horizontal orbit correction; middle: vertical orbit correction; right: skew quadrupole correction.

### 3.1.6 Supporting structures

***SSS units***

The magnets in the short straight sections (SSS) - quadrupoles and sextupoles - are supported by a common girder where they are pre-aligned and fixed, to guarantee the stability of their relative positioning and ease the global alignment of the beam line elements in the tunnel. The pre-assembly of the SSS magnets is also necessary to install a common vacuum chamber across them and minimise the number of vacuum interconnections.

There are three versions of the Short Straight Sections (SSS), depending on whether they contain no sextupoles, one sextupole, or two. Two configurations are under consideration for the arc half-cell layout. In the first option, the lengths of the SSS girder and dipoles are adapted to the number of sextupoles in the arc half-cell, resulting in three different variations of dipole and SSS lengths. In the second option, all SSS girders are manufactured at a uniform length, corresponding to the longest version capable of housing two sextupoles. When fewer than two sextupoles are required, a short dipole of equivalent length replaces the missing sextupole.

The second option offers the advantage of standardising SSS and dipole lengths across the arcs, along with their support structures. This approach also provides greater flexibility in redistributing SSSs if optical adjustments are needed during different operational phases of the machine. However, it introduces the drawback of requiring additional interconnections for SSSs that do not contain two sextupoles, compared to the first option, leading to a slight reduction in the dipole filling factor.

Both options are technically feasible and will be evaluated in detail during the pre-TDR phase of the project, taking into account functionality, machine performance, and cost.

***Dipole units***

Similar considerations apply to the dipole assemblies as to the SSS units. They will be pre-assembled at the surface, incorporating their vacuum chambers and synchrotron radiation (SR) shielding blocks, before being transported, installed, and aligned in the tunnel as individual units.

The dipole supporting scheme must ensure both precise alignment functionality and the minimisation of magnet sag due to its weight and the additional mass of shielding blocks surrounding the SR absorbers, which are distributed approximately every five to six metres. This aspect is particularly critical, as the dipole magnets have a flat and elongated aspect ratio, resulting in relatively low mechanical inertia that makes them susceptible to deformation. Consequently, the supporting scheme must be designed to align with the longitudinal distribution of the SR shielding along the dipoles, ensuring structural stability and optimal performance.

More information on the supporting structures can be found in Section 3.10.

## 3.2 Vacuum system and electron cloud mitigation

### 3.2.1 Introduction

The FCC-ee vacuum system design is rather challenging due to its sheer size and the fact that it must cope with a synchrotron radiation (SR) environment that changes significantly depending on the beam energy.

The Z machine at 45.6 GeV is characterised by an SR spectrum with a critical energy of only 21 keV, while the highest energy $t\bar{t}$ machine at 182.5 GeV has a critical energy of 1.35 MeV. The Z machine has a very large beam current, around 1.4 A, leading to a significant photon-stimulated desorption (PSD) rate. Since nearly all SR photons are absorbed in a very thin layer of the vacuum chamber walls, they contribute locally to the PSD dynamic gas load.

The $t\bar{t}$ SR fan, on the other hand, penetrates deeply into the material of the vacuum chamber where it generates high-energy Compton showers, giant dipole resonance, and particle creation, necessitating a thorough analysis of the related radiation field in the tunnel. Efficient solutions must be implemented to mitigate detrimental effects on accelerator and tunnel components, particularly shielding cables and electronics in the tunnel, which has proven to be a challenging task.

The analysis and modelling carried out so far by the vacuum group have taken into account concerns raised by other groups, such as:

- Machine physics group (e.g., beam-stay clear aperture, impedance contributions of all vacuum components)
- Magnet group (e.g., integration of vacuum chamber cross section and SR shielding within the magnets' geometries)
- FLUKA team (e.g., SR shielding, radiation protection, and related R2E issues)
- Tunnel integration and logistics (e.g., total power dissipation in air and cooling water in the tunnel arcs)
- Cost estimation and machine installation planning

The beam parameters are chosen to maintain an SR power of 50 MW per beam across all beam energies. Consequently, the linear SR power in the arcs is approximately 650 W/m, regardless of beam energy.

This approach would also help contain high-energy Compton-scattered secondaries once the beam energy is increased to 182.5 GeV later in the experimental programme. Additionally, the associated synchrotron radiation (SR) power is concentrated on the absorbers rather than being distributed throughout the entire vacuum chamber, improving thermal management and overall system efficiency.

The design of the FCC-ee machines, including the full-energy booster, have provided valuable insights into the challenges ahead. In particular, the selection of suitable materials for the vacuum chamber and the adoption of a lumped SR absorber design have emerged as effective solutions to several critical issues related to vacuum performance and machine physics.

The following sections outline the proposed design and the analysis conducted thus far.

### 3.2.2 Pressure Requirements

The FCC-ee is a powerful synchrotron radiation (SR) source, at all beam energies. From the point of view of the photon-stimulated desorption (PSD) gas load, the Z machine is the toughest one to deal with, since it has the largest beam current. Basically, any photon in the Z SR spectrum below its critical energy (i.e., 91% of the total SR photon flux) can generate photoelectrons which are emitted within a very thin layer of the internal surface of the vacuum chamber.

It is therefore necessary to maximise the pumping efficiency and minimise the PSD gas load, while finding ways to speed up the vacuum conditioning as much as possible, so that vacuum quickly becomes good enough to store nominal currents at the Z energy. The photoelectrons are very efficient at desorbing molecules inside the vacuum and contribute to the electron cloud in the positron ring.

The pressure requirements, which stem mainly from beam-gas scattering arguments, mean that the vacuum lifetime, $\tau_{\text{vac}}$, should be much longer than the lifetimes of all other effects, such as Bhabha scattering, collisions burn up, beam-thermal photon scattering, etc.

Extensive experience for similar SR-dominated storage rings and colliders, e.g., SuperKEKB and light sources, shows that an average pressure in the low $10^{-9}$ mbar range, nitrogen-equivalent, should give a $\tau_{\mathrm{vac}}$ of several tens of hours.

### 3.2.3 Physical Aperture Requirements

Based on geometric impedance arguments, a circular cross section of the vacuum chamber was requested at the beginning of the FCC-ee study. The internal diameter compatible with the design of the magnets and other machine components and diagnostics had initially been set at 70 mm. Later in the study, a reduction of the inner diameter (ID) of the magnet yokes became necessary, and the ID of the vacuum chamber was reduced to 60 mm.

The application of non-evaporable getter (NEG) coating makes this ID reduction irrelevant for vacuum since NEG coatings generate a rather uniform distributed pumping, contrary to a lumped pumping design, which would need maximisation of the vacuum chamber conductance. A reduction from 70 to 60 mm would mean a 37% loss of conductance loss (ratio of IDs cubed). All design and modelling are therefore based on a 60 mm ID vacuum chamber.

The vacuum chamber is made of oxygen-free silver-bearing (OFS) copper and can be extruded up to 12 m. The cross section is shown in Fig. 3.5. The chamber has two appendages, called winglets, in the horizontal plane to accommodate short, localised synchrotron radiation absorbers (SRAs) and pumping slots. These do not protrude inside the circular part of the chamber.

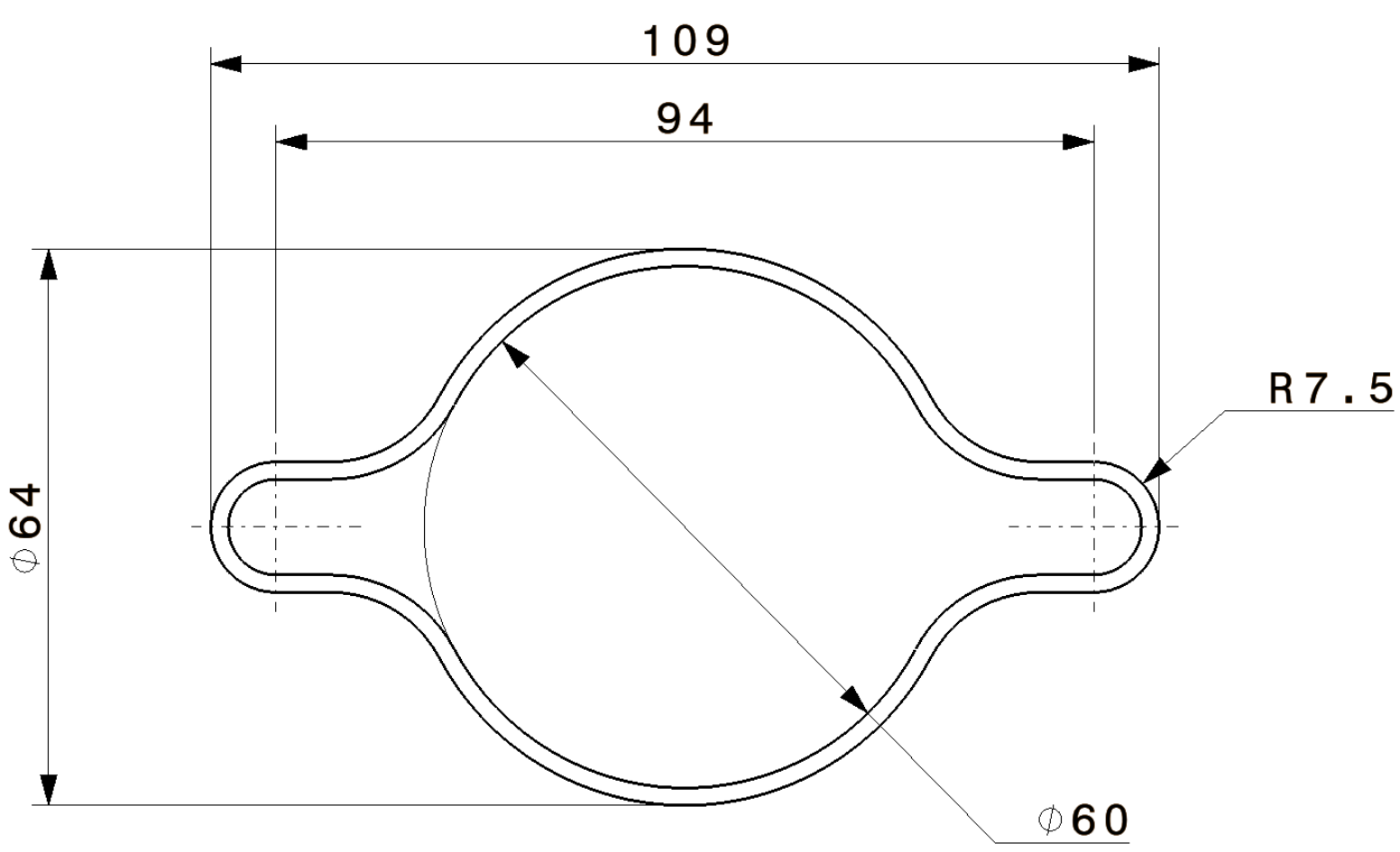

Fig. 3.5: Cross section of the FCC-ee arc vacuum chamber with winglets.

### 3.2.4 Pumping System

NEG-coating will be applied throughout the vacuum chamber. There have been tests at a KEK Photon Factory beamline of the effectiveness of 'thin' NEG-coatings. It was found that the NEG-coating is still capable of being fully activated after 10 cycles of saturation and re-activation [239] for thicknesses down to 200 nm. For comparison, the standard NEG-coating thickness of the long straight sections of the LHC is 2 µm. The reduction to 200 nm is compatible with requirements of the resistive-wall impedance.

One more advantage of the application of NEG-coating is that ∼90% of its PSD outgassing comprises $H_2$ , and the favourable effect on the beam-gas scattering of a dominant $H_2$ becomes important.

The modelling and analysis carried out throughout the FCC study period have demonstrated that the application of NEG-coating to 100% of the vacuum chamber's internal surface will drastically reduce the need for lumped pumps, therefore minimising the number and length of the expensive control and power cables for such pumps.

However, a few lumped pumps are needed every 30-50 m to pump non-getterable gases (methane and inert gases such as He, Ar and Kr), although $CH_4$ is effectively pumped by beam ionisation as soon as $e^-$ or $e^+$ are injected. The lumped pumps which have been identified as suitable are integrated NEG-ion pumps. The readout of the 12 l/s ion pump will be used to get an estimate of the pressure at the pump's location by conversion of its ion-pump current. This methodology is both effective and current, enabling the early detection of developing leaks, which are typically identified through changes in pressure, particularly under static vacuum conditions where no beam is stored. This is a crucial operational consideration, as locating leaks in a machine as large as the FCC-ee can be challenging and time-consuming, as previously experienced during LEP operation. By facilitating early leak detection, this approach helps to minimise machine downtime and improve overall operational efficiency.

### 3.2.5 NEG-Coating

As mentioned, detailed measurements have proved that a 'thin' NEG-coating can be envisaged for the FCC-ee collider rings. A thickness of 200 nm is deemed ideal [240].

The major drawback of the NEG-coating solution is that a reliable, radiation-resistant, in-situ bake-out system is needed. Prototyping of a solution based on a thin titanium track sandwiched between two plasma-sprayed ceramic insulation layers (see Fig. 3.6), gave a rather uniform vacuum chamber temperature profile and reasonable electric power needs. The latter mainly depends on the efficiency of the radiation-resistant thermal insulation layer, which is under study now. Cost-efficient optimisation of this heating system is underway.

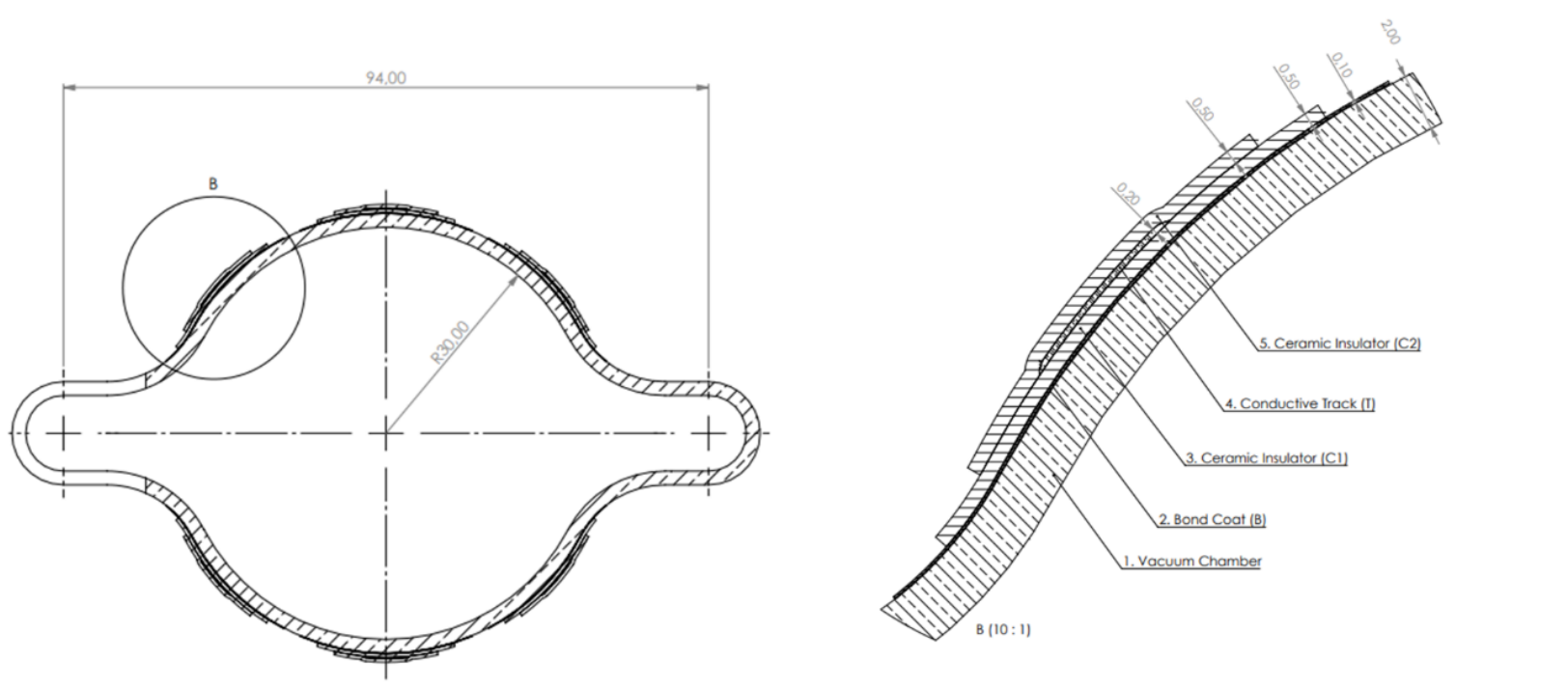

Fig. 3.6: Bake-out system based on cold-sprayed conductive tracks.

The detailed design of horizontal NEG-coating benches has not yet been done. However, experience with a similar coating setup developed for the HL-LHC stand-alone magnets provides confidence that such a system can be implemented. The proposed approach involves a horizontally moving carriage equipped with a permanent magnet magnetron trap, which would travel along the axis of the chambers, up to 12 m in length, depositing NEG coating in localised sections of approximately 20 cm at a time.

Transfer of this CERN-developed technology to industry for mass production for the two 91 km rings will be necessary.

### 3.2.6 Vacuum Profiles

Extensive modelling of the pressure profile along a representative length of the FCC-ee arcs has been carried out. The comparison between a NEG-coating-based solution and one based on lumped pumps shows that the former is much more efficient at reducing the average pressure along the arcs, and it conditions a long beam-gas scattering lifetime faster than the latter. The benefits of using the SRA and the NEG coatings can be seen in Figs. 3.7 and 3.8.

Extensive data from existing light sources, some with 100% NEG-coating (MAX-IV and SIRIUS) and some without NEG-coating, makes it clear that NEG-coating allows an almost immediate availability of long vacuum lifetimes, at least for the Z energy machine, which is the most demanding in vacuum terms.

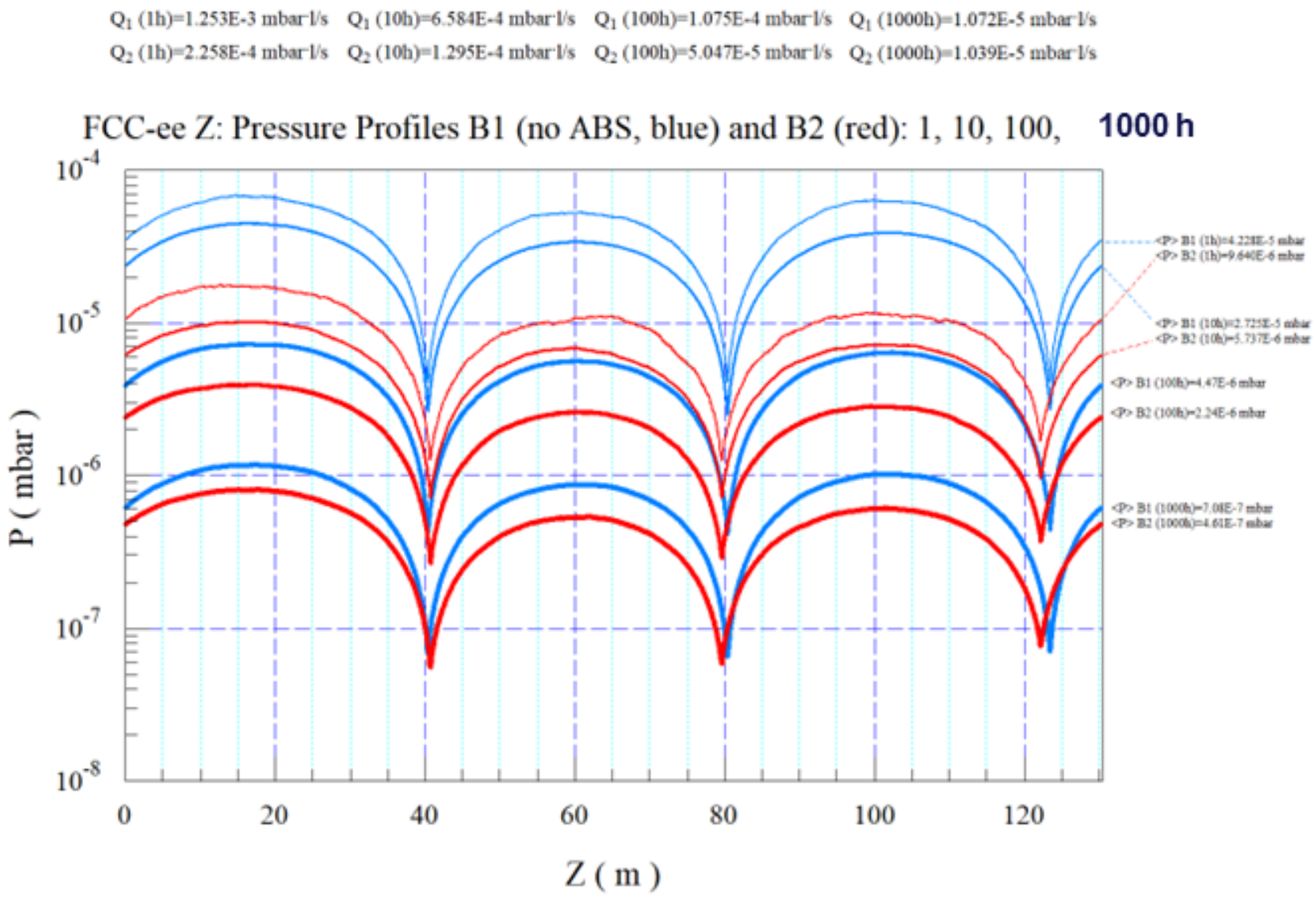

Fig. 3.7: The pressure profiles assuming no NEG-coating are shown for different vacuum conditioning beam doses of 1, 10, 100, and 1000 Ah. One of the two beams has lumped SRAs, the other one does not. It is evident that for each beam dose, the pressure profile for the case with SRAs gives a lower pressure and, therefore, an advantage in terms of vacuum commissioning time.

### 3.2.7 Synchrotron Radiation Absorbers (SRA)

The current design of the SRA is displayed in Fig. 3.9. It is a 390 mm long copper component welded into a dedicated aperture in the vacuum chamber. It needs to be actively cooled by specific water-cooling circuits integrated in the absorber.

The channels have a twisted tape design, see Fig. 3.10, that significantly improves cooling capacity by enhancing turbulence mechanisms.

The synchrotron radiation absorbers (SRA) are produced using laser powder bed fusion (LPBF) additive manufacturing technology. This approach offers the advantage of enabling the fabrication of intricate internal structures, such as the complex twisted-tape design, which would be more time-consuming and less refined if manufactured using traditional methods. Copper materials produced through 3D printing are currently being qualified for ultra-high vacuum (UHV) applications. A green laser machine is employed to deposit copper alloy layer by layer, ensuring negligible degradation of its mechanical, thermal, and electrical properties. New machines with lower energy requirements and smaller beam wavelengths are entering the market, promising a more cost-effective and energy-efficient solution.

A promising candidate for the absorber material is the copper alloy CuCrZr, which is undergoing detailed thermal and mechanical testing to assess its performance and suitability for this application.

Both analytical and numerical analyses are being conducted to support the design of the synchrotron radiation absorber, particularly in relation to the twisted-tape cooling channels within it. The design is based on implicit formulations by Manglik and Bergles [241], which show strong agreement with experimental observations.

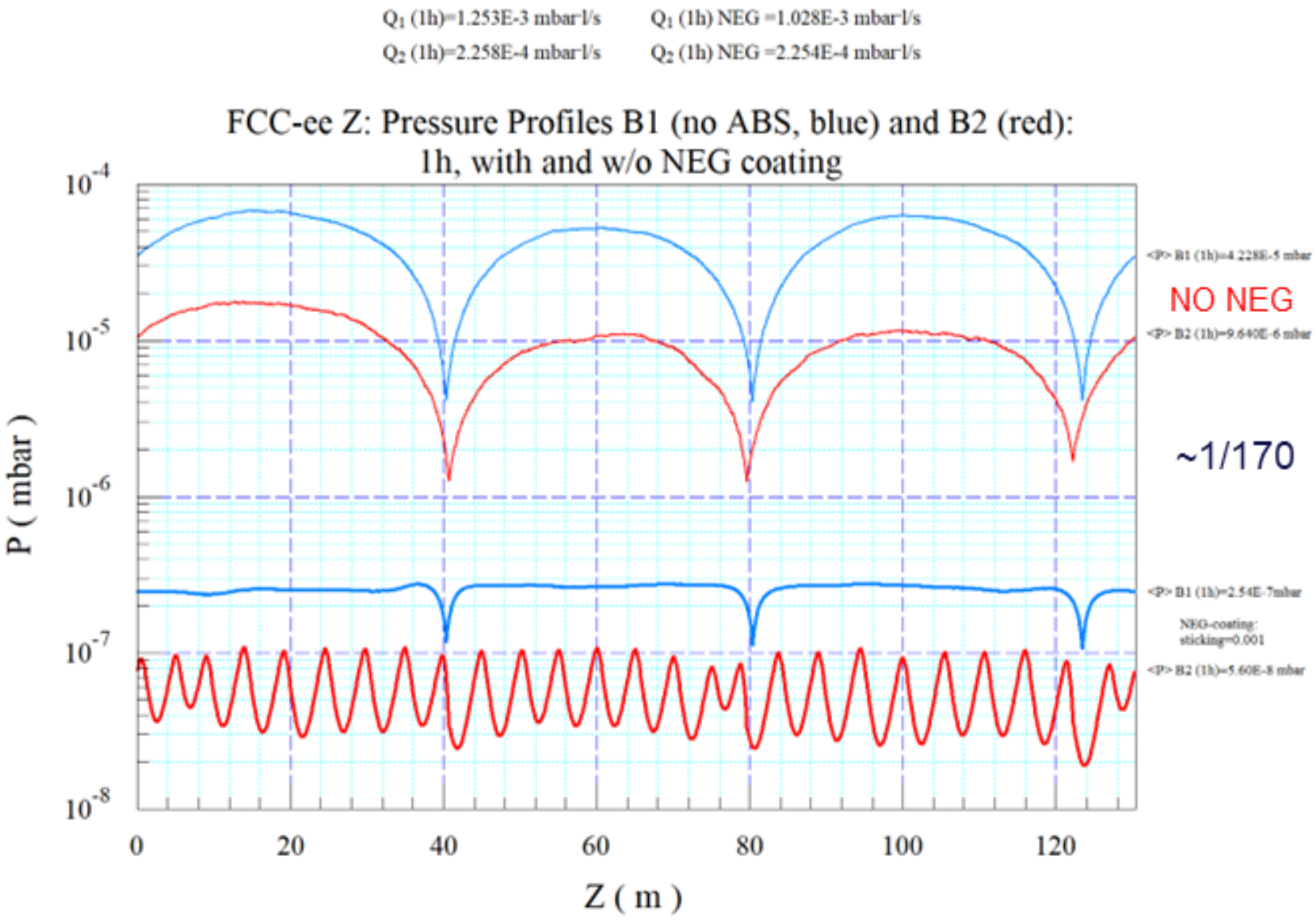

Fig. 3.8: The pressure profiles for the two rings, with and without SRAs, are shown here for the 1 Ah beam conditioning dose, i.e., the initial condition of the rings. In this case, the two sets of curves allow the comparison of adding NEG-coating to the two rings. The corresponding pressure curves drop by approximately two orders of magnitude, greatly shortening the vacuum commissioning time.

The geometric and fluid dynamics parameters for the internal cooling channels are shown in Table 3.6.

Table 3.6: Table of parameters and data for each absorber channel.

| **Parameter** | **Data per absorber channel** |
|---|---|
| $d$ | 7 mm |
| $H$ | 0.5 mm |
| $v$ | 30 mm |
| $Re$ | 3 m/s |
| $m$ | $2.1 \times 10^4$ |
| $h$ | 115 g/s |
| $\Delta P$ | 0.2 bar |

The absorbers intercept a high heat load from SR on their slanted surface. The temperature needs to be maintained below safety levels to ensure mechanical integrity and avoid water phase transitions. However, this also leads to an increased pressure drop that requires a design optimisation. The heat intercepted by the absorbers is not the same around the collider ring. The current worst-case scenario indicates a total power of 4471 W and a peak power density of 74.6 W/mm$^2$. SR strikes a ($< 2$ mm wide) strip horizontally across the middle of the absorber surface, see Fig. 3.11. This refers to the 91 GeV centre-of-mass machine. For the $t\bar{t}$ machine configuration, the total power remains unchanged; however, the power density is expected to increase, and the strip width to decrease. Due to the Compton effect, the power density transitions from a surface phenomenon to a volumetric distribution, potentially resulting in an equivalent surface density.

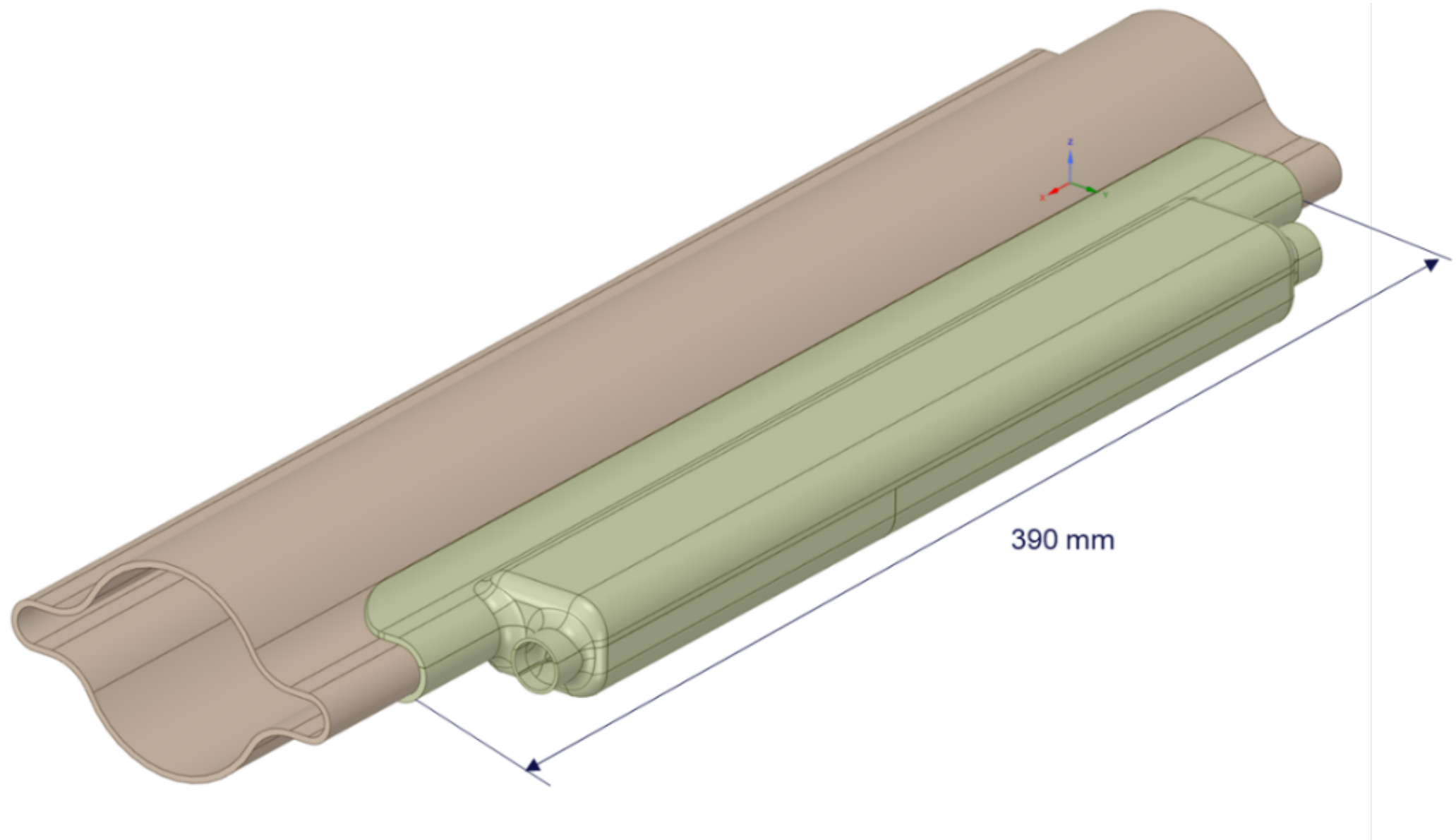

(a)

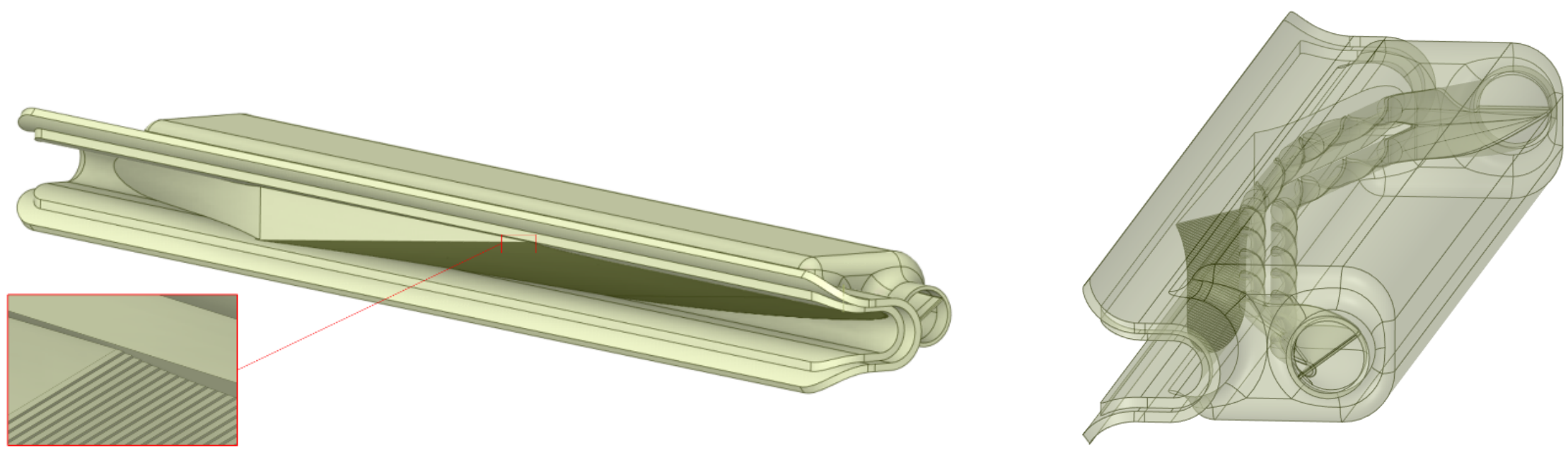

(b) (c)

Fig. 3.9: (a) Synchrotron radiation absorber integrated in the vacuum chamber, (b) sawtooth profile highlighted, and (c) cooling channels with twisted tape.

To accurately determine the volumetric power distribution, detailed simulations using FLUKA are required, as Synrad lacks the capability to model volumetric effects.

The synchrotron radiation (SR) power density peak is not constant but decreases with each successive sawtooth, as shown in Fig. 3.12. For the thermo-mechanical analysis, the SR power density corresponding to the highest peak (from the first tooth) is used. This conservative assumption accounts for fewer sawtooth impacts by SR than would occur in reality as the total power remains the same, i.e., 4471 W.

The temperature distribution of the absorber due to SR power is shown in Fig. 3.13. The maximum temperature is 219°C.

***Sawtooth profile***

Introducing a sawtooth profile on the face of the SRA modifies the photon reflection and, thereby, it potentially reduces the fraction of photons which, after reflections, impinge on the central circular part of the vacuum chamber. Two cases, with and without a sawtooth profile on the SRA faces, have been compared. the images in Fig. 3.14 show the result, side by side, and the selection of facets relevant to the photoelectron emission. The sawtooth case (on the left) decreases the amount of SR photons absorbed by the circular part of the dipole vacuum chamber segment (985 cm long) by 81%.

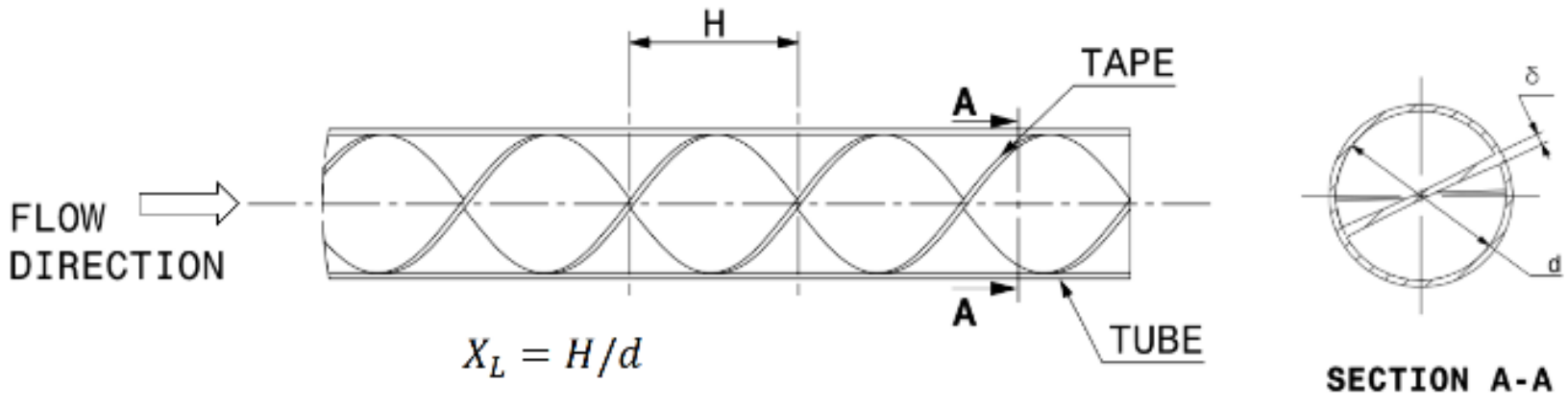

Fig. 3.10: Twisted-tape design of the absorber channels.

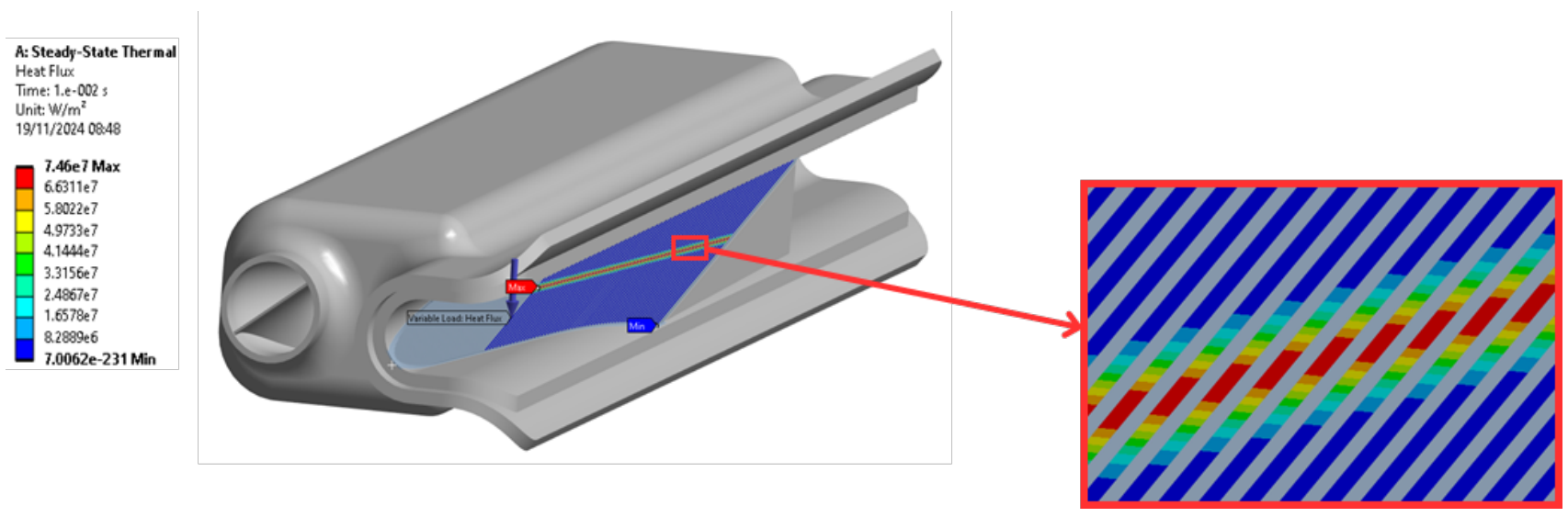

Fig. 3.11: Synchrotron radiation strip in the absorber. The sawtooth structure alternates impacts, with every other side of the tooth being struck, as highlighted in the red box.

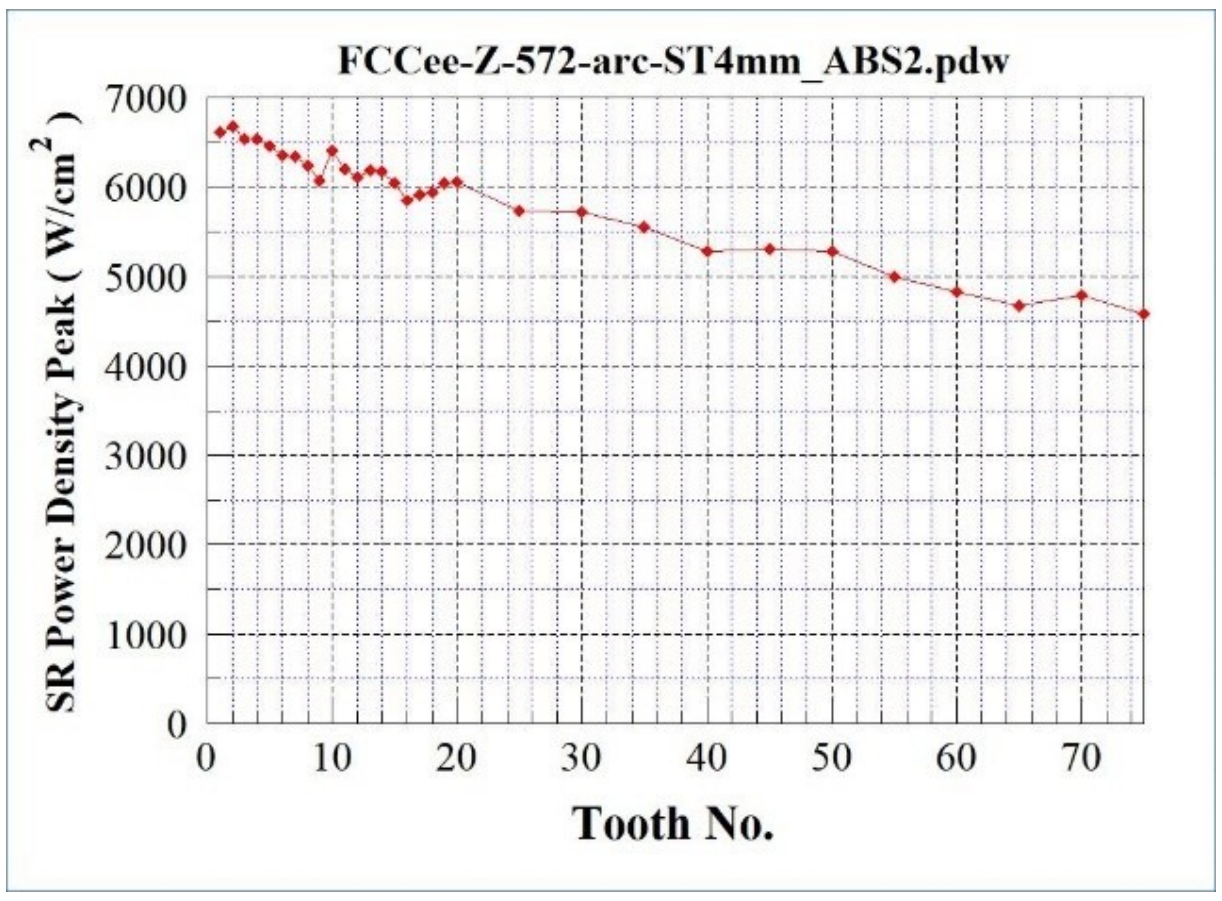

Fig. 3.12: The SR power density peak along the tooth of the SRA.

It can be seen that the sawtooth design removes the two reflected photon spots (in green-blue colour) near the axis of the vacuum chamber, on the lower part of it.

Without sawtooth, the two spots would generate primary photoelectron, which would be trapped along the vertical field lines along the dipole magnet and perturb the stored beam.

The trajectory of the beam is shown by the curved yellow line. Although these simulations had only 2 SRA with sawtooth out of the 26 of the whole arc model ( 130 m-long), it is concluded that the introduction of the sawtooth profile on the inclined surface of the SRA would be beneficial for a reduction of the primary photoelectrons. More refined simulations will be carried out shortly.

An optimisation for the position of the cooling channels was performed in a 2D approximation

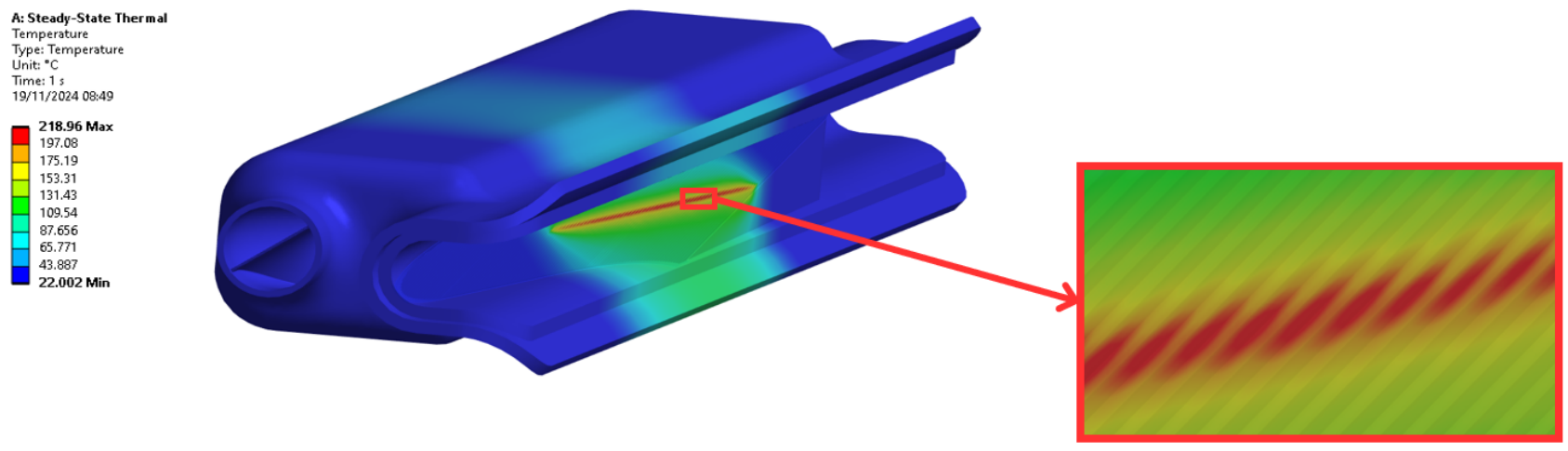

Fig. 3.13: Temperature distribution of the SRA with 2 cooling channels.

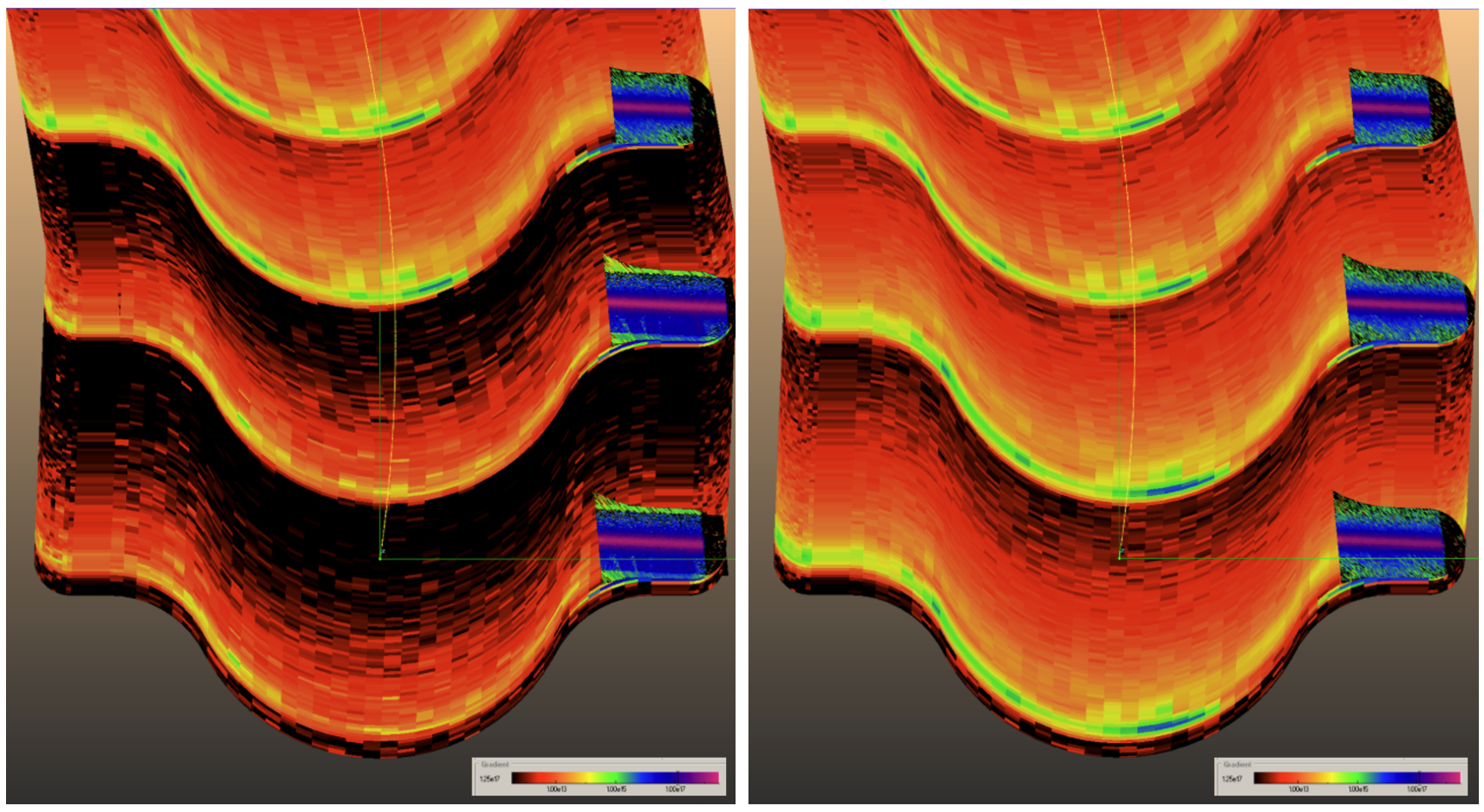

(a) Case with 2 SRA with sawtooth profiles

(b) Case with "regular" SRA, one inclined surface

Fig. 3.14: The effect of sawtooth on synchrotron radiation power density. (a) Case with 2 SRA and sawtooth profiles and (b) Case with 'regular' SRA and one inclined surface

with the aim of lowering the temperature of the internal cooling channels. To this end, the SR power distribution in a section perpendicular to the horizontal mid-plane was divided into small regions equal to the width of the sawtooth, as illustrated in Fig. 3.15.

A third cooling channel is beneficial to lower the temperature of the internal channels and, in general, of the whole absorber, as shown in Fig. 3.16. The temperature of the internal surface of the cooling channels, presented in Fig. 3.17, always stays below 70°C.

The Von Mises stress due to SR heat load is shown in Fig. 3.18. The maximum stress is 285 MPa on the sawtooth receiving the highest SR power density. These values are below the elastic limit of the copper alloy and are deemed safe. However, the material properties of the copper alloy must be measured experimentally to validate the results of the thermal and mechanical simulations.

The cooling layout for the vacuum chamber in the half arc cell is illustrated in Fig. 3.19. A single cooling channel serves the vacuum chamber for the magnets on the girder (quadrupoles and sextupoles), which then splits into two channels for the dipole magnets.

The cooling channel on the opposite side of the absorbers ensures an even temperature distribution around the absorber area, minimising differential deformations. To enhance cooling efficiency and achieve a more uniform outlet temperature and to equilibrate pressure drop, the two cooling circuits cross over between the dipoles before recombining at the water outlet.

The maximum pressure difference between the inlet and outlet of the cooling line should be around 2-3 bar, and the inlet water temperature is around 27°C (EN/CV inputs). Considering a conservative case accounting for three absorbers per vacuum chamber (12 m long) and the total resistive losses [242] (150

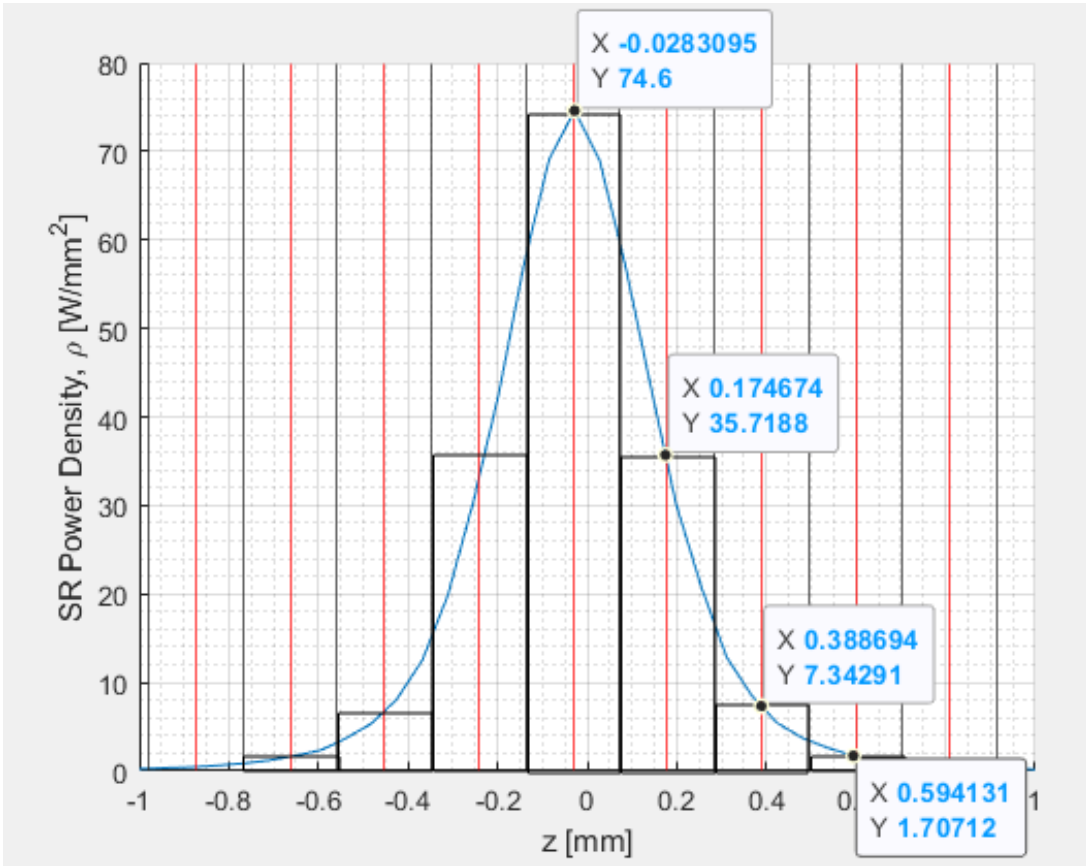

Fig. 3.15: Power density distribution in the absorber, plotted in a section perpendicular to the mid-plane. The distribution is divided into seven regions, each 0.23 mm wide, with constant power density.

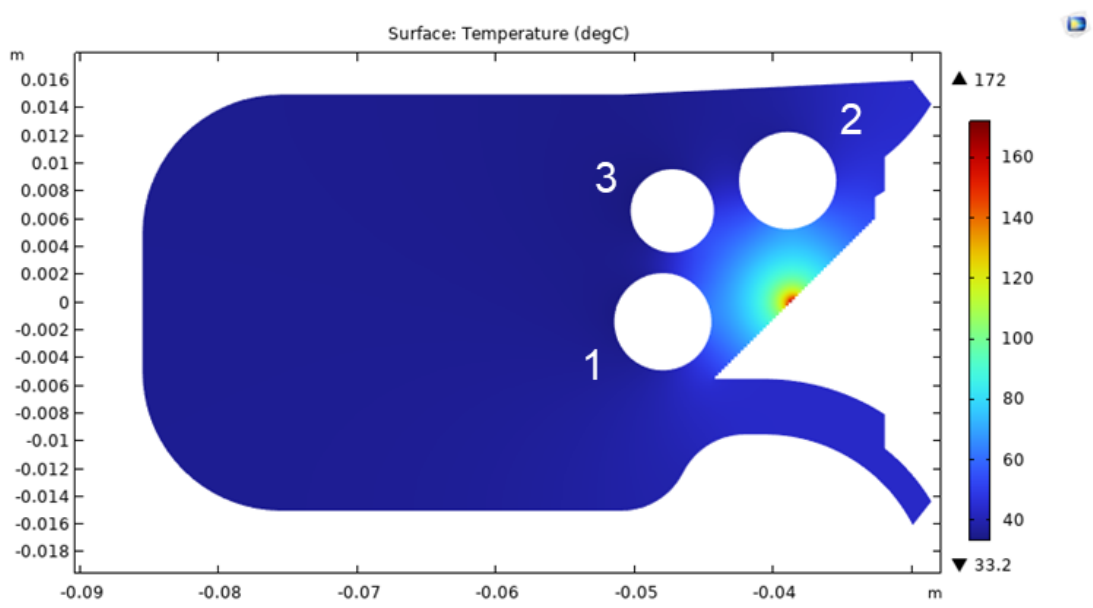

Fig. 3.16: Cross section of the SRA temperature distribution with 3 cooling channels.

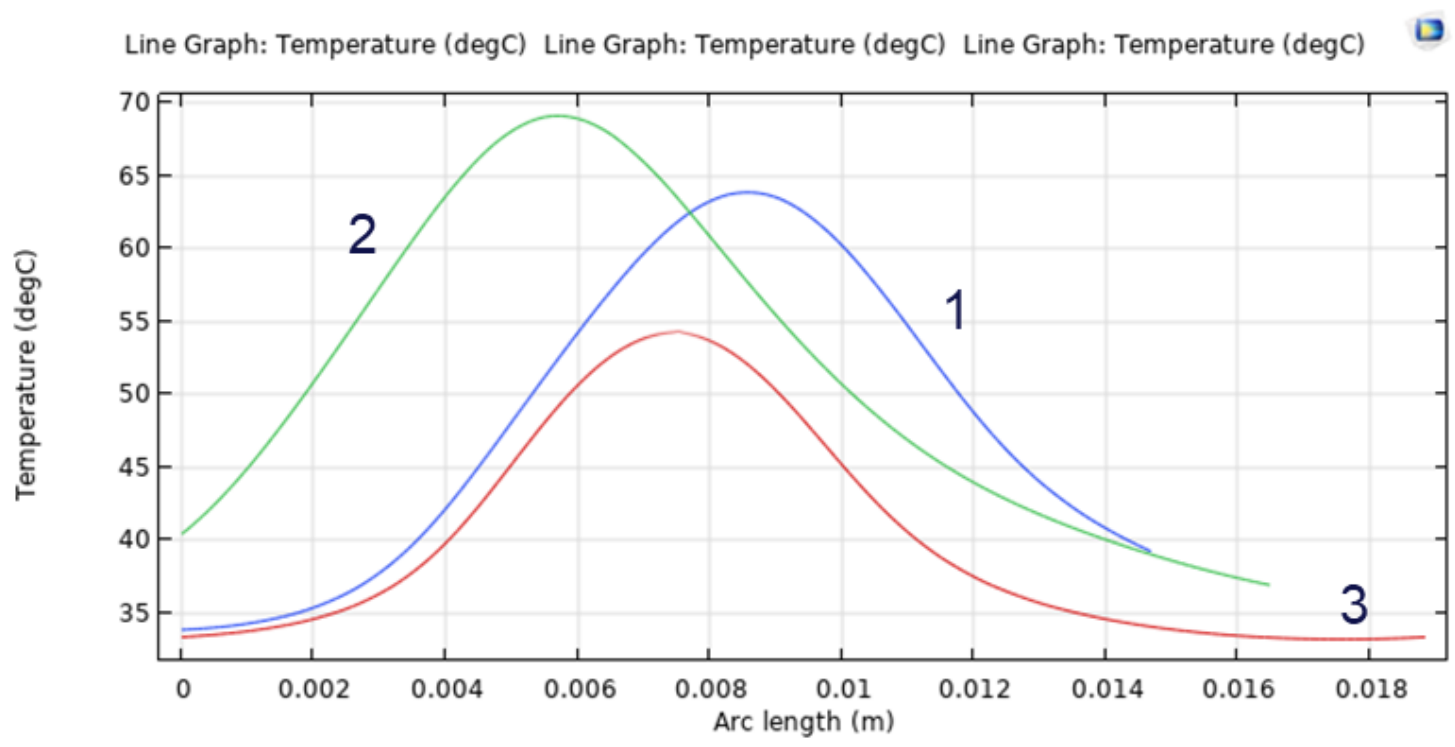

Fig. 3.17: Temperature of the internal surface of the cooling channels of the SRA. These are numbered as in Fig. 3.16.

W/m), total power of about 31 kW is generated per half arc cell, i.e., SRA = (4.5 kW × 6) + impedance losses = (150 W/m×28 m). The flow rate needed for the SRA would be 115 g/s×3 =345 g/s. Considering the parallel configuration of the two cooling tubes, the combined mass flow rate would be 690 g/s, leading to an average temperature difference of about 11°C between the inlet and outlet.

The pressure drop along the parallel smooth channel along the 12 m long vacuum chamber is 0.25 bar for an internal diameter of 14 mm, without considering bends. The cooling channel on the absorber side incurs a pressure drop of 0.2 bar per absorber, resulting in a total pressure drop of 0.6 bar for all three absorbers. The theoretical pressure drop in the dipole would then be 0.85 bar. Considering all the fittings and bends, a pressure drop of 1.5/2 bars from the inlet to the outlet of the half arc cell is expected.

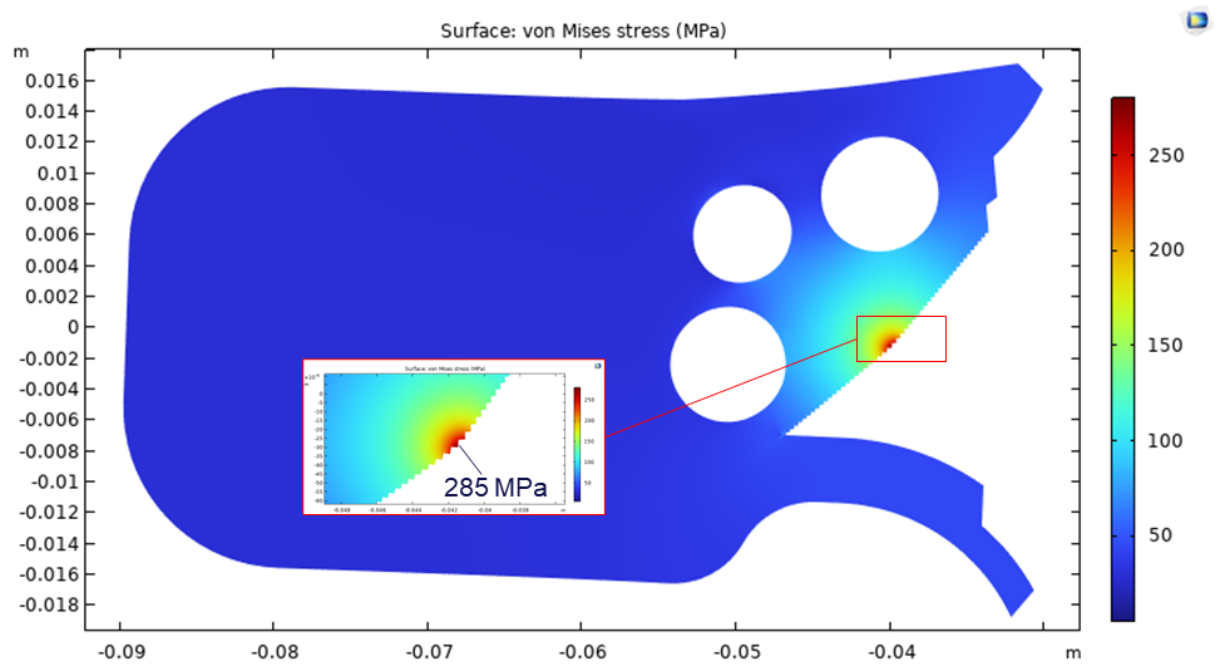

Fig. 3.18: Von Mises stress distribution of the SRA.

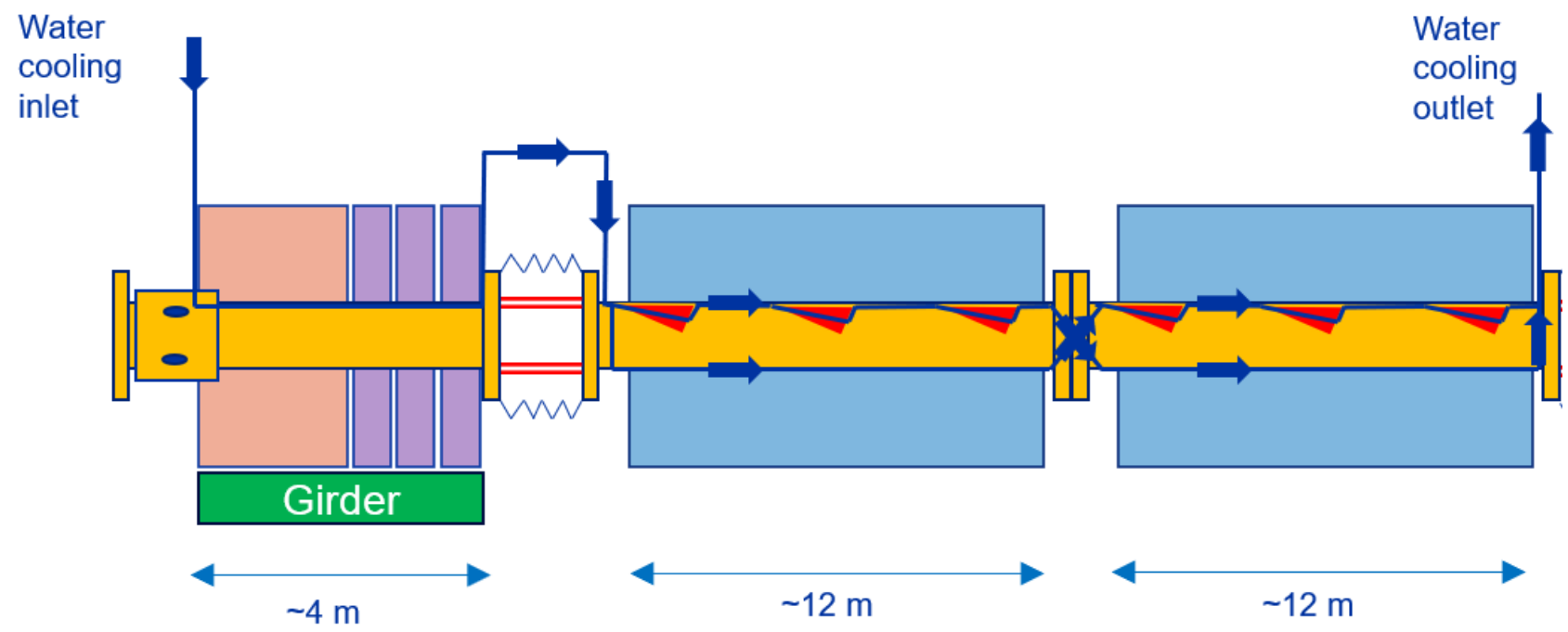

Fig. 3.19: The vacuum chamber cooling layout for the arc half cell.

Interconnection modules, shown in Fig. 3.20, are implemented to cope with the thermal expansion of the vacuum chambers during bakeout, NEG activation thermal cycles, and beam operation. These modules ensure that any resulting transversal misalignments of magnets and BPMs stay within the mechanical alignment tolerances. To reduce the quantity of these critical assemblies, interconnection modules are placed only on either side of the quadrupole-sextupole magnet girder. This configuration minimises the mechanical coupling between the dipole chambers and the quadrupole chamber integrating the BPM. An axial stroke of 65 mm is required, and a 3 mm transverse offset has been chosen as a design requirement.

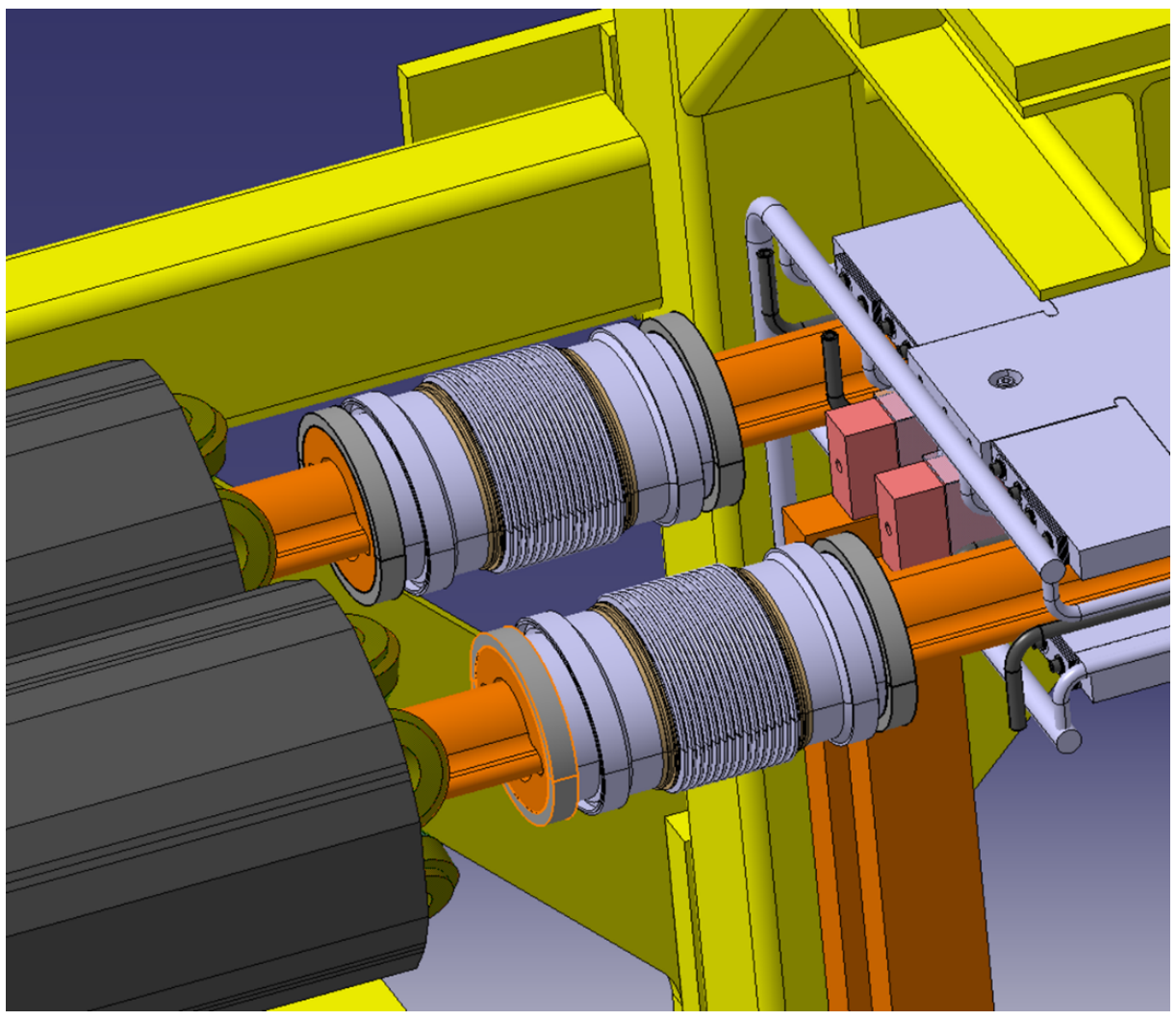

Fig. 3.20: Interconnection modules.

The interconnection modules are dismountable and are based on an external vacuum enclosure and a smooth internal RF transition to ensure electrical continuity for the image current, avoiding higher order modes. The vacuum enclosure integrates a thin-walled hydroformed bellows expansion joint which has an axial and lateral stiffness of about 10 N/mm. A minimum length of 150 mm is needed for the bellows. Oval flanges are integrated on the module and chamber extremities and the leak tight connections are achieved by shape memory alloy rings and soft gaskets. Significant beam induced heat loads are generated in the module (a few 100s W) and an appropriate robust RF transition will be designed. Various options are considered for the bellows shielding. The first one is based on a deformable RF bridge. The concept is based on a thin bridge attached to the adjacent chamber extremities, stretched and almost straight in operating conditions. This solution is used in LHC and HL-LHC. The second solution extends the technical solution used at superKEK and is based on a comb design. A third one is being developed at CERN. In this solution, the shielding of the bellows is achieved by robust sliding RF fingers. A set of machined copper parts is used to ensure a smooth transition between the vacuum chamber cross-section and the oval RF shielding.

## 3.3 Radiation shielding

### 3.3.1 Dipole shielding

The emission of synchrotron radiation by the stored electron and positron beams in the collider can have a significant impact on machine components and other equipment in the tunnel (see Section 1.9). A dedicated shielding must be installed on the dipoles, tightly enclosing the synchrotron radiation absorbers described in Section 3.2. A preliminary conceptual shielding design for the collider arcs is shown in Fig. 3.21. A first optimisation of the shielding geometry has been performed, but further iterations are needed in the technical design phase. Several hundred kilograms of shielding material is needed for each synchrotron radiation absorber in order to reduce the ionising dose in the tunnel sufficiently. The design of the shielding and its integration entails many technical challenges, which require detailed engineering studies. This section highlights some of the main design considerations.

The selection of the shielding material is a trade-off between shielding efficiency, raw material costs, material availability in the industry, engineering aspects (fabrication, machining), and radiological (operational and waste) considerations. High-density materials such as tungsten heavy alloys (17–$18.5\,\mathrm{g/cm^3}$) are very efficient in absorbing photons. However, the costs are prohibitive when considering the large number of shielding units required. Lead-based alloys are less dense and, therefore, require a larger shielding volume but are available at a fraction of the cost. The current baseline material for the dipole shielding is based on a lead-antimony alloy, which is commonly used as shielding material for photons. The antimony content is needed for structural reasons; it hardens the material, increasing its mechanical stability and machinability. Lead-antimony has a good corrosion resistance, which is an important factor in high-radiation environments. While lead alloys with different antimony content are available in the industry, an antimony fraction between 6 and 8% is considered sufficient for dipole shielding. The resulting material density of such a binary alloy ($10.88\,\mathrm{g/cm^3}$) is only slightly reduced compared to pure lead ($11.35\,\mathrm{g/cm^3}$). The shielding might require a casing or coating to enable safe handling during the installation or during maintenance work. Although lead antimony seems to be a good candidate material, the final material selection for the shielding will be confirmed in the technical design phase.

The anticipated activation of the shielding material is a key factor in its life cycle and requires a thorough assessment. Radionuclide production within the shielding is primarily driven by neutrons generated from photo-nuclear interactions, particularly in the synchrotron radiation absorbers. Initial studies indicate that neutron production due to synchrotron radiation is expected to be minimal up to ZH operation, as photon energies remain largely below the ($\gamma$,n) threshold. However, a significant neutron flux is expected during $\mathrm{t\bar{t}}$ operation, as the synchrotron photon spectrum extends beyond the giant dipole resonance of most materials. A preliminary radiological assessment of antimonial lead as a shielding material suggests that it can be considered non-radioactive after a cool-down period of one to two years following $t\bar{t}$ operation, with residual activity initially dominated by the antimony content. Further studies are required to evaluate potential contributions from other radiation sources, particularly beam-gas scattering, which can lead to shielding activation at all beam energies due to the emission of Bremsstrahlung photons with significantly higher energies than synchrotron radiation.

The current shielding configuration represents only a preliminary conceptual design and must be refined into a realistic technical solution using state-of-the-art engineering practices. Given that the collider ring requires more than 20 000 shielding units, a careful balance between design optimisation and cost efficiency is essential. Assuming 20 synchrotron radiation absorbers per FODO cell for ZH and $t\bar{t}$ operation, and an estimated shielding weight of 400 kg per absorber, the total shielding material required for the arcs amounts to approximately, 10 400 tons, with an additional 1000 tons needed for the experiment insertions. The sheer scale of the raw material required introduces a significant risk factor in terms of procurement, necessitating a well-defined strategic sourcing plan. Furthermore, a full life cycle assessment must be conducted to explore potential reuse options for the shielding material following the decommissioning of the FCC-ee. The material could either be reintegrated into the market or repurposed for use in other CERN facilities.

Integrating the shielding is a complex task and requires a coordinated design effort for the systems concerned (shielding, vacuum chambers, synchrotron radiation absorbers, and magnets). Direct contact between shielding and vacuum system components must be avoided, since the shielding would act as a heat sink during the bake-out of the vacuum chambers. This requires a careful assessment of tolerances and alignment requirements and needs a detailed study of the assembly procedure for all components. Considering the weight of the shielding (about 4 tons per 20 m dipole), the mechanical design of the dipole and its supporting scheme needs to be reassessed in order to the cope with the additional load generated by the shielding. Given the complexity of the shielding integration and assembly, a staged shielding approach for different operation modes is not favoured.

Another important aspect of the shielding design is the heat load management and the associated structural stability at higher temperature. The shielding absorbs about 20% of the synchrotron radiation power during ZH and $t\bar{t}$ operation. (see Table 1.20).

This amounts to 20 MW for the full ring, or 2.5 MW per arc, which has to be dissipated by a dedicated cooling circuit embedded in the horizontal shielding inserts. For a single shielding element, the power load can reach up to several hundreds of Watt and could give rise to higher than tolerable peak temperature (especially due to the low melting point of Pb-alloys). Detailed energy deposition and thermo-mechanical simulations and associated system engineering considerations are essential for the design optimisation of the shielding and routing of service systems, the dimensioning of the circuit, and the definition of the corresponding infrastructure requirements.

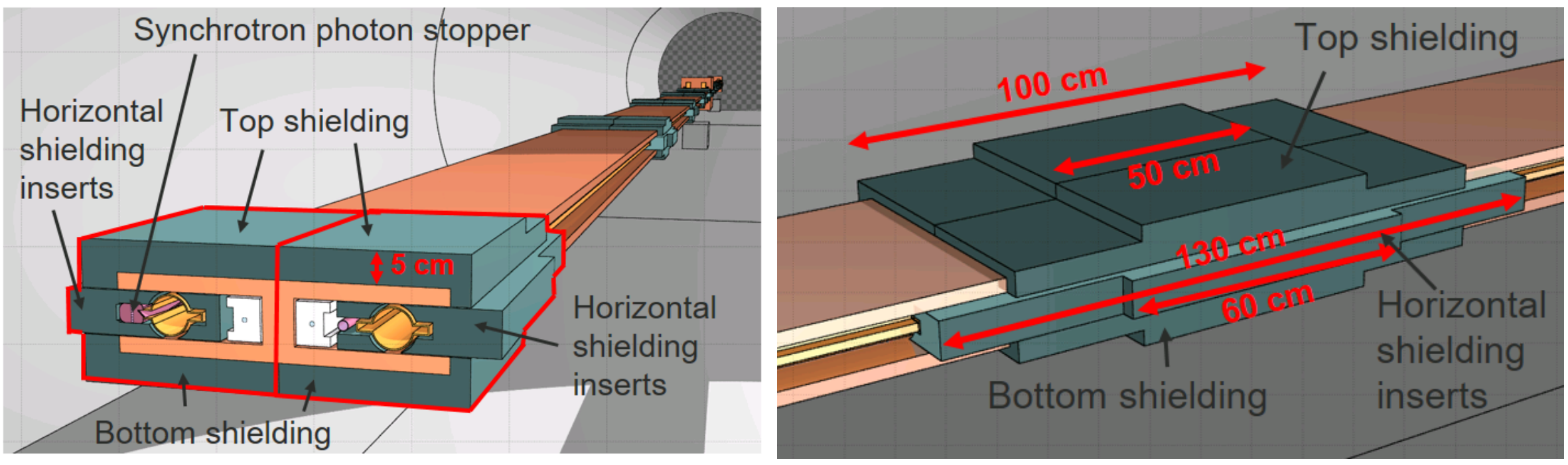

Fig. 3.21: Preliminary conceptual radiation shielding design for the collider dipoles in the FCC-ee arcs. The shielding is assumed to be made of antimonial lead and has a weight of about 400 kg per synchrotron radiation absorber.

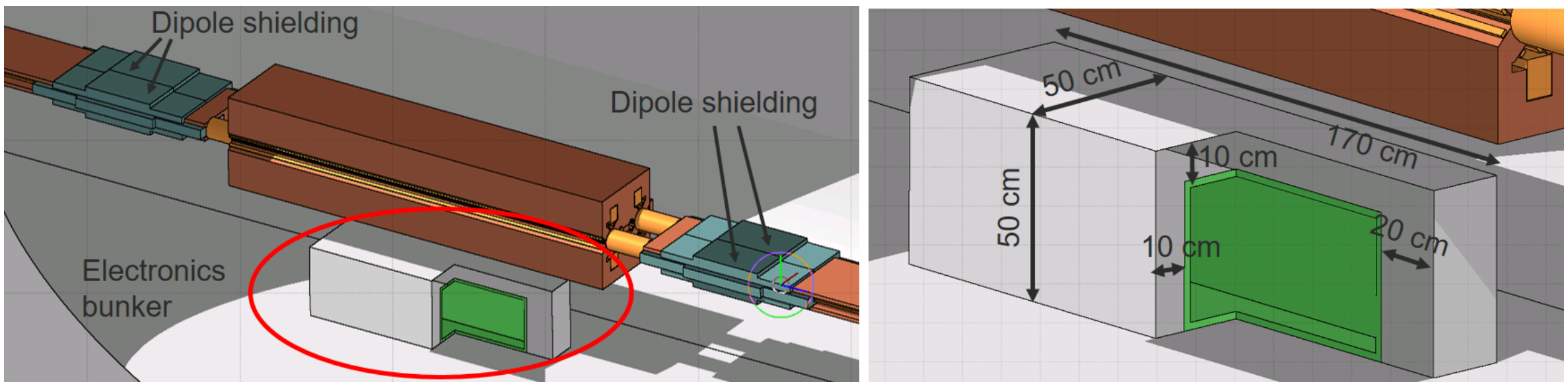

Fig. 3.22: Possible electronics bunker (red circle) near the lattice quadrupoles in the FCC-ee arcs. In this very first design, the bunker is assumed to be made of concrete (gray walls) and borated polyethylene (green sheets on the inside). The dimensions shown have to be revised once the number and size of the electronics racks have been confirmed.

### 3.3.2 Electronics bunker

A conceptual design of a possible electronics bunker has been devised (see Fig. 3.22), in order to reduce the radiation levels for electronics. In this very first design, the bunker is assumed to be made of 10–20 $\mathrm{cm}$-thick concrete walls, which are covered on the inside by borated polyethylene sheets. The latter are needed for moderating and capturing neutrons. The concrete walls can also be replaced by other materials, which can affect the required wall thickness. The actual bunker size will depend on the final space requirements for electronics racks. The outer dimensions shown in Fig. 3.22 need to be adapted once a complete inventory of the required racks has been established. It is presently assumed that one such bunker is needed per arc quadrupole in the collider, which amounts to more than 2800 units. The bunkers might also be needed in the insertion regions.

The technical implementation of such a bunker remains to be studied. In particular, the integration of the bunker inside or near the quadrupole girder has to be assessed. Another important aspect is the accessibility of electronics in case of interventions. The racks need to be accessible without the need of heavy lifting equipment in case an electronics card has to be exchanged. Furthermore, a suitable ventilation system has to be designed for the bunkers to extract the heat generated by the systems inside the bunker and to have precise temperature control.

## 3.4 Radio frequency system layout, configurations and parameters

### 3.4.1 Introduction

The superconducting radio frequency (SRF) system of FCC-ee [13] accelerates two beams of particles circulating in opposite directions. The system is designed to provide 50 MW of RF power in continuous wave (CW) to each beam in order to compensate synchrotron radiation (SR) losses. Beam currents and required RF voltages for four operating points are summarised in Table 3.7. At the Z operating mode the beam current of 1.3 A is very high and the total RF voltage is only about 100 MV. When switching to the W and Higgs operating points, the beam currents are reduced by one order of magnitude and are 135 mA and 26.7 mA respectively, while the total RF voltage is ten times higher, at 1.05 and 2.1 GV . In the $\mathrm{t\bar{t}}$ mode the beam energy is significantly increased (182.5 GeV). The beam current and RF voltage vary again by one order of magnitude and are 5 mA and 11.3 GV making the accelerator a very high gradient machine. The collider RF system requires a transverse feedback system to cure coupled bunch transverse instabilities, as described in Section 1.4. Strip line kickers operating at a multiple of the bunch repetition frequency are an obvious choice. For the collider, the function of depolariser and transverse feedback can be combined as outlined in Section 1.7.

The SRF system will be located in a single straight section at point PH, while the one for the booster will be grouped in point PL. The requirement to locate all the RF at a single location rather than being more distributed around the ring is driven by the need for extremely precise centre-of-mass energy calibration at the Z. This localisation also has the benefit of consolidating the cryogenics, electrical distribution, and RF maintenance. The following sections describe the main aspects of the RF system and important changes since the CDR and mid-term report.

Table 3.7: Main RF-related FCC-ee parameters.

| Modes | Energy [GeV] | Current [mA] | RF Voltage [GV] |
|---|---|---|---|
| Z | 45.6 | 1292 | 0.089 |
| WW | 80.0 | 135 | 1.049 |
| ZH | 120.0 | 26.8 | 2.098 |
| $t\bar{t}$ | 182.5 | 5 | 11.300 |

### 3.4.2 Baseline FCC-ee collider SRF system layout

Table 3.8: Evolution of FCC-ee collider RF system layout.

<table>
<tr><th>Operating point</th><th></th><th>Z</th><th>WW</th><th>ZH</th><th>$t\bar{t}$</th></tr>
<tr><td colspan="6">Conceptual design report</td></tr>
<tr><td>RF frequency</td><td>[MHz]</td><td colspan="3">400</td><td>400/800</td></tr>
<tr><td>Common RF system for two beams</td><td></td><td colspan="3">no</td><td>yes</td></tr>
<tr><td>Number of cavities</td><td></td><td colspan="2">104</td><td>272</td><td>272/372</td></tr>
<tr><td>Number of cells per cavity</td><td></td><td>1</td><td colspan="2">4</td><td>4/5</td></tr>
<tr><td>RF power per cavity</td><td>[kW]</td><td colspan="2">962</td><td>368</td><td>149/155</td></tr>
<tr><td colspan="6">Feasibility study mid-term report</td></tr>
<tr><td>RF frequency</td><td>[MHz]</td><td colspan="3">400</td><td>400/800</td></tr>
<tr><td>Common RF system for two beams</td><td></td><td colspan="2">no</td><td colspan="2">yes</td></tr>
<tr><td>Number of cavities</td><td></td><td>112</td><td>264</td><td>264</td><td>264/488</td></tr>
<tr><td>Number of cells per cavity</td><td></td><td>1</td><td colspan="2">2</td><td>2/5</td></tr>
<tr><td>RF power per cavity</td><td>[kW]</td><td>901</td><td>378</td><td>382</td><td>78/163</td></tr>
<tr><td colspan="6">Feasibility study final report</td></tr>
<tr><td>RF frequency</td><td>[MHz]</td><td colspan="3">400</td><td>400/800</td></tr>
<tr><td>Common RF system for two beams</td><td></td><td colspan="2">no</td><td colspan="2">yes</td></tr>
<tr><td>Number of cavities</td><td></td><td colspan="3">264</td><td>264/408</td></tr>
<tr><td>Number of cells per cavity</td><td></td><td colspan="3">2</td><td>2/6</td></tr>
<tr><td>RF power per cavity</td><td>[kW]</td><td colspan="3">380</td><td>78/195</td></tr>
</table>

Due to the large span in beam current and RF voltage of the four FCC-ee operating modes, it is very challenging to design a unique RF system suitable for all the operating points. Nevertheless, several steps were performed towards a more compact and efficient solution.

The main parameters of the RF system described in CDR [13] are summarised in Table 3.8. For the Z operating point only, 1-cell elliptical cavities at 400 MHz with low shunt impedance were foreseen, with a cavity RF shape carefully optimised to minimise the higher-order mode (HOM) power. The 4-cell cavities were considered for WW and ZH operating points due to significantly higher RF voltages and lower beam currents. Thanks to the small number of bunches required for the highest energy operating point ($\mathrm{t\bar{t}}$), it was assumed that there is a common RF system for both electron and positron beams that reuses all 400 MHz 4-cell cavities complemented by more compact 800 MHz 5-cell cavities [243].

The main RF system changes in the mid-term report concerned WW and ZH operating points. The first considerations of 2-cell cavity scenario were described in Ref. [244] and highlighted a strong HOM damping efficiency compared to the 4-cell designs while still permitting a moderate accelerating gradient. On that basis and due to the increased RF voltage requirements for the WW operating point, 2-cell cavities were chosen, and a common RF system for the ZH operating point was also adopted. This led to an increased number of high-gradient 800 MHz 5-cell cavities for the $\mathrm{t\bar{t}}$ operating point.

Two additional modifications led to the present baseline RF system. The 2-cell cavities were also adopted for the Z operating point requiring the reverse phase operation (RPO) mode [245] discussed in the following section. In addition, 6-cell cavities at 800 MHz have been proposed instead of 5-cell cavities at the $\mathrm{t\bar{t}}$ operating point, leading to a significant reduction of the total number of cryomodules thus a reduction of the investment costs. The main RF parameters for the collider are summarised in Table 3.9.

Table 3.9: Main RF parameters of the FCC collider.

| Parameters | Z | W | ZH | $\mathrm{t\bar{t}}$ | |
|---|---|---|---|---|---|
| Common RF system for two beams | no | no | yes | yes | |
| RPO | yes | no | no | no | |
| Total RF voltage [MV] | 89 | 1049 | 2098 | 2098 | 9202 |
| Beam current [mA] | 1283 | 135 | 53.6 | 10 | |
| RF frequency [MHz] | | 400.79 | | 400.79 | 801.58 |
| Operating temperature [K] | | 4.5 | | 4.5 | 2 |
| Number of cells per cavity | | 2 | | 2 | 6 |
| Quality factor $Q_0$ | | $2.7 \times 10^9$ | | $2.7 \times 10^9$ | $3 \times 10^{10}$ |
| Cavity voltage [MV] | | 7.95 | | 7.95 | 22.5 |
| Accelerating gradient $E_{\mathrm{acc}}$ [MV/m] | | 10.6 | | 10.6 | 20.1 |
| RF power per cavity [kW] | | 380 | | 78 | 195 |
| Coupling factor $Q_L$ | | $9.2 \times 10^5$ | | $4.5 \times 10^6$ | $4.1 \times 10^6$ |
| Number of cryomodules | | 66 | | 66 | 102 |
| Number of cavities | | 264 | | 264 | 408 |

### 3.4.3 Reverse Phase Operation

The optimal cavity detuning, $\Delta f_{\mathrm{opt}} = f_0 - f_{\mathrm{RF}}$, and optimal quality factor, $Q_{L,\mathrm{opt}}$, are commonly used to minimise RF power requirements in high-current synchrotrons. Assuming 132 2-cell cavities for the Z operating point, the lowest RF voltage per cavity, $V_{\mathrm{cav}} \approx 0.7\,\mathrm{MV}$ results in $\Delta f_{\mathrm{opt}} \approx -70\,\mathrm{kHz}$ and $Q_{L,\mathrm{opt}} \approx 5 \times 10^3$. Both are extremely difficult to achieve because of the enhancement of longitudinal coupled-bunch instabilities due to fundamental mode (FM) modulations of bunch-by-bunch parameters due to transient beam loading, as well as increased critical fields in a fundamental power coupler (FPC). An alternative approach, the reverse phase operation (RPO) mode [245] was studied for the FCC RF system to overcome these challenges partially. The RPO was originally developed and tested in KEKB for various scenarios [245–247] and was adopted as a baseline solution for the electron storage ring of the Electron-Ion Collider (EIC) [248]. It is based on introducing groups of focusing and defocusing cavities. They are de-phased with respect to the reference phase and provide accelerating voltage at the beam phase, adding up to the required total RF voltage as illustrated in Fig. 3.23. In this case, $V_{\mathrm{cav}}$ can be chosen to be the same for the WW and ZH operating points, resulting in a common optimal quality factor for three modes given by:

$$Q_{L,\text{opt}} = \frac{V_{\text{cav}}^2 N_{\text{cav}}}{2P_{\text{SR}}(R/Q)}. \tag{3.1}$$

Here $N_{\text{cav}}$ is the total number of cavities, $P_{\text{SR}}$ is the synchrotron radiation power, and $(R/Q)$ is the ratio of the shunt impedance to the quality factor of the cavity FM expressed in circuit ohm. Although this simplifies the fundamental power coupler design, the first drawback is that the total RF voltage can only be changed in discrete steps of $N_{\text{foc}} - N_{\text{defoc}}$, (see, e.g., Ref. [249]),

$$V_{\text{RF}} = \frac{U_0}{e}\sqrt{1 + \left(1 - \frac{e^2 N_{\text{cav}}^2 V_{\text{cav}}^2}{U_0^2}\right)\frac{(N_{\text{foc}} - N_{\text{defoc}})^2}{N_{\text{cav}}^2}}, \tag{3.2}$$

where $U_0$ is the energy loss per turn due to synchrotron radiation. Therefore, the RF voltage has been increased from 79 to 89 MV, which assumes $N_{\text{foc}} = 71$ and $N_{\text{defoc}} = 61$. The RPO mode needs to be used for the Z operating point of the collider and all cycles of the high-energy booster (see Section 6.3).

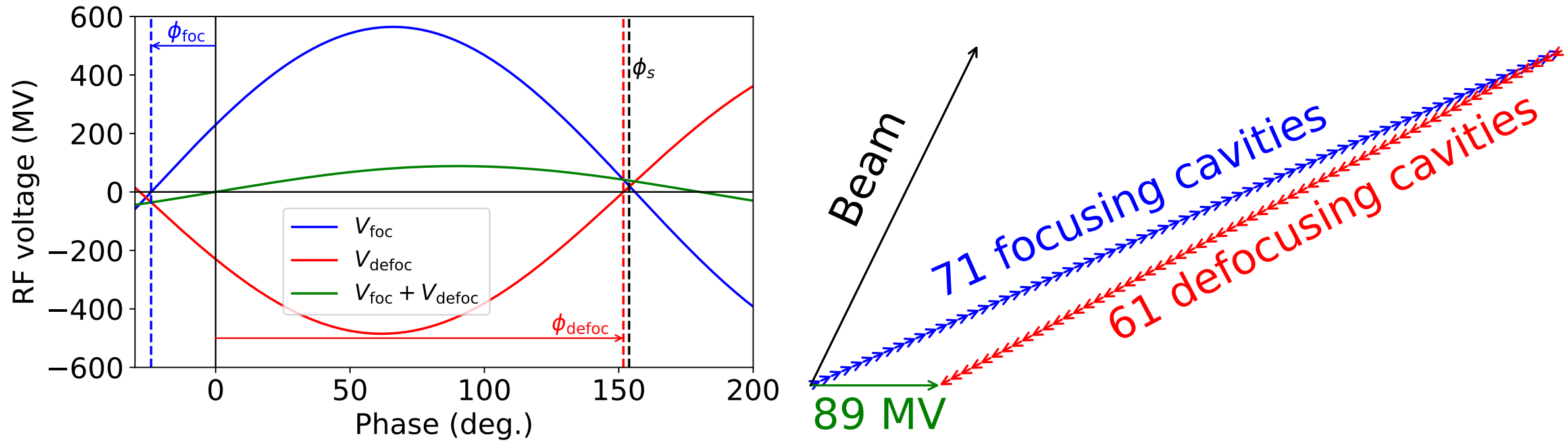

Fig. 3.23: RF waves (left) and phasors (right) for the RPO mode.

### 3.4.4 Instabilities due to fundamental mode (FM) and transient beam loading

Cavity detuning leads to impedance asymmetry around the RF frequency, which can drive the coupled bunch instability. Direct RF feedback [250] (Fig. 3.24, left) is assumed to be implemented in the low-level (LL) RF system to reduce the effective impedance 'seen by the beam'. The closed loop impedance is

$$Z_{\text{cl}}(\omega) = \frac{Z(\omega)}{1 + G_{\text{FB}} Z(\omega) e^{-i\tau_{\text{delay}}\omega + i\phi_{\text{adj}}}}, \tag{3.3}$$

where $G_{\text{FB}}$ is the feedback gain, $\tau_{\text{delay}}$ is the overall loop delay assumed to be 700 ns (similar to the LHC RF system [251]), and $\phi_{\text{adj}}$ is the phase adjustment required to correctly set the feedback negative at the detuned cavity resonant frequency. The flat response is achieved for $1/G_{\text{FB}} = 2(R/Q)\omega_{\text{rf}}\tau_{\text{delay}}$. The instability growth rates were computed assuming a uniformly filled ring for the Z, WW, and ZH operating points (Fig. 3.24, right). The synchrotron radiation damping is sufficient to suppress any coupled-bunch modes driven by fundamental cavity impedance for WW and ZH operating points, whereas direct RF feedback is necessary for stability at the Z operation point.

To evaluate the transient beam loading, the small-signal (Pedersen) model [252] was extended to the RPO case. The filling scheme assumed in the mid-term report considered 20 trains of 560-bunches with the 25-ns bunch spacing resulting in about 1.2 µs gaps between bunch trains. In this case, the modulation of the effective RF voltage reaches about 50% for the baseline $V_{\text{RF}} = 89$ MV (Fig. 3.25, left) resulting in the synchrotron tune spread of 30% (Fig. 3.25, right). This spread is not acceptable for transverse beam stability due to limited space in the tune diagram (Section 1.4.3), and therefore, two mitigation schemes were proposed: a higher RF voltage or a shorter gap length.

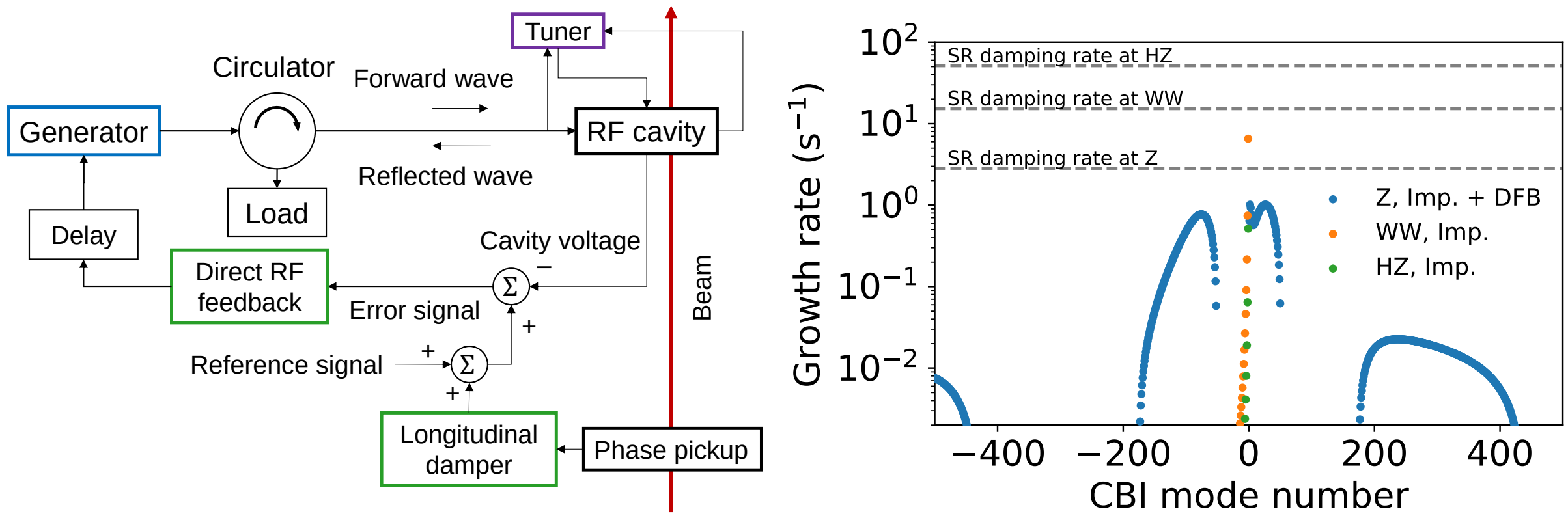

Fig. 3.24: Simplified block diagram of the LLRF system for the FCC-ee 400 MHz RF system (left) and Growth rates of longitudinal coupled bunch instabilities due to fundamental cavity impedance (right).

After a modification of the injection and extraction system layouts (Section 1.8), a new filling scheme of 40 trains of 280-bunches with 0.6 µs gaps was adopted, and a stable working point was found for all bunches (see Fig. 1.17). The impact of low-intensity pilot bunches was also verified, and a uniform filling of all gaps is recommended to avoid an additional 1% increase in the synchrotron frequency spread.

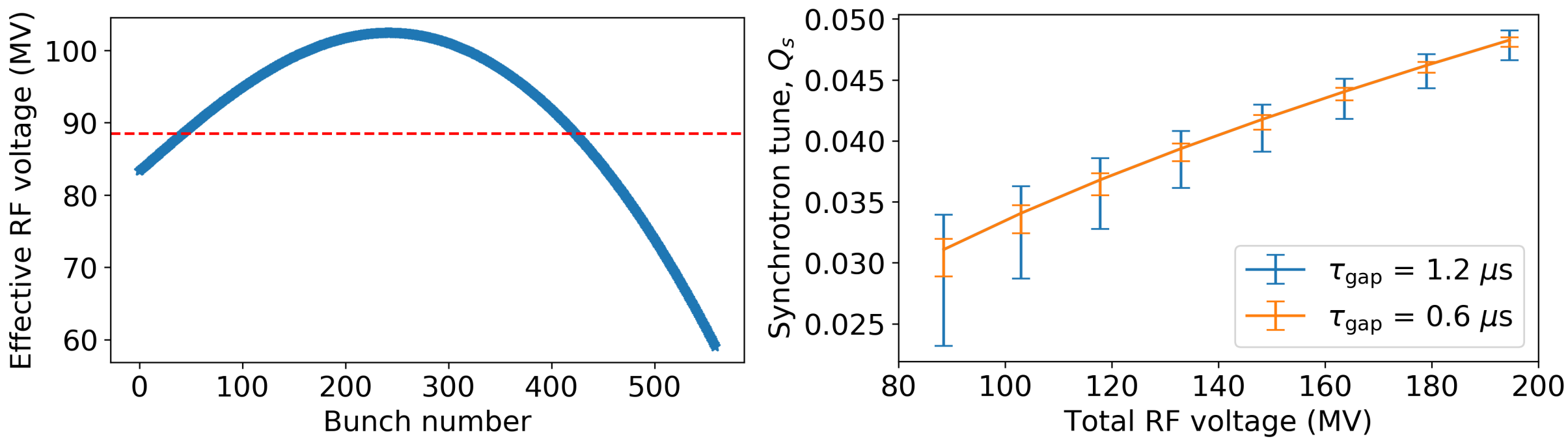

Fig. 3.25: Left: bunch-by-bunch modulation of the effective RF voltage with RPO for 20, 580-bunch trains filling scheme (one train shown) as a function of the bunch number. Right: synchrotron tune spreads as a function of total RF voltage.

### 3.4.5 Analysis of RF system trip

A time-domain model based on [253] was developed to evaluate beam-cavity interaction in the event of an RF system failure. It solves differential equations describing the cavity-generator and LLRF building blocks (Fig. 3.24, left) coupled with the longitudinal equations of motion for centroids of all bunches similar to Ref. [254]. The two highest current FCC operating points were considered.

#### *Z operating point*

It is assumed that the LLRF system is capable of detecting the event of a single RF cavity (or RF amplifier, LLRF system, etc.) trip within one turn (Fig. 3.26, top left, green trace). It then adapts the reference signals of the remaining cavities (blue and orange traces) to compensate for missing RF voltage within the following turn thanks to a strong direct RF feedback. In that case, the RF voltage of the tripped cavity overshoots for a short time by about 6% of the nominal RF voltage and then settles slightly below the nominal value. At the same time, the RF power (Fig. 3.26, top right) is modulated at the synchrotron frequency due to the synchrotron oscillations excited of different bunches (Fig. 3.26, bottom). Although the peak RF exceeds the nominal value by about 40% (shaded blue and orange areas), the total average power of all remaining cavities increases by less than 10% (solid black line). In this scenario, the RF power system adapts the high-voltage set point to efficiently cope with the corresponding overshoot (Section 3.4.12). The amplitude of bunch oscillations remains within a few ps, which is significantly smaller than the rms bunch length. The beam, however, becomes unstable due to the increase of the impedance around the FM (the direct RF feedback is no longer active). In particular, coupled-bunch modes -2 or 2 become unstable depending on the type of cavity tripped. To suppress these instabilities, a longitudinal damper system can be employed (Fig. 3.24, left) using the remaining RF cavities as the kicker cavities and providing the shortest damping time of the order of two synchrotron periods [255, 256]. The algorithm needs to be adapted for the RPO case, and the potential increase of RF power transients should be further studied. After that, a recovery scenario can be evaluated. For completeness, the results of simultaneous trips of focusing and defocusing cavities were evaluated. In that case, the RF power modulations are unacceptably high (more than 70%) and the beams must be dumped. Note that coupled-bunch instabilities due to the impedance of two tripped cavities of the same type can not be suppressed by the longitudinal damper.

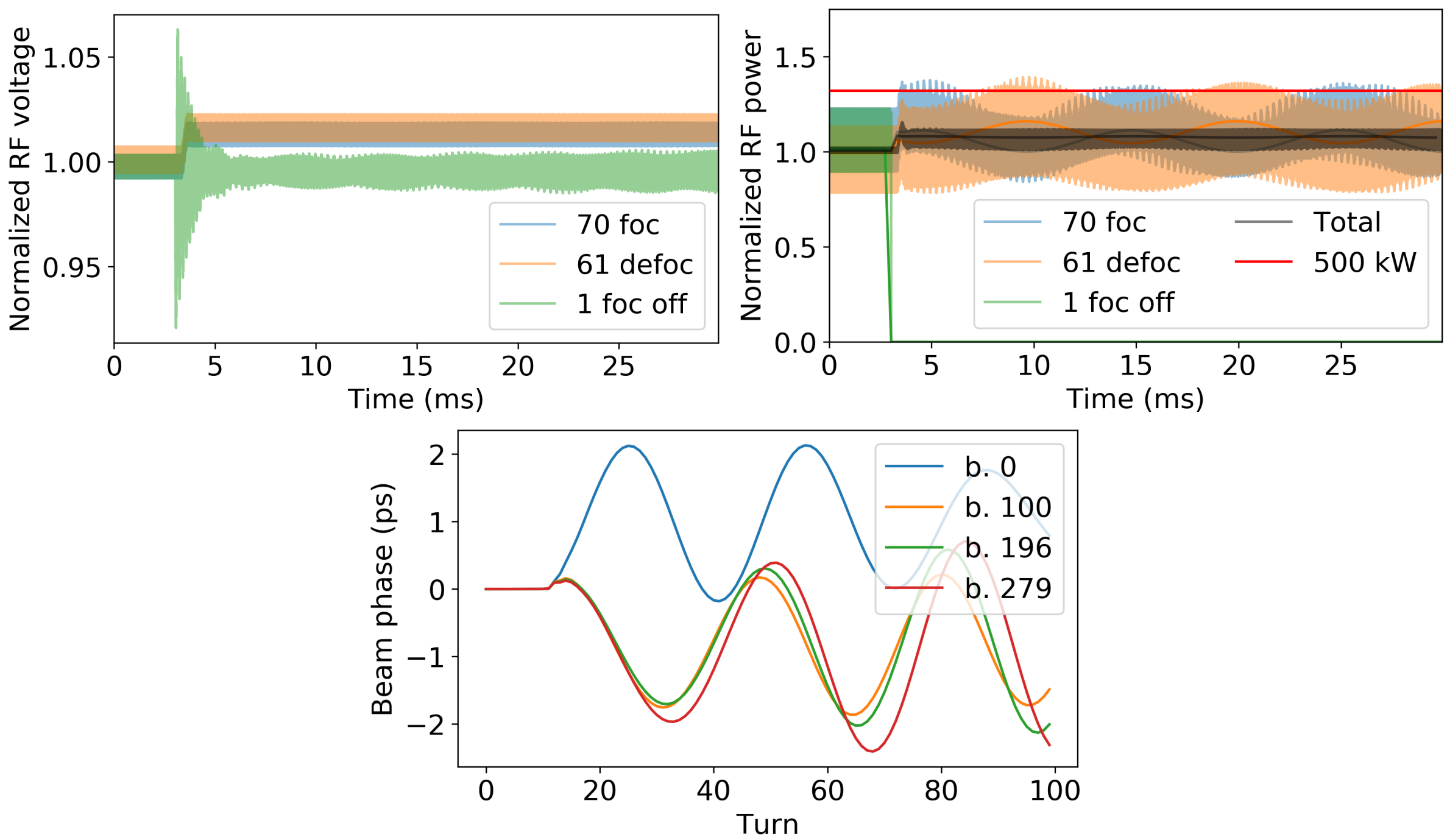

Fig. 3.26: Transients in the event of a single focusing RF cavity trip for Z operating point. Top left: evolution of RF voltages normalised by the nominal value of 7.95 MV as functions of time. Top right: evolution of the RF power normalised by the nominal value of 380 kW for the focusing, defocusing, and tripped RF cavities. Bottom: synchrotron oscillations of different bunches in a train.

#### *WW operating point*

A similar analysis was performed for the scenario with all cavities being in phase for the W operating point. A tripped RF system cannot drive a coupled-bunch instability due to a lower beam current, higher energy, and stronger synchrotron radiation (see Fig. 3.24). An example of simultaneous trips of six RF systems is shown in Fig. 3.27. It is the worst-case scenario as the probability of this event is rather low and sequential trips can appear instead. Applying a similar strategy of detecting missing RF cavities and adapting the total RF voltage, the power transients can be kept below 30% while the peak RF voltage can increase by 15% with respect to the nominal value. This requires increasing the time of RF voltage adaptation by another turn at a cost of about 10 ps amplitude of the bunch oscillations (Fig. 3.27, bottom), which becomes comparable with the 18 ps rms bunch length. No coupled-bunch instability is expected in this case due to sufficiently fast synchrotron radiation damping.

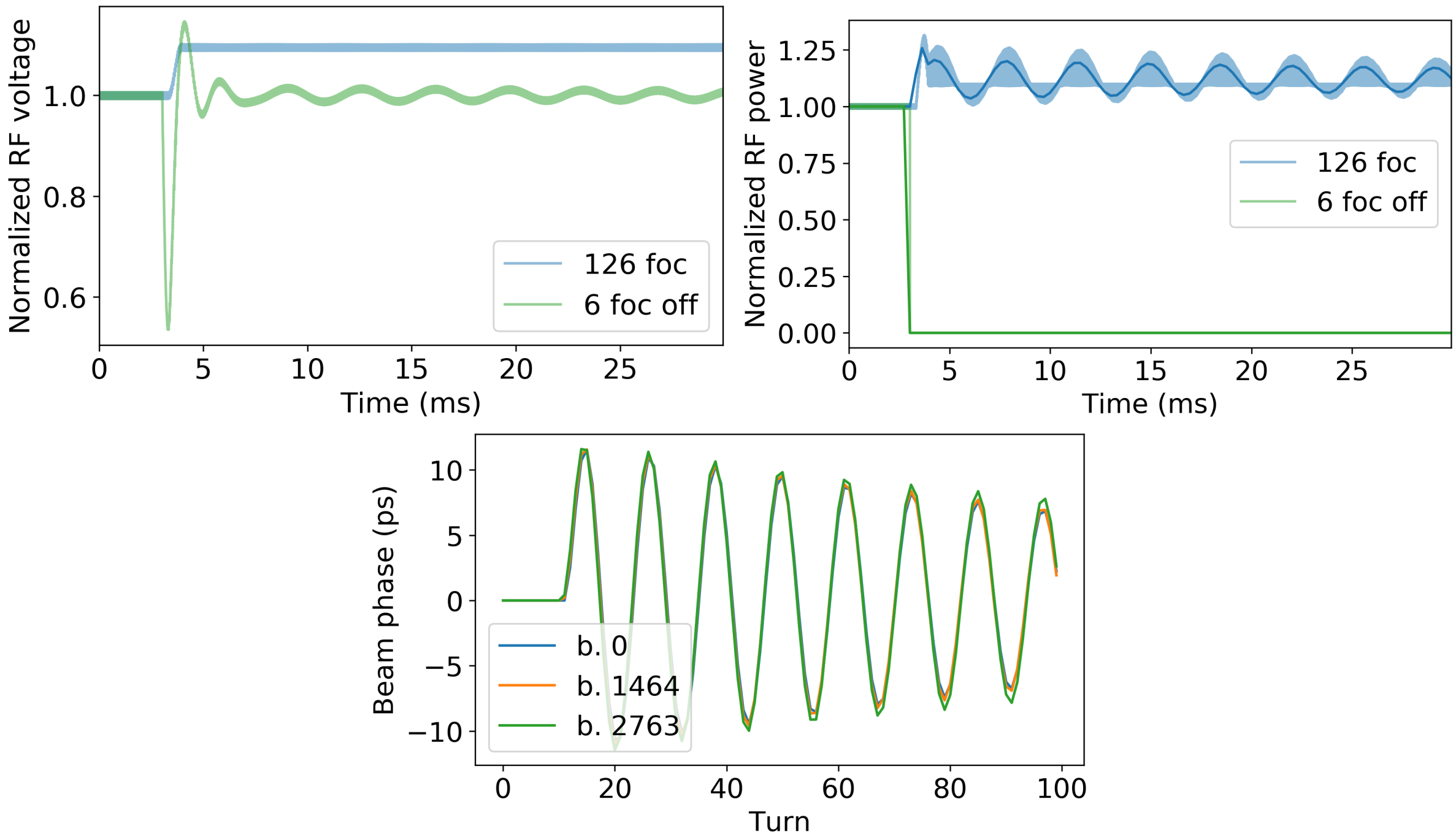

Fig. 3.27: Transients in the case of six RF cavity trips for the WW operating point. Top left: evolution of RF voltages normalised by the nominal value of 7.95 MV as a function of time. Top right: evolution of the RF power normalised by the nominal value of 380 kW for the focusing and tripped RF cavities. Bottom: synchrotron oscillations of different bunches in a train.

### 3.4.6 RF synchronisation aspects

The ring tunnel of the FCC-ee must also be compatible with the FCC-hh accelerator to be installed in the same infrastructure once the lepton physics programme has been completed. Therefore, the tunnel circumference, which will obviously remain an unchangeable parameter throughout the entire lifetime of the FCC, must be carefully chosen. Flexibility in terms of beam parameters and transfer schemes is key to not exclude any possibilities even in the far future.

For the FCC-ee, there are only a few restrictions as long as the booster and collider rings have exactly the same circumference. RF frequencies for the ultra-relativistic leptons are moreover constant and synchrotron radiation naturally damps longitudinal oscillations of the bunches.

The choice of circumference is more constrained for FCC-hh, since protons will either be transferred from a high-energy booster (HEB) in the existing SPS or LHC tunnel. In this injector the hadrons must be accelerated with beam control loops to mitigate common-mode dipole oscillations. These beam-derived corrections change the RF frequency during acceleration and hence modify the azimuthal position of the bunches at the arrival at the flat-top. This requires a cogging process to move them to the desired azimuth prior to the bunch-to-bucket transfer. Additionally, following acceleration and synchronisation in the HEB, multiple transfers of batches of about 80 bunches to the collider cannot be avoided because of the need to keep the stored energy of the injected beam at an acceptable level for protection devices.

The first analysis to identify an FCC circumference compatible with the HEB in the SPS or LHC tunnel can be found in Ref. [257]. Once synchronised to the same RF frequency, two circular accelerators can be modelled as cogwheels. For a circumference ratio of $C_2/C_1 = h_2/h_1 = n_1/n_2$, where $h$ are the harmonic numbers, the bunches in both rings are at the same azimuth only every $n_1$ turns of the ring with the circumference $C_1$, which corresponds to $n_2$ turns in ring with $C_2$. This is also the basic periodicity with which the beam transfer can take place. In the CDR, a 97 750 $\mathrm{m}$ circumference ring was chosen as a baseline solution with the corresponding ratios with the HEB $C_{\mathrm{FCC}}/C_{\mathrm{LHC}} = 11/3$ and $C_{\mathrm{FCC}}/C_{\mathrm{SPS}} = 99/7$. The evolution of the placement required an additional detailed study [258]. For the present FCC baseline circumference of 90 658.2 m the ratios with the HEB are $C_{\mathrm{FCC}}/C_{\mathrm{LHC}} = 1010/297$ and $C_{\mathrm{FCC}}/C_{\mathrm{SPS}} = 1010/77$. This means that beam can only be transferred every 1010 turns in the injector, corresponding to about 90 $\mathrm{ms}$ for the HEB in the LHC tunnel. While FCC circumferences of 179.5 $\mathrm{m}$ shorter or longer would be ideal from the RF point of view, with much smaller numerators and denominators in the circumference ratios, and provide important flexibility at the transfer, they are excluded based on the integration in the tunnel.

The baseline circumference ratio of the HEB in the LHC tunnel of $1010/297 \simeq 3.40067$ is extremely close to $17/5 = 3.4$ [259]. A promising alternative in view of FCC-hh would therefore be an 18 m shorter FCC circumference of 90 640.2 m, hence exactly 3.4 times the circumference of the LHC. It would allow the injection of hadrons every 17 turns of the HEB, corresponding to only 5 turns in the collider. Such low numbers of turns, with an almost instantaneous transfer, enable non-adiabatic RF manipulations, like bunch rotation to compress or stretch bunches for the injection into the collider, in combination with multiple transfers to limit the maximum stored energy for reasons of machine protection. These manipulations are excluded with the present baseline scenario. To keep the baseline harmonic number of $h_{\mathrm{FCC}} = 121\,200 = 2^4 \cdot 3 \cdot 5^2 \cdot 101$ for the full flexibility with bunch spacings, the RF frequency of the FCC-ee would have to be increased by only 80 kHz, from 400.79 $\mathrm{MHz}$ to 400.87 $\mathrm{MHz}$. This is small enough to stay within the frequency range of the operational cavity tuning systems. The impact of a circumference change as small as 18 $\mathrm{m}$ on the tunnel integration would be minor. Table 3.10 compares the baseline scheme with the proposed alternative circumference.

Table 3.10: Summary of baseline and alternative tunnel circumferences. The length difference with respect to the baseline circumference is indicated by $C_{\mathrm{FCC}}$.

| $h_{\mathrm{FCC}}$ | $C_{\mathrm{FCC}}$ [m] | $\Delta C_{\mathrm{FCC}}$ [m] | $\frac{h_{\mathrm{FCC}}}{h_{\mathrm{LHC}}}$ | $\frac{h_{\mathrm{FCC}}}{h_{\mathrm{SPS}}}$ | $f_{\mathrm{RF}}$ [MHz] | Comment |
|---|---|---|---|---|---|---|
| 120 960 | 90 478.6 | -179.5 | $112/33$ | $144/11$ | 400.8 | Ideal for RF, too short for tunnel integration |
| 110 160 | | | | | 364.4 | Option for FCC-hh |
| **121 200** | **90 640.2** | **-17.95** | $\mathbf{17/5}$ | $\mathbf{459/35}$ | **400.9** | **Proposal for FCC-ee, ideal for RF with minimal change** |
| 122 400 | | | | | 404.8 | Option for FCC-ee |
| 146 880 | | | | | 485.8 | Option for FCC-hh |
| **121 200** | **90 658.2** | **–** | $\mathbf{1010/297}$ | $\mathbf{1010/77}$ | **400.8** | **Baseline** |
| 121 440 | 90 837.7 | +179.5 | $92/27$ | $92/7$ | 400.8 | Ideal for RF, too long for tunnel integration |

### 3.4.7 Technological choices for the SRF cavities

The technological choices proposed for the two types of SRF cavities of the FCC-ee collider are driven by the experience of particle accelerators operating at similar parameters. The accelerating gradients are fixed to $10\,\mathrm{MV/m}$ at 400 MHz and $20\,\mathrm{MV/m}$ at 800 MHz. These accelerating gradients must be reached in operation and be very reliable. Thus, a 20% margin on the accelerating field $E_{\mathrm{acc}}$ and the unloaded quality factor $Q_0$ is added between the values during qualification tests in a vertical cryostat and operation in the machine.

It is indeed important to remember that performance degradation occurs between cavity qualification in the vertical cryostat and cavity performance after cryomodule assembly. This is a well-known phenomenon in the SRF domain. Taking into account an additional margin for reliable operation, performance targets on $E_{\mathrm{acc}}$ and $Q_0$ have been specified for all steps of the cavity lifetime (see Table 3.11).

Table 3.11: Main RF performances targets of the 400 MHz and 800 MHz cavities.

| Cavity configuration | | bare | dressed | cryomodule | operation |
|---|---|---|---|---|---|
| Orientation in test | | vertical | vertical | horizontal | horizontal |
| Added elements | | | helium tank<br>HOM couplers | FPC<br>tuner<br>shieldings | |
| 400 MHz Cavity | $E_{\mathrm{acc}}$ [MV/m] | 13 | 12.4 | 11.8 | 10.6 |
| | $Q_0$ | $3.3 \times 10^9$ | $3.15 \times 10^9$ | $3 \times 10^9$ | $2.7 \times 10^9$ |
| 800 MHz Cavity | $E_{\mathrm{acc}}$ [MV/m] | 24.8 | 23.6 | 22.5 | 20.25 |
| | $Q_0$ | $3.8 \times 10^{10}$ | $3.65 \times 10^{10}$ | $3.5 \times 10^{10}$ | $3 \times 10^{10}$ |

At the Z, WW and ZH operating points, the 2-cell cavity technological design is inspired by the 400 MHz LHC cavities accelerating a proton beam of about 0.5 A (1 A for HL-LHC) and operating at 300 kW RF power in CW [260]. The cavity is heavily damped thanks to four coaxial couplers placed very close to the accelerating cell. The cavity is made of copper and is coated with a superconducting niobium thin film, operating at the temperature of 4.5 K. Its very large aperture of 300 mm diameter allows the propagation of most of the high-frequency modes induced by the beam. The Nb/Cu version of the 352 MHz - 5 mA multicell cavities used in LEP is also a good reference for the design of the 2-cell FCC-ee cavity [261] where some experience can be gained to push the accelerating gradient to 10 MV/m and above, while keeping the HOM damping properties very efficient.

Bulk niobium technology is very suitable for the 800 MHz 6-cell cavities used at the $\mathrm{t\bar{t}}$ energy. Similar 5-cell cavities of this type have already been developed for high-intensity proton accelerators like the SNS linac at ORNL in Oak Ridge [262], the SPL project at CERN [263], the ESS accelerator in Lund [264], and the PIP-II linac at Fermilab [265]. It is also important to remember that a prototype FCC-ee 5-cell bare cavity was manufactured and successfully tested in a vertical cryostat by the Jefferson Laboratory in 2018 [266].

### 3.4.8 RF design of elliptical cavities

A multi-objective optimisation was performed for the 2-cell 400 MHz cavity, targeting both the FM properties, such as peak surface electric and magnetic fields, and the longitudinal impedance of the HOMs. At the Z operating point, the longitudinal impedance of the 0-mode in the fundamental passband becomes critical for ensuring longitudinal stability, as this mode only couples to the power coupler with a loaded quality factor of approximately $10^6$. To address this, the $(R/Q)$ of the 0-mode was incorporated into the optimisation problem. As a result, the half-cell lengths were reduced from 187 mm to 180 mm, lowering the $(R/Q)$ of the 0-mode to below 0.01 $\Omega$. This modification was achieved while maintaining the FM figures of merit and HOM performance metrics at levels comparable to the previous design. The design of the 6-cell 800 MHz cavity is based on the 5-cell design reported in the mid-term report, with an additional mid-cell added to the structure. The mid-cells were optimised to reduce peak surface fields, enabling a higher accelerating gradient ($E_{\mathrm{acc}}$) required for $\mathrm{t\bar{t}}$ operation. Additionally, the end-cell was designed to facilitate the damping of potentially harmful trapped modes. The parametrised model of a cavity-cell, along with the shapes adopted for the 2-cell 400 MHz and 6-cell 800 MHz cavities, are shown in Fig. 3.29. The corresponding geometrical parameters are given in Table 3.12 and some important figures of merit for the cavities are presented in Table 3.13.

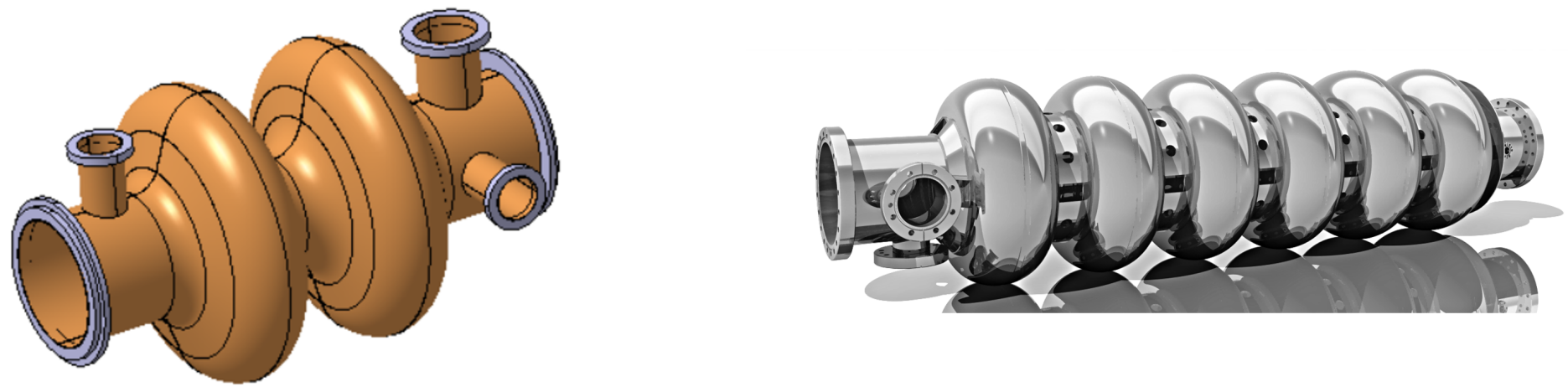

Fig. 3.28: Artist's view of the 2-cell 400 MHz (left) and 6-cell 800 MHz (right) SRF cavities.

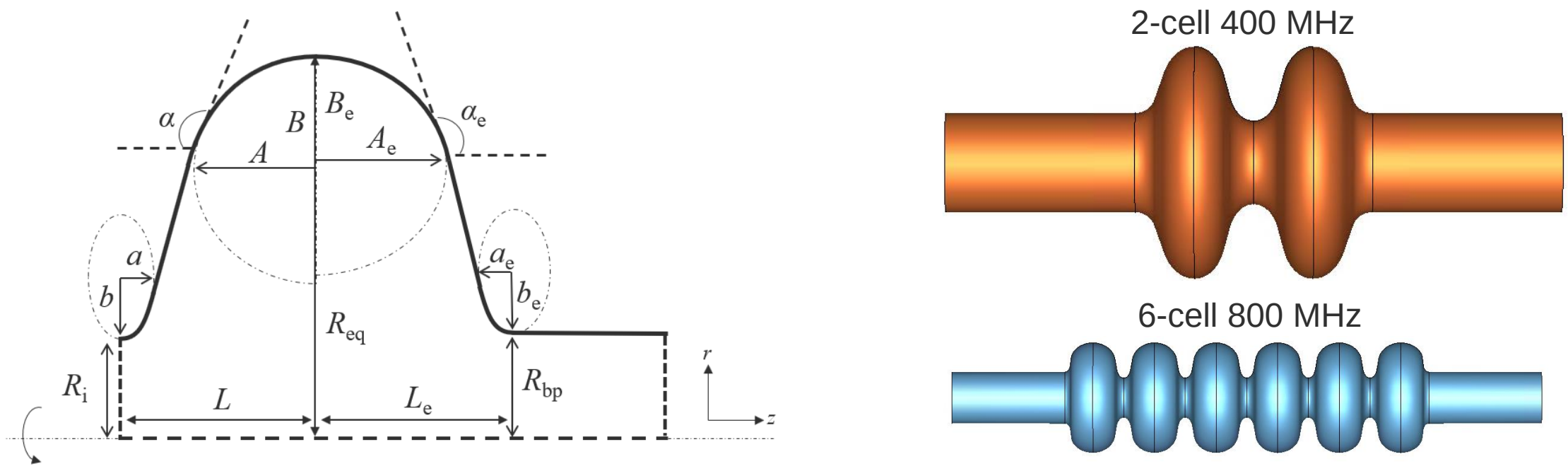

Fig. 3.29: Parametrised model of the cavity's end-cell (left) and the shapes of the two FCC-ee cavities (right). The inner side of the end-cell has the same shape as the middle cells in the 6-cell cavity.

Table 3.12: Geometric parameters of the FCC cavities.

| | $A/A_e$ [mm] | $B/B_e$ [mm] | $a/a_e$ [mm] | $b/b_e$ [mm] |
|---|---|---|---|---|
| **2-cell 400 MHz** | 94.06 / 102.81 | 127.81 / 137.97 | 77.19 / 61.11 | 109.06 / 58.94 |
| **6-cell 800 MHz** | 67.72 / 66.5 | 57.45 / 51.0 | 21.75 / 17.0 | 35.6 / 23.0 |
| | $R_i/R_{bp}$ [mm] | $L/L_e$ [mm] | $R_{eq}$ [mm] | $\alpha/\alpha_e$ [°] |
| **2-cell 400 MHz** | 125.58 / 150 | 180 / 180 | 351.041 | 105.3 / 109.6 |
| **6-cell 800 MHz** | 60.0 / 78.0 | 93.5 / 85.77 | 166.591 | 100.0 / 96.9 |

Figure 3.30 presents the impedance spectrum of each cavity type up to 3.4 GHz without any damping features. For the 2-cell 400 MHz cavity, no critical modes with high longitudinal impedance ($Z_{\parallel}$) are identified. However, dipole modes near 530 MHz are trapped, leading to excessive transverse impedance above the coupled bunch instability threshold. To mitigate this, two hook-type couplers are installed per cavity, reducing transverse impedance to approximately 60 $\mathrm{k\Omega/m}$. A transverse feedback system with a moderate damping rate (about 50 turns) can ensure stability at the Z operating point. The hook coupler geometries are shown in Fig. 3.31(a). For the 6-cell 800 MHz cavity, two DQW-type HOM couplers, as shown in Fig. 3.31(b), suffice to meet stability requirements for $t\bar{t}$ working point.

Table 3.13: Some figures of merit for the FCC-ee cavities, with the LHC cavity as a reference for comparison.

| | LHC (reference) | FCC 2-cell | FCC 6-cell |
|---|---|---|---|
| $f$ [MHz] | 400.79 | 400.79 | 801.58 |
| $(R/Q)_{\text{linac}}$ [$\Omega$] | 88.1 | 182.7 | 630.4 |
| $G$ [$\Omega$] | 252 | 232.7 | 272.8 |
| $E_{\text{pk}}/E_{\text{acc}}$ [-] | 2.3 | 2.0 | 2.04 |
| $B_{\text{pk}}/E_{\text{acc}}$ [mT/MV/m] | 5.1 | 5.33 | 4.31 |
| $k_{\parallel}$ [V/pC] | 0.13 | 0.26 | 3.58 |
| | ($\sigma_z$ = 14.6 mm) | ($\sigma_z$ = 14.6 mm) | ($\sigma_z$ = 2.32 mm) |

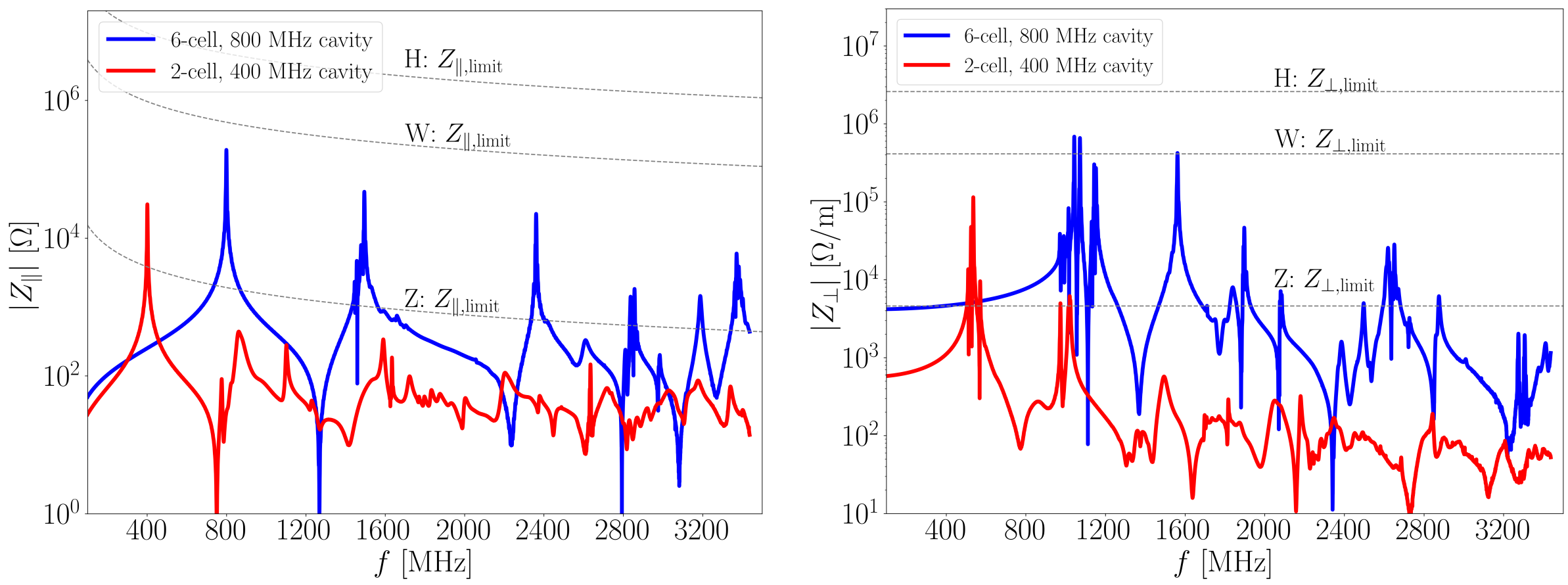

Fig. 3.30: Longitudinal (left) and transverse (right) impedance of the two cavity types. The wakefield simulations are conducted on the cavities without any damping coupler. Consequently, due to the truncation of the wake potential, the impedance peaks are not fully resolved.

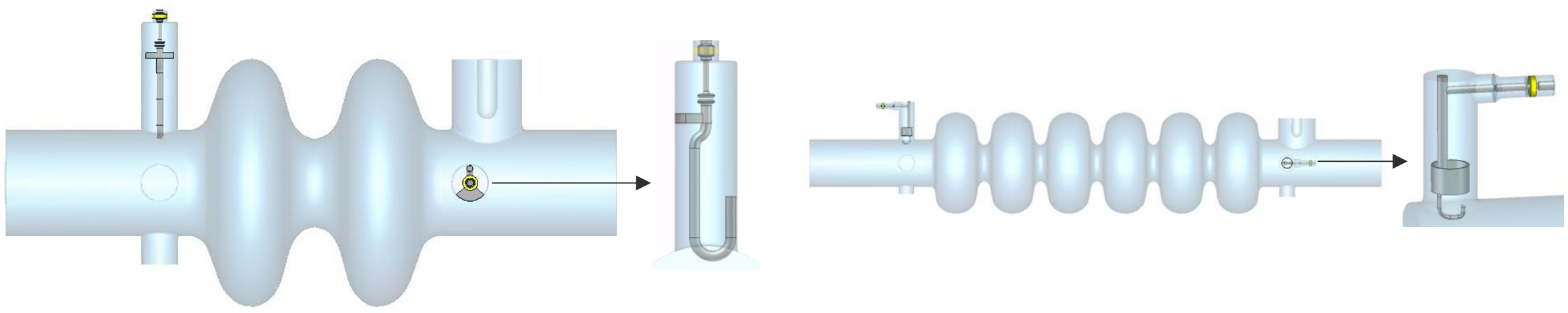

(a) (b)

Fig. 3.31: (a) Two hook-type couplers are used to damp the trapped dipole modes in the 2-cell 400 MHz cavity. (b) DQW-type HOM couplers are employed for HOM damping in the 6-cell 800 MHz cavity.

### 3.4.9 HOM power calculation

A short bunch length of a few millimetres can excite high-frequency HOMs in the cavities, reaching up to tens of GHz. These high-frequency modes must be extracted from the cryomodules and dissipated in air-cooled or water-cooled RF loads. Several types of RF extractors have been studied, including rectangular waveguides, ridged waveguides, beam line absorbers and coaxial lines. A configuration with two coaxial waveguides oriented at 90° and connected to the beam pipe is chosen for its simplicity, its compactness and its ability to handle high RF power. Since the 2-cell cavities are designed to avoid trapped modes with large longitudinal impedance in them, there is no need to place the HOM power extractors close to the cavity. Consequently, the coaxial lines, which also lack an FM rejection mechanism, must be positioned far from the cavity on the beam pipe to prevent the extraction of FM energy while absorbing HOM energy. Another advantage of this configuration is that it eliminates the creation of a transverse kick caused by the two couplers interacting with the FM field, as the influence of coaxial lines on the FM is minimal. However, this design choice comes at the cost of requiring a larger distance between the cavities to accommodate these couplers on the beam pipe.

A similar coaxial line inter-cavity damping concept was considered for the 800 MHz cavity. Due to space limitations and the need to shorten 800 MHz cryomodules, studies are ongoing to eliminate the coaxial line for the 800 MHz system. This appears feasible, with the two additional HOM ports near the cavity utilised for HOM couplers to damp other dangerous HOMs mainly required for the booster cavities which have lower beam instability thresholds. Since neither the coaxial lines nor the HOM couplers have broadband transmission capabilities up to tens of GHz, room-temperature high-power beam line absorbers (BLA) are necessary between cryomodules to absorb HOM power above the beam pipe cut-off frequency for both 400 MHz and 800 MHz cryomodules.

Figure 3.32(a) shows the HOM power distribution for the Z operating point. Each coaxial line should handle several kilowatts of HOM power, with up to 7 kW expected for the coaxial extractors located in the middle of the cryomodule. Additional margins must be considered, as some HOMs can generate up to 10 kW of additional HOM power if the beam spectral line aligns with a high-impedance peak. Depending on the mode excited and its coupling to the coaxial lines, this power can propagate into different couplers. Therefore, to ensure reliability, the coaxial line extractors will be designed to handle power levels between 15 kW and 20 kW to accommodate such worst-case scenarios. Figure 3.32(b) illustrates the power distribution for collider cavities in $\mathrm{t\bar{t}}$ operation without using inter-cavity coaxial lines. In this scenario, each DQW coupler absorbs, on average, 0.1 kW of HOM power, with most of the power propagating out of the cryomodule which has to be damped by inter-cavity BLAs.

It is important to highlight that, in the case of the bunch length without collisions (referred to as SR for the synchrotron radiation) there is a significant increase in the HOM power at the Z working point compared to when beams are in collision, where beamstrahlung (BS) increases the energy spread and bunch length. Figure 3.33 illustrates the loss factor of the four-cavity module with tapers and without couplers for both the BS and SR bunch lengths. The resulting HOM power at the Z operating point is calculated using the formula $P_{\mathrm{HOM}} = k_{\parallel,\mathrm{HOM}} Q I$, where $k_{\parallel,\mathrm{HOM}}$ represents the longitudinal loss factor of HOMs, $Q$ denotes the bunch charge, and $I$ stands for the average beam current. Figure 3.33 also presents the HOM power levels, showing an increase by approximately a factor of 4.4 when transitioning from the BS to the SR bunch length. A significant portion of HOM power for the SR bunch length is caused by 300 mm to 100 mm diameter tapers at the cryomodule ends. Removing these tapers reduces HOM power by 100.4 kW (SR) and 13.2 kW (BS). Replacing them with 300 mm to 160 mm diameter tapers reduces HOM power by 52.3 kW (SR) and 6.8 kW (BS).

### 3.4.10 Fundamental Power Couplers

A new family of fundamental power couplers (FPC) has to be developed for FCC-ee which all must operate reliably in CW mode. For the 400 MHz 2-cell cavities, the reverse-phase operation scheme reduces the input power requirement for the Z working point from 0.9 MW to approximately 400 kW, matching the requirements for the WW and H working points at the same $Q_L$ level of $9.2 \times 10^5$. However, additional margins must be considered for the RF power modulations and failure scenarios, where one or more cavities may trip, requiring the remaining cavities to compensate. An adjustable coupler is needed to cover a $Q_L$ range from $9 \times 10^5$ to $4.5 \times 10^6$, as required for the $\mathrm{t\bar{t}}$ working point, to minimise the input power to $\approx 80$ kW (Table 3.9).

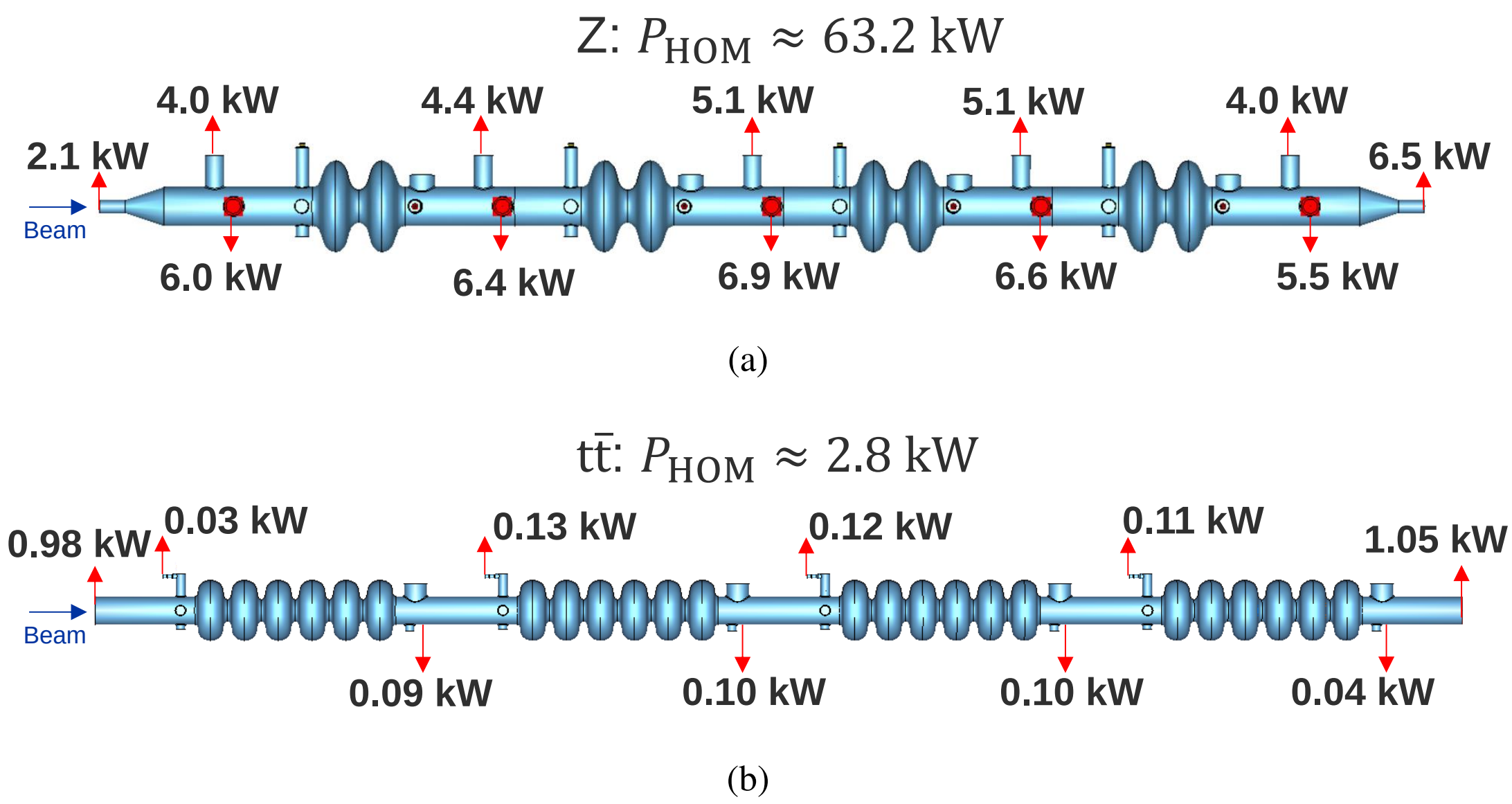

Fig. 3.32: (a) The figure displays an approximation of the HOM power distribution at the Z operating point in a cryomodule configuration consisting of four 2-cell 400 MHz cavities, each equipped with two hook-type couplers for trapped dipole mode damping, and two coaxial lines on the beam pipe for HOM power extraction. The total HOM power propagating through all eight hook-type couplers is approximately 0.7 kW. (b) An approximation of the HOM power distribution in a cryomodule configuration with four 800 MHz cavities at $\mathrm{t\bar{t}}$ operation. Each cavity has two DQW HOM couplers. Most of the HOM power propagates out through the beam pipes, requiring beamline absorbers between modules. In both cases, BS bunch length is considered for HOM power calculation.

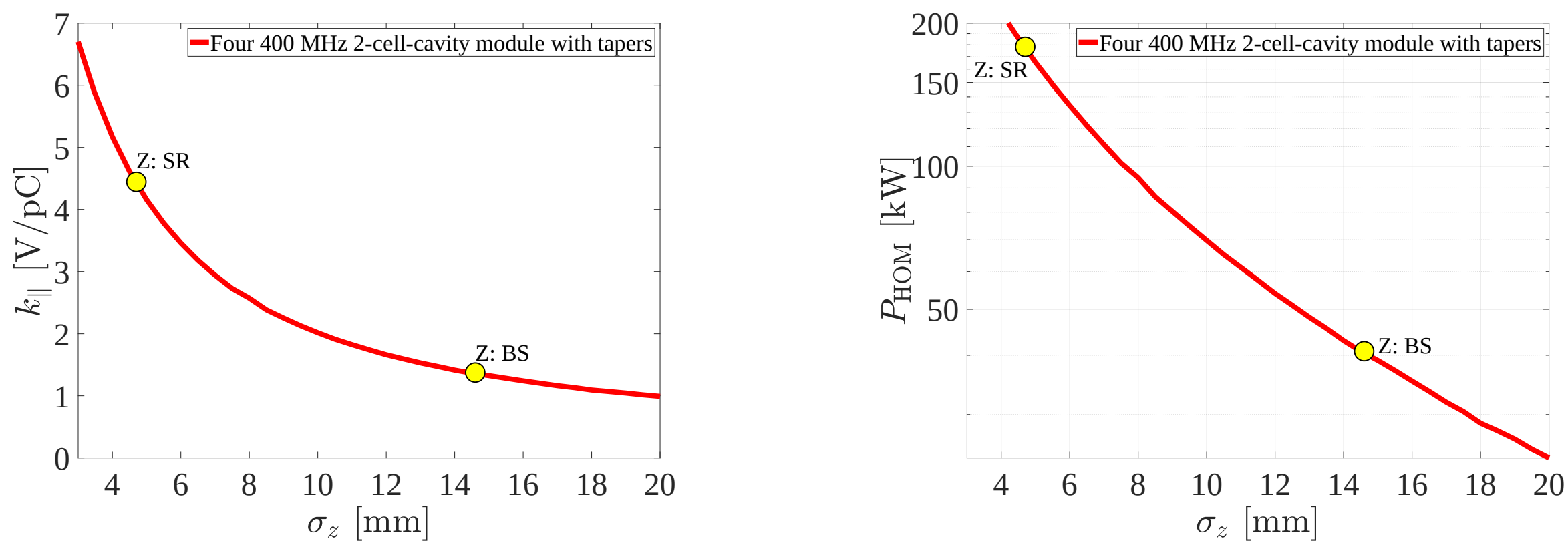

Fig. 3.33: The figures show the loss factor (left) and HOM power (right) at the Z working point for 2D-axisymmetric modules consisting of four 2-cell cavities with tapers at both ends and no couplers. Circle markers highlight the loss factor and HOM power corresponding to the BS and SR bunch lengths at the Z working points.

At 800 MHz, the couplers installed on the 6-cell cavities for the $\mathrm{t\bar{t}}$ require $Q_{L,\mathrm{opt}} = 4.2 \times 10^6$ to operate at $\approx 200$ kW. The same concept of the adjustable FPC is foreseen to unify the coupler designs with the collider and booster 800 MHz RF systems. The $Q_L$ range is from $4.2 \times 10^6$ to $2.7 \times 10^7$, which is a compromise between the tuning capabilities and RF power overhead (see Section 6.3).

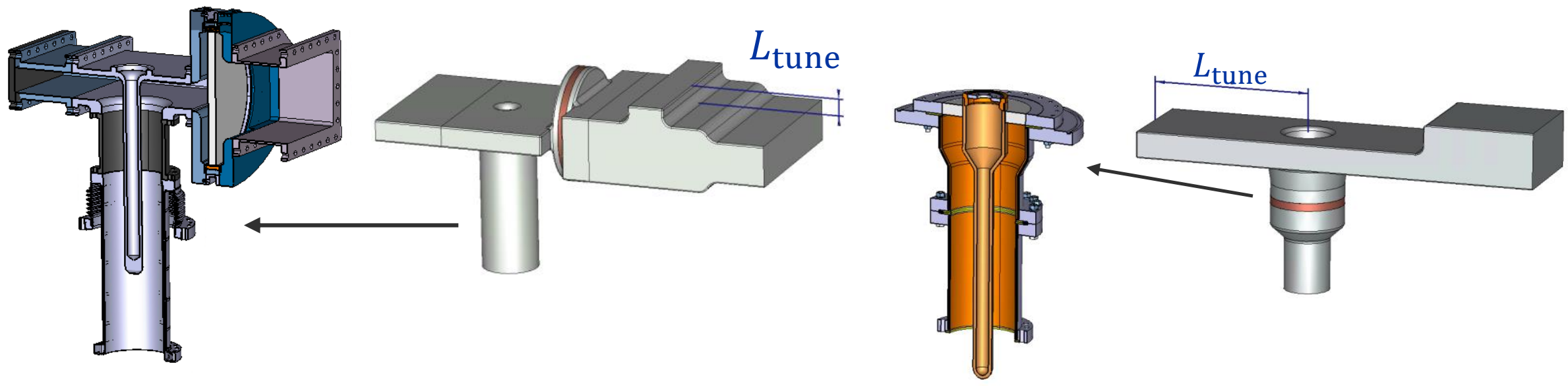

(a) (b)

Fig. 3.34: The RF models of the 400 MHz(a) and 800 MHz(b) power couplers, along with their mechanical representation concepts. $L_{\mathrm{tune}}$ is used to adjust $Q_{\mathrm{L}}$ when transitioning between working points by altering a waveguide plate on the air side of the coupler.

At 400 MHz, with a nominal power of approximately 400 kW, the proposed design concept incorporates an alumina ceramic disc window positioned within the rectangular waveguide, following an approach similar to that developed for Linac4. This configuration includes a step transition to a vertically mounted coaxial coupler, which interfaces with both the cavity and the surrounding cryomodule. The placement of the ultra-high vacuum (UHV) boundary within the rectangular waveguide presents logistic challenges for assembly, as the insertion into the outer vacuum vessel is intended to occur outside the clean room. However, technical solutions involving localised enclosures are being explored to address these constraints.

At 800 MHz, for a nominal power of 200 kW, the design adopts a more conventional annular disc window within the coaxial section of the coupler. Integration with the conceptual design of the 800 MHz cryomodule necessitates an innovative approach to the structural support system, ensuring that the horizontally mounted coupler is adequately supported while avoiding hyperstatic constraints that could impede thermal expansion. Engineering solutions to address these challenges are currently under development. The RF models of both couplers, along with their respective mechanical design concepts, are presented in Fig. 3.34.

### 3.4.11 SRF cryomodules

Two types of cryomodules are planned to accommodate the 400 MHz and 800 MHz cavity strings, covering the four operating modes of FCC-ee. The first type is designed for operation at 4.5 K and will house four 2-cell 400 MHz cavities, with a total length of 11.24 m. For the 6-cell 800 MHz cavities, the cryomodule will be designed to operate at 2 K, with an expected total length of 10.25 m.

The RF design of the cavity string has been refined through iterative development to optimise several key aspects: improving the integration of mechanical elements such as bellows and flanges within the string, enhancing the overall integration of the string within the cryomodule, including the orientation of rigid components such as fundamental power couplers (FPC) and coaxial extractors, and ensuring compactness and efficient longitudinal integration within the FCC tunnel.

The 400 MHz cryomodule conceptual design includes an isostatic supporting system for the cavities that allows free thermal contraction of the different elements, the weight of the FPC and the vacuum forces are intended to be discharged on the vacuum vessel. Helium tank, tuner and reinforcement structures are still to be engineered. In the conceptual design, the space occupation has been identified by rescaling the correspondent elements of the LHC cryomodule. In Fig. 3.35 it can be seen that there is limited clearance to integrate the actively cooled thermal shield and magnetic shield, not present in the LHC cryomodule. The cryomodule outer diameter is constrained by the length of the FPC outer conductor, actively cooled with supercritical helium at 4.5 K. There is a margin to reduce this value towards a more compact cryomodule design once the FPC maximum power and RF losses for the Z working point are finalised.

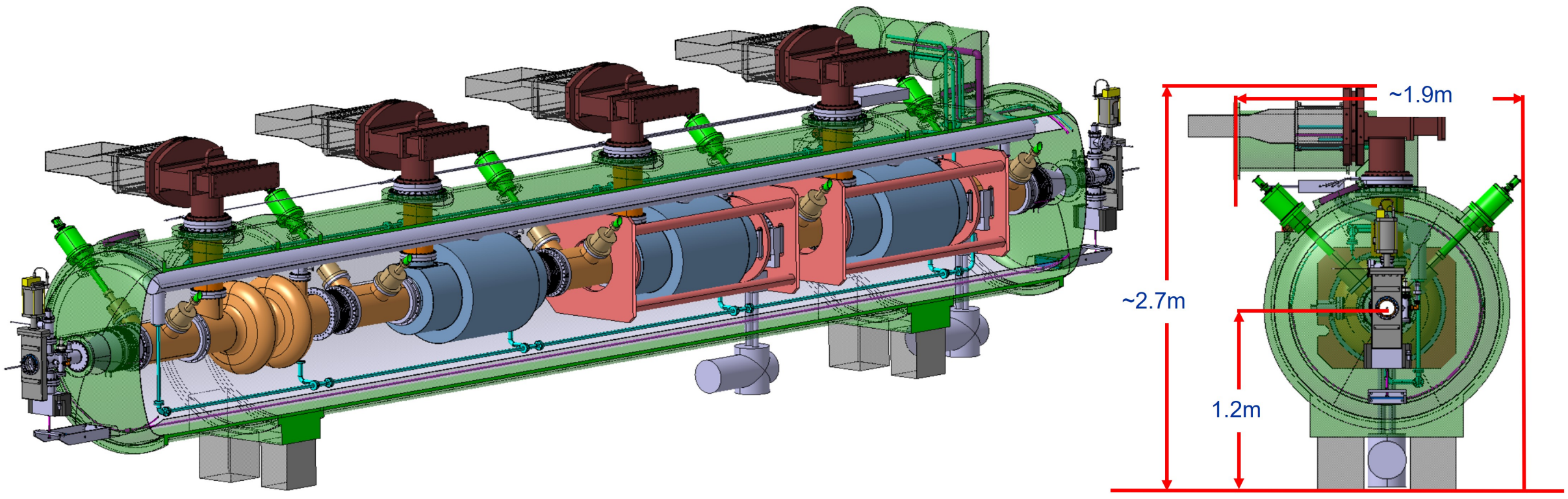

Fig. 3.35: Lateral and front view of the 400 MHz cryomodule.

The current cryomodule design and assembly sequence are based on a cylindrical vacuum vessel, which is the preferred choice for maintaining structural integrity under vacuum forces. This design is commonly used in machines with a large number of cryomodules, such as XFEL and LCLS-II. A cylindrical vacuum vessel also presents a cost-effective solution in several respects. First, it simplifies material procurement, as it allows the use of low-carbon steel tubular products. Second, it facilitates an industrialised assembly process, following established practices from previous machines, where the cavity string slides into the vacuum vessel without requiring the vessel itself to be cleaned to cleanroom standards at any point.

However, the new fundamental power coupler (FPC) design, which incorporates a ceramic window in the waveguide, necessitates a revision of the assembly process. The first integration approach, which is compatible with the current cryomodule conceptual design, involves inserting the coupler antenna and ceramic window in a second step—after the cavity train has been placed inside the vacuum vessel. This would be achieved using a local cleanroom or a glovebox. The local cleanroom approach has been successfully employed for FPC maintenance in other facilities, such as ESS, XFEL, and SNS, but it poses a potential risk of cavity contamination. Therefore, both the local cleanroom concept and the associated assembly procedure must be designed and thoroughly validated with prototype testing before finalising the current cryomodule design.

If this integration method proves too complex or incurs excessive R&D costs, an alternative design is under consideration. This would involve a vacuum vessel composed of two separate sections, as used in other projects such as Crab cavities and SPL. In this configuration, the top part of the vessel would be brought into the cleanroom along with the cavity string, enabling the FPC assembly to take place in a fully controlled clean environment. While this solution minimises the risk of cavity contamination, it introduces significantly greater complexity and increases the cost of the vacuum vessel.

The design of the HOM for the 400 MHz cryomodule is determined by the highest power extraction requirements at Z operating point, the coaxial extractors in between cavities should be dimensioned to extract up to 6 kW-10 kW each. Thus, the coaxial extractors shall be rigid connectors ($P_{\rm HOM} \times 6$ > 1 kW) reducing the freedom in the port positioning across the vacuum vessel, with complications in matching the routing and the assembly and maintenance needs. For the absorption of the HOM power leaking through the beam pipe, it is foreseen the integration of warm BLA in the cryomodule interconnection regions. Examples of water-cooled BLA have been tested for EIC, and their impact on the machine reliability and cavity contamination is now under assessment.

The 800 MHz cryomodule conceptual design, Fig. 3.36, is based on the PIP-II HB650 cryomodule, with the strong back support for the cavity string, and has been developed in collaboration with FNAL.

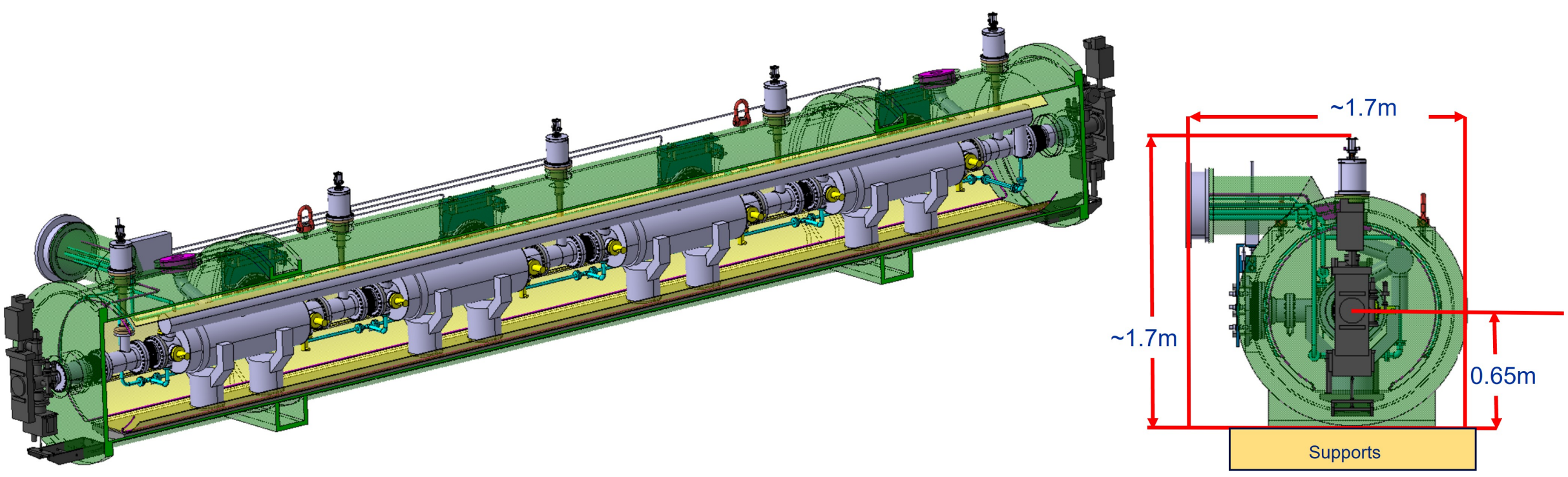

Fig. 3.36: Lateral and front view of the 800 MHz cryomodule.

The parameters that require adjustment of the current PIP-II design are: (i) the power for the FPC, (ii) the presence of HOM extractors, (iii) the requirement of the design to be compatible with both segmented and continuous architecture. The FPC of the HB650 cryomodule, designed for a peak power of 65 kW, has a bellows in the outer conductor to decouple thermo-mechanically the waveguides from the outer conductor, avoiding transmitting cantilever forces to the cavity string. The bellows must be eliminated from the FCC FPC design, given the requirement of 250 kW of forward RF power to be transmitted. Active cooling is preferred for the cooling of the FPC outer conductor, using supercritical helium at 4.5 K to be able to adjust the cooling capacity by adjusting the helium flow rate in the different working points and phases of machine operation. The requirements for the HOM extractors for the 800 MHz cavity string are still under definition due to the necessity of introducing BLAs. The current design is compatible with a continuous machine architecture, although if the power to be extracted from the BLAs is too high to be dumped in the cryogenic lines (XFEL design for 100 W) the requirement of warm BLA would push towards a segmented architecture for the 800 MHz cryomodule, as curently defined for the 400 MHz. The conceptual design of both cryomodules is well advanced, in both cases the FPC remains the most critical component to design and integrate and may require modifications to the current models. The current design allows some space contingency: the next step will focus on the engineering and dimensioning of the components inside the cryomodules.

To address the definition of the RF cryogenic capacity needs, heat load budgets have been calculated for both cryomodules. The values of static heat loads in Table 3.14 include a 50% margin to account for the preliminary level of maturity of the two design concepts. This margin will likely decrease when engineering the cryomodule internal components. The dynamic heat loads only account for the power dissipated by the cavity at the nominal values of $Q_0$ and $E_{\rm acc}$. The heat loads from power coupler, HOMs and non-superconducting elements of the beamline are not included. It is noted that the values of $Q_0$ and $E_{\rm acc}$ are target values of the R&D processes (Table 3.11). At present, there are no elliptical 2-cell and 6-cell cavities with the stated performance at 400 MHz and 800 MHz, respectively, the specifications are based on the rescaling of performance obtained at different frequencies, with different surface treatment recipes. The stated performances rely on the success of the R&D programmes.

Table 3.14: Cryomodules heat loads budget.

| | 400 MHz | 800 MHz |
|---|---|---|
| Max. heat load to thermal shield per CM [W] | 327 | 180 |
| Max. static heat load to 4.5 K / 2 K bath per CM [W] | 197 | 56 |
| Max. dynamic heat load to 4.5 K / 2 K bath per CM [W] | 619.2 | 130.2 |

For the dynamic loads, it was assumed that at least a 20% operational margin should be added to the nominal value to ensure machine operation with up to 10% faulty or unpowered cavities, and voltage needs being compensated by increasing the gradient of the remaining cavities. For both cryomodules preliminary cryogenic schemes have been defined, together with the dimensions of the cryomodule inner pipes. The cavities are immersed in a saturated helium bath, the helium tanks are connected on top through a 2-phase tube, from which vapour pumping and pressure control ensures temperature control of the helium bath. Cavity cool-down is done from a lower point through a valve-controlled line at 4.5 K. Steady-state operation filling is ensured by a valve directly feeding the 2-phase line. For the 800 MHz, Fig. 3.37, the valve is a JT subcooling the liquid to 2 K. Helium level control in the 2-phase is ensured by liquid level measurement (LT) in a phase separator, connected to the bottom supply line, as in the helium tanks, so that the level can be measured in one point only (communicating vessels).

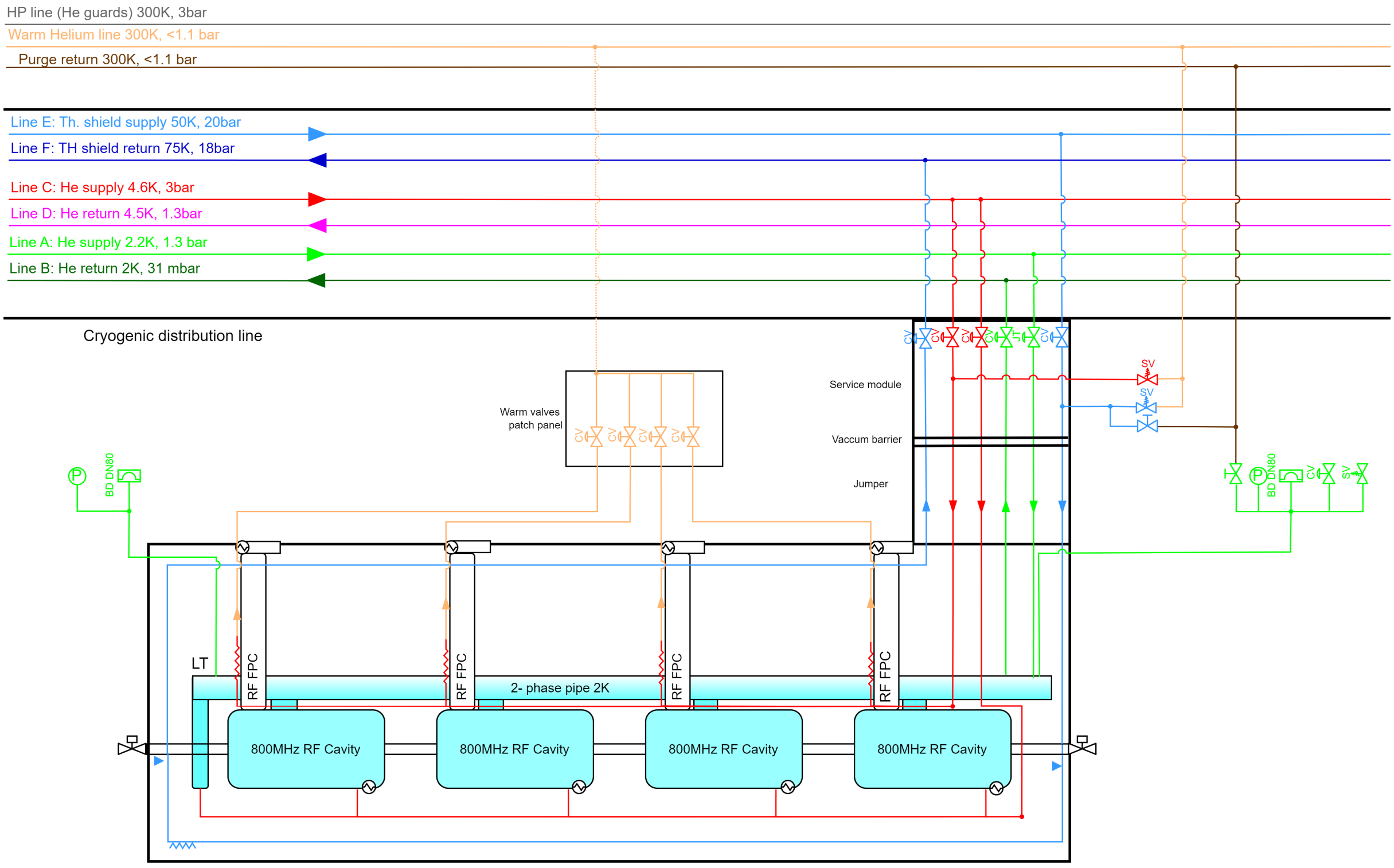

Fig. 3.37: Cryogenic scheme for the 800 MHz cryomodule.

An actively cooled thermal shield at 50 K is needed to intercept the conduction heat from the thermal bridges between ambient temperature and helium bath, such as the cold-to-warm transitions in the beam tubes, cavities supporting system, etc. The static and dynamic heat loads linked to the FPC external conductor are gas-cooled through a double-walled heat exchanger; cold supercritical helium is injected at the coldest end of the FPC and is extracted at room temperature at the vacuum vessel interface. Temperature-based mass flow regulation on each FPC is possible with warm valves outside the CM. Finally, two burst disk exhaust lines protect the cavity circuit against overpressure in the worst-case event of accidental venting of the cavity vacuum with air. For the less critical case of insulation vacuum break, only one of the two burst disks should be enough to evacuate the mass flow rate of helium vapour. The scenarios in which the cryogenic supply is interrupted, or the machine is subjected to a power cut, are considered abnormal operating conditions which could happen frequently, not an emergency scenario. The limited mass flow will be recovered through the cryomodule return line – the return valve default position is open in case of a power cut. If there are problems with the return line valve, the helium mass flow will be released in the tunnel through a pressure relief valve, which ensures leak-tight reclosing. It is important to mention that this is a preliminary schematic, several components still need to be included for cryomodule operation, for example the instrumentation, the valves and/or lines for individual cryomodule warm-up/cool-down,etc.

### 3.4.12 High power RF system

The high-power RF systems will comprise electrovacuum amplifiers and solid-state power amplifiers (SSPA), chosen to meet the RF power level required at different operating points and frequencies. For the FCC collider, the efficiency of electric power conversion from the grid to RF was considered a priority, together with cost-effectiveness and footprint minimisation. In the last three years, a novel concept of the compact (3 m high) low voltage (<60kV) two-stage (TS) multi-beam klystron (MKB) [267] was pioneered within CERN's High Efficiency klystron Project. Such a tube can generate CW RF power up to 1.2 MW at 400 MHz with very high efficiency – above 85%. TS MBK design was completed and is ready for prototyping in industry. TS MBK layout and the tube performance are shown in Fig. 3.38. In the CDR and MTR, it was assumed that for the Z operating point, one tube would feed one cavity and then be reused for other operating points by splitting the RF output power. The novel TS MBK technology can be considered as an optimal choice for various high-energy colliders like CLIC, ILC and Muon Collider.

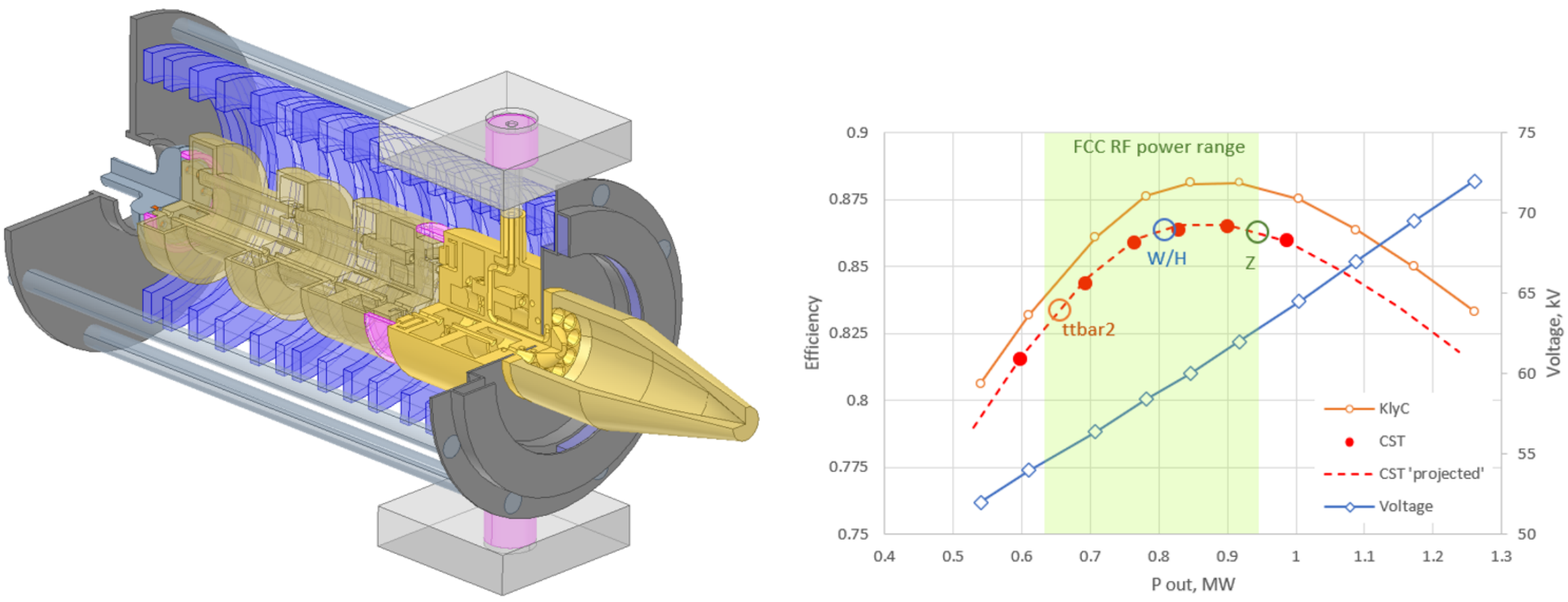

Fig. 3.38: Artist's view of TS MBK is shown on the left. Simulated RF power/efficiency performance of the TS MBK operated at different High Voltage levels is shown on the right.

The present baseline FCC collider layout with a common 2-cell 400 MHz SRF cavity used for Z, WW, and ZH operating points reduces the maximum RF power needed for a single cavity from 1 MW to below 0.5 MW. In addition, the RPO mode leads to different RF power transient modulations for focusing and defocusing cavities due to gaps in the beam filling schemes (Fig. 3.39). Thus, the single 0.5 MW RF power source feeding one cavity was the solution that satisfied the requirements for all beam energies.

Following these changes, the TS MBK design was scaled down in RF power from 1 MW to 0.5 MW, providing high efficiency as before with the original tube. However, analysis of the klystron operation with required RF power transient modulation regulated by the LLRF system brought the operational efficiency of the tube down from 85% to 68%. The reason is that the highest efficiency is delivered by a klystron when it operates in saturation. Without changing the operating high voltage on a fraction of microsecond scale (not yet technically feasible), efficiency will be linearly degraded by reducing the RF power. CEPC adopted a possible solution using the depressed collector (5 stages) to recover the energy back into HV modulator [268]. At this point, klystrons will operate with an average efficiency of about 65%, but the overall RF system efficiency can be pushed up to 85%. This approach is not commonly used in industrial high-power klystrons, and it will result in a significant increase in the tube complexity and cost. Moreover, the transient nature of the recuperated beam power will require specialised electronics to return the energy to the DC modulator.

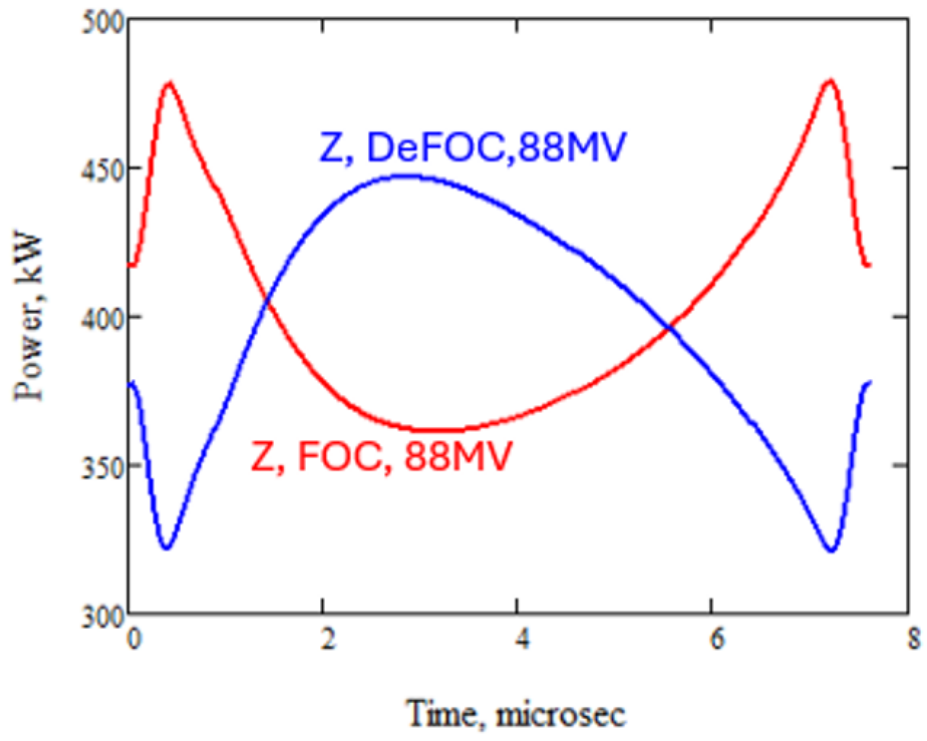

Fig. 3.39: Transient RF power modulation profiles for two types of accelerating cavities (Z-pole).

A gridded tube, like an inductive output tube (IOT), is another class of electrovacuum device that is renown for their ability to operate efficiently in a wide range of output RF power levels regulated by the input RF signals. For example, a 1.2 MW, 0.7 GHz multi-beam (10 beams) IOT was successfully prototyped in industry for ESS almost a decade ago [269], proving that powerful MB IOT is a mature technology. However, the practical efficiency in these tubes is limited to 70-75%. The extended IOT version, called Tristron (a hybrid of triode and klystron) was proposed in the late 1960s, but it has never been commercialised. A tristron comprises an additional idler cavity located before the output cavity. This cavity dramatically improves bunching quality and increases efficiency. A tristron was studied and optimised at CERN as a candidate for FCC RF power source [270]. The final MB (10 beams) tristron design at 400 MHz showed excellent performance in RF power range from 300 kW to 600 kW with efficiency exceeding 90%. A snapshot of particle dynamics in the tristron and simulated RF power/efficiency performance at different operating voltages are shown in Fig. 3.40. Compared to TS MBK, the tristron provides a much more compact and cost-effective solution.

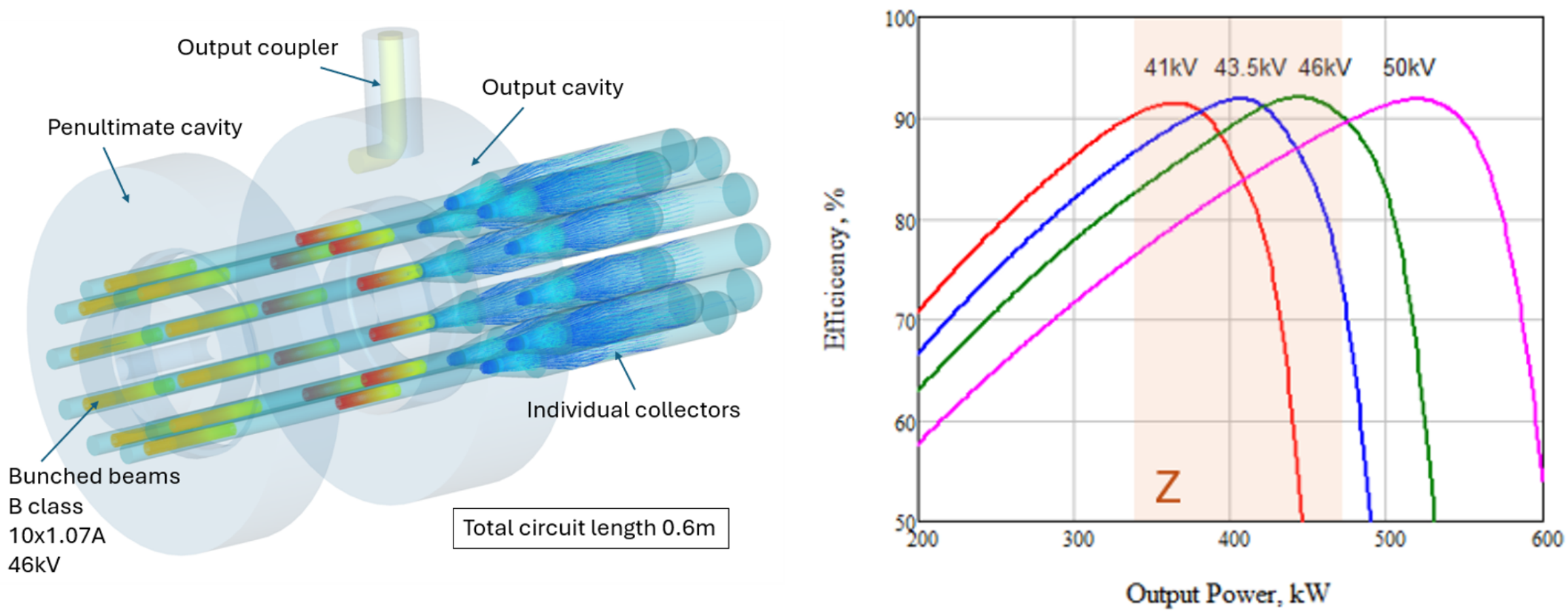

Fig. 3.40: Left: A snapshot of particle dynamics in the tristron simulated in 3D PIC CST. Right: RF power/efficiency performance at different operating voltages.

The RF power modulation waveform shown in Fig. 3.40 can be convoluted with the tube performance at a fixed voltage for the operational efficiency of the tristron at the Z operating point. , The tristron transient efficiency simulated for the operating voltage of 46 kV is shown in Fig. 3.41. Despite the very broad range of the RF power modulation required, a remarkably high average efficiency of 88.7% can be obtained.

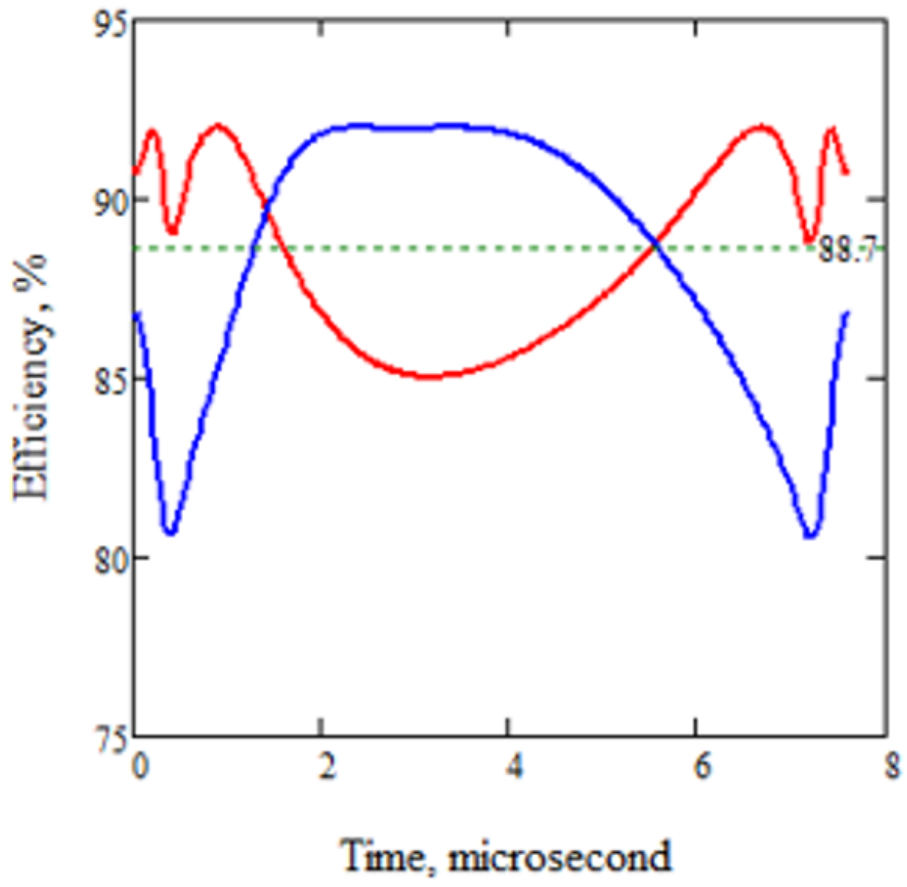

Fig. 3.41: Transient RF efficiency of the tristron when operated at the Z-pole.

The failure scenario of a single cavity trip was also studied. In this case, there must be a dramatic change of the RF power modulation waveform, so that the peak RF power required can temporarily exceed 520 kW (Fig. 3.26). The proposed recovery scenario is based on the slow (200-300 µs) ramp-up of the tristron operating voltage from 46 kV to about 50 kV. At this new set point, the operating efficiency will be slightly reduced: from 88.7% to 85.2%. For the WW and ZH poles the RF power transient sweep required will stay within 1%, thus, tristron will operate with an efficiency exceeding 90%, while similar performance degradation is expected in failure scenarios. Finally, if a 1-cell 400 MHz cavity for the Z operating point is re-considered in the future, two tristrons can be combined through the 3-dB hybrid and deliver 1 MW RF power. The booster and $t\bar{t}$-pole will be operated with 800 MHz RF power sources. Direct scaling of 400 MHz of tristron to the higher frequency and reduced RF power (200 kW) is a straightforward task, it will be initiated upon completion of the technical design of the 400 MHz tristron. This 800 MHz tube will feed a different number of the cavities depending on its location: one tube per cavity $t\bar{t}$ collider and one tube per four cavities for the Z, WW and ZH poles and the booster.

A preliminary 3D mechanical model of the tristron is shown in Fig. 3.42. The tristron development programme is separated into two phases. The first one is a technology demonstrator based on a retrofit and upgrade of the existing ESS MB IOT. This work will be done in a collaboration of CERN, ESS and Thales. It is expected that the project can be completed on a short timescale of 12-18 months, providing ESS with an additional option for the efficiency upgrades of their RF system in the future. The second phase is technical design and prototyping of a 400 MHz tristron for FCC. During this phase attention will be focused on system optimisation and cost reduction. It is antcipated that a tristron prototype will be built and tested 30-36 months after the formal project assignment between CERN and Thales.

Tristron technology will cover all the power and frequency ranges for FCC collider and booster at different energies. However, $t\bar{t}$ mode booster only requires 13 kW at 800 MHz to feed the cavity. It is then possible to split RF power into 10 cavities from the 200 kW tristron which is operated efficiently at a lower RF power of about 130 kW. However, the RF distribution system could be too bulky and expensive. As an alternative, a 15 kW SSPA can be employed. There is no specific development of such an amplifier at CERN and most of the development is in industry. Recently it was reported that L-band 6 kW GaN/SiC single transistors operated at 100 V are commercialised [271], providing drain efficiency of about 80%. So far these chips are operated at a low duty cycle (10%), but predictions are for further performance improvements in the coming years.

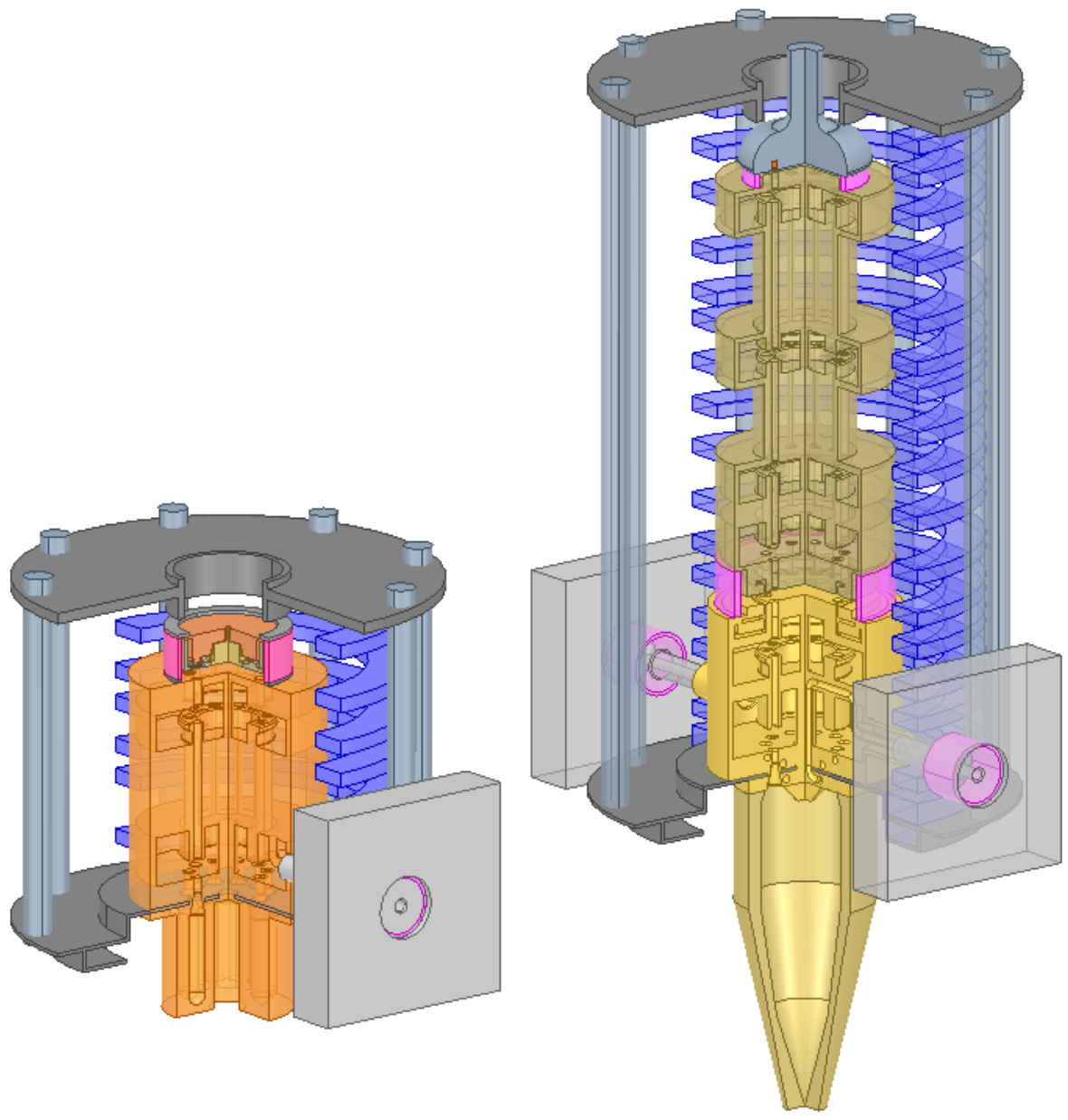

Fig. 3.42: Tristron 3D view (left) compared to TS klystron (right).

The RF power sources are grouped by RF units, each one being individually powered by a power converter located at the surface. Fig. 3.43 summarises the distribution schemes for the collider and booster at the different beam energy levels.

### 3.4.13 RF system integration in the tunnel

The integration studies of the RF system in the accelerator tunnel and in the klystron gallery were considered as high priority and were completed to verify that: (i) the cryomodules and the dedicated services fit, with minor exceptions, in a tunnel with 5.5 m diameter, both for the collider and the booster ring (ii) the RF equipment necessary for the $\mathrm{t\bar{t}}$ mode working point fit in the 2032 m of the long straight section. The civil engineering layout, transport constraints, cable trays and general services, as designed for the arcs, were input to the study. The focus was the integration of the cryomodule, cryogenic distribution line and waveguide connection from the FPC to the klystron gallery. All the different working points were considered to ensure feasible transitions from Z to $\mathrm{t\bar{t}}$ and coherence in the beam positions between the collider and booster rings, Fig. 3.44.

The cross section in Fig. 3.45 presents the 400 MHz cryomodule in point PH connected to the cryogenic distribution line (QRL) through the jumper. Some space was allocated along the QRL for the service module containing the cryogenic valves and a local heat exchanger (if needed), in the sector with the 800 MHz cryomodules at 2 K, in case the current baseline with a centralised heat exchanger is not viable.

The cryogenic services required for the $\mathrm{t\bar{t}}$ working point will be installed at the beginning of machine operation, including all the service modules and jumpers required to feed all the cryomodules. A platform has been included on top of the QRL for the access and maintenance of the FPC and waveguides. Access to both sides of the cryomodule is ensured for the maintenance of the HOM coupler connectors and tuner in case of need. In point PH the 800 MHz cryomodules sit in the shadow of the 400 MHz cryomodules, thus integration is less critical. The cross section in Fig. 3.46 presents the 800 MHz cryomodule in point PL, the same cryomodule design will be compatible with installation on the ground, for the collider, and installation on an elevated platform for the booster, for this reason the FPC is horizontally oriented.

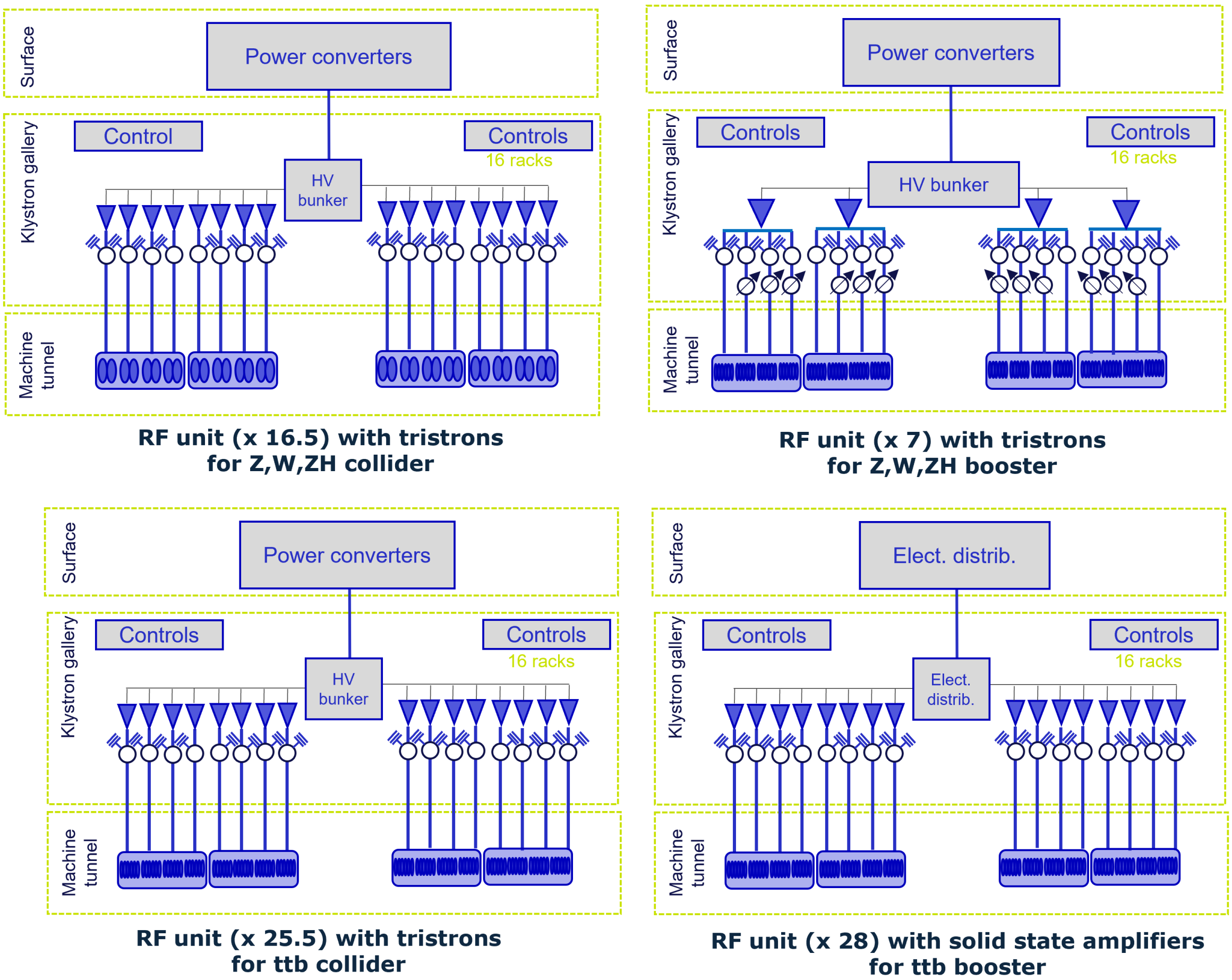

Fig. 3.43: High power RF distribution layouts for the collider and booster.

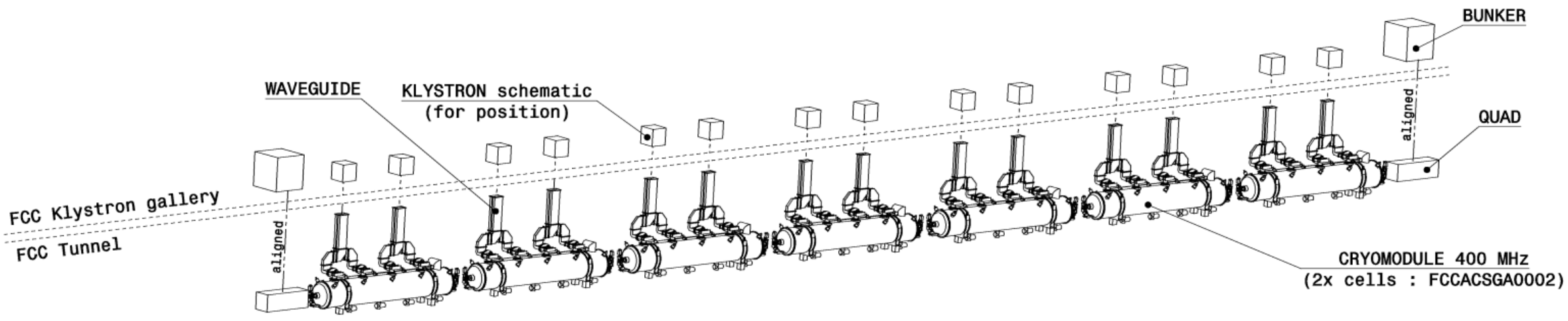

Fig. 3.44: RF longitudinal integration of the 400 MHz cryomodules in point PH, and waveguide system layout from the FPCs to the klystron gallery

The clashes highlighted in dashed lines can be resolved with the adaptation of the general services, pavement and false ceiling to the requirement of the RF straight section, different to what is required in the arcs. The beam height in the RF straight section should be increased, from the current value of 980 mm in the arcs, to 1200 mm to accommodate the current design of the 400 MHz cryomodule to avoid the clash between the cryomodule supports and the pavement. Several solutions are under evaluation, in case the diameter of the cryomodule cannot be reduced, a slope in the tunnel should be introduced to lower the pavement in only the collider straight sections. After the revision of all the civil engineering and the services, it will also be possible to refine the integration work by adding more details, i.e., patch panels for warm valves, water pumps for the FPC inner conductor cooling, HOM load dumpers and BLA with the dedicated water cooled circuits, etc. In the following design stage, the integration work will be refined to adjust the position of the non-accelerated beam, necessary for Z and WW working points, to avoid it being in the transport/passage area.

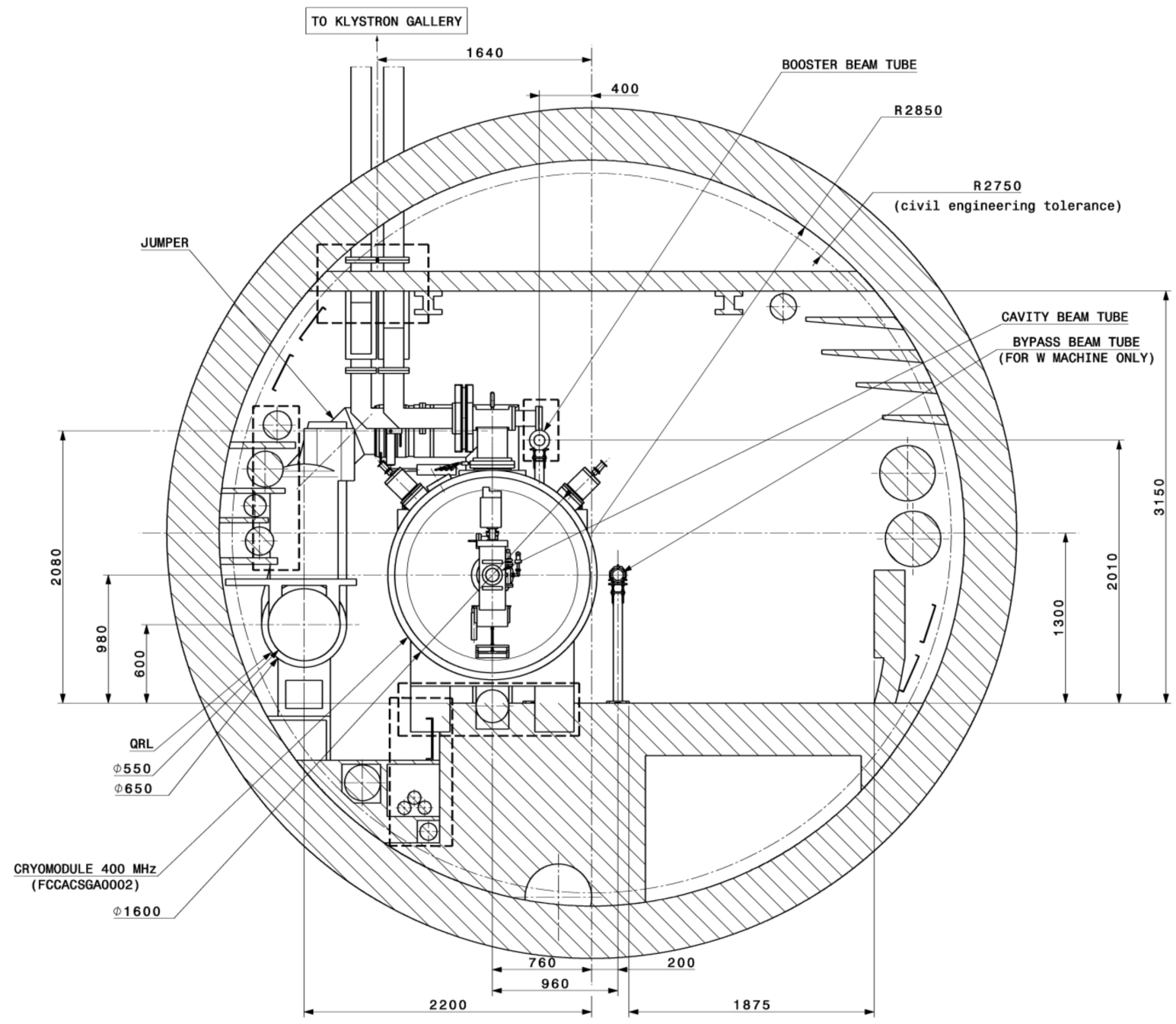

Fig. 3.45: 400 MHz cryomodule, integration in point PH (collider). Non-accelerated beam on the right of the cryomodule is necessary for Z and WW working points only.

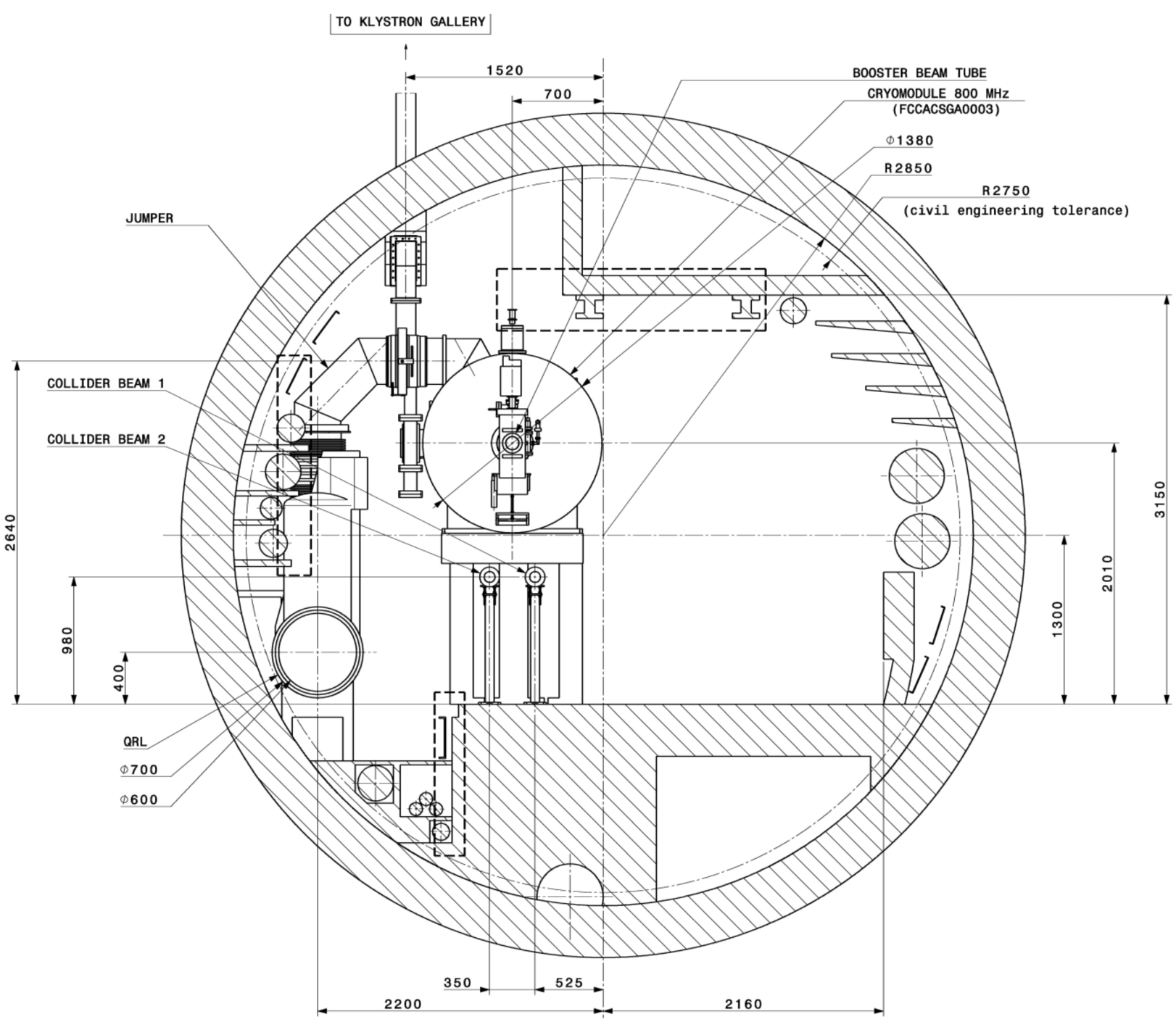

Fig. 3.46: 800 MHz cryomodule, integration in point PL (booster).

The collider beam positions and spacing in the RF straight section at the various working points will be achieved by a beam optics system for beam separation and recombination (see Fig. 3.47). It is important to underline that, in point PH both collider beams need steering, the position of the incoming beam needs to be adjusted compared to the arcs, otherwise, the cryomodule would be centred in the machine tunnel, strongly protruding in the transport/passage area. It has been possible to maintain the beam spacing in the collider at point PL at 350 mm, as in the arcs.

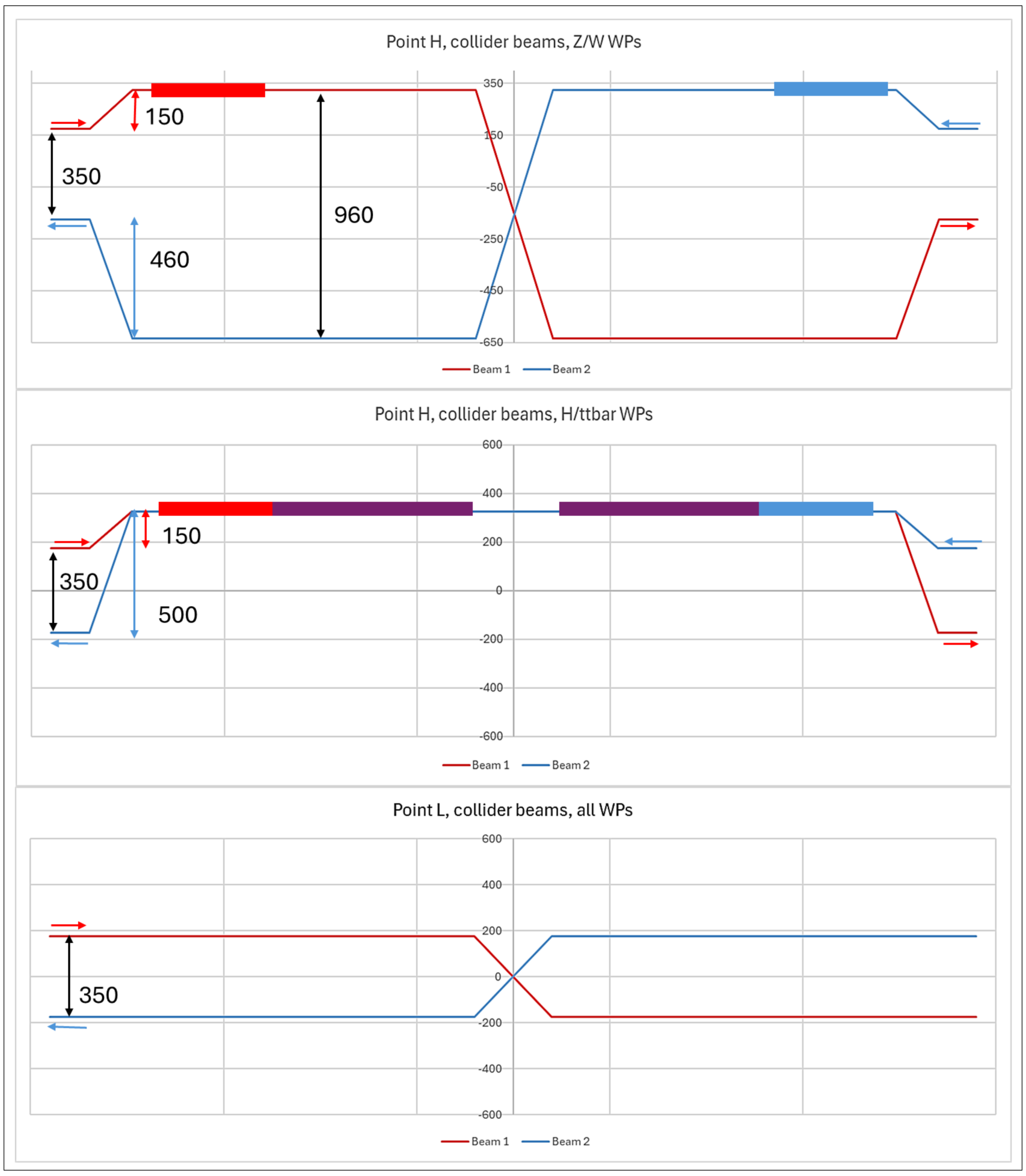

Fig. 3.47: Beam location required at the arc-RF interconnection region, in point PH (top and centre) and in point PL (bottom).

The longitudinal integration of the RF equipment along the long straight section was also considered, in parallel with the disposition of klystrons and bunkers in the klystron gallery. A vertical arrangement of the booster and main rings is chosen to ease the installation of the waveguide lines in the ducts. Thanks to a concrete chicane located at the duct entrance, the level of X-rays in the klystron gallery will be limited, thus allowing personnel access to the klystron gallery during commissioning and operation. Seven 400 MHz cryomodules and eight 800 MHz cryomodules are placed between two consecutive quadrupole magnets, whose location vertically corresponds to the location of the bunkers in the klystron gallery. This choice for the longitudinal integration maximises compactness and accounts for the constraint that every bunker can have a maximum of four klystrons on each side. The disposition of the RF equipment, at the $t\bar{t}$ working point, on each side of the IP point is summarised by the scheme in Fig. 3.48.

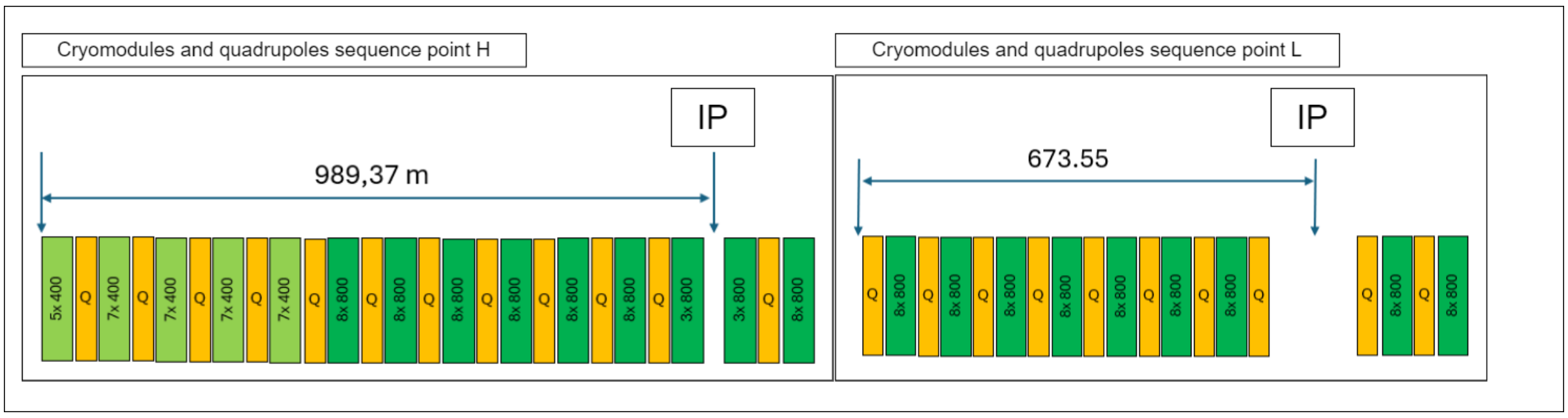

Fig. 3.48: Scheme illustrating the distribution of cryomodules and quadrupoles in point PH (left) and point PL (right).

### 3.4.14 R&D on SRF

Considering the timeline of the FCC-ee project, there is sufficient opportunity to conduct R&D aimed at enhancing the performance of SRF cavities. This would enable the optimisation of both the size and cost of the overall SRF systems [272].

The 400 MHz cavities built with the Nb/Cu technology require advanced engineering techniques for the copper substrates. The design and fabrication of two bulk seamless 400 MHz copper cavities are ongoing (see Fig. 3.49), allowing the ultimate SRF performance with this geometry and Nb/Cu copper technology to be assessed without the effect of welds in the equatorial area.

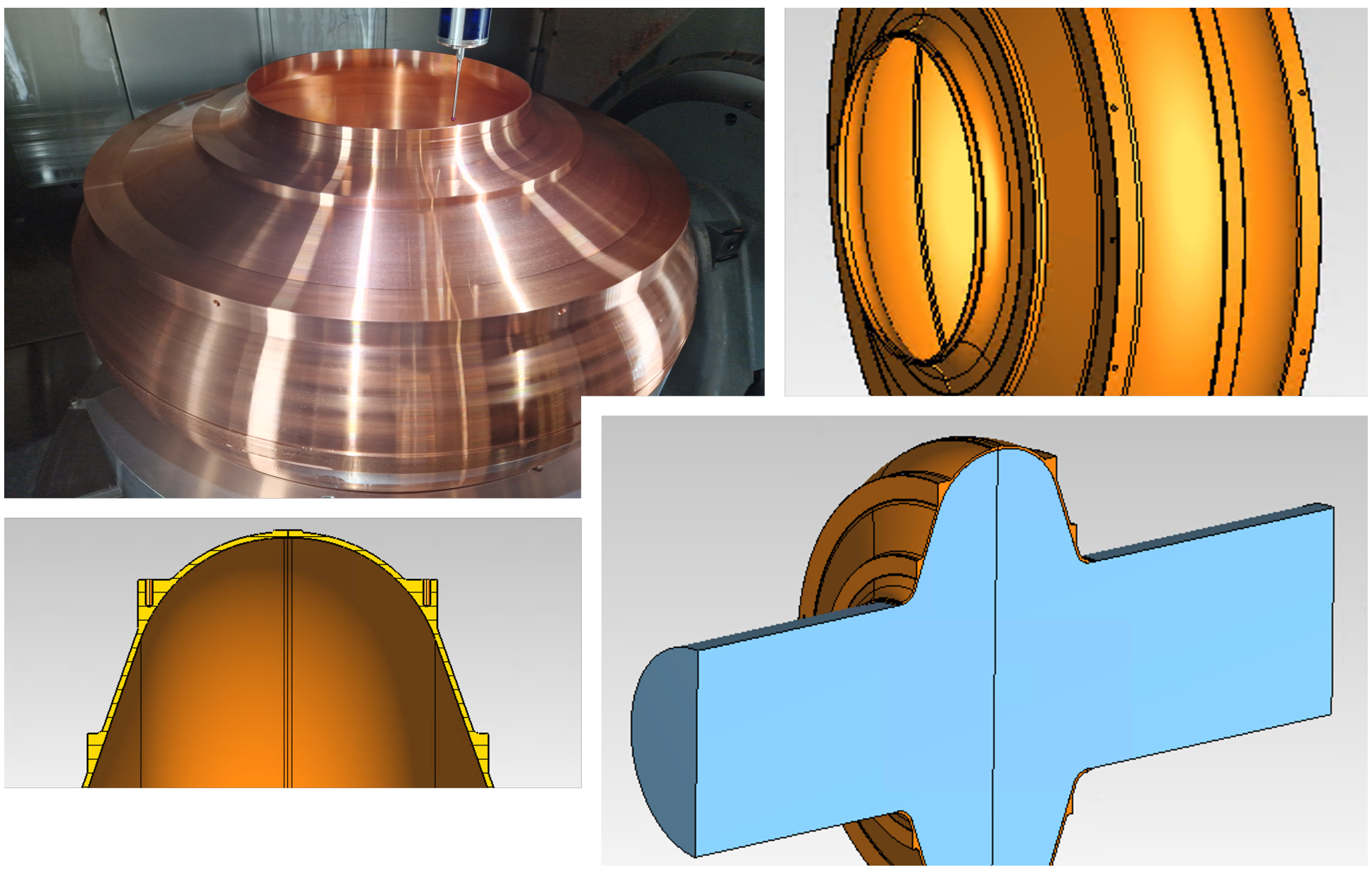

Fig. 3.49: Design and fabrication of the first 400 MHz 1-cell cavity machined from bulk copper.

Hydroforming technology, which is highly attractive for series production, is being explored in collaboration with KEK. The development of internal welding techniques is also being pursued, in particular, the welding of the RF ports on the cavity beam tubes.

Electropolishing is the best technique for preparing the surface of copper substrates for low surface roughness. The electropolishing setup has been available at CERN since 2022 and was successfully benchmarked on smaller cavities. Some R&D is still needed to optimise the polishing parameters on larger cavities, but a first successful attempt was on an LHC type 1-cell 400 MHz cavity with a simplified geometry (without RF ports - see Fig. 3.50). The most promising coating approach of niobium on copper is high-power impulse magnetron sputtering (HiPIMS), which has shown excellent results on several 1.3 GHz 1-cell cavities. It was performed on a 400 MHz LHC type cavity in 2023 and the test results at 4.5 K and 1.7 K were reported in the mid-term report.

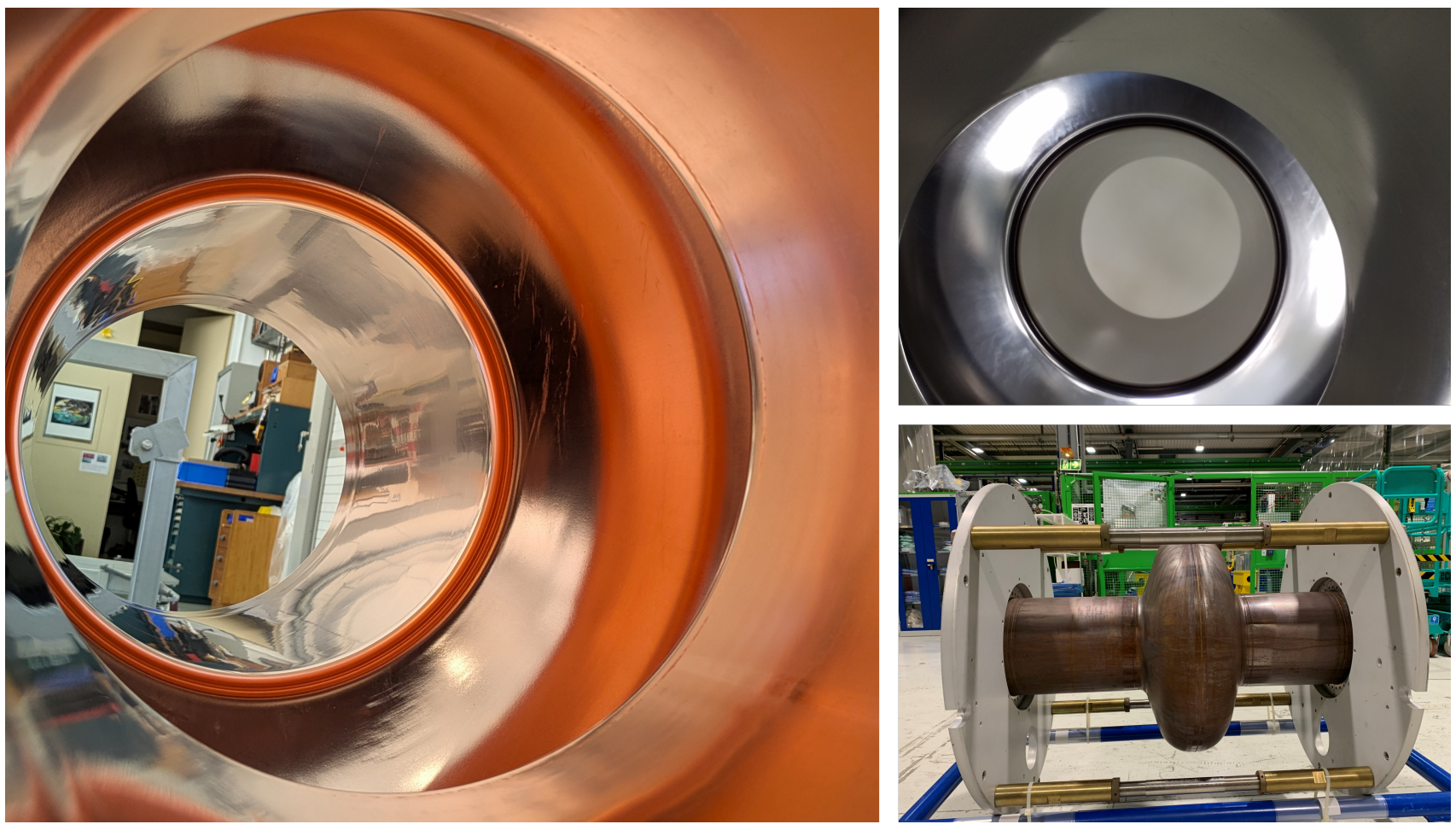

Fig. 3.50: Electropolishing and niobium coating of a simplified LHC 400 MHz cavity.

For the 800 MHz cavities built in bulk niobium, the 5-cell bare cavity developed by JLAB on a similar RF design shape has already demonstrated that an accelerating gradient of 30 MV/m is achievable with a quality factor of $Q_0 = 3.0 \times 10^{10}$ . To reach a $Q_0 = 3.8 \times 10^{10}$ or higher values, a dedicated R&D programme in collaboration with FNAL [273] is on going. It will be based on a combination of nitrogen doping and mid-temperature bake-out, inherited from the 5-cell 650 MHz cavity R&D performed within the PIP-II project.

An innovative concept of the slotted waveguide elliptical cavity (SWELL) is being developed [274–276] in parallel to the baseline study. The design consists of elliptical cavities where longitudinal waveguide slots crossing perpendicularly to the RF surface are added to damp transverse HOMs. Thanks to this approach, the cavity is seamless by its nature and can be built by sectors, which is very appropriate for precise manufacturing techniques. This quadrant-based configuration allows direct access to the RF surface when separated, thus facilitating the surface preparation, surface inspection and thin film deposition (with any kind of superconducting material). The cavity is also robust against frequency detuning by Lorentz forces or microphonics, and allows cryogenic cooldown with a significantly reduced volume of liquid helium.

RF design efforts led to 400 MHz 2-cell and 800 MHz 6-cell SWELL cavities as alternatives with almost one order of magnitude better transverse impedance damping. A prototype of a simplified SWELL version of a 1-cell 1.3 GHz elliptical cavity has been fabricated for a feasibility demonstration of this new concept (see Fig. 3.51). The cavity was installed in a vertical cryostat at CERN in August 2024. The $Q_0$ factor was measured at 2.2 K using a self excited loop (SEL) digital LLRF system and the standard decay time procedure. After tens of hours of RF conditioning of low-field multipacting barriers, the accelerating field has been gradually increased. At $E_{acc} = 1.1$ MV/m, the cavity quenched repetitively, most likely due to the presence of a surface defect observed on one quadrant before cavity assembly. The cavity reached a quality factor of $Q_0 = 1.0 \times 10^{10}$ at 1 MV/m with a very good reproducibility. This corresponds to a residual surface resistance of 20 $\mathrm{n\Omega}$ as shown on the plot in Fig. 3.51, which is a remarkable result for such a complex RF structure. It demonstrates, in particular, that no RF leakage occurs through the slots and in the contact surfaces of the quadrants. Additional niobium re-coatings, surface treatments and cold RF tests are planned in the next months to study and fully qualify this very attractive and innovative concept.

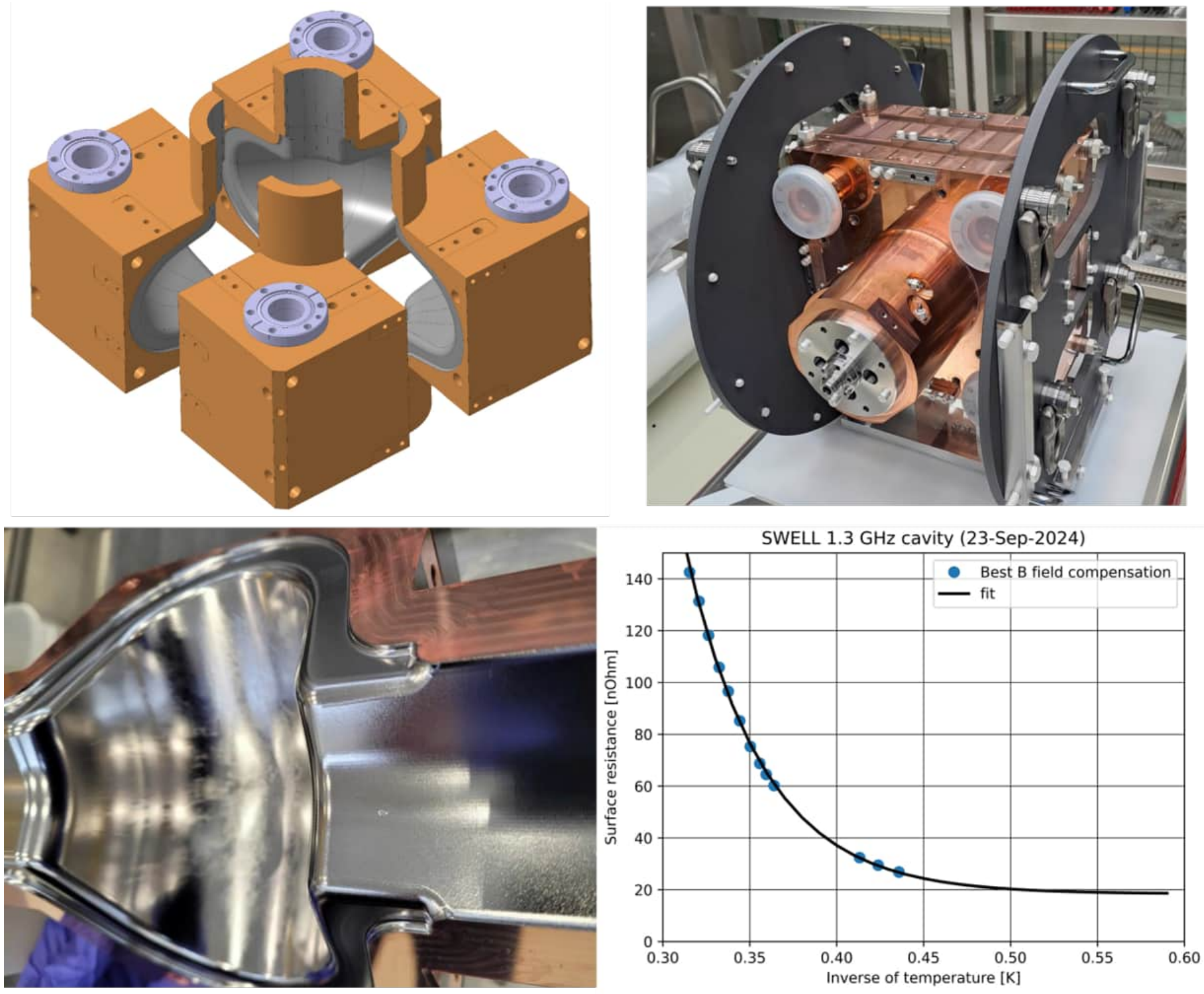

Fig. 3.51: SWELL cavity prototype design, fabrication and test.

## 3.5 Survey and alignment systems

The survey and alignment systems and strategy are based on the steps detailed in this section. Machine and detector components undergo fiducialisation and assembly measurements at the surface to ensure precise alignment once installed in the tunnel. This process begins with marking the floor to indicate the designated positions for the jacks that will support each component. After placement, these jacks are surveyed to confirm their correct positioning.

Following this, absolute alignment is carried out relative to the underground geodetic network, establishing a stable reference for all components. Once the components are interconnected, a process known as smoothing is performed to refine their relative alignment, ensuring seamless integration.

The final and most time-consuming phase of alignment involves maintaining the initial precision over time. This aspect is particularly critical for a newly constructed tunnel, such as that of FCC-ee, where long-term stability must be carefully monitored and preserved.

### 3.5.1 Alignment tolerances for the FCC-ee machine

The initial mechanical alignment tolerances assumed for the FCC-ee arcs are described below. It is assumed that adjacent arc quadrupoles and sextupoles are pre-aligned at 50 µm accuracy with respect to a common $\sim$ 6 m long girder. In the tunnel, the alignment tolerances from girder to girder or between girders are 200 µm over 50 m and 500 µm over 200 m. In the interaction regions (IR), transverse misalignment errors for quadrupole and sextupole reference axes are taken to be $\pm$100 µm ($1\sigma$) ($\pm$ 250 µm ($1\sigma$) longitudinally, and $\pm$0.25 mrad in roll). Current requirements regarding the alignment are of 30 µm for the final focusing quadrupoles; the LumiCal will need to be aligned at 50 µm, and the screening and compensation solenoids at 100 µm (all values referring to $1\sigma$).

The figures presented above represent misalignment errors at the level of the reference axis, incorporating both the fiducialisation process and the subsequent adjustment steps. To meet the specified alignment requirements, position determination and adjustment solutions will need to achieve precision and accuracy at least three times higher than the stated tolerances.

With regard to fiducialisation, the required measurement tolerances are well within reach for synchrotrons operating under stable environmental conditions with rigid support structures. However, for FCC-ee, the process must be automated due to the large number of girders and components that require fiducialisation and pre-alignment.

Furthermore, in a newly constructed tunnel such as that of FCC-ee, where stable and unstable areas have yet to be identified, a significantly greater number of components will require (re)alignment compared to the LHC. The implementation of automated solutions will be essential in achieving and maintaining the necessary alignment standards over the machine's operational lifetime.

### 3.5.2 Alignment tolerances of FCC-ee experiments

The alignment accuracy for the assembly of the experiment is assumed to be similar to those of the LHC experiments, i.e., 0.5 mm with respect to the machine geometry. The positioning and stability tolerances and requirements of each sub-system of the FCC-ee experiments need to be identified in order to estimate the means of adjustment.

### 3.5.3 Theoretical data

The spatial position and orientation data for the beamline elements including the FCC detectors need to be extracted from the beam optics calculations such as BEATCH/MADX beamline definition files. Additional parameters necessary for geodetic metrology can be derived from the layout drawings of the detectors.

### 3.5.4 Metrology

In addition to the metrological controls of detector and machine components throughout the manufacturing and assembly process, the final transfer measurement of the component axis, called fiducialisation, is a major step in the metrologic controls and is mandatory for later alignment. This job is carried out for each piece of equipment whose precise alignment parameters are to be determined. The proposed techniques are similar to those proposed for the Compact LInear Collider (CLIC), i.e., laser tracker and close-range photogrammetry. For smaller components, coordinate measuring machines (CMM) and new technologies such as frequency scanning interferometry (FSI) may be used, depending on the accuracy requested.

For larger components, the effects of thermal expansion must be carefully accounted for and compensated to ensure precise alignment. Metrology measurements should be conducted in a controlled environment to maximise accuracy. On-site measurements in the tunnel or experiment caverns should be minimised, as they inherently introduce uncertainties that can reduce the overall precision of the alignment process.

The feasibility of metrology measurements within the tunnel or caverns must be thoroughly assessed, as factors such as environmental conditions, spatial constraints, supporting systems, and concurrent activities can impact measurement quality and accuracy.

The placement, number, and distribution of alignment targets (fiducials) on the components play a critical role and must be determined based on survey requirements, selected alignment technologies, and the specific constraints of the experiment caverns and accelerator tunnel. Additionally, the supporting system must adhere to alignment specifications and constraints while following established survey guidelines [277].

Depending on the object, the parameters of the fiducialisation are stored in the survey reports or a database and can originate from metrology reports, geometrical quality control measurements or from a specific fiducialisation operation. The external fiducials are supposed to stay visible and accessible for the various measurement operations during the lifetime of the component.

### 3.5.5 Alignment of accelerator components

***Introduction***

The alignment of the beam components of future CERN accelerators will be achieved with various techniques depending on the alignment tolerance (mainly depending on component type and physics requirements), the sector concerned (long straight sections, arcs, transfer lines, injection/extraction zones, etc.), the geometric dimension of the components, and the phase of the project (initial installation of the machine, a single isolated component alignment, voluntary displacement, smoothing campaign).

The radiation level will also influence the instrumentation and methods chosen for measurement and alignment.

***Marking the beam line and supporting system on the floor***

Using the MAD-X sequence files transmitted by the physicists in charge of beam optics, and prior to the installation of the first component in the tunnel, the vertical projection of the beam point assembly and the position of specific component supports such as jacks must be traced on the ground. This will be based on the geodetic reference network of the tunnel. This tedious work, previously done manually, should be automated in the future (see below). The ground marking prepares the floor for the installation of the equipment support systems and later the components.

***Position control of supporting systems***

Following the installation of the supporting systems of the largest and most voluminous components (mainly jacks), their position must be verified to ensure safe installation and to detect any positioning errors prior to the critical component installation step. The positioning accuracy of the supports can be easily achieved with well-known and standard 3D measurement systems.

***Component pre-alignment***

The pre-alignment phase in the tunnel (not to be confused with the pre-alignment of components on girders before installation in the tunnel) consists of the initial alignment of the components after their installation. This activity is performed using the absolute reference frame, i.e., based on the geodetic reference network of the tunnel.

During this phase, the tolerance required is slightly less stringent than in the relative alignment stage of the machine (smoothing phase), allowing the use of 3D polar measurement techniques such as total stations and laser trackers.

A similar approach will be used when aligning an isolated component, for instance, when replacing a unit with a spare. In such cases, the new component is typically positioned relative to the nearest reference component to ensure continuity in alignment.

Once all components have been pre-aligned, they can be mechanically interconnected, ensuring a stable and precise overall assembly.

***Machine smoothing***

Smoothing consists of refining the relative position of neighbouring components to meet relative alignment tolerances and avoid offset between adjacent components.

The first step is to precisely measure the vertical and radial position of the whole machine or section concerned. In the following step, components outside the alignment tolerances are identified and will be displaced.

The initial machine smoothing after installation is distinguished from maintenance smoothing, which is repeated regularly to compensate for displacements due to mechanical constraints, geological movements or support instability. The frequency of periodic maintenance smoothing depends mainly on physics requirements, ground motion and time available during the technical stops.

It is a relatively time-consuming activity. Thus, during machine operation, where maintenance time is limited to yearly technical stops, it is not possible to measure the entire machine. Smoothing is then restricted to some sensitive portions of the machine.

Generally, the acquisition techniques depend on the machine plane to be measured (vertical, radial) and on the accuracy required. For radial measurements, wire offset measurements are favoured, while direct levelling is used for vertical position determination.

Outside the arcs and within much smaller volumes (a few dozen metres at most), 3D polar techniques can also be used for machine alignment.

The most restrictive areas in terms of precision will be equipped with permanent and automatic measurement systems. In addition, most of the machines could, in principle, be measured by the remote alignment system to be developed, which is described below.

***Transfer lines***

The transfer lines are different from the main machine in terms of geometry (continuously changing slope and roll), topology (fewer components and greater spacing) and the alignment needs as the beam passes only once. The alignment techniques will need to be compatible with these different constraints and in particular offset measurements and direct optical levelling are much more difficult to realise.

***Notion of primary and secondary components***

Since the machine components are of all sizes and weights, their stability in time is also different. The largest and heavier components are more stable. These are usually magnets, such as dipoles, which have an impact on the trajectory of particles and are considered as 'primary' elements. The other components are qualified as 'secondary' elements.

From a practical point of view, the precise alignment of primary components is generally performed during the smoothing campaign (see above), while secondary components are mainly aligned with respect to the primary elements in a second phase.

For the FCC-ee the alignment tolerances of the dipoles are quite loose, e.g., at the level of 1 mm transversely. The relative alignment of quadrupole magnets, sextupole magnets, and beam-position monitors (BPMs) is more sensitive. These elements, forming the arc 'short straight sections', should be prealigned on common girders prior to installation in the tunnel, as it is being done for modern light sources. The closest of these girders, which are about 25 or 50 m apart in the tunnel, should then be aligned with respect to each other.

***Impact of component type on alignment***

There are several categories of components, each with specific alignment tolerances. Magnetic elements, which actively influence the beam, typically require more precise alignment in at least one plane—whether roll angle, horizontal, or vertical. This is particularly critical for dipole and quadrupole magnets, where small misalignments can significantly affect beam dynamics.

Beam diagnostic equipment forms another group of components with distinct alignment constraints. Some, such as beam position monitors (BPMs), require precise knowledge of their position relative to the primary beamline. This can be achieved using beam-based alignment techniques, ensuring optimal accuracy in beam monitoring and control.

The weight and dimensions of the components play a crucial role in determining the appropriate supporting system and alignment strategy. Stability and safety considerations necessitate the use of isostatic supporting systems, which minimise structural deformations. Additionally, the mechanical design of the components must ensure that deformations remain negligible. The length of the components, along with associated lever arm effects, is a key factor influencing both alignment precision and interconnections within the system.

### 3.5.6 Interaction regions including MDI

The alignment of the MDI is described further in Section 1.6.

The initial alignment in the interaction region, surrounding the MDI ($\pm$900 m around the IP) will be performed with the rest of the collider.

Position monitoring is planned for the components for the incoming beam in this region, using a combination of hydrostatic levelling systems (HLS), FSI distance measurements, wire positioning systems (WPS), and inclinometers, resulting in a configuration similar to the HL-LHC. Knowledge gained during CLIC studies (see [278] and [279]) could also be applied if needed. In addition, innovative systems, such as the structured laser beam (SLB) [280], are being studied for use in this region.

### 3.5.7 Remote alignment and monitoring

Along the straight sections, for example, in the IR areas, the same alignment solutions as those proposed for the implementation phase of the CLIC project can be used [138]. The solutions developed for CLIC will certainly meet the FCC requirements. The transverse position of components will be measured with regard to a long-range straight alignment reference over several hundreds of metres. Alignment sensors measuring the transverse offsets with respect to an alignment reference will be installed on the components. Overlapping references will be used over very long distances, allowing a very accurate determination of one common straight reference. Two types of alignment references can be considered: a structured laser beam (SLB), or a stretched wire. Both systems will have to be combined with a hydrostatic levelling system (HLS) which will provide relative vertical references.

The SLB can be defined as a pseudo non-diffractive optical beam, with a very bright central core, sharp boundaries, a minimal divergence and a theoretically infinite range (tested on 1 km). All of its properties are under evaluation, and its application to the transverse alignment of components and, more particularly, to the FCC-ee is the object of two PhD theses. The R&D on the SLB has just started. Preliminary studies indicate that maintaining the straightness of the SLB along its trajectory requires it to be enclosed within a vacuum pipe. However, due to the specific properties of the beam [281], the required pipe diameter is smaller compared to those used in optical-based alignment systems.

Stretched wires used as straight references are not at a development stage. They have been in use, combined with WPS, for many years in the LHC for the continuous determination of the position of the low beta quadrupoles. The same combination will be used for a full remote alignment system of the main components of the long straight sections around IP1 and IP5 for the HL-LHC project. (see Refs. [282] and [283]). Their main drawback comes from the limited length of cables between each sensor and its remote electronics/acquisition system (maximum 120 m). Cables will be a great concern in the FCC-ee tunnel, from the integration/space available and radiation perspective. One R&D development to overcome this issue would be the development of WPS sensors based on FSI measurements (replacing the current capacitive technology). A 'chained' optical fibre configuration could be developed that would take far less space in the tunnel and could be compatible with the high level of radiation expected during the machine's operation.

Extrapolating such solutions to the arcs will require additional R&D as there is currently no solution allowing the permanent monitoring of the position of components in the arcs. Alignment solutions have been developed for straight portions, not circular ones and should be adapted, either using the spe-cific properties of SLB, or developing a new configuration of alignment references based on 'broken lines' of wires.

Position adjustment is an activity that will be challenging to automate. Since 3D adjustments of components are required, continuous monitoring is essential when using actuators to perform these adjustments in all three dimensions. This ensures that the bellows at the extremities of the component can accommodate the displacement without excessive stress or deformation. Additionally, for longer components, even small movements can create significant lever arm effects, leading to unintended shifts at the opposite end. Real-time monitoring throughout the adjustment process is, therefore, necessary to maintain alignment precision and prevent mechanical strain.

### 3.5.8 Survey robot options

Given the size of the collider, many alignment steps will have to be automated, from the 3D scans to the initial alignment. A few examples of robot options are provided below. 3D scans could be performed by a 'dog robot', provided that there are permanent targets which define the underground network and allow the geo-referencing of scans. For the step consisting of marking the beam axis and the jack position on the tunnel floor, increasing numbers of floor marking robots piloted by tacheometers are now available on the market. An automatic inspection of the position of components pre-aligned on girders could be performed with high accuracy by a robot in the tunnel once the girder is at its final location. In all cases, adaptations to the specific tunnel configuration and the underground geodetic network will be necessary. These may include increasing the density of the network and incorporating permanent targets that various types of instruments can measure to enhance alignment precision.

### 3.5.9 Impact of beam-based alignment on alignment system requirements

The FCC-ee design assumes one BPM close to each quadrupole. Beam-based alignment (BBA) will determine the BPM position offsets (reflecting both mechanical and electrical errors) with respect to the magnetic centre of the nearby quadrupole and sextupole magnets, and it will help steer the beam through the magnetic centre of these elements by using orbit correctors along with dipolar and (skew) quadrupolar trim coils to minimise feed-down effects. Instead of relying on magnetic trims, Beam-Based Alignment (BBA) could also be carried out using movers. While movers are not included in the current baseline scenario for the arcs, they could be a viable solution for the strong sextupole magnets in the experiment insertions. If movers were to be implemented, their displacement range would determine whether additional measurement systems are required to assess potential risks and ensure safe operation. Since BBA must be performed regularly to monitor and correct possible alignment drifts, the choice of adjustment mechanism should also consider long-term stability and operational efficiency.

### 3.5.10 Experiments

In the FCC-ee experiments, the geodetic metrology survey work will be performed in different steps: design, manufacturing construction, assembly and alignment of the detector elements. This means:

- Participating in the project at an early stage, e.g., when collecting the geometrical parameters, alignment needs and discussing the integration of the survey needs in the design of the infrastructure, tools or detector elements.
- Providing the necessary geometric data for adjustment and control of the assembly infrastructures.
- Providing the position and orientation information related to the FCC-ee detector assembly and tests, i.e., the geometric information:
  - for the detector assembly tooling alignment,
  - for the geometrical follow-up and adjustment of the detector elements during assembly,
  - for the positioning of the detector elements for tests.

– Establishing, measuring, computing and maintaining geodetic networks or coordinate systems and defining the parameters linking them when needed.
– Providing surveyed position and orientation information to locate the detector in the CERN Coordinate System (CCS) and to link it to the accelerator geometry.
– Providing, when necessary, metrology measurements for the fiducialisation of the detector elements as well as of module assemblies, as described in the metrology section above.
– Providing, when and where necessary, geometric control and validation measurements for detector elements as well as of module assemblies.
– Providing the control of the position and the alignment of elements and module assemblies in the experiment area.
– Participating in any upgrade projects at an early stage to ensure high-quality geometric information during the lifetime of the detector.
– Providing stability measurements of cavern walls and floor on demand.

This covers both the theoretical and the practical aspects of the geodetic metrology work for the FCC-ee project. The theoretical positions of the beamline elements are provided by the accelerator optics team.

In-field measurements should be performed using suitable survey instrumentation and methods such as total stations, optical levelling, photogrammetry, 3D laser scanners or laser trackers.

### 3.5.11 Link of machine geometry to the experiment area

The very first geometric link between the tunnel geometry and the experiment cavern is expected to pass directly from the tunnel to the cavern while the access and visibility exist. In later stages, the geometric link between the accelerator and the experiment area will be performed through the survey galleries using geodetic methods and dedicated monitoring systems as described in the MDI section above.

***Marking out***

With respect to the geodetic network, reference marks representing the projected beam line and the elements to be aligned can be painted on the floor and walls by the survey team. These marks help with the installation of the services and beamline elements. Everybody working in their vicinity must ensure that these marks remain visible. The marks and annotations required have to be defined in collaboration with technical coordination.

***Geometric quality control measurement***

The survey team should provide on-demand, where and when necessary, the geodetic measurements and analysis for the geometrical and dimensional control of prototype and production elements.

***Positioning***

Prior to the installation of the detector elements, their supports are installed by others and pre-adjusted to their nominal position by the survey team. These survey interventions can also happen before the support installation if shims are required to overcome local floor deformations.

Once the detector elements are installed, their initial positioning will be carried out with respect to the geodetic network of the area. Precise survey methods and instrumentation have to be used, such as laser trackers, total stations and direct levelling.

In collaboration with the technical coordination of the experiment, the development and installation of simple alignment systems for the positioning of the detectors after maintenance or equivalent could be included.

### 3.5.12 As-built survey, 3D scans and measurements

The FCC-ee tunnel will have a diameter of 5.5 m. All services and cable paths will have to be optimised inside. Furthermore, the FCC project has a rather long lifetime: it is planned to dismantle the components of FCC-ee and replace them with a superconducting machine for the FCC-hh after 15–20 years of operation. It will be necessary to perform 3D scans of the empty tunnel as soon as possible, before any services are installed and subsequently perform 3D scans at well-defined milestones: all services in place before the installation of components, all components in place, etc. The geo-referenced 3D point clouds obtained during the installation process will be highly valuable for verifying that components are positioned according to their theoretical locations and for identifying potential interferences between systems. Additionally, these data will play a crucial role in facilitating the future installation of new equipment once the FCC-ee is in operation, ensuring seamless integration and minimising disruptions.

It is recommended to execute equivalent as-built measurements using 3D scans of the civil engineering structures in the experiment caverns, followed by as-built measurements of the infrastructure installed and, finally, the experiments in order to provide 3D documentation of the experiment areas of the FCC-ee and to save time during installation and future upgrade works. The survey team should pre-process these measurements to give the geo-referenced point clouds and their 3D coordinates in the CCS, or a defined experiment coordinate system, to the integration team of the FCC-ee experiments.

By utilising 3D scans, a comprehensive digitisation strategy can be developed, incorporating data-to-cloud solutions for remote visualisation. This will enable access to detailed historical documentation, such as girder assembly records, while also supporting the implementation of digital twin technology. A digital twin would allow real-time anomaly detection and advanced simulations, including studies on the impact of temperature variations and other environmental factors.

### 3.5.13 Software

The Survey Database plays a critical role in storing all the parameters needed for surveyors to align and determine the position of accelerator components in the LHC era. This database stores the calibrated component geometry (fiducialisation) as well as their theoretical positions within the global CERN Coordinate System (CSS). It also contains an element's voluntary displacement (bump) and its measured offset to the theoretical position at a given date. Furthermore, it stores all the measurements acquired and other crucial information, such as calibration parameters of the instruments and sensors.

For the FCC-ee, the survey database will require similar capabilities, including the ability to handle an ever-increasing number of accelerator elements followed in 4D (3D position plus time). Additionally, future concepts such as girders housing elements and so-called cells will need to be accounted for in the database.

The survey database will need to seamlessly interface with the digital twin framework and any Building Information Model (BIM) solutions implemented for the construction and maintenance of the FCC-ee accelerator infrastructure.

To support planned monitoring systems, the development of custom databases or integration into a broader CERN-standard database will be essential. These databases must efficiently manage an ever-growing number of sensor parameters—both existing and newly developed—to ensure reliable data processing for computational models. Additionally, automated, robot-based, and as-built measurement data will need to be incorporated, either within a dedicated system or as part of CERN's global database infrastructure.

Moreover, due to the inherent complexity of the data structure, dedicated APIs should be developed to enable secure and efficient access to internal data for various CERN applications, including survey-related tools. These interfaces will be designed using standard IT technologies, ensuring interoperability with other systems and facilitating seamless data exchange across different platforms.

Given the sheer volume of data and the survey processes, special software will be needed for the geodesy, alignment, and survey processes. Commercial software may have serious limitations due to the specificity of the alignment processes that involve combining several methods of observation, geodesy aspects, and non-widely used sensors and instruments.

Even though a significant level of automation is planned, in-field intervention of surveyors to observe, check, and, in some cases, move an accelerator element will happen. As such, specific computing interfaces will be needed to acquire geometrical observations from the survey instrument. In-house developed solutions will provide the geometric data in a reliable and adapted way. New processes will require either the evolution of the current software such as SMART [284] and TSUNAMI [285] or the development of entirely new software adapted to the FCC-ee alignment strategy. Most likely, surveyors in the field will use a combination of existing commercial software that satisfies typical needs and follows standard procedures, with in-house software that meets CERN and FCC-ee specific workflows and instrumentation. Common rules, formats and standards should, therefore, be defined and implemented; this strategy will involve strong and long-term industrial partnerships.

Monitoring systems will also be put in place, and acquiring data from these systems will require special development to be integrated into future standards at CERN according to the evolution of the control system IT-infrastructure (communication middleware and low-level framework controlling the accelerator equipment). Post-processing the permanently acquired data is a crucial step. It involves a variety of tools available to surveyors or machine-based routines to compute the positions and their related statistics for numerous geometric elements such as points, planes, lines, or circles. In configurations where the monitoring system must provide real-time positions, for example, when synchronised with the accelerator's timing system, optimisation strategies should be found and implemented in the global calculation process. Collaborations with academic partners will be a key to success in developing new algorithms and achieving the expected performance.

The main computation shell, LGC [286], will need to evolve or be re-developed to handle new mathematical models describing the behaviour of new instruments or sensors. It will need to be a single point entry from across most of the survey, alignment, and geodesy projects. It will need to be able to handle most of the survey processes and combine several sources of data in a reliable, fast, and integrated way.

Special post-processing algorithms and dedicated routines, such as smoothing algorithms or best-fit processes for various geometric primitives will need to be available. Automated processes intended for fiducialisation, robot-based measurements or any other geometrical checks may also require dedicated post-processing tools and algorithms.

In summary, the feasibility study has provided valuable insights into the surveying, alignment, and geodesy requirements for the FCC, highlighting the need for further analytical refinement and dedicated development to establish a robust software environment and adequate database. While many existing solutions can be adapted, some aspects of the FCC-ee project will require the development of specialised tools or the evolution of current software. Particular attention must be given to data storage and management strategies to ensure long-term accuracy, reliability, and interoperability with other systems.

## 3.6 Beam intercepting devices

The FCC-ee complex, including the collider, booster and injectors, will require several beam intercepting devices to be installed in the various accelerators, covering a large variety of functionalities. At this point, the following devices have been identified:

- Positron target (injector)
- Extraction dumps & spoilers (collider and booster)
- Beamstrahlung (photon) dump
- Betatron and momentum collimators

- Synchrotron radiation (SR) collimators
- Interaction region (IR) masks and machine detector interface (MDI) devices
- Injection protection devices (collider and booster)
- Extraction protection devices (collider and booster)
- Beam stoppers
- Slits/scrapers
- Collimators (booster)

A detailed study and design process are required to ensure that all these devices meet their specific operational requirements. So far, three key components—collimators, the positron target, and photon dumps—have been identified as presenting significant technological challenges. Addressing these challenges necessitates dedicated R&D efforts and prototyping to develop viable solutions.

As the functional requirements of additional components become fully defined, further challenges are expected to emerge. In particular, injection protection devices likely need to intercept a low-emittance beam with high-energy density, a condition that could potentially damage or destroy conventional materials. Therefore, advanced material research and innovative engineering solutions are critical.

Similarly, some of the interaction region (IR) and machine-detector interface (MDI) masks are expected to experience substantial energy deposition, leading to significant thermo-mechanical loads. Their design must adopt suitable materials and structural configurations capable of withstanding these conditions. In extreme cases, sacrificial protection devices may need to be incorporated to safeguard critical elements and ensure long-term system integrity.

### 3.6.1 Lepton dumps and spoilers

Specific dumps are necessary to safely absorb the beams from the collider and booster. There are two dumps for this purpose, one for positrons and one for electrons. Each dump will receive the beams from both the collider and the booster. Preliminary estimates indicate that spoilers upstream of the dumps are not required, as the beam will be sufficiently diluted that the dump can manage the thermo-mechanical loads safely.

Nevertheless, if, after detailed studies, spoilers are deemed necessary, they have already been studied. The principle would be to use passive graphite cylinders some hundreds of metres upstream of each dump [287].

The design concept that is planned to be used for the dumps is similar to the current LHC dumps [288], i.e. a combination of different graphite and CfC grades, enclosed in a metallic, cylindrical vessel (Fig. 3.52). Also, the dimensions are expected to be similar to those of the LHC dumps (diameter in the order of 400-700 mm $\times$ length $\sim$5 m).

### 3.6.2 Beamstrahlung radiation dumps

High-intensity beamstrahlung (BS) radiation is expected to be generated at the interaction points due to the synchrotron radiation emitted during the collision in the electromagnetic field of the opposing beam. Dedicated devices are required to safely absorb the power carried by this beam, which can reach hundreds of kilowatts. One BS dump is required on each side of each interaction point; hence, a total of eight dumps are needed for the entire collider. One of the most efficient materials for absorbing this type of radiation is lead. Moreover, in order to avoid thermal-stresses and unpredictable change of phase during operation, liquid lead, circulated in a closed circuit, has been chosen as the baseline (Figs. 3.53 and 3.54). Moreover, due to the strong dependence of the photon beam power to beam separation, a robust system capable of absorbing rapid power excursions is required. Research and development activity for the optimisation of this device is ongoing, and a robust prototyping activity is envisaged in the next phase of the project.

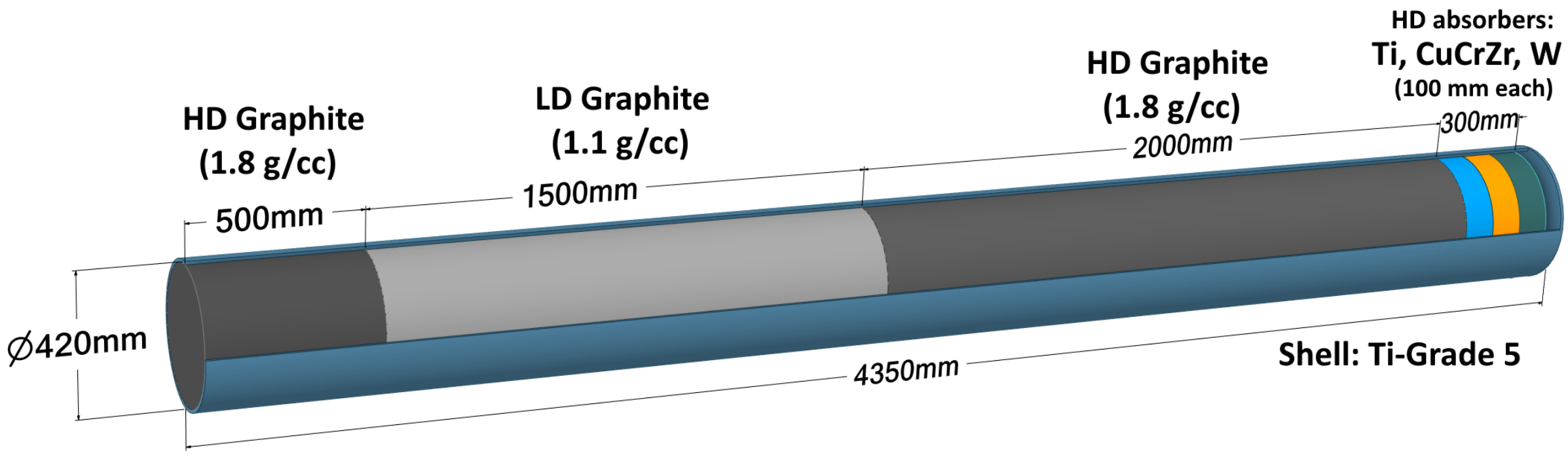

Fig. 3.52: Schematic diagram of the FCC-ee dump block design, showing the material for the dump core and the vessel.

In addition, a robust shielding enclosure needs to be installed around the liquid lead system so that activation of the cavern and limitations to personnel access to the area are avoided. An alternative design using gas-cooled graphite discs is also being considered. Both design options are to be studied and prototyped to have a robust design for such a device.

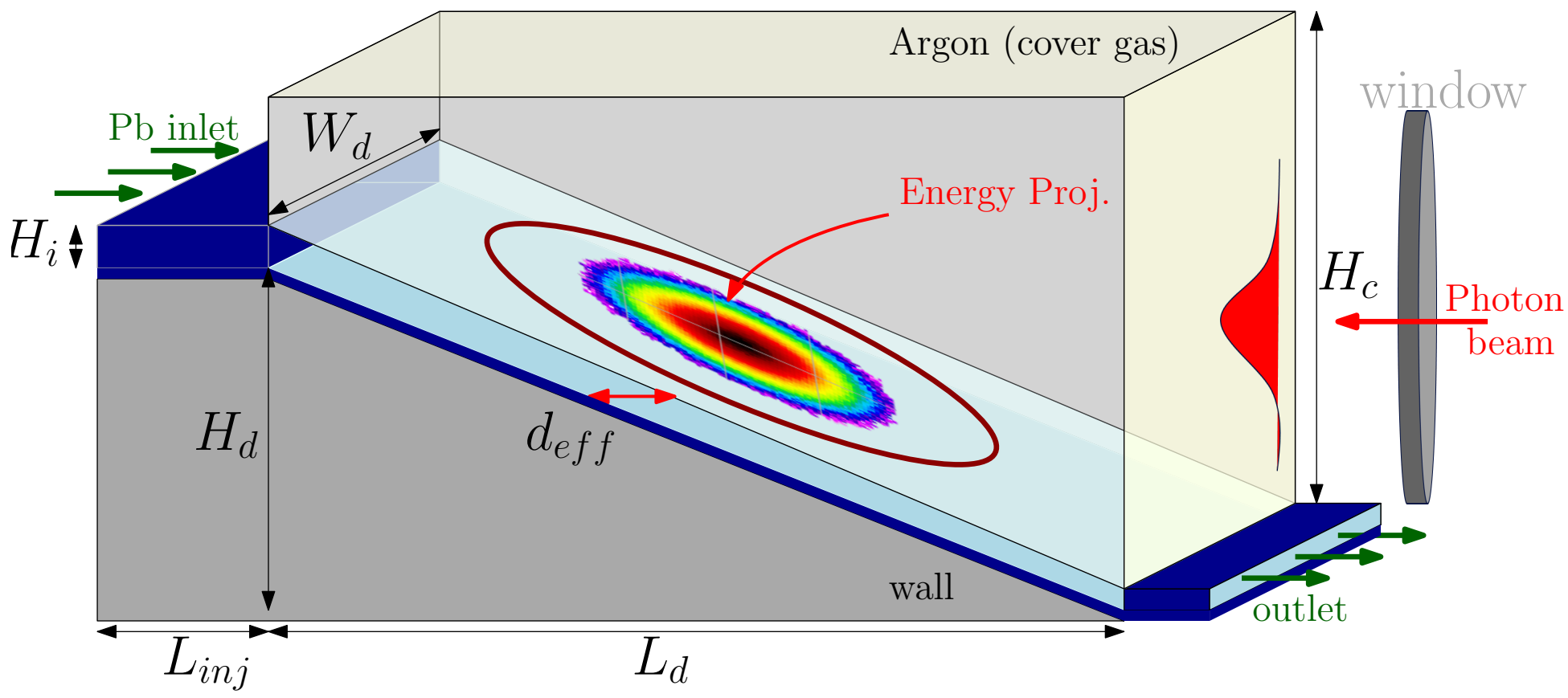

Fig. 3.53: Schematic of the FCC-ee beamstrahlung dump.

### 3.6.3 Betatron and momentum collimators

The design of these devices will depend on the operating requirements and the accident scenarios that they may need to withstand. Since these devices are located in proximity (and even interact with) the beam, special constraints are imposed so that their impedance is as low as possible.

The selected materials and geometries must preserve beam quality while ensuring that the devices can withstand beam impacts under off-normal conditions, including potential accident scenarios.

A total of 58 units will be required, covering all installations at point PF, within the experiment insertions, and the shower absorbers. The design will incorporate key improvements based on operational experience from the LHC and SuperKEKB collimation systems. Openable tanks will be implemented to facilitate maintenance and replacement, while optimised bellows movement will enhance mechanical flexibility and reduce wear. The use of high-performance absorbing materials will ensure the system can withstand sustained beam loads without degradation, and enhanced radiation resistance will contribute to long-term durability. Additionally, reliable actuation mechanisms will be developed to enable precise and repeatable positioning. These design choices aim to maximise system robustness, ease of maintenance, and overall operational reliability.

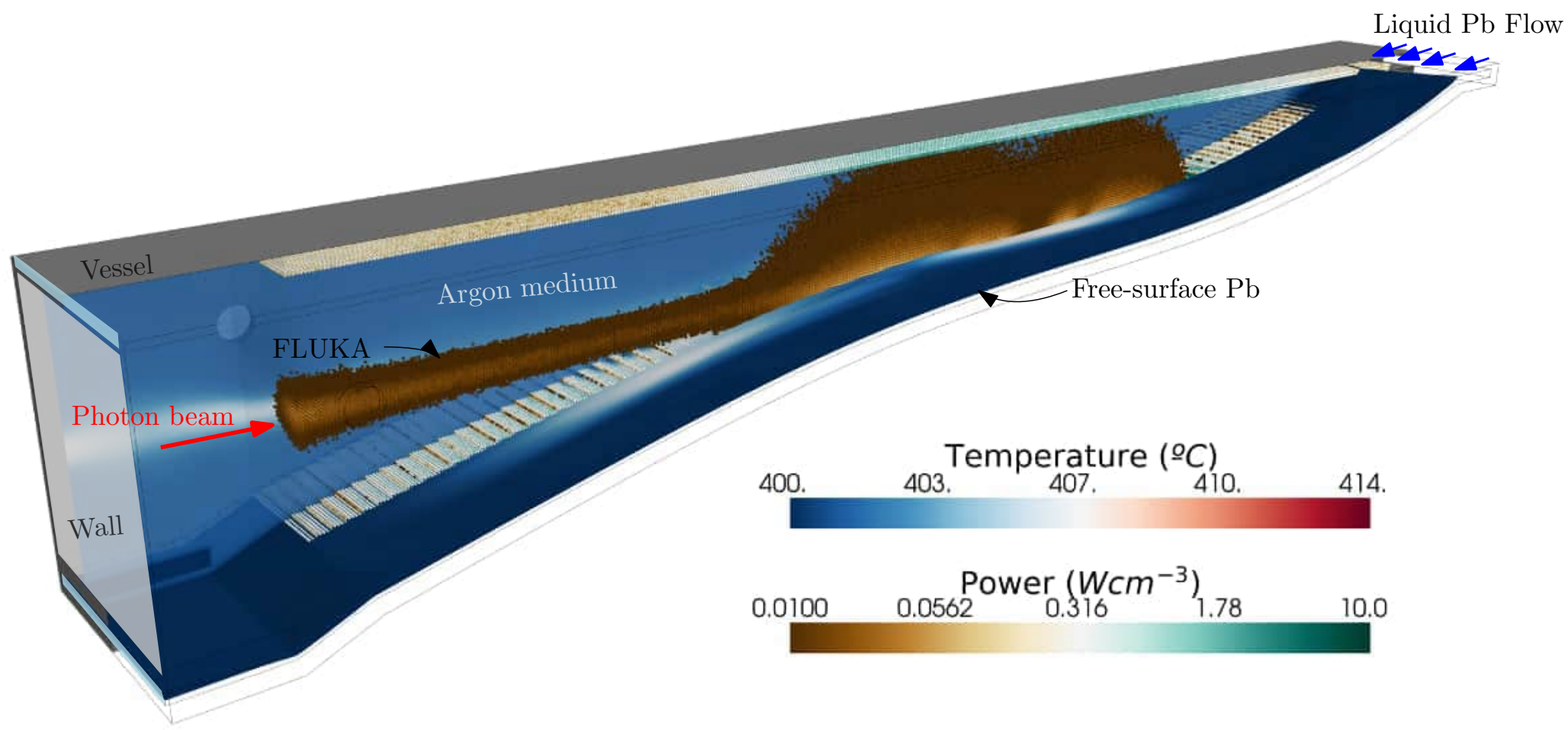

Fig. 3.54: Preliminary simulation of liquid-lead beeamstrahlung dump.

### 3.6.4 Synchrotron radiation collimators

These devices have a different function to the betatron and momentum collimators. However, the design considerations and constraints are expected to be similar due to their proximity to the beam. The materials and technologies selected for the hardware design will be based on the functional requirements. A total of 48 units will be required for operation.

### 3.6.5 Interaction region masks and absorbers

As their names suggest, these devices absorb secondary particle showers generated by upstream collimators and to protect sensitive hardware in the interaction regions.

The specific design of these components may vary depending on their exact function and installation layout. They may incorporate either fixed or movable absorbing elements, depending on operational requirements. Given the high-energy environment, high-density materials such as tungsten heavy alloys are preferred for their superior absorption properties.

A total of 16 of these devices will be needed with two per beam positioned upstream of each interaction point.

### 3.6.6 Injection protection devices

These devices protect the machine from any physical damage in case of injection failure (e.g., beam on the wrong trajectory). These devices are expected to have a similar function to the TDIS for the (HL-)LHC machine (see Fig. 3.55).

The materials and design to be selected must fulfil the functional requirements whilst guaranteeing robust protection of the machine elements that may be exposed to accidental injection failures. One device per injection point will be required (hence, a total of 2).

### 3.6.7 Extraction protection devices

These devices are responsible for protecting the machine from any physical damage in case of extraction failure scenarios (e.g., a beam at the wrong trajectory). These devices are expected to hold a similar function as the TCDS/TCDQ system for the (HL-)LHC machine.

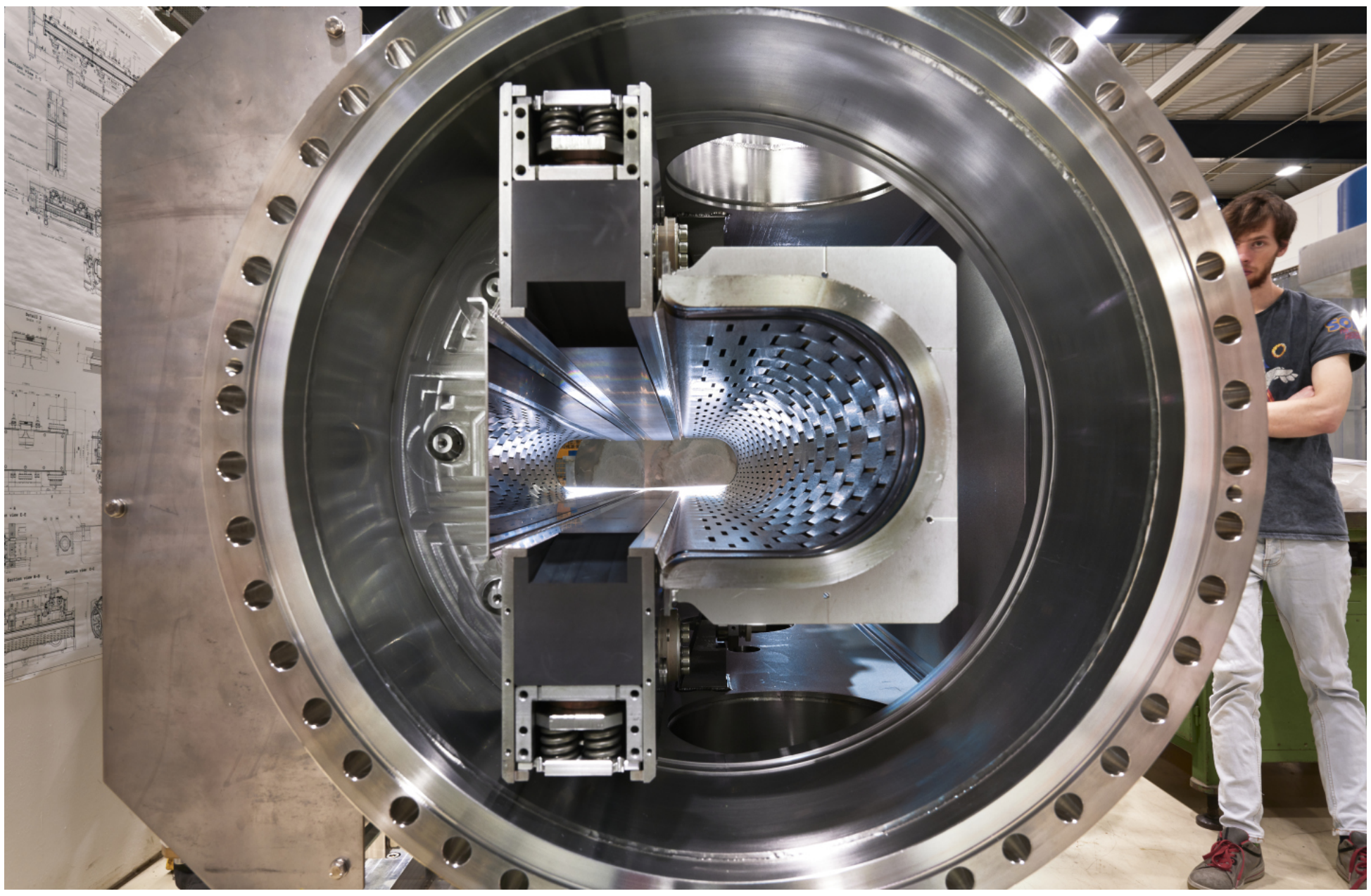

Fig. 3.55: Front image of the (HL-)LHC injection protection device, the TDIS. It shows the two graphitic absorbers, kept in place by a TZM back stiffener and TiGr5 clamps. The RF screen for the circulating beam is visible on the right side.

The selected materials and design must meet the functional requirements while providing reliable protection for machine elements that could be exposed to accidental injection failures. To ensure adequate safeguarding, one device will be required per extraction point, resulting in a total of two units.

## 3.7 Beam transfer systems and separators

### 3.7.1 Beam transfer

The concepts for the beam transfer systems of the collider and their associated requirements are discussed Section 1.8. This section focuses on a few systems but a comprehensive assessment of the technical choices and specifications can be found in [178].

#### *Thin septum*

For collider injection, a very thin septum with an apparent blade thickness of up to 2.8 mm is required (see Table 1.18). The long pulse length of at least 304 µs prevents using an eddy-current, so the proposed topology is a direct drive, under vacuum, septum. Figure 3.56 shows the mechanical concept for the septum, with the thin blade carrying the drive current between the injected and circulating beam, shown in magenta. A circular perforated shield carries the image current of the circulating beam to minimise the impedance.

During injection, the distance between the blade and the circulating beam becomes very small. The design will need to consider both synchrotron radiation and direct impact from the beam halo, with the possible need for a specific mask to protect the thin blade of the septum.

#### *Kicker systems*

For the collider dump and injection systems, the kicker requirements are listed Tables 1.18 and 1.19. Benefiting from the design of a completely new accelerator complex, the same kicker hardware design was selected for both systems. A ferrite-loaded lumped-inductance kicker magnet operating at relatively low current and voltage is selected. This choice of an out-of-vacuum magnet enhances reliability, reduces costs, and simplifies maintenance, making it more favourable compared to other technologies, such as striplines, considered during the feasibility study. Additionally, this magnet topology was previously implemented in LEP, where it demonstrated reliable performance in a lepton collider environment.

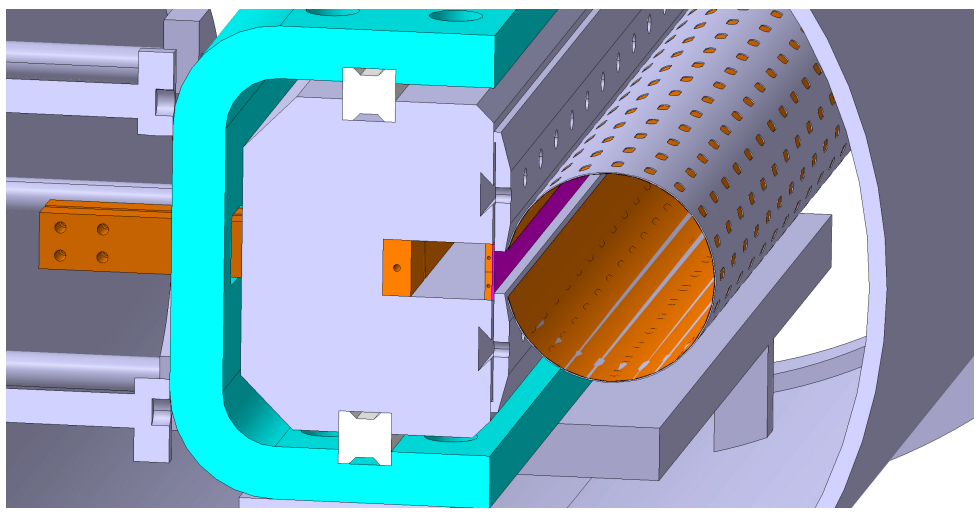

Fig. 3.56: Mechanical design concept for the collider thin, under vacuum, injection septum.

A primary constraint for the system presently considered is to rise and fall within the time between trains, within less than 600 ns. This limits the maximum length of each kicker to 0.7 m and requires a total of 16 kickers for the injection and 18 for the dump. Every magnet will be connected to its generator with a maximum cable length of approximately 250 m. Additionally, harmonising the systems across the FCC machine can lead to significant cost savings.

For the collider injection, two generator options are possible: a pulse forming network (PFN)-based pulse generator or a Marx generator, with the latter being the preferred choice as it maintains a lower voltage. To compensate for the long flattop duration, a significantly large capacitor is required, necessitating adequate space allocation in the service gallery. The magnets will operate in short-circuit mode for the dump system since there is no fall time requirement.

### 3.7.2 Separators

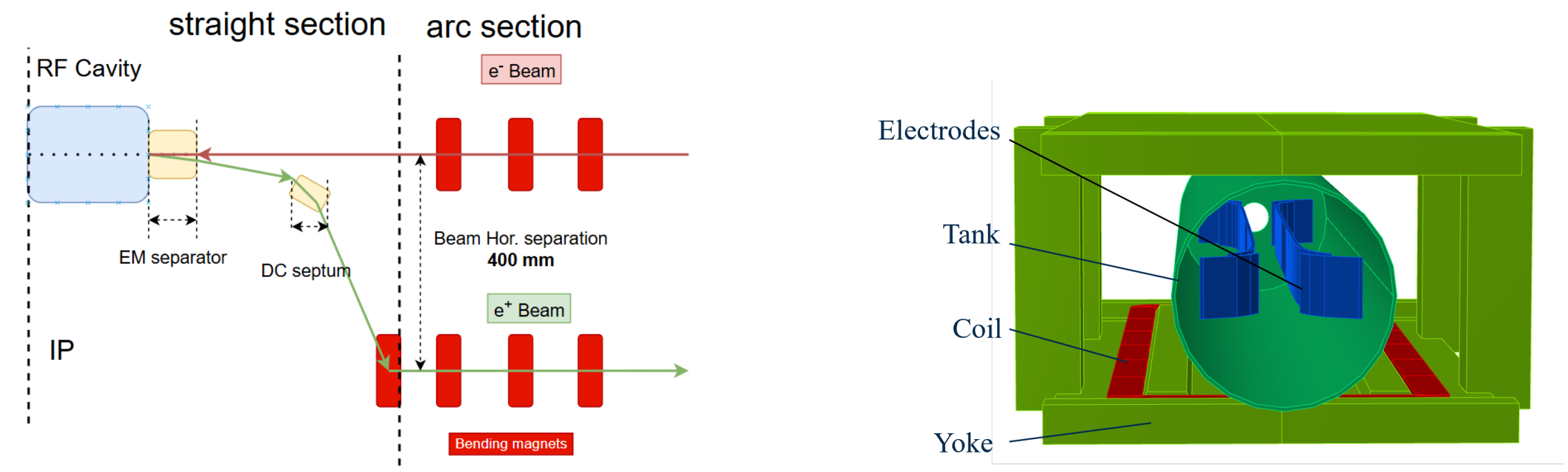

(a) (b)

Fig. 3.57: Schematic representation of the separator system function in PF (a) and the concept of an EM separator combining electric and magnetic fields (b).

An electromagnetic (EM) separator is required at the entry and exit of the RF section to merge and separate beams travelling in opposite directions (see Fig. 3.57a). Operating for H and $t\bar{t}$ modes, they allow both beams to circulate on the same trajectory in the cavity.

To achieve this, perpendicular electric and magnetic fields are used to deflect the outgoing beam while allowing the incoming beam to maintain a straight trajectory, preventing synchrotron radiation from being directed towards the RF section. A key challenge lies in generating the extremely low mag-netic fields required to counteract the electric field in one direction, as remanent magnetisation can disrupt the delicate field balance. The current design concept [289], which explores shaping the end fields and potentially implementing an air-coiled dipole magnet, is under active investigation to ensure that the required field overlap is both technically feasible and stable (see Fig. 3.57b).

Additional design considerations include beam impedance, thermal loads, and the risk of high-voltage sparks. To minimise impedance, the integration of ground electrodes along the length of the separator has been proposed. However, synchrotron radiation striking these electrodes must be accounted for in the cooling strategy. The potential for stray particles and electrical discharges further necessitates a design that allows either collective or independent power regulation of the separators.

Maintaining proper overlap of the electric and magnetic fields is another key challenge. Online measurements of the magnetic field or beam position could be incorporated into fast feedback loops, enabling real-time corrections to ensure field balance during operation. While the current design does not include active cooling for the electrodes, energy deposition due to RF heating and synchrotron radiation must be carefully assessed. Although active cooling remains a viable option, its implementation would increase both cost and system complexity [290].

Important uncertainties remain regarding the high voltage breakdown probabilities in the presence of synchrotron radiation. An R&D programme has been initiated to quantify this relation, but the present concept addresses it by limiting the maximum electric field to $1.46\,\mathrm{MV\,m^{-1}}$. Electric breakdown during beam operation will also need to be considered for interlocking and machine protection strategies, as the sudden loss of electric field will cause strong betatron oscillations of the circulating beam.

Dedicated DC septa magnets are needed towards the arc from this separator to further separate the trajectories up to the nominal arc separations. Two groups with thin and thick septa will operate at a relatively high field of up to $0.244\,\mathrm{T}$ for $\mathrm{t\bar{t}}$ mode. The resulting high power of the synchrotron radiation generated will need specialised absorbers to protect other nearby components and downstream septa.

The combined system of EM separator and DC septa is required for the higher energy modes where the collider cavities are shared between the two beams. The present concept results from the initial studies conducted [289, 291] and a more detailed review of the overall implementation and operation of the separator system will be needed in the technical design phase.

## 3.8 Powering systems

### 3.8.1 Collider Magnet Powering Systems

***Global optimisation of the magnet powering systems***

Evaluating cost-effective, efficient and reliable powering solutions is critical to assessing the feasibility of the FCC project. This requires balancing factors such as current density, number of turns, magnets in series, power converters location, energy storage systems and redundancy, while complying with constraints like footprints, power losses, and machine performance.

For instance, increasing cable current density lowers capital cost by allowing smaller cross-section cables but raises operational costs due to higher power losses. Identifying a current density that minimises total costs, both capital and operational, is crucial. Tunnel and alcove space must also be considered: Is there room for larger cables, or should space prioritise high-current cables ? Could shorter cables reduce expenses further ?

A comprehensive optimisation model addresses these complexities. This global model evaluates interconnected systems/circuits configurations, machine placements, and constraints, providing actionable insights to guide design decisions. Sub-models for magnets, cables, trays, power converters, alcoves, and cooling systems were developed in collaboration with feasibility study teams. Figure 3.58 illustrates their interconnections, which are critical for global optimisation.

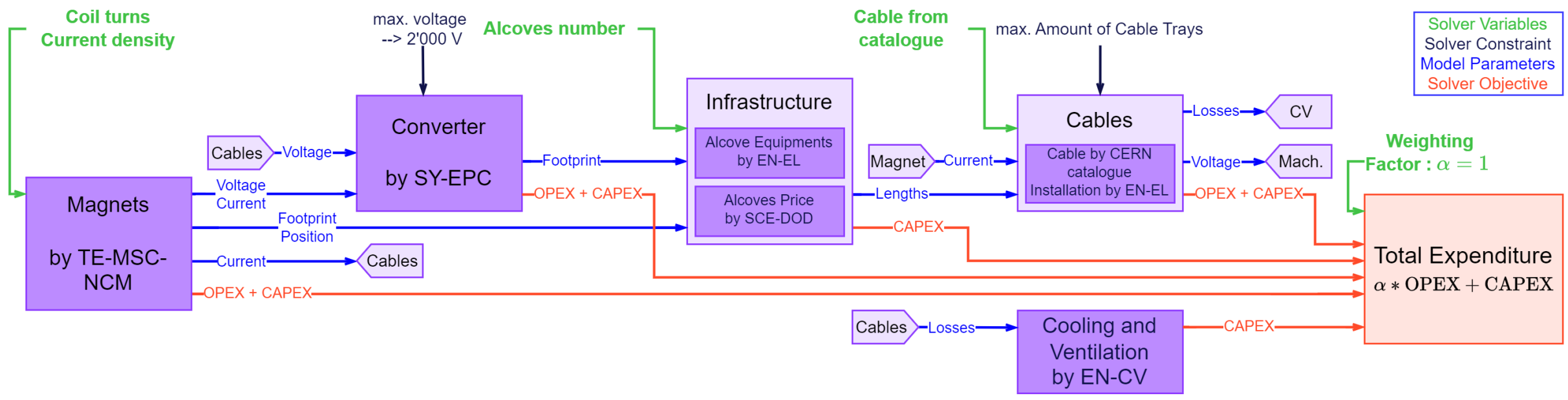

Fig. 3.58: Interconnected sub-models within the global optimisation of the magnet powering systems.

The main output, total expenditure (TOTEX), combines capital (CAPEX) and operational (OPEX) expenditures :

$$\mathrm{TOTEX} = \mathrm{CAPEX} + \alpha \cdot \mathrm{OPEX} \tag{3.4}$$

Where TOTEX [CHF] is the total expenditure; CAPEX [CHF] is the capital expenditure; OPEX [CHF] is the operational expenditure; $\alpha$ is the weighting factor, a parameter to balance the relative importance of CAPEX and OPEX.

The global model also outputs other important parameters such as power consumption, power losses, material masses and carbon footprint. Key inputs include the number of alcoves, cable and magnet current density, magnet turns and circuit configurations. Constraints include maximum voltage drop and available cable trays.

An evolutionary optimisation algorithm, suited for multi-variable, non-linear and non-continuous systems, iteratively determines the input parameters that minimise TOTEX while respecting constraints.

The model guided choices for magnet and cable parameters to identify cost-effective solutions. The magnet parameters were refined by the magnet team to meet additional constraints and forms the baseline of this report.

### *Collider Magnet Circuits*

Magnet powering is performed by power converters in alcoves located either at the end of the straight section or in the machine tunnel, called *Big Electrical Alcove* and *Small Electrical Alcove* respectively. The installation of power converters in the machine tunnel is not considered due to space and radiation constraints.

The location of the power converters is dictated by several factors, including the granularity of control required, the maximum voltage tolerance of the cable insulation and the resulting impact on expenditure for cables and converters. The granularity of control required by the optic layout :

- Collider dipoles, as well as collider quadrupole focusing and defocusing magnets, can each be powered in series within their respective family.
- Collider sextupoles need to be powered in groups of two or four (depending on the energy level).
- Collider tapering magnets can be powered in groups of up to four consecutive magnets.
- Collider correctors need to be powered individually.

The power converters for the collider dipoles and quadrupoles are installed in the big electrical alcoves at the end of the straight section.

In contrast, all other converters are located in the small electrical alcoves of the tunnel, as they power fewer magnets in series. Table 3.15 presents the quantity of magnets and circuits and the powering parameters.

Magnet specifications and quantities for the *straight section* and the *injection* is not yet defined, estimations were made.

For the Booster details, see Table 6.7 in Section 6.7.1

Table 3.15: Collider Magnet Powering Circuits.

| | Magnet Quantity | Circuit Quantity | Peak Current (A) | Peak Voltage (V) |
|---|---|---|---|---|
| Dipole | 5680 | 16 | 3665 | 420 |
| Quadrupole (F and D) | 2836 | 32 | 366 | 1914 |
| Sextupole (F and D) | 4672 | 1168 | 178 | 284 |
| Dipole Tapering | 5680 | 710 | 7 | 148 |
| Quadrupole Tapering | 5672 | 709 | 9 | 150 |
| Horizontal Corrector | 2824 | 2824 | 8 | 47 |
| Vertical Corrector | 2824 | 2824 | 14 | 48 |
| Skew Quadrupole | 2824 | 2824 | 10 | 44 |
| †*Straight Section* | *n/a* | *1666* | *n/a* | *n/a* |
| †*Injection* | *n/a* | *n/a* | *n/a* | *n/a* |
| **Collider subtotal** | **34 698** | **12 773** | - | - |
| Booster Subtotal | 21 068 | 7917 | - | - |
| **Total** | **55 766** | **20 690** | - | - |

†Magnet specifications not yet fully defined or nonexistent, values extrapolated

***Precision of power converters***

Converter precision class is used to encapsulate metrics such as accuracy, reproducibility, stability, resolution and tracking. Depending on the precision class, the cost and complexity of the power converters varies significantly. Table 3.16 shows the precision classes used for CERN power converters.

Currently power converters offering class 4 precision have been considered, but the current precision requirements for the power converters remain unspecified. These requirements are essential to determine the precision class and, consequently, the cost implications of the converters.
Standard precision converters (class 7) are readily available from manufacturers, offering a cost-effective solution for most applications. However, a step higher in precision involves additional costs due to more stringent stability and resolution requirements, which remain manageable within standard practices. In contrast, the highest precision converters, often required for the most demanding applications, are not only difficult to manufacture but also come with substantial cost increases. Defining the precision class required is thus critical to accurately estimate the capital expenditure and to identify whether standard or custom-designed converters will be necessary.

Table 3.16: HL-LHC Precision class and metrics.

| | Class | | | | | | | |
|---|---|---|---|---|---|---|---|---|
| | **0** | **1** | **2** | **3** | **4** | **5** | **6** | **7** |
| Resolution [ppm] | 0.5 | 0.5 | 1.0 | 1 | 1 | 1 | 1 | 1 |
| Initial uncertainty after cal. [2x RMS ppm] | 2.0 | 2.0 | 3.0 | 7 | 10 | 50 | 100 | 200 |
| Linearity [max. ppm] | 2.0 | 2.0 | 5.0 | 8 | 9 | 20 | 50 | 100 |
| Stability during fill (12 h) [2x RMS ppm] | 1.0 | 2.6 | 15.5 | 33 | 39 | 50 | 100 | 200 |
| Short term stability (29 min) [2x RMS ppm] | 0.2 | 0.4 | 1.2 | 2 | 5 | 10 | 20 | 50 |
| Noise ($< 500$ Hz) [2x RMS ppm] | 3.0 | 5.0 | 7.0 | 15 | 19 | 50 | 100 | 200 |
| Fill to fill repeatability [2x RMS ppm] | 0.7 | 1.6 | 14.5 | 32 | 38 | 60 | 100 | 200 |
| Long term fill to fill stability [2x RMS ppm] | 9.5 | 9.5 | 26.5 | 56 | 64 | 200 | 500 | 1000 |

#### *Availability of Power Converters*

Please see Section 2.4.4.

### 3.8.2 RF Powering

The RF systems of the FCC-ee machine will be distributed between point PH, which will host the collider RF, and point PL, which will house the booster RF. The RF infrastructure will evolve throughout the operational lifecycle of FCC-ee, adapting to the different modes of operation. The primary RF amplifiers will be tristrons (or potentially klystrons), while the booster in the $t\bar{t}$ phase will utilise solid-state amplifiers (or possibly IOTs).

The tristrons will require a high-voltage DC supply, which will be provided by the RF main power converter system. This system is planned to be located on the surface and will require a 40 kV AC supply. In contrast, the solid-state amplifiers for the booster during $t\bar{t}$ operation will be installed underground in the klystron galleries and will operate with a low-voltage AC supply.

The electrical infrastructure must be designed to accommodate all modes of operation, ensuring sufficient capacity for the $t\bar{t}$ phase, which has the highest power demand. Given the substantial energy requirements of the RF systems, a dedicated branch of the machine's electrical network will be allocated to their supply, ensuring stable and efficient power distribution.

#### *Point PH Layout*

The number of RF cavities and RF amplifiers will increase with the beam energy, see Fig. 3.59. There will be updates of the machine for each operating mode. The strategy is to increase the number of cavities and RF amplifiers within the operating year. The technical infrastructure to cover all modes of operation will be installed from the beginning, see Fig. 3.60.

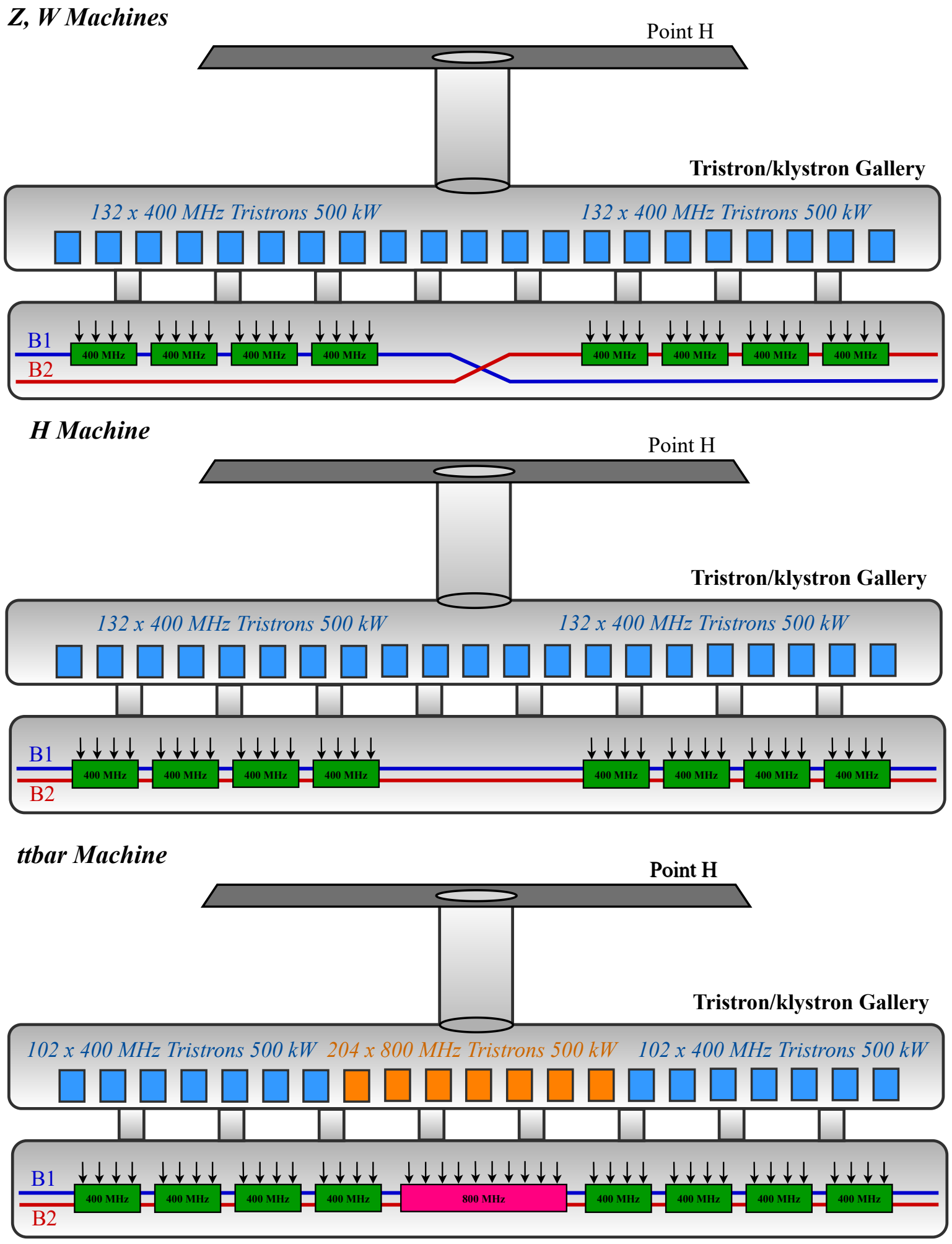

Fig. 3.59: Collider RF amplifiers as a function of operating mode.

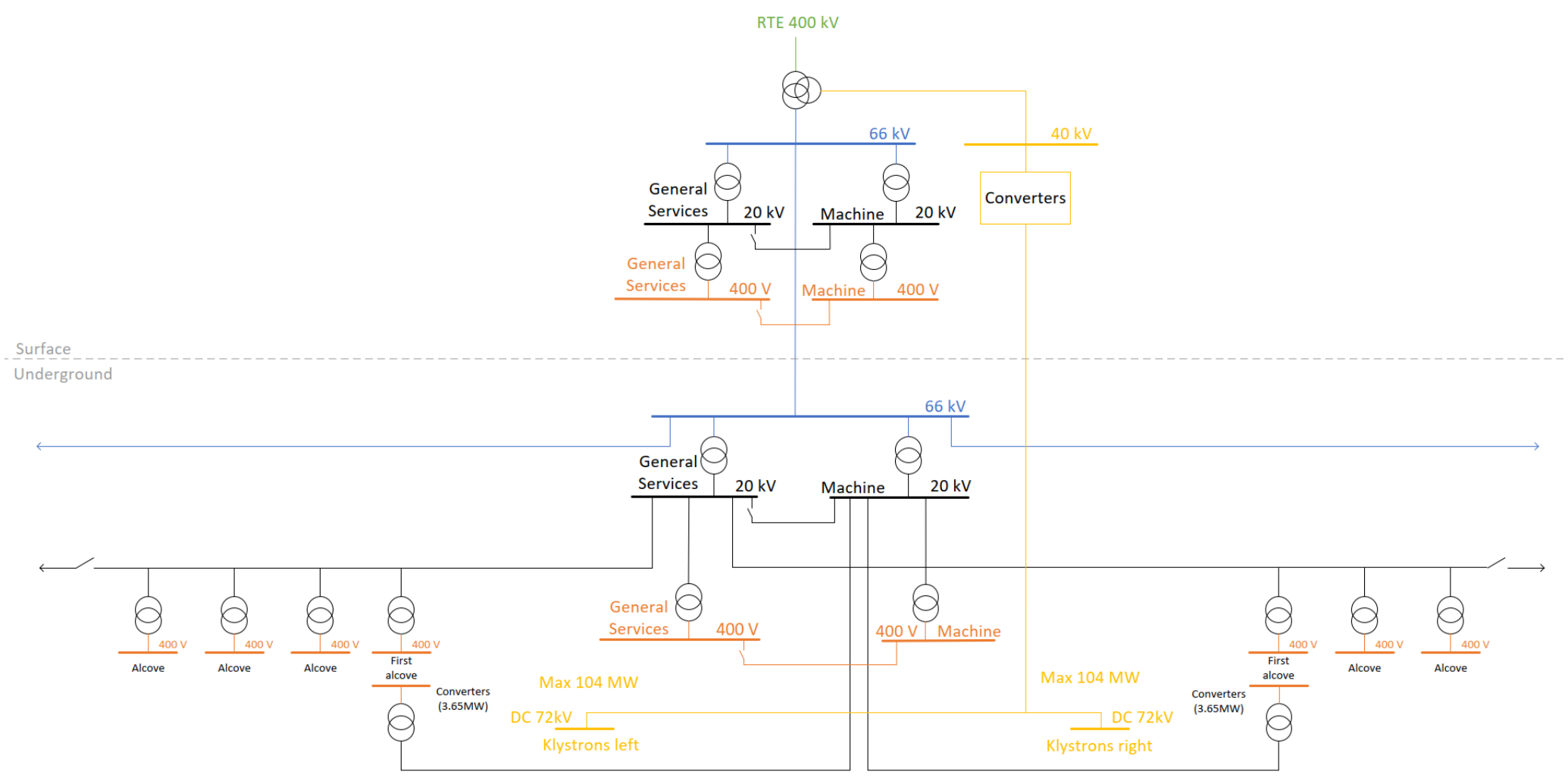

Fig. 3.60: Concept of electrical distribution at point PH.

### *Point PL layout*

Like the collider, the booster will be upgraded with more cavities and RF amplifiers as the operating mode advances, see Fig. 3.61. The technical infrastructure will also be installed from the beginning to cover all operating modes, see Fig. 3.62.

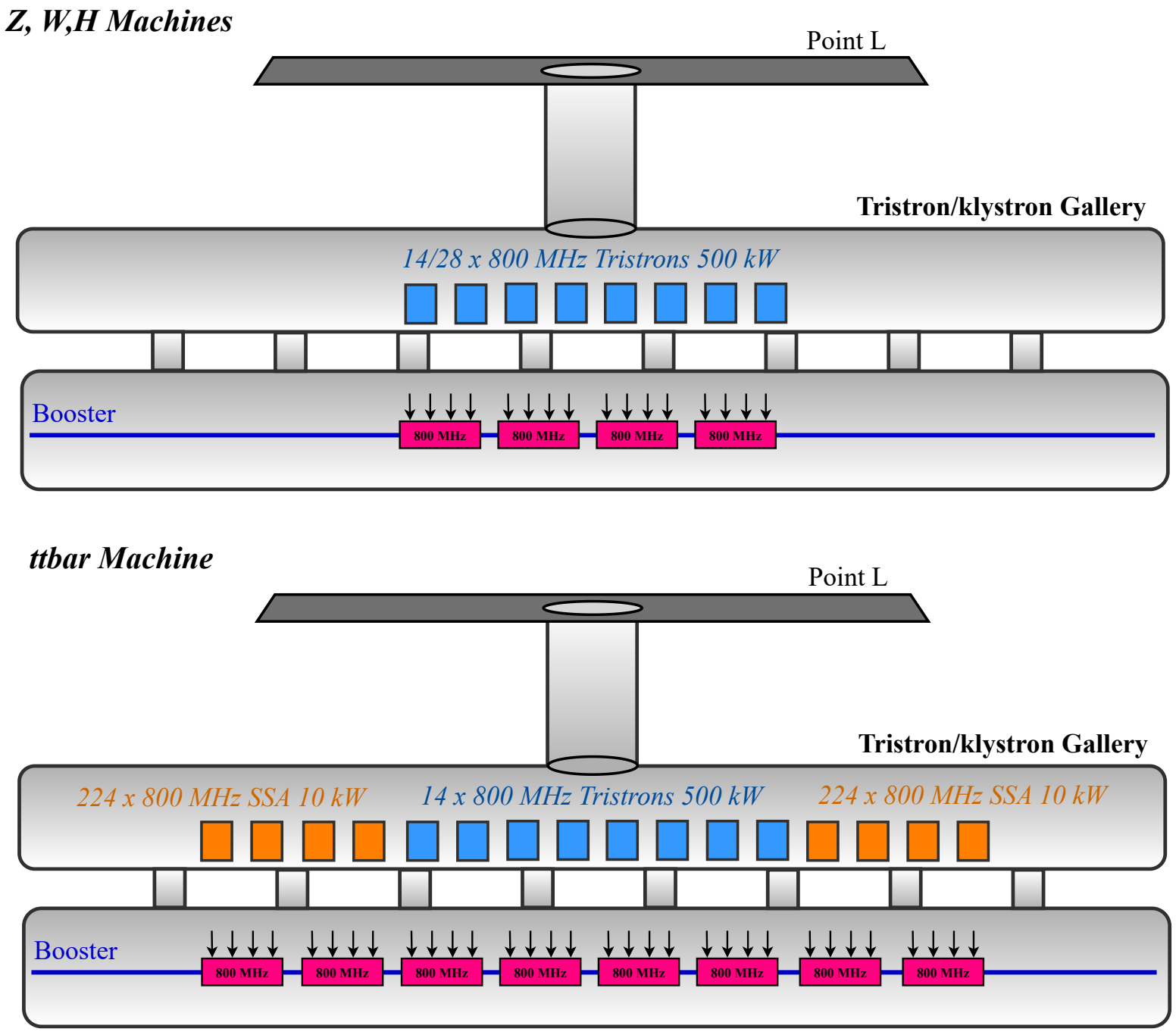

Fig. 3.61: Booster RF amplifiers in function of operation mode.

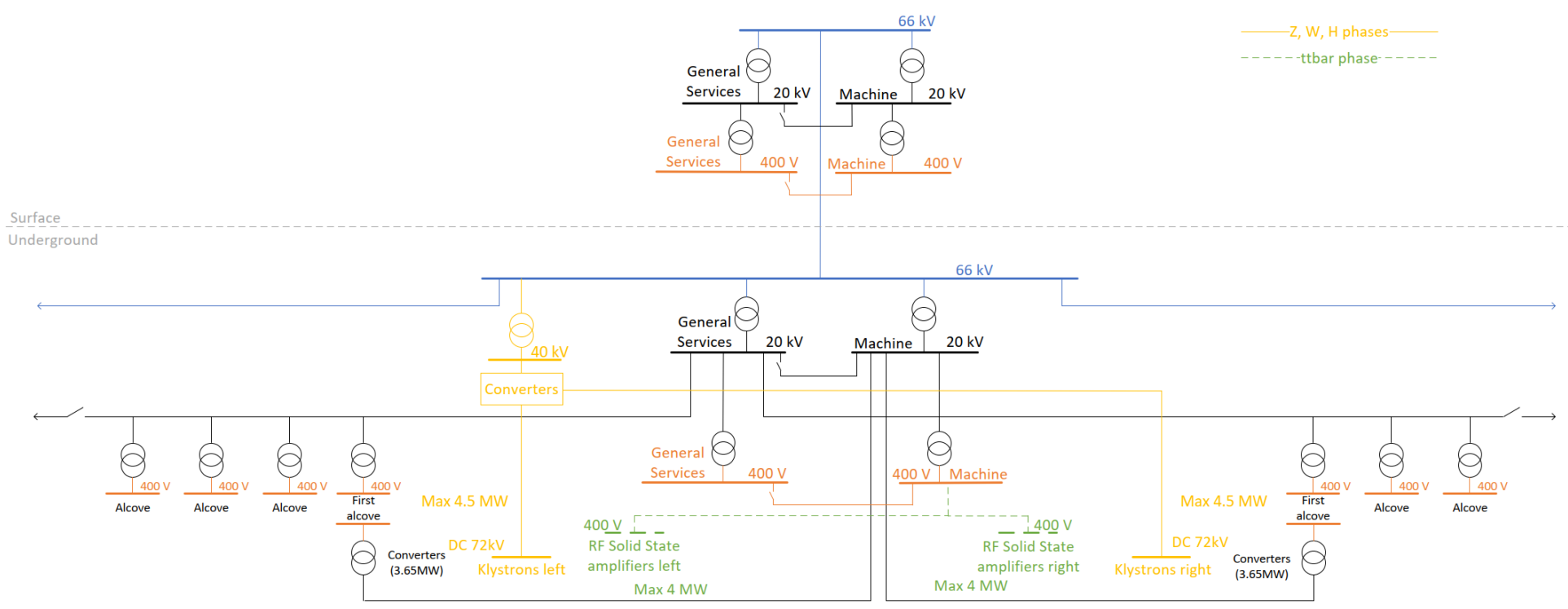

Fig. 3.62: Concept of Electrical distribution layout at point PL.

### *RF high-voltage main power converters*

The powering solution for the tristrons (or klystrons in case they are used) of the collider still needs to be optimised. However, a single and centralised high-voltage power converter installed on the surface has many advantages. Figure 3.63 illustrates the principle of this case.

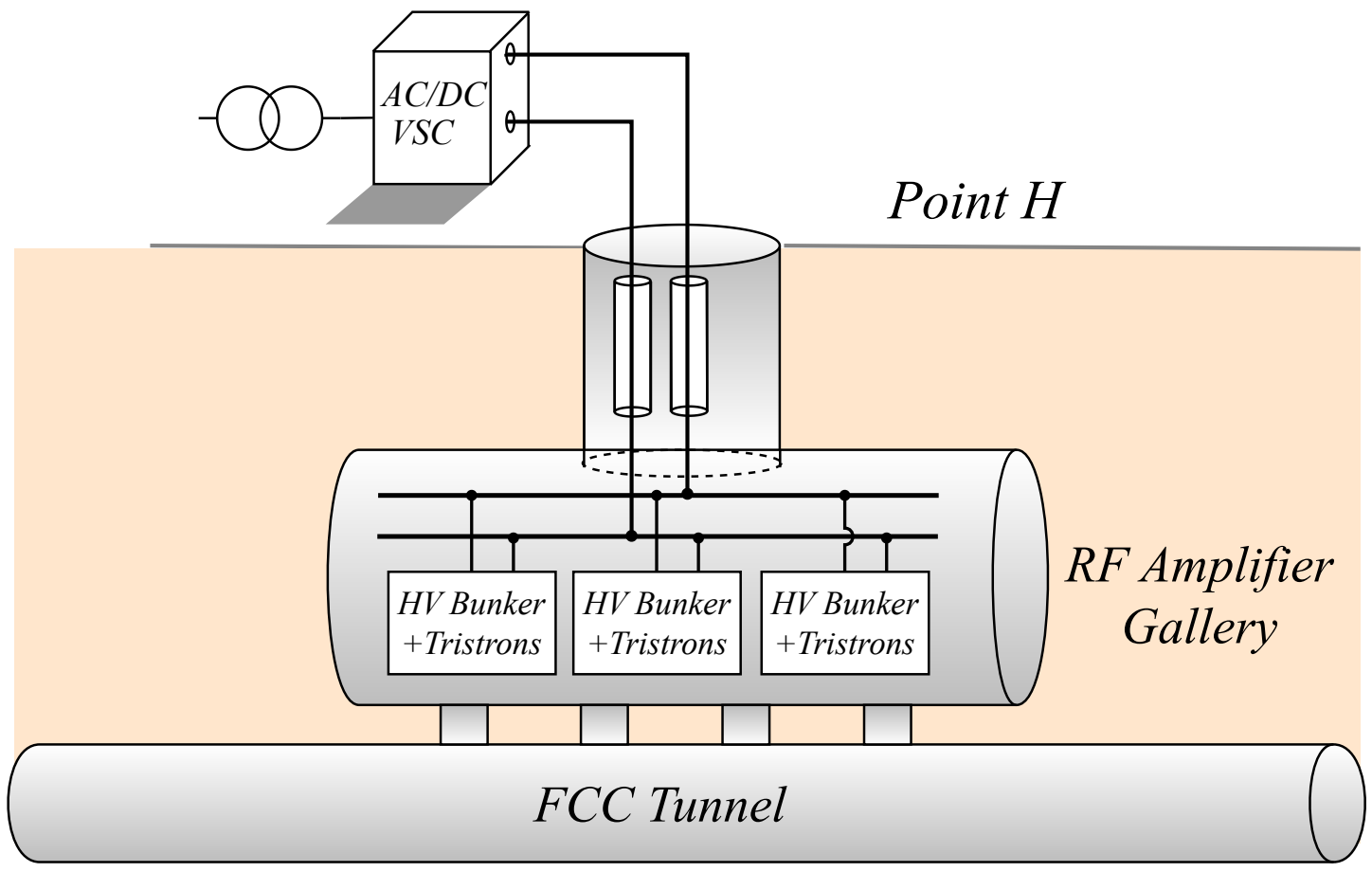

Fig. 3.63: The principle of the integration of FCCee RF powering.

Even with a centralised power converter solution, some RF powering equipment needs to be installed in the klystron gallery (HV tank in Fig. 3.63):

- Filtering capacitors in the proximity of the tristrons for DC filtering and DC bus decoupling.
- If needed, a small power converter can be used to trim the voltage on each tristron (powering fine tune / klystron perveance drifts in case they are used, etc.)
- A crowbar protection system (probably a series HV switch) to disconnect a klystron in case of a fault.
- Small power converters for the tristron (or klystron) filament heater and solenoid.

As a first rough estimation, a volume of 1 $m^3$ is to be reserved in the RF amplifier gallery for each tristron. The centralised RF high voltage power converter can be based on the so-called MMC (Modular Multilevel Converter), offering very high efficiency (97-98%) and high modularity. These types of converters are already deployed (e.g., for high-voltage DC transmission systems) at even higher voltages and power. An illustrative draft layout of the 150 MW / 60 kV RF power converter is shown in Fig. 3.64. The estimated total surface area for the collider RF power converter is 50 m$\times$30 m = 1500 $m^2$.

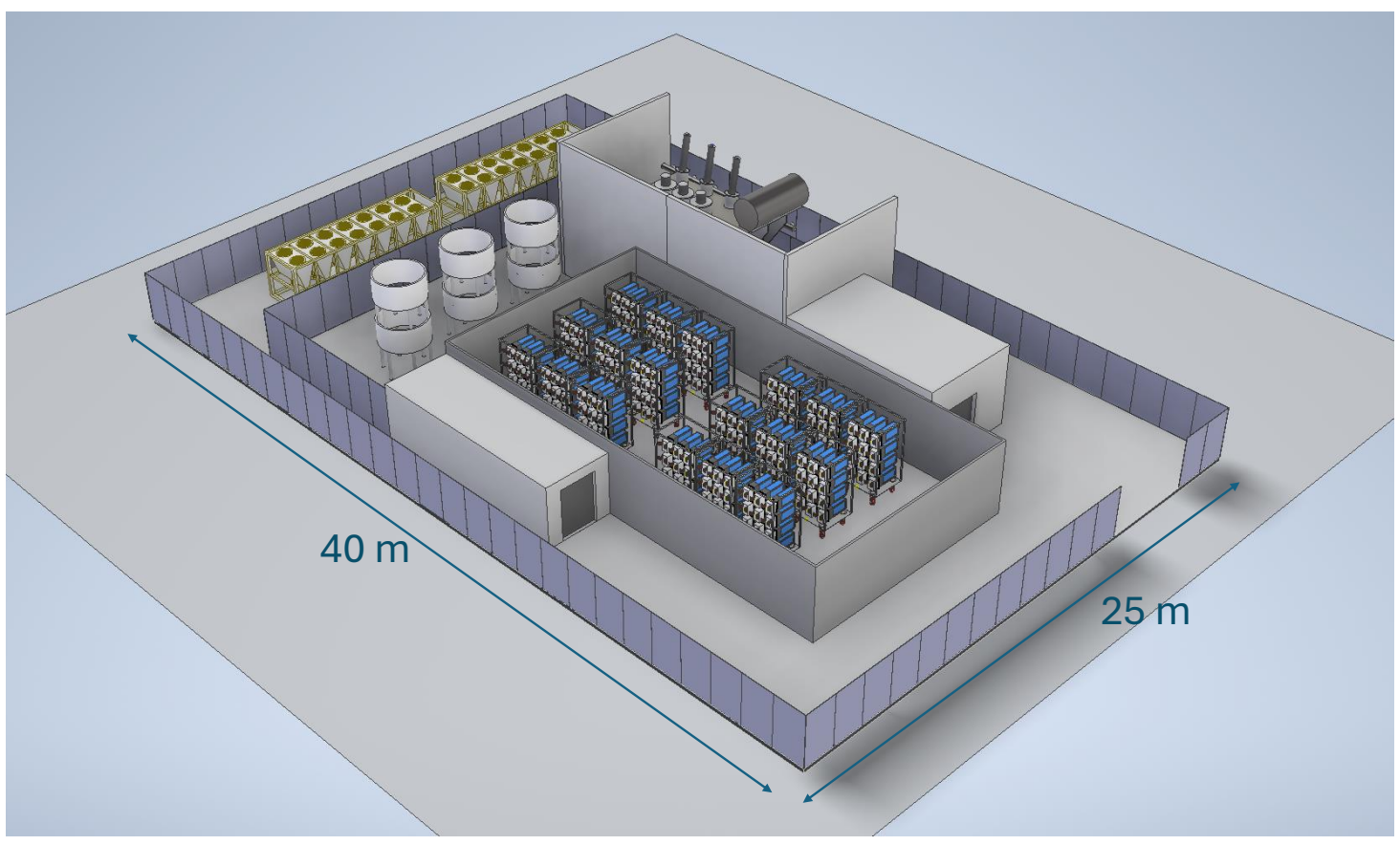

Fig. 3.64: Centralised RF high-voltage power converter.

There can be a similar approach for the booster but with a smaller power level of around 10 MW.

## 3.9 Beam diagnostics

The FCC-ee beam instrumentation (BI) will be composed of a very large number of devices that will provide a means to measure relevant beam properties across the FCC complex (injector complex, booster and main ring). Table 3.17 shows the overview of the number and type of BI systems needed in the complex. A feasibility study was only carried out for specific systems in the main ring that are considered the main challenges for the FCC BI. The section includes the requirements and design solutions for such systems.

Table 3.17: Number of BI systems in the FCC-ee complex

| | **Dump line** | **Main ring** | **Booster ring** | **Inj. lines** | **HE linac** | **Comm. linac** | **$e^-$ linac** | **$e^+$ DR** | **$e^+$ TL** | **$e^+$ linac** | **TOTAL** |
|---|---|---|---|---|---|---|---|---|---|---|---|
| Beam position | | | | | | | | | | | |
| Quad BPM | 20 | 5800 | 2944 | 420 | 82 | 35 | 11 | 258 | 30 | 17 | **9617** |
| Special BPM | | 20 | 5 | | | | | 4 | | | **29** |
| Collimator BPM | | 66 | | 5 | | | | | | | **71** |
| Beam loss | | | | | | | | | | | |
| Fast BLM channels | 34 | 152 | 126 | 200 | | | | | | | **512** |
| Arc BLM channels | | 17616 | 8808 | | | | | | | | **26424** |
| Arc BLM crates | | 1468 | | | | | | | | | **1468** |
| Beam intensity | | | | | | | | | | | |
| Fast BCT & WCM | 3 | 4 | 2 | 1 | 2 | 2 | 2 | 1 | 4 | 2 | **23** |
| DC BCT | | 4 | 2 | | | | | 1 | | | **7** |
| Transverse profile | | | | | | | | | | | |
| Imaging screen | 6 | 6 | 2 | 20 | 2 | 2 | 2 | 2 | 4 | 2 | **48** |
| SR-based | | 4 | 1 | | | | | 1 | | | **6** |
| Laser Wire Scanner | | 2 | | | | | | | | | **2** |
| Longitudinal profile | | | | | | | | | | | |
| b/b(EO/streak) | | 2 | 1 | | 1 | 1 | 1 | 1 | 1 | 2 | **10** |
| Luminosity / collision rate | | | | | | | | | | | |
| Beamstrahlung | | 8 | | | | | | | | | **8** |

### 3.9.1 Beam position diagnostics

The R&D for the FCC-ee BPM systems is focused on the design, integration and alignment of button-type BPM pickups for the arc and the interaction regions (IR) of the main rings, as well as R&D of the related signal processing electronics. Given the size of the system and its importance for FCC operation, the reliability and potential repair strategies should be considered at all stages of the system design.

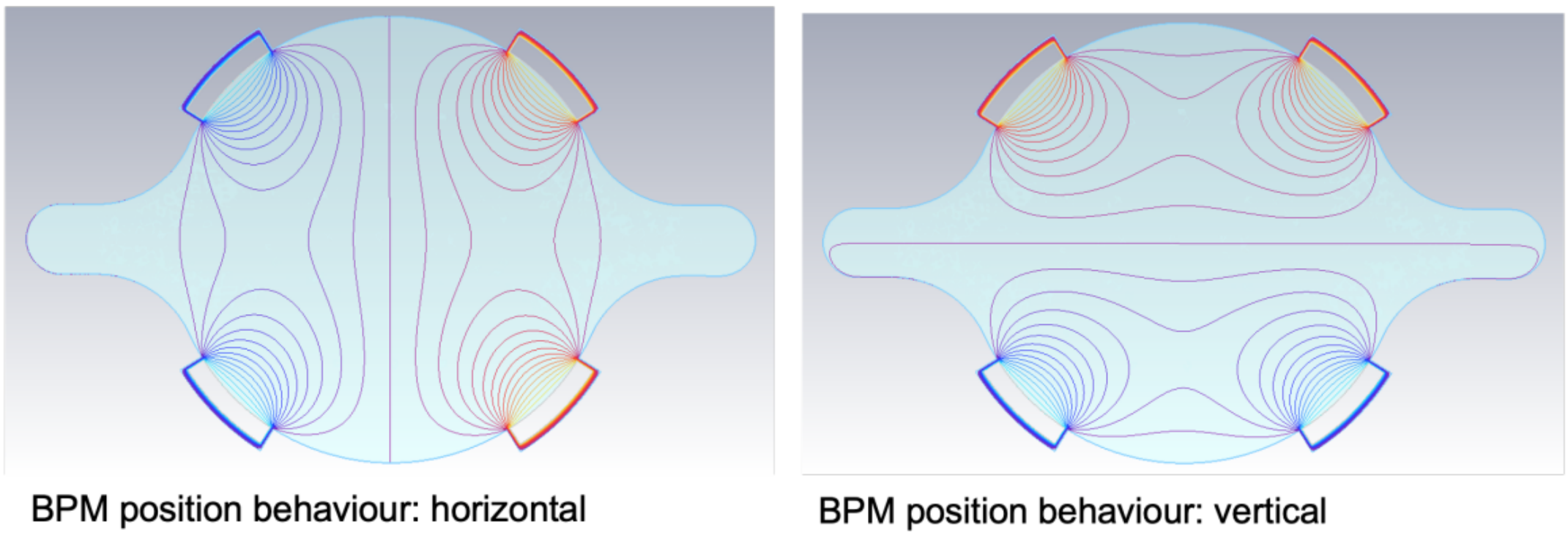

Fig. 3.65: Electrode arrangement of an FCC-ee main ring button BPM, along with lines of constant beam horizontal (left) and vertical (right) displacement.

Figure 3.65 illustrates the skewed arrangement of the BPM button electrodes to avoid the exposure of the synchrotron light fan.

The BPM system, with the button pickups as the signal source located at every quadrupole, has to meet several challenging requirements, such as a turn-by-turn resolution of 10 µm and an orbit resolution better than 1 µm, with the relative accuracy and the alignment tolerances in the same range. A bunch-by-bunch measurement capability is required for the nominal FCC-ee bunch spacing of 25 ns. The BPMs will also serve several feedback applications, thus the layout, segmentation, and data transmission need to be studied to minimise the latency of the BPM readings.

To tune and optimise the luminosity in the interaction points (IP), BPMs will be symmetrically located on each IP side. One pair of BPMs is planned in the combined beam vacuum chamber near the luminosity calorimeter to serve an IP luminosity feedback system. Another 3 - 5 BPMs accommodated in the IR cryostat with the beams in separate chambers will be mounted next to the segmented superconducting quadrupoles. These IR BPMs and associated signal cabling need to be particularly reliable, as their locations will be unreachable after the final assembly.

Requirements and details of the BPM system are still being discussed. Current studies based on electromagnetic simulations are targeted to find an optimal button electrode arrangement and size, ensuring at the same time that there are sufficient signal levels to achieve the required resolution and the BPM beam impedance within the limit assigned to the BPM system for all FCC-ee operational modes [292]. Preliminary studies have been performed [293, 294], and the longitudinal beam coupling impedance of 4000 BPM pickups with conical button electrodes was compared to other impedance contributing elements, such as the resistive wall (RW), bellows, RF cavities and taper section. So far, initial studies have shown that the BPM impedance budget of 40.1 V/pC should not be very difficult to fulfil and more challenging could be thermal aspects related to beam heating and resulting changes of the BPM mechanical dimensions and material properties [13].

In many aspects, the performance required from the BPM systems for the FCC-ee and current synchrotron light sources are quite similar, and therefore, technical solutions used in such systems could be used as a reference. However, the size of the FCC-ee BPM systems, the radiation levels, and tunnel temperature gradients are by far less favourable, therefore new design, production, installation and maintenance challenges will have to be addressed.

Early estimates for both the mechanical alignment of the BPMs required with respect to the quadrupoles and the alignment's long-term stability are of the order of $100\,\mu\mathrm{m}$. Solutions used in synchrotron light sources suggest that the BPMs may have to be rigidly attached to their corresponding quadrupoles to meet this alignment requirement. The relative positions of the BPM and quadrupole geometrical axes would need to be measured with even greater accuracy. As this requirement could influence the design of the quadrupoles and alignment strategies, aspects related to BPM alignment should be studied at an early stage of the arc cell design. Given the importance of the BPM alignment for the FCC-ee commissioning and subsequent operation, the solutions chosen should be confirmed by measurements performed on an arc cell prototype.

Contrary to light sources, there will be high levels of ionising radiation in the FCC-ee tunnel. The current experience with BPM electronics shows that they can be built from commercial off-the-shelf (COTS) components carefully selected and qualified with radiation tests when the integrated radiation dose over the whole lifetime of the project stays below some 1 kGy. Any higher radiation doses may require designing dedicated integration circuits, which would require significant financial and workforce resources, especially since the CERN BI group currently has no experience in designing radiation-tolerant BPM electronics.

Temperatures in the FCC-ee tunnel will have significant gradients and variations related to changing operational conditions, unlike in light sources. To achieve the beam position measurement accuracy required, it may be necessary for the BPM temperature drifts to be reduced by active cooling to ensure that the BPM dimensions stay within the necessary margins. Similarly, the BPM electronics may also require active cooling to stabilise their temperatures to a level that ensures the long-term accuracy required of the system. Such aspects will be studied in more detail once the temperature variations and gradients in the machine tunnel are better quantified, along with the temperature sensitivity of the prototypes of the BPM electronics.

Beam signals from FCC-ee arc BPM electrodes will not be able to be sent to the signal processing electronics over long coaxial cables due to very limited space in the tunnel cable trays. Instead, the beam signals must be treated close to the BPMs, so that the resulting beam data can be sent over optical fibres to an alcove. At this initial stage of the system design, it is expected that there will be a mini-rack located near each BPM pair (one BPM per beam, installed close to each arc quadrupole). This mini-rack would accommodate BPM electronics, receive beam signals from the BPMs over short coaxial cables, and transmit results over optical fibres to an alcove. Racks to accommodate computers to process and concentrate beam position data at rates adequate for sending it to the surface will be located in the alcoves.

Currently, the design effort is focusing on the arc BPMs, as they are most important for the design of the arc cell, which is a crucial part of the machine design. It is expected that the booster BPMs and their electronics will have a similar design.

The IR BPMs have stricter requirements for the measurement resolution and accuracy, therefore their mechanical design will be addressed separately. Also, the IR BPM processing electronics will need to fulfil very challenging requirements. It would be favourable to accommodate the IR BPM electronics in places free from high levels of ionising radiation so that the electronics can be based on the best commercial components available on the global market. Otherwise, it might be very difficult to design and produce radiation-tolerant versions.

### 3.9.2 Transverse diagnostics

Transverse profile measurements must be non-invasive during regular operation, whilst invasive measurements may be considered in the initial stages of commissioning or dedicated low-intensity fills.

Monitoring the relative evolution of the transverse emittance is the main priority for transverse diagnostics, and the precision target is set to 2% in emittance, corresponding to 1% in beam size. Absolute accuracy is less critical and not strictly necessary from the very beginning of the operation. The accuracy target is set at $\pm 15\%$ in emittance. These precision and accuracy targets apply to both transverse planes, irrespective of the machine's operation mode.

Most machine operations require a relatively slow monitoring of the evolution of the average beam emittance. Measuring the average emittance at a frequency of at least 1 Hz is desirable, as it can provide input for feedback systems aimed at optimising luminosity. However, faster diagnostics enabling bunch-by-bunch measurements become necessary for studying effects like single bunch instabilities or detecting any emittance patterns in the filling scheme. Bunch-by-bunch capabilities are not required from all instruments, but at least one system per beam must provide such measurements. Bunch-by-bunch measurements can be acquired by integrating tens or hundreds of revolutions to enhance the signal as needed. These measurements can occur either through a continuous bunch scan or with on-demand measurements. In both cases, the full-ring scan should not exceed a few minutes to ensure sufficient beam stability. Single-bunch measurements recorded in a single-turn may be investigated but are not a priority.

As is customary in high-energy lepton machines, the transverse diagnostics for FCC-ee will primarily rely on synchrotron radiation (SR). To achieve the resolution needed for picometre-level transverse emittance, diagnostics techniques operating in the x-ray domain become necessary.

The baseline locations for the SR diagnostics are downstream of the two major experiment interaction points (PA and PG), just after the wiggler straight section. This location features a series of weak dipoles that are suitable candidates for the radiation source, with the photons emitted efficiently separated by the strong dipoles downstream that transport the beam towards the arc. The light will propagate to the instrumentation in a dedicated extraction line. Extraction lines approximately $100\,\mathrm{m}$ long will be required to transversely separate the light from the main beam and intercept the radiation at a distance of $0.5\,\mathrm{m}$.

Placing the diagnostics outside the arcs offers several advantages, including reduced radiation levels, improved accessibility to the equipment, and more flexibility in adapting the machine design to meet diagnostic requirements. Moreover, in the baseline GHC optics model, the location downstream of the interaction points is favourable due to relatively high betatron functions. These produce vertical beam sizes exceeding $40\,\mu\mathrm{m}$, thereby relaxing the resolution demands on the instrumentation.

The radiation energy range for diagnostics is expected to remain between 20 keV and 80 keV, well within the capabilities of existing diagnostic systems at modern light sources. To address variations in the synchrotron radiation spectrum across different beam energies, segmenting one of the existing dipoles into a sequence of tunable bends has been proposed. This approach minimises changes between operational modes while remaining transparent to the machine optics as long as the global bending angle of the dipole is preserved.

A multilayer monochromator will be installed as the first element of the photon beamline, positioned immediately after the vacuum exit window. Beyond its primary function of providing quasi-monochromatic light for specific diagnostic techniques, this monochromator also bends the photon beam, reducing the transverse space occupied by the diagnostics line. Additionally, it absorbs the majority of synchrotron radiation power emitted outside of the spectral range of interest, thereby reducing the thermal load on the detector.

Due to their simplicity and robustness, X-ray pinhole cameras are identified as the baseline for transverse diagnostics. These instruments will fulfil the requirement for precise average emittance monitoring. Bunch-by-bunch measurements can also be achieved with a fast gated detector. The optimised pinhole width for a typical photon energy of $20\,\mathrm{keV}$ is $60\,\mu\mathrm{m}$, which can be manufactured using the same technologies as modern-day light sources. This configuration provides a theoretical resolution of $35\,\mu\mathrm{m}$, comparable to the (vertical) beam sizes of interest, though not ideal for absolute beam size measurements.

Techniques based on synchrotron radiation interferometry (SRI) can improve resolution and compensate for the limited accuracy of pinhole cameras when the beam size approaches its resolution limit. Diagnostics exploiting the Heterodyne Near-Field Speckles (HNFS) technique [295] is being developed as a potential SRI candidate, and Young's double-slit interferometers have also been studied for FCC-ee. All interferometry-based devices could be installed on the same beamline as the pinhole camera, and the configurations could be switched as needed.

Alongside SR-based techniques, scintillating screens for beam observation (BTVs) will be installed in the transfer and dump lines. BTVs will also be placed in the main ring for beam detection during the initial stages of machine commissioning.

### 3.9.3 Beam loss monitoring

For circulating machines, the main functionality of a beam loss system is to:

– detect beam losses fast enough to protect accelerator components by triggering a beam extraction before the equipment is damaged and to
– provide regular measurements of the amount and location of beam losses along the accelerator to optimise machine operation and assess its performance.

However, the design of the beam loss monitoring (BLM) system depends on the knowledge of beam losses that are expected to be detected. Beam losses could be classified as accidental (or also known as irregular beam losses) and unavoidable (also known as regular beam losses). Regular beam losses are those that can be minimised but not completely avoided; they correspond to beam debris from collimation cleaning, physics collision debris, Touschek scattering along the accelerator or Coulomb scattering from residual gas interactions, among others. Therefore, continuous monitoring of beam losses with a distributed BLM system is required. Irregular beam losses can appear as a result of equipment failures, beam instabilities, micro-particles interacting with the beam or aperture restrictions, i.e., an object inserted in the way of the beam. These are, to some extent, avoidable losses, but a beam loss system has to detect and define the levels allowed for such losses to protect the accelerator.

Stored energies in the FCC-ee collider ring are expected to be lower than present existing machines like the LHC or its high luminosity upgrade (HL-LHC). However, as the beam size becomes very small, particularly for the vertical plane, the expected beam energy density is comparable to HL-LHC parameters. Assuming a beta function in the arc of $50\,\mathrm{m}$ (on average), beam sizes are expected to be of the order of microns, and the beam energy deposition per surface can go up to $11\,082\ \mathrm{MJ/mm^2}$ in the most critical case of the Z-pole configuration, see Table 3.18.

Ionisation chambers have collection times of the order of µs for electrons and $\mathrm{ms}$ for ions, and although they perform very stably under radiation and could be very sensitive, they are not ideal for bunch-by-bunch measurements. Instead, solid-state detectors have risen and decay times of the order of $\mathrm{ns}$, which will make them the most suitable candidates for fast loss detection. A new beam loss detection technology under study is the use of large-core silica fibres to convert the secondary charged particles from beam impacts into Cherenkov light. This process is instantaneous. The signal is then transported to both fibre ends and read-out by photo-sensors. The photo-sensor choice defines the time response of the device. Regular photo-multipliers, can be fast, in the order of $1-2\ \mathrm{ns}$ rise and decay times. Silicon photo-multipliers, are of the same magnitude although a bit slower, in the order of $10-20\ \mathrm{ns}$.

Final specifications on minimum and maximum beam losses to be measured and the required timing resolution will define the final choice of technology. In the meantime, R&D is needed to evaluate the performance of such devices under the $50\ \mathrm{MW}$ synchrotron radiation levels per beam expected in the tunnel.

Table 3.18: Collider beam parameters.

| | **HL-LHC** | **Z** | **W** | **ZH** | **$t\bar{t}$** |
|---|---|---|---|---|---|
| Beam Energy [GeV] | 7000 | 45.6 | 80 | 120 | 182.5 |
| Particle | p | | $e^+e^-$ | | |
| No. bunches | 2748 | 11200 | 1780 | 440 | 60 |
| Beam current [mA] | 1090 | 1270 | 137 | 26.7 | 4.9 |
| Bunch Intensity [$10^{11}$] | 2.2 | 2.14 | 1.45 | 1.15 | 1.55 |
| Stored beam energy [MJ] | 678 | 17.5 | 3.3 | 0.97 | 0.3 |
| Geom. emittance hor. $\varepsilon_x$ [nm] | 0.34 | 0.71 | 2.17 | 0.71 | 1.59 |
| Geom. emittance ver. $\varepsilon_y$ [pm] | 340 | 1.9 | 2.2 | 1.4 | 1.6 |
| Hor. beam size $\beta_x = 50$ m [µm] | 130 | 188 | 329 | 188 | 282 |
| Ver. beam size $\beta_y = 50$ m [µm] | 130 | 9.7 | 10.5 | 8.4 | 8.9 |
| Energy density [MJ/mm$^2$] | 40 118 | 11 082 | 682 | 645 | 87.2 |

#### ***Fast beam loss monitors***

With unprecedented stored beam energies for an electron collider, up to 17.5 MJ in the Z-pole operations, the electron/positron beams are highly destructive. A collimation system will be set in place in order to reduce the backgrounds to the experiments as well as to protect the machine from unavoidable losses. There is one system in PF for global halo collimation (betatron and off-momentum) and a second system around the colliding points. Beam loss monitors need to be installed at each collimator, with the aim of measuring turn-by-turn losses or faster, if possible. With the presently proposed collimation system, 27 collimators will be installed per beam, and thus 54 beam loss monitors covering both beams in the collider ring.

Regarding the synchrotron radiation collimation system, presently, six collimators and two masks per beam upstream of the interaction points are planned. To monitor the losses in these locations, a total of 16 beam loss monitors per collision point will be needed (corresponding to 64 beam loss monitors in the collider ring).

The injection and extraction to the booster and collider rings will also need beam loss monitoring, in particular at the location of collimators, beam masks or absorbers, kicker magnets and septum magnets. Table 3.19 shows the present estimate of beam loss monitors to cover the collider ring collimation and the injection/extraction needs for both rings.

The injectors to the Booster will need additional beam loss monitoring, and a first estimate of an additional 200 monitors is being considered.

#### ***Arc cell beam loss monitors***

Potential beam losses occurring in the arc cells of the collider and booster rings need to be continuously monitored. Beam loss monitors are typically located where losses are expected. In the case of the arc cells that correspond to near the arc quadrupoles, the aperture available is smaller. On the other hand, it is not predictable where micro-particles will interact with the beam.

A generic beam loss system in the arc cell should be able to cover a distance of several metres and reconstruct the main beam loss location. The implementation of this beam loss system is still under R&D, but a minimum configuration with 3 monitors or channels distributed along that distance could provide the location of the loss by triangulation. This number has been used to estimate the number of beam loss channels and read-out tunnel electronics needed to cover both the collider and the booster rings. Figure 3.66 shows a schema of the configuration of a generic beam loss system in the FCC-ee arc cell.

Table 3.19: Fast beam loss monitors for Colliding and Booster ring, excluding injectors.

| Purpose | Colliding ring | Booster ring | Total |
|---|---|---|---|
| Collimation Halo (PF) | 27/beam | | 54 |
| Collimation SR (PA/PD/PG/PJ) | 32/beam | | 64 |
| Booster injection region | | 10/beam | 20 |
| Booster to Collider extraction | | 15/beam | 30 |
| Booster to Dump extraction | | 15/beam | 30 |
| Booster to Dump transfer line | | (12+5)/beam | 34 |
| Booster to Collider transfer line | | 6/beam | 12 |
| Collider injection region | 2/beam | | 4 |
| Collider to dump extraction | 15/beam | | 30 |
| Collider dump line | (12+5)/beam | | 34 |
| **Total Injectors** | | | **200** |
| **Total Ring** | **93/beam** | **63/beam** | **312** |

The tunnel read-out electronics will be installed under the arc quadrupoles. A total of 18 BLM channels per arc cell can be read-out with a single crate. This includes the beam loss monitors for the collider and the booster rings. Table 3.20 shows the number of channels in both rings together with the number of crates needed in the tunnel. R&D is needed to develop radiation-tolerant electronics that are able to withstand the FCC-ee radiation level, which comes mainly from synchrotron radiation.

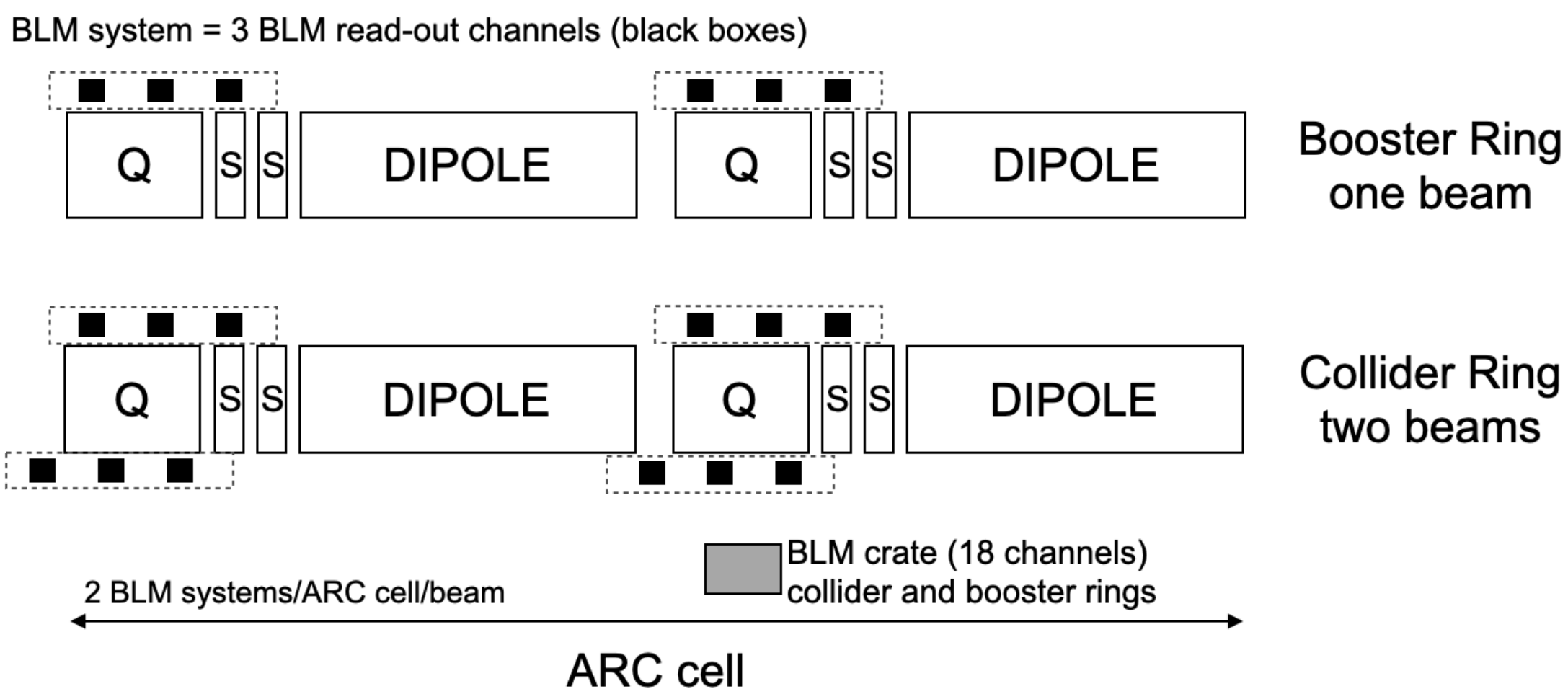

Fig. 3.66: Proposal of beam loss system in the arc cells, for Booster ring (on top) and Collider ring (on bottom), sharing the same read-out crate.

### 3.9.4 Longitudinal diagnostics

Longitudinal profile monitoring is needed to assess the effect of beamstrahlung, which will increase bunch lengths depending on the charge balance between colliding bunches. A bunch-by-bunch measurement of the bunch length is deemed crucial to determine the collision condition and the energy spread. Turn-by-turn measurement is not needed as the evolution of the bunch length will be in the order of the longitudinal damping time (i.e., from 1320 turns at Z to 20 at $t\bar{t}$). The precision of the order of 1% between bunches (both beams) is needed, while absolute accuracy is not strictly needed (still undefined). Considering a minimum bunch length of 1.91 mm or 6.4 ps ($t\bar{t}$) bunch length measurements must have a precision of better than 60 fs. A bunch-by-bunch profile measurement on selected bunches is required during commissioning and when requested to check the quality of top-up injected bunches, where fresh bunches at 10% nominal population are injected onto circulating bunches.

Table 3.20: Arc beam loss monitors.

| | Collider | Booster | Total |
|---|---|---|---|
| BLM channels per quadrupole | 6 | 3 | 9 |
| **BLM channels in the ring** | **17 616** | **8808** | **26 428** |
| BLM crates per arc cell | 1 | 0 | 1 |
| **BLM crates in the ring** | **1468** | **0** | **1468** |

As part of this feasibility study, electro-optical spectral decoding (EOSD) is one of the candidate techniques that could fulfil these requirements. The Karlsruhe Institute of Technology (KIT) is working on the design of an EOSD system adapted to the FCC-ee environment and parameters that allow single-shot longitudinal measurement with sub-picosecond resolution and a repetition rate in the MHz range [296]. Simulations have been carried out to investigate its performance under FCC-ee conditions that present significant challenges for EOSD measurements, especially the long bunches during Z operation with $\sigma$ ranging from 4.7 to 15.5 mm and the high bunch charge. A prototype was tested in the CLEAR facility at CERN in 2024. Further studies will be needed to optimise thermal management of the EO pick-up system under very high repetition rates, investigate the placement of monitor and laser/detection system and optimise the resolution to achieve 50 fs resolution.

Cherenkov diffraction radiation (ChDR) is an alternative to EOSD for longitudinal diagnostics [297, 298]. ChDR is generated as a charged particle passes in the vicinity of a dielectric material with a velocity which exceeds the speed of light in the given dielectric material. As the radiation is emitted at the well-known Cherenkov angle, the extraction of the signal is simplified in comparison to diagnostic techniques based on synchrotron radiation, which is emitted at a very small angle. For that reason, the incoherent part of the ChDR spectrum is a promising candidate for measuring bunch length in the FCC-ee. However, two analytical models [299, 300] predict a very different photon yield with very little experimental data [301, 302]. To assess the potential of incoherent ChDR for FCC-ee, a photon counting experiment is being prepared to measure the photon yield in the visible spectrum. The ATF2 beamline at KEK [303] is a suitable candidate for these tests, as it provides high particle energy and charge and a small beam size, which allows a beam-radiator distance in the sub-mm range, maximising the photon yield obtained.

### 3.9.5 Luminosity monitoring diagnostics for IP tuning

At each interaction point where the opposing beam is encountered, beamstrahlung emitted during the beam-beam collision results in an intense flux of photons on both sides of the IP, that are extracted through a dedicated 500 m long dilution channel terminated by a liquid lead dump. Specific instrumentation dedicated to monitoring the beamstrahlung profile position and its intensity is under development. Since the position of the photon peak is sensitive to the beam-beam transverse distance in the nanometre range as well as to residual optical aberrations at the IP, these signals are a valuable input for the IP tuning; see Fig. 3.67. The sensitivity range spans almost linearly over 200 nm and then saturates. This beamstrahlung information is used, in particular, for precision IP tuning scans.

A beam television screen (so-called BTV) is under development with the goal of imaging the beamstrahlung photon profile. The scintillator technology used for this device should be optimised for providing the best signal-to-noise ratio (i.e., synchrotron radiation vs beamstrahlung radiation) while keeping the thermal load from the ultra-intense flux of photons acceptable. Considering the worst-case scenario, a hollow screen is being simulated such that the profile position is tracked by fitting the tails only. The result obtained with a gap of 30 mm (equivalent to 1 sigma of the profile) is presented by the red curve in Fig. 3.67. It accurately follows the curve obtained by fitting the whole distribution (blue curve). If a screen imaging technique cannot be employed due to the heat load, the ionisation of rest gas is a possible alternative means of measuring the beamstrahlung profile [304].

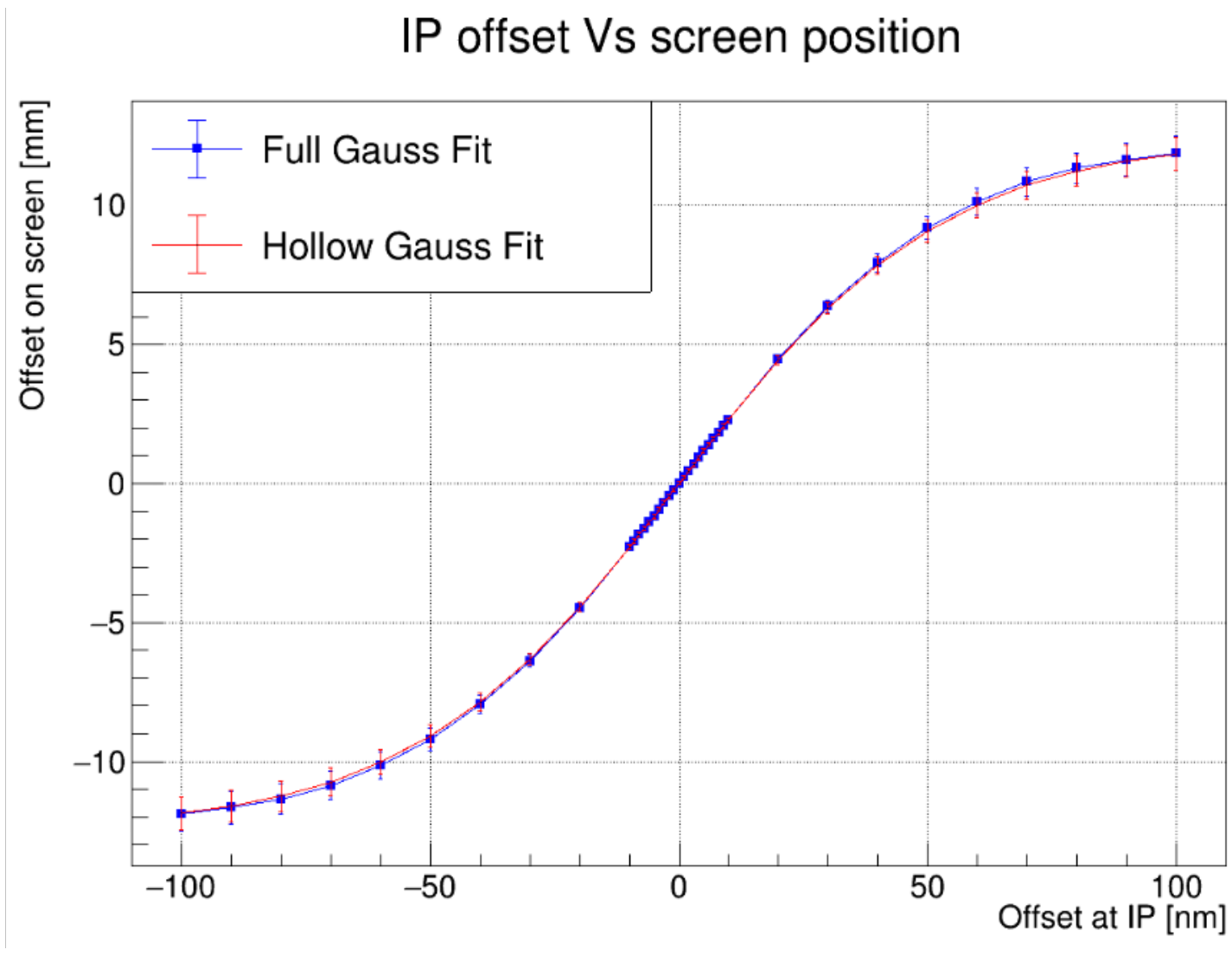

Fig. 3.67: Position of the beamstrahlung peak on the BTV screen installed at 400 m from the interaction point. In blue is a Gaussian fit applied to a full-screen image. The Gaussian tails fit to a hollow screen image are shown in red. The tail fit is based on the use of a 30 mm aperture.

Another process of interest in measuring luminosity is the rate of radiative Bhabha scattering. Since the resulting off-energy leptons mainly hit the beam-pipe about 150 m from the interaction point, a fast beam loss monitor installed at that location would allow monitoring bunch-by-bunch collision rate, complementing the beamstrahlung profile imaging technique. The latter might not be able to resolve individual bunch crossings. An example of a similar instrument is the BRAN developed for the LHC [305].

### 3.9.6 Inverse Compton polarimeter for energy calibration

A beam energy measurement based on resonant spin depolarisation will be implemented [306], providing precise collision energy calibration for the Z and WW modes. This technique performs regular transverse beam excitation tune scans on a dedicated pilot bunch to identify the spin resonant depolarisation frequency, which then provides the collision energy with great precision. The subject is covered in detail in the EPOL section of this feasibility report (Section 1.7). Here, the need for the inverse Compton polarimeter (IPC), which will perform the bunch polarisation measurement, is highlighted. The vacuum chamber for this 100 m long instrument must comply with the impedance budget of the machine, and a design study has been initiated. The simulation of the detector made good progress in building a digital twin of the future instrument. The EPOL section of this document gives more insight into the whole energy calibration procedure and the associated instrumentation.

## 3.10 Arc region: integration and supporting systems

The current FCC-ee layout has a circumference of around $90\,\mathrm{km}$, and about $85\,\%$ (i.e., $77\,\mathrm{km}$) are taken by the arcs. Arcs are made of a sequence of FODO half-cells, about 3000 for the high-energy optics, and consist of a short straight section (SSS) with quadrupole, sextupoles, beam instrumentation and correctors, followed by a long dipole length that can be achieved with a series of two or three interconnected magnets. This section describes the optimisation of the arc-supporting structures to maximise their performance, easing the installation and maintenance while minimising cost. The section also reports the state of advancement of the construction of the arc half-cell mock-up.

### 3.10.1 Collider-Booster placement configurations

The placement of the booster and the collider must be optimised in the radial and vertical directions of the tunnel cross-section, with a hard constraint of a maximum tunnel inner diameter of $5.5\,\mathrm{m}$. The longitudinal positioning of the SSS of the collider relative to the position of the SSS of the booster also requires optimisation. An optimised configuration of the tunnel cross-section, as described in Ref. [307], with respect to that presented in the CDR [13] has the booster placed on top of the collider. This frees more space for the services, especially the cooling and ventilation piping, the space reserved for transport vehicles, and the alignment system (Fig. 3.68). In addition to being more compact, the vertical configuration provides further advantages:

- Permits a smaller tunnel diameter in the RF sections.
- Same basement configuration as FCC-hh.
- Easier access for handling and removal of booster magnets and SSS (when leaving enough vertical clearance to allow installing/uninstalling both machines once the supports are already in position).
- Better from a radiation point of view, as the highest dose is generated on the outer side of the tunnel, which could be detrimental to the booster ring in case of a horizontal placement [308].

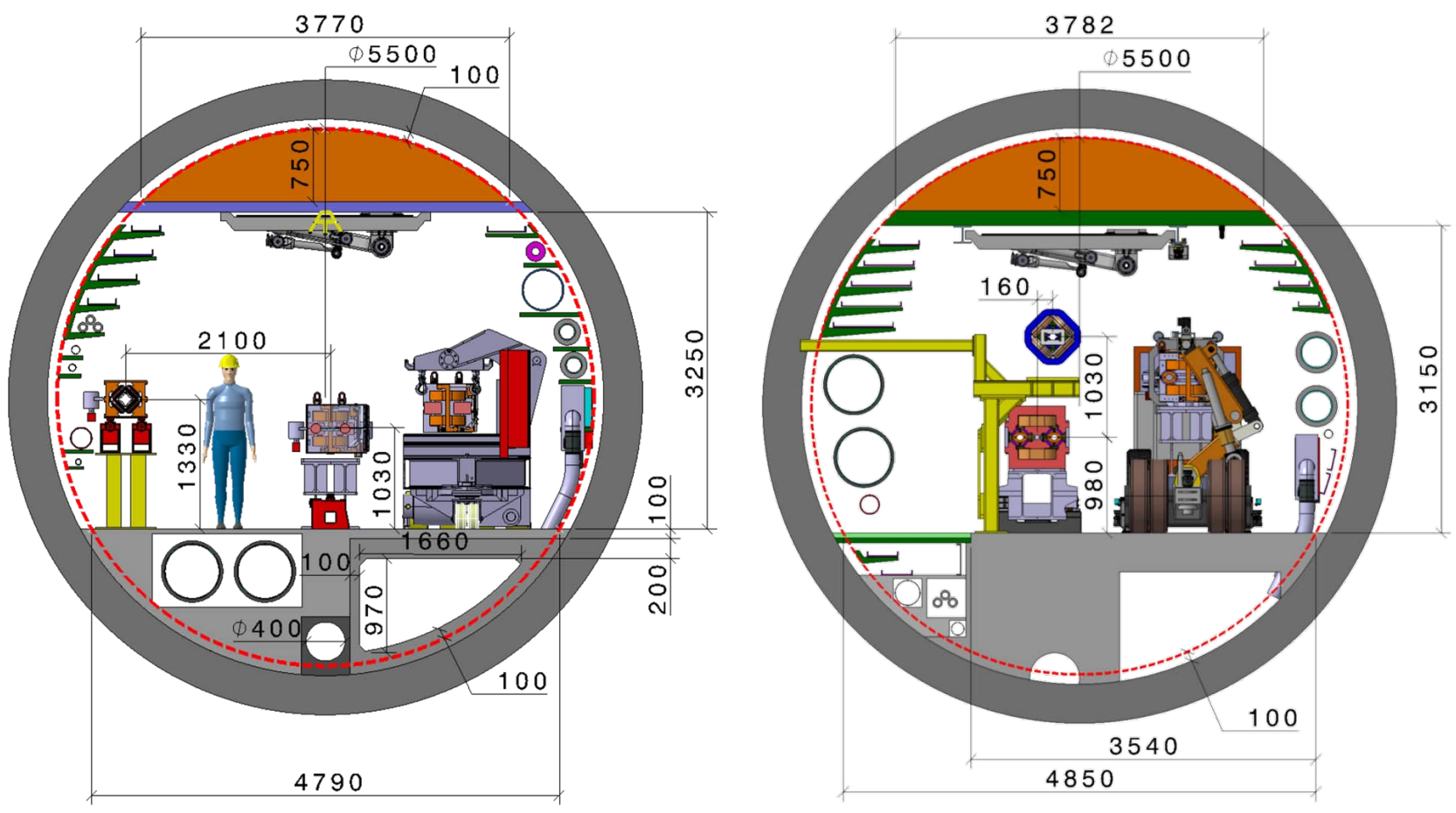

Fig. 3.68: Configurations for the relative placement between booster and collider. Left: horizontal configuration. Right: vertical configuration.

However, the vertical placement of the booster could raise dynamic stability challenges: due to the longer lever arm between the ground and the magnet, the booster could oscillate further, exceeding the tight dynamic positioning tolerances, particularly in the SSS region. For this reason, a significant effort in design and simulations has been made to improve the supporting system and maximise the stability of the two accelerators.

Given that the booster will be positioned above the collider, it is essential to maintain sufficient clearance with the collider underneath. This could be done by longitudinally shifting the SSS of the collider with respect to the SSS of the booster, keeping the cell periodicity (i.e., the azimuthal distance between the booster SSS and the collider SSS is maintained constant along the arc). Since the SSS is the bulkiest section of the arc for both machines, the proposed modification allows the SSS of the collider to be longitudinally positioned corresponding to the location of the smaller and more compact booster dipoles, and vice versa (example for ZH/$\mathrm{t\bar{t}}$ phases with the Quad-Sext-Sext configuration of the collider Fig. 3.69).

Further work will be required during a next phase of the project to optimise the interface and spacing between the collider, the booster, and their support structures. Sufficient clearance must be ensured to facilitate handling procedures, allowing for the installation or removal of individual magnets from each accelerator without interfering with the magnets of the other machine. This must be achieved while keeping the support structures in place, avoiding unnecessary disassembly and minimising operational disruptions.

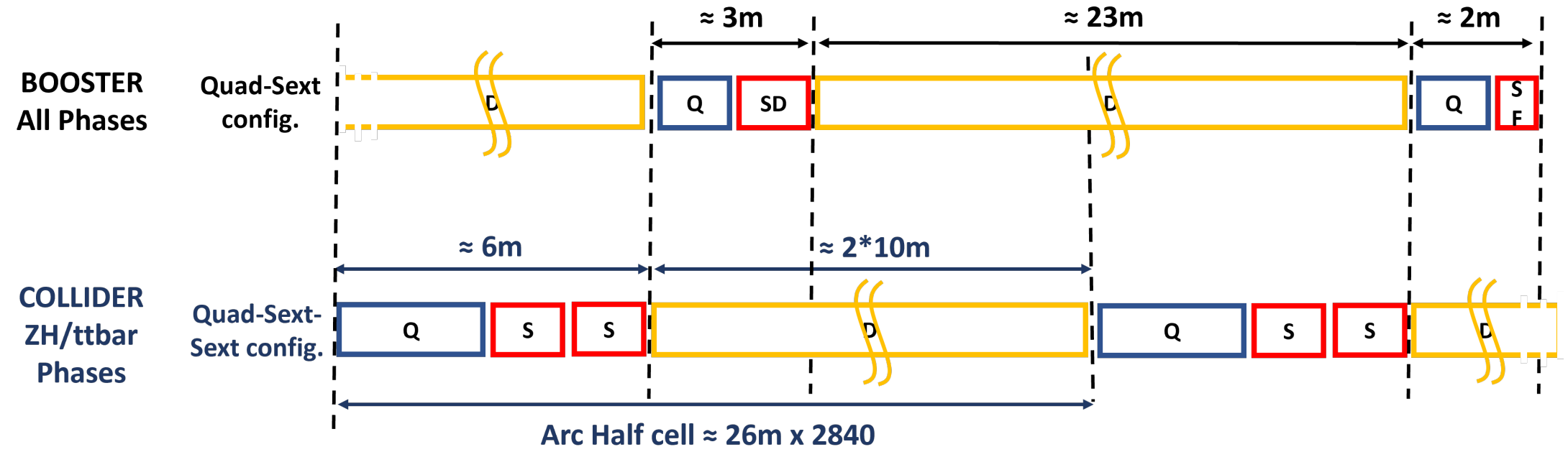

Fig. 3.69: Azimuthal shift between booster and collider - collider optics V24.3_GHC and booster optics V24_FODO.

### 3.10.2 Arc cell configurations of the collider

The configuration of the collider arc cell changes from the low to the high energy phase. The number of quadrupoles per unit length is doubled between the Z/WW and ZH/$\mathrm{t\bar{t}}$ phases as the length of the FODO cells is halved: the length of the half cell is $52\,\mathrm{m}$ for Z/WW phases and $26\,\mathrm{m}$ for the ZH/$\mathrm{t\bar{t}}$ phases. As can be seen in Fig. 3.70, there are two configurations for the Z/WW phase: 568 short straight sections with one quadrupole and one sextupole and 856 short straight sections with a single quadrupole. Similarly, there are two configurations for the ZH/$\mathrm{t\bar{t}}$ phase: 1136 short straight sections with one quadrupole and two sextupoles, and 1704 short straight sections with a single quadrupole.

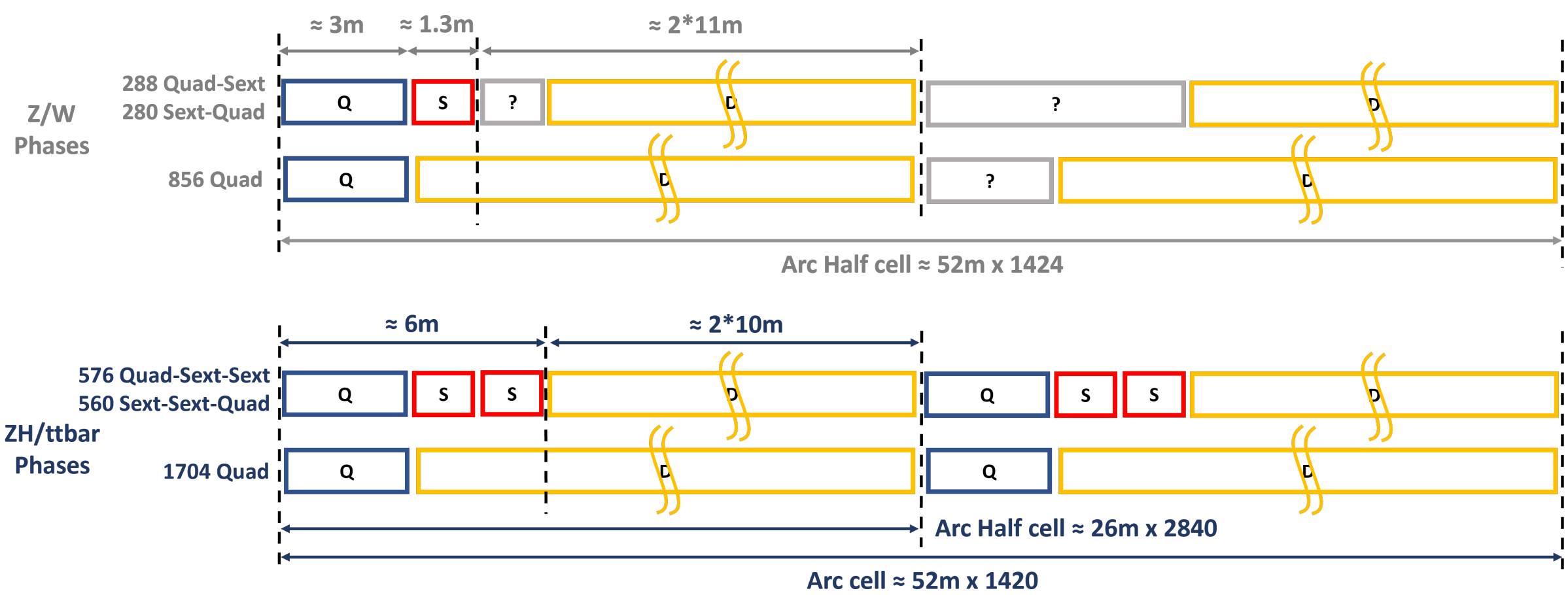

Fig. 3.70: Arc cell configurations for the collider, for the Z/WW phases and the ZH/$\mathrm{t\bar{t}}$ phases - optics V24.3_GHC.

Among these different SSS configurations, the Quad-Sext-Sext configuration has been studied in detail and will be installed in the mock-up. This SSS configuration is the bulkiest and the most challenging for its integration. It is also the most complex in terms of static and dynamic stability.

The optics of the collider is still evolving. However, the reference optics for the first phase of the mock-up was frozen at version V24.3_GHC by the end of 2024, to allow installation in the first half of 2025. To document this baseline solution, two drawings were produced, one of the cross-section (Fig. 3.81) and one longitudinal view representing an arc half-cell (Fig. 3.82), for the reference baseline optics V24.3_GHC. These two drawings, which constitute the baseline configuration for the mock-up, will be versioned following the mock-up study and will evolve in the coming years.

Further work is needed to define the strategy for the modification of the FCC-ee between the Z/WW and ZH/$t\bar{t}$ phases. At first glance, a mechanical reconfiguration of the collider will be necessary between these phases. To understand this reconfiguration, consider an ensemble of five FODO cells at Z/WW phase, corresponding to 10 cells at ZH/$t\bar{t}$, this forms a pattern that repeats 17.5 times in one arc, and over the eight arcs of the entire collider. In these five cells at Z/WW (i.e., 10 cells at ZH/$t\bar{t}$), there are 20 slots where SSS can be installed (10 cells at ZH/$t\bar{t}$ multiplied by 2 SSS per cell). Over these 20 slots, moving from the Z/WW to ZH/$t\bar{t}$ phase will involve:

– 2 slots are not changing
– 2 slots are for single quadrupoles at Z/WW phase, which change polarity at ZH/$t\bar{t}$ phase.
– 2 slots are for single quadrupoles at Z/WW phase, which will change slot location at ZH/$t\bar{t}$ phase.
– 6 empty slots at Z/WW phase will receive single quadrupole at ZH/$t\bar{t}$ phase.
– 4 slots are for quadrupole - sextupole at Z/WW phase, which will be removed at ZH/$t\bar{t}$ phase.
– 4 empty slots at Z/WW phase will receive quadrupole - sextupole - sextupole at ZH/$t\bar{t}$ phase.
– 4 occupied slots at Z/WW phase will receive quadrupole - sextupole - sextupole at ZH/$t\bar{t}$ phase.

The mechanical reconfiguration of the collider between the Z/WW phase and the ZH/$t\bar{t}$ phase presents many inconveniences and challenges. The number of SSS that would need to be dismounted, moved or installed is very large, requiring a large number of activities in the tunnel, transport to and from the access points and the surface, as well as breaking the vacuum of the collider beam pipes and proceeding with full realignment of the collider. Another solution to alleviate most of these issues requires the installation of all SSS girders from the start, that is, installing the full contingent of elements required for the ZH/$t\bar{t}$ operation, while less than half of these would be used for the Z/WW phase of operation. A careful optimisation of the power converter circuits for the magnets in the arcs would allow a very fast (order of one hour) switch between the two modes of operation. This solution is being actively studied to avoid the mechanical and operational challenges linked to a collider reconfiguration.

### 3.10.3 Optimisation of the collider supporting structure - static and dynamic analyses

***Overview and principles***

Figure 3.71 shows the 3D model of an arc half-cell, with a focus on the short straight section of the collider and booster. The supporting system of the booster shown in this picture is one of the possible solutions under study, compared to the conceptual configuration reported in Section 8.2.7.

The principle for assembling, installing, aligning and maintaining the elements in the SSS on their supporting system has been extensively analysed.

Using girders to support the common elements in the SSS offers significant practical advantages. The SSS elements, magnets, and vacuum chambers can be pre-assembled and pre-aligned on a girder in a workshop at the surface with the proper tools and environment. The entire SSS module (girder with quadrupole, sextupoles, alignment fiducials and vacuum system) can then be transported as a single object to the tunnel, optimising the transport and maintenance operations. Hot spares of SSS modules can be stored and rapidly prepared for installation in case of a major fault. Repair can then be done in the correct conditions at the surface. Finally, a pre-alignment of elements on the girder at the surface reduces the time required for the positioning and final alignment of the girder in the tunnel.

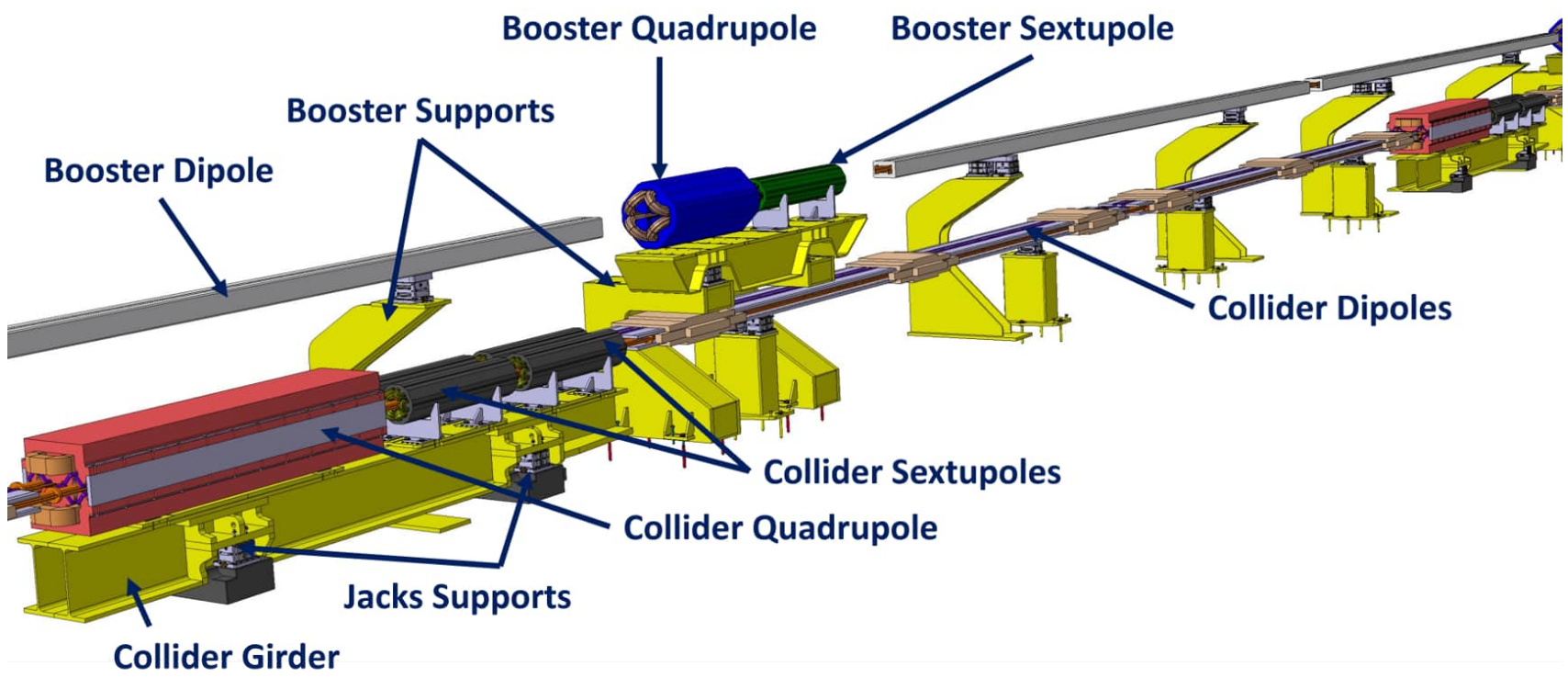

Fig. 3.71: CAD model of an arc half-cell, with focus on the Short Straight Section.

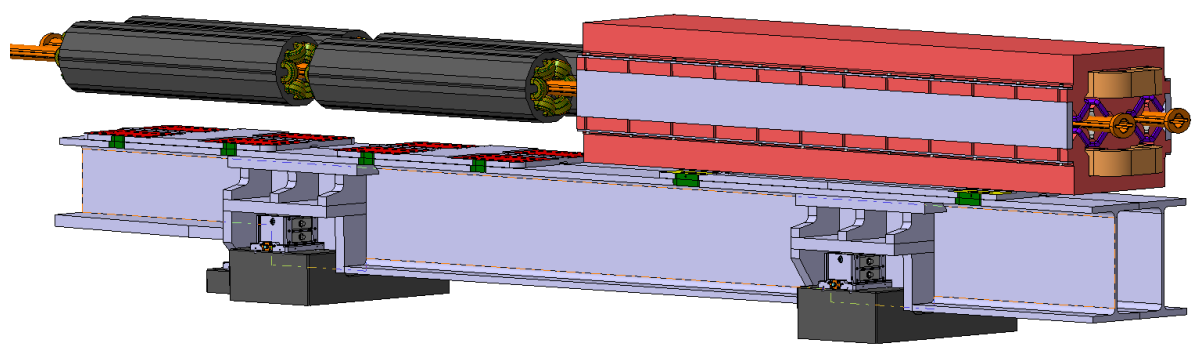

Fig. 3.72: CAD model of the optimised actual steel girder.

However, a potential disadvantage of a girder is that it usually requires more space vertically, than, for example, a supporting system with standard jacks for each magnet of the SSS. This means that the vertical position of the accelerators would be higher, with possibly a detrimental effect on the dynamic stability at the level of the beam axis. It is, therefore, important to maximise the girder stability by optimising its design and materials.

***Specifications***

An initial estimation of the acceptable vibrations in the SSS was defined in 2022, and is reported in Table 3.21. This specification did not distinguish between vertical and lateral motion and between booster and collider.

To consolidate the tolerance estimates, the sensitivity to vibrations of the GHC optics at the Z operating point was evaluated assuming a tolerance on the beam oscillation amplitude at the collision point of less than 5% relative to the collision point beam size [309]. Due to the small emittance ratio, vertical oscillations are more than an order of magnitude more critical than horizontal oscillations. Assuming that an orbit feedback will efficiently damp beam oscillations with frequencies below 1 Hz, the integrated RMS motion for frequencies higher than 1 Hz is ideally around 10 (100) nm for the vertical (horizontal) plane. If the beam orbit feedback bandwidth can be extended above 1 Hz, it would be possible to set the targets for the vertical (horizontal) plane to 20 (200) nm, in line with Table 3.21. It must be noted that the sensitivity may evolve in the future with the optics and the machine layout. For the low-beta regions, the tolerances are an order of magnitude tighter. These considerations are summarised in Table 3.22, the key frequency of interest is 1 Hz, and tolerances are given at the quadrupole (SSS) magnetic axis.

Table 3.21: Proposed dynamic stability requirements in the arcs, presented at FCC IS workshop for the arcs [310] in 2022.

| Frequency range | Tolerance | Correlation |
|---|---|---|
| 0.01 Hz < f < 1 Hz | 1 µm | 10 km |
| 0.01 Hz < f < 1 Hz | 100 nm | none |
| 1 Hz < f < 10 Hz | 20 nm | none |
| 10 Hz < f < 100 Hz | 5 nm | none |
| 100 Hz < f | 1 nm | none |

Table 3.22: Updated dynamic stability requirements in the arcs at the level of the quadrupole magnetic axis.

| | Tolerance at 1 Hz frequency |
|---|---|
| Collider vertical direction | 20 nm |
| Collider lateral direction | 200 nm |
| Booster vertical direction | 40 nm |
| Booster lateral direction | 400 nm |

***Historical background***

It is interesting to compare the current stability specifications with what was studied in other projects or achieved in past CERN machines, such as the Large Hadron Collider (LHC/HL-LHC) and the future Compact LInear Collider (CLIC). For the LHC/HL-LHC quadrupoles [311]:

– In standard operation, the root mean square (RMS) should be < 5 µm at 1 Hz;
– Beam instabilities can be provoked if the RMS is between 5 µm and 20 µm at 1 Hz;
– A beam dump is usually needed for an RMS > 20 µm at 1 Hz.

On the other hand, the future Compact LInear Collider (CLIC) has significantly more stringent specifications: given its very small beam sizes, even minor oscillations of one quadrupole reduce the luminosity. It has been estimated that in the vertical direction the RMS must be below 1 nm at 1 Hz and similarly, it must be below 5 nm at 1 Hz in the lateral direction to ensure sufficient performance [312]. A study has demonstrated that these specifications can be achieved, albeit using active stabilisation based on piezo-actuators combined with a stiff and optimised design of the quadrupole support [313].

Hence, the FCC arc specifications are closer to those of a linear accelerator (CLIC) than those of existing circular accelerators (LHC/HL-LHC), which illustrates their challenging aspects.

***Numerical methodology***

It is possible to calculate the vibrations expected for a given configuration of the SSS numerically and compare it with the specification above. A finite element method (FEM) has been defined and is described in the following steps:

1. Definition of a baseline model (support and simplified magnets);
2. Carry out static analysis of the system → to assess the structural resistance and the static stability of the support;
3. Carry out modal analysis → to assess the rigidity of the system, study the dynamic characteristics of the system;

4. Identify the transfer function of the support from the ground to the magnet axis → identify the critical modes of the supports, carry out a comparative study between different supports;
5. Conduct random vibration analysis in response to ground motion → study the impact of the support on the RMS at the magnetic centre, and obtain an estimate of this RMS;
6. Add vibrational cross-talk between the booster and the collider, add excitation forces dependent on the support environment such as pumps, water pipes, ventilation, etc.;
7. Compare the results to the specifications.

For the moment, this methodology is being applied from step 1 to step 5. The vibrational cross-talk between the booster and the collider, and how to account for it in the simulations, is currently being studied [314]. Further investigations are required to determine and include excitation forces.

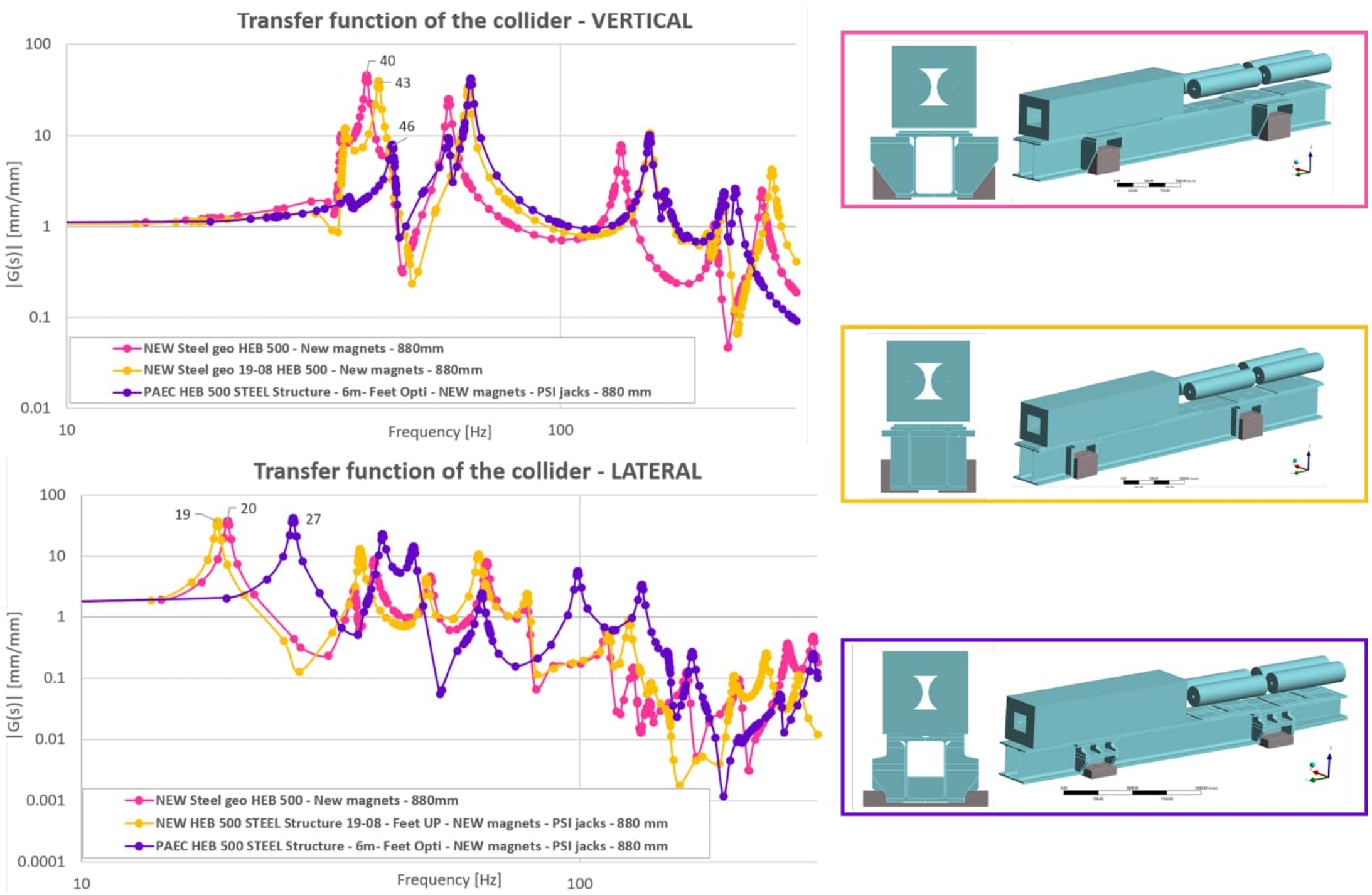

Fig. 3.73: Transfer function comparison for different girder geometries in the vertical and lateral directions.

An example of a transfer function comparison for different geometries is shown in Fig. 3.73. The first graph displays the transfer functions obtained in the vertical direction and the second in the lateral direction. Each plot represents the amplification or reduction of the input oscillation through the system over a given range of frequencies. The transfer functions provide insight into how the system behaves and reacts over a range of frequencies. The higher the natural frequencies and rigid body frequencies, the more rigid the system is considered to be. Comparative studies can be carried out to analyse the impact on stability of the geometry, materials, position, rigidity, number of feet, etc. and to determine the most suitable geometry.

Knowing the spectrum of the ground motion over the relevant range of frequencies, as well as the transfer function of the supporting structure between the ground and the centre of the magnet, it is possible to estimate the displacement of the mechanical axis of the magnet in response to the ground motion. As values of the estimated movement of the ground in the FCC tunnel are not yet available, the power spectral density (PSD) of the ground motion measured in the LHC tunnel [315] has been used instead as an initial input. When performing a random vibration analysis of the system with such a ground PSD as an input for the calculation, the PSD of the magnetic axis is obtained as an output, and the integrated root mean square (RMS) displacements at the level of the axis can then be computed. The methodology is detailed in Fig. 3.74.

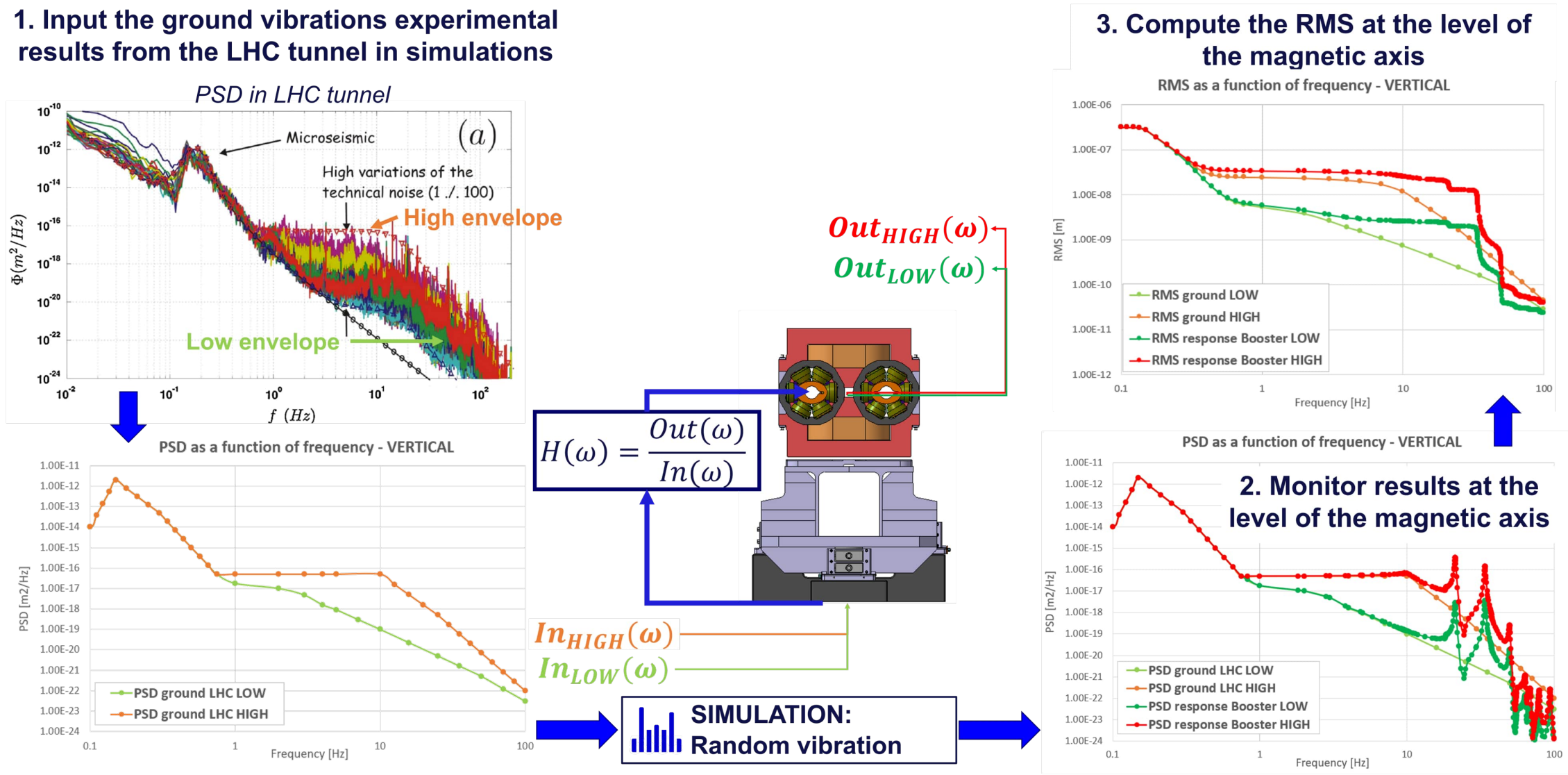

Fig. 3.74: Methodology to assess the stability of the supporting systems.

### *Experimental benchmarking*

It is important to note that the method described above is sensitive to several parameters that cannot be precisely estimated at this stage of the study; similarly, the simulations contain approximations, such as simplified magnets and interfaces (ground to support, support to magnets, etc.). It is thus of paramount importance to experimentally benchmark and tune the simulations to acquire confidence in the numerical model. A simple 2.5 m-long Short Straight Section demonstrator (see Fig. 3.75) was therefore assembled to allow an understanding of how the different elements of the SSS affect the system stability. This SSS consists of feet/jacks supporting a granite girder, on which the prototype quadrupole built by TE-MSC during the CDR phase [316] was installed.

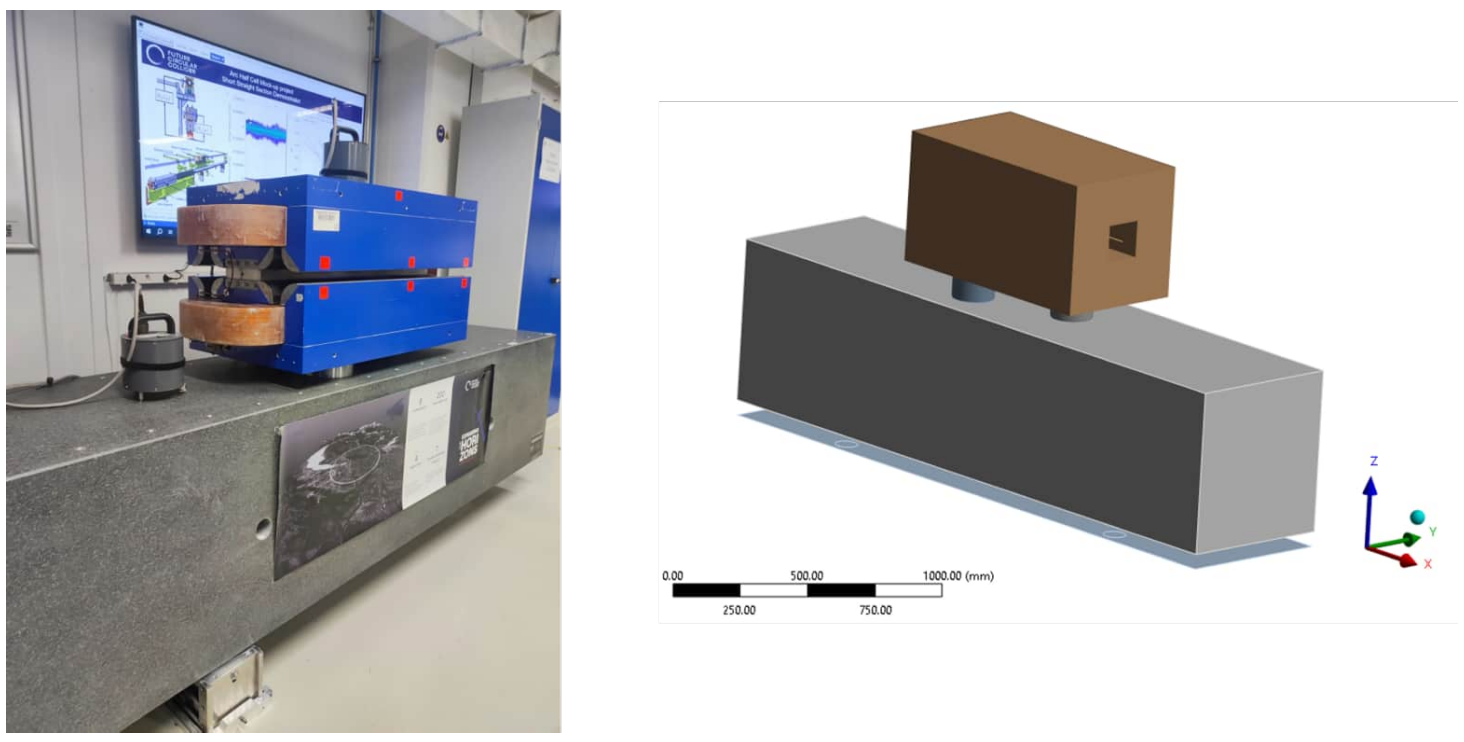

Fig. 3.75: 2.5 m long short straight section demonstrator and its equivalent in simulation.

A multistep experimental characterisation of each element and, successively, of the full assembly, in terms of modal analysis and transfer function, was then performed. These measurements were carried out at CERN in the Mechanical Measurement Laboratory of EN-MME. The experimental results are compared with the numerical analyses to gradually refine the simulations and to determine the dynamic stability of the different elements more accurately. The different steps and their key results are summarised below.

1. **Characterisation of the prototype quadrupole:** the vibrational natural frequencies of the prototype quadrupole, even after extrapolation for the 2.9 m-long quadrupole, are quite high, around 100 Hz, and typical of a very stiff system. It will thus only marginally affect the vibrational behaviour of the SSS. Particular attention should be given to the interfaces between quadrupole and girder, as these interfaces, depending on their stiffness, can be the source of rigid body modes, generally at low frequencies, which have a major impact on stability. For more details, see Ref. [311].
2. **Characterisation of the** 2.5 m **long girder:** the natural frequencies of the girder are above 400 Hz, whereas the rigid body modes are below 100 Hz and even fall below 20 Hz for the first. The results show that the frequencies associated to the girder rigid body modes are determined by the vibrational modes of the feet/jacks. The girder is therefore a rigid component, but the feet/jacks have a significant effect on its stability, which is why it is important to work on the design of the feet/jacks that will be used to support the girder. For more details, see [311].
3. **Characterisation of the** 2.5 m **long girder with the prototype quadrupole:** the third step involved characterising the 2.5 m long girder with three jacks and the 1 m long quadrupole prototype on top. This configuration was chosen because it is the closest to that which is most likely to be adopted in the FCC-ee tunnel. The key results are presented below and details can be found in Ref. [311].

To summarise, the measurements demonstrated that individual subcomponents such as the quadrupole and the girder are relatively stiff compared to the supports on which they are placed, i.e., the jacks and the interface between magnets and girder.

Table 3.23 displays the integrated RMS of the 2.5 m long girder with the prototype quadrupole, at the level of the ground, girder and quadrupole, in both vertical and lateral directions. This table also compares experimental measurements with data obtained from simulations.

Table 3.23: Experimental and simulation results for the 2.5 m long granite girder with the prototype quadrupole: Integrated Root Mean Square at 1 Hz at the level of the floor, the girder and the quadrupole in the vertical and lateral direction.

| Integrated RMS values at 1 Hz | Vertical direction EXP. | Vertical direction SIMU | Lateral direction EXP. | Lateral direction SIMU |
|---|---|---|---|---|
| On the ground | 5 nm | 5 nm | 7 nm | 7 nm |
| On the girder | 12 nm | 10 nm | 162 nm | 124 nm |
| On the quadrupole | 13 nm | 15 nm | 309 nm | 220 nm |

Direct comparison of the values presented in Table 3.23 with the specifications presented in Table 3.22 shows that with the 2.5 m long demonstrator, the specifications have been reached or even exceeded. However, the real SSS will be 6 m long instead of 2.5 m, which will have a detrimental effect on stability.

Given the consistency found between the experimental results and the data obtained via simulation, a first extrapolation was carried out: simulating a 6 m long granite girder with the simplified quadrupole and sextupoles (see Fig. 3.76). Table 3.24 presents the modal results following extrapolation. Given the increase in mass and length of the system, there is a clear decrease in all mode frequencies: the system becomes less rigid. The refinement process on the simulations is still in progress to determine the integrated RMS for the 6 m length of the short straight section.

#### ***Conclusions and highlights***

The experimental measurements facilitate the understanding of how the different elements of the SSS affect the stability of the system and indicate an order of magnitude of the displacements expected in the frequency range of interest for FCC-ee. Individual subcomponents such as the quadrupole and the girder are relatively stiff (vibrational natural frequencies are in the order of a few hundred Hz) compared to the supports on which they are placed, i.e., the jacks and interfaces girder/magnets. From the dynamic stability point of view, the latter thus requires more work, during the next project phase.

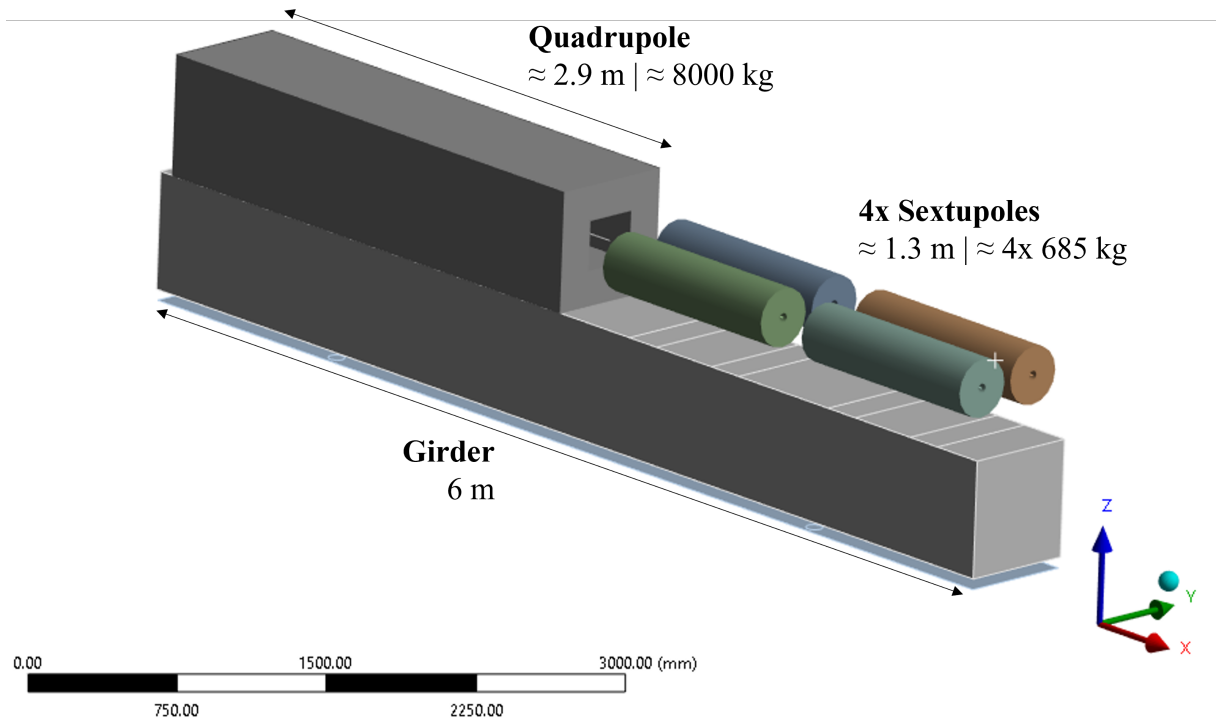

Fig. 3.76: 6 m long short straight section in simulation composed of the granite girder, the simplified quadrupole and sextupoles.

Table 3.24: Simulation results of the modal analysis of the 2.5 m long granite girder with the prototype quadrupole vs. the 6 m long granite girder with the quadrupole and sextupoles

| Modes | Frequency for 2.5 m-long config. (simu. and exp. results) | Frequency for 6 m-long config. (simu.) | Mode shape identification |
|---|---|---|---|
| S1 | 14 Hz | 6 Hz | Tilting mode around X |
| S2 | 30 Hz | 16 Hz | Tilting mode around Y |
| S3 | 34 Hz | 12 Hz | Tilting mode around Z |
| S4 | 47 Hz | 26 Hz | Rotation mode around X |
| S5 | 55 Hz | 30 Hz | Up and down mode |

The measurements of integrated RMS on the 2.5 m long demonstrator show vibrations comparable to the current specifications. However, simulations performed so far on the 6 m length of the SSS demonstrate that, with the current configuration of the systems, the specification will be exceeded. Moreover, for the moment, the study considers only the ground motion and does not account for additional external vibration sources (such as pumps, cooling systems, and ventilation) or the vibration cross-talk between the booster and collider structures through the ground. These additional sources will have a detrimental effect on the stability of the system, and they require further work.

The specifications appear challenging, and considerable work is required to optimise the system in terms of stability. This stability challenge must be taken into account by all stakeholders who will have equipment in the tunnel (pumps, ventilation, cooling, tunnel cross-section etc.) that will have a direct impact on the stability of the supporting structure.

### 3.10.4 Integration studies magnet/vacuum system

The design of key components, including magnets, the vacuum system, alignment mechanisms, and beam instrumentation, is still in the early stages for both the collider and the booster. To advance this work, a dedicated design study was initiated with the primary objective of refining the collider's integration model, with a particular focus on system interfaces. This study has proven to be crucial not only for optimising the integration layout but also for improving the optical design, as discussed in the following sections.

As part of the study, several critical aspects were analyzed, including the integration of the vacuum absorber cooling system, heating jackets, dipole-to-dipole and dipole-to-quadrupole interconnections, magnet busbar insulation, vacuum shape-memory alloy (SMA) flange tooling, and overall accessibility to interconnections. More details on this work can be found in Refs. [317–319]. The present discussion, however, focuses on one of the key outcomes of the study: the separation of the $e^+/e^-$ beams in the collider.

In the CDR [13] and, still, at the beginning of the Arc Half-Cell Mock-up Project, the optics design of the collider had a 30 cm radial spacing between the $e^+$ and the $e^-$ beams. In light of the results of the interfaces design study, this distance has been increased to 35 cm. The main reasons are the following:

1. Integration of vacuum synchrotron radiation absorber cooling circuit, heating jackets and dipole busbar insulation.
   (a) 300 mm $e^+/e^-$ separation results in jacket/busbar interference
   (b) 350 mm $e^+/e^-$ separation allows >25 mm radial clearance between the two systems, a part of which will host the connection fittings of the cooling water supply tubes of the SR absorbers
2. Dipole to dipole interconnection (see Fig. 3.77).
   (a) 300 mm $e^+/e^-$ separation results in a clash between vacuum chamber flanges and busbar insulation
   (b) 350 mm $e^+/e^-$ separation guarantees 24.5 mm clearance between vacuum chamber flanges and insulation, as well as a continuous dipole busbar interconnections (no need of jumpers)
3. Vacuum chamber interconnection (see Fig. 3.78).
   (a) 350 mm $e^+/e^-$ separation provides more space for SMA flange tooling and general access to interconnections
   (b) ∼127 mm between flange-to-flange insulation (assuming ∼10 mm insulation)
   (c) Enough space for BPM positioning (3× vacuum chamber outer diameter from SMA flange connection.)

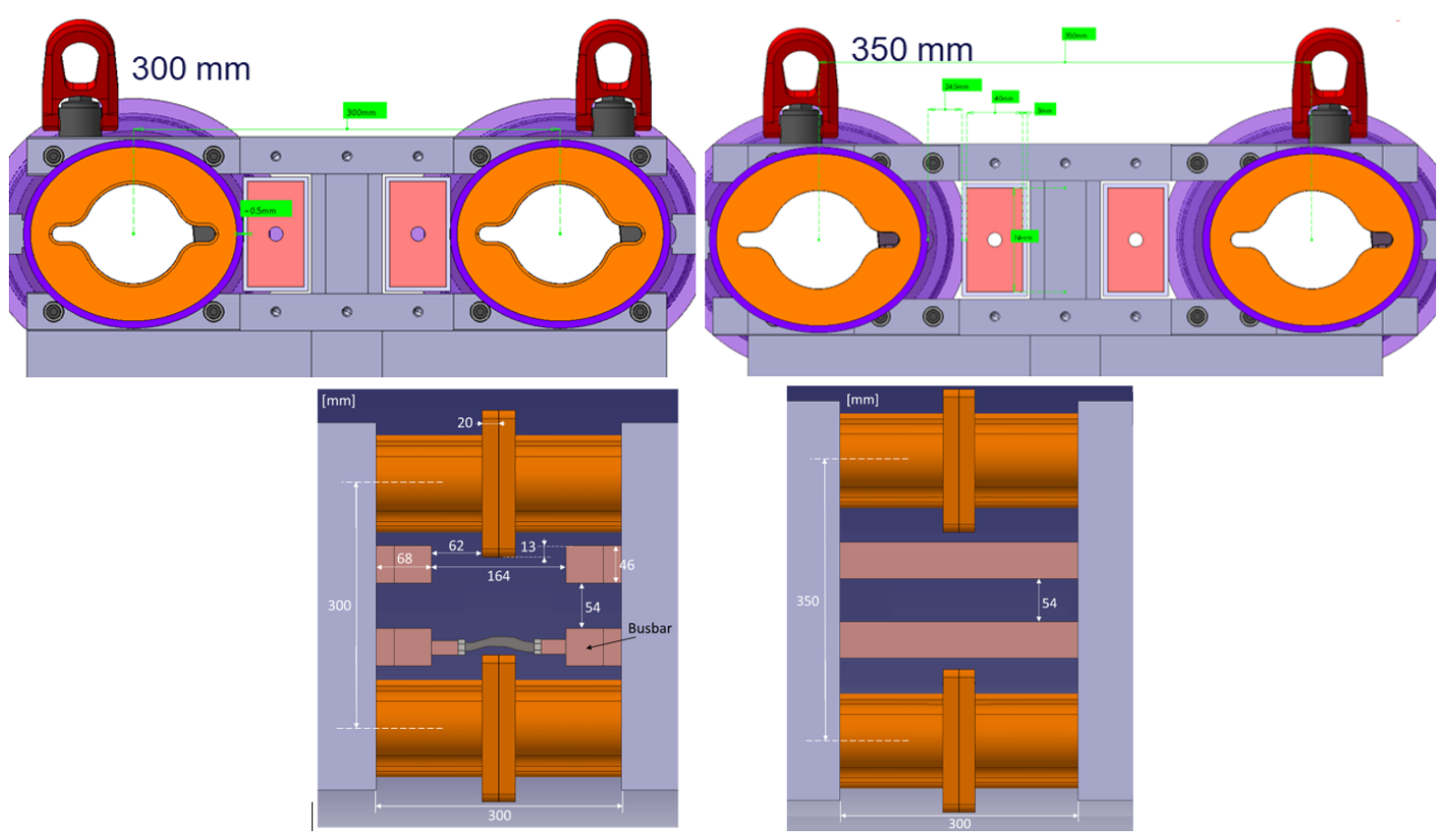

Fig. 3.77: Relative position of vacuum flanges and busbar insulation. Top and bottom Left: 300 mm $e^+/e^-$ separation. Top and bottom Right: 350 mm $e^+/e^-$ separation.

### 3.10.5 Arc half-cell mock-up: concept verification

A mock-up of the arc half-cell is being built at CERN, on the Meyrin site and specifically in building 355/358. Among its objectives, a distinction can be made between short-term (2025) and long-term ones:

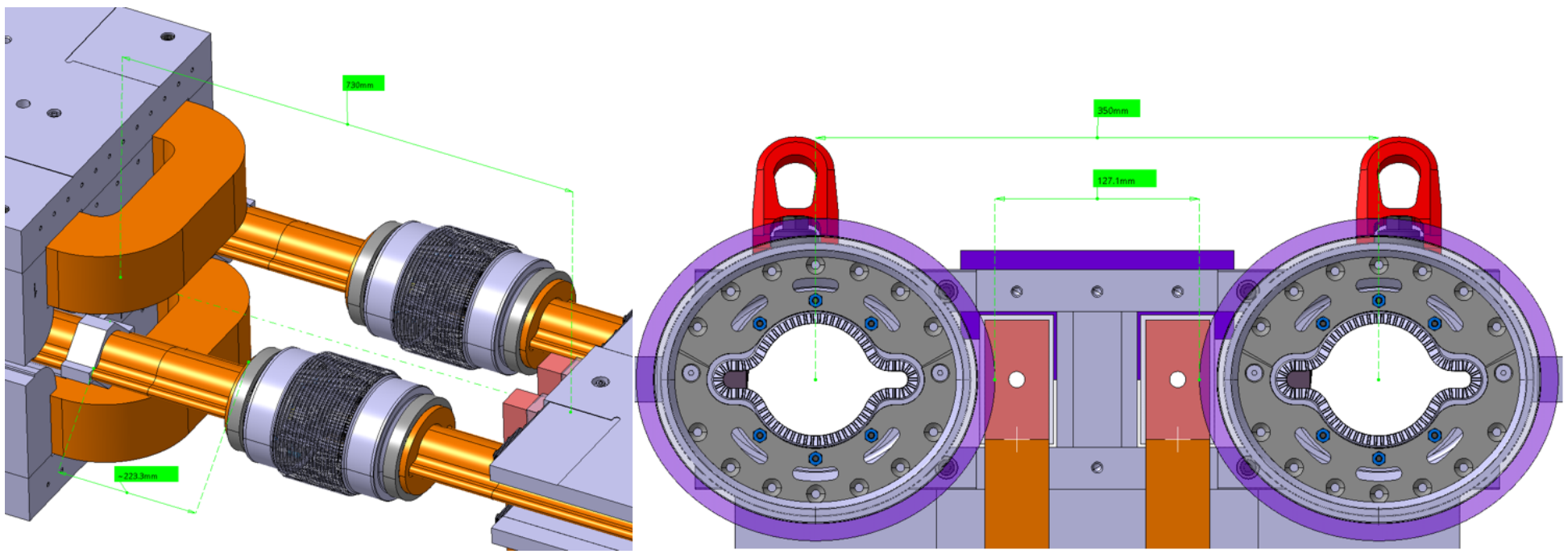

Fig. 3.78: 350 mm $e^+/e^-$ separation, vacuum interconnections and SMA flanges.

- **Short term objective:** a mock-up allowing the testing of the integration of simplified elements within a short time-frame (detailed study of the integration of elements, analysis of access and compatibility with safety requirements, test of alignment strategy and mechanical stability, etc.). This mock-up will also be used for outreach, it is a 'visual' demonstrator for the stakeholders of the FCC.
- **Long term objective:** an evolving mock-up allowing equipment groups to install and test their equipment. The mock-up could, in the future, house the full-size/weight functional elements.

The mock-up will be 1:1 scale, meaning that the 5.5 m of the tunnel, and the half-cell length at high machine energy ( 30 m), will be accurately reproduced. The only area that will not be reproduced in the mock-up is the trench under the ground level.

As shown in Fig. 3.79, the mock-up structure will be made of steel, featuring an arch every 3 metres, reinforced by transverse beams. Wooden or aluminium panels will be attached to the structure and painted to replicate the appearance of concrete accurately. The main advantage of the steel structure is that it can be easily dismounted and remounted if needed, as well as updated and adapted to the evolution of the machine arc region.

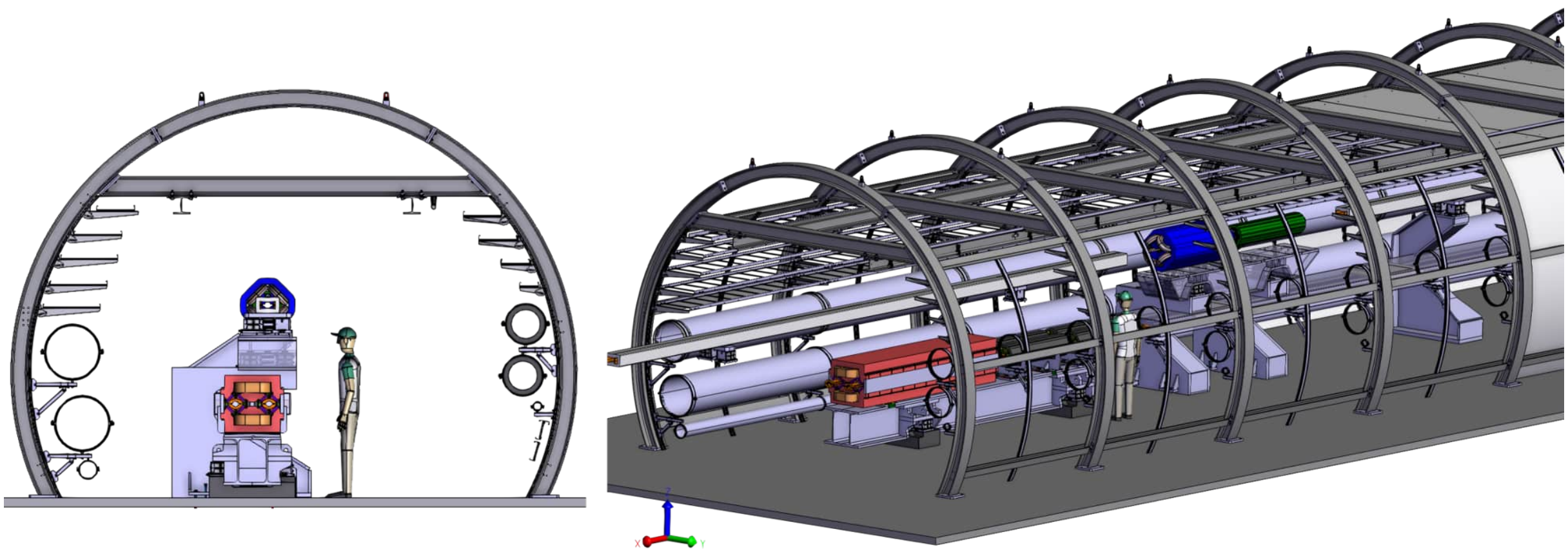

Fig. 3.79: CAD of the mock-up structure in development.

For the 2025 installation, the initial mock-up will consist of a combination of real and dummy elements, with the flexibility to evolve over time. To facilitate integration testing and optimisation, the magnets will be constructed from wood, while actual support structures, including jacks, girders, and pillars, will be installed. The arc envelope is designed to support the full weight of all real components from the outset, ensuring that no structural modifications will be required when transitioning from dummy to real elements.

A prototype of the remote maintenance and inspection system (RMIS) robot will be installed to evaluate various operational scenarios, including routine maintenance, emergency procedures, and leak detection. Additionally, a prototype of the fire door and its partition will be deployed to assess its positioning, integration, installation process, and interaction with the RMIS. Following these initial installations, all essential services—including cooling and ventilation systems, electrical infrastructure, and IT networks—will be implemented.

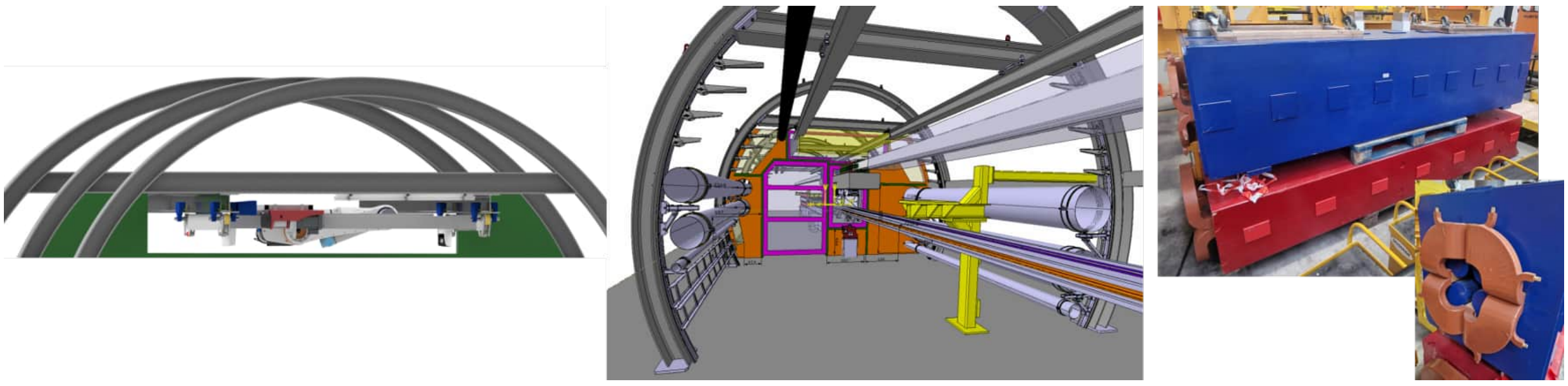

Fig. 3.80: CAD of the integration of the RMIS, CAD of the integration of the fire door and partition and services, example of wooden magnets.

### 3.10.6 Next steps

As mentioned above, a significant amount of work still needs to be done before reaching a satisfactory solution for the integration and stability of the arc half-cell elements. In particular, the following main activities have been identified:

- Install the first version of the mock-up, and test integration aspects.
- Build an arc configuration compatible with the installation/uninstallation/handling procedures defined by the Handling Engineering team (see Section 8.6).
- Develop and commission the mixed reality system in the mock-up region.
- Continue the study and optimisation of the supports (jacks, girders, supporting structures) from the stability and cost point of view, also via prototypes and dedicated experimental measurements in the mock-up (e.g., ground-induced vibrations on the 6 m SSS, random-vibrations effects).
- Perform thermomechanical simulations to evaluate, for example, the thermal deformations induced by the tunnel temperature variation on the elements of the girder, also validating the proposed air flow cooling system.
- Update and upgrade the mock-up main systems. In particular, the initial dummy structures (wooden magnets, vacuum chambers) were replaced with the first prototypes produced by the equipment groups while also adapting the mock-up to the changes of the arc region.

## 3.11 Machine protection hard- and software systems

This section describes the systems, hardware and software, closely related to the machine protection of the accelerator. They are the systems which presently fall under the responsibility of the TE-MPE group.

### 3.11.1 Interlock systems

Interlock systems are essential for the safe operation of the FCC-ee. The following interlock systems have been identified for the FCC-ee collider and booster.

- Warm magnet interlock controller (WIC). The system protects the warm magnets and the synchrotron radiation absorbers from over-heating because of missing (water) cooling and/or running at too high a load (magnet current or synchrotron radiation load). In case of a power converter fault, it can also act as an interface between the power converters and the beam interlock system. The system is connected to thermal switches, flow meters of the water-cooling system, the power converters and the beam interlock system.

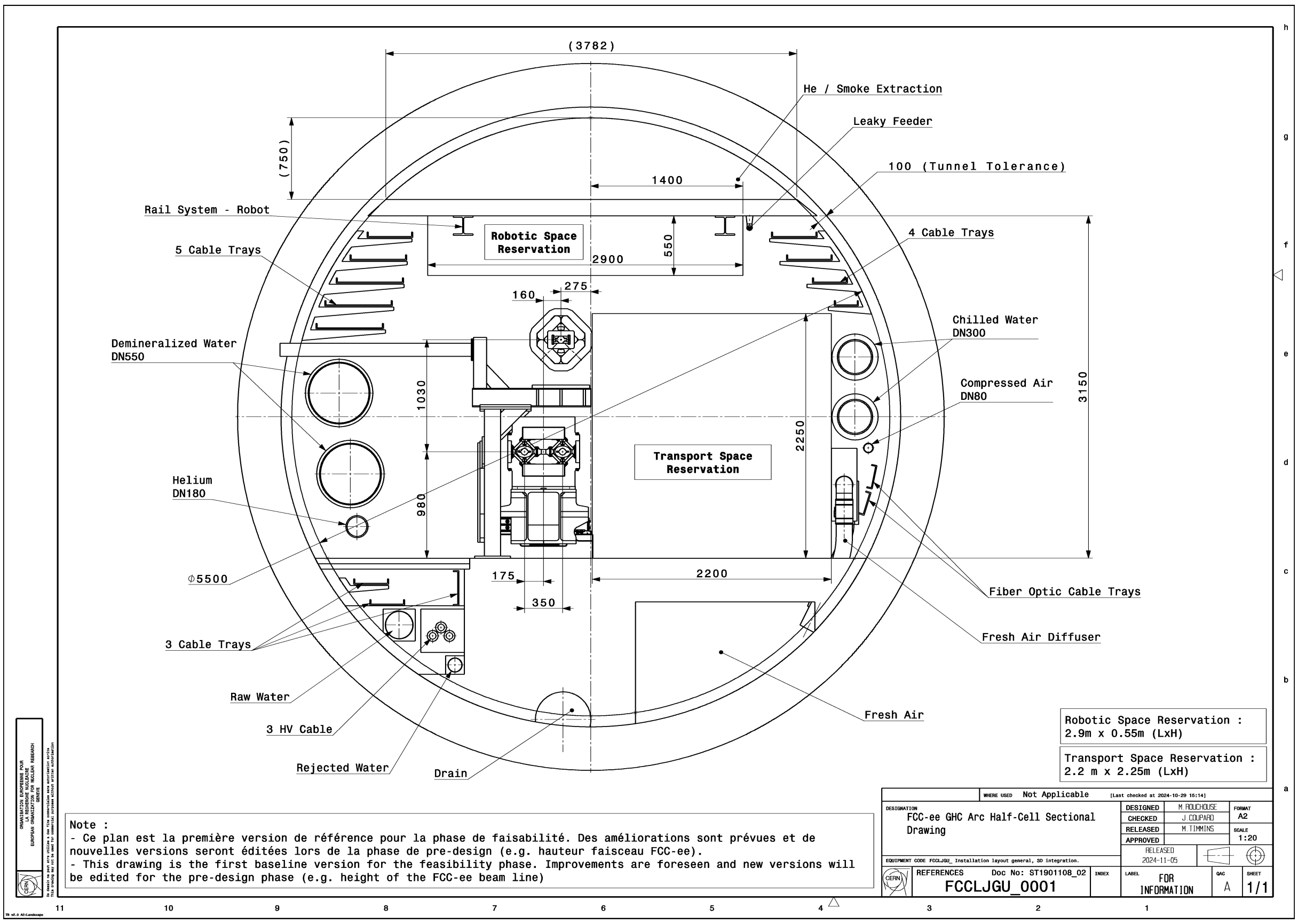

Fig. 3.81: FCC-ee GHC Arc Half-Cell sectional drawing - EDMS 3180552 [320]

A diagnostic system that identifies which magnet or synchrotron radiation absorber segment has an over-temperature needs to be developed. The system should not require individual cables to be installed to these magnets; such a system does not exist at the moment. In the existing machines, individual cables are installed, but this would be too expensive for a machine as large as the FCC-ee.

– Beam Interlock System (BIS). This is the core of the machine protection system and connects the many Users (Power converters, RF, Beam Loss Monitors, Interlock Systems etc.) to the beam dumping system. There are dedicated BIS for the injection and extraction systems, which can inhibit injection or extraction.

  The BIS is based on customised electronics and needs to be designed and developed based on the failure cases and reaction times derived by detailed machine protection studies for the main ring, the full energy booster-ring and the transfer lines. The current LHC system does not account for any especially fast interlock channels. Potential designs of extra fast interlock channels have to be studied and the electronics required designed and developed in case the machine protection studies show that reaction times in the microsecond range are required.

– Safe Machine Parameters (SMP). The SMP is closely linked to the BIS. It provides flags to the different BIS to allow masking of certain interlock channels when operating with low beam intensities and/or energies (needed for setting up the accelerator). It also provides beam parameters to many different systems like the beam loss monitors and the experiments, all in a highly reliable

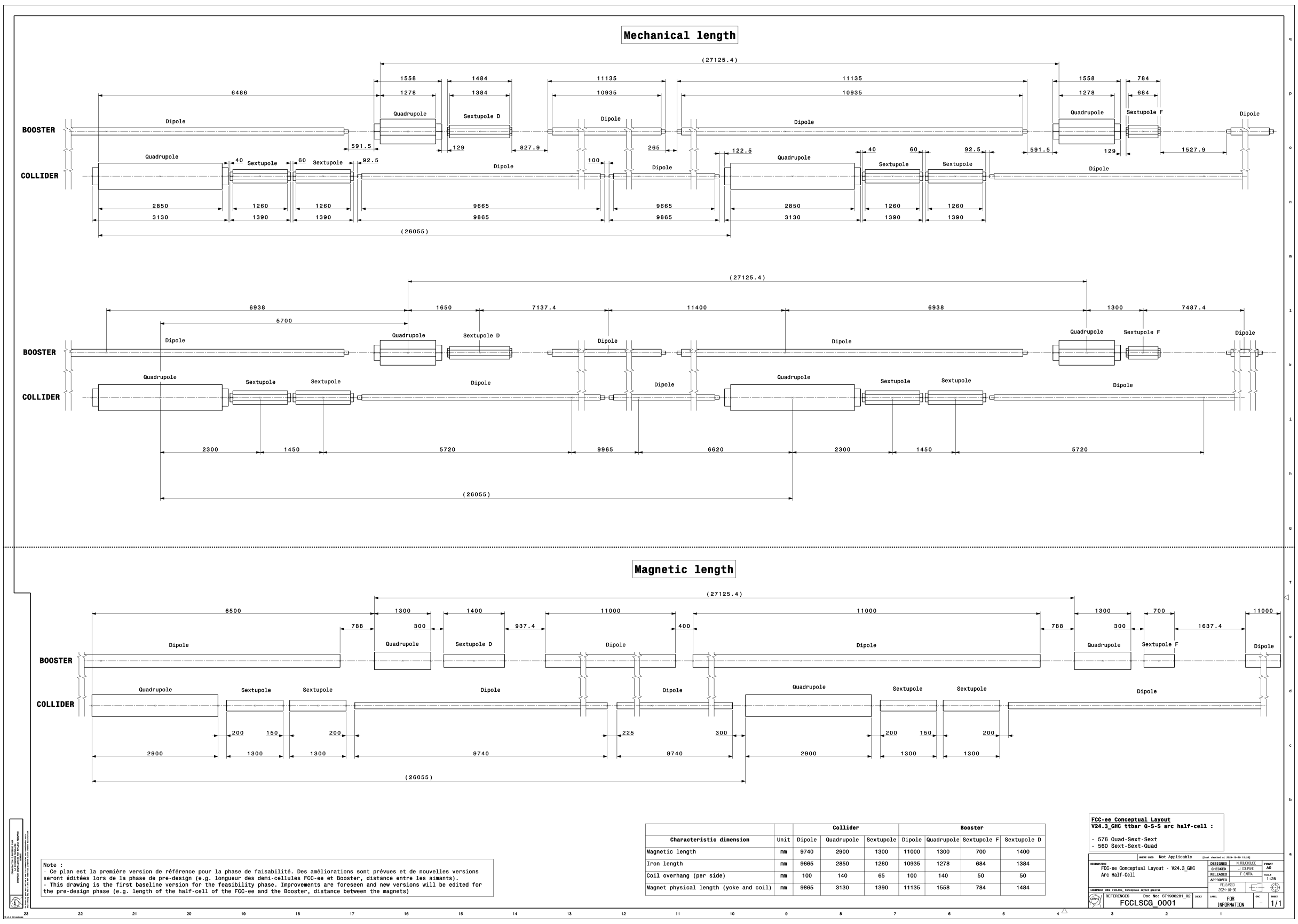

Fig. 3.82: FCC-ee Conceptual layout Arc Half-Cell - V24.3_GHC Q-S-S configuration - EDMS 3180559 [321]

environment.

– Fast Magnet Current Change Monitor (FMCM). For some normal conducting magnet systems the reaction time from the power converters to request a beam request in case of powering failures can be too long. For those systems a dedicated system can be installed. The FMCM has a very fast reaction time to request a beam dump in case of fast magnet current changes (in general a power converter trip).
– Powering Interlock Controller (PIC). This system interfaces the protection of the super conducting magnets (Quench Detection System) to the Beam Interlock System and requests a beam dump in case of any problems related to the superconducting magnets. For this reason it also interfaces to the many systems to which the superconducting magnets are connected: power converters, cryogenics, UPS etc.

### 3.11.2 Protection systems for superconducting magnets

The FCC-ee will be dominated by normal conducting magnets and circuits. However, superconducting circuits of the FCC-ee (e.g., the circuits of the final focusing magnets at the IPs) will require dedicated protection systems to ensure their safe operation.

– Quench detection and data acquisition system (QDS). The QDS is made of high-precision, fast, dedicated electronic boards for the detection of quenches in superconducting magnets, busbars, links and HTS current leads. It interfaces all superconducting magnet and circuit protection systems (energy extraction system, local protection units, coupling loss induced quench systems), the powering interlock controllers and the beam interlock system. If a quench is detected, the QDS will send high-resolution data to the post-mortem system, which is crucial to validate the correct behaviour of the circuit and the protection system before releasing them for re-powering.

- Energy extraction systems (EE): These are clusters of racks with high-current switches, controls and extraction resistors. If there is a power failure or quench of a superconducting magnet, the EE systems are activated and they ensure the timely and safe extraction of the energy stored in the circuit to the extraction resistor. The EE systems interface the QDS and the PIC and send high-resolution data to the post-mortem system.
- Local protection units. These are quench heater power supplies, to protect superconducting magnets in case of a quench. They consist mainly of an energy storage system based on capacitor banks, which discharge their energy into quench heater strips installed on or in the coils of the magnets if there is a quench. They interface the QDS and the instrumentation feed boxes (IFS).
- Coupling loss induced quench system (CLIQ). These are special double racks size systems, to protect superconducting circuits in case of a quench or powering failure. They consist mainly of an energy storage system based on capacitor banks, which discharge their energy into the magnet's inductance, creating a high di/dt in the circuit. They interface the QDS and the Instrumentation Feed Boxes (IFS).
- Cold diodes for circuit protection. These are radiation hard by-pass diodes required to divert the circuit current around the quenching magnet. They are specially designed together with industry and need to be qualified for the integrated radiation dose expected, and the 1 MeV equivalent neutron fluence levels.
- Warm diodes for circuit protection. These are warm by-pass diodes to ensure a discharge path of the oscillating currents induced by the activation of the CLIQ systems around the power converter of a circuit. They are usually procured from industry.
- Electrical quality assurance (ElQA). This is the dedicated high precision, high voltage (mobile) hard- and software for the ElQA testing of all superconducting circuit elements during production, reception and commissioning.
- Proximity equipment (current lead heaters). These consist of regulators and power transformers with support structures and controls racks. They are required to heat the normal conducting end of the HTS current leads to avoid the creation of condensation and ice. They are usually custom designs based on commercial components.
- Instrumentation feed boxes (IFS). These boxes interface cables for discharge and instrumentation from the cold mass of superconducting magnets and links. They consist of dedicated PCBs for interfacing the QDS, cryogenic and protection equipment (CLIQ, Local protection units) and are installed on top of the instrumentation flanges of the cryo-assemblies.

### 3.11.3 Software systems

Software systems play an important role for interlocking, efficient and safe validation and safe operation of the FCC-ee.

- Software Interlock System (SIS). The SIS completes the hardware interlock system. It generally reads the equipment status over software and is connected to the BIS to request a beam dump when needed. It is less reliable and significantly slower in reaction than a full hardware system however it is more flexible. This system is presently developed mainly by the BE-CSS group. However, as it is important and related to machine protection, while very likely not listed in any other category, it has been added to the TE-MPE list of items.
- Post mortem systems. The safety of the accelerators depends on the correct functioning of the complete chain of the machine protection system and the in-depth understanding of the reason for beam dump requests and powering failures. The post-mortem system receives and reliably stores high-resolution data from all relevant systems and performs automatic analyses of these data. It interfaces with the SIS and can block beam operation if the analysis criteria are not met.

- Accelerator testing (AccTesting) & analysis tools for commissioning and operation. These are software systems running on top of the CERN IT infrastructure for scheduling, automatically performing and evaluating hardware commissioning tests of the accelerator systems. They are essential for efficient and safe hardware commissioning and for validating the conformity and correct operation of the various hardware systems, many of which are related to the different electrical circuits and to machine protection.
- Fault tracking and machine availability. These are software tools for the simulation and the tracking of system and accelerator faults. They are very important during the design and operation of the accelerators and their different hardware systems, especially those related to the safety of the accelerator. The machine availability needs to be studied and tracked to optimise performance and determine where investments are best placed to improve performance.

#### 3.11.4 Conclusion on machine protection hard- and software systems

New techniques and concepts will need to be developed for the WIC system to identify the faults' location, without having to pull individual cables to the sensors. In case required, special extra fast channels of the BIS will need new technologies. No specific new developments have been identified for all other systems described in this chapter.

### 3.12 An alternative arc magnet design

FCC is also pursuing an alternative design for the magnets in the arc short straight sections, which could bring the following advantages compared to the baseline design:

1. Lower power consumption and lower weight.
2. More flexibility in optics design.
3. An improved filling factor.
4. State-of-the-art technology with increased societal impact.

The lower power consumption is achieved by replacing the main quadrupoles and sextupoles of the arcs with superconducting ones. In the baseline scheme, about 80 MW of electrical power is consumed at top energy by the collider main magnets, the majority of which goes to power the quadrupoles and sextupoles. An estimated further 14 MW is consumed by the cooling and ventilation systems for these magnets. By employing superconducting quadrupoles and sextupoles, one can dispense with ohmic losses at the expense of cooling power, which is estimated to be only a fraction of the ohmic load.

Replacing the normal conducting sextupoles and quadrupoles with magnets based on high-temperature superconductors (HTS) would result in the following gains:

- Magnets can be nested, increasing the dipole filling factor;
- Magnets for the electron and positron beams are independently powered, giving greater flexibility;
- Power consumption for the relevant systems is significantly reduced;
- The use of novel materials and techniques (HTS conductors) increases the relevance of FCC to society and its sustainability credentials.

The idea has led to the approval of two projects, named *FCCee-HTS4* and *FCCee-CPES*, that have been financed through the CHART programme, a Swiss mission-oriented research network focused on technology development for the FCC.

HTS4 aims to investigate the nested magnet idea using HTS conductors. The end goal of the project, a metre-class prototype, is supported by subscale sextupole demonstrators manufactured at CERN and PSI. The first two demonstrators investigate the options of using a wax-impregnated canted-cosine-theta (CCT) coils, based on insulated HTS tape, and a partial-insulation based cosine-theta (CT) configuration. The prototype will consist of a nested quadrupole-sextupole configuration.

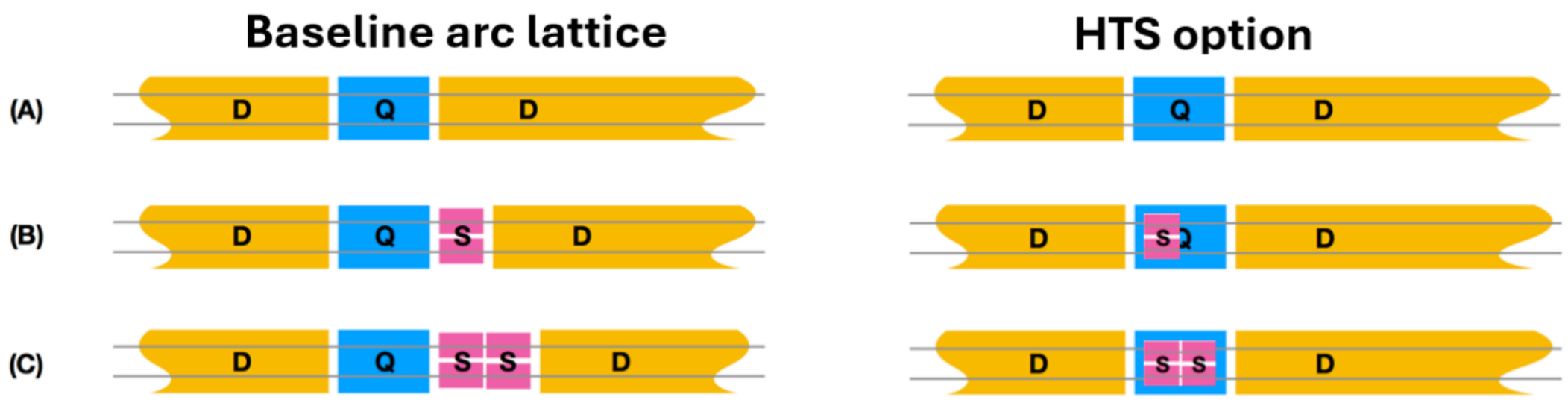

Fig. 3.83: The three different types of short straight sections of the baseline design (left). On the right, the single type using nested HTS magnets.

The minimum length of the short straight section (SSS), 3.5 m, is dictated by the minimum length of the quadrupoles, below which synchrotron radiation issues arise. This corresponds to a total magnetic length of 3 m. To make the magnets as compact as possible, the sextupole can be nested in the same axial space as the quadrupole.

Two cooling strategies are being considered. Both involve a dry, cryogen-free magnet. The first consists of a cryogenic distribution line connected to several cryoplants located on the surface. The distribution line services the HTS modules via Neon-based heat exchangers. This arrangement allows simple installation and replacement of magnets.

A second option is to give each magnet its individual cooling system in the form of multiple redundant cryocoolers. By utilising high-reliability single-stage coldheads, it is expected that such a configuration can achieve high overall system reliability and availability.

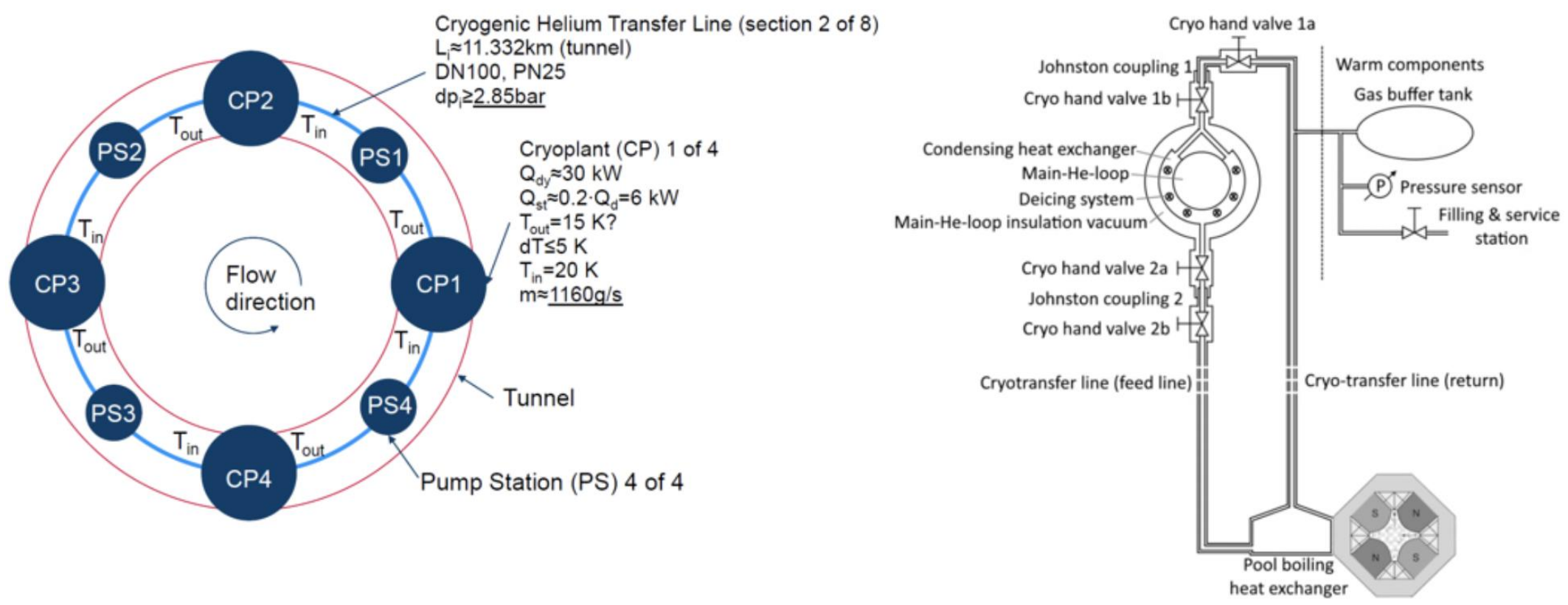

Fig. 3.84: Concept of cooling method based on cryogenic distribution line (left), and heat exchangers between the line and the HTS SSS's (right). Sketches courtesy of J. Bessler, E. Rosenthal, Forschungszentrum Jülich.

The optimum operating temperature of an HTS-based short straight section (SSS) is found by balancing the operational costs (dominated by electricity use for cooling) with capital costs (dominated by HTS conductor). For the cryocooler-based cooling option, 40 K seems to be a sweet spot for a wide range of energy and conductor prices. It is expected that results from a conceptual design study of the distributed cooling line would yield an overall lower total cost-of-ownership option, with the optimum at a lower temperature.

The sextupole demonstrators (Fig.3.85) are designed to generate a gradient of 1000 T/m$^2$ (the FCC design calls for a gradient of 820 T/m$^2$) at a current of 250 A.

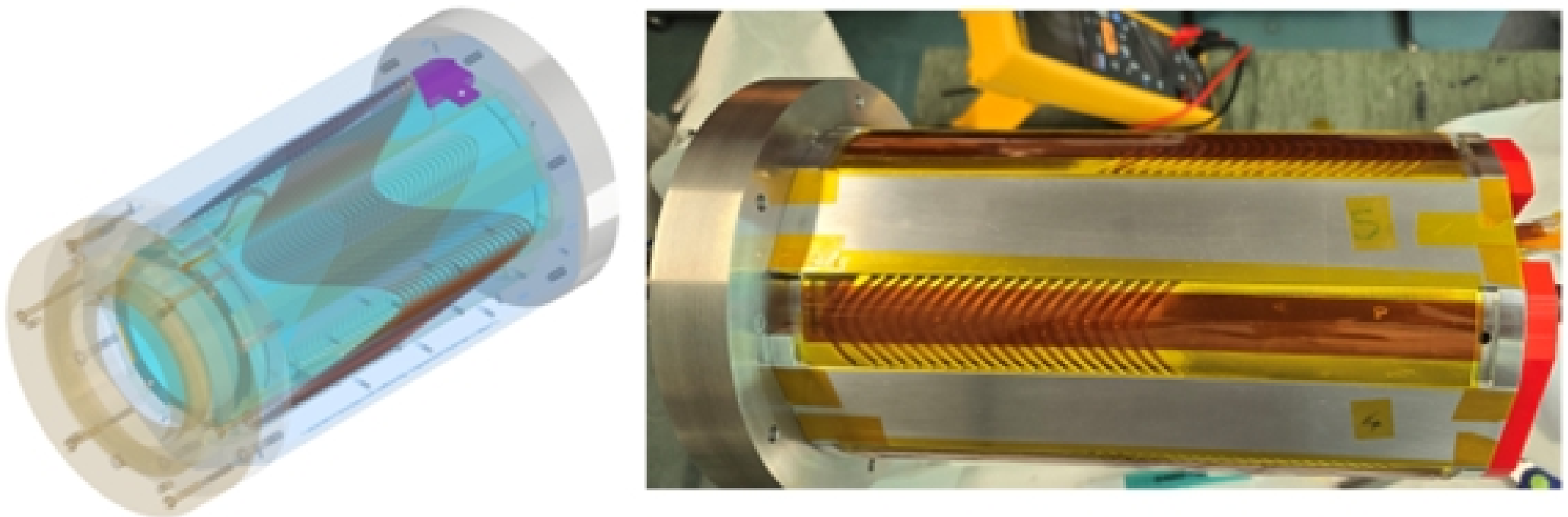

Fig. 3.85: The HTS4 sextupole demonstrator: left: CAD design; right: the magnet after winding, before the assembly of its aluminium sleeve.

The improvement expected in the dipole filling factor over the baseline solution is 7%. This would translate to 7% higher luminosity at the same beam power or one year of running less in the FCC-ee 14-year programme for the same physics output. This also translates to 7% less RF voltage needed at top running, and thus the number of 800 MHz cavities required is reduced.

The nesting of quadrupoles and sextupoles paves the way for nesting a dipole component as well see, for instance, Ref. [322], increasing the filling factor to 120% of the baseline value. Such an inclusion in the HTS SSS is possible.

HTS4 is also pursuing technologies of partially insulated HTS tapes. Partially insulated coils have distinct advantages in robustness and effective current density, as they reduce the amount of stabilising copper required in case of a quench. Developing such a technology is important for the future of HTS accelerator magnet technology and a coating station (Fig. 3.86) has been built at PSI to coat HTS tapes with a partially insulated layer.

The CPES project at ETHZ designed a highly efficient power supply suitable for powering accelerator magnets while operating inside the cryostat at cryogenic temperatures. In this way, the large heat load associated with the high-current conduction-cooled leads required to transfer the current from room temperature to cryogenic conductions can be avoided. This is a key component in making cryogen-free magnets energy efficient.

CPES built a demonstrator (Fig. 3.87) with five full-bridge phase modules and up to 100 A output. Test results of this module at 100 A are extrapolated to a heat load of 4.4 W at 250 A. In comparison, a pair of high-current conduction-cooled current leads would have 23 W of losses in the cryostat. For cryogenic power supplies to be viable in an accelerator environment, additional studies on component- and system-wide reliability and radiation hardness are required.

For the cryo-cooled approach, equipment needs to be placed in the collider tunnel (cryocooler and associated electronics) or indeed inside the SSS cryostat (power supply). Therefore a radiation study has been performed to ensure that equipment will continue functioning during the whole lifetime of the FCC. This study, that uses somewhat more shielding around the SR photon stoppers, shows that doses of 1 kGy or less are possible for the equipment concerned (to be validated in the next phase).

The possibility of using HTS superconducting magnets operated at or below 40 K and replacing the baseline approach of iron-based non-superconducting magnets seems promising but also involves some risks (mainly having to do with costs and the fact that this is a radically different approach) that need to be understood. The HTS4 prototype will be tested in early 2026, at which point the technological readiness level as well as integration into the FCC-ee tunnel and timeline will be reviewed.

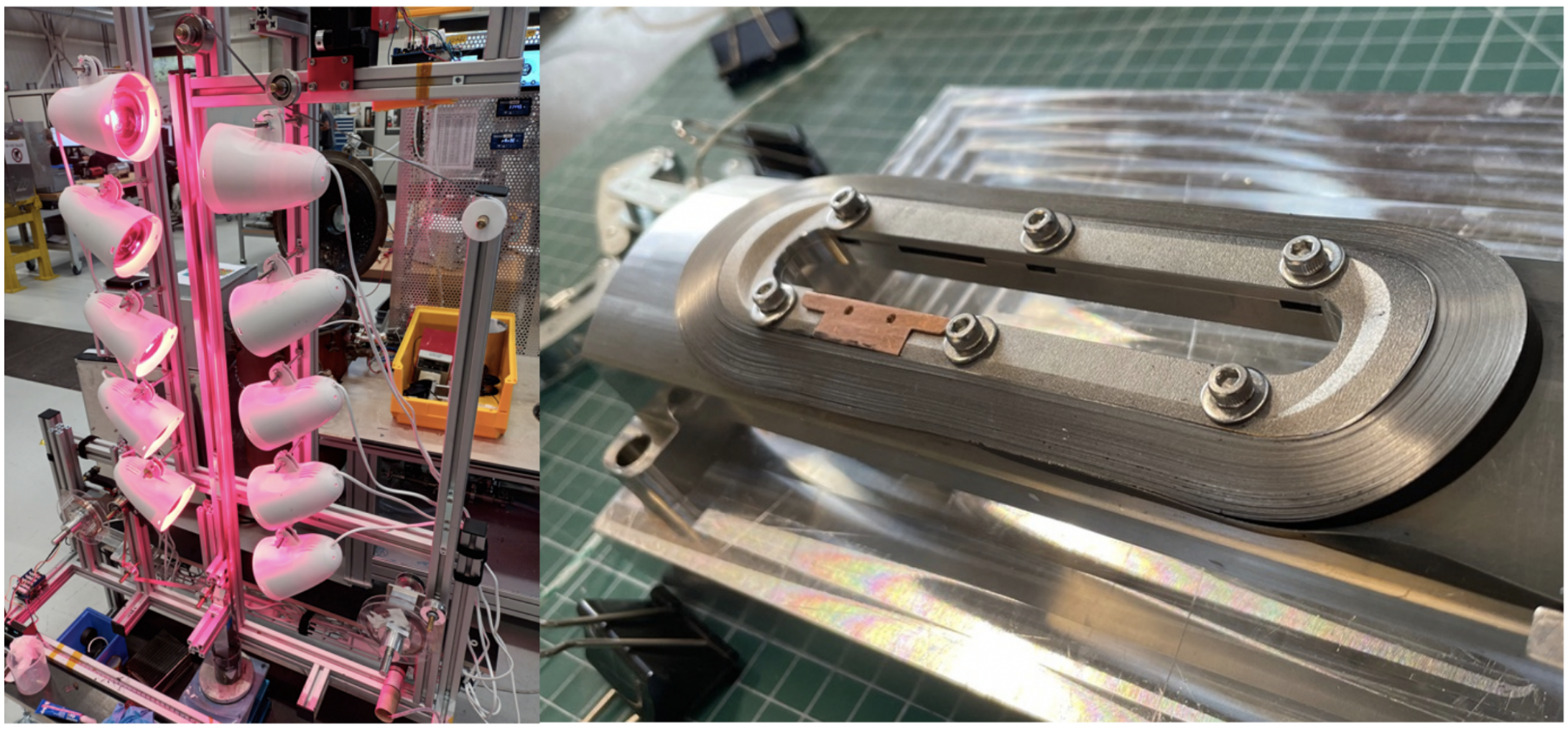

Fig. 3.86: Left: The coating station, with the uncoated tape-spool on the left and coated tape spool on the right. The coating occurs near the bottom of the setup, after which the tape is heated to be dried before re-spooling. Right: one CT sextupole coil wound using 4 mm coated dummy tape.

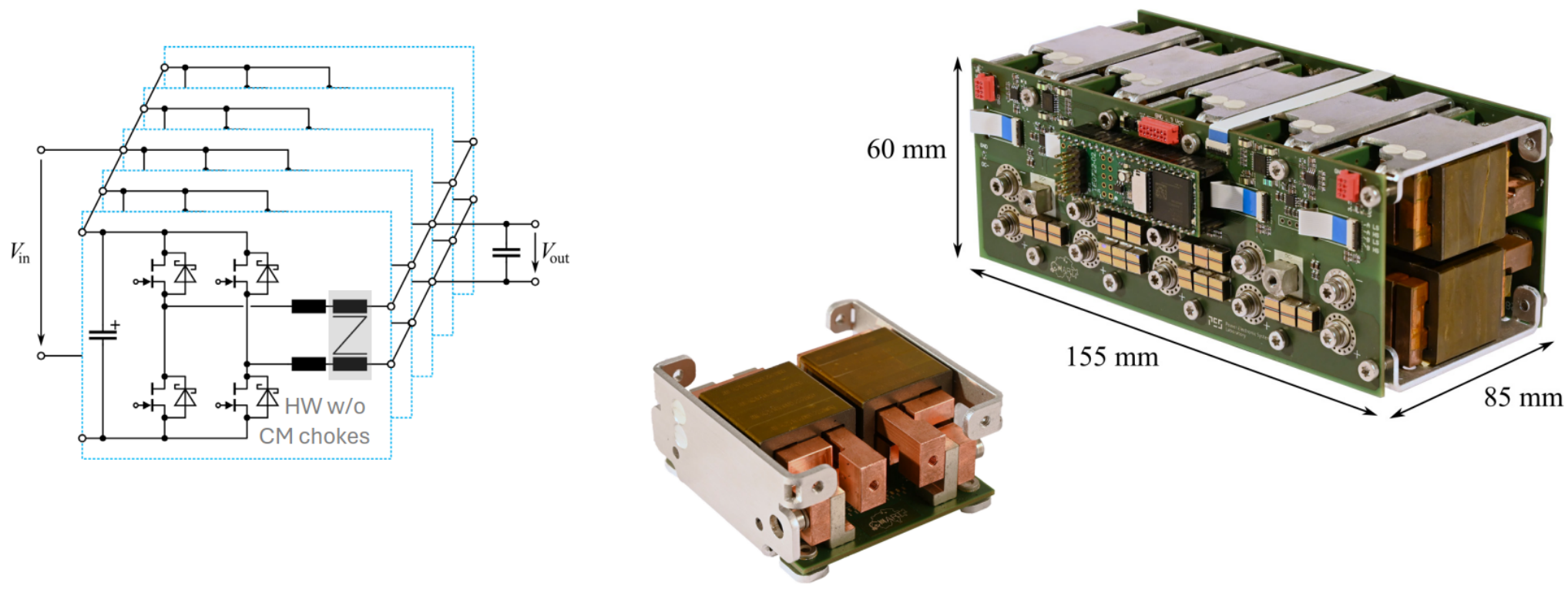

Fig. 3.87: The CPES demonstrator comprises five full-bridge phase modules using Gallium Nitride (GaN) transistors.

### 3.13 Dismantling FCC-ee

In order to make way for the FCC-hh, the FCC-ee machine and infrastructure will have to be removed. This section presents a first look at the strategy and process for dismantling FCC-ee and identifies the main cost drivers as well as logistics and planning constraints. Given that the details of the machine and its components are still being refined, it is clear that the considerations here can only be approximate but can be refined as the project progresses. The dismantling study drew heavily on the experience of dismantling LEP [323].

### 3.13.1 Basic strategy

***Objectives***

The objective is to dismantle the FCC-ee machine and its specific infrastructure in order to leave the underground areas in a state ready for the installation of the FCC-hh. Since the regular arcs account for 80% of the FCC circumference and 90% of the mass, their clearing will dominate the overall dismantling process and drive the logistic constraints.

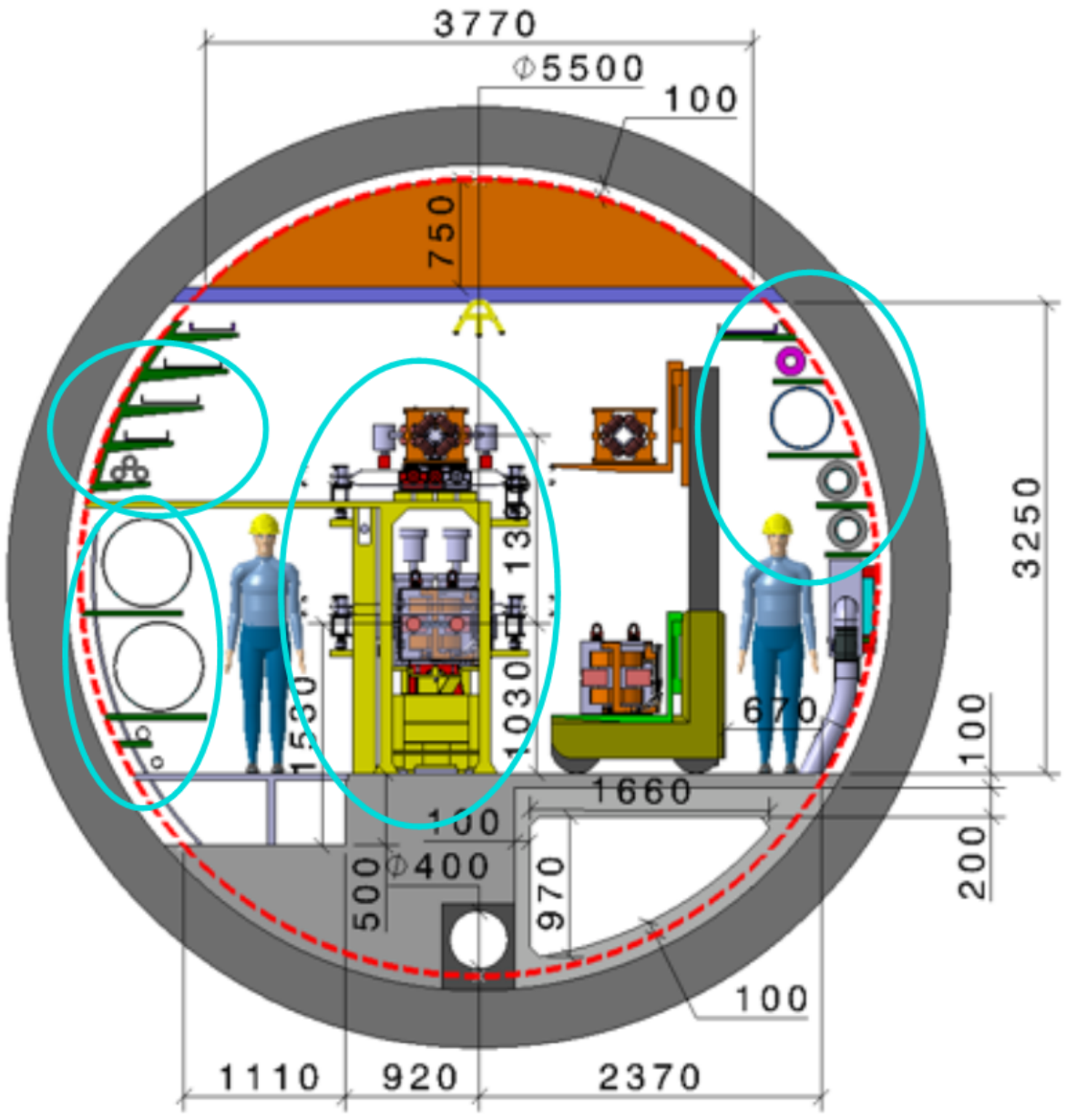

Fig. 3.88: Cross-section of the current FCC-ee tunnel integration in the regular arcs. The main areas to be dismantled are circled in blue.

Figure 3.88 shows the cross-section in the regular arc of the FCC-ee. The areas circled in light blue represent the principal areas that will be dismantled in the machine tunnel. The equipment consists of the collider and booster rings along with their metallic support and alignment structures, control and DC powering cables and piping for a variety of cooling circuits. It is assumed that the main ventilation system as well as safety systems, AC power cables and fibre optic cables will not require dismantling. In addition to the main machine tunnel there are alcoves, located at about every $\sim$1.5 km along the circumference of the arc, housing local equipment, like low voltage distribution and repeaters. Much of this equipment will not be dismantled. However, the alcoves will also house the power converters for some of the FCC-ee magnets. The removal of this equipment, along with the associated DC cabling represents a significant workload and logistic challenge.

***Assumptions for the study***

A number of assumptions were made for the study and these are listed below (in no particular order) along with a brief explanation of the reason for the assumption and its impact.

- FCC-ee operation will end with $\sim$5 years of physics at the top energy (360 GeV). This is the most constraining scenario from a radiological point of view as operation at this energy will produce the highest levels of induced activity in some machine elements. It is assumed that the arc magnets and vacuum chambers of the collider and booster, as well as the photon stoppers will become activated. However, the materials chosen for these components will be selected so that the levels of activity are minimised, and it is expected that the activity will decay to below release levels within a few years. In the straight sections, parts of the RF modules and individual components (such as beam dumps and collimators) will also become activated. The levels will be determined from FLUKA simulations as well as measurements in-situ (see Section 3.13.3) during and after operation. The remainder of the machine elements will be treated as conventional.

- To determine the timescale of dismantling, it is assumed that the work will be done during two 8-hour shifts per day, working five days a week. Such a schedule assumes that equipment maintenance can be carried out overnight and during weekends.
- The machine shafts at all 8 points will be used to extract the equipment (although not necessarily at the same time). The dismantling of the experiments will be via the experiment cavern shaft(s) and kept separate from the machine dismantling.
- As the bottleneck for dismantling is likely to be in the underground areas, the logistics of the disposal of the equipment once at the surface are considered in less detail here. The main consideration given to the surface sites is to make some estimates of the needs in terms of temporary storage.

***Equipment to preserve***

Most of the equipment removed from the machine will be treated as waste. After radiological checks, it will be disposed of via the appropriate pathway. However, some of the equipment is of high value and has potential for reuse. The major system in this class is the RF, with a total installed voltage of over 23 GV, and a construction cost well in excess of 1 BCHF. Table 3.25 summarises the RF system installed for the $t\bar{t}$ run.

Table 3.25: RF system installed for $t\bar{t}$ running (180 GeV/beam).

| **Frequency** [MHz] | **Machine** | **Number of cryomodules** | **Total Volume** [m$^3$] |
|---|---|---|---|
| 400 | Collider | 70 | 2519 |
| 800 | Collider | 100 | 4320 |
| 800 | Booster | 124 | 5357 |
| | **Sum** | 294 | 13196 |

Preserving the cryomodules in a way that would allow their re-use in another facility implies storing them in appropriate conditions, with the cavities themselves either under vacuum or filled with an inert gas. Assuming that cryomodules can be stacked on top of each other to a maximum of 3 modules, the storage space needed would amount to a floor surface area of 4400 m$^2$ (not including access space). In addition to the cryomodules, there is the RF power system, consisting of klystrons, circulators, waveguides and power supplies. With 2 klystrons per module a total of 140, 400 MHz klystrons and 288, 800 MHz klystrons would need to be stored.

For this study, it is assumed that all RF related equipment apart from the waveguides is kept and stored, although in reality a much more detailed study will be required to determine which parts of the system to preserve and which to discard. Besides the RF system, there may be other pieces of equipment to be kept, but these are likely to be smaller, individual items (e.g., specific beam instruments) which are all installed in the long straight sections. None of the machine elements in the arcs will be preserved.

It is also assumed that the radioactivity of the arc components (magnets and vacuum vessels) will take a few years to decay, and they will require interim storage for this period.

### 3.13.2 Regulatory framework

The regulatory framework concerning the elimination of radioactive waste is currently defined by the tripartite agreement between CERN and the Host States as explained in Ref. [324]. The agreement allows CERN to eliminate radioactive waste in the two Host States by using the most technically and economically advantageous pathways in both countries. This principle allows CERN to optimise its radioactive waste elimination by choosing the most appropriate solution corresponding to the type of waste.

Concerning the elimination of radioactive waste produced by CERN experiments, the principle is that the collaborating institute remains the owner of the equipment they have provided (even if radioactive) and that they take it back unless otherwise agreed.

The cost estimate for the disposal of radioactive waste from CERN's facilities is based on an inventory indicating the amount and radiotoxicity of present and future radioactive waste, i.e. waste stored at CERN and waste that will be produced by preventive and corrective maintenance or by the upgrade of CERN's facilities or experiments. The estimate of future waste will not include an estimate of waste produced in case of the decommissioning of CERN's facilities until the decommissioning has been approved.

### 3.13.3 Radiological estimations and zoning

The radiological zoning is a key driving factor for the logistics of dismantling the machine. The aim of the process is to identify all components and structures which are likely to become radioactive. This will be based on knowledge of the beam's behaviour and its interaction with matter and the history of operation. The classification of a zone determines the precautions and procedures which have to be applied. The synchrotron radiation emitted by the high-energy leptons will strike the localised absorbers in the vacuum chamber and will induce radioactivity. Other areas expected to become radioactive are beam scrapers, collimators, shields, beamstrahlung, main beam dumps and, to some extent, the magnets themselves. The vacuum vessels and magnets and, in particular, the absorbers, which are placed every 4-5 m around most of the machine circumference, obviously present the largest volume of radioactive material. Simulations show that the level of activity in the activated components will be very weak (TFA) and appropriate precautions will be taken in the dismantling process. Current studies indicate that the magnets and vacuum vessels of both booster and collider will not become activated until the $\mathrm{t\bar{t}}$ run. During dismantling, the most active components will be removed first unless they are in an area which can be closed off until the rest of the machine has been removed. This could apply, for example, to the beam dump where the tunnel containing the dump transfer line and radioactive dump block, could be sealed off.

Predictions of the levels of activity are made by simulations using tools such as FLUKA and these will form the basis of the initial zoning. Unexpected beam losses may lead to additional localised radioactivity, and this information will be incorporated in the overall picture. The operational zoning for dismantling will be confirmed by radiation measurements carried out after the definitive stop of the machine and before any dismantling is started.

### 3.13.4 Dismantling process

It is assumed that the dismantling can be done from all 8 points. However, it is unlikely that all 8 points will be used at the same time. The activities will be split into a series of 'trains' that will move progressively through the arcs. Each train will consist of the personnel, equipment, and transport vehicles needed to remove a specific set of equipment or carry out specific actions. Each train will have to complete its work before the next train starts. However, some optimisation may be possible. The order of the activities is listed below:

– Radiological measurements and zoning – measurements of radioactive activity. Final verification

of the zoning of the arcs.
- Where essential systems and services (fire protection, communications, safety systems etc.) will be interrupted or removed, compensatory measures will have to be installed before the start of dismantling activities.
- Cryogenic systems will have to have the helium evacuated and be warmed up before the start of dismantling equipment in the affected areas.
- Cooling circuits will have to be emptied.
- Electrical safety: all elements to be removed have to be made electrically safe with an electrical lockout, followed by a definitive separation from the supply.
- Radiation survey measurements of the equipment and its surroundings (checking for anomalies and contamination risks).
- All of the equipment removed from the machine will undergo a triage measurement to determine its further destination: release, detailed free release measurement or further storage as a radioactive item. There will have to be space and dedicated measurement equipment for this free release measurement.
- Removal of radioactive components like collimators, which will be more highly activated.
- Removal of booster magnets and vacuum chambers as activated materials.
- Dismantling of the metallic structures supporting the booster as the last part of booster machine element removal.
- Removal of the collider vacuum chambers and magnetic elements together with the photon stoppers.
- Dismantling of the remaining machine support structures (can probably be done in parallel with the removal of the collider).
- Removal of straight section elements and equipment.
- Removal of unused piping.
- De-cabling.
- Final cleaning and floor/walls repair/repainting as necessary.

***Critical logistics paths for dismantling***

As already mentioned, this analysis is based on dismantling two half-arcs from each point. Since the main components to remove will be the arc magnets, this defines the scale and sets the limits of the dismantling process.

*Tunnel transport of arc components*

The average distance travelled will be 3 km each way from the loading location to the extraction point. Based on a vehicle speed of 10 kph loaded and 20 kph empty, the average travelling time for a round trip will be $\sim$30 minutes. Clearly, journeys to the beginning of the arc will be short, and those to mid-arc will be longer, but the average time/distance was used in the calculations. Dismantling will start on one side with the closest magnets and on the other at the 6 km mid-arc extremity. To be added to the travelling time are the loading/unloading times: 45 minutes for loading (135 minutes for the 3-dipole trains) and around 30 minutes for unloading.

Taking into account that there will be two 3-dipole trains for every three SSS trains, each side of the arc would be capable of delivering 6 or 7 loads to be lifted within a double 8-hour shift. Therefore, it is assumed that only one transport vehicle will be needed on each side in the tunnel to deliver enough magnets to saturate the crane capacity (see below).

*Arc equipment dismantling*

As well as the transport team, each train will have a team at the worksite to cut or dismantle the equipment and prepare it for transport. It is assumed that enough time will be available between each transport to prepare the next load.

*Lifts and cranes*

The bottleneck in the extraction process is the crane lifting time. The depth of the shafts varies from 180 m (PD) to 400 m (PF) giving an average shaft depth of 240 m. For heavier loads, assuming a lifting speed of around 8 m/minute [325], a single hoist would take around 35 minutes to complete. Allowing 30 minutes for loading and 30 minutes for unloading, an average single round trip to the surface (240 m) will take ~1.8 hours. This sets an upper limit on the number of hoists during a double 8-hour shift in the range 6 to 10, according to the depth of the shaft.

### 3.13.5 Material quantities

***Arcs***

Table 3.26: Contents of a half-cell in the regular arc

| Description | Quantity | Length [m] | Approx. Weight [kg] |
|---|---|---|---|
| Booster dipoles (2 sections) | 4 | 5.55 | 2500 |
| Collider main dipoles (2 sections) | 4 | 5.55 | 5000 |
| Booster short straight section | 1 | 2 | 4200 |
| Collider short straight section | 1 | 6.3 | ~11 000 |
| Metallic support structures | 1 | | 10 000 |

Details of the quantities concerned have been addressed in a separate document [326], and only the main conclusions are presented here. These numbers have been derived based on the current state of the design, which will certainly evolve but should not affect the conclusions by more than ±20%. Table 3.26 summarises the contents of the regular half cells; the weight of booster magnets has been assumed to be the same as those of the collider ring. For the support structures, an overall weight of 10 tonnes per half-cell is assumed.

Excluding the busbars, total material weight is 118.35 ktonnes for the arcs and 121.25 ktonnes for the whole machine if the straight sections are included. If one assumes that copper cored cables are used, there will be around 2.4 ktonnes for the whole machine (0.8 ktonnes, if aluminium is used). Based on cabling of the sextupole families using 120 $mm^2$ section copper-cored cables, their weight will be 1.6 ktonnes for the whole main ring. A similar value for the booster sextupole powering can be assumed and in both cases the weight will be reduced by a factor 3 if aluminium is used. Each half-cell will also contain other equipment, which is cabled to the nearest alcove (orbit correctors, beam position monitors, beam loss monitors, interlock cables and miscellaneous signals). It has been estimated that there will be ~31 000 km of these cables. It is assumed that the other cables for such things as AC power, fibre optics and the access system will not be removed as they will serve FCC-hh.

The collider and booster dipole magnets can be transported and lifted stacked in groups of 3 on the transport trailer. The collider QSS girders are the heaviest elements and will be evacuated one at a time. There will be 368 lifts for each of the quadrupole girders (QSS, QS and Q) and booster SSS and 245 lifts for the collider dipoles and a similar number for the booster dipoles (see Table 3.27).

Table 3.27: Hoist inventory for one extraction point, assuming dipoles can be transported three at a time, working for a minimum duration of 28, 5-day weeks with 2 shifts per day.

| **Element** | **Quantity** | **No. Hoists** | **No. per week per crane** |
|---|---|---|---|
| Booster dipoles | 736 | 245 | 11 (×3) |
| Booster SSS | 368 | 368 | 16 |
| Collider dipoles | 736 | 245 | 11 (×3) |
| Collider SSS | 368 | 368 | 16 |
| **Total** | | **1226** | |

### *LSS*

The long straight sections of the collider and booster contain a further 520 quadrupoles and 200 sextupoles, and they have a total weight ∼2900 tonnes.

### *RF*

There will be 170 cryomodules installed in the collider at PH and 124 in the booster at PL. These will be powered by the klystrons in the gallery above the tunnel and linked to the modules by the waveguides.

### *Cryogenics*

The cryogenics equipment is located in the klystron galleries, the long straight sections, the service caverns, shafts and surface buildings. The equipment to be removed the comprises control equipment, distribution lines, cold boxes and storage vessels. Sensitive elements like instrumentation and valves will be removed first. The cryogenic equipment associated with the machine-detector-interface, MDI, has not been specified at the time of writing and its dismantling has not been considered. However, it is not expected to have a significant impact on the overall timescale or cost.

It is assumed that both cryoplants at PH will be dismantled, although one may be kept for FCC-hh.

### *Alcoves*

Detailed estimates of the equipment in the alcoves which has to be removed are given in Ref. [326]. Table 3.28 presents an overview of this equipment and the associated cabling.

Table 3.28: Summary of electrical equipment to be dismantled from the various alcoves.

| | | **Cables** | | **Converters** | |
|---|---|---|---|---|---|
| | **Number of alcoves** | **Number per alcove** | **Length per alcove** [m] | **Number of converters** | **Number of racks** |
| Big alcoves | 16 | 460 | 274 666 | 228 | 273 |
| Small alcoves | 40 | 880 | 429 148 | 440 | 238 |
| **Grand Total** | **56** | **42 560** | **21 560 578** | **21 248** | **13 888** |

### 3.13.6 Timescales

***Arcs***

Given the average 1.8 hours needed for a round trip of the crane hook, the time required to perform the 1226 hoists corresponds to about 28 weeks working two shifts per day and five days per week.

***LSS***

It is assumed that the machine elements in the LSS between the IP and the arc can be dismantled in $\sim$4 weeks.

***RF***

The rate of removal of RF cryomodules is limited by the time it takes to carefully dismantle and transport them, and it is expected to take around the same time as the installation. The removal of klystrons and waveguides should be $\sim$50% faster than installation and can be done in the shadow of the cryomodule dismantling. The rate of dismantling cryomodules is expected to be around 16 stations per month, working two shifts, 5 days per week. Working in PH and PL in parallel, the time envelope determined by the dismantling of the 170 cryomodules in PH will be $<$12 months.

***Cryogenics***

The sensitive elements in the service caverns and tunnel will be removed first and the cold boxes will be purged and sealed before removal. Once connecting pipes have been removed, the boxes can be lifted and removed. At the surface, as many elements as possible will be carefully removed for potential reuse in FCC-hh. Dismantling at the surface can only start once all cryogens have been stored and underground elements have been disconnected.

It has been shown in Ref. [324] that if PH and PL are done in parallel, the cryogenics system can be dismantled in $<$1 year allowing around 6 weeks for the initial preparatory stages.

***Overall timescale***

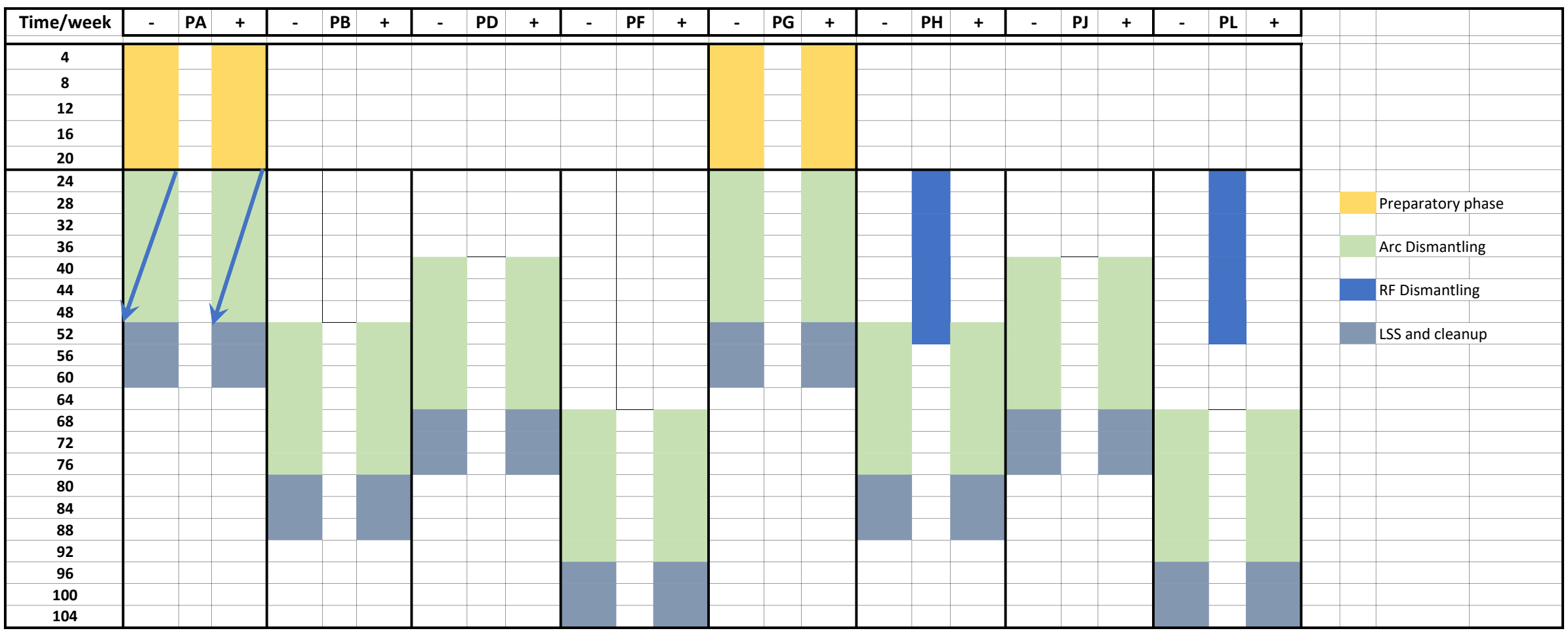

Fig. 3.89: Schematic representation of a possible timeline of machine dismantling based on 2 shifts working, 5 days per week. The arrows on PA indicate the direction of progress through the arc. The preparatory phase is only shown for the first two arcs.

The overall duration is governed by the rate at which the main arc components can be removed from the tunnel. In the following, it is assumed that the overhead crane will be working continuously 5 days per week and 16 hours per day.

To be added to 28 weeks for arc equipment removal are the times needed for initial activities such as radiological zoning, electrical safety, implementation of compensatory systems for safety and services, cryogenics warmup etc. as well as the dismantling of the straight sections and the de-cabling and de-piping activities which follow. It has been assumed that the LSS dismantling and cleanup will require around 3 months after the arc has been cleared. Careful overall planning should allow much of the work to be performed in the shadow of the overall dismantling of the machine elements in other areas. Smaller loads can be brought to the surface using the lifts, which have a 3 t capacity.

Based on 5 days, two shifts working, and these assumptions, a possible timeline for dismantling is shown in Fig. 3.89. The direction of the trains' progression is indicated on PA, showing that one train starts from mid-arc and the other at the IP end. The preparatory phase is only indicated for PA and PG as it will be in parallel with other activities for the other points. The total time required for this scenario is $\sim$2 years. Adding two shifts and working at weekends would reduce it to a total of about 1.4 years.

### 3.13.7 Surface storage and logistics

Once the equipment reaches the surface, it will have to be transported from the shaft head to a temporary storage where it can be re-checked for radioactivity before being prepared for onward transport. At this stage, traceability data will also need to be generated for each load. The preparations may include further dismantling and sorting of the material before loading into (new) specific transport containers. Cranes will be needed for the new storage facilities.

After $\mathrm{t\bar{t}}$ running, the arc magnets and vacuum chambers will remain radioactive for a period of up to a few years. They should be stored on-site for the decay period before entering the disposal pathway. The space requirement for this would amount to $\sim$20 000 $\mathrm{m^2}$ if they are piled four high. This interim storage could be in the form of tent-like structures.

A tent-type structure which has no foundations but which could be heated to keep it above freezing should also suffice for temporary surface storage (triage, etc.).

To set the scale on the size of the covered storage and preparation area, it can be assumed that it should be capable of holding the material which comes out of the machine in 1 week, and there should be sufficient space around each element to allow any dismantling/conditioning to take place. Based on the numbers given in Table 3.27 and that dipoles will require a footprint of 11 m$\times$3 m (to allow access), a covered area of 1500 $\mathrm{m^2}$ would be required for the 5-working day, two-shift scenario. To be added to this will be space for loading and unloading the equipment. The installation of a temporary industrial tent/hanger of an appropriate size at each surface point will, therefore, be required. It should be noted that during dismantling, equipment for FCC-hh will be arriving at CERN for assembly and testing, which will put an additional strain on the space available on CERN. Site security is an important issue, given the value of the materials being temporarily stored at surface sites, as well as the sensitive nature of low-level radioactive elements. Additional security will, therefore, be necessary.

For equipment requiring long term storage in controlled conditions, a more substantial building will be required. This type of building typically has large access doors, thermal insulation and a small capacity crane. These lightweight buildings are limited to a span of around 20 m so it may be necessary to construct a number of them.

### 3.13.8 Dismantling budget items

As much as possible of the work will be done by CERN staff and their regular support contractors so that they are familiar with the equipment being dismantled. However, they will not be able to do all the work and therefore additional contractors will be required. These contracts will also cover things such as operating the traceability system, removing cables and pipework, and ensuring safety supervision.

Details of the costs for the various labour components have been given in the report of the FCC-ee dismantling study [324].

Estimates of the cost of infrastructure changes and additional equipment, summarised in Table 3.29, have also been included in Ref. [324].

Table 3.29: Summary of equipment and infrastructure changes required for dismantling

| **Description** |
|---|
| New temporary storage facilities ($8\times1500\,\mathrm{m}^2$) |
| New storage facilities for radioactive decay ($20\,000\,\mathrm{m}^2$) |
| New long term storage facilities ($5000\,\mathrm{m}^2$) |
| Transport and handling materials (incl. cranes, vehicles, fuel etc.) |
| Tooling for dismantling |
| Transport containers |
| Equipment maintenance/consolidation (vehicles, cranes…) |
| Equipment for traceability, safety, signage etc. |
| Infrastructure modifications (access/safety systems, surface areas etc. ) |

### 3.13.9 Pathways for FCC-ee components after removal

As mentioned above, high-value equipment that can be reused will be stored in storage. Materials that are not classified as radioactive (i.e., conventional waste) can be sold for recycling, and the remainder will be treated as waste. Institutes in member states will be offered the possibility of receiving equipment which is not needed by CERN and that they can use at their facilities. Radioactive waste will be disposed, using the Host State authorised facilities. The cost of this will be minimised by the appropriate choice of materials for components which will be activated, thereby limiting the volume of waste classified as radioactive. Both long-term and temporary storage will be required at CERN: the temporary areas will serve as buffer zones for equipment to be re-cycled and as intermediate storage for materials requiring processing before disposal. Long-term storage will be for equipment of high value which can be reused.

### 3.13.10 Dismantling of the experiments

The FCC-ee detector systems resemble LEP detectors in the sense that they are optimised for electroweak precision physics and precision measurements of the Higgs sector in a similar energy range. The overall size of the detectors is comparable to the LEP experiments. The CLIC-like Detector (CLD), for instance, has a height of 12 m and a length of 10.6 m, compared to DELPHI, which was $\sim$10 m in both length and diameter. The typical mass of a LEP experiment was approximately 3000 tons (barrel + endcaps), concentrated in the instrumented iron of the return yoke. Depending on the size and weight of the solenoid magnets, their lowering/lifting may require dedicated lifting equipment exceeding the capacity of the overhead cranes in the assembly halls on top of the caverns. The FCC-ee caverns are equipped with two independent lifts capable of evacuating 300 people within 30 minutes. The occupancy of the caverns is therefore not expected to be a limiting factor for the dismantling schedule. A crucial difference between LEP and FCC-ee may be the radioactive activation of detector (and accelerator) hardware originating from the up to $10^5$ times higher instantaneous luminosity.

#### *Dismantling of the LEP experiments*

Traditionally, the detector systems and specific infrastructure are owned by the international collaborations that designed and built them, with CERN as a member of the collaborations. It is thus also the collaborations that are in charge of the preparation and implementation of the dismantling. Dismantling of FCC-ee experiments is expected to follow the LEP approach, which is outlined below.

The dismantling of the LEP experiments had to be integrated into the global LHC project plan and tightly coordinated with the accelerator dismantling. CERN, as the host lab, was in charge of the experiment sites, logistics, radiological and general safety, planning, and coordination. The year before the start of dismantling was dedicated to the preparation of the site, e.g., the mechanical workshop, setting up storage area for sub-detectors and zones for the safe storage of activated materials, zones for temporary storage of high-value waste, such as copper cables and pipes.

The dismantling teams of the four LEP experiments typically consisted of 10-15 persons. A small core of experienced CERN technical staff (2-4 persons) with in-depth knowledge of the detector, its infrastructure and the experiment site was reinforced by handling and electro-mechanical support personnel (about 5 persons). The international collaborations assumed their responsibility and sent expert teams that, ideally, had been involved in the installation process of a sub-detector approximately 12 years earlier.

A radiological zoning analysis based on detailed modelling of the detector and its environment was performed well before the start of the dismantling. The main steps of dismantling were:

- Making the areas safe (flushing detector gases, cutting electrical power, coolant supplies, removal of beam pipe and calibration sources)
- Cutting and removing cables and pipework
- Extraction of sub-detectors in the reverse order of installation
- Removal of remaining components and materials

The overall dismantling budget of the 4 experiments was about 3 MCHF (in year 2000 prices). About 1 MCHF was contributed by CERN, mainly used for financing of additional handling personnel, rental of cranes and special transports. Each of the 4 scientific collaborations foresaw a budget of about 0.5 MCHF. This covered the additional personnel for the systematic radioactivity monitoring, additional personnel for the dismantling operations, purchase of special tools and infrastructure modifications/preparations. The sales of the dismantled material brought a nett income of about 100 kCHF per experiment. A big fraction of the sales price of the return yokes (corresponding to about 10 000 tons of steel) was offset by the cost of cutting and transport (special heavy-load transports).

The unprecedented luminosity of the FCC-ee storage ring may lead to significantly higher activation of the detector hardware and its infrastructure. Hardware close to the beam pipe, like the luminometers, will be activated the most.

***Conclusions from LEP dismantling***

Preliminary considerations suggest that the dismantling of the FCC-ee experiments can follow the same approach that was used for the four LEP experiments in 2001. The estimates of time and resources required appear still valid. The degree of activation and the related classification of certain parts of the detector need to be assessed with dedicated simulations based on realistic detector models and operational scenarios. The amount of material declared as TFA may have an impact on the planning and cost of the dismantling.

### 3.13.11 Recycling of materials

It can be anticipated that there will be some offset in the costs from the sales of materials for recycling and some specialist equipment. In the case of LEP dismantling, the sales revenue corresponded to $\sim$15% of the total cost of dismantling. The LEP dipole magnets, which formed the bulk of the materials, were made from steel embedded in concrete, and at the time, it was not financially viable to sell them for recycling. However, the FCC-ee magnets will be made of steel, weigh a total of ∼120 ketones, and will form the bulk of the recyclable materials to be removed. Copper and aluminium cables/busbars, as well as stainless steel pipework, will constitute further materials to be sold for re-cycling. Heavy metals (like lead and tungsten) will potentially be used for shielding, which will be available for recycling. The vacuum chambers in FCC-ee will be fabricated from high-purity copper and will, therefore, be of high value for recycling. There will be more copper and/or aluminium in the busbars and cables. A total of 3000 tonnes of copper or 1000 tonnes of aluminium will be required for magnet powering, further boosting the income from sales.

***Planning***

Figure 3.89 presents a possible timeline for the dismantling process based on an initial preparatory phase of 20 weeks followed by 40 weeks of dismantling activities. The schema presented assumes that up to 4 arcs can be dismantled simultaneously. Such a scheme, however, will require additional transport equipment and larger facilities for triage, storage, and processing of the equipment than if only two arcs were dismantled simultaneously. If the LSS could be done in parallel to an arc at another point and the sequence was, for example, to complete dismantling from PA and PG (28 weeks), then start PD and PJ and so on, the overall duration would increase to around 144 weeks, but the total labour costs would remain the same. The gain would come from a reduction in the quantity of transport equipment required and fewer infrastructure requirements.

### 3.13.12 Conclusions

The general scheme for dismantling up to 8 half arcs through 4 access shafts concurrently has been analysed and is feasible given the current design of the FCC-ee and its infrastructure. If it were necessary to only work through two shafts concurrently due e.g., to bottlenecks in the extraction channels, the total cost would not change very much, but the overall duration would increase from around two to three years.

The collaborations will fund the dismantling of the experiments, but some support will be required from CERN staff. Once the design has been completed and the specifications of the materials are known, it should be possible to estimate the cost of the elimination of radioactive waste.

# Chapter 4

# FCC-ee booster design and performance

## 4.1 Optics design and Beam dynamics

### 4.1.1 Booster parameters

The booster is located in the same tunnel as the collider. To simplify its mechanical integration, the length of one arc cell is set to $52\,\mathrm{m}$, matching the short cell length in the collider. More precisely, the arc periodicity corresponds to five short cells due to the non-interleaved sextupole scheme.

However, the booster layout (see Fig. 4.1) differs in several aspects from the collider layout. These differences are highlighted in Fig. 4.2 and summarised below:

– The booster beam runs at a radial offset of $8\,\mathrm{m}$ on the outer side relative to the interaction point not to interfere with the experimental detector.
– The booster beam pipe is positioned at the top level of the tunnel, with a vertical offset of $1030\,\mathrm{mm}$ and a horizontal offset of $161\,\mathrm{mm}$ towards the inner side. This offset ensures that the booster and collider maintain exactly the same circumference. Additionally, the booster cells have a slight longitudinal offset to enhance mechanical stability and facilitate tunnel integration. Due to minor differences in the arc cell patterns of the booster and collider, the distance between their reference trajectories oscillates by approximately $\pm 4\,\mathrm{mm}$. This orbit variation is also influenced by the dipole arrangements in the arcs.
– The RF cavities in the booster are positioned in the long straight section H, whereas in the collider, they are located in section L.

The parameters for the booster are summarised in Table 4.1.

### 4.1.2 Main constraints

At injection energy, Transverse Coupled Bunch Instabilities (TCBIs) may limit the total beam current, while the Transverse Mode Coupling Instability (TMCI) may determine the maximum allowed bunch charge. Both effects could, in principle, be mitigated by a larger momentum compaction factor.

For the $\mathrm{ZH}/\mathrm{t\bar{t}}$ modes, TCBI is not a concern due to the significantly lower average current. However, TMCI remains a limiting factor because of the large bunch charge ($4\,\mathrm{nC}$) required during collider filling. Studies on the filling scheme have shown that a bunch charge of $4\,\mathrm{nC}$ is not essential for all operations. A reduced bunch charge of $1.6\,\mathrm{nC}$, which remains above the minimum required to mitigate bootstrapping instability, has been identified as a viable alternative. This reduction has a negligible impact on the collider filling time for the $\mathrm{ZH}/\mathrm{t\bar{t}}$ modes.

For the Z/W modes, collective effects studies have demonstrated that the beam pipe is the main driver of TMCI. Two approaches were investigated to accommodate a bunch charge of $4\,\mathrm{nC}$: increasing the momentum compaction or reducing the transverse impedance. Increasing the momentum compaction would lead to a larger equilibrium emittance, necessitate different optics for different operational modes, and require additional magnet families to operate the booster at different working points, similar to the collider. Conversely, reducing the transverse impedance implies the need to minimise the contribution from the resistive beam pipe.

The proposed solution is to use a beam pipe made from copper or copper-coated stainless steel and to enlarge the inner diameter of the beam pipe from $50\,\mathrm{mm}$ to $60\,\mathrm{mm}$. Studies have shown that further increasing the diameter beyond 60 mm would result in a significant rise in magnet costs. A 60 mm diameter provides a reasonable compromise between the need for larger and heavier magnets and achieving lower transverse impedance.

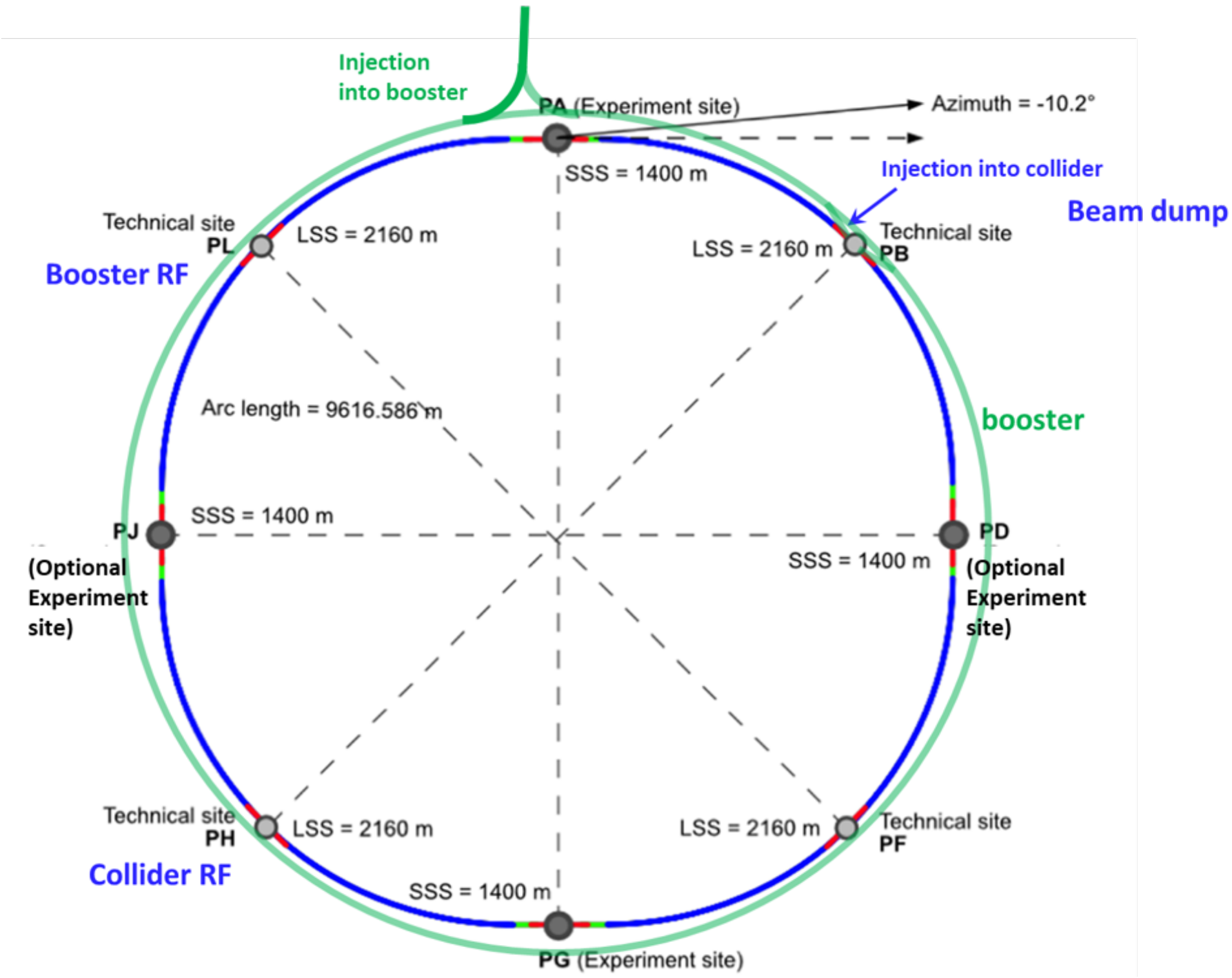

Fig. 4.1: Layout of the booster and collider. The booster shares the tunnel with the collider. The RF cavities of the booster are located in point L.

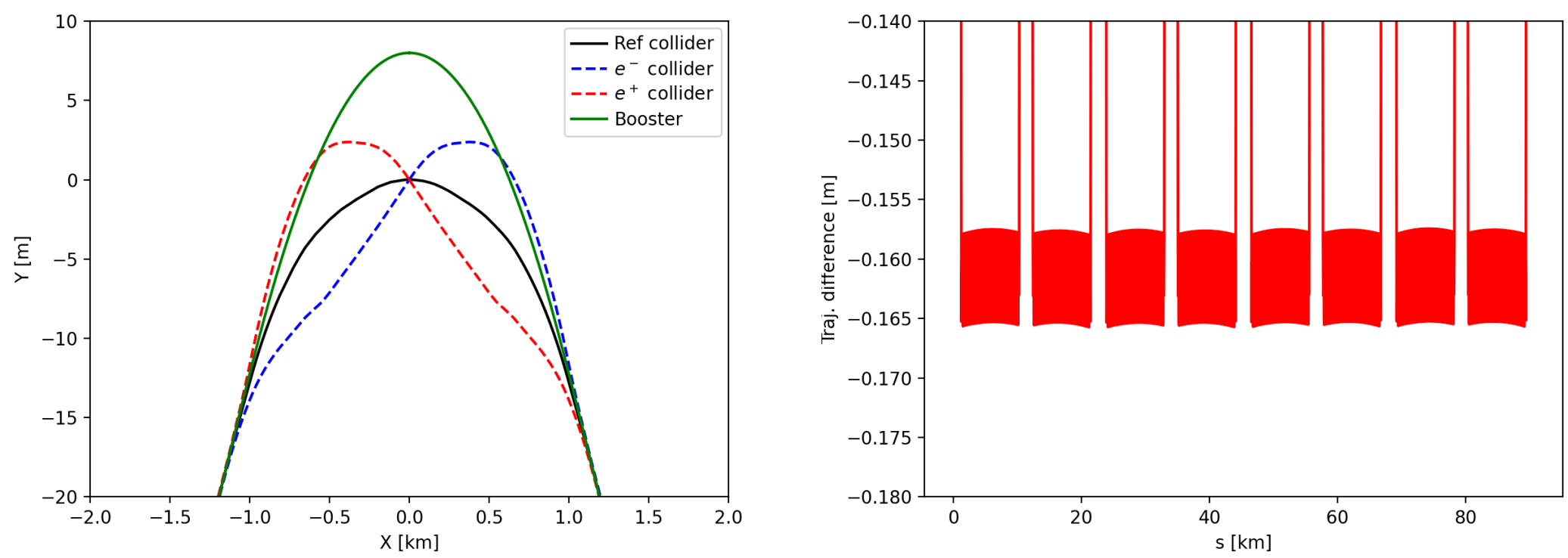

Fig. 4.2: Layout of the booster and collider in the section A, left. The distance between the reference axis of the booster and collider arcs is shown on the right picture.

Finally, reducing the stored current in the booster by a factor of ten significantly alleviates the constraints associated with TCBI, reducing the need for large momentum compaction. Consequently, the current strategy is to adopt the same optics for all operation modes while increasing the inner diameter of the vacuum chamber.

Ensuring beam stability in the booster requires chromaticity correction, which necessitates the use of sextupoles. The CDR compared three different sextupole schemes, where sextupoles separated by a

Table 4.1: Preliminary key parameters of the high-energy booster of FCC-ee. We consider here a linac of 20 GeV as a pre-injector and a high-energy damping ring.

| Running mode | | Z | WW | ZH | $t\bar{t}$ |
|---|---|---|---|---|---|
| Circumference | [km] | | 90.65871376 | | |
| Injection energy | [GeV] | | 20 | | |
| Extraction energy | [GeV] | 45.6 | 80 | 120 | 182.5 |
| Number booster ramps per cycle | | 10 | 2 | 1 | 1 |
| Number of stored bunches | | 1120 | 928 | 300 | 64 |
| Particle number/bunch (filling)† | [$10^{10}$] | 2.725 | 1.268 | 1.268 | 1.268 |
| Particle number/bunch (top-up)† | [$10^{10}$] | 2.725 | 1.035 | 1.268 | 1.125 |
| Collider top-up interval | [s] | 43.405 | 14.772 | 11.286 | 10.446 |
| RF frequency | [MHz] | | 800 | | |
| Lattice version | | | V24_FODO | | |
| Momentum compaction | | | $7.12 \times 10^{-6}$ | | |
| Coupling | | | $2 \times 10^{-2}$ | | |
| Injection emittances (norm.) | [µm] | | $20 \times 2$ | | |
| Extraction horizontal equilibrium emittance | [nm] | 0.087 | 0.27 | 0.61 | 1.4 |
| Extraction vertical equilibrium emittance | [pm] | 1.75 | 5.37 | 12.1 | 28.0 |
| Injection Energy loss / turn | [MeV] | | 1.34 | | |
| Extraction Energy loss / turn | [MeV] | 36.1 | 342 | 1730 | 9270 |
| Injection bunch length | [mm] | | 4 | | |
| Extraction bunch length | [mm] | 2.43 | 2.56 | 2.26 | 1.98 |
| Injection RMS energy spread | [$10^{-3}$] | | 1 | | |
| Extraction RMS energy spread | [$10^{-3}$] | 0.38 | 0.67 | 1.01 | 1.53 |
| Injection Maximum relative energy acceptance | [%] | | 3 | | |
| Extraction Maximum relative energy acceptance | [%] | 1 | 1.01 | 1.51 | 2.29 |
| Injection RF voltage | [MV] | | 50.1 | | |
| Extraction RF voltage | [MV] | 57.2 | 402 | 1960 | 10200 |
| Filling time | [s] | 2.8 | 2.315 | 3 | 0.64 |
| Up-Ramp time | [s] | 0.706 | 0.857 | 1.429 | 2.321 |
| Flat top | [s] | 0.1 | 0.1 | 0.1 | 0.1 |
| Down-Ramp time | [s] | 0.334 | 0.689 | 1.148 | 1.866 |
| Total cycling time | [s] | 39.4 | 7.922 | 5.68 | 4.927 |

† The required particle numbers in the booster assume an injection efficiency into the collider of 80% as specified in 2.3.9

phase advance of $\pi$ form a family:

- Classical sextupole scheme. After every quadrupole a sextupole magnet is installed leading to a maximum number of sextupoles. In this case there are two families per plane.
- Interleaved sextupole scheme, consisting of paris of sextupoles separated by $180^{\text{deg}}$ betatron phase advance . The sextupoles are considered to mainly, or only, act in one plane, and the pair for one plance is interlaced with a pair of the other plane.
- Completely non-interleaved sextupole scheme. In order to optimise the cancellation of the sextupole geometric effect, only linear elements are installed between two sextupoles forming a pair, and there is no interference between the pairs affecting one or the other plane.

The non-interleaved sextupole scheme was found to give the best dynamic aperture and momentum acceptance. Thus, this scheme has been retained.

To further increase the momentum acceptance, we have also added some transparency conditions on the dispersion suppressors and insertions to match the chromatic functions:

- The phase advance between the sextupoles in the dispersion suppressor is $\pi$ in both planes to maximize the geometric aberration cancellation.
- The angles of some dipoles in the dispersion suppressors have been matched to cancel the second-order dispersion. However, the total angle of the dispersion suppressor stays unchanged.
- The phase advance between the end of the upstream arc (or the beginning of the upstream dispersion suppressor) and the beginning of the downstream arc (or the start of the downstream dispersion suppressor) is equal to the phase advance of one arc cell plus an integer.
- Although the insertions are not perfectly symmetric, we have preferred to keep some waist conditions in the middle. In other terms, we ask to have $\alpha_x = \alpha_y = D'_x = 0$ at the middle of the insertion. We also ask to keep the chromatic derivative of $\alpha$ and the second-order dispersion derivative near 0: $A_x \approx A_y \approx dD'x_/d\delta \approx 0$.
- We match the Twiss parameters but also the Montague functions $A_x, A_y, B_x, B_y$ and second-order dispersion $dD_x/d\delta$ and $dD'_x/d\delta$ at the entrance of the right arc.

### 4.1.3 FODO lattice

The starting configuration for the design of a booster cell is a sequence of five FODO cells with 90 degree phase advance, and each approximately $52\,\mathrm{m}$ long.

The booster cell must follow the geometry of the collider determined by the structure for the collider short 90/90 FODO cell of the GHC optics. This condition requires the basic unit length of five booster arc cells to be $260.554\,\mathrm{m}$.

To maximise the cancellation of geometric aberrations introduced by the arc sextupoles, a phase advance of $\pi$ in both planes between the sextupoles of a pair is required, imposing four constraints. Additionally, the global tune of the machine is fine-tuned by adjusting the arc cells, adding two more constraints. As a result, the arc cell tune is $1.25 \pm \epsilon_{x,y}$ with $\epsilon_{x,y} \ll 1$. To accommodate these constraints, the arc cell requires six quadrupole families to control the phase advances between sextupole pairs and the global tune, along with two sextupole families to regulate the global chromaticity. The optical and chromatic functions of the arc cell, based on a FODO lattice, are shown in Fig. 4.3 (top).

The dispersion suppressor consists of 1.5 FODO arc cells followed by two additional FODO cells of the same type as the arcs, but with only one dipole per half-cell instead of two (see Fig. 4.3 (middle)). The dipole lengths have been optimised to minimise second-order dispersion at the end of the dispersion suppressor. The dispersion naturally damps along this section, requiring minor quadrupole adjustments in the dispersion suppressor to cancel the dispersion, correct second-order dispersion, and ensure a phase advance of $180^\circ$ between the last sextupole in the arcs and the entrance of the dispersion suppressor. This phase advance constraint allows the possible later insertion of an additional sextupole to further correct geometric aberrations.

The dispersion suppressors and insertions have been optimised according to transparency conditions. The optical functions and second-order chromatic functions for one quarter of the booster, based on a FODO lattice, are shown in Fig. 4.3 (bottom).

### 4.1.4 Alternative: HFD lattice

The Hybrid FODO (HFD) lattice is an alternative to the FODO cell. The main motivation of this cell is to reduce the anharmonicity and second-order chromaticity of the FODO cell to enlarge the momentum acceptance and dynamic aperture. The main difference between the HFD and FODO cells are:

- The phase advances between the two sextupoles of the same pair are near $\pi$ (and not exactly $\pi$).

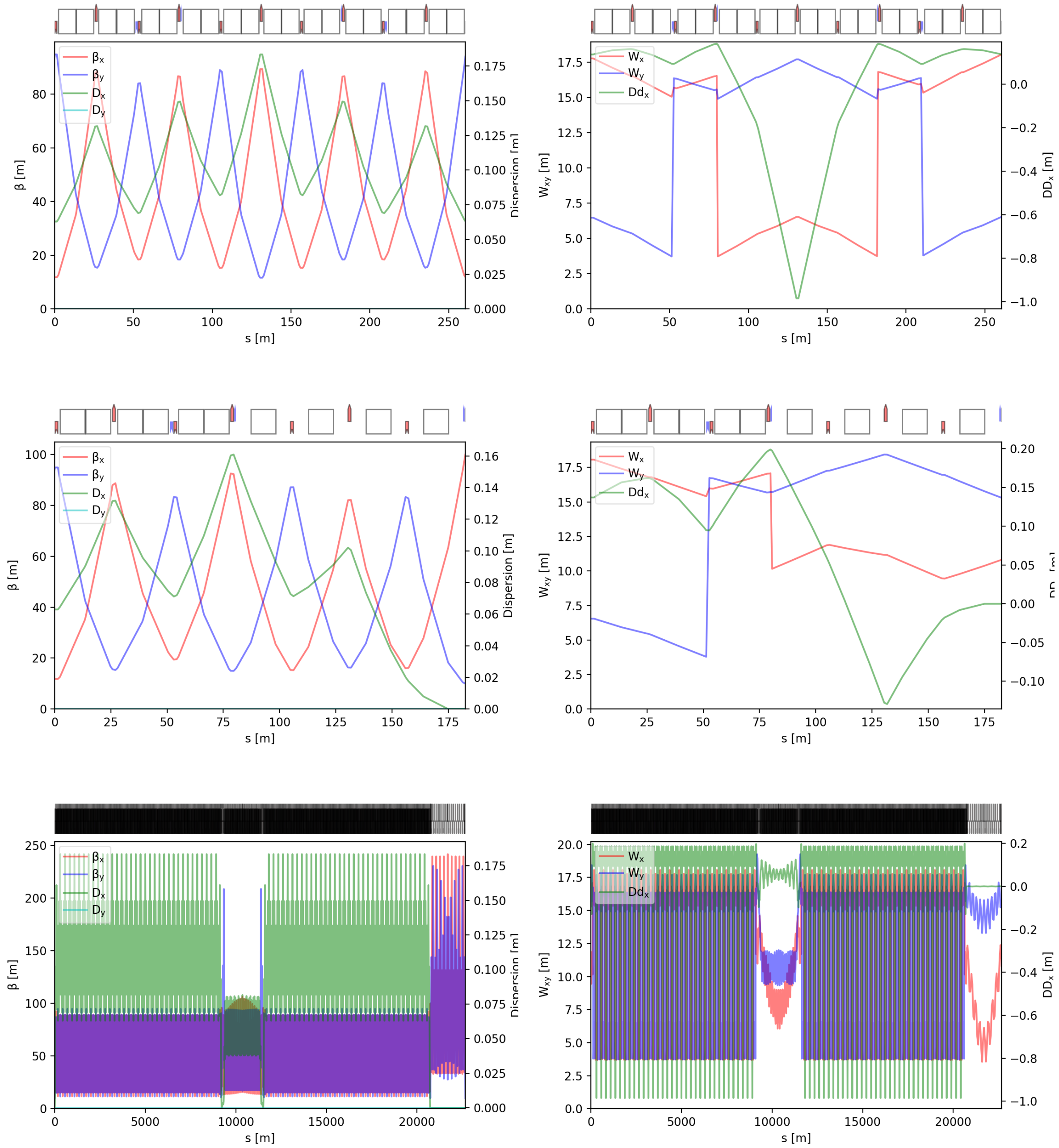

Fig. 4.3: Optical functions (left) and second-order chromatic functions (right) in the arc cells (top), in the dispersion suppressor before section B (middle), or in one quarter of the booster (bottom) for the booster baseline optics with FODO arc cells.

- The dipoles do not have the same length, which enable a modulation of the distance between the quadrupoles. The total length of the dipoles stay the same, to keep the same curvature radius and thus radiated power per turn.
- The horizontal and vertical tunes are quite different. In the case of the FODO cell, the tunes are about 1.25 in both planes against 1.25/1.15 for the HFD cell.

The optical functions and second-order chromatic functions in the arc cell and in one-quarter of the booster based on an HFD lattice are shown in Fig. 4.4.

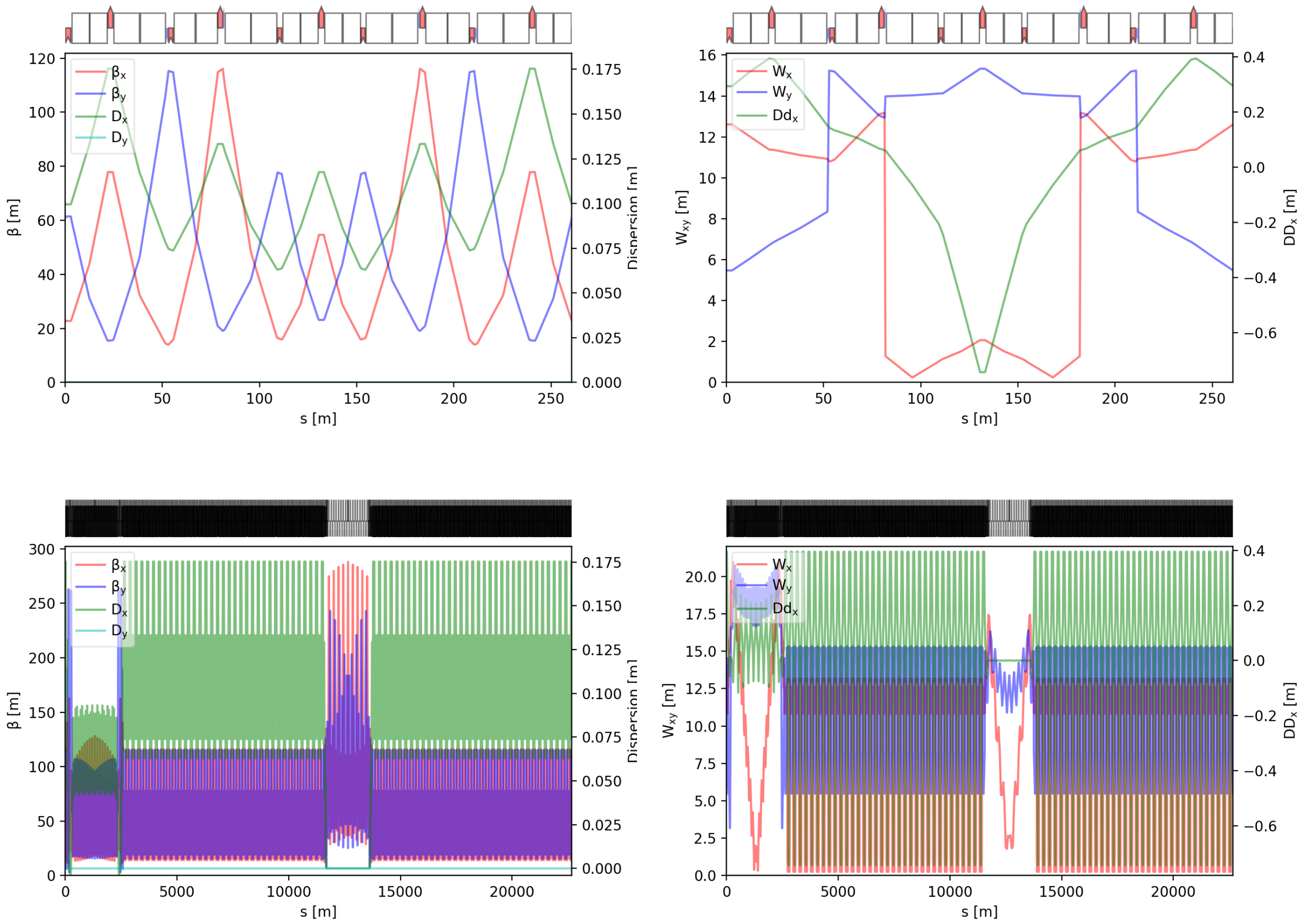

Fig. 4.4: Optical functions (left) and second-order chromatic functions (right) in the arc cells (top) and in one-quarter of the booster (bottom) with the version with HFD arc cells.

### 4.1.5 Dynamic aperture and momentum acceptance

One of the main concerns is to ensure a sufficiently large dynamic aperture and momentum aperture to keep the beam losses small at injection. The dynamic aperture without errors is calculated by scanning the initial particle positions that are stable on the $x$ and $y$ axes after 1000 turns, without taking into account synchrotron radiation damping. Different initial energy offsets between $-2\,\%$ and $+2\,\%$ are also considered checking the momentum aperture. Figure 4.5 compares the horizontal and vertical stable regions as a function of the energy offset, illustrating the dynamic aperture of both the FODO and Hybrid FODO lattices.

In both cases, the margin is sufficient to ensure the stability of the injected beam. The target dynamic aperture of $15\,\sigma$ is indicated by the dashed red line in Fig. 4.5.

The corresponding momentum aperture for the FODO lattice exceeds $\pm 0.75\,\%$ in the horizontal plane and $\pm 1\,\%$ in the vertical plane, equivalent to $5\,\sigma_\delta$ where $\sigma_\delta$ denotes the rms relative momentum spread of the injected beam. The HFD lattice exhibits a larger momentum aperture in the horizontal plane. However, this difference is partly thanks to the absence of an injection optics in one of the insertions. This injection optics disrupts the superperiodicity of the FODO lattice. Injection has not yet been developed for the HFD lattice.

The impact of eddy currents on the dynamic and momentum aperture has also been investigated. The cyan dashed lines in Fig. 4.5 represent tracking results for the FODO lattice, including a systematic $b_3$ field error component applied to each dipole in the arcs.

The expected sextupole gradient in the dipoles due to eddy current effects is $-0.015\,\mathrm{T\,m^{-2}}$, cor-responding to an integrated $b_3$ of approximately $-0.0025\,\mathrm{m}^{-2}$ at injection energy. To assess its impact, the dynamic aperture was evaluated for ten different integrated $b_3$ values over a range of $0.01\,\mathrm{m}^{-2}$.

The phase variation that corresponds to the $\Delta p/p$ is shown on the top horizontal axis of Fig. 4.5 and is calculated following section 3.4 of Ref [327]. The impact of the linear and random non-linear field errors on the stability region and a possible mitigation strategy are the next steps of investigation.

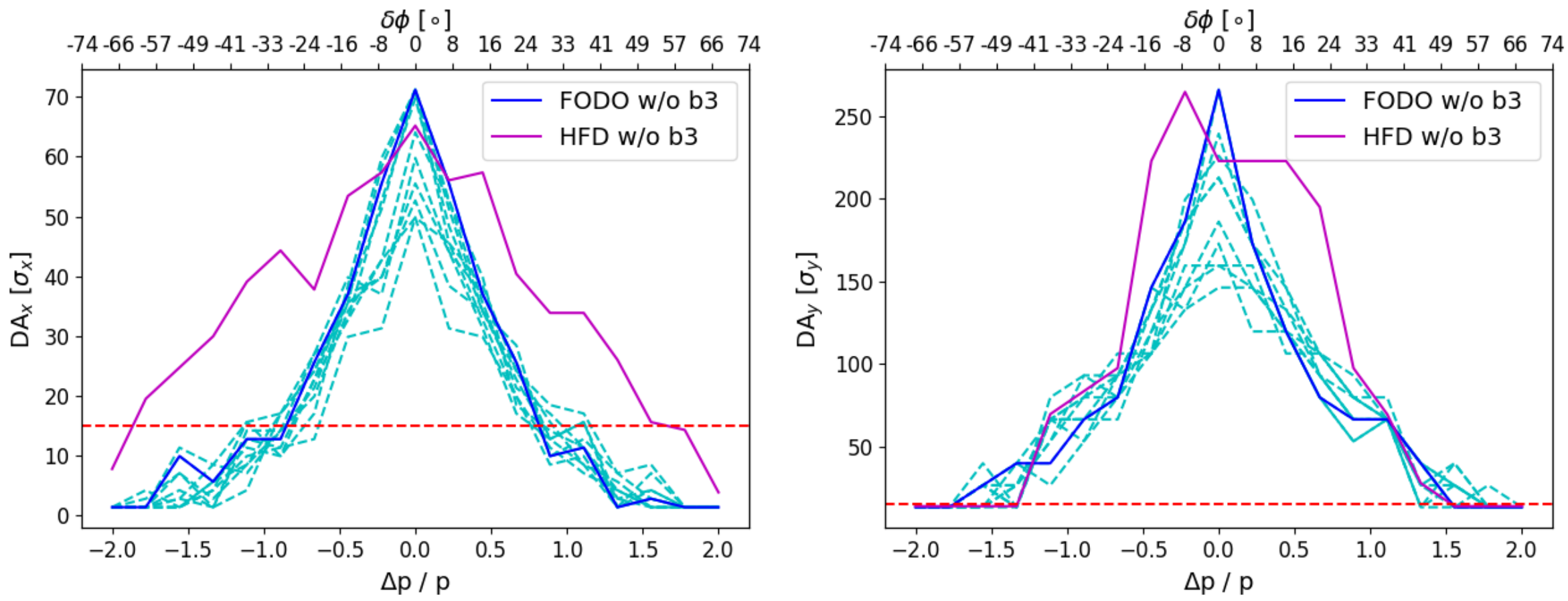

Fig. 4.5: Horizontal (left) and vertical (right) Dynamic Aperture at injection as a function of momentum deviation for the FODO and the HFD optics. The normalised horizontal and vertical emittances used to compute the beam sizes are $20\,\mu\mathrm{m}$ and $2\,\mu\mathrm{m}$, respectively. The dashed red line is the target value of $15\,\sigma$ transverse dynamic aperture, the cyan dashed line show the effect of several values of the systematic $b_3$ component due to eddy-current in the main dipole, as described in the text.

### 4.1.6 Tapering

The energy lost per turn due to synchrotron radiation scales with the fourth power of the beam energy. Consequently, the relative energy loss per turn scales with the third power of the energy. As a result, the relative energy lost per turn increases from $6.7 \times 10^{-5}$ at injection to $5 \times 10^{-2}$ at extraction for the $\mathrm{t\bar{t}}$ operation.

In consequence, the beam energy varies around the ring with a maximum just after the RF section and a minimum, when the beam arrives again at the RF section one turn later. The magnetic fields have to be adjusted to ensure that the beam remains close to the center of the vacuum chamber.

Since the arcs are individually powered, the dipoles in different arcs can be supplied with different currents, allowing for tailored magnetic field adjustments. Assuming a single power supply per arc, the maximum beam deflection in the booster can be reduced to $553\,\mu\mathrm{m}$ (see Fig. 4.6). Further improvement can be achieved using horizontal dipole correctors positioned near the focusing quadrupoles. In this case, the maximum orbit variation is reduced to $55\,\mu\mathrm{m}$, while the maximum integrated strength of the dipole corrector remains at $6.3\,\mathrm{mT\,m}$, which is below the specified dipole strength listed in Table 5.4.

These results indicate that a combination of varying the dipole field for each arc and employing dipole correctors is sufficient to maintain the orbit within the required tolerances. However, this study could be further refined by increasing the number of sectors within the dipole arcs to achieve a finer magnetic field distribution.

Additionally, while this analysis has focused on orbit correction, synchrotron radiation losses also induce beta-beating and impact the equilibrium emittance. Further studies should investigate beta-beating correction through arc rematching with the interaction region.

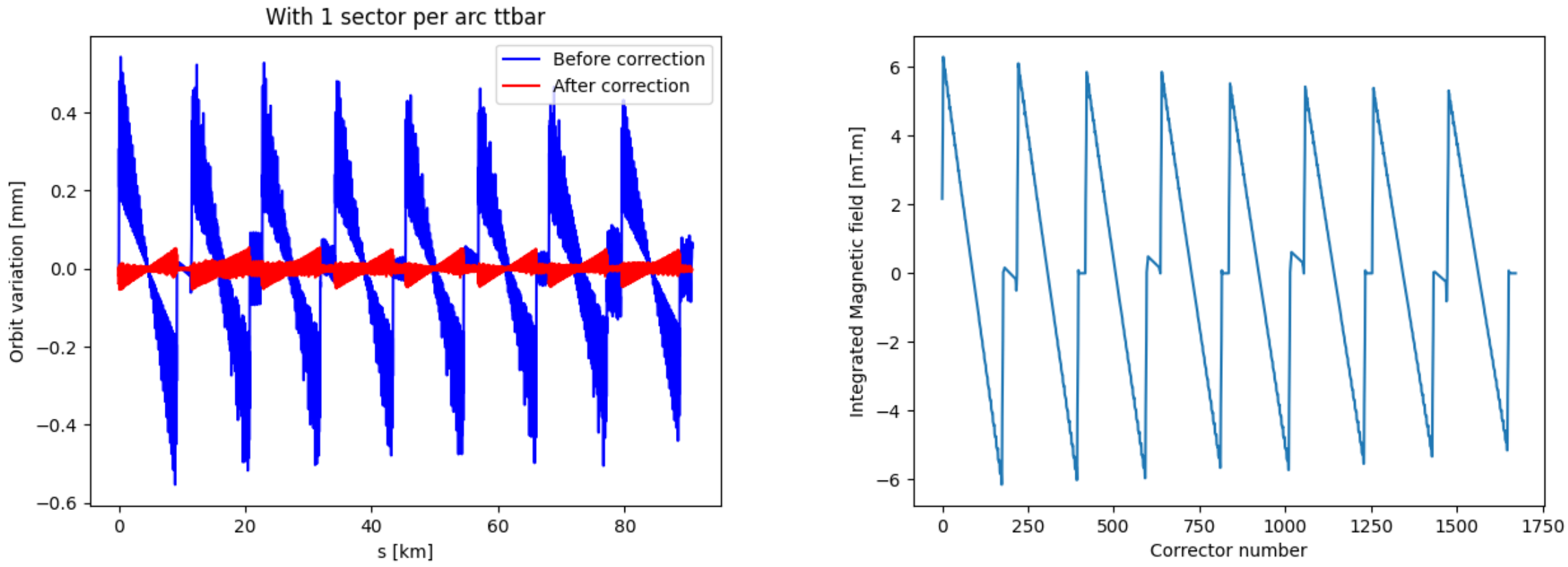

Fig. 4.6: Left: Beam orbit around the booster at the extraction energy for the $\mathrm{t\bar{t}}$ mode taking synchrotron radiation losses into account, and comparing the cases of a different magnetic field in each arc with the horizontal correctors off (in blue) or on (in red). Right: Integrated field of the horizontal dipole correctors to minimise the orbit with synchrotron radiation.

## 4.2 Collective effects

### 4.2.1 Baseline assumptions for collective effects studies

The high-energy booster may be subject to impedance-induced instabilities, which depend on various factors and can manifest as either short-range or long-range effects. Some of these factors, such as the number of bunches and the number of particles per bunch, are linked to the cycling and operational mode. In contrast, others, like the momentum compaction factor, are determined by the optics design.

At this stage of the study, only resistive wall contributions and RF cavities have been taken into account. While these do not represent the full spectrum of impedance effects, they are dominant and play a crucial role in defining key physical parameters of the beam pipe, such as its diameter and material. These parameters, in turn, influence other critical subsystems, including the vacuum system and magnet designs.

### 4.2.2 Mismatched beams at injection

The high-energy booster serves as the intermediary between the high-energy LINAC and the main ring. Its ability to deliver a beam that meets the required specifications depends on multiple factors, including the properties of the injected beam from the high-energy LINAC. While the booster design demonstrates robustness to a range of transverse beam parameters at injection energy, the longitudinal parameters can introduce significant mismatches and microwave instabilities. A study was conducted to ensure that the injected beam does not induce instabilities. The injector complex is capable of providing beams with specific bunch lengths and energy dispersion. A two-dimensional parametric scan was performed to further assess injection energy stability. The longitudinal mismatch $\xi_z$ has been quantified in terms of:

$$\xi_z = \sigma_z^{\mathrm{eq}} / \sigma_z^{\mathrm{inj}} \,, \tag{4.1}$$

where $\sigma_z^{\mathrm{inj}}$ is the bunch length at injection and $\sigma_z^{\mathrm{eq}}$ is the bunch length at equilibrium. Figure 4.7 shows that the baseline bunch lengths and energy dispersion requirements from the LINAC provide a significant safety margin regarding longitudinal mismatch and microwave instabilities. Figure 4.7 also rules out a configuration where an energy compressor would not be used after the high-energy LINAC, resulting in an injection energy dispersion of $0.25\,\%$ and a bunch length of $1\,\mathrm{mm}$. The figure shows that the present baseline injection parameters from the high-energy LINAC, with $\sigma_z = 4\,\mathrm{mm}$ and $\sigma_e = 0.1\,\%$, allow a significant margin to avoid mismatch and microwave instabilities at injection energy.

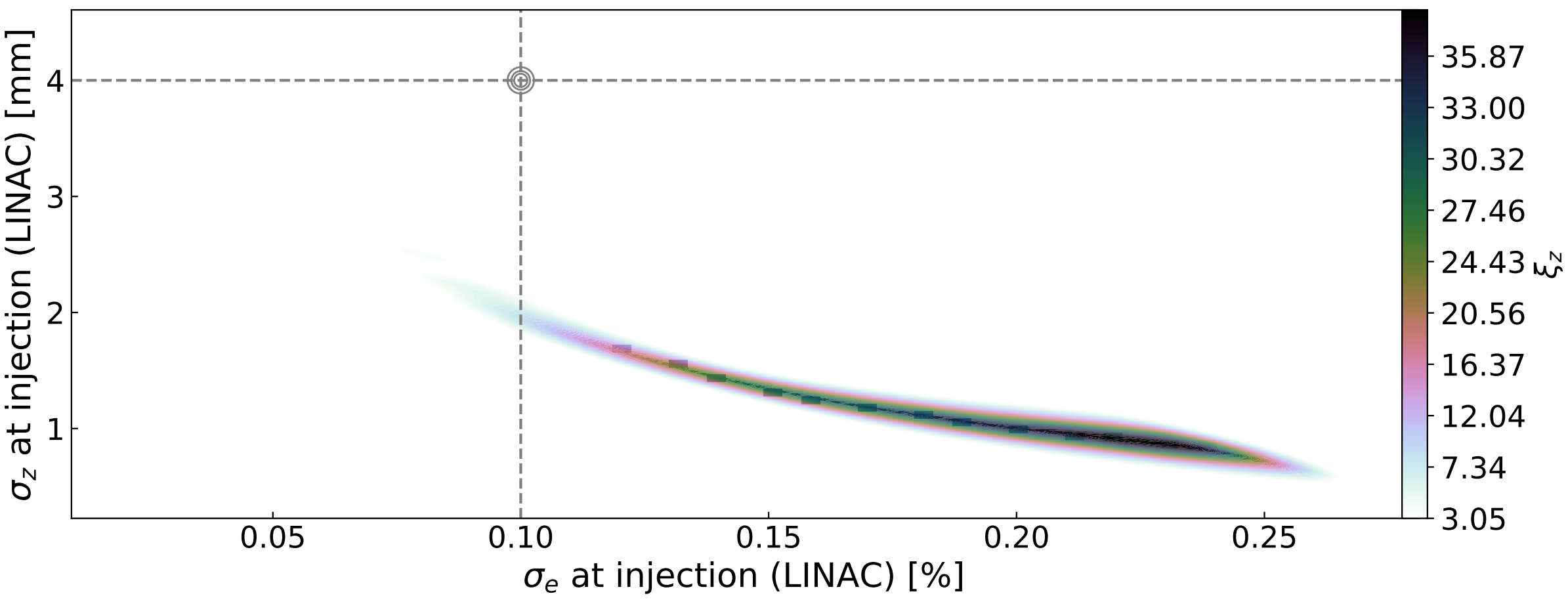

Fig. 4.7: Kernel density estimate of the longitudinal mismatch as a function of bunch length and energy dispersion at injection. The energy considered is $20\,\mathrm{GeV}$, and the different booster parameters considered are those of the Z-pole operation (see Table 4.1). The target circle shows the present baseline injection parameters to the high-energy booster.

Another effect being studied is the result of a transverse jitter of up to $1\,\sigma$ in the vertical and horizontal planes. Present particle tracking studies do not show any effect at equilibrium. However, amplitude detuning has not yet been taken into account, and these results need to be confirmed with more realistic simulation parameters.

### 4.2.3 Coupled bunch instabilities

***Transverse coupled bunch instabilities due to resistive wall impedance***

Transverse resistive wall wakefields contain long range components. These can lead to transverse coupled-bunch instabilities that are destructive to the beam. Given the following assumptions, one can estimate analytically the transverse growth rate due to coupled bunches: 1. Equally spaced Gaussian bunches 2. Only coherent bunch modes 3. Only the most prominent radial mode in the longitudinal azimuthal mode.

With these assumptions, the transverse growth rate $\tau_\perp$ [328] can be expressed by:

$$\tau_\perp \sim \frac{N_p \cdot N_b}{4\pi \cdot Q_{x,y} \cdot E} \cdot \mathrm{Re}(Z_\perp) \cdot \mathcal{G}(Q_{x,y}, \sigma_z, \sigma_e) \tag{4.2}$$

where $N_p$ is the number of particles per bunch, $N_b$ is the number of bunches, $Q_{x,y}$ is the transverse tune, $E$ is the energy, $Z_\perp$ is the transverse impedance, and $\mathcal{G}$ a factor which can be approximated by 1 in our case. With the Z operation mode being the worst-case scenario, one can estimate the growth rate as a function of the mode numbers. By normalising the modes, one can compare the growth rate for two different booster configurations, namely from 2023 and 2024, respectively. Figure 4.8 shows that the growth rate has been reduced in the new baseline design compared to the previous one. The most drastic change is at the Z-pole operation, which represents the worst-case scenario. For this case, a reduction is observed from $374\,\mathrm{s}^{-1}$, i.e., (8.7 turns)$^{-1}$ in the 2023 design to $10\,\mathrm{s}^{-1}$, i.e., (310 turns)$^{-1}$ in the 2024 design. This is due to several important changes, including the increase of the beam pipe diameter from $50\,\mathrm{mm}$ to $60\,\mathrm{mm}$ and the reduction of the number of bunches from 15 880 to 1120 bunches. While the new growth rates are still faster than the transverse synchrotron damping ($\sim$ 30 000 turns), they reduce the constraints on the dampers needed to mitigate such coupled-bunch instabilities.

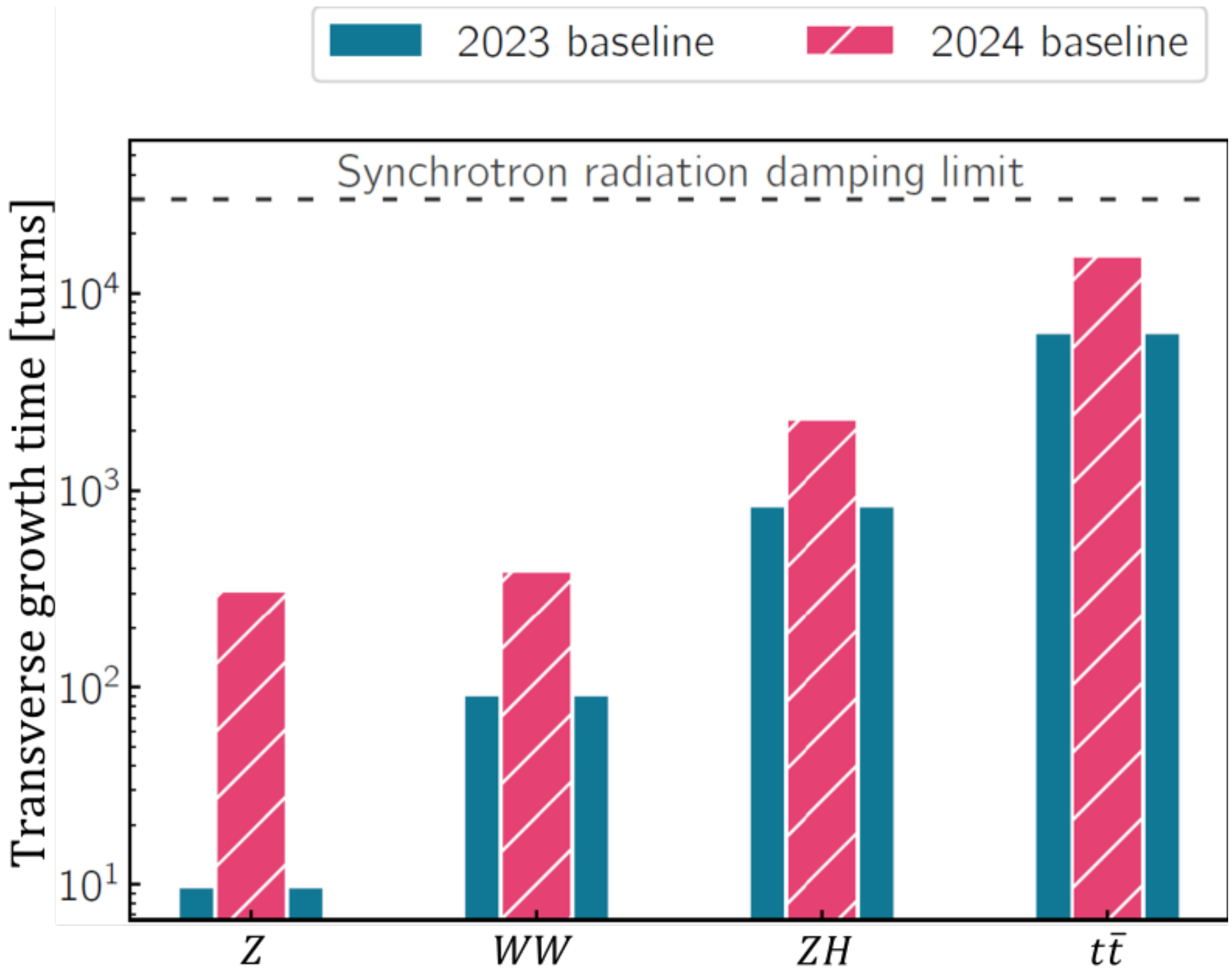

Fig. 4.8: Resistive wall transverse impedance induced transverse growth time constants as a function of the mode number for the 2024 (hashed-red) and 2023 baseline (plain-blue).

***Longitudinal coupled bunch instabilities due to cavities high order modes***

The higher order modes of the RF cavities are the main drivers of the longitudinal coupled bunch instabilities. The strategy to keep the RF system below the stability limit of longitudinal coupled bunch instabilities is detailed in Section 6.3.3.

### 4.2.4 Single Bunch instabilities

Single bunch instabilities that may arise in the high-energy booster and hinder achieving technical targets are of two types: microwave instabilities (MI) and transverse mode coupling instabilities (TMCI) (see [328]). The bunch population thresholds for these instabilities can be expressed as:

$$N_{p,th}^{TMCI} = \frac{Q_{x,y} \cdot Q_s \cdot E \cdot \sigma_z}{\mathrm{Im}\{Z_\perp\} \cdot e \cdot c}, \quad N_{p,th}^{MI} \propto \frac{n \cdot \alpha_c \cdot E \cdot \sigma_e \cdot \sigma_z}{|Z_\parallel|}, \tag{4.3}$$

where $Q_{x,y}$ is the transverse tune, $Q_s$ is the synchrotron tune, $E$ is the energy, $\sigma_z$ is the bunch length, $Z_\perp$ is the transverse impedance, $\alpha_c$ is the momentum compaction factor, and $Z_\parallel$ is the longitudinal impedance.

Although these two effects are interconnected, transverse mode coupling instabilities are particularly critical due to their potential to cause destructive transverse exponential growth in the beam. In contrast, microwave instabilities primarily result in longitudinal emittance growth, which, given the injected beam parameters, remains within the limits required for extraction to the main ring.

The design of the high-energy booster requires balancing multiple dominant effects, as outlined in Eq. (4.3). Among these, resistive wall contributions are the primary source of impedance effects, while the momentum compaction factor plays a crucial role in determining the optimal optical design.

The following section examines the influence of these parameters on transverse mode coupling instabilities.

### *Resistive wall contribution*

The large circumference of the booster ring makes the resistive wall (RW) impedance the dominant contributor to collective effects. In this study, the VACI SUITE [329] was used to analyse different materials and beam pipe radii. The resulting impedance calculations were directly incorporated into XSUITE [175] and PYHEADTAIL [330] for beam dynamics simulations. These tools enabled precise modelling of beam behaviour and its interaction with the surrounding environment, ensuring accurate predictions and optimisations for the beam pipe design.

A detailed study of beam pipe materials has been done for three cases: a copper beam pipe (Cu), a stainless steel pipe (SS), and a stainless steel pipe with an internal copper coating (SS-Cu). Various copper coating thicknesses were evaluated, and a 1 mm copper layer was identified as the optimal configuration. The different material scenarios (Cu, SS, and SS-Cu) are illustrated in Fig. 4.9. The choice of materials played a crucial role in controlling RW impedance, directly impacting overall beam stability and performance.

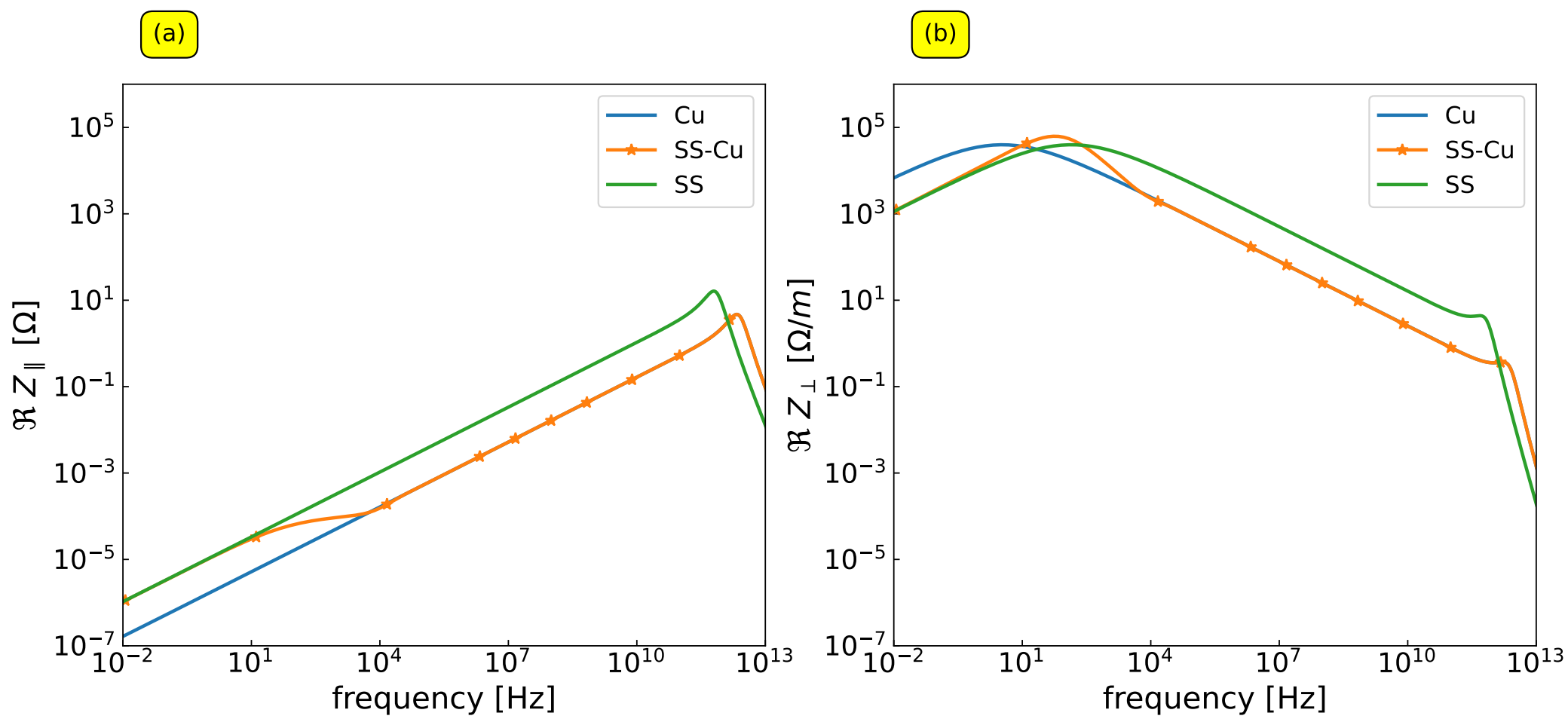

Fig. 4.9: (a) Longitudinal impedance and (b) Dipolar impedance for a pipe with R=30 mm and L=1 m, for three different scenarios (Cu, SS, and SS-Cu).

Given that the beam pipe has a round geometry, no detuning (quadrupolar) impedance was observed, even in the case of the coated pipe, which featured a narrow strip of missing copper on the inner surface. While the absence of a continuous copper layer might have been expected to introduce a detuning impedance, the VACI results indicated that the effect of this strip was negligible. Consequently, the final design adopts the SS-Cu configuration with a 1 mm copper coating, providing an optimal balance between material properties and impedance control.

### *Beam-pipe radius and material*

Two materials were considered for the beam vacuum chamber: stainless steel and copper. Stainless steel offers advantages such as lower initial cost and reduced eddy currents induced by magnetic field ramps during acceleration. However, it comes with increased longitudinal and transverse impedance (see Fig. 4.9). Given the low magnetic fields involved, eddy currents are not a concern, allowing for a 1 mm copper coating: beyond approximately 2 kHz, the skin depth ensures the beam interacts predominantly with the copper layer.

The impact on beam dynamics was assessed using the PyHEADTAIL[1] tracking code [330].

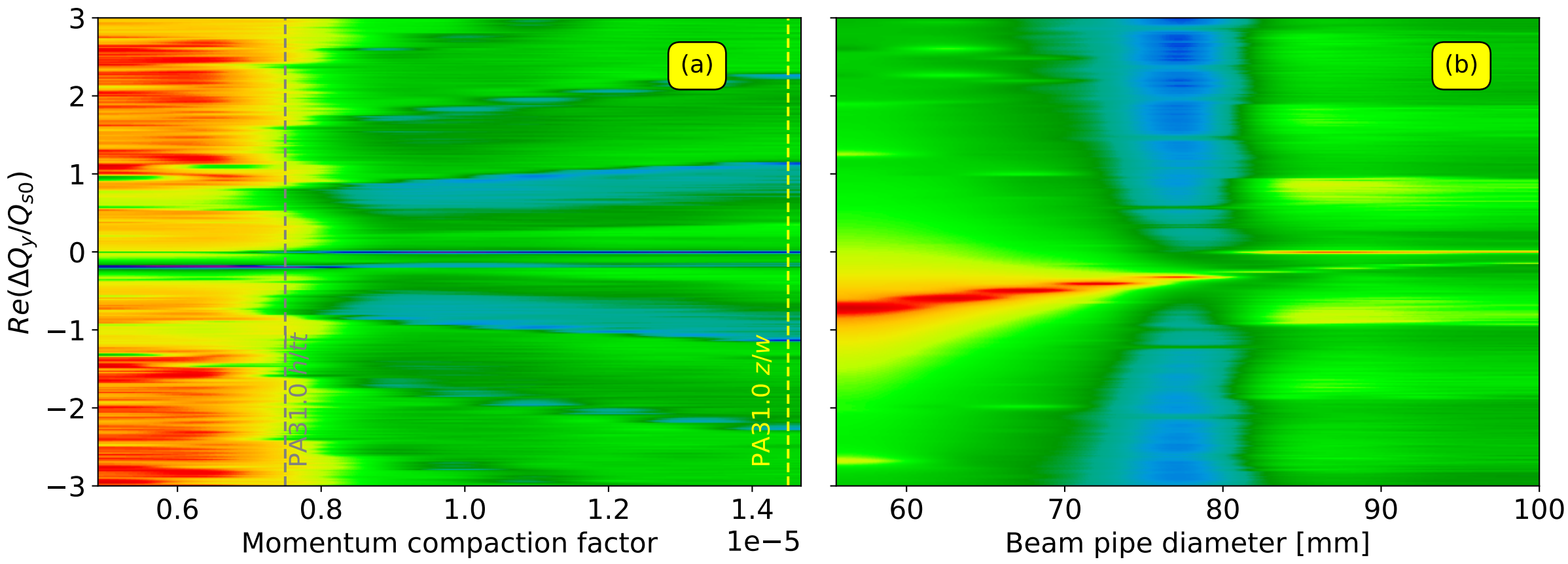

Fig. 4.10: Real part of the tune shift of the first azimuthal transverse coherent oscillation modes normalised by the synchrotron tune as a function of the momentum compaction factor (left) and the beam-pipe diameter (right) for the Z-operation mode. For (a), a copper beam-pipe of 50 mm diameter and a bunch population of $2.5 \times 10^{10}$ particles are considered. For (b), a stainless steel beam-pipe with a bunch population of $2.5 \times 10^{10}$ particles and a momentum compaction $\alpha_c = 7.34 \times 10^{-6}$ are considered.

A comparison was conducted on a circular beam pipe made of either copper or stainless steel, each with a diameter of 50 mm, using the `PA31.0` baseline design (2023 CDR baseline) for the Z-pole operation. A substantial increase in bunch length and longitudinal emittance was observed with a stainless steel vacuum chamber of 50 mm diameter, necessitating an increase in the inner diameter of the beam pipe. To address this, an optimisation study was conducted to determine the ideal beam pipe diameter for stainless steel. Modal analysis, following the methodology in Ref. [61], established a minimum of 70 mm, with a potential increase up to 100 mm to accommodate impedance margins.

#### *Momentum Compaction Variation*

For a given bunch of particles, variations in the relative path length due to changes in momentum can significantly affect transverse mode coupling instability (TMCI) bunch population thresholds. This is quantified by the variation of the momentum compaction. Figure 4.10 (a) illustrates that the 2023 optics for $\mathrm{t\bar{t}}$ operation with $\alpha_c = 7.34 \times 10^{-6}$ and a 50 mm diameter copper beam pipe resides in an unstable region. By contrast, the 2023 Z mode optics ($\alpha_c = 1.49 \times 10^{-5}$), with nearly twice the momentum compaction, falls within a stable zone.

#### *Bunch Population Variation*

A high-energy booster design featuring a 50 mm diameter beam pipe, whether made of copper or stainless steel, coupled with an optics design featuring a momentum compaction of $\alpha_c = 7.34 \times 10^{-6}$, proves impractical. Conversely, adopting a single optics design instead of separate values for different operation modes offers benefits. A compromise was reached, maintaining a single momentum compaction value ($\alpha_c = 7.12 \times 10^{-6}$) while increasing the beam pipe diameter from 50 mm to 60 mm and opting for copper as the beam pipe material. Figure 4.11 demonstrates that this new baseline design increases the TMCI threshold bunch population from $2.5 \times 10^{10}$ to $5.7 \times 10^{10}$. Additionally, incorporating a 300-turn transverse damper would further enhance the threshold and provide additional safety margins. These results, however, require validation with a more comprehensive machine impedance model.

[1]https://github.com/PyCOMPLETE/PyHEADTAIL

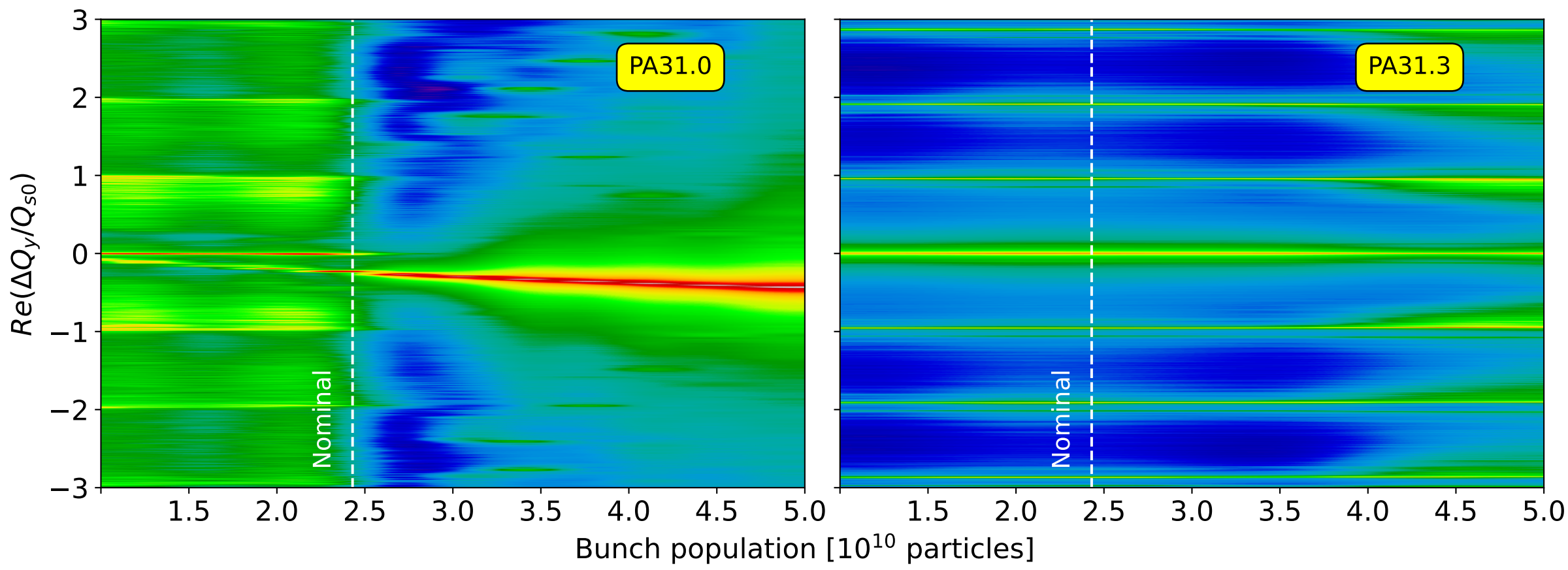

Fig. 4.11: Real part of the tune shift of the first azimuthal transverse coherent oscillation modes normalised by the synchrotron tune as a function of bunch population for the `PA31.0` 2023 CDR baseline (left) and the `PA31.3` 2024 baseline design (right).

## 4.3 Radiation environment

Given the significant power dissipated by synchrotron radiation in FCC-ee, appropriate mitigation measures must be implemented to prevent radiation-induced equipment failures and degradation of machine performance. Although the radiated power is considerably higher in the collider than in the booster, the booster's contribution must still be carefully evaluated, as no dedicated photon stoppers are foreseen, unlike in the collider ring.

This section examines the ionising dose generated by synchrotron photon emission in the booster arcs and assesses the shielding efficiency of the dipole yokes. Other radiation effects, such as single-event effects in electronics or radiation-induced corrosion, fall outside the scope of this section and will be addressed in future studies. Additionally, other radiation sources in the booster, such as beam-gas scattering, must also be considered. However, the dose studies presented here provide an initial assessment of the shielding requirements for the booster.

### 4.3.1 Synchrotron radiation emission during a booster cycle

The primary source of radiation in the FCC-ee arc tunnel is synchrotron radiation emitted in the collider ring. By design, this radiation is limited to $100\,\mathrm{MW}$ across all operating modes ($50\,\mathrm{MW/beam}$). In contrast, the average synchrotron power emitted by the booster is significantly lower due to several factors: the lower stored beam intensity, the strong dependence of synchrotron radiation power on beam energy ($\propto E^4$), the relatively short duration of booster cycles, and the intervals without beam between cycles.

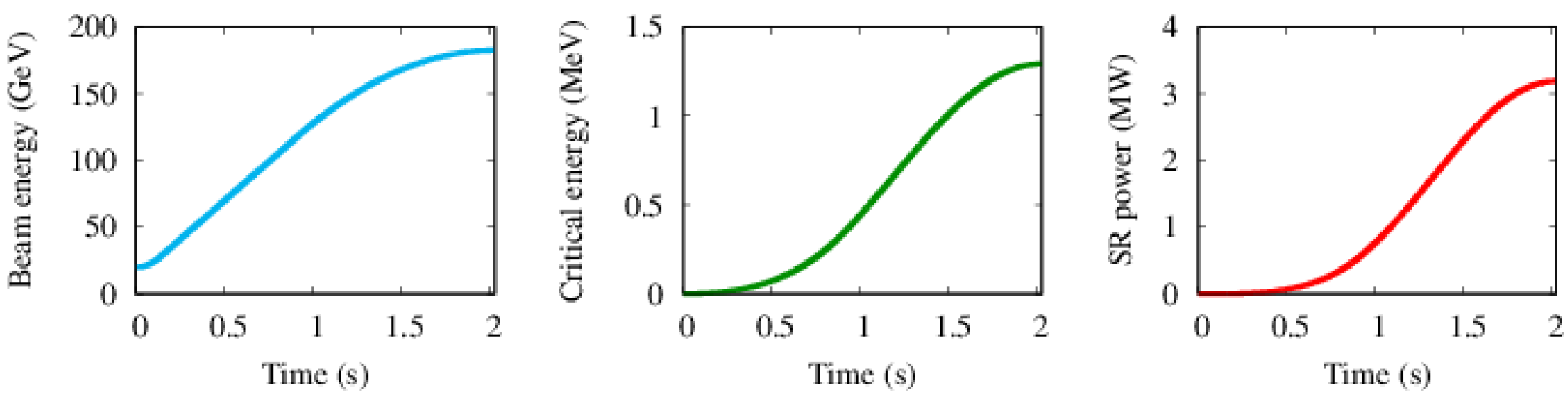

Fig. 4.12: Time evolution of the beam energy (left), critical energy of the synchrotron photon spectrum (centre) and emitted synchrotron radiation power (right) for a booster ramp from $20\,\mathrm{GeV}$ to $182.5\,\mathrm{GeV}$ ($\mathrm{t\bar{t}}$ operation). The flat bottom and flat top plateaus are not shown.

For top-up injection, the booster train intensity is expected to be lower than for full collider refills, as the bootstrap bunch charge can vary between $0\,\%$ and $100\,\%$. However, top-up cycles are anticipated to have a greater impact on cumulative radiation effects because they will be executed more often.

The evolution of the radiated power and photon spectra during a booster cycle depends on the ramp function. Figure 4.12 shows the time dependence of the beam energy, critical energy and emitted power during a ramp from $20\,\mathrm{GeV}$ to $182.5\,\mathrm{GeV}$ in $\mathrm{t\bar{t}}$ operation. The figure assumes a bunch train of 64 bunches, with a bunch intensity of $10^{10}\,\mathrm{e}^{\pm}$, which is the maximum top-up intensity in $\mathrm{t\bar{t}}$ cycles (see Table 4.1). At the end of the ramp, the power reaches about $3\,\mathrm{MW}$, while the average power during the $\sim$2 s-long ramp is about $1\,\mathrm{MW}$. The presently foreseen repetition rate of booster cycles for $\mathrm{t\bar{t}}$ operation is $10.4\,\mathrm{s}$, alternating between electrons and positrons. Taking into account the time without beam between cycles, the average power is less than $0.2\,\mathrm{MW}$, i.e., more than 500 times lower than in the collider. This estimate can still slightly change depending on the need of a flattop plateau before beam extraction. The peak power and top-up interval is similar for ZH operation, but the radiation leakage from the magnets is expected to be less compared to $\mathrm{t\bar{t}}$ due to the lower critical energy at flattop. For the other beam modes (Z and WW), the top-up intervals are longer and the peak power and critical energies are lower. They are, hence, expected to be less relevant for cumulative radiation effects than the two higher-energy modes.

### 4.3.2 Booster contribution to the ionising dose in the arcs

The synchrotron radiation absorbers in the collider ring need to be surrounded by heavy shielding in order to sufficiently reduce the ionising dose in the tunnel (see Section 1.9). It is, therefore, highly desirable that the booster contribution to the ionising dose remains as low as reasonably achievable. Contrary to the collider, the synchrotron photons emitted in the booster impact directly on the vacuum chamber. The chamber walls are too thin to fully shield the photons and secondary particles, but the H-shaped yokes of the booster dipoles still provide a significant attenuation of the secondary radiation. The shielding efficiency required of the booster dipole yokes represents an important prerequisite for the magnet design. In order to study the necessary yoke thickness, the radiation leakage from the magnets and the resulting dose in the tunnel was estimated by means of FLUKA Monte Carlo simulations [126, 127, 331]. In this very first study, the booster was modelled as a series of $11\,\mathrm{m}$-long dipoles with $30\,\mathrm{cm}$ drift spaces. The studies will have to be repeated in the future with a more realistic simulation model that includes quadrupoles and sextupoles; nevertheless, the results obtained provide a first estimate of the booster-induced radiation environment.

Figure 4.13 compares the booster and collider contributions to the annual dose in the arc tunnel for $\mathrm{t\bar{t}}$ operation. The simulation assumes 185 days of operation, with 75% machine uptime. The results for the collider assume a $400\,\mathrm{kg}$ lead-alloy shielding around the photon stoppers, with a thickness of $10\,\mathrm{cm}$ on the internal and external side of the magnets (see Section 1.9). For the booster, dipole iron yokes with thicknesses of $2\,\mathrm{cm}$ and $4\,\mathrm{cm}$ were simulated. As a conservative assumption, each booster cycle was modelled at maximum top-up intensity, corresponding to the highest bootstrap bunch charge. The results indicate that, with the current collider shielding, the dose contribution from the collider remains dominant.

At the location of the upper cable trays on the walls (above $2\,\mathrm{m}$ from the floor), the radiation dose from collider operation reaches $2-10\,\mathrm{kGy/year}$, while the booster contributes up to approximately $2\,\mathrm{kGy/year}$ when using $2\,\mathrm{cm}$-thick dipole yokes. To maintain the dose at the cable trays below $10\,\mathrm{kGy/year}$, the booster's contribution cannot be entirely neglected. Reducing booster-induced radiation at the walls to below $1\,\mathrm{kGy/year}$ in $\mathrm{t\bar{t}}$ operation can be achieved by increasing the booster dipole yoke thickness to $4\,\mathrm{cm}$.

The results presented indicate that the booster's contribution to overall dose levels in the tunnel is non-negligible and must be considered alongside collider shielding. Further studies are required for both the collider and the booster to optimise the overall shielding strategy.

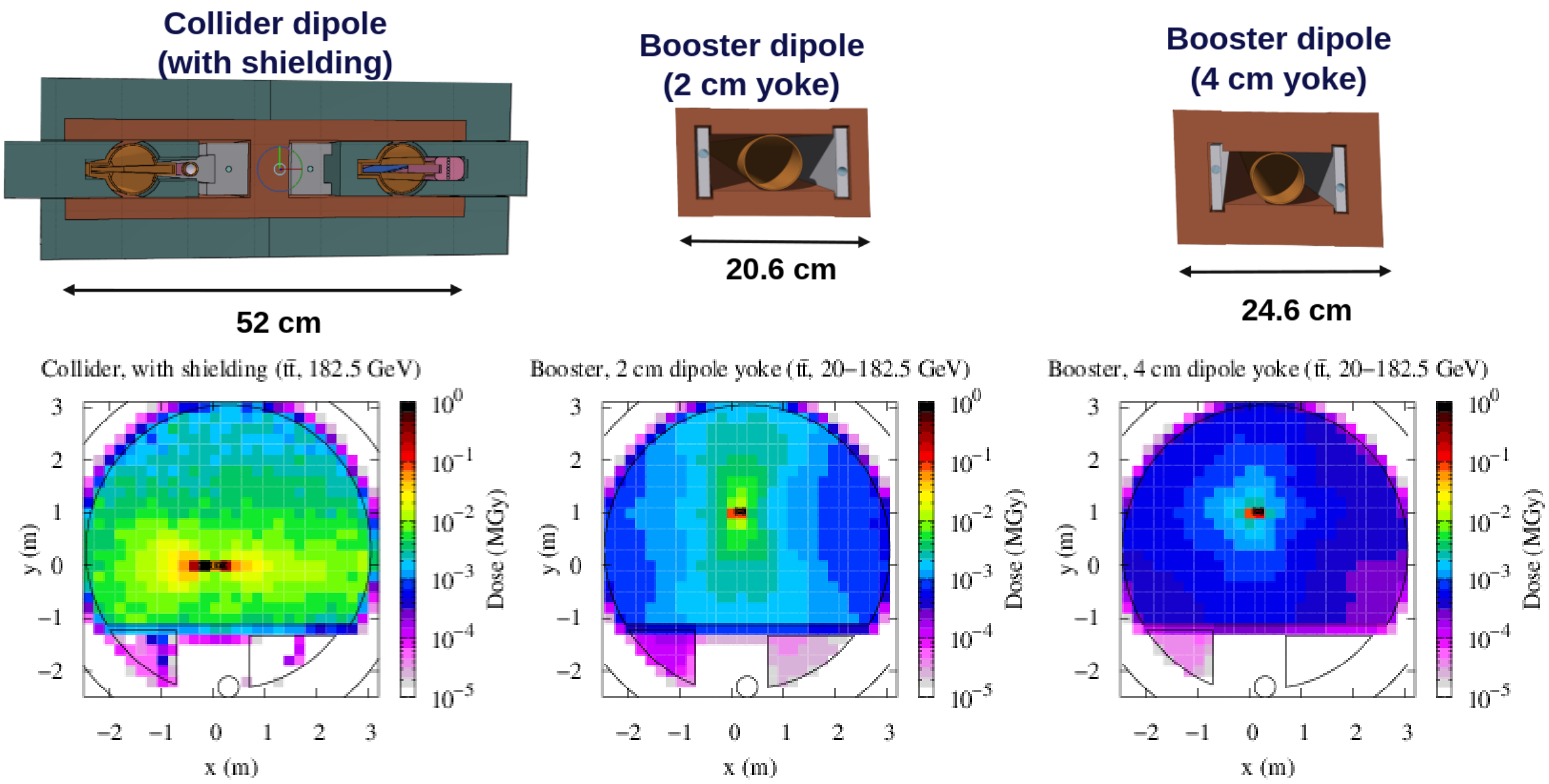

Fig. 4.13: Annual ionising dose in the arc tunnel during $t\bar{t}$ operation, showing separately the contributions of the collider (left) and the booster (middle: with a $2\,\mathrm{cm}$-thick yoke, right: with a $4\,\mathrm{cm}$-thick yoke). The collider simulations assume a preliminary lead-based shielding around discrete photon stoppers (see Section 1.9). The studies were carried out using the FLUKA code. The FLUKA magnet models are shown at the top of the figure.

## 4.4 Injection and extraction

### 4.4.1 Injection

The beam is accelerated to $20\,\mathrm{GeV}$ in the high energy linac and transported from the injector complex to the collider complex close to point PA, through the injection transfer lines shown with bright green colour in Fig. 4.1. The injection transfer-line tunnels merge with the collider tunnel on either side of the PA experiment straight section, approximately $600\,\mathrm{m}$ from the IP. This layout is symmetric around the interaction point (IP), and since the booster optics is also symmetric, the injection scheme described below for the clockwise direction applies equally to the counter-clockwise injection on the opposite side of the straight section.

At the junction with the collider tunnel, the current baseline design features an angle of approximately $150\,\mathrm{mrad}$ between the injection and booster beamlines. While the precise geometry of the injection line has not yet been finalised, the present concept envisions using approximately $10\,\mathrm{m}$ of dipoles to achieve a total deflection of $125\,\mathrm{mrad}$, followed by a septum system providing an additional $25\,\mathrm{mrad}$ deflection to align with the booster trajectory [332].

The design of the booster injection has primarily focused on the booster FODO lattice and on the injection elements. The detailed trajectory and optical matching from the injection line will be addressed at a later stage, with no significant challenges identified so far.

The beam produced by the injector complex is composed of up to 4 bunches separated by $25\,\mathrm{ns}$ per batch, whose parameters are detailed in Table 4.1. The booster accumulates the injected batches at a maximum rate of $100\,\mathrm{Hz}$ and for up to several seconds. The baseline uses a fast injection scheme, allowing consecutive batches to be injected between circulating ones. Therefore, the injection kicker has to provide a flattop of up to $80\,\mathrm{ns}$.

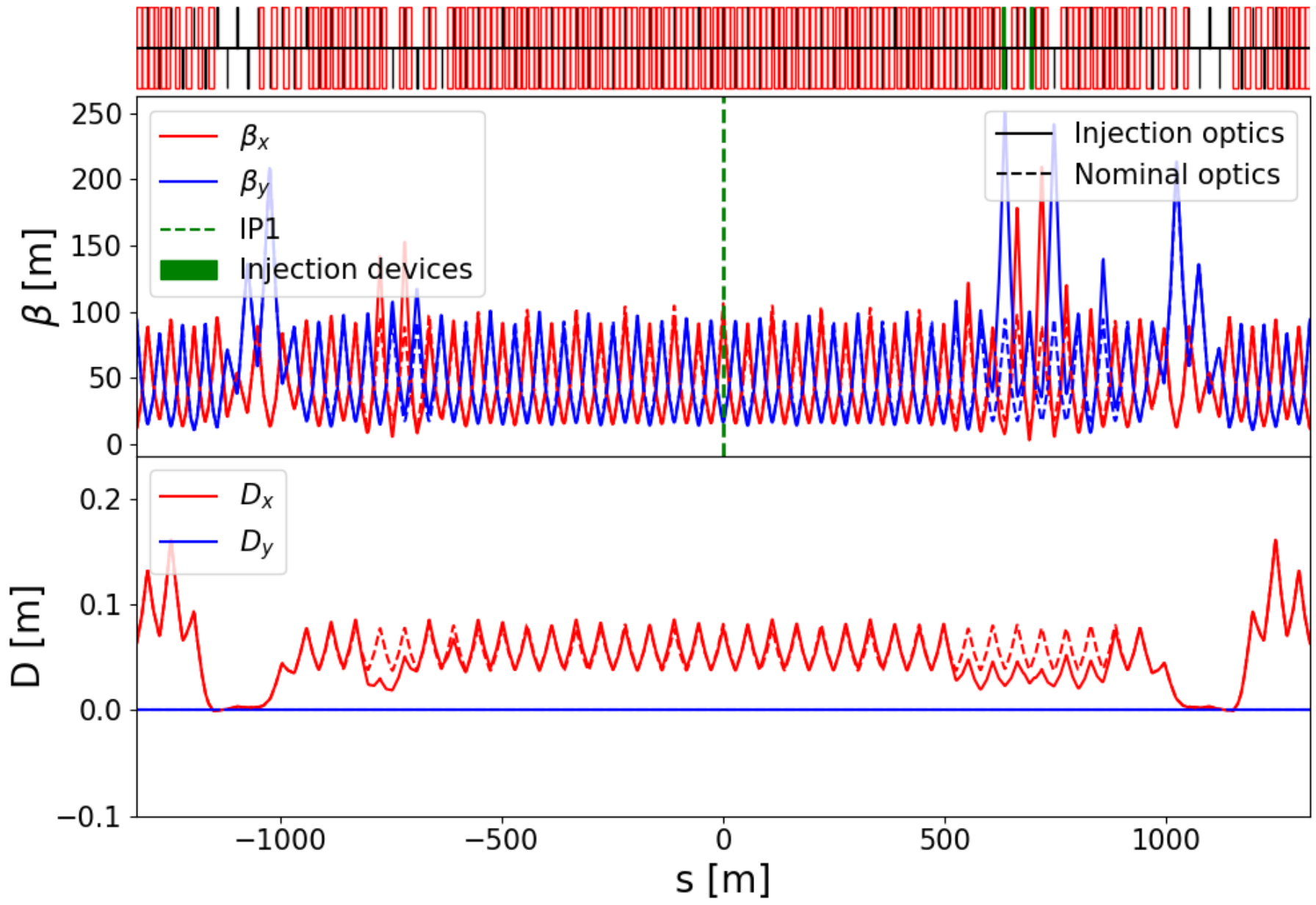

Fig. 4.14: Magnets layout, apertures, beam paths with envelopes and Twiss functions for the booster ring along the PA straight section with the injection optics on the positron injection side and the nominal optics on the electron injection side. The dipole position is shown in red, and the quadrupoles are depicted in black.

Taking advantage of the low field of the dipole magnets in the injection region close to an IP, where the average bending radius is twice as large as in the arcs (19.8 km), the current injection scheme removes selected dipoles and redistributes their deflection among the adjacent ones. In total, six dipole magnets are removed on each side of the PA section, with their deflection transferred to six neighbouring dipoles. This redistribution does not affect the reference trajectory outside each injection region and results in a negligible impact on the ring circumference (28 µm). The dipoles are removed symmetrically with respect to the FODO lattice to minimise optical perturbations, particularly in the dispersion function. Some minor effects on the local dispersion remain, as seen around the electron injection point in Fig. 4.14.

The second key requirement of the injection scheme pertains to the optical functions and phase advances in the vertical plane at the injection devices. A high beta function at both the kicker and the septa, combined with a phase advance close to $90^\circ$, enhances the kicker's efficiency and maximises the separation between the circulating and injected beams at the septum. To achieve this while maintaining symmetry in the lattice around the local FODO structure, two additional quadrupoles (`inj0` and `inj1`) and independent powering of $2 \times 11$ existing quadrupoles are introduced.

The resulting injection optics, illustrated for the positron injection side in Fig. 4.14, features a vertical beta function of up to 250 m at the septum. While this optics modifies the phase advance and the local sextupole correction scheme, a global correction allows both effects to be compensated [333]. Once the injection process is completed, at the start of the energy ramp the optics can adiabatically be reverted to the nominal configuration for the remainder of the booster cycle, thereby minimising the impact of the injection optics on the booster beam dynamics.

The placement of the injection elements and the threading of the injected beam through the lattice elements can be seen in the close-up of the injection region in Fig. 4.15. Following the injected envelope, a side channel in the horizontal plane will need to be installed between the poles of the quadrupole `065`. This seems feasible with the present quadrupole design in the grey part of the aperture, between 31.5 mm

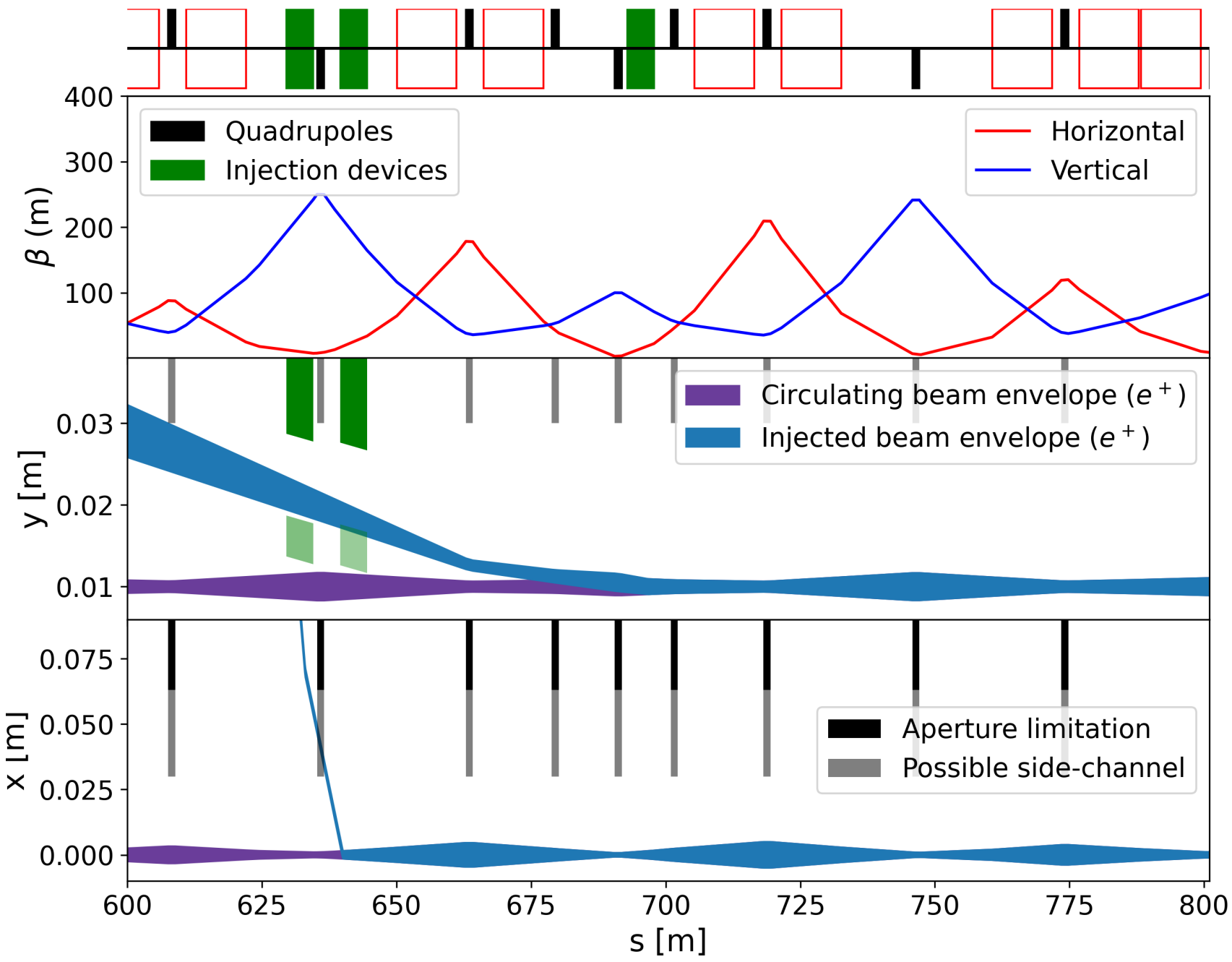

Fig. 4.15: Layout, apertures, Twiss functions, beam paths and envelopes in both planes around the clockwise injection point of the booster. Injected and circulating envelopes are defined by the $\pm 15\,\sigma$ beam size and the injected beam parameters.

and $65\,\mathrm{mm}$ [334].

In the vertical plane, a closed orbit bump of $10\,\mathrm{mm}$ at the septum is foreseen to bring the circulating trajectory closer to the injected one in order to reduce the required kicker strength. This bump is taken into account only as a constant vertical orbit shift in Fig. 4.15, but properly modeled in beam dynamics simulations. The inside edge of the septum is located $11.5\,\mathrm{mm}$ from the centre of the beam pipe and the injected beam is placed closer to the blade with an envelope of $\pm 3\,\mathrm{mm}$ to be compared with the blade thickness of $5\,\mathrm{mm}$ and the septum vertical aperture of $10\,\mathrm{mm}$. At the septum, the vertical slope of the injected beam is $260\,\mathrm{\mu rad}$ and reduced to $86\,\mathrm{\mu rad}$ after the quadrupole `066`. Downstream, a fast kicker provides an angle of $90\,\mathrm{\mu rad}$.

The injection optics features optimised beta function and phase advance, but one may notice that the vertical beta function at the kicker remains around $100\,\mathrm{m}$. The low vertical beta function at the kicker allows relaxing the requirements on the kicker pulse flatness [178]. In practice, an active damper of the injection oscillations is required but at this stage, only injection system and ring optics are considered.

Another source of injection offset may come from the high energy linac with up to $1\,\sigma$ jitter in each transverse plane [335]. In the longitudinal plane, the maximum relative energy jitter is $3 \times 10^{-3}$ (see Table 7.13). Such a jitter should not impact the aperture requirement at injection but active damping may need to be considered to avoid emittance growth.

This baseline injection scheme relies on a set of hardware requirements summarised in Table 4.2, considered achievable using existing technologies. A comprehensive discussion on the hardware choices is provided in Section 6. Further investigations are needed to advance this concept towards a technical design, focusing on error analysis and their impact, as well as the operational and performance optimisation of the scheme to maintain a reliable injection despite the various drifts that may occur.

Table 4.2: Summary of the booster injection scheme hardware requirements

| | Kicker | Septum |
|---|---|---|
| Beam energy (GeV) | 20 | |
| Deflection angle per system (mrad) | 0.09 | 4.5 |
| Maximum repetition frequency (Hz) | 100 | DC |
| Rise/Fall time (ns) | 25† | – |
| flattop time (ns) | 80 | – |
| Blade thickness (mm) | – | 7 |
| Aperture (H×V mm) | – | 5×10 |
| Longitudinal available space (m) | 5.5 | 20 |

†A 25 ns rise or fall time is not needed with the present filing schemes described in Section 2.3.8.

### 4.4.2 Extraction

The booster beam is extracted in the technical straight section in PB (see Fig. 4.1) towards the booster-to-collider beam transfer line. The structure of the circulating beam depends on the operation mode and the required filling scheme of both booster and collider as well as collective effects considerations in both synchrotrons [336]. However, the present concept is based on the most stringent requirements of a fast extraction of a full turn of 304 µs, with a rise time of the kicker system within the planned minimum collider gap between trains of 600 ns [2].

Unlike the experiment sections, the PB technical straight section of the booster is physically straight and consists solely of quadrupoles. Due to the ample space available between these quadrupoles, no modifications to the layout are required to accommodate the extraction elements.

To facilitate the extraction of both electron and positron beams, a symmetric design is preferred. Since the collider injection system is positioned in the second half of the straight section (see Section 1.8.1), the booster extraction system is placed at the centre of the section. Figure 4.16 illustrates the layout along the entire straight section, where the extraction scheme is centrally located and symmetrically configured to extract both beams towards the section's exit.

The extraction scheme operates entirely in the horizontal plane, with both the kickers and septa providing a horizontal deflection.

The nominal optics follows a FODO structure without dispersion and features a strong beating over a period of two cells (see Fig. 4.16). The large beta function and near 90° phase advance per cell already match the typical requirements of a fast extraction system. However, the optics was adjusted to relax the requirement on the kicker pulse flatness to limit the extracted beam offset to $1\,\sigma$. This extraction optics uses 9 independent power supplies, but no additional magnets will be needed. The injection optics is required only for a short time before extraction without perturbing the rest of the booster cycle.

The extraction layout is fully symmetric for electron and positron beams, with the kickers placed at the precise centre of the straight section on either side of the focusing quadrupole 022 (see Fig. 4.17a). Although the present concept considers dedicated kickers for electron and positron beam extraction, this placement at the centre of the straight section could allow the reuse of the same kickers for the extraction of both beams.

The circulating beam is moved closer to the extraction septum using a closed-orbit bump of 10 mm, although it is presently only modelled as a shift of the reference trajectory in Fig. 4.17a. The kicker system provides a total angle of $0.2\,\mathrm{mrad}$ to ensure the required clearance at the septum blade located upstream of the following focusing magnet. In this scheme, the septum blade has a thickness of $8\,\mathrm{mm}$, with its inside edge located $20\,\mathrm{mm}$ from the reference position of the booster beam. Downstream of the septum, the extracted beam envelopes pass through the quadrupole between its poles, within a side channel that will need to be installed—similar to the booster injection setup (Section 4.4.1).

[2]For the baseline bootstrapping injection and filling scheme described in Section 2.3.8, the gaps in the booster are significantly larger than 600 ns.

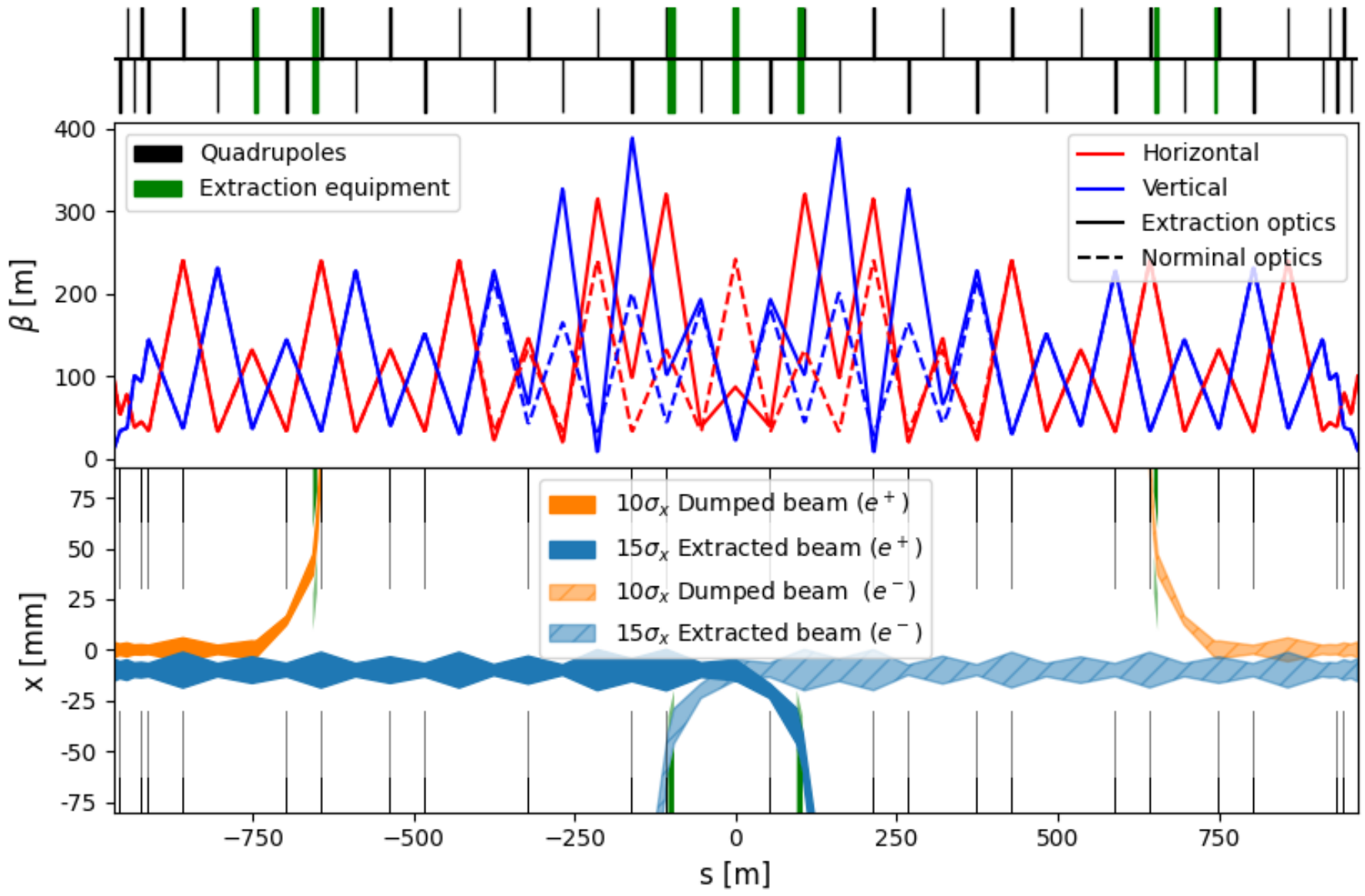

Fig. 4.16: Layout, apertures, beam envelopes, and Twiss functions for the booster ring along the PB straight section, showing the extracted and stored beam. The envelopes are computed using the largest equilibrium beam parameters (reached for $\mathrm{t\bar{t}}$-mode).

Table 4.3: Summary of the booster extraction scheme hardware requirements

| | Kicker | Septum |
|---|---|---|
| Beam energy (GeV) | 45.6 – 182.5 | |
| Deflection angle per system (mrad) | 0.2 | 2 |
| Maximum repetition frequency (Hz) | 0.3 | 0.3 |
| Rise/fall time (μs) | 1.1 | N/A |
| flattop time (μs) | 304 | N/A |
| Blade thickness (mm) | N/A | 8 |
| Aperture (H×V mm) | N/A | 18×10 |
| Longitudinal available space (m) | 15 | 15 |

The hardware requirements for this extraction scheme are summarised in Table 4.3. This concept is based on realistic hardware specifications achievable with current technologies. A detailed discussion of the hardware choices is provided in Section 6. Comprehensive error studies and failure mode analyses are required to advance the present concept toward a technical design.

Passive absorbers must be installed after the extraction kickers to protect the septum and the most exposed downstream components from potential failures that could result in miskicked bunches striking the booster aperture. Likewise, a system of masks should be implemented in the transfer lines to the collider to intercept particles with large oscillations, preventing them from reaching the collider aperture and shielding the transfer line itself.

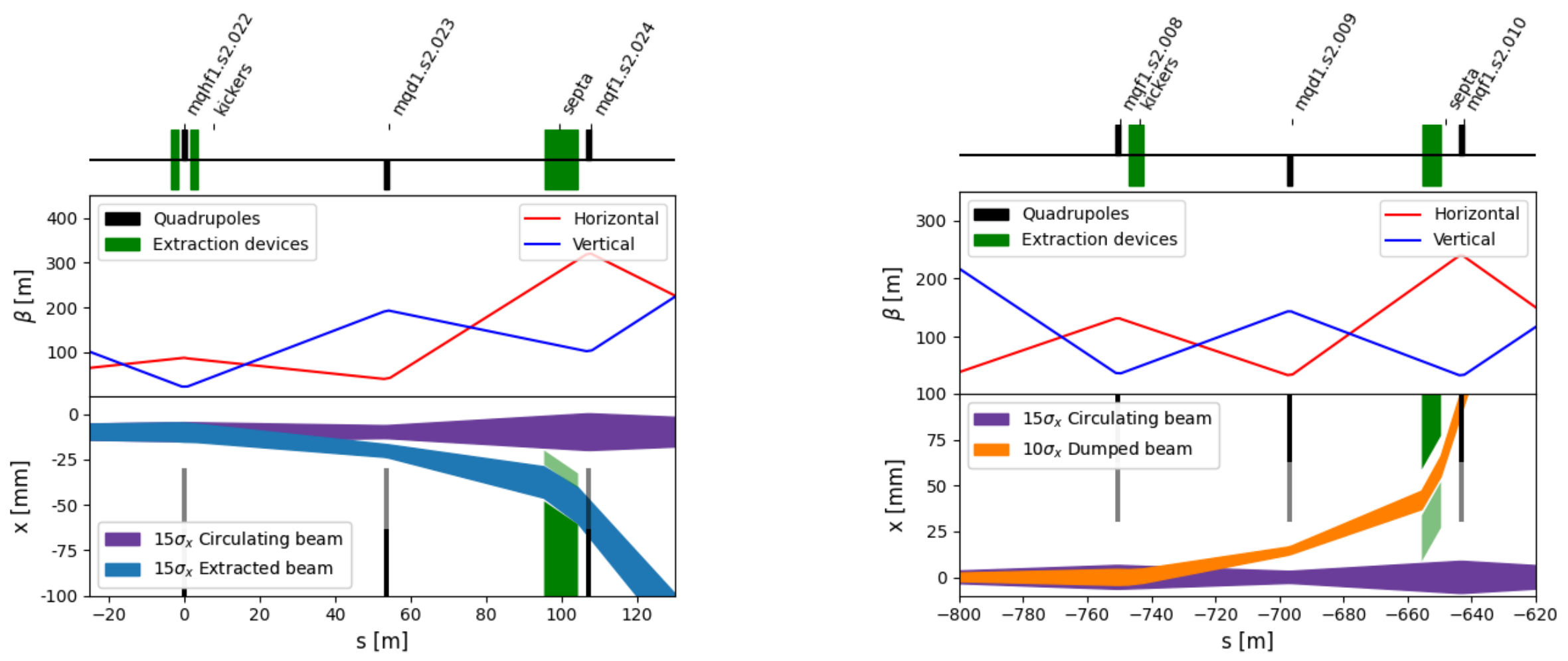

(a) (b)

Fig. 4.17: Magnet layout, Twiss functions, beam paths and envelopes in the horizontal plane for the booster extraction (a) and dump (b). The envelopes are computed using the largest equilibrium beam parameters (reached for $t\bar{t}$-mode).

### 4.4.3 Dump

The booster transfer dump system closely resembles the collider dump system described in Section 1.8.1. It is positioned at approximately the same longitudinal location (see Fig. 4.16) and, like the collider system, extracts the beam at the entrance of the straight section towards its centre. The system must rise within the minimum possible gap between trains of $600\,\mathrm{ns}$ (or larger, for the present filling schemes) and extract the entire beam in a single turn. This necessitates a fast extraction scheme with a flattop duration of $304\,\mu\mathrm{s}$, with both the kicker and septa designed to provide horizontal deflection.

Unlike injection or extraction systems, a dump may be required anytime. As a result, the system must be capable of safely extracting the circulating beam within a few turns. This requirement means that the concept cannot depend on manipulating the optics or the closed orbit before the kicker system is triggered. Consequently, the optics in the extraction region remains unchanged, and the reference position of the circulating beam stays at the centre of the vacuum chamber.

The kicker system provides a slightly larger deflection angle of $0.3\,\mathrm{mrad}$ compared to the extraction scheme. Further downstream, the septum is positioned upstream of the next focusing quadrupole (see Fig. 4.17b) to take advantage of the approximately $90^\circ$ phase advance per cell and the larger beta function. At the septum, the dumped beam is shifted by a total of $43\,\mathrm{mm}$, allowing for a septum blade thickness of $25\,\mathrm{mm}$ and placing its inner edge $4\,\mathrm{mm}$ from the ring reference position. The total septum deflection is $5\,\mathrm{mrad}$, and the downstream quadrupole will need to be adapted to accommodate a side channel for the dumped beam.

Since the circulating beam energy varies continuously throughout the booster cycle, both the septum power supply and the kicker charging system must adjust to match the beam energy, from injection at $20\,\mathrm{GeV}$ up to the maximum energy of $182.5\,\mathrm{GeV}$. Additionally, specialised control systems will be required to trigger a dump if any of these elements fails to follow their expected voltage or current profiles.

The transport line to the dump consists solely of a drift tube, without dipoles or quadrupoles. Small correctors may be considered at a later stage to fine-tune the beam trajectory. With a total distance of $1200\,\mathrm{m}$ from the dump kicker to the dump, the beam naturally expands to a $1\,\sigma$ size of $3.5\,\mathrm{mm}$ in the horizontal plane and $1\,\mathrm{mm}$ in the vertical plane. These beam sizes are calculated for the smallest possible beam parameters, corresponding to the design equilibrium emittance in Z-mode.

For dump system, reliability is a critical factor, and one particular aspect must be addressed from the earliest stages of development. The total deflection provided by the kicker system is typically distributed across multiple independent magnets and generators, but the failure of at least one component is always a possibility. Therefore, it is essential to ensure that beam extraction and transport to the dump remain functional even in the event of a missing kicker.

In the current design, a total of nine kickers are used, meaning that the loss of a single unit results in an overall deflection reduction of $11\,\%$. The present concept guarantees that an extraction kick variation of up to $\pm 20\,\%$ keeps the dumped beam envelope at the dump location within $\pm 150\,\mathrm{mm}$, which is considered acceptable for the dump's size and design.

Additionally, to accommodate a $\pm 20\,\%$ variation in kick strength, the horizontal aperture of the septum must exceed $30\,\mathrm{mm}$.

Table 4.4: Summary of the booster dump scheme hardware requirements

| | Kicker | Septum |
|---|---|---|
| Beam energy (GeV) | 20 – 182.5 | |
| Deflection angle per system (mrad) | 0.3 | 5 |
| Maximum repetition frequency (Hz) | 0.3 | 0.3 |
| Rise/fall time (μs) | ~1.5 | N/A |
| flattop time (μs) | 304 | N/A |
| Blade thickness (mm) | N/A | 25 |
| Aperture (H×V mm) | N/A | 30×10 |
| Longitudinal available space (m) | 20 | 20 |

The hardware parameters required for this dump scheme are summarised Table 4.4 and a detailed discussion on the technology choice of the hardware system can be found in Section 6.4.

To develop this dump concept towards a technical design, a comprehensive review of the failure cases and mitigation measures, as well as possible errors in both the circulating beam and the dump line, will need to be carried out. Similar considerations as those presented in the previous section hold for protecting the booster aperture in case of failure of the dump extraction elements.

## 4.5 Ongoing studies and possible upgrades

Since the CDR, the booster has faced several major changes:

- The layout has changed to follow the different changes on the collider layout (see Section 4.1.1: the circumference has decreased; the length of the insertions has changed to go from a scheme with 2 IPs to 4 IPs; the booster is located on top of the collider with a transverse offset in the arcs; the location of the RF cavities has changed.
- The baseline optics is based on FODO cells (Section 4.1.3. Maintaining the same optics for the different operation modes may reduce the costs. An alternative, based on hybrid cells (Section 4.1.4) has been developed and compared. Improving the matching conditions, including second-order considerations, has enlarged the momentum acceptance and dynamic aperture.
- The injection and extraction systems were developed. The placement of the transfer lines from the pre-injector complex to the booster was optimised (Section 4.4.1). The extraction and dump lines are located in section B (see Section4.4.2). A baseline of the injection and extraction sections exists and this has been integrated in the booster FODO lattice. The baseline injection scheme requires a slight modification of the PA section by changing a few dipoles, adding a couple of quadrupoles and by adding 22 independent power supplies for existing quadrupoles. The injection and extraction sections are able to transport the required envelopes of the circulating and injected/extracted beams. The machine protection at extraction is highly challenging due to the high stored energy and has pushed the reduction of the total number of bunches at Z and WW operation. The hardware required for the injection, extraction, and dump have been listed.

- The collective effects (see Section 4.2) have been studied for the updated optics and considered different sources of instabilities: mismatching at injection, microwave, transverse mode coupling, transverse and longitudinal coupled bunches. The studies have shown that present baseline injection parameters from the high-energy LINAC allow a significant margin to avoid mismatch and microwave instabilities at injection energy (see Section 4.2.2). Studies on the coupled bunch instabilities have shown that the new parameters (especially the reduced number of stored bunched at Z-operation and the enlarged beam pipe) increase the threshold (see Section 4.2.3). Nevertheless, a transverse damper (although with relaxed damping time) is still mandatory. Single-bunch instabilities were investigated for copper and stainless steel beam pipes of several diameters as well as for different values of momentum compaction. The analysis has led to the choice of a copper beam pipe with a diameter of $60\,\mathrm{mm}$ as the baseline. With such parameters and with only the beam pipe as an impedance contributor, the beam is stable for all operating modes with the baseline optics.
- The synchrotron radiation power emitted from the booster during one cycle has been evaluated (see Section 4.3.1) to estimate the ionising dose in tunnel (see Section 4.3.2). The studies have shown that the contribution of the booster to the overall dose levels in the tunnel has to be studied in conjunction with the collider shielding. A recommendation is to increase the thickness of the booster dipole yokes from $20\,\mathrm{mm}$ to $40\,\mathrm{mm}$ for better shielding efficiency.

There has been a lot of progress, and many changes have been made to the booster design since the CDR. The current booster performance fulfills most of the requirements. Several additional studies remain to be done for a complete validation of the booster design. Among the different topics to be addressed are the following :

- The placement of injection elements for the vertical injection scheme and threading of the injected beam through the lattice elements is particularly challenging. It is necessary to undergo detailed studies on the feasibility of the magnets and proposed channels for the injected beam. Some technical solutions are proposed for the septa and kickers and could be developed towards a technical design but could also benefit from further R&D to apply newer technologies. The apertures and strengths used in the extraction and dump baseline concept are capable of transporting the envelopes of the circulating and extracted beams, but a more detailed study, including kicker technology choices and filling scheme requirements, will be needed to finalise the requirements and converge on the technology choices. Despite the reduced beam intensity in the booster, such beams remain extremely critical and potentially destructive. Additional measures must be considered to ensure the safety and integrity of all machine components in case of failure during the extraction and transfer processes. It is necessary to develop a dedicated system to monitor the position of the beam in the extraction and dump regions, to evaluate the required reaction time to detect anomalies and safely dispose of the beam, and to have interlocks to avoid drifts beyond well-defined limits.
- According to studies carried out so far, collective instabilities will not be a imitation with nominal beams in the HEB with the foreseen larger Cu coated chamber and a transverse damper for TCBI. Nevertheless, the studies have to be refined by taking all impedance sources into account, and the instability thresholds need to be confirmed by self-consistent simulations, which include possibly compounding effects, like intrabeam scattering and synchrotron radiation.
- To further validate the booster dipole design, the next step is to produce and test a short prototype. The optimisation of the magnets will be pursued in order to include in the process: (1) the yoke thickness of the dipole to provide additional radiation shielding, (2) the operational temperature of the magnets to improve the efficiency of waste heat recovery, (3) the industrialisation and design for manufacture and assembly, and (4) detailed integration with the other technical systems.

- The results presented demonstrate that the contribution of the booster to the overall dose levels in the tunnel must be studied in conjunction with the collider shielding. In the next steps, it will be necessary to perform further investigations in order to optimise the overall shielding strategy.

# Chapter 5

# FCC-ee booster operation concept

## 5.1 Operation and performance

### 5.1.1 Filling scheme

Following the 2023 mid-term review of the FCC Feasibility Study, a revised proposal for the injectors introduces a Linac that produces and accelerates four pulses with a 25 ns spacing. The repetition frequency is set to 100 Hz for the Z and WW operation modes, and 50 Hz for the ZH and $t\bar{t}$ modes. The damping ring now operates at 2.86 GeV and accommodates both electrons and positrons. The evolution of the damping ring (DR) bunch pattern is sketched in Fig. 5.1 for the old (left) and new (right) injector scheme.

This revised approach is now the new baseline, with the updated collider filling scheme fully detailed in Section 2.3.7. The booster-filling scheme has been adapted to follow this new collider-filling pattern.

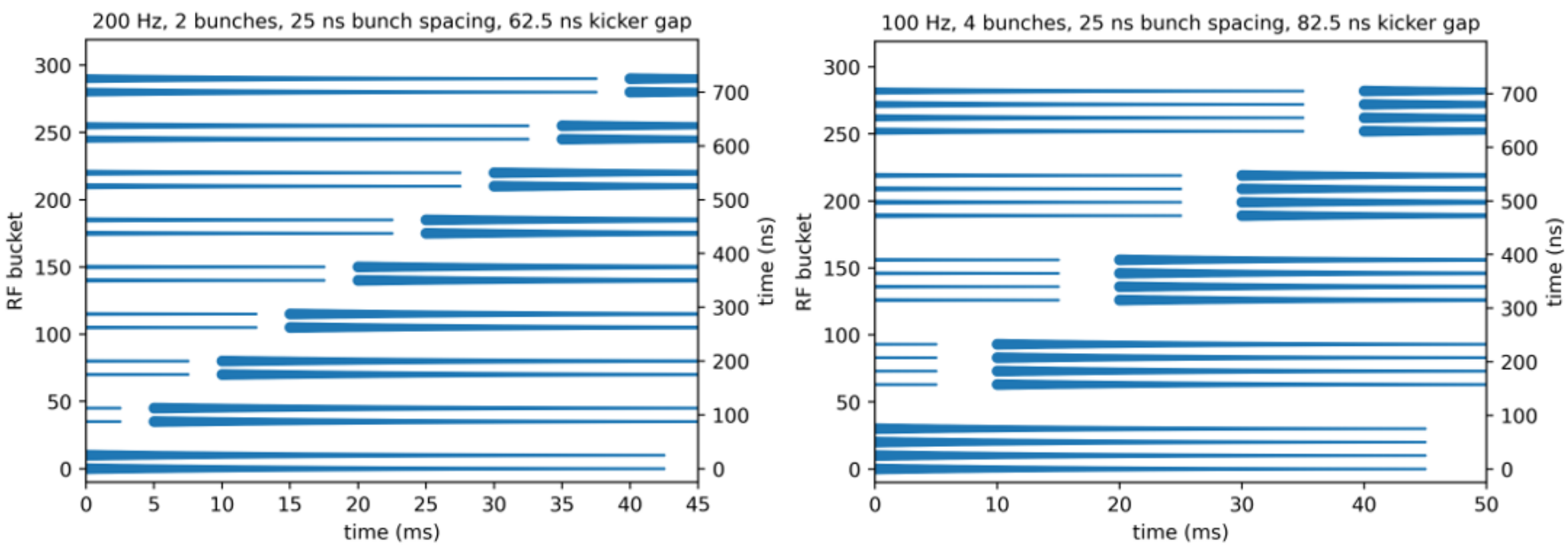

Fig. 5.1: Pulse structure in the damping ring before injection into the linac of 20 GeV for an operating frequency of 200 Hz (left) or 100 Hz (right).

For the Z mode, the new scheme implies that each booster cycle provides beam to only one tenth of the collider bunch positions. The smaller number of bunches in each booster cycle has several advantages:

– Machine protection considerations only allow for about 1/10th of all bunches, i.e., 1120 (each with max. 1/10th of nominal collider bunch intensity) injected into collider at once at Z energy.
– The shorter booster injection plateau relaxes the vacuum quality required for tolerable emittance growth from rest gas collisions and reduces the impact of intra-beam scattering (IBS) and, thus, emittance growth on early injected bunches.
– It allows the bunches to be distributed around the booster circumference. This relaxes the fast beam ion instability and tune shifts for the electron beam, as well as the RF power required for beam loading compensation. This distribution can also be exploited to optimise the filling of the collider to mitigate e-cloud for positrons.

However, the main drawback is the increase in the number of ramps, which in turn affects the time required for the top-up of all bunches (the effective cycle length), as illustrated in Fig. 5.2. In the new scheme, the goal is to inject up to 1/10th of all collider bunches per booster cycle. c A consequence of the new scheme that either the accelerating ramp in the booster must be shortened (for operational margin), as assumed in Table 5.1, or the full collider top-up duration length be slightly increased (e.g., the total ramping time in the Z mode rising from 1.0 s to 1.14 s).

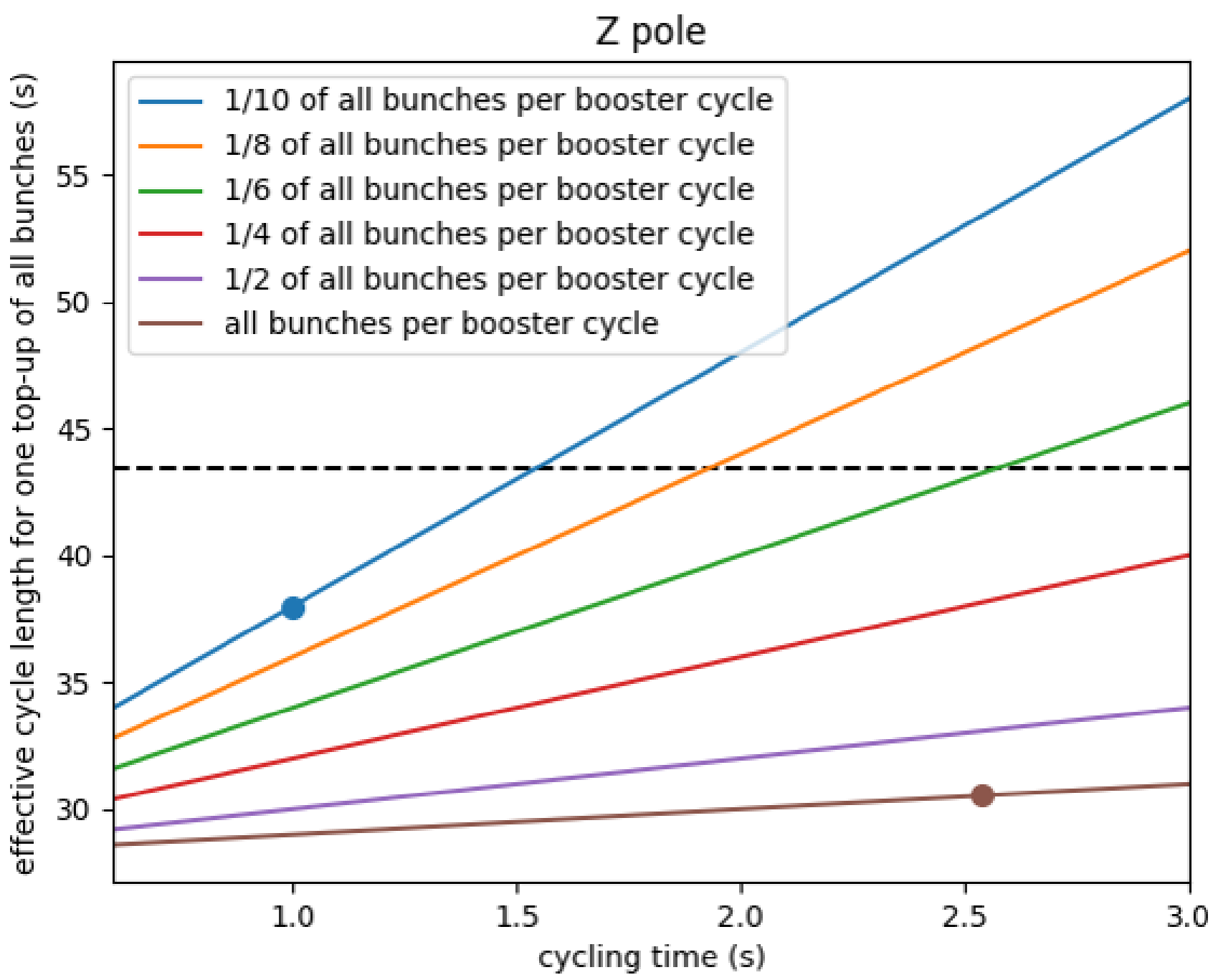

Fig. 5.2: Effective cycle time as a function of the ramping time and the number of bunches transferred per booster cycle. The brown dot corresponds to the scheme in presented in the mid-term review and the blue dot is for the new scheme.

An example of the filling pattern in the booster is shown in Fig. 5.3. It is worth noting that for Z operation, it is possible to choose which subset of bunches is injected from the booster to the collider. Such a scheme provides flexibility in the accumulation phase of the collider for the creation of gaps in the bunch train and possibly mitigates the electron cloud effect.

The three operation modes run with a smaller number of bunches and lower bunch intensity compared to Z. The low bunch intensity needed from the injector ($< 1 \times 10^{10}$) resolves transverse coupled mode instabilities (TMCI) limitations in the booster (see Section 4.2). The three operation modes allow longer abort gaps, partly occupied by non-colliding bunches for energy calibration. There is still a margin for reducing the injection rate (i.e., pausing between cycles) for energy saving in top-up mode or increased booster cycle length (e.g., if beam injection in the collider needs to be done piecemeal). For WW operation, the transfer of half of the bunches at each injection is considered to limit the length of the injection plateau. For the ZH and $t\bar{t}$ operation, one requirement for the filling scheme is to always have only one beam present in the common RF section at a time. The number of bunches is small enough to run the injector complex at 50 Hz with two bunches.

### 5.1.2 Emittance evolution

The transverse damping time is respectively 9 s, 0.763 s, 0.141 s, 0.0419 s and 0.0119 s at the injection energy of 20 GeV and at the extraction energy of the 4 operating modes, respectively. The direct consequence is that the damping time at Z operation is of the same order of magnitude as the total ramping

Table 5.1: Filling parameters

| | | Z | WW | ZH | $t\bar{t}$ |
|---|---|---|---|---|---|
| Linac repetition rate | [Hz] | 100 | 100 | 50 | 50 |
| Bunches per Linac pulse | | 4 | 4 | 2 | 2 |
| Linac bunch spacing | [ns] | 25 | 25 | 25 | 25 |
| Booster accumulation time | [s] | 2.8 | 2.32 | 3.0 | 0.64 |
| Booster total ramping time | [s] | 1.0 | 1.6 | 2.6 | 4.3 |
| Booster cycle length | [s] | 3.8 | 3.92 | 5.6 | 4.94 |
| Bunches per booster cycle | | 1120 | 928 | 300 | 64 |
| Number of bunches in collider | | 11 200 | 1856 | 300 | 64 |
| Max. bunch intensity injected in collider | $10^{10}$ | 2.725 | 1.268 | 1.268 | 1.268 |
| Nominal bunch intensity in collider | $10^{10}$ | 21.5 | 13.8 | 16.9 | 14.8 |
| Allowable charge imbalance | [%] | 5 | 3 | 3 | 3 |
| Beam lifetime: lumi 4 IPs, (q,BS,lattice)/4 | [s] | 916 | 517 | 428 | 497 |

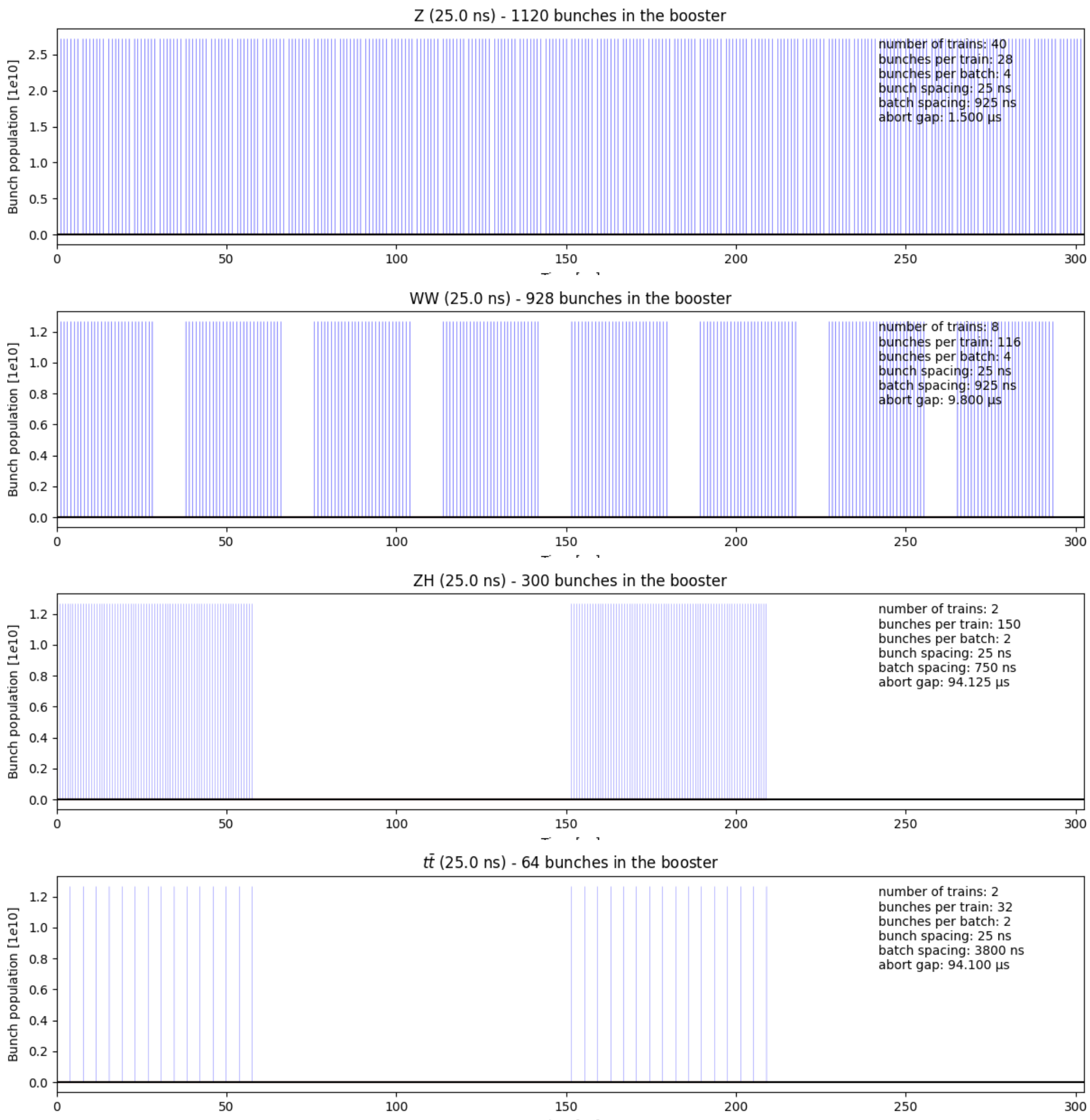

Fig. 5.3: Filling pattern for the different operation modes (from top to bottom): Z, WW, ZH, and $t\bar{t}$. For the Z mode, 4 bunch batches are separated by longer gaps, which is not visible on this scale.

time given in Table 5.1 whereas the damping time is small for the other modes. Therefore, the emittance and energy spread at the extraction at the Z mode will depend on the initial injection parameters, whereas the equilibrium emittance will be reached for the other modes. For this reason, only the Z mode has been considered for assessing whether the extracted emittance remains within the collider injection tolerances. This study examines two injection scenarios:

**Case 1 (Linac alone for $20\,\mathrm{GeV}$ electrons)** $\varepsilon_x = 10\,\mu\mathrm{m}$ , $\varepsilon_y = 10\,\mu\mathrm{m}$ , $\sigma_\delta = 1 \times 10^{-3}$.

**Case 2 (Baseline: Damping ring)** $\varepsilon_x = 20\,\mu\mathrm{m}$ , $\varepsilon_y = 2\,\mu\mathrm{m}$ , $\sigma_\delta = 1 \times 10^{-3}$

The equilibrium emittance in the collider for Z operation, $\epsilon_{x,\mathrm{RMS}}$, is $0.71\,\mathrm{nm} \times \epsilon_{y,\mathrm{RMS}} = 1.9\,\mathrm{pm} \times \sigma_\delta = 1.09 \times 10^{-3}$. A first study on the injection into the collider suggests that the collider can accept up to 5 times larger vertical emittance. That is why the target for the emittance at extraction, $\epsilon_{x,\mathrm{target}}$, is $0.71\,\mathrm{nm} \times \epsilon_{y,\mathrm{target}} = 9.4\,\mathrm{pm}$. However, since the equilibrium horizontal emittance in the booster is lower than in the collider for the Z and WW modes, the expected horizontal emittance at extraction is about $0.12\,\mathrm{pm}$ (see Fig. 5.4).

The baseline energy ramp has a parabolic increase from $0 – 80\,\mathrm{GeV\,s^{-1}}$, a linear ramp up to a maximum energy higher than the extraction one (to speed up the radiation damping), a decrease back to the extraction energy, a flat-top, and finally a ramp-down of the magnets. The current total time of the booster ramp is $1.14\,\mathrm{s}$, near the target of $1\,\mathrm{s}$ and can be shared between a ramp-up of $0.706\,\mathrm{s}$, a flat-top of $0.1\,\mathrm{s}$, and a ramp-down of $0.334\,\mathrm{s}$. It is worth noting that the maximum slope for the magnetic field in the dipole in the ramp-up is the same as in the ramp-down. However, it is likely that it will be possible to have a faster ramp-down since no beam is circulating during this time. The stability constraints and beam loading in the RF cavities are not a concern during the ramp-down. Under these conditions, it is possible to reach a total ramp time of $1\,\mathrm{s}$ with a ramp-down of $0.194\,\mathrm{s}$.

The evolutions of the beam energy, the radiated power, the minimum RF voltage, the horizontal and vertical RMS emittance of the energy spread are given in Fig. 5.4 for the baseline. The maximum energy delivered by the cavities is $90.6\,\mathrm{MeV/turn}$. In case 1, the final RMS vertical emittance is $30\,\mathrm{pm}$, well above the target of $9.4\,\mathrm{pm}$. In case 2, the final vertical RMS emittance is $6.12\,\mathrm{pm}$, which is above the equilibrium emittance in the collider but within the injection acceptance.

If the ramp does not include an energy overshoot but maintains the same up-ramp, flat-top, and down-ramp durations with a parabolic profile, the maximum energy delivered by the cavities is reduced to $51.7\,\mathrm{MeV/turn}$. In this scenario, the final RMS vertical emittance reaches $55.4\,\mathrm{pm}$ and $11.2\,\mathrm{pm}$ in both cases—significantly exceeding the target value of $9.4\,\mathrm{pm}$. This highlights the clear advantage of increasing the energy to accelerate damping and achieve lower emittance values. Several mitigation strategies have been evaluated to enhance damping during the ramp and achieve a smaller vertical emittance while maintaining a total ramping time of $1.14\,\mathrm{s}$. In all cases, accelerating damping requires an increase in radiated power, which in turn necessitates a higher cavity voltage.

Three solutions are proposed to achieve a final smaller vertical emittance. The first possibility is to increase the time allocated to the ramp-up by decreasing the time of the ramp-down. For instance, if the ramp-down time is reduced by $170\,\mathrm{ms}$, the ramp-up time increases to $876\,\mathrm{ms}$. This also enables reaching a higher maximum energy, thereby accelerating damping due to increased power consumption. The maximum energy delivered by the cavities is then $130.0\,\mathrm{MeV/turn}$. The final vertical RMS emittance is $9.27\,\mathrm{pm}$ and $1.99\,\mathrm{pm}$ for the two cases, respectively.

The second possibility is to achieve a higher energy by increasing the slope of the magnetic field from $80 – 100\,\mathrm{GeV\,s^{-1}}$, for instance. In this case, the maximum energy delivered by the cavities is $158.5\,\mathrm{MeV/turn}$. The final vertical RMS emittance is $9.1\,\mathrm{pm}$ and $1.96\,\mathrm{pm}$ for the two cases, respectively. The third proposal is to insert a wiggler in one of the straight sections.

For these studies, the following parameters were assumed: a number of wigglers, $n_W = 2$, each $4.925\,\mathrm{m}$ long, with $n_P = 43$ poles, each $L_P = 95\,\mathrm{mm}$ long, a magnetic field in the gap of $B_W = 1\,\mathrm{T}$, and a gap between the poles of $20\,\mathrm{mm}$. The wiggler parameters are preliminary and require further refinement. The aim in the framework of this study was to obtain an order-of-magnitude estimate of what is needed.

The parameters of the wiggler were calculated to go from $\mathcal{I}_2 = 0.59\,\mathrm{mm}^{-1}$ to $\mathcal{I}_2 = 2.36\,\mathrm{mm}^{-1}$ and from $\mathcal{I}_3 = 0.057\,\mathrm{mm}^{-2}$ to $\mathcal{I}_3 = 27\,\mathrm{mm}^{-2}$ at an injection energy of $20\,\mathrm{GeV}$. The evolution of the beam energy, voltage, and beam emittance with an additional wiggler is given in Fig. 5.4. The emittance evolution takes into account that $\mathcal{I}_\in$ and $\mathcal{I}_\ni$ vary in time since it is assumed that there is a constant magnetic field in the wiggler. The maximum energy delivered by the cavities is then $122.5\,\mathrm{MeV/turn}$. The final RMS vertical emittance for the baseline case is $1.84\,\mathrm{pm}$. Having a larger $I_3$ significantly enlarges the final energy spread to $1.9 \times 10^{-3}$, which is above the requirements. An optimisation of the wiggler parameters should enable a smaller final energy spread to be achieved.

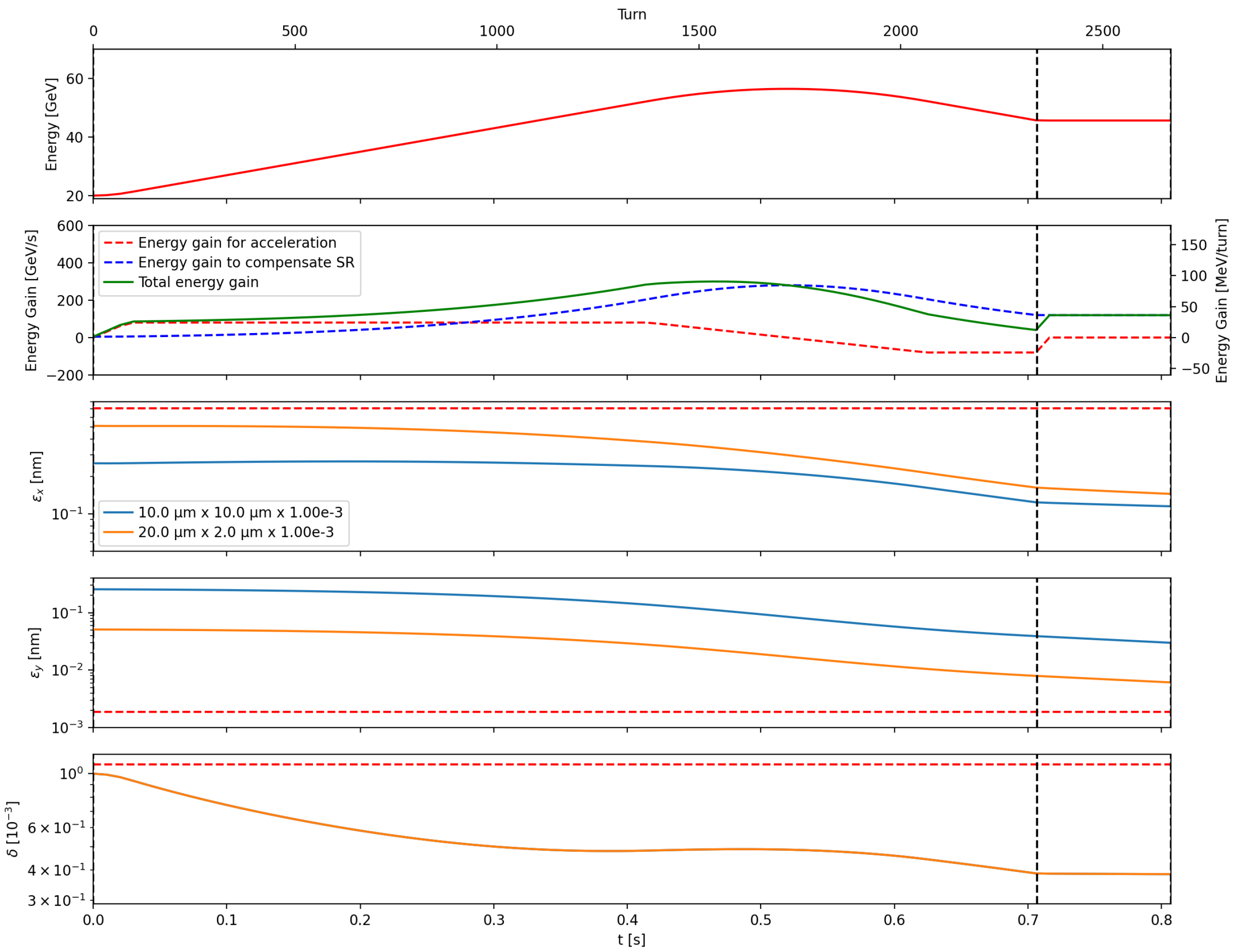

Fig. 5.4: Evolution of the beam energy (first line), radiated power and RF voltage required (second line), the horizontal RMS emittance (third line), vertical RMS emittance (fourth line), and RMS energy spread (last line) as a function of time for Z mode operation.

In summary, achieving the required final vertical emittance is only possible with the baseline scenario if two conditions are met: using a high-energy damping ring and maintaining a normalised vertical emittance of $2\,\mathrm{pm}$ throughout the linac. However, at the price of more RF power, it is possible to reach the target vertical emittance with a ramp of $1\,\mathrm{s}$ if it is possible to combine a shorter ramp-down with a wiggler. Increasing the acceleration rate to reach higher energies more quickly can be beneficial, but caution is needed to avoid excessively high energies, where the radiated power will increase significantly, leading to high power consumption.

### 5.1.3 Ramping strategy

Following an accumulation time, the booster will ramp the beam energy from the injection energy ($20\,\mathrm{GeV}$) up to extraction energy ($45.6\,\mathrm{GeV}$ for Z, up to $182.5\,\mathrm{GeV}$ for $\mathrm{t\bar{t}}$). The optimisation strat-egy has to ensure beam stability and reach the target emittances.

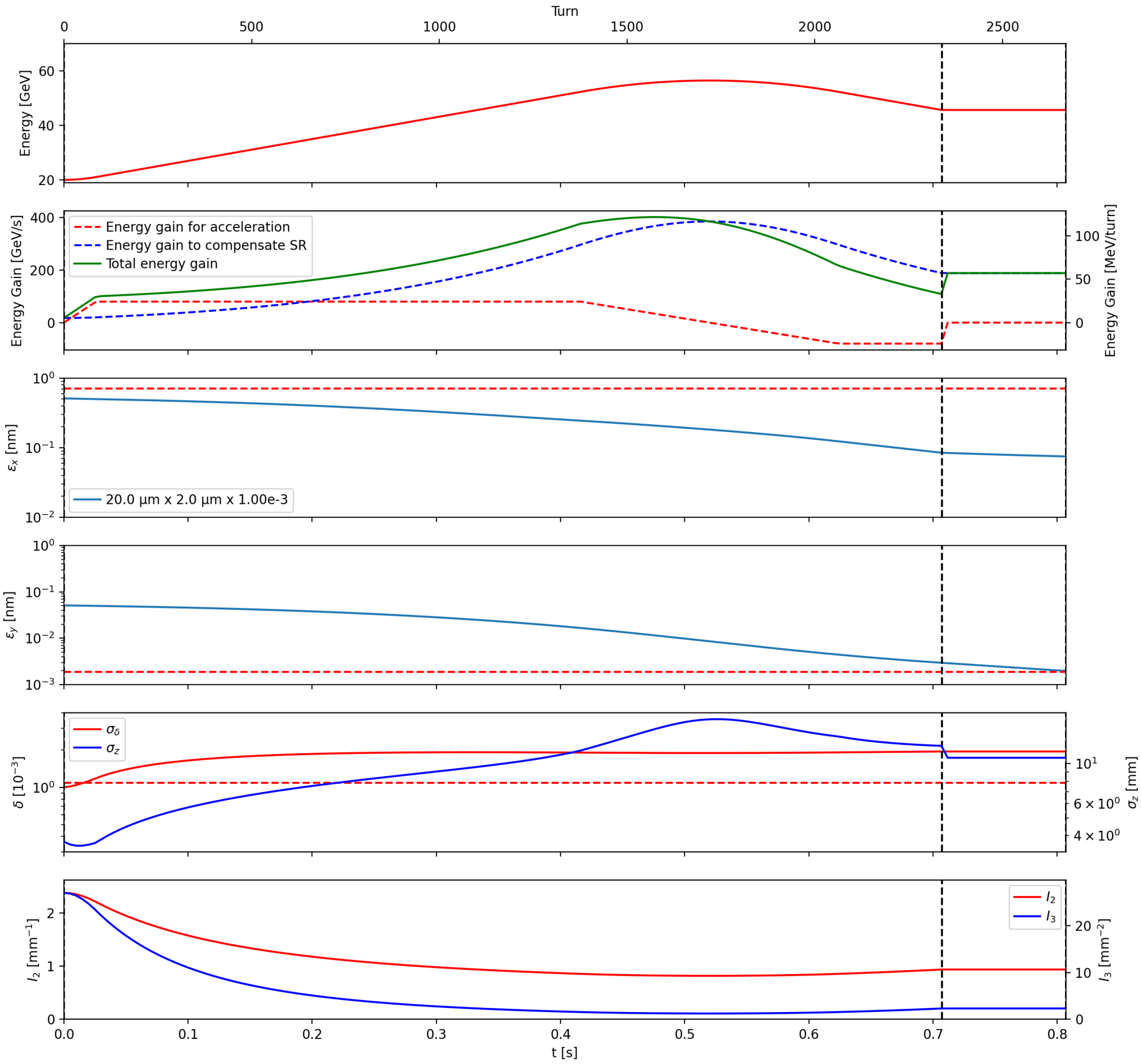

Fig. 5.5: Evolution of the beam energy (first line), the radiated power and required RF voltage (second line), the horizontal RMS emittance (third line), vertical RMS emittance (fourth line), RMS energy spread (fifth line), and synchrotron integrals $\mathcal{I}_{\in}$ and $\mathcal{I}_{\ni}$ as a function of time, for Z mode operation with a wiggler installed in one of the straight sections.

- The start of the ramp shall be adiabatic to avoid shaking the bunches.
- The energy gain per turn is limited by eddy currents in the magnets. A conservative maximum ramp rate of $80\,\mathrm{GeV/s}$ is considered.
- When approaching high energies, the energy gain is dominated by the energy provided to compensate for losses due to synchrotron radiation. The extraction energy needs to be reached adiabatically.

The tracking simulation suite BLonD [337] was used to design the energy and RF voltage ramps for all four modes. The next sections give details of the energy ramps of the two extreme modes (the high current mode $Z$ and the high energy mode $\mathrm{t\bar{t}}$).

### *Z mode ramp*

To achieve the target emittances required by the collider (see Section 5.1.2), the ramping strategy includes an energy overshoot up to $53\,\mathrm{GeV}$. Since energy loss due to synchrotron radiation scales with the fourth power of the energy, the 8 GeV excess must be compensated by a high RF voltage.

For designing the voltage ramp, the phase-space longitudinal stability regions are considered at three key points: the flat bottom, the overshoot maximum, and the flat top, as shown in Fig. 5.6. To maintain a sufficiently small filling fraction (the ratio of bunch emittance to stability area), an RF voltage of at least 85 MV is required.

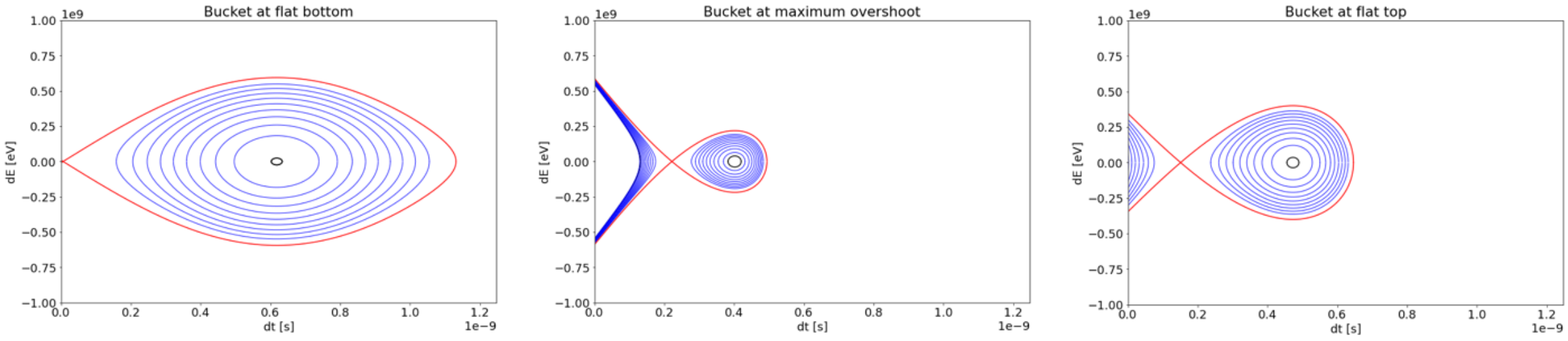

Fig. 5.6: Longitudinal phase space stability regions delimited by the separatrix (red curve). Blue lines show curves of constant energy. Left panel corresponds to flat bottom, with an energy $E = 20$ GeV and voltage $V_{\mathrm{RF}} = 50.1$ MV. Right panel corresponds to flat top, with an energy $E = 45.6$ GeV and voltage $V_{\mathrm{RF}} = 57.2$ MV. The middle panel corresponds to the maximum of the energy overshoot, $E = 53$ GeV, with voltage $V_{\mathrm{RF}} = 85$ MV.

A doubly parabolic voltage ramp up to 85 MV is proposed, see Fig. 5.7. The first 1 % voltage increase is parabolic to avoid shaking the beam; then there is a steep linear part, and finally a parabolic curve for the remaining 20 % of the voltage increase. From 85 MV, the voltage is ramped down to 57.2 MV with a similar strategy, parabolic for the initial 20 %, then linear, then parabolic for the remaining 40 %, before a 0.1 s flat top. The proposed ramp, based on considerations of longitudinal dynamics, will be fine-tuned according to hardware considerations.

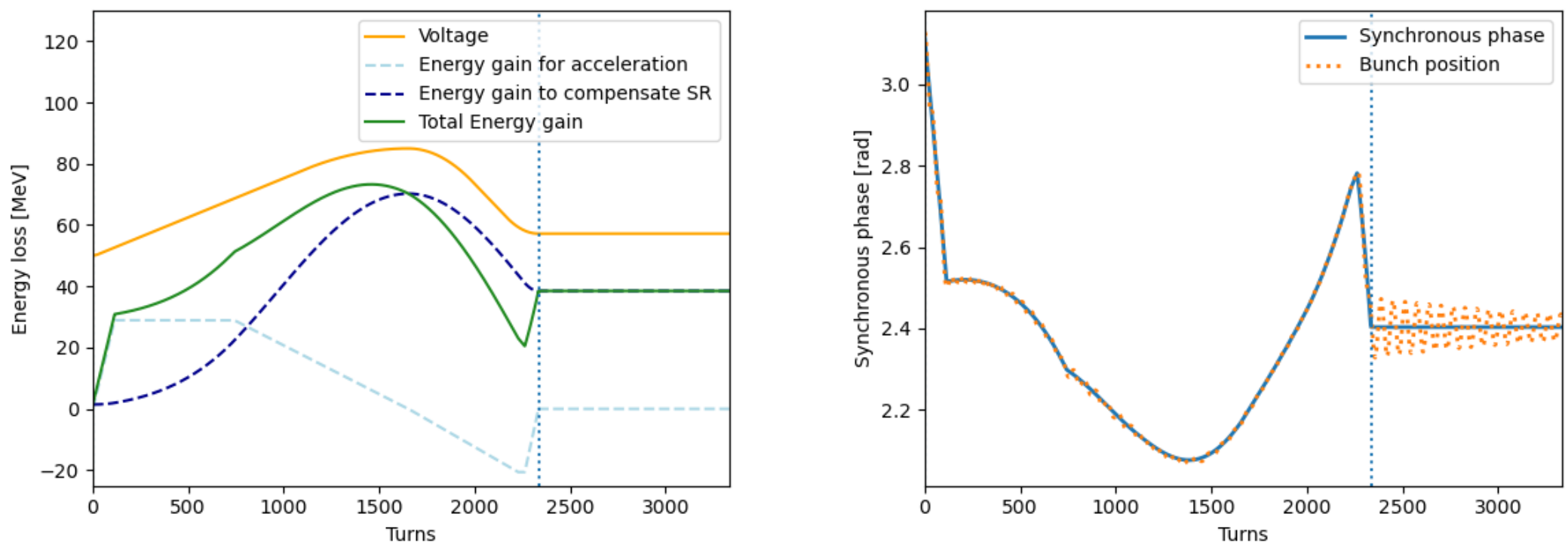

Fig. 5.7: Left: Energy and voltage ramping strategies for the Z mode. The total energy gain is plotted as a full green line; it is the sum of the energy gain needed for acceleration (light-blue dashed line) and the energy required to compensate for synchrotron radiation losses (dark blue dashed line). The voltage is shown as a full orange line. Right: Synchronous phase, theoretical (full blue line) against simulated bunch position (dotted orange line).

### *$t\bar{t}$ mode ramp*

The $\mathrm{t\bar{t}}$ mode ramp lasts 2.03 s, corresponding to 6713 turns in the booster ring, followed by a 0.1 s flat top. The energy increases nine-fold, from 20 GeV to 182.5 GeV. At flat top, the energy loss due to synchrotron radiation is substantial (over 10 GeV/turn). The top energy has to be reached as slowly as possible. The proposed energy and voltage ramps are shown in Fig. 5.8. A preliminary solution with a linear voltage increase from 50.1 MV up to 11.533 GV is proposed.

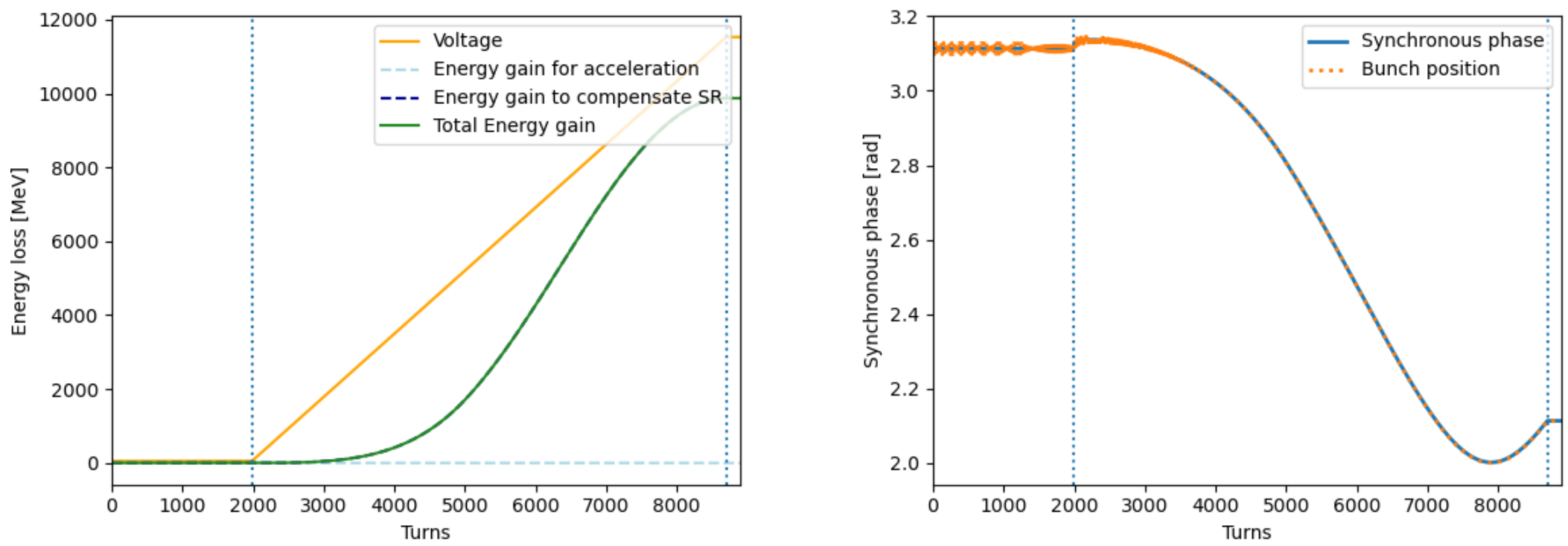

Fig. 5.8: Left: Energy and voltage ramping strategies for the $t\bar{t}$ mode. The total energy gain is plotted as a full green line; it is the sum of the energy gain needed for acceleration (light-blue dashed line) and the energy required to compensate for synchrotron radiation losses (dark blue dashed line). The voltage is shown as a full orange line. Right: Synchronous phase, theoretical (full blue line) against simulated bunch position (dotted orange line).

## 5.2 Requirements

### 5.2.1 Magnets

The magnet design is based on the FODO lattice described in Section 4.1.3. The main cell of the arcs uses 1 dipole family, 6 quadrupoles circuits (to tune the phase advances between the sextupoles and also the global tune of the cell), and 2 sextupoles families (to correct the chromaticity in both planes).

Each quadrupole is combined with one dipole corrector. The correction plane is directly linked to the polarity of the quadrupole (horizontal if positive and vertical if negative). More details on the correction scheme in the booster are given in Section 5.2.6. The requirements for the main magnets of the booster are summarised in Table 5.2.

Table 5.2: Main magnet requirements for the booster.

| | Dipole | Quadrupole | Sextupole Focusing | Sextupole Defocusing | Corrector |
|---|---|---|---|---|---|
| Total number in lattice ... | 6164 | 3346 | 576 | 560 | 3346 |
| ... of which in arcs | 5536 | 2768 | 576 | 560 | 2768 |
| Aperture [mm] | 65 | 65 | 65 | 65 | 65 |
| Length [m] | 11 | 1.3 | 0.7 | 1.4 | $< 0.3$ |
| Max strength, arc at $t\bar{t}$ | 58.9 mT | 28.7 T m$^{-1}$ | 1147 T m$^{-2}$ | 1219 T m$^{-2}$ | 20 mT m |
| Min strength, arc at 20 GeV | 6.45 mT | 2.8 T m$^{-1}$ | 126 T m$^{-2}$ | 134 T m$^{-2}$ | 20 mT m |

### 5.2.2 Injection

To assess the injection requirements for the booster, the 6-D dynamic aperture was calculated at injection over 1000 turns. The calculation was performed by tracking a grid of macro-particles with initial actions ranging from $1-50\,\sigma$ in $1\,\sigma$ steps, considering four different angles in the $x$-$y$ plane: $0^\circ$, $30^\circ$, $60^\circ$ and $90^\circ$.

Macro-particles were injected at the injection kicker with an initial time offset between $-250$–$250\,\mathrm{ps}$ and a relative energy spread between -2% and +2%. The maximum action for which a macro-particle remains unlost after 1000 turns is shown in Fig. 5.9, with the $15\,\sigma$ limit indicated in red.

If the RMS energy spread at injection is $10^{-3}$ and the RMS bunch size is $4\,\mathrm{mm}$, the corresponding requirements are a maximum time jitter of $50\,\mathrm{ps}$ and a relative energy spread tolerance of $5 \times 10^{-3}$. It is worth noting that the injection transfer line to the booster requires a relative energy difference error of $3 \times 10^{-3}$. To summarise, the time jitter for the injection complex should be $50\,\mathrm{ps}$ and the relative energy error should be less than $3 \times 10^{-3}$.

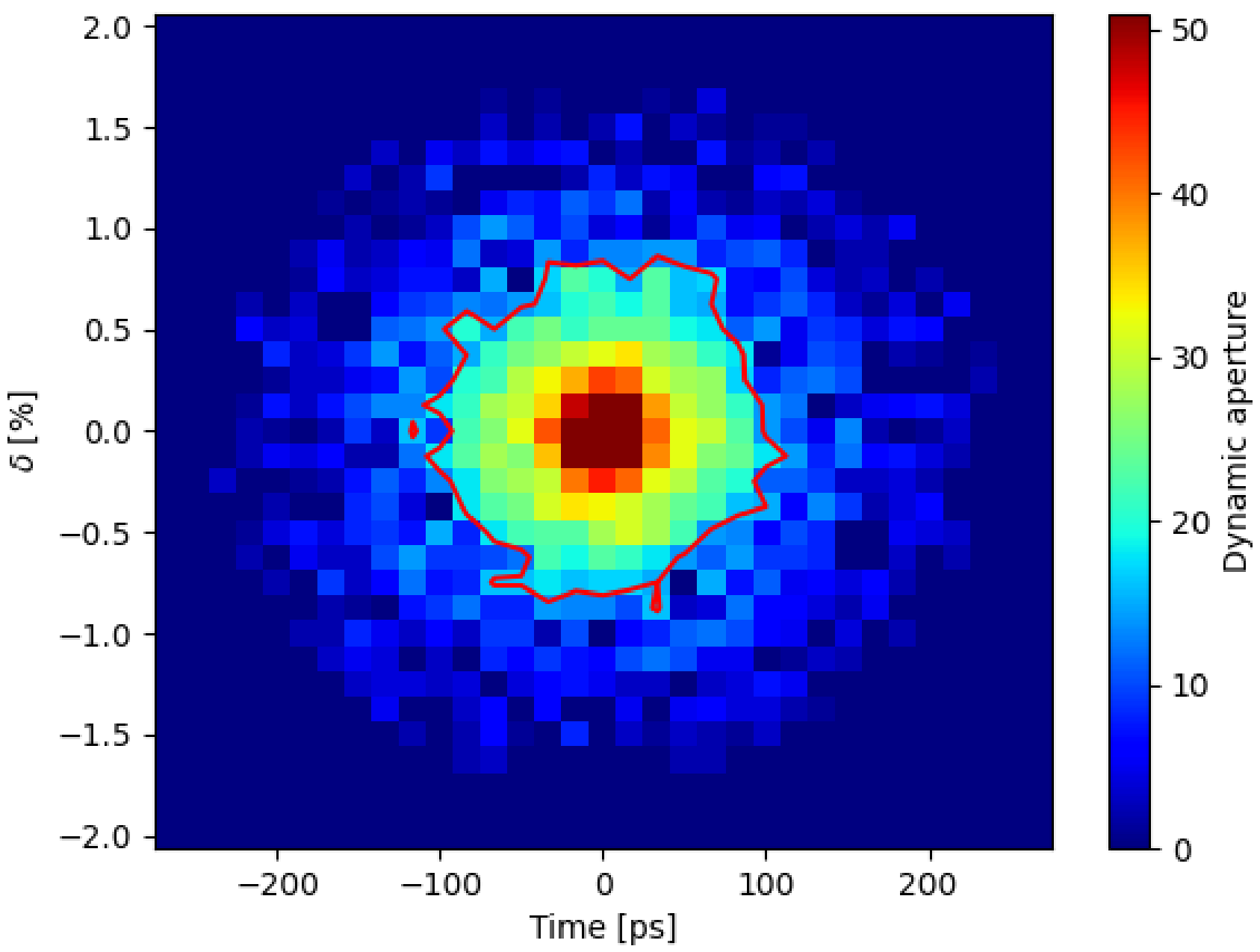

Fig. 5.9: 6-D dynamic aperture, in units of rms beam size ($\sigma$), at injection into the booster for 1000 turns as a function of the initial time delay and relative energy difference. The injection emittances are $20\,\mu\mathrm{m} \times 2\,\mu\mathrm{m}$. The red line gives the contour for a dynamic aperture of $15\,\sigma$.

### 5.2.3 RF staging

The high-energy booster aims to accumulate and accelerate the bunches before injection into the two collider rings. The RF system will be installed in point PL.

In the first stage, the RF system must operate at the Z, WW or ZH operating points without any hardware modification, allowing fast switching between the three different modes as is done in the collider. This is achieved by installing 112 cavities which corresponds to 28 cryomodules with 4 cavities per cryomodule. Due to the large range in RF voltage between injection and extraction energy, reverse phase operation (RPO) [245] is planned (more details are given in Sections 3.4.3 and 6.3).

The second step will consist of adding 332 cavities (84 cryomodules) to have a total of 448 cavities to run at the $\mathrm{t\bar{t}}$ energy, corresponding to a total of 112 cryomodules. This important installation sequence will be performed during a year of shutdown after the last run at the Z, WW, ZH operating points and before the start of the $\mathrm{t\bar{t}}$ run.

The same 6-cell $800\,\mathrm{MHz}$ elliptical cavities as designed for the collider $\mathrm{t\bar{t}}$ mode will be used. As a baseline, the cavity manufacturing technology is the standard bulk niobium technology which has a limitation restricting the usable accelerating gradient to $20\,\mathrm{MV/m}$ in operation, assuming that the cavities are qualified in a vertical cryostat at a $20\,\%$ higher gradient and $Q_0$ (see Table 3.11).

### 5.2.4 Vacuum

Collective effects arising from the accumulation of oppositely charged particles in the beam can lead to various undesirable consequences. As these effects are highly sensitive to bunch spacing, they are a primary concern at the Z operating point, where the collider will run with a large number of closely spaced bunches.

In particular, electron cloud build-up and photoelectron emission during operation with positrons and ion accumulation during operation with electrons may lead to beam instabilities, emittance growth and other detrimental effects in the high-energy booster. These effects can most effectively be mitigated by limiting the production of electrons and ions in the beam chamber environment, which sets requirements on the properties of the vacuum chamber surface as well as the vacuum level in the machine.

#### *Ion accumulation and instabilities*

Beam-induced gas ionisation gives rise to electrons and ions along the beam path. The positive ions are attracted by the electron beam field and may be trapped in oscillation along the bunch train. Trapped ions accumulate over the passage of a bunch train and can seed a coupled-bunch instability known as the fast beam-ion instability [338, 339]. Ions are trapped between bunches if their molecular mass number, $A$, exceeds a critical mass number, $A_c$, which in the linear approximation can be defined as [99]

$$A_c \equiv \frac{N_b r_p c\, \Delta t_{\mathrm{sep}}}{2\sigma_y(\sigma_x + \sigma_y)}\ , \tag{5.1}$$

where $N_b$ is the bunch intensity, $\Delta t_{\mathrm{sep}}$ the bunch spacing and $r_p$ is the classical proton radius. Since the trapping mass is inversely proportional to the beam size, the most critical conditions for ion trapping and instabilities occur immediately after injection, when the beam sizes are at their largest, and relax as the emittance decreases during the booster cycle.

The following parameters are assumed: a bunch spacing of $25\,\mathrm{ns}$, as in the collider filling scheme, the critical mass number $A_c \approx 1$, the transverse beta functions $\beta_{x,y} \simeq 40\,\mathrm{m}$ and the horizontal dispersion $D_x \simeq 0.125\,\mathrm{m}$. In this case, even hydrogen gas, $H_2$ ($A = 2$), would be trapped around the bunch train and avoiding instabilities in such conditions would set unrealistic constraints on the acceptable vacuum levels, especially for an unbaked vacuum without active pumping from a NEG or similar surface. Following the recent updates after the mid-term review, the proposed strategy for Z mode involves transferring only one-tenth of the total number of bunches at a time from the booster to the collider. This allows the bunches to be distributed around the booster circumference, thereby increasing the bunch spacing.

If each group of four bunches from the linac is separated by the maximum possible distance of approximately $1\,\mu\mathrm{s}$, the ion trapping mass between successive 4-bunch trains can reach $A_c \approx 35$, significantly mitigating constraints. In this scenario, ions from the lightest and most abundant gases would be trapped along the 4-bunch trains but would dissipate during the gaps before the next group, making them unlikely to cause detrimental effects. However, heavier gas species such as $CO_2$ could still be trapped from train to train, and their potential impact on vacuum level requirements will need to be studied further.

#### *Electron cloud*

Electron clouds can be created by secondary electron emission through a beam-induced multipacting process, an accumulation of photoelectrons, or a combination thereof. Electron cloud formation through multipacting depends strongly on the secondary electron yield (SEY) of the beam chamber surface, defined as the ratio between the emitted and the impinging electron currents. The risk of electron cloud build-up for different values of the SEY has been assessed by simulating the process of electron cloud formation in the beam chamber at injection and extraction energy using the PYECLOUD code [73]. Assuming the nominal collider filling pattern with uniform trains of bunches spaced by $25\,\mathrm{ns}$, electron cloud build-up in drift spaces, as well as in the main dipolar, quadrupolar and sextupolar fields is suppressed if the SEY is kept at the value of 1.5 or below. Here, the strongest constraints come from the sextupole magnets at extraction energy, whereas the requirement in other elements is more relaxed. Such values of the SEY are expected to be readily achievable after beam-induced conditioning for warm copper surfaces [340]. In addition, these requirements can be significantly further alleviated if the flexibility available in the booster filling pattern for spacing the bunches over the entire ring, is employed.

Photoelectrons, produced through photoemission from the chamber walls due to the synchrotron radiation emitted by the circulating beam, can enhance the electron cloud build-up process, and in very large quantities can induce electron cloud effects even in the absence of beam-induced multipacting. The amount of photoelectrons emitted is determined by the photoelectron yield (PY) of the beam chamber surface, defined as the ratio between the emitted photoelectrons and the number of impinging photons, along with the number and distribution of synchrotron radiation photons in the beam chamber. While build-up studies for the different photoemission levels have not been finalised for the booster, studies for the collider can be used (see Section 1.4.4) for a first estimate of the acceptable level in the booster. For the collider, build-up and ray-tracing studies imply that the photon absorbers need an efficiency well above 90% to ensure beam stability. Since the beam current in the booster is around a factor 100 lower than in the collider, the photon flux is lower by a similar factor and therefore effectively fulfils the collider constraint. In addition, since electron multipacting is expected to be well suppressed in the booster, the risk of photoelectrons enhancing the multipacting to a worrying degree is low. Therefore, the effect of the photoemission is not expected to have a major impact on the beam quality. Detailed studies are pending.

### 5.2.5 Machine protection systems

***Beam intensity limitations for beam transfers***

Ensuring robust machine protection is a key priority when operating with the high-intensity and small-emittance beams of the Z mode. While the stored beam intensity in the booster is lower than in the collider, uncontrolled beam losses could still pose a risk to equipment. In particular, beam loss incidents may occur during the beam transfer process due to potential kicker failures or timing errors.

Given the large number of booster-to-collider transfers required for top-up injection, the reliability of the hardware systems involved is paramount. To further enhance safety, protection absorbers will be strategically placed in critical areas, including the booster extraction region, the transfer lines, and the collider injection region. These absorbers will ensure that any mis-steered beams are safely intercepted, preventing damage to the accelerator systems.

Similar protection measures are successfully implemented in other high-intensity accelerators, such as the SPS and LHC at CERN, providing a well-established foundation for ensuring safe operation.

The maximum acceptable load on protection absorbers poses a limitation for the beam intensity, which can be safely transferred from the booster to the FCC-ee collider. Some of the most robust absorber materials for protection absorbers used nowadays at CERN include isotropic graphite and Carbon/Carbon composites. Material tests with $440\,\mathrm{GeV}$ proton beams in the CERN HiRadMat facility showed that such materials can resist energy deposition densities as high as $5\,\mathrm{kJ/g}$ without sustaining any damage ($5\,\mathrm{kJ/g}$ corresponds to a peak temperature of roughly $3000\,^{\circ}\mathrm{C}$). The maximum acceptable energy density may even be higher, but firm limits can only be established once higher-intensity beams become available in HiRadMat or other facilities.

In order to provide a first order estimate of the expected intensity limitation for booster-to-collider beam transfers in the Z mode, a generic beam loss scenario is considered, where a mis-steered $45.6\,\mathrm{GeV}$ bunch train is intercepted by a graphite or Carbon/Carbon absorber with a density of $1.8\,\mathrm{g/cm^3}$. The energy deposition in the absorber block is calculated with the FLUKA radiation transport code. It is assumed that all bunches impact on the same spot and have an emittance of $\epsilon_{x,\text{target}} = 0.71\,\text{nm}$ in the horizontal plane and $\epsilon_{y,\text{target}} = 9.4\,\text{pm}$ in the vertical plane; these values are the target emittances for booster extraction defined in Section 5.1.2. Assuming the bunch train corresponds to around 1% of collider intensity (1200 bunches with a bunch intensity of $2.14\times10^{10} - 2.68\times10^{10}\,\text{e}^{-/+}$), the simulations suggest that the maximum energy density in in the block can reach $5-6\,\text{kJ/g}$ if $\beta_{x,y} = 500\,\text{m}$ at the absorber location, and $3-4\,\text{kJ/g}$ if $\beta_{x,y} = 1\,\text{km}$ (in both cases the possible contribution of the dispersion to the beam spot size was neglected). These results show that a transfer of 1% of the collider intensity might be acceptable in the Z mode if large $\beta$-functions can be achieved at protection absorbers. On the other hand, transferring a train equivalent to 10% of the collider intensity, as originally envisaged in the mid-term report, is considered too high and poses a non-negligible risk for machine protection.

The presented numbers provide an initial estimate of intensity limitations, but a more comprehensive assessment will be necessary through detailed thermo-mechanical studies. These studies will enable a thorough evaluation of material response to energy deposition and temperature gradients induced by particle showers.

Additionally, refining the acceptable load on absorber materials will require dedicated beam impact tests to ensure their resilience under operational conditions. The final intensity limits will also be influenced by the beam's emittance and optics.

Furthermore, a systematic analysis of potential failure scenarios during beam transfer will be essential. This will allow more precise estimates of energy deposition in protection absorbers, ensuring their effectiveness in safeguarding machine components.

***Other machine protection aspects for the booster***

Accidental beam losses can occur not only during the transfer from the booster to the collider but also within the booster ring itself, due to factors such as beam instabilities, interactions with dust particles, or hardware failures. A thorough assessment of potential failure modes and their associated time scales is essential for defining the machine protection architecture of the booster.

Given the destructive potential of the booster beams at Z mode—even at just 1% of the collider's intensity—it is likely that a minimal collimation system will be required to protect the machine throughout the booster cycle. At top energy in Z mode operation (45.6 GeV), the stored beam energy in the booster is comparable to that of SuperKEKB, where collimator jaws suffered damage from the beams. This underscores the importance of machine protection in the booster, with the robustness of collimators being a key area of study, similar to the protection absorbers discussed previously.

For safe operation, a beam monitoring and loss detection system will also likely be necessary in the booster. The specific requirements for booster beam instrumentation, such as beam loss monitors, beam position monitors, and beam current monitors, will be determined based on the failure modes identified. In cases of excessive beam loss, a rapid extraction system must be in place to remove the beam in a single turn. As outlined in Section 4.4, the current extraction system design assumes that the booster shares a beam dump with the collider. The required reaction time for triggering a beam abort remains to be defined but is expected to be within a few turns.

Beyond beam losses, the impact of synchrotron radiation-induced heating must also be carefully assessed. While the synchrotron power emitted in the booster ring is significantly lower than in the collider (see Section 4.3), power deposition on equipment remains non-negligible. Detailed energy deposition studies will be necessary to ensure that sensitive components are adequately protected from heat generated by synchrotron photons.

### 5.2.6 Correction strategy

The schematic workflow of the correction strategy is shown in Fig. 5.10 and is presented here for the FODO lattice described in Section 4.1.3. The emittance tuning is divided in two parts: one with the sextupoles turned off (or very low strength) and the other with the sextupoles ramped to full strength in 3 steps. The momentum aperture of the nominal lattice without errors reduces to $\pm 0.3\%$ when sextupoles are switched off. For the moment, this value is considered acceptable since the high-energy linac can provide dedicated single bunches with an energy spread of $0.05\%$, during commissioning. Further studies are required in the next phase, including the errors in the momentum aperture evaluation. It is assumed that all the orbit correctors of the booster are individually powered, and they are placed at each quadrupole together with the BPM (i.e., 1400 dipole correctors per plane as given in Table 5.2). Firstly, the sextupoles are off. The orbit correction is applied segment-by-segment (SbS), i.e., in this case, arc by arc, which is similar to the LHC commissioning [341]. After the SbS, two iterations of orbit correction (using the singular value decompositions method) are made on all arcs and in line in order to reduce the residual orbit.

This is sufficient to reduce the orbit enough to locate the closed orbit. Following this step, multiple iterations of orbit correction in the ring are performed until the residual RMS orbit falls below the analytical value [342]. Finally, the sextupoles are set to $33\%$ of their nominal strength, and a final iteration of orbit correction in the ring is carried out.

Four iterations of orbit correction, coupling resonant driving terms (RDTs), horizontal and vertical dispersion correction, phase advance adjustments, and tune corrections are performed at this stage.

A total of 560 normal quadrupole correctors are distributed within the `F2D2` main quadrupole family along the eight arcs. In comparison, 568 skew quadrupole correctors are positioned at the sextupoles in the eight arcs.

This procedure is then repeated with the sextupoles set to $66\%$, of their nominal strength before being performed at full sextupole strength.

Various case scenarios were studied for 100 seeds. The alignment and field errors used are reported in Table 5.3. Assuming 200 µm misalignment from one girder to the other (about 25 m distance) and 50 µm misalignment for the elements placed on top of the same girder (BPM, Quadrupole, and Sextupole).

Table 5.3: Summary of the different error types and their values.

| Error type (Gaussian RMS ) | Value | Unit |
|---|---|---|
| MB relative field error | $10^{-3}$ | |
| MB main dipole roll error | 300 | µrad |
| MQ offset (with respect to the girder) | 50 | µm |
| MQ roll | 100 | µrad |
| MS offset (with respect to the girder) | 50 | µm |
| BPM offset (with respect to the girder) | 50 | µm |
| Girder-to-girder offset | 200 | µm |

It is worth noting that all the errors applied on the elements are randomly Gaussian distributed within $\pm 3$ RMS.

### 5.2.7 Residual orbit and corrector strength required

Figure 5.11 shows the RMS values of the residual orbit for the 100 configurations of errors studied with the correction procedure described in the previous section. The dashed red lines on the distributions represent $\pm 3$ times the RMS calculated analytically. The right panel of Fig. 5.11 shows the distribution of the RMS orbit corrector strength for the same 100 machine configurations.

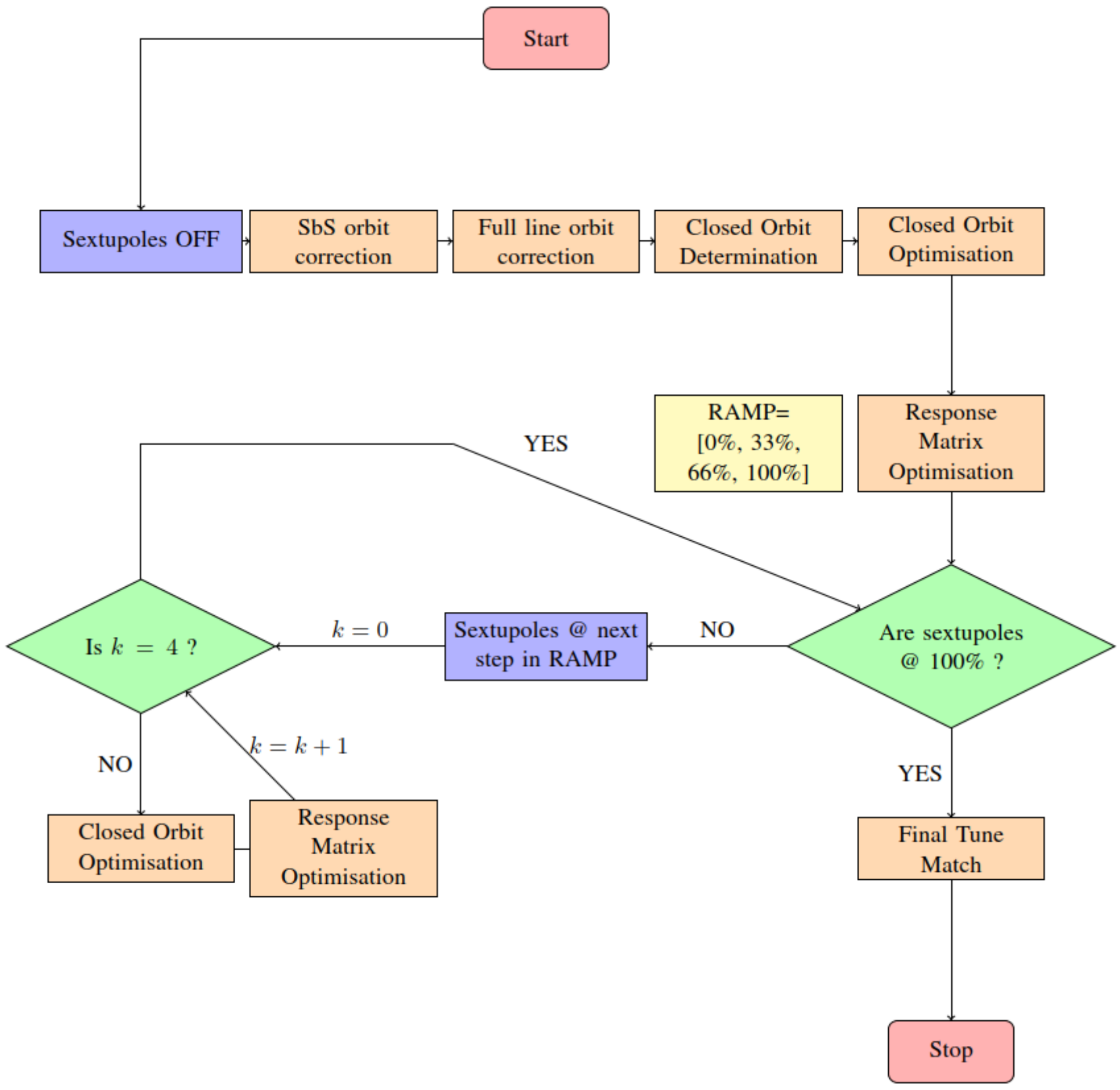

Fig. 5.10: Workflow chart of the tuning scheme of the booster.

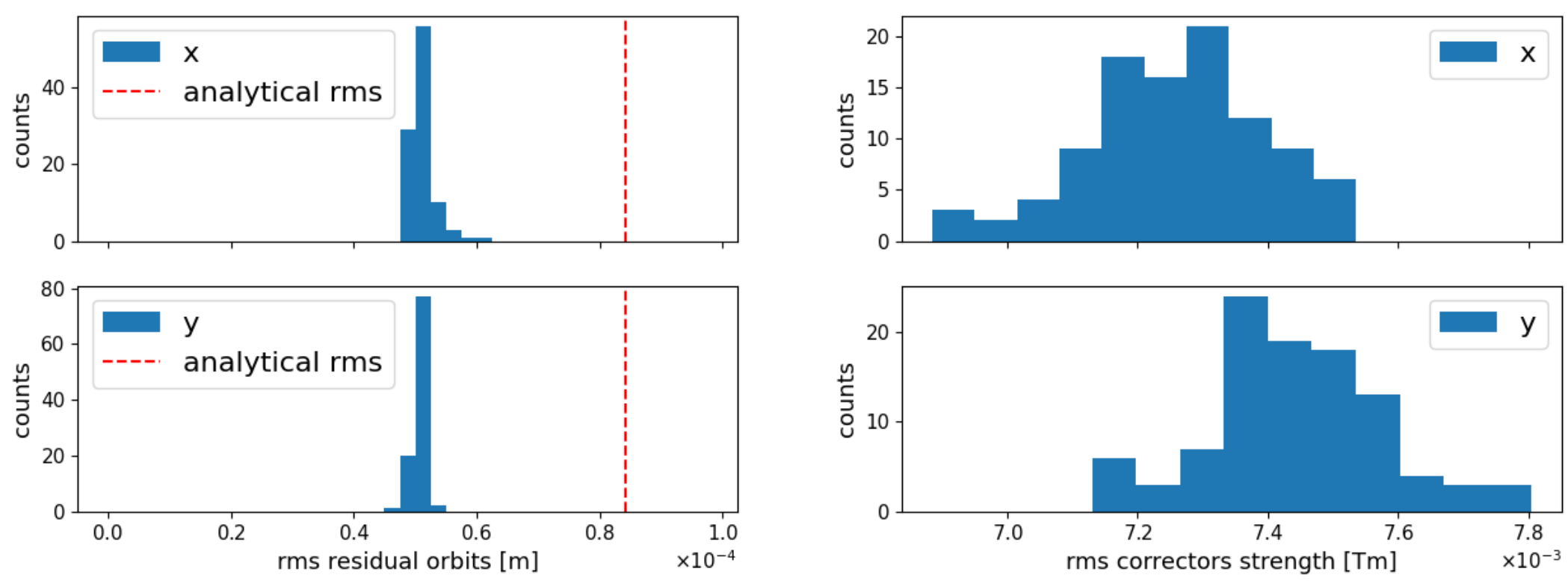

Fig. 5.11: RMS values of the residual orbit (left) and of the corresponding corrector strengths (right) for the different error configurations described in the text.

The RMS values of the correctors' strength for the same 100 different configurations are shown in Fig. 5.12. As for the residual orbit both optics have similar RMS corrector strength values. The average values of the RMS residual orbit and of the maximum corrector strength are presented in Table 5.4.

### 5.2.8 Emittance evaluation

The final equilibrium emittance for the 100 converging machine configurations is evaluated at a beam energy of $45.6\,\mathrm{GeV}$. The distribution of the equilibrium emittance after orbit correction and before the correction of the coupling RDTs, dispersions, phase advances and tunes are shown in orange in Fig. 5.13.

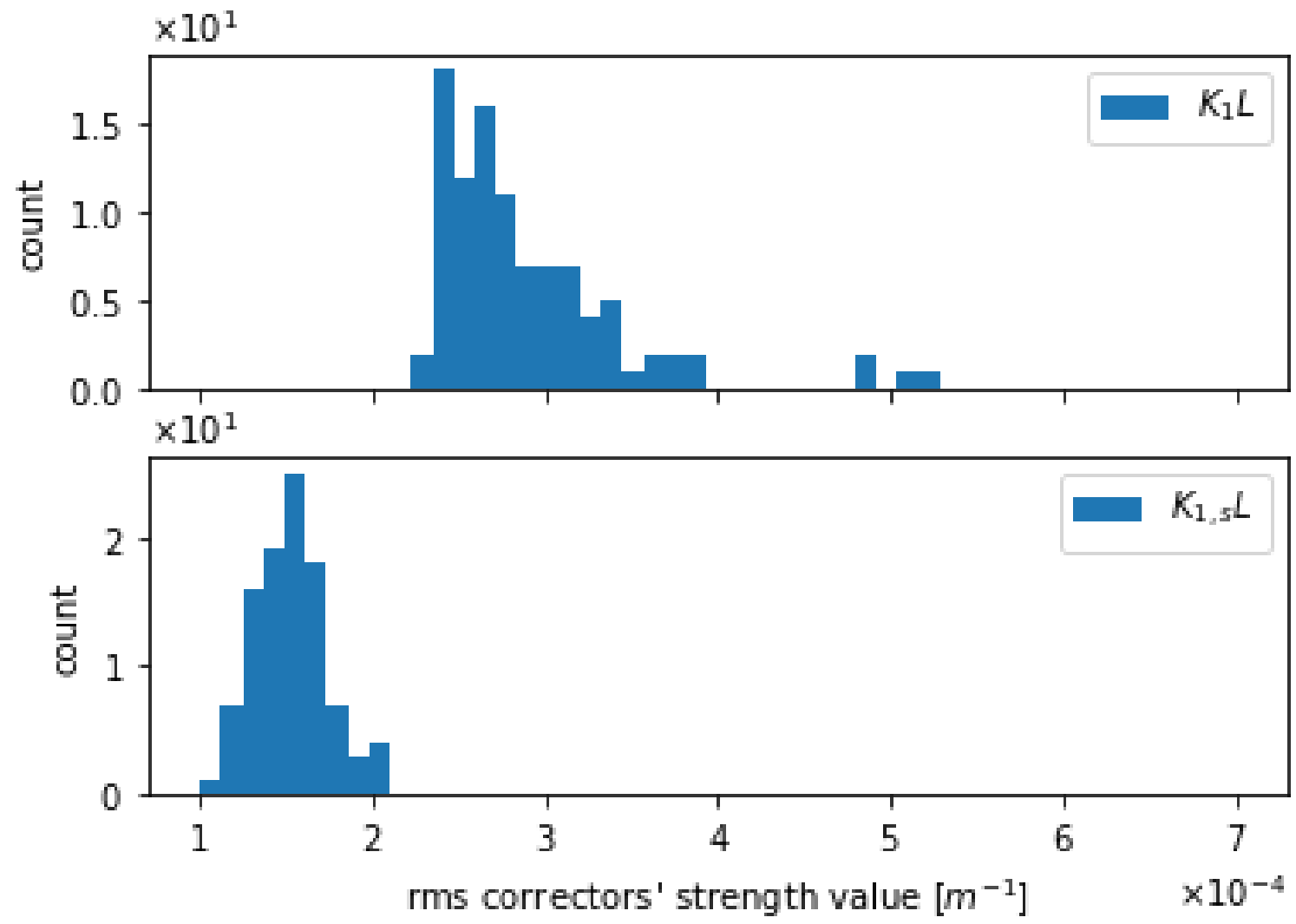

Fig. 5.12: RMS values of the normal and skew quadrupole correctors strength for the 100 configurations of errors, as described in the text.

Table 5.4: Average RMS residual orbit and maximum RMS corrector strength values after correction for the 100 configurations based on Table 5.3.

| | | Plane | | 3×RMS Analytic | 3×RMS Seeds |
|---|---|---|---|---|---|
| Residual orbit | [μm] | x | | 253 | 181 |
| | | y | | 262 | 160 |
| Orbit Corrector | [mT.m] | x | $t\bar{t}$ | 20 | 23 |
| | | | inj | 2.2 | 2.5 |
| | | y | $t\bar{t}$ | 21 | 24 |
| | | | inj | 2.2 | 2.6 |
| Integrated Quadrupole Corrector | [$T.m^{-1}.m$] | Normal | $t\bar{t}$ | - | 0.32 |
| | | | inj | - | 0.03 |
| | | Skew | $t\bar{t}$ | - | 0.13 |
| | | | inj | - | 0.02 |

After the four iterations of orbit, coupling RDTs, dispersions, phase advances and tunes corrections the equilibrium emittance distributions are centred around the target values at Z energy.

### 5.2.9 Perspectives

Future studies can focus on reducing the number of independent correctors to simplify the correction process and enhance computational efficiency.

To achieve this, the autocorrelation of the trim (resp. skew) correctors, as well as the linear and non-linear correlations between correctors around the ring, will be analysed.

Finally, chromaticity correction will be implemented, and the dynamic aperture will be evaluated while accounting for various errors and their associated corrections.

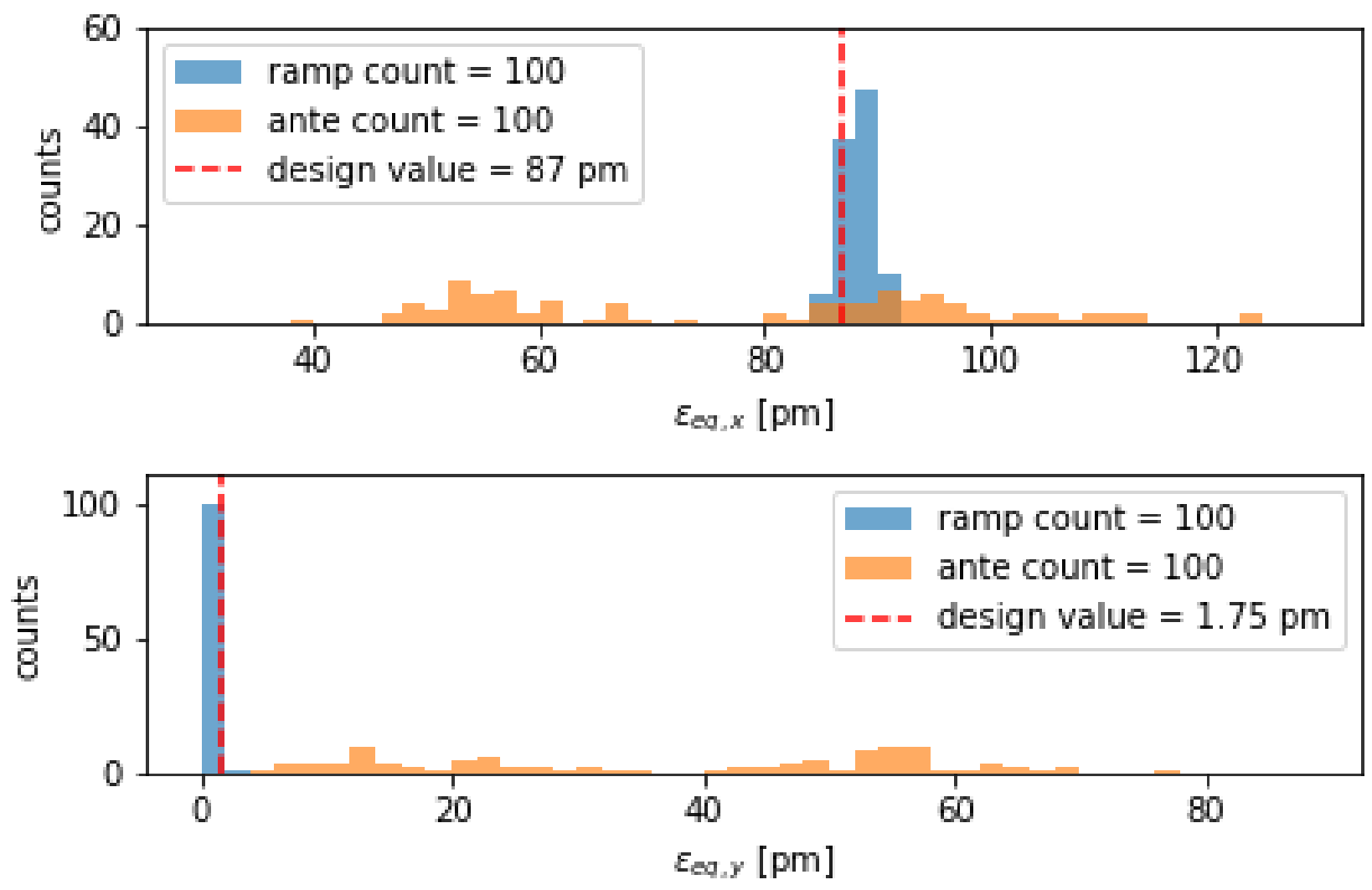

Fig. 5.13: Distribution of the equilibrium emittance in the horizontal (top) and vertical (bottom) planes at Z energy before (ante) and after (post) the four iterations of orbit and emittance correction described in Section 5.2.6.

## 5.3 Availability

This section details results from the enhanced Monte Carlo simulation environment for FCC-ee availability described in Section 2.4, focusing on systems specific to the booster ring.

### 5.3.1 Contributing Systems

The same general framework for availability approximation was applied as per the collider systems, described in Section 2.4.1. The specifics of this process used for each booster system are described below. Only faults leading to downtime in the representative system were considered, thereby assuming a similar degree of redundancy for each basic component family as exists currently in the working accelerator. Details of subsystems, scaling numbers, and any additional system-level redundancy implemented are provided in Table 5.5. Fault data was taken from CERN's Accelerator Fault Tracking (AFT) database [235] and is specific to the LHC physics operation 2015-2024, unless otherwise stated.

1. **Beam Instrumentation**: Beam instrumentation faults are given beam position monitor (BPM), beam loss monitor (BLM) and "other" categories. These were scaled according to the relative number of monitors in the booster compared to the LHC.
2. **Beam Losses**: Downtime due to Unidentified Falling Objects (UFOs) and beam instabilities. Dumps in the former category were scaled with the length of the beam pipe, assuming similar amounts of dust and beam interaction as in the LHC. Dumps due to beam instabilities were applied without scaling. The LHC sees relatively long downtimes due to beam loss effects that cause quenches in nearby superconducting magnets, which is not representative of FCC-ee. Therefore, downtime due to quench recovery was subtracted to calculate the MTTR. FCC-ee dumps due to beam losses then appear frequently, with a short recovery time.
3. **Extraction**: Describes systems required for beam abort. These include controls and kicker hardware to divert the beam into the dump block. They appear twice in the LHC (one for each circulating beam), as for the booster, so they are applied without scaling. The dump blocks are excluded as these are shared with the collider rings.
4. **Injection Systems**: This includes kickers, septa, and control systems needed to inject and extract

the beam under regular operation. The LHC has two injection systems (one for each beam). The booster has two injection systems from the injector complex and two extraction systems for the collider.

5. **Machine Protection & Interlocks**: Scaling for LHC subcategories is approximated from the relative number of hardware instances foreseen in the FCC-ee.
6. **Magnets**: Failure data from 3 532 normal-conducting magnets around the CERN complex was applied to 12 500 magnets in the booster.
7. **Power Converters**: Failure data was taken from two families of power converters, each grouped according to similar MTBF, to represent powering hardware for various magnet types. Only faults leading to downtime in the LHC were considered, thereby assuming a similar degree of redundancy for each magnet group as currently exists in the LHC. Due to the large number of corrector magnets in the latest optics configuration, an MTBF of 9 days is observed in some categories.
8. **Radio Frequency (RF)**: The LHC has 16 accelerating superconducting RF cavities. The number in FCC-ee varies with energy mode. All cavities are horizontally tested to a margin 10 % above their nominal voltage so, theoretically, nominal beam energy can be preserved if no more than 10 % of the cavities are unavailable. The effect of losing cavities on beam stability and hardware protection has only been modelled in the collider so far, where it was determined that redundancy is severely limited in $Z$ and $W$ modes due to beam loading. In the booster, the transients required due to energy ramping complicate beam stability calculations, and it is presently unclear whether the beam could be preserved in the event of cavity loss. Pending further study, no redundancy was assumed in the lower energy (higher beam current) modes for the booster. In $ZH$ and $t\bar{t}$, 10 % is given as per the collider cavity circuits.
9. **Transverse Damper**: The booster will have one transverse damper system. The failure rate is scaled down from the two dampers in the LHC.
10. **Vacuum**: LHC vacuum faults are categorised according to the failed component (pump, gauge, valve, controller). These are then scaled according to the relative number of components in the booster vacuum system.

### 5.3.2 Inherent Resilience to Shorter Fault Types

In the event of an outage in the booster and injector complex, stable beams can be maintained in the main collider rings for a lifetime of 10-15 minutes, depending on energy mode (see Table 2.2). If top-up injection can be restored at this time, normal physics can resume. This significantly reduces the FCC-ee's sensitivity to short-duration fault types in the booster and injector complex and leads to an inherent advantage for availability.

### 5.3.3 Simulation Results

The breakdown of unavailability and lost luminosity contribution from each booster system is shown in Fig. 5.14. Contributions from the injector complex are also provided for illustration. For comparison with collider and technical infrastructure systems, see Fig. 2.22.

The RF is the largest contributor to unavailability and lost luminosity in Z and WW modes, largely because it does not benefit from redundancy as it does in the collider. Power converters suffer from the same challenges as identified for the collider in Section 2.4.1. Beam losses, although occurring at the same rate per metre of beam pipe as the collider, are much less significant for lost luminosity as their recovery time is relatively short.

### 5.3.4 R&D Opportunities

Several R&D opportunities are identified to improve availability in the booster ring:

Table 5.5: Parameters used to simulate availability of systems and subsystems in the FCC-ee Booster.

| **System** | **Subsystem** | **Representative System** Machine | # | **FCCee** # | **Redundancy*** # (%) | **Location** AP | **Unit MTBF** days | **Group MTBF** days |
|---|---|---|---|---|---|---|---|---|
| Beam Instruments | BPM | LHC | 1000 | 3369 | 0 (0) | all | 20 671 | 6.1 |
| | BLM | | 3600 | 200 | 0 (0) | all | 36 632 | 183.3 |
| | Other | | 1 | 1 | 0 (0) | all | | 24.5 |
| Beam Losses | UFOs | LHC | 54 | 91 | 0 (0) | all | | 2.4 |
| | Beam Instability | | 1 | 1 | 0 (0) | all | | 4.3 |
| Extraction | Controls | LHC | 2 | 2 | 0 (0) | PB | 36 | 18.2 |
| | Hardware | | 2 | 2 | 0 (0) | PB | 65 | 32.7 |
| | Other | | 2 | 2 | 0 (0) | PB | 71 | 35.6 |
| Injection Systems | MKI | LHC | 2 | 4 | 0 (0) | PB | 20 | 5 |
| | TDI | | 2 | 4 | 0 (0) | PB | 71 | 17.8 |
| Machine Protection | FMCM | LHC | 1 | 10 | 0 (0) | all | | 39.2 |
| | BIS | | 1 | 2 | 0 (0) | all | | 32.7 |
| | SMP | | 1 | 1 | 0 (0) | all | | 87.1 |
| | PIC | | 1 | 0.2 | 0 (0) | all | | 261.4 |
| | WIC | | 1 | 6 | 0 (0) | all | | 21.8 |
| Magnets | Normal-conducting | CERN | 3532 | 12 500 | 0 (0) | all | $71 \times 10^6$ | 5681 |
| Power Converters | Dipole | LHC | 194 | 16 | 0 (0) | all | 3672 | 229.5 |
| | Quadrupole | | 194 | 32 | 0 (0) | all | 3672 | 114.8 |
| | Sextupole Focusing | | 194 | 32 | 0 (0) | all | 3672 | 114.8 |
| | Sextupole Defocusing | | 194 | 32 | 0 (0) | all | 3672 | 114.8 |
| | Dipole Tapering | | 1032 | 346 | 0 (0) | all | 15 485 | 44.8 |
| | Quadrupole Tapering | | 1032 | 346 | 0 (0) | all | 15 485 | 44.8 |
| | Horizontal Corrector | | 1032 | 1672 | 0 (0) | all | 15 485 | 9.3 |
| | Vertical Corrector | | 1032 | 1672 | 0 (0) | all | 15 485 | 9.3 |
| | Quadrupole Corrector | | 1032 | 1384 | 0 (0) | all | 15 485 | 11.2 |
| | Skew Quadrupole | | 1032 | 1384 | 0 (0) | all | 15 485 | 11.2 |
| Radio Frequency | Cavity Circuit Z | LHC | 16 | 112 | 0 (0) | PL | 1462 | 0.54 |
| | Cavity Circuit W | | 16 | 112 | 0 (0) | PL | 1462 | 0.54 |
| | Cavity Circuit ZH | | 16 | 112 | 11 (10) | PL | 1462 | 314.6 |
| | Cavity Circuit $t\bar{t}$ | | 16 | 600 | 60 (10) | PL | 1462 | 465.2 |
| Transverse Damper | | LHC | 2 | 1 | 0 (0) | PL | 65.4 | 65.4 |
| Vacuum | Pumps | LHC | 891 | 8360 | 0 (0) | all | 349 448 | 41.8 |
| | Gauges | | 1052 | 904 | 0 (0) | all | 274 997 | 304.2 |
| | Valves | | 323 | 226 | 0 (0) | all | 84 434 | 373.6 |
| | Controllers | | 789 | 2328 | 0 (0) | all | 55 915 | 20.8 |

* System-level redundancy in addition to that already implemented in the component family of the representative system.

### *Radio Frequency*

Presently, it is unclear whether redundancy could be implemented in the booster RF cavities as the transients due to ramping lead to additional complications. However, the beam current within the booster is significantly lower than in the collider, suggesting beam loading effects should be less problematic. Implementation of redundancy in this system could significantly benefit the achieved integrated luminosity.

### *Power Converters*

Proposed solutions are treated in Section 2.4.4. Collaboration with the optics working group is required to consider how to preserve the beam if combinations of magnets fail. A system-level redundancy ap-proach could make significant gains in this case, besides reducing the number of families for the corrector magnets.

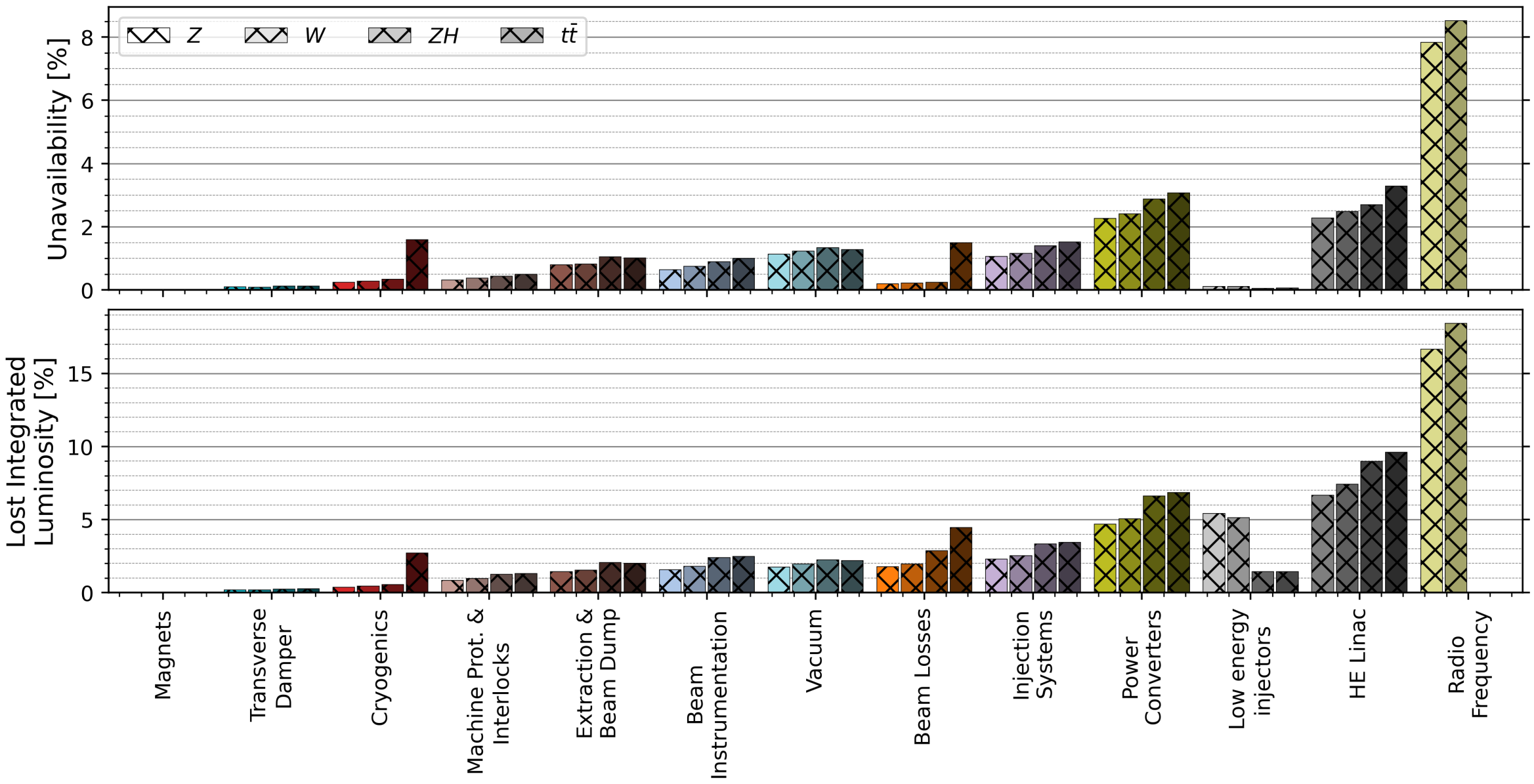

Fig. 5.14: Unavailability and lost luminosity contribution of each system in the booster. Systems are ordered according to $Z$ mode lost luminosity contribution.

#### *Exploiting Resilience to Short Faults*

Additional opportunities exist to exploit the booster's natural resilience to shorter fault types. A lengthy turnaround time may be avoided if a failed component can be brought back online before the beams in the main collider expire. To this end, there are two parallel approaches:

1. If the collider beam lifetime without top-up injection can be made longer, systems will have a better opportunity to recover before a beam abort is necessary. This requires intricate optimisation of the beam optics under various possible failure mechanisms; however, this could greatly alleviate the design task on the collider technical systems.
2. Measures to bring the recovery time of failed systems to below 10 minutes could be extremely powerful for the booster. For example, the automatic reset of recurring RF trips or the installation of redundant or back-up modules that can quickly be brought online to replace failed components.

## 5.4 Conclusion

The results presented in this section demonstrate the feasibility of the Future Circular Collider (FCC) design, outlining the operational strategies, technical requirements, and performance benchmarks necessary for its successful implementation. Through detailed analysis, the revised filling scheme, ramping strategy, and emittance evolution have been optimised to enhance collider efficiency while addressing challenges related to beam stability and machine protection. The study has also highlighted key advances in magnet technology, RF staging, vacuum control, and correction strategies to mitigate instabilities and improve overall reliability. Furthermore, availability assessments and R&D opportunities underscore the importance of system redundancy and resilience to minimise downtime and maximise integrated luminosity. Moving forward, continued refinements and targeted technological developments will be essential to ensuring the FCC meets its scientific and operational goals, paving the way for groundbreaking discoveries in high-energy physics.

# Chapter 6

# FCC-ee booster technical systems

## 6.1 Main magnets

The proposed booster magnet system meets the requirements of the booster FODO lattice V24 described in Section 4.1.3. The regular FODO lattice of the booster ring contains 2768 arc half-cells composed of a short straight section (SSS) and two dipoles. There are three lengths of arc half-cell depending on the composition of the short straight section (SSS): one quadrupole and either no sextupole, or a focusing sextupole, or a defocusing sextupole. The arc-half cells are organised into a periodic structure of 5 cells every $260.554\,\mathrm{m}$, matching the collider. The positrons and electrons circulate in opposite directions in the booster so the magnet polarity is the same for both filling cycles.

In this version, all arc dipoles are identical. The dipoles in the dispersion suppressors are slightly longer and require different field levels but share a common cross-section with the arc dipoles. There are six main quadrupole circuits in the arcs, with strengths ranging from $24.9 - 26.6\,\mathrm{T\,m^{-1}}$. Since the strengths of the arc quadrupoles are similar, a single magnet design is proposed for all quadrupoles, with slight variations in powering. In the dispersion suppressors, strengths of up to $28.7\,\mathrm{T\,m^{-1}}$ are required. The current proposal is to use similar magnets, with additional magnetomotive force provided by trim circuits.

For the sextupoles, the dispersion in the defocusing sextupoles is half that of the focusing sextupoles. Consequently, the defocusing sextupoles require twice the integrated gradient. This has been achieved by using a common cross-section for both sextupoles while making the defocusing sextupoles twice as long. The sextupole strengths in the dispersion suppressor region are lower than those in the main arcs. The current proposal is to use the same magnets with a current bypass.

The requirements of the arc magnets are summarised in Table 6.1. A summary of the magnet cycles is given in Table 4.1. The relative field error dipole-to-dipole is required to be less than $1 \times 10^{-3}$. The relative harmonic field error is required to be better than $1 \times 10^{-4}$ on a reference radius of $10\,\mathrm{mm}$ for all magnets. The following section describes the technical solutions and performance of the main magnets of the arcs: dipoles, quadrupoles, and sextupoles.

Table 6.1: Magnet requirements for the booster FODO lattice V24.

| | Dipole | Quadrupole * | Sextupole | |
|---|---|---|---|---|
| | | | Focus | Defocus |
| Total number in lattice. . . | 6146 | 3346 | 576 | 560 |
| . . . of which in arcs | 5536 | 2728 | 528 | 528 |
| Aperture [mm] | 65 | 65 | 65 | 65 |
| Length, arc [m] | 11 | 1.3 | 0.7 | 1.4 |
| Max strength†, arc at $t\bar{t}$ ext. | $58.9\,\mathrm{mT}$ | $26.6\,\mathrm{T\,m^{-1}}$ | $1147\,\mathrm{T\,m^{-2}}$ | $1219\,\mathrm{T\,m^{-2}}$ |
| Min strength†, arc at 20 GeV in. | $6.45\,\mathrm{mT}$ | $2.7\,\mathrm{T\,m^{-1}}$ | $126\,\mathrm{T\,m^{-2}}$ | $134\,\mathrm{T\,m^{-2}}$ |

* The ratio of min:max quadrupole strength does not match the beam rigidity as there are multiple families.
† Sextupole strength given as $B''$, $B'' = 2S$.

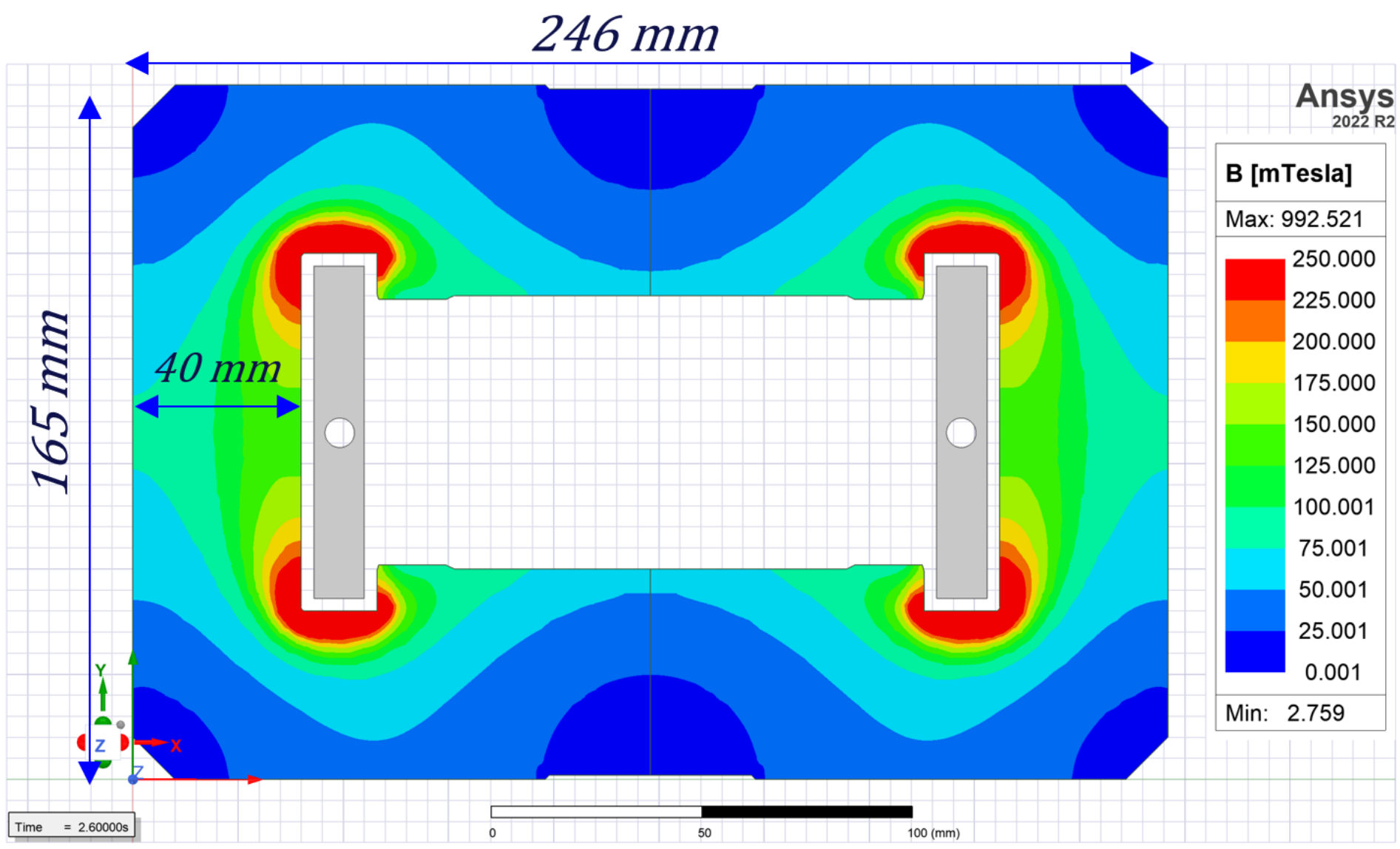

Fig. 6.1: Field map of the booster dipole at $t\bar{t}$ energy.

### 6.1.1 Dipoles

The cross section of the proposed booster dipole is shown in Fig. 6.1, and its key parameters are summarised in Table 6.2. The proposed dipole is an H topology with a laminated steel yoke and directly cooled aluminium busbars, the following paragraphs will discuss each of these key design aspects.

Three topologies were initially considered: C, O, and H. A C topology is attractive from an integration perspective as it gives easy access to vacuum components. However, early studies showed that a C topology was not able to effectively shield the earth's magnetic field and led to relative field errors of $3 \times 10^{-4}$ at injection energy ($6.5\,\mathrm{mT}$, 20 GeV) and therefore an O or H topology is necessary. It should be noted that the beam may be impacted by the Earth's field in areas not shielded by the magnets. An H topology is advantageous over an O topology as the poles can be used to optimise the field quality across the full cycle. The H topology was further optimised for field quality across the full range of operations. This optimisation led to a back leg that is wider than strictly necessary for magnetic return, $40\,\mathrm{mm}$ with a peak field around $150\,\mathrm{mT}$. A lower back leg peak field reduces the remanent magnetisation of the steel, yielding a magnetic performance benefit. Additionally, the extra material removes the need for dedicated synchrotron radiation shielding and, therefore, represents a holistic design optimum. The yoke is laminated to reduce the eddy current during the ramp. The material chosen for the prototype is M270-50A, a $0.5\,\mathrm{mm}$ thick electrical steel with low coercivity that is readily available due to its prevalence in industrial applications. The lamination thickness is conservative, 1/15 of the skin depth, mitigating problems from yoke eddy currents.

Busbars have been used instead of coils for two key reasons: they allow end-to-end magnet interconnection, reducing transmission losses, and they simplify magnet construction. Aluminium is preferred over copper as it is cheaper for a given resistance per unit length. This choice is facilitated by the low field requirement, the magnet envelope is still reasonable even with the larger busbar section. The ground insulation will be an air gap set by $3\,\mathrm{mm}$ thick ceramic spacers. This solution is radiation hard and withstands voltages well in excess of the required $1\,\mathrm{kV}$ to ground. The root mean squared (RMS) current density is low, $1.4\,\mathrm{A\,mm^{-2}}$ for the $t\bar{t}$ cycle. This current density has been shown to optimise the combined total lifetime cost of the magnets and technical infrastructure. While a lower current density would allow an air-cooled solution, it is preferable to dissipate heat from the tunnel using water rather than air. Therefore, the magnets will be water-cooled.

Table 6.2: General parameters of the arc dipole magnets.

| Parameter | Unit | Value |
|---|---|---|
| Strength, $B$, $20 - 182.5\,\mathrm{GeV}$ | $\mathrm{mT}$ | 6.5 - 58.9 |
| Aperture (horizontal × vertical) | $\mathrm{mm}^2$ | $130 \times 65$ |
| Magnetic length | $\mathrm{m}$ | 11 |
| Outer envelope | $\mathrm{mm}^2$ | $246 \times 165$ |
| Peak current | A | 3065 |
| Magnet resistance | $\mathrm{m}\Omega$ | 0.70 |
| Magnet inductance | µH | 44 |
| Peak voltage, magnet | V | 2.2 |
| Conductor (Aluminium) | $\mathrm{mm}^2$ | $12 \times 79$, ⌀7 |
| Turns | | 1 |
| RMS current density, $t\bar{t}$ | $\mathrm{A\,mm}^{-2}$ | 1.4 |
| Magnet active mass | kg | 2452 |
| Busbar active mass (Al) | kg | 54 |
| Yoke active mass | kg | 2398 |

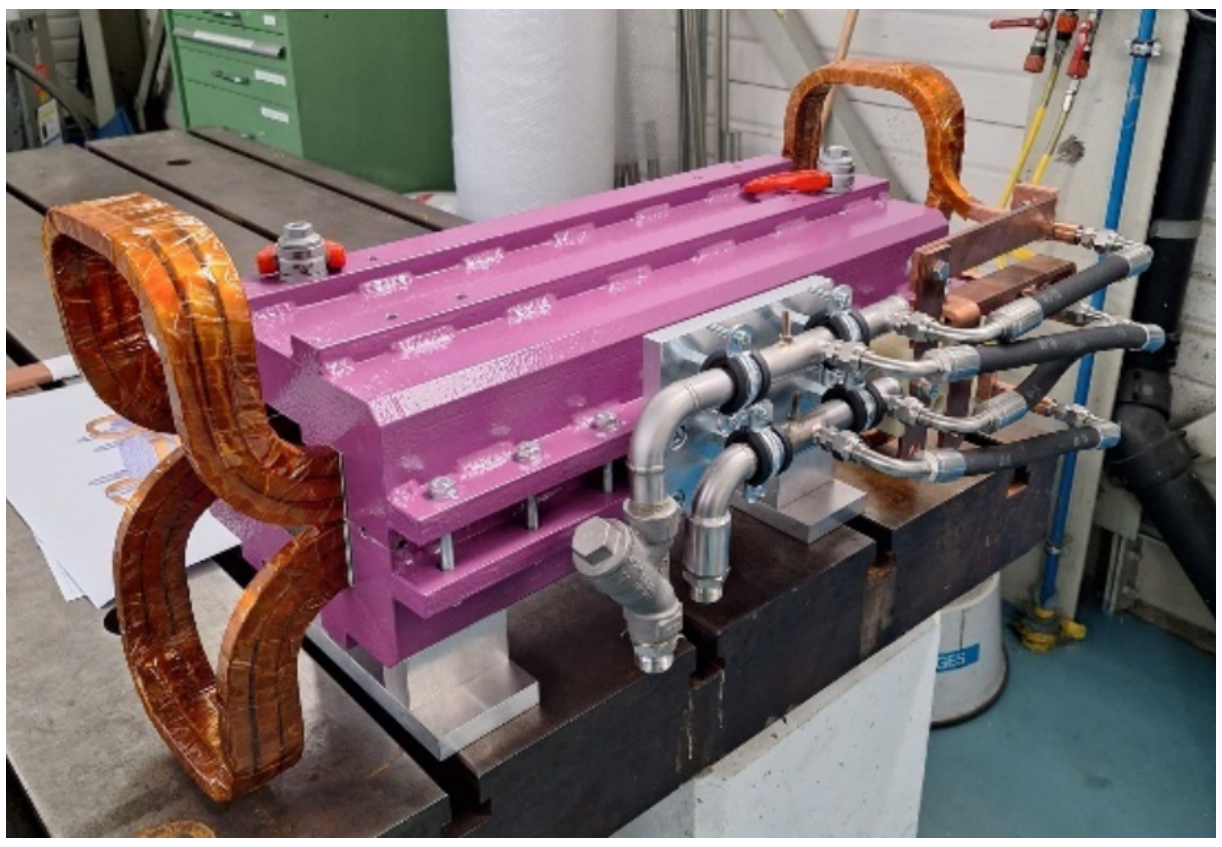

Fig. 6.2: The short prototype booster dipole built and tested to demonstrate the feasibility of low field levels. In this prototype, the busbars are substituted with a coil for compatibility with test facilities.

Other water-cooled systems in the tunnel are made of copper. To avoid the risk of galvanic corrosion or the need for multiple cooling circuits, the inner surface of the cooling tubes will be made of either stainless steel or copper. The baseline proposal is to use a mechanical assembly of a tube and a busbar; however, co-extrusion and electroplating are also being considered. This will be studied in detail during the next phase.

***Test of short prototype dipole magnet***

The main field of the booster dipole at injection is lower than previous machines at CERN, less than half of both the LEP ($22\,\mathrm{mT}$) and the SPS as LEP injector ($15.7\,\mathrm{mT}$). Additionally, hysteresis effects are complex to model as they depend on material characteristics and previous magnetic states. Therefore, a $0.5\,\mathrm{m}$ long prototype booster magnet has been built to demonstrate the feasibility of the low field levels with respect to the remanent effect in the iron yoke. The short prototype produced is shown in Fig. 6.2.

The allowed field harmonics measured during magnetic cycles from $6.5 - 58.9\,\mathrm{mT}$ are presented in Fig. 6.3. The total field harmonics are below $1 \times 10^{-4}$ relative field error up to a radius of $17\,\mathrm{mm}$. The results show that the magnet stabilises quickly after only a few pre-cycles, with only minor, non-systematic field distortions observed throughout the cycle.

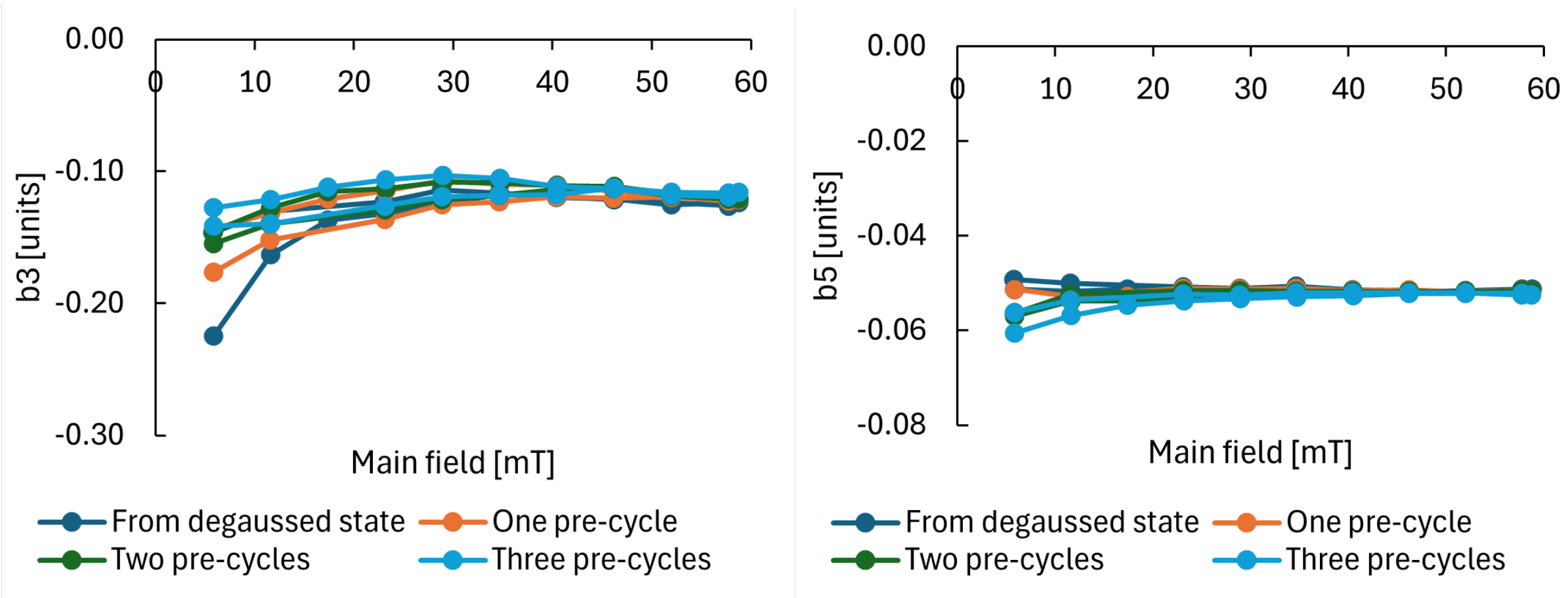

(a) Sextupolar component. (b) Decapolar component.

Fig. 6.3: 2D field harmonics in the short prototype booster dipole at the $t\bar{t}$ field levels. Values quoted at R = $10\,\mathrm{mm}$ in units of $1 \times 10^{-4}$ [343].

For practical reasons, testing has so far been conducted in a quasi-static manner, meaning it does not account for eddy fields from either the copper vacuum chamber or the magnet itself. Since the vacuum pipe is circular, the eddy fields have been calculated using analytical formulae and cross-checked with numerical modelling, showing good agreement. The eddy currents in the vacuum tube will generate a sextupole component of $15\,\mathrm{mT/m^2}$, corresponding to 0.15% of the total sextupole strength of the machine at injection. Due to the relatively long ramp times, the modelled eddy field harmonics from the magnet itself are $0.1 \times 10^{-4}$ relative to the main field at injection.

The transfer function of the short prototype booster dipole is shown in Fig. 6.4. The remanent magnetisation of the yoke contributes around $60\,\mu\mathrm{T}$, or 1%, of the main field at injection. From validated models, it can be seen that the remanent magnetisation is locally linear with material coercivity. It is therefore readily concluded that the coercivity of the yoke must be controlled to at least 10% between magnets to be within the currently allowed dipole-to-dipole relative field error of $1 \times 10^{-3}$. This consistency can be readily achieved by the use of a simple control on the steel supply[1] and, potentially, shuffling. It seems likely that a magnetic length correction with end shims will be needed. However, this is not a foregone conclusion and remains to be studied in detail. The transfer function is inversely proportional to the aperture, so a tolerance of $1 \times 10^{-3}$ in the main field implies a $65\,\mu\mathrm{m}$ aperture tolerance within range for a stamped and stacked yoke. Material controls, shuffling, and length correction were successfully implemented with the SPS dipoles, which have a similar total weight of steel as the booster dipoles, though they are less numerous.

### 6.1.2 Quadrupoles

The parameters of the proposed quadrupole magnet are summarised in Table 6.3. The field map for the quadrupole magnet at $t\bar{t}$ energy is given in Fig. 6.5. To minimise the length of the short straight sections, thereby increasing the dipole filling factor and reducing the synchrotron radiation losses, the quadrupole has been designed with a relatively high gradient and a pole tip field around $0.9\,\mathrm{T}$. As the requirements are more typical, the quadrupole design is correspondingly more conventional than that of the dipole. The coils will be simple racetrack-shaped windings made from hollow conductors, with copper used for compactness. The yoke will be laminated, as the booster operates in a cycled mode. The vacuum chamber diameter is constrained by impedance requirements, resulting in an aperture that is large relative to the good field region. Consequently, a relative harmonic distortion below $1 \times 10^{-4}$ is achievable.

[1] The value of 10% is an upper bound of the variation seen even between completely different suppliers for a given electrical steel grade.

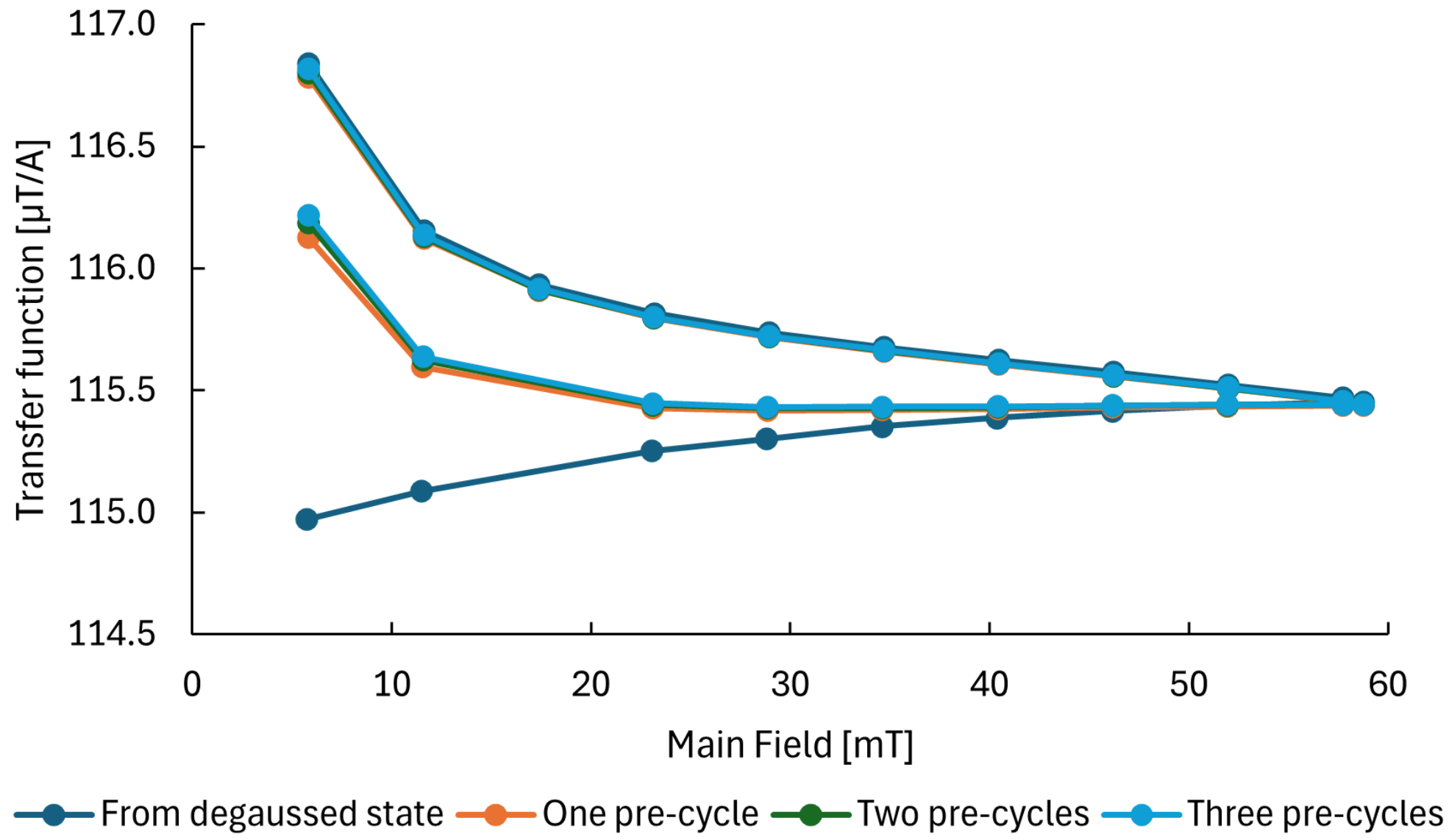

Fig. 6.4: The centered transfer function of the short prototype booster dipole at the $\mathrm{t\bar{t}}$ field levels [343].

Table 6.3: General parameters for the quadrupole magnet.

| Parameter | Unit | Value |
|---|---|---|
| Strength, main arc at $\mathrm{t\bar{t}}$ ext. | $\mathrm{T\,m^{-1}}$ | 26.6 |
| Ultimate strength, disp. sup. at $\mathrm{t\bar{t}}$ ext. | $\mathrm{T\,m^{-1}}$ | 28.7 |
| Aperture diameter | mm | 65 |
| Length | m | 1.3 |
| Outer envelope | mm | ⌀584 |
| Current, arc at $\mathrm{t\bar{t}}$ ext. | A | 939 |
| Additional MMF for disp. sup. at $\mathrm{t\bar{t}}$ ext. | A | 1350 |
| Magnet resistance | mΩ | 11.4 |
| Magnet inductance | mH | 13.3 |
| Peak voltage magnet | V | 15.5 |
| Conductor (copper) | $\mathrm{mm^2}$ | $14.3 \times 18.25$, ⌀2.8 |
| Turns | | 13 |
| RMS current density, main arc at $\mathrm{t\bar{t}}$ | $\mathrm{A\,mm^{-2}}$ | 1.5 |
| Temperature rise at 6 bar | °C | 11.2 |
| Magnet active mass | kg | 2070 |
| Coil active mass | kg | 368 |
| Yoke active mass | kg | 1702 |
| Coil overhang | mm | 140 |

The cross section of the quadrupole has been optimised to minimise the total lifetime cost of both the magnet itself and the technical infrastructure. This optimisation began with a parametric cross-section, which was then refined in two stages. First, the geometry inboard of the coil was adjusted to maximise the gradient delivered per ampere-turn. Next, a global optimisation was performed to determine the number of turns, current density, and back leg thickness (see Section 3.8.1). These three parameters—number of turns, current density, and back leg thickness—are key trade-offs between capital expenditure (CapEx), operational expenditure (OpEx), and infrastructure requirements and costs.

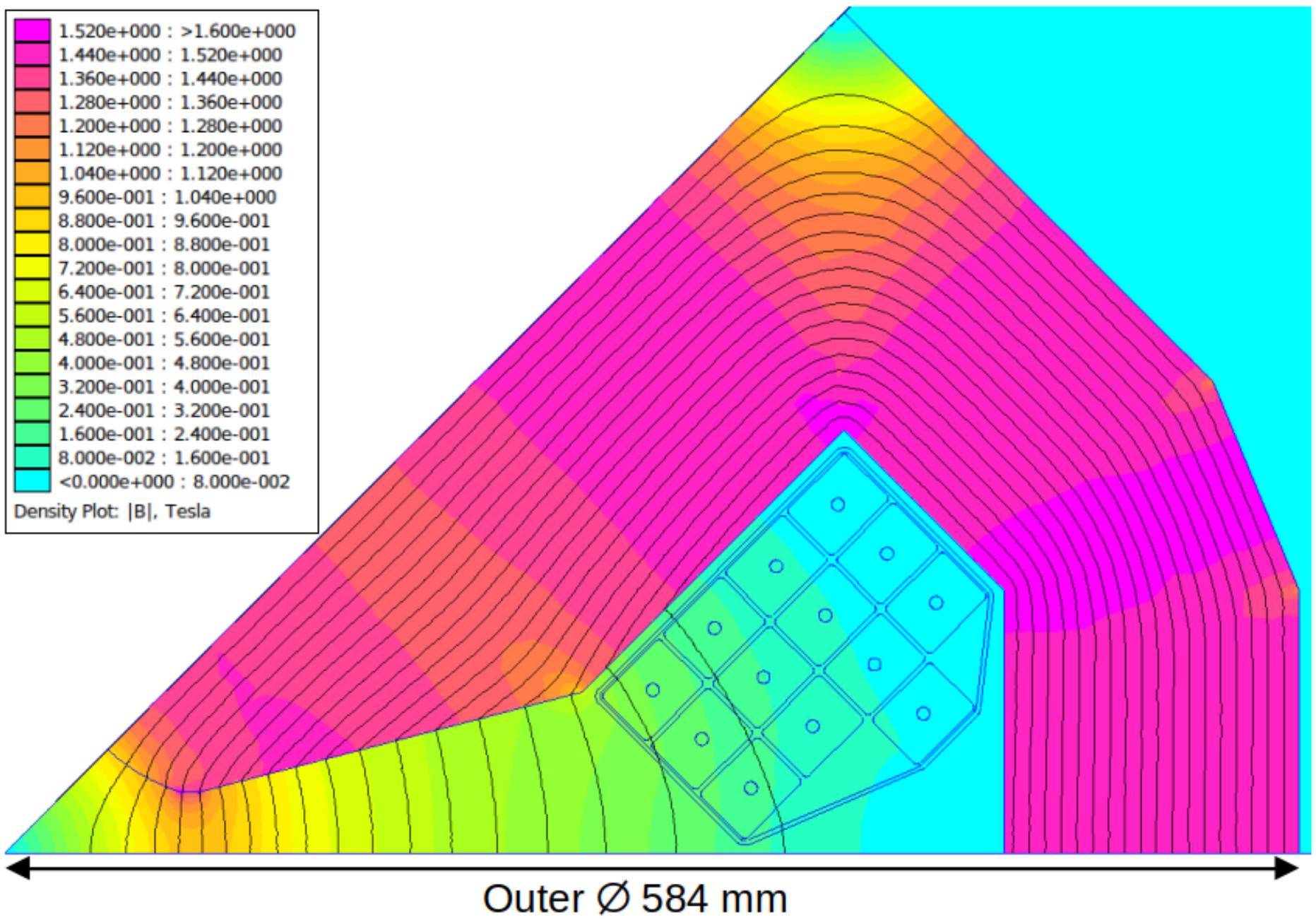

Fig. 6.5: Field map of the booster quadrupole magnet at $26.6\,\mathrm{T\,m^{-1}}$, main arc at $t\bar{t}$ extraction.

To facilitate such a complex optimisation, a parametric code was used to generate design space maps, allowing for rapid interpolation during function calls. These maps plotted stored magnetic energy, magnetic efficiency, and yoke cross-section as functions of current density and pole width. A design could then be specified based on three interpolated variables and the number of turns. The required ampere-turns were determined from efficiency and bulk current density, while the cooling hole diameter was solved using Newton-Raphson for a given temperature rise and pressure. The resistance was calculated based on the cooling hole diameter and filling factor, and the inductance was derived from magnetic energy and the number of turns. Finally, the mass was determined from the yoke cross-section and the coil. The recommended design was then manually reviewed to finalise the cross-section.

The magnet's current density is relatively low, at $1.5\,\mathrm{A\,mm^{-2}}$ RMS in the main arcs during the $t\bar{t}$ cycle, demonstrating that despite the fixed magnet lifetime, investing in larger coils remains beneficial to reduce power consumption and electrical infrastructure requirements.

There are 24 quadrupoles in the dispersion region that require ultimate strength. At this field level, the yoke begins to saturate, causing efficiency to drop from 96% at nominal to 93% at ultimate. A bespoke solution for these quadrupoles may be proposed at a later stage.

### 6.1.3 Sextupoles

The parameters of the proposed sextupole magnet are summarised in Table 6.4. The field map for the sextupole magnet at $t\bar{t}$ energy is given in Fig. 6.6. The sextupole is again relatively strong with a pole tip field of $0.7\,\mathrm{T}$, and again a conventional design is proposed, with material and design choices similar to those for the quadrupole. The total harmonic distortion is below $1 \times 10^{-4}$ relative to the main field.

Table 6.4: General parameters for the sextupole magnet.

| Parameter | Unit | Sextupole | |
|---|---|---|---|
| | | Defocus | Focus |
| Strength, $B''$ | $\mathrm{T\,m^{-2}}$ | 1219 | 1147 |
| Aperture diameter | mm | 65 | |
| Length | m | 1.4 | 0.7 |
| Outer envelope | mm | ⌀305 | |
| Peak current | A | 595 | 556 |
| Magnet resistance | mΩ | 51 | 27 |
| Magnet inductance | mH | 9.3 | 4.7 |
| Peak voltage magnet | V | 32 | 16 |
| Conductor (copper) | $\mathrm{mm^2}$ | 8.4 × 8.4, ⌀2.5 | |
| Turns | | 10 | |
| RMS current density at $t\bar{t}$ | $\mathrm{A\,mm^{-2}}$ | 3.7 | 3.5 |
| Temperature rise (6 bar) | °C | 14 | 4.6 |
| Magnet active mass | kg | 614 | 312 |
| Coil active mass | kg | 105 | 57 |
| Yoke active mass | kg | 509 | 255 |
| Coil overhang | mm | 50 | |

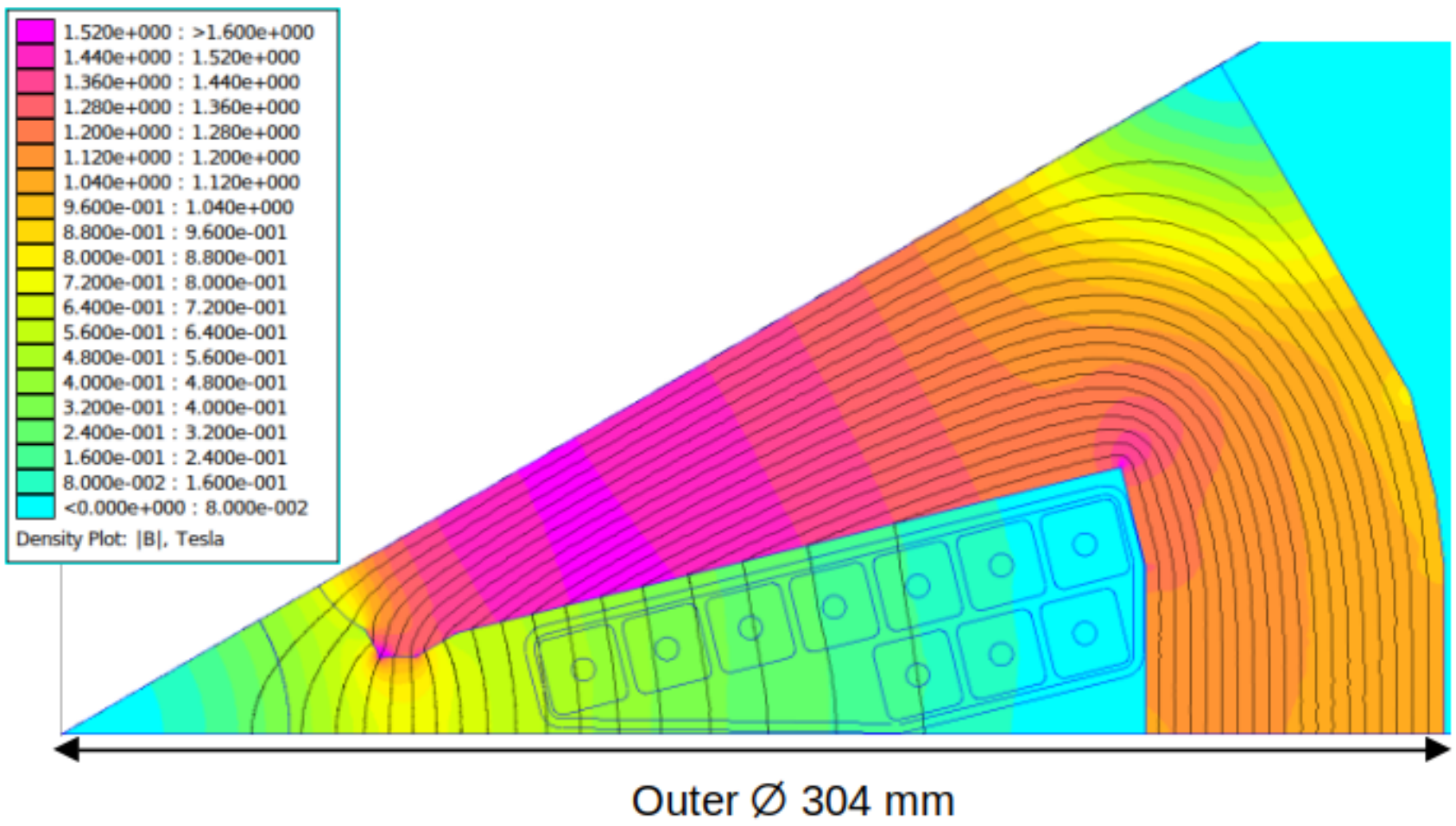

Fig. 6.6: Field map of the booster sextupole magnet at $t\bar{t}$ energy, $B'' = 1219\,\mathrm{T\,m^{-2}}$.

### 6.1.4 Summary

In conclusion, the headline magnet requirements of v24 FODO optics are technically feasible including low dipole field and lens strengths. For reference, the power consumption of the proposed magnets are summarised in Table 8.15. There is a large design space of feasible magnets, giving an opportunity for holistic design and a cost optimised solution. Moving into the next phase the industrialisation, design for manufacture and assembly, and detailed integration with the other technical systems will be performed.

## 6.2 Booster vacuum system

The booster chamber design is based on a round seamless tube in OFE copper. The inner diameter is 60 mm and the thickness is 1.5 mm, see Fig. 6.7. A copper tube (size yet to be determined) is spot welded on it for water cooling. No lump photon absorber is planned. There is no need for a bakeout thermal cycle, and therefore, non-shielded bellows are used to compensate for mechanical and alignment tolerances. The chamber is neither coated nor baked.

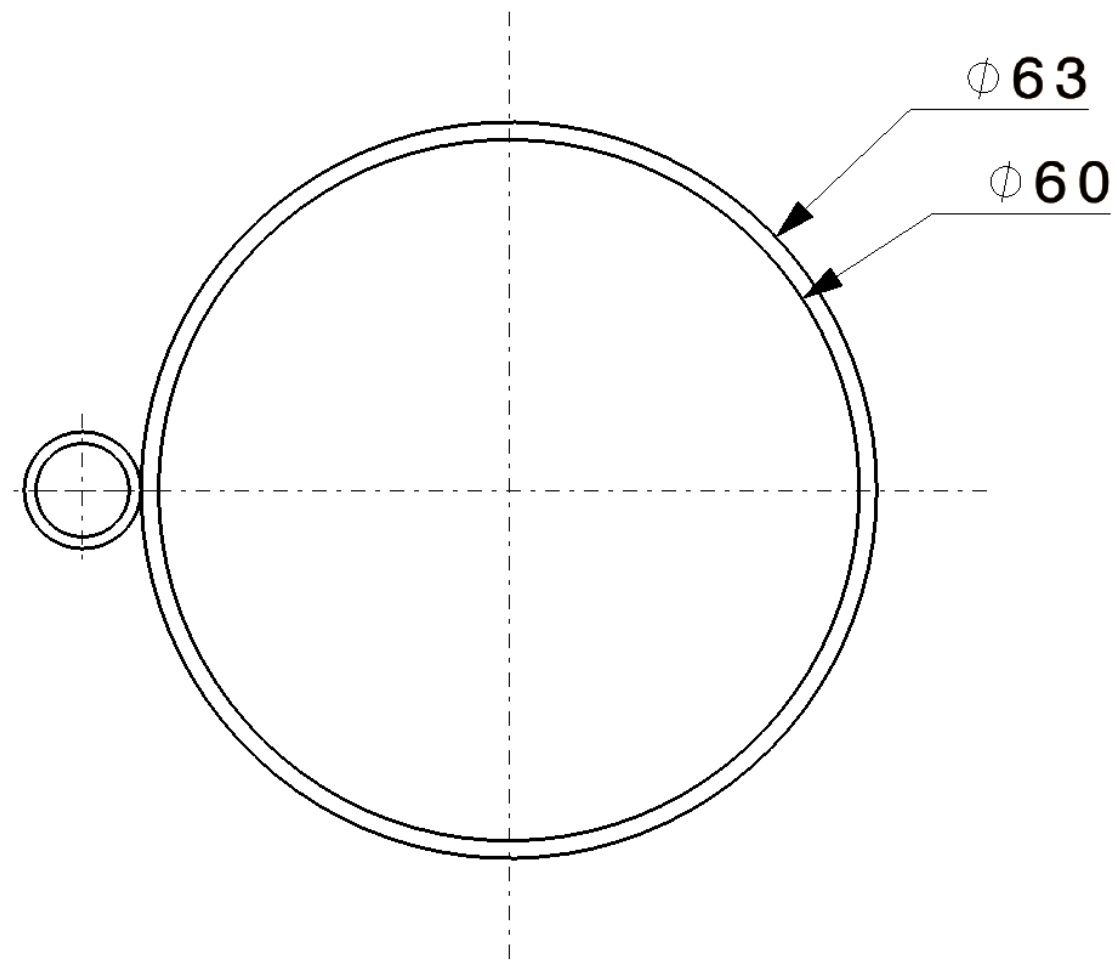

Fig. 6.7: Cross section of the FCC-ee booster vacuum chamber.

### 6.2.1 Input parameters

To determine the pressure levels in the FCC-ee booster, the end-of-2024 state of the design has been used as a baseline, as summarised below: The vacuum chamber results in 26.6 l.m/s specific conductance (a value similar to the two storage rings).

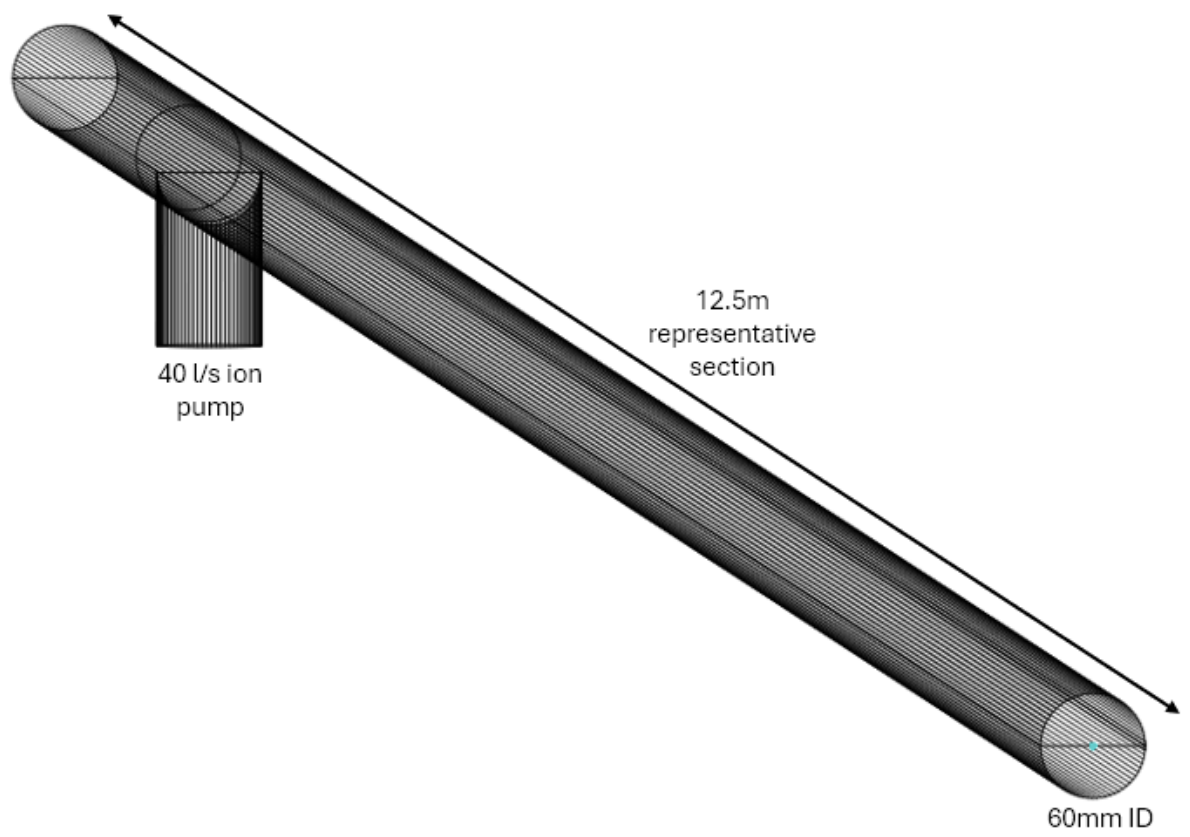

Fig. 6.8: Representative section of the vacuum chamber in the FCC-ee booster, modelled over 12.5 m.

It is assumed that one pump is placed every 12.5 m around the ring. Consequently, a 12.5 m section with periodic boundary corrections has been modelled (see Fig. 6.8). The section includes a single ion pump, which is simulated to operate at a pumping speed of 40 l/s, accounting for the conductance limitation of the orifice, despite its rated speed of 60 l/s.

In the following discussion, a distinction is made between static (background) gas, which is always present in the system, and dynamic (time-dependent) gas, which arises during operation. Since their calculation methods differ, they are addressed in separate sections. In both cases, the gas levels are estimated after 100 hours of conditioning.

### 6.2.2 Static pressure

For unbaked metals, the dominant gas to desorb is water vapour. Empirical tests show [344] that the outgassing rate is inversely proportional to the pumping time and can be estimated as:

$$Q_{H_2O} \approx \frac{3 \times 10^{-9}}{t} \quad (\text{mbar.l/s/cm}^2) \tag{6.1}$$

which corresponds to a specific outgassing rate of $3 \times 10^{-11}$ mbar.l/s/cm$^2$ after 100 hours.

A MOLFLOW simulation was performed on the model of the representative section for this outgassing from the walls and the 40 l/s pumping speed on the pumping port.

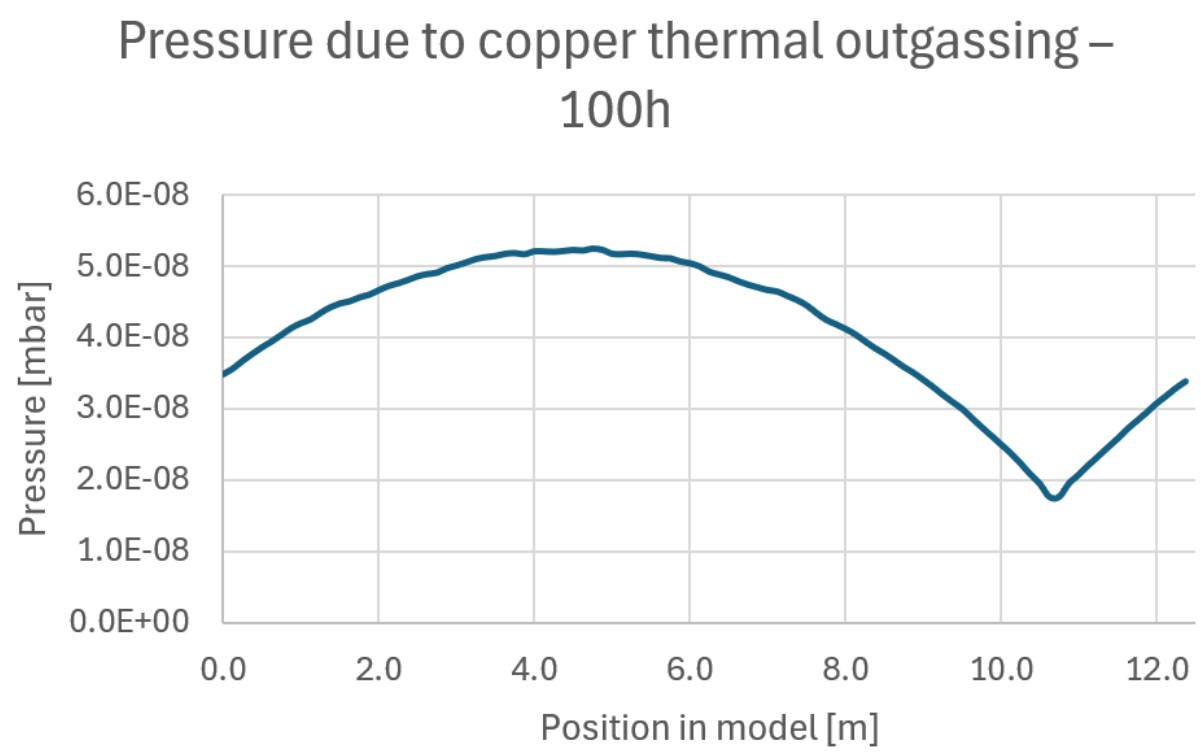

Fig. 6.9: Static (background) pressure after 100 h of conditioning in a 12.5 m long representative section.

The resulting pressure profile, in Fig. 6.9, shows an average pressure in the $3 \times 10^{-8}$ mbar range, with the lowest pressure at the pump location, as expected.

### 6.2.3 Dynamic vacuum

High-energy photons are created by synchrotron radiation (SR) when the booster is operating. Estimating the dynamic vacuum requires calculating the amount of SR, calculating the resultant outgassing through photon-stimulated desorption (PSD), and then performing a vacuum simulation. As the booster is a cyclic machine, all quantities are time-dependent.

The SR photon flux scales linearly with the beam energy and current:

$$\text{Flux [photons/sec]} = 8.08 \times 10^{17} \times E_{\text{beam}}[\text{GeV}] \times I_{\text{beam}}[\text{mA}] \tag{6.2}$$

The booster's operation parameters in the Z mode are as follows:

- 1120 bunches per cycle.
- $2.5 \times 10^{10}$ electrons per bunch.
- Revolution frequency: 3.3 kHz.
- Current: 14.8 mA.
- Operation cycle:
  - 0-2.8 s: Accumulation (linear increase from 0 to 14.8 mA at 20 GeV).
  - 2.8-3.6 s: Ramp-up (energy increase from 20 GeV to 45.6 GeV).
  - 3.6-3.8 s: Ramp-down (cycling magnets, no beam).

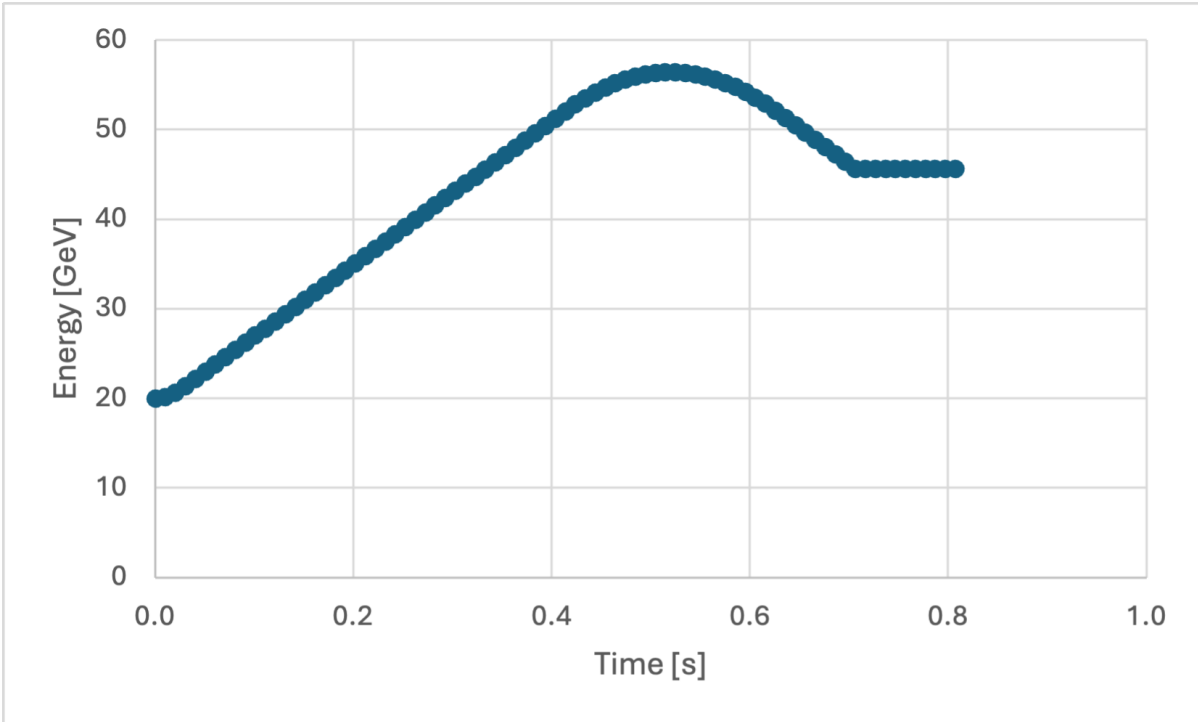

Fig. 6.10: Energy ramp-up from 20 to 45.6 GeV with an overshoot.

### 6.2.4 SynRad simulations

To estimate the SR, the formula can be made more precise by applying the so-called SR_factor, which specifies what portion of the SR spectrum has an energy over 4 eV. The reason is that photons below that energy (equivalent to the work function for photo-electron generation on most metal alloys) are not powerful enough to desorb molecules from the surface.

The Monte Carlo simulation code SynRad was used to model and simulate four magnets in a periodic setup (D1-D2-D3-D4, then D1 again, see Fig. 6.11). Each magnet is 11 m long, and the magnetic field scales linearly with the energy, from 6.5 mT at 20 GeV.

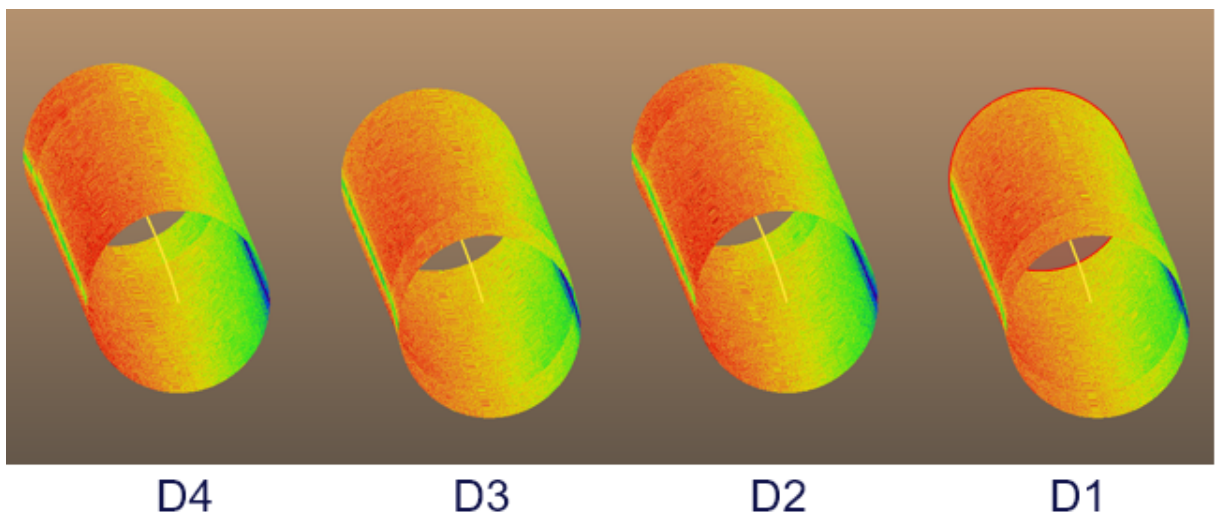

Fig. 6.11: SynRad model representing the four booster dipoles creating SR, in a periodic setup.

SynRad does not support time-dependent beam parameters, however, it is possible to obtain the average SR factor by simulating 26 distinct energies equally spaced from 20.6 GeV to 45.6 GeV and taking the average (see Fig. 6.12).

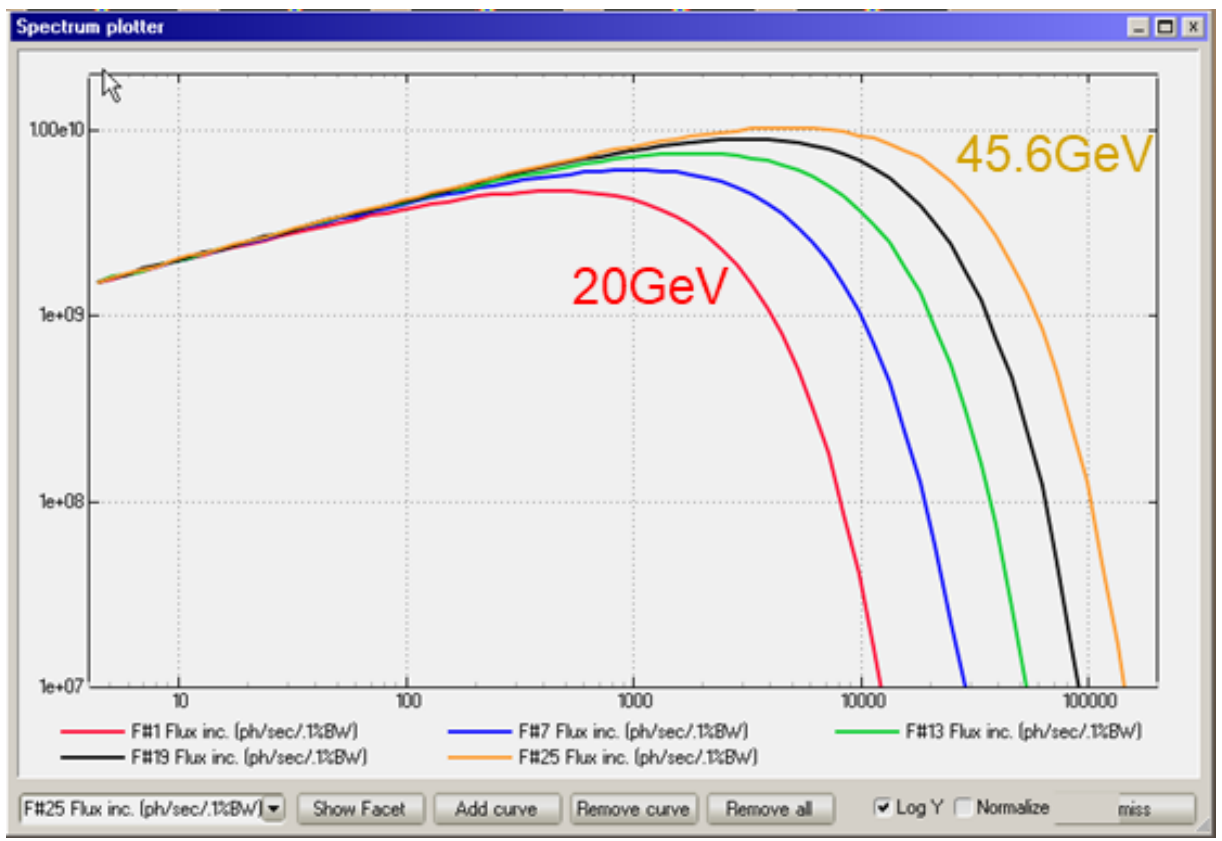

Fig. 6.12: Part of SR spectrum above 4 eV for beam energies at 20, 26, 32, 38 and 45.6 GeV.

Averaging the flux for the 26 energies at 14.83 mA, SYNRAD estimates the flux to be $2.24 \times 10^{17}$ photons/s. Taking only the average SR factor of 88% from SYNRAD, and substituting in (6.2):

$$\text{Flux} = 8.08 \times 10^{17} \times 32.8 \times 14.83 \times 0.88 \times \frac{44}{64486} = 2.36 \times 10^{17} \text{photons/s} \tag{6.3}$$

where 32.8 GeV is the average energy during the ramp, and 64 486 m is the dipoles' magnetic length. The SYNRAD and the theoretical results are thus very close.

The results of the 44 m length modelled in SYNRAD (for calculating the SR) to the 12.5 m length in MOLFLOW (for calculating the vacuum). The booster comprises more than just dipoles. It is estimated that the total radiation from the dipoles is distributed over a 90 km arc length :

$$\text{Flux [12.5 m]} = 8.08 \times 10^{17} \times E \times I \times 0.88 \times \frac{12.5\,\text{m}}{90\,\text{km}} \tag{6.4}$$

The following formula is used to convert the flux to outgassing:

$$Q[\text{molecules/sec}] = \text{Flux} * \eta \tag{6.5}$$

where $\eta$, the molecular yield, describes how many photons on average are desorbed by an absorbed SR photon. This depends on the material and decreases over time as the surface is conditioned. The duration of this conditioning is estimated to be 100 h, based on data from an experiment performed at KEK [345] where 1.2 m long uncoated stainless steel vacuum chambers were irradiated with SR.

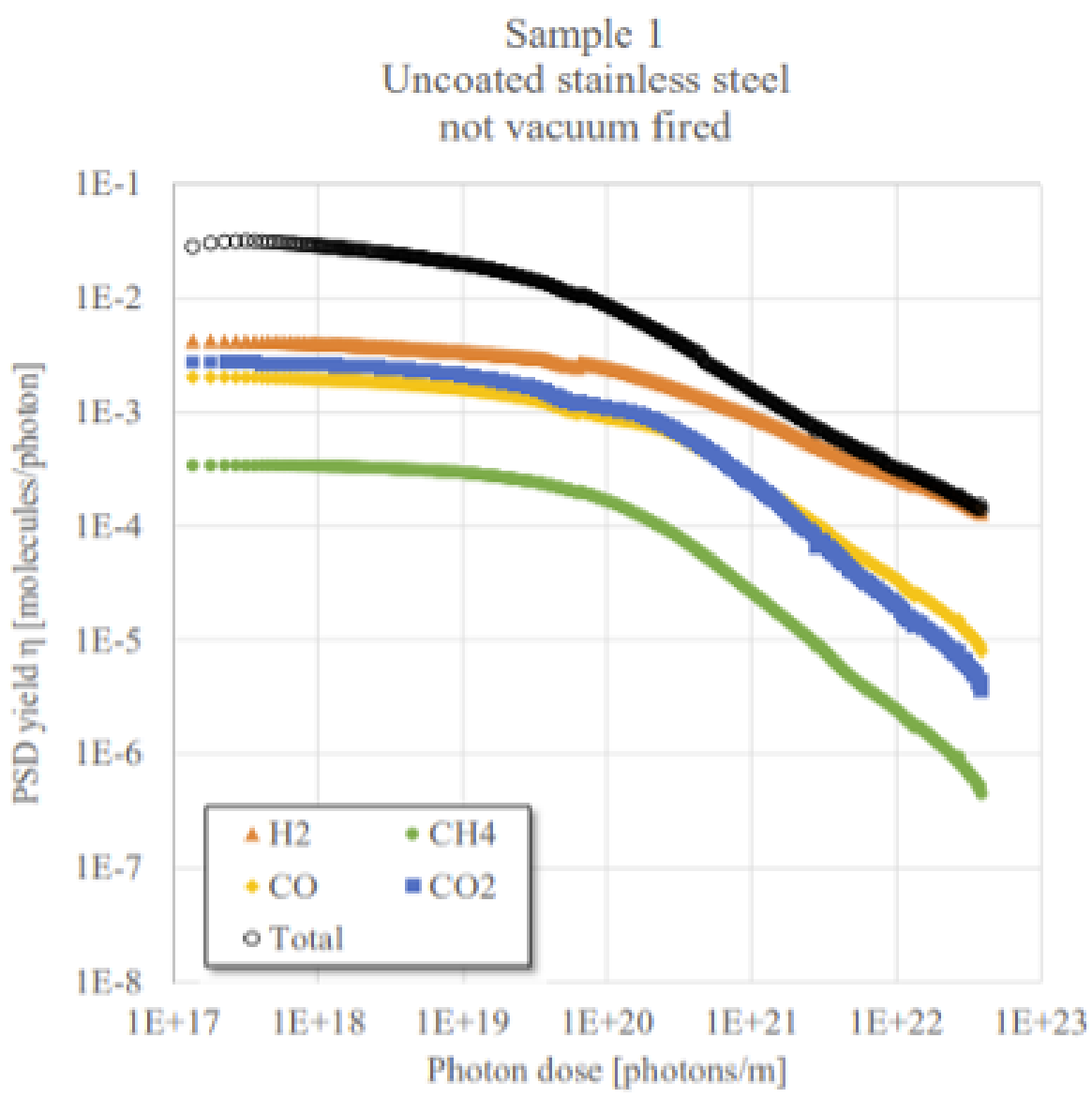

Fig. 6.13: Photon stimulated desorption yield of uncoated stainless steel.

As shown in Fig. 6.13, the dominant gas species are typically $H_2$, CO, $CO_2$, $CH_4$, ordered by fraction (plus $H_2O$, if unbaked, which can initially be the predominant gas species–but the sample in the figure was baked). In this case, CO was considered because it has an atomic number much higher than that of hydrogen, and therefore its effect in terms of beam bremsstrahlung (BS) is approximately 56 times bigger than for $H_2$. The dependence for BS given by $Z_{\text{eff}}(Z_{\text{eff}} + 1)$ , where $Z_{\text{eff}}$ is the weighted effective atomic number for C and O.

Calculating with an average cycle current of 8 mA, 100 h of operation corresponds to a photon density of $1 \times 10^{21}$ photons/m, resulting in a CO yield of $3 \times 10^{-4}$ mol/photon. As a result, the outgassing is:

$$Q = 8.08 \times 10^{17} \times E \times I \times 0.88 \times \frac{12.5}{90\,000} \times k_B T \times 10 \tag{6.6}$$

where $E$ is the energy in GeV, $I$ is the beam current in mA, $T$ the temperature in K and the $k_B T$ term originates from the ideal gas equation, allowing conversion from molecules/s to Pa·m$^3$/s, and the factor of 10 is the conversion from Pa·m$^3$/s to mbar·l/s. A temperature of 300 K was assumed.

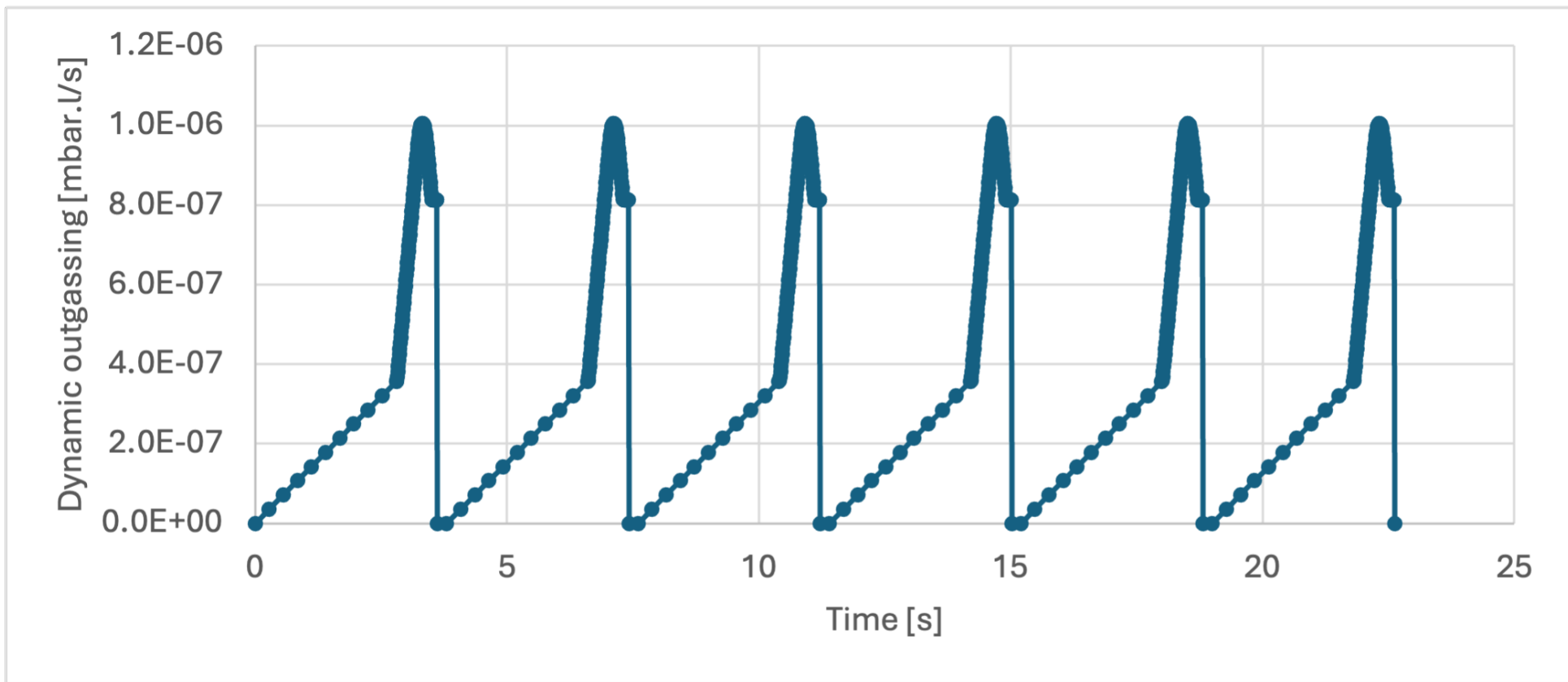

Fig. 6.14: Dynamic outgassing caused by PSD, over 6 booster cycles on a 12.5 m modelled region.

### Dynamic vacuum simulations

The result of the equation (visualised in Fig. 6.14) was imported into MOLFLOW as a time-dependent outgassing parameter and applied it to the side wall of the modelled region, as the location of primary incidence of SR.

As the residence time of molecules in the booster is not negligible, the dynamic pressure builds up after approximately 3 cycles when starting the booster from a static state, as can be seen in Fig. 6.15.

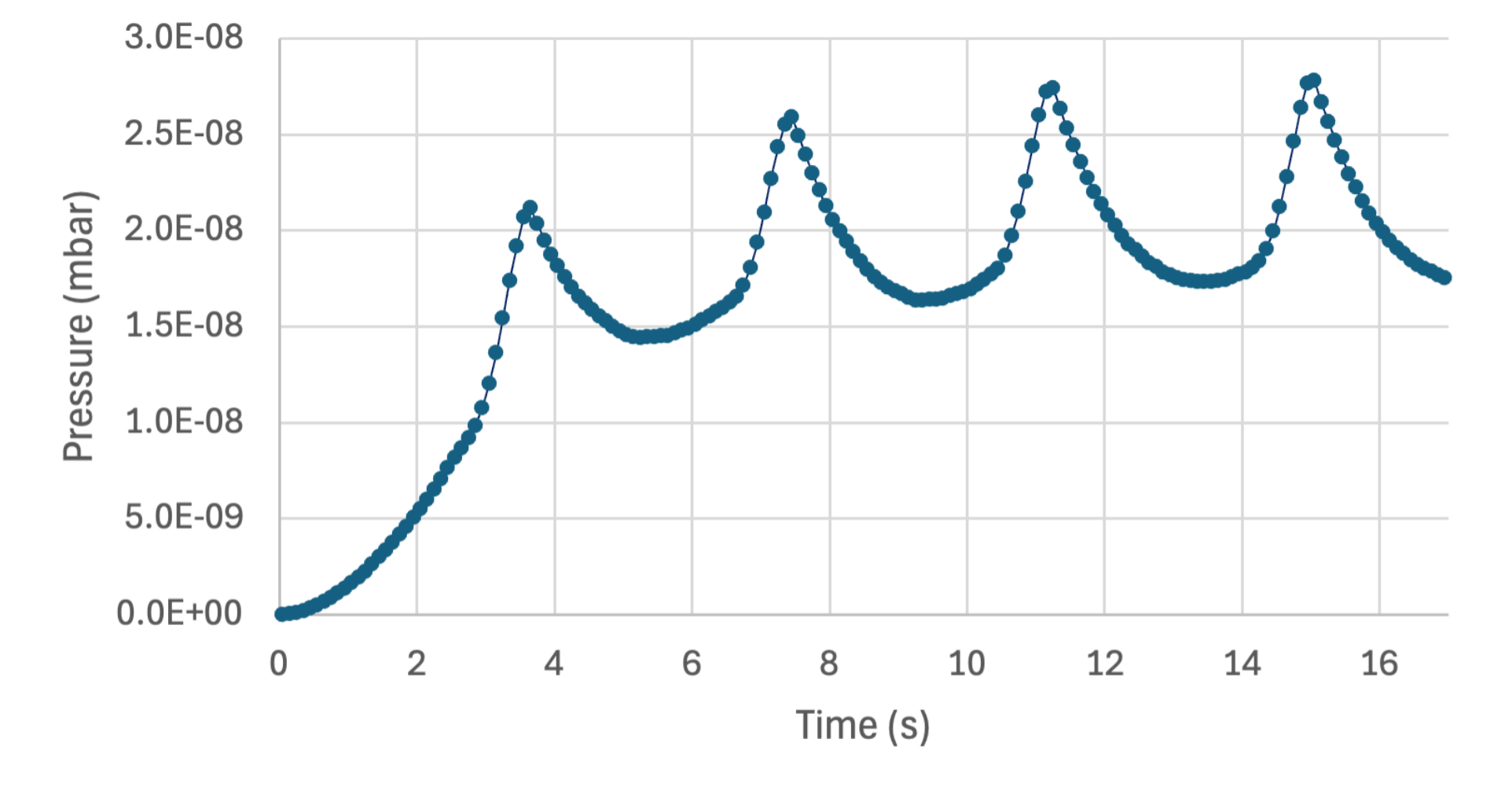

Fig. 6.15: Dynamic pressure, averaged over the 12.5 m modelled region, after starting the booster.

The average pressure, after reaching a periodic state, cycles between $1.7 \times 10^{-8}$ mbar and $2.8 \times 10^{-8}$ mbar. The peak pressure was chosen to have a conservative estimate in the following stages.

### 6.2.5 Comparison and conclusion

The pressures from the two sources, and their sum are plotted in Fig. 6.16. While the static pressure, due to thermal outgassing, is higher after 100 h, it also conditions faster:

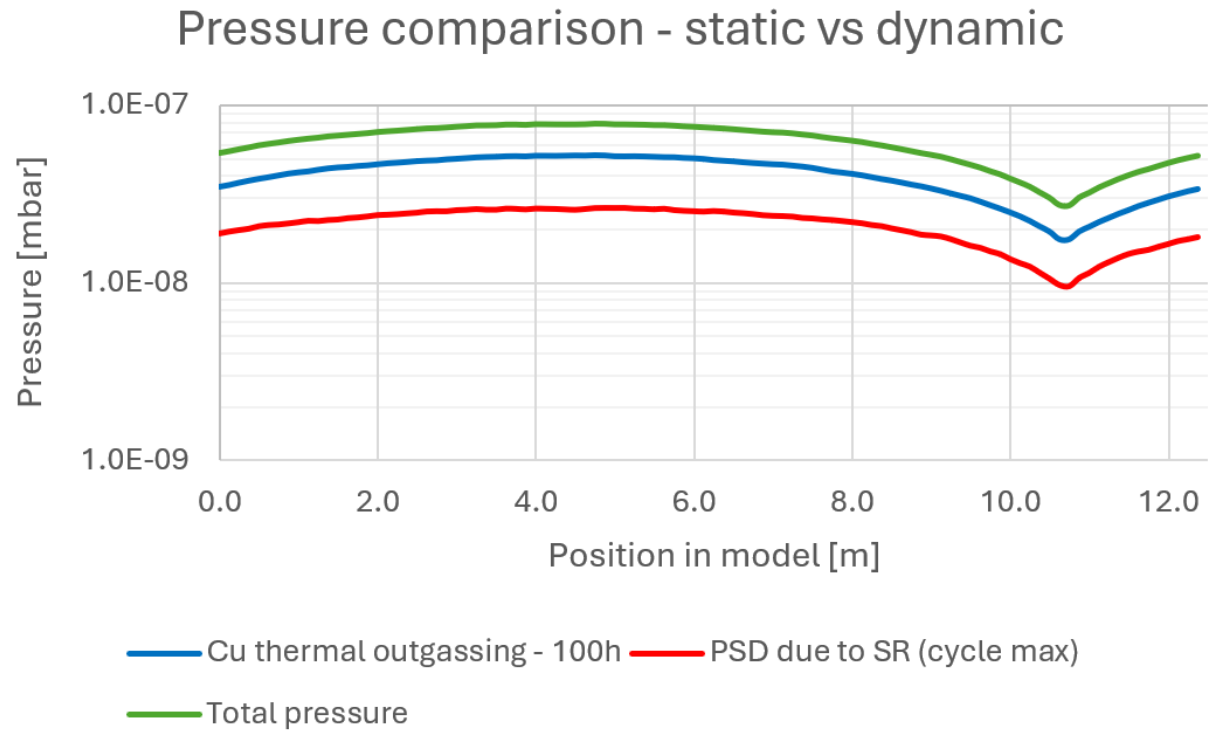

Fig. 6.16: Comparison of static and peak dynamic pressures after 100 h of conditioning.

- Thermal outgassing decreases with $t^{-1}$ [344]
- Photon stimulated desorption of stainless-steel decreases with $t^{-0.6}$ [345]

After $100\,\mathrm{h}$, the total pressure at the peak of a booster cycle is in the order of $4 \times 10^{-8}$ mbar. The latest beam-gas calculations estimate that the booster pressure needs to be below 30 nTorr [346], which is the expected level.

It would be preferable to have an in-situ bakeout system installed, and consequently NEG-coating of the vacuum chambers, as this would reduce the number of external pumps and their cabling by a factor of about 5. Additionally, the NEG-coated solution would guarantee very fast reconditioning of any vacuum sector needing venting during operation.

## 6.3 Radio frequency system layout, configurations, and parameters

### 6.3.1 Introduction

The high-energy booster SRF system accelerates the electron and positron bunches before injection into the two collider rings. A booster transverse feedback system is required to cure coupled-bunch transverse instabilities, as specified in Section 4.2.1 as well as damping injection oscillations. Both RF systems will be installed in point PL.

### 6.3.2 Parameter choices for the SRF system

The present baseline RF system requires flexibility for switching between Z, WW, and ZH operating points. Then it needs to be upgraded for the highest energy $\mathrm{t\bar{t}}$ operating point. The baseline solution assumes the same 6-cell $800\,\mathrm{MHz}$ elliptical cavities as in the collider. The main RF-related global parameters are summarised in Table 6.5. Note that the indicated maximum synchrotron radiation power includes the energy overshoot for the Z operating point without wigglers. To avoid hardware modification

Table 6.5: RF-related parameters for FCC-ee high-energy booster.

| Operating point | | Z | WW | ZH | $\mathrm{t\bar{t}}$ |
|---|---|---|---|---|---|
| Maximum beam current* | [mA] | 16.2 | 6.2 | 2.0 | 0.4 |
| Extraction Energy loss / turn | [MeV] | 36.1 | 342 | 1730 | 9270 |
| Injection RF voltage | [MV] | 50.1 | | | |
| Extraction RF voltage | [MV] | 57.2 | 402 | 1960 | 10200 |
| Maximum synchrotron radiation power | [MW] | 1.07 | 2.13 | 3.49 | 3.99 |

* Including 80% injection efficiency in the FCC-ee collider.

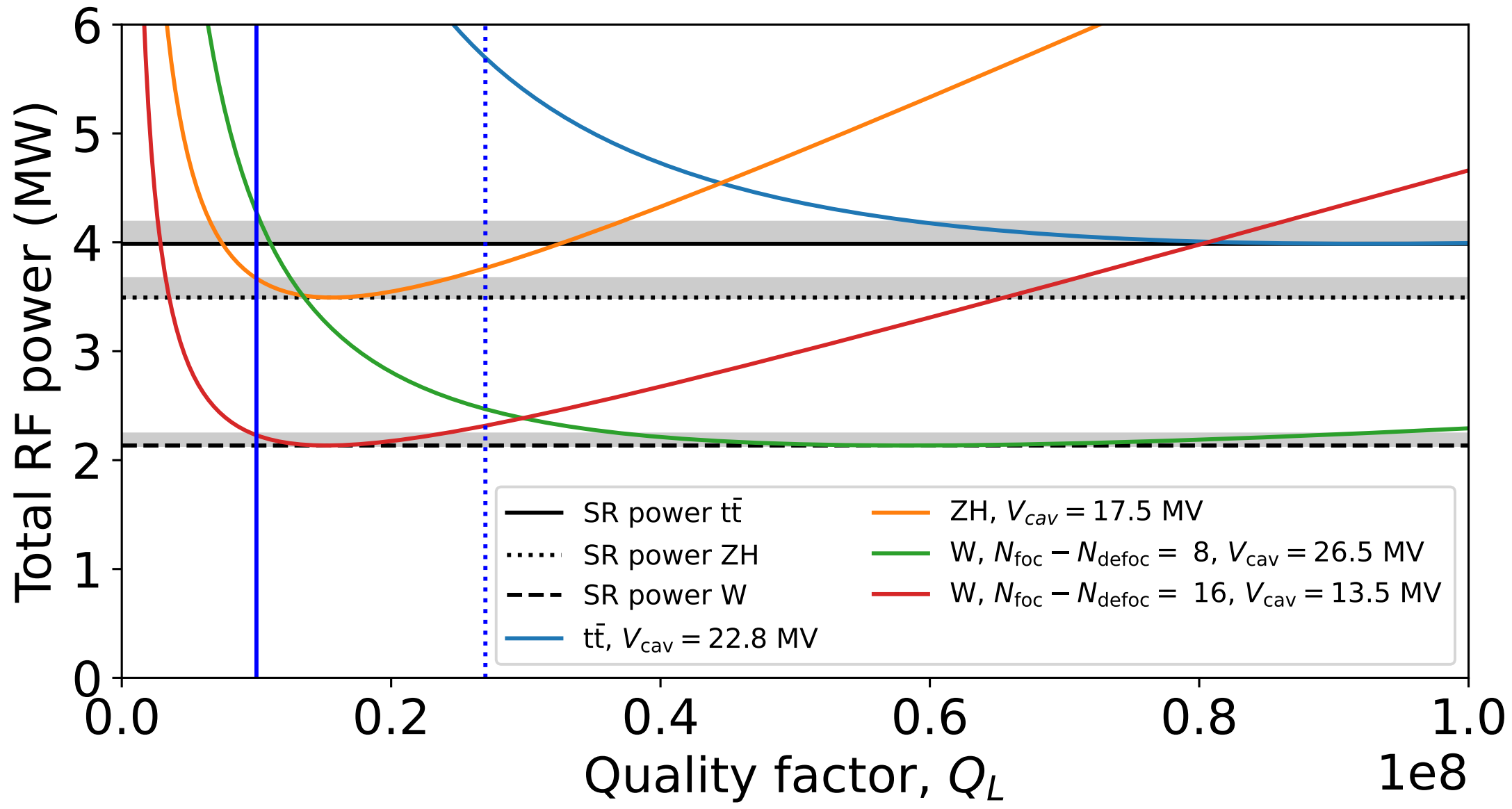

Fig. 6.17: RF power requirements as a function of the cavity quality factor. Vertical lines indicate the chosen quality factors for the first (solid) and second (dotted) stages of HEB SRF systems.

in the first stage, a common quality factor that minimises RF power requirements needs to be chosen. The RPO scheme must be deployed at the injection energy. At higher energies, it remains active for the Z and WW operating points, while the transition to the normal phase operation mode has to be done for the ZH and $t\bar{t}$ ramps to achieve the maximum total RF voltage. It is assumed that the first 112 cavities share a single RF power source per four cavities. Therefore, the difference between the number of focusing and defocusing cavities, $N_{\text{foc}} - N_{\text{defoc}}$, can be changed in steps of eight cavities. Several scenarios for the WW operating points and corresponding RF power requirements were analysed (Fig. 6.17). Choosing $N_{\text{foc}} - N_{\text{defoc}} = 16$ and $Q_L = 10^7$ (solid blue vertical line) seems to be a good compromise that leads to a small RF power overshoot for both WW and ZH operating points and keeps the cavity bandwidth (half-height full-width) $\Delta f = f_{\text{RF}}/Q_L \approx 80\,\text{Hz}$. Switching to the $t\bar{t}$ operating point, $Q_L$ needs to be changed to reduce RF power requirement, which can be done by adjusting waveguide length as described in Section 3.4.10. The optimal $Q_L = 9.2 \times 10^7$ results in an extremely small $\Delta f \approx 9\,\text{Hz}$ which might be challenging to control in an accelerator environment. Therefore, the $Q_L = 2.7 \times 10^7$ ($\Delta f = 30\,\text{Hz}$) is chosen with a drawback of 43% more RF power than SR power at 182.5 GeV. The range $Q_L = 4.2 \times 10^6 - 2.7 \times 10^7$ assumed to be covered by the common 800 MHz FPC design for collider and booster.

***RF power requirements at injection energy***

The RF power requirements during the injection process depend on the LLRF system details and the beam loading compensation scheme in use. Extremely slow SR damping rate and very high impedance of the fundamental mode require a direct RF feedback system to suppress the longitudinal coupled-bunch instability. The expected growth rate for the Z operating point at the end of the filling at the injection energy is shown in Fig. 6.18. The stability can be improved by enhancement of the SR damping rate using a wiggler as described in Section 5.1.2. The dynamic model, similar to the one of the collider (Section 3.4), was applied to evaluate RF power transients during injection (Fig. 6.19, left). The peak RF power increases significantly even after injection of the first four bunches due to the fast and strong reaction of the direct RF feedback, so its gain must be reduced. This can be acceptable since the instability growth rates shown in Fig. 6.18 are computed for the maximum beam current. The RF power can be kept acceptable at the 50 kW limit if cavities are gradually detuned as the beam current increases (Fig. 6.19, right). At any moment the filling schemes should be maintained as uniform as possible to reduce modulations of beam and RF cavity parameters.

Table 6.6: Main RF parameters of the FCC booster.

| | Z | W | ZH | $t\bar{t}$ |
|---|---|---|---|---|
| RPO at extraction | yes | yes | no | no |
| RF frequency [MHz] | 801.58 | | | |
| Operating temperature [K] | 2 | | | |
| Number of cells per cavity | 6 | | | |
| Quality factor $Q_0$ | $3 \times 10^{10}$ | | | |
| Cavity voltage at extraction [MV] | 5.6 | 13.5 | 17.5 | 22.8 |
| $E_{acc}$ [MV/m] | 4.9 | 12.0 | 15.6 | 20.3 |
| Max. RF power per cavity [kW] | 42 | | 8.9 / 12.7 | |
| Coupling factor $Q_L$ | $1 \times 10^7$ | | $9.2 \times 10^7$ / $2.7 \times 10^7$ | |
| Number of cryomodules | 28 | | 112 | |
| Number of cavities | 112 | | 448 | |

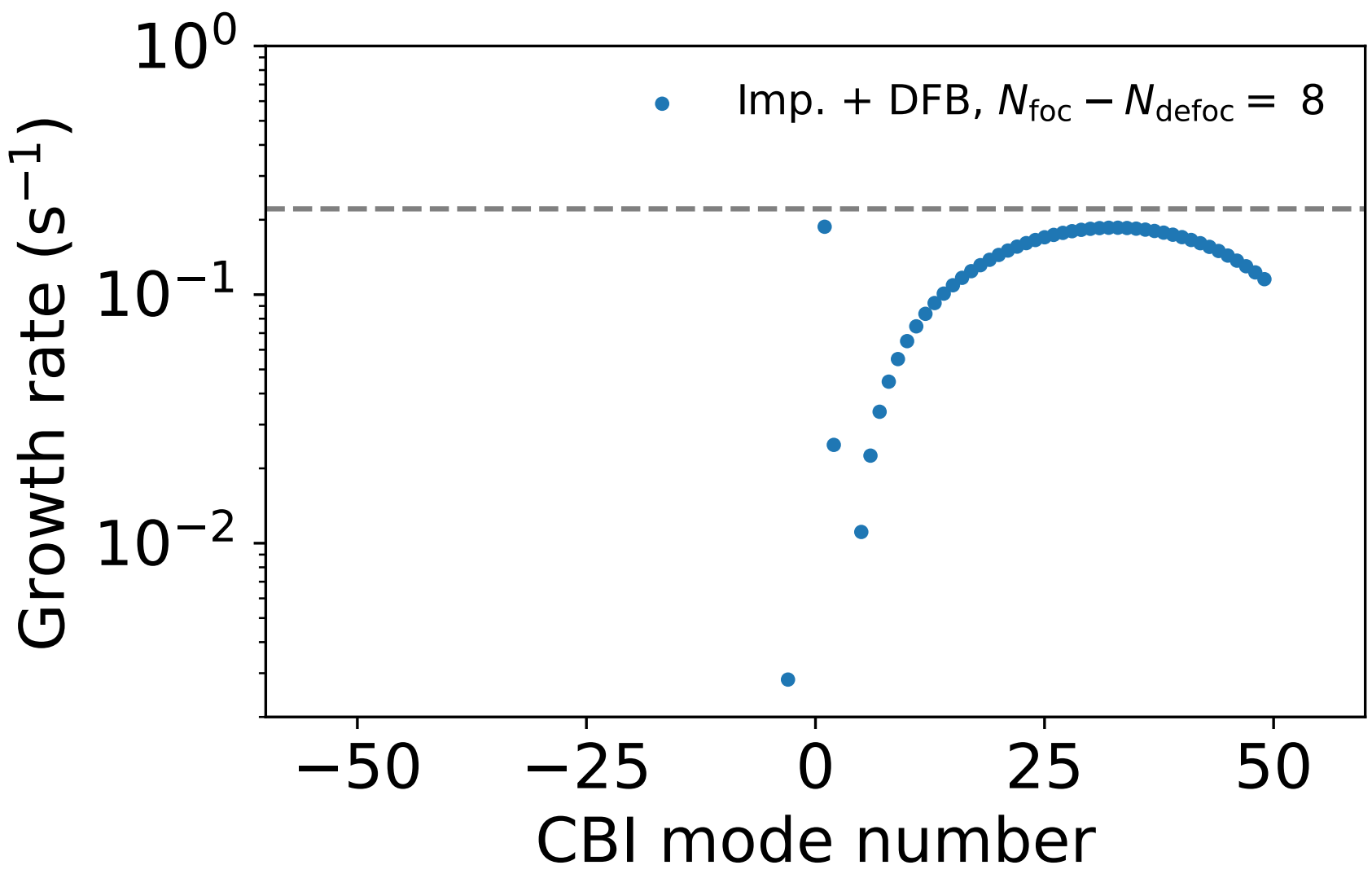

Fig. 6.18: Growth rates of longitudinal coupled-bunch instabilities driven by the fundamental cavity mode for the Z operating point.

The development of a frequency tuning system able to operate in a pulsed regime with a resolution at the Hz level will be mandatory. It can be based on the elastic axial deformation of the cells by an electromechanical system equipped with a stepping motor and a piezo-electric actuator. A new type of tuning technology can also be considered, for example, a non-mechanical tuner using ferroelectric materials [347, 348].

### 6.3.3 Longitudinal coupled-bunch instabilities due to RF cavity HOMs

The low beam injection energy, combined with long longitudinal and transverse damping times, reduces the beam stability limit at the booster injection phase. The simultaneous installation of all 112 cavities brings the total beam impedance close to the stability threshold. To mitigate this, a transverse feedback system and wigglers can help improve stability.

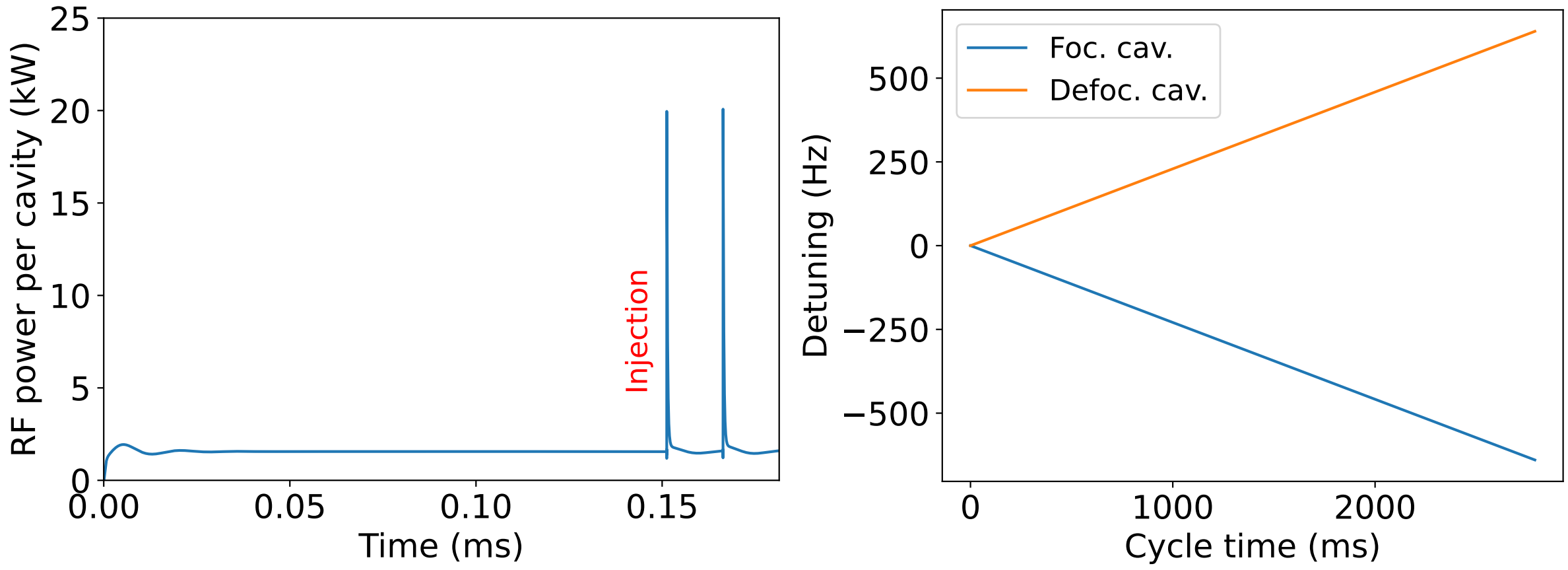

Fig. 6.19: Left: evolution of RF power before and after injection of the first four bunches at the Z operating point with parameters: $N_{\text{foc}} - N_{\text{defoc}} = 8$, the feedback gain is 20% of the optimal value. Right: evolution of cavity detuning at 20 GeV for Z operating point.

From a cavity design perspective, modifying one of the end cells can improve the damping of the mode with the highest longitudinal impedance. Additionally, frequency spread in HOMs, caused by manufacturing imperfections, leads to resonance at different frequencies, which helps distribute and reduce peak impedance. A perturbation analysis, following the approach in Ref. [349], was conducted on the 6-cell cavity to assess its effect on the total longitudinal impedance of the cavity chain.

Figure 6.20(a) shows the longitudinal impedance of the lower-order modes within the fundamental mode (FM) passband for all 112 cavities, where the cavity geometries have been perturbed, and the total longitudinal impedance has been calculated accounting for these perturbations. Since these five modes lie within or below the FM passband, they couple primarily to the power coupler rather than the HOM couplers. Their coupling to the fundamental power coupler (FPC) depends on the target $Q_L$ of the FM. However, for two different $Q_L$ values, the frequency spread of the eigenmodes ensures that the longitudinal impedance peak remains below the Z-booster stability threshold. For the HOM with the highest longitudinal impedance, an alternative end-cell design (V2) was investigated as a potential improvement over the current design (V1) to reduce mode trapping. Additionally, implementing cell sorting during assembly, as described in Ref. [349], can help prevent HOMs from being trapped at 2.37 GHz, as shown in Fig.6.20(b). The combined effects of the transverse feedback system, increased beam energy loss from wigglers, the asymmetric end-cell design and frequency spread due to manufacturing imperfections should help keep the system below the stability limit, although with only a small safety margin.

## 6.4 Beam intercepting devices (halo collimators, beam dump)

As for any large accelerator, beam intercepting devices will be needed in the booster. At this stage the studies are not mature enough to define the actual requirements; however, it is estimated that some collimators will be needed, as well as extraction and injection protection devices, which will need to be able to withstand partial or full impacts of the high energy beams operated in the booster. Even though the total energy potentially deposited by the booster beams will be lower than that in the collider the materials used in the beam intercepting devices will also be submitted to significant loads. In addition, the other requirements (e.g., UHV compatibility, low impedance) will also make the design of these devices challenging. Another important device required by the booster is a beam dump. As described in Section 1.8.1, the collider dumps will be designed also to absorb the beams extracted from the booster. Additional requirements of beam intercepting devices, not foreseen at this stage, may be identified during the functional studies of the booster.

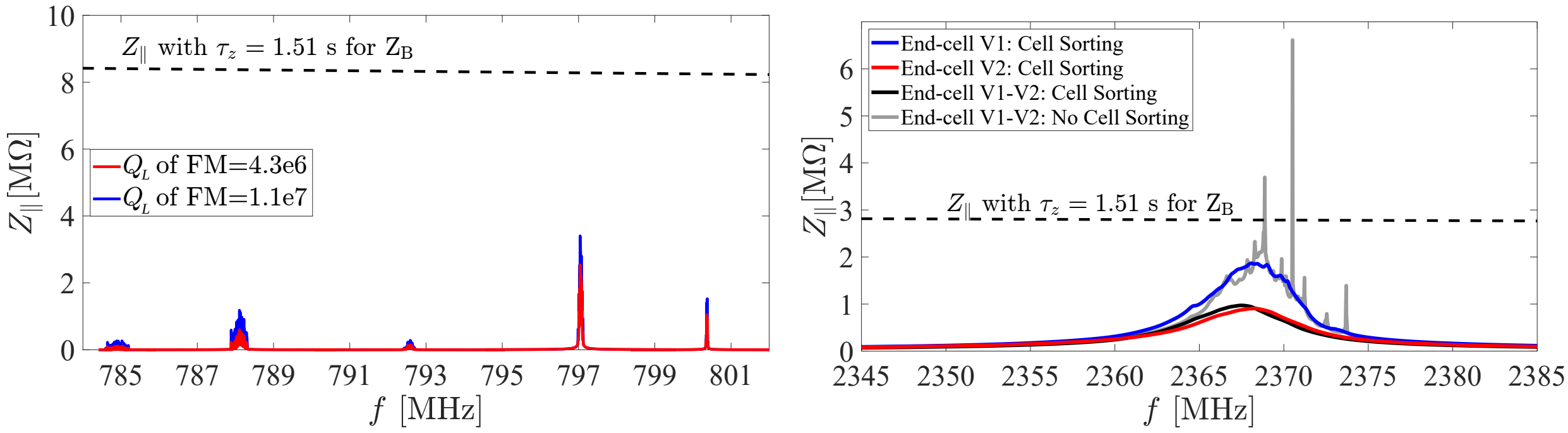

(a) (b)

Fig. 6.20: (a) Longitudinal beam coupling impedance at the FM passband for 112 cavities under geometrical perturbations, compared with the beam stability limit determined by synchrotron radiation. The impedance is calculated by considering the average $(R/Q)$ of the modes and the $Q_L$ of the modes through their coupling with the FPC. (b) Longitudinal impedance for the 2.368 GHz mode, corresponding to the HOM with the highest longitudinal impedance. The impedance for 112 cavities, with no HOM coupler and damping occurring only through the open boundary condition, is calculated under geometrical uncertainties and compared to the stability limit (including a wiggler). Asymmetric end-cells and cell sorting during cavity assembly can reduce the impedance by a small margin below the stability limit.

## 6.5 Beam transfer systems

### 6.5.1 Septa

***Injection***

As discussed in Section 4.4.1, the beam is transferred from the injector complex and injected into the booster on either side of the PA experiment straight section. The hardware requirements for the present injection concept are listed in Table 4.2. The operation mode of each system will see injection at up to 100 Hz for up to ∼30 s, followed by ∼30 s without beam.

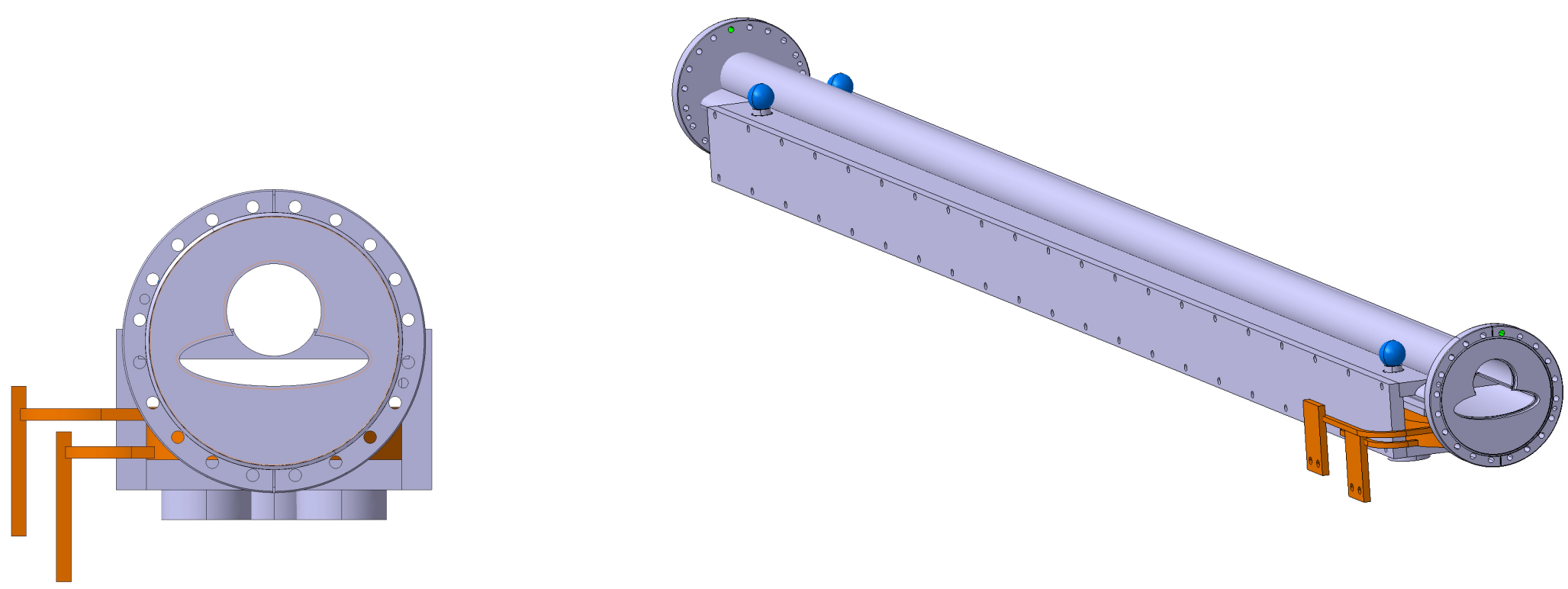

(a) (b)

Fig. 6.21: Mechanical drawing of the Lambertson septum for booster injection in a cross-section side view (a) and isometric view (b).

For injection into the booster, several septa topologies have been studied, and a Lambertson septum design has also been selected for the present concept [178]. To keep the apparent septum thickness to a minimum, the coil is located outside vacuum, but the second pole, including the septum, the magnet gap and the orbiting beam area are all under vacuum, see Fig. 6.21.

#### *Extraction*

The septa used for extraction towards the collider will only extract one energy for each operation mode, and the list of requirements from the extraction design is summarised in Table 4.3. The leak field of the septum is less critical than for the collider injection septa since they only have to be powered at extraction and are not required to track the beam energy like the booster dump septa. This opens the possibility of using a pulsed device, which is less challenging with respect to thermal dissipation. Another option that was explored to achieve a more sustainable solution was the use of a permanent magnet. This latter option, however, would lead to an apparent septum thickness that exceeds the requirements.

The present baseline consists of 2 pulsed septa outside vacuum. The pulse length is chosen to be relatively long to allow a flattop of $>304\,\mu s$ to allow extraction of the full circulating beam as well as to provide sufficient time for eddy currents in the extraction vacuum chamber to decay, preventing the generation of a lower quality field. To limit design effort and benefit from economies of scale, this magnet is chosen to be identical to the collider thick injection septum, albeit operating at a different current. To optimise the cost, the two magnets of each system are connected electrically in series.

#### *Dump*

The booster dump septa shown in Fig. 6.22 are the same design as the collider dump septa described in Section 4.4.3. However, the following operational differences have an impact on the final design of the booster dump septa. First, the field of the dump septa in the booster must follow the energy of the particles in the booster during the ramp, making it a ramped device rather than a purely DC-operated device. Second, the dynamic range of the booster dump septa is larger, as the device will be used across the energy range from booster injection energy level to top energy in the $\mathrm{t\bar{t}}$ mode.

Due to the first reason, the system will include a link to the Beam Energy Tracking System. Because of the further increased dynamic range with respect to the collider dump septa, further development of the proposed low-power septa topology is needed to ensure the field homogeneity and leak field levels remain within specification throughout the dynamic range. For both extractions, the two septa will be powered electrically in series, i.e., one power converter per extraction.

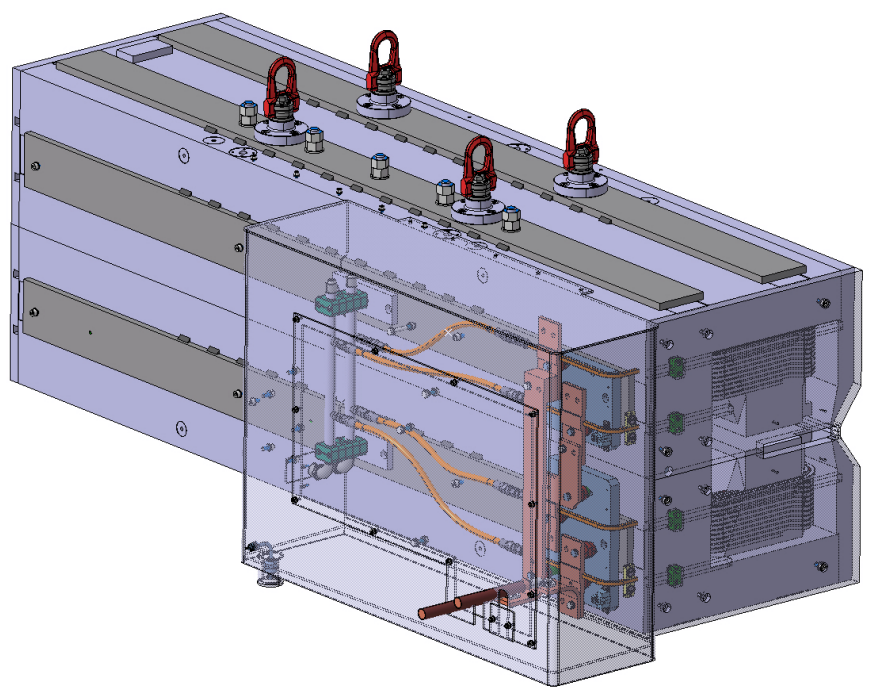

Fig. 6.22: Mechanical drawing of the booster dump septum without its vacuum chamber.

### 6.5.2 Kickers

***Injection***

A stripline design has been chosen for the injection kicker system to meet the requirements listed in Table 4.2 and in particular the fast rise and fall times of $25\,\mathrm{ns}$. Two systems are used, each consisting of a single device.

To meet the field homogeneity requirements, the stripline must be carefully designed. Numerical field simulations were carried out comparing several cross sections, finally resulting in a half-moon shaped electrode design offering the best field homogeneity parameters. To minimise the stripline's impact on the circulating beam, it is essential to achieve the desired characteristic impedance in the even mode while maintaining the odd-mode characteristic impedance as close as possible to $50\,\Omega$ to ensure the required field quality.

Further optimisation studies on impedance matching will be conducted, such as considering a termination network on the load side of the stripline. This termination method also requires a resistor between the electrodes, which presents some challenges. An in-vacuum design is difficult to implement due to concerns about power losses and vacuum compatibility, whereas an outside-vacuum design would be prone to parasitic inductance. In addition, mechanical design optimisation studies should commence soon to ensure the design is capable of withstanding synchrotron radiation.

For the generator side, the short flattop duration of $80\,\mathrm{ns}$ allows the use of either an inductive adder or very short pulse forming lines (PFL). Separate generators deliver both positive and negative driving pulses to the stripline via $50\,\Omega$ coaxial cables.

***Extraction***

The extraction of electron and positron beams from the booster is achieved through two systems, each consisting of 14 kicker magnets and following the extraction scheme requirements listed in Table 4.3. These systems employ the same lumped inductance topology used for collider injection and dump kickers, leveraging the benefits of economies of scale. Each magnet could either be driven by a pulse-forming network or a Marx generator and is operated in short-circuit mode. The Marx generator consists of a main high-voltage stage and a low-voltage stage. Both stages are composed of multiple units: the stages in the high-voltage section are triggered throughout the entire pulse duration, while the low-voltage stages are triggered sequentially to compensate for voltage droop along the pulse. A matching resistor can be included on the generator side to address the mismatch between the magnet and the transfer cable. In order to provide the required flattop quality, the length of the cable connecting the magnet to the generator is limited to $100\,\mathrm{m}$.

The cable length between each extraction kicker and its generator is defined by the pulse requirements. Thus, the location of the service galleries is affected by the required cable lengths. The required cable length is $100\,\mathrm{m}$ which includes the bending radius ($1\,\mathrm{m}$) and the handling path ($15\,\mathrm{m}$).

***Dump***

The booster beam dump system will be very similar to the collider dump (see Section 4.4.3), only requiring adjustments for placement and control. The same magnet modules and generator topologies will be used. Compared to the collider dump, which operates at a fixed energy, beam energy tracking is required. This feature can be achieved using commercially available power supplies.

***Controls***

An inductive adder (IA) pulse generator may be considered for some systems but would be a new technology for CERN, requiring a thorough analysis of its impact on controls. The IA introduces challenges such as managing the $100\,\mathrm{Hz}$ repetition rate, which limits the time for tasks like post-operational checks.

Radiation to electronics (R2E) factors, including high-energy hadron flux and accumulated radiation dose, also need careful consideration to ensure the system's robustness.

Similarly, the Marx generator, another novel technology, requires a detailed evaluation of the impact of its control system and offers potential synergies with collider injection systems to enhance efficiency. A generator based on the LBDS design is planned for the dump kickers, balancing reliability with R2E considerations. An active or hybrid fail-safe retriggering architecture, akin to the SBDS system, is preferred to avoid complexities like signal reflections associated with passive systems.

## 6.6 Beam Instrumentation

In the course of the feasibility study, the type of beam diagnostics has been identified for the booster, with an estimation of the number of instruments needed as outlined in Table 3.17. Similar to the collider ring, beam position monitors (BPMs) and beam loss monitors (BLMs) form the vast majority of the installed systems. Presently, the overall concept for such systems is that they will share the acquisition scheme adopted for the main ring BPMs and BLMs, where signals will be acquired and digitised in shielded racks installed in the tunnel in each arc half cell.

Current (DC and bunched) measurement, transverse and longitudinal profile measurement systems will also be present and capable of measuring throughout the energy ramp. This will form one of the main challenges of the next phase where a preliminary design for the booster BI systems will be produced, particularly for monitors based on synchrotron radiation due to the changing spectrum as a function of the beam energy.

## 6.7 Powering system

### 6.7.1 Booster Magnet Powering Systems

***Global optimisation of the Magnet Powering Systems***

Please see Section 3.8.1.

***Booster Magnet Circuits***

Magnet powering is performed by power converters in alcoves located either at the end of the straight section or in the machine tunnel, called Big Electrical Alcove and Small Electrical Alcove, respectively The installation of power converters in the Machine Tunnel is not possible due to space and radiation constraints.

The location of the power converters is dictated by several factors, including the granularity of control required, the maximum voltage tolerance of the cable insulation and the resulting impact on expenditure for cables and converters. The granularity of control is determined by the optics layout:

– Booster dipoles, as well as booster quadrupole focusing and defocusing magnets, can be powered in series within their respective family.
– Focusing and defocusing booster sextupoles can be powered by half-octant.

The power converters for the booster dipoles, quadrupoles, and sextupoles are installed in the big alcoves at the end of the straight section. All other converters are located in the small alcoves of the tunnel, as they power fewer magnets in series. Table 6.7 presents the quantity of magnets and circuits and the powering parameters.

Magnet specifications and quantities for the tapering, correctors, dispersion suppressor, straight section and injection are not yet defined and therefore estimations were made. A different powering scheme is needed for the dispersion suppressor which uses the same magnets as in the arcs, see Section 6.1.

For the collider details, see Table 3.15 in Section 3.8.1.

Table 6.7: Booster magnet powering circuits quantities and parameters

| | Magnet Quantity | Circuit Quantity | Peak Current (A) | Peak Voltage (V) |
|---|---|---|---|---|
| Dipole | 5536 | 16 | 3065 | 851 |
| Quadrupole (F and D) | 2768 | 32 | 939 | 1998 |
| Sextupole Focusing | 576 | 16 | 525 | 1002 |
| Sextupole Defocusing | 560 | 16 | 595 | 1492 |
| * *Dipole Tapering* | 2768 | 346 | 10 | 396 |
| * *Quadrupole Tapering* | 2768 | 346 | 10 | 382 |
| * *Horizontal Corrector* | 1672 | 1672 | 20 | 68 |
| * *Vertical Corrector* | 1672 | 1672 | 20 | 69 |
| * *Quadrupole Corrector* | 1384 | 1384 | 20 | 65 |
| * *Skew Quadrupole* | 1384 | 1384 | 20 | 65 |
| * *Straight Section* | n/a | 1033 | n/a | n/a |
| * *Injection* | n/a | n/a | n/a | n/a |
| **Booster Subtotal** | **21 068** | **7917** | - | - |
| Collider subtotal | 34 698 | 12 773 | - | - |
| **Total** | **55 766** | **20 690** | - | - |

* Magnet specifications not yet fully defined or non-existent, values extrapolated.

***Precision of Power Converters***

Please see Section 3.8.1.

***Availability of Power Converters***

Please see Section 5.3.4.

## 6.8 Arc region: integration and supporting systems

This section complements Section 3.10, which describes the integration and development of supporting systems in the collider arcs. Arcs are made of a sequence of FODO half-cells, about 3000 for the high-energy optics, and consist of a short straight section (SSS) with quadrupole, sextupoles, beam diagnostics and correctors, followed by a long dipole length that can be achieved with a series of 2 or 3 interconnected magnets. This section describes the optimisation of the arc-supporting structures to maximise their performance, easing the installation and maintenance while also minimising cost. This study is being conducted within the scope of the FCC-ee Arc Half Cell Mock-up Project, and of the FCC-ee design study. It is planned to study elements of the arc cell in a full-scale mock-up (detailed in Section 3.10.5).

The analysis of the relative placement between the collider and the booster is not detailed in this section, as it has been covered in Section 3.10.1.

### 6.8.1 Arc cell configurations of the booster

The configuration of the booster arc cell does not change from the low to the high energy mode. In fact, the length of the arc half-cell remains 26 m for all the phases. As can be seen in Fig. 6.23, there are two configurations: 1136 half-cells with 1 quadrupole and 1 sextupole, and 1704 half-cells with 1 quadrupole. It should be noted that in Fig. 6.23, the drift distances are not represented. In addition, for the booster, the focusing and defocusing sextupoles are not of the same length: $1.4\,\mathrm{m}$ for the defocusing sextupole and $0.7\,\mathrm{m}$ for the focusing sextupole.

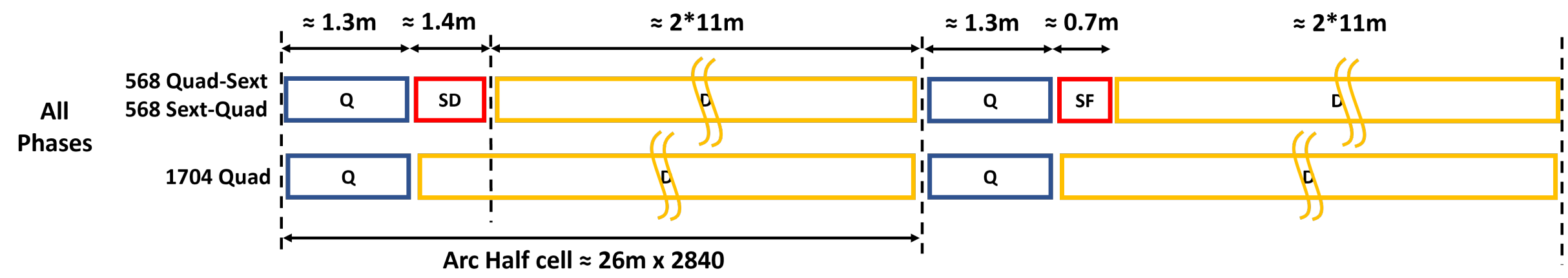

Fig. 6.23: Arc cell configurations for the booster, for all the phases - optic V24-FODO.

Among the different configurations, the FCC-ee Quad-Sext (defocusing) type has been studied in detail and will be installed in the mock-up. This SSS configuration is the bulkiest and, therefore, the most challenging in terms of integration. It is also the most complex in terms of static and dynamic stability.

### 6.8.2 Optimisation of the booster supporting structure - static and dynamic analyses

***Overview and principles***

As discussed in Section 3.10.3, the goal is to optimise the design of the supporting structure by finding a balance between static and dynamic structural stability, integration constraints, and cost, while ensuring compliance with all safety requirements. This optimisation process is still ongoing, with alternative geometries currently under evaluation.

The design of the booster supporting structures for the SSS, which is currently under study, is illustrated in Fig.6.24. This structure consists of two cantilever supports on which a girder is installed. Following the same principle as the collider's SSS, the use of girders to support the common elements in the SSS provides significant practical advantages. The SSS elements, including magnets and vacuum chambers, can be pre-assembled and pre-aligned on a girder in a clean room outside the tunnel, using the appropriate tools and environment. The module can then be transported as a single unit to the tunnel, streamlining transport and maintenance operations (for more details, see Section3.10.3).

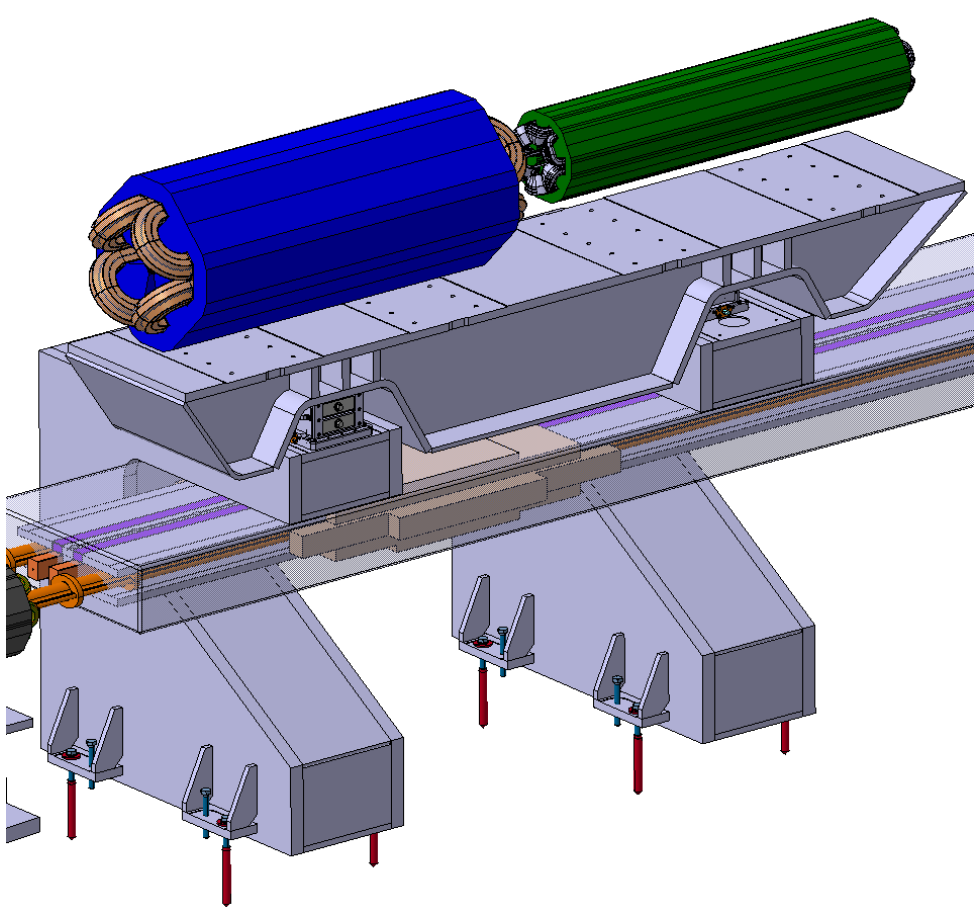

Fig. 6.24: CAD model of a possible optimised version of the booster SSS.

***Specifications***

An initial estimation of the acceptable vibrations in the SSS was defined in 2022, and is reported in Table 6.8. This specification did not distinguish between vertical and lateral motion, and between booster and collider.

To consolidate the tolerance estimates, the sensitivity to vibrations of the GHC optics at the Z operating point was evaluated assuming a tolerance on the beam oscillation amplitude at the collision point of less than 5% relative to the collision point beam size [309]. Due to the small emittance ratio, vertical oscillations are more than an order of magnitude more critical than horizontal oscillations. Assuming that an orbit feedback system will efficiently damp beam oscillations with frequencies below 1 Hz, the integrated RMS motion for frequencies higher than 1 Hz is ideally around 10 (100) nm for the vertical (horizontal) plane. If the beam orbit feedback bandwidth can be extended above 1 Hz, setting the targets for the vertical (horizontal) plane to 20 (200) nm, in line with Table 6.8 is possible. It must be noted that the sensitivity may evolve in the future with the optics and the machine layout. For the low-beta regions, the tolerances are an order of magnitude tighter.

These considerations are summarised in Table 6.9. The key frequency of interest is 1 Hz, and in this frequency range:

- The collider quadrupole acceptable vertical displacement is 20 nm.
- Laterally, the acceptable displacement of the collider quadrupole is higher by roughly one order of magnitude. Currently, a value of 200 nm is tentatively assumed.
- It is very likely that the quadrupoles in the booster can accept larger displacements than in the collider. It is, however, not evident to define how much the requirements can be relaxed. An increase of the tolerance by a factor of two in vertical and lateral directions is currently tentatively assumed, resulting in values of 40 nm and 400 nm, respectively.

Table 6.8: Dynamic stability requirements in the arcs proposed in 2022%, presented by T. Raubenheimer at FCC IS workshop for the arcs [310].

| Frequency range | Tolerance | Correlation |
|---|---|---|
| 0.01 Hz < f < 1 Hz | 1 µm | 10 km |
| 0.01 Hz < f < 1 Hz | 100 nm | none |
| 1 Hz < f < 10 Hz | 20 nm | none |
| 10 Hz < f < 100 Hz | 5 nm | none |
| 100 Hz < f | 1 nm | none |

Table 6.9: Updated dynamic stability requirements in the arcs at the level of the magnetic axis.

| | Tolerance at 1 Hz frequency |
|---|---|
| Collider vertical direction | 20 nm |
| Collider lateral direction | 200 nm |
| Booster vertical direction | 40 nm |
| Booster lateral direction | 400 nm |

***Historical background***

It is interesting to compare the current stability specifications with what was studied in other projects or achieved in past CERN machines, such as the Large Hadron Collider (LHC/HL-LHC) and the future Compact LInear Collider (CLIC). Concerning the LHC/HL-LHC quadrupoles:

– In standard operation the root mean square (RMS) value should be < 5 µm at 1 Hz;
– Beam instabilities can be provoked if the RMS is between 5 µm and 20 µm at 1 Hz;
– A beam dump is usually triggered for an RMS > 20 µm at 1 Hz.

On the other hand, the future Compact LInear Collider (CLIC) has significantly more stringent specifications: given its very small beam sizes, even minor oscillations of one quadrupole reduce the luminosity. It has been estimated that in the vertical direction, the RMS must be below 1 nm at 1 Hz and similarly, it must be below 5 nm at 1 Hz laterally to ensure sufficient performance [312]. A study has demonstrated that these specifications can be achieved, albeit using active stabilisation based on piezo-actuators combined with a stiff and optimised design of the quadrupole support [313].

Hence, the FCC arc specifications are closer to those of a linear accelerator (CLIC) than those of existing circular accelerators (LHC/HL-LHC), which illustrates their challenging nature.

***Numerical methodology***

As explained and detailed in Section 3.10.3, the supporting structures of the collider and the booster are being optimised in terms of static and dynamic stability. For this scope, a finite element method (FEM) procedure has been defined (see Section 3.10.3).

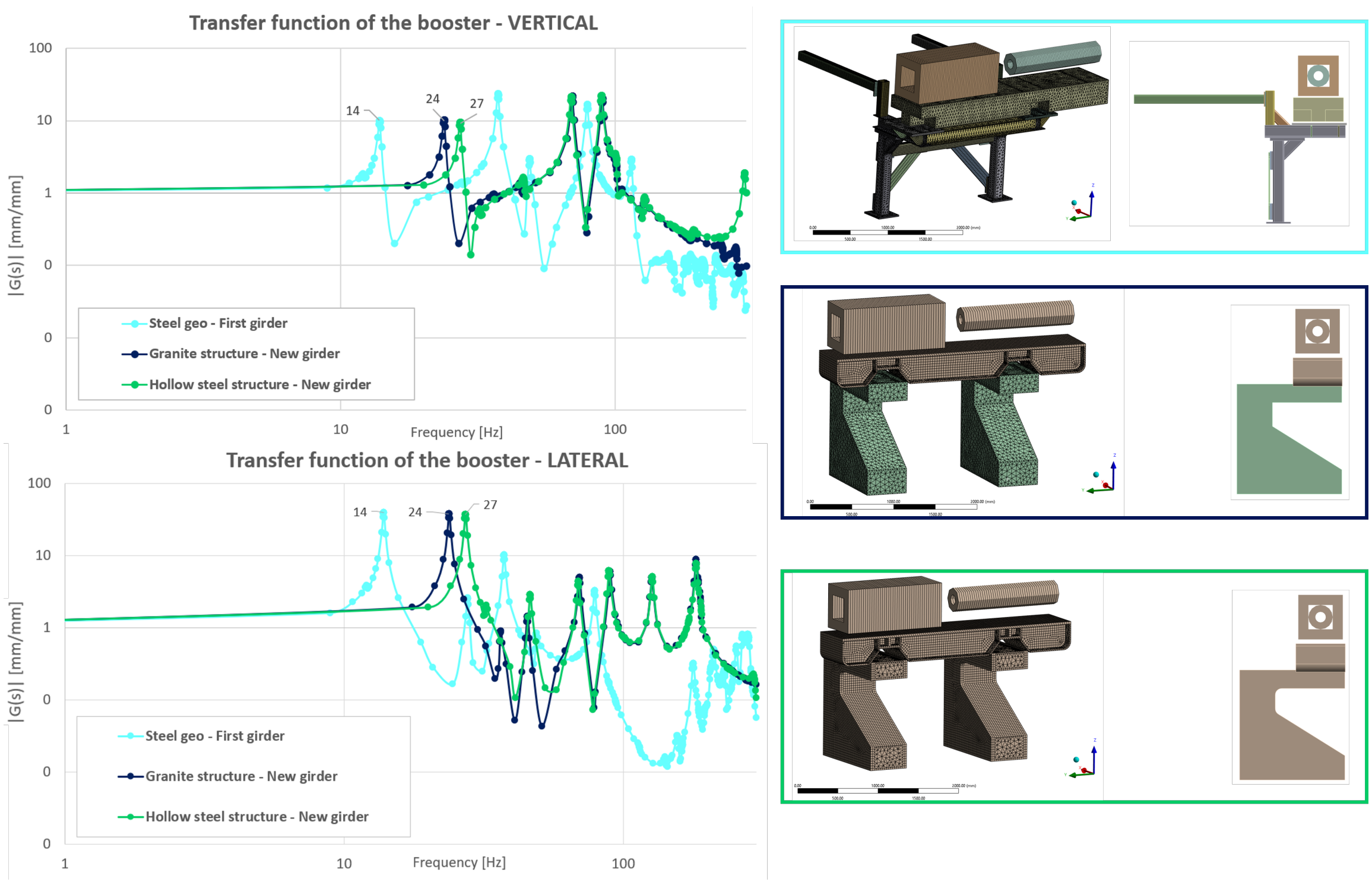

Fig. 6.25: Transfer function comparison for different supporting structure geometries in the vertical and lateral directions.

An example of a transfer function comparison for different geometries of the booster supports is shown in Fig. 6.25. Two graphs are presented: the first displays the transfer functions obtained vertically, and the second the transfer function obtained laterally. The transfer functions provide insight on how the system behaves and reacts over a range of frequencies. The greater the natural vibrational and rigid body frequencies, the more rigid the system is considered to be. Comparative studies can be carried out to analyse the impact on stability of the geometry, materials, position, rigidity and number of feet, etc. and to determine the most suitable geometry.

Knowing the spectrum of the ground motion over the relevant range of frequencies, as well as the transfer function of the supporting structure between the ground and the centre of the magnet, it is possible to estimate the displacement of the magnet mechanical axis in response to the ground motion. Since the values of the estimated movement of the ground in the FCC tunnel are not yet available, the power spectral density (PSD) of the ground motion measured in the LHC tunnel [315] has been used as an initial input instead. When performing a random vibration analysis of the system with such ground PSD as an input for the calculation, the PSD of the magnetic axis is obtained as an output, and then the integrated root mean square (RMS) displacements at the level of the axis can be computed. The methodology is explained in Fig. 3.74.

***Experimental benchmarking***

It is important to note that the method described above is sensitive to several parameters that cannot be precisely estimated at this stage of the project. Similarly, the simulations involve certain approximations, such as simplified representations of magnets and interfaces (e.g., ground-support and support-magnet interactions). Therefore, it is crucial to experimentally benchmark and refine the simulations to build confidence in the numerical model.

To better understand how the different elements of the SSS influence system stability, a simple $2.5\,\mathrm{m}$ long short straight section demonstrator was assembled (see Section 3.10.3). The design of this demonstrator closely resembles what will be installed in the collider. However, during the next phase, the plan is to extend these measurements to the booster's SSS as well.

## 6.9 Machine protection

Machine protection hardware and software will be required. These protection elements are very similar to those described for the collider and are described in more detail in Section 3.11. As there are no superconducting magnets foreseen in the booster, there will be no such protection systems.

# Chapter 7

# FCC-ee injector complex

## 7.1 Injector overview

The FCC-ee injector complex must provide the electron and positron bunch trains for alternating bootstrapping injection during, both, top-up and filling-from-scratch operations. Given the charge per bucket and lifetime for this operational mode, an alternating injection of a train of positrons and electrons approximately every few tens of seconds is required to ensure the correct balance between positron and electron bunch charges during collider operation. It includes separate linacs for electrons and positrons up to a beam energy of 2.86 GeV – the electron linac (e-Linac) and the positron linac (p-Linac), respectively. Figure 7.1 shows the basic layout of the injector complex schematically. Following the positron and electron linacs, both species are injected into the damping ring (DR) for emittance reduction. The layout also includes the high-energy (HE) linac, which boosts the beam energy from 2.86 GeV up to 20 GeV in order to inject beams directly into the booster ring (BR). The baseline for the positron source is based on a conventional scheme using electrons from the e-Linac impinging on a tungsten target. This approach allows all linacs to operate at 100 Hz with 4 bunches per pulse to meet the collider ring filling specification for the most demanding Z running mode.

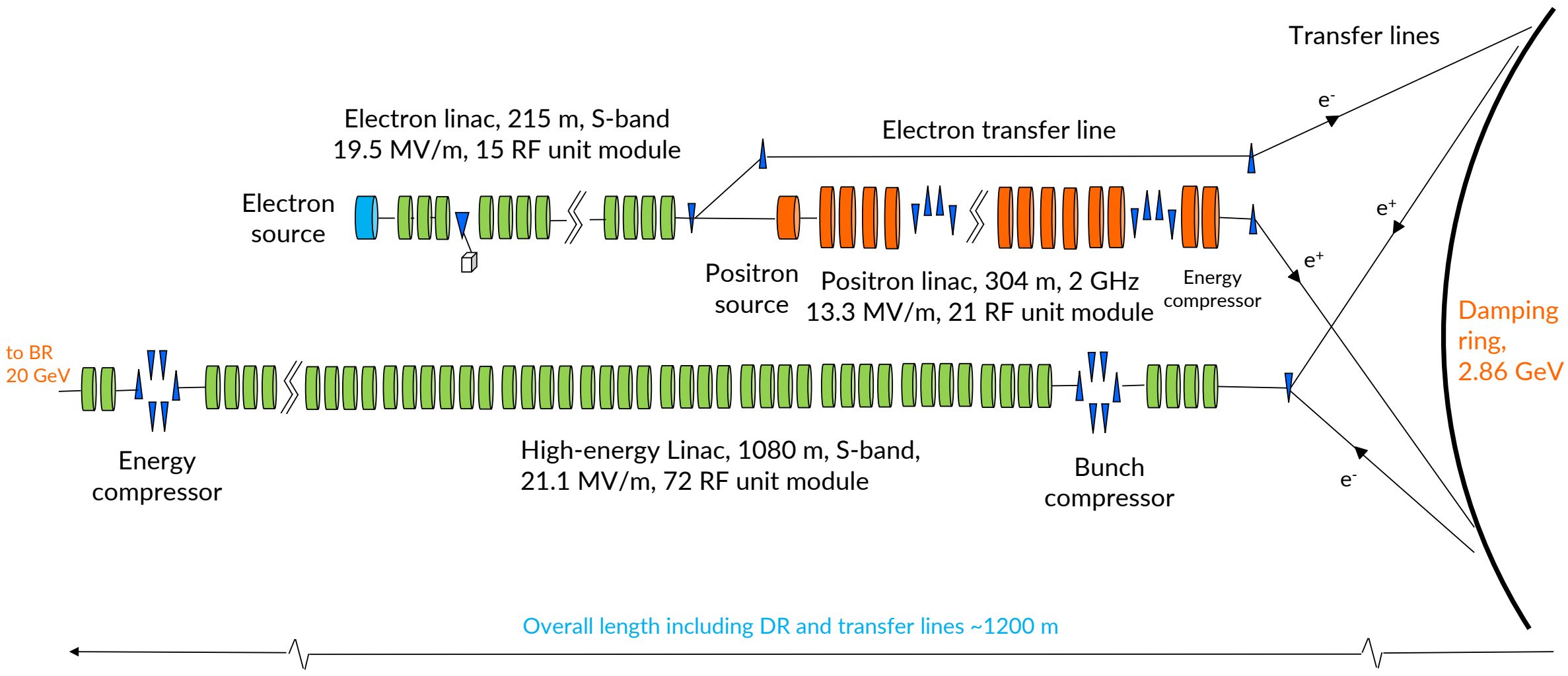

Fig. 7.1: Baseline layout of the pre-injector complex, including the high-energy (HE) linac.

Operating at a 100 Hz repetition rate with four bunches per RF pulse, the linac system incorporates beam-loading compensation and long-range wakefield suppression for enhanced stability. Although the injector complex is now longer, this new layout improves reliability, featuring a damping ring at a higher energy of 2.86 GeV and eliminates the need for a common linac, which would otherwise require doubling the repetition rate. This revised concept provides a more efficient and sustainable solution aligned with performance and operational goals.

Table 7.1 lists the collider and booster parameters used as specifications for the injector design. It is worth emphasising that the injector must operate continuously due to the short beam lifetime and the strict requirement that the charge imbalance between the electron and positron beams in the collider remain within a narrow range of 3-5%. This constraint requires a precise and uninterrupted injection process to maintain beam-beam stability. For example, assuming a beam duration of about 1000 seconds, the injector must alternate between injecting electrons and positrons at intervals of about 50 seconds. This requirement imposes a significant operational challenge, particularly in the Z-mode, where any interruption in injector functionality could severely impact the collider performance. Consequently, the reliability and availability of the injector are of critical importance, as any downtime could compromise the overall efficiency of the collider. This issue represents a critical limitation presented in more detail in Section 7.9, where potential risks and mitigation strategies associated with injector failures are discussed in more detail.

Table 7.1: Collider and booster parameters used as specifications for the injector design. Bunch charge is the maximum bunch charge to be injected into the collider ring. Emittance, bunch length and energy spread are the specifications at the injection into the booster ring.

| **Running mode** | **Z** | **W** | **ZH** | $\mathbf{t\bar{t}}$ | **Unit** |
|---|---|---|---|---|---|
| Number bunches in collider | 11200 | 1856 | 300 | 64 | |
| Nominal bunch charge in collider | 34.40 | 22.08 | 27.04 | 23.68 | nC |
| Allowable charge imbalance | 5 | 3 | 3 | 3 | % |
| Beam lifetime, lumi 4 IPs (q, BS, lattice)/4 | 916 | 517 | 428 | 497 | s |
| Trains/Bunches per booster cycle | 40×280 | 8×232 | 2×150 | 2×32 | |
| Max injected bunch charge | 3.43 | 3.43 | 1.60 | 1.60 | nC |
| Number of bunches | 4 | 4 | 2 | 2 | |
| Linac rep. rate | 100 | 100 | 50 | 50 | Hz |
| Bunch spacing | 25 | | | | ns |
| Beam energy at BR | 20 | | | | GeV |
| Norm. emittance (x, y) (rms) (BR) | <20,2 | | | | mm mrad |
| Bunch length (rms) (BR) | ∼4 | | | | mm |
| Energy spread (rms) (BR) | ∼0.1 | | | | % |

The top-up operation for each operating mode requires the charge of the individual bunch in the train to vary from a few tens of pC to about 4 nC per injection, depending on the charge imbalance of electrons and positrons of the individual bucket in the collider. This requirement results from the different lifetimes of the individual bunches in the collider rings, which will also determine the filling pattern for each injection. Regarding emittance, the specifications for injection into the BR require a beam with a flat normalised emittance of 20 mm·mrad and 2 mm·mrad in the horizontal and vertical planes, respectively, to ensure a shorter cycle in the booster itself. This specification has an impact on both electron source and DR parameters. In particular, the electron source must guarantee this emittance even during the required charge variation for top-up operation, and this question has an impact on the optimisation of the photo-cathode RF gun.

In order to achieve independent design specifications for the linacs in the injector from those of the BR, an energy compressor in the transfer line from the HE linac to the BR is planned. This arrangement is depicted in the left part of Fig. 7.1. By adopting this approach, the design of the linacs can converge towards a solution for the beam length and energy spread at the linac end specified in Table 7.1, without considering more complex layouts that include compression and/or decompression of the beam along the different linacs.

Table 7.2 provides a comprehensive summary of the key parameters of the electron and positron beams as they progress through the injector and booster up to the point of injection into the collider in Z-mode. This table serves as a crucial reference, consolidating the relevant beam characteristics at each stage of the acceleration and manipulation processes. By collecting these parameters in a structured manner, Table 7.2 played a key role in coordinating the various simulation studies conducted on the different injector subsystems. It allowed the results to be systematically compared and validated, ensuring consistency between simulations. Furthermore, this collection was crucial to verify that the beam parameters remain within the specifications required for successful injection in both the booster and the collider.

The findings from the injector study have also been documented in four detailed scientific reports submitted to CHART [350–353]. These reports serve as a comprehensive record of the research conducted, encompassing various aspects of the injector's design, performance, and operational constraints. Additionally, these reports include an extensive list of references that provide further context and background to the study. This bibliography features a wide range of scientific publications, including peer-reviewed journal articles, conference proceedings, and contributions presented at workshops and international conferences.

The following sections provide a summary of the key findings and achievements resulting from the efforts to address the recommendations outlined in the mid-term review. These results highlight the progress made in optimising the injector's design, performance, and integration within the overall system. Additionally, this discussion includes an overview of the proposed location of the injector on the CERN site. The positioning of the injector is a critical aspect, as it directly influences factors such as beam transport efficiency, infrastructure requirements, and operational feasibility.

Table 7.2: Electron and positron bunch parameters along the injector and booster up to the injection into the collider for the Z-mode. Some parameters still have to be calculated (tbc). LE=Low-energy, HE=High energy, DR=Damping ring.

| | **Beam Energy** | **Bunch charge** | **Transm.** | **Bunch length (rms)** | **Rel. energy spread (rms)** | **Norm. emit. (rms) H/V** |
|---|---|---|---|---|---|---|
| | **[GeV]** | **[nC]** | **[%]** | **[mm]** | **1E-3** | **[mm mrad]** |
| LE Linac injection | 0.2 | 3.79 | | 1 | 5 | 3/3 |
| LE Linac exit | 2.86 | 3.75 | 0.99 | 1 | 6 | 3.3/3/3 |
| Positron source target | 0.045 | 26.53 | 7.07 | 1.34 | >100 | 21 000/20 000 |
| Positron capture exit | 0.185 | 14.81 | 0.56 | 9 | >100 | 13 000/12 000 |
| Positron linac injection | 0.263 | 12.06 | 0.81 | 8 | 140 | 13 000/12 000 |
| Positron linac exit | 2.86 | 10.73 | 0.89 | 2.8 | 8.7 | 13 000/12 000 |
| Energy Compressor | 2.86 | 10.09 | 0.94 | 2.8 | 8.7 | 13 000/12 000 |
| DR injection | 2.86 | 5.04 | 0.5 | tbc | tbc | 13 000/12 000 |
| DR extraction | 2.86 | 4.99 | 0.99 | 4.8 | 0.72 | 10/1 |
| LE Linac injection | 0.2 | 5.20 | | 1 | 5 | 3/3 |
| LE Linac exit | 2.86 | 5.15 | 0.99 | 1 | 6 | 3.3/3.3 |
| Transfer line | 2.86 | 5.09 | 0.99 | 1 | 6 | 3.3/3.3 |
| DR injection | 2.86 | 5.04 | 0.99 | 1 | 7 | 5/5 |
| DR exit | 2.86 | 4.99 | 0.99 | 4.8 | 0.72 | 10/1 |
| Bunch Compressor | 2.86 | 4.94 | 0.99 | 1 | 7 | 10/1 |
| HE Linac injection | 2.86 | 4.89 | 0.99 | 1 | 7 | 12/1 |
| HE Linac exit | 20 | 4.84 | 0.99 | 1 | 6.1 | 16/1.6 |
| Energy Compressor | 20 | 4.80 | 0.99 | 4 | 1 | 16/1.6 |
| Transfer line | 20 | 4.56 | 0.95 | 4 | 1 | 16/1.6 |
| Booster injection | 20 | 4.33 | 0.95 | 4 | 1 | 20/2 |
| Booster extraction | 45.6 | 4.29 | 0.99 | 2.43 | 0.38 | 10.71/0.89 |
| Collider injection | 45.6 | 3.43 | 0.8 | 2.43 | tbc | 10.71/0.89 |

## 7.2 Electron source

The baseline configuration of the injector complex, illustrated in Fig. 7.2, includes a single electron source for producing both the nominal electron beam and the driver electron beam for positron production. The electron source is composed of a photo-cathode 2.6 cell RF photo gun followed by three RF accelerating structures reaching the beam energy of approximately 200 MeV. The main requirement for the electron source is to generate four bunches with a charge of 5 nC each, keeping the normalised emittance below 4 mm·mrad in order to have a margin in any emittance growth along the linacs and transfer lines between linacs.

Extensive simulations have been conducted on the electron source and downstream linacs. Uniform and truncated Gaussian initial distributions have been studied, and Table 7.3 lists the optimised beam parameters that were achieved at the end of the electron source. These parameters and the simulated distributions have been used as input for the design and simulations of the subsequent linacs. In particular, investigations into the yield of positron production indicate that 5 nC electron bunches are sufficient to achieve the desired positron bunch charge.

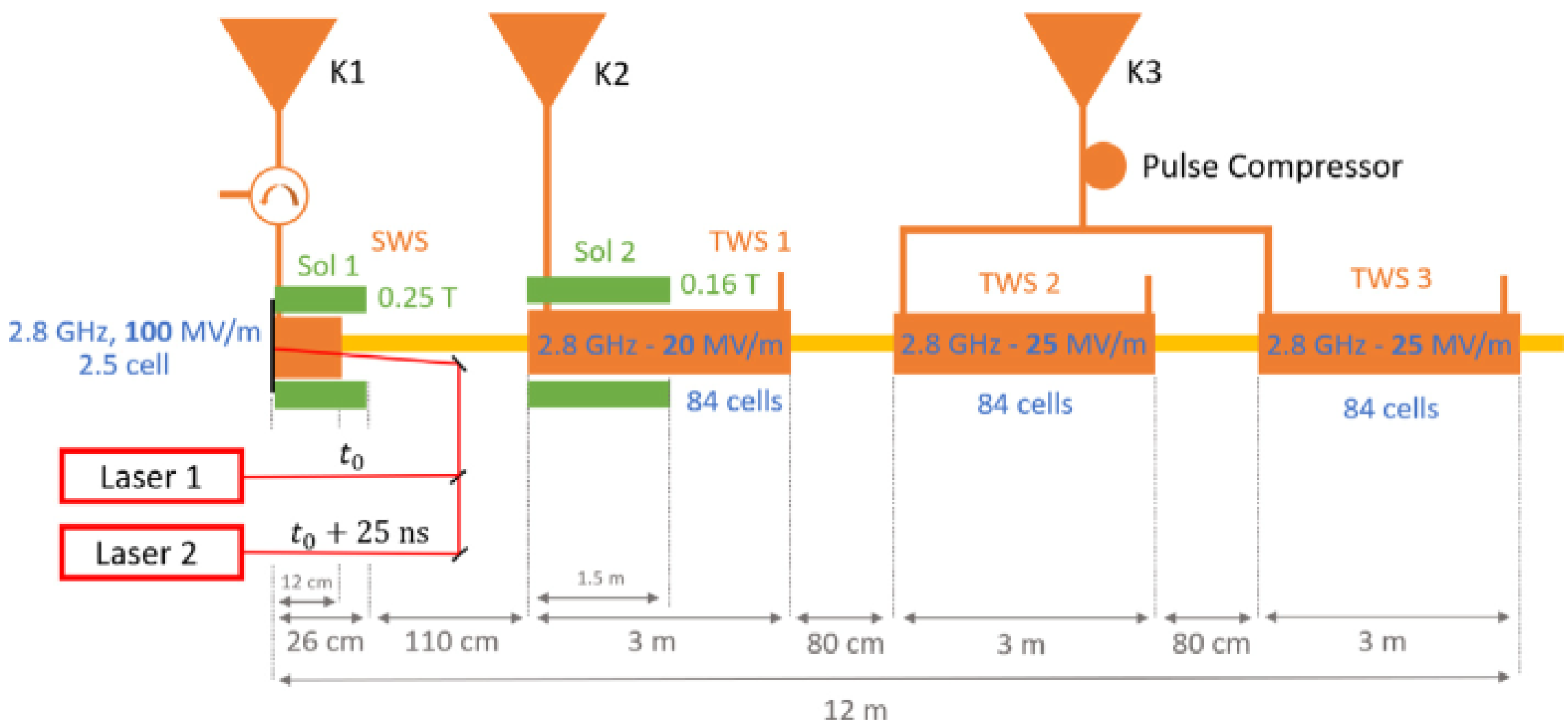

Fig. 7.2: Schematic layout of the 200 MeV pre-injector consisting of an RF-gun and 3 accelerating structures. Here the option with two redundant laser systems is shown.

Table 7.3: Electron source beam parameters at 200 MeV with a bunch charge of 5 nC.

| Parameter | Uniform Distribution | Gaussian Distribution |
|---|---|---|
| Transverse Emittance [mm·mrad] | 2 | 3 |
| Energy Spread rms [%] | 0.4 | 0.25 |
| Bunch Length rms (mm) | 0.98 | 1.3 |

One of the most challenging aspects for the electron source is the top-up mode for the collider. In this operational mode, the bunches circulating in the collider rings will be topped up with charge, compensating for the charge decrease during collisions. Therefore, the injector has to deliver varying bunch charges in the range of 10-100% for each bunch. Nominal operation consists of a four bunch per RF pulse scheme, with a spacing of 25 ns. In theory, a single laser pulse could be split in 4 and then individually manipulated, but using up to 4 lasers would allow more flexibility to change the charge independently of a few hundred pC up to 5 nC. Several lasers may also be needed to achieve the required availability (see Section 7.9). The source and the linac will work with a 100 Hz repetition rate, leaving only 10 ms between pulses to adjust laser, RF or magnet parameters. A detailed beam dynamics study has been started to determine which parameters need to be changed for different bunch charges to deliver beams as similar as possible. The best option would be to leave RF and magnet parameters constant and only change the laser spot size and intensity on the cathode so that the charge density stays as similar as possible. The spot size could be manipulated fast enough using a controllable mirror array. Simulations show that, in this case, the beam parameter variation at the exit of the electron source is not too big (see Fig. 7.3). More details on the electron source study can be found in Ref. [354].

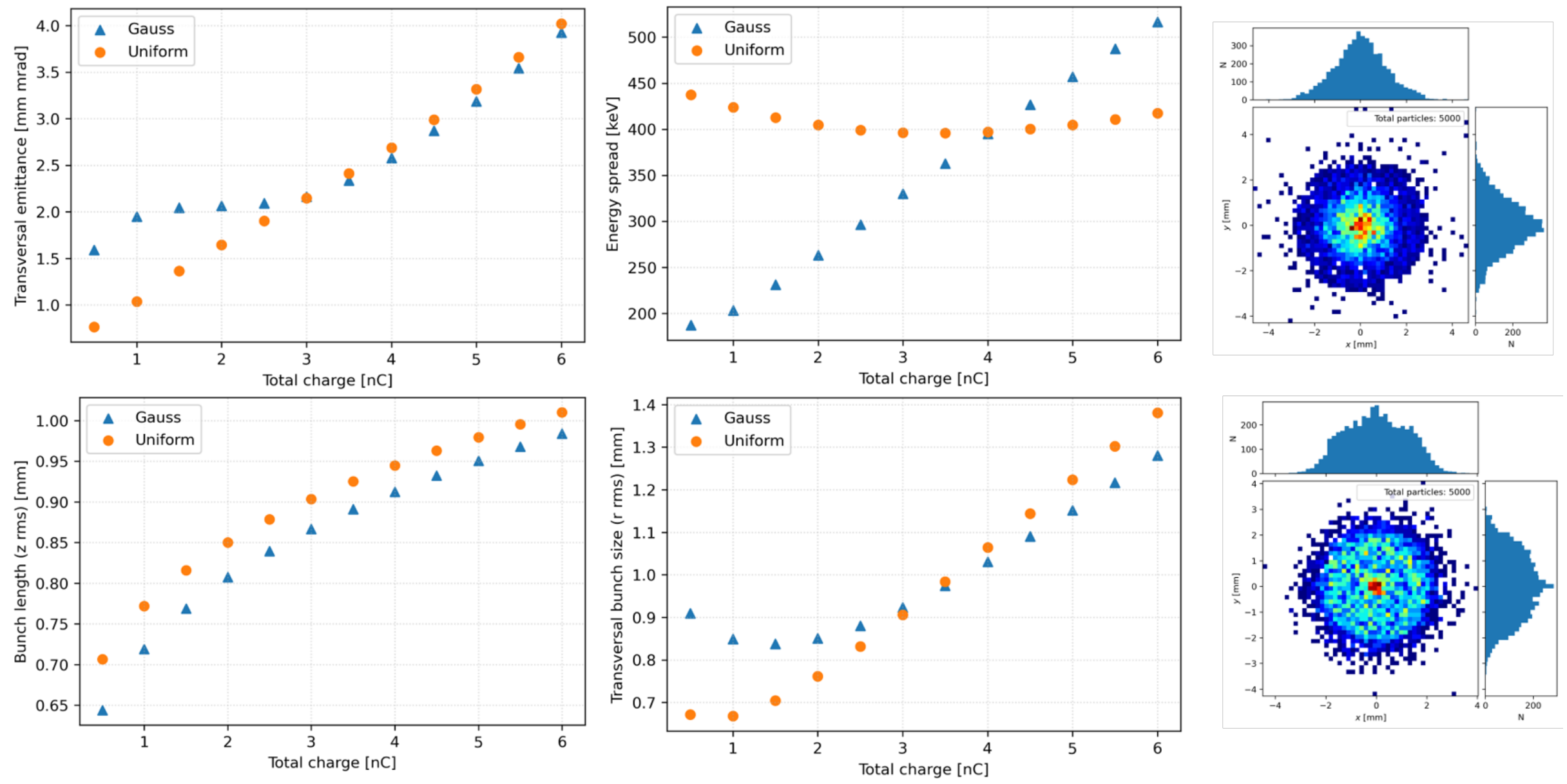

Fig. 7.3: Beam parameter variation as a function of bunch charge simulating the top-up operation of the electron source and pre-injector.

## 7.3 Electron linac

The electron linac accelerates electron beam of up to 4 bunches of 5 nC each from the energy of about 200 MeV to 2.86 GeV at the repetition rate of 100 Hz for Z-mode, which presents the most critical case for the linacs. The electron linac is located right after the electron source, as shown in Fig. 7.1. In the electron production mode, the beam goes to the DR, and in the positron production mode, it goes to the target.

Beam dynamics simulations have been done to study both longitudinal and transverse beam dynamics in the electron linac using the tracking code RF-TRACK [355]. The main purpose was to define the specifications for RF structures in terms of iris aperture, working frequency, structure length, gradient, and lattice parameters, like quadrupole separation, kind of lattice, and phase advance. The following beam parameters were used in the simulations: bunch length 1 mm, relative energy spread 0.25% and emittance 3.2 mm·mrad at the start of the electron linac for the case of the 5 nC electron bunch.

The linac comprises a FODO lattice with 90 degrees phase advance per cell with one quadrupole and one BPM per RF structure, with a total cell length of 7.5 m. The RF structures are operated on-crest. This allows the accelerating efficiency to be maximised, and the beam quality degradation to be minimised.

Tracking simulations were conducted in terms of emittance growth to evaluate the robustness of the linac design against static misalignments. Gaussian-distributed misalignments were assumed for the RF accelerating structures, quadrupoles, and BPMs; the corresponding rms values are summarised in

Table 7.4. A BPM resolution of 10 µm was also included in the computation of some of the steering corrections applied.

Based on these simulations, an RF structure aperture corresponding to $a/\lambda = 0.15$ was determined. To mitigate the effects of misalignments, a combination of one-to-one correction and dispersion-free steering (DFS) was applied sequentially, incorporating the randomly distributed errors. The minimum RF aperture corresponding to $< a > /\lambda = 0.15$ was determined by compromising the growth of transverse emittance from static effects with the efficiency of the RF structure. Finally, a maximum emittance growth of 0.3 mm·mrad was obtained at the end of the linac for 98 % or more of the simulation seeds assuming the optimised RF parameters (on-crest operating phase and aperture corresponding to $a/\lambda = 0.15$).

Table 7.4: RMS of the Gaussian random distributions assumed for the misalignments of the lattice elements for the static effect simulations.

| **Element** | **Value** [µm] |
|---|---|
| Quadrupoles | 50 |
| RF accelerating structures | 100 |
| BPM | 30 |

Dynamic effects were also investigated, focusing specifically on the amplification of the incoming beam transverse jitter, whether in position, angle or a combination of both. To simulate transverse beam jitter, the transverse phase space was painted following a circular path at the entrance of the linac. The jitter amplification (JA) was then calculated as the square root of the ratio between the areas of the transverse phase space at the entrance and at the exit of the section. The final jitter, in either position or angle, was determined by multiplying the incoming jitter by the computed amplification. This method provides a quantitative measure of the robustness of the linac design to incoming beam transverse jitter and enables optimisation of parameters such as the RF structure aperture, the lattice phase advance per cell, and the spacing between quadrupoles linked in this design to the length of the RF structure. Both single- and multi-bunch effects were analysed following a different procedure. In the case of the single-bunch the JA was computed all along the linac varying the above-mentioned parameters. As an example, Fig. 7.4 shows the dependence of the JA on the RF structure aperture. The aperture corresponding to $a/\lambda = 0.15$ produces a maximum final JA of approximately 1.2. Assuming an initial orbit jitter of 0.12 $\sigma$ (where $\sigma$ is the transverse beam size), like for example, that measured in AWAKE for similar

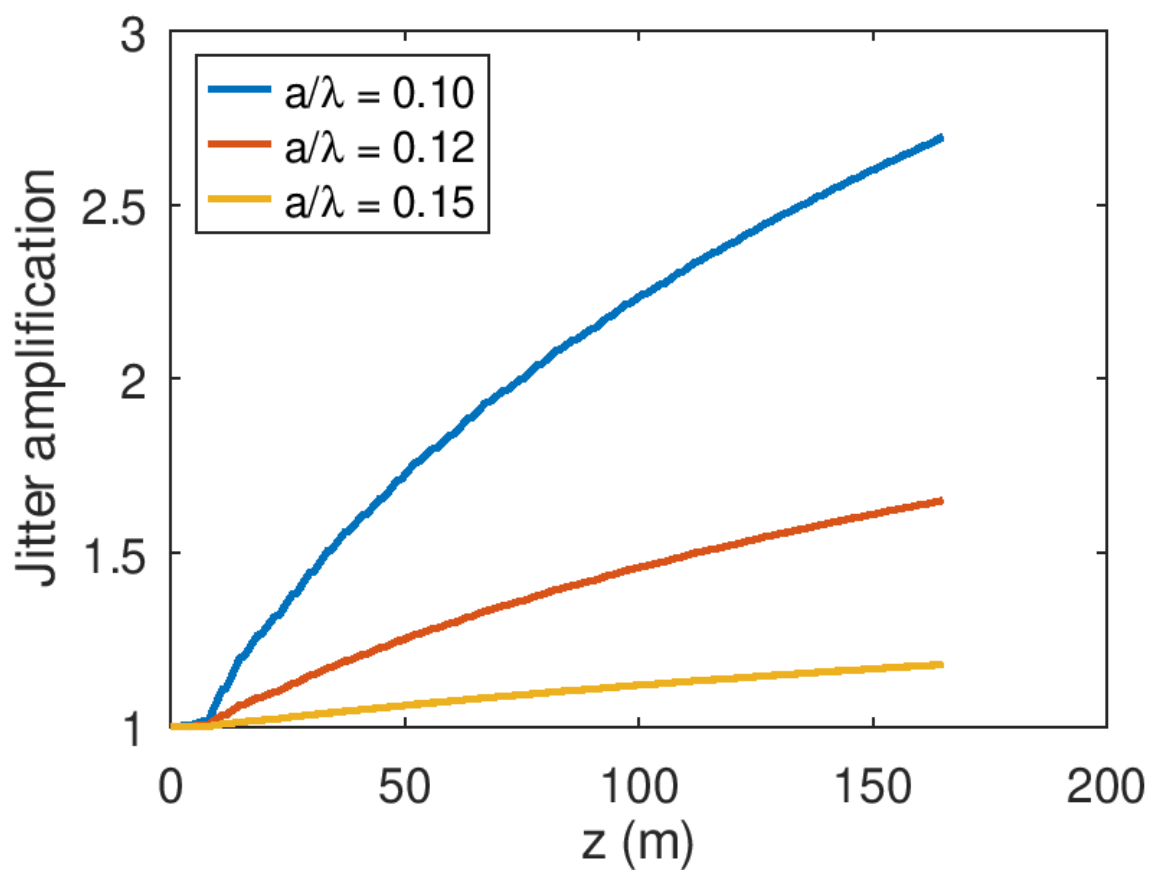

Fig. 7.4: Single-bunch jitter amplification along the electron linac at different RF apertures.

beam parameters, results in a jitter of $0.14\ \sigma$.

The JA was also used to study multi-bunch dynamic effects, but assuming a variable kick imparted by the first bunch on the second bunch. Given the amount of jitter that can be accepted, the maximum tolerable kick was determined and used to determine the specification for the HOM dipole suppression in the RF structure. The results are shown in Fig. 7.5, demonstrating that the kick on the following bunch of 0.1 V/pC/mm/m corresponds to a JA of 1.02 at the exit of the linac, which is significantly smaller than the single-bunch JA. From these simulations, the total JA, calculated as a product of single- and multi-bunch JAs, is 1.22. The corresponding total jitter is smaller than $0.15\ \sigma$, assuming $0.12\ \sigma$ as the incoming jitter. This value has a negligible impact on the positron production, and is expected to be in the acceptance of the DR. The parameters obtained from the beam dynamics studies are summarised in Table 7.5 for convenience.

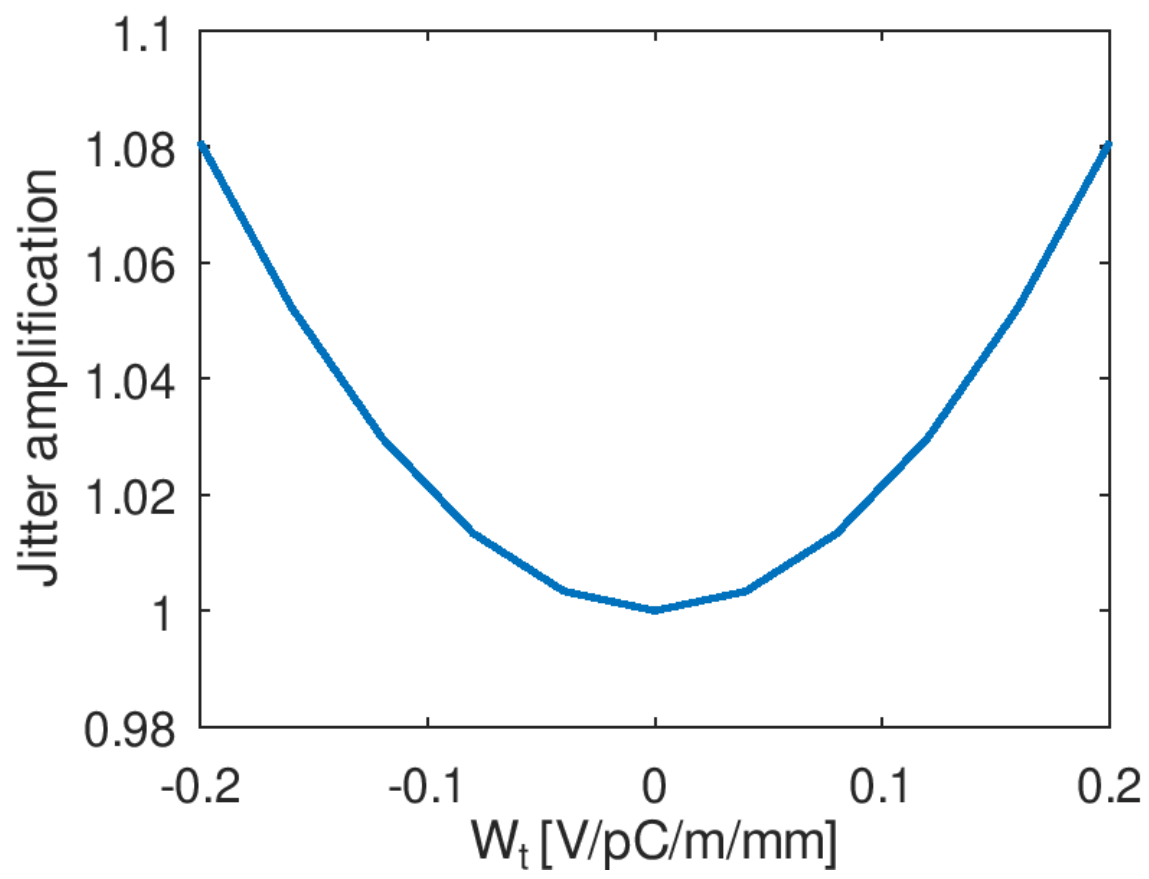

Fig. 7.5: Multi-bunch JA at the exit of the electron linac versus transverse wakefield kick imposed by the first to the following bunch.

Table 7.5: Summary of the optimised RF and lattice parameters based on the beam dynamics studies.

| **Parameter** | **Value** |
|---|---|
| Mean RF structure aperture (mm) | 16.1 ($a/\lambda = 0.15$) |
| RF structure length (m) | 3 |
| RF structure operating phase | on-crest |
| Phase advance/cell (degrees) | 90 |
| Number of BPM/RF structure | 1 |
| Number of quadrupoles/RF structure | 1 |
| Distance between the quadrupoles (m) | 3.75 |

### 7.4 Positron source and linac

The production of positrons is always an extremely important topic for any electron-positron collider, especially for future colliders like the FCC-ee, which are designed to operate at extreme parameters. For the FCC-ee, a high-yield positron source is essential to provide the low-emittance positron beam with sufficient intensity to reduce the injection time into the collider. Specifically, at Z-pole operation, a positron bunch intensity of $2.14 \times 10^{10}$ particles is required at injection into the collider rings. The positron rate for the FCC-ee is twice that achieved at the SLC at SLAC, while remaining an order of magnitude lower than the values typically proposed for linear collider projects [356]. At the injector level, the primary requirement for the positron source is to deliver a positron bunch charge of $5\,\mathrm{nC}$, which must be accepted into the damping ring (DR), as indicated in Table 7.2. Based on the available experience of designing and operating previous or current positron sources, a safety margin of 2.56 has been applied to the FCC-ee positron source design, requiring the delivery of a total positron bunch intensity of $12.8\,\mathrm{nC}$ at the injection into the DR.

### 7.4.1 Positron production and capture system

A conventional positron source using 2.86 GeV electrons impinging on a $15\,\mathrm{mm}$ thick tungsten target is the basis for FCC-ee positron production. The bremsstrahlung radiation of the electrons in the field of the target nuclei is converted in $e^+e^-$ pairs. The target thickness has been optimised to maximise the number of positrons produced at the target exit. This conventional production method has been successfully employed in all the $e^+e^-$ colliders (ADA, ACO, DCI, SPEAR, ADONE, LEP, and also for the first linear collider SLC).

The capture section includes an Adiabatic Matching Device (AMD) [357], followed by a capture linac embedded in a DC solenoidal magnetic field to accelerate the positron beam to about $170\,\mathrm{MeV}$. At the end of the capture linac, positron and electron bunches are separated using a chicane at $170\,\mathrm{MeV}$ and the solenoid focusing is used up to a positron energy of $930\,\mathrm{MeV}$. After the matching section at $930\,\mathrm{MeV}$, the positron beam passes through quadrupole focusing and is accelerated up to the DR energy $2.86\,\mathrm{GeV}$. An energy compressor system (ECS) is used before the DR to increase the number of positrons within the DR energy acceptance. The DR is an important part of the positron source design as its dynamic aperture, longitudinal and transverse acceptance parameters define the final performance of the positron source. The baseline design of the DR is described in Section 7.5.

Two AMD designs were investigated during the Feasibility Study: one employing a flux concentrator (FC) based on pulsed magnet technology (currently used in the SuperKEKB collider [358]) and another using a superconducting (SC) solenoid based on high-temperature superconducting (HTS) materials. The latter, an innovative approach for positron sources, will also be tested in the PSI Positron Production ($P^3$) experiment at SwissFEL [359], which has been designed as a demonstrator for the FCC-ee positron source technologies (see Section 7.4.6).

For the classical FC-based approach, several models were evaluated for the FCC-ee positron source design. These included the one designed by BINP for the FCC-ee and ILC projects, the FC developed by KEK for the ILC project, and the FC currently used in the SuperKEKB collider. Due to the conceptual and mechanical constraints of the FC, the peak of the magnetic field is located downstream the target and as a result, the available field on the target is reduced to $\sim$3.5 T/$\sim$1.1 T (for the BINP/SuperKEKB designs respectively) manifesting a significant drop in capture efficiency. Moreover, the presence of a high transverse magnetic field component (with strong domination of dipole harmonic) makes the trajectories of positrons strongly distorted. As a result, the positron beam receives an offset in the vertical and horizontal planes. This must be mitigated for the positron beam transport in the capture line. Compared to the FC systems used in the SuperKEKB or BINP designs, the FCC-ee requires a higher repetition rate (up to $100\,\mathrm{Hz}$), ideally with stronger magnetic fields and larger apertures. These requirements pose substantial technological and engineering challenges, particularly for high-power sources.

To address these challenges, an SC solenoid based on HTS technology was proposed for positron capture. The HTS solenoid offers several advantages over the FC design. It provides a significantly higher magnetic field at the target exit surface, a larger aperture and greater flexibility in target positioning, as the target can be placed inside the magnet bore. The axial symmetry of the solenoid ensures zero transverse magnetic fields at the magnet axis, eliminating beam distortion issues encountered with the FC option. The comparison of the field profiles for the FC and HTS solenoid designs, as used in the FCC-ee positron source studies, is shown in Fig. 7.6. Based on these considerations and simulation results, the AMD employing the HTS solenoid was selected as the baseline for the FCC-ee positron source.

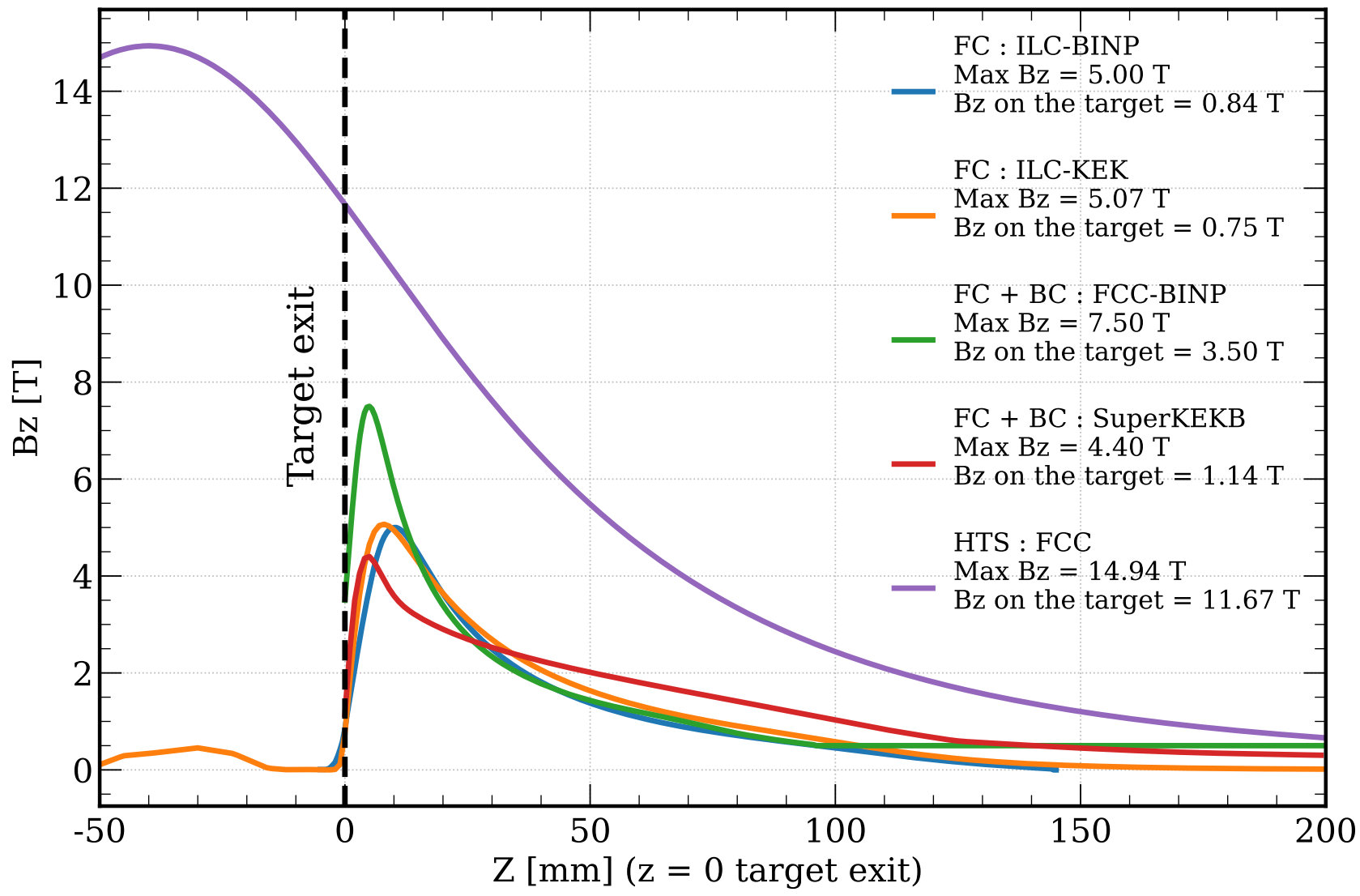

Fig. 7.6: Magnetic field profile of the AMD implemented in the form of the FC and HTS solenoid magnet. BC refers to the bridge coils, which are solenoid magnets surrounding the FC. The following FC models were analysed: BINP design for the ILC, KEK design for the ILC, BINP design for FCC-ee, and KEK design for SuperKEKB. A dashed line indicates the target exit surface.

The capture linac consists of six 3 m long, travelling wave (TW) 2 GHz RF structures with large iris apertures (2a = 60 mm), designed to provide enhanced transverse acceptance for positrons. The baseline design assumes an average RF gradient of 13.3 MV/m. Each accelerating structure is embedded within ten solenoid magnets, forming a solenoidal magnetic channel with a field strength of approximately 0.5 T, which efficiently guides the positron beam through the capture linac aperture. An additional solenoid magnet, referred to as the tuning solenoid, is placed between the AMD and the first accelerating structure. This solenoid increases the magnetic field experienced by the positron beam prior to entering the capture linac, further improving beam focusing and capture efficiency. The RF phases of the capture linac were optimised using the Xopt package [360] to maximise the final positron yield accepted by the DR. Figure 7.7 presents the key simulation results for the capture section, including the separator chicane and the first two accelerating structures of the positron linac. At the beginning of the capture section, a 33% drop in capture efficiency is observed, primarily due to the transverse acceptance of the accelerating structures. This is followed by a smaller 9% reduction in efficiency after the separator chicane.

### 7.4.2 Radiation load studies for target and capture system

The interaction of the electron drive beam with the positron production target gives rise to an intense flux of secondary particles. Only a fraction of the original drive beam energy contributes to the final positron yield, while most of the power is dissipated in the target, the AMD and the downstream capture linac. The resulting thermal load and cumulative radiation damage in the different components require a careful assessment in the engineering design process. This concerns in particular, the design of the production target and the optimisation of shielding components, which protect sensitive components like the SC coils of the AMD.

In order to quantify the impact of secondary radiation fields, radiation transport studies were carried out with the FLUKA Monte Carlo code. The FLUKA geometry model is illustrated in Fig. 7.8. The target was modelled as a stationary tungsten disk with a thickness of 15 mm. The target is surrounded by cylindrical tungsten shielding (13 cm long and 2 cm thick walls), which reduces the heat load and radiation damage in the AMD. The inner radius of the HTS coils of the AMD, placed in a cryostat, was 6.1 cm. A second tungsten shielding with a tapered aperture was assumed to be located between the AMD and the tuning solenoid, and a third between the tuning solenoid and the first RF structure of the capture linac.

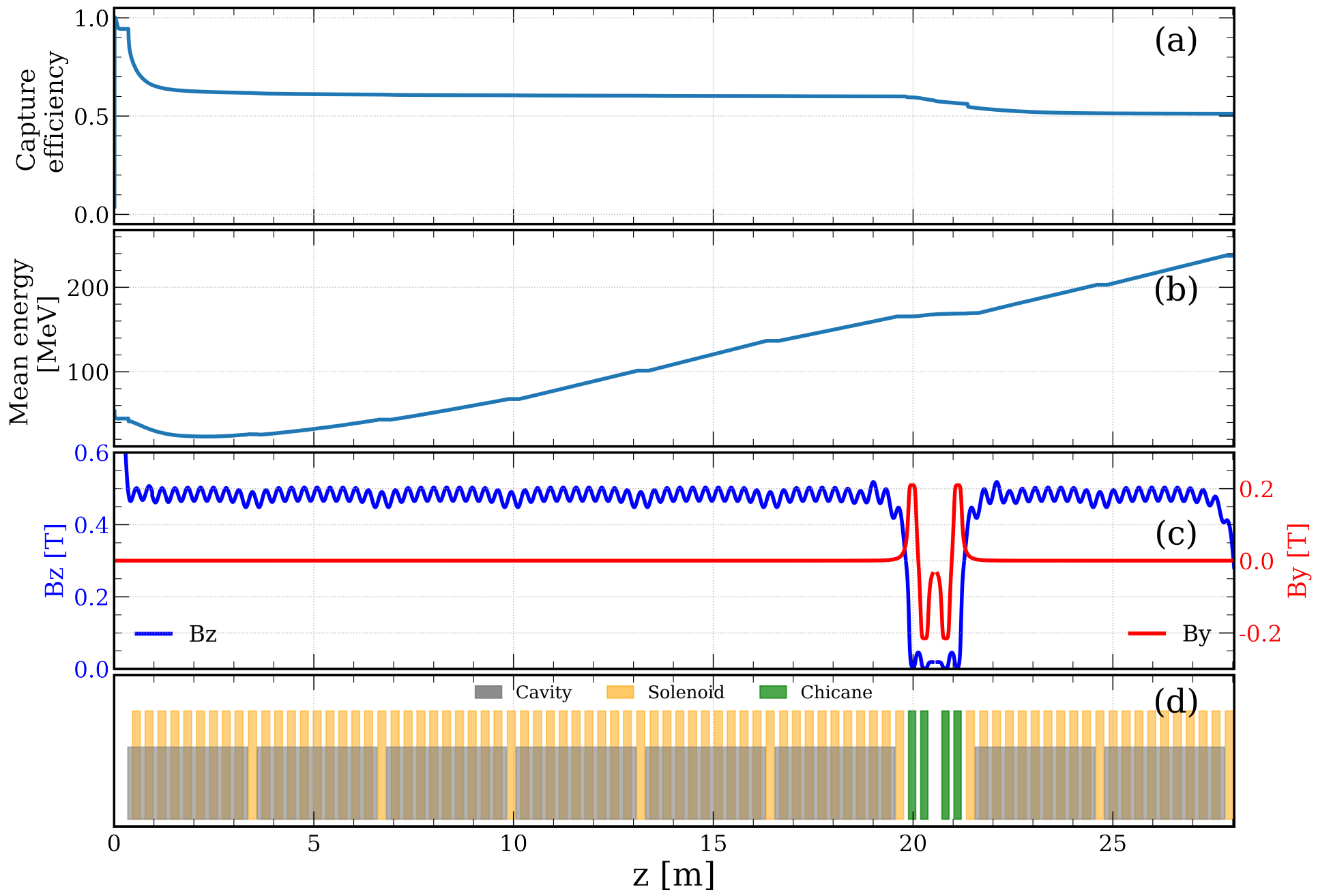

Fig. 7.7: Simulation results of the positron capture section, starting from the target exit surface. The plots illustrate: (a) the evolution of positron capture efficiency, (b) the positron beam energy, (c) the magnetic field profile along the capture section, and (d) the schematic layout of the capture section.

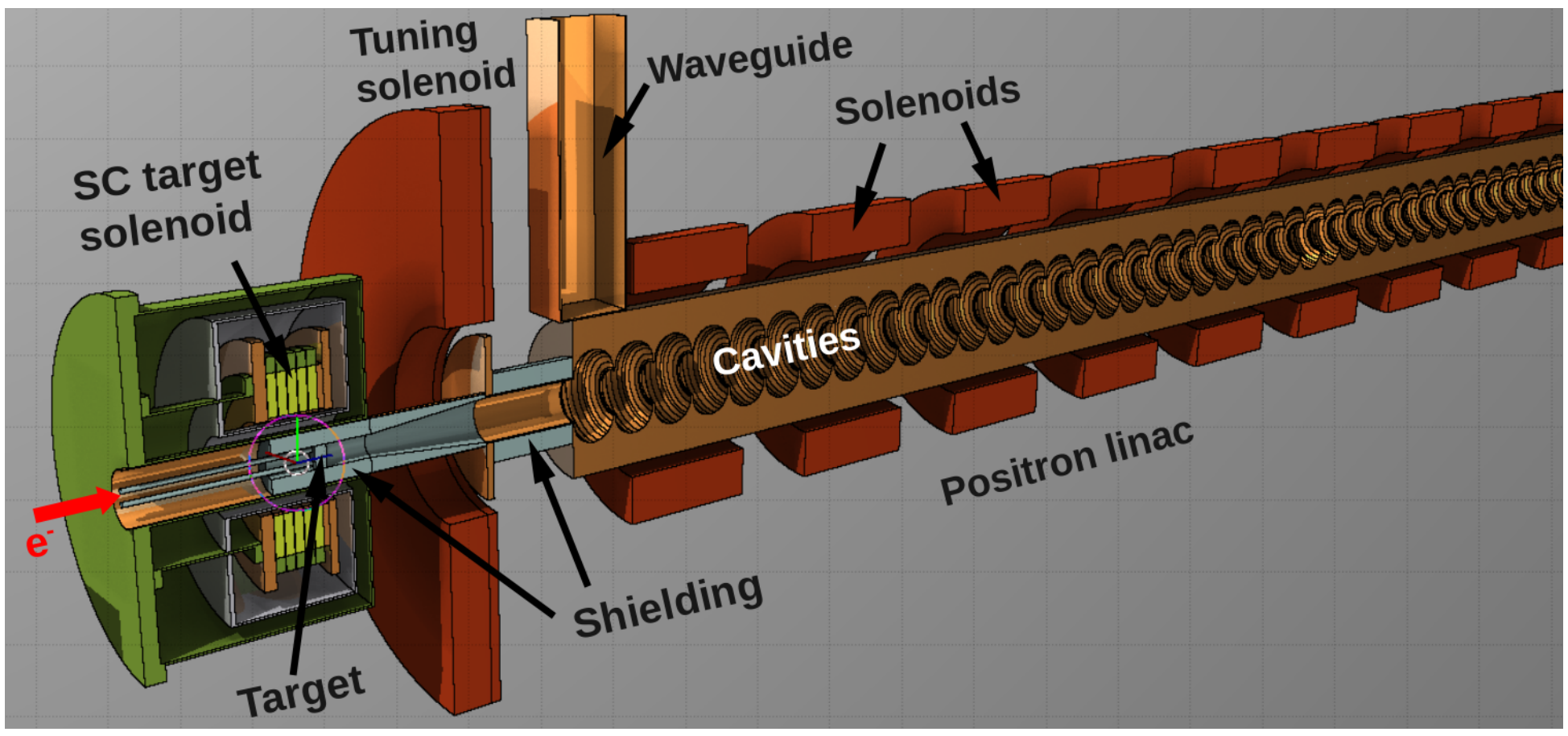

Fig. 7.8: FLUKA geometry model of the positron production target, the AMD with SC solenoid, the tuning solenoid, and the downstream positron capture linac.

The radiation load studies assumed a 2.86 GeV electron drive beam with four bunches, a bunch charge of $2.37 \times 10^{10}$ $e^-$, and a repetition frequency of 100 Hz, which results in an average drive beam power of 4.3 kW (Z-pole). The simulated power density distribution in the target, shielding, AMD and the first linac cells is illustrated in Fig. 7.9.

With such electron beam parameters, the target and surrounding shielding absorb around 1.3 kW,

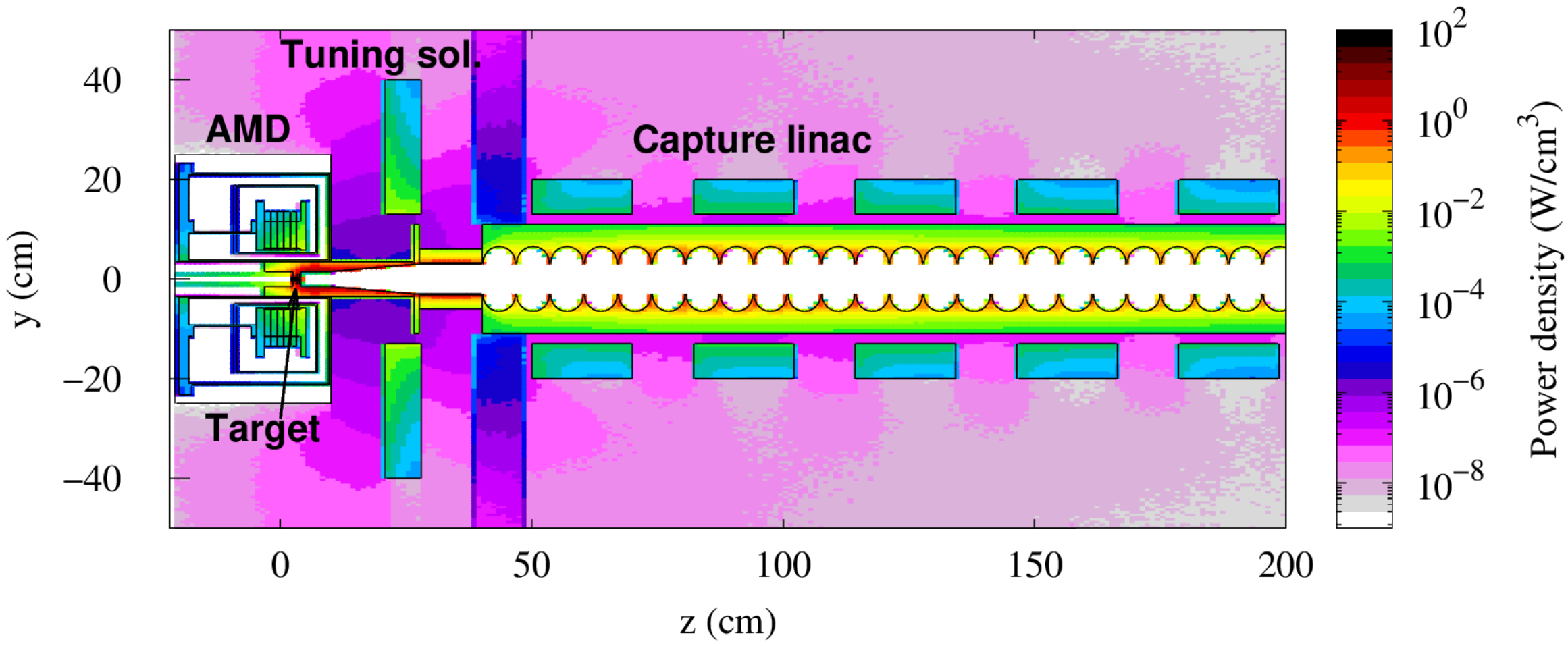

Fig. 7.9: Power density (Z-pole) in the target, AMD, shielding, tuning solenoid and the first RF cells of the capture linac. The peak power density in the target reaches $10\,\mathrm{kW/cm^3}$ (out of the scale of the plot).

which poses a challenge for the target design; possible engineering solutions and cooling options are discussed in the next section. Another issue is the atomic dislocations in the target, which are mostly concentrated along the beam axis. Assuming 185 days of operation per year and a duty factor of around $80\%$, the radiation transport simulations show that the displacement damage in the target can reach a peak value of 1–2 $\mathrm{DPA/year}$ for FCC-ee operation at the Z-pole. Possible mitigation measures need to be determined, e.g., the design of a remote handling system, which would enable regular replacement of the target assembly.

The power deposition in the AMD, including cryostat, HTS solenoid and support structures, is only about $10\,\mathrm{W}$, which demonstrates the effectiveness of the shielding. In particular, the power density in the HTS coils remains below $10\,\mathrm{mW/cm^3}$, which is considered acceptable and is safely below the quench level of the solenoid. The simulations also show that the cumulative displacement damage in the HTS tapes is less than $1 \times 10^{-4}\,\mathrm{DPA/year}$ (Z-pole), which is not expected to degrade their properties. Furthermore, the total radiation dose to the coils reaches about $6\,\mathrm{MGy/year}$. Since no organic insulation materials are used in the coils, such dose values should not pose a problem for the HTS solenoid, but require further assessment. If needed, a slight increase in the shielding thickness can reduce the dose further.

The capture linac is assumed to consist of six RF structures with 44 cells, which are surrounded by solenoids. About $0.6\,\mathrm{kW}$ are deposited in the shielding between the AMD and the capture linac. At the same time, most of the remaining power, i.e., about half of the power originally carried by the electron drive beam, is lost in the linac, mainly in the first structure. The tungsten shielding between the AMD and the linac protects the front face of the linac, but cannot intercept the most energetic secondary particles near the beam axis, which are then lost on the cavity walls. The highest radiation-induced power deposition in a single RF cell is about $70\,\mathrm{W}$, but decreases to about $15\,\mathrm{W}$ at the end of the first RF structure. The thermal load due to RF wall losses is estimated to reach similar (or even higher) values as the average radiation-induced power deposition per cell. With an adequate cooling design, the radiation load in the linac is expected to be manageable. The solenoids around the RF structure are assumed to be normal conducting. The integrated dose in the solenoids is estimated to be about $1\,\mathrm{MGy/year}$ for operation at the Z-pole, which is considered acceptable. However, the peak dose in the upstream tuning solenoid reaches $5\,\mathrm{MGy/year}$, which must be considered when choosing insulation materials. A better shielding of the tuning solenoid might be needed.

### 7.4.3 Design and integration of the positron source target

The current baseline design is based on a fixed target made of polycrystalline tungsten (W) with a thickness of 15 mm. The selection of tungsten as a material for the target is due to its high atomic number and its remarkable thermo-mechanical properties at high temperatures. However, to properly dissipate the thermal power produced by the beam impact, a thermal management strategy must be included in the design. For this purpose, a pressurised water cooling circuit is added in the target, as shown in Fig. 7.10a. This consists of a pair of embedded tantalum pipes that transport water from an upstream source and circulate through a 180° elbow inside the tungsten core. This setup will allow the beam-impacted region to properly transfer the 1.26 kW deposited on the target and its shielding and avoid the direct contact of water with bare tungsten. The power density distribution obtained from Monte Carlo simulations is shown in Fig. 7.10b, where the peak value is 10.4 kW/cm$^3$ and it takes place along the primary beam axis (z-axis), close to the exit face of the target.

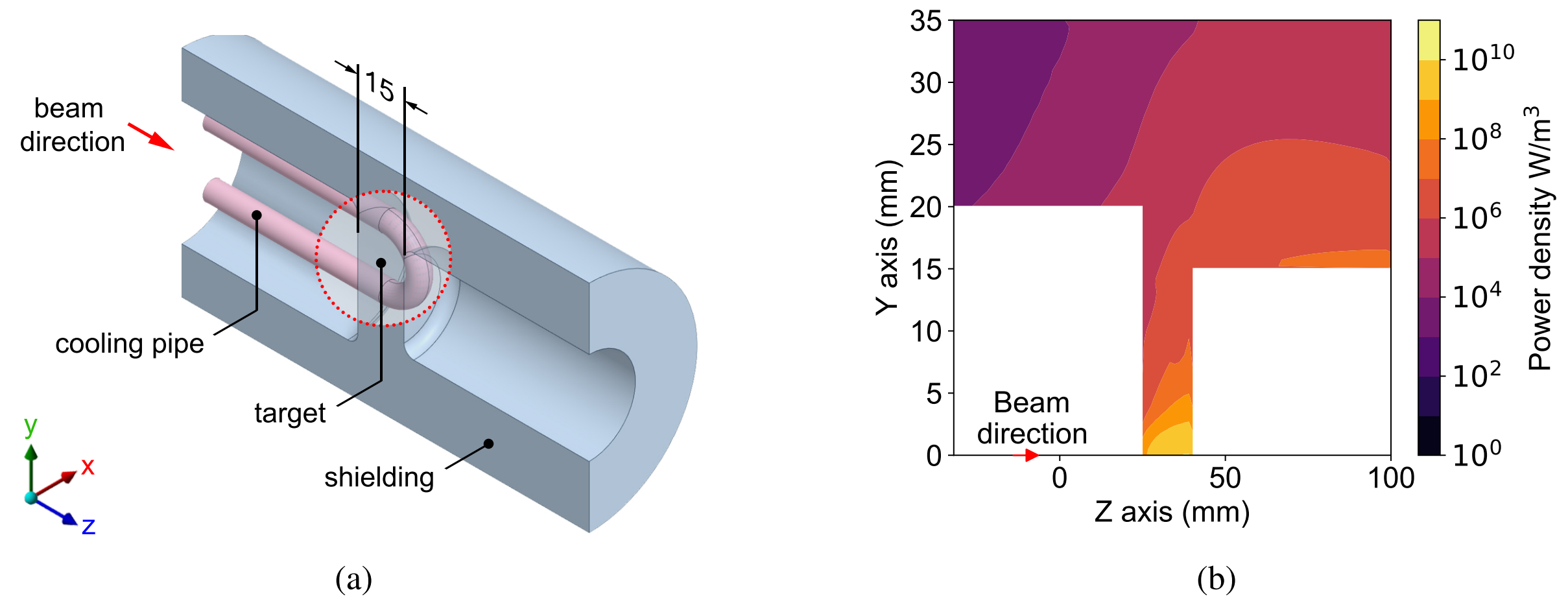

Fig. 7.10: FCC-ee e$^+$ source target. (a) Baseline design geometry: the detailed zone shows the embedded tantalum cooling pipes. Only one half of the geometry is shown because it is symmetrical with respect to the x-axis and (b) power density deposition map obtained from FLUKA.

Figure 7.11 shows the steady-state thermo-mechanical results. Note that the location of maximum temperature (P1) is not coincident with the position of maximum equivalent thermal stresses (P2). While P1 is along the beam axis beneath the exit surface, P2 is located at the exit surface at a height above P1. The maximum temperature at P1 is 284°C. This means that the target is working below the ductile-to-brittle transition temperature (DBTT) for tungsten[1]. In terms of stresses at P1, the 99 MPa registered on the design are due the constrained material surrounding the target. On the other hand, P2 reaches 166°C with a maximum equivalent stress of 138 MPa located at the surface level and produced due to the strong thermal gradient. The resulting stress values are below the yield stress at the associated temperatures. The thermal fatigue analysis performed using the Universal Slope method [362] showed that the target is capable of withstanding the extreme service conditions and coping with the expected lifetime of the device, set to 155.4 days/year, which corresponds to $1.34\times10^9$ thermal cycles[2] with a duty factor of 0.84 [363]. Then, the number of thermal cycles is obtained by including the primary beam frequency of 100 Hz.

From the integration standpoint, the positron source target is a subsystem and its interaction with the required infrastructure is based on the mechanical layout of the P$^3$ experiment. The current configuration is being used to study the space requirements during the injector complex design. Figure 7.12 shows the tunnel cross section with dimensions of 4×4 m to host the expected services around the beam intercepting device. With the aim of allowing the installation and replacement of the target inside the HTS solenoid, a drift space of around 1.5 m is included. In addition, an overhead travelling crane with a capacity of 500 kg to handle the target and the surrounding shielding is under consideration. In parallel, a more detailed study of radiation protection where the use of mobile shielding is an option is being carried out. At the same time, an evaluation of the utilities required (e.g., cooling, cabling, handling equipment) needs to be developed. Further integration studies are ongoing to define the requirements in terms of civil engineering. The results of these studies will be used in the model for general integration of the injector complex.

[1]For the results presented in this document, the DBTT was set to 400°C, based on the behaviour of tungsten at high strain-rate loading conditions reported in [361].

[2]The target lifetime is estimated assuming an operation cycle of 185 days/year [13]

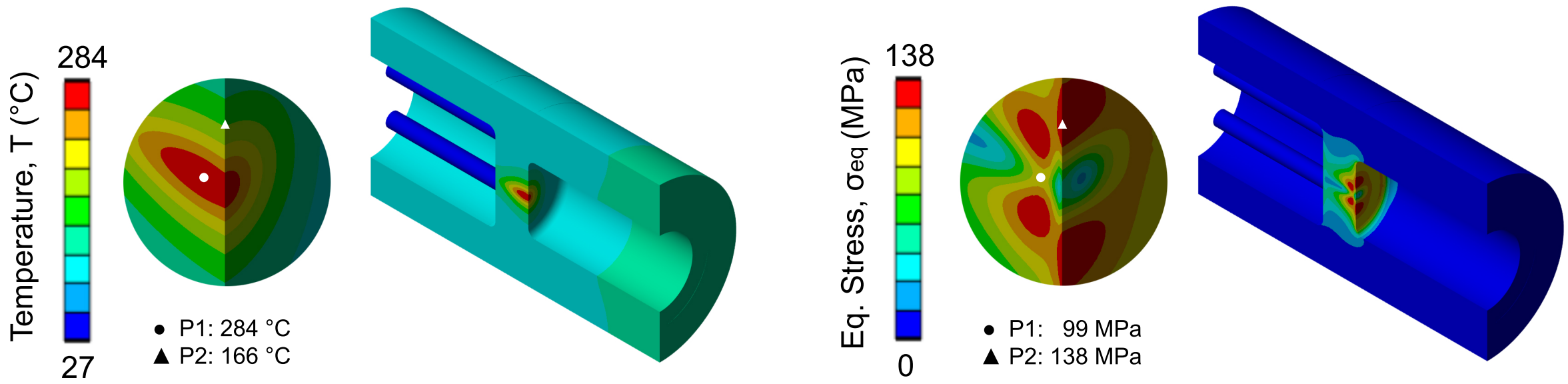

Fig. 7.11: Steady-state thermo-mechanical results: temperature (left) and equivalent stress (right) distributions for the baseline W target. Maximum temperature and stress points are marked with a circle and triangle, respectively.

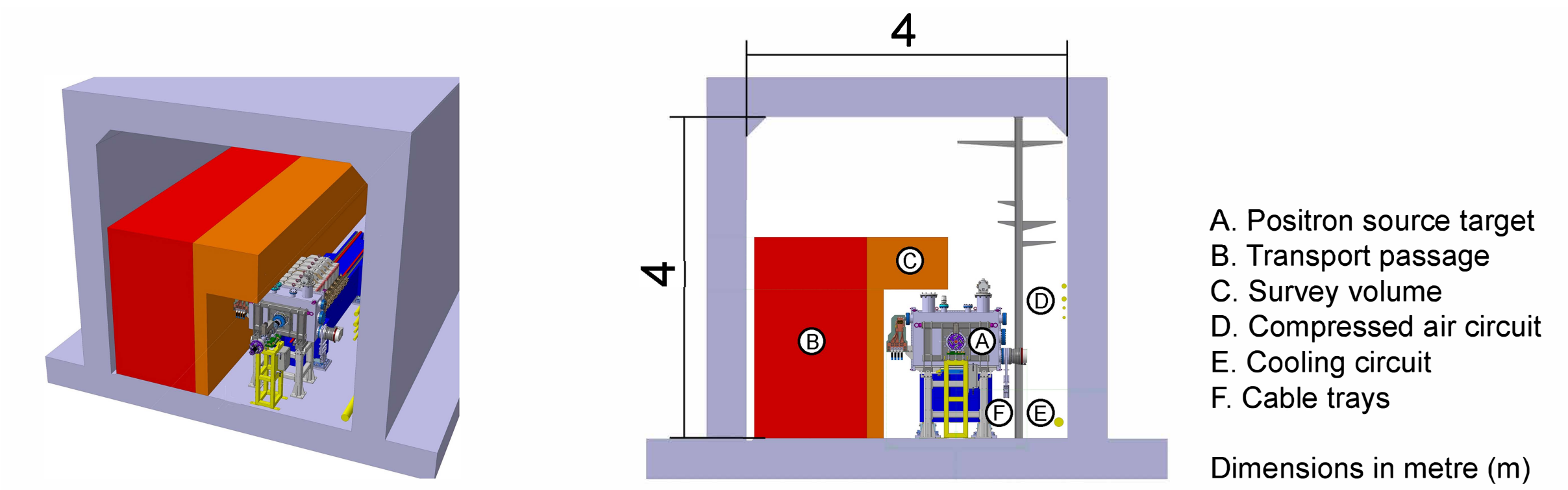

Fig. 7.12: Current layout of the FCC-ee injector complex cavern: tunnel overview (left) and expected utilities around the positron source target (right).

### 7.4.4 Positron linac

The positron linac (p-Linac) begins with a chicane equipped with a collimator (beam stopper) at its centre to remove electrons and photons co-propagating with the positron beam. The p-Linac is divided into two sections, each with distinct layouts. In Section 1, the layout is the same as that of the capture linac, where the accelerating structures are surrounded by solenoids. In contrast, Section 2 employs a simple FODO lattice, with each FODO cell containing two accelerating structures and a phase advance of 76.35°, optimised for minimum positron beam size. The accelerating structures and solenoids used in both sections are identical to those in the capture linac. The separator chicane is situated between the capture linac and Section 1, featuring a symmetric layout of four dipole magnets with identical designs but different current settings. A schematic diagram of the chicane, along with the collimator/beam stopper, is presented in Fig. 7.13a.

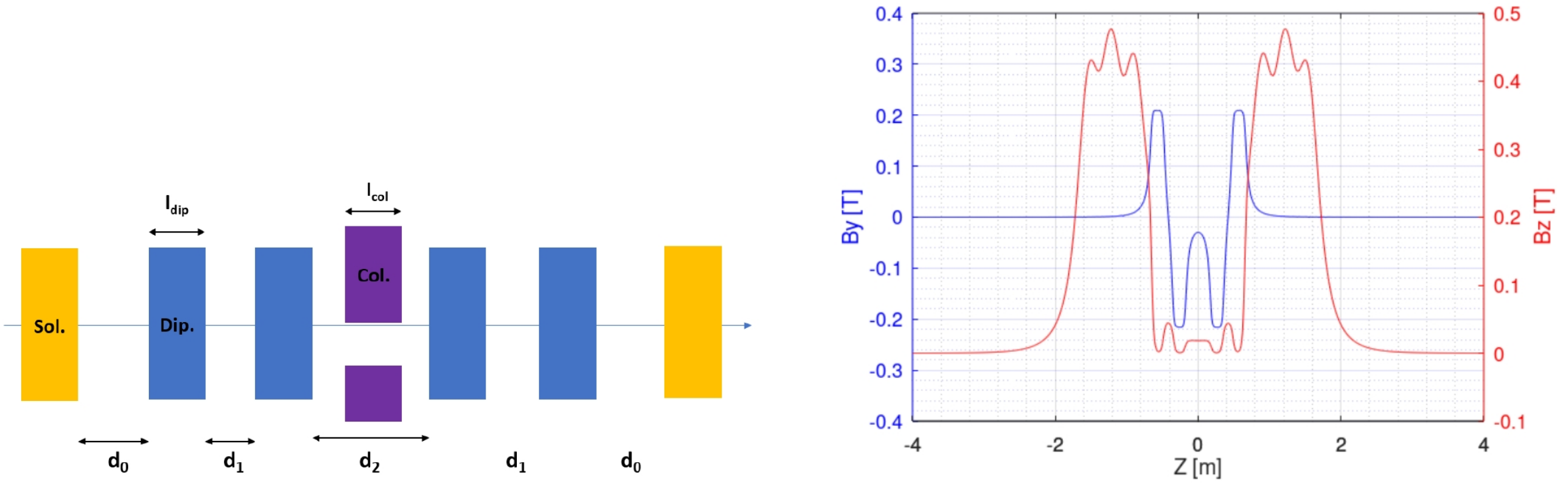

(a) (b)

Fig. 7.13: Separator chicane design. (a) Schematic layout of the chicane and collimator used to stop the electron and photon beams. Sol. refers to Solenoid magnet; Dip. refers to Dipole magnet ($l_{\text{dip}}$ = 180 mm) and Col. refers to Collimator ($l_{\text{col}}$ = 120 mm) with $d_0 = d_1 = 125$ mm and $d_2 = 350$ mm. (b) On-axis magnetic field of the chicane, including three neighbouring solenoids on each side.

The dipole yoke in the chicane has a length of 180 mm and a vertical aperture of 70 mm. The beam pipe within the dipoles assumes a rectangular aperture of $\Delta x = 150$ mm and $\Delta y = 50$ mm. The collimator is horizontally offset by $-35$ mm. To account for field crosstalk between the chicane and the upstream and downstream solenoids, a 3D magnetic field simulation was performed using MAXWELL3D. This simulation includes three solenoids upstream and three downstream of the chicane. The resulting on-axis magnetic field is shown in Fig. 7.13b.

Section 1 of the p-Linac contains 20 accelerating structures, with an average RF gradient of 13.3 MV/m. The RF phase is set to -10° off peak, optimised to maximise the final positron yield. At the end of this section, the average energy of the positron beam around the bunch core is approximately 932 MeV. Section 2 contains 52 accelerating structures, with an average RF gradient of 12.8 MV/m. Here, the RF phase is adjusted to 5° off crest for optimal positron yield.

Beam tracking throughout the p-Linac was simulated using RF-TRACK, including short-range wakefield and space charge effects. To estimate the effective positron yield accepted by the DR, particle selection was applied using energy and time cuts. For all results presented here, an energy window of ± 2% around 2.86 GeV (i.e., 2.86 GeV ± 57.2 MeV) and a time window of ± 10 mm/c were used. In the next phase of the FCC, positron tracking simulations in the positron linac followed by the injection in the DR should be carried out to have a more realistic estimate of the positron yield accepted.

### 7.4.5 Simulation Results and Final Performance

Comprehensive start-to-end tracking simulations were conducted to evaluate the performance of the FCC-ee positron source. The final longitudinal phase space of positrons at the end of the p-Linac is shown in Fig. 7.14. Approximately 99% of the positrons reaching the end of the p-Linac are accelerated in the first main RF bucket. Furthermore, about 88% of these positrons fall within the assumed DR acceptance window, demonstrating the effectiveness of the positron source and linac design. The evolution of the positron yield along the longitudinal axis, from the target exit to the end of the p-Linac, is depicted in Fig. 7.15.

Table 7.6 summarises the key parameters and simulation results for the baseline design of the positron source and linac. Beam dynamics simulations confirm that the proposed design ensures reliable positron production, achieving a final accepted yield of 3 $N_{e^+}/N_{e^-}$. This meets the requirements set by the FCC-ee (Z-pole) with a safety margin of 2.56. To date, no critical issues have been identified that would prevent the use of a superconducting solenoid as the AMD, along with the proposed capture system and positron linac, in the baseline design. For comparison, in a similar layout using an FC-based capture system (SuperKEKB model), the accepted positron yield is approximately 1.5 $N_{e^+}/N_{e^-}$. This lower yield would likely result in use of a thermionic gun to deliver the required electron drive beam bunch charge ($\geq 5\,\mathrm{nC}$), significantly increasing the target power load.

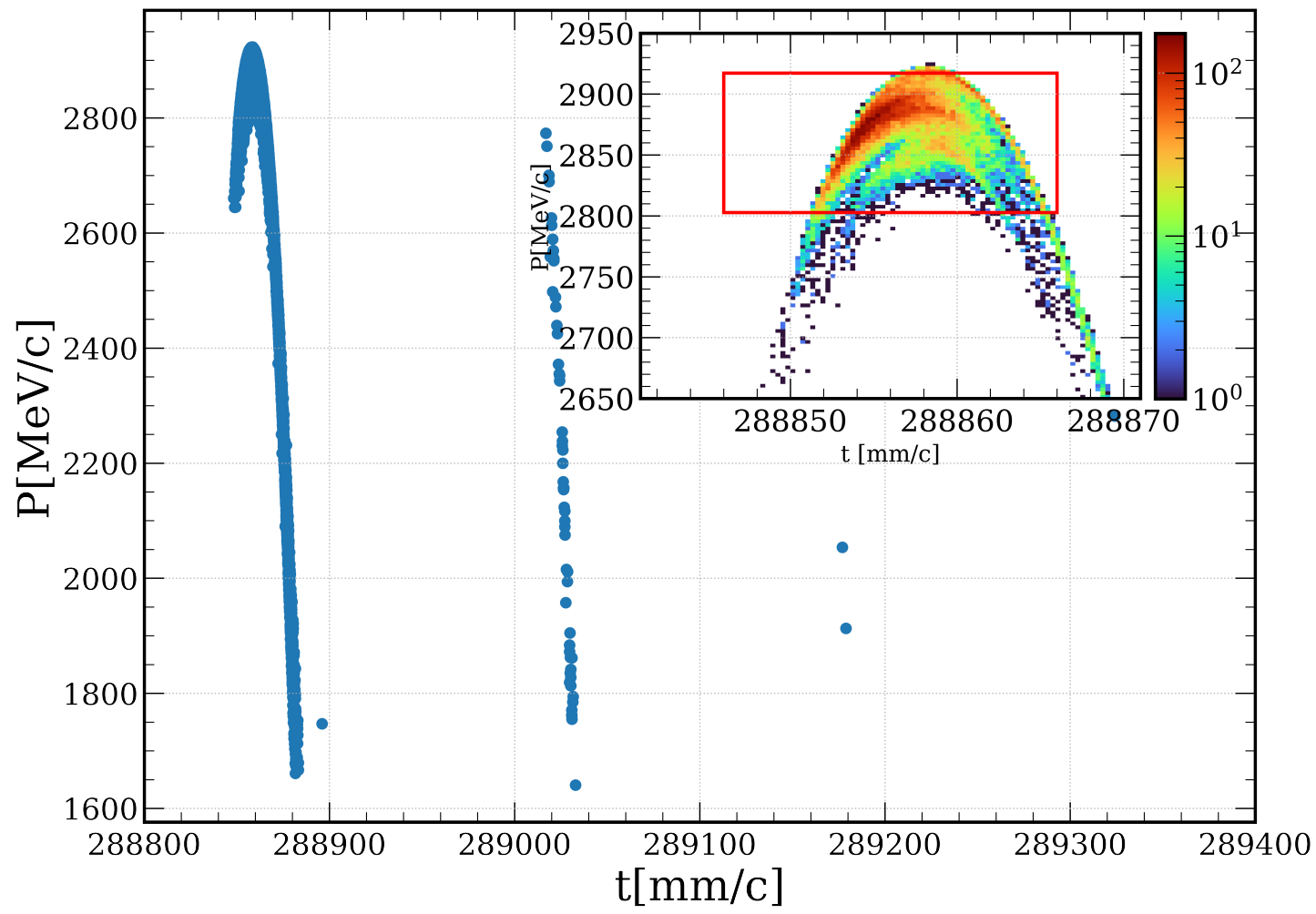

Fig. 7.14: Longitudinal phase space at the end of the positron linac. The inset provides a zoomed view of the main RF bucket. The red region represents the energy and time cut window used to estimate the fraction of positrons accepted by the damping ring.

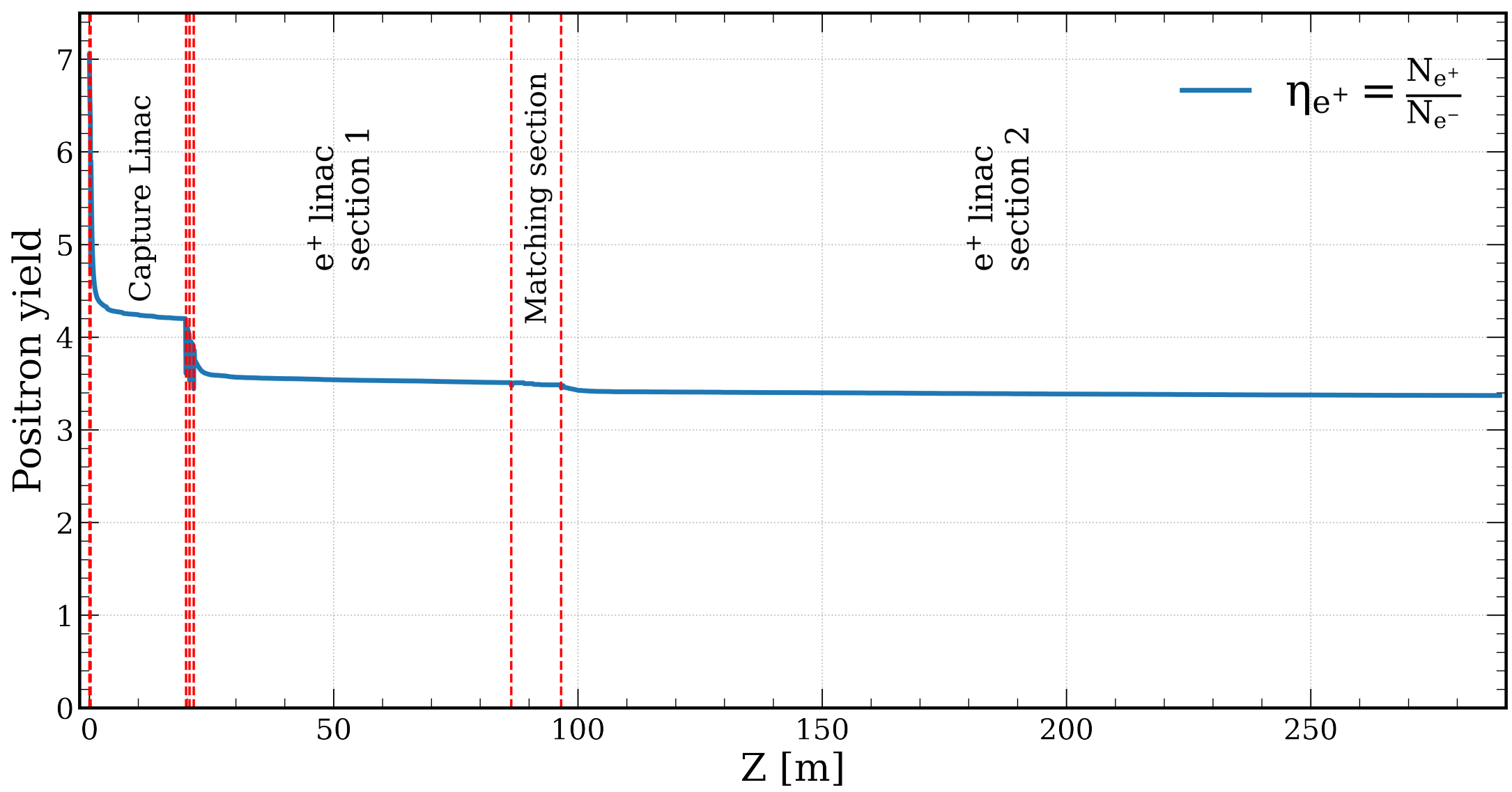

Fig. 7.15: Evolution of the positron yield along the longitudinal axis, from the target exit to the end of the positron linac. The dashed lines indicate the boundaries of the different sections of the positron pre-injector up to the end of the positron linac.

Simulations incorporating imperfections in the positron source systems were also conducted. The imperfections considered, from the target to the end of the p-Linac, are summarised in Table 7.7, with RMS values reported. The impact of these imperfections is found to be negligible. The average positron yield accepted by the DR decreases by only 1.3%, while the average transverse emittances increase by 0.4% horizontally and 0.8% vertically. These results indicate the robustness of the proposed design.

Table 7.6: Parameters and simulation results for the baseline design of the positron source and linac, delivering a positron bunch charge of 5 nC accepted into the damping ring, including safety margins. CS (capture system), CL (capture linac), p-Linac (positron linac), DR (damping ring).

| **Parameter** | **Value** | **Unit** |
|---|---|---|
| **$e^-$ Drive Beam** | | |
| Beam energy | 2.86 | GeV |
| Repetition rate | 100 | Hz |
| Number of bunches per pulse | 4 | |
| Bunch charge | 3.8 | nC |
| Bunch length (rms) | 1 | mm |
| Beam size (rms) | 1 | mm |
| Beam power | 4.3 | kW |
| **Target** | | |
| Thickness | 15 | mm |
| Production rate | 7.07 | $N_{e^+}/N_{e^-}$ |
| Deposited power | 1 | kW |
| PEDD | 5.8 | J/g |
| **Capture System** | | |
| AMD peak field (@Target) | 15 (12) | T |
| Solenoid strength | 0.5 | T |
| AMD/CS aperture | 60 | mm |
| Average energy @CL | 173 | MeV |
| Positron yield @CS (before chicane) | 4.2 | $N_{e^+}/N_{e^-}$ |
| **Positron Linac** | | |
| Positron yield @p-Linac | 3.4 | $N_{e^+}/N_{e^-}$ |
| Positron yield accepted @DR | 3.0 | $N_{e^+}/N_{e^-}$ |
| Average energy | 2.87 | GeV |
| Bunch length (rms) | 2.84 | mm |
| Energy spread (rms) | 0.87 | % |
| Spot size x/y (rms) | 5.28 / 2.78 | mm |
| Normalized emittance x/y (rms) | 13.1 / 13.0 | mm·rad |
| Geometric emittance x/y (rms) | 2.36 / 2.32 | mm·rad |

Table 7.7: Summary of imperfections considered from the target to the end of the positron linac, with RMS error values reported for each parameter.

| Imperfection | Unit | Value |
|---|---|---|
| Transverse position error | µm | 100 |
| Transverse angular error (solenoids and dipoles) | µrad | 200 |
| Transverse angular error (other elements) | µrad | 100 |
| Magnetic strength error | % | 0.1 |
| RF gradient error | % | 1 |
| RF phase error | ° | 0.1 |
| Beam position error | µm | 100 |
| Beam divergence error | µrad | 100 |

### 7.4.6 PSI Positron Production ($P^3$) Project

The PSI Positron Production ($P^3$) experiment is a demonstrator for the positron source and the goal is to design and install such a demonstrator in the SwissFEL facility, and experimentally validate a range of novel techniques that, according to simulations, have proven potential to increase the positron yield by one order of magnitude with respect to the state of the art [359]. The $P^3$ project is driven by the high luminosity requirements of the FCC-ee collider ring and its results will be one of the key outcomes of the feasibility study concerning the injector. The remarkable positron capture capabilities of $P^3$ are enabled to a great extent, by the usage of a high-temperature superconducting (HTS) solenoid around the target region, as well as a novel standing-wave solution for the RF cavities that provides a large iris aperture.

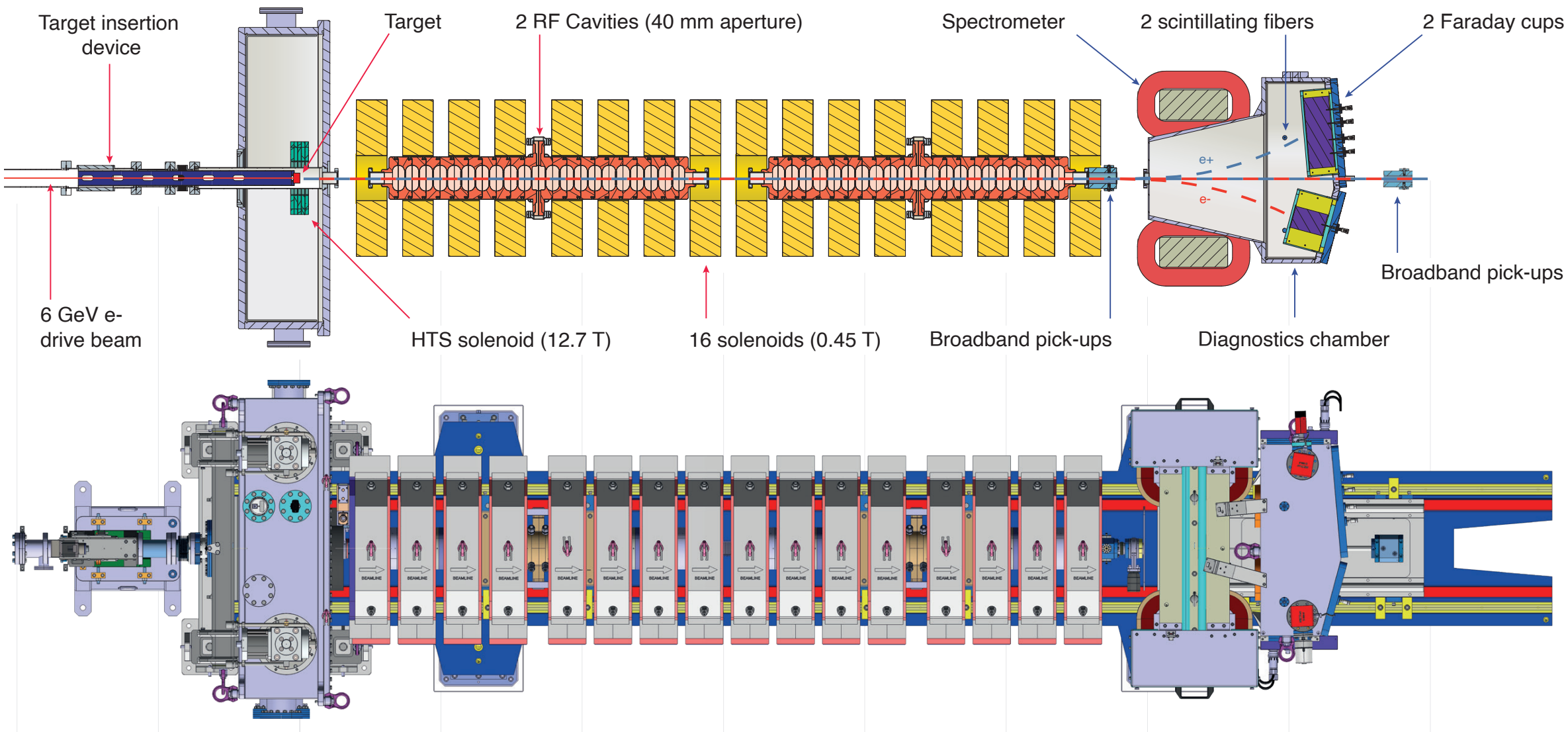

Fig. 7.16: Overview of the technology for the $P^3$ experiment.

Figure 7.16 presents a technical drawing of the experiment, illustrating all the technologies developed or under development at PSI and CERN. These technologies have been carefully designed and optimised based on an in-depth study involving advanced beam dynamics simulations. This detailed simulation work made it possible to model and predict the behaviour of particle beams under various conditions, thereby refining the experimental setup and ensuring the optimal performance of each component. The figure also illustrates the integration of these technologies, highlighting key aspects such as the integration of the target in the cryostat that houses the HTS coils, the RF structures surrounded by normal conducting solenoids, and the diagnostics chamber that will allow measurement of the charge and energy spectrum of the positrons generated.

The procurement and assembly of most accelerator and diagnostic components are progressing on schedule, ensuring the timely completion of the project milestones. A key achievement has been the successful demonstration of the HTS solenoid at PSI, which achieved magnetic fields up to 18 T, a significant step forward for this advanced component.

Figure 7.17 shows photos of the $P^3$ component production process, including the HTS coils and their cryostat, the broadband pickups, and the first accelerating RF structure. Overall, the $P^3$ experiment is making steady progress, with installation work at SwissFEL proceeding smoothly during scheduled shutdown periods, which take place three times a year. Essential infrastructure components, such as segments of the extraction line and the high-voltage klystron-modulator system, are being installed in the tunnel according to plan. The primary installation phase is expected to conclude by the end of 2025, paving the way for the experiment to begin positron operations in 2026. It is worth noting that the SwissFEL facility is an ideal location for hosting the $P^3$ experiment, as it can deliver an electron beam energy of up to 6 GeV, matching the maximum drive beam energy required for the FCC-ee positron source. Currently, two beamlines (Aramis and Athos) are operational at SwissFEL, and the accelerator tunnel has reserved space for a future third beamline (Porthos). This layout provides sufficient room for the temporary installation of the $P^3$ experiment and switchyard. Figure 7.18 illustrates the current design of the $P^3$ experiment in its planned final configuration within the SwissFEL tunnel.

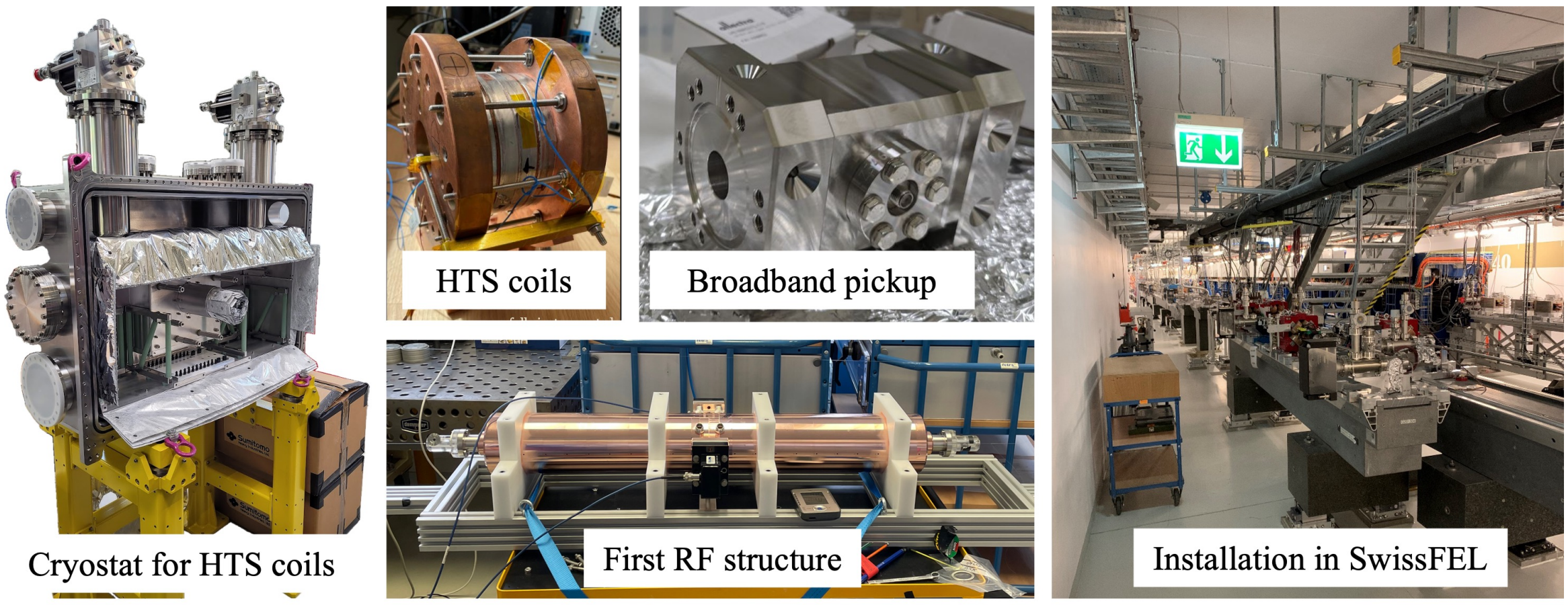

Fig. 7.17: $P^3$ components production and installation in SwissFEL.

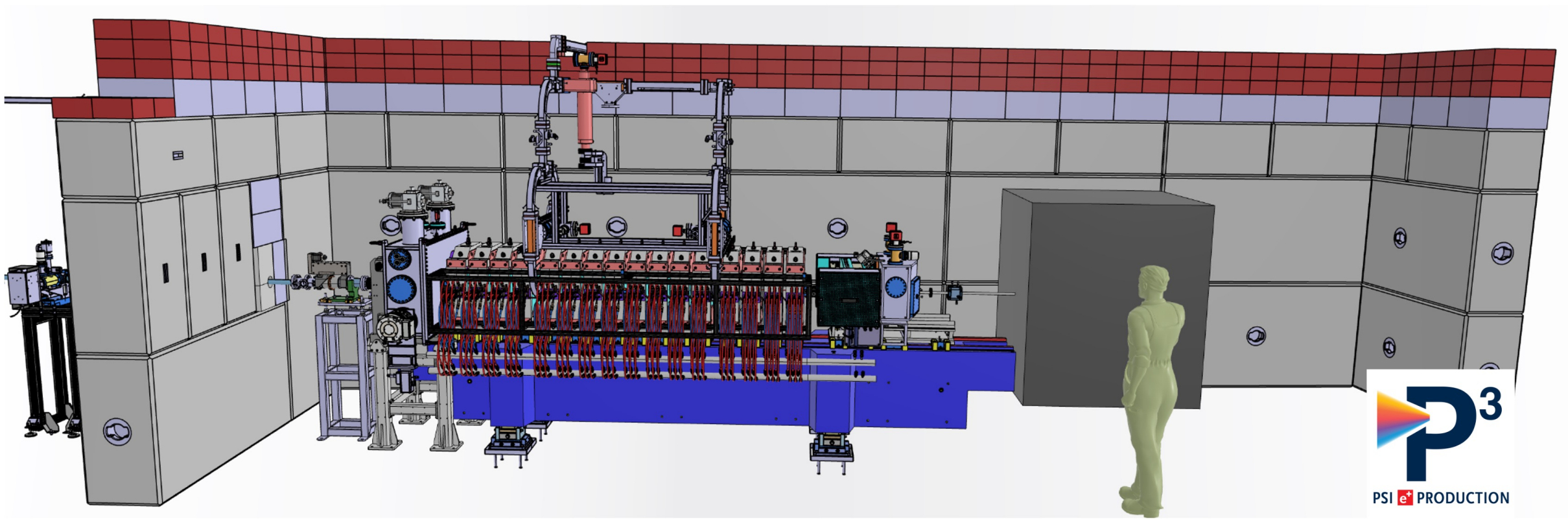

Fig. 7.18: The PSI Positron Production ($P^3$) Experiment.

### 7.5 Damping ring and bunch compressor

The new optimised FCC-ee injector layout imposed a review of the intrinsic structure of damping ring (DR) Transfer Lines (TLs). The presence of two independent linacs for electron and positron beams, and the elimination of the common linac, naturally led to increased DR energy which, to avoid spin resonances, was set at the value of 2.86 GeV.

The main concept driving the DR and TLs design consists of achieving an overall efficiency of the order of 80 % in transporting electron and positron beams from the respective linacs, through the DR for emittance cooling, to the end of the TLs conveying extracted beams toward the collider booster. Electron and positron linacs produce beam pulses at 100 Hz, each pulse consists of 4 bunches spaced by 25 nsec, each bunch stores a variable charge intensity up to 5 nC. The DR is mainly needed to reduce the emittance of the incoming positron beam by more than three orders of magnitude, from $2.36\times10^{-6}$ m·rad to about $1.8\times10^{-9}$ m·rad.

However, in the latest injector layout, the DR will also be used for electron beam cooling to cure possible emittance dilution induced by misalignments and space-charge effects. Several options have been studied and evaluated for the DR arcs, such as using multi-bend cells, and FODO cells. One of the options studied is based on a 6-fold symmetry ring, and has multi-bend arc cells. This design is presented as the main option in the following section. An alternative design of the DR has also been studied. A FODO cell is chosen and this alternative DR design, consists of three arcs and three straight sections that locate damping wiggler magnets, the RF cavity and injection/extraction equipment. The ring is about 384 m long, and its energy is 2.86 GeV. The injected beam emittance could be reduced to the required emittance value of 1.76 nm·rad, horizontal damping time is 6.4 ms and the energy loss per turn is 1.13 MeV. The total length of the damping wiggler magnets is about 36.45 m. They have a 2 T magnetic field and are distributed in the three straight sections. It could be possible to lower the magnetic field of the damping wiggler to 1.5 T for optimising the phase advance for minimum emittance (around 135°). However, this may cause even more challenging dynamic aperture optimisation.

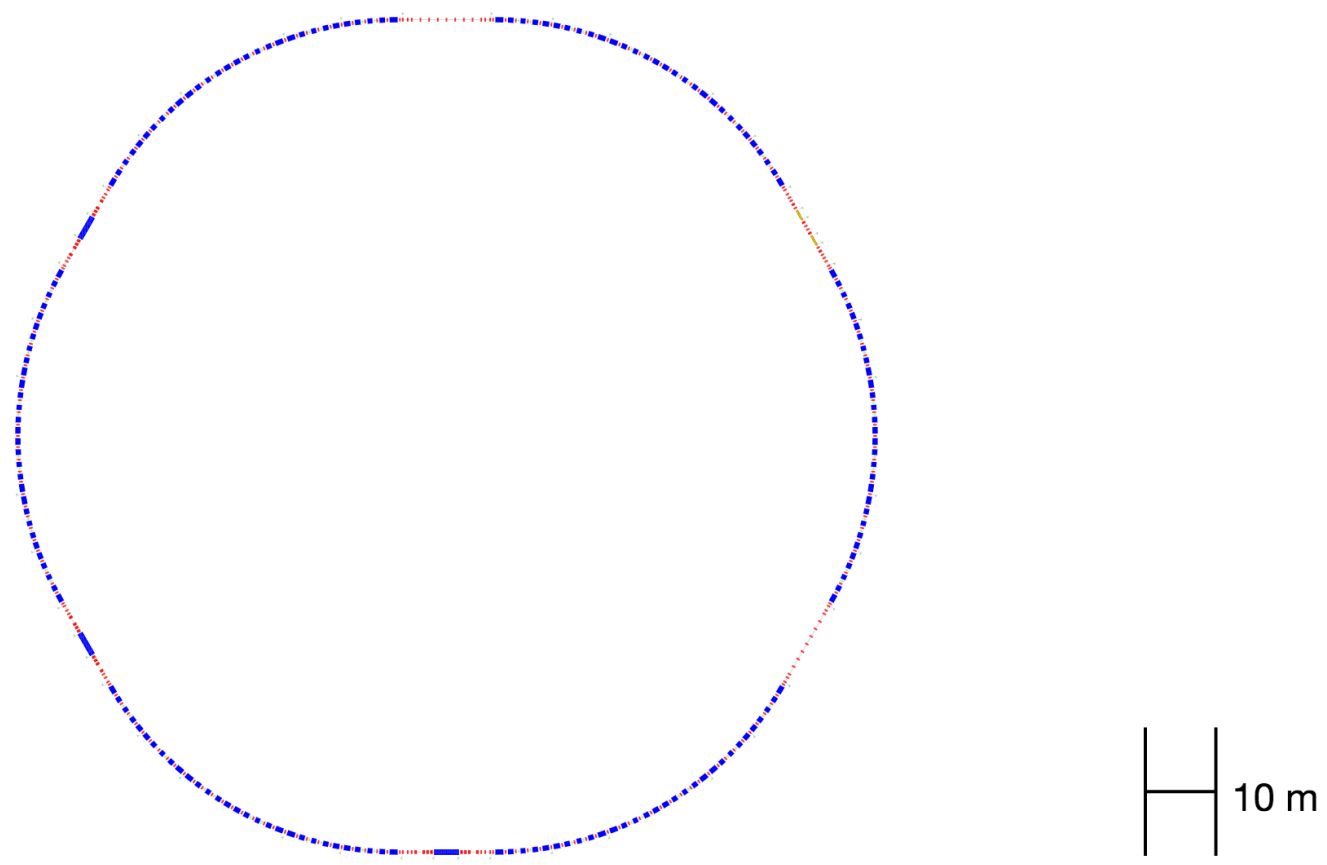

Fig. 7.19: Damping Ring layout. The six-fold symmetry allows having straight sections dedicated for different equipment: RF cavity, injection/extraction septa and kickers and wigglers (to reduce damping time and equilibrium emittance).

### 7.5.1 New damping ring design

The new DR lattice features a six-fold symmetry, as shown in Fig. 7.19. It consists of six arc cells connected by six straight sessions. Each straight session is used to host three wiggler magnet insertions, one RF cavity module, and two independent injection/extraction sections. Injection and extraction will be implemented in the same branch for the two-particle species in order to avoid changing the polarities of the DR magnets, thus ensuring fast and reliable operation modes for both electron and positron. The injection will be performed using an on-axis scheme.

Arc cells are based on an achromatic multi-bend optics, symmetric with respect to the cell centre. Each half cell provides 30° deflection angle, using 15 bends of five different types. This approach allows keeping the maximum excursion of the horizontal dispersion, optimising damping time, and shaping the $H_5$ function along the cell to achieve low emittance. A further reduction of the damping time is obtained using three 3.5 m long wiggler magnets, each with a moderate magnetic field intensity of 1.8 T. Straight sections are based on the FODO structure and modified according to their function.

The DR optics is presented in Fig. 7.20, it features moderate betatron oscillation amplitudes achieved with a relatively weak focusing lattice producing low chromaticity per cell and, consequently, wide on- and off-momentum dynamic aperture. Limiting betatron oscillation amplitudes in the transverse planes below a maximum value of the order of 10 m is also beneficial in keeping the ring requirement in terms of stay-clear aperture under control, which is crucial especially for DR operation with positron beam. A complete list of the DR parameters is presented in Table 7.8.

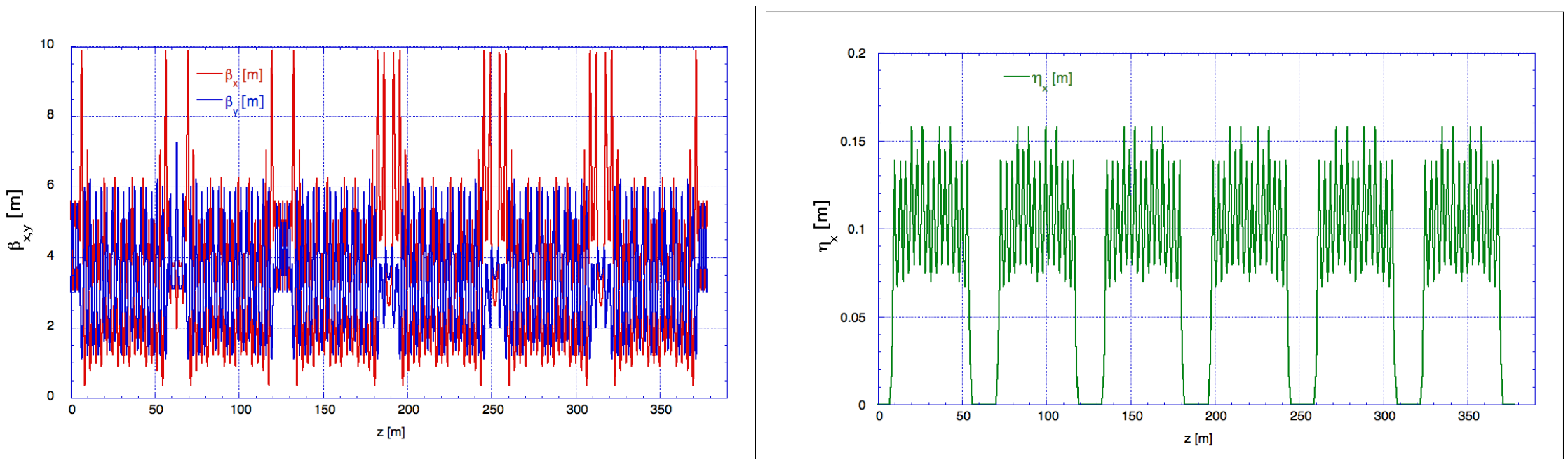

Fig. 7.20: Damping Ring optics: betatron amplitude (left) and dispersion (right).

Table 7.8: Damping ring parameters.

| Parameters | Value |
|---|---|
| Energy [GeV] | 2.86 |
| Circumference [m] | 373.46 |
| Arc Cell | multi-bend |
| Lattice shape | six-fold symmetry |
| Nat. emittance [nm rad] (WGL on/off) | 1.3 / 2.3 |
| Bunch Length [mm] | 5.1 |
| Damping time $\tau_{x,y}$ (WGL on/off) [ms] | 16.9 / 29.4 |
| Nat. Chromaticity (x/y) | -38.2/-28.3 |
| Nat. energy spread (WGL on/off) [$10^{-4}$] | 7.1 / 5.2 |
| Betatron amplitude max (x/y) [m] | 9.66 / 6.49 |
| Betatron amplitude min (x/y) [m] | 0.5 / 1.1 |
| Tune ($Q_x$,$Q_y$) | 27.8707 / 22.3728 |
| Momentum compaction (WGL on/off) [$10^{-3}$] | 1.55 / 1.57 |
| Revolution period [µs] | 1.2457 |
| Dipole #, length [m], field [T] | 180 , 0.7  1.13, 0.34  0.39 |
| Wiggler #, length [m], field [T] | 3, 3.5 , 1.8 |
| Cavity #, length, voltage [MV] | 1.5, 4 |
| Max. # Bunch stored, Bunch Curr. [mA] | 40 /  4 |
| Store time | 5 $\tau_y$ |
| Energy loss per turn (WGL on/off) [keV] | 422.2 / 246.7 |
| SR power loss wiggler [kW] | 27.83 |
| Kicker rise time [ns] | 50 |

### 7.5.2 Timing

The maximum number of trains that could be simultaneously stored in the damping ring depends on the revolution period ($T_{per}$), the train length ($\Delta t$) and the kicker pulse rise time ($t_K$)

$$n_{train} = \frac{T_{per}}{\Delta t + t_K}$$

The storing time ($T_{store}$) for each train of bunches it is fixed by the requirements imposed from the next steps of the injector chain: the positron vertical emittance must be damped from 2.34 mm·mrad ($\epsilon^y_{inj}$) to 0.18 nm·rad ($\epsilon^y_{ext}$);

$$\epsilon_y(t) \sim \epsilon^y_{inj} \mathrm{e}^{-\frac{2t}{\tau_y}}$$

The damping required implies:

$$T_{store} = -\frac{\tau_y}{2} \ln \frac{\epsilon^y_{ext}}{\epsilon^y_{inj}} \simeq 5\tau_y$$

Using the values in Table 7.8 for the revolution period and the linac repetition rate of 100 Hz ($T_{RepRate}$ = 10 ms) gives:

$$\tau_y \leq \frac{n_{train} T_{RepRate}}{5} \simeq 20 \text{ ms}$$

with $n_{train} \simeq 10$ that corresponds to $T_{per} \geq$1.25 µs.

### 7.5.3 Bunch compressor

The bunch compressor is designed to reduce the bunch length of the beam originating from the damping ring, from an initial range of 4–5 mm to a final length of $\sim$1 mm. Figure 7.21 (right) illustrates the schematic layout of the bunch compressor, which consists of a magnetic chicane formed by four C-shaped bending magnets. Each magnet has a magnetic length of 1.9 m and a maximum magnetic field strength of 1 T. The momentum compaction factor $R_{56}$ is -0.336 m, and the bending angle of each magnet is 11 degrees, resulting in a maximum dispersion of 0.98 m. The distance between dipoles 1 (3) and 2 (4) is 2.9 m, while the separation between dipoles 2 and 3 is 2 m. To induce the necessary energy chirp for compression, two RF structures, providing a maximum accelerating voltage of 122 MV, are employed. Additionally, four RF structures are utilised to partially remove the residual chirp in order to meet the specifications of the high-energy linac (HE-linac). These structures operate with a maximum accelerating voltage of 256 MV. Beam dynamics simulations were performed using the ELEGANT code for a bunch with a maximum charge of 5 nC. Figure7.21 (left) shows the longitudinal phase space after the compression and de-chirping. The results indicate a residual energy spread of 0.7% after de-chirping, a compression factor of 5.75, and a reduction in the bunch length from 4.6 mm to 0.8 mm. The horizontal emittance growth is approximately 20%, while the vertical emittance growth remains below 1%.

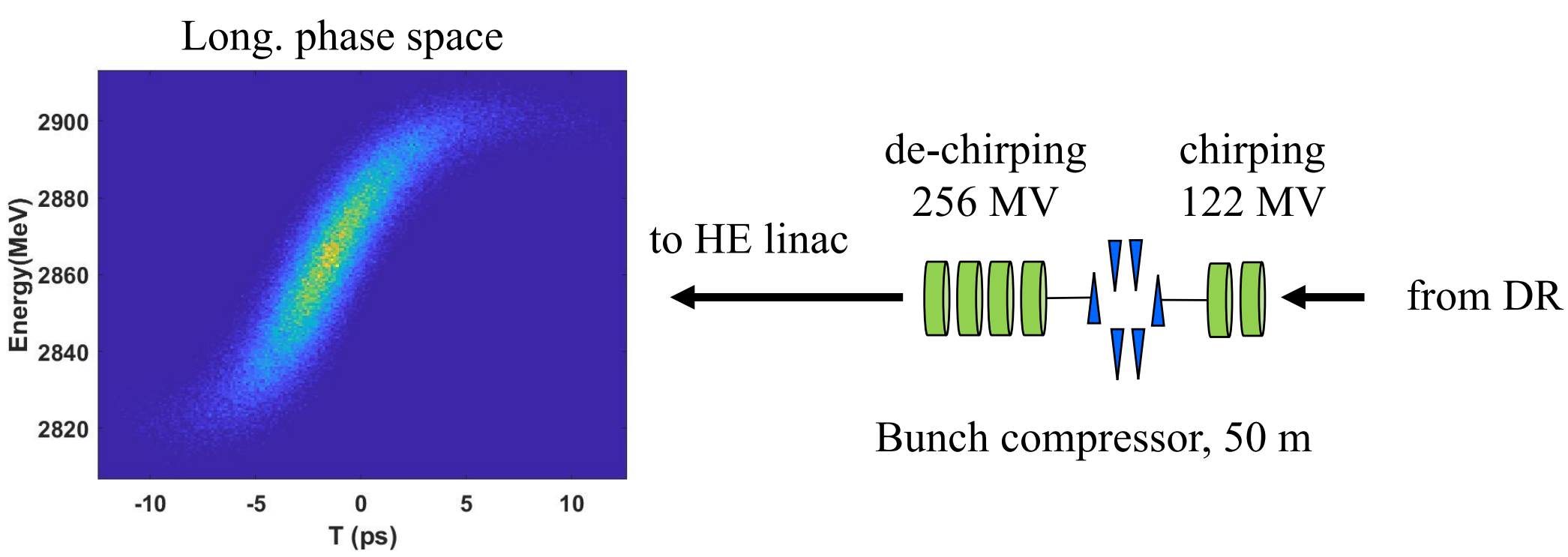

Fig. 7.21: Bunch compressor at 2.86 GeV placed between the DR and the HE-linac. Left: longitudinal phase space after the compression and de-chirping. Right: schematic layout and overall length.

## 7.6 High energy linac and Energy Compressor

### 7.6.1 High energy linac

The high energy (HE) linac accelerates electron and positron beams from the exit of the chicane downstream of the DR at 2.86 GeV energy up to the transfer line towards the booster ring injection at 20 GeV energy.

Analogous to the e-linac, beam dynamics simulations have been done to study both longitudinal and transverse beam dynamics in the HE-linac. The following beam parameters were used in the simulations: bunch length 1 mm, emittance 1 mm·mrad and 10 mm·mrad in the vertical and horizontal plane, respectively at the start of the HE-linac for a 5 nC electron bunch. The lattice is the same as the electron linac: a FODO lattice with 90° phase advance per cell, one quadrupole, and one BPM per RF structure. The RF structures operate at the peak of the RF electric field (on-crest) to maximise accelerating efficiency and minimise beam quality degradation. The beam tracking simulations described in this document were performed using RF-TRACK [355].

To evaluate the robustness of the linac design against static misalignments in terms of emittance growth, the same Gaussian-distributed misalignments assumed for the electron linac and reported in Table 7.4 and the BPM resolution of 10 µm were considered.

To mitigate the effects of misalignments, a combination of one-to-one correction and dispersion-free steering (DFS) was applied sequentially, incorporating the randomly distributed errors. In this case, the focus is on the vertical plane, which is much more critical due to the smaller margin between the starting value and the value accepted by the booster. Finally, a maximum emittance growth of 0.6 mm·mrad in the vertical plane was obtained at the end of the linac for 98% or more of the simulation seeds, giving a final emittance of 1.6 mm·mrad below the maximum 2 mm·mrad required at the booster injection. A similar emittance increase is expected in the horizontal plane, which would give a final emittance smaller than 15 mm·mrad, still satisfying the booster requirement (maximum emittance smaller than 20 mm·mrad) in the horizontal plane. These results were achieved by varying the RF structure aperture and optimising the minimum number of sections (resulting in 8 for the HE-linac) in which the linac must be divided to avoid propagating an error in the correction over a long distance. These simulations determined a minimum aperture of the RF structure corresponding to $a/\lambda = 0.12$.

The dynamic effects were investigated using the same approach as the electron linac. The JA was determined by tracking simulations varying the length of the RF structure, impacting the distance among the quadrupoles, the phase advance per cell, and the RF structure aperture. Figure 7.22 shows, for example, the JA along the HE-linac assuming a constant RF structure length of 3 m (corresponding to a distance among the quadrupoles of 3.75 m) for several RF structure apertures. In particular, for an RF aperture corresponding to $a/\lambda = 0.12$, a final JA of 1.01 was obtained for 3 m RF structure length. The multi-bunch effects were investigated by imposing a kick from the first to the following bunch, analogous to what was done for the e-linac. The results are shown in Fig. 7.23. The RF structure has been designed to provide a long-range wakefield corresponding to a maximum kick equal to 0.11 V/pC/m/mm for the 5 nC bunch charge. This corresponds to a multi-bunch JA of 1.02.

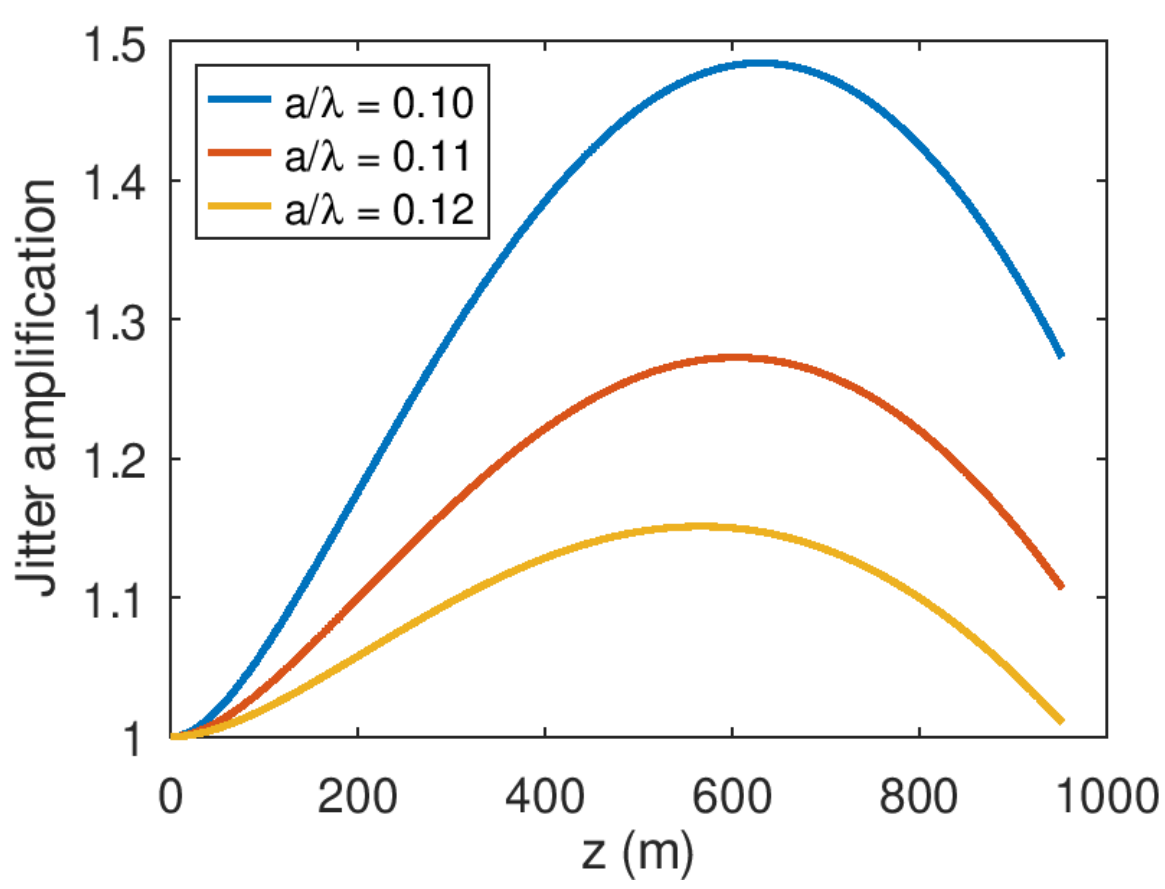

Fig. 7.22: Jitter amplification along the HE-linac for several RF structure apertures. The RF structures are operated on-crest, with a quadrupole spacing of 3.75 m (corresponding to the RF structure length of 3 m).

Single- and multi-bunch JAs give a total JA of 1.03. This allows a very large transverse jitter coming from the upstream sections to be tolerated. The latter is expected to be much smaller than the one at the exit of the gun section, because of the damping in the DR (which is expected to reduce the incoming jitter), and the kickers in current use (which introduce an angle jitter). Table 7.9 summarises the most important lattice and RF structure parameters determined by the beam dynamics simulations. In summary, the RF structure aperture, length, and lattice of the HE-linac fulfil the specifications determined by the transfer line and booster in terms of static and dynamic effects.

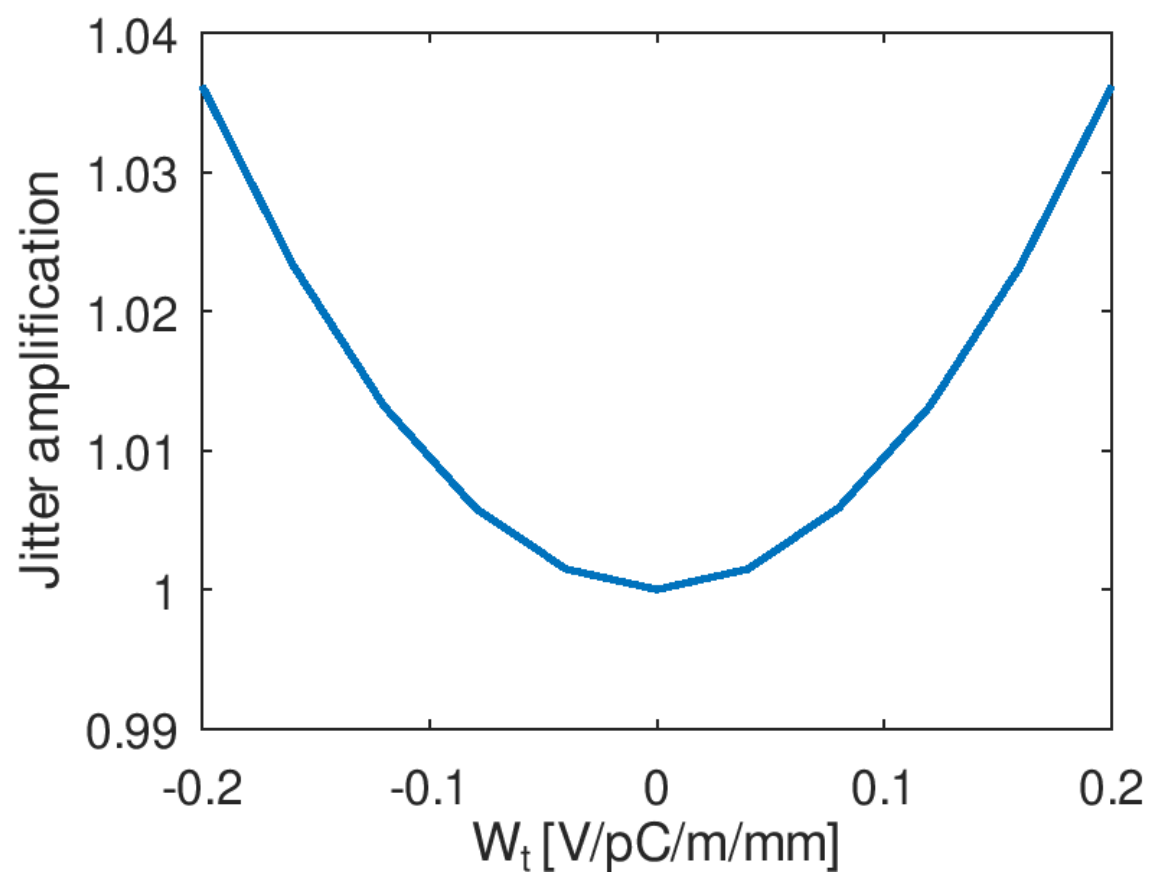

Fig. 7.23: Multi-bunch jitter amplification at the end of the HE-linac.

Table 7.9: Summary of the optimised RF and lattice parameters based on the beam dynamics studies.

| **Parameter** | **Value** |
|---|---|
| Mean RF structure aperture (mm) | 13.9 ($a/\lambda = 0.12$) |
| RF structure length (m) | 3 |
| RF structure operating phase | on-crest |
| Phase advance/cell (degrees) | 90 |
| Number of BPM/RF structure | 1 |
| Number of quadrupoles/RF structure | 1 |
| Distance between the quadrupoles (m) | 3.75 |

### 7.6.2 Energy compressor

The Energy Compressor (EC) [364] at the end of the HE-linac is essential to achieve optimal performance along the HE-linac and, at the same time, the beam parameters required by the transfer line and the booster ring. This kind of system, previously utilised in accelerators such as SuperKEK-B to enhance positron capture efficiency in the damping ring (DR), consists of a magnetic chicane followed by RF structures operated at zero-crossing phase. Using the EC presents several advantages:

- Possibility to operate the RF structures of the HE-linac on-crest. This is advantageous for the mitigation of emittance growth and for accelerating efficiency.
- It minimises the single-bunch energy spread due to the RF curvature and the longitudinal short-range wakefields.
- It reduces the bunch-to-bunch energy variation due to a very large variation from several nC down to nearly 0 nC bunch charge and associated beam loading effects when using a ‘golden’ RF pulse during top-up mode operation.
- It reduces the overall bunch-to-bunch energy jitter.
- It manipulates the beam longitudinal phase space to match the booster requirements.

The EC also has some drawbacks:

– It requires more hardware and more space.
– It converts energy jitter and offset to arrival time jitter and offset, which must be within the tolerance of the downstream sections.

While the transfer line and booster requirements can be met without this system, its absence would lead to reduced performance along the HE-linac.

The EC may be utilised exclusively for single-bunch effects, as well as for both single- and multi-bunch effects. In the first case, the $R_{56}$ of the chicane and the integrated voltage of the RF structures are determined by the target bunch length and energy spread, given the incoming chirp. The simulations also incorporate a residual chirp resulting from compression downstream of the DR, estimated at 0.7% for the 5 nC bunch charge. Table 7.10 summarises the results obtained for a maximum and minimum charge of 5 nC and 5 pC, respectively. In this configuration, increasing the integrated voltage can reduce the single-bunch energy spread by a factor of 3 to 4 while maintaining the bunch length. The booster injection does not currently require this, but it is stressed that the system can, in principle, provide it.

Table 7.10: Parameters of the high energy EC assuming the same machine settings for the maximum 5 nC and a minimum 5 pC single-bunch charge. 'Initial' indicates that the parameter is computed at the entrance of the EC chicane and 'final' at the exit of the downstream RF modules. The voltage is 410 MV.

| **Parameter** | **Q = 5 nC** | **Q = 5 pC** |
|---|---|---|
| Initial rms bunch length (mm) | 1.00 | 1.00 |
| Initial single bunch $\Delta$E/E (%) | 0.56 | 0.20 |
| Final rms bunch length (mm) | 4.00 | 1.43 |
| Final single bunch $\Delta$E/E (%) | 0.10 | 0.12 |

Although bunch-to-bunch beam loading compensation can be managed through a low-level RF (LLRF) system, this system must have the capability to modify the RF phase and/or amplitude for each of the infinite combinations of bunch charges in the timescale of the bunch separation (presently 25 ns) for four bunches accelerated during the same RF pulse. Ongoing studies are therefore investigating whether the EC fulfils this task by leveraging a single constant golden pulse optimised by the LLRF, which remains the same for all the possible charge combinations. In this context, the bunch length was identified as a third degree of freedom to fine-tune the single-bunch properties. The bunch length selected is 0.8 mm rms instead of the 1 mm assumed so far. This modification is not detrimental to the other aspects of the design since a shorter bunch would even be advantageous for static and dynamic effects at the price of a slightly increased single-bunch energy spread of 0.05% at the EC entrance.

Like the previous case, the simulations incorporate the residual chirp resulting from compression downstream of the DR, estimated to be 0.7% for all bunches, and the beam loading effect as a function of the bunch charge calculated assuming the golden pulse discussed in this document, see Fig. 7.30. The optimal $R_{56}$ of the chicane is about 0.55 m and the voltage of the downstream RF structures is equal to 620 MV. This corresponds to a total EC length of less than 90 m including matching sections upstream and downstream of the chicane, the chicane itself, made of normal conducting dipoles, and three RF modules. Table 7.11 summarises the results for the maximum single-bunch charge of 5 nC. The results of the low-charge scenario simulations without modifying any machine parameters optimised for the high-charge case are shown in Table 7.12. The shorter single-bunch length compared to the higher charge mode appears to be acceptable for the booster, as lower charges are less susceptible to instabilities.

In conclusion, the EC enables the operation of the HE-linac at settings optimised for beam dy-namics and RF structures while ensuring that the target parameters for the downstream lines are matched at its output. The energy jitter and offset caused by substantial bunch charge variations ranging from a few nC to nearly zero are converted into time delays and jitter, which are better tolerated by the transfer line and the booster ring compared to energy variations. Table 7.13 summarises the expected range of parameter variation for the final bunch. Some minor effects, like the variation of the bunch length and the residual chirp from the bunch compressor before HE-linac for the 5 pC charge, are presently neglected.

Table 7.11: Parameters of the EC for the maximum 5 nC bunch charge, assuming a target single-bunch energy spread of approximately 0.1%. $B_j$ corresponds to $j^{th}$ bunch along the train. The extra time delay is computed by subtracting an increasing integer number of RF structures periods. The voltage is 620 MV.

| | $B_1$ | $B_2$ | $B_3$ | $B_4$ |
|---|---|---|---|---|
| Initial rms bunch length (ps) | 0.80 | 0.80 | 0.80 | 0.80 |
| Initial single bunch ΔE/E (%) | 0.61 | 0.61 | 0.61 | 0.61 |
| Initial offset centroid ΔE/E from $B_1$ (%) | 0 | -0.31 | -0.59 | -0.93 |
| Final rms bunch length (mm) | 4.06 | 4.07 | 4.09 | 4.10 |
| Final single bunch ΔE/E (%) | 0.11 | 0.11 | 0.10 | 0.09 |
| Final offset centroid ΔE/E from $B_1$ (%) | 0 | -0.01 | -0.03 | -0.05 |
| Final centroid Δt from $B_1$ (ps) | 0 | 5.7 | 11.0 | 17.3 |

Table 7.12: Parameters of the EC for a 5 pC bunch charge, assuming the machine parameters optimised for the 5 nC bunch charge (RF structures phasing and strength of the dipoles set using the 5 nC bunch as a reference). The relative energy spread, energy offset, and extra time delay are computed with respect to the $B_1$ of the 5 nC bunch charge.

| | $B_1$ | $B_2$ | $B_3$ | $B_4$ |
|---|---|---|---|---|
| Initial rms bunch length (ps) | 0.80 | 0.80 | 0.80 | 0.80 |
| Initial single bunch ΔE/E (%) | 0.14 | 0.14 | 0.14 | 0.14 |
| Initial offset centroid ΔE/E from $B_1$ at 5 nC (%) | 1.62 | 1.97 | 2.30 | 2.54 |
| Final rms bunch length (mm) | 1.04 | 1.04 | 1.03 | 1.03 |
| Final single bunch ΔE/E (%) | 0.13 | 0.13 | 0.12 | 0.12 |
| Final offset centroid ΔE/E reference $B_1$ at 5 nC (%) | 0.14 | 0.21 | 0.29 | 0.36 |
| Final centroid Δt from $B_1$ at 5 nC (ps) | -29.0 | -35.2 | -41.0 | -44.9 |

Table 7.13: Spread of the beam parameters at the EC exit for the different charges (assuming the same bunch charge along the four bunches), and considering the maximum single-bunch variation from 5 nC down to 5 pC. The peak-to-peak values are taken for the $B_4$, which are the maxima.

| | Initial | Final |
|---|---|---|
| Single-bunch rms energy spread @ 5 nC (%) | 0.61 | 0.103±0.009 |
| Single-bunch rms bunch length @ 5 nC (ps) | 0.80 | 4.08±0.02 |
| Single-bunch rms energy spread @ 5 pC (%) | 0.14 | 0.125±0.006 |
| Single-bunch rms bunch length @ 5 pC (ps) | 0.80 | 1.035±0.006 |
| Peak-to-peak centroid energy offset variation from 0-5 nC (%) | ±1.74 | ±0.21 |
| Peak-to-peak centroid Δt variation from 0-5 nC Δt (ps) | 0 | ±31 |

Others are overestimated, such as the single-bunch beam loading, which is included both in the golden pulse calculation (solely the fundamental mode) and in the tracking code. Even with these assumptions, the HE-linac and EC design fulfils the requirements of the downstream sections. Further studies are underway to fine-tune the results.

## 7.7 Transfer lines from HE-linac to Booster

### 7.7.1 Transfer line geometry

The transfer line geometry has changed significantly between MTR and this report. The transfer lines at the MTR stage passed close to the SPS to allow synergy with hadron transfer lines for FCC-hh and for the physics programme in the SPS. The design suggested here provides a direct connection from the output of the HE-linac on the surface at the CERN Prévessin site, see P1 in Fig. 7.24 to the two extremities of the collider tunnel straight section of PA, see P8 and P10, respectively. This design is driven mainly by civil engineering constraints related to scheduling and shaft availability. It also features a symmetric line design for electrons and positrons, keeping the beam dynamics impact of synchrotron radiation in the final bending sections equal for both species. The location of the injector complex allows a very efficient connection between HE-linac and the CERN North Area (NA); see beamlines close to P2 in Fig. 7.24. It remains to be discussed if a beam transfer in the reverse direction from the booster back to, e.g., the NA or a direct line from the HE-linac to the SPS is required and to design the transfer lines accordingly.

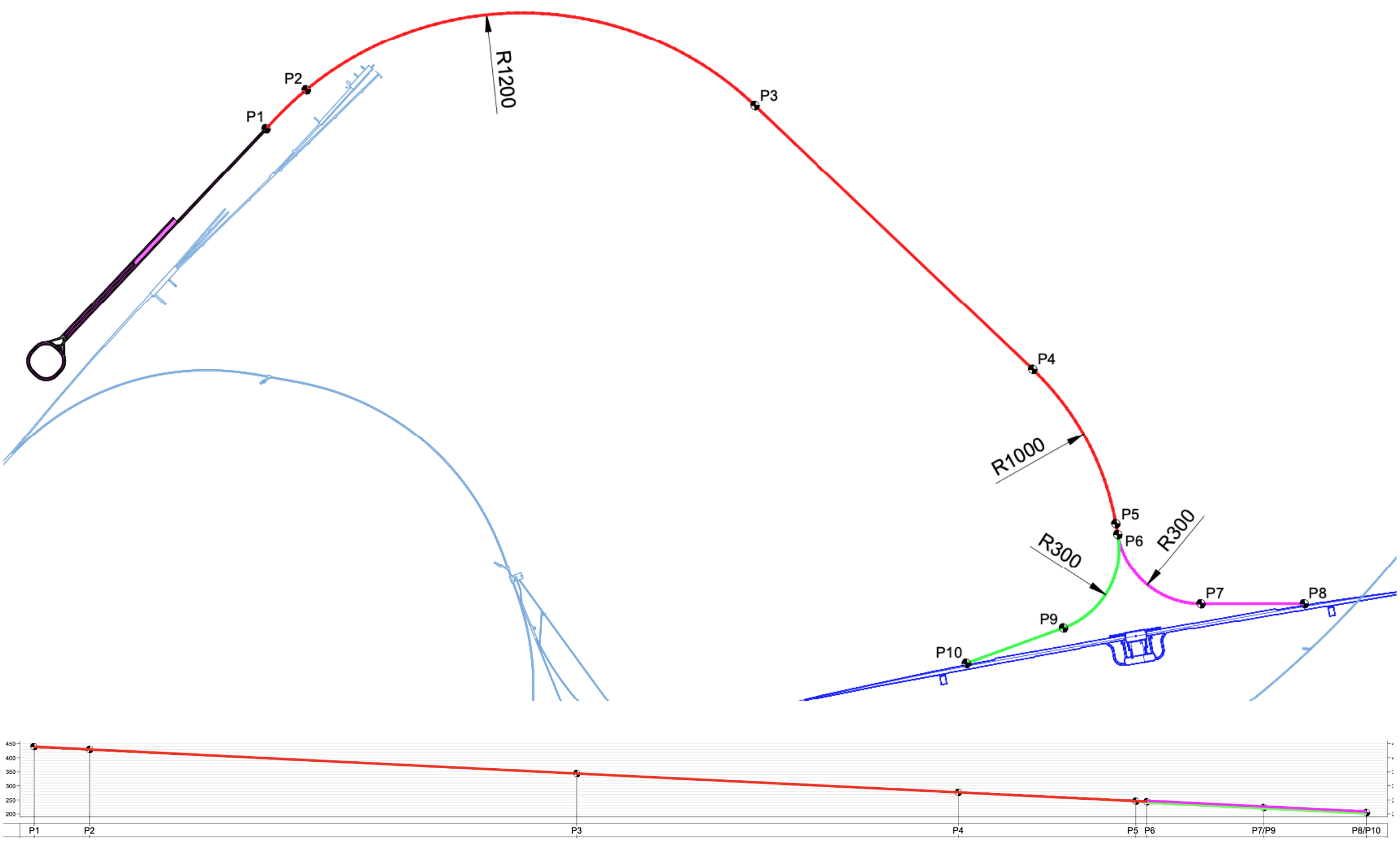

Fig. 7.24: Lepton transfer lines from the injector complex on the surface to the collider tunnel.

### 7.7.2 Cell design and beam dynamics

As shown in Fig. 7.24, the high energy linac to booster transport has been designed with a constant negative vertical slope of $-3^{\circ}$ in the CERN Coordinate System (CSS), combined with horizontal bending in sectors P2-P3, P4-P5, P7-P8 and P9-P10. The slope will be applied and removed at the first and last cells of the line, respectively. To maintain it along the line, the local bending plane must be progressively tilted following the expressions described in Ref. [365]. Figure 7.25 confirms that such a procedure yields the correct beam trajectory. This *twisting reference frame* will introduce a coupling between horizontal and vertical motion. Skew quadrupoles may be needed to control such a coupling, although they have not been included in this initial iteration.

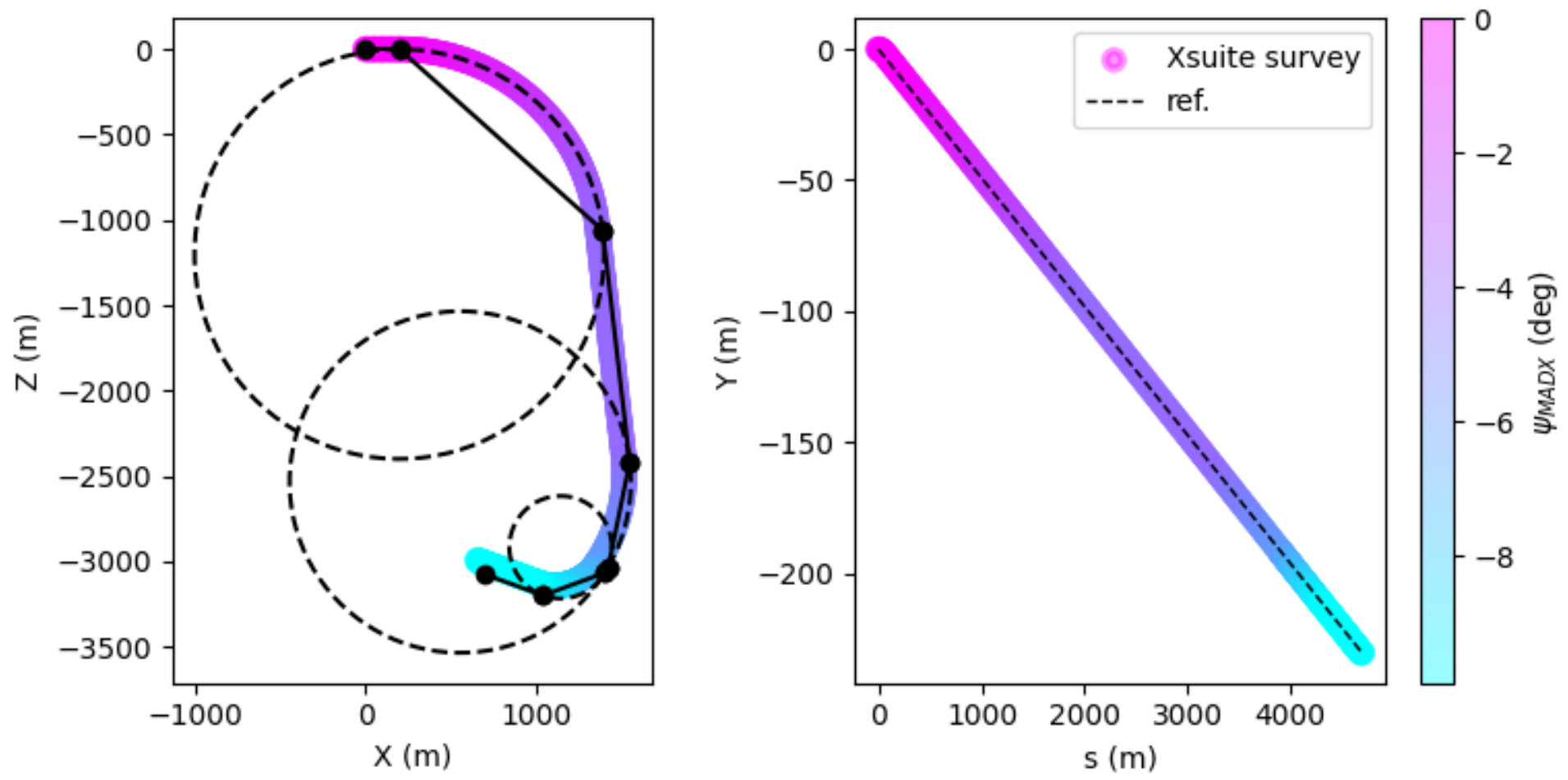

Fig. 7.25: XSUITE survey output and reference trajectory provided by Civil Engineering. The colour bar shows the local roll angle $\psi_{MADX}$, which is necessary to maintain a constant vertical slope in the global frame.

Additionally, the beam transport from the high-energy linac to the booster must be designed to stay within the 'beam quality' budget specified in Table 7.14, which will drive the lattice design choices.

Table 7.14: Summary of Sector Parameters.

| **Sector** | **Bunch len.** (mm) | **RMS dp/p** ($10^{-3}$) | $\epsilon_{x,N}$ (µm) | $\epsilon_{y,N}$ (µm) |
|---|---|---|---|---|
| (...) | | | | |
| HE Linac | 1 | 7.5 | 16 | 1.6 |
| E. Compressor | 4 | 1.0 | 16 | 1.6 |
| Transfer | TBD | TBD | TBD | TBD |
| Booster Inj. | 4 | 1 | 20 | 2 |
| (...) | | | | |

FODO cells have been chosen to design the transfer line due to their simplicity. Later iterations may explore other types of cells, which explicitly aim to minimise the impact of synchrotron radiation (such as theoretical-minimum-emittance cells). The chosen phase advance is 135° as it is close to the optimum for minimising emittance blow-up [366], while remaining an easy number to work with when designing orthogonal correction schemes. Figure 7.26 shows the horizontal emittance blow-up as a function of the cell length and the dipole fill factor for the sectors P2-to-P3[3] and P6-to-P7 (or equivalently P9-to-P10). The required quadrupole gradients, bending fields and apertures are also shown. The aperture required has been calculated with the formula:

$$A_{x,y} = n_\sigma * \sqrt{\beta_{x,y} * \frac{\epsilon_{norm,x,y}}{\gamma}} + D_{x,y} * (2 * \frac{\delta_p}{p} + \delta_{jitter}) + traj * \sqrt{\frac{\beta_{x,y}}{\beta_{max}}} + align \tag{7.1}$$

[3] P4-to-P5 has very similar behaviour to P2-to-P3.

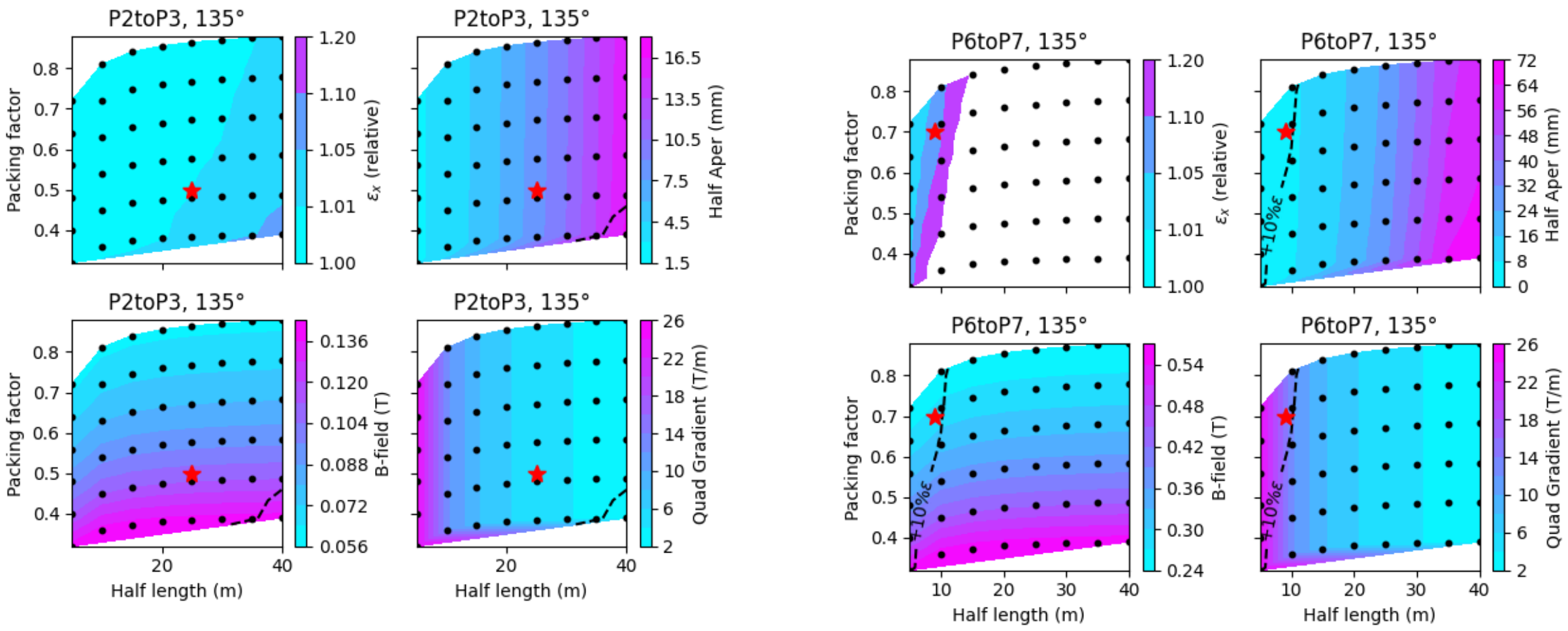

Fig. 7.26: Horizontal emittance blow-up, bending field, quadrupole gradient and aperture as a function of cell half length and dipole fill factor.

where $n_\sigma = 6$, $\epsilon_{norm,x/y} = 20/2\,\mu\text{m}$, the $1\sigma$ momentum spread $\frac{\delta_p}{p}$ is $0.1\%$ ($\pm 2\,\sigma$ taken into account and dispersion contribution conservatively added linearly). The trajectory variation is assumed to be $\pm 2\,\text{mm}$ at the locations of maximum betatron functions, the momentum jitter $\delta_{jitter}$ is assumed to be a maximum of $\pm 0.3\,\%$, and there are $\pm 2\,\text{mm}$ taken into account for alignment errors.

Based on the parametric scans above, two cells have been chosen to build the line: a long FODO for the shared part of the transport (P1-to-P6) and a short FODO for the separate parts of the transport (P6-to-P8, P6-to-P10). The latter is required due to the tight bending radius of $R = 300\,\text{m}$, which leads to significant synchrotron radiation. Table 7.15 lists the parameters for each FODO cell. The 135° phase advance is close to the optimum for minimising emittance blow-up [366], while remaining an easy number to work with when designing orthogonal correction schemes.

Table 7.15: FODO cell parameters.

| **Attribute** | **Long FODO (P1-to-P6)** | **Short FODO (P6-to-P8/P10)** |
|---|---|---|
| Cell length (m) | 50 | 18 |
| Dipole fill factor | 0.5 | 0.67 |
| Dipole field (mT) | 130 | 330 |
| Dipole length (m) | 6 | 6 |
| Dipoles per half cell | 2 | 1 |
| Quadrupole length (m) | 1 | 1 |
| Quadrupole gradient (T/m) | 5 | 14 |
| Cell phase advance (deg) | 135 | 135 |

The schematics and optics functions for both cell configurations are shown in Fig. 7.27. The number of magnets required has been documented in Ref. [367]. For the beginning and end of the line, the bending magnets will bend vertically to introduce (and then remove) the -3° slope present along the entire line descending from the HE-linac to the collider tunnel. In those cells, the dispersion will be vertical but similar in amplitude and shape to the one shown in Fig. 7.27. Under such a configuration, tracking simulations show that both the $dp/p$ blow-up and the bunch length increase remain under 5%. However, further investigations might be needed when including matching sections and errors, which have not been considered in this report.

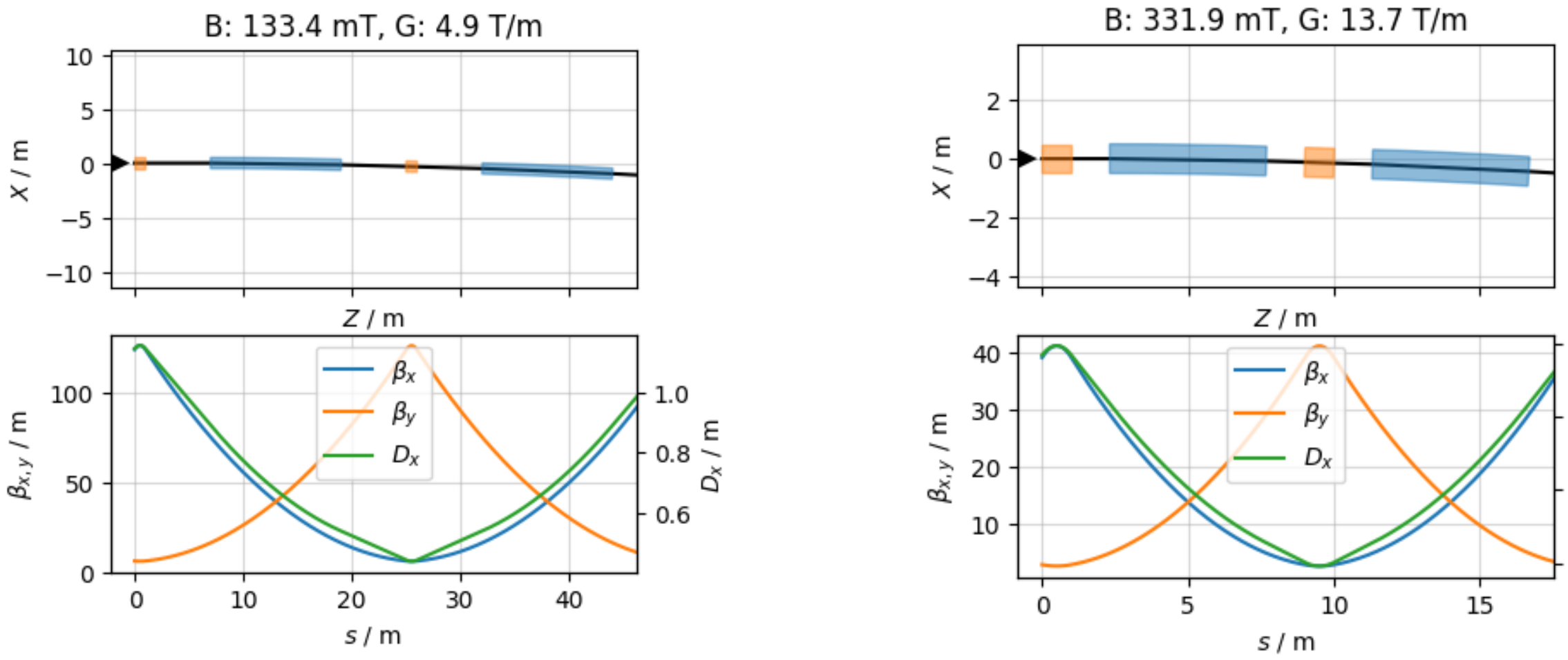

Fig. 7.27: Magnet layout and optics for long and short FODOs (quads in orange and dipoles in blue). Only horizontal dispersion is shown, but a similar vertical dispersion is expected for the cells where the $-3\,\mathrm{deg}$. the vertical slope is applied.

### 7.7.3 Magnet technology and technical infrastructure needs

The magnet system of the transfer lines is specified in Table 7.16. The constant transfer energy of 2 0 GeV opens the door to permanent magnet technology, thus both technologies, electro- and permanent magnets were studied and documented in Ref. [368]. It can be concluded that both magnet technologies are technically feasible. The installation and running cost for the permanent magnet option is well below the electromagnet option. The technical infrastructure needs are also much reduced, however the transfer lines as a whole, contribute less than 10% of the injector complex power requirements. The total power requirements of the transfer lines amount to about 2 MW, compared to about 25 MW for the full injector complex. The permanent magnet option would require about a factor 10 less power for e.g., corrector magnets and ventilation. Operational flexibility is the main aspect to consider when choosing between magnet technologies. If the are beam stability issues in the booster, a transfer energy increase can be beneficial. On the other hand, if the booster could accept a lower energy beam at injection, reducing the transfer energy reduces the power consumption of the whole injector complex which is driven by the HE-linac. The main field of permanent magnets can be tuned by about $\pm 20\%$ by increasing or decreasing a leakage field via mechanical shunts. The shunt modification and subsequent magnetic field measurement require several weeks. Another argument for operational flexibility is the duty cycle of the injector complex, which varies between 5 and 75% depending on the operation mode. In particular, in the low-duty modes, the injector beam can be used for experiments beyond FCC physics and shot-to-shot variability of beam parameters, including the transfer energy, open the door to a very diverse physics programme as discussed in Ref. [369]. At this stage, electromagnets have been chosen as the baseline technology due to the higher operational flexibility for the FCC and any science programme beyond the FCC.

## 7.8 RF system for linacs

### 7.8.1 RF accelerating structure

The RF design of the accelerating structures FCC injector complex is optimised to deliver efficient beam acceleration in all 3 linacs: the electron (e-)linac , the positron (p-)linac and the high-energy (HE)-linac.

Table 7.16: Summary of magnet specifications, assuming polarity switching and the FODO parameters from Table 7.15. For ease of manufacturing and costing, the 6 m dipoles have been split into 1 m segments. The number of dipoles quoted refers to the number of 1 m segments.

| | **Unit** | **Quadrupoles** | **Dipoles** | **Correctors** |
|---|---|---|---|---|
| Total number | | 338 | 286x6=1716 | 224 |
| # magnets in common line | | 162 | 192x6=1152 | 108 |
| Length | m | 1 | 1 | tbd |
| Aperture (diameter) | mm | 30 | 30 | 30 |
| Gradient | T/m | 5-15 | - | - |
| Field | mT | - | 150-400 | tbd |
| Deflection | μrad | | | O(10) |
| Field homogeneity | | O($10^{-3}$) | O($10^{-3}$) | tbd |
| Polarity switching time | s | O(1) | O(1) | |

Table 7.17: Accelerating structure parameters for e-, p- and HE-linacs.

| | e-linac | p-linac | HE-linac | Unit |
|---|---|---|---|---|
| Frequency | 2.8 | 2 | 2.8 | GHz |
| Length | 3 | 3 | 3 | m |
| Average aperture | $0.15\lambda$ | $0.2\lambda$ | $0.12\lambda$ | |
| Cell aperture: first/last | 17.13/14.99 | 30/30 | 14.85/10.85 | mm |
| Iris thickness: first/last | 10.4/13.7 | 14.3/20.0 | 2.84/4.04 | mm |
| $V_g/c$: first/last | 3.14/1.38 | 2.58/1.92 | 3.92/1.25 | % |
| $r/Q$: first/last | 3.28/3.67 | 1.49/1.52 | 3.63/4.38 | kΩ/m |
| $Q$: first/last | 14599/13668 | 20977/19102 | 16571/16039 | |
| Filling time | 486 | 447 | 460 | ns |
| SLED coupling | 15 | 17 | 15 | |
| $R_{\text{sh,eff}}$ (4 bunches) | 81.69 | 36 | 95.65 | MΩ/m |
| Repetition rate | 100 | 100 | 100 | Hz |
| Klystron power per structure | 14.2 | 14.2 | 14.2 | MW |
| Average power per structure | 3.76 | 3.68 | 3.72 | kW |
| Bunch charge | 5 | 15 | 5 | nC |
| Average loaded gradient (4 bunches) | 19.50 | 13.31 | 21.06 | MV/m |
| $E_{s,\text{max}}$ (instant.) | 77 | 55 | 73 | MV/m |
| $S_{c,\text{max}}$ (instant.) | 453 | 298 | 501 | mW/mm$^2$ |

While each linac has some unique operational parameters, they share several common design features: all have a RF structure of total length of 3 m, are configured to accelerate 4 bunches with a separation of 25 ns, and operate with a repetition rate of 100 Hz. The HE- and e-linacs are designed to operate at a frequency of 2.8 GHz, with a bunch charge of 5 nC, while the p-linac operates at a lower frequency of 2 GHz to accommodate its higher bunch charge of 15 nC. This difference reflects the tailored optimisation of each linac to the specific beam dynamics requirements described above. The RF structures have tapered geometries, which effectively balance beam transport with short- and long-range wakefield suppression requirements, $W_t < 0.1$ V/pC/mm/m and a value specific to each linac of $< a > /\lambda$. The detailed parameters and design specifications for all three linacs are summarised in Table 7.17. Similar design principles are applied to all 3 linacs, providing a uniform and scalable approach across all structures. The RF design of the HE-linac, is described as an example, in detail below.

The parametric sweep of iris aperture and thickness was utilised to identify structures that satisfy the long-range transverse wakefield constraint $W_t < 0.1$ V/pC/mm/m for a given $< a > /\lambda = 0.12$ related to the short-range wakefields, ensuring stable beam propagation. The long-range transverse wakefields were calculated using a lookup table, which employs frequency-domain parameters of the 20 lowest higher-order modes (HOMs).

To achieve more realistic and accurate wakefield predictions and to benchmark the lookup table calculations, the ECHO2D time-domain solver was used for the structure of the final choice. This approach is more accurate, accounting for an infinite number of HOMs and the coupling between cells. The envelope-of-the-envelope approach was applied to the wakefield data from both ECHO2D and the lookup table to rigorously assess worst-case scenarios and long-range wakefield effects. The results are shown in Fig. 7.28.

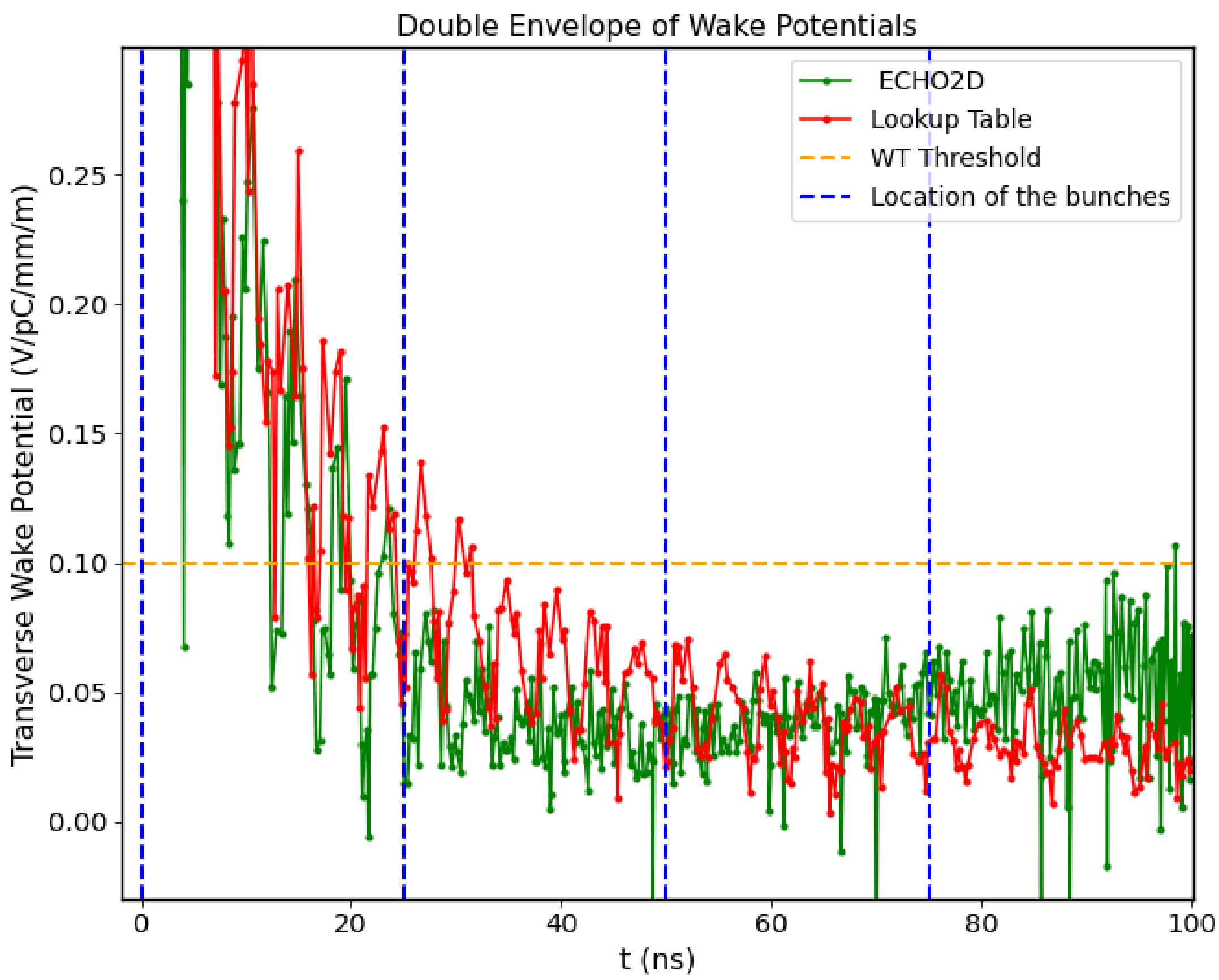

Fig. 7.28: : Envelope-of-the-envelope of transverse wakefield potentials for the HE-linac structure, calculated over time using both the lookup table and ECHO2D methods

In the design of travelling wave accelerating structures, minimising bunch-to-bunch energy spread is a critical objective for ensuring high beam quality and stable accelerator performance. In operation during the collider filling with nominal bunch charge, the input RF power pulse shape is optimised to minimise bunch-to-bunch energy spread, taking into account the beam loading effect. As shown in the red trace in Fig. 7.29, the input RF power is modulated using a step-like amplitude modulation. This approach ensures a constant average loaded gradient for all four bunches. The gradients for unloaded and loaded conditions are represented by the blue and orange traces, respectively. The four red dots indicate the temporal positions of the bunches. In Fig. 7.30(a), the unloaded and loaded gradients are shown in more detail for comparison near the position of the 4 bunches, showing a maximum energy difference of 2.4% between unloaded and loaded gradients on the fourth bunch, highlighting the beam loading effect in HE-linac. However, in top-up operation, where bunch charges vary from 0 to 100% among the four bunches, a single optimal input RF pulse must accommodate both extremes. The ‘golden pulse’ is introduced to address this by averaging the optimised RF pulses for unloaded and loaded voltages. While it does not individually minimise energy variation to ∼0-level, it provides a balanced compromise, reducing energy spread across all possible charge variations. Figure 7.30(b) demonstrates the golden pulse in top-up operation, reducing energy spread to +1.1% and -1.1% for unloaded (4 bunches at 0 bunch charge, small red dots) and loaded cases (4 bunches at nominal intensity of 5 nC, large red dots), respectively. This result underscores the golden pulse's ability to balance energy spread across bunches, ensuring stable performance even under varying charge conditions in top-up mode.

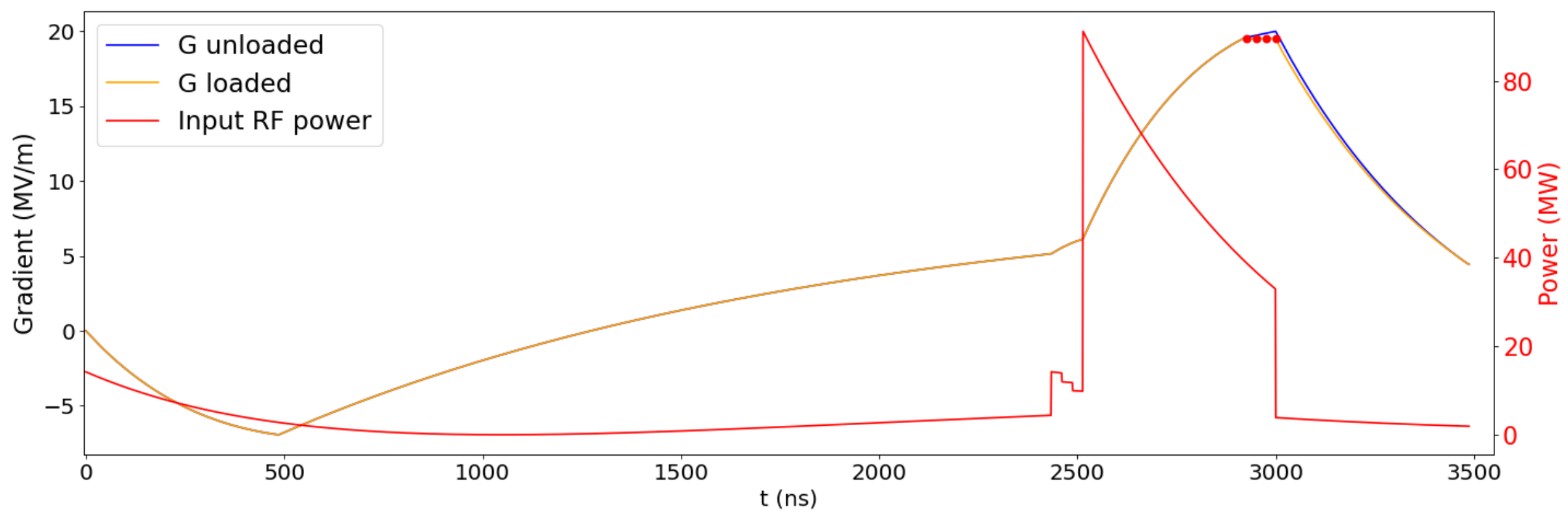

Fig. 7.29: Input RF power and average gradients versus time.

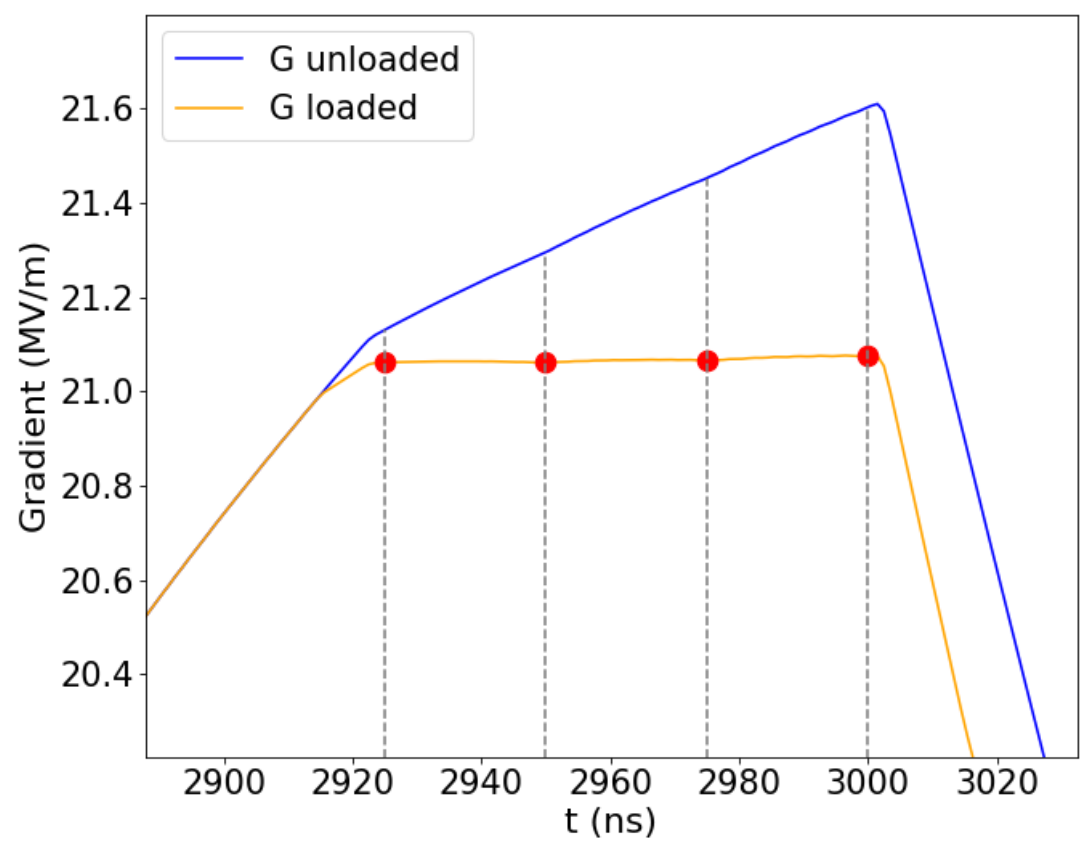

(a) Nominal charge condition.

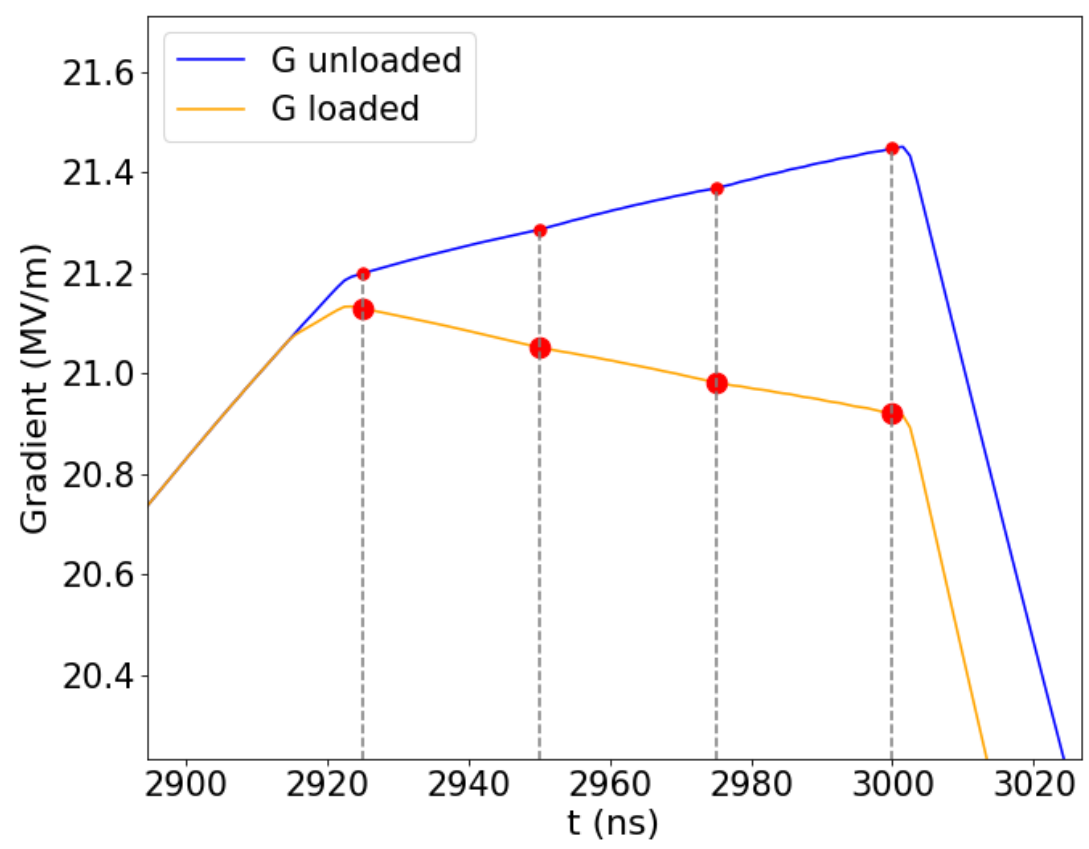

(b) Top-up operation.

Fig. 7.30: Bunch-to-bunch energy minimisation for the nominal bunch charge (a) and top-up operation (b).

### 7.8.2 RF module and RF power

Each RF module of the electron and high-energy linacs consists of one high voltage modulator (HV), one klystron operating at 2806 MHz, one pulse compressor (Barrel-Open Cavity or SLED type) and four accelerating structures with one quadrupole, one corrector magnet and one BPM between each structure. Its total length is 15 m. The choice of an RF module consisting of four structures results from cost and power consumption optimisation. Figure 7.31 shows the schematic layout of such an RF module.

The positron linac has two types of RF modules. The first type equips the end part of the capture linac and the S1 linac. It is about 13 m long, since instead of 4 quadrupoles, it has 40 solenoids, i.e., 10 solenoids per structure for beam focusing. A schematic layout of this type of module is shown in

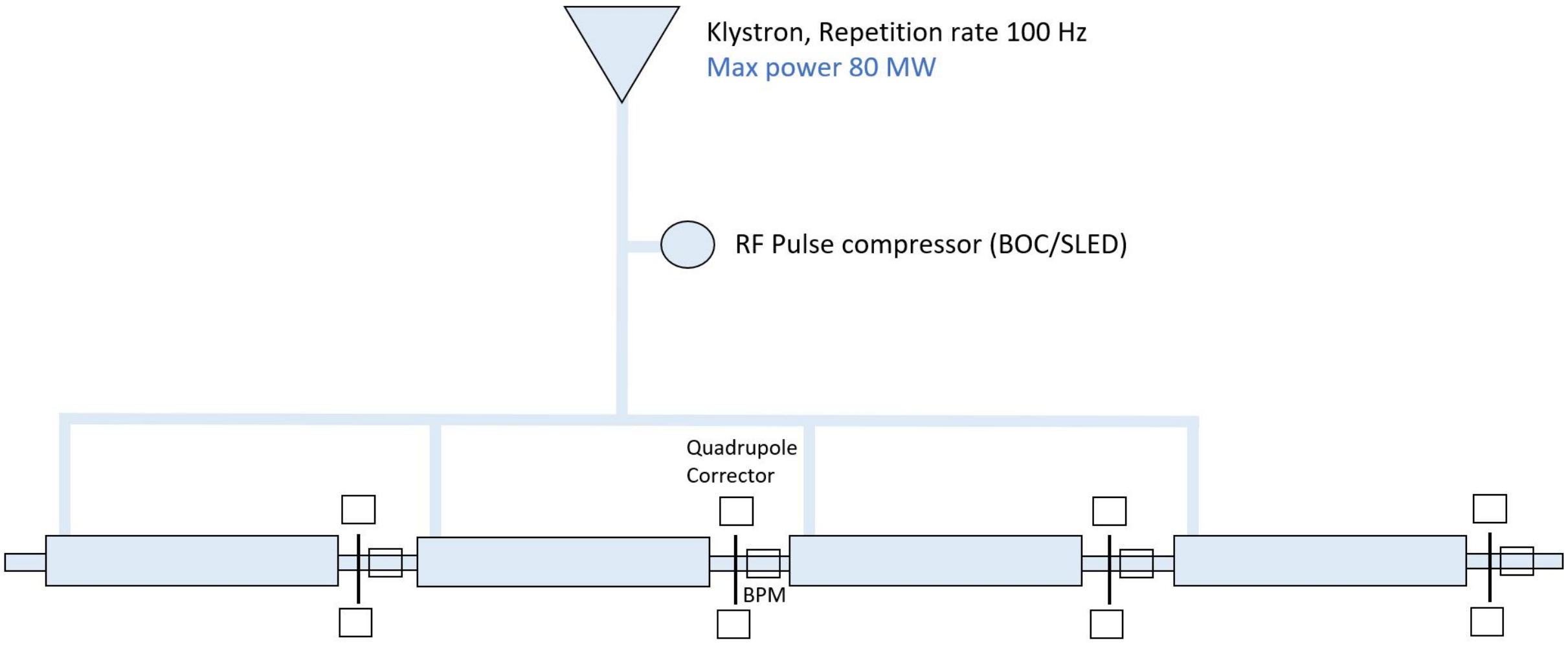

Fig. 7.31: Schematic layout of an RF module for the electron, S2 positron and high-energy linacs.

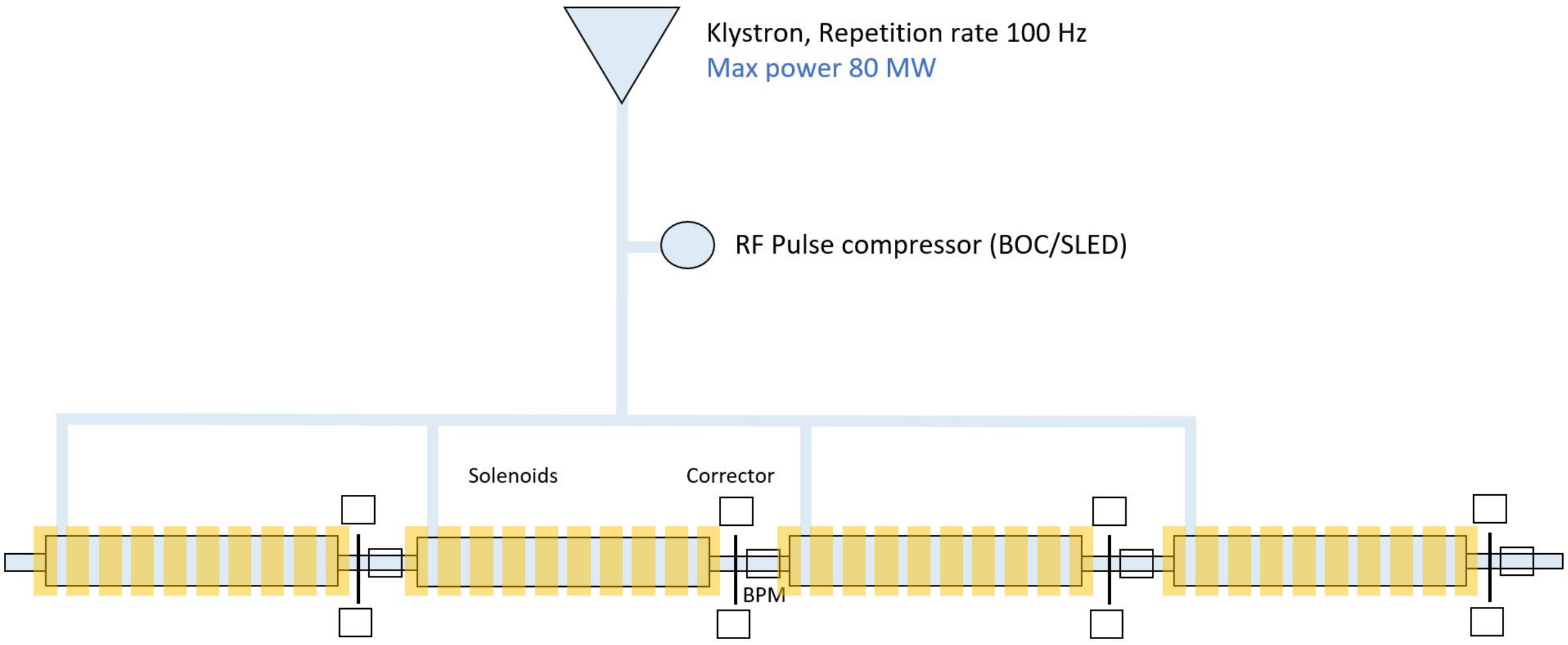

Fig. 7.32: Schematic layout of the RF module for the positron capture and S1 linacs.

Fig. 7.32. The second type of RF module in the p-linac is similar to the electron and high-energy linac module (see Fig. 7.31). The klystron operates at 2004 MHz in the p-linac.

For the electron and the high-energy linacs, the peak power specification of each klystron is 80 MW. Its repetition rate and RF pulse length are 100 Hz and 3 µs, respectively. For the positron linac, the power specification of the klystron and its repetition rate are also 80 MW and 100 Hz, respectively, but its RF pulse length is 5 µs. Although no klystron exists at the specific frequencies of 2806 MHz and 2004 MHz, many klystrons have demonstrated reliable performance at 2856 MHz with such a power specification and RF pulse length. The development of a conventional klystron at 2806 MHz with a 100 Hz repetition rate is straightforward by retuning its cavities and redesigning its collector. As for the design of a 2004 MHz klystron with such specifications, it would be a completely new development but presents no technical challenges. Several companies have indeed been contacted to assess the feasibility of such conventional klystrons and confirmed that such specifications are achievable. The RF efficiency of such klystrons would be 42%. Their parameters are summarised in Table 7.18.

Both types of klystron operate at 80% of their peak power specification, i.e., at 64 MW. For the electron and high-energy linacs, it is assumed that the RF waveguide system comprises S-band WR284 waveguides and is 25 m long. Taking into account the waveguide losses, the klystron power per structure is then 14.2 MW before pulse compression. For the positron linac, the RF waveguide system comprises the less lossy L-band WR510 waveguides. With a 25 m long waveguide system, the klystron power per structure is 15.4 MW before pulse compression. The specifications of each RF module for all linacs are summarised in Table 7.19.

Table 7.18: Klystron parameters.

| **Linac** | **Freq.** [MHz] | **Peak power specification** [MW] | **Rep. rate** [Hz] | **RF pulse length** [μs] | **Duty factor** [$10^{-3}$] | **Average power** [kW] | **Number required** |
|---|---|---|---|---|---|---|---|
| e-Linac | 2806 | 80 | 100 | 3 | 0.3 | 24 | 15 |
| p-Linac | 2004 | 80 | 100 | 5 | 0.5 | 40 | 21 |
| HE-Linac | 2806 | 80 | 100 | 3 | 0.3 | 24 | 72 |

Table 7.19: RF module summary table for the injector linacs.

| | e-Linac | p-Linac | HE-Linac | Unit |
|---|---|---|---|---|
| RF frequency | 2.8 | 2.0 | 2.8 | GHz |
| Repetition rate | 100 | 100 | 100 | Hz |
| Modulator max. peak power | 190 | 190 | 190 | MW |
| Modulator max. average power | 114 | 152 | 114 | kW |
| Klystron max. RF power | 80 | 80 | 80 | MW |
| Klystron RF pulse length | 3 | 5 | 3 | μs |
| Structures per klystron | 4 | 4 | 4 | |
| Klystron power per structure | 14.2 | 15.4 | 14.2 | MW |
| Average loaded gradient | 19.5 | 13.3 | 21.1 | MV/m |
| Number of rf modules | 1+14 | 1+6+14 | 72 | |
| Module Length | 15 | 13, 14.8 | 15 | m |
| Length of all modules | 215 | 304 | 1080 | m |

In recent years, research and development have been conducted on pulsed high-efficiency klystrons, and it is conceivable that more of these klystrons will be available in the near future. A comparison of the plug power consumption for all linacs assuming 42% and 70% klystron efficiencies is shown in Table 7.20. The advantage of operating with high-efficiency klystrons is remarkable for the electron and high-energy linacs but less so for the positron linac since the power consumption of the structure solenoids in the capture and S1 linacs is a substantial fraction of the positron linac total power consumption.

Table 7.20: Comparison of the plug power consumption for the injector linacs assuming 42 % or 70 % klystron efficiencies.

| Linac | Plug power [MW] | |
|---|---|---|
| Klystron efficiency [%] | 42 | 70 |
| e-Linac | 2 | 1.2 |
| p-Linac | 8 | 6.5 |
| HE-Linac | 9.5 | 6 |

## 7.9 Availability

This section details results from the enhanced Monte Carlo simulation environment for FCC-ee availability described in Section 2.4, focusing on systems specific to the injector complex.

### 7.9.1 Contributing Systems

For availability modelling, the injector chain is divided into two top-level accelerators. The statistical failure rate and blocking duration from subsystems in each accelerator are taken from two representative machines with similar performance characteristics. These are simulated in the injector complex of the FCC-ee to gauge the effect of equivalent availability on global physics performance.

1. **Lower Energy Injectors:** The electron and positron source, linac and damping ring up to an energy 2.86 GeV are modelled from fault data from the SuperKEK-B $e^-/e^+$ source, injector linac and damping ring between 1998-2023, detailed in Ref. [370].
2. **High Energy (HE) Linac:** with an energy up to 20 GeV is modelled using fault data from the Linac Coherent Light Source (LCLS) at SLAC [371].

Subsystems within each accelerator have been re-categorised for illustration to preserve naming conventions in the CERN complex. These include the relevant supporting technical infrastructure for each accelerator. Only faults leading to downtime in each accelerator were considered, thereby assuming a similar degree of redundancy in the FCC-ee as exists in practice for each representative subsystem.

### 7.9.2 Inherent resilience to shorter fault types

In the event of an outage in the booster and injector complex, stable beams can be maintained in the main collider rings for a lifetime of 10-15 minutes, depending on energy mode (see Table 2.2). If top-up injection can be restored in this time, normal physics can resume. This significantly reduces the FCC-ee's sensitivity to short-duration fault types in the injector complex and leads to an inherent advantage in availability. The contribution for lost luminosity of subsystems in the injector complex is therefore biased towards those with longer-duration fault types.

### 7.9.3 Results

The overall contribution to downtime of each accelerator in the injector complex is compared with the booster systems in Fig. 5.14. The lost luminosity contributions in the Z mode from the low energy injectors and HE linac are significant, second only to the booster RF system. The availability of the injector complex will be a challenge for overall FCC-ee performance.

#### *Lower Energy Injectors*

The contribution to unavailability and lost luminosity from each subsystem in the low energy injectors is shown in Fig. 7.33. These only contribute to downtime, as there is high redundancy in the relevant systems, meaning that faults generally block the beam for only short durations. The majority of faults are due to RF breakdowns, which have a large contribution to lost luminosity in Z and W modes due to the lengthy turnaround time needed to recover from dumps.

#### *High Energy (HE) Linac*

The contribution to unavailability and lost luminosity from systems in the HE linac are shown in Figure 7.34. RF breakdowns are again the most problematic, representing contributions from S-band klystrons, high-power sub-boosters, waveguides and modulators. The control system is also a high contributor, composed of micros and CAMAC crates as well as the MCC/VMS computing and timing systems. The access system is represented by the personnel protection system (PPS). Contribution from the gun laser system is also significant.

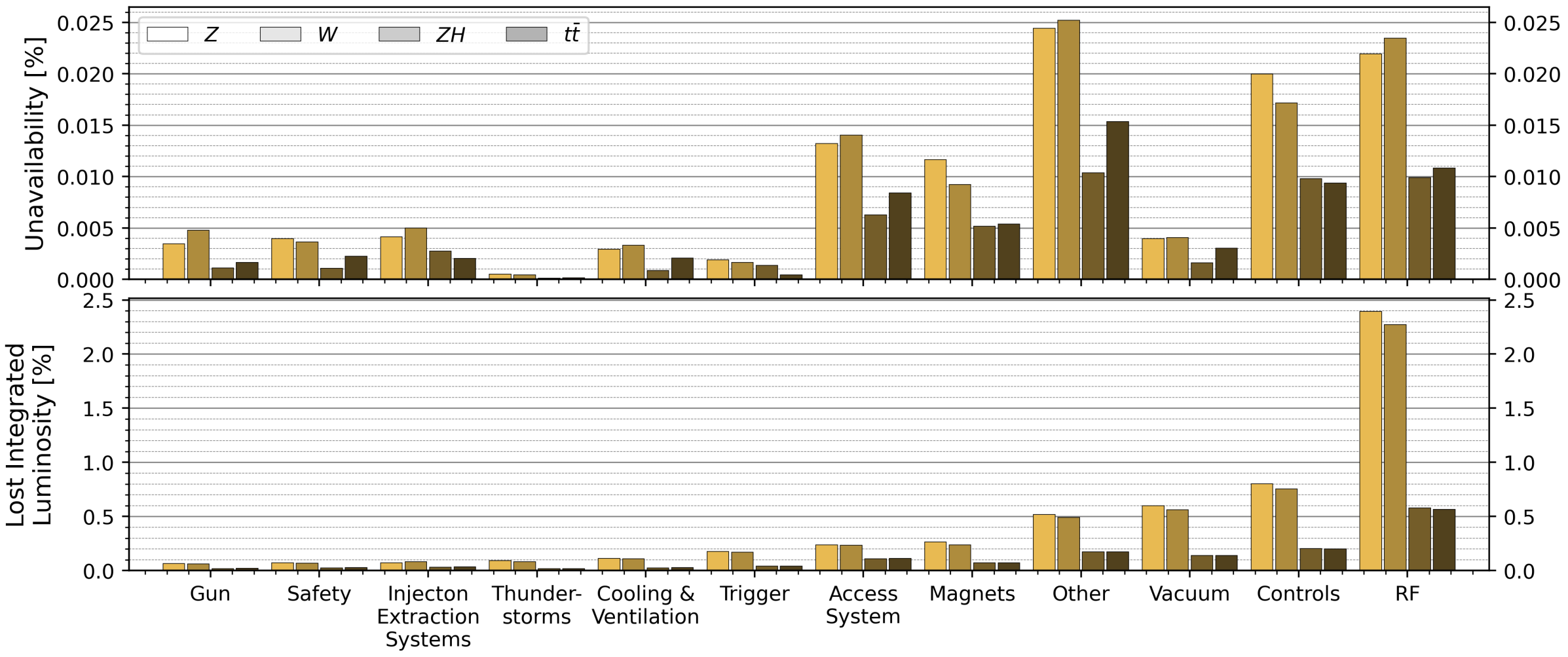

Fig. 7.33: FCC-ee unavailability and lost luminosity contribution of the low energy injectors (up to 6 GeV) determined by applying fault statistics from equivalent equipment at SuperKEK-B [370]. Systems are ordered according to Z mode lost luminosity contribution.

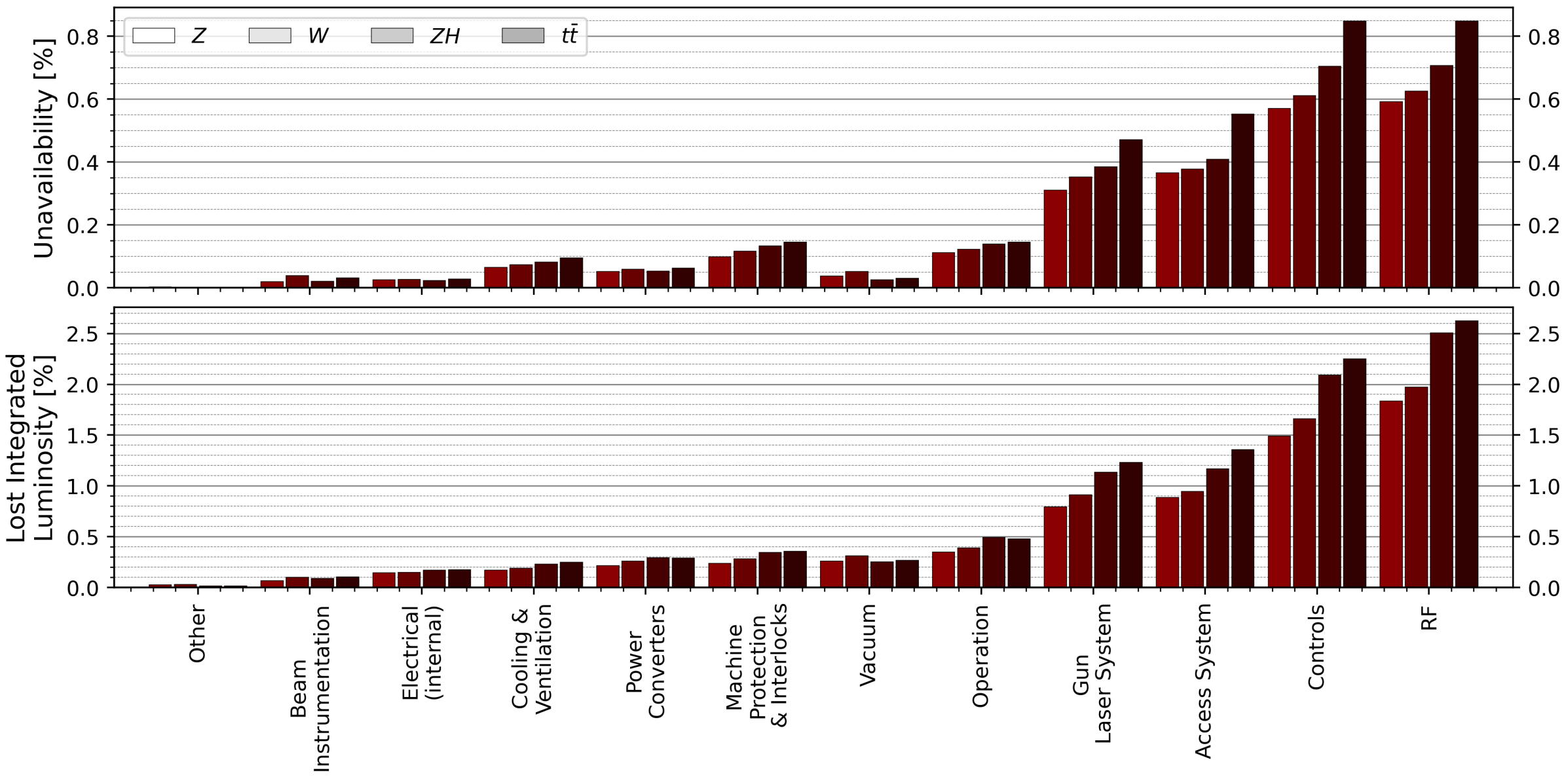

Fig. 7.34: FCC-ee unavailability and lost luminosity contribution of the HE Linac determined by applying fault statistics from the Linac Coherent Light Source (LCLS) at SLAC [371]. Systems are ordered according to Z mode lost luminosity contribution.

### 7.9.4 R&D Opportunities

Opportunities exist for exploiting the injector complex's natural resilience to shorter fault types. If a failed component can be brought back online before the beams in the main collider expire, a lengthy turnaround time may be avoided. This mirrors the opportunities already discussed for the booster in Section 5.3.4.

The short beam lifetime is especially significant for Z mode operation, where the charge imbalance between electron and positron beams in the collider must be less than 5%. Assuming a lifetime of 10-15 minutes, the injection of electrons and positrons must alternate for top-up every 50 seconds. This places significant design constraints on the injector hardware and has profound implications for reliability and availability, as even short interruptions could impact the collider's performance. This requirement limits the breakdown (BD) rate in the injector's accelerating structures. The time required to bring an accelerating structure back into operation after a BD is on the order of a minute. This poses a problem with the injector's availability. An analysis of BDs in the SwissFEL linac is conducted in Ref. [372]. During regular user operation, the RF modules failure rate began to reduce thanks to RF conditioning incidental to nominal operation. This incidental conditioning followed a power law trend that saw the BD rate decrease by over three orders of magnitude in the first three years of user operation, from $10^{-6}$ to less than $10^{-9}$ BD per pulse per metre or about two BD per day in the entire linac. Nonetheless, RF BDs continue to account for about half of all broken interlocks in SwissFEL.

Given the number of RF structures planned for the FCC injector and scaling, the results obtained for SwissFEL, translate into a reduction of the BD rate from about 850 events per day to about eight events per day after the first four years of operation. The proposed solution, which is to be further investigated, to mitigate these events is to use some additional RF modules during these BD events. The low probability of having more than two events simultaneously led to the adoption of one additional RF module for the positron and electron linacs and two additional RF modules in the HE linac. Thanks to this approach, the injector's availability due to the BDs could be greater than 99% after a few years of operation.

## 7.10 Civil engineering

### 7.10.1 Injector complex

The preferred location for the FCC-ee injector complex is on the existing CERN Prévessin site. Several options for the location of the high-energy linac were explored, with the preferred choice being near the northwest edge of the Prévessin site. A more detailed placement study considering also environmental aspects will be performed in the next phase before confirming the location. The envisaged layout of the facility is shown in Fig. 7.35.

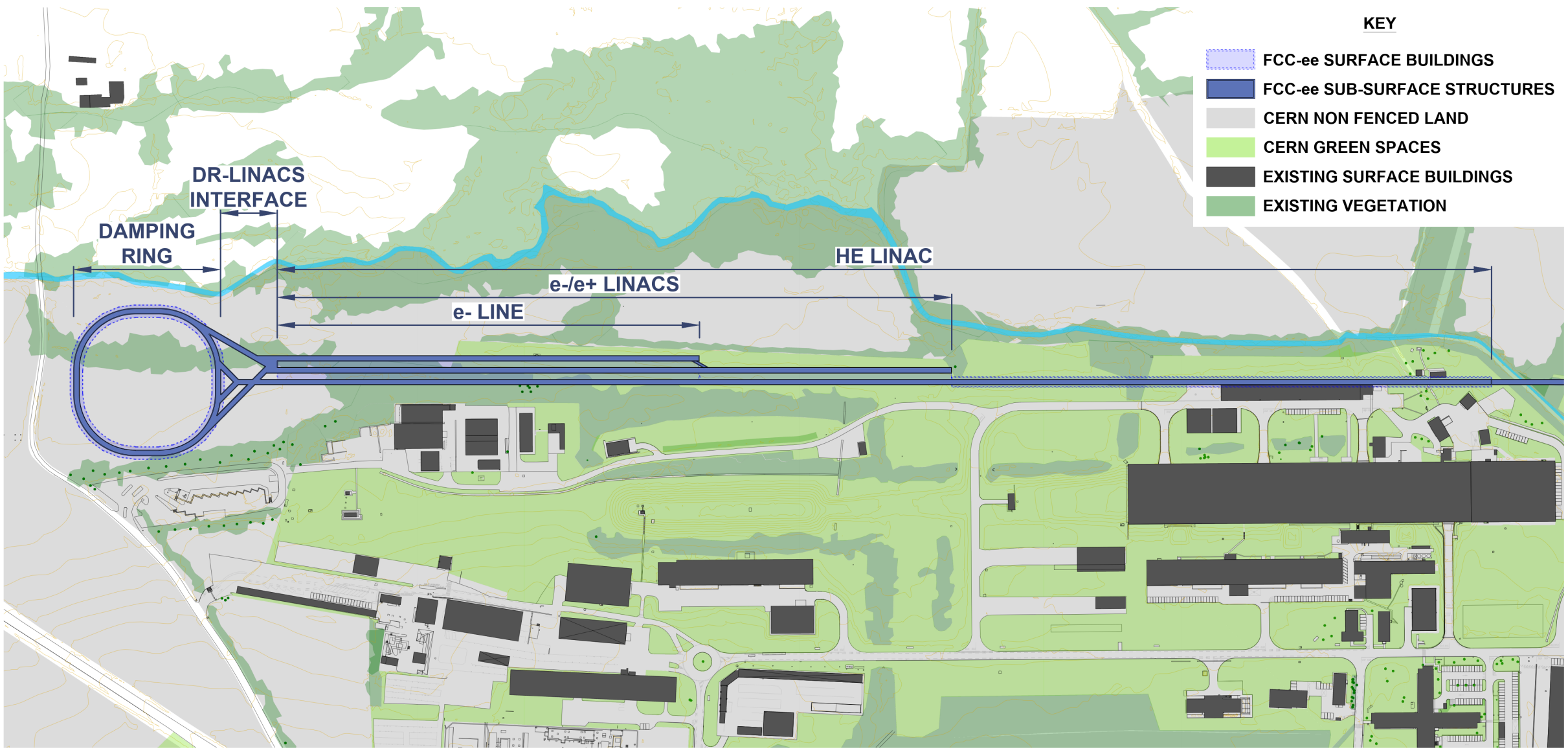

Fig. 7.35: Plan view of high energy linac civil engineering.

The civil engineering for the high-energy linac will consist of four main elements, namely: the high-energy linac tunnel, the associated $e^+$ and $e^-$ tunnels, the damping ring and the transfer tunnels/structures to connect the damping ring to the rest of the facility. All these structures will consist of reinforced concrete buried structures extending over a length of about 1.2 km. For shielding purposes, these structures will be built in a series of trenches up to 15 m below the ground level. These so-called ‘cut-and-cover’ tunnels will be backfilled using the excavated material from the trench formation, and then a surface hall will be constructed above the tunnels. A series of 6 m long vertical mini-shafts of about 0.8 m diameter will connect the tunnels to the surface hall for connecting the klystrons and to allow for the passage of services and cables. A schematic section of this arrangement is shown in Fig. 7.36.

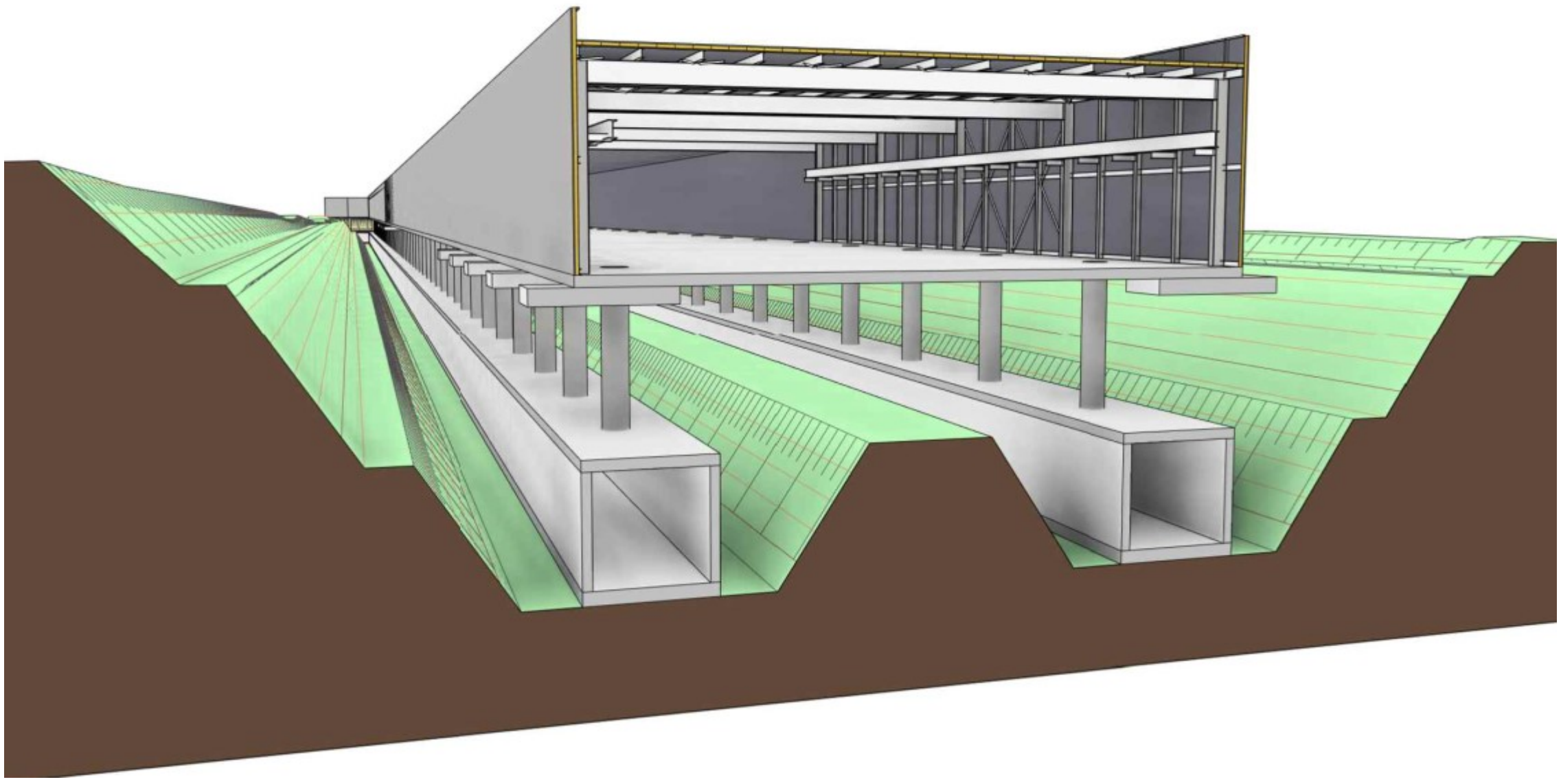

Fig. 7.36: Cross-section of high energy linac civil engineering.

Compared to the other options studied for the high energy linac, the preferred site has a number of significant advantages including:

- The proposed location is quite flat with a total difference in ground level along the length of the HE-linac of about 5 m
- The proposed location has minimal existing infrastructure. A small number of underground networks will need to be rerouted before construction, and two or three older structures on the Prévessin site will be carefully dismantled or relocated as needed.
- The location and orientation are well suited for a potential future connection of beam lines to the North Area of Prévessin, where fixed-target physics research is conducted.

### 7.10.2 Transfer Tunnel

The beam lines from the high energy linac will be connected to the FCC main accelerator tunnel through a transfer tunnel. This transfer tunnel will consist of a single tunnel of internal diameter 3 m and total length of 5 560 m. As this tunnel approaches PA of the FCC, the tunnel will bifurcate to provide a clockwise and anti-clockwise link to the FCC. These two link tunnels will be symmetrical and each 747 m in length. This arrangement is shown schematically in Fig. 7.37 and a plan view is given in Fig. 7.38.

The main characteristics of the transfer tunnel are:

- Three horizontal curves with a minimum radius of 300 m are required
- The tunnel will have a slope of 4.9%.

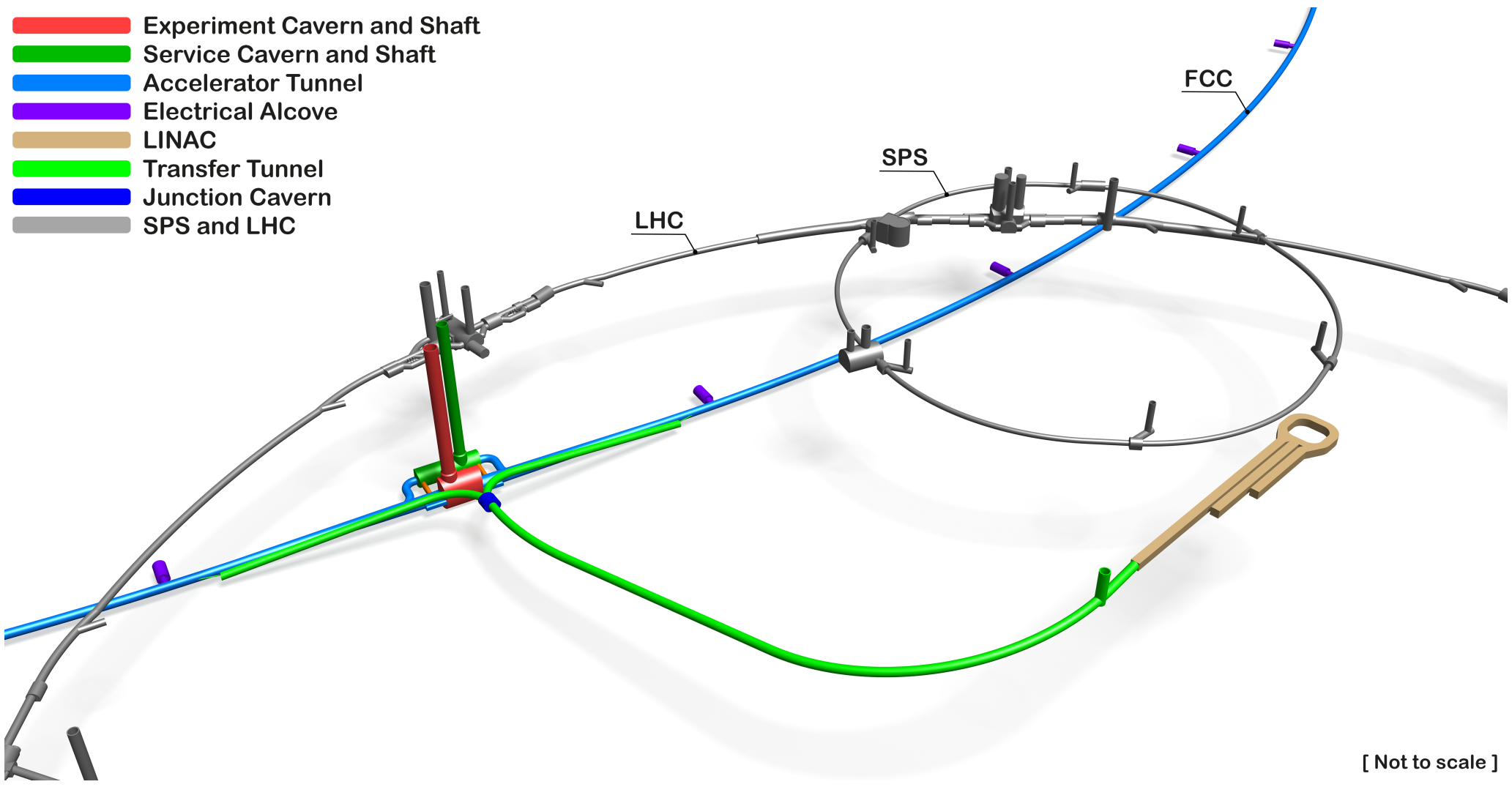

Fig. 7.37: Schematic view of transfer tunnels.

- Connection of the tunnel to the high energy linac will be made on the CERN Prévessin site at the end of the high energy linac.
- Connections to the FCC main tunnel would be created with 0.8 m diameter drilled horizontal passage for the clockwise and anti-clockwise beam lines. A small personnel access connection would also be created between the FCC and transfer tunnels for emergency access and ventilation. These connections would be constructed in the widened section of the accelerator tunnel either side of PA, where the span of the accelerator tunnel is 14.5 m.
- A temporary shaft located within the existing Prévessin site would be built to provide access for the construction of the transfer tunnel. This shaft would have a depth of 30 m and an internal diameter of about 10 m to suit the contractor's plant and equipment. The shaft may not be needed for the operation of the FCC. The transfer tunnel will be excavated from this shaft using either a tunnel boring machine or roadheader or a combination of the two.
- A junction cavern of 30 m length and 9 m span will be necessary to accommodate the bifurcation of the tunnel for the clockwise and anti-clockwise injection into the FCC machine.

Geological conditions for tunnelling this small-diameter tunnel are expected to be favourable, with molasse strata serving as the tunnelling medium from the access shaft down to the FCC accelerator tunnel. The connections to the FCC tunnel will be made at a depth of approximately 200 m below ground level. Additional confidence in the feasibility of construction comes from CERN's extensive experience in underground civil engineering projects in this region, including the SPS, LHC, and CNGS, all of which were successfully constructed in the same general area. If a tunnel boring machine (TBM) is used for excavation, reinforced concrete segments will likely be employed for support, similar to the main FCC accelerator tunnel. Alternatively, if a roadheader is used, a two-phase approach will be followed for tunnel support, consisting of an initial lining with rock bolts and shotcrete, followed by a secondary cast in-situ reinforced concrete lining.

## 7.11 Technical infrastructure

The new injector complex will be integrated in CERN's Prévessin site, requiring additional energy and services. Initially constructed for the SPS accelerator and its fixed-target physics programme, the Prévessin site hosts the SPS North Area and serves as a critical hub for CERN's technical infrastructure. This includes the 400 kV power lines from RTE (France's Transmission System Operator) that supply power to the entire CERN complex, the CERN Control Centre (CCC) responsible for operating all CERN accelerators, the new data centre, and the Ethernet backbone hub. These infrastructures are regularly upgraded to support the evolving demands of CERN's programmes. The new injector complex will become a significant addition to the existing layout. The injector complex consists of five main areas: the e-linac, the p-linac, the damping ring, the HE-linac, and the transfer line to the booster. Each area includes an accelerator tunnel and its associated surface building. Dedicated technical infrastructures will be established to meet the needs of these areas, which can operate independently. For electricity and cooling, there will be a shared infrastructure. A new electrical substation will connect to the existing main substation to supply power to the complex. Cooling towers, grouped in a dedicated building near the end of the HE linac, will dissipate heat generated by the five areas. Each area will also have its own ventilation systems.

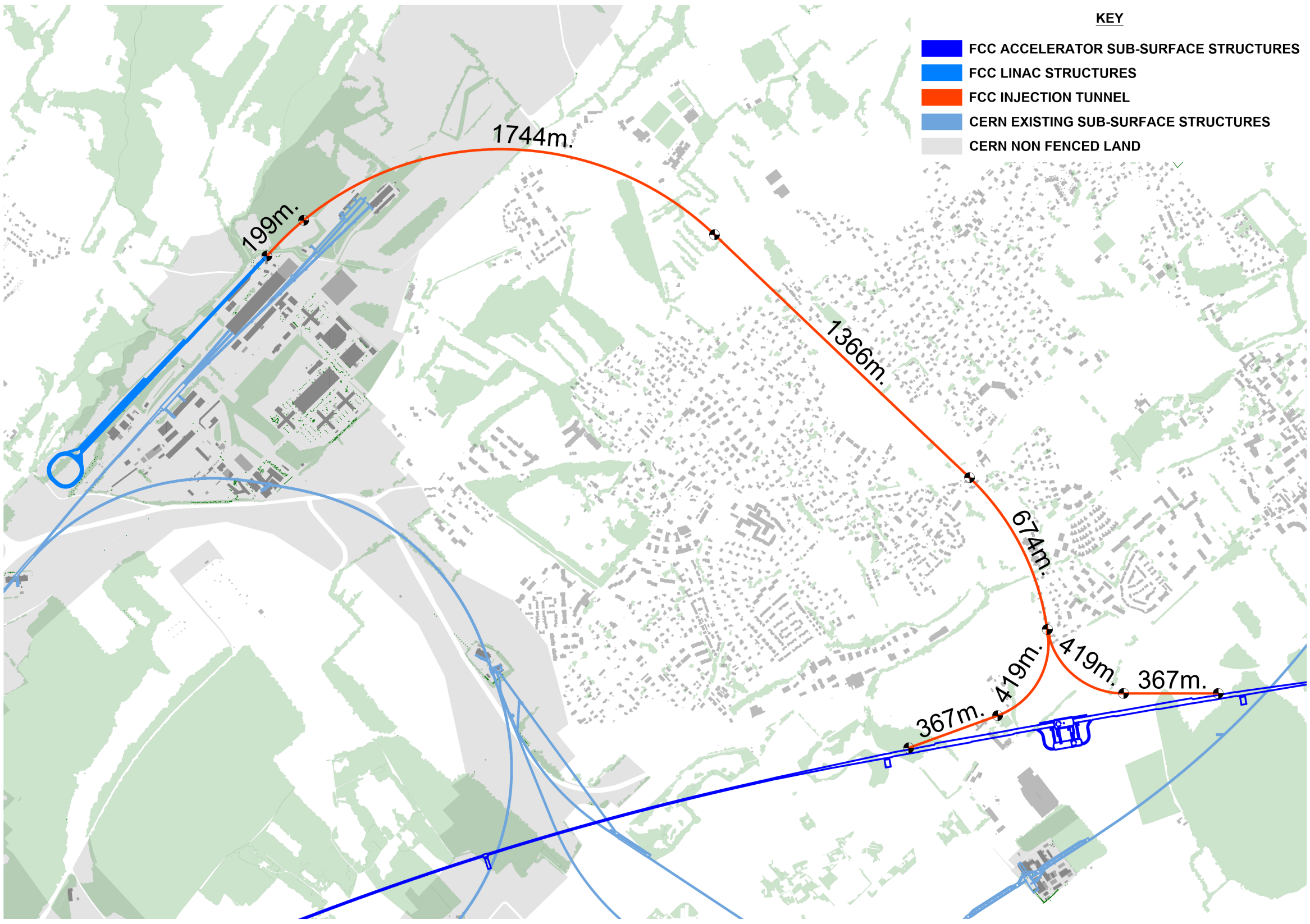

Fig. 7.38: Plan view of transfer tunnels.

The tunnels will house accelerator system components such as RF cavities, magnets, instrumentation, and beam pipes. The surface buildings will accommodate RF klystrons, high voltage modulators, magnet power supplies, vacuum control systems, and beam instrumentation electronics. These buildings will require power distribution, water cooling, ventilation, control systems, and safety features such as fire detection and alarm systems. Services like electricity, cooling, and control will be distributed from the tunnels to the surface buildings.

Additional technical infrastructures will support the complex, including access systems, smoke extraction, environmental monitoring, radiation protection systems, handling and transport systems, and communication systems. These systems will be integrated into the surface buildings above the tunnels. The injector complex can be operated remotely from the existing CERN Control Centre. The estimated main accelerator electrical loads for the injector complex are presented in Table 7.21. In addition to the main loads, the power demand of the other systems is presented in Table 7.22. The injector complex will require a main electrical substation with a capacity of 40 MVA, and its annual energy consumption is estimated at 150 GWh. The cooling towers will have a capacity of approximately 25 MW, with an annual water consumption of 200 000 $m^3$. This water will be sourced from the existing distribution system, which uses water from Lake Geneva. From a technical infrastructure perspective, integrating the injector complex into the Prévessin site is feasible without affecting SPS operations.

Table 7.21: Main electrical loads from the accelerators of the injector complex.

| Accelerator loads | Area | Power rating |
|---|---|---|
| RF | e-linac | 2 MW |
| RF | p-linac | 4 MW |
| Magnets | p-linac | 4 MW |
| RF | HE-linac | 9.5 MW |
| Magnets | Damping ring | 2.5-4 MW |
| Magnets | Transfer line | 2 MW |
| **Total** | | **24-25.5 MW** |

Table 7.22: Additional loads of the injector complex.

| Other loads | Types | Power rating |
|---|---|---|
| Accelerator systems | Auxiliary magnets, vacuum, instrumentation, control… | 2 MW |
| General services | AC plugs, lighting,safety systems… | 2 MW |
| Cooling and ventilation | Pumps, motors, fan coils… | 3 MW |
| **Total** | | 7 MW |

### 7.12 Ongoing studies and possible upgrades

As the study advances there is strong support from Switzerland and PSI to continue and strengthen the CHART collaboration. The injector project team, consisting of PSI and CERN members and external partners, is preparing a comprehensive proposal to be included in the CHART programme for the financial period 2025-2028, complete with detailed resource and cost estimates, which should be completed in early 2025. In the next phase, several critical areas will require focused attention. The updated baseline design includes a 2.86 GeV damping ring, which must be integrated into the forthcoming technical design study.

The current baseline design assumes an RF frequency in the linacs compatible with the main rings to retain operational flexibility. However, the limited availability of suitable power sources on the market presents a challenge, prompting efforts to adopt a commercial S-band frequency for the RF system in the linacs. Additionally, the positron linac will undergo optimisation to shift from the current 2 GHz setup to a more widely available S-band frequency. Reliability and availability of the injector complex will be key factors in the injector's design, particularly for continuous 'top-up' operation, which requires a low-gradient injector configuration with an experimentally validated RF breakdown rate. Assessing the impact of short interruptions caused by RF breakdowns in top-up operations will also be critical for ensuring stable performance.

The involvement of international partners will remain crucial as the injector study progresses to-wards the technical design and construction stages. A comprehensive technical design report could follow the feasibility study phase, building on the SwissFEL linac technology for RF accelerating structures and related upgrade projects. This report will also benefit from the expertise of international partners who have experience in synchrotron light and positron source operations. Its primary objective will be to establish detailed specifications for the accelerator and technical infrastructure requirements necessary to initiate civil engineering design. These specifications will ensure that all system and subsystem requirements are addressed, making the initial construction phase and mass production of critical components smoother.

A comprehensive TDR for the accelerator and the associated technical infrastructure will be completed should the project move to its next phase. This will be pivotal, enabling further engineering design efforts and finalising all necessary specifications for civil engineering work. This phase is essential to confirm that all design elements undergo rigorous examination and are fully integrated into the project, ensuring a seamless transition to the construction phase. A prototyping phase will be critical in this context. During this phase, RF accelerator structures will undergo exhaustive testing to validate their performance and reliability. Additionally, based on a photo-cathode RF gun, the electron source concept will be demonstrated to support top-up operation for injection into the collider ring reliably. This demonstration is essential to confirm that the chosen technology can meet operational requirements and integrate smoothly into the overall system.

Furthermore, the $P^3$ positron production experiment at PSI will be completed and commissioned, with subsequent development focusing on a hybrid scheme expected to reduce the power dissipated on the target while keeping the efficiency of positron production. This development phase will refine the specifications for the positron source, ensuring that all components meet the operational criteria for integration into the injector complex.

# Chapter 8

# Technical infrastructure for FCCs

## 8.1 Requirements and design considerations

The infrastructure requirements and the proposed technical solutions for FCC-ee have been defined for the PA31-1.0 layout, which, as described above, features eight points with associated surface areas. Four of the points are experiment points, and four are technical points. The technical infrastructure systems described here are electricity and energy management, cooling and ventilation, integration, geodesy and survey, cryogenic systems, transport and logistics, and computing infrastructures and robotics. The present study focuses on the technical infrastructure related to FCC-ee, but wherever design choices are made which may impact FCC-hh (in particular, the space requirements and dimensions), constraints from and possible synergies with the hadron collider are taken into account.

### 8.1.1 Assignment of points

Four of the eight FCC points are designated for experiments: PA, PD, PG, and PJ. Among these, PA and PG are considered as large experimental areas for FCC-hh, whereas PD and PJ are optional sites that may accommodate smaller hadron collider detectors requiring less infrastructure. All four points can accommodate the much smaller experiments of the FCC-ee collider.

The remaining four points are technical. Point PB, the only one located in Switzerland, has been chosen to house the beam dump system. Points PL and PH have been selected to house the booster and collider RF, respectively. Point PF was found less appropriate to house the RF due to difficult access conditions (the access shaft is displaced laterally w.r.t. the FCC ring by several hundred metres) [373], but it could house the collimation system.

The assignment of points PL and PH was made based on constraints from accelerator physics (preference to concentrate the collider RF in one point only), constraints at the surface areas and infrastructure constraints (accessibility, resources, electrical network connection and powering). Since point PL features a relatively complicated surface site with many constraints, it was decided to install the booster RF 800 MHz cryomodules at this point. The Booster RF requires much less infrastructure than the collider RF and can more easily be integrated. Point PH was designed to house the collider RF (400 and 800 MHz cryomodules), which requires substantial infrastructure.

### 8.1.2 Electricity supply

Electrical power will be provided by the French network and fed into the FCC at three points (PA, PH, and PD). Further distribution is then accomplished via the FCC tunnel. The electrical infrastructure is designed to support all FCC-ee configurations seamlessly, eliminating the need for additional substations between different operational stages. The proposed setup also includes the possibility of operating the machine without one sub-station (PA or PD). The PH substation, which has the main RF loads (200 MW), is mandatory for beam operation. This setup will ease the maintenance and repairs as all substations will be connected together through the HV network. RTE (Réseau de Transport d'Électricité) has launched a study on how to connect the FCC to the French network. The requirements of the FCC project are not considered significant by RTE, and will not impact the French electrical network operation.

### 8.1.3 Cooling and ventilation

The potential sources of cooling water in the area are lake Geneva and the Rhône and Arve rivers. As a baseline, the required cooling water will be taken from Lake Geneva via the existing infrastructure, which has enough capacity to supply the FCC. Pipework within the tunnel will connect the remaining points to PA, PD, or PJ to ensure their respective water supply. The drainage flow rates, corresponds to around $30\,m^3/h$ in the RF points and a maximum of $11\ m^3/h$ in the remaining points.

A semi-transverse ventilation scheme has been adopted for the FCC tunnel, with air supplied through a dedicated duct running along the entire sector and extracted either via the tunnel itself or through an emergency extraction duct. To ensure continuous airflow in case of a duct failure, air is supplied to each sector from both endpoints, with the same configuration adopted for the extraction system.

### 8.1.4 Cryogenic systems

The cryogenic system design of FCC-ee follows the requirements of the superconducting devices, which are distributed throughout the machine. Whilst the technical points PB and PF do not feature any cryogenic needs, the technical points PH and PL are associated with the radiofrequency (RF) system, which needs to be operated at 2 K and at 4.5 K. These RF points have the largest cryogenic installations due to their high concentration of heat loads. Additionally, each experiment point – PA, PD, PG, and PJ – requires a dedicated cryoplant to support the cooling needs of the detector magnet and the magnets in the machine detector interface (MDI) region.

### 8.1.5 Transport and logistics

In the first phase of the analysis, an inventory was established detailing the type, quantity, dimensions, and weight of the components to be transported. Simultaneously, the number of personnel requiring transport to their workplace in the tunnel was determined, and evacuation studies were conducted.

Based on these input data, two vehicle concepts have been studied: the first for the transport of the equipment, including the accelerator components, and a second for the transport of personnel throughout the installation, operation and dismantling phases.

The vehicle for transporting materials will be electric, fed by batteries. Charging stations will be installed all along the tunnel. The design comprises a trailer carrying the material, pulled by a tractor hosting the batteries. The trailer is equipped with two robotic arms capable of handling and manoeuvring the components up to their final location.

As the market already offers valid solutions, the personnel vehicle will be based on an existing system. This will then be customised to meet the specific requirements and constraints of the FCC tunnel. The personnel vehicles will be electric and powered by batteries. The personnel transport will consist of a mixture of individual and scheduled transport systems. Its characteristics will be defined to guarantee the evacuation of the people present in a sector at a safe time.

For safety reasons, vehicle overtaking needs to be possible at any location in the tunnel, putting a constraint on the width of the vehicles.

In addition to technical solutions for material and personnel transport, a logistics study is being conducted. This study develops logistics scenarios based on various boundary conditions, such as the number of shafts available for material transfer and the number of vehicles used for underground transport.

For each scenario, key performance indicators (KPIs) are generated, including the number of magnets that can be installed per day and the total installation time required for the accelerator. Potential bottlenecks in material flow are also identified. These indicators serve as an objective tool for evaluating and comparing different scenarios, enabling the selection of the most suitable approach for the overall project.

### 8.1.6 Geodesy

The size of the FCC requires a refined and extended geodetic infrastructure, including reference frames and a gravity field model. The geodetic infrastructure must be compatible with each phase of the project. This comprises the generation of a primary surface geodetic network. To this end, Swisstopo and IGN are increasing the density of their national geodetic networks. In order to meet the vertical alignment accuracy, a refined model of the gravity field is being prepared.

### 8.1.7 Communications, computing and data services

Three different infrastructures will be deployed to enable data, telephone and radio networks. They are all built on top of a fibre optics infrastructure. Regarding the IT infrastructure, the FCC will feature two redundant data centres on the Meyrin and Prévessin sites.

### 8.1.8 Surface areas

Based on the design choices above, and input from the infrastructure and equipment groups, the needs for infrastructure and buildings at the eight surface points have been compiled. The tables summarise the key elements (element, location, construction type, size, capacity) by domain (safety, electricity, and energy management, cooling and ventilation, geodesy and survey, cryogenic systems, transport and logistics, computing infrastructures). The contents of these inventory tables are the basis for the layout of the surface points [374].

### 8.1.9 Underground areas

Integration of the underground areas, including the tunnel, main caverns and service caverns, shafts and alcoves, has been performed. Equipment sensitive to radiation will be placed in alcoves. There are seven alcoves per arc. The alcoves are linked to transport lay-by zones, which allow transport vehicles to park, pass and overtake each other.

In addition to the alcoves, there are service caverns at the eight points. The current civil engineering design for the service caverns at technical areas is 60 m long, 25 m wide and 15 m high. These caverns will each be divided into three levels to provide additional floor area on the upper levels. The service caverns are mainly occupied by power supplies, cooling and ventilation equipment as well as cryogenics equipment.

## 8.2 3D Integration Studies

The 3D integration studies of the FCC-ee have been detailed for all the areas of the underground tunnels and caverns, fitting the requirements from the work packages (e.g., civil engineering, infrastructure, general services, transport, safety, optics, accelerators requirement, etc.). This section presents the results of the configuration and layout of the accelerators and infrastructure systems that were studied for the feasibility study report. This is an evolution of the previous studies made for the conceptual design report. [13, 307]. The 3D integration studies take into account the configurations of both the FCC-ee and the FCC-hh to ensure space compatibility.

### 8.2.1 General layout of the FCC

The FCC is a 90.6 km circumference tunnel including eight access points equally spaced above the accelerator tunnel and currently named as point PA, PB, PD, PF, PG, PH, PJ and PL (see Fig. 8.1). The points PA, PD, PG and PJ will host the detectors in dedicated civil engineering areas. They are mentioned as large experiment points for points PA and PG and small experiment points for points PD and PJ. The points PB, PF, PH and PL are mentioned as technical points hosting different accelerators part.

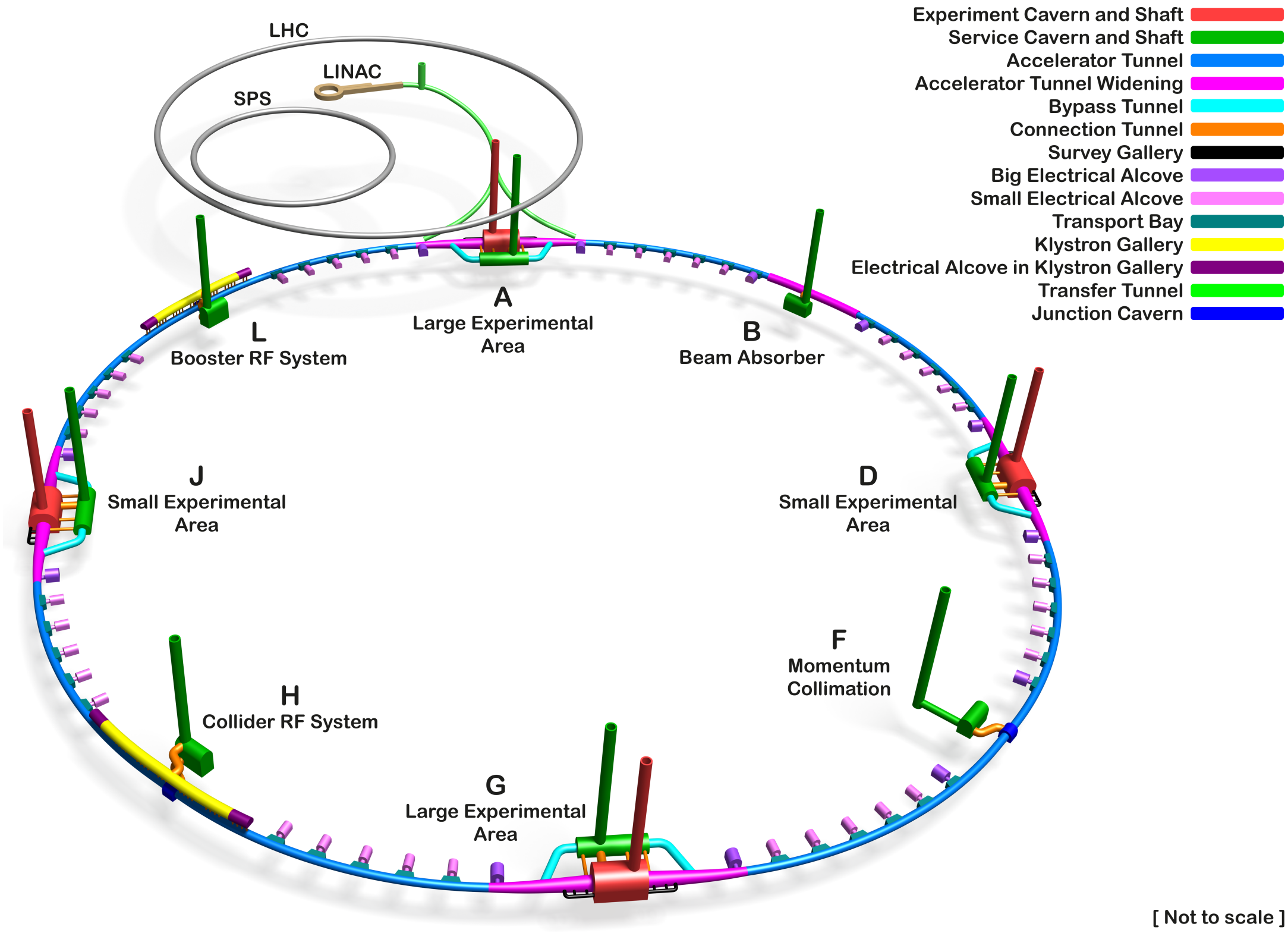

Fig. 8.1: FCC-ee schematic view

For the FCC-ee machine,

– The point PB will host the extraction system from the booster, transfer line and injection system to the collider, and the extraction systems and transfer lines to the dumps for both the collider and the booster,
– The point PF will host the collimation systems (betatron and momentum),
– The point PH will host the collider RF system,
– The point PL will host the booster RF system.

For the FCC-hh machine,

– The point PB will host the collider injection lines and the extraction lines to the dumps,
– The point PF will host the betatron collimation system,
– The point PH will host the momentum collimation system,
– The point PL will host the RF system.

### 8.2.2 Integration of point PA and point PG

Point PA and point PG will serve as large experiment points, hosting detectors in the cavern in both the FCC-ee and the FCC-hh phases of operation. The cavern size and extent meet the needs of the detectors for both FCC-ee and FCC-hh. In addition, for the FCC-ee, the long straight sections (LSS) on either side of point PA and point PG will host the beamstrahlung dumps and the polarimeter system (see Fig. 8.2 and Fig. 8.3). The LSS of point PA will also host the end of the transfer line from the injector linac, into the booster.

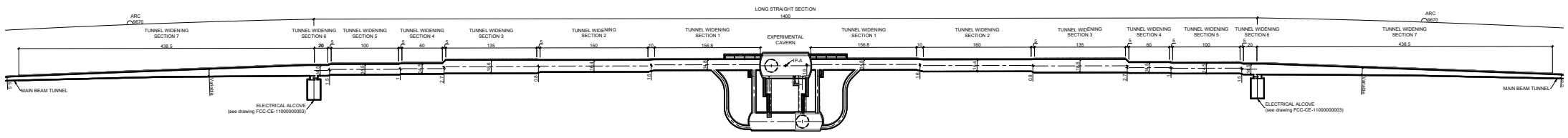

Fig. 8.2: FCC underground - civil engineering in point PA.

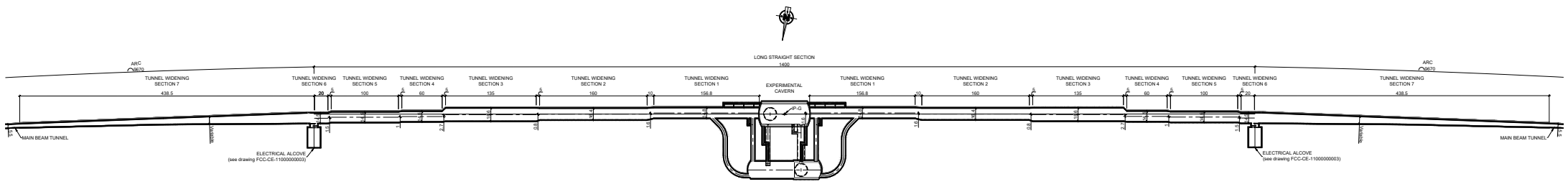

Fig. 8.3: FCC underground - civil engineering in point PG.

The figures included in this section are the results of the 3D integration studies for the point PA and the point PG.

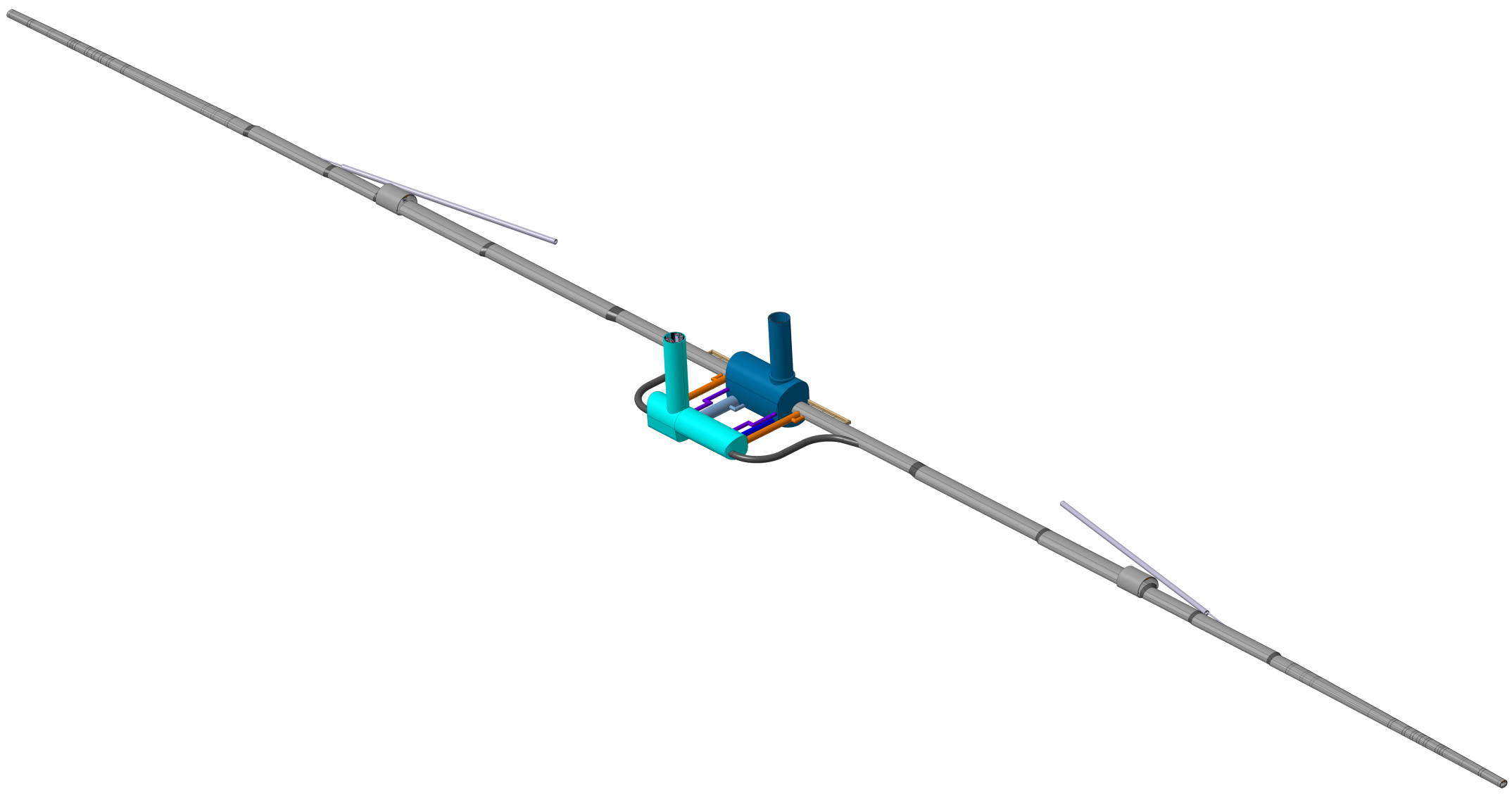

Fig. 8.4: FCC-ee point PA and point PG - general overview.

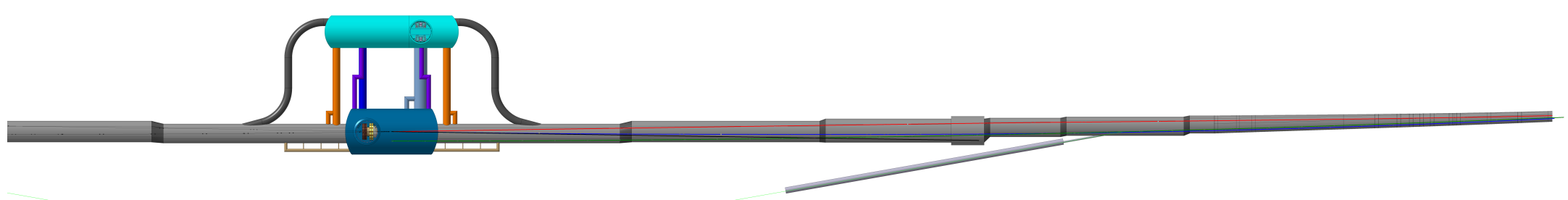

Fig. 8.5: FCC-ee point PA - top overview with optics.

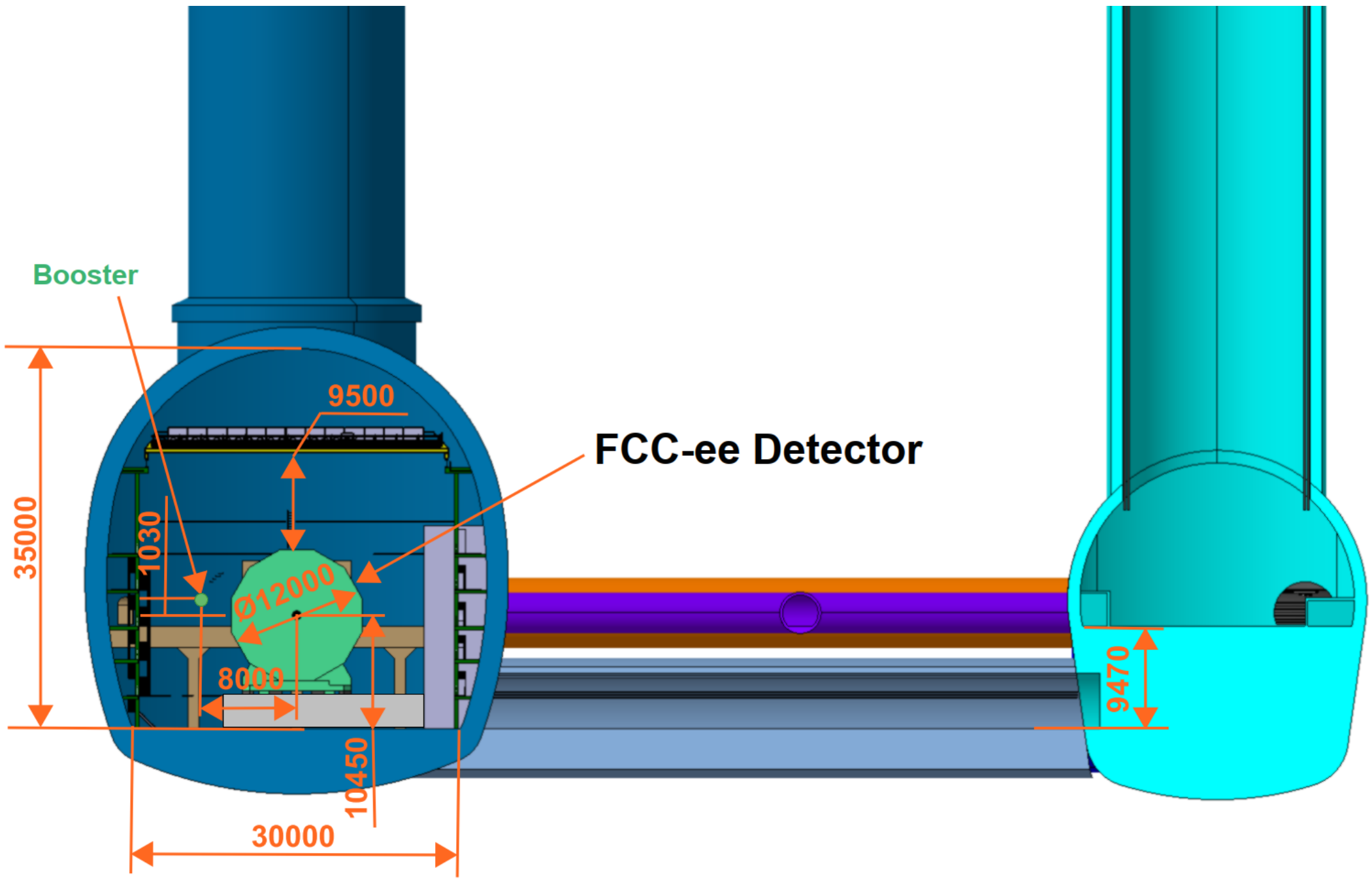

Fig. 8.6: FCC-ee point PA - experiment cavern cross-section.

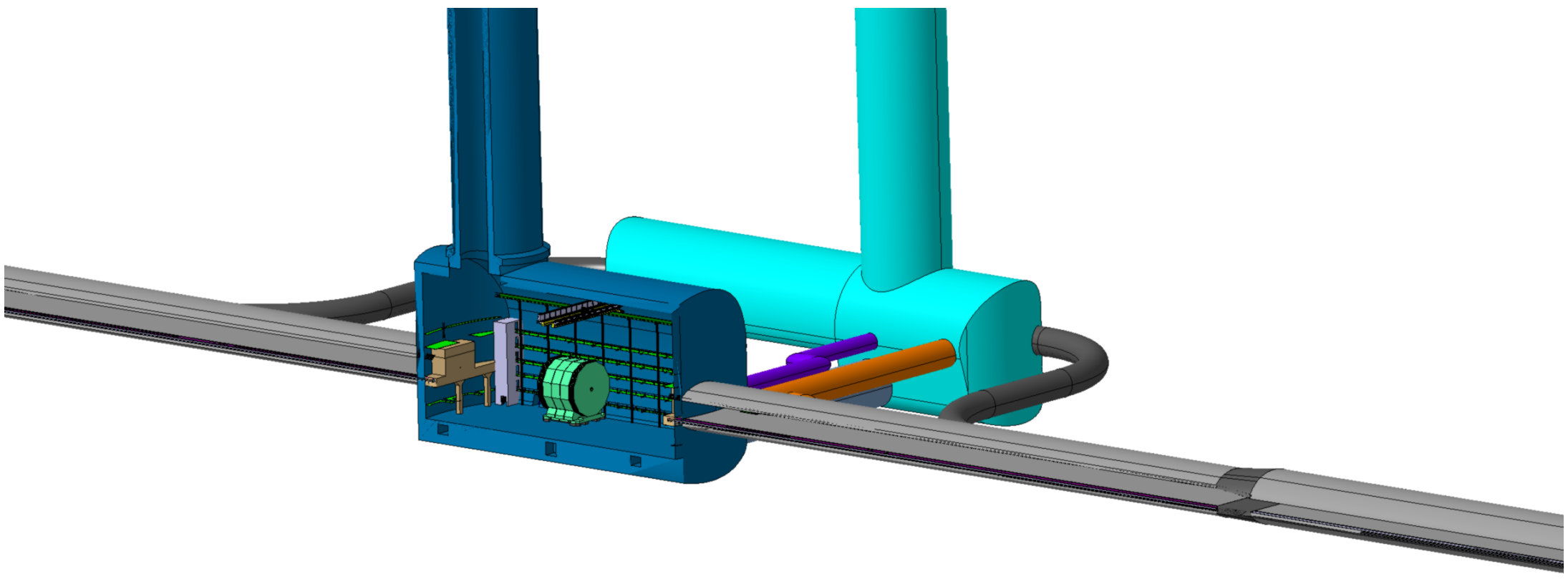

Fig. 8.7: FCC-ee point PA - experiment cavern iso view.

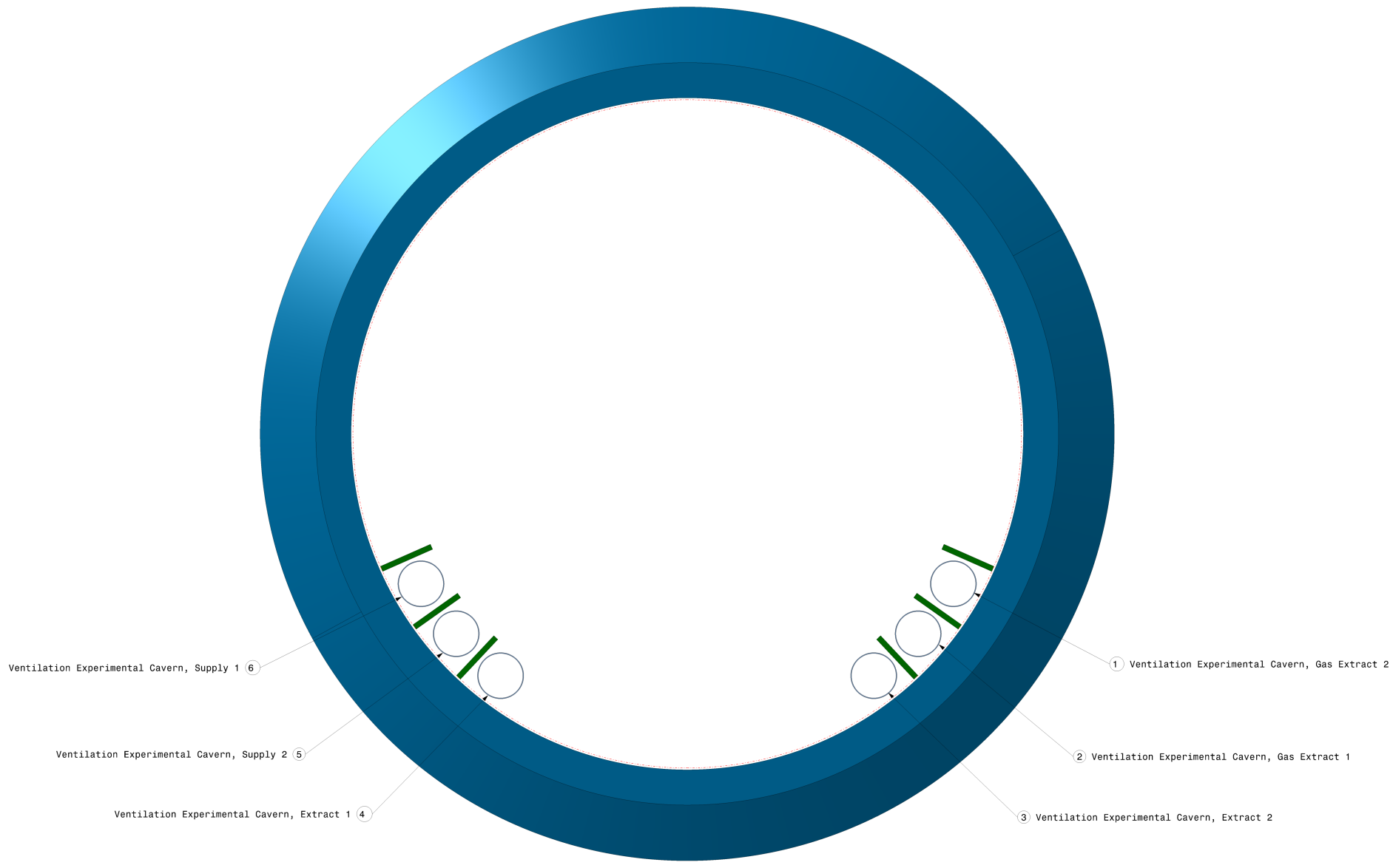

Fig. 8.8: FCC-ee point PA - shaft of the experiment cavern.

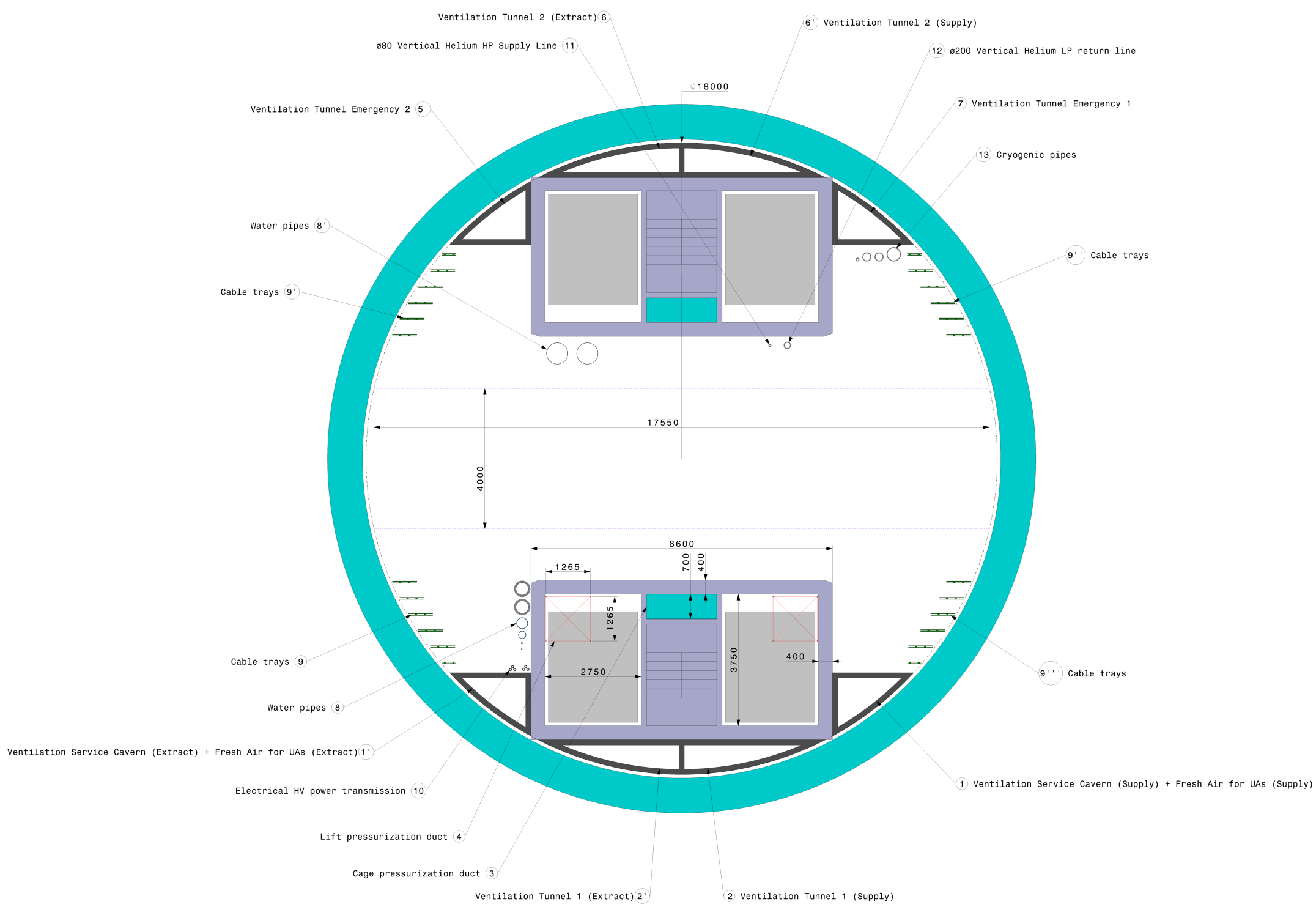

Fig. 8.9: FCC-ee point PA - shaft of the service cavern.

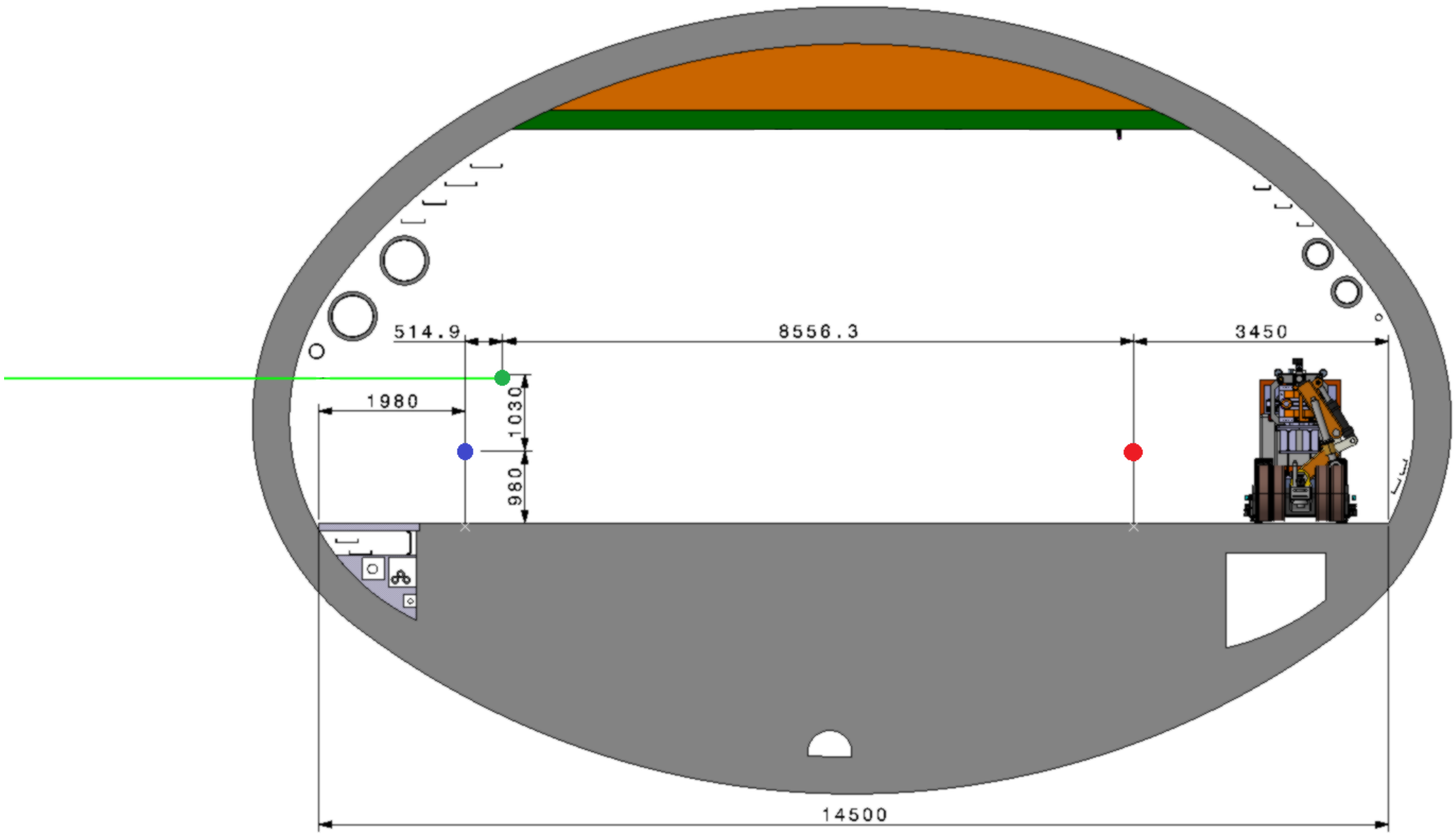

Fig. 8.10: FCC-ee point PA - tunnel cross-section with booster.

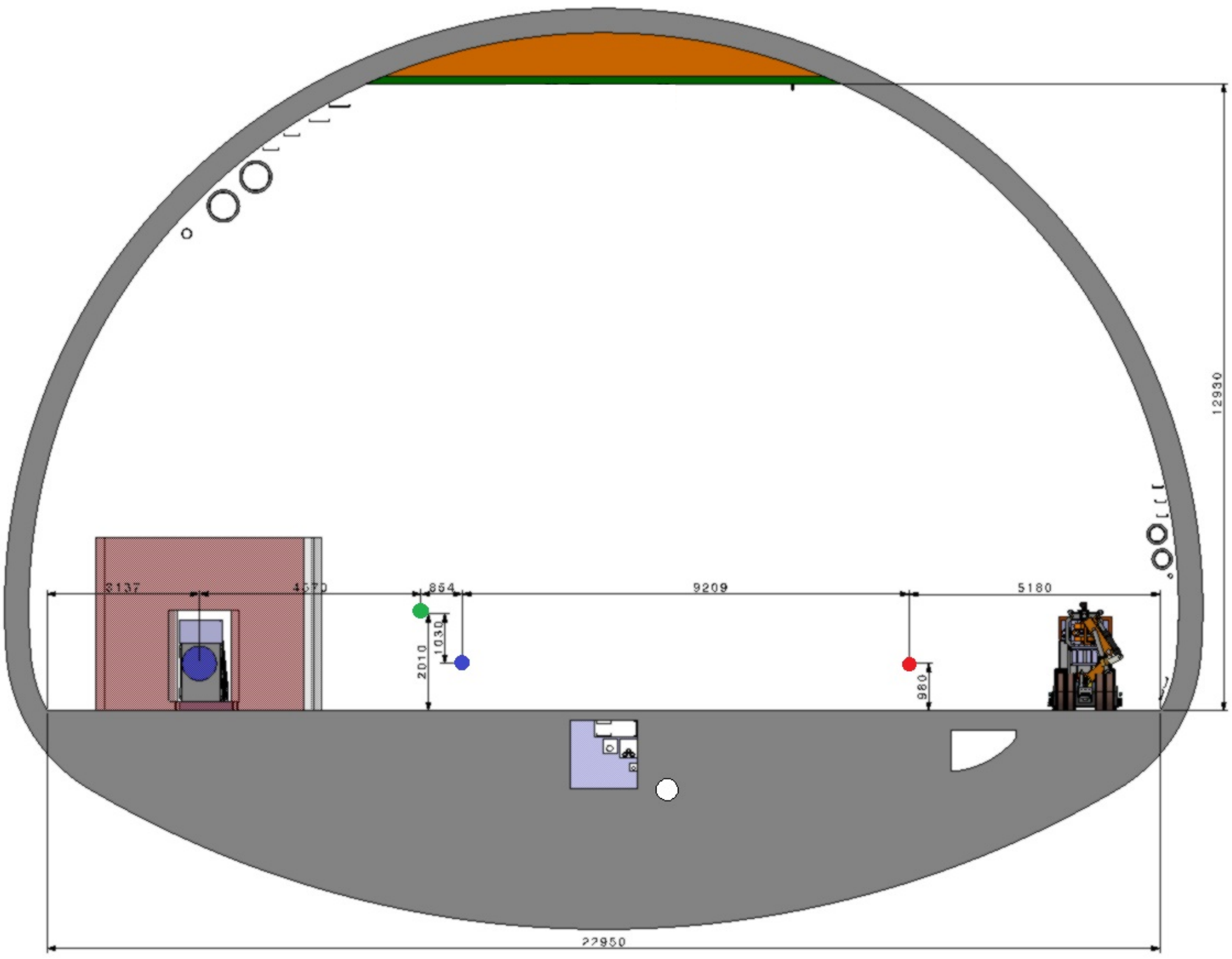

Fig. 8.11: FCC-ee point PA - tunnel cross-section with dump.

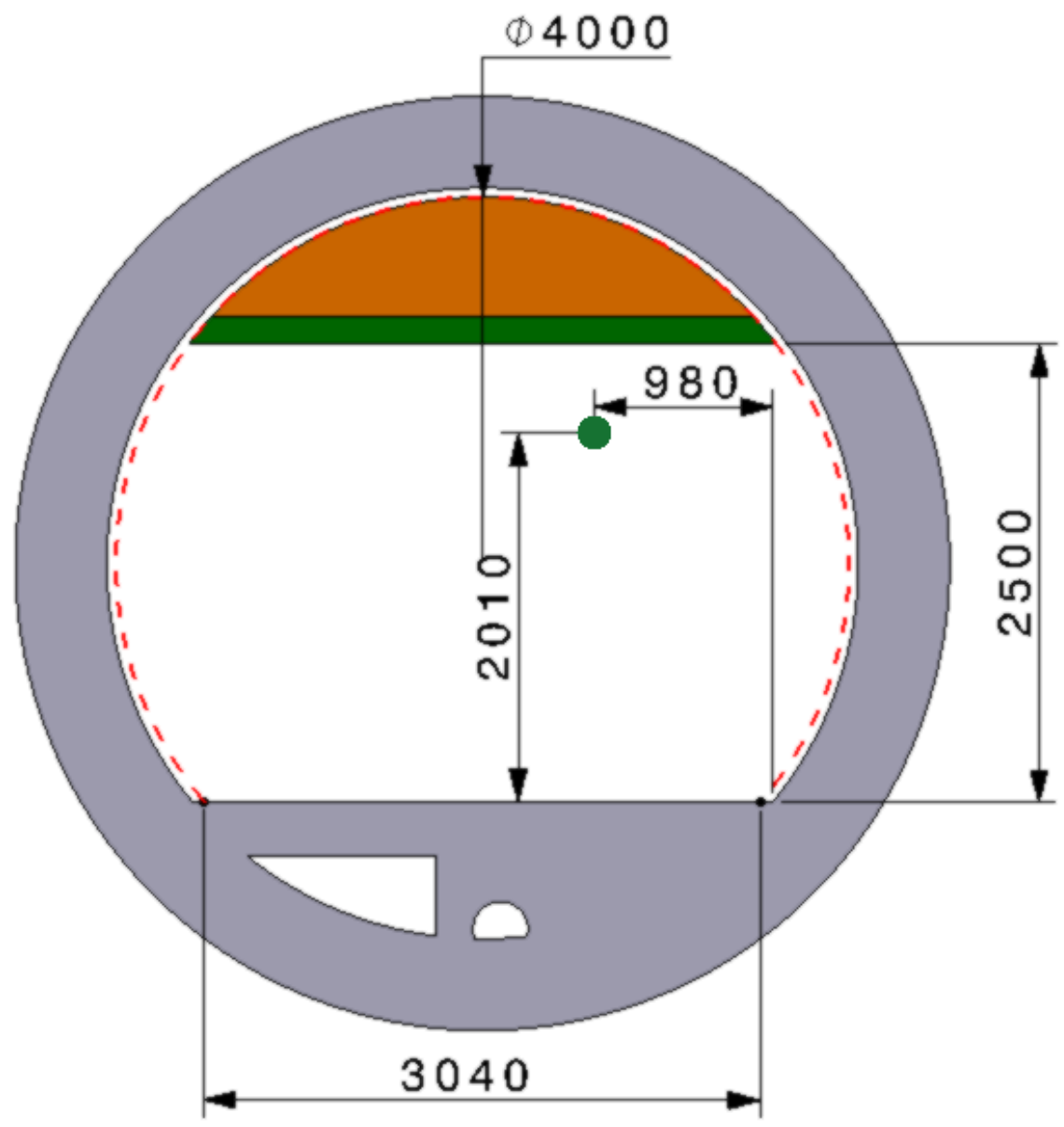

Fig. 8.12: FCC-ee point PA - booster injection tunnel.

### 8.2.3 Integration of point PB

Point PB will house the FCC-ee beam dump and injection from the booster into the collider. In the next phase of operation, FCC-hh will house the beam dump system. The civil engineering volumes meet the needs of FCC-ee (see Fig. 8.13).

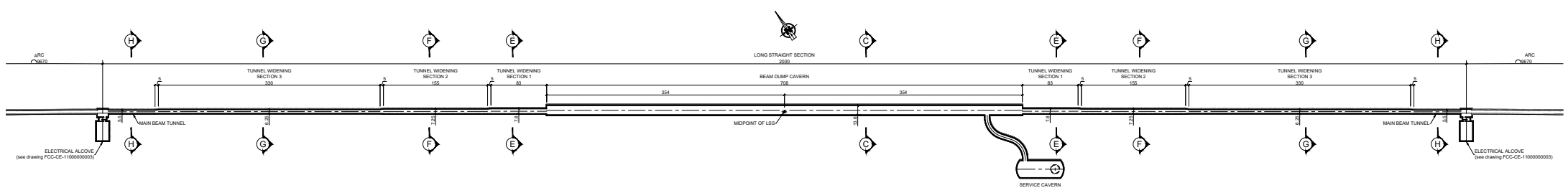

Fig. 8.13: FCC underground - civil engineering in point PB

The figures included in this section are the results of the 3D integration studies for the point PB.

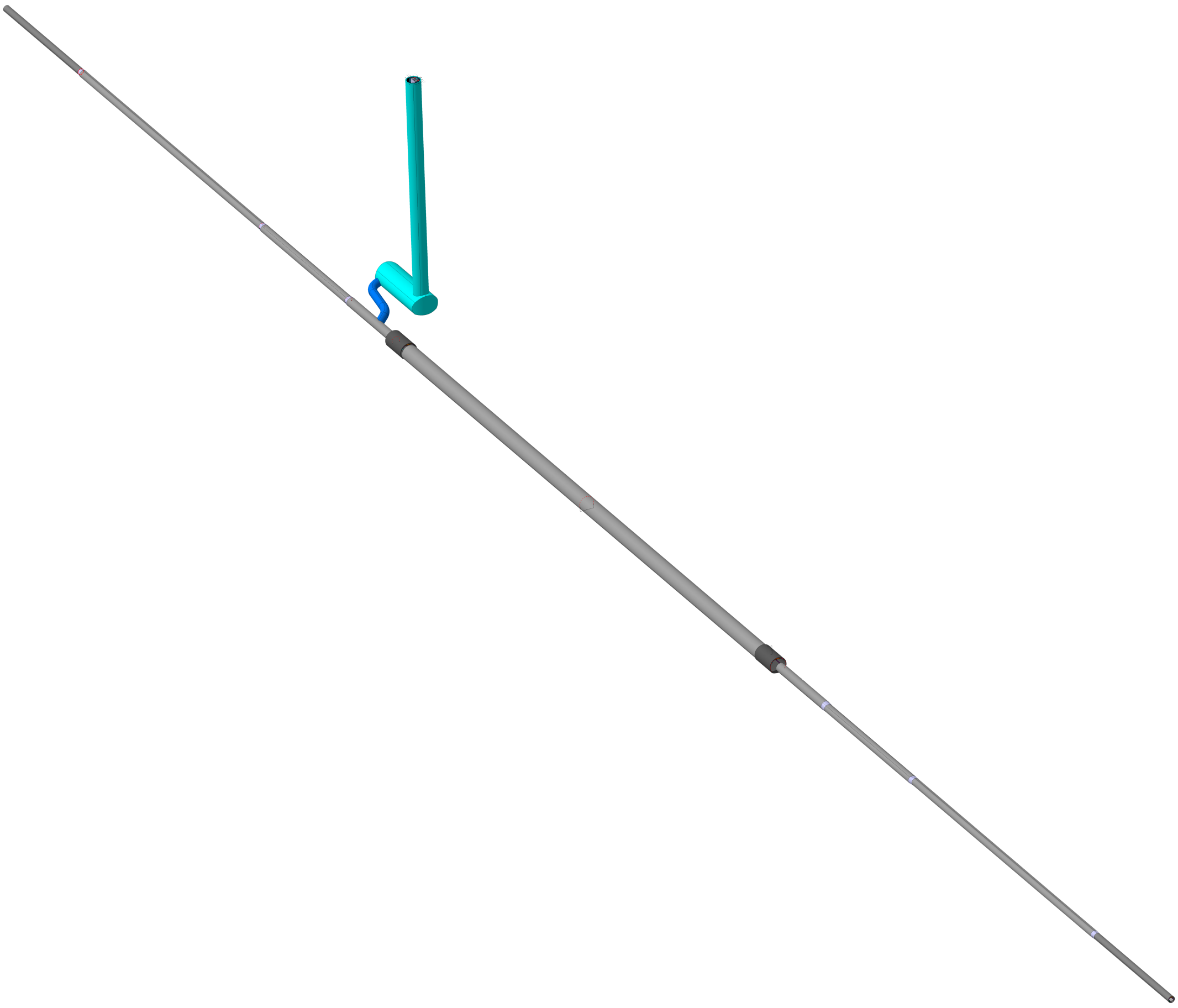

Fig. 8.14: FCC-ee Point PB - general overview.

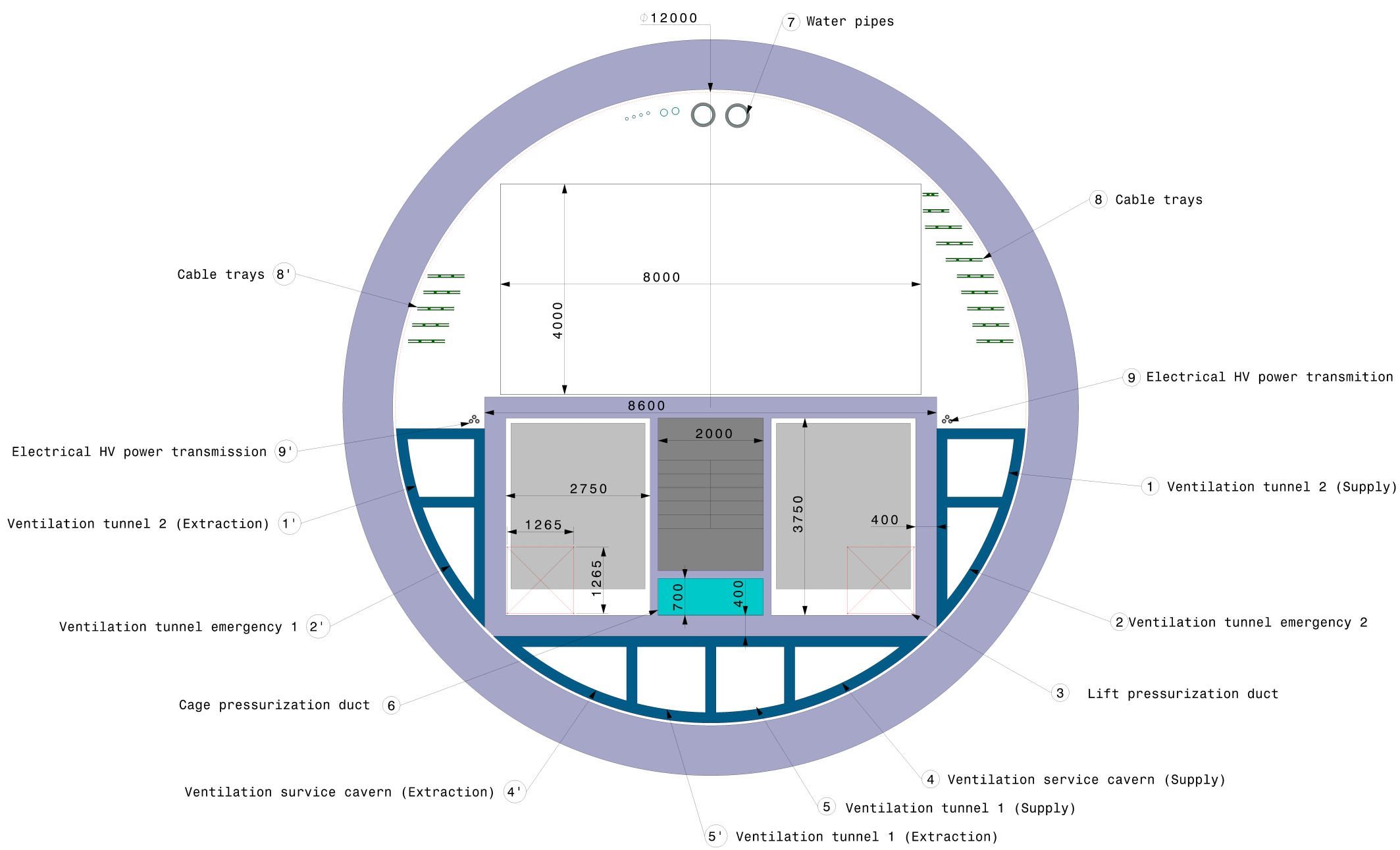

Fig. 8.15: FCC-ee Point PB - shaft of the service cavern.

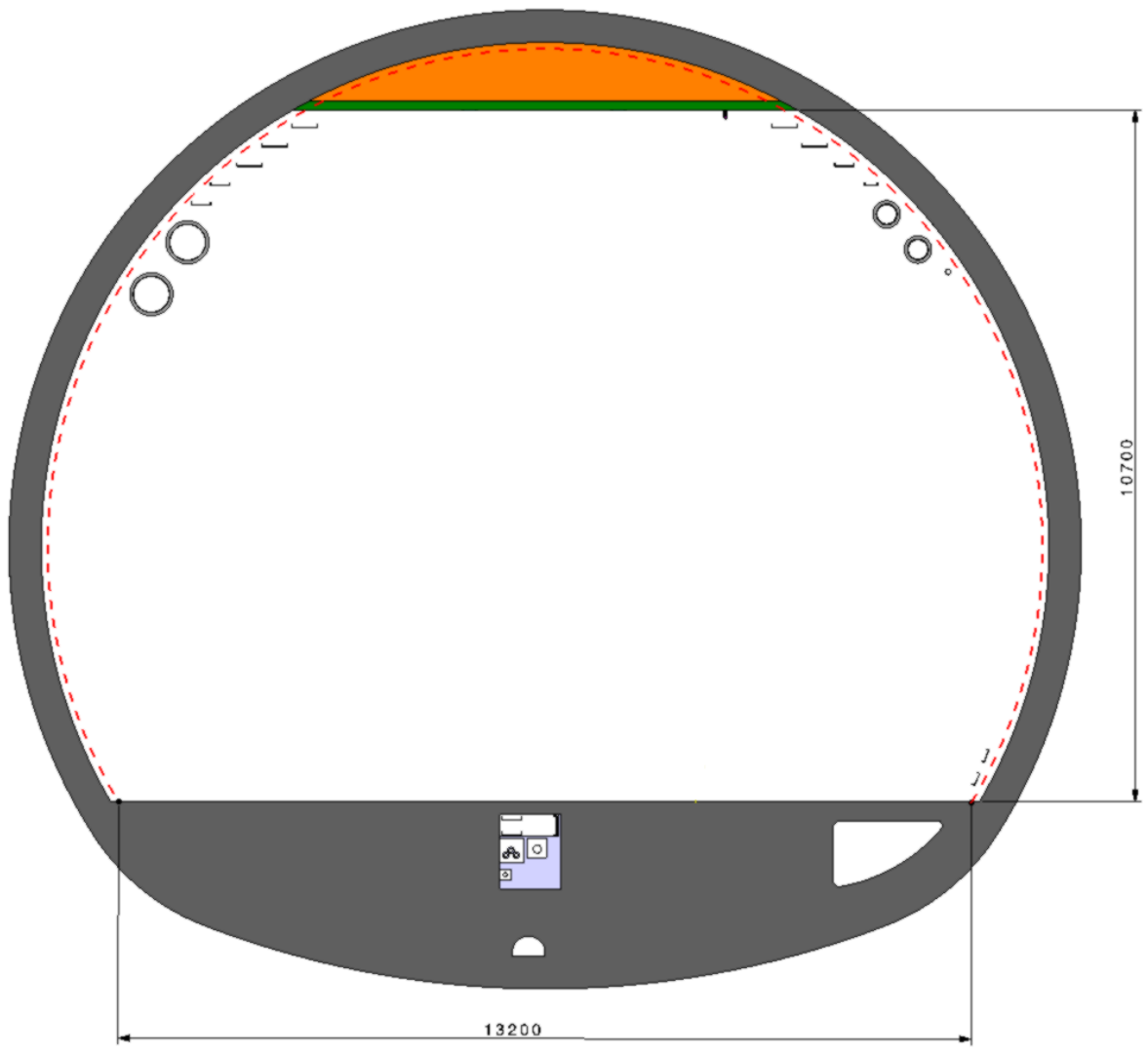

Fig. 8.16: FCC-ee Point PB - tunnel including the dump.

### 8.2.4 Integration of point PD and point PJ

Point PD and point PJ will serve as small experiment points, hosting detectors in the cavern in both the FCC-ee and the FCC-hh phases of operation (see Fig. 8.17 and Fig. 8.18). The cavern size and extent meet the needs of the detectors for both FCC-ee and FCC-hh. In addition, for the FCC-ee, the long straight section (LSS) on either side of point PD and point PJ will host the beamstrahlung dumps.

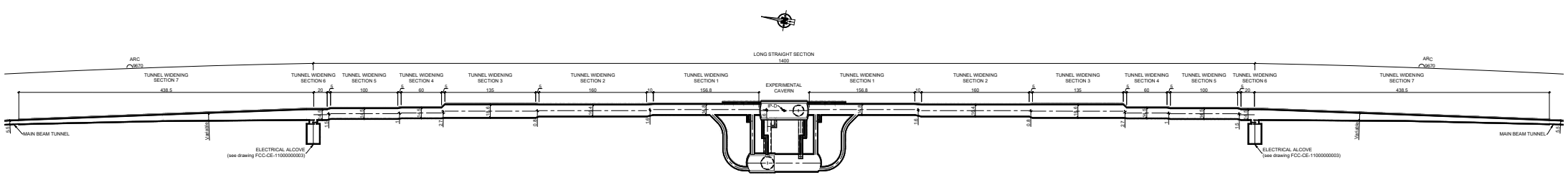

Fig. 8.17: FCC underground - civil engineering in point PD.

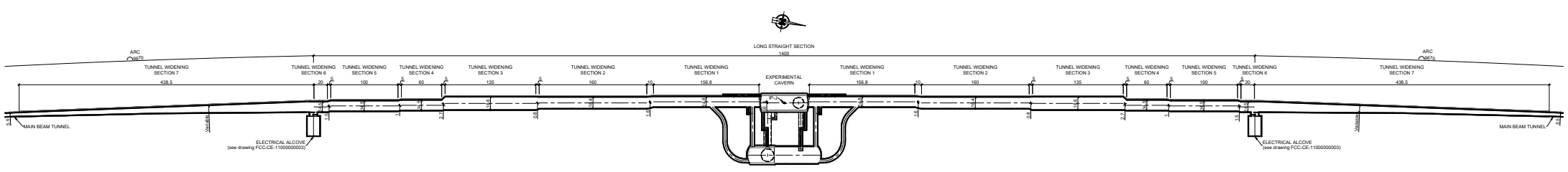

Fig. 8.18: FCC underground - civil engineering in point PJ.

The figures included in this section are the results of the 3D integration studies for point PD and point PJ.

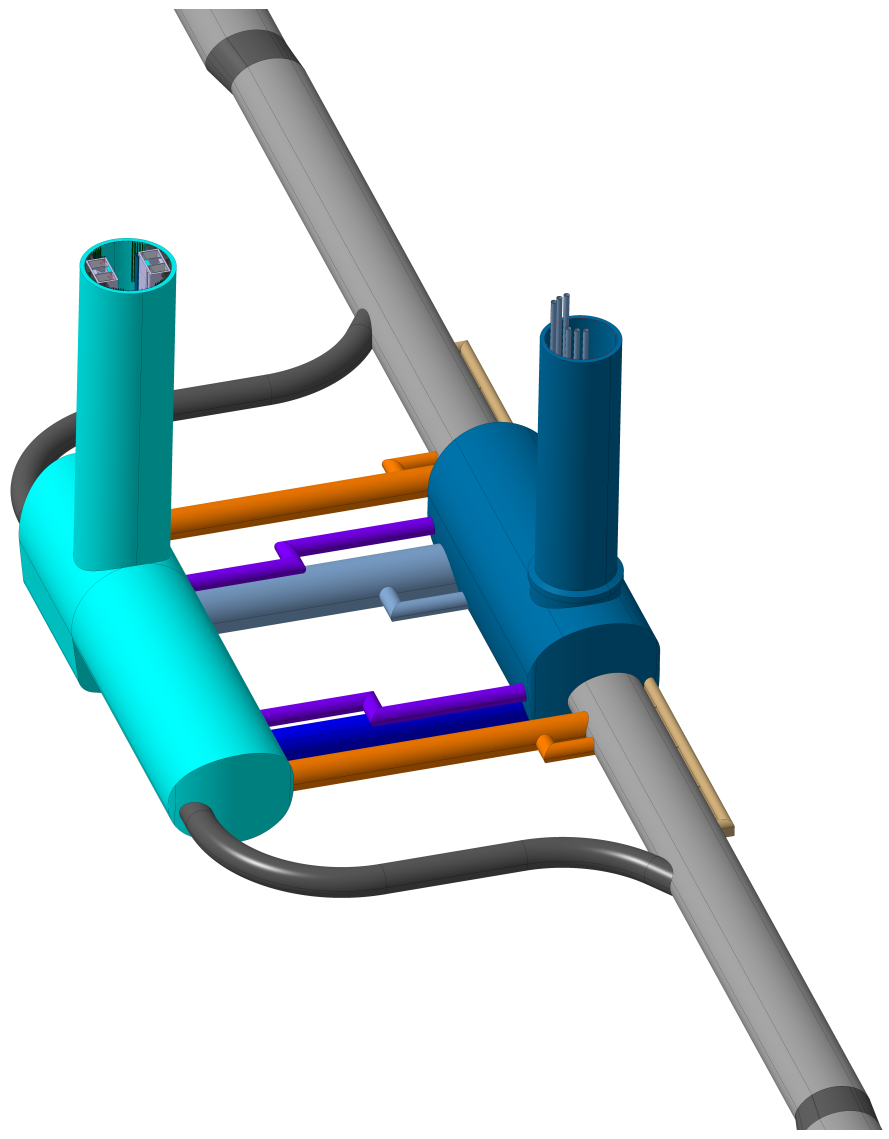

Fig. 8.19: FCC-ee point PD and PJ - general overview.

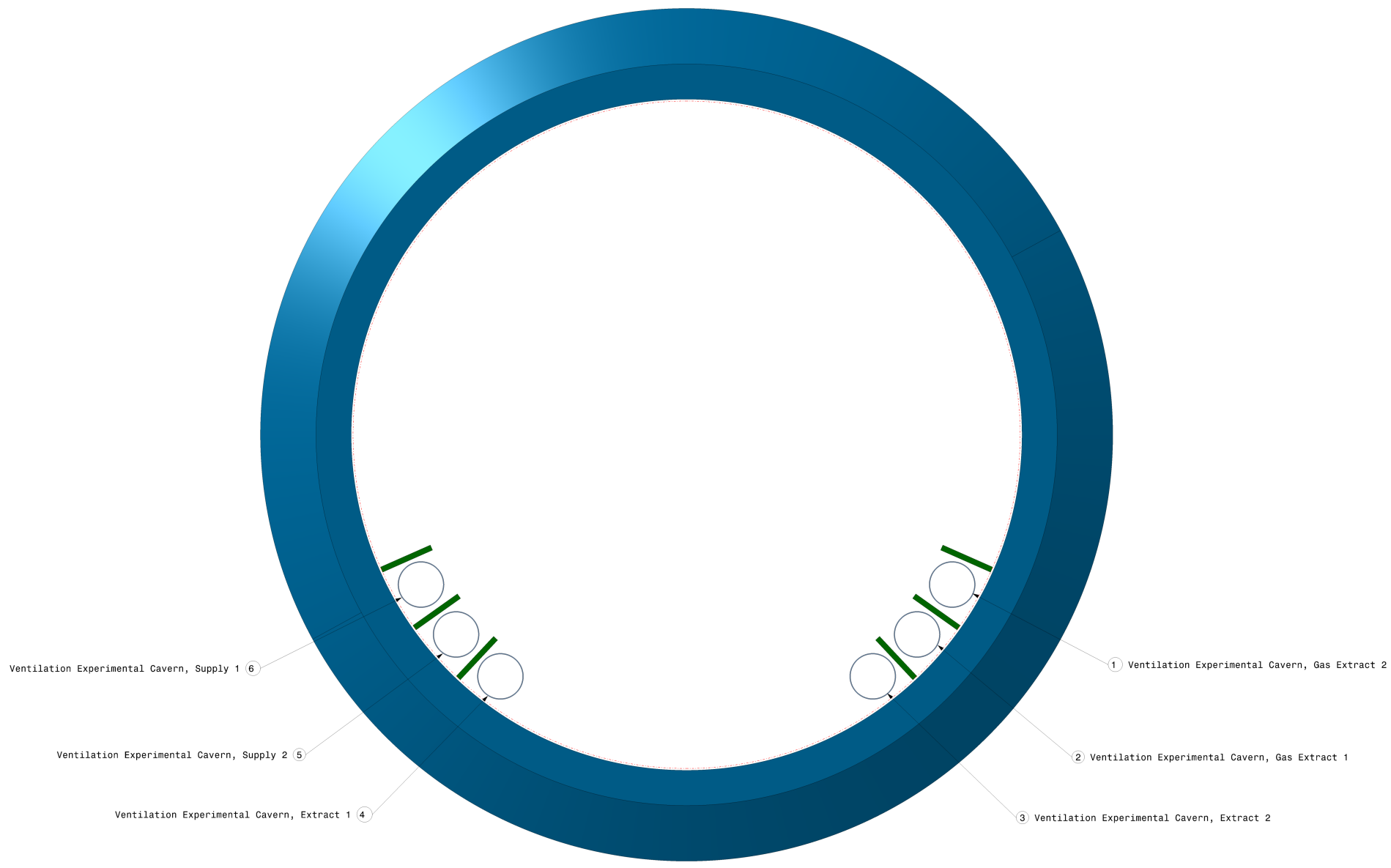

Fig. 8.20: FCC-ee point PD and PJ - shaft of the experiment cavern.

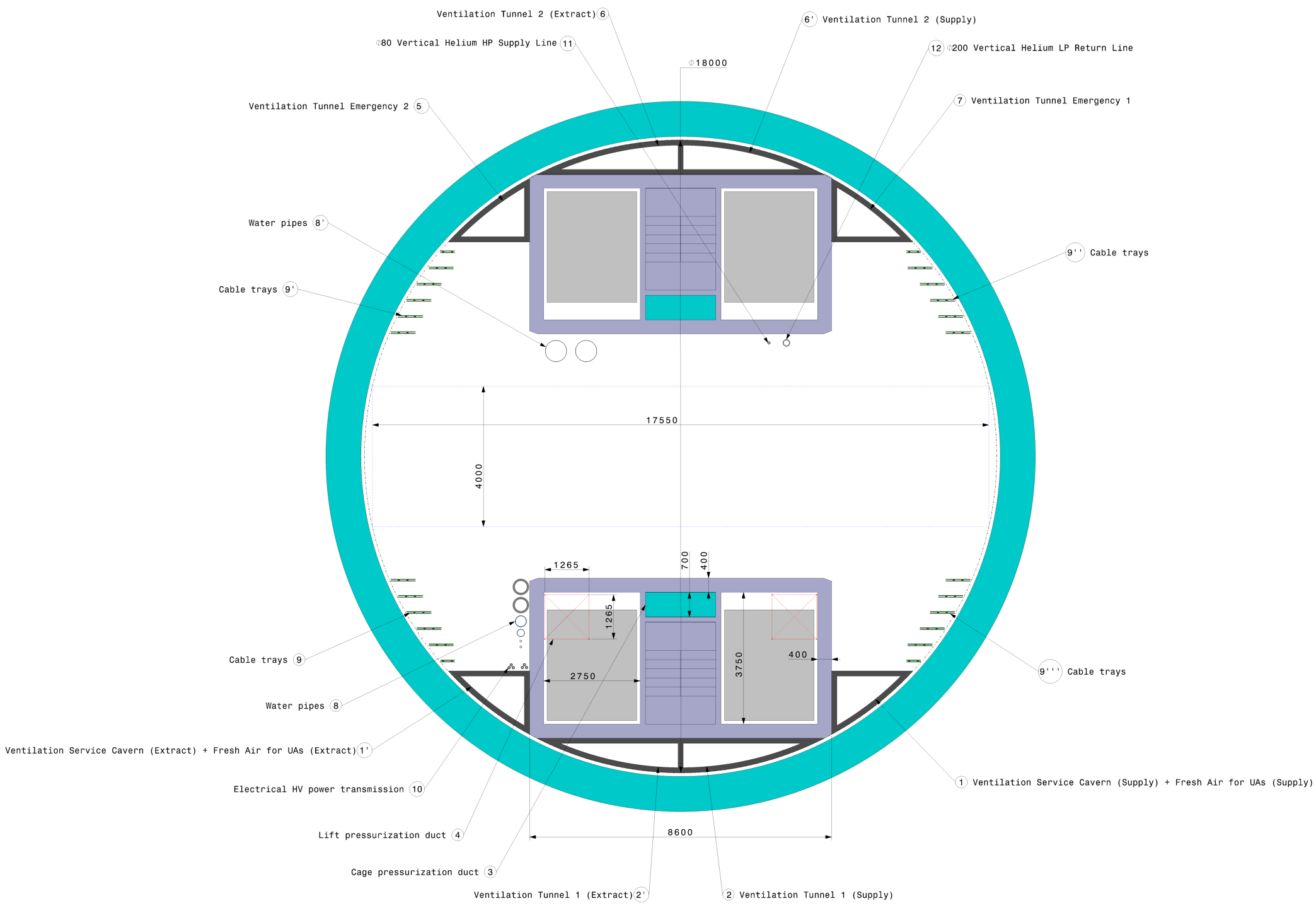

Fig. 8.21: FCC-ee point PD and PJ - shaft of the service cavern.

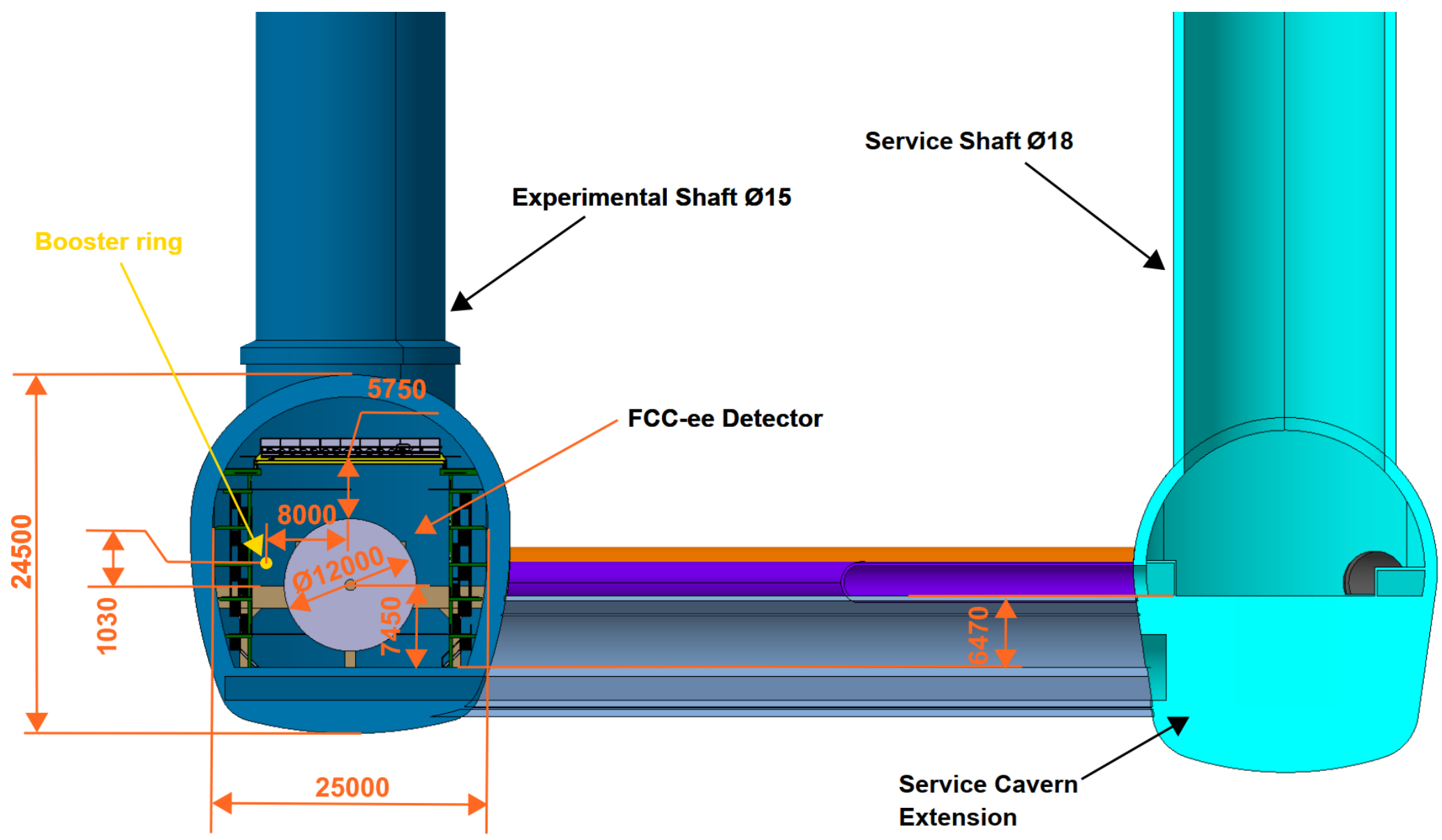

Fig. 8.22: FCC-ee point PD and PJ - experiment cavern cross-section.

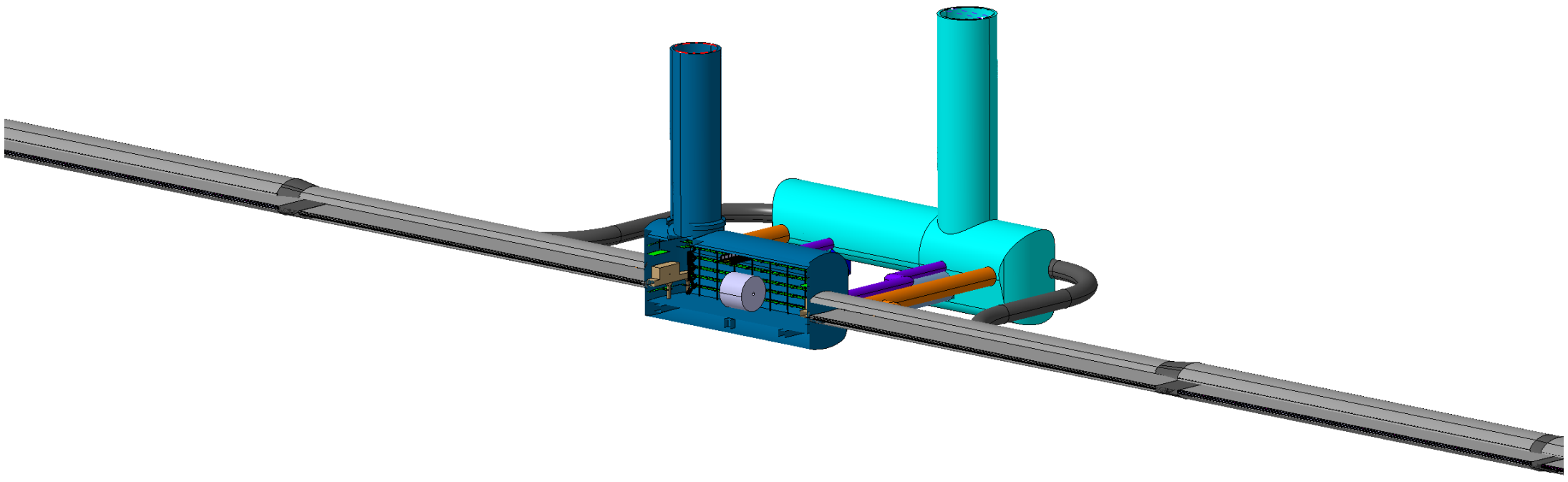

Fig. 8.23: FCC-ee point PD and PJ - experiment cavern iso view.

### 8.2.5 Integration of point PF

Point PF will house FCC-ee betatron and momentum collimation systems. In a next phase of operation, FCC-hh will house the beam dump systems. The civil engineering volume meet the needs of FCC-ee (see Fig. 8.24).

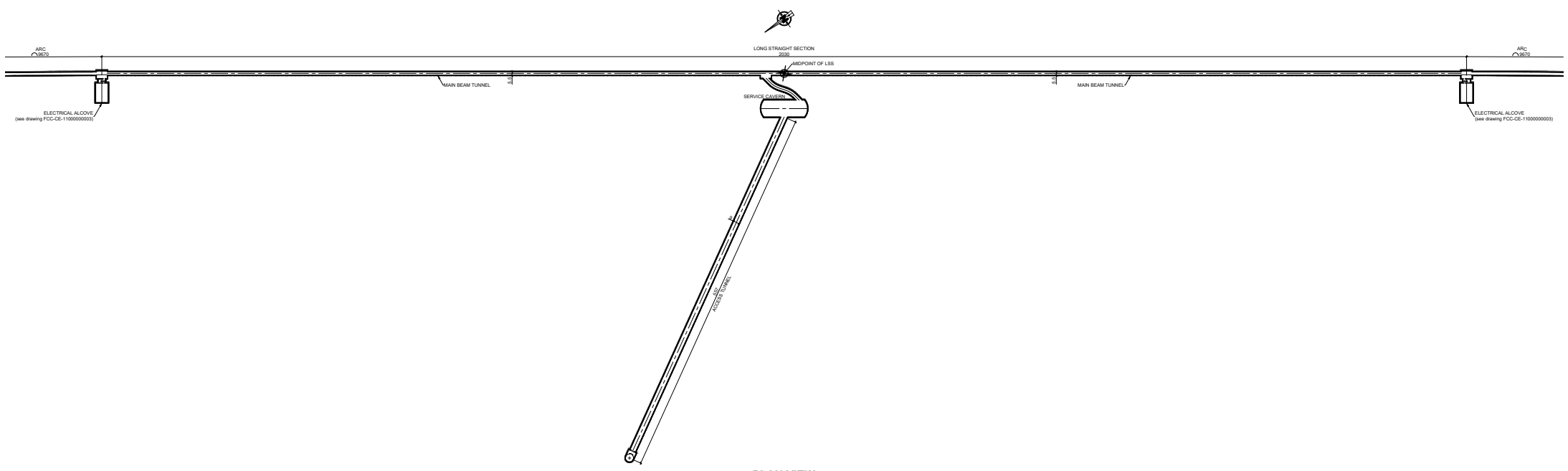

Fig. 8.24: FCC underground - civil engineering in point PF.

The figures included in this section are the results of the 3D integration studies for the point PF.

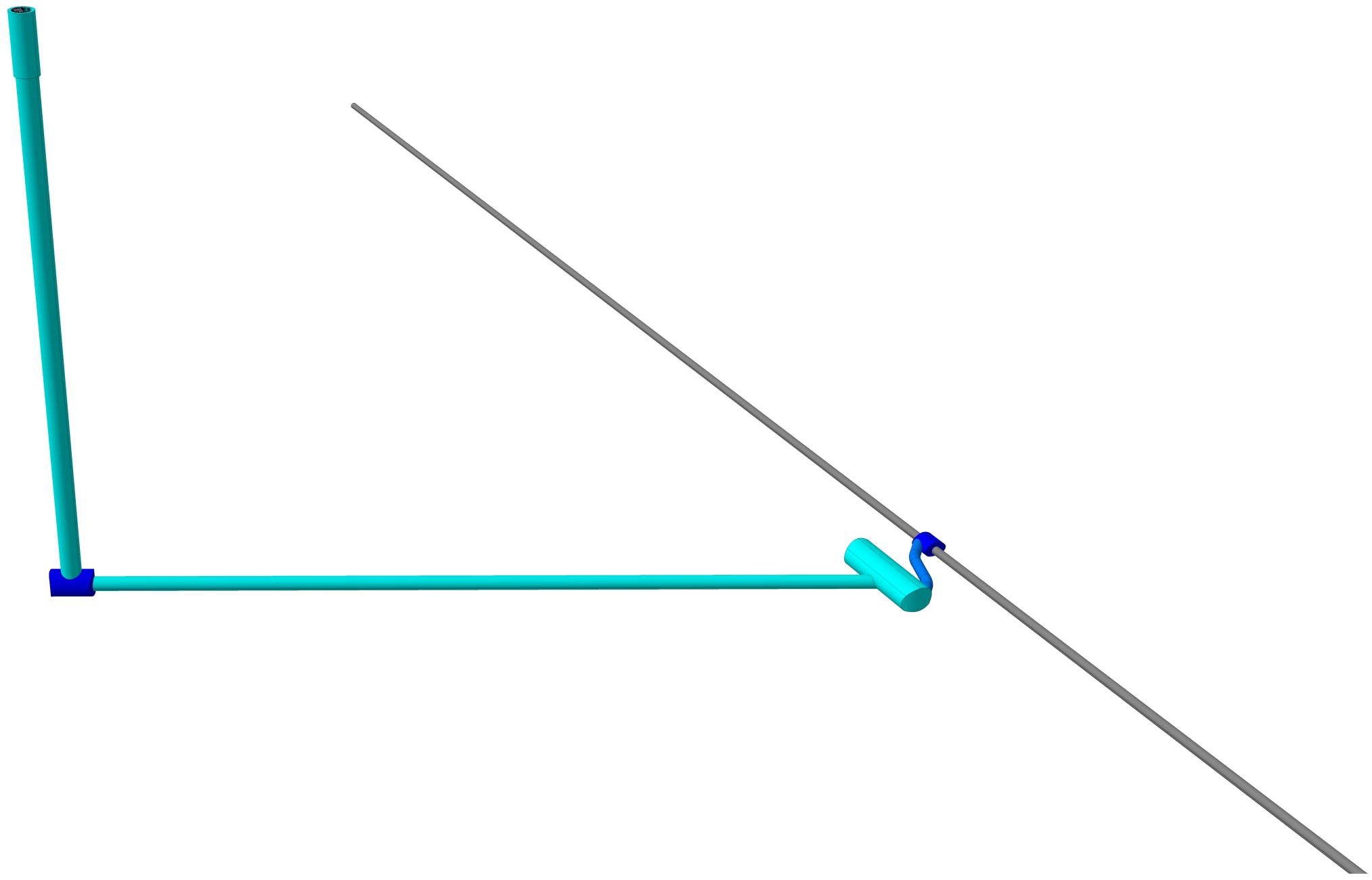

Fig. 8.25: FCC-ee point PF - general overview.

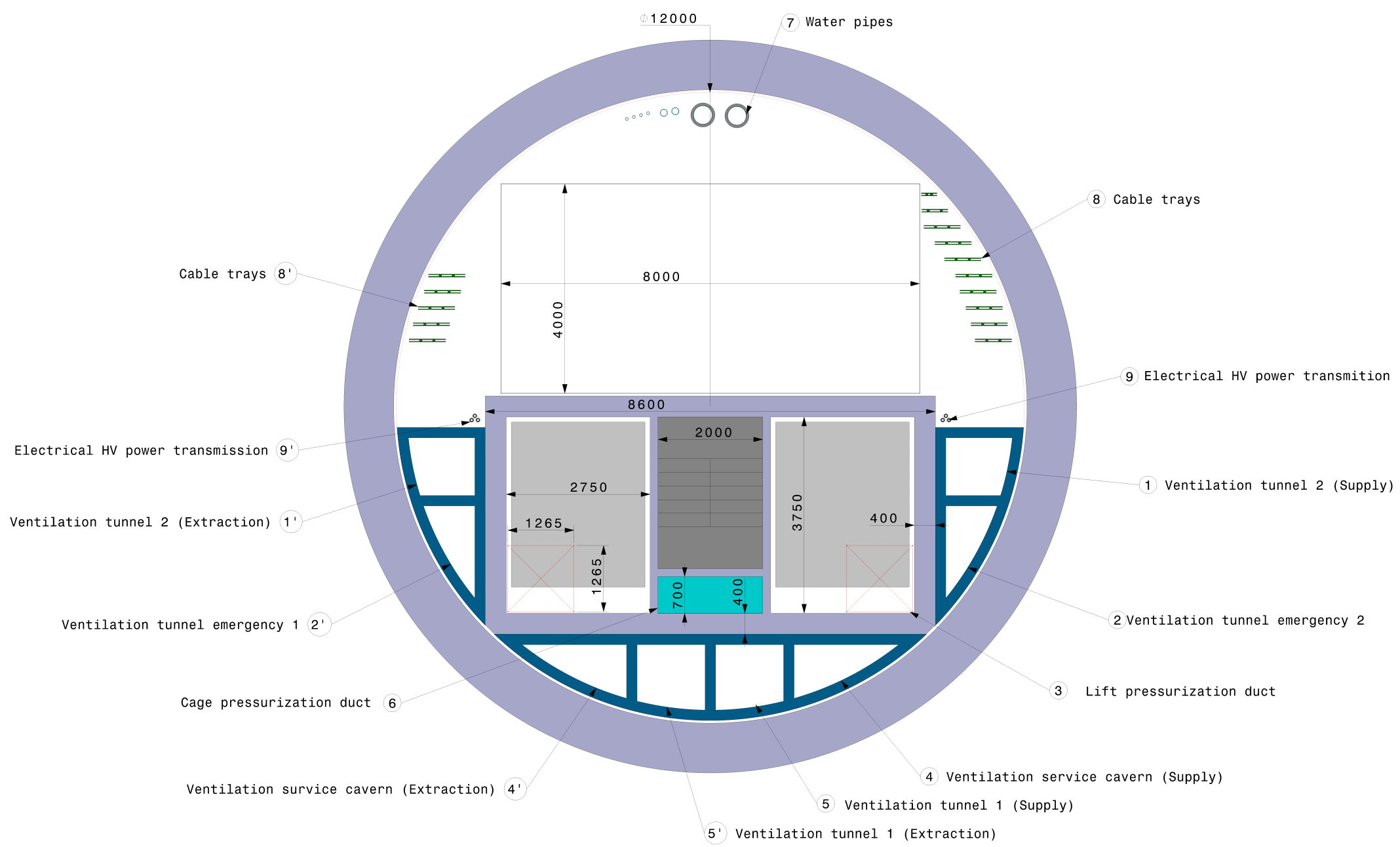

Fig. 8.26: FCC-ee point PF - shaft of the service cavern.

### 8.2.6 Integration of point PH and point PL (RF Systems)

The RF systems of the FCC-ee will be located in point PH for the collider and point PL for the booster (see Fig. 8.27 and Fig. 8.28).

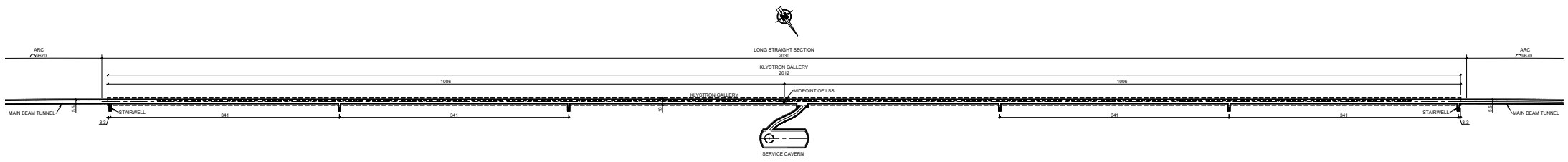

Fig. 8.27: FCC underground - civil engineering in point PH

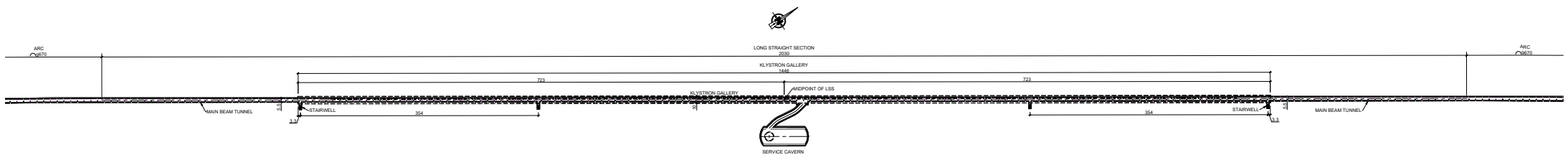

Fig. 8.28: FCC underground - civil engineering in point PL

***Configuration and layout of collider RF systems - point PH***

The figures included in this section are the results of the 3D integration studies for point PH.

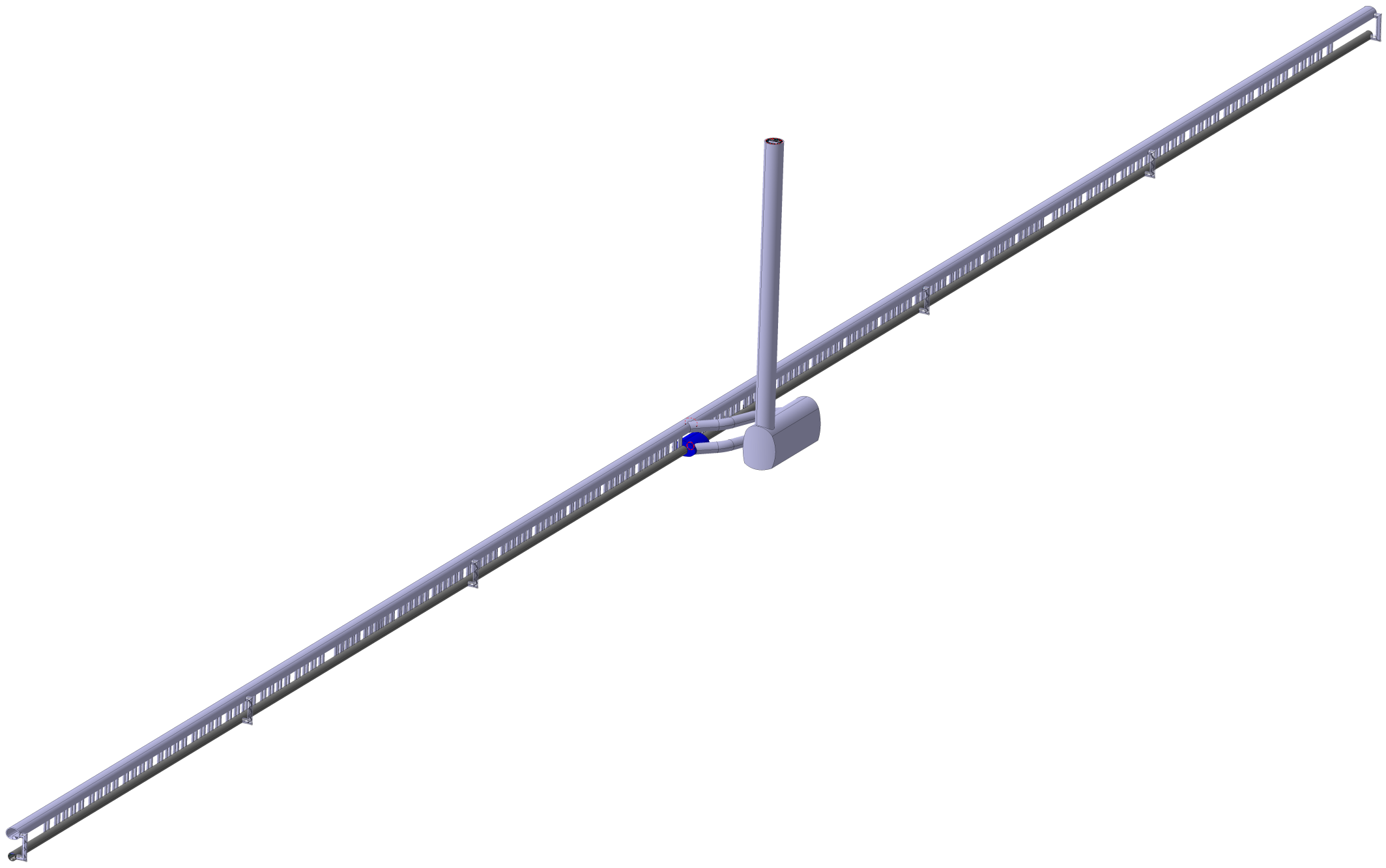

Fig. 8.29: FCC-ee point PH - general overview.

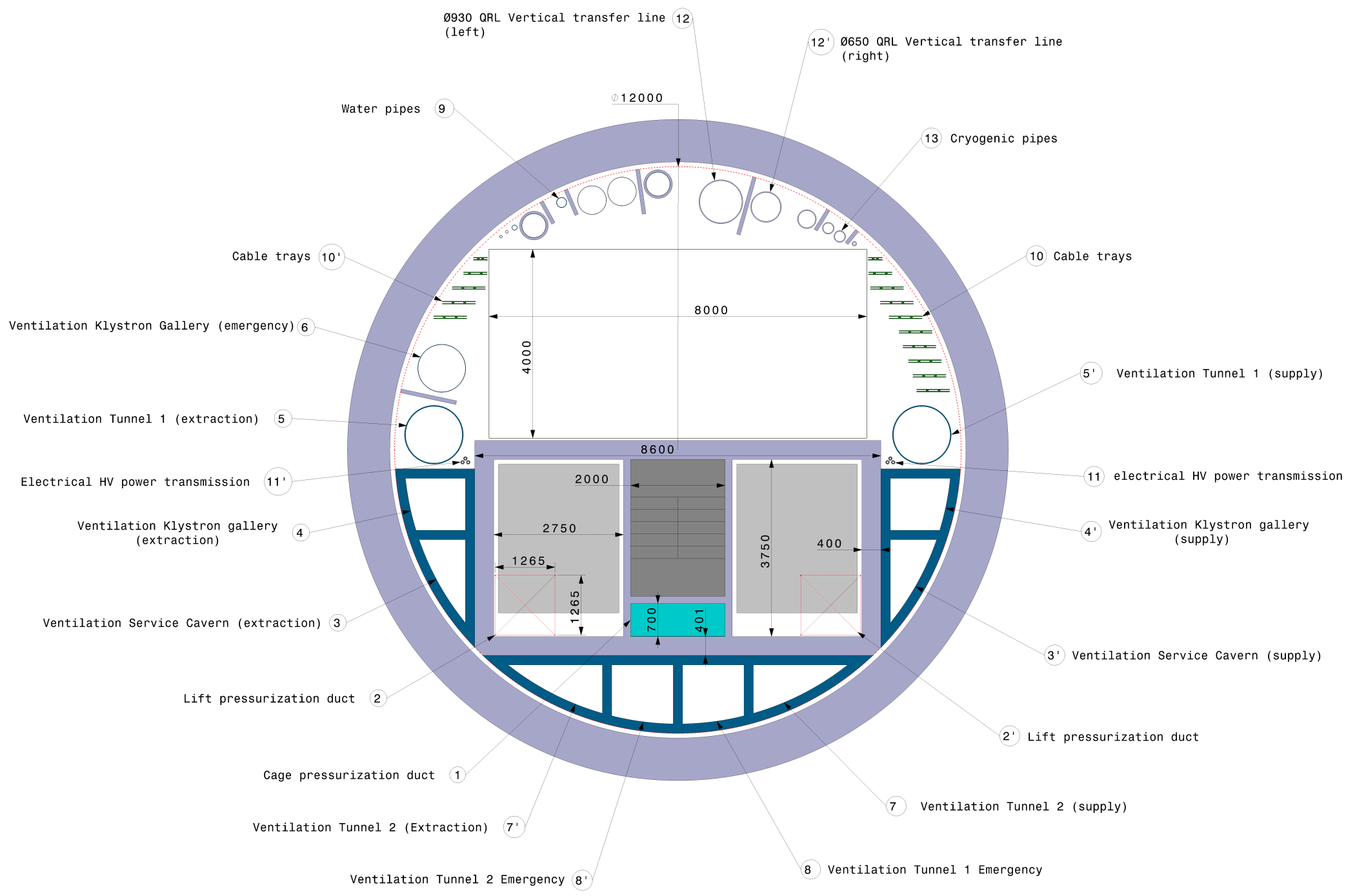

Fig. 8.30: FCC-ee point PH - shaft of the service cavern.

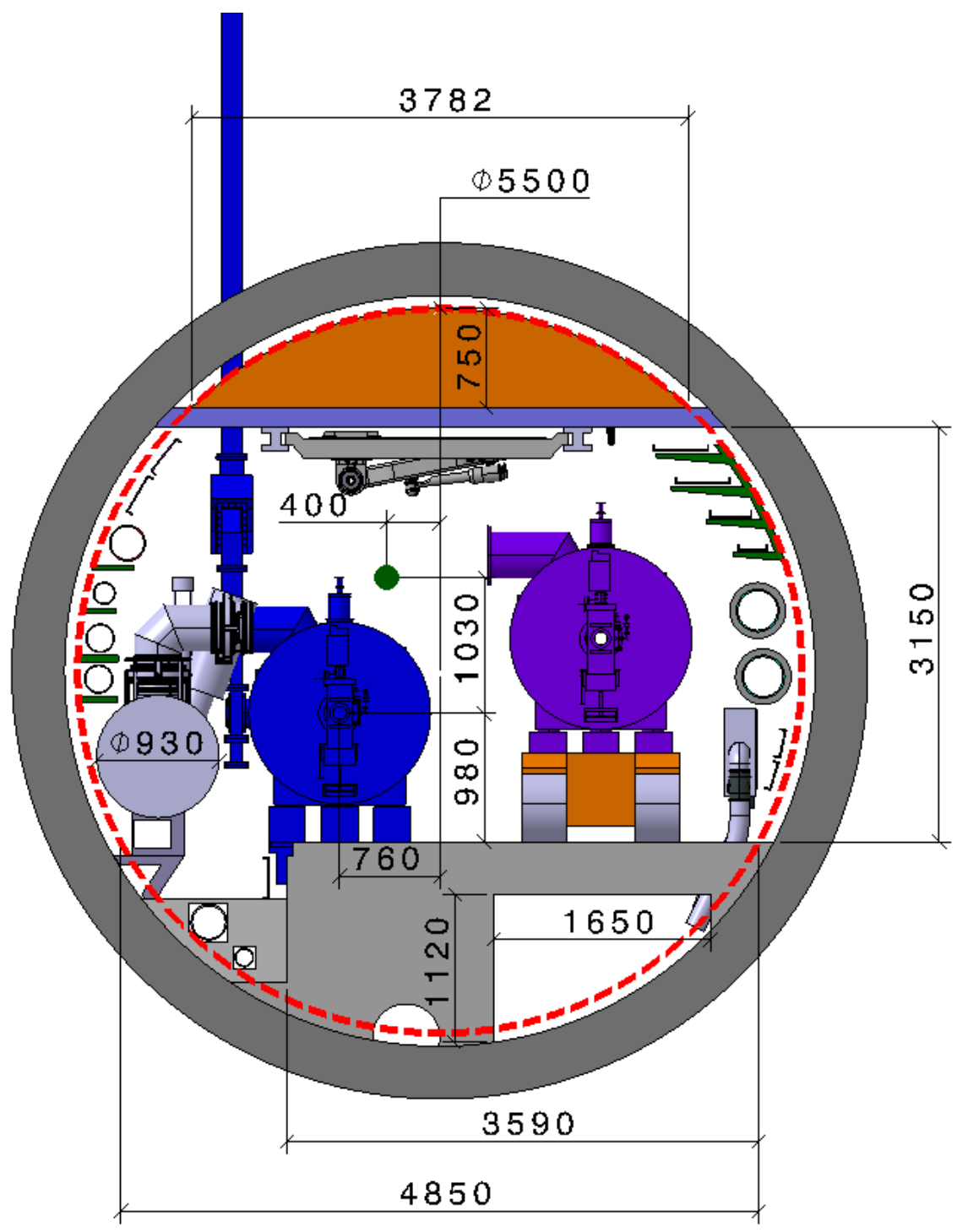

Fig. 8.31: FCC-ee point PH - tunnel cross-section with FCC 800 MHz cavity.

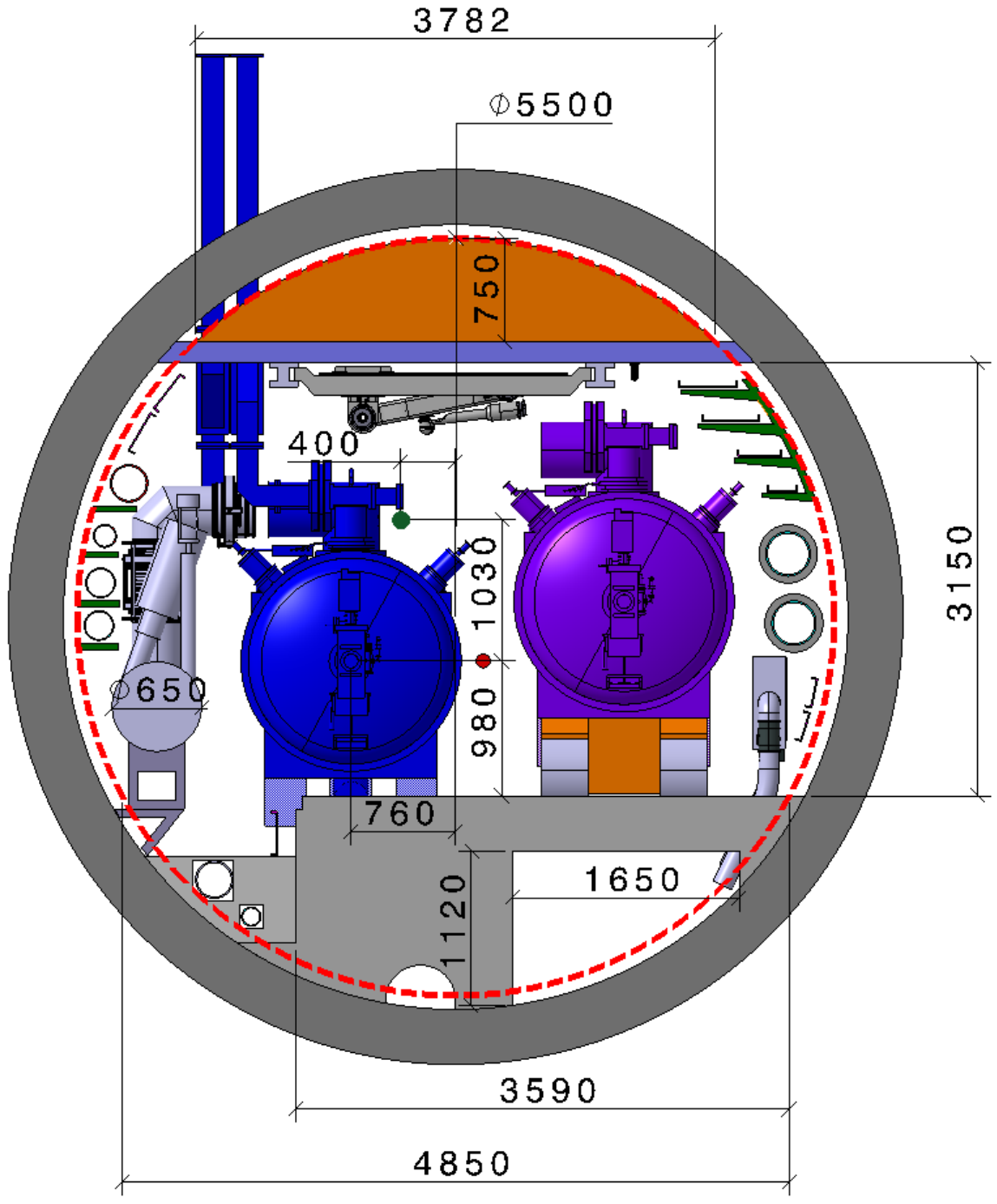

Fig. 8.32: FCC-ee point PH - tunnel cross-section with FCC 400 MHz cavity.

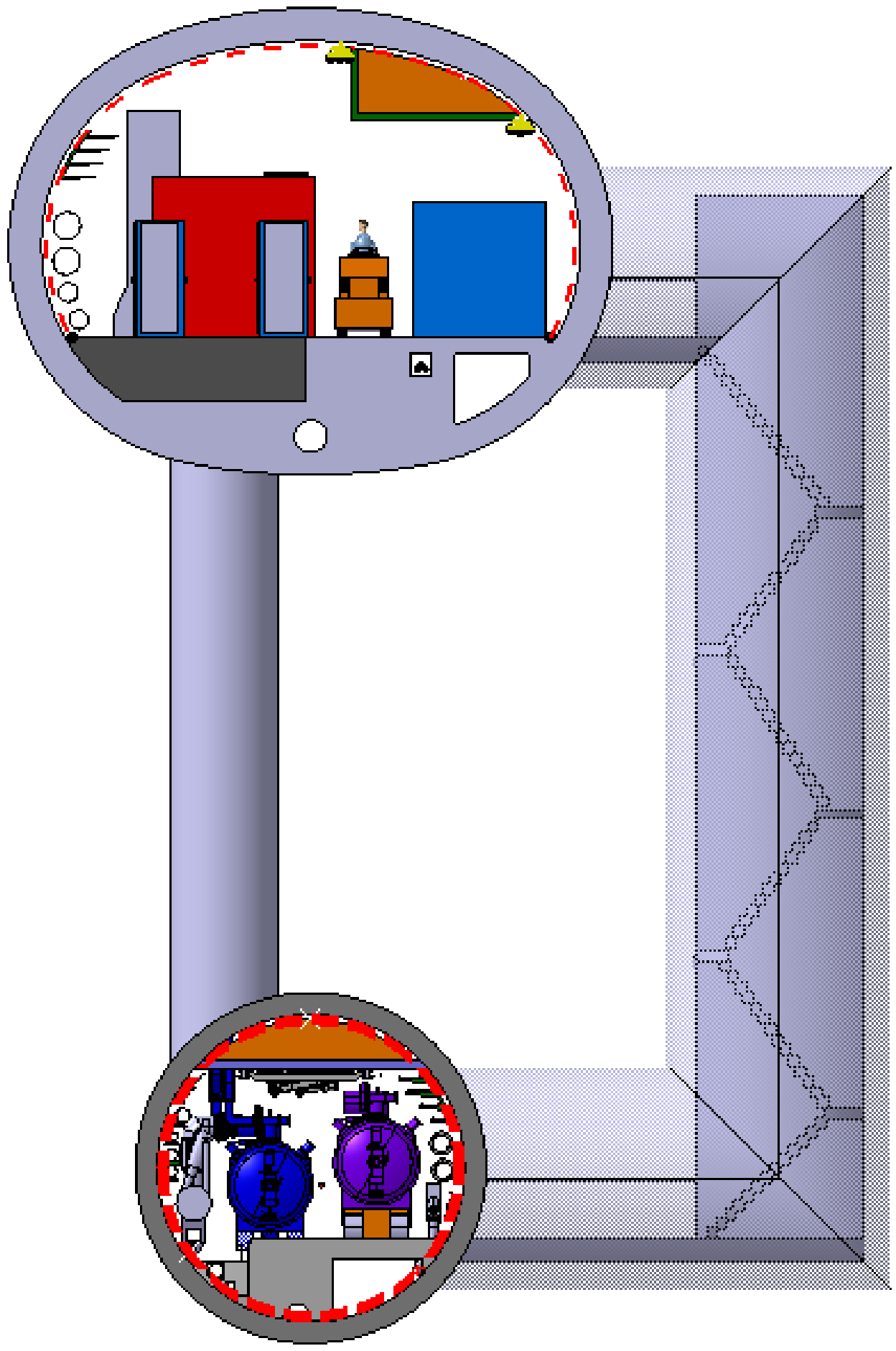

Fig. 8.33: FCC-ee point PH - tunnel and gallery cross-section.

For radiation safety, the distance between the klystron gallery and the machine tunnel is 10 m (Fig. 8.33, Fig. 8.37) and shielding is installed in the klystron gallery. The vertical cores house the waveguides and electrical cables. For fire protection there are partition walls every 400 m in the klystron gallery, and a smoke extraction system on the ceiling. There is a stairway connection between the klystron gallery and the machine tunnel every 280 m.

#### *Configuration and layout of booster RF systems - point PL*

The figures included in this section are the results of the 3D integration studies for point PL.

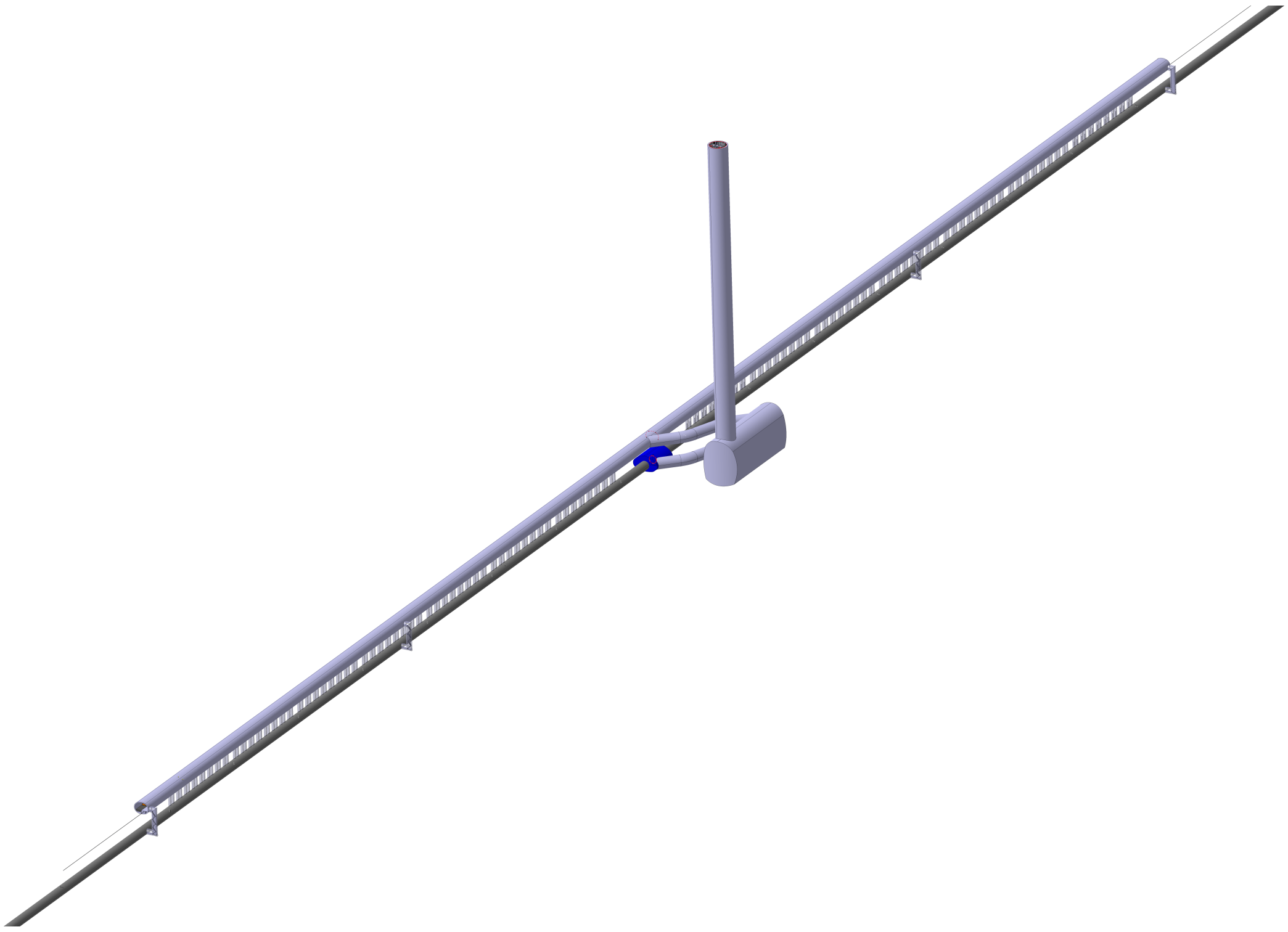

Fig. 8.34: FCC-ee point PL - general overview.

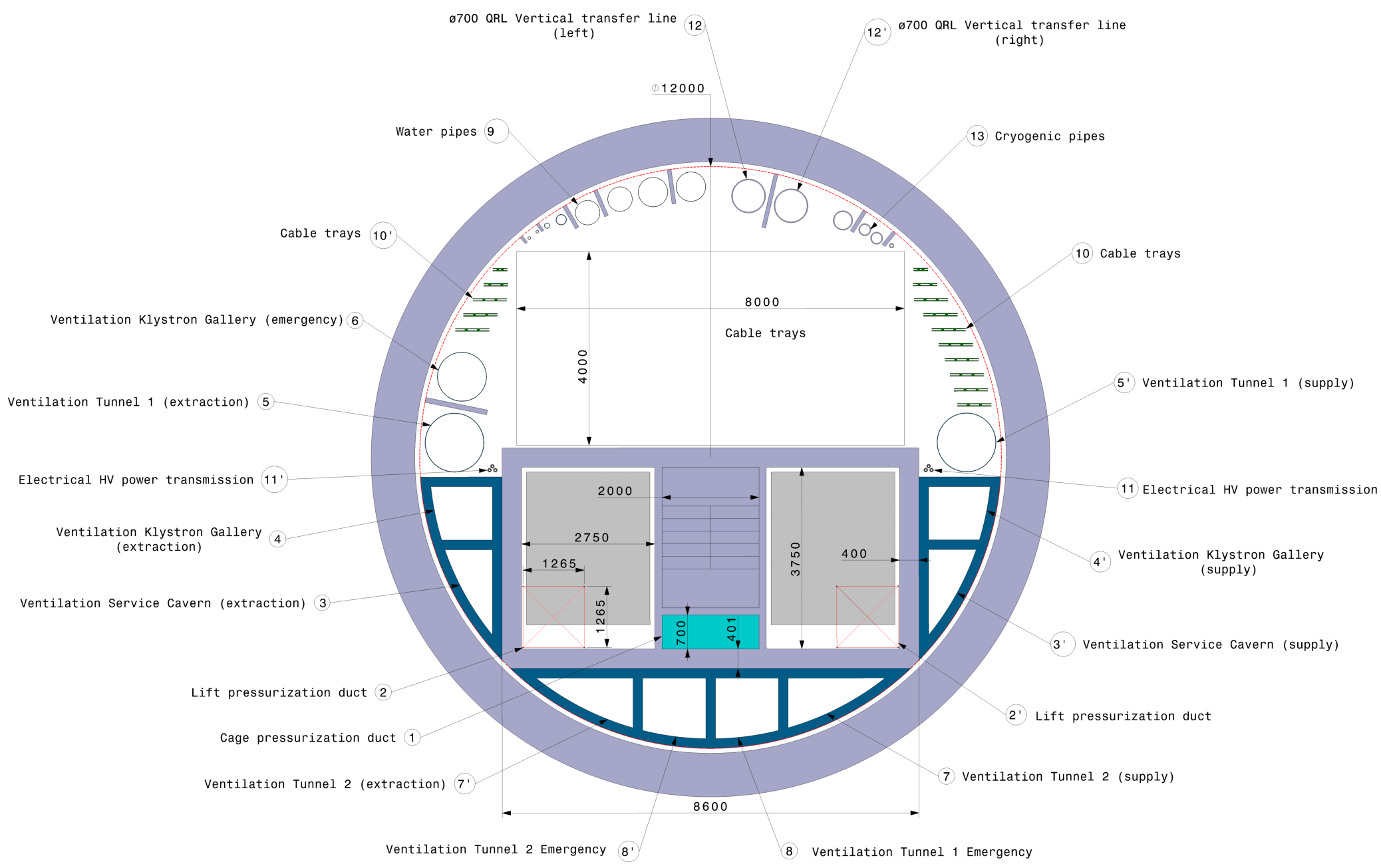

Fig. 8.35: FCC-ee point PL - shaft of the service cavern.

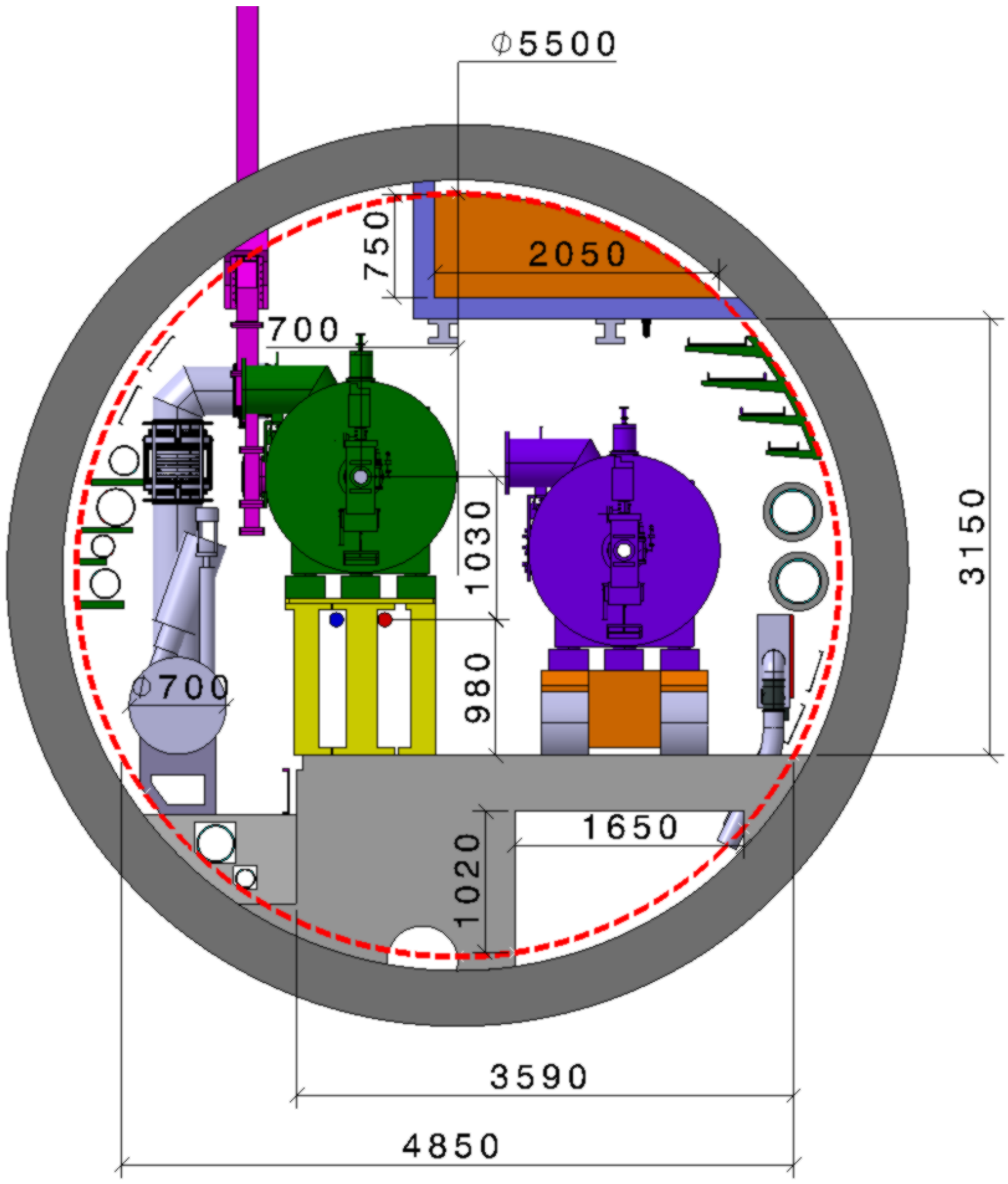

Fig. 8.36: FCC-ee point PL - tunnel cross-section with booster 800 MHz cavity.

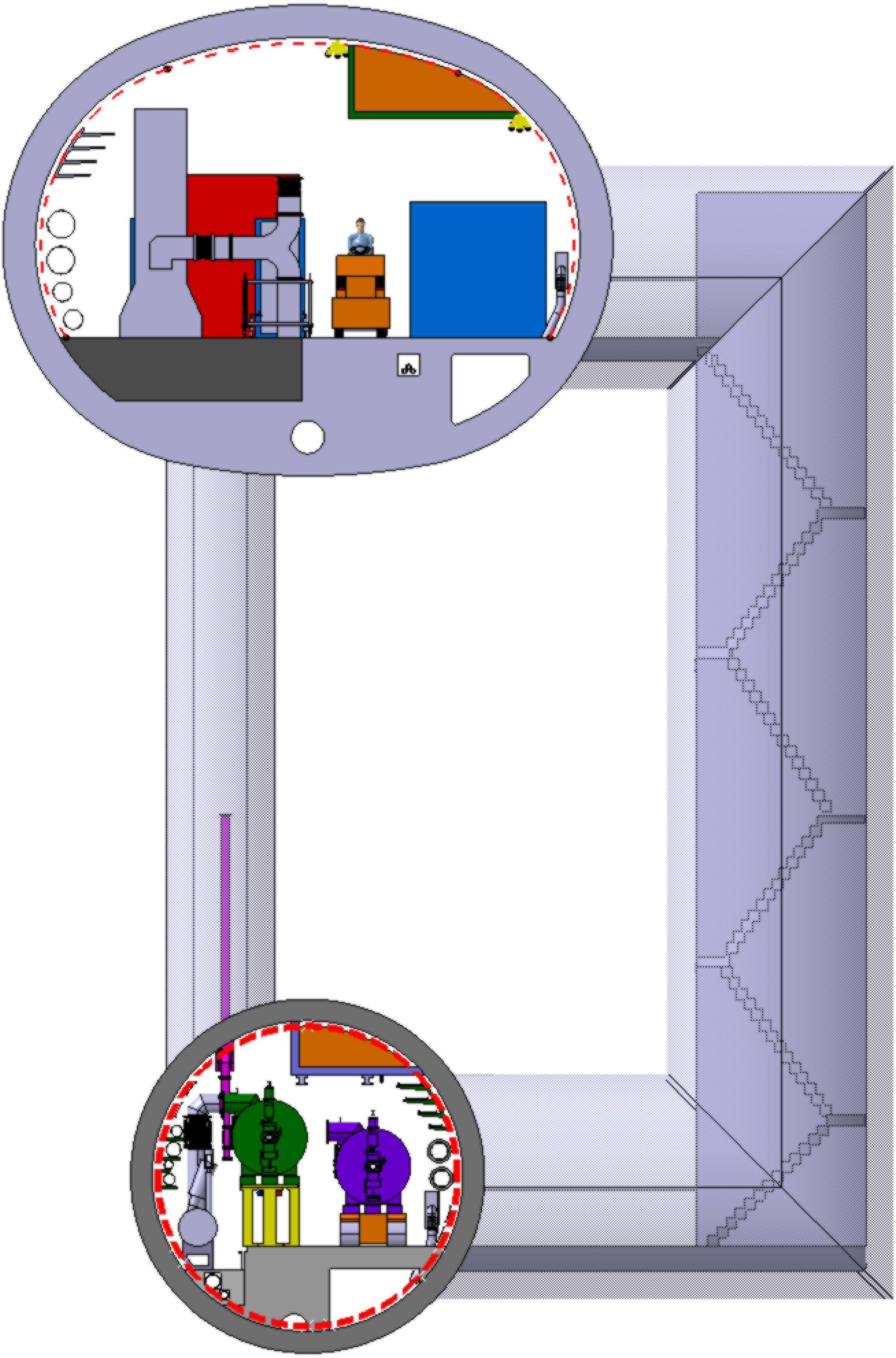

Fig. 8.37: FCC-ee point PL - tunnel and gallery cross-section.

### 8.2.7 Integration of the Arcs

The machine tunnels will house the FCC-ee accelerator and, in the following phase of operation, the FCC-hh accelerator. The civil engineering volume meet both accelerators needs.

Since the publication of the conceptual design report [13], the collider and booster have been lowered to give more space above the beam lines for safety requirements. After collecting updated requirements from all the stakeholders, the space requirements were reviewed with respect to installation and maintenance accessibility. A robot was added to the ceiling of the tunnel for better control of installation and maintenance of both magnets and support services.

The graphics included in this section are the results of the 3D integration studies for the arcs based on one of the baseline scenarios of a half-cell layout (see Fig. 3.81 and Fig. 3.82). The optimisation of the arc-supporting structures to maximise their performance, easing the installation and maintenance while minimising cost is facilitated by the FCC-ee Arc Half Cell Mock-up (see Section 3.10).

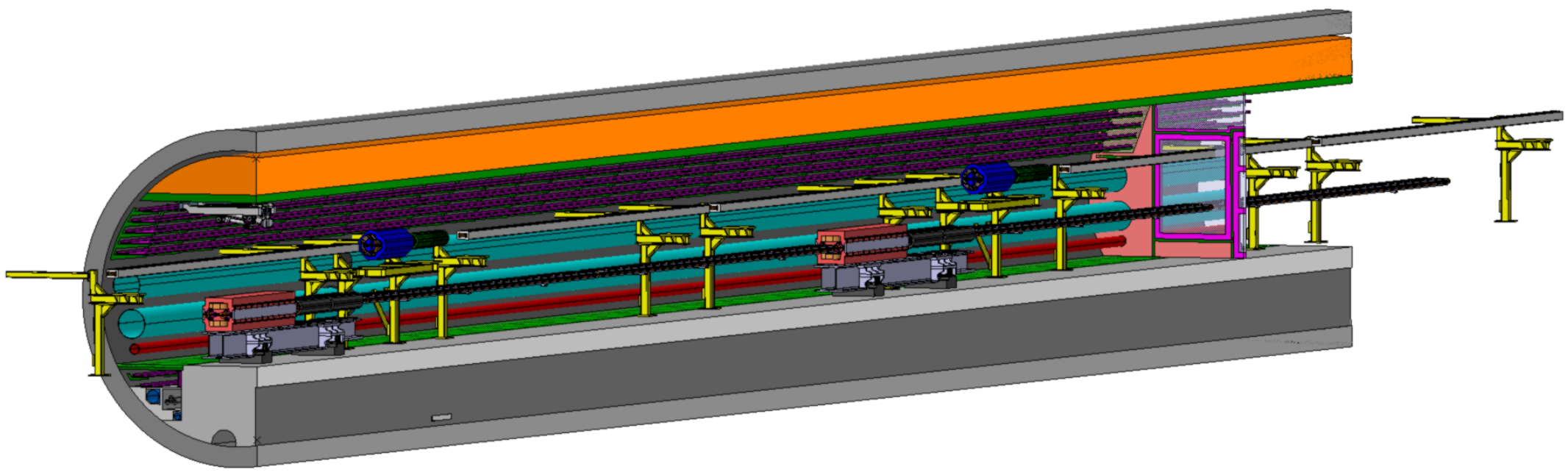

Fig. 8.38: FCC-ee layout in the arcs - longitudinal view.

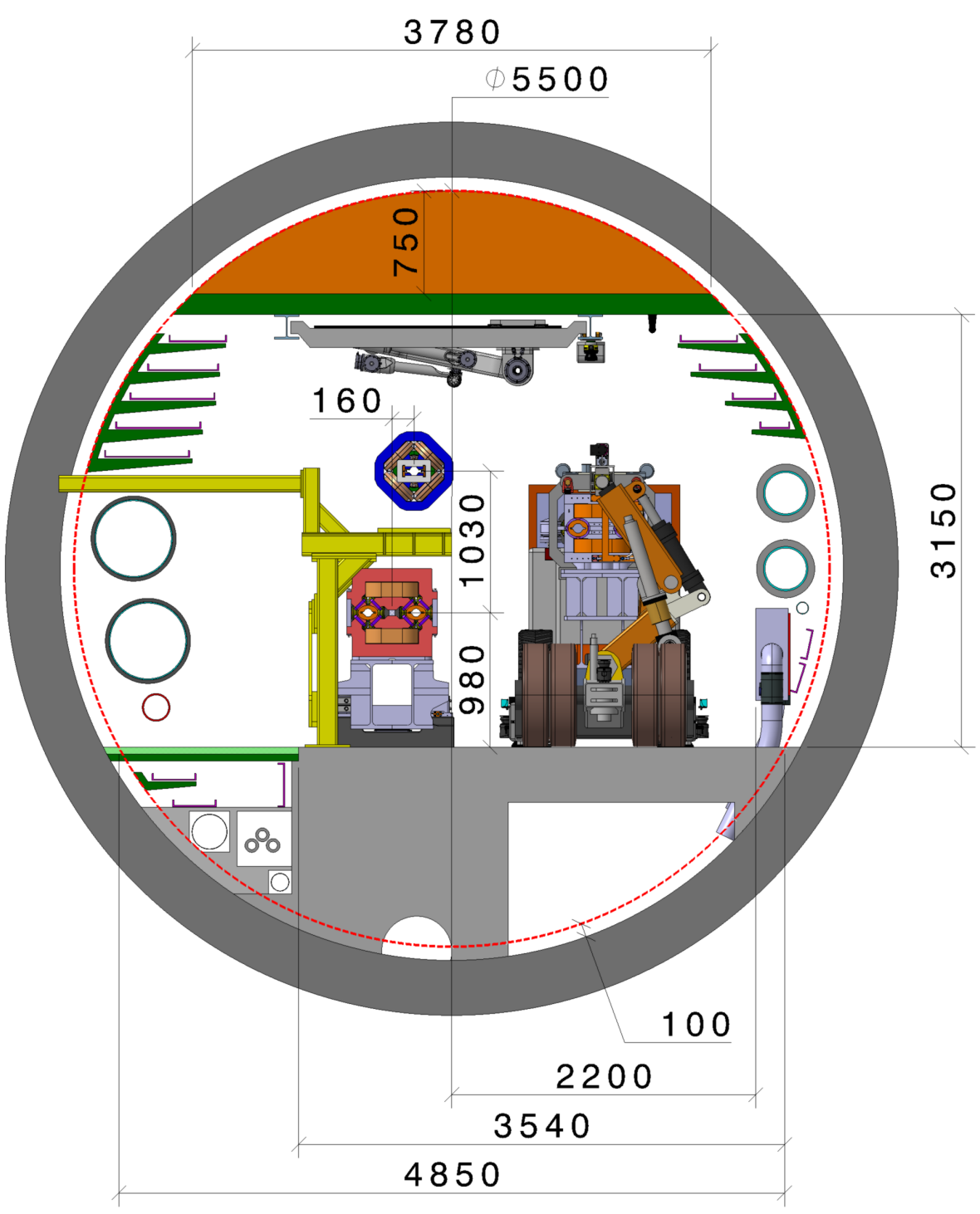

Fig. 8.39: FCC-ee layout in the arcs - cross-section view.

### 8.2.8 Integration of the Alcoves

In addition to the technical and experiment points, there will be 56 alcoves distributed around the ring (7 per arc). These alcoves will mainly house electronics and power supplies which need to be shielded from ionising radiation. The alcoves integrate the electrical equipment to supply a sector of the FCC, including systems installed in the alcoves themselves, plus the tunnel infrastructure like lighting, general services and a secured network. The entrance of the alcoves will also serve as a parking area for the transport vehicles (see Figs. 8.40 and 8.41).

The drawings included in this section are the results of the 3D integration studies for the alcoves.

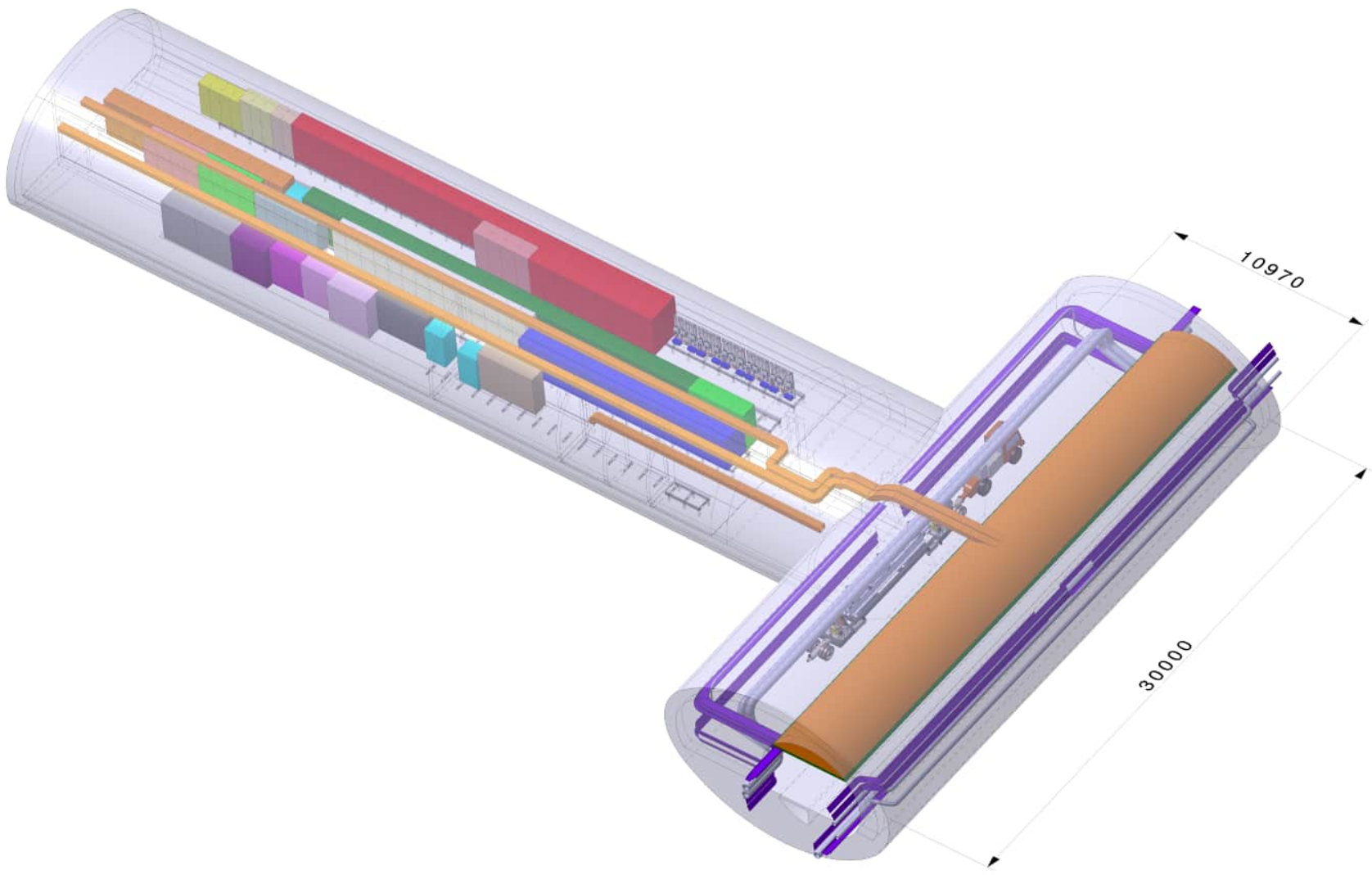

Fig. 8.40: FCC-ee alcove in the arcs with big parking area.

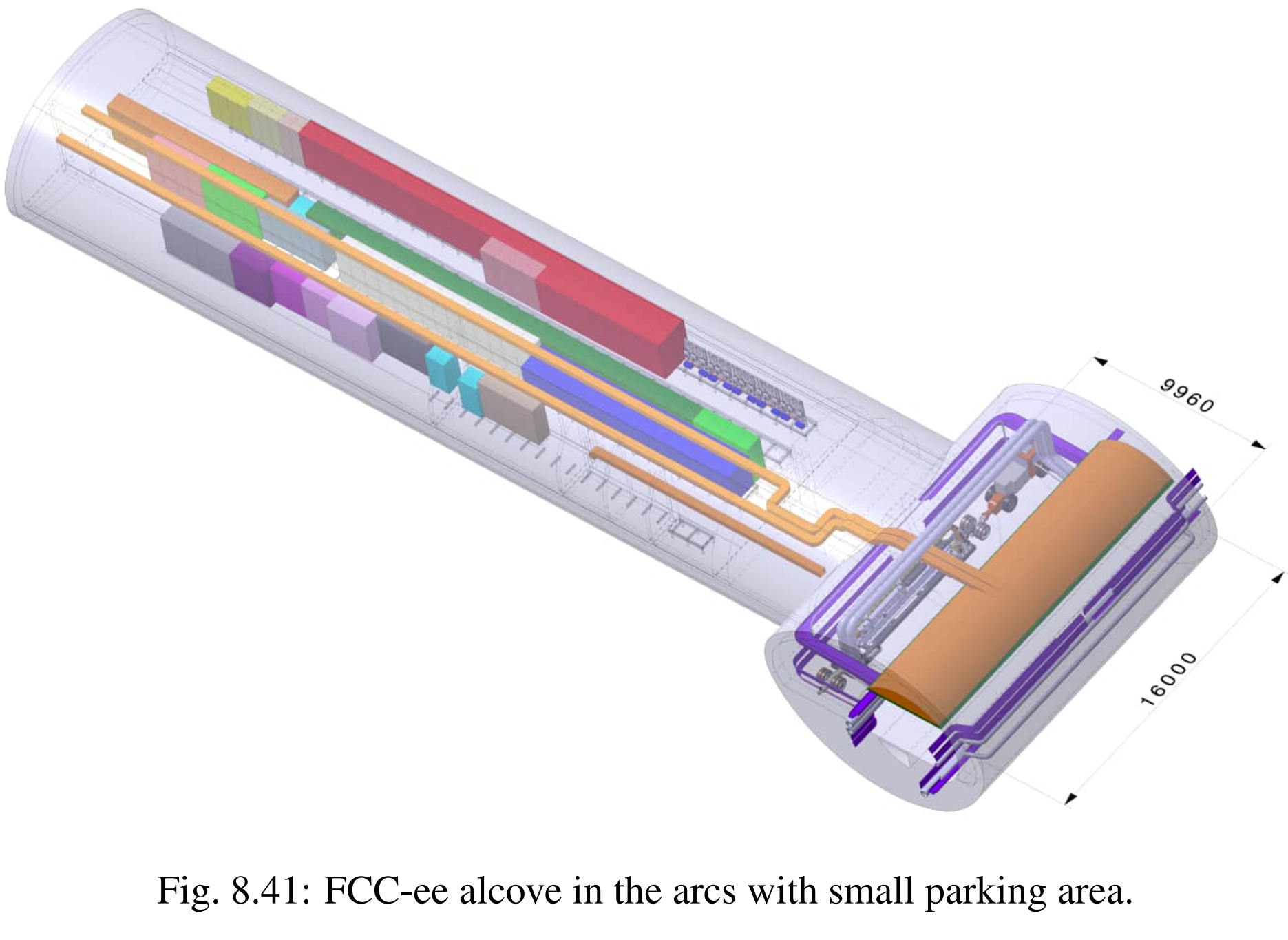

Fig. 8.41: FCC-ee alcove in the arcs with small parking area.

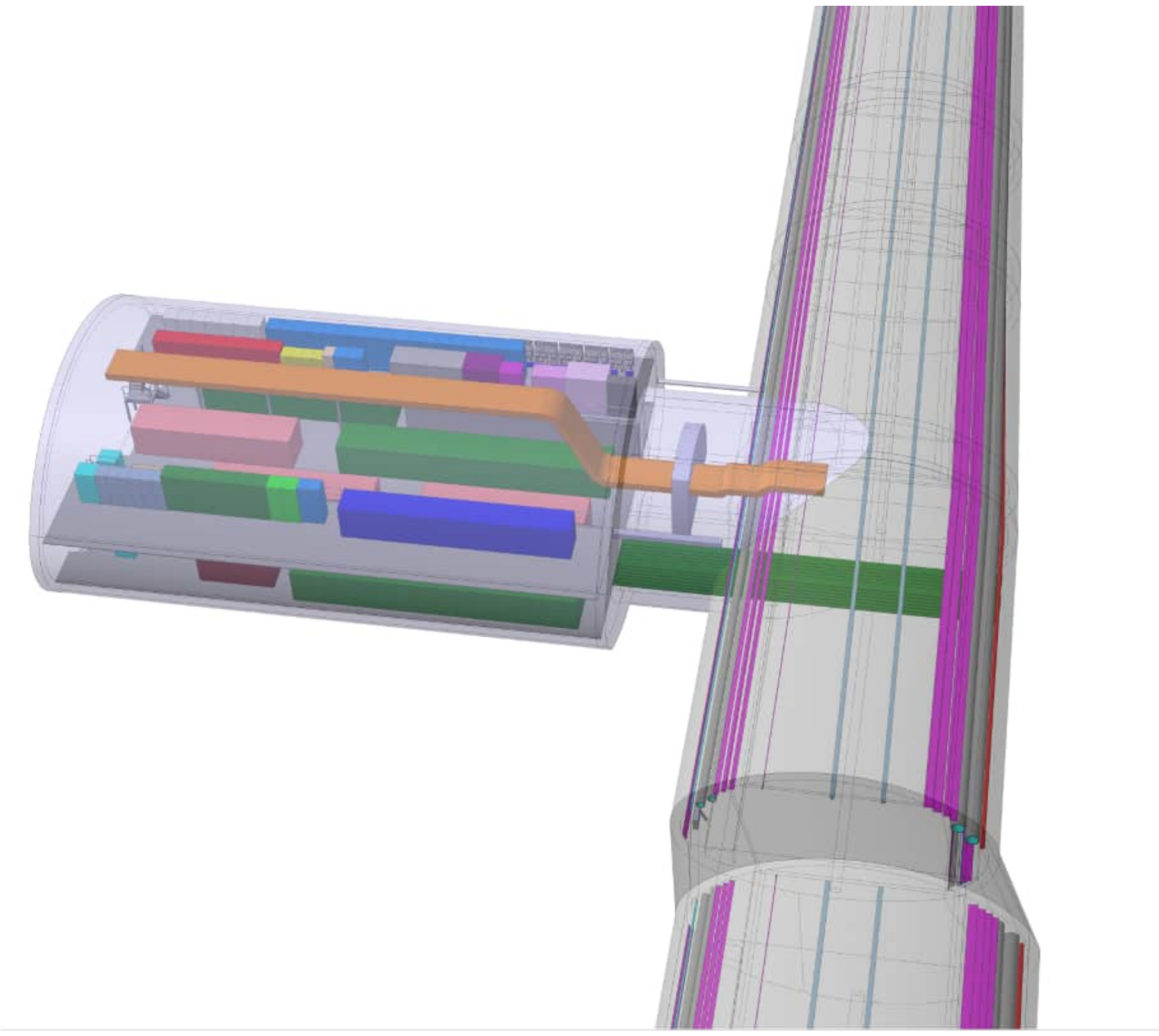

Fig. 8.42: FCC-ee alcove at point PA and PG.

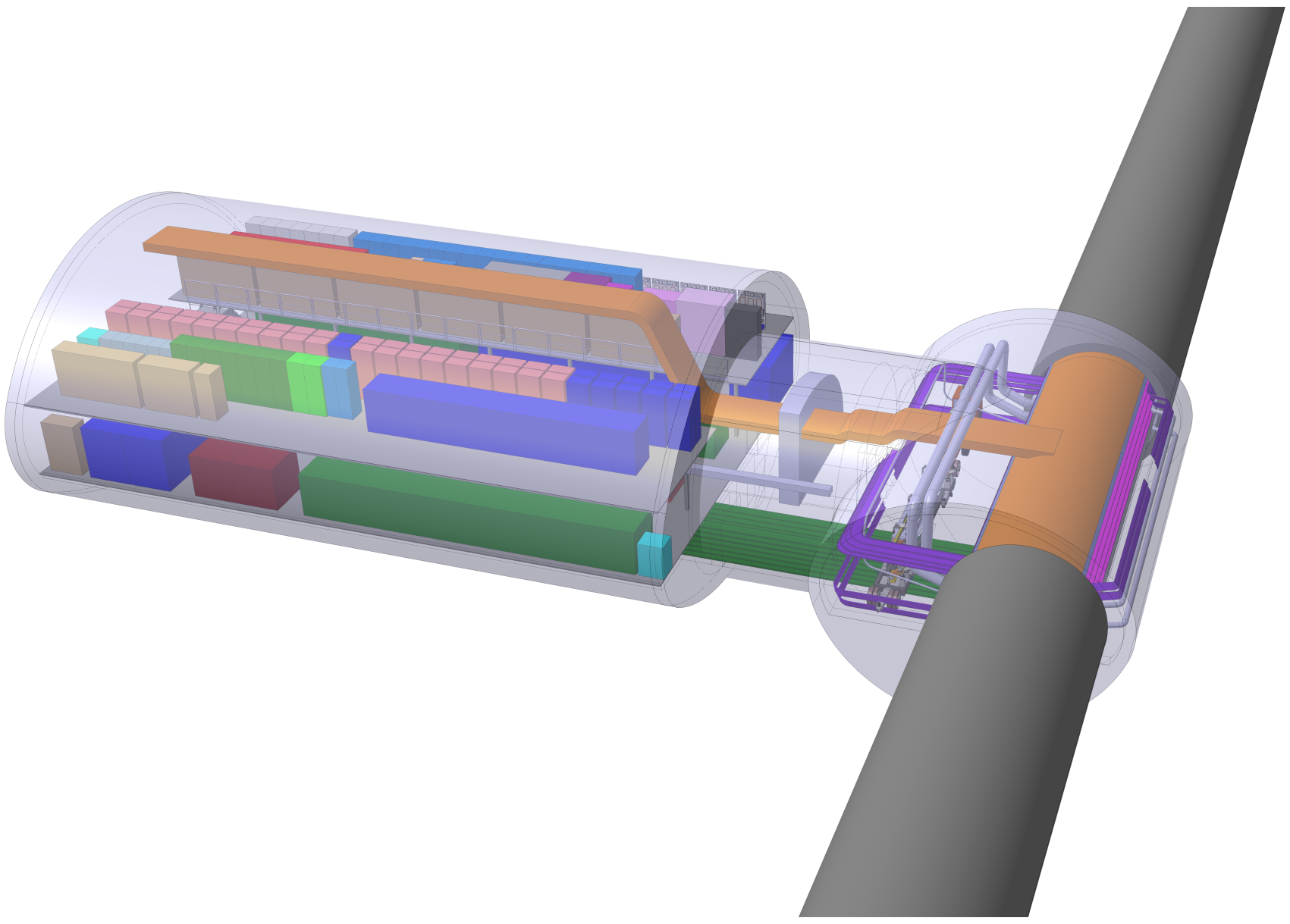

Fig. 8.43: FCC-ee alcove at point PB.

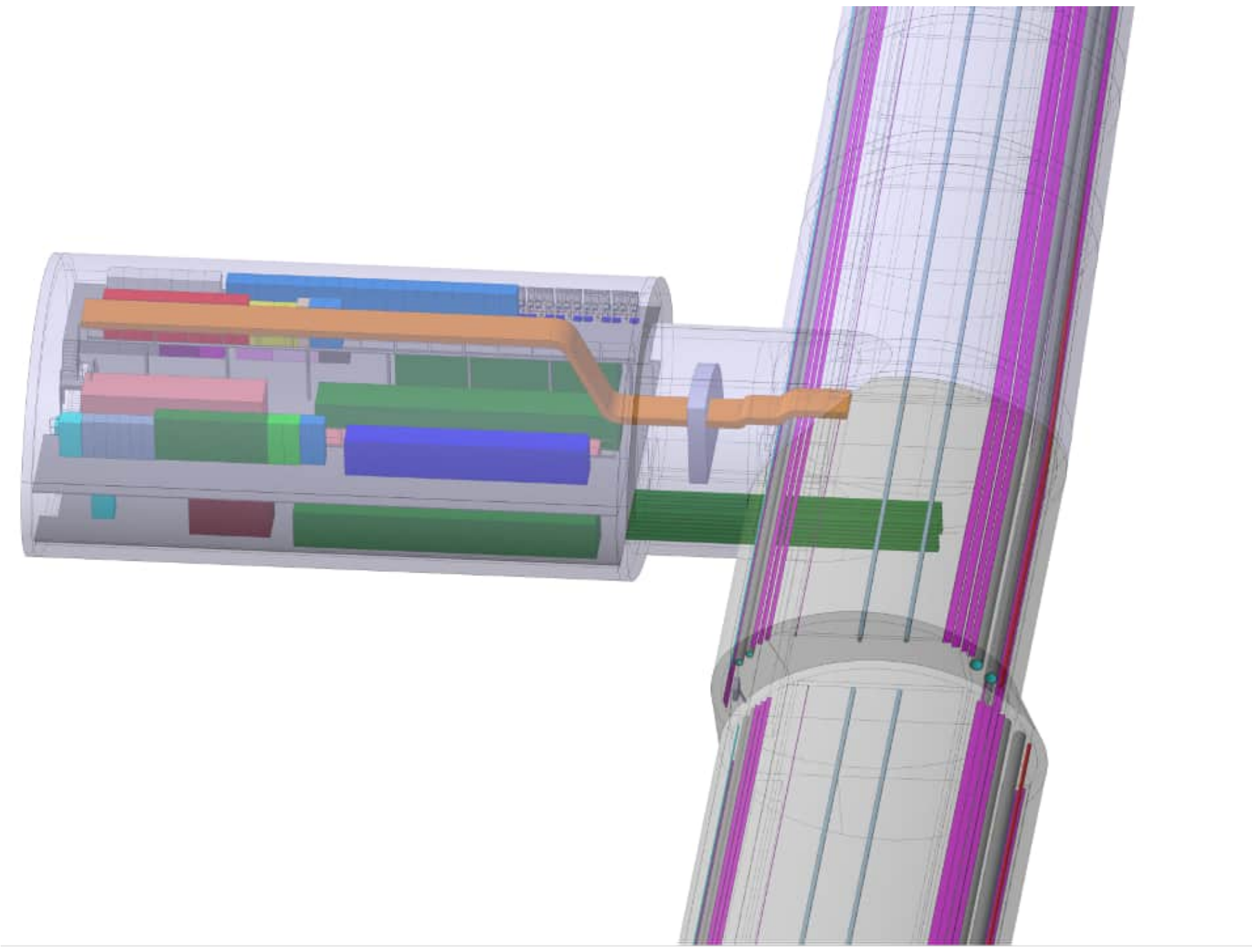

Fig. 8.44: FCC-ee alcove at points PD and PJ.

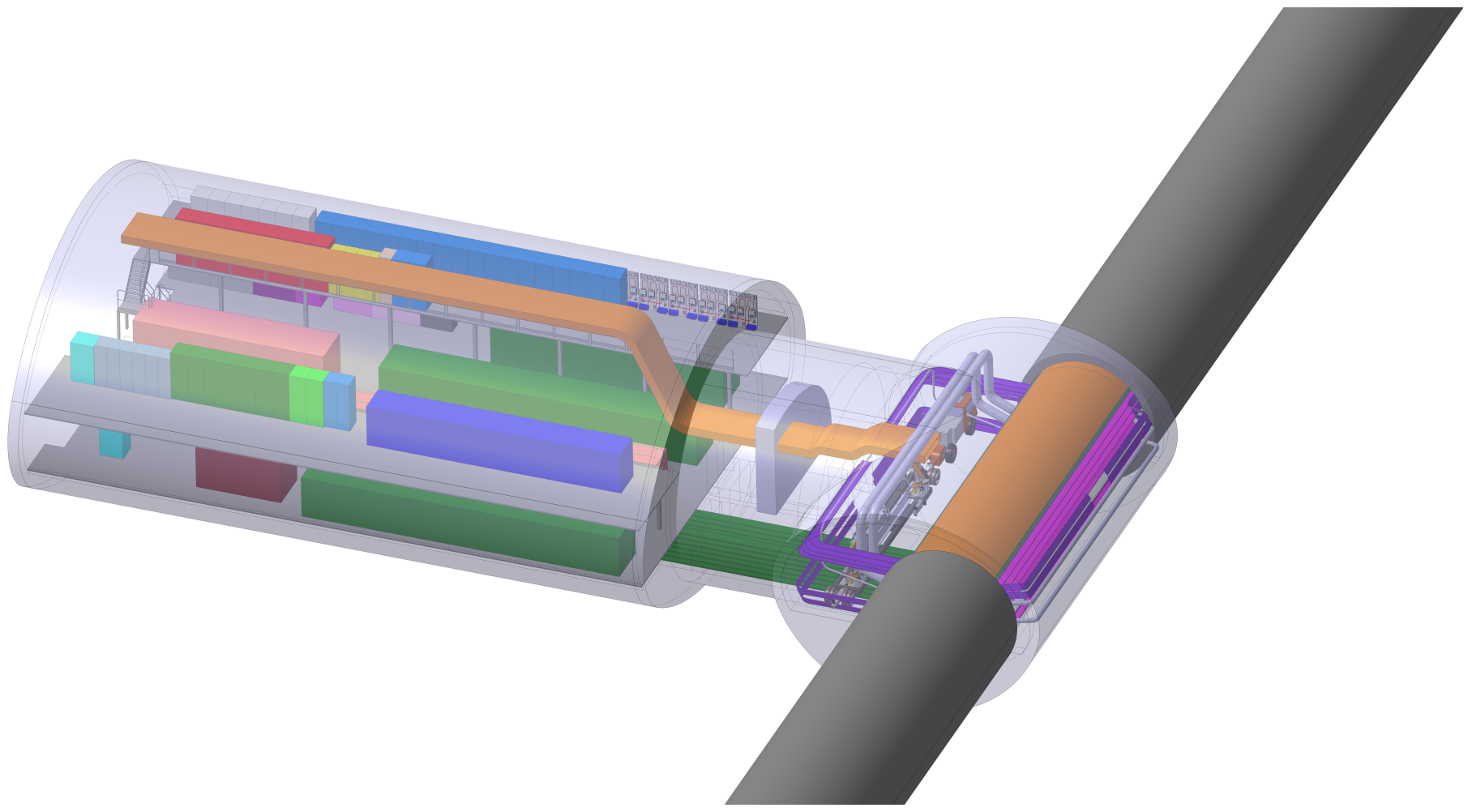

Fig. 8.45: FCC-ee alcove at point PF

The integration of power converters in the alcoves and the routing of cables in the machine tunnel present significant challenges due to space constraints and the high number of circuits involved. Power converters are installed in both small and big alcoves. Table 8.1 provides an overview of the number of converters in the alcoves and the number of cables in the cable trays.

There are five cables trays of varying sizes located above the booster supporting structure. The layout of the cables in the cable tray must respect the normal cabling rules. The proposed radiation shielding must reduce the radiation level to be low enough for the cables to survive for the entire lifetime of operation.

Table 8.1: Quantities of converter racks in the alcoves and cables in cable trays at the exit of an alcove.

| Magnet | Quantity of Racks in | | Cables | |
|---|---|---|---|---|
| | Big Alcove | Small Alcove | Size ($mm^2$) | Quantity |
| Collider Dipole | 27 | - | 1x500 | 2 |
| Collider Quadrupole (F and D) | 46 | - | 1x300 | 4 |
| Collider Sextupole (F and D) | 42 | 84 | 1x70 | 28 |
| Collider Dipole Tapering | 1 | 2 | 1x6 | 16 |
| Collider Quadrupole Tapering | 1 | 2 | 1x10 | 16 |
| Collider Horizontal Corrector | 2 | 3 | 1x10 | 60 |
| Collider Vertical Corrector | 2 | 3 | 1x16 | 60 |
| Collider Skew Quadrupole | 2 | 3 | 1x10 | 60 |
| **Collider Straight Section* | *18* | - | - | - |
| **Collider Injection* | *n/a* | *n/a* | *n/a* | *n/a* |
| Booster Dipole | 35 | - | 1x500 | 2 |
| Booster Quadrupole (F and D) | 86 | - | 1x185 | 8 |
| Booster Sextupole Focusing | 4 | - | 1x240 | 2 |
| Booster Sextupole Defocusing | 11 | - | 1x185 | 4 |
| **Booster Dipole Tapering* | *1* | *2* | *1x4* | *8* |
| **Booster Quadrupole Tapering* | *1* | *2* | *1x6* | *8* |
| **Booster Horizontal Corrector* | *3* | *6* | *1x25* | *36* |
| **Booster Vertical Corrector* | *3* | *6* | *1x25* | *36* |
| **Booster Quadrupole Corrector* | *3* | *5* | *1x16* | *30* |
| **Booster Skew Quadrupole* | *3* | *5* | *1x16* | *30* |
| **Booster Straight Section* | *23* | - | - | - |
| **Booster Injection* | *n/a* | - | *n/a* | *n/a* |
| **Total** | **314** | **123** | - | ***410*** |

*Magnet specifications not yet fully defined or inexistent – values extrapolated

### 8.2.9 Next phase for the integration studies

The integration studies have been made with the pre-design of volumes and will evolve with detailed technical design in the next phase of the FCC studies. Optimisation of the 3D integration studies with respect to the requirements from work packages will continue for all the areas following the identified technical and space needs:

- Optimisation of the large experiment points PA and PG [375] [376] and small experiment points PD and PJ [377], integrating more detailed detectors design in the experiment caverns,
- Optimisation of the technical point PB [378], integrating more detailed mechanical design of the accelerators and transfer lines,
- Optimisation of the technical point PF [379], integrating more detailed mechanical design of the collimation systems,
- Optimisation of the technical point PH and PL [380] [381], integrating the more detailed mechanical design of the radiofrequency and cryogenic systems and enlarging the space for transport and safety in the passage part of the tunnel,
- Optimisation of the arc half-cell [382], including the optimisation of the arc-supporting structures studied by the FCC-ee Arc Half Cell Mock-up (Section 3.10) to maximise their performance, ease the installation and maintenance while minimising cost.

– Optimisation of the alcoves [383], updating the control and powering racks layout with respect to technical specification evolutions.

## 8.3 Cooling and ventilation

### 8.3.1 Introduction

The cooling systems for the FCC mainly concern several water systems for the machine and its infrastructure, including accelerator and detector equipment (electronic racks, cryogenic plants, water-cooled magnets, RF equipment, etc.). Raw water will be used to fill the cooling circuits in some firefighting systems and as make-up water for the cooling towers. Chilled water is needed to extract the heat load from air in ventilation systems. Demineralised water would be employed to cool sensitive equipment like magnets, synchrotron radiation absorbers, and power converters.

The ventilation systems for the FCC should ensure the necessary conditions of humidity and temperature in the tunnel during operation, providing a supply of fresh air for personnel working within the facilities and heating for the working environment. The extraction systems must be capable of purging the air in the tunnel before access is allowed, as well as removing smoke and gases during an emergency.

Other systems comprise several and varied applications, some unrelated to cooling and ventilation. The compressed air systems for the FCC would provide compressed air to end clients requiring it, including the dampers employed in the FCC tunnel. Effluent water from the cooling towers and filters would be concentrated in salts and returned to a water treatment plant. Drinking water would be needed for sanitary purposes. Finally, a drainage network would evacuate the water drained from the surface facilities and underground areas.

***Cooling systems***

*Water supply*

The FCC accelerator would need to be supplied with water at the eight surface points. The main water consumers are the cooling towers, but water is also needed for the fire-fighting systems, the demineralised water production plant and all the cooling plants in the surface buildings as well as in the tunnels and caverns.

As a baseline, raw water is supplied from point PA to the rest of the accelerator, as represented in Fig. 8.46. Lake Geneva, located near PA, also serves as a water source for other accelerators within the CERN complex. Pipes running through the FCC tunnel sectors distribute water to one or more contiguous points: one branch supplies water from point PA to points PB, PD, PF, and PG, while another branch supplies water from point PA to points PL, PJ, and PH. Table 8.2 presents the approximate maximum raw water flow rates required, along with the effluent water generated at the cooling towers and filters.

An alternative design to this baseline could include the Rhône and Arve rivers as potential water sources for the FCC. In that case, point PA would feed points PL and PB with water from Lake Geneva, point PD would feed points PF and PG with water from the Arve river and point PJ would feed point PH with water from the Rhone river. This would effectively decrease the piping and pumping system sizes.

Firefighting water is available across each surface site, as well as in the service and experiment caverns. An additional flow rate of 240 $m^3/h$ per point is allocated for firefighting purposes. Each surface site is equipped with water tanks capable of storing the estimated water supply needed for one hour.

In addition to the cooling towers and firefighting systems, the demineralised and chilled water circuits also require periodic refilling, with an estimated maximum demand of 0.5 $m^3/h$.

Water transfer between points is managed through a pumping station at point PA and booster pumps positioned at the bottom of the shaft at point PF, as illustrated in Fig. 8.47, which also depicts the approximate surface and underground elevations, along with shaft depths.

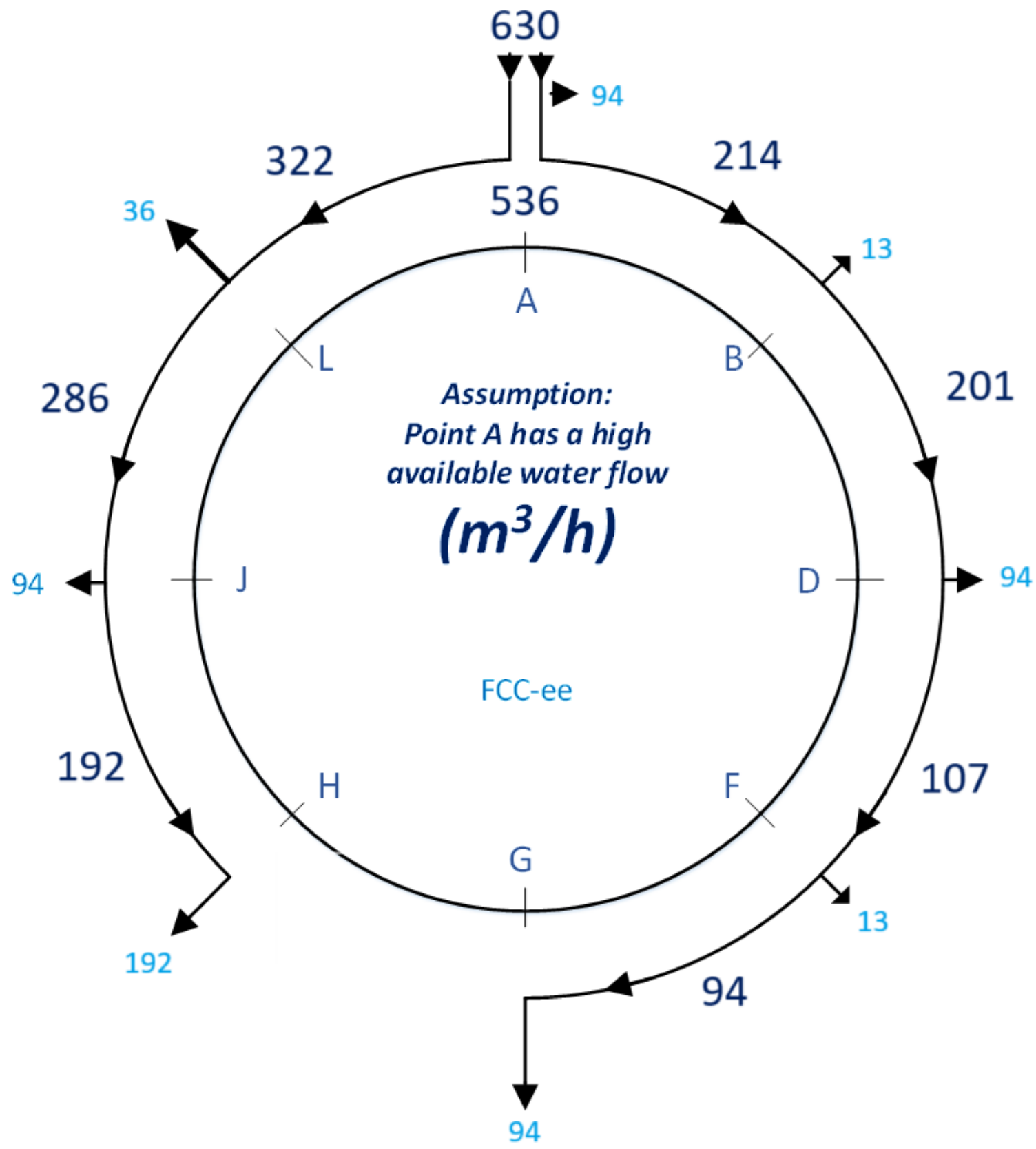

Fig. 8.46: Distribution of raw water coming from point PA, Lake Geneva.

Table 8.2: Maximum make-up and effluent water of cooling towers ($m^3$/h).

| Point | PA | PB | PD | PF | PG | PH | PJ | PL |
|---|---|---|---|---|---|---|---|---|
| Make-up water for cooling towers | 94 | 13 | 94 | 13 | 94 | 192 | 94 | 36 |
| Effluent water from cooling towers and filters | 13 | 2 | 13 | 2 | 13 | 26 | 13 | 5 |

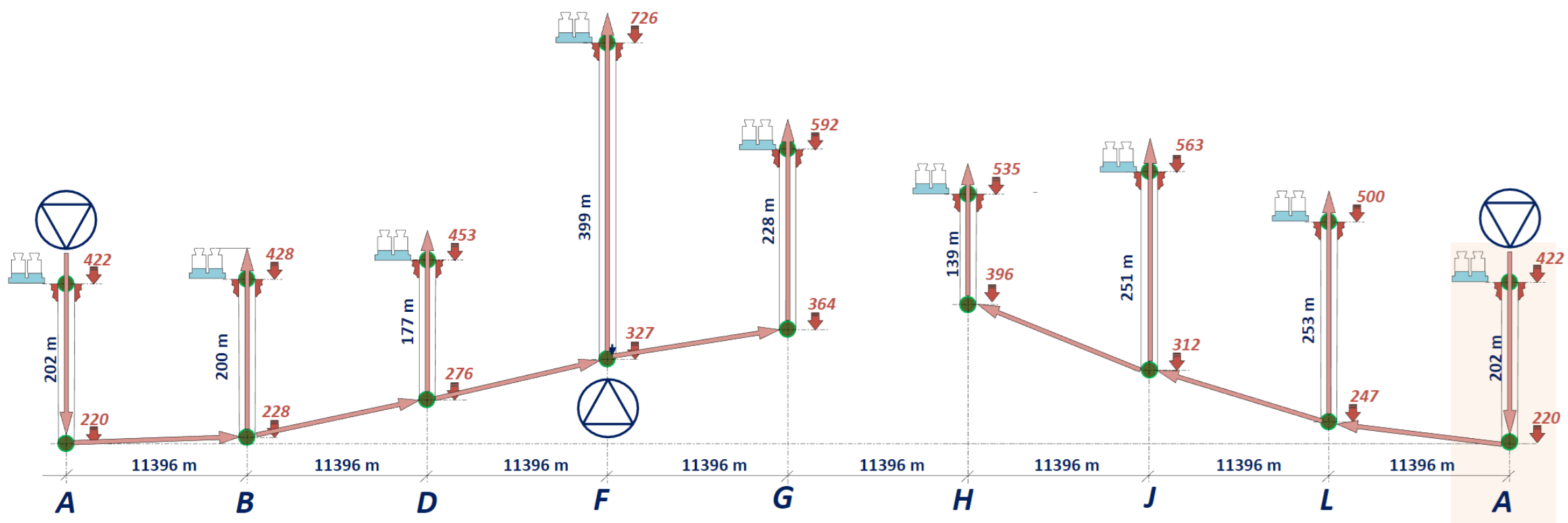

Fig. 8.47: Pumping stations and booster pumps of the raw water system.

***Primary cooling circuits.***

The cooling towers will remove most of the heat generated by the accelerator equipment, the detectors and the technical areas. There will be a set of field erected, open wet cooling towers per surface site.

The evaporative cooling towers are of the mechanical-induced draft counter-flow type. Ventilation fans are positioned at the top of the cells, ensuring easy access for maintenance. Each tower consists of multiple cells with identical cooling capacity, with one cell per tower reserved for backup purposes (N+1 redundancy). The towers are designed to operate with water treated with anti-corrosion and biocide chemicals. The cooling towers comply with the following parameters and conditions:

- Outside air wet bulb temperature: 21°C.
- Cooling tower inlet temperature: 40°C.
- Cooling tower outlet maximum temperature: 25°C.

The cooling power to be installed at each point has been determined based on the users' cooling requirements. Certain equipment, particularly cryogenic systems, will be cooled directly by the primary circuit, while other equipment will be connected to the primary systems via heat exchangers (secondary circuits). In most cases, the secondary circuits will operate in a closed loop using demineralised water.

Table 8.3 gives an overview of the cooling requirements by type of equipment: each row represents a different type of equipment, except for the 'Underground' row, which groups all equipment cooled from the same primary circuit and is located underground. In addition, Table 8.4 provides the cooling tower complex capacity per point, including the backup cells. Figure 8.48 shows the cooling tower complexes per point, including the backup cells.

Table 8.3: Cooling demands (MW) of the primary circuit.

| **Point** | **PA** | **PB** | **PD** | **PF** | **PG** | **PH** | **PJ** | **PL** |
|---|---|---|---|---|---|---|---|---|
| Cryogenics | | | | | | 34.0 | | 10.0 |
| Experiments | 0.5 | | 0.5 | | 0.5 | | 0.5 | |
| General Services | 2.0 | 2.0 | 2.0 | 2.0 | 2.0 | 2.0 | 2.0 | 2.0 |
| Power Converters | | | | | | 4.5 | | 0.1 |
| Chilled Water | 5.8 | 5.2 | 5.8 | 5.2 | 5.8 | 11.3 | 5.8 | 5.8 |
| Underground | 42.5 | 1.0 | 42.5 | 1.0 | 42.5 | 48.9 | 42.5 | 2.7 |
| **Total power required** | 50.8 | 8.2 | 50.8 | 8.2 | 50.8 | 100.7 | 50.8 | 20.6 |

Table 8.4: Cooling capacity (MW) of the primary circuit.

| **Point** | **PA** | **PB** | **PD** | **PF** | **PG** | **PH** | **PJ** | **PL** |
|---|---|---|---|---|---|---|---|---|
| Number of cells per cooling tower | 6 | 2 | 6 | 2 | 6 | 8 | 6 | 3 |
| Cooling capacity per cell | 10.0 | 10.0 | 10.0 | 10.0 | 10.0 | 15.0 | 10.0 | 11.0 |
| **Total cooling capacity** | 60.0 | 20.0 | 60.0 | 20.0 | 60.0 | 120.0 | 60.0 | 33.0 |

The cooling towers operate continuously in open air with the ambient conditions present in Geneva and its surrounding area, with only a short stoppage each year to carry out maintenance. The baseline strategy implemented to reduce water consumption in the cooling tower is the recycling of its blowdown water. To do so, its conductivity is first reduced to values of around 20 µS/cm. The resulting water is then mixed with normal raw water, which is eventually used as a low-conductivity make-up water for the cooling towers.

An alternative to the baseline design is the adoption of hybrid cooling tower technology, which combines the advantages of both wet and dry cooling methods to enhance performance and reduce the environmental impact. In this system, the wet cooling tower - similar to the baseline design - uses water to absorb heat from the process, releasing it through evaporation. Dry air-water heat exchangers supplement this. The airflows from the wet and dry sections are mixed before reaching the fan stack, helping to reduce or eliminate the visible water vapour plume typically observed in wet cooling towers, particularly under certain weather conditions. However, hybrid cooling towers require a larger surface area and entail higher costs.

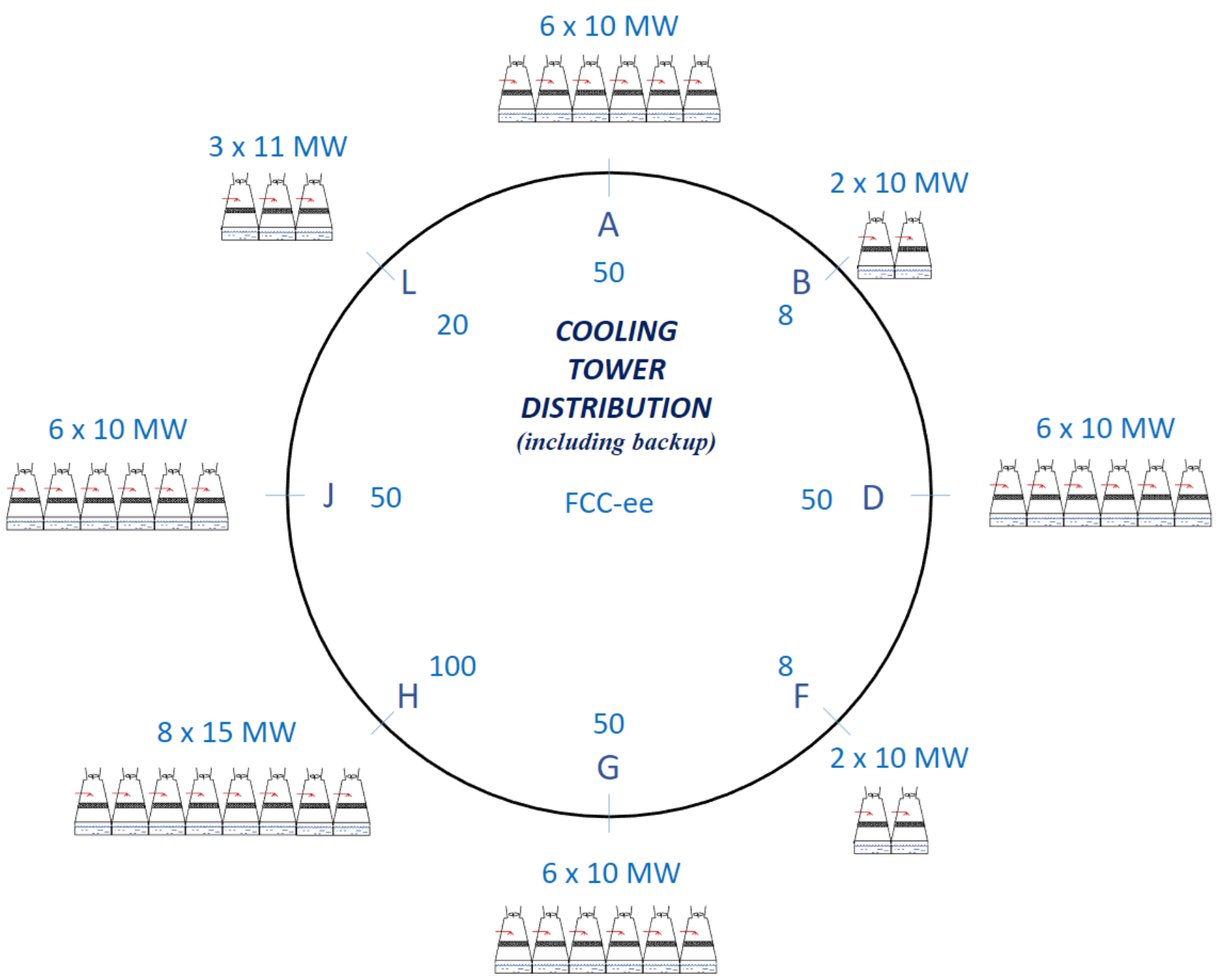

Fig. 8.48: Cooling tower complex for FCC-ee, including the backup cells.

***Demineralised water cooling circuits.***

The secondary circuits are connected to the primary system through heat exchangers. In most cases, the secondary circuits use demineralised water in a closed loop. Demineralised water cooling circuits are grouped according to the typology of the equipment to be cooled and to the equipment pressure ratings. Since the underground areas are up to 400 m below ground level (case of point PF), it is necessary to install an underground cooling station in the service cavern at all points. Here, heat exchangers separate the circuit coming from the surface (with a static pressure of up to 40 bar) from the underground distribution circuit. For operability and maintenance purposes, both surface and underground cooling stations are accessible during accelerator running. Figure 8.49 shows the demineralised water distribution in the FCC tunnel, the water being supplied at the experiment points. The types of underground equipment cooled by demineralised water are listed in Table 8.5. The magnets, alcoves and SR absorbers rows correspond to equipment located in the tunnel. The other four columns refer to equipment located close to the access points.

The heat loads from the accelerator's radio frequency system come directly from the RF equipment. Furthermore, the cryogenic installation, which cools part of the equipment to cryogenic temperatures, is cooled by primary water, not demineralised water. The first fill of the demineralised water circuits can be done by a CERN-owned on-truck demineralised water production station and leave the

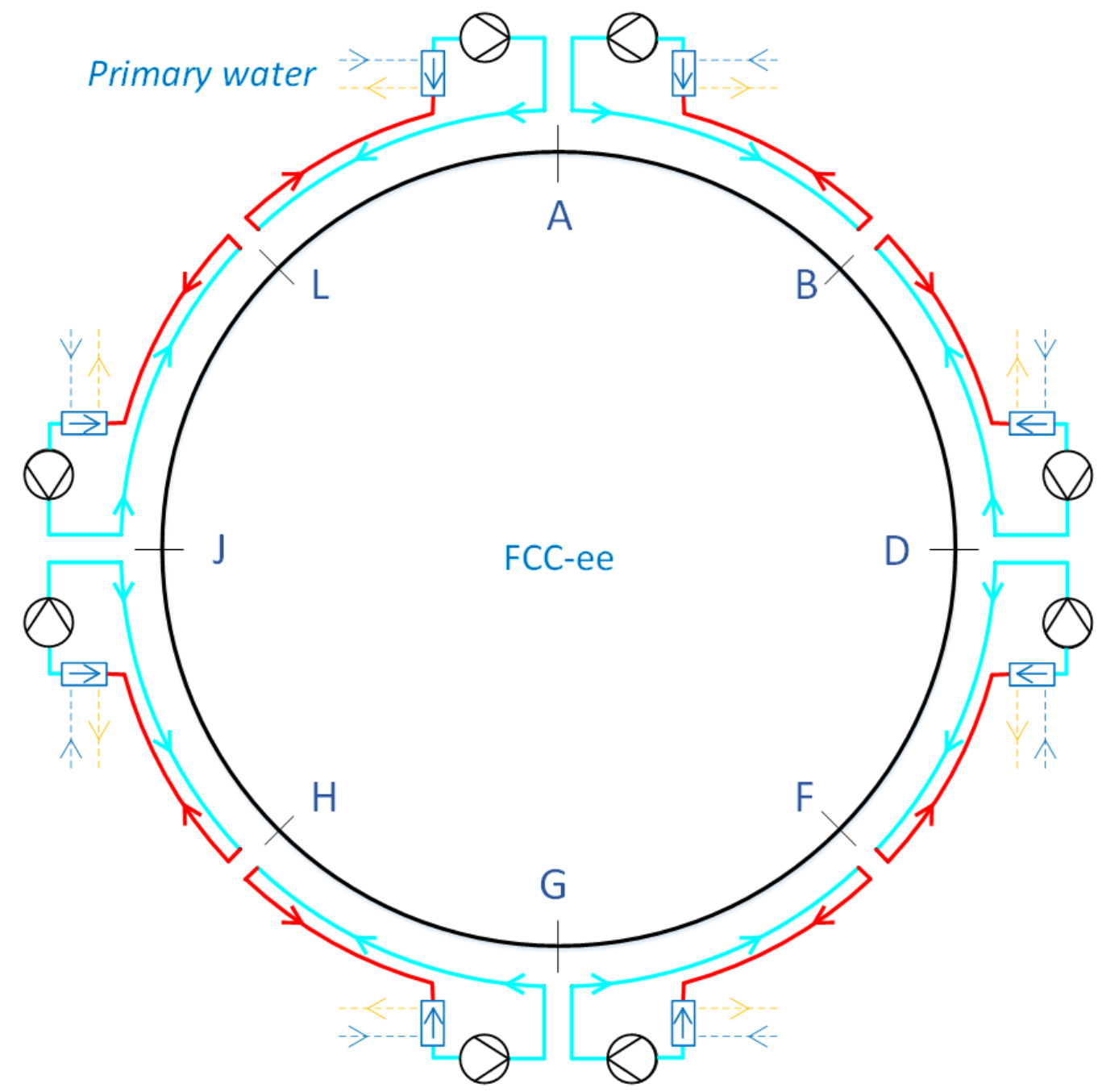

Fig. 8.49: Demineralised water distribution in the FCC tunnel.

Table 8.5: Demineralised water cooling demands (MW) per underground user.

| **Point / Sector** | **PA / PL-PA PA-PB** | **PB** | **PD / PB-PD PD-PF** | **PF** | **PG / PH-PJ PJ-PL** | **PH** | **PJ / PH-PJ PJ-PL** | **PL** |
|---|---|---|---|---|---|---|---|---|
| Magnets | 2×7.1 | | 2×7.1 | | 2×7.1 | | 2×7.1 | |
| Alcoves | 2×0.9 | | 2×0.9 | | 2×0.9 | | 2×0.9 | |
| SR Absorbers | 2×12.5 | | 2×12.5 | | 2×12.5 | | 2×12.5 | |
| Experiment Area | 0.5 | | 0.5 | | 0.5 | | 0.5 | |
| Power Converters | 1.0 | 1.0 | 1.0 | 1.0 | 1.0 | 1.0 | 1.0 | 1.0 |
| RF | | | | | | 45.7 | | 0.7 |
| Cryogenics* | 0.01* | | 0.01* | | 0.01* | 2.2* | 0.01* | 1.0* |
| **Total power** | 42.5 | 1.0 | 42.5 | 1.0 | 42.5 | 48.9 | 42.5 | 2.7 |

* Cryogenic equipment uses primary water quality in its secondary circuit.

topping up of the circuits to smaller local demineralisers at each point. Alternatives include renting an on-truck production station or designing the local demineralisers to carry out the first fill too. In this case, the production station would be underutilised, operating at reduced capacity most of the time.

***Chilled water cooling circuits.***

Chilled water at 6°C is needed for air dehumidification and cooling purposes and for the cooling of electrical equipment with specific temperature requirements.

The water is cooled by industrial air-cooled or water-cooled chillers. The maximum heat load to be removed by chilled water per point is given in Table 8.6, together with the flow needed for a temperature difference of 6 K between the supply and return chilled water temperature. The number of chillers and power per chiller is shown in Table 8.7: all of them have similar cooling demands except for point PH, which has a higher demand due to its hosting the most power-intensive RF section.

The distribution of chilled water in the tunnel is designed as shown in Fig. 8.50. The selection of refrigerant for the chillers will be guided by both technological compatibility and environmental sustainability, with careful consideration of ongoing research developments. This decision will be made at a later stage, ensuring it reflects the latest advances and best practices. Additionally, the potential integration of mixed water production at 12°C, which is not currently planned, will be assessed based on user requirements and implemented if deemed necessary to optimise efficiency and performance.

Table 8.6: Chilled water cooling demands (kW) and supply water flow rate ($m^3$/h) per point.

| **Point** | **PA** | **PB** | **PD** | **PF** | **PG** | **PH** | **PJ** | **PL** |
|---|---|---|---|---|---|---|---|---|
| Cooling power | 4895 | 4322 | 4875 | 4322 | 4895 | 9367 | 4875 | 4757 |
| Flow rate | 703 | 620 | 700 | 620 | 703 | 1345 | 700 | 683 |

Table 8.7: Chilled water cooling capacity (kW) per point.

| **Point** | **PA** | **PB** | **PD** | **PF** | **PG** | **PH** | **PJ** | **PL** |
|---|---|---|---|---|---|---|---|---|
| Number of chillers | 6 | 6 | 6 | 6 | 6 | 7 | 6 | 6 |
| Cooling power of chiller | 1000 | 900 | 1000 | 900 | 1000 | 1800 | 1000 | 1000 |
| **Total cooling power** | 6000 | 5400 | 6000 | 5400 | 6000 | 12 600 | 6000 | 6000 |

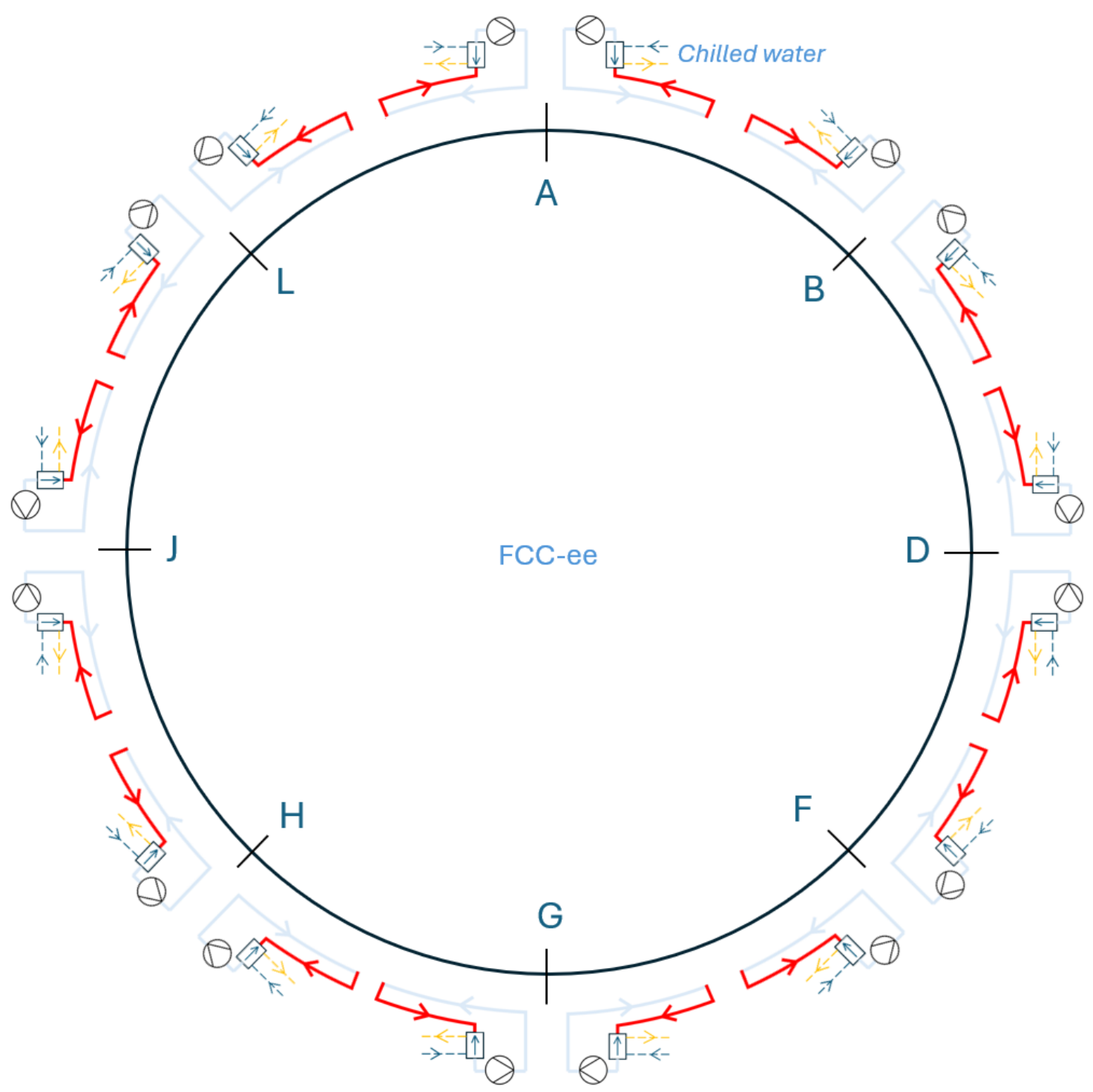

Fig. 8.50: Chilled water distribution in the FCC tunnel.

### 8.3.2 Ventilation systems

***Design principles and heat loads***

The FCC ventilation systems supply a sufficient amount of fresh air to meet the air quality and thermal requirements: the ambient temperature is suitable for the accelerator and auxiliary technical equipment. In addition, the air supplied is dehumidified to prevent condensation on equipment and structures.

In the current design, the underground areas are generally ventilated by air-handling units located on the surface, accessible at all times. Redundant units (N + 1) are planned everywhere to avoid affecting accelerator operation in case of breakdown. In addition, fancoils provide local cooling wherever necessary. Tables 8.8, 8.9 and 8.10 present the air heat loads in all surface sites, in the underground areas and for each sector of the tunnel, respectively.

Table 8.8: Heat loads on air (kW) on the surface, per point.

| Point | PA | PB | PD | PF | PG | PH | PJ | PL |
|---|---|---|---|---|---|---|---|---|
| Cryogenics* | 13 | | 13 | | 13 | 1400* | 13 | 402* |
| Experiment Areas | 50 | | 40 | | 50 | | 40 | |
| General Services | 500 | 500 | 500 | 500 | 500 | 500 | 500 | 500 |
| Power Converters* | | | | | | 2250* | | 35* |
| Shaft pressurisation | 300 | 150 | 300 | 150 | 300 | 150 | 300 | 150 |
| Fresh air for Underground | 150 | 50 | 150 | 50 | 150 | 150 | 150 | 150 |
| **Total heat load to Chilled water** | 1013 | 700 | 1003 | 700 | 1013 | 800 | 1003 | 800 |

* Cryogenics and power converters heat loads that are extracted without chilled water.

Table 8.9: Heat loads on air (kW) underground, per point.

| Point | PA | PB | PD | PF | PG | PH | PJ | PL |
|---|---|---|---|---|---|---|---|---|
| Cryogenics | 10 | | 10 | | 10 | 145 | 10 | 60 |
| RF | | | | | | 4600 | | 75 |
| Experiment Areas | 50 | | 40 | | 50 | | 40 | |
| Power Converters | 220 | 220 | 220 | 220 | 220 | 220 | 220 | 220 |
| CV Zone Underground | 200 | | 200 | | 200 | 200 | 200 | 200 |
| **Total heat load to Chilled water** | 480 | 220 | 470 | 220 | 480 | 5165 | 470 | 555 |

Table 8.10: Heat loads on air (kW) for each arc of the tunnel.

| Magnets | Cables* | Synchrotron radiation absorbers | Alcoves | Total heat load to Chilled water |
|---|---|---|---|---|
| 352 | 2500* | 250 | 300 | 3402 |

* Heat load from the cables extracted without chilled water.

***Tunnel ventilation.***

For the FCC tunnel, a semi-transverse ventilation scheme has been adopted. The air is supplied through a specific duct running throughout the sector and extracted either through the tunnel itself or by an emergency extraction duct. Air is supplied to each sector from both endpoints to ensure air supply even in case of a duct failure; the same configuration has been adopted for the extraction (Fig. 8.51).

The air supply duct runs in the concrete floor slab and supplies air to the tunnel about every 100 m via diffusers at floor level. A closed circular segment in the upper part of the tunnel is used for emergency extraction. The structure consists of 70 mm thick steel panels, secured to the tunnel lining using post-drilled anchors, and is equipped with passive fire protection on both sides. Inlet diffusers and extraction grills are strategically offset to ensure optimal air distribution within the tunnel and to prevent direct airflow shortcuts between supply and extraction points.

Fire-resistant dampers are installed at every connection between diffusers and extraction grills, enhancing ventilation control in the event of a fire or helium release. These dampers help manage airflow within the affected tunnel compartment, improving safety and containment.

To further control the spread of smoke and helium gas, each tunnel sector is divided into 28 compartments, separated by fixed fireproof panels and automatic doors. This design prevents smoke propagation in case of fire and mitigates the spread of helium gas in the event of an accidental leak from accelerator equipment.

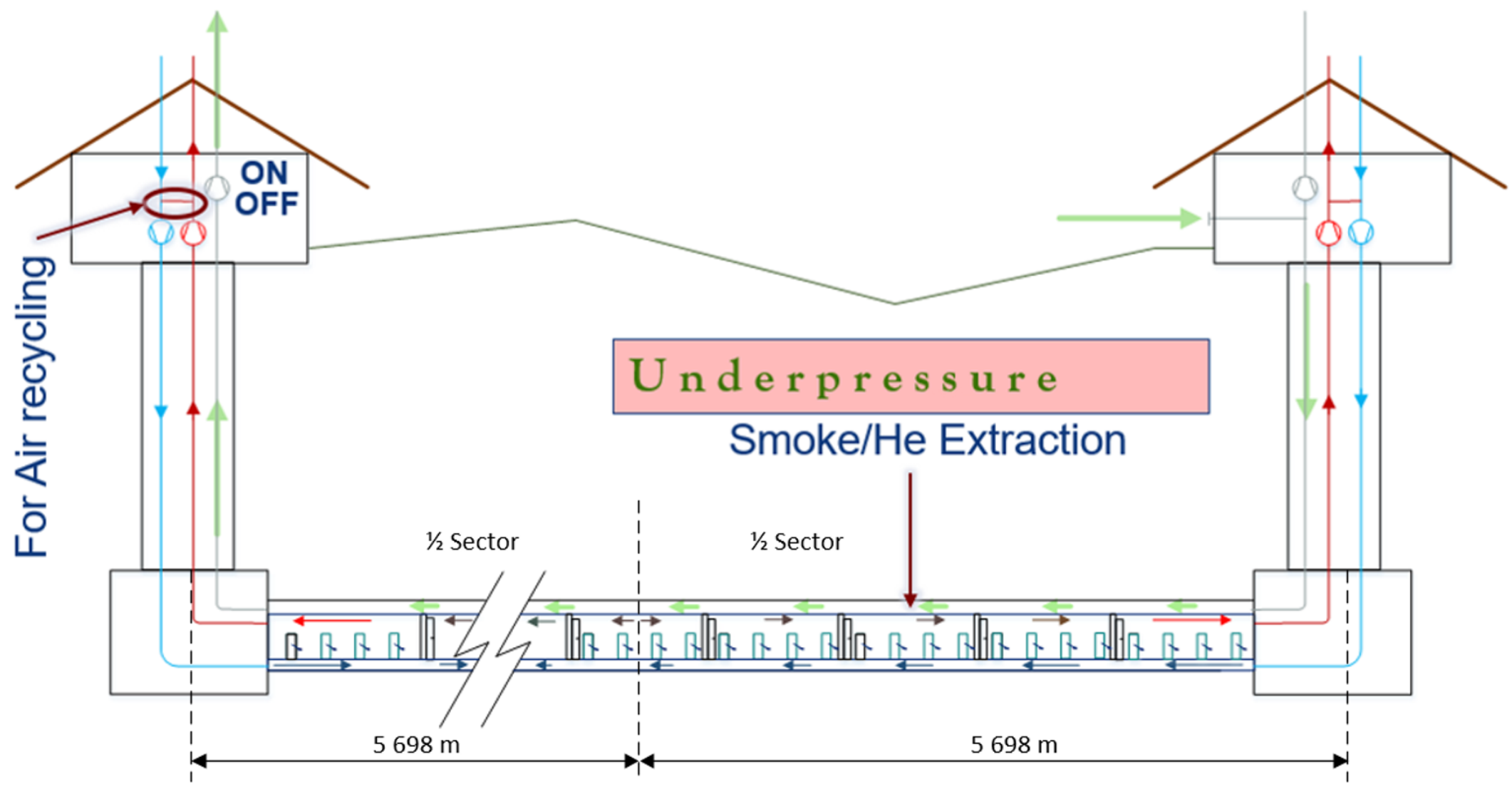

Fig. 8.51: Operation of the ventilation elements in one sector of the tunnel during normal operation. Smoke and helium extraction in green, general extraction in red and air supply in blue.

Under normal conditions, there are four different working modes for tunnel ventilation.

– *Run mode*: the accelerator is in operation, and most of the electrical systems are powered; a high heat load is transferred to the air in this mode. In addition,n radiation protection aspects are taken into account.
– *Access mode*: technical personnel can work in the tunnel and therefore occupational safety requirements are considered.
– *Economy mode*: a minimum airflow is employed in situations without personnel access, reducing the airflow. In this mode, the dew point of the air is controlled, but wider ranges of dry temperature are accepted to save energy.

– *Flushing mode*: used to renew the air underground completely and takes place when moving from Run to Access modes. The extracted volumes are filtered before release.

Table 8.11: Airflow conditions for the tunnel ventilation.

| | **Air Supply** | **Regular Extraction** | **Emergency Extraction** |
|---|---|---|---|
| Run and Access mode | 2×27 000 $m^3$/h per sector. | 2×27 000 $m^3$/h per sector. | Standby under pressure operation, no extraction. |
| Flushing mode | 100 000 $m^3$/h per sector. | 100 000 $m^3$/h per sector. | Standby under pressure operation, no extraction. |
| Emergency conditions | 10 000 $m^3$/h in affected compartments (max. 2). 3500 $m^3$/h in adjacent compartments (max. 2). 2160 $m^3$/h in the rest of the compartments. | Max. 54 000 $m^3$/h Sharing between the two shafts depends on affected compartment location. | 10 000 $m^3$/h in affected compartments (max. 2). 3500 $m^3$/h in adjacent compartments (max. 2). |

The airflow requirements under normal and emergency conditions are presented in Table 8.11. When the accelerator is in operation with full heat loads, the maximum dry temperature is limited to 32°C and the maximum dew point is limited to 12°C, inside the tunnel. In this mode, the nominal supply temperature is of 17°C. In flushing mode, the minimum supply temperature is 15°C. If the external temperature is below 0°C, the airflow will be reduced to maintain the 15°C supply temperature. For the access mode, the temperature inside the tunnel is set between a minimum of 18°C and a maximum of 26°C, while respecting the regulations regarding air renewal. In all operating modes, environmental sustainability and energy efficiency are key priorities. As a general principle, the ventilation system is designed to minimise electrical consumption by incorporating free cooling and air recycling strategies.

In run and access modes, airflow is supplied through an inlet duct integrated into the tunnel floor slab. In flushing mode, however, air is not delivered via this duct but is distributed longitudinally across the entire tunnel cross-section.

An alternative approach to tunnel ventilation would be to extend the longitudinal airflow concept of flushing mode to both run and access modes, resulting in a fully longitudinal ventilation scheme. This solution would simplify system controls, enhance robustness, and free up space required for the integration of the inlet duct. However, before implementation, the system's performance under degraded conditions and emergency scenarios must be carefully assessed to ensure full compliance with safety standards.

***Experiment areas ventilation.***

The ventilation system for the experiment areas (experiment caverns at points PA, PD, PG, and PJ) consists of a ventilation system on the surface with a supply and extraction unit and the possibility of recycling 100% of the air or the option to supply a controlled percentage of fresh air. The ventilation principle for the experiment caverns is represented in Fig. 8.52. No additional local cooling systems are planned.

The nominal airflow conditions are presented in Table 8.12. During operation, the experiment areas will have a temperature distribution of 18/32°C from floor to ceiling. The same parameters as for the FCC tunnel are applicable in flushing and access modes.

Table 8.12: Airflow conditions for the experiment areas ventilation.

| | **Air Supply** | **Regular Extraction** | **Gas Extraction** |
|---|---|---|---|
| Run and Access mode | 50 000 $m^3$/h to 70 000 $m^3$/h | 50 000 $m^3$/h to 70 000 $m^3$/h | Switched off |
| Flushing mode | 50 000 $m^3$/h to 70 000 $m^3$/h | 50 000 $m^3$/h to 70 000 $m^3$/h | Switched off |
| Fire emergency (gas extraction) | Up to 70 000 $m^3$/h | Up to 70 000 $m^3$/h | Switched off |
| Gas emergency (gas extraction) | 20 000 $m^3$/h | Switched off | 20 000 $m^3$/h |

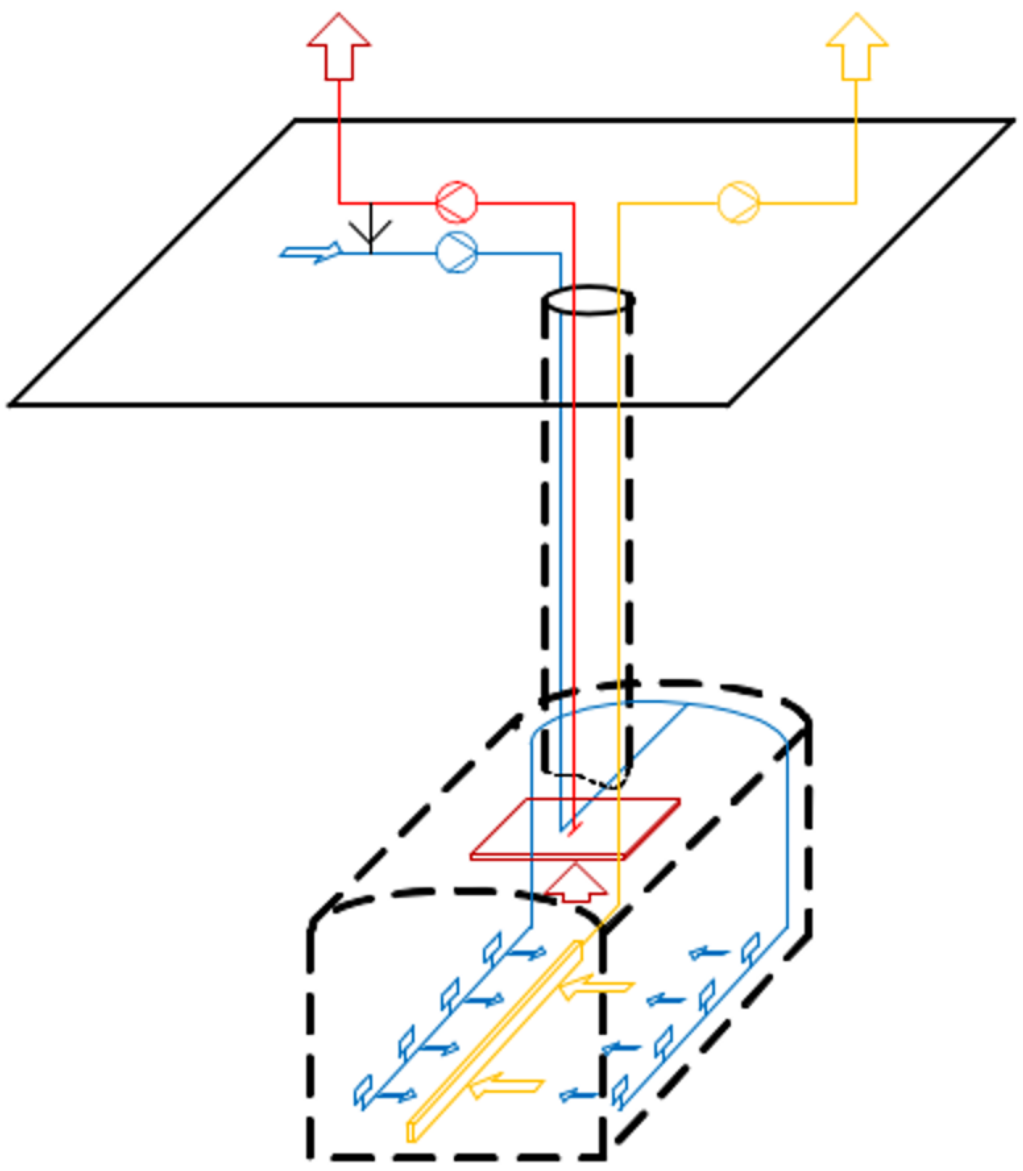

Fig. 8.52: Ventilation systems of the experiment cavern. Smoke and helium extraction in orange, general extraction in red and air supply in blue.

***RF areas ventilation.***

The RF system is cooled mainly by the demineralised water circuits in the underground klystron galleries of points PH and PL. A considerable heat load nevertheless remains to be extracted: 4600 kW in point PH and 75 kW in point PL.

The ventilation design has main ventilation from the surface, with a fresh air supply, plus local cooling with fan coils cooled with chilled water. Table 8.13 presents the airflow requirements for each of the modes; flushing is not needed because contaminants are not present in the klystron gallery equipment.

***Technical areas ventilation.***

Dedicated ventilation systems serve the technical zones around each point and the connecting galleries between areas. These areas comprise the service caverns and connecting galleries at all points.

The design of each ventilation system is different because the general demands and heat loads differ greatly between the points. A difference is made between the technical areas in experiment points (points PA, PD, PG, PJ), the technical areas in technical points (points PB and PF), and the technical areas in RF points (points PH and PL). The ventilation solutions proposed under normal and emergency conditions are presented in Table 8.14. Finally, other ventilation systems pressurise airlocks to create pressure cascades for safety purposes.

Table 8.13: Airflow conditions for the RF areas ventilation.

| | **Air Supply** | **Regular Extraction** | **Emergency Extraction** |
|---|---|---|---|
| Run and Access mode | 30 000 $m^3$/h | 30 000 $m^3$/h | Switched off |
| Emergency conditions | <30 000 $m^3$/h | <30 000 $m^3$/h | 10 000 $m^3$/h in affected compartments (max. 2). 3500 $m^3$/h in adjacent compartments (max. 2). |

Table 8.14: Airflow conditions for ventilation of the technical areas.

| | **Air supply** | **Extraction** | **Emergency Extraction** |
|---|---|---|---|
| Technical Areas in Experiment points | Fresh air for UAs, 20 000 $m^3$/h. Ventilation for Service Cavern, 40 000 $m^3$/h. | Extraction through the Service Cavern, 60 000 $m^3$/h. | Emergency extraction through the Ventilation system for Service Cavern, 60 000 $m^3$/h. |
| Technical Areas in Technical points | Ventilation for Service Cavern, 40 000 $m^3$/h. | Extraction through the Service Cavern, 40 000 $m^3$/h. | Emergency extraction through the Ventilation system for Service Cavern, 40 000 $m^3$/h. |
| Technical Areas in RF points | Ventilation for Service Cavern, 40 000 $m^3$/h. | Extraction through the Service Cavern, 40 000 $m^3$/h. | Emergency extraction through the Ventilation system for Service Cavern, 40 000 $m^3$/h. |

***Surface buildings ventilation***

Each surface building will be ventilated by a dedicated air-handling unit. Where the building size requires it, it is planned to have several units in the same building, each of them taking care of a part of the building. At present, it is not considered necessary to have redundant units in these buildings. Should this be needed, it can easily be implemented. All surface buildings will be equipped with a mechanical system on the roof to extract smoke, designed and certified for operation at 400°C for a minimum period of 2 h.

### 8.3.3 Other systems

***Drinking water***

Drinking water will be used by personnel. It is planned that this will be provided by the local water network at each point.

***Reject water and sumps***

Discharging effluent water into local disposal networks or small water bodies presents challenges due to its high salinity levels. To address this, a centralised treatment system is proposed, in which all effluent water is transported through the tunnel to a treatment plant at point PA.

An alternative approach would involve local treatment of effluent water at each point. Implementing minimum or zero liquid discharge (ZLD) technologies could significantly reduce the volume of effluent by concentrating the salts into solid pellets. However, this solution would introduce higher costs and increased surface area requirements due to the complexity of these treatment systems.

***Compressed air***

The compressed air for all equipment and actuators will be provided by compressed air stations located on the surface at each point. These will supply both surface and underground areas. A level of redundancy of N +1 is planned to ensure the reliability and maintainability of the plant.

***Sumps and clear water***

Two separate pump systems to lift clear water and sewage will be installed underground at each point. They will be connected to the point's local drainage network. All underground equipment (tunnel and caverns) must have redundancy in order to avoid affecting operation in case of a breakdown. The sump pumping capacities for the tunnel at the points are 30 $m^3$/h, and the experiment caverns have an additional 30 $m^3$/h of sump pumping capacity. Alarms for 'high level' and 'level too high' will be implemented in all basins.

The key parameters of the clear water, including temperature and pH, will be systematically monitored prior to discharge. If the clear water fails to meet the required quality standards or poses a risk of environmental contamination, appropriate mitigation measures will be implemented. These measures may include the activation of retention basins at designated discharge points to ensure compliance with environmental regulations.

## 8.4 Power consumption and electricity distribution

### 8.4.1 Energy consumption of the FCC-ee machine

***Overall power Demand***

Since FCC-ee is a large-scale accelerator, accurately identifying and understanding its electrical loads is essential for designing a robust electricity infrastructure and assessing overall energy consumption. The accelerator's main electrical loads include the radiofrequency (RF) systems, magnets, cryogenic systems, cooling and ventilation, experiments, and general services, which cover all general-purpose energy needs. The total power demand varies depending on the operation mode of the machine (Z, WW, H, and $\mathrm{t\bar{t}}$).

The collider radiofrequency system represents the largest electrical load, requiring 146 MW across all operation modes. It consistently supplies 50 MW per beam to compensate for synchrotron radiation losses. The global efficiency of the RF chain is 68%, with the goal of having klystron amplifiers operating at 80% efficiency. The power demand for the magnets and their powering chains depends on the beam energy, starting from 6 MW in Z mode, increasing to 39 MW in H mode, and reaching 89 MW in $t\bar{t}$ mode.

The cryogenic power demand remains relatively low at 13 MW up to H mode but becomes significant at 35 MW in $t\bar{t}$ mode. The cooling and ventilation systems require between 25 MW and 33 MW, while the power demand of the experiments and data centres is estimated to be 10 MW for all four experiments, with an additional 4 MW allocated for local data centres, given that detailed technical designs are not yet available. Finally, general services are expected to require up to 26 MW, based on scaling from LHC consumption.

These estimates will form the basis for the development of the FCC-ee's electrical infrastructure, ensuring it can accommodate the varying power demands while optimising efficiency and sustainability. Table 8.15 shows the estimate of the maximum power demand for the four different modes of operation.

Table 8.15: Power demand by technical system, in MW.

| | **Z** | **W** | **H** | **$t\bar{t}$** |
|---|---|---|---|---|
| Beam energy, GeV | 45.6 | 80 | 120 | 182.5 |
| Collider radiofrequency | 146 | 146 | 146 | 146 |
| Booster radiofrequency | 2 | 2 | 2 | 2 |
| Collider cryogenics | 1.2 | 11.5 | 11.5 | 27.6 |
| Booster cryogenics | 0.35 | 0.8 | 1.5 | 7.4 |
| Cooling and ventilation | 25 | 26 | 28 | 33 |
| Collider magnets | 6 | 17 | 39 | 89 |
| Booster magnets | 1 | 3 | 5 | 11 |
| 4 experiments, PA, PD, PG, PJ | 10 | 10 | 10 | 10 |
| 4 datacenters, PA, PD, PG, PJ | 4 | 4 | 4 | 4 |
| General services | 26 | 26 | 26 | 26 |
| Total power during beam operation | 222 | 247 | 273 | 357 |

*Operational model*

The energy consumption is estimated based on the accelerator operational model. The power demand depends on the period of the year and the state of the machine. The machine schedule defines the different periods during the year and the power demand varies from the minimum during the shutdown to the maximum during beam operation.

The schedule consists of six distinct periods: shutdown, commissioning, physics operation, short downtime (without access to the machine), technical stops (or long downtime), and machine developments. The power demand calculated for each period is shown in Table 8.16.

Table 8.16: Power demand by operation mode, in MW.

| **Beam energy mode** | **Z** | **W** | **H** | **$t\bar{t}$** |
|---|---|---|---|---|
| Shutdown | 30 | 33 | 34 | 41 |
| Technical stop | 67 | 78 | 81 | 108 |
| Short downtime | 74 | 98 | 125 | 209 |
| Commissioning | 144 | 163 | 177 | 233 |
| Machine Development | 96 | 121 | 147 | 231 |
| Beam operation | 222 | 247 | 273 | 357 |

The machine schedule comprises 120 days of shutdown, 30 days of commissioning, 20 days of machine development, 10 days of technical stops and 185 days for physics, see Table 8.17.

The machine availability is expected to be at least 80%. Unlike other accelerators, FCC-ee does not lose time for acceleration, meaning that its operational efficiency is determined solely by technical downtime and the associated time required for beam refilling and recovery. Based on empirical data from similar machines, the effective efficiency factor is estimated to be approximately equal to the hardware availability (80%), with an additional 5% reduction to account for beam recovery after a failure.

Of the 185 days allocated for physics operation, an estimated 46 days are lost due to downtime, leaving 139 days for colliding beams.

The downtime is categorised into three types: long stops due to major failures, such as the loss of cryogenic conditions or RF cavity conditioning; downtime with machine access, allowing repairs and maintenance; and downtime between cycles, which includes beam dumps and refilling that do not require machine access.

Table 8.17: Distribution of operating modes in time.

| **Period Type** | **Duration** |
|---|---|
| Shutdown | 120 days |
| Commissioning | 30 days |
| Physics Operation | 185 days |
| Beam colliding | 139 days |
| Downtime | 46 days |
| Machine development | 20 days |
| Technical Stops | 10 days |

***Power demand by points***

As the distribution of the power demand is not uniform across the points (see Table 8.18), the installed power and its position need to be known to be able to design the electrical infrastructure. The load survey performed in the first phase of the study provided the following localised power requirements:

- Surface at each point: between 6.1 and 42.2 MW,
- Underground at each point: between 2.1 and 215 MW,
- Tunnel between points: around 20.2 MW.

This first analysis allowed the creation of a load mapping of the FCC-ee machine for the design of its electrical network, see Fig. 8.53.

Table 8.18: Power demand by point, $t\bar{t}$ mode.

| **Point** | **Type** | **Max Power [MW]** |
|---|---|---|
| Point PA | Experiment | 23 |
| Point PB | | 20 |
| Point PD | Experiment | 23 |
| Point PF | | 20 |
| Point PG | Experiment | 23 |
| Point PH | Collider RF | 194 |
| Point PJ | Experiment | 23 |
| Point PL | Booster RF | 30 |

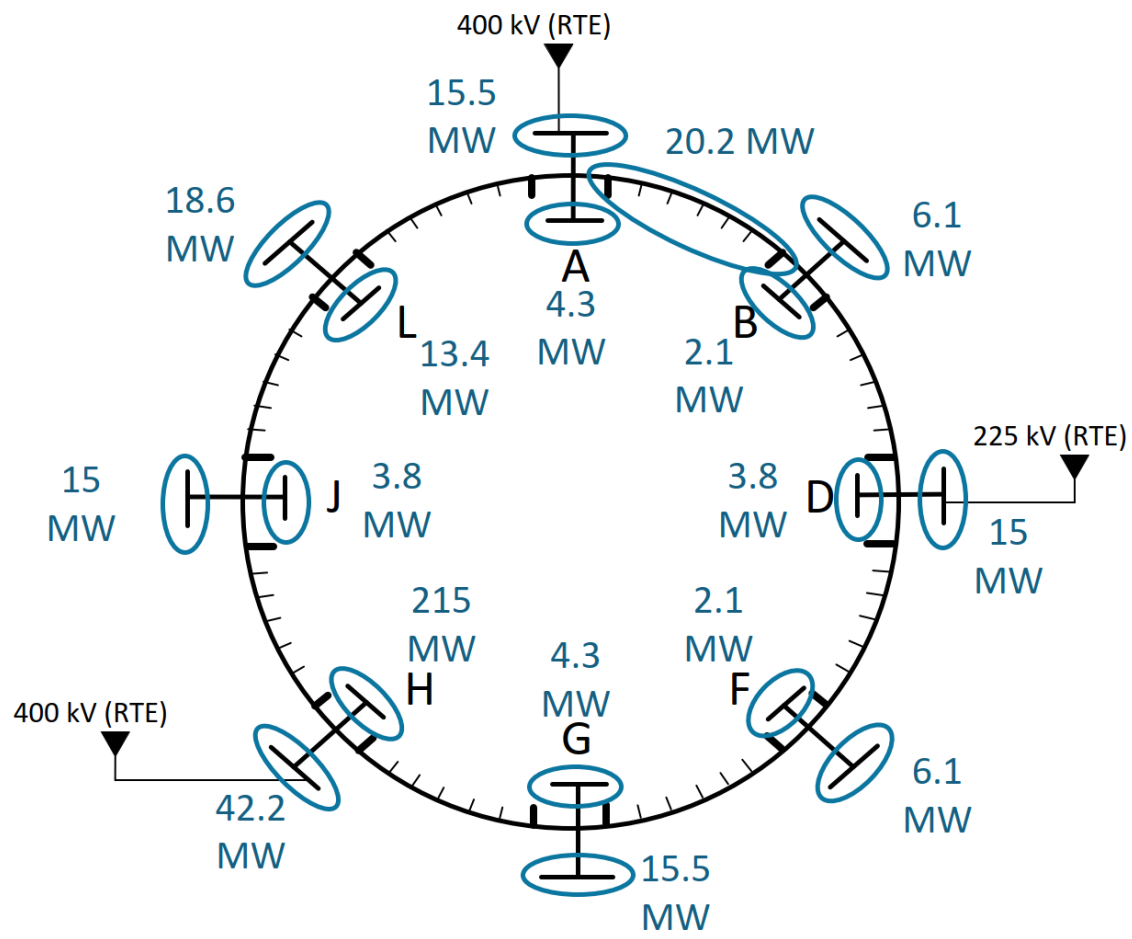

Fig. 8.53: Load mapping of the FCC-ee machine.

#### *Energy consumption*

With the schedule and the expected power demand for the various periods, the annual energy consumption can be calculated, see Tables 8.19 and 8.20 for Z and $t\bar{t}$ operation respectively. The energy consumption varies from 1.07 TWh/year for Z mode operation to 1.77 TWh/year for $t\bar{t}$ mode operation. For comparison, the complete CERN complex energy consumption in 2024 was 1.3 TWh/year. This yearly consumption is expected to increase after the completion of the HL-LHC project.

Table 8.19: Energy Consumption, Z mode.

| **Z mode** | **Duration, Days** | **Power, MW** | **Energy, GWh** |
|---|---|---|---|
| Beam colliding | 139 | 222 | 740 |
| Downtime | 46 | 67 | 75 |
| Commissioning | 30 | 144 | 103 |
| Machine development | 20 | 96 | 46 |
| Technical Stops | 10 | 67 | 16 |
| Shutdown | 120 | 30 | 86 |
| Total Z mode operation | | | 1070 |

Table 8.20: Energy consumption, $t\bar{t}$ mode.

| **TT mode** | **Duration, Days** | **Power, MW** | **Energy, GWh** |
|---|---|---|---|
| Beam colliding | 139 | 357 | 1200 |
| Downtime | 46 | 108 | 157 |
| Commissioning | 30 | 233 | 168 |
| Machine development | 20 | 231 | 111 |
| Technical Stops | 10 | 108 | 26 |
| Shutdown | 120 | 41 | 117 |
| Total $t\bar{t}$ mode operation | | | 1770 |

### *Grid Connections*

To provide power to the machine which is spread over many points, the strategy is to connect three points to the high-voltage European grid and to create an internal transmission network through the tunnel to distribute it to the other points. These connections will be rated at the same level of 220 MVA at each point. These points are PH, PA, and PD (see Fig. 8.54) and will be connected to the French grid, operated by RTE. RTE confirmed that, beside a minimum additional infrastructure to connect the new FCC points, no further upgrades or power stations are required in the existing French grid.

PH needs a dedicated substation for the collider RF systems. PA can be powered through an existing CERN substation. PD is needed to cover the other parts of the machine, for redundancy with PA and for the future operation of FCC-hh.

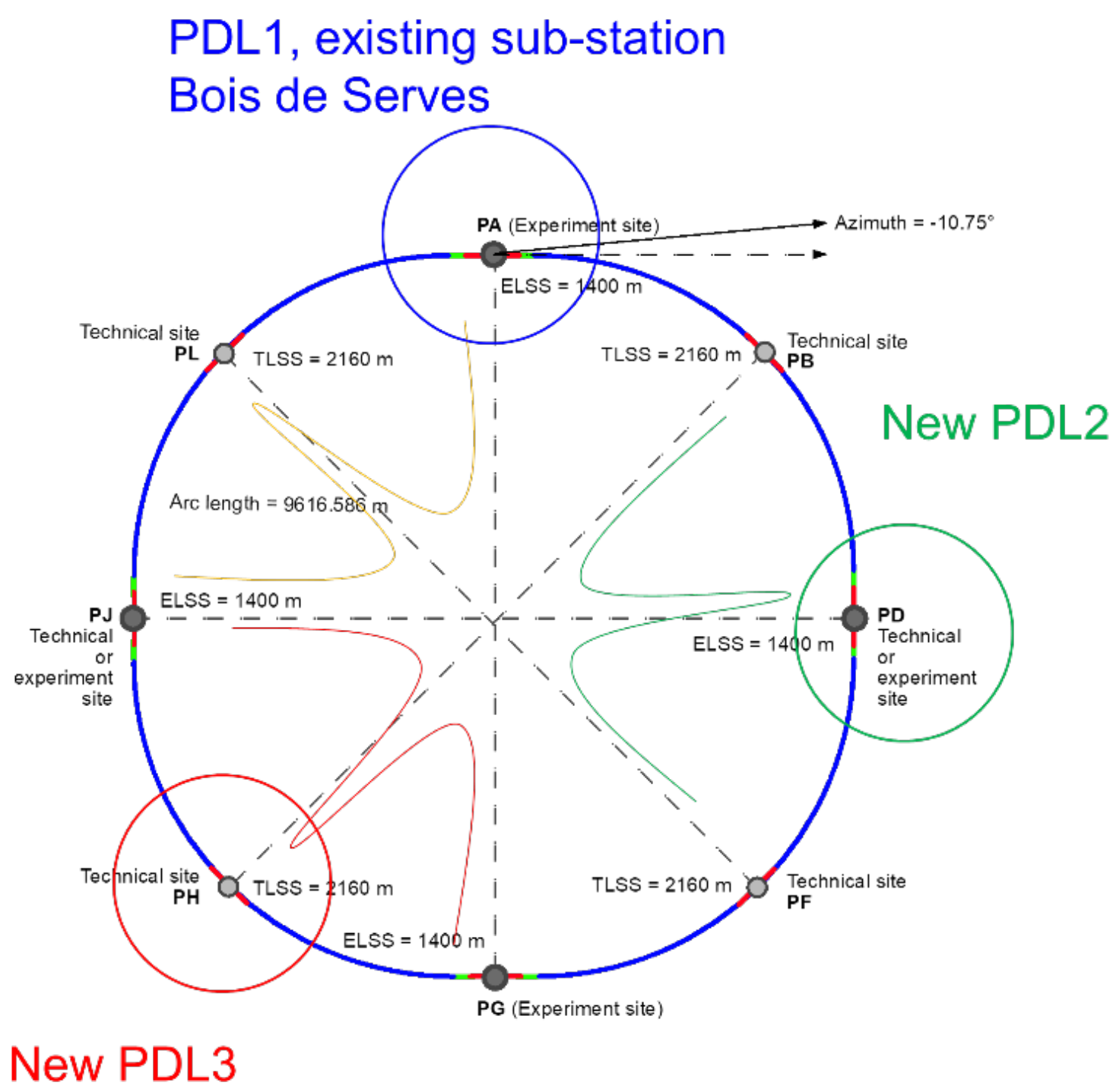

Fig. 8.54: High Voltage grid connections, PDL as delivery points.

In addition to that, all the points will be connected at much lower power to the local distribution grids: SIG (Services Industriels de Genève) for the point in Switzerland (PB), and ENEDIS for points located in France, except for point PG which will be connected to Energie et Services de Seyssel. These connections are needed for the civil engineering works, and later they will be used as backup sources. The voltage level of these connections will be at medium voltage (20 kV in France, 18 kV in Switzerland), and they will be in general rated at 14 MW, to cover the TBM loads in the civil engineering works phase.

## 8.4.2 FCC electrical grid

### *Transmission network*

The FCC can be powered from the French national grid, operated by RTE (France's Transmission System Operator, Réseau de Transport d'Électricité), through three access points (PA, PD, and PH) at two possible voltage levels (400 kV or 225 kV). Power will be distributed to the other access points via an internal high-voltage (HV) transmission network, which will be implemented using HV cables installed within the accelerator tunnel. These cables will link all access points, ensuring power transmission from one point to the adjacent ones.

Following verification with RTE, it has been confirmed that operating two or three high-voltage supplies of the FCC in parallel will not be permitted by the grid operator, as this would introduce the risk of uncontrolled or unwanted power transfers between connection points.

As a result, the option of operating the HV grid of the FCC as a closed loop, initially considered feasible, has been discarded at this stage of the study. Instead, the three high-voltage supplies will be operated in an antenna configuration, with each connection radially supplying other points in the accelerator. In the normal configuration, three distinct sectors have been defined, each powered by a single RTE connection: PA will supply PB, PJ, and PL; PD will supply PF and PG; while PH will operate standalone, as the RF systems at this location represent the largest power load of the machine.

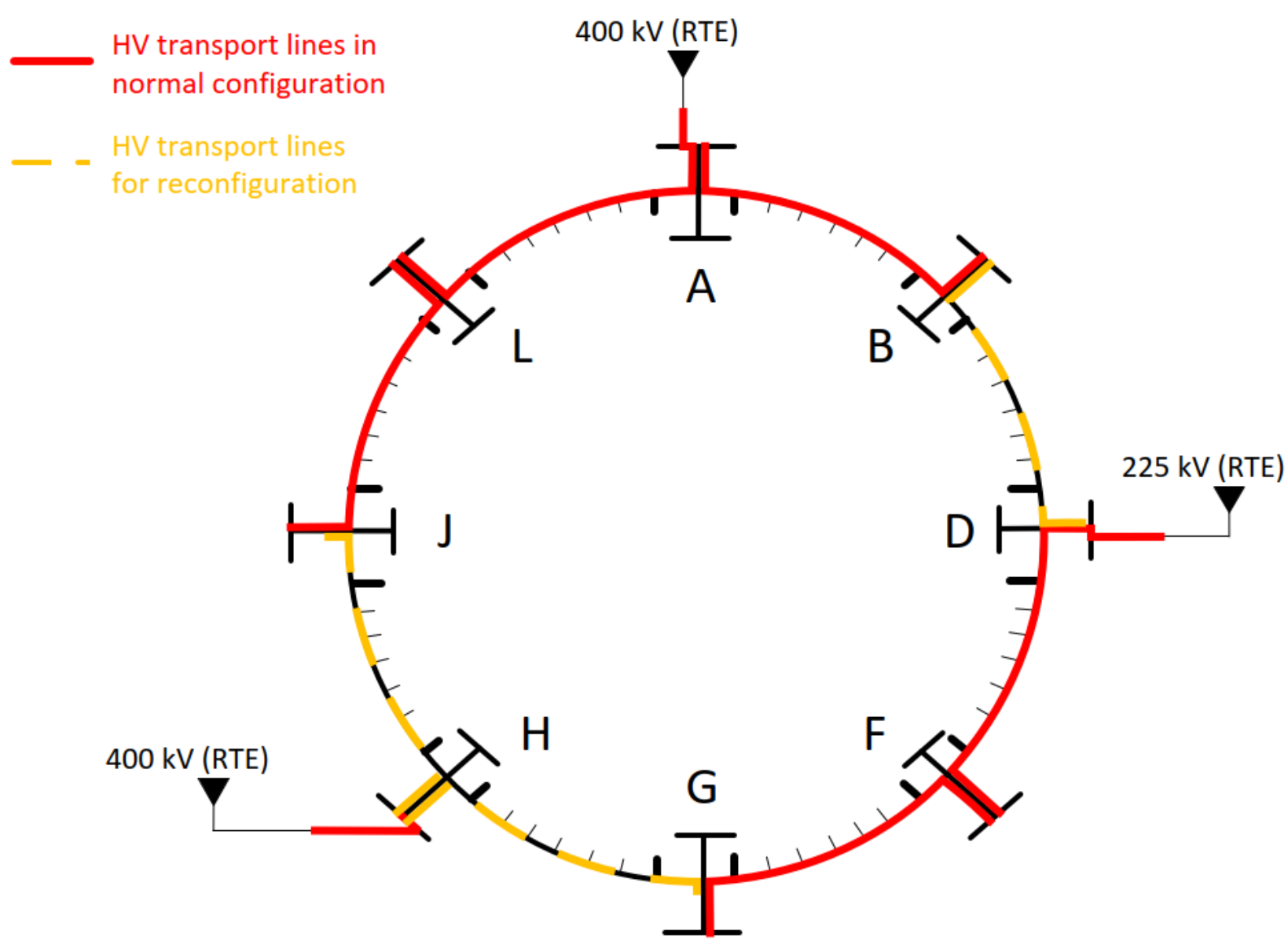

Fig. 8.55: General single line diagram of the transmission network of FCC.

Even if the HV cables between PB and PD, between PG and PH, and between PH and PG will not be loaded during normal operation, they can be used as a backup supply in case part of the HV network is unavailable. The possible continuity of service based on the [N-1] principle has been analysed by simulating different failures in the network and reconfigurations to overcome these failures have been determined. The only scenario in which the machine will cease operation without the possibility of network reconfiguration is the loss of supply from PH, as the RF loads at this location are too high to be compensated by other power sources. Apart from this case, the network is designed to allow reconfiguration, ensuring that the accelerator can continue normal operation in the event of a failure of one HV source or one HV branch. The list of possible HV network failures that the system can withstand, along with the corresponding reconfiguration strategies, is summarised in Table 8.21.

Table 8.21: Transmission network configuration modes to overcome failures.

| | PA - PB | PB - PD | PD - PF | PF - PG | PG - PH | PH - PJ | PJ - PL | PL - PA |
|---|---|---|---|---|---|---|---|---|
| Normal configuration | ON | OFF | ON | ON | OFF | OFF | ON | ON |
| Loss of PA | ON | ON | ON | OFF | ON | ON | OFF | ON |
| Loss of PD | ON | ON | ON | OFF | ON | ON | OFF | ON |
| Loss of PA - PB | OFF | ON | ON | ON | OFF | OFF | ON | ON |
| Loss of PD - PF | ON | OFF | OFF | ON | ON | OFF | ON | ON |
| Loss of PF - PG | ON | OFF | ON | OFF | ON | OFF | ON | ON |
| Loss of PJ - PL | ON | OFF | ON | ON | OFF | ON | OFF | ON |
| Loss of PL - PA | ON | OFF | ON | ON | OFF | ON | ON | OFF |

This analysis also provides the minimum criteria to dimension the sources and the lines of the HV network, in addition to the load requirements. Finally, to also cope with the objective of keeping the same HV grid for the future FCC-hh, it was decided that all the HV cables will be sized equally to a nominal rating of 115 MVA, to allow the infrastructure to be compatible with the FCC-hh load.

*Selection of the voltage level, the location and the technology of the HV transmission network*

The selection of the operating voltage of the transmission network has been based on a study aimed to optimise infrastructure and operational costs. As a result, the voltage will be stepped down from 400 or 225 kV, to a lower level, see Fig. 8.55.

For the radiofrequency (RF) systems, the voltage will be stepped down to the required level for the main RF power conversion system, which is currently planned to be 40 kV. While adopting 40 kV for the entire transmission network would offer some benefits in terms of material uniformity, an initial analysis clearly indicated that this voltage level is not optimal due to the large cable cross-sections required, leading to high space occupancy in the tunnel and significant power losses. For these reasons, 40 kV was discarded from the beginning of the study.

To determine the most suitable internal transmission voltage, the analysis considered three reference voltages, selected from standard values used by European Grid Operators: 63 kV, 90 kV, and 132 kV. The study was conducted for these three voltage levels, allowing extrapolation to evaluate performance trends across different voltage ranges. For each voltage level, two options have been considered : i) substations installed on the surface or underground (in the technical galleries of the accelerator); ii) air-insulated substations (AIS) or gas-insulated substations (GIS).

The following parameters were included in the analysis, with the aim of evaluating the impact of the different alternatives on all the main aspects of the total lifetime cost of the network:

- OPEX[1] of the HV lines: electrical losses of the lines depending on their operation over the full FCC machine (based on the operating models of the phases of FCC-ee, and on a simplified estimation of the same models for FCC-hh).
- CAPEX[2] of the HV lines: initial cost of the cables, the accessories and the installation works required.
- OPEX of the HV substations: electrical losses of the transformers depending on their operation over the full FCC machine programme and the cost of the maintenance of the substations.

[1]Operational expenditure
[2]Capital expenditure

– CAPEX of the HV substations: initial cost of the substations, and equivalent financial value of their space occupation.

The values of CAPEX and OPEX of each case have been assessed based on detailed studies of the substations and the lines at the different voltage levels, but also with estimated financial values of the space occupied and energy cost.

This comprehensive analysis gave the total lifetime cost for each scenario, with the variation trends illustrated in Fig. 8.56. The study concludes that selecting a 63 kV transmission voltage and using GIS substations installed on the surface are the optimal choices within the range analysed. Additionally, manufacturers are demonstrating a clear trend toward reducing the greenhouse gases used in GIS mixtures, significantly improving the environmental sustainability of this technology. The final technical solution was developed based on these findings.

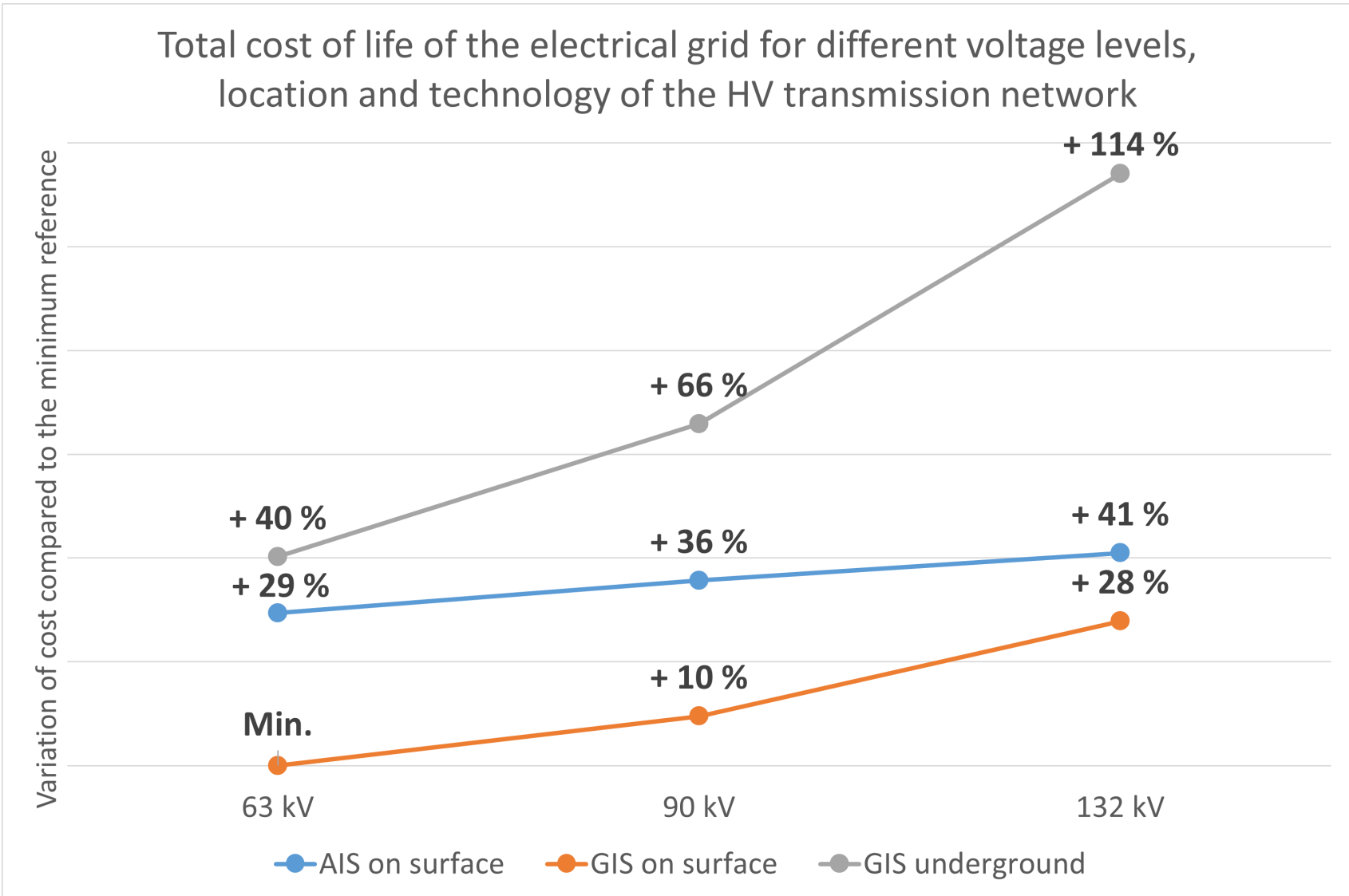

Fig. 8.56: Trend of the variation of the total lifetime cost of the electrical grid for different voltage levels, locations and technologies of the HV transmission network.

This analysis will be kept running in the future phases of the FCC study, as some parameters have a significant impact and can change the conclusions.

*Main high voltage substations*

The design of the HV substations has been developed according to the results of the analysis shown on Fig. 8.56.

A 400/63 or 225/63 kV GIS substation will be installed with a main stepdown transformer of 220 MVA in the three points connected to the European grid. The main switchgears and auxiliary systems will be housed inside a dedicated building, while the transformer will be installed nearby, protected by firewalls and equipped with all necessary safety systems to manage potential oil spills. These include a containment pit, oil separator, retention basin, and fire protection measures to ensure safe operation and environmental compliance. It is worth noting that in PH an additional bay will be available in the 400 kV switchgear, to supply an additional 150 MVA 400/40 kV transformer that will be dedicated to the power conversion system for the FCC-ee RF system.

A 63 kV GIS substation will be installed at all the other access points, as shown in the single line diagram of the transmission network in Fig. 8.58a. Also, in this case, the main switchgear will be housed with the auxiliary systems in a building. At each point, the 63 kV GIS substation will supply two 40 MVA 63/20 kV transformers that will be the sources of the distribution networks for each point.

In each access point connected to the grid, the total surface area of the HV substation will be approximately 10 000 $m^2$; in each of the other points, it will be approximately 4000 $m^2$, these requirements are already included in those considered by the civil engineering team for the surface occupation and the layouts of the different sites.

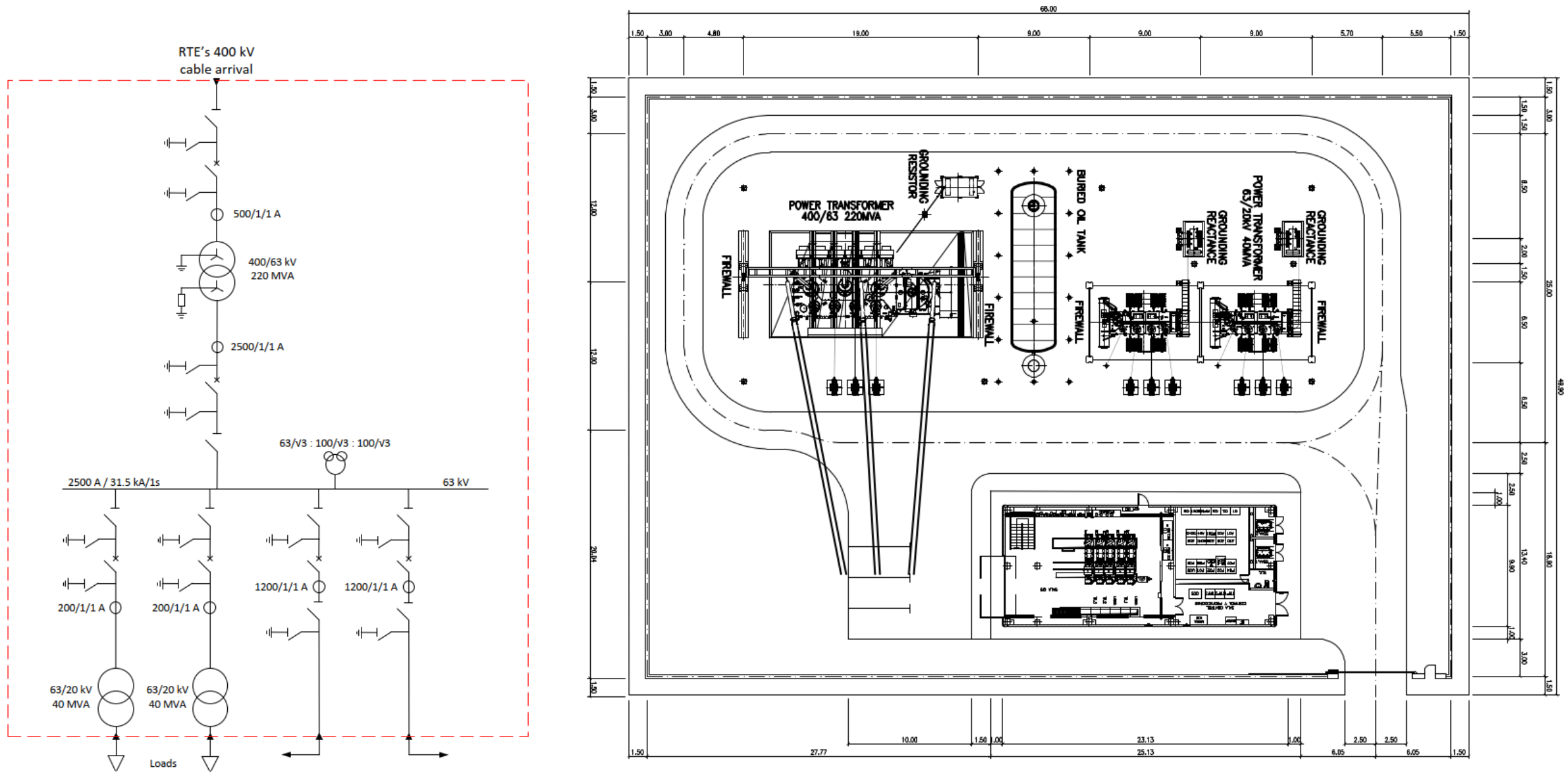

(a) Single line diagram.

(b) Layout.

Fig. 8.57: Typical 400/63 kV substation

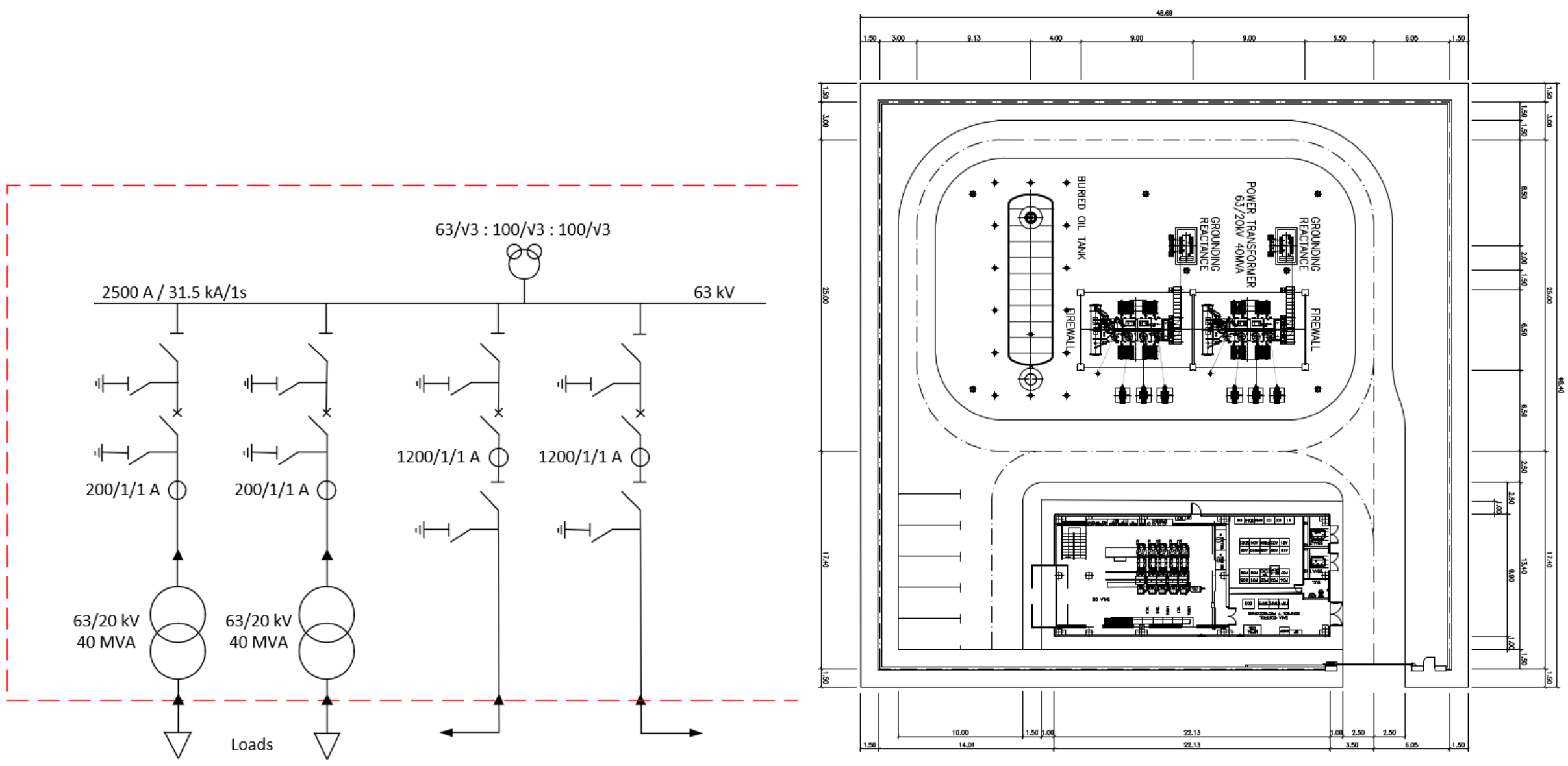

(a) Single line diagram.

(b) Layout.

Fig. 8.58: Typical 63 kV substation

*Installation of HV cables in the accelerator tunnel*

As the installation of HV transmission networks rated at tens or hundreds of MVA is not a standard solution adopted in the underground accelerator facilities (e.g., in LHC today only cables of up to 18 kV and rated at 15 MVA are installed for the electrical distribution), a detailed and specific feasibility analysis has been performed, in order to confirm the absence of showstoppers and to provide requirements for civil engineering.

The final report of this analysis confirms the feasibility of installing such cables in the underground areas of the accelerator. It provides models and datasheets of cables with their related losses, different installation methods with their constraints and the required space and civil engineering works to install and house the cables. It also includes a cost estimate and a tentative schedule for the installation. All the details can be found in Ref. [384], but a summary is presented in Table 8.22.

Table 8.22: Summary table of the requirements of the HV lines.

| | **63 kV** | **132 kV** |
|---|---|---|
| Proposed cable | $3\times$ 1200 mm$^2$,<br>Aluminium cable | $3\times$ 630 mm$^2$,<br>Aluminium cable |
| Proposed space occupation in the tunnel | Concrete duct of 520×500 mm. | |
| Proposed quantity of junction chambers | 13 junction chambers (including 1 earthing junction chamber) | |
| Proposed space occupation of junction chambers | With earthing of screens:<br>duct of 1500×500 mm<br>with a length of 9.5 m. | With earthing of screens:<br>duct of 1800×500 mm<br>with a length of 12 m. |
| Preferred installation method | Within PEHD ducts with the so-called cable train pulling method. | |
| Cost of installation | Lower cost | + 4% compared to 63 kV |
| Schedule estimation of installation (per line between two points) | 160 days | 170 days |
| Level of magnetic field in the tunnel | See magnetic field graphs on Fig. 8.59. | |

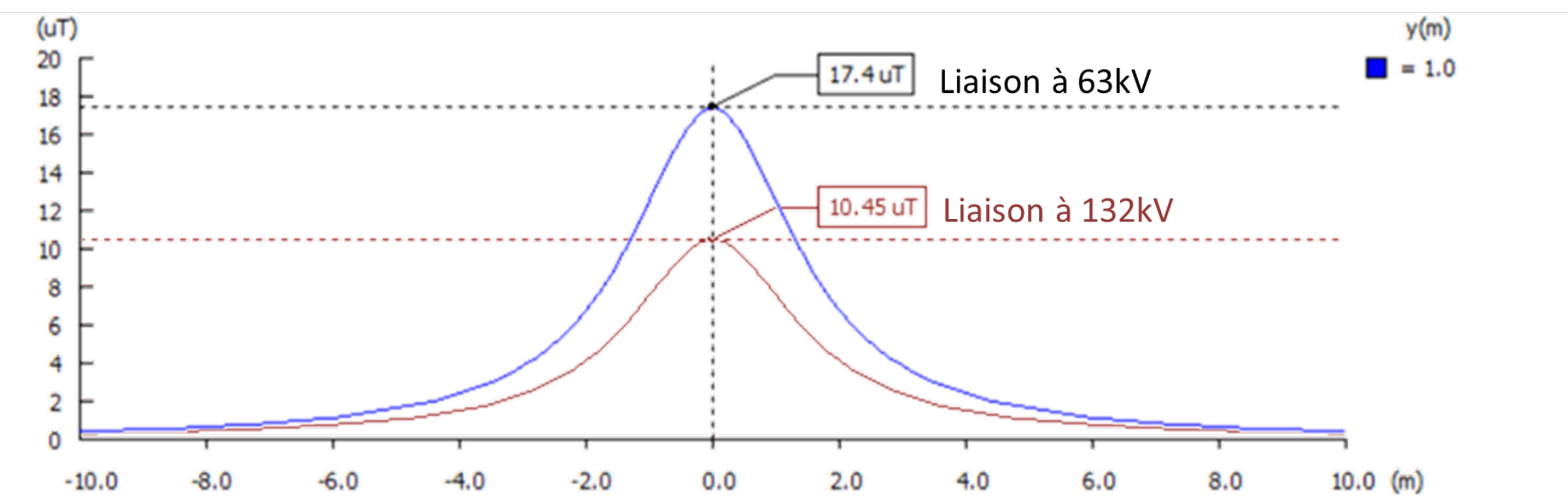

Fig. 8.59: Comparison of magnetic field simulations for cables (63 kV in blue, 132 kV in red).

Two important points deserve to be mentioned: the cable runs in a concrete duct of 520×500 mm in the machine tunnel requiring a specific volume, as can be seen in Fig. 8.61; and the cables along a tunnel sector need junction chambers every approximately 1-1.5 km, requiring a significant space in the tunnel of approximately 1200×500×16 000(length) mm as shown in Fig. 8.62. This second point is important, as it was decided to create the junction chambers corresponding to each alcove (in the current baseline, every 1.6 km ), housing them in the transport parking areas to ease their construction and future accessibility.

Réf. : 202405281055 ind A

| Prysmian Group | Fiche Technique Commerciale | MI |
|---|---|---|

Câble 1200 mm² AL 63 kV PEA - GSC

| Voltage | Isolation | Ecran | Icc |
|---|---|---|---|
| 36 / 63 / 72.5 (kV) | PR | ALU | 8 kA-0,5s |

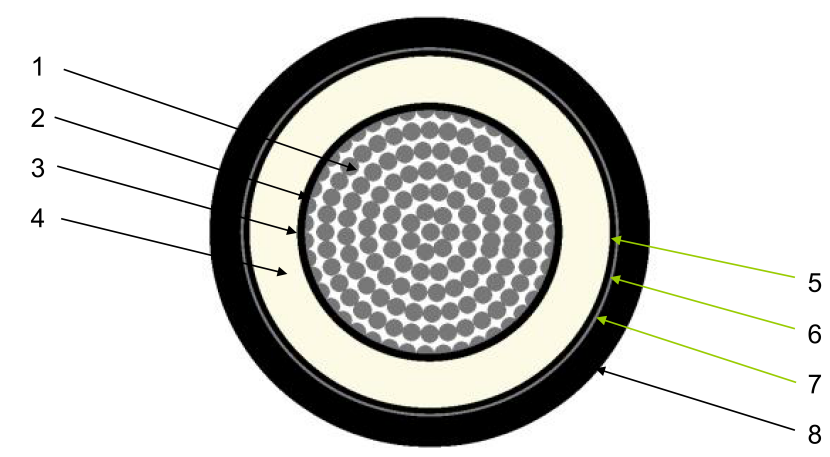

| Item | Description | Ep. Nominale (mm) | Détails | Diam. (2) Fabrication (mm) |
|---|---|---|---|---|
| 1 | Conducteur | - | Rond aluminium | *41.5* |
| 2 | Rubanage | - | Ruban semi-conducteur | -- |
| 3 | Ecran SC interne | 1.3 | Polymère semi-conducteur | -- |
| 4 | Isolation | 7.8 | PR | *64* |
| 5 | Ecran SC externe | 1 | Polymère semi-conducteur | -- |
| 6 | Rubannage | - | Ruban S-C hygroscopique | -- |
| 7 | Gaine métallique | 0.5 | Ruban Al posé en long à recouvrement et contrecollé | -- |
| 8 | Gaine extérieure (1) | 5 | PE + GSC | *79* |
| | | | Masse linéique indicative (kg/m) : | *7.3* |

| | |
|---|---|
| Effort de tirage maximal | 4320 kg |
| Rayon de courbure pendant le tirage | 3.16 m |
| Rayon de courbure en installation fixe | 1.58 m |

**(1) Gaine extérieure (PE) avec une couche de semi-conducteur extrudé (GSC)**

Fig. 8.60: Specification of the proposed 63 kV cable.

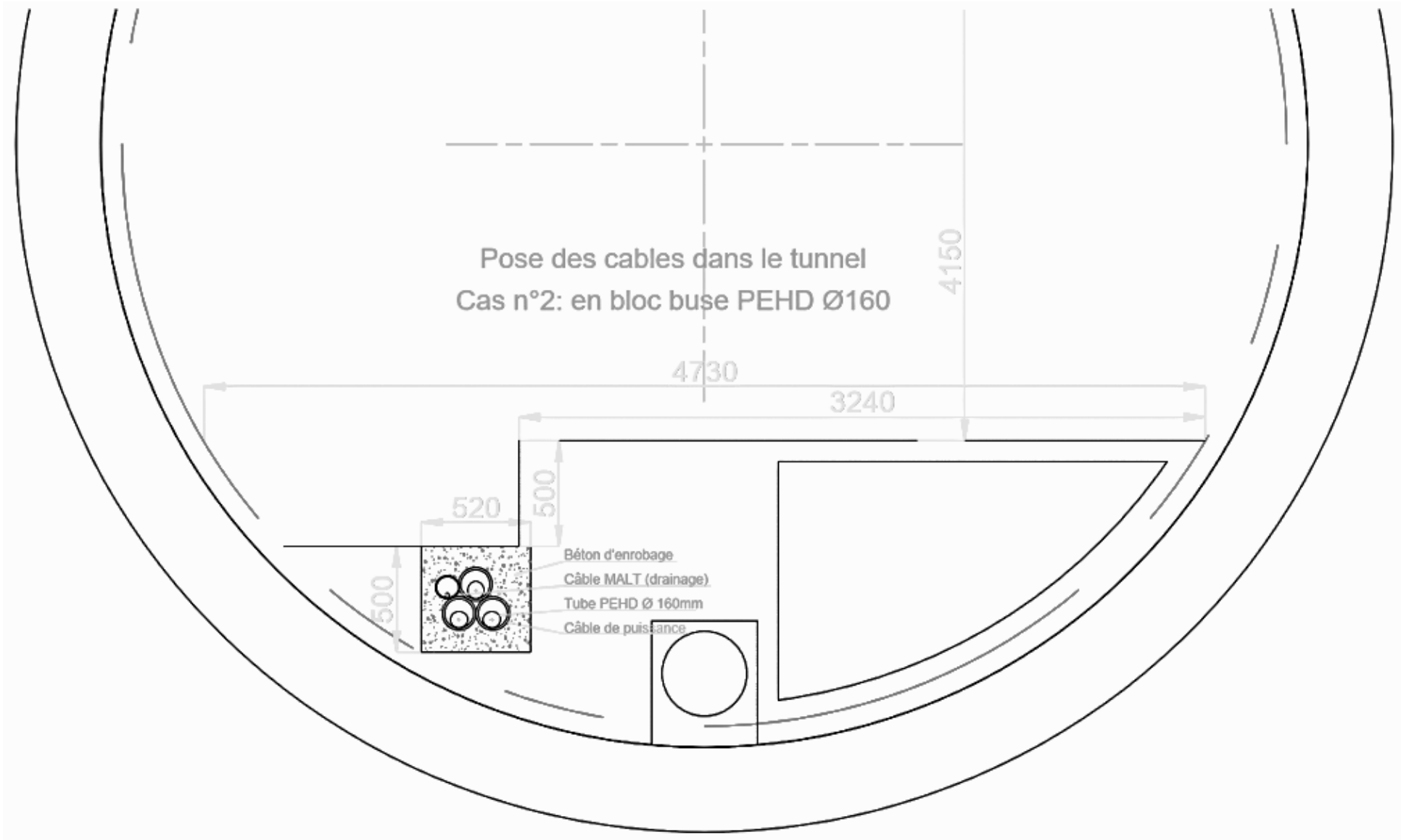

Fig. 8.61: Proposed space occupation of the cables in the machine tunnel.

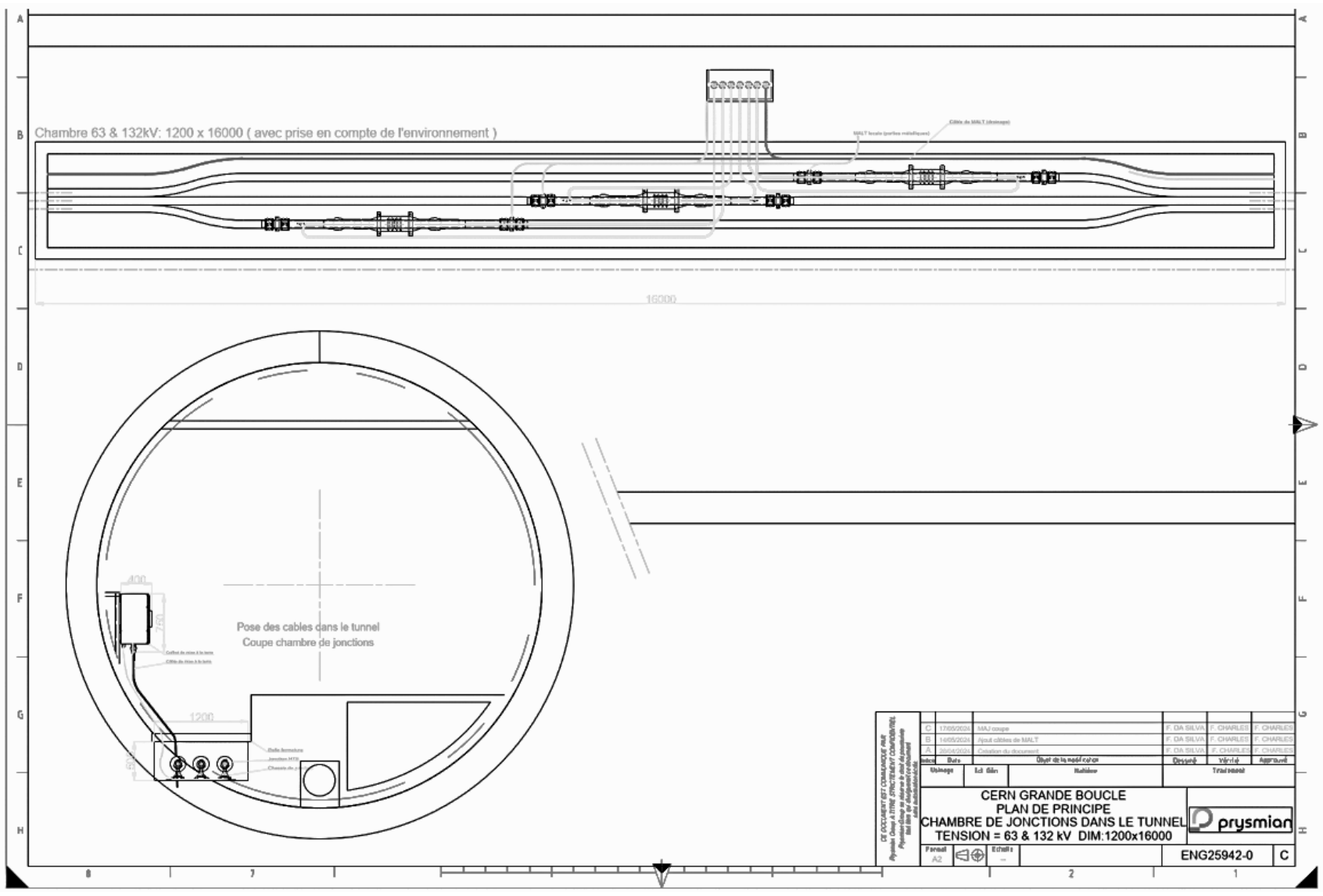

Fig. 8.62: Proposed space occupation of the junction chambers in the machine tunnel.

***Distribution networks***

The power is locally dispatched to the electrical loads of the surface and underground facilities by the medium and low voltage distribution networks available at each point.

The various loads can be classified based on their function and criticality, and each category has a dedicated distribution network with specific features, as shown in Table 8.23. The functional separation of the networks was one of the criteria in the study of the distribution at each point. This was to ensure users' power quality (in particular, to minimise potential disturbances from power electronics) and to enhance reliability, operability and maintainability of the grid.

Table 8.23: Network types and characteristics.

| Network Type | Loads type (non-exhaustive list) | Power unavailability duration in case of degraded scenario |
|---|---|---|
| Machine | Power converters, RF, cooling pumps, fan motors, etc. | Until return of main supply |
| General Services | Lighting, outlets | Until return of main or secondary supply |
| Secured | Personnel safety-related loads (lighting, pumps, elevators) | 10-30 s |
| Uninterruptible | Personnel safety (evacuation and anti-panic lighting, fire-fighting system, oxygen deficiency, evacuation)<br>Machine safety (sensitive processing and monitoring, beam loss, beam monitoring, machine protection) | Interruptions are not allowed; continuous service is mandatory |

The distribution network concept is the same for all points. The HV transmission network supplies two 63/20 kV transformers to create two main types of networks at 20 kV: General Services (GS) and Machine. The power is dispatched through two medium voltage substations (one in a surface building, the other in the underground service cavern) to 20/0.4 kV transformers that supply the low voltage switchboards in charge of local distributing to the terminal loads.

The GS and machine networks can be coupled at the 20 kV substations to allow a backup supply: the size of the main 20 kV transformers, busbars and links (primary and backup) has been set to allow full reconfiguration and nominal operation in case one element of the network is unavailable as a result of maintenance or failure. For these reasons, and also to cope with the load forecast of the FCC-hh, the two main transformers of each point have been sized at 40 MVA, and the main 20 kV links at 20 MVA.

An example is shown in the single line diagram of the distribution network of point PA shown in Fig. 8.63, and the layout of the 20 kV substations on the surface at the same point is shown on Fig. 8.64.

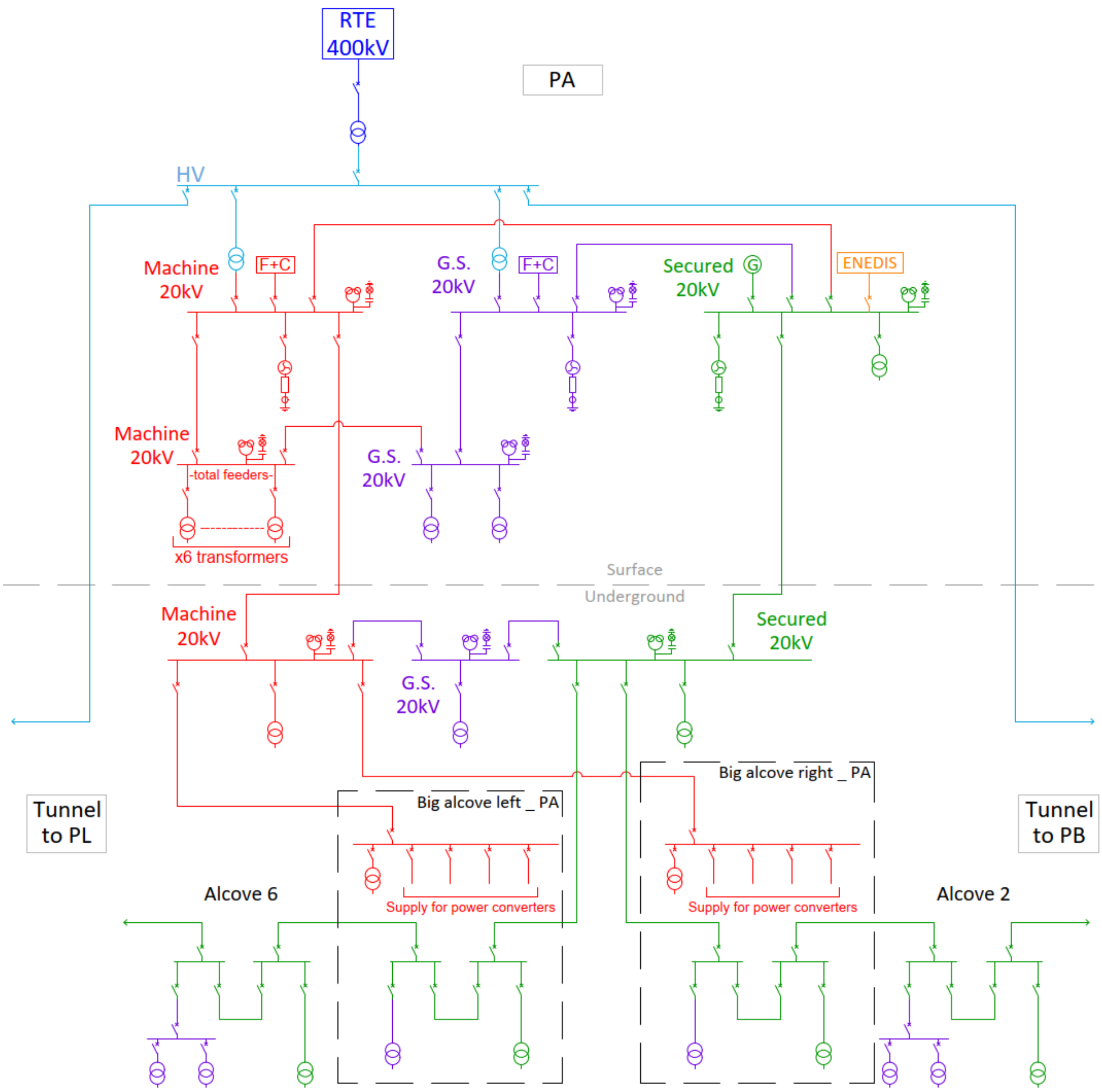

Fig. 8.63: Conceptual electrical distribution for point PA.

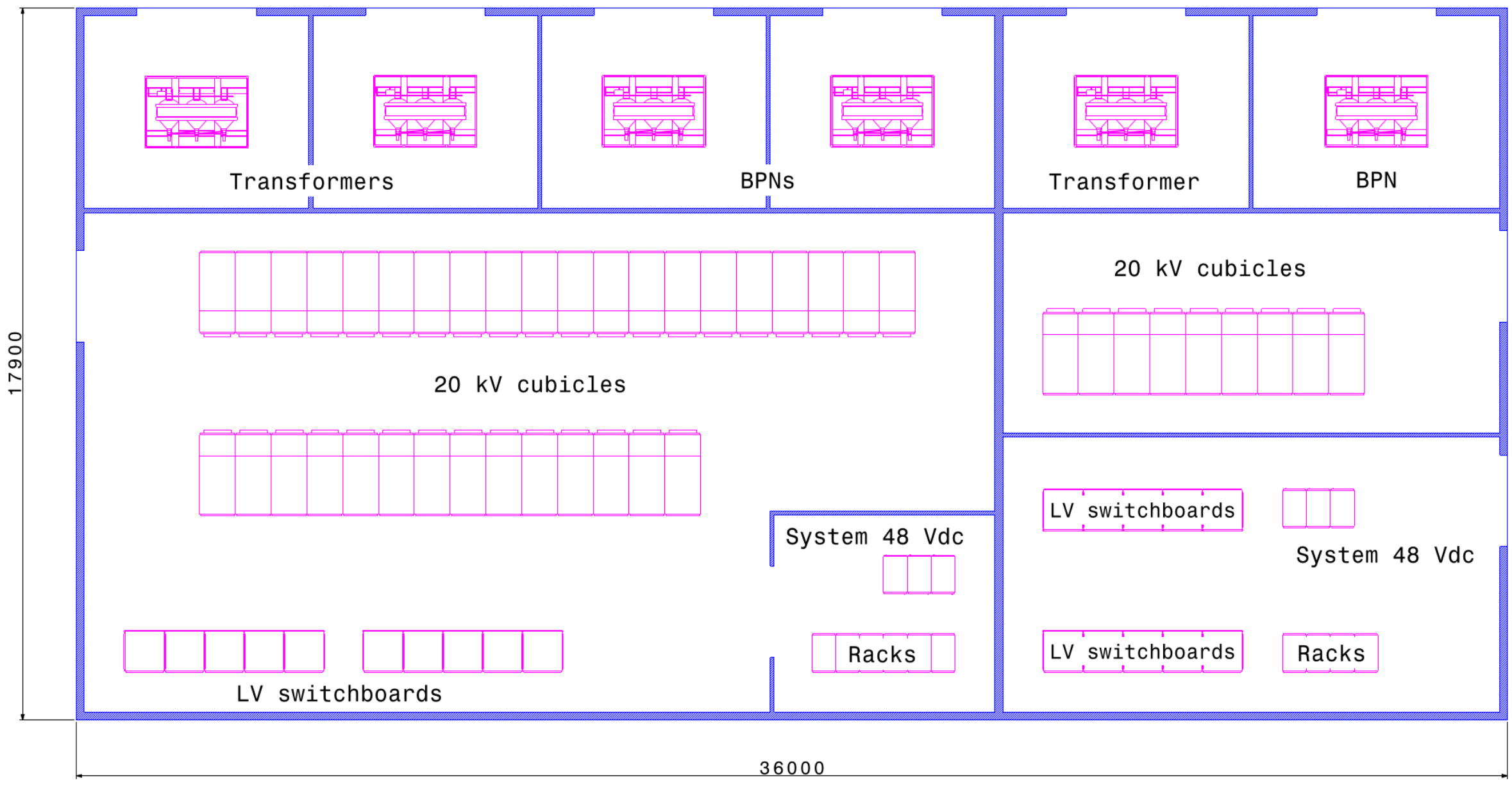

Fig. 8.64: Layout of the 20 kV substation of point PA.

***Secured and uninterruptible networks***

Based on the load categories of Table 8.23, there is a secured network with backup sources available at each point to power critical loads within a few seconds of the main supply becoming unavailable. The first priority backup sources have an emergency power station installed on the surface close to the substation (the baseline is a diesel generator, alternatives like hydrogen-based groups are under evaluation). As an alternative, there will be a connection to the medium voltage utility deployed during the construction phase which will be kept available.

The backup sources will be at 20 kV and will be connected to the surface substation, on a busbar where only critical loads are hosted; from there, a cable running down the shaft will connect to the equivalent in the underground substation. From the underground substation, the secured network will be distributed to each alcove through a 20 kV line, and each secured load in the tunnel will be powered from its closest alcove.

The distribution from the secured substation in the tunnel must be done at medium voltage, as stepping down to low voltage would mean having an excessive voltage drop in the circuits running along the sector. In this sense, a possibility for optimisation was identified during the study, and to avoid duplicating the 20 kV cables in the tunnel (one for the normal network, one for the secured power), it is proposed to use the same link deployed for the general services, applying a load shedding logic in case of power outage. The aim is to minimise the space occupation and ease integration in this tight area.

In all the substations (surface, underground, alcoves) there will be a coupling between the critical loads 20 kV busbar and the general services busbar; in normal operation, the coupling will be closed, and the link between the substations will serve both critical and not-critical loads. In case of a failure, maintenance or any other unavailability of the main source, the electrical network will be supplied using the backup power sources: in this case, the couplings between the secured network and the general services network will be automatically disconnected, and in the tunnel the same 20 kV link will only keep the loads related to the safety of people live.

The secured network of one access point covers all the facilities of the point itself at medium voltage and up to half a sector left and right in the tunnel. To enhance the reliability of the system, there is an MV coupling to the last alcoves of each half sector: this stays open during normal operations. If a part of the network connected to one point is not operational, all the secured network busbars can be powered from the adjacent points through the tunnels by closing this coupling.

The secured network described so far is specifically designed to support electrical loads related to the safety of personnel. If a similar level of backup is required for systems associated with the accelerator or experiments, a parallel infrastructure with equivalent characteristics will need to be developed and implemented. For loads that cannot accept a brief interruption of few seconds, whether they are safety-related or critical to operation, an uninterruptible power supply (UPS) system supported by batteries will be made available. This system will be strategically installed in buildings, service caverns and alcoves, to ensure a continuous power supply with the necessary autonomy. This infrastructure will serve critical loads of systems like vacuum, machine protection, cryogenics, RF, control, IT.

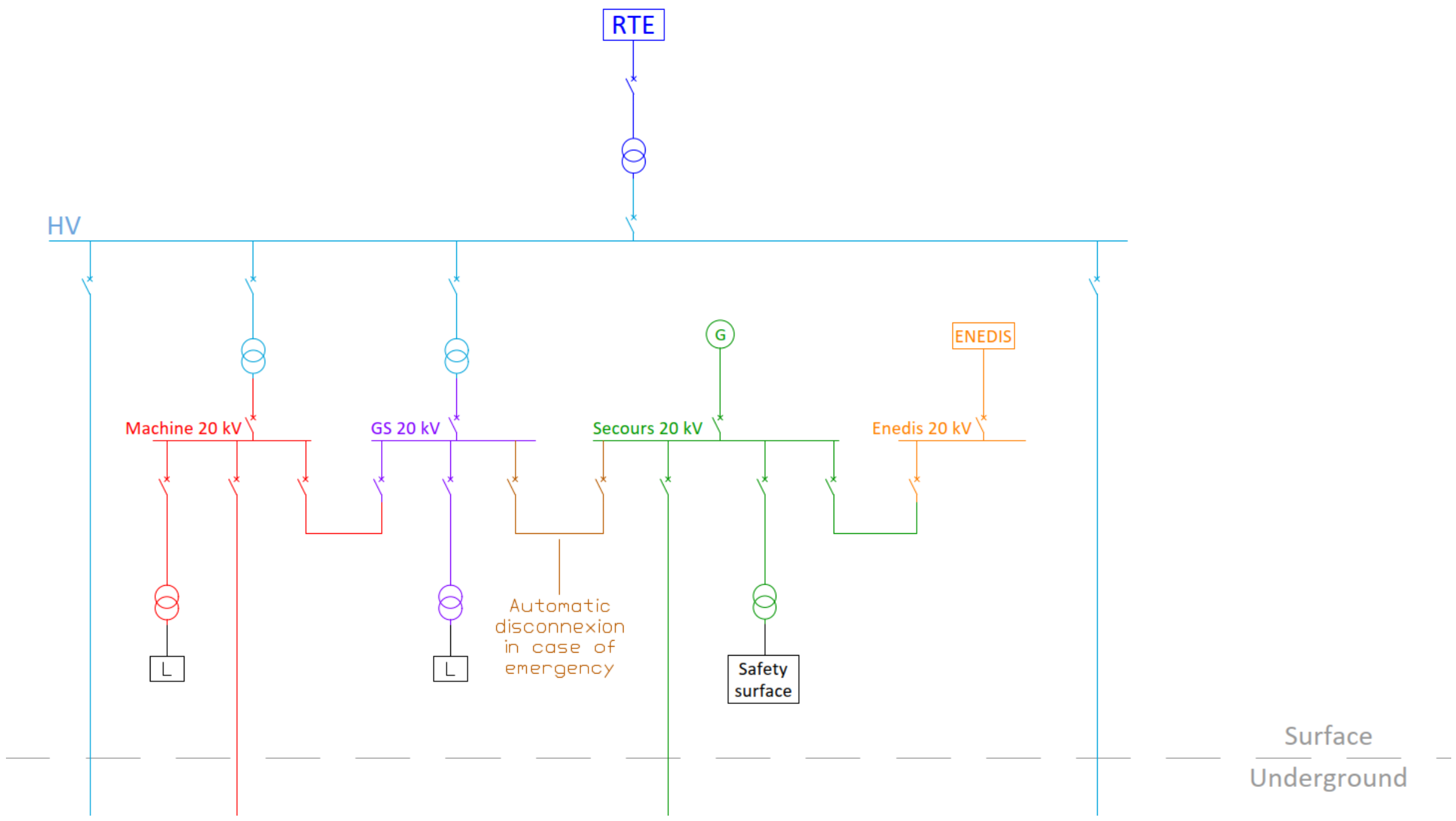

(a) On surface,

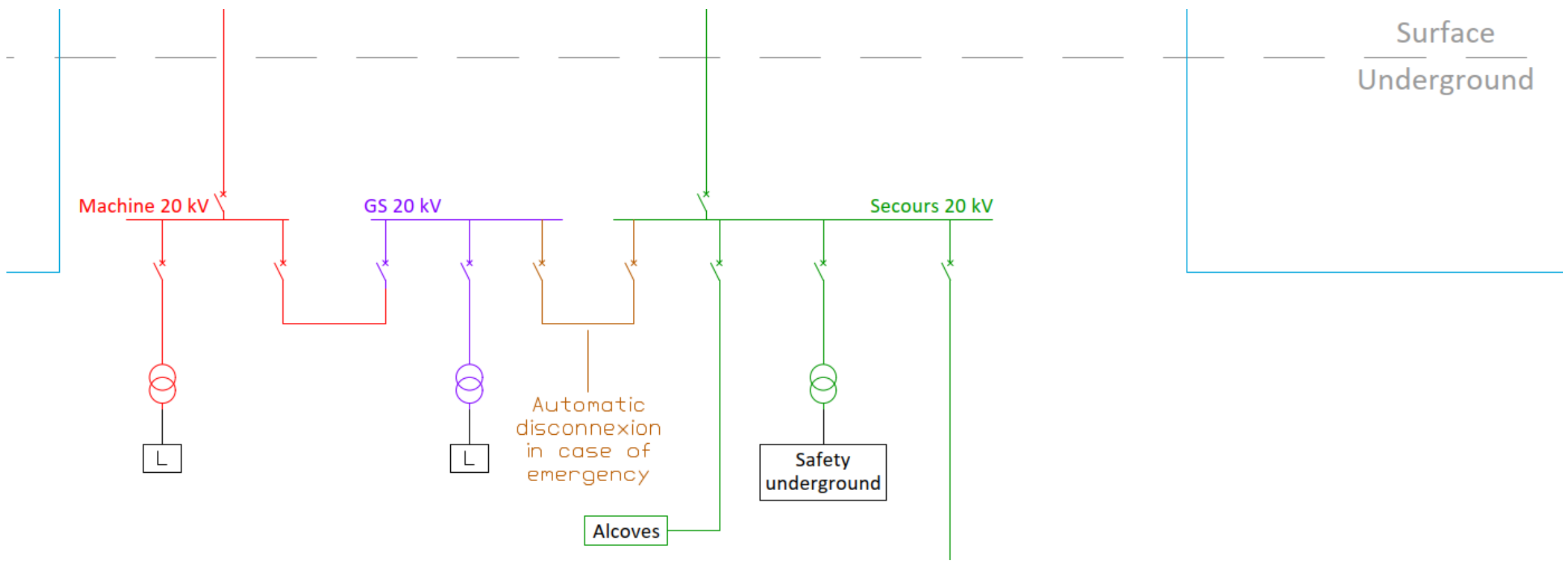

(b) underground,

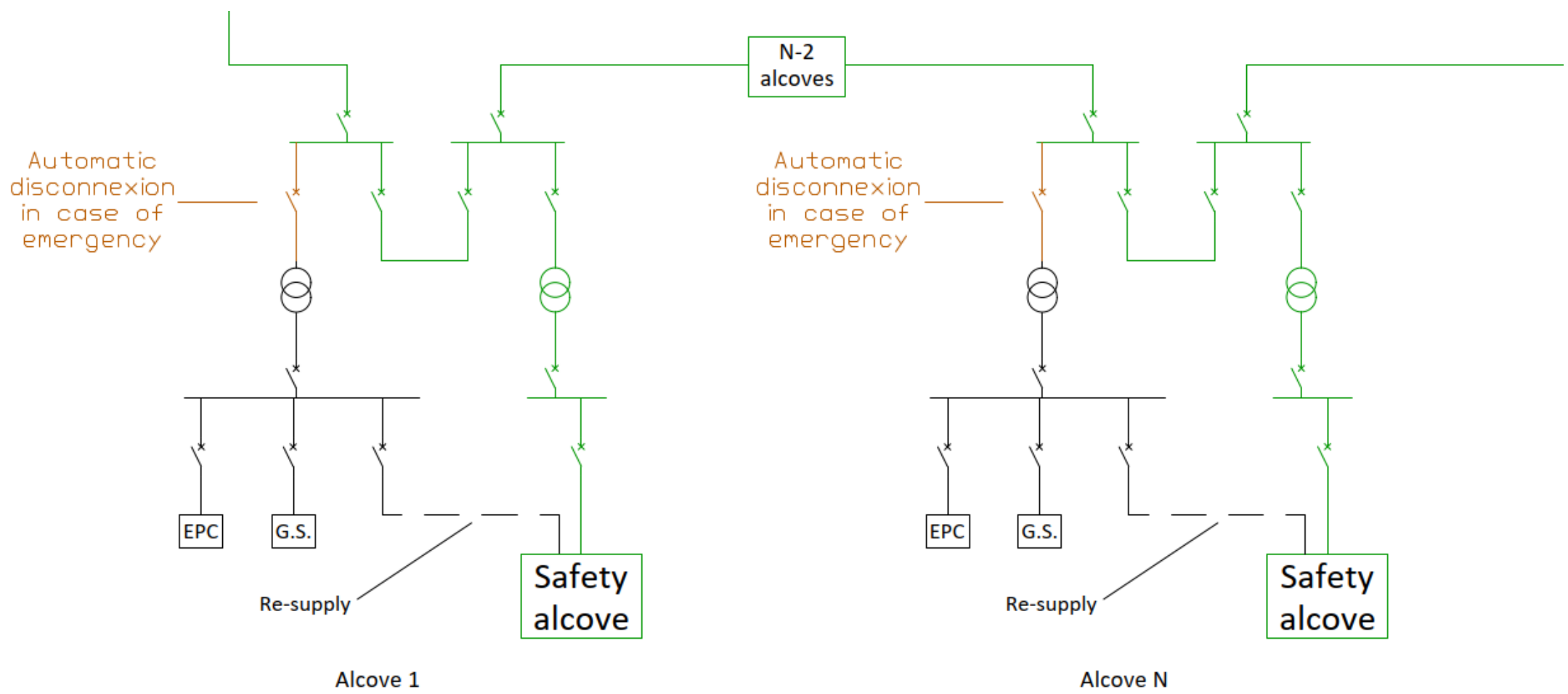

(c) in the tunnel.

Fig. 8.65: Conceptual electrical distribution for the secured network.

### 8.4.3 RF Powering

The RF system requires a power source capable of delivering a stable, high-quality DC high voltage. In existing implementations, such as the LHC and other accelerators, this is typically achieved by supplying a small group of RF amplifiers with a power converter operating in the MW range. However, for FCC-ee, this approach has been deemed economically unfeasible due to the large number of power converters and the extensive surface infrastructure that would be required at point PH. To address this challenge, a centralised power conversion strategy has been adopted, utilising a modular multilevel converter (MMC) to efficiently meet the system's power requirements.

Given that the RF system's power demand exceeds 100 MW, a thyristor-based converter was considered but ultimately found to be impractical due to power quality issues. If such a converter were implemented, a large-scale reactive power and harmonic compensation system would be required to maintain acceptable power quality. In contrast, the MMC solution offers significant advantages: it requires minimal or no harmonic filtering on the AC side and provides flexible reactive power control. This capability could potentially be leveraged to partially compensate for the reactive power consumption of the accelerator, further optimising energy efficiency.

### 8.4.4 Arc powering and alcoves

Power is distributed through the alcoves along the tunnel from the underground 20 kV substations. The alcoves serve multiple functions: they house equipment that must be protected from radiation, as well as local distribution systems for the arc, including accelerator systems, cooling and ventilation, safety systems, lighting, and power boxes. Additionally, they accommodate the power converters responsible for supplying the magnets in the arc.

The current baseline design includes seven alcoves per arc, with two larger alcoves positioned at the ends of the LSS (long straight section) of each point. These larger alcoves will host the power converters for the main arc magnets (dipoles and quadrupoles), as they require more space. The remaining alcoves will be spaced every 1600 m, each covering a sector extending 800 m on either side, distributing power accordingly.

The machine and general service networks will be separately available only in the first alcoves at the LSS ends, as the power converters in these locations have significantly higher loads and must be powered using dedicated transformers connected to a separate machine network. However, for the other alcoves, where power requirements for converters and other accelerator systems are much lower, it has been decided to supply them only from the general service network to avoid installing two separate 20 kV links in the tunnel.

To achieve this optimisation, the functional separation criteria outlined in Table 8.23 has been suspended for the alcoves. Instead, the choice is based on the assumption that the electrical parameters of the installed systems, particularly the power converters, will comply with specific requirements for connection to the network, including resilience to voltage and frequency variations and low total harmonic distortion (THD).

A similar synergy has been adopted for the safety-related loads and the secured network in the alcoves. These will be supplied through the same 20 kV link running through the tunnel, using a load-shedding logic to prioritise general services when necessary.

Following these technical choices, one 20 kV cable will run through the tunnel to the first smaller alcove from the alcoves at the end of the LSS at each access point, connecting it and continuing to the next, up to the end of the arc at the next access point (see Fig. 8.66. In the nominal configuration, each access point will supply the alcoves of half sectors to the left and to the right, with the one in the middle having an open coupling between the two sides. Closing this coupling and having the 20 kV cable running from one point to the other, will allow re-supplying all the alcoves from only one point if the other is unavailable.

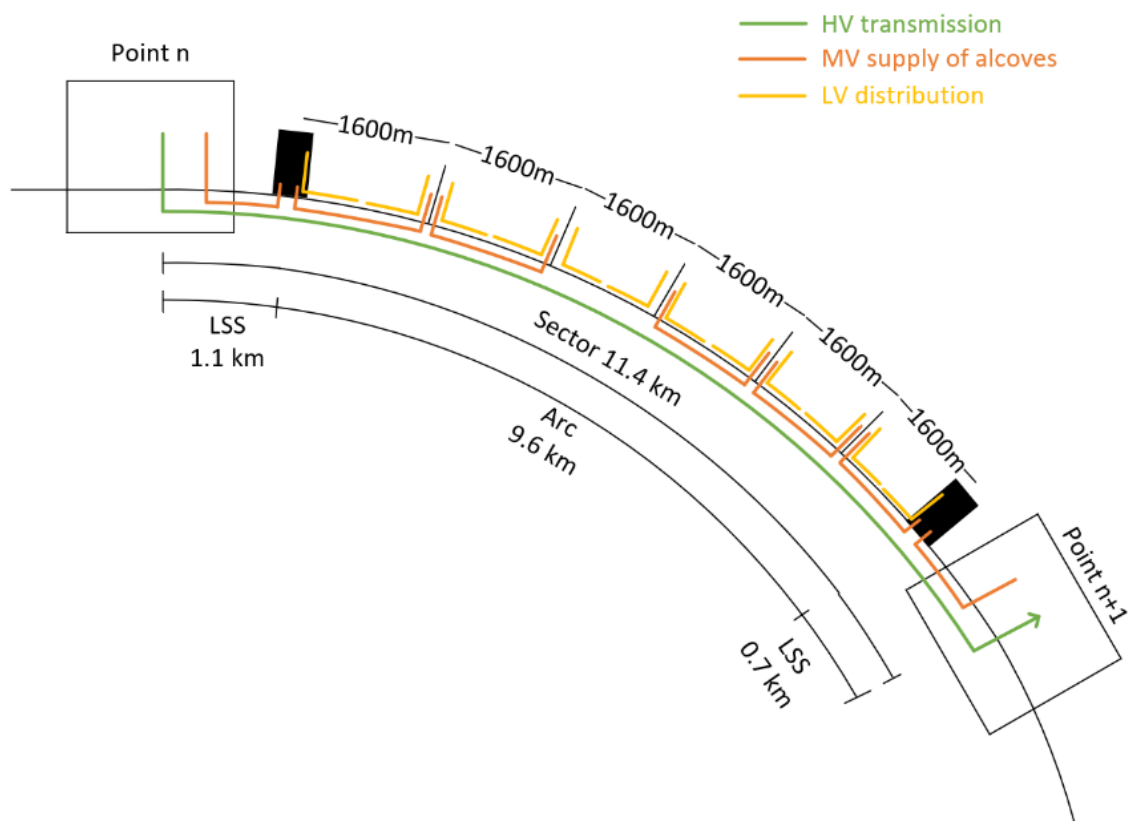

Fig. 8.66: Arc and alcoves powering layout concept.

### *Electrical equipment in the alcoves*

The power distribution of the alcoves has been studied and sized based on the load requirements identified for the FCC-ee arc. The 20 kV cable coming from the arc will arrive at a switchgear divided into two busbars for secured and other loads. The switchgear will be connected to 20/0.4 kV transformers powering various low voltage switchboards, from where all the circuits going to the alcove and the tunnel are supplied. A redundant UPS will provide the required backup to create an uninterruptible network for the critical loads, and a redundant power supply will be dedicated to the emergency lighting.

In addition to the power converters, a certain number of racks will be installed, supplied through ad-hoc busbar trunking systems, and hosting the systems required for the accelerator and the services in a sector. Today the number of these racks is estimated at 13 units but could evolve in accordance with the evolution of the systems' requirements. Figure 8.67 shows the typical single-line diagram of the power distribution of an alcove.

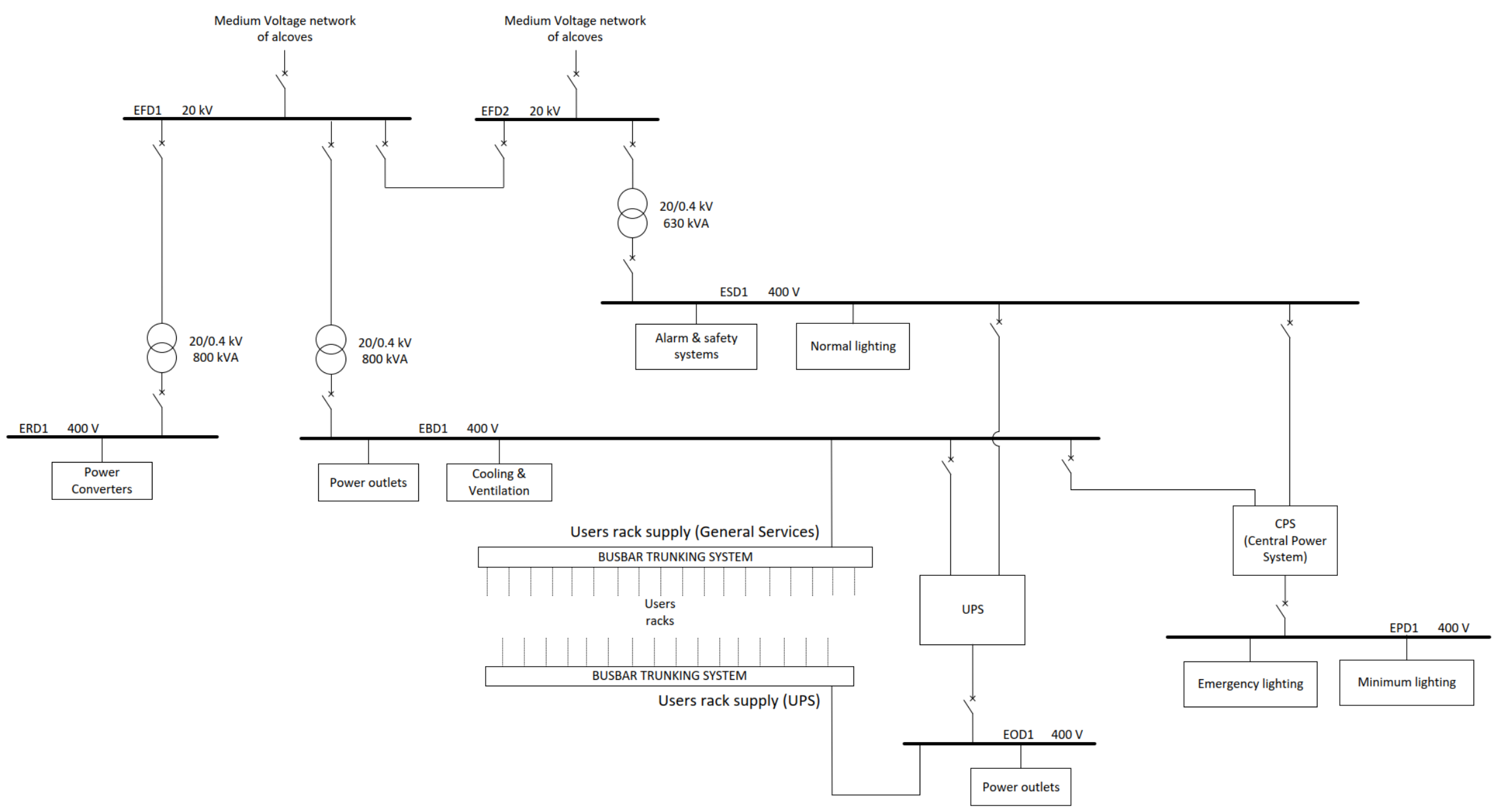

Fig. 8.67: Typical single line diagram of the power distribution of an alcove.

A preliminary study has been conducted for the so-called big alcoves at the end of the LSS, con-cerning the powering of the power converters and of the loads of the half sector. The principles that have been applied are the same as used for the other alcoves, and the single line diagram is shown on Fig. 8.68.

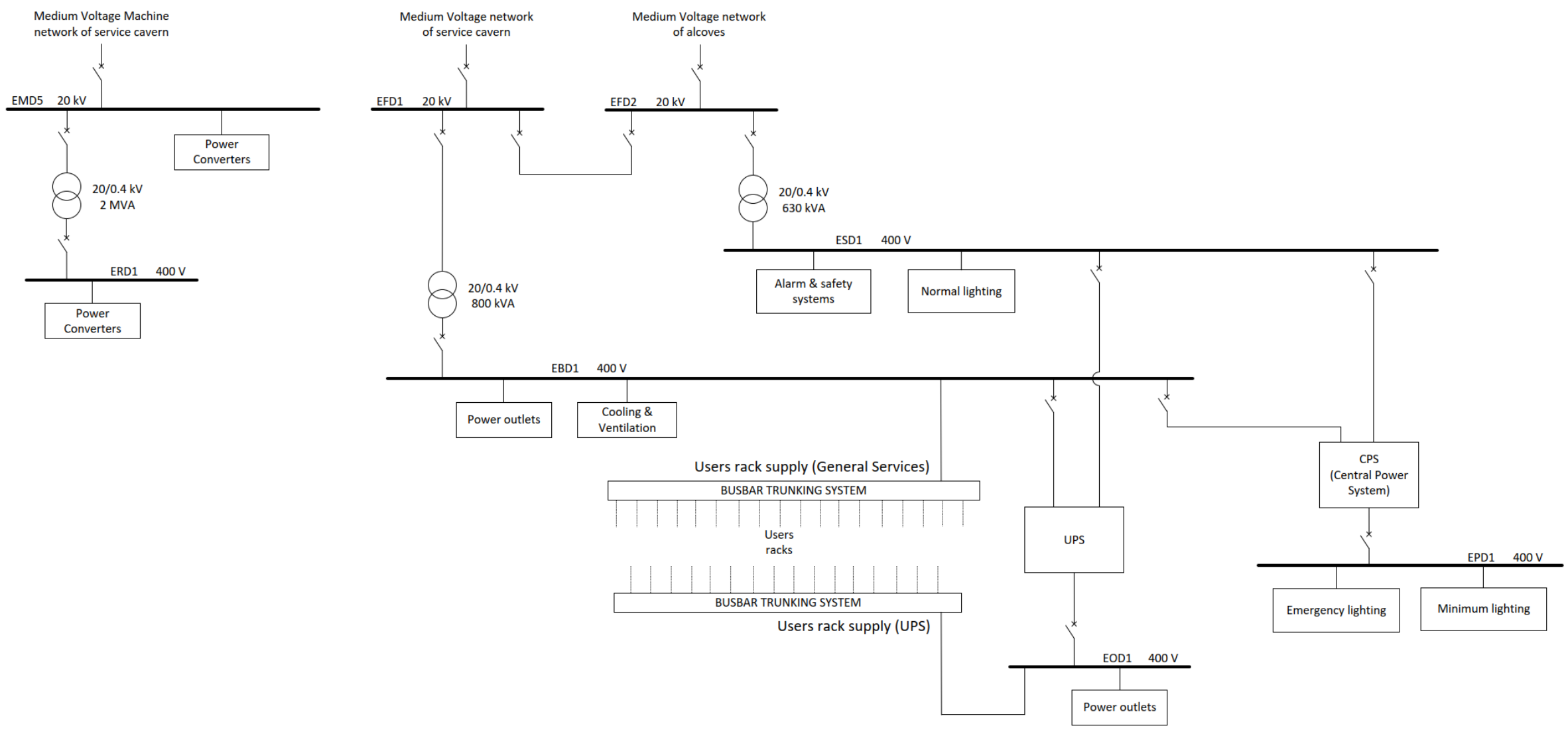

Fig. 8.68: Typical single line diagram of the power distribution of a big alcove at the end of LSS.

*Electrical equipment in the arc*

The list of the equipment and circuits for the powering of the arc includes:

- One power outlet box on the general service network every 50 m, powered from only the closest alcove (maximum length of the circuit: 800 m), without redundancy.
- One power outlet box on UPS network every 50 m, with secured redundant powering from both alcoves to the right and left (maximum length of the circuit: 1600 m).
- One normal light every 5 m, powered from only the general service network of the closest alcove (maximum length of the circuit: 800 m), with a circuit connected in parallel to the secured switchboard of the same alcove supplying 1 light out of 5 (to create a set of minimum lighting every 20 m).

In addition, the list of the electrical equipment related to safety in the arc includes:

- One electrical emergency button (AU) every 50 m.
- One emergency light every 14 m, with secured redundant powering from both alcoves on the right and left (maximum length of the circuit: 1600 m).

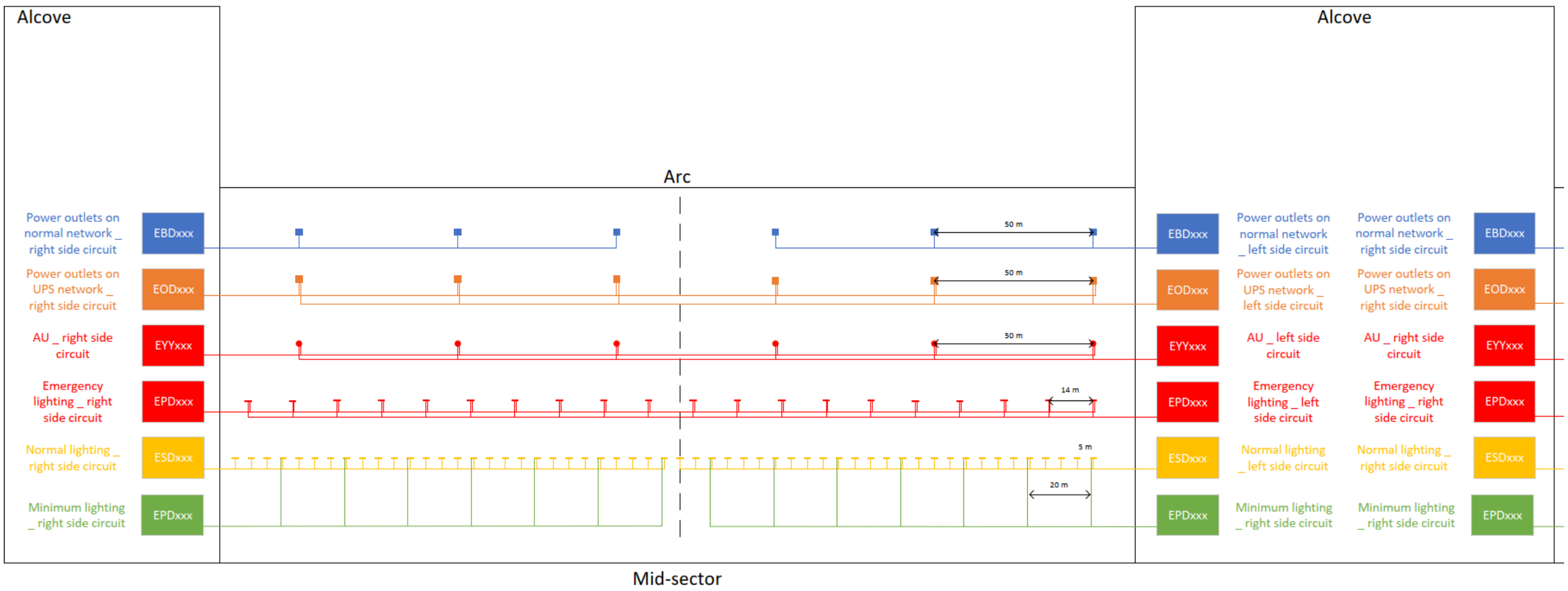

Fig. 8.69: Electrical distribution in the arc.

The list of the cables composing these circuits is given in Table 8.24.

Table 8.24: Estimation of the power cables in the arc.

| **User** | **Load** | **Cable tray** | **Cable type** | **Size (mm)** | **Number of lines in the tunnel** |
|---|---|---|---|---|---|
| **Powering** | MV power of alcoves | MV | $3\times(1\times400\,\text{mm}^2)$ + $(1\times120\,\text{mm}^2)$ Cu | $3\times$d=50 + d=21 | 1 |
| | Power outlets on GS network | LV | $3\times(1\times150\,\text{mm}^2)$ + $(1\times95\,\text{mm}^2)$ Cu | $3\times$d=21 + d=18 | 1 |
| | Power outlets on UPS | LV | $3\times(1\times150\,\text{mm}^2)$ + $(1\times95\,\text{mm}^2)$ Cu | $3\times$d=21 + d=18 | 2 |
| | Classic lighting | LV | $(5\times10\,\text{mm}^2)$ Cu | d=22 | 1 |
| | Emergency lighting | Safety | $(5\times2.5\,\text{mm}^2)$ Cu, flat | 24x6 | 2 |
| | Emergency stops link (AU) | Safety | $14\times(2\times1\,\text{mm}^2)$ Cu | d=21 | 2 |

The quantification of these cables is of great importance in the assessment of the space required for the cable trays and other containment systems (e.g., cablofil, duct) that need to be installed in the arc.

For the same reason, the signal and control cables and the optical fibres used by other systems and installed in the arc have also been estimated. This process started from the various system's owners when available, or from extrapolation from previous accelerator projects at CERN when the information was not available. The list of these cables is given in Table 8.25.

The collection of the powering and cables' requirements in the tunnel will be constantly updated in the next stages of the study to consider every possible modified or new request and to adapt the technical solution accordingly. Furthermore, the use of optical fibres should be prioritised over copper cables to minimise space occupancy when developing users' cable requirements.

Table 8.25: Estimation of the control and signal cables, and optical fibres crossing the arc.

| User | Load | Cable tray | Cable type | Size (mm) | Number of lines in the tunnel |
|---|---|---|---|---|---|
| **Fibre optics** | Backbone | FO | 3 cables (×24 fibres) | d=25 | 1 |
| | Underground | FO | 9 cables (×24 fibres) | d=25 | 2 |
| | BI (only BPMs) | FO | 13 cables (×12 fibres) +8 cables (×24 fibres) | d=25 | 4 |
| | Sensing | FO | 2 cables (×6 fibres) | d=25 | 1 |
| **Vacuum** | ion pumps | Signals | 1×0.63(HV) + $2\times0.25\,mm^2$ Cu | d=10.7 | 28 |
| | NEG (power) | Signals | $3\times2.5\,mm^2$ Cu | d=13 | 16 |
| | Penning | Signals | 3-axis 0.8/8.4 mm Cu | d=10.3 | 12 |
| | Pirani | Signals | $1\times(4\times1\,mm^2)$ Cu | d=6.5 | 12 |
| | BA power | Signals | $6\times(2\times0.75\,mm^2)$ Cu | d=14.5 | 6 |
| | BA collector | Signals | 3-axis 0.5/5.7 mm Cu | d=7 | 6 |
| | Sector valve | Signals | $6\times(2\times0.75\,mm^2)$ Cu | d=14.5 | 6 |
| | Profibus | Signals | $2\times(1\times0.35\,mm^2)$ Cu | d=8 | 1 |
| **Access & Alarms** | Sector doors | Safety | 13× 2×0.5 | d=17 | 1 |
| | Fire doors - magnet | Safety | 2×1.5 | d=10.5 | 3 |
| | Fire doors - position contacts | Safety | 2×1 | d=5 | 3 |
| | Fire doors - flashing lights | Safety | 4×1.5 | d=10.5 | 3 |
| | Call points - break-the-glass | Safety | 2×1 | d=5 | 2 |
| | Call points - telephones | Safety | 1× 4×0.6) | d=7.4 | 2 |
| | Evacuation - voice alarm | Safety | 2×2.5 | d=13 | 2 |
| **Cooling & Ventilation** | Fancoils | Signals | $1\times70\,mm^2$ | d=16.6 | 16 |
| | Dampers | Signals | $1\times35\,mm^2$ | d=12.1 | 21 |
| | Valves & other equipments | Signals | $1\times35\,mm^2$ | d=12.1 | 3 |
| **Radiation protection** | Radiation detectors ($t\bar{t}$ phase only) | Safety | 2×CEH50 + $2\times(2\times0.22\,mm^2)$ | d=9.5 | 2 |
| **Magnet Protection** | WIC fieldbus | Signals | $1\times50\,mm^2$ | d=14 | 4 |
| | BIS fibres | FO | 1 duct | d=25 | 1 |
| **Power Converters** | DC cables | | | | |
| **Others** | Unknown | | | | |

The information summarised in Table 8.24 and Table 8.25 allows the selection and dimensioning of the cable containment systems to be installed (mainly cable trays). This is one of the most important factors for the determination of the layout and cross section of the arc tunnel. The number and dimensions of cable trays are:

- Twelve 500 mm wide and 60 mm high cable trays for: DC cables (five cable trays considered), MV power cables, LV power cables, signal cables and optical fibres.
- Two 200 mm wide, 60 mm high and fire-resistant cable trays for safety systems.
- One 520×500 mm concrete duct for the HV cables.

### 8.4.5 Control of FCC grid

***AC Network Control Using Unified Power Flow Controllers***

Unified power flow controllers (UPFCs) are power-electronic-based network controllers capable of performing voltage control, power flow control, reactive power compensation, and protection against network perturbations. The versatility of these systems makes them highly suitable for controlling and stabilising the FCC AC transmission network, offering an alternative to systems such as static var compensators (SVCs) or DC networks.

A UPFC consists of two back-to-back converters, each connected to the network via a transformer. One converter is connected in parallel, enabling reactive power or voltage control at the associated network node. The other converter is connected in series, allowing active and reactive power flow control, voltage control, and voltage dip mitigation by injecting a specific voltage between the network and the loads. Figure 8.70 illustrates the basic structure of the UPFC.

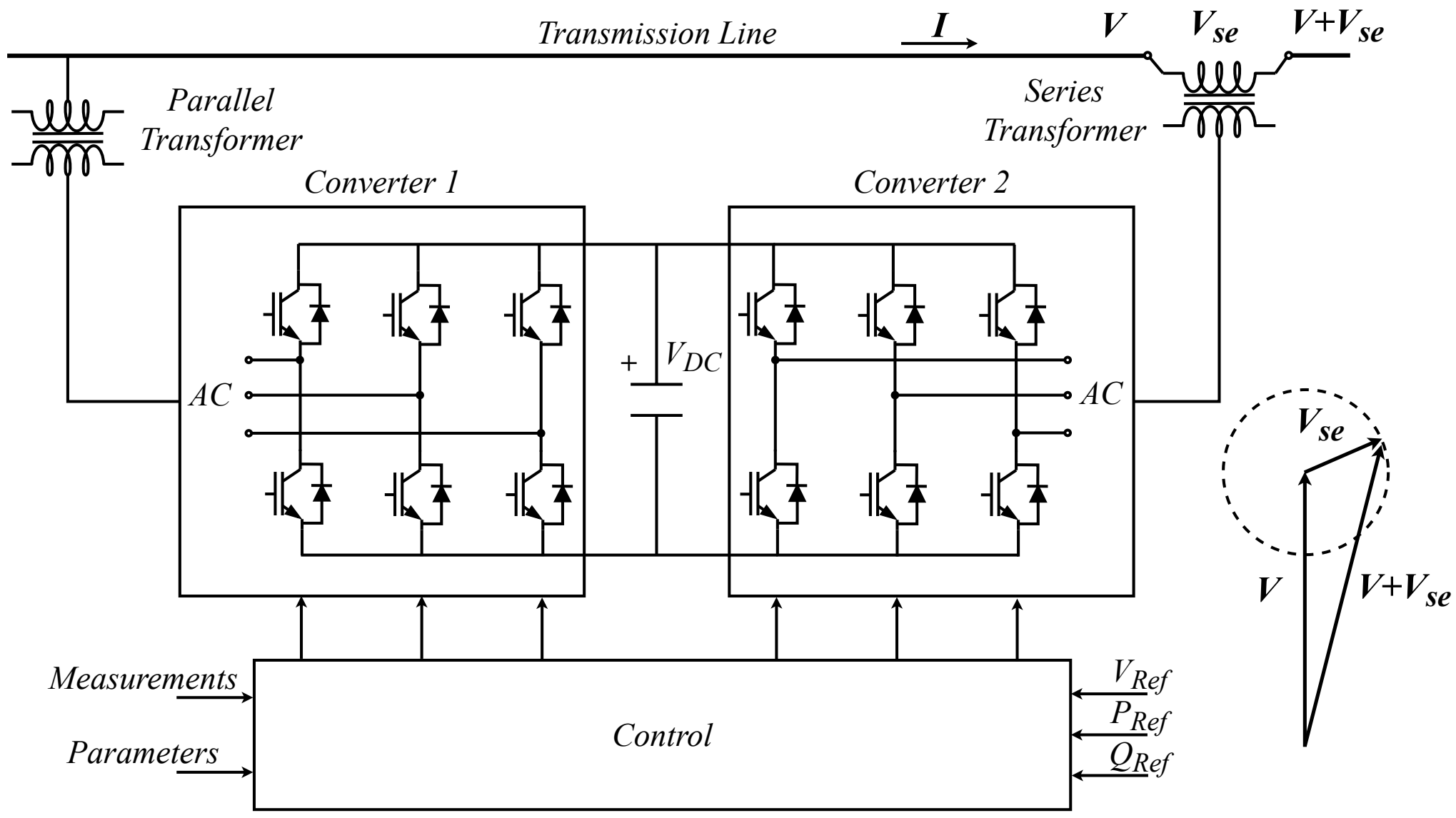

Fig. 8.70: Diagram of a unified power flow controller (UPFC) used to enhance controllability of the FCC network.

The application of UPFCs in the FCC network can be tailored to specific protection and reactive power compensation requirements. In the high-voltage transmission ring, UPFCs can be installed at each network connection point to manage the accelerator's reactive power using the parallel converter. At the same time, the series converter provides protection against voltage dips and prevents undesired power flows, particularly under closed-loop operation.

At the medium-voltage level, UPFCs enable voltage control and protect access points from voltage dips. To optimise equipment efficiency, protection can be focused on critical sections of the network by connecting the series transformer to selected feeders. This approach creates a dedicated machine network, effectively decoupling it from general services, following the model already employed in CERN's existing infrastructure.

Compared with existing compensating equipment such as SVCs, UPFCs offer enhanced functionality. While they perform the same tasks of reactive power compensation and voltage control, UPFCs are more robust to network perturbations, exhibit faster response times, and, through the series converter, provide load protection against transients - an advanced feature not currently available in SVCs.

At this stage of the project, the exact compensating requirements are not fully defined. However, it is anticipated that the reactive compensation needs for each access point will be approximately 10 MVAr. Regarding voltage dip compensation, most of the recorded voltage dips at CERN exhibit a depth of less than 30% of the nominal voltage. With this figure in mind, and assuming a power rating of 10 MVA for the loads to be protected at each access point, the UPFC must be designed to inject up to 3 MVA during such events. Based on these considerations, the total installed UPFC power per access point can be estimated at approximately 15 MVA.

#### *DC Distribution Alternative*

Recent developments in power electronics have enabled the use of DC for power transmission and distribution. Compared with conventional AC distribution, DC networks offer lower transmission losses, as only active power flows through the cables and the skin effect is absent. Additionally, DC networks provide precise power flow control and mitigate perturbations originating from the AC network.

The use of DC to transmit power from network connection points to accelerator loads is under consideration. In this transmission scheme, loads are supplied by high-voltage converters with energy buffers that decouple them from network dynamics. This configuration enables the local generation or absorption of reactive power as required by the loads and protects them from network perturbations. To implement these features, the network configuration shown in Fig. 8.71 is proposed.

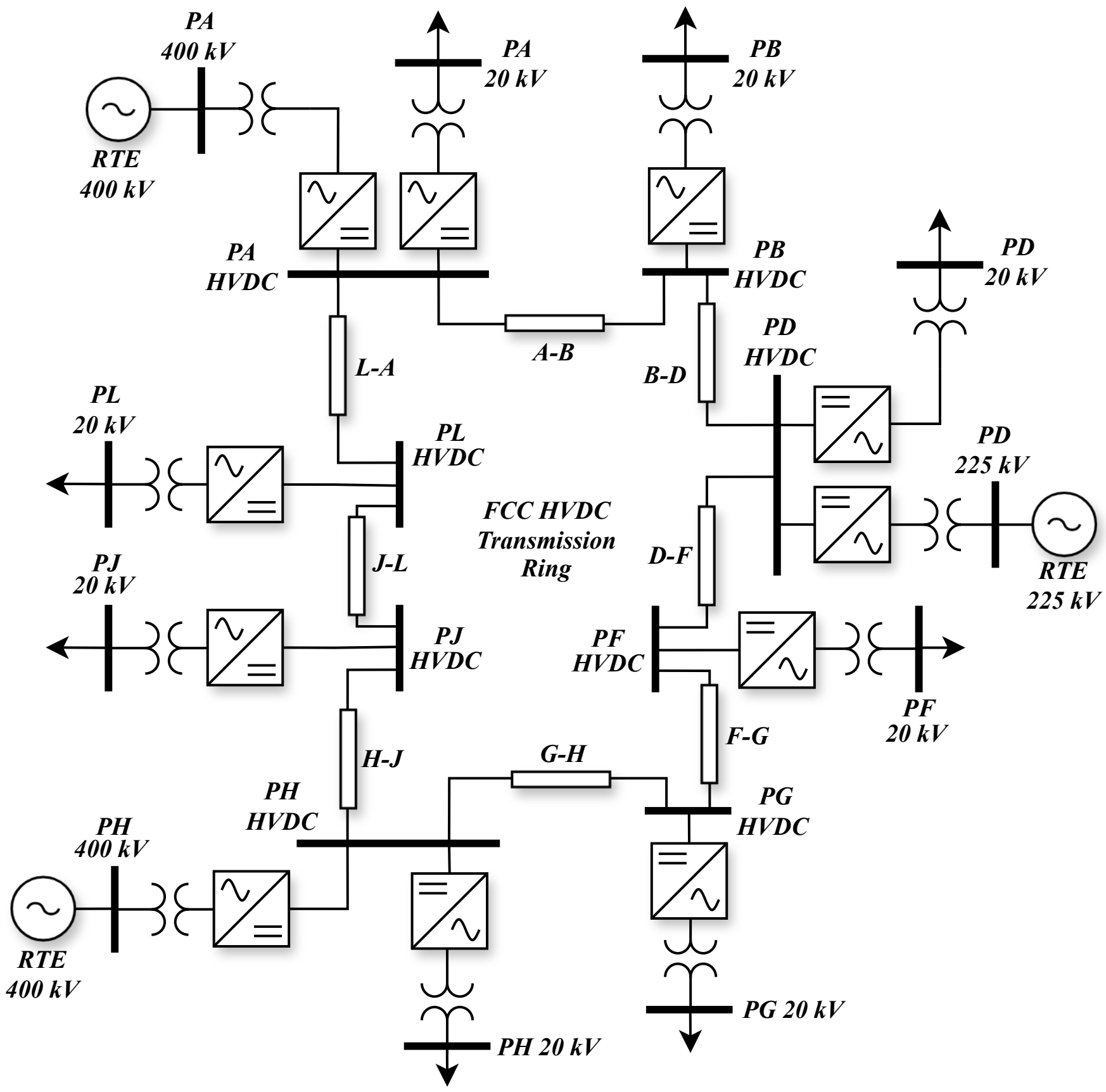

Fig. 8.71: High Voltage DC network used as an alternative to conventional AC distribution along the accelerator ring.

It consists of eleven power converters: three operating as rectifiers and eight as inverters. A key feature of this configuration is the closed-loop operation of the transmission ring, which guarantees precise control of power flows. This operational mode enables adjustments to the power drawn from the main connection points, optimising flow and ensuring resilience in the event of a feeder outage. In such scenarios, any point on the accelerator can still be supplied by at least two feeders located on opposite sides of the ring.

### 8.4.6 Energy storage systems

The use of energy storage systems (ESS) is being evaluated for the FCCee and FCChh. The primary objectives of ESS are:

- Providing energy to critical systems (e.g., safety or cryogenics) during electrical outages lasting from minutes to hours (some of these systems currently rely on diesel generators).
- Storing lower-cost energy from renewable sources (local or otherwise) for later use.
- Temporarily storing the energy of FCC-hh magnets during the deceleration phase of the acceleration cycle and returning it during the next cycle.

Technologies such as batteries, supercapacitors, flywheels, electrolysers, fuel cells (hydrogen-based), and gravitational energy storage systems have been preliminarily analysed and compared for the above-mentioned applications.

A collaboration has been established to explore the integration of hydrogen as an energy storage system (ESS) vector to support one or more functions within the accelerator complex. Hydrogen-based systems are particularly well-suited for both short and long-term energy storage, offering the advantage of decoupling energy storage capacity - determined by the hydrogen mass - from charging speed, defined by electrolyser capacity, and power output, which is set by the fuel cell capacity.

The heat generated by electrolysers and fuel cells is available at high temperatures, which can enhance the usability of waste heat. For instance, this heat can be used to increase the temperature of water used for cooling various equipment, thereby improving the efficiency and transportability of waste heat over longer distances. Additionally, the heat recovered can be leveraged to drive absorption cooling cycles, contributing to the cooling needs of magnets, power converters, tunnels, and other infrastructure.

In principle, hydrogen-based ESS technology offers high flexibility and adaptability across a wide range of applications. However, it has some drawbacks, including relatively low conversion efficiency and the requirement for robust hydrogen storage systems. Nevertheless, the co-generation of usable heat and cooling can partially offset these efficiency limitations, making hydrogen a promising candidate for integration into the FCC infrastructure. Studies on the feasibility and implementation of hydrogen ESS within the FCC are ongoing and will be further intensified in the future.

For the FCC-hh superconducting magnets, preliminary assessments suggest that batteries appear to be the most suitable ESS technology for recovering energy during the ramp-down phase of the cycle and re-injecting it into the magnets during the next acceleration cycle.

## 8.5 Cryogenic systems

Cryogenic infrastructure is essential for various components of the accelerator. The experiment points PA, PD, PG, and PJ will each be equipped with two cryoplants: one dedicated to the detector magnet (detector cryoplant) and another to the machine detector interface (MDI) region magnets (interaction region (IR) cryoplant). These systems ensure the required low-temperature environments for the sensitive components in the experiment regions.

While the technical points PB and PF do not require cryogenic refrigeration, the technical points PH and PL are associated with the radiofrequency (RF) system, which necessitates operation at 2 K and 4.5 K to efficiently accelerate particles to the required energies. Each technical point features a long straight section (LSS) of 2032 m, symmetrically distributed around the interaction point (IP). The RF system, composed of a string of superconducting cavities and klystrons, is housed within these LSS regions and requires dedicated cryogenic refrigeration to maintain the necessary low-temperature conditions for efficient operation.

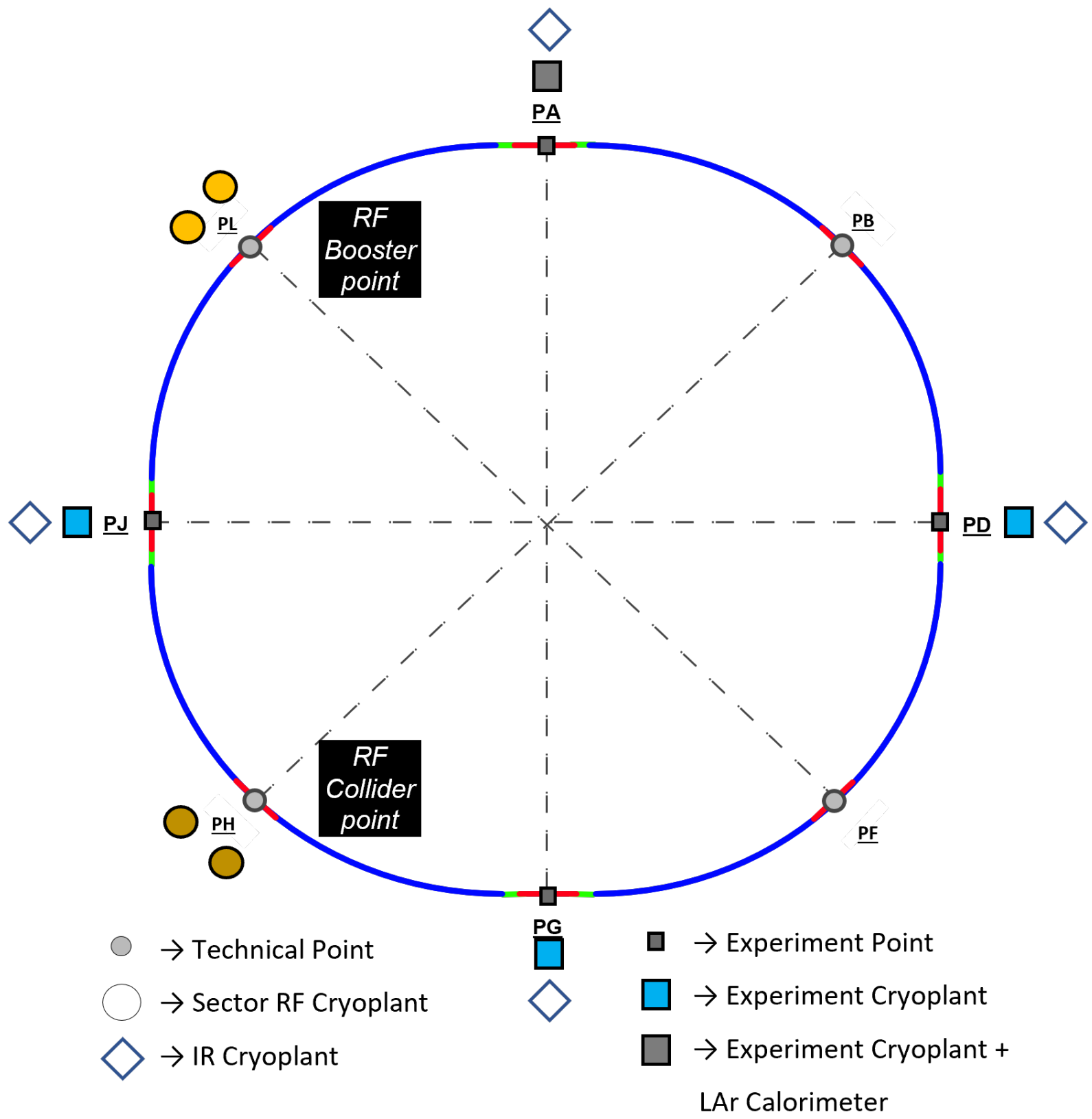

Fig. 8.72: FCC-ee collider cryogenic system layout for $\mathrm{t\bar{t}}$ operation.

Installation of the cryogenics for the cooling of the superconducting RF systems will follow the same strategy defined for the cryomodules to be operated in the technical points PH and PL.

A new RF layout was proposed in June 2024 with the aim of simplifying the overall installation of the SRF system. As described in Section 3.4.3, it consists of installing and, from the beginning, operating all the 2-cell 400 MHz cavities for the Z, WW and H mode, with each cavity powered at 500 kW in continuous wave (CW) mode. In other words, all the cryomodules necessary for the operation of FCC-ee at the energies of Z, WW and H levels will be installed at once and operated with the heat loads corresponding to the Z energy. This is valid for both 400 MHz and 800 MHz cryomodules.

This new layout significantly simplifies the design and staging of the cryogenic system. There will only be two stages: one for the Z, WW and H operation, with cryoplants operated to cover heat loads at H energy, and another one for $\mathrm{t\bar{t}}$ operation, with the same cryogenics scenario. This major modification drove the update of the cryogenic system with respect to the CDR, and the corresponding results can be found in the following sections.

### 8.5.1 Cryogenics for superconducting RF systems

#### *Layout and architecture*

The general cryogenic layout of the FCC-ee RF system is shown in Fig. 8.73. Cryoplant locations are indicated using small circles near the experiment and technical points.

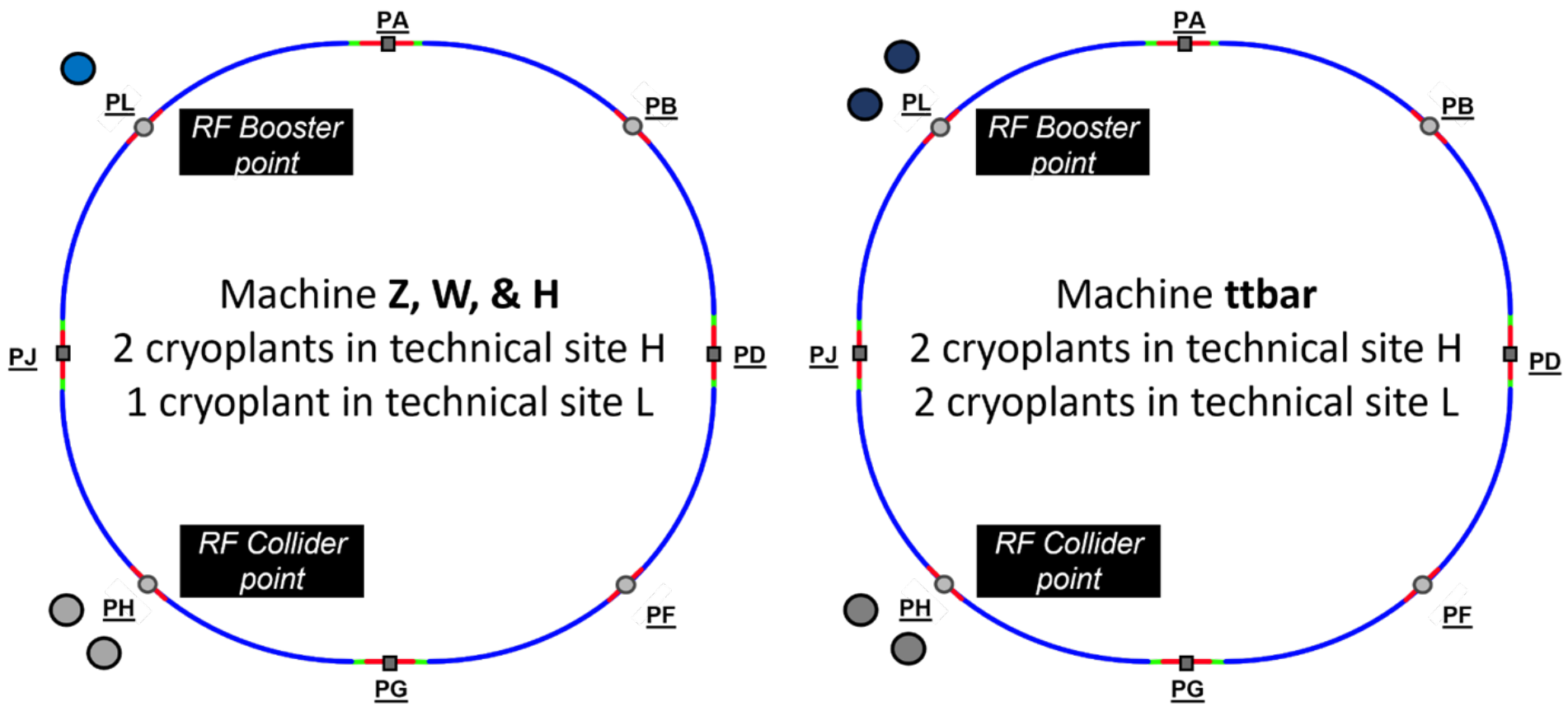

Fig. 8.73: General cryogenic layout for the technical points.

Aiming to optimise the integration of the machine, the RF systems of the collider and the booster have been split between two points. Point PH contains the collider RF, while point PL contains the booster's. This separation is also optimised from the cryogenic point of view as the booster and the collider present different heat load profiles. The booster ramps up the energy of the particles from injection energy to collider energy while the collider maintains a constant energy. Both the dynamic and static loads vary greatly. Moreover, the booster is composed solely of bulk Niobium 800 MHz cavities operating at 2 K, whilst the collider is mainly composed of coated elliptical 400 MHz cavities operating at 4.5 K. For the collider at point PH, there is an exception at the $\mathrm{t\bar{t}}$ stage, where 800 MHz cavities are installed to further increase the accelerating gradient achieved with the 400 MHz cavities.

Figures 8.74 and 8.75 show the cryogenic plant architectures for point PH and PL respectively. For certain stages that contain two cryoplants, the cryogenic plant architecture includes an interconnection box (QUI) that couples the refrigeration equipment to the cryogenic distribution line. The interconnection box facilitates a level of redundancy amongst the refrigeration plants and eases maintenance procedures. The other elements that are shown in the figures are the warm compressor station (WCS) and the upper cold box (UCB), located on the surface. Also included is the lower cold box (LCB), located in the cavern. This split box architecture has a temperature cut of 40 K between the two cold boxes, which reduces the hydrostatic losses incurred due to the depth of the tunnel.

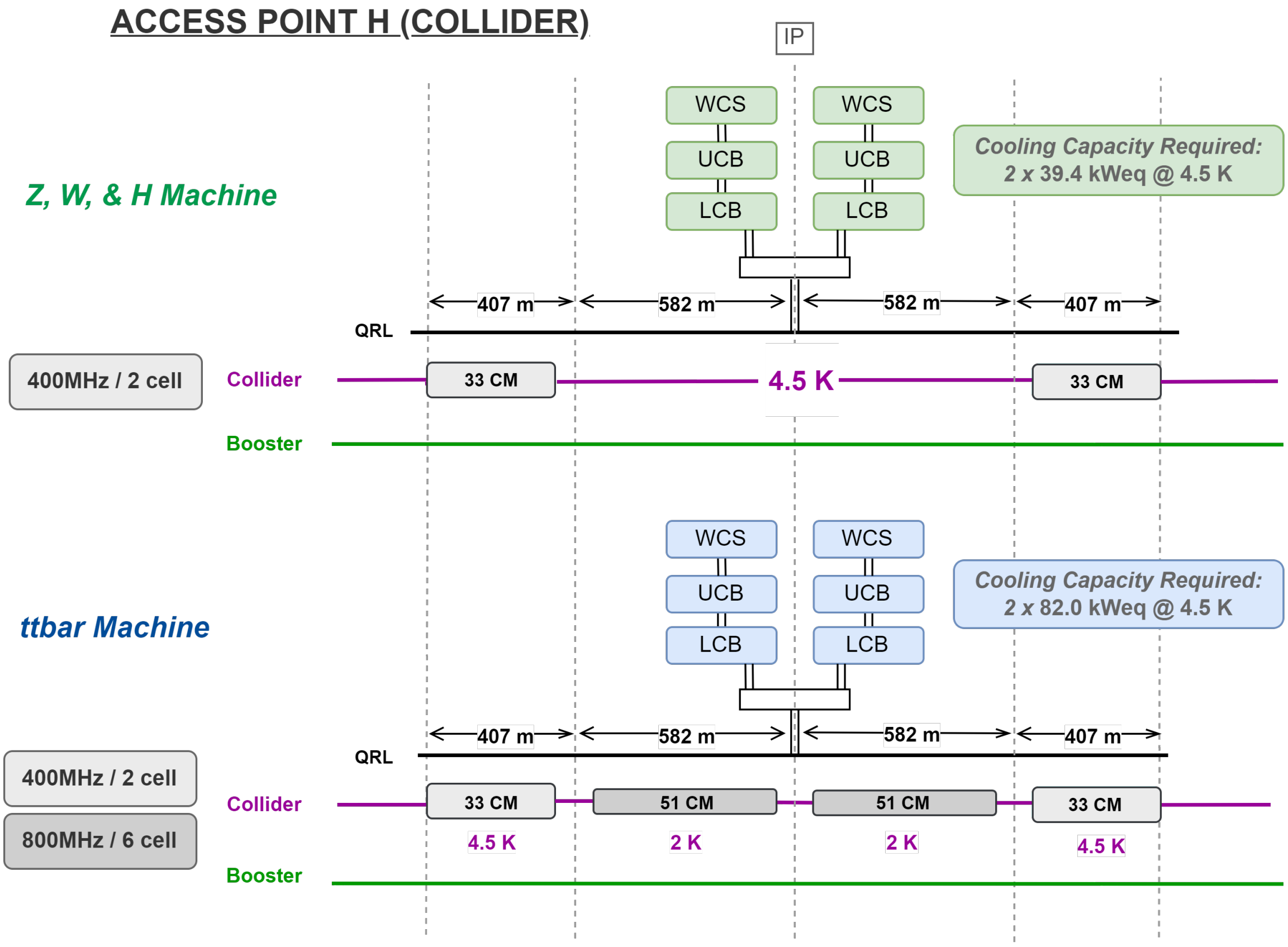

Fig. 8.74: Cryogenic plant architecture at point PH (RF collider).

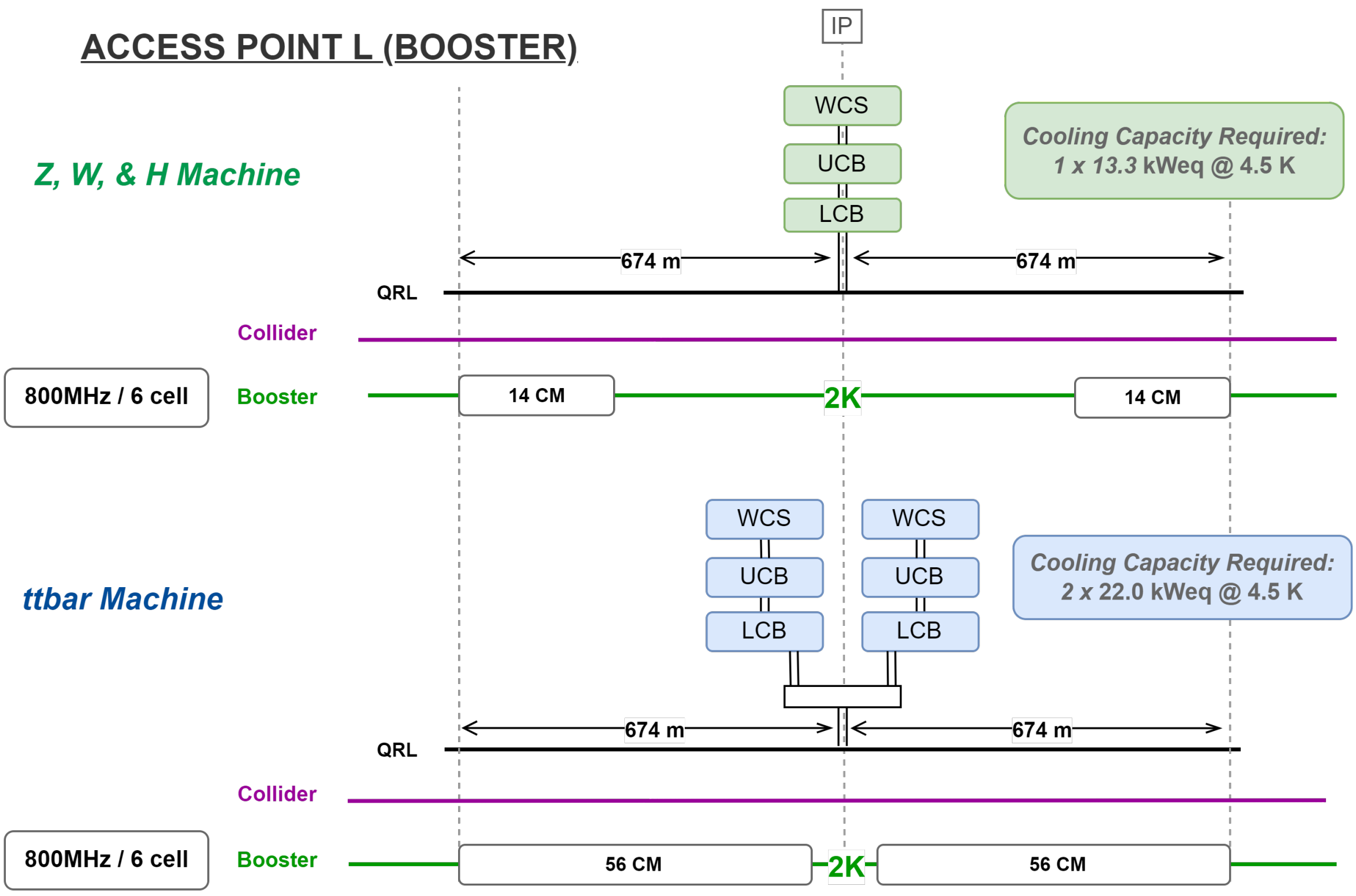

Fig. 8.75: Cryogenic plant architecture at point PL (RF booster).

***Temperature levels***

In view of the high thermodynamic cost of refrigeration at 2 K and 4.5 K, the thermal design of cryogenic components aims to intercept the largest fraction of heat loads at higher temperatures, hence the use of multiple staged temperature levels. These are:

- 50 K – 75 K for thermal shield as the first major heat intercept, sheltering the cavity cold mass from the bulk of heat inleaks from the environment.
- 4.5 K normal saturated helium for cooling 400 MHz superconducting cavities.
- 2 K saturated superfluid helium for cooling the 800 MHz superconducting cavities.

The cryomodules (CM) and cryogenic distribution line (QRL) combine several low temperature insulation and heat interception techniques which will have to be implemented on an industrial scale. These techniques include low-conduction support systems made of non-metallic fibreglass/epoxy composite, low impedance thermal contacts under vacuum for heat intercepts and multi-layer reflective insulation for wrapping the cold surface.

***Heat loads***

Inward static heat leaks (inleaks) arise from the ambient temperature environment and depend on the cryomodule design. The current design adopts a fully segmented architecture for both the booster and the collider, where each cryomodule has an independent connection to the cryogenic distribution line. Additionally, this distribution line is positioned externally to the cryomodule. The thermal calculations for the cryomodules are based on the thermal performance data of similar cryogenic assemblies.

The heat loads in the RF cryomodules are dissipated in the cavity baths at $4.5\,\mathrm{K}$ and $2\,\mathrm{K}$. These heat loads are determined by both the thermo-mechanical design of the cryomodule and the heat dissipation from the cavities during operation.

Table 8.26 presents the heat load values for point PH, while Table 8.27 provides the corresponding values for point PL. The reported values (excluding margins) originate directly from the SRF group. For the booster, these values are expected to increase, as the final RF cavity voltage and duty cycle have not yet been determined. The heat loads given in Table 8.27 for the booster assume a duty cycle of 15%. However, duty cycle values across different working points will range between 15% and 90% and still require finalisation.

Additionally, the reported heat loads do not account for all margins or the dynamic heat loads resulting from dissipation in the fundamental power coupler and higher-order modes (HOMs). These values are valid for an installed quality factor of $3.0 \times 10^{10}$ for the $800\,\mathrm{MHz}$ cavities and $2.7 \times 10^{9}$ for the $400\,\mathrm{MHz}$ cavities.

Notably, at the $\mathrm{t\bar{t}}$ stage for both point PH (collider) and point PL (booster) there will be 800 MHz / 2 K cavities. These 800 MHz cavities have different heat loads at the different points. While the static heat loads are similar, the dynamic losses are very different with 27.1 $\mathrm{W/cav_{collider}}$ and 3 $\mathrm{W/cav_{booster}}$. The dynamic losses are very different since the collider operates in continuous waves with 27.1 $\mathrm{W/cav_{collider}}$ while the booster dynamic losses are dependent on the duty cycle. This value is likely to increase following the expected increase of the booster duty cycle.

Table 8.26: Collider heat loads at point PH (as of October'24).

| **FCC-PH (collider)** | **Z, WW, & H** | **$t\bar{t}$** | |
|---|---|---|---|
| Freq [MHz] | 400 | 400 | 800 |
| Temperature [ K] | 4.5 | 4.5 | 2.0 |
| No. of cryomodules | 66 | 66 | 102 |
| No. of cavities (×4) | 264 | 264 | 408 |
| Static losses to the thermal shield / CM (no margins) [W] | 218 | 218 | 120 |
| Static losses to the helium bath / CM (no margin) [W] | 131 | 131 | 37.3 |
| Dynamic losses / cavity (no margins) [W] | 128 | 129 | 27.1 |
| Dynamic losses / CM (no margins) [W] | 512 | 516 | 108.5 |
| Total static losses to the thermal shield (with uncertainty factor) [kW] | 21.6 | 21.6 | 18.4 |
| Total static losses to the helium bath (with uncertainty factor) [kW] | 13 | 13 | 5.7 |
| fTotal dynamic losses (with overcapacity factor) [kW] | 50.7 | 51.1 | 16.6 |

Table 8.27: Booster heat loads at point PL with an assumed duty cycle of 15% (as of October'24).

| **FCC-PL (Booster)** | **Z, WW, & H** | **$t\bar{t}$** |
|---|---|---|
| Freq [MHz] | 800 | 800 |
| Temperature [ K] | 2.0 | 2.0 |
| No. of cryomodules | 28 | 112 |
| No. of cavities (×4) | 112 | 448 |
| Static losses to the thermal shield / CM (no margins) [W] | 103 | 103 |
| Static losses to the helium bath / CM (no margin) [W] | 37 | 37 |
| Dynamic losses / cavity (no margins) [W] | 3 | 3 |
| Dynamic losses / CM (no margins) [W] | 12 | 12 |
| Total static losses to the thermal shield (with uncertainty factor) [kW] | 4.3 | 17.3 |
| Total static losses to the helium bath (with uncertainty factor) [kW] | 1.6 | 6.2 |
| Total dynamic losses (with overcapacity factor) [kW] | 0.5 | 2.0 |

Dimensioning the cryoplants requires adding margins to the raw heat loads, by defining and applying two factors, the uncertainty and the overcapacity. The uncertainty, evolving during the project lifetime, is set to 50% for the feasibility study and covers the design uncertainties of the cryo facility. It is applied to the static heat loads. The overcapacity, stable during the project lifetime, is also set to 50% for the feasibility study, and ensures nominal performance of the cryo facility by covering specific risks such as reduced performance induced by ageing and operational flexibility. The overcapacity is applied to both the dynamic heat loads and the static heat loads after the uncertainty factor. Figure 8.76 shows a visual representation of how the margins were applied.

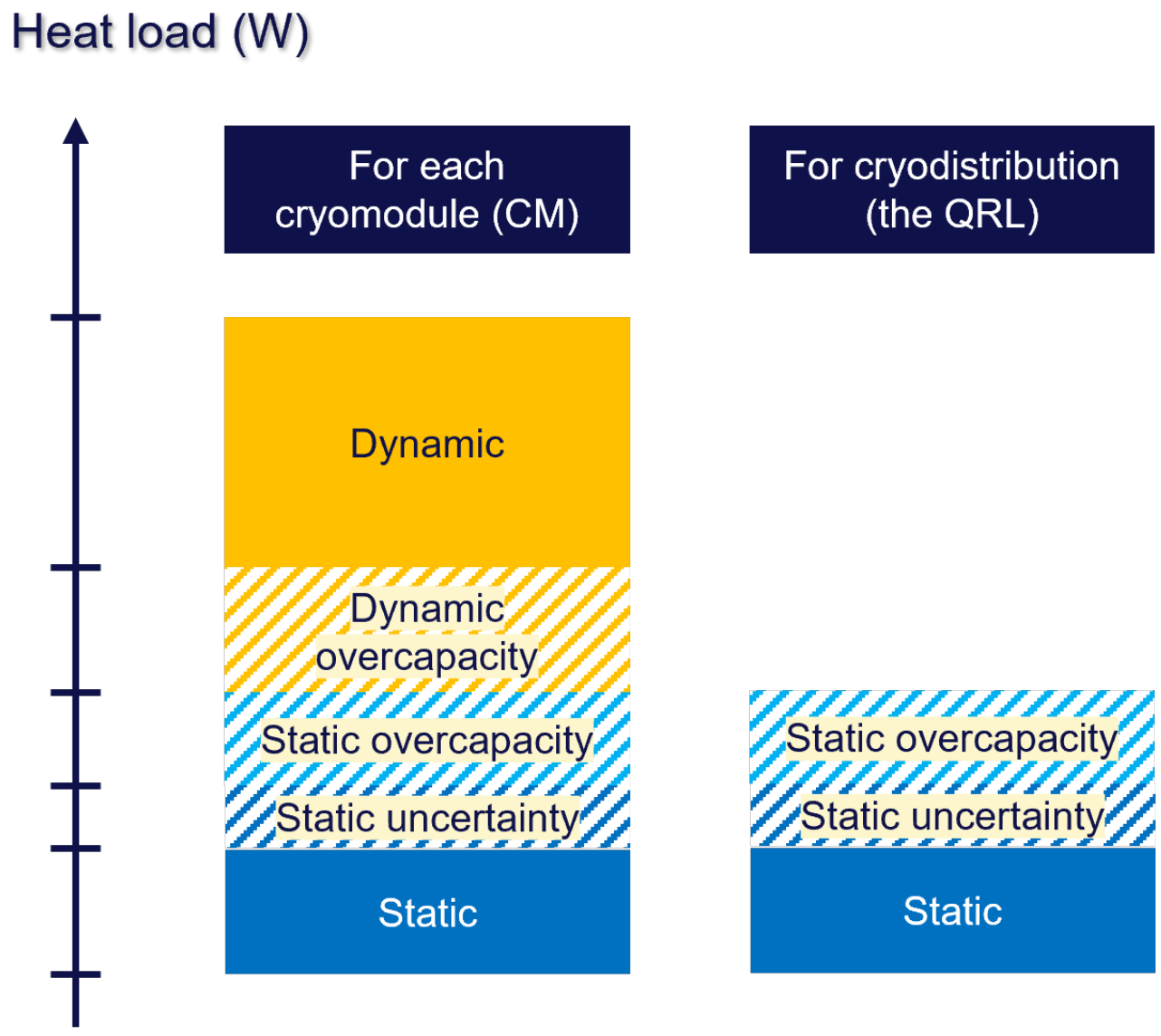

Fig. 8.76: Uncertainty and overcapacity margins on the raw heat loads

### 8.5.2 Cooling scheme and cryogenic distribution

The cooling scheme for the cryomodules is represented in Fig. 8.77. The 4.5 K cavity cold masses are immersed in saturated helium baths, which are supplied by line C. The saturation pressure is maintained by line D, which recovers the evaporated vapour. The 2 K cavity cold masses are immersed in saturated helium baths, which are supplied by line A. The low saturation pressure is maintained by pumping the vapour through line B. Each cryomodule has a dedicated thermal shield and heat intercept circuit cooled in parallel between line E and F.

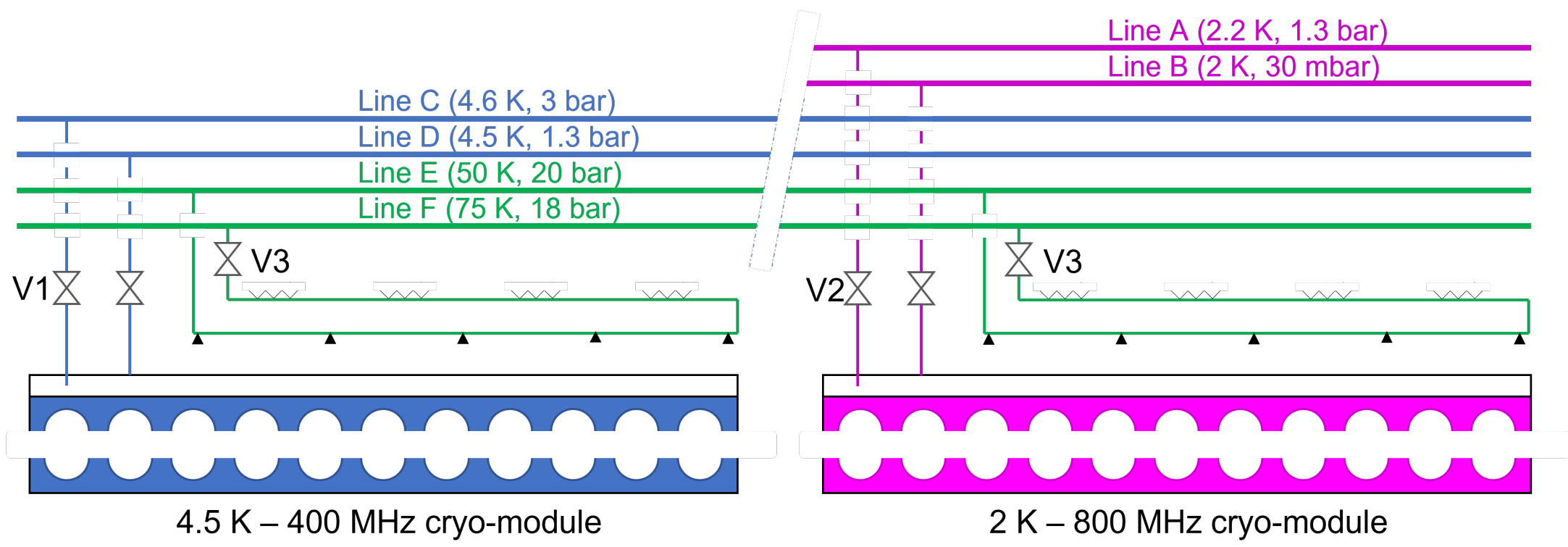

Fig. 8.77: Cryogenic flow scheme for the cryomodules. Here the cryogenic plant is located to the right of the 800 MHz section.

Table 8.28 gives the size of the main cryogenic distribution system components. The cryogenic distribution line (QRL) was designed for the final $t\bar{t}$ stage, making it oversized for the Z, W, and H stages. This design choice was made to lower installation complexity and costs. A visual cross-sectional view of the QRL can be seen in Fig. 8.78. A front view of the lengths of the QRL along with other connecting components in the service cavern and LSS sections can be seen in Fig. 8.80 and Fig. 8.79.

Table 8.28: Dimensions of the main cryogenic distribution line components.

| **Component** | **Diameter [mm] Point PH** | **Diameter [mm] Point PL** |
|---|---|---|
| Line A: 1.3 bar, 2.2 K | 80 | 60 |
| Line B: 30 mbar, 2.0 K | 345 | 265 |
| Line C: 3 bar, 4.6 K | 120 | 20 |
| Line D: 1.3 bar, 4.5 K | 200 | 20 |
| Line E: 20 bar, 50 K | 80 | 60 |
| Line F: 18 bar, 75 K | 80 | 60 |
| Vacuum jacket (400 MHz) | 550* | - |
| Vacuum jacket (800 MHz) | 830* | 600* |

* +100 mm for bellows and flanges.

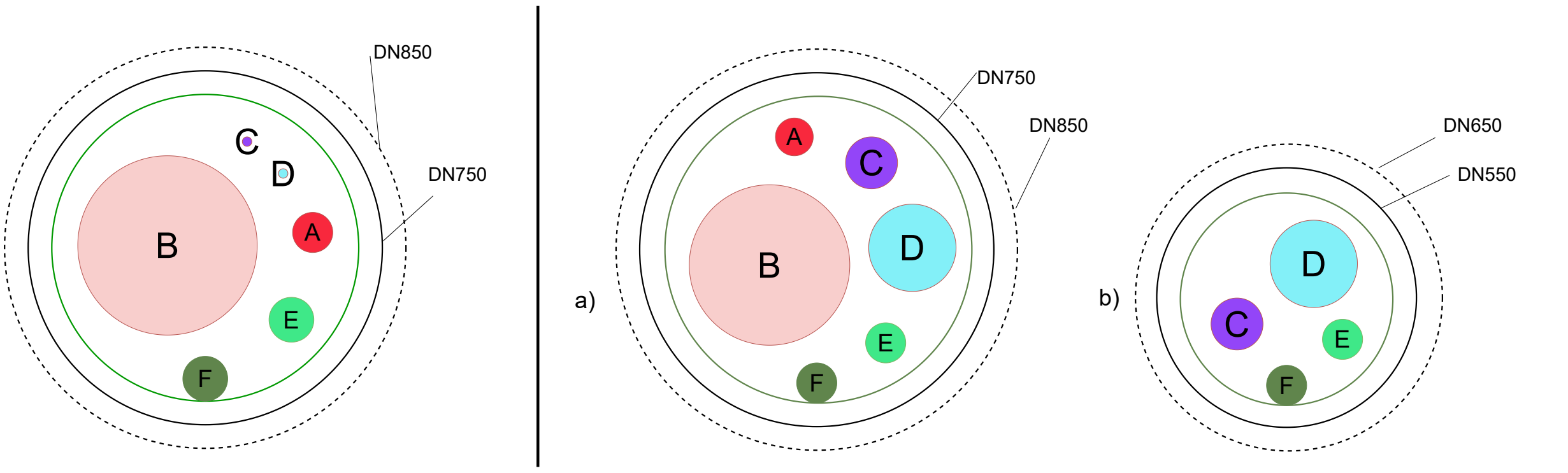

Fig. 8.78: Cross sections of point PL (left) and point PH (right) cryogenic distribution line. a) for the cooling of 800 MHz cryomodules. b) for the cooling of 400 MHz cryomodules.

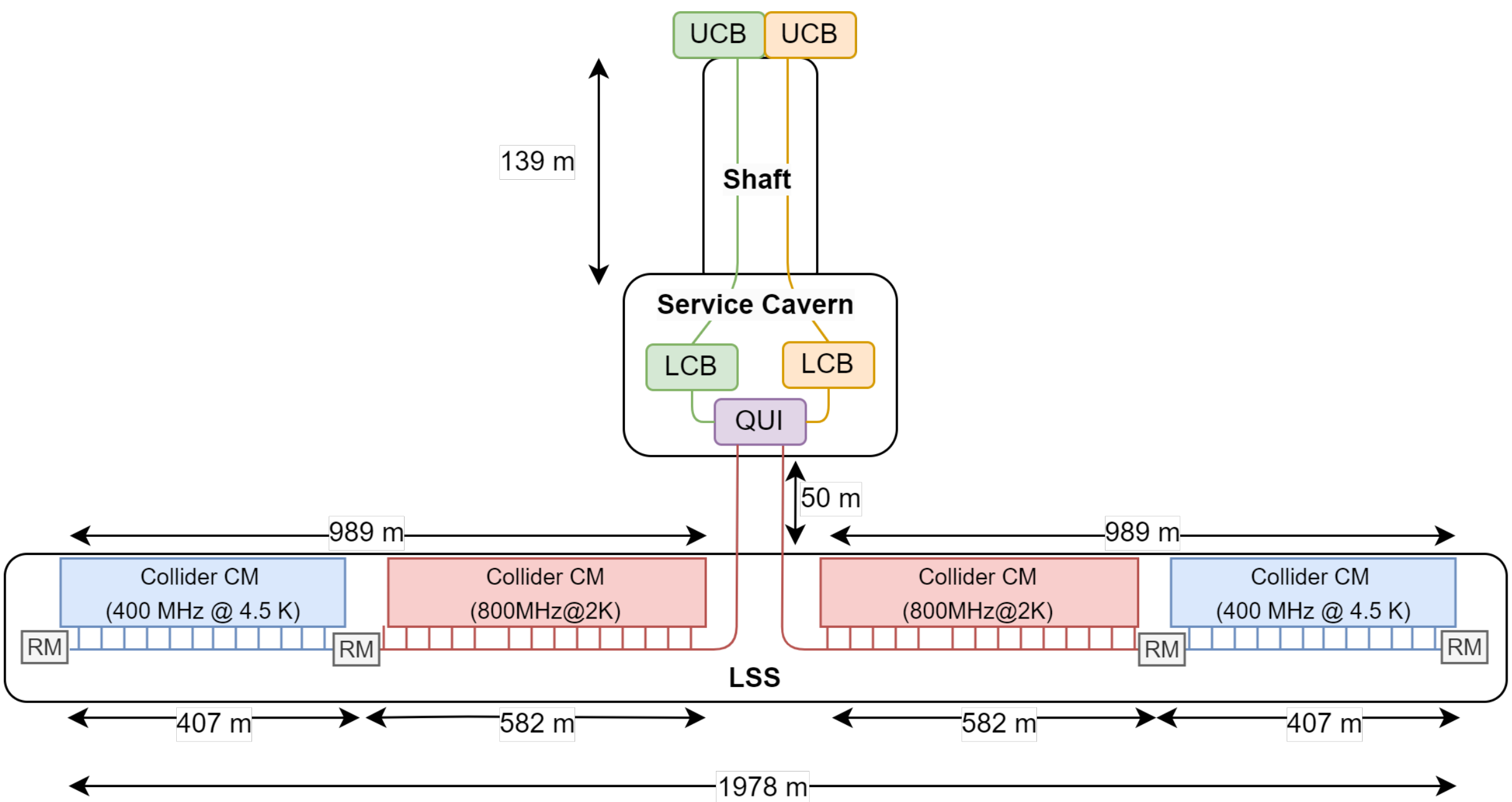

Fig. 8.79: Collider (point PH) cryogenic plant distribution architecture at the $t\bar{t}$ stage.

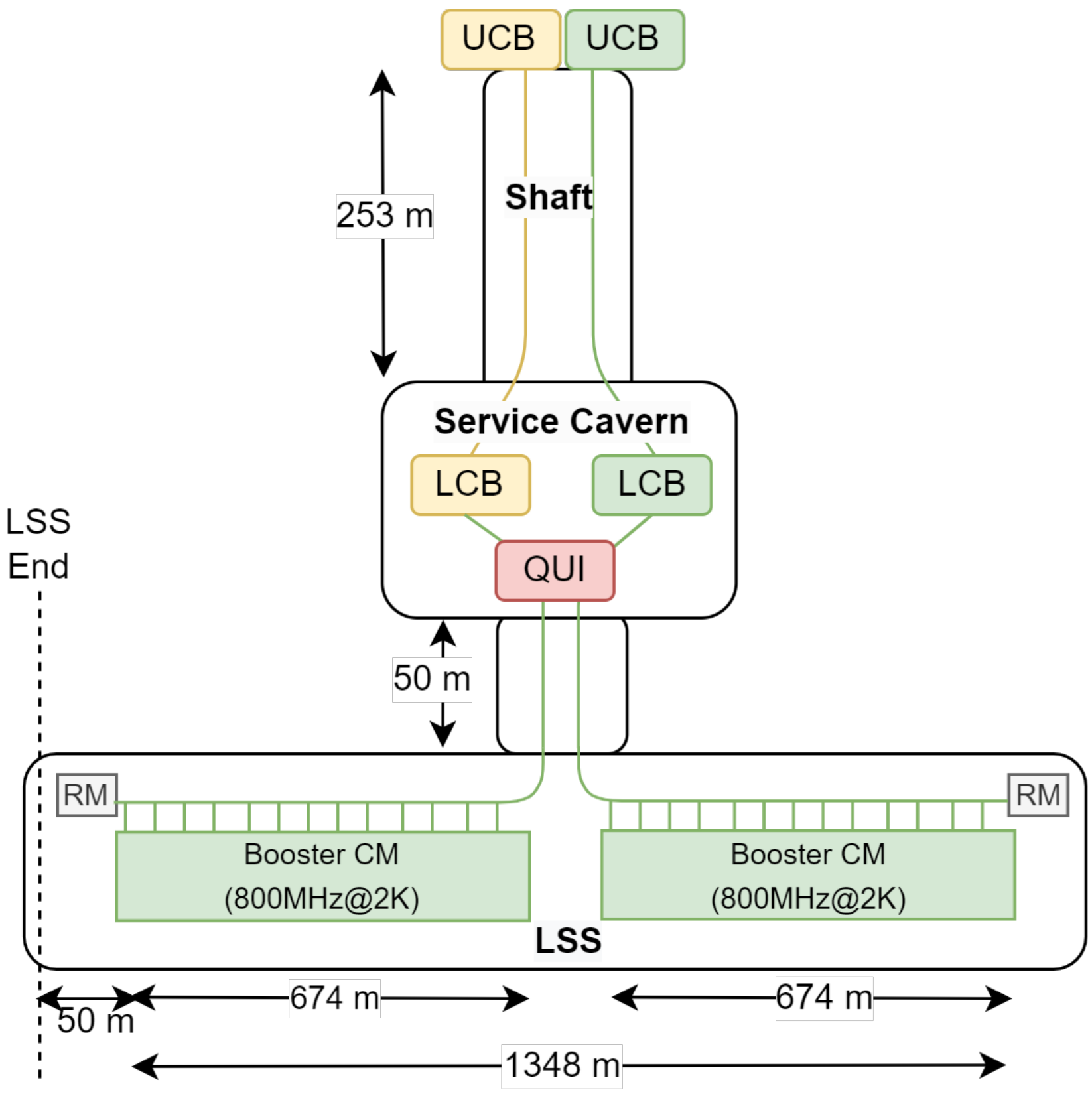

Fig. 8.80: Booster (point PL) cryogenic plant distribution architecture at the $t\bar{t}$ stage.

***2 K system optimisation***

Refrigeration of the RF cavities in a 2 K saturated bath involves low-density helium (He) with process pressures in the 30 mbar range. A series of compressors are required to achieve these pressure levels. Whenever high mass flow rates of helium need to be treated, performing a large fraction of the compression at low temperatures allows operating with a higher density, effectively limiting the size of the compressors. During $t\bar{t}$ operation of FCC-ee, about 550 g/s or 250 g/s [385] per train will need to be pumped at points PH and PL, respectively. In both cases, a cold compressor system (CCS), composed of a series of centrifugal compressors, will be required to provide the compression back to semi-atmospheric levels ($\approx$ 400 mbar). From previous studies, it is considered that a CCS of 12 kW at 1.8 K is feasible [386, 387]. This translates into a mass flow rate of about 500 g/s with one train of four compressors at a suction pressure of 16 mbar. Addressing FCC-ee needs, the suction pressure of such a CCS can be increased as the saturation temperature desired is 2 K. This leads to slightly higher densities and, hence, allows operation with slightly more than 500 g/s mass flow rates.

The lower the design suction pressure of a CCS is, the more compression stages may be required to reach semi-atmospheric levels, increasing the complexity of the system further. Since the cavities are cooled by a saturated He-II bath [388], the furthermost 2 K cryomodule will need to be at the saturation pressure of helium at 2 K, that is, at 31 mbar. The pressure drop along the distribution line to that cryomodule needs to account for this and must be minimised so that the suction pressure at the inlet of the CCS is kept as high as possible. The design goal is to limit the total pressure losses such that the suction pressure of the CCS stays above 22 mbar. Out of the 9 mbar of available pressure drop, 7 mbar is left as a margin for heat exchangers, elbows and valves, and 2 mbar is set as a design criterion for sizing the very low pressure (VLP) return line or line B in Fig. 8.81. The choice of 2 mbar provides a good compromise for the CCS suction pressure and the overall size of the VLP line. Limiting the pressure drop further could hinder the already difficult distribution line integration in the 5.5 m diameter tunnel, where the space is very limited due to the presence of both booster and collider machines.

Ensuring a thermodynamically well-adapted architecture is essential to reduce the high cost associated with such a system. Several architectures have been assessed and compared so far, with the goal of finding the optimum choice when considering both energy and integration constraints. Figure 8.81 shows the two main options that are being considered.

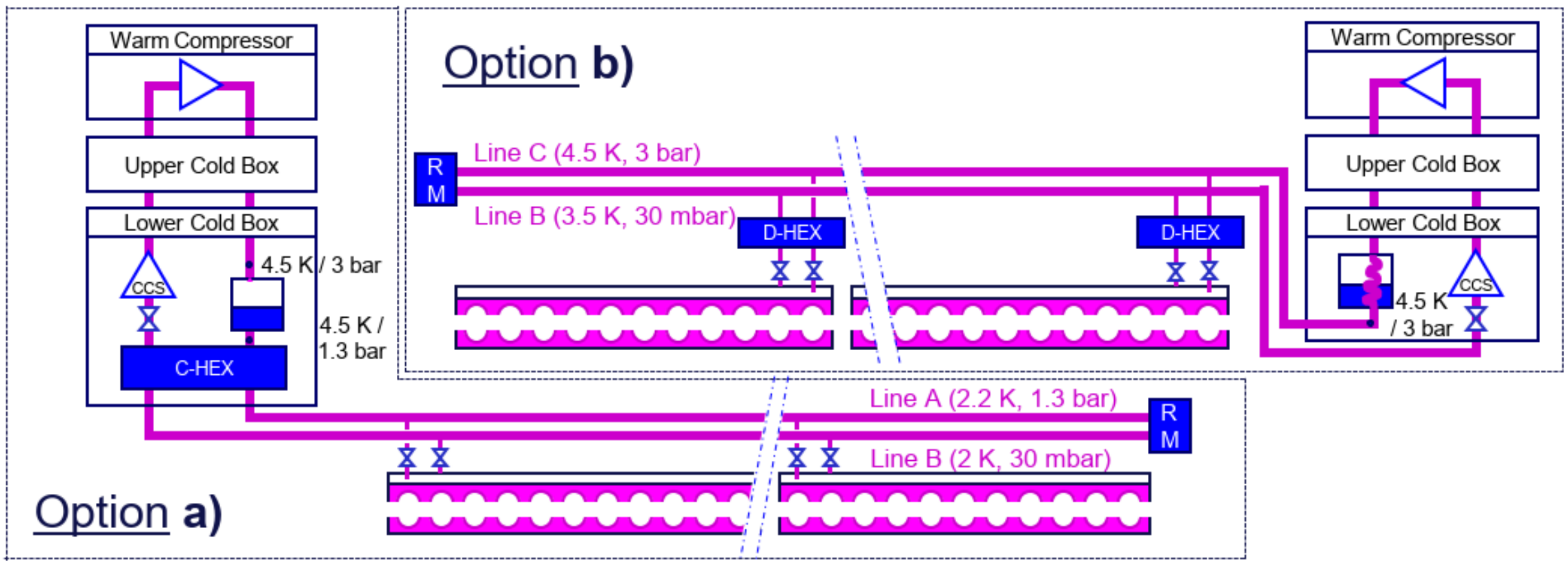

Fig. 8.81: FCC-ee 2 K system distribution architecture options studied, a) and b). CCS being the Cold Compressor System, RM the Return Module, C-HEX the Centralised Heat Exchanger and D-HEX the Distributed Heat Exchanger.

The current calculations for the cryoplants and QRL are based on the preliminary architecture choice of the C-HEX design. An energetic optimisation was performed [389] to justify the preliminary architecture choice. The conclusions of the study showed that choosing a D-HEX based architecture means only a 1.3% increase in total cryoplant size and energy consumption of FCC-ee. The main gain of the C-HEX option is the smaller size of the VLP return line, as well as the central location of one single HEX, reducing the number of components in the tunnel and easing the overall integration. This conclusion is subject to achieving high-efficiency on a C-HEX twice as big as the current state of the art. The current calculations for the cryoplants and QRL are based on the C-HEX design.

### *Helium Recovery System*

The latest available cryomodule inventory values are 55 kg of He at 2 K per 800 MHz cryomodule, and 116 kg of He at 4.5 K per 450 MHz cryomodule [388]. Because the SRF cavities are low-pressure rated devices, the risk of inventory loss in a non-nominal scenario is high. Without a recovery system the pressure inside the helium tank of the cryomodules would start building up, in the case of non-nominal operation, eventually reaching the set pressure of the pressure relief valve first, then ultimately the burst disc. The latter is the main safety component preventing the cavity from sustaining any mechanical damage. The scenarios that could lead to such a situation are:

- **Scenario 1:** Isolated cryomodule(s) from the cryoplant due to a malfunctioning valve.
- **Scenario 2:** Loss of the full sector cooling capacity (e.g., due to a power outage).
- **Scenario 3:** Beam vacuum break.
- **Scenario 4:** Insulation vacuum break.

Scenarios 3 and 4 generate very high mass flow rates due to the large heat load that appears after a vacuum break. Therefore, these are to be addressed with the pressure relief devices alone. The helium recovery system (HRS) will cover Scenario 1 and 2, which consist of only inherent static heat loads. The concept proposed for such a system is depicted in Fig.8.82 and has been presented in detail in [389].

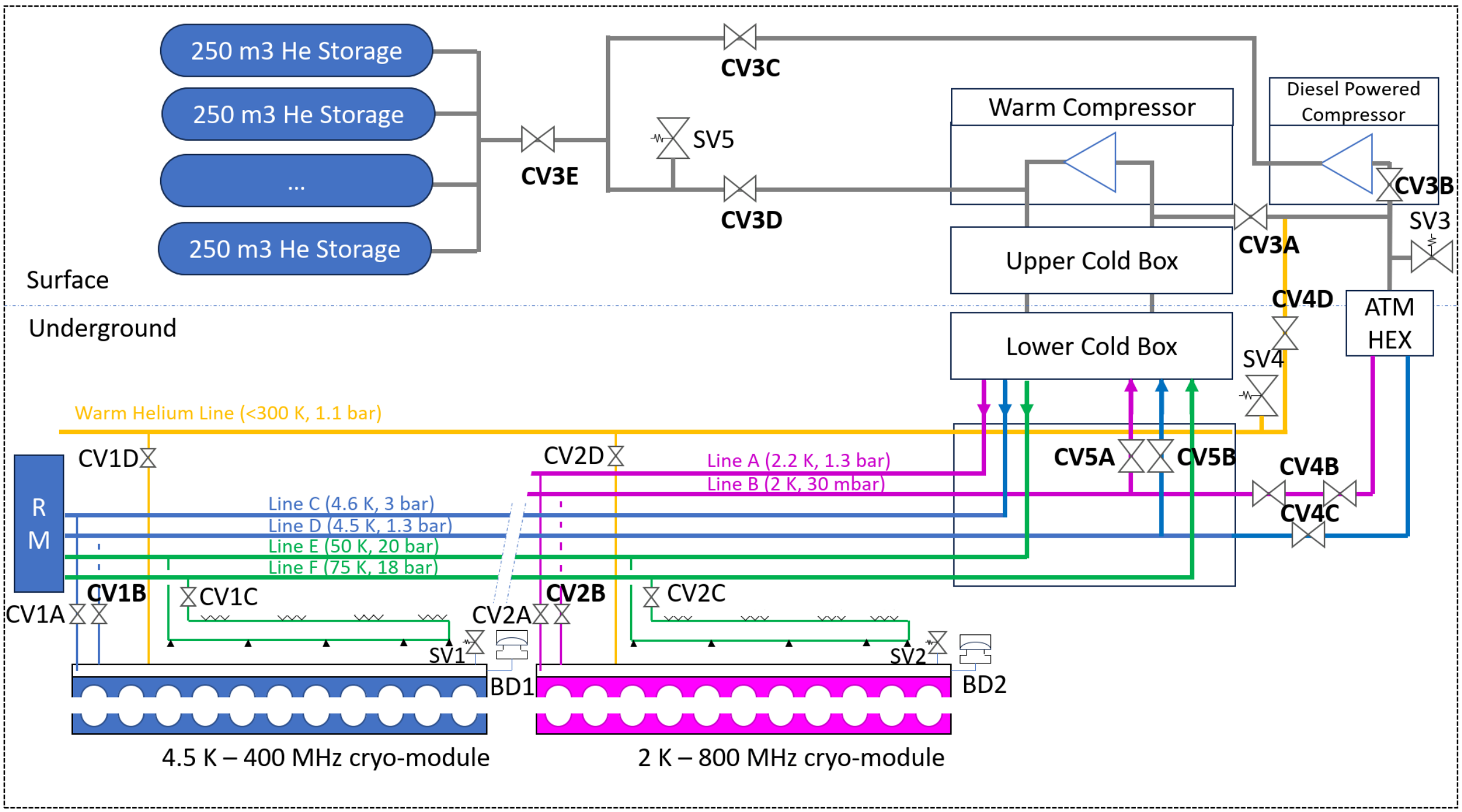

Fig. 8.82: FCC-ee proposed helium recovery system (HRS) concept.

### 8.5.3 Cryogenic plants

Table 8.29 and Table 8.30 below give the nominal cooling capacity per cryogenic plant required at the various temperature levels for the technical points, including an overcapacity margin factor of 1.5.

Table 8.29: Point PH cryoplants and their staging, including an overcapacity margin factor of 1.5.

| | **Z, WW, & H** | **$t\bar{t}$** |
|---|---|---|
| Type | C1 | C1+ (Upgrade of C1) |
| Schema | See Fig. 8.83 (a) | See Fig. 8.83 (b) |
| 50-75 K, kW per cryoplant | 23.7 | 37.5 |
| 4.5 K, kW per cryoplant | 36.4 | 36.6 |
| 2.0 K, kW per cryoplant | - | 13.0 |
| Nominal equivalent cooling capacity at 4.5 K per cryoplant [$kW_{eq}$] | 39.4 | 82.0 |
| Number of cryoplants required | 2 | 2 |

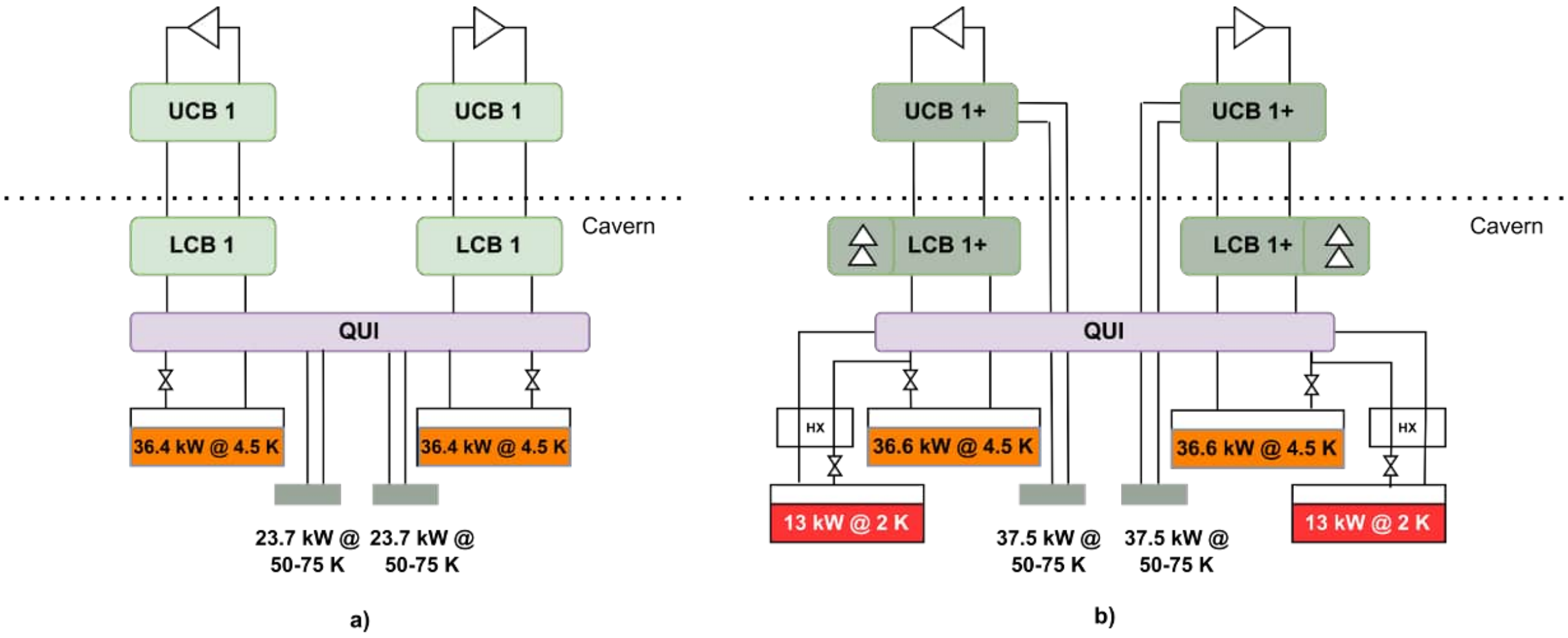

Fig. 8.83: Point PH cryoplant architecture at Z, WW and H (a), and $t\bar{t}$ (b).

Table 8.30: Point PL Cryoplants and their staging, including an overcapacity margin factor of 1.5.

| | Z, WW, & H | $t\bar{t}$ |
|---|---|---|
| Type | B1 | B2 |
| Schema | See Fig. 8.84 (a) | See Fig. 8.84 (b) |
| 50-75 K, kW per cryoplant | 14.4 | 16.9 |
| 4.5 K, kW per cryoplant | - | - |
| 2.0 K, kW per cryoplant | 3.6 | 6.1 |
| Nominal equivalent cooling capacity at 4.5 K per cryoplant [$kW_{eq}$] | 13.3 | 22.0 |
| Number of cryoplants required | 1 | 2 |

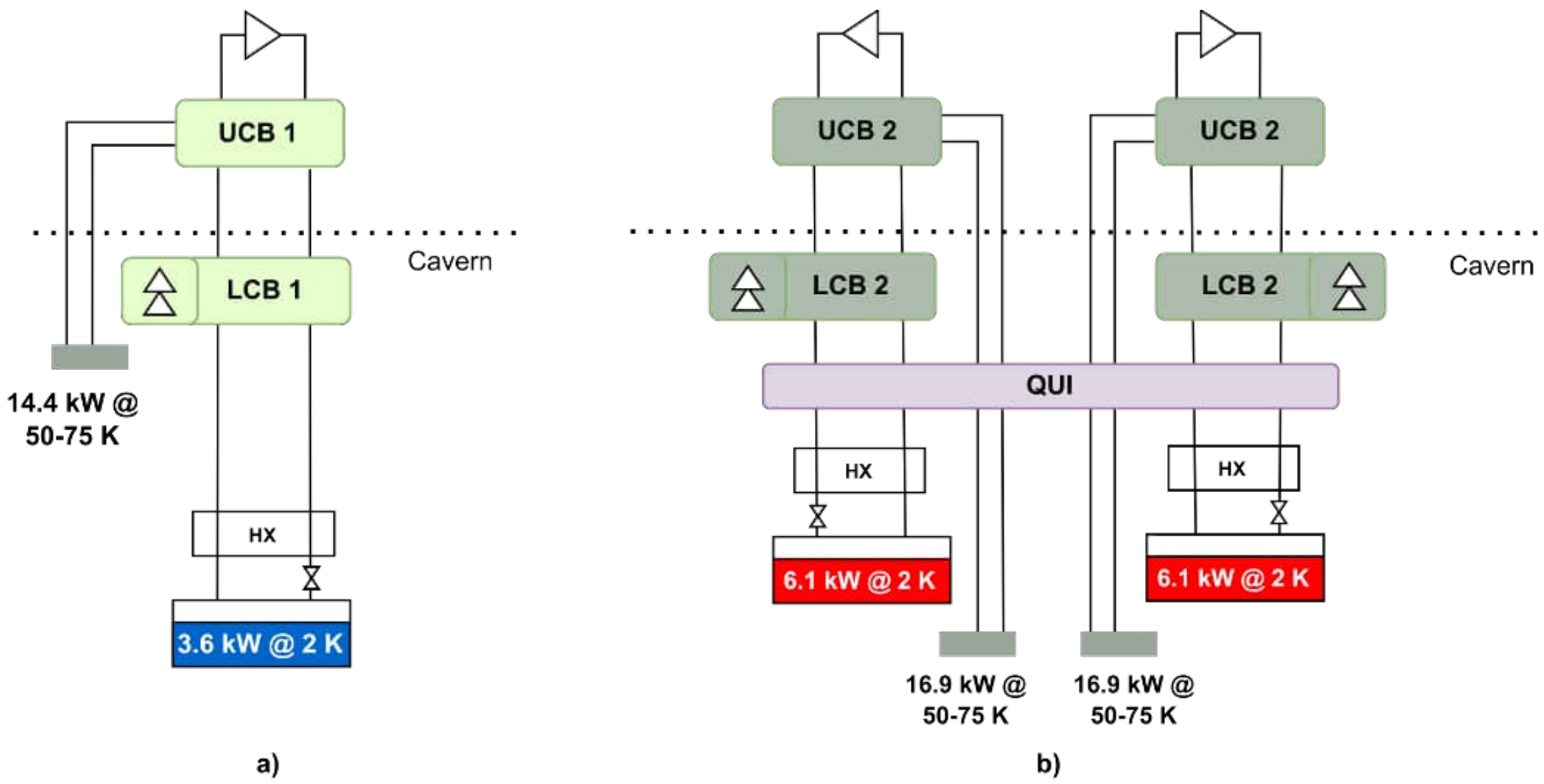

Fig. 8.84: Point PL cryoplant architecture at Z, WW and H (a), and $t\bar{t}$ (b).

### 8.5.4 Cryogen inventory and storage

The cryogenics system for the RF points will require helium and nitrogen. Nitrogen will only be needed for the regeneration of absorbers and dryer beds. Consequently, one standard 50 $m^3$ liquid nitrogen ($LN_2$) reservoir is planned for each point PH and point PL. The helium inventory is mainly driven by the cryomodule cold mass baths and the cryogenic distribution system. The combined helium inventory of point PH (collider) and point PL (booster) is 17.4 tons for the Z, WW, and H stage and 32.7 tons for the $t\bar{t}$ stage. The storage will be provided using 250 $m^3$ medium-pressure (MP, 18 bar) storage tanks, the total number of storage tanks required can be found in Table 8.31.

Table 8.31: Inventory of helium and its storage for the FCC-ee RF points.

| **Machine** | **Z, WW & H** | $t\bar{t}$ |
|---|---|---|
| Cryomodules (t) | 9.2 | 19.4 |
| Distribution (t) | 5.0 | 6.0 |
| Cryoplant (t) | 3.2 | 7.2 |
| **Total (t)** | 17.4 | 32.7 |
| **No. of 250 $m^3$ MP storage tanks at point PH (collider)** | 21 | 34 |
| **No. of 250 $m^3$ MP storage tanks at point PL (booster)** | 4 | 12 |
| **Total No. of storage tanks (+1 auxiliary tank)** | 26 | 47 |

### 8.5.5 Economic mode for energy savings

One of the particularities of the FCC-ee is the large difference between the static and the dynamic heat loads that the cryogenic system must handle. The dynamic heat loads represent between 22% and 68% of the total heat loads depending on the machine stage and RF point. See Table 8.32 below, where details are given.

Table 8.32: Static and dynamic heat load sharing for the different FCC-ee phases.

| **FCC-ee Machine** | **Z, WW & H** | | $t\bar{t}$ | |
|---|---|---|---|---|
| Point | PH | PL | PH | PL |
| Static heat loads at 4.5 K [$kW_{eq}$] | 24.8 | 10.4 | 54.8 | 32.5 |
| Dynamic heat loads at 4.5 K [$kW_{eq}$] | 54.0 | 2.9 | 109.2 | 11.6 |
| Total heat loads at 4.5 K [$kW_{eq}$] | 78.8 | 13.3 | 164.0 | 44.1 |
| Dynamic heat load percentage | 68% | 22% | 67% | 26% |

Due to the proportion of large dynamic heat loads and the requirement to maintain the RF cavities at cryogenic temperatures (<5 K) throughout the years between the long shutdowns (LS), the cryogenic systems must be designed in such a way that the electrical and water consumptions are optimised throughout the year.

The cryogenic system must deliver full power during the physics, commissioning, and machine development (MD) periods. During these periods the RF cavities need to be fully operational at top energy. During the end-of-year technical stops (YETS) and the technical stops (TS), all RF cavities (including the 800 MHz cavities at 2 K) will be switched off and maintained at 4.5 K using an economic mode of the cryogenic plant. During this economic mode, there will be only static heat loads from the RF cavities. In the current evaluation of the FCC-ee yearly operation [4], around 36% of the time could be operated with such a cryogenic economic mode as depicted in Fig. 8.85.

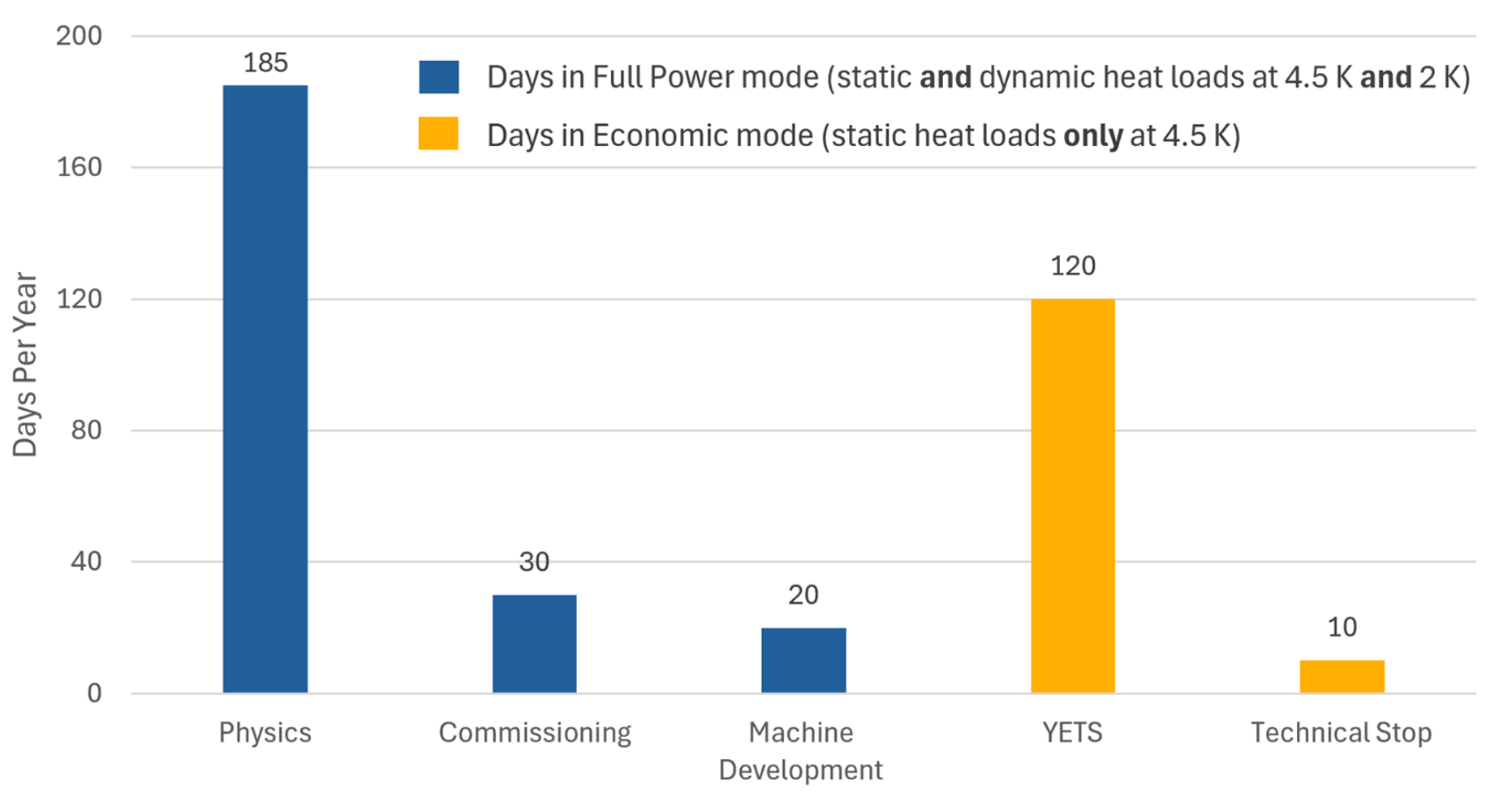

Fig. 8.85: FCC-ee yearly cryogenic operation modes between long shutdown periods.

To allow these energy savings, the cryoplants must be designed taking into account these different operation modes. The coefficients of performance (COP) of the cryoplants must be optimised for the full power mode and the economic mode, where a reduced cryogenic power will be delivered. To achieve this objective in partnership with the cryoplant manufacturers, different solutions will be studied and developed. For instance, specific piping and some additional cryogenic equipment will be needed, such as smaller warm compressors with variable frequency drives and specific turbines that could be operated efficiently during this economic mode. A COP-1 of 220 $W_{elec}$/W at 4.5 K for the full power mode and a COP-1 of 250 $W_{elec}$/W at 4.5 K for the economic mode were considered for this report.

Because of the economic mode, the cryogenic system will consume less power during the YETS and TS periods. The energy consumption per year for the FCC-ee cryogenic system is shown in Table 8.33. The table includes both scenarios with and without the implementation of an economic mode. By using the economic mode, electrical energy savings of up to 26% can be achieved. Note that the cooling water system dedicated to cryogenics will be also alleviated in about the same proportion.

Table 8.33: Cryogenic system electrical energy consumption per year.

| **PL and PH plants** | **Z, WW & H** | **$t\bar{t}$** |
|---|---|---|
| Full power [MW elec] | 20.3 | 45.8 |
| Eco power [MW elec] | 7.6 | 12.9 |
| Elec energy/year [GWh] no eco mode | 177.4 | 401.0 |
| Elec energy/year [GWh] with eco mode | 138.1 | 298.3 |
| **Electrical energy savings per year** | 22% | 26% |

### 8.5.6 Cryogenics for experiments and MDI region

***Concept***

The experiment points PA, PD, PG and PJ each have a detector cryoplant that covers the needs of the detector magnet, as well as a second IR cryoplant for the machine detector interface (MDI) region magnets.

Figure 8.86 shows the loads expected for the detector and IR cryoplants. In yellow is the 2 T detector solenoid, which will be cooled by a dedicated detector cryoplant. The other loads, represented in blue, will be covered by the IR cryoplant. These loads remain under study, but could include the MDI magnets, which contain focusing quadrupoles (QC**), compensating and screening solenoids, as well as the crab sextupoles (SY**). One of the detectors will also include a liquid argon calorimeter and its associated liquid nitrogen cryoplant.

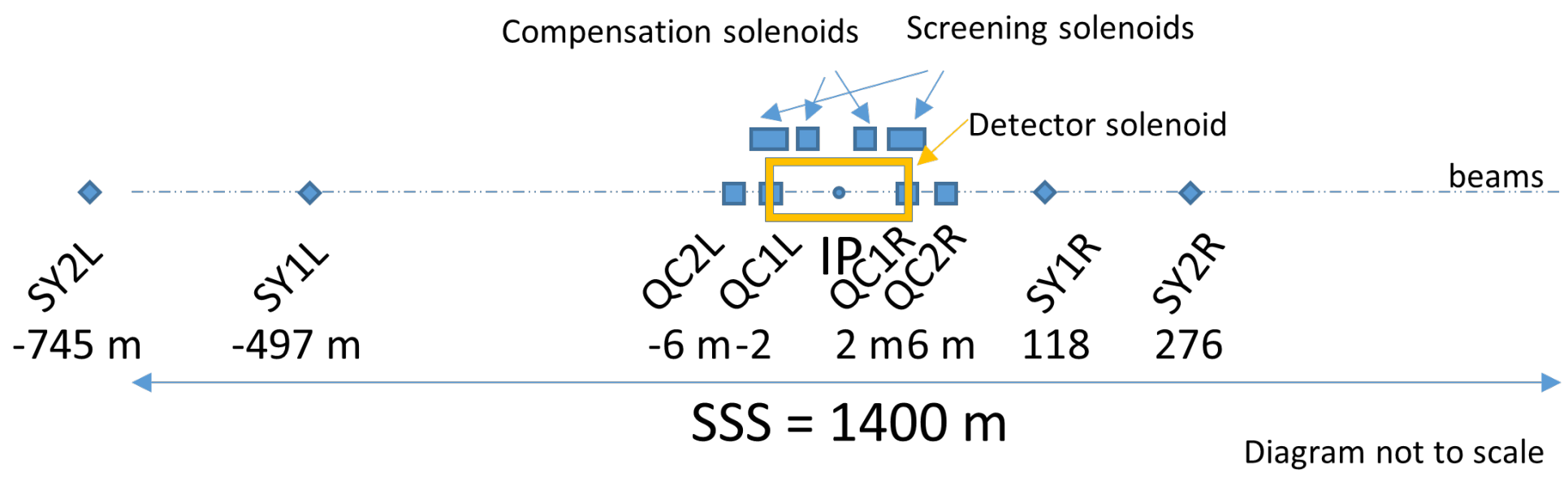

Fig. 8.86: Interaction Region superconducting electrical loads.

***Basic parameters***

The current working scenario has the following cryoplants:

- Four cryoplants for the detector solenoids. CMS-like 1.5 $kW_{eq}$ at 4.5 K plants are considered as an envelope case following the latest FCC experiment sites civil engineering and technical infrastructure review
- Four cryoplants for the IR area magnets. Further details on heat loads and the mechanical design of the cryostat will define the cryoplant size and architecture.
- One nitrogen liquefier for the detector containing the liquid argon calorimeter. An ATLAS-like 20 kW at 80 K plant is considered at this stage.

### 8.5.7 Cryogenics for alternative solutions

An alternative solution currently under study aims to exchange all the short straight section (SSS) warm magnets (arc quadrupoles, arc sextupoles and correctors) with superconducting ones based on ReBCO HTS tapes. The main goal of this alternative solution is power reduction. It also envisages nesting the quadrupoles and the sextupoles. Saving space and relaxing the RF system requirements and costs. The proposal involves 2900 distributed cryostats, each being 3.5 m long, affecting some 11% of the entire machine. The operating temperature is currently defined at 40 K [390]. The adoption of this solution would have a large impact on the amount of cryoplants and distribution lines that would be needed.

## 8.6 Transport

### 8.6.1 Equipment transport requirements

The transport of equipment within the underground facilities is a key activity during the installation phase. Given the current stage of the study, some of the data collected remain partial. Therefore, assumptions have been made based on similar existing projects. The identified requirements have been consolidated in the document Transport Requirements [391].

The types of items considered in the preliminary transport study include:

- Accelerator components (magnets, supports, beamstrahlung dump system);
- Power converters;
- Electrical equipment;
- Cryogenic equipment;

– Cooling and ventilation equipment.

Among these, the collider and booster ring components have the greatest impact on the design of the transport vehicles. To keep underground installation time as short as possible, certain components, such as quadrupoles and sextupoles, will be pre-assembled on a supporting structure at the surface and then transported as a single unit (QSS unit), as illustrated in Fig. 8.87.

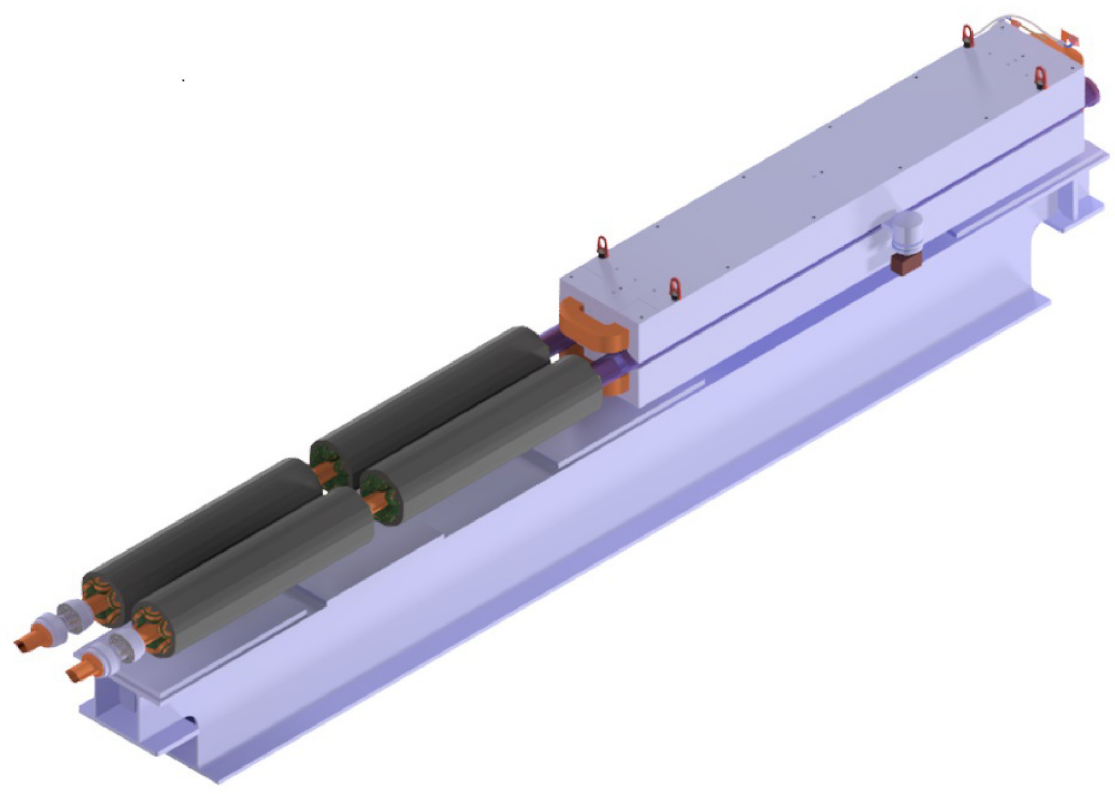

Fig. 8.87: Quadrupole with two sextupoles and the supporting girder (QSS unit)

The weight of the QSS unit is expected to be close to 14.5 t for a length of 7 m, which represents the maximum load that the vehicle will lift. The dipoles will be transported stacked in groups of three, with an overall weight close to 16 t and a length of approximately 12 m. The beamstrahlung dump system is also significant in terms of weight and dimensions, a specific preliminary study has been conducted on this topic and is summarised in the present report Section 8.6.1 . The other elements considered are more standard (mainly pipes, racks and cable reels) and will be transported by electric tractors and trailers as is usual for CERN accelerators.

***Overhead cranes and shafts***

Overhead cranes will be installed in almost every surface building and within the service caverns to facilitate handling operations during both the installation and operation phases.

This type of equipment is generally preferred over floor-based handling machines, such as forklifts or telehandlers, as it offers several advantages. In terms of safety, the suspension point is located above the object being handled, reducing the risk of accidental collisions or tipping. From an operational perspective, overhead cranes provide greater positioning precision, ensuring more controlled and efficient handling. The use of overhead cranes eliminates the need for wide aisles for manoeuvring and high floor load capacities, leading to potential cost savings in infrastructure design and implementation.

Table 8.34 contains the list of the overhead cranes foreseen to be installed in the facilities of the experiment and technical points.

The overhead cranes will comply with the relevant European Directives (currently the Directives 2006/42/EC [392], 2014/30/EU [393] and 2014/35/EU [394]) and the CERN Safety Rules. According to the expected frequency of use (intensive during the installation phase, very low during the operation phase) and the load spectra (the maximum load will rarely be lifted), they will be considered as light/medium duty overhead cranes and classified as A4-M4 according to the FEM (European Materials Handling Federation) guide 1.001.

Table 8.34: List of overhead cranes.

| Building | Location | Capacity [t] |
|---|---|---|
| Assembly hall SX$^{\dagger}$ | Surface | 120 |
| Head-Shaft building SD$^{\dagger}$ | | 75 |
| Head-Shaft building SD$^{\ddagger}$ | | 25 |
| Tunnel and service areas ventilation building SU | | 7.5 |
| Experiment ventilation building SUX$^{\dagger}$ | | 7.5 |
| Cooling plant SF | | 3.5 |
| Power converters building SR | | 5 |
| Compression station SH$^{\dagger}$ | | 20 |
| Service cavern | Underground | 20 |

$^{\dagger}$ Only in the experiment points

$^{\ddagger}$ Only in the technical points

The hoist design will incorporate specific features to ensure that the crane is suitable for handling fragile components with high precision in positioning. The hoist will be equipped with two ropes, ensuring that the hook is lifted and lowered without horizontal drift, thereby minimising the risk of unintended movement. Additionally, all motors will be driven by inverters, allowing for smooth acceleration and precise control during lifting and lowering operations.

The speed values will be set in the following ranges:

– Lifting speed: 8 to 10 m/min;
– Cross-travel speed: 12 to 15 m/min;
– Long-travel speed: 15 to 20 m/min.

Each building layout will ensure safe access to the cranes during preventive and corrective maintenance (e.g., a walkway at the side of the railway). Access to any position where the crane may stop in case of breakdown will be possible.

***Shaft cranes***

The overhead cranes installed in the assembly halls and in the shaft-head buildings will require a bespoke design due to their particular lifting heights, as shown in Table 8.35. The overhead cranes in the SD buildings of the experiment points have a capacity of 75 t in order to allow the handling of the FCC-hh magnets in the future. This will avoid expensive modifications of the cranes and buildings arising from increasing the capacity from 25 t to 75 t.

Table 8.35: Characteristics of shaft cranes.

| Overhead crane | Sites | Capacity [t] | Lifting height [m] |
|---|---|---|---|
| Assembly hall crane | PA, PD, PG, PJ | 120 | 253 |
| Head-shaft building crane type A | PB, PH, PL | 25 | 253 |
| Head-shaft building crane type B | PF | 25 | 400 |
| Head-shaft building crane type C | PA, PD, PG, PJ | 75 | 253 |

To minimise the overall dimensions of the hoist while accommodating the required rope length, the cranes will be equipped with two trolleys, each capable of lifting approximately 55% of the crane's total capacity. When handling a full-capacity load, both trolleys will operate together, using a spreader beam (see Fig. 8.88).

Each trolley will house two independent hoisting units (motor – gearbox – rope drum) that support the hook block. Under normal operating conditions, both hoists will function simultaneously, with the rope winding evenly across both drums.

This configuration provides mechanical redundancy in the event of a hoist component failure. If any component of one hoist malfunctions and prevents normal operation, the other hoist will still be able to complete the lifting operation. In such a scenario, the rope will be wound on the functioning drum in a double-layer configuration, ensuring continued operation and safety.

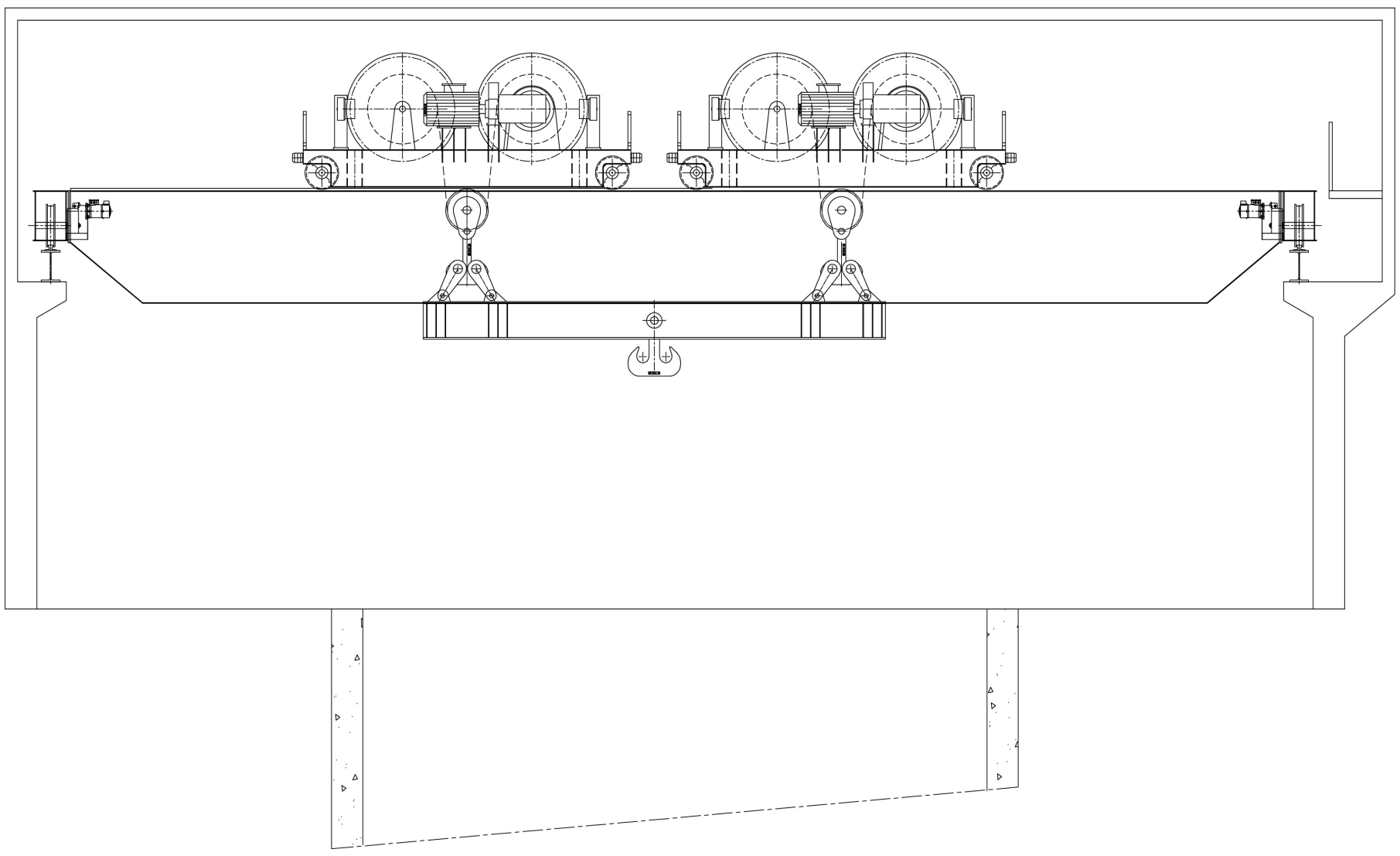

Fig. 8.88: Concept of shaft crane.

The design of the shaft cranes will also include the following specific features:

- Emergency brake on each rope drum to avoid dropping the load should any element of the hoisting drive chain fail;
- FEM classification of the hoist gearboxes: M8
- Hoisting speed at full load: 15 m/min - hoisting speed without load: 30 m/min
- Laser sensors, encoders and a programmable logic controller (PLC) to control the hook position at any moment.

The cranes will also be equipped with a dedicated monorail, installed below one girder, hosting a traversing trolley on which a platform is suspended. The platform allows personnel access inside the shaft in both the construction and operation phases to install or maintain the technical infrastructure (e.g., ventilation ducts).

In the coming five years, CERN will also test a new technology for the lifting ropes, based on textile ropes instead of steel ropes, which reduces the overall weight suspended from the structure of the crane and, therefore, optimising the overall crane design and cost.

***Underground transport vehicles***

The diversity of magnets to be transported creates a number of different requirements for the transport and handling technology in the tunnel. The basic principle will consist of two types of trailers, which are specially equipped with the technology required for the particular magnet transport and magnet handling.

A detailed study is accessible in Ref. [395], based on QSS units of 11 t and dipoles of 3.7 t. In 2024 the design of the magnet has been updated with QSS units of 14.5 t and dipoles of 5 t, however, this change does not have a significant influence on the design of the vehicle, the main outcome being that the outer width may increase by 200 mm maximum, which still fits inside the transport volume.

The underground transport vehicles will be composed of convoys, each consisting of two tractors that can move in both directions and a specialised trailer designed for transporting and handling magnets within the regular arcs. There are two main categories of magnets and assemblies to be transported, both for the collider and the booster, each with specific weight distribution characteristics requiring adapted trailers. The dipole magnets are long and relatively light, with a unit weight of 5 t and 12 m long, whereas the QSS units are shorter and heavier, weighing up to 14.5 t and 7 m long.

The tractors and all actuators on the trailers will be fully electric, eliminating exhaust gas emissions and reducing noise in the tunnel during magnet assembly. The tractors and trailers will be powered by battery systems, requiring the installation of charging stations in the service caverns to allow recharging during vehicle downtime periods, such as magnet loading and maintenance. The battery-powered approach is preferred compared to a conductor rail system, as it provides greater flexibility, requires no additional tunnel infrastructure, and enhances safety by eliminating exposed live electrical components. Additionally, battery technology is expected to improve in capacity and reliability in the coming years, further strengthening this choice.

The specialised trailers will be loaded with magnets at the base of the service cavern shafts, which will be equipped with the necessary infrastructure for lowering them into the tunnel. Autonomous driving will be implemented for the journey from the shaft/loading area to the installation points in the tunnel, minimising personnel requirements and improving operational efficiency.

To accommodate vehicle movements, enlargements are planned at every alcove, following a fixed pattern determined by the alcove locations. These enlargements will allow vehicles to pass or overtake when necessary, and can also serve as evacuation points in case of an emergency. However, the baseline approach for magnet transport is to operate overnight without personnel present in the tunnel, thereby avoiding co-activity with workers. While interventions by personnel during magnet transport have not been entirely ruled out, additional compensatory safety measures would need to be implemented to ensure proper evacuation procedures in such cases.

The handling technique relies on three-axis movement. First, the components are lifted vertically from the trailer. Once at the required height, a movable arm equipped with gripping technology places the components onto jacks. Finally, a fine longitudinal adjustment can be made using the gripper, ensuring precise positioning of the magnets.

*Transport of quadrupoles and sextupoles*

The QSS unit represents the biggest weight requirement for the equipment. The special trailer (shown in Fig. 8.89) which has the capability to carry the QSS unit is equipped with a damping system that protects the load from small bumps in the surface of the carriageway during travel through the tunnel. Hydraulic cylinders (alternatively electric cylinders) will be used by the handling system to generate the necessary force to move the load to its final destination. The hydraulic cylinders perform the main movements while placing the steel girder on top of alignment blocks.

Motors on the gripping system will allow the final fine adjustment longitudinally, due to the limitation in accuracy of the positioning of the whole convoy (as shown in Fig. 8.90).

*Transport of dipoles*

The dipoles are nearly 12 m long and are the longest items to be transported; they require dedicated equipment for their transport and handling. The trailer which is planned for carrying the dipoles has the capability to carry three dipoles at one time, thereby increasing transport efficiency (see Fig. 8.91).

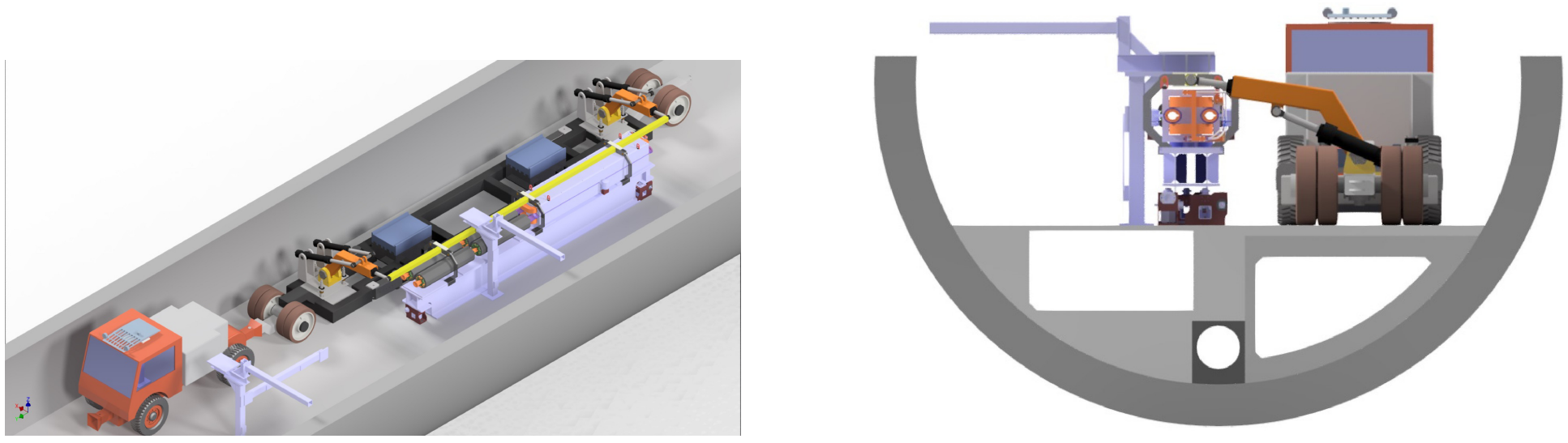

(a) (b)

Fig. 8.89: Transport and installation of a QSS unit.

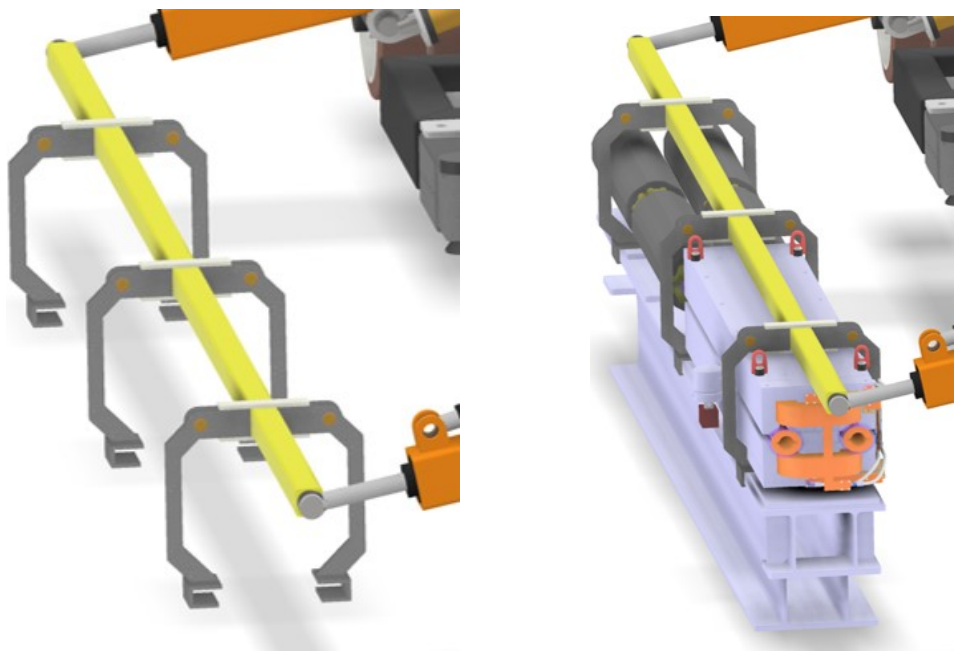

(a) (b)

Fig. 8.90: Detail of the gripping system.

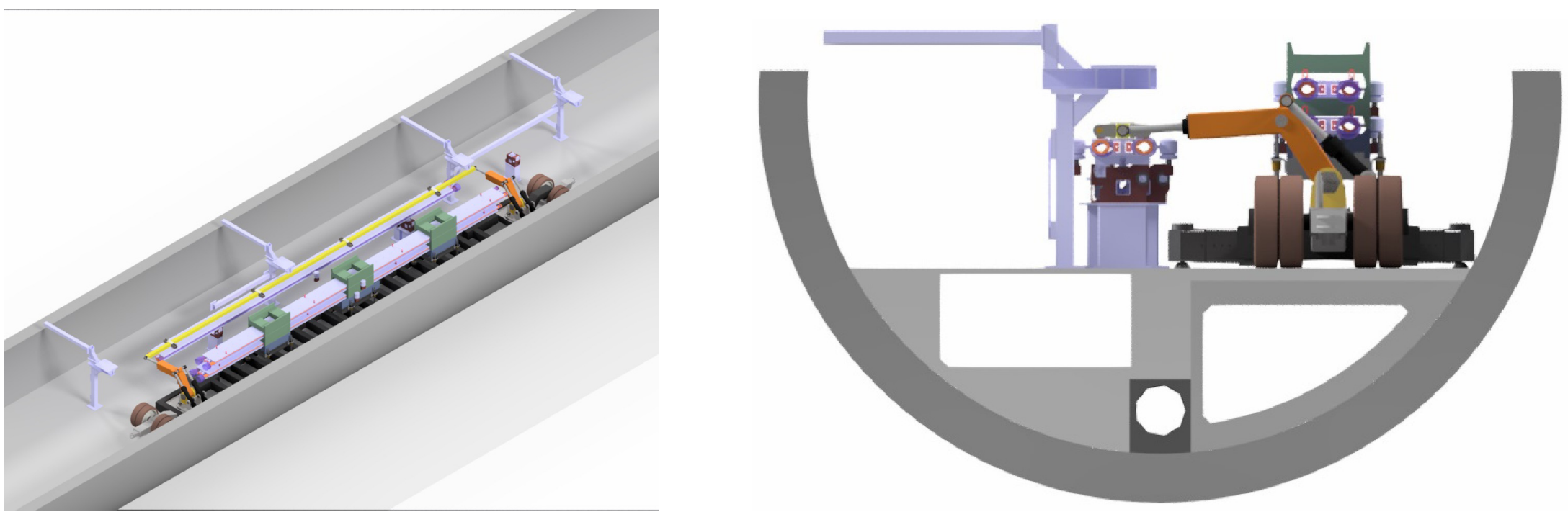

(a) (b)

Fig. 8.91: Transport and installation of 3 dipoles.

To enable the transport of three dipoles as one load, the dipoles will be stored in a special rack. The rack can be raised and lowered by the spindle lifting gear so that the handling system can always pick up the dipoles and place them in their final destination from the one position.

*Transport of booster ring components*

The components of the booster ring are mounted above the collider ring. The trailers for the dipoles can install magnets on both the collider and the booster, the same strategy applies for trailers dedicated to the QSS units.

### *Lifts*

The lifts will comply with the relevant European Directives (currently 2014/33/EU [396]) as well as CERN Safety Rules and EN standards. The lifts will use state-of-the-art technologies like those used in high-rise buildings, which already cover heights of up to 450 m. Two lifts will be installed in each of the four shaft-head buildings at the technical points, while four lifts will be installed in each of the four shaft-head buildings at the experiment points, within the service shafts. Lifts are the only authorised means of personnel access to and from the underground areas, making their reliability and safety critical. To enhance safety, the lift shaft concrete modules will be over-pressured, preventing the ingress of contaminants or smoke in case of an emergency.

The lifts will be powered by the secure power network, ensuring they remain fully operational even in the event of a failure of the standard electrical network. To further improve safety, operational efficiency, and maintenance availability, each lift shaft module will be equipped with two lifts, providing redundancy and easing access and maintenance operations.

The lift capacity and speed have been determined based on the results of evacuation simulations [397]. A maximum cycle time of 4 minutes has been defined, covering the entire process from people entering the lift, travelling from underground to the surface, and exiting the lift. This calculation assumes a shaft depth of 400 m and a lift speed of 4 m/s.

During the LHC programme, approximately 80% of the non-machine components were transported using lifts, suggesting that a similar proportion will apply to FCC. After evaluating different options, lifts with a 3 t capacity were chosen as the baseline solution, as they offer the best cost-to-capacity ratio while meeting the project's operational and logistic requirements.

The main characteristics of a lift are:

– Speed : 4 m/s;
– Capacity : 3000 kg/38 persons;
– Shaft height : (up to) 400 m;
– Car (length × width × height) : 2700 mm × 1900 mm × 2700 mm;
– Door (width × height) : 1900 mm × 2700 mm;
– Shaft width : 2750 mm;
– Shaft length : 3750 mm;
– Headroom : 7700 mm;
– Pit depth : 5900 mm.

### *Preliminary detailed handling studies in LSS sections in PA and PB*

The underground areas identified as 'Machine tunnel widening' areas in Fig. 8.92, are zones where handling processes will be studied in detail since the configuration of the machine elements is different from the regular arcs. The input from the beam optics induces tunnels with specific configuration and handling challenges.

Studies based on the preliminary design of the beamstrahlung dump system, located 500 m away from the interaction point on each side of each experiment cavern, have been performed to determine the possible handling process and equipment to install it. The same conceptual study was performed for PB, hosting the beam dump system of the collider and the booster. Details can be found in Ref. [398]. The handling and installation of the beamstrahlung dump system is feasible using a mobile crane and a standard electric tractor pulling a trailer. This avoids the installation of an overhead crane above each system, allows the reuse of the mobile crane for the installation of several systems, and helps reduce the size of the cavern required to install this system (see Fig. 8.93).

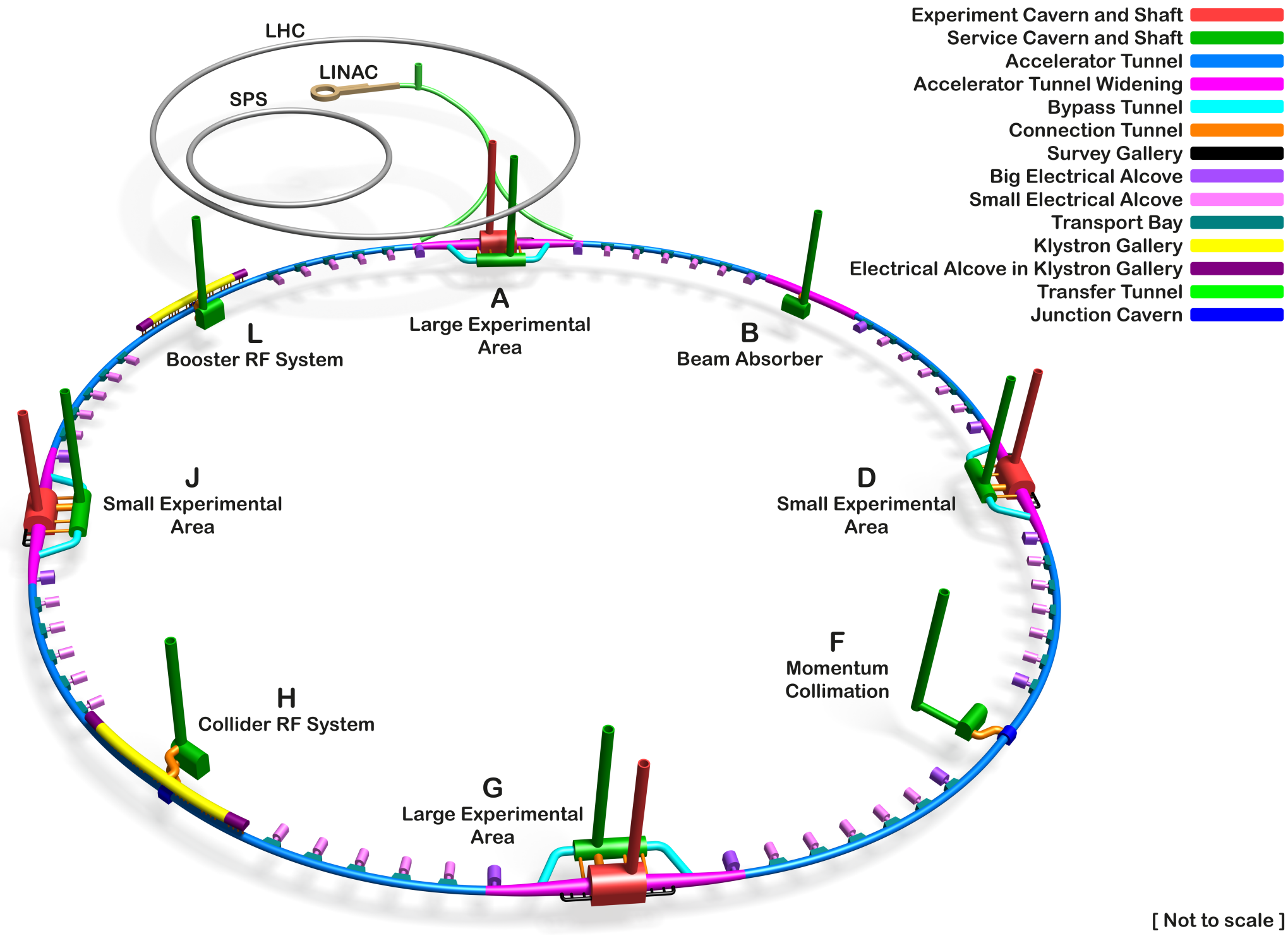

Fig. 8.92: FCC-ee Layout including machine tunnel widening areas.

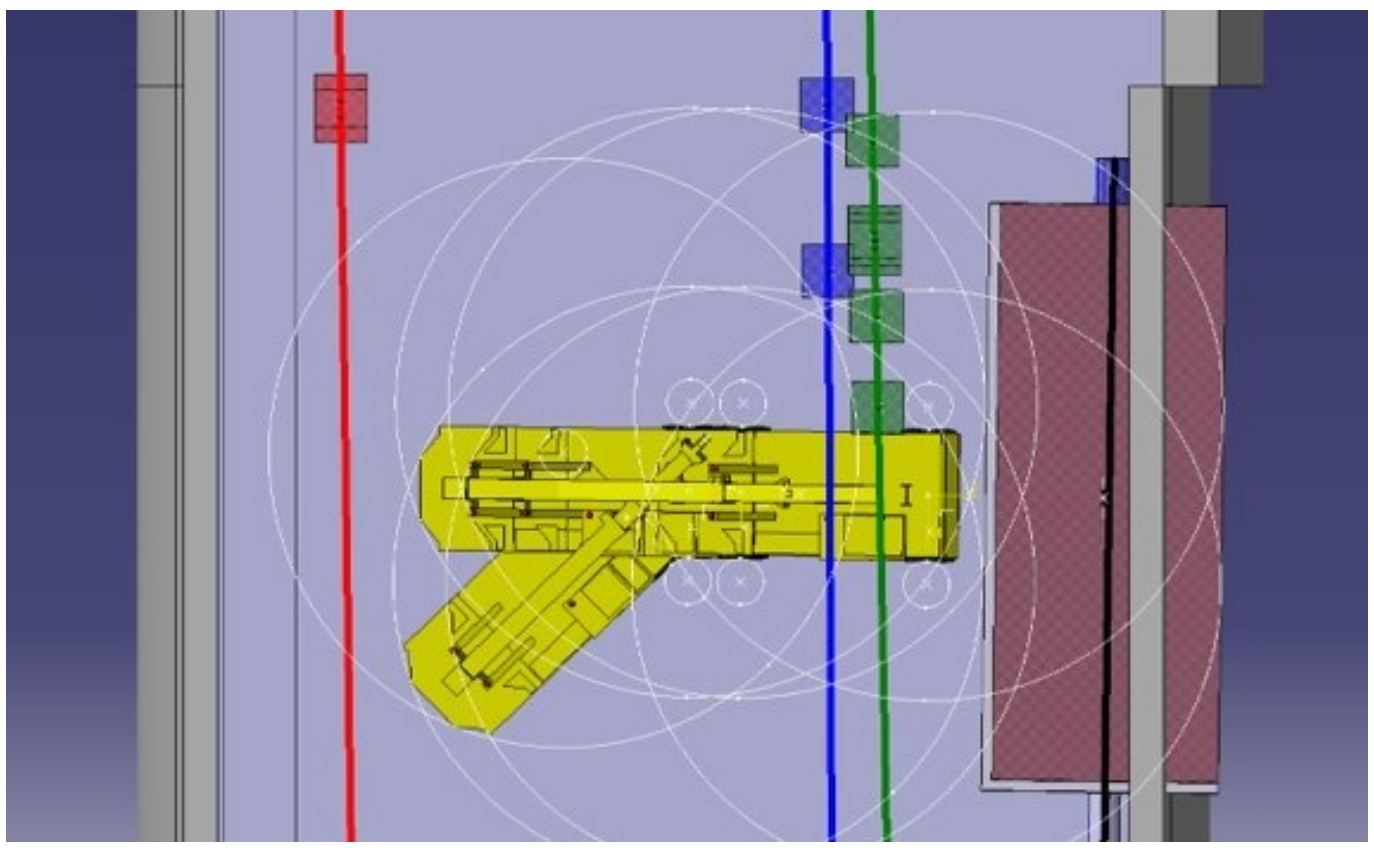

Fig. 8.93: Handling study for the beamstrahlung dump system.

Due to the limited space between beam lines in the area, the handling and installation of components in PB will be done by overhead cranes above the beam lines and beam dumps. Specific overhead cranes with a capacity of 20 t will be installed above each beam dump system to allow their installation and maintenance, as their weight is significantly higher than the elements of the beam lines estimated to be up to 5 t (see Fig. 8.94).

Once the final design of the machine components in this area is available, a detailed handling study will be performed in order to determine the most optimised handling strategy both for the installation and operation phase.

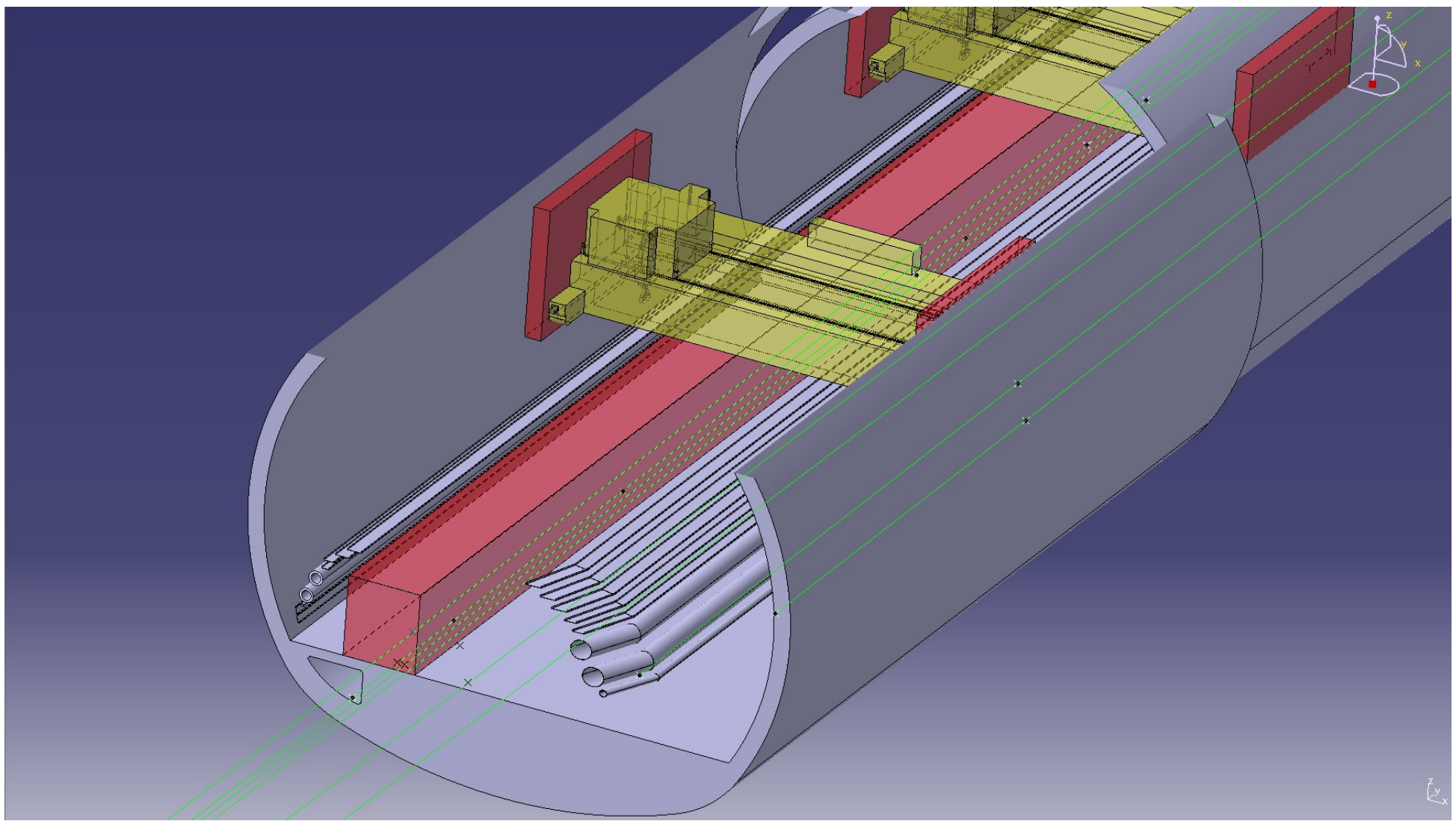

Fig. 8.94: Handling study in point PB, beam lines are highlighted in green.

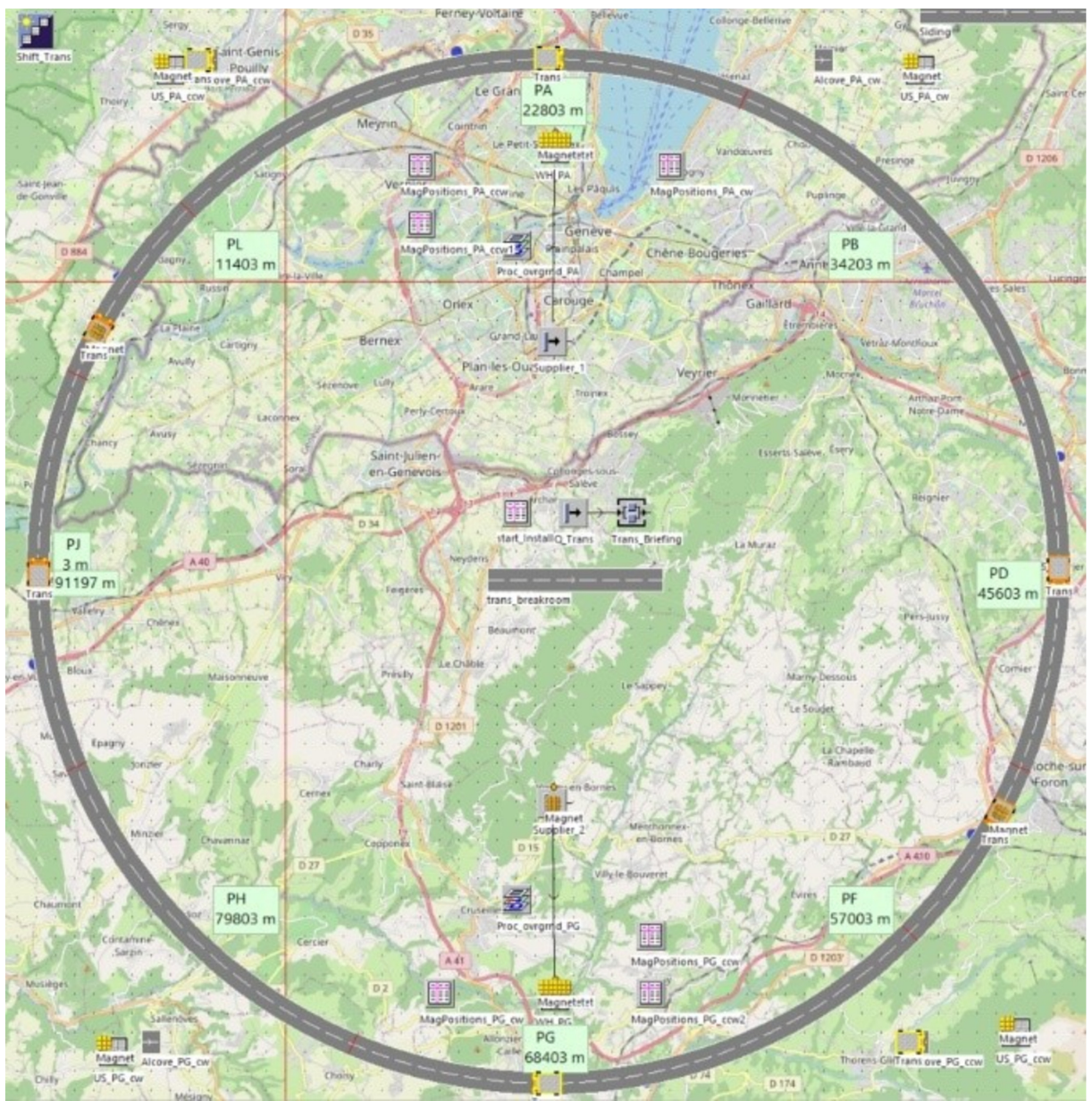

Fig. 8.95: Tunnel model developed for logistics studies.

### 8.6.2 Logistics

***Material flow simulation***

Logistics is of great importance for the construction, assembly, and operation of the FCC. An event-discrete material flow simulation study has been conducted to integrate all the dynamic interdependencies to create a robust schedule, identify bottlenecks and potential improvements, and estimate the resources required. Figure 8.95 shows the tunnel model which was developed. In addition, the corresponding necessary process flows and dependencies were defined and abstracted so that they could be integrated into the digital model. The simulation was run using Tecnomatix Plant Simulation©.

The underlying input parameters and handling strategy are described in Ref. [395]. Several sets of simulations were developed, allowing the identification of the optimised scenario: the dipole magnets will be handled through the experiments point service shafts, the QSS units will be handled through the technical points service shafts. This allows the workload of the shaft cranes to be balanced and optimisation of the overall arc transport time.

The main final outcome is that the effort required to handle the dipole magnets from experiment points to their final location in the arcs is 1200 man-hours or 75 days working in two shifts. The effort required to handle the QSS assemblies from the technical points is 1392 man-hours or 87 days working in two shifts. This result is compatible with the original assumption of 100 days duration and provides contingency in case of failure or logistic chain discontinuity. The workload of the shaft cranes allows the installation of two arcs from the same surface point to be performed in parallel, if the final schedule requires doing so. At the beginning of the simulation, it is assumed that the tunnel civil engineering has been completed and the general services technical infrastructure (cooling and ventilation, cabling, supporting jacks etc.) is installed so that the tunnel is ready for magnet transport. The logical flow of the programmed process can be seen in Fig. 8.96.

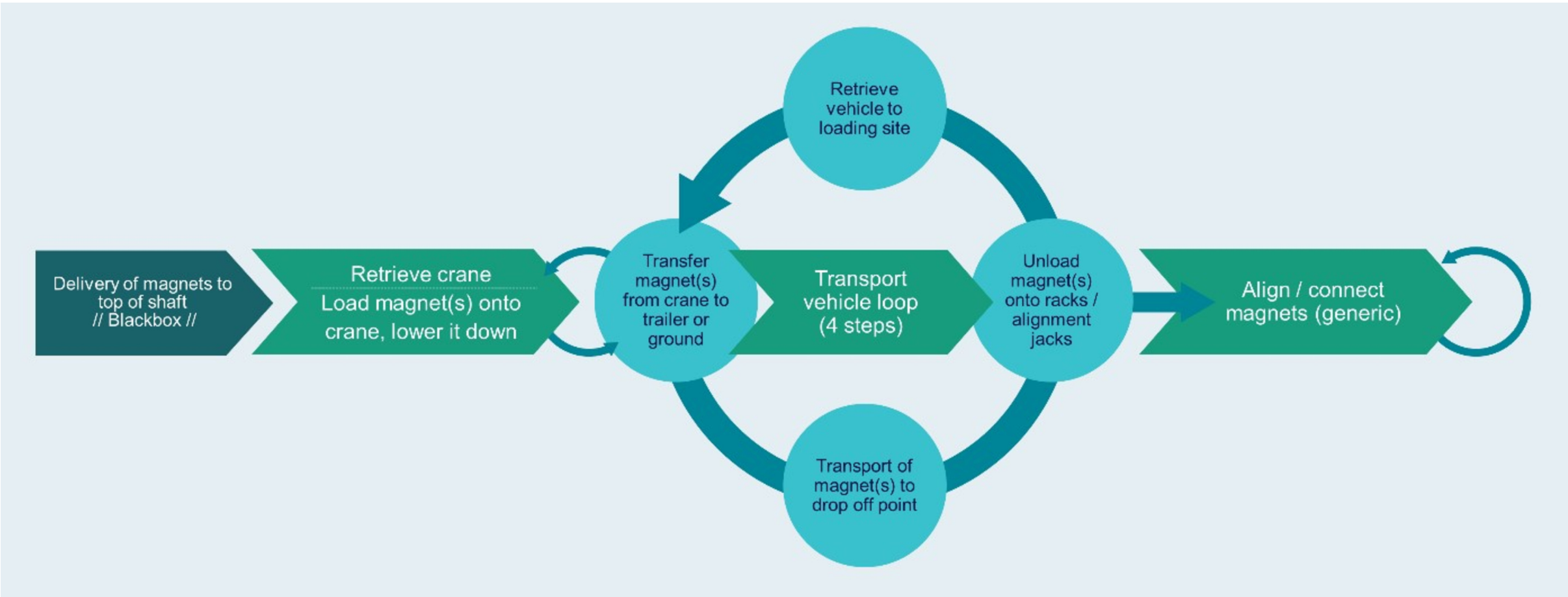

Fig. 8.96: Programmed logistic process.

The quadrupoles and sextupoles are pre-aligned on the girder and assembled into transport units. As described earlier, three dipoles can be grouped and transported simultaneously as a dipole pack. The magnet transport vehicle is designed to carry one QSS unit or one dipole pack at a time. During magnet transport operations, no other traffic will be present in the tunnel to ensure safety and efficiency.

To enable continuous transport operations, a variable number of magnets can be stored underground at the bottom of the shafts. Simulations indicate that a buffer of three QSS units or three dipole packs in the underground service cavern allows smooth operation. Expanding this buffer to 16 dipole packs and 8 QSS units would provide a one-day operational margin, mitigating potential disruptions in the surface logistics chain.

Magnet transports will not be scheduled to pass through areas where installation teams are actively aligning and connecting the magnets, as this approach optimises transport times. While co-activity of simultaneous transport and installation could be possible by reducing convoy speed to enhance safety, it is not included in the baseline plan due to its impact on efficiency.

The collider ring and booster ring can be installed simultaneously during the same installation phase, thanks to the flexibility of the trailers. On the surface sites, storage areas are planned near each shaft, with a capacity of two days (equivalent to 32 dipole packs or 16 QSS units) to facilitate direct crane loading.

It is important to note that the time required for aligning and connecting the magnets is still an estimate at this stage. As a result, any conclusions drawn based on this parameter should be interpreted with caution.

***Magnet production flow simulations***

As detailed in Ref. [399], the anticipated manufacturing timelines for dipoles, quadrupoles, and sextupoles are aligned with the current FCC-ee installation schedule. By targeting an average production throughput of 18 magnets per day, manufacturing and installation can proceed in parallel, ensuring timely deliveries while allowing sufficient time for R&D and procurement activities prior to full-scale production. This assessment confirms that the planned manufacturing rates are adequate to support the on-time installation phase, demonstrating the overall feasibility of the FCC-ee magnet production strategy.

***Surface transport of 240 t transformers and 60 t magnets***

The transport of equipment exceeding 24 t in a single unit requires an exceptional convoy and, in some cases, modifications to public roads to ensure feasibility. The heaviest piece of equipment identified in the project is the high-voltage transformers to be installed at sites PA, PD, and PH. A study [400] has confirmed that these transformers can be transported to site PH via four different routes with exceptional convoy procedures and limited modifications to public roads.

The second heaviest equipment in the project consists of the cryo-dipoles for the FCC-hh machine, which have according to today's estimates a unit weight of approximately 60 t and a length of 15 m. A separate study [401] demonstrated the feasibility of transporting these magnets to their final destination along the last kilometres leading to sites PA, PD, PG, and PJ, where service shafts are sufficiently wide to accommodate the handling of 15 m-long magnets.

***Overall surface road transport volumes for FCC-ee installation***

The scenario involves logistical planning for the transport of equipment on public roads. A key consideration is limiting the potential impact of daily transport operations. The most transport-intensive phase will be the magnet installation, during which the daily traffic per site is expected to remain below 10 trucks per day under the current schedule. This corresponds to an average of approximately one truck per hour, ensuring a steady, manageable flow of transport. Further details on transport logistics and planning can be found in Ref. [402].

***Handling for Experiments***

Handling activities in the experiment caverns will be conceptually similar to those currently carried out for the LHC Experiments. Detector components will be lowered to the cavern from the assembly hall through the shaft directly connecting the two facilities; this will be done with the overhead cranes installed in the surface hall. The components will then be handled with two overhead cranes installed inside the experiment cavern; each of them will have a capacity of 20 t. These overhead cranes will have the same characteristics as described in Section 8.6.1 plus the additional features listed below:

– the hoist gearbox will include a differential unit which allows the installation of a second hoist motor; this motor will be used for operations at very low speed (of the order of 0.3 m/min) which is usually requested for the precise assembly of detector components;
– an emergency brake on the rope drum to avoid dropping the load if any element of the hoisting drive chain fails.

Personnel access to the various parts of the detectors will be facilitated by mobile elevating working platforms (scissor and boom lifts).

### 8.6.3 Personnel transport

***Requirements for Personnel Transport Vehicle***

Throughout all stages of the life cycle, personnel will need to be transported from the vertical shafts to their designated workplaces. Following the handover of the shafts and tunnel from civil engineering, the installation of general infrastructure, such as electrical cabling, cooling, and ventilation, will take place, followed by the placement and connection of magnets. Various specialists will be required for these installation tasks, with the number of personnel varying depending on the specific phase of the project.

Once the collider is in operation, there will be technical and unplanned stops during which maintenance teams will enter the tunnel for repairs, inspections, and system upgrades. The safety requirements for personnel transport are outlined in Section 9.4.1, and further details can be found in Ref. [395].

The maximum distance between two shafts is 11 km, meaning that when some shafts are closed for access, the longest round-trip distance for personnel transport could be 22 km. To ensure efficient transport and minimise travel time, vehicles should be capable of speeds up to 30 km/h.

Fire doors installed in the tunnel define the maximum permissible vehicle size. Personnel transport vehicles should be able to pass each other to avoid blockages during evacuations and provide operational flexibility. The maximum driveway width is 2.2 m, and the maximum allowable vehicle height is 2.25 m. This allows vehicles to have a maximum width of 0.8 m, ensuring a 0.2 m clearance between vehicles as they pass, as well as between the vehicles and tunnel walls or machine equipment (see Fig. 8.97).

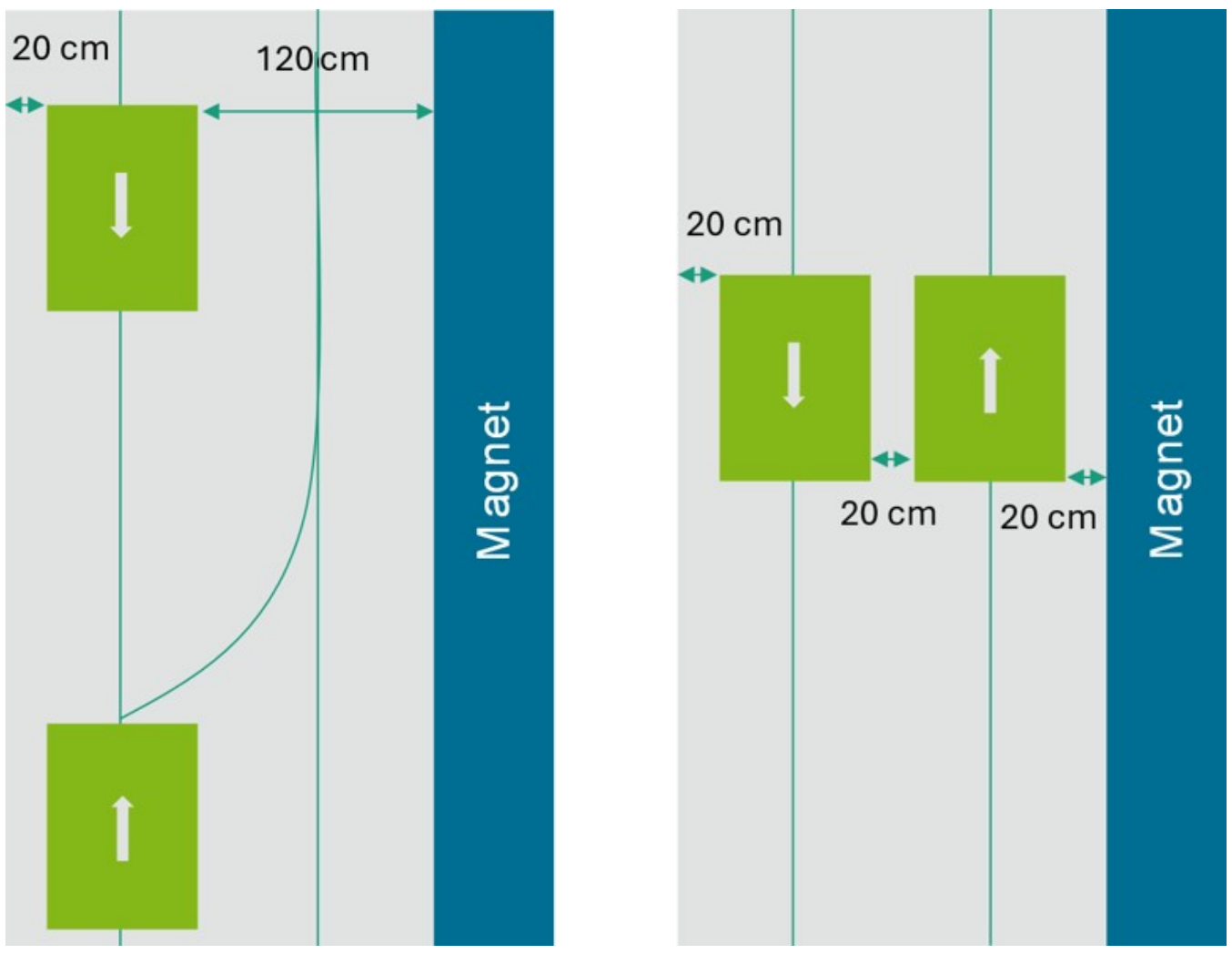

(a) (b)

Fig. 8.97: Concepts of vehicle circulation in the regular arc.

Based on these requirements, a design of a vehicle for four people has been established (see Fig. 8.98), the other constraint being the length of the vehicle to ensure maneuverability inside the tunnel (see Fig. 8.99).

The vehicles should be capable of autonomous driving using contour navigation, minimising the need for additional infrastructure. However, manual steering should remain an option for emergencies, maintenance, and operational flexibility. The vehicles will be battery-powered, with an estimated range of approximately 200 km based on an available volume of 216 L. This autonomy is sufficient, allowing for over six hours of continuous operation at full speed or nine round trips along the maximum distance of 22 km. During peak times, up to 200 people may be present in a single sector, requiring 50 vehicles per arc, with designated parking positions in each alcove (see Fig. 8.99).

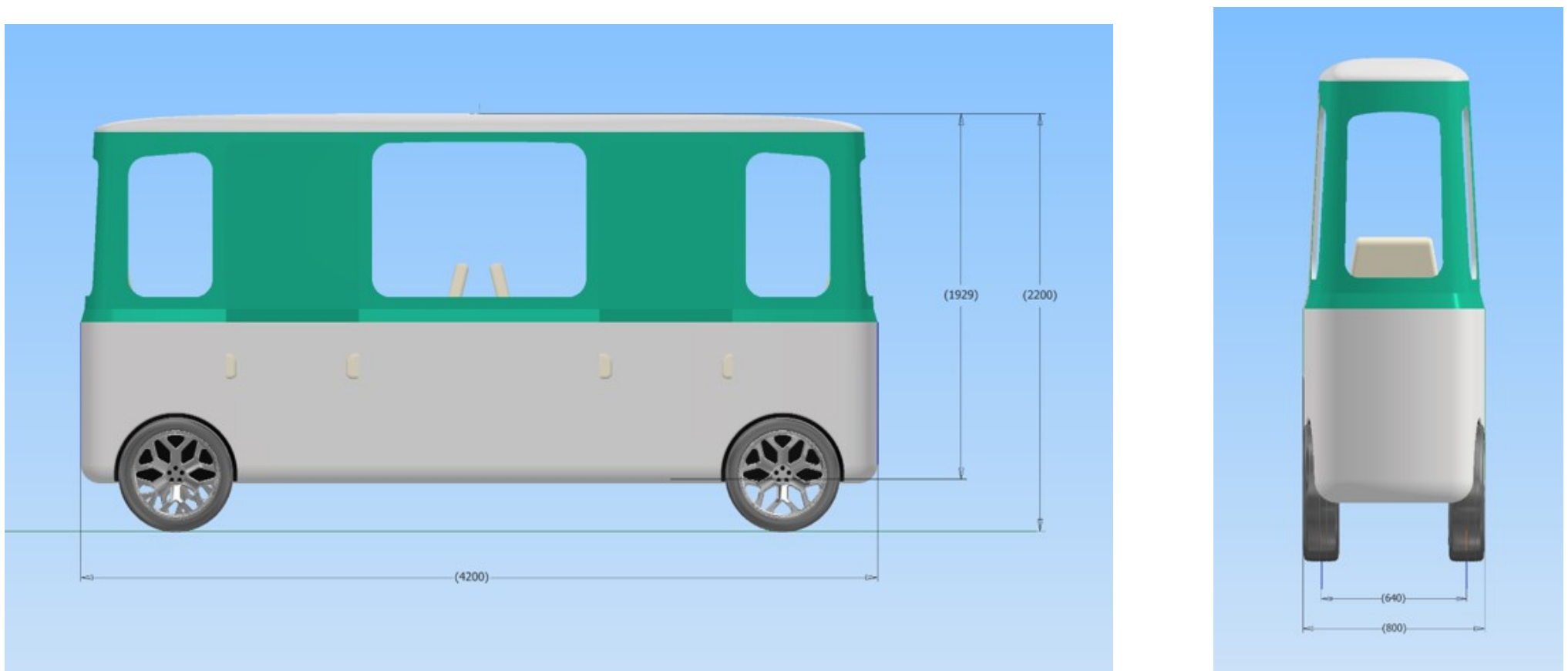

(a) (b)

Fig. 8.98: Dimensions of the conceptual design of the personnel transport vehicle.

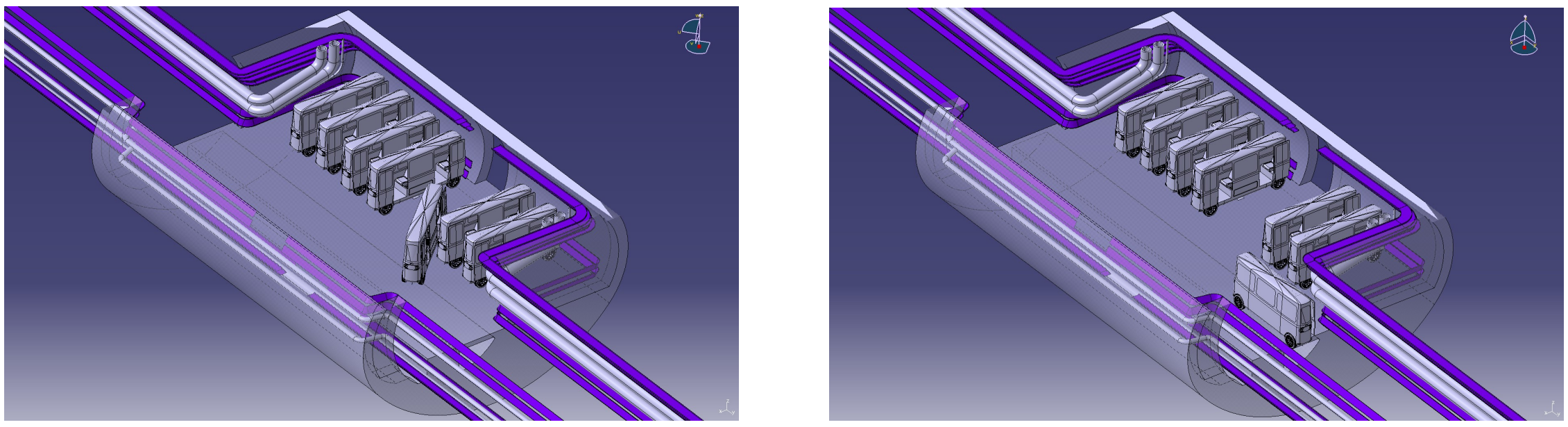

(a) (b)

Fig. 8.99: Illustration of manoeuvre of personnel transport vehicles in and out parking place

*Operation concept*

The concept of operation requires a distinction between the installation and operation phases of the accelerator, as each has distinct conditions. During installation, large quantities of materials must be transported to specific work sites, following a clear workflow. In contrast, the operation phase involves maintenance and other tasks distributed throughout the entire tunnel.

The main challenge during the installation phase is the simultaneous transport of materials and personnel within the tunnel. To minimise interference, the most effective approach is to prevent intersections of material flow. The baseline strategy for the magnet installation phase - the most transport-intensive phase - is to schedule material transport operations overnight while installation work takes place during the day. This method ensures better coordination and reduces congestion in the tunnel.

This night-time transport strategy can also be selectively applied to other phases where material movement is expected to be significant, such as the delivery of cables and pipes before their installation. However, in some cases, the presence of personnel near moving vehicles will be unavoidable during installation activities.

To enhance safety, personnel transport vehicles will be equipped with collision avoidance systems to detect and prevent accidents involving workers and obstacles within the transport zone. These systems will account for potential hazards, including toolboxes, materials, or personnel, ensuring safe and efficient tunnel operations (see Fig. 8.100).

The speed of the vehicles will be reduced in zones where activity is detected, ensuring safe coex-istence of movement and ongoing work. For both the installation and operation phases, two scenarios for personnel transport can be defined:

- Groups of people use a vehicle assigned to them, after drop-off the vehicle will park in the closest alcove during the whole stay in the tunnel;
- Groups of people use a vehicle to get to their place of work, the vehicle then moves somewhere else to pick up the next team.

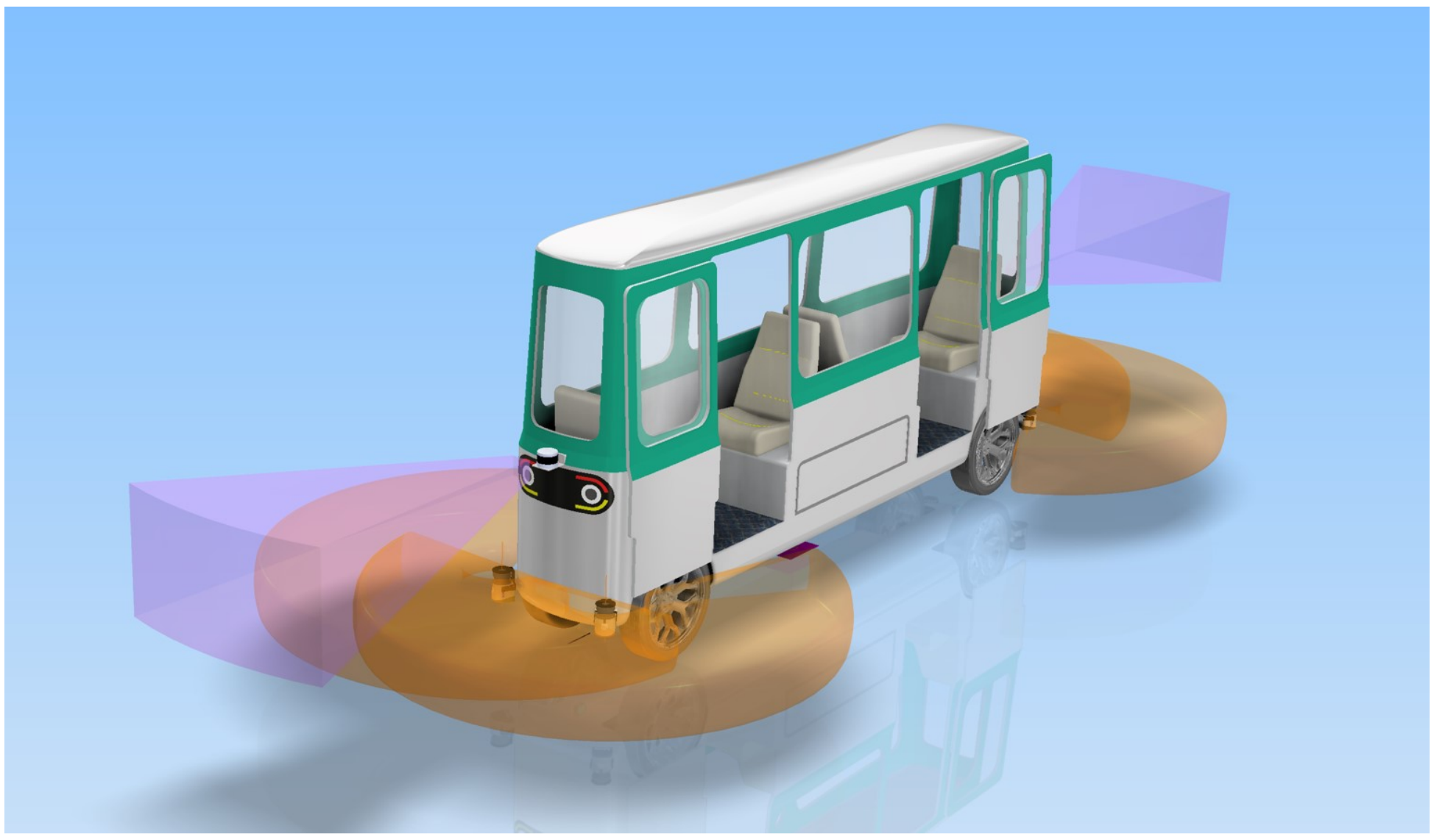

Fig. 8.100: Obstacle detection systems.

The two scenarios will be used depending on the type of work to be performed. The first scenario will be typically the one used for punctual work in several locations, the second being more dedicated to teams working in a fixed place.

*Evacuation concept*

The personnel transport vehicles will be parked in the lay-by zones of the alcoves. The maximum number of workers per arc is limited to 200, meaning that up to 50 vehicles may be required per arc, ensuring that each worker always has an assigned seat in a vehicle for evacuation purposes. Figure 8.99 illustrates how seven vehicles can be parked in the small lay-by zones, providing a storage capacity of 49 vehicles per arc. Additionally, 20 vehicles can be parked in the service caverns at each point, offering contingency parking spaces to support traffic management.

The current installation schedule foresees work being carried out simultaneously in all eight arcs, requiring a maximum of 400 personnel transport vehicles across the entire accelerator. However, further studies on detailed installation scenarios may help optimise this number, particularly if it is possible to reduce the number of personnel present in a single arc during certain phases. Such an optimisation would require a real-time personnel and vehicle tracking system, ensuring that each worker always has a seat available in an evacuation scenario.

*Current state of investigation*

Ongoing studies have led to the design of a vehicle that meets all dimensional, manoeuvrability, and autonomy requirements. The main challenges associated with these vehicles are their autonomous driving capabilities and the management of the fleet, both for daily operations and emergency evacuations.

CERN has prior experience with autonomous driving technologies. In addition, the industry has made significant advances in autonomous vehicle technology, with some manufacturers offering autonomous forklifts capable of operating safely alongside personnel. Modern factories, such as for example car manufacturing plants, are today typically deploying autonomous vehicles alongside human personnel. The challenge will be to integrate this autonomous functionality into a custom-designed personnel transport vehicle.

Fleet management presents another key challenge, as it requires integrating data from multiple systems, including safety systems, personnel tracking, and vehicle localisation. This integration will allow a centralised system to issue real-time instructions to each vehicle, ensuring efficient operations under normal conditions and coordinated response in case of evacuation. The development of this system will be addressed in a later phase.

## 8.7 Communications, computing and data services

The following sections outline the key components required for the computing infrastructure of the FCC, based on the Future Circular Collider Conceptual Design Report, Volume 3 [10]. These sections detail the equipment and infrastructure necessary to support data, voice, and radio communications, as well as the broader computing infrastructure. The document has been developed drawing on experience from previous colliders, particularly the LHC, to ensure a robust and efficient design.

The set of services and users was introduced in Ref. [10]. For completeness, it is depicted in Fig. 8.101.

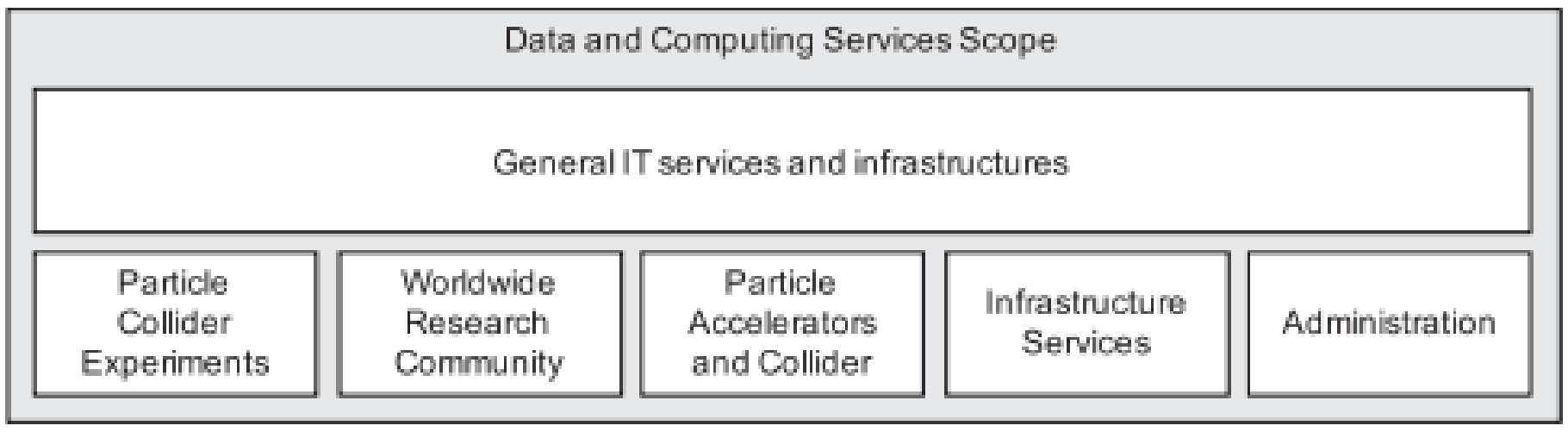

Fig. 8.101: Users and services of communication services.

### 8.7.1 Communication Services

The FCC will require the same communication services portfolio as provided at the LHC. These services include:

- The site-wide campus network, including Wi-Fi coverage,
- The high-performance data centre network,
- The dedicated technical network supporting accelerator operations and control networks for experiments,
- CERN's fixed and mobile telephony services,
- The TETRA digital radio service for the Fire Brigade and site guards, and
- Support for IoT devices, including a LoRaWAN infrastructure.
- WhiteRabbit network for timing distribution and synchronisation of all accelerator machines.
- Access control and video surveillance systems
- CERN Safety Alarms Monitoring (CSAM) system

To support the diverse services and technologies required for communications, three independent infrastructures will be deployed within the FCC facilities to enable data, telephony, and radio networks. While these three communication systems operate on separate network equipment, they will all rely on a shared fibre infrastructure that will interconnect the FCC's surface and underground facilities with the rest of CERN.

It is important to note that the choice of fibre infrastructure will have a significant impact on the equipment and associated costs required to implement data, telephony, and radio services.

The following sections present the fibre infrastructure solutions currently under consideration for the FCC, along with the requirements for data, telephony, and radio networks.

### 8.7.2 Fibre network deployment

Three different possibilities are being evaluated:

- Deploy the cables on the surface. This might be very complicated, given the size of the FCC and the distances between the points. Laying the cable on the surface will require agreements with multiple entities.
- Deploy the cables underground using the tunnel. The main issue with this approach is the level of radiation that will affect the cable. Presently, there are two types of fibre optic channels: the standard one and a more radiation tolerant version. The price difference is very significant; therefore, placing standard cables and shielding them behind concrete might be more economical.
- Rely on existing telecommunication operators. This approach has issues regarding the management of the network. The planning of interventions will also be more cumbersome. Moreover, there is the risk of locking in with a single operator.

Another key aspect under study is the network topology. In the LHC, a star topology is used, where each point has direct and redundant access to the CERN Computer Centre. This design provides a resilient and efficient architecture, simplifying traffic exchange at the data centre, which serves as the central hub for all infrastructures, including the internet, WLCG, experiments, and GPN. While this star topology remains the preferred design for the FCC, the final decision will depend on the optical fibre deployment strategy and whether the fibre infrastructure will be owned by CERN or an external entity.

Additionally, the connections between the alcoves and access points are being carefully designed. In the LHC, each alcove is directly connected to the two access points of its sector. However, in the FCC, with seven alcoves per sector, an alternative approach is being considered: a daisy-chain topology, where each alcove is connected only to its two nearest alcoves or access points. This solution would significantly reduce the number of fibres required but would come at the cost of reduced resilience in case of network failures.

### 8.7.3 Telephony and Radio Networks

This section covers the voice and radio communication services requirements for FCC's facilities and experiments. The 'red phones' currently installed in the LHC are required, and a radiation resistance study should be carried out. The fixed telephony [403] is currently based on IP phones on the surface and the CERNPhone application [404].

In the LHC, CERN IT provides a mobile voice and data services [405], a TETRA [406] radio network for the fire brigade, and a LoRaWAN network [407] for battery-powered wireless sensors to the accelerator chain and the experiments via a radiating cable. WiFi is only available on the surface, the alcoves and certain areas of the experiments, such as the WiFi access points, are not radiation-free, and the WiFi frequencies are too high to be injected in the radiating cable.

The deployment of the radiating cable is essential to ensure the availability of mobile and TETRA services in the accelerators and experiment areas. Since the FCC environment may expose the cables to higher radiation levels than those currently supported, a study is underway to assess their long-term durability. The cable must be installed in a way that provides a direct line-of-sight for service users—such as personnel, robots, and sensors—while also optimising its placement to extend its lifespan.

By the time the FCC could become operational, telecommunications technologies will have evolved further. Current trends suggest a convergence of mobile telephony, TETRA, and IoT services into future mobile network generations (5G and beyond). While the specific technologies may change, the core services - including mobile data, voice communication, IoT connectivity, and safety functions - will remain fundamental.

From a deployment point of view, there are three points to be assessed:

- Interconnection:
  - Fibres between one or several central points, today Meyrin and Prévessin, and all the FCC access points.
  - Fibres between each FCC access point and its adjacent alcoves.
- Radio equipment:
  - Mobile/TETRA/LoRaWAN radio emitters: 1 rack per access point, 1 per alcove and probably 1 extra for each large experiment.
  - If safety is a concern for these services (as is the case today for TETRA), the racks should have a secure supply with backup from diesel generator sets and/or UPS.
- Antennae to emit the radio signals:
  - Radiating cable in all the tunnels and experiments..

### 8.7.4 Data network

The data network design of the FCC will follow the same approach as the LHC, as described in the LHC Computing TDR [408]. It is based on a tiered structure, with CERN being the Tier-0, several remote sites, or Tier-1s, connect directly to CERN and store, analyse and redistribute the data to other Tier-2 institutes. Connectivity from CERN to Tier-1's and other WLCG members is established through the national research and education networks, and the FCC will not impact its structure, apart from the foreseeable increase in capacity and members.

The networks at CERN are composed of several independent infrastructures. Following today's LHC design, the FCC network will need to offer user connectivity to:

- General Purpose Network (GPN): CERN's Campus Network offering Internet access, wired and wireless connectivity to users.
- Technical Network (TN): Control network critical for the management and operation of the accelerators.
- FCC Computing Grid (equivalent to today's LCG): High-bandwidth network connecting the server farms in the FCC experiments to the datacentres for data storage at the Tier-0 and communication with the Tier-1 and Tier-2 centres.

The FCC network will be composed of passive and active equipment. The passive equipment, providing the physical media to support data communication (mostly fibre and UTP), will include around 260 starpoints with racks, patch panels, patches, and the fibre infrastructure to connect each starpoint to the upper element in the topology, either FCC points or the datacentres. The starpoints will also house the active network equipment.

The active equipment, enabling data communication, will mainly include switches, wireless access points and routers. Based on today's LHC scale and user density, 300 and 280 switches will be needed for TN and GPN respectively. Around 600 access points will be needed to provide wireless coverage in the surface buildings, underground facilities and alcoves. To interconnect all FCC facilities to both TN and GPN networks, 34 routers will be required. While today the GPN and TN infrastructures are completely separated and use different passive and active equipment, some economies could be made in the future by sharing the equipment and using virtualisation to implement the separation.

As mentioned in Section 8.7.2, the fibre network deployment will have an important impact on the way the network topology is built and how alcoves, buildings and FCC points will be interconnected.

- Starpoints in the alcoves should, ideally, be connected to their closest FCC points to provide redundant TN and GPN access. Fibres could be direct to the points or patched between alcoves.
- Starpoints in the surface buildings and underground facilities will use internal fibres to connect the wired and wireless network to GPN and TN.
- FCC points will need redundant fibre connectivity to the datacentres to have access to GPN, TN and, in the case of experiments, to LCG.

To ensure redundancy and fault tolerance of the FCC network, the backbone equipment interconnecting all FCC points will be distributed between the datacentres in Building 513 (Meyrin) and Building 775 (Prévessin). The Second Network Hub (Building 773) in Prévessin will also offer possibilities for increased resiliency.

### 8.7.5 IT infrastructure

As described in the previous section, the FCC will combine computing resources distributed all over the world. CERN will be the Tier-0. By the time the FCC starts, CERN will have two datacentres: one in Meyrin and a second one in Prévessin. This redundancy will help with business continuity and disaster recovery. The Prévessin datacentre is currently being built [409]. During 2025, it will provide 12 megawatts of computing resources. It will use the latest cooling technologies, and it will recuperate heat for other buildings. The power usage effectiveness (PUE) is expected to be 1.1. For comparison, the current PUE of the Meyrin datacentre is 1.5.

The LHC experiments follow the CERN Open Data Policy [410], which ensures a consistent approach towards the openness and preservation of experiment data. The experiments at the FCC will continue in this direction, ensuring the use cases of re-interpretation and re-analysis of physics results are available to the public. The Data Preservation and Long Term Analysis in High Energy Physics Project (DPHEP) [411] identified four levels of data (published results, outreach and education, reconstructed and raw data). This goes in the direction of Open Science, which applies to multiple areas of the FCC and this Feasibility Study.

Given the distributed nature of the collaboration, cyber-security plays a critical role in the whole infrastructure. It is of vital importance that all the components are secured from the first day and that they can withstand the constant cyber attacks that target scientific infrastructures. Ensuring security, data resilience and protection, and long-term data accessibility requires dedicated organisation from the beginning of the project.

## 8.8 Robotics

The fourth industrial revolution drives automation and data interconnection across industries, including space, warehouses, and harsh environments. Industry 4.0's pillars are IoT, Wireless Sensors, Cloud Computing, AI, ML, and Robotics. Robots are vital for tasks humans avoid due to danger, size constraints, or extreme environments. CERN has developed robots for accelerator maintenance, reducing risk and enhancing uptime. Envisioning advances over two decades, robots could revolutionise FCC tunnel interventions, replacing manual work or risky interventions.

### 8.8.1 Robotic impact

Robotics can enhance machine availability by improving maintainability through both corrective and preventive maintenance, as well as by increasing reliability with predictive maintenance. Additionally, robotics improves operational safety by reducing the need for workers to perform hazardous tasks and improves emergency safety with fast interventions by readily available robotic systems in the underground facilities, able to provide situational awareness and support rescue teams [412].

The potential impact of robotics on the availability can be demonstrated by looking at the luminosity of the particle accelerator, a metric that can be approximated as the volume of physics data it generates within a given time frame. To reach the luminosity targets [13] of the FCC, the machine availability must reach a minimum of 80 %, see Section 8.10. Recent studies, based on expert interviews and historical LHC data, have shown that this target can only be achieved by a 15-fold increase of the mean time between failures of certain critical systems. Introducing a readily available robotic system in the accelerator and thus reducing the drive time to intervention locations, allows to relax this constraint by a 10-fold, see Section 8.10.

As part of the safety concept for the FCC, robotics will enhance infrastructure and personnel safety during emergency scenarios within the 91 km long FCC tunnel. A three step emergency response strategy including robotics has been proposed by the CFRS [413]:

1. Response by trained workers on site
2. Emergency response robots
   (a) Situation awareness
   (b) First intervention (fire fighting, search and rescue)
3. Professional human responders
   (a) Verify situation
   (b) Second intervention (finalise situation awareness and fire fighting, specific damage control)

### 8.8.2 Level of robotic automation

Automation has the potential to be a primary driver of cost reduction across all stages of the accelerator's lifecycle, particularly in sustaining high availability throughout years of operation - directly influencing the organisation's data output and, consequently, its value and profit. It is therefore essential to determine the optimal level of automation that maximises benefits, while identifying the point beyond which additional automation leads to diminishing economic returns. The level of automation throughout the accelerator life cycle has to be identified in upcoming studies.

### 8.8.3 Standards, conventions, and guidelines towards efficient robotic automation

It is essential to emphasise that effective and economical robotics depends on incorporating automation requirements from the earliest design stages. Hardware and software components must adhere to well-defined standards, norms, and conventions, ensuring seamless integration of robotic automation. In the near future, a central point of contact or service needs to provide guidance, support, and oversight on implementing the conventions and guidelines for infrastructure and intervention procedures.

### 8.8.4 Robotic R&D required

The initial focus of the robotic development will be on a system with the greatest potential to enhance availability and safety, targeting the largest area of the FCC complex—the main tunnel. Previous studies [414] have identified a rail-based robotic system installed on the ceiling as the most efficient and robust solution for the accelerator tunnel. However, robotic systems covering other areas are expected to further enhance machine availability and safety. The necessary developments for a rail-based robotic system for the FCC tunnel area can be split into four phases:

1. **Definition of procedures and conventions**
   (a) Defining required tasks and intervention procedures
   (b) Defining remote maintenance code of practice
2. **Integration**
   (a) Define rail placement in regular tunnel cross section and parking locations
   (b) Design radiation safe spaces for parking and robot maintenance
   (c) Tool manager
   (d) Design of automated hatches in fire doors
3. **Technology R&D**
   (a) Infrastructure (Energy management, logistics)
   (b) Locomotion
   (c) Manipulator
   (d) Control (Motion, grasping)
   (e) Human-robot interfaces (Tele-operation, proprioception, haptics, collaboration)
   (f) Recovery scenarios and emergency interventions
   (g) Tool manager
   (h) Localisation,mapping & perception
   (i) Cognition
4. **Proof of concept**
   (a) Inspection and measurements in full autonomous mode
   (b) Teleoperation mode
   (c) Collaborative mode
   (d) Dipole alignment
   (e) Vacuum leak detection
   (f) Reconnaissance

The plan establishes progressive levels of capability:

1. **BASIC**: Full operation with limitations in automation and safety, requiring substantial tele-operation.
2. **BASIC+**: Emergency intervention capabilities added.
3. **MEDIUM**: Tool Manager, improved Human-Robot Interface, and higher efficiency.
4. **ADVANCED**: Semi-autonomy achieved, enabling collision avoidance and human-robot collaboration.
5. **ADVANCED+**: Full autonomy attained, introducing cognition and unforeseeable interventions.

This roadmap outlines the journey towards a comprehensive robotic solution for FCC. It addresses various technical challenges while highlighting the potential for enhanced safety, efficiency, and required innovation in maintenance operations.

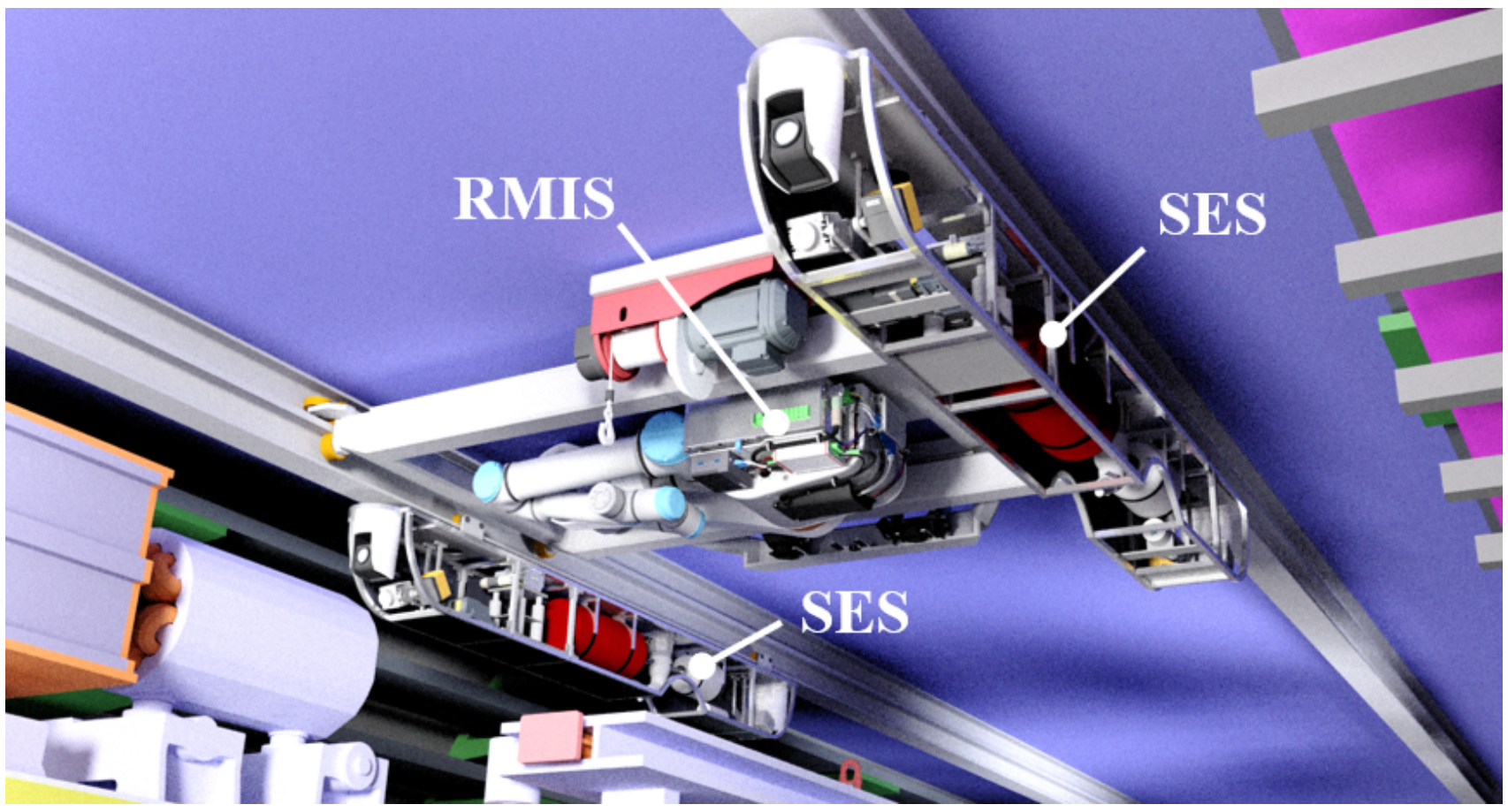

Fig. 8.102: The current baseline of the rail-based FCC robotic system with one RMIS and two SES systems.

### 8.8.5 Development status

The study on a rail-based robotic system for the FCC tunnel area departed from the first phase of the implementation plan outlined in the previous section. This rail-based robotic system, referred to as FCC robotic system, consists of two subsystems: the Remote Maintenance and Inspection System (RMIS) and the Surveillance and Emergency Shuttle (SES). The RMIS is designed to enhance machine availability and operational safety, while the SES focuses on improving emergency safety.

In Phase 1, requirements were gathered from stakeholders to enhance machine availability and operational safety [415], as well as emergency safety [416]. The requirements for emergency safety have reached a high level of maturity, while those for machine availability and operational safety are still in progress, as they are closely tied to the infrastructure design. Additionally, a draft for a Remote Maintenance Code of Practice has been developed, outlining conventions for infrastructure design and guidelines for intervention procedures [417].

In Phase 2, the integration of the rails beneath the ceiling of the accelerator has been finalised. A designated area within the tunnel's cross-section has been allocated for the robotic system, and radiation-shielded parking locations in the service caverns have been identified.

In Phase 3, a prototype of the 11-degree-of-freedom Remote Maintenance and Inspection System (RMIS) has been developed for proof-of-concept experiments in the FCC tunnel mock-up, with installation scheduled for 2025. The design of the Surveillance and Emergency Shuttle (SES) remains at the conceptual stage. The current baseline designs for both the RMIS and SES are illustrated in Fig. 8.102.

Phase 4, which begins in end of 2025, will focus on tests with functional infrastructure in the FCC tunnel mock-up.

## 8.9 Geodesy

Since the FCC-ee will extend well beyond the current CERN site, spanning both Switzerland and France across areas with diverse topographical and geological characteristics, an enhanced and extended geodetic infrastructure will be required. This will include an evolution of existing reference frames and an updated gravity field model to ensure precise positioning throughout the project.

A robust geodetic foundation will support the planning, construction, alignment, and operation of the FCC-ee, accommodating different levels of accuracy. These range from the initial large-scale placement studies to the final sub-millimetric precision alignment, which will continue to be refined throughout the accelerator's lifetime. The geodetic infrastructure will be adaptable to meet the needs of each project phase.

During the preparatory phase, a primary geodetic network will be established as early as possible to provide a stable reference frame for all subsequent activities, including civil engineering. Once the FCC-ee layout is finalised and construction begins, the geodetic network will be extended underground through the shafts, forming a dedicated underground geodetic network to ensure precise alignment of the infrastructure.

This section describes the core components of the geodetic infrastructure, while the alignment strategy for the accelerator components and detectors is detailed in Section 3.5.

### 8.9.1 Geodesy

***Definition of the coordinate reference systems for the FCC-ee***

A key component of the geodetic infrastructure will be a static coordinate reference system, established through the CERN Terrestrial Reference Frame (CTRF), along with a kinematic model (CKM) that represents the temporal evolution of the CTRF's reference points. This system will enable connections between CERN's existing reference frames and international, national, and local reference frames, ensuring compatibility and long-term stability. Additionally, it will serve as a foundation for analysing crustal deformations in the region.

For civil engineering works, a compound Coordinate Reference System (CRS) will be used, consisting of a projected CRS for horizontal positioning and a vertical CRS for gravity-related height determination. The horizontal coordinates will be defined through a CERN Projected Frame (CPF), while the Vertical Reference Frame (CVF) will provide height information consistent with the gravitational field.

For alignment purposes, the CERN Coordinate System (CCS) - which is currently used for all CERN machines - will remain in use. This will ensure a consistent and reliable link between the FCC infrastructure and the existing CERN facilities.

Since the FCC is a cross-border project, existing geo-referenced data such as geological maps, digital terrain models, and aerial imagery are expressed in different geodetic horizontal and vertical datums. To ensure compatibility, these datasets will be harmonised and transformed into the CTRF. Transformation models and their associated uncertainties will be carefully computed to maintain consistency across all datasets. Additionally, dedicated geodetic transformation software will be developed and made available to stakeholders to facilitate seamless integration.

Figure 8.103 shows the connection amongst the different parts of the reference system infrastructures. A detailed description of the coordinate reference systems is provided in Ref. [418].

***Implementation of the surface geodetic network***

Since it will be the reference for all survey and civil engineering work, a primary surface geodetic network (P-SGN) will have to be created as soon as possible to implement the CTRF (see [419]). During the period of the FCC Feasibility Study, IGN and Swisstopo increased the density of their national geodetic network (Réseau de Base Français and Points fixes planimétriques, respectively) over the FCC area and built new geodetic pillars and installed a new continuously operating global navigation satellite system (GNSS) reference station at locations suitable for the FCC-ee (see Fig. 8.104). The coordinates of the P-SGN will be determined and tied in the latest implementation of the European Terrestrial Reference Frame using simultaneous GNSS observations, ensuring absolute accuracy of 3 to 5 mm.

During the tunnel construction, the P-SGN would increase its density with auxiliary points, and a portal network would be created at each shaft to define the orientation of the tunnel. A surface levelling network will also be created, linking the eight surface sites and access shafts.

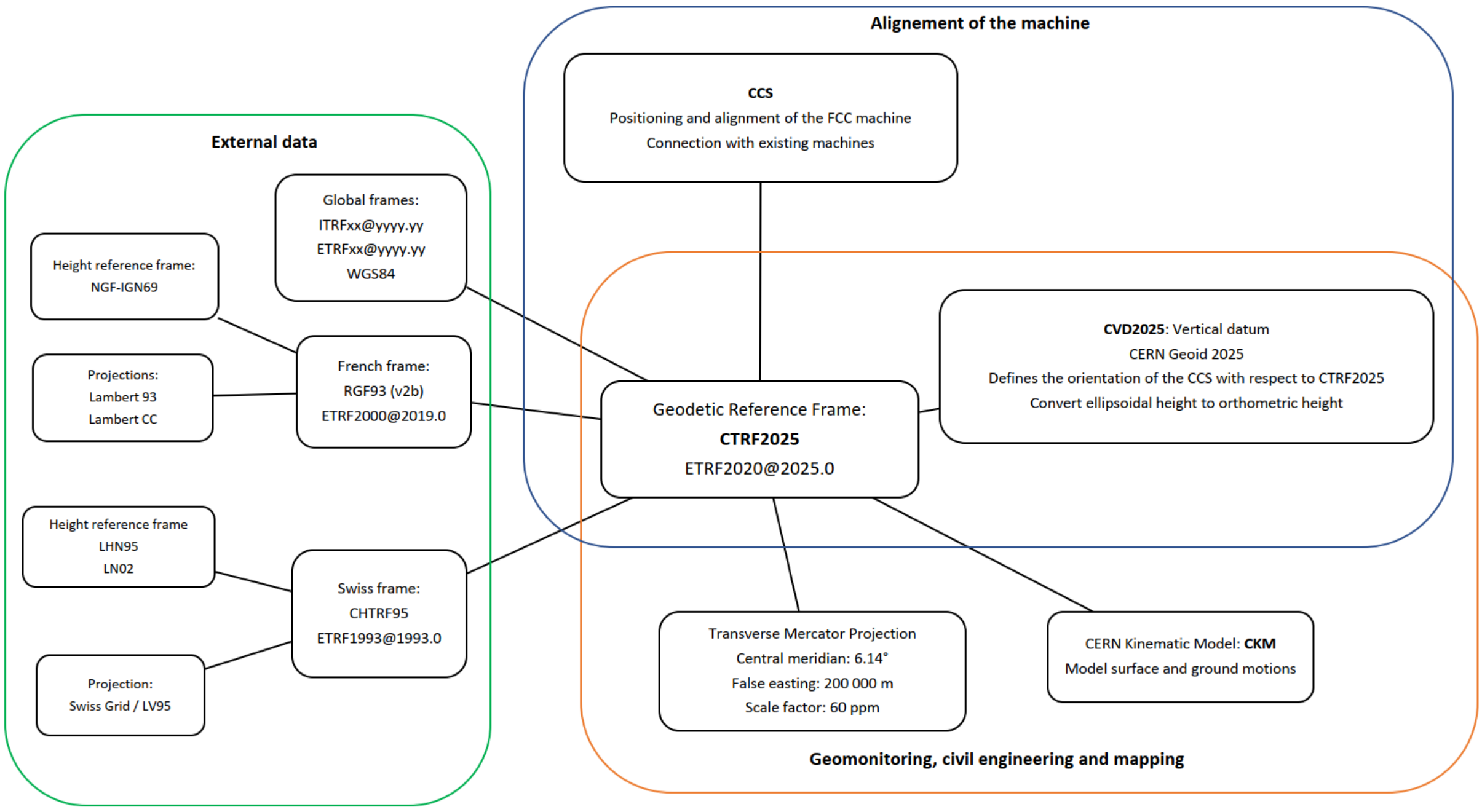

Fig. 8.103: Graphical outline of the coordinate reference systems for the FCC-ee.

### *Gravity field model*

To meet the vertical alignment accuracy requirements and to overcome the differences between the French and the Swiss altimetric systems, the local variations of the gravity field must be known or modelled with high accuracy and resolution (i.e., at a very short wavelength).

A centimetric (1 cm) accuracy could be achieved for the civil engineering and tunnelling work and as a basis for computing the gravity field model at the tunnel level. The latter is required to align the machine in an Euclidean plane. The computation of these models will require additional R&D.

A control profile, composed of GNSS-levelling and astrogeodetic observations, has already been determined to control the different geoid solutions that will be computed [420].

### *Underground geodetic network*

Once the tunnel shafts have been excavated, the coordinate reference system will be transferred underground. At this stage, the coordinates of reference markers regularly spaced on the floor and on the wall of the tunnel will be computed (see Fig. 8.105).

The underground geodetic network will be the reference for all automatic or manual alignment activities. Systems and instruments detecting the movements of the reference markers will be developed and installed to permanently or periodically monitor the stability of the reference network.

The development of the optimum methodology for the determination and monitoring of the underground geodetic network will require additional R&D efforts.

A geodetic reference frame that includes the nominal beamline must be established for each of the FCC-ee experiment caverns. To achieve this, geodetic reference points will be installed within the experiment caverns, typically implemented using wall brackets, nests, permanent tripods, or ground inserts equipped with CERN standard survey reference sockets. The entire geodetic network will be defined within the CERN Coordinate System (CCS) as part of the broader underground geodetic network. The network will be measured using high-precision instrumentation, including laser trackers, total stations, and direct levelling techniques. To maintain accuracy and stability, the geodetic networks of the ex-periment caverns will be regularly updated, ensuring that alignment remains consistent throughout the lifetime of the FCC-ee.

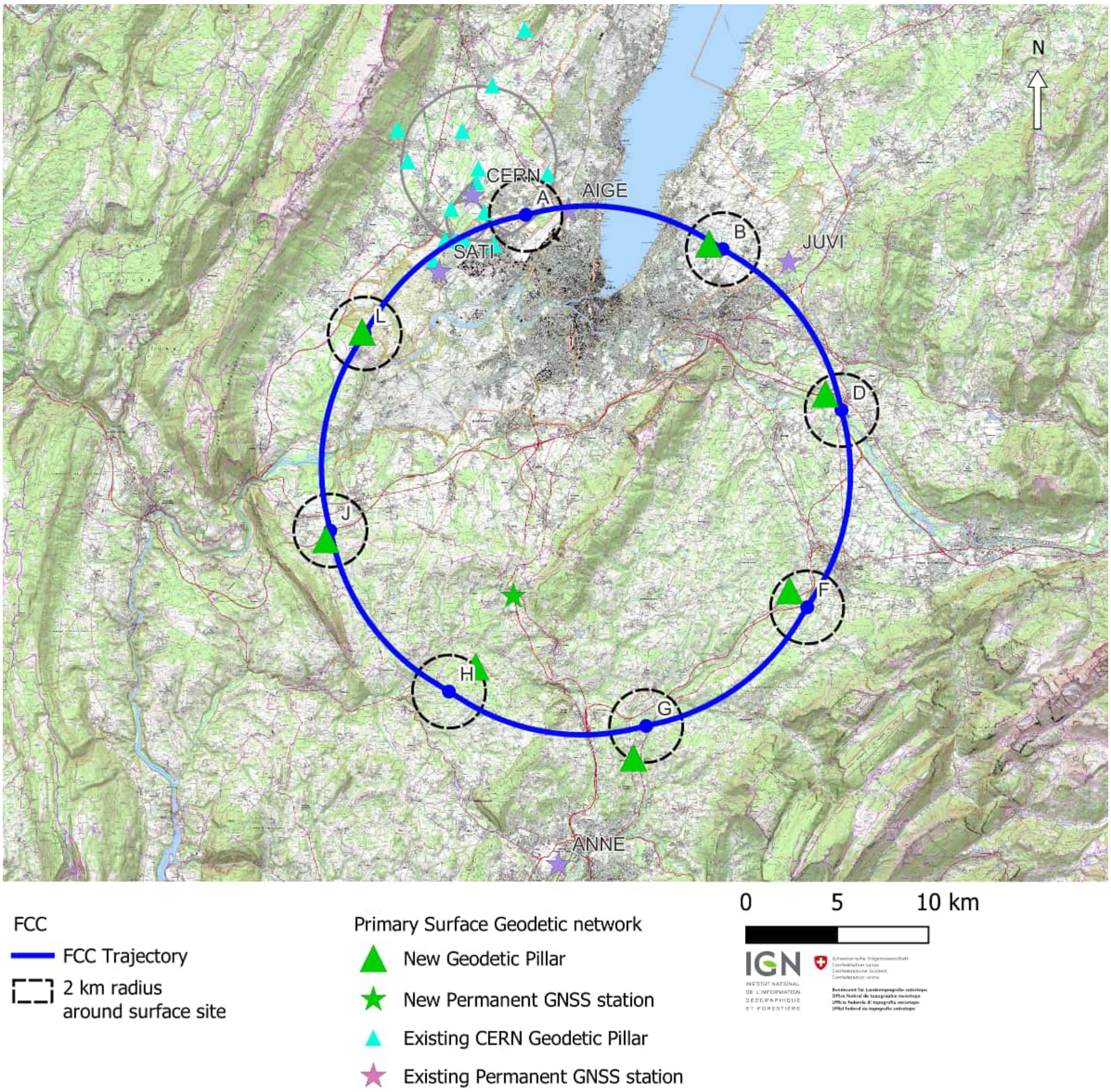

Fig. 8.104: Primary Surface Geodetic Network for the Future Circular Collider.

***Calibration, checking and testing of the geodetic instruments***

To meet surveying and alignment requirements, geodetic instruments, sensors, and tools used during the construction, installation and operation of the FCC must be regularly tested. A concept for calibration, checking, and testing (CCT) of geodetic instruments for the FCC has been developed (see Ref. [421]). A mix of in-house activities and facilities as well as outsourcing to external service providers would cover the needs. Dedicated spaces must be allocated at strategic locations for checking instruments like total stations, levels, laser trackers, 3D scanners prior usage. External service providers can be in charge of yearly maintenance and fixing identified defects.

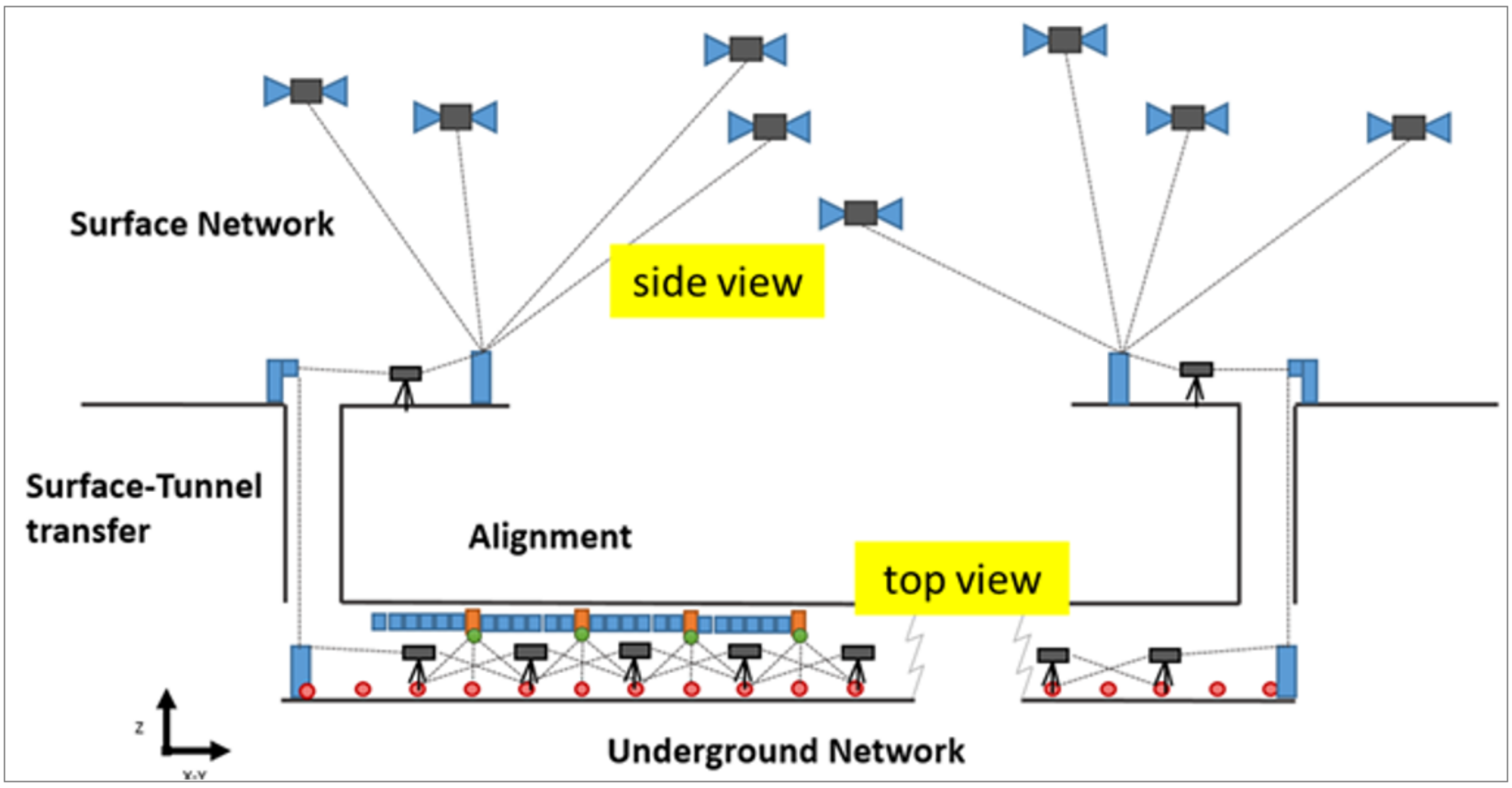

Fig. 8.105: Schematic representation of the coordinate transfer.

## 8.10 Availability

This section details results from the enhanced Monte Carlo simulation environment for FCC-ee availability described in Section 2.4, focusing on systems specific to the Technical Infrastructure serving the collider and booster. Infrastructure serving the injector complex is detailed in Section 7.9.

### 8.10.1 Contributing Systems

The same general framework for availability approximation was applied to the collider systems, described in Section 2.4.1. Specifics of this process used for each technical infrastructure system are described in the following paragraphs. Only faults leading to downtime in the representative system were considered, thereby assuming a similar degree of redundancy for each basic component family as currently exists in the working accelerator. Details of subsystems and scaling numbers are provided in Tables 8.36 and 8.37. Fault data was taken from CERN's Accelerator Fault Tracking (AFT) database [235] and is specific to the LHC physics operation 2015-2024, unless otherwise stated.

1. **Cooling & Ventilation:** Fault rate was scaled using the number of critical circuits at each access point, which is similar to the LHC (160 in LHC, 168 in FCC-ee). This does not consider the overall volume served by these critical circuits, which is significantly larger in the FCC-ee.
2. **Electrical Network:** The FCC-ee is connected to the PA, PD and PH external grid. The distribution between PA and PD is connected such that one can power the other in the event of failure with approximately 15 minutes of changeover time following beam dump. At PH, there is no backup reconfiguration option due to the high power demand in the collider RF system. Any electrical network failure at PH will, therefore, block the beam until failed hardware is restored. Two types of faults are categorised:
   (a) External network perturbations are a high contributor to downtime in the LHC, as they can trigger child faults in sensitive hardware across the complex. Failure rate and downtime from this category are applied without scaling to each of the three access points, assuming the same resilience to glitches as the LHC.
   (b) Internal network faults occur due to hardware failure in the electrical distribution designed and managed by CERN. This is scaled according to the number of substations at each location (PA: 4, PD: 3, PH: 1). This excludes the parallel power line feeding the collider RF at PH, which has not yet been modelled. A study to optimise the reliability of this parallel line against cost factors will take place in the pre-TDR phase.

3. **Cryogenics:** The FCC-ee requires only two cryoplants to serve the collider and booster superconducting RF systems at PH and PL, respectively. In the absence of a detailed design to indicate the number of components, the failure rate at each point is assumed to scale with the reference 4.5 K power level relative to one LHC cropland at 144 kW. This is constant for lower energy modes (Z, W, ZH) at 80 kW (collider) and 9 kW (booster). During the long shutdown prior to $t\bar{t}$ operation, each cryoplant is upgraded to 168 kW and 53 kW, respectively.

### 8.10.2 Simulation Results

The breakdown of unavailability and lost luminosity contribution from each technical infrastructure system is shown in Fig. 8.106. The electrical network has the largest contribution to lost luminosity in Z, WW, and ZH operation due to its more frequent and shorter duration fault types. Cryogenics is the most significant contributor to unavailability and lost luminosity in $t\bar{t}$ mode due to the energy upgrade required for additional cavities in the superconducting RF system.

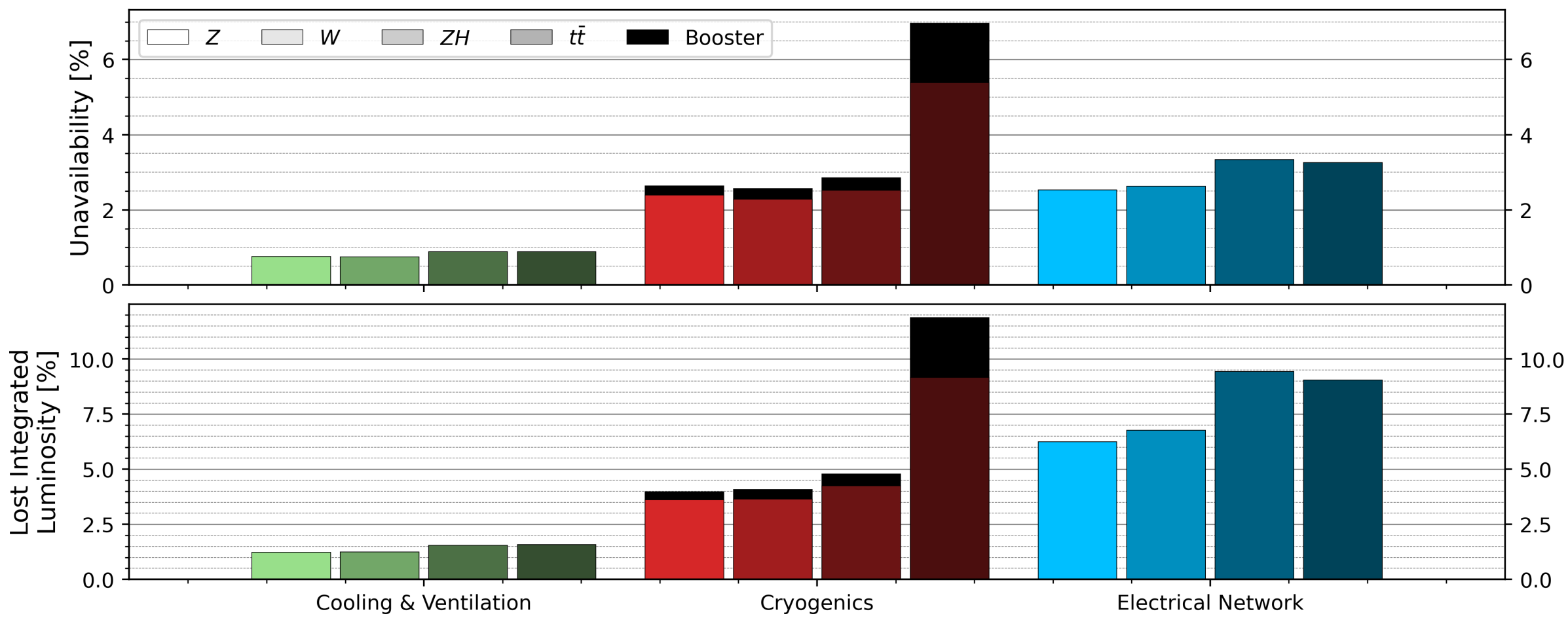

Fig. 8.106: Contribution of technical infrastructure subsystems to unavailability and lost luminosity. Systems are ordered according to Z mode lost luminosity contribution.

### 8.10.3 R&D Opportunities

Several areas for improvement are identified:

#### *Internal Electrical Network Faults*

The internal electrical network significantly contributes to lost luminosity despite a 15-minute changeover redundancy between PA and PD. Designs must incorporate a more robust redundant connection between these two points. An automatic switch could significantly reduce the changeover time. However, the largest gains would be made by avoiding beam dumps altogether, as this eliminates lost luminosity due to turnaround time in the initial operation phases (see Section 2.3.2).

Reliability must get particular attention at PH, as this cannot be reconfigured to draw power from another access point. Further, this means the parallel power line feeding the collider RF must be designed to a particularly high-reliability target.

Table 8.36: Parameters used to simulate availability of systems and subsystems in the collider Technical Infrastructure.

| System | Subsystem | LHC<br># | FCCee<br>Z,WW,ZH / $t\bar{t}$<br># | Location | Group MTBF<br>Z,WW,ZH / $t\bar{t}$<br>days |
|---|---|---|---|---|---|
| Cooling & Ventilation | Critical circuits: | | | | |
| | PA | 200 | 30 | PA | 111.3 |
| | PB | 200 | 20 | PB | 166.9 |
| | PD | 200 | 30 | PD | 111.3 |
| | PF | 200 | 20 | PF | 166.9 |
| | PG | 200 | 30 | PG | 111.3 |
| | PH | 200 | 25 | PH | 133.6 |
| | PJ | 200 | 30 | PJ | 111.3 |
| | PL | 200 | 25 | PL | 133.6 |
| Electrical Network | Internal Network: | | | | |
| | PA | 1 | 4 | PA | 3.6 |
| | PD | 1 | 3 | PD | 4.8 |
| | PH | 1 | 1 | PH | 14.3 |
| | External Glitch | 1 | 3 | PA, PD, PH | 7.7 |
| Cryogenics (collider) | Tunnel: | | | | |
| | Instrumentation | 144 | 80 / 168 | PH | 70.6 / 33.6 |
| | PLC | 144 | 80 / 168 | PH | 352.9 / 168.1 |
| | Production: | | | | |
| | Temperature | 144 | 80 / 168 | PH | 47.1 / 22.4 |
| | Controls | 144 | 80 / 168 | PH | 470.6 / 224.1 |
| | Instrumentation | 144 | 80 / 168 | PH | 74.3 / 35.4 |
| | PLC | 144 | 80 / 168 | PH | 235.3 / 112.0 |
| | Vacuum | 144 | 80 / 168 | PH | 352.9 / 168.1 |
| | Operation: | | | | |
| | PNO-SAM | 144 | 80 / 168 | PH | 352.9 / 168.1 |
| | PNO-BSCR | 144 | 80 / 168 | PH | 235.3 / 112.0 |
| | PNO-REF | 144 | 80 / 168 | PH | 282.3 / 134.5 |
| | PNO-DFB | 144 | 80 / 168 | PH | 17.9 / 8.5 |
| | PNO-HF | 144 | 80 / 168 | PH | 282.3 / 134.5 |
| | Other | 144 | 80 / 168 | PH | 117.6 / 56.0 |
| | Specific Operation | 144 | 80 / 168 | PH | 176.5 / 84.0 |
| | Users | 144 | 80 / 168 | PH | 282.3 / 134.5 |

***External Electrical Perturbations***

The simulation assumes that the same resilience to external network perturbations will occur in the FCC-ee as currently appears in the LHC. This is a coarse assumption, as the resulting downtime is most commonly due to child faults in sensitive accelerator systems, e.g., high-precision power converters, RF, etc.; the number of which could be significantly higher in FCC-ee. Learning from LHC experience, all FCC-ee systems must be designed according to an established standard for resilience to electrical glitches.

Table 8.37: Parameters used to simulate availability of systems and subsystems in the booster.

| System | Subsystem | LHC<br>#| FCCee<br>Z,WW,ZH / $t\bar{t}$<br>#| Location | Group MTBF<br>Z,WW,ZH / $t\bar{t}$<br>days |
|---|---|---|---|---|---|
| Cryogenics (booster) | Tunnel: | | | | |
| | Instrumentation | 144 | 9 / 53 | PL | 627.4/ 106.5 |
| | PLC | 144 | 9 / 53 | PL | 3137 / 532.7 |
| | Production: | | | | |
| | Temperature | 144 | 9 / 53 | PL | 418.3/ 71.0 |
| | Controls | 144 | 9 / 53 | PL | 4183 / 710.3 |
| | Instrumentation | 144 | 9 / 53 | PL | 660.5 / 112.2 |
| | PLC | 144 | 9 / 53 | PL | 2091 / 355.2 |
| | Vacuum | 144 | 9 / 53 | PL | 3137 / 532.7 |
| | Operation: | | | | |
| | PNO-SAM | 144 | 9 / 53 | PL | 3137 / 532.7 |
| | PNO-BSCR | 144 | 9 / 53 | PL | 2091 / 355.2 |
| | PNO-REF | 144 | 9 / 53 | PL | 2510 / 426.2 |
| | PNO-DFB | 144 | 9 / 53 | PL | 158.8/ 27.0 |
| | PNO-HF | 144 | 9 / 53 | PL | 2510 / 426.2 |
| | Other | 144 | 9 / 53 | PL | 1046/ 177.6 |
| | Specific Operation | 144 | 9 / 53 | PL | 1569/ 266.4 |
| | Users | 144 | 9 / 53 | PL | 2510 / 426.2 |

***Cryogenics***

Despite the significantly smaller cryogenic load compared to LHC, this is the highest contributor to downtime from the technical infrastructure systems. As designs develop, special consideration must be given to ensure availability.

### 8.10.4 Outlook

Technical infrastructure availability is of particular importance in the next phase as many of these systems have significant footprints on the civil engineering layout, e.g., the size of alcoves and caverns, routing and location of cables and equipment, radiation protection, access, etc. The coming years must see a thorough reliability/availability analysis of these high priority systems to ensure performance ahead of civil engineering procurements.

# Chapter 9

# FCC safety concepts

## 9.1 Introduction

This chapter outlines the safety concept for FCC. Given the intricate nature of particle accelerators, which involve high-energy beams, radio frequency cavities, cryogenic systems, and sophisticated electromagnetic equipment, a comprehensive safety strategy is essential to ensure operational integrity.

Key safety considerations include identifying the scope of the concept, its objectives, as well as the underlying impact of domains such as deep underground siting, radiation protection, structural integrity, fire and oxygen deficiency, and emergency response protocols. Integrating safety measures into the FCC design has an impact on the layout of the facilities. This chapter underlines the specific strategies and technologies employed, demonstrating our commitment to creating a safe and efficient research environment.

### 9.1.1 Safety considerations

The geographical distribution of surface sites and the time required to reach each site by road, the challenges linked to the underground working conditions, and the diverse composition of technical equipment at surface sites require the systematic development of an integrated safety concept.

The results presented in this chapter are based on work that assumes the *baseline* layout and parameters outlined in this report. If future phases of the study introduce deviations from the *baseline*, their impact on the safety concept will have to be assessed. Certain safety systems form the foundation of the safety concept; therefore, any modifications to these systems may necessitate the development of an entirely new safety concept (see Section 9.3.4).

## 9.2 Safety goals & objectives

### 9.2.1 Regulatory framework

CERN, an intergovernmental organisation established under international law by its Convention (1 July 1953, amended 16 June 1972), is headquartered in Geneva and operates across the Franco-Swiss border.

Under its regulatory framework, CERN establishes safety rules as necessary for its functioning. The CERN Safety Policy [422] defines the overarching principles governing safety at CERN and is complemented by CERN-specific rules tailored to its operational needs. Where specific rules are not established, national laws apply within their respective territories. This framework ensures that the rules are uniform throughout the site and adapted to the specific (particularly technical) requirements of the organisation.

CERN collaborates with the Host States under domain-specific treaties and tripartite agreements, e.g., in efforts to minimise environmental impact and uphold best practices in radiation protection. While the Host States facilitate its operations, CERN remains committed to limiting its impact on their territories and ensuring its activities do not compromise their security.

This regulatory framework reflects the current perspective and serves as the foundation for developing this safety concept.

### 9.2.2 CERN safety policy

CERN has a safety policy [422] in place that defines the safety objectives for every project and activity in order of priority:

1. Life safety: ensure the best possible protection according to CERN's Safety Rules in health and safety matters of all persons, irrespective of their status, participating in the organisation's activities or presence on its site, as well as of the population living in the vicinity of its installations.
2. Environment protection: limit the impact of the organisation's activities on the environment.
3. Asset protection and business continuity: guarantee the use of best practices in matters of safety.

The FCC feasibility study covers all these objectives, although the scope of the safety concept, developed at this stage, focuses on the life safety objective. The environmental protection aspects are covered in a dedicated stand-alone document, which is equally referenced in the third volume of the Feasibility Study Report.

Although the safety concept described focuses on the FCC-ee accelerators (injector, booster, and collider), possible incompatibilities with the civil engineering layout for FCC-hh are highlighted and integrated based on the requirements for the feasibility study. Modifications to the technical infrastructure needed for FCC-hh but which are not yet included in the design will be integrated in the safety concept when the infrastructure is being designed.

The transfer tunnels from the injector chain are not included in this concept, since safety aspects can only be studied for those elements once a design baseline is developed. The impact on safety of this part of the infrastructure would therefore be analysed in the next phase of the project.

### 9.2.3 Lifecycle phases

The Safety concept covers the following lifecycle phases:

- Construction phase.
- Installation phase.
- Operation phase, including maintenance periods and repair activities.

Each phase is characterised by different safety aspects, associated with the safety systems deployed and available, and with the time occupants are exposed to those conditions. The development of the safety concept described in this section is based on analysis of the *operation phase* of the FCC.

### 9.2.4 Safety Organisation & Management Plan

A safety management plan is required for ensuring the safe operation of such a large research infrastructure. This plan will not be limited to the respect of safety objectives but also safeguard the long-term viability of the facility by fostering a culture of safety, compliance, and continuous improvement. Through risk assessment, planning, and the establishment of clear procedures, a management plan ensures that safety is prioritised in all phases of the infrastructure.

The safety regulation on the 'Responsibilities and organisational structure in matters of safety at CERN' (SR-SO) [423] is based on the the CERN Safety Policy [422].

Current practice at CERN foresees that the appointed project leader (PL) is responsible for safety within the project [423]. The PL may decide to appoint a project safety officer (PSO) with the mandate to support the project leader in meeting the obligations in matters of safety [424]. At the end of the project phase, safety responsibilities are transferred from the project leader to the organic units in charge of the operation.

## 9.3 Planning for safety

### 9.3.1 Hazard Register

A *hazard register* is the result of systematic identification of activities, equipment, and substances with their associated hazards for workers, the public, or the environment. A systematic inventory of hazards serves as a support not to overlook any danger and unifies the terminology among different assessors. Such lists are available from different occupational safety organisms (see for example Ref. [425]). As they target manufacturing and services, they must be modified for particle accelerators and for research infrastructures in general to include hazards unique to these environments. Table 9.1 shows the headlines of a hazard register for particle accelerators. Ref. [426] provides a hazard list, tailored to particle accelerators. The adaptation of the hazard list to the local workplace by adding or suppressing hazardous equipment, activities, or substances precedes establishing the register.

The focus of the hazard register depends on the lifecycle phase. In the planning stage, it identifies hazards which can be controlled by design, for example, by adopting standards. In this early stage, organisational and psychosocial hazards and risks have not yet been assessed. In the installation and operational phase, workplace hazards become more important, for example, related to the organisation of work or physiological constraints.

Table 9.1: Hazard domains for a hazard register, with examples.

| **Hazard Domain** | **Examples** |
|---|---|
| External | Earthquake, climate, malicious action, cyber criminality |
| Physical | Temperature, noise, electromagnetic fields |
| Ionising Radiation | Particle beam, stray radiation, activation |
| Non-ionising radiation | Static magnetic fields, UV light, microwaves, lasers, RF |
| Noxious substances | Chemically or biologically harmful substances |
| Fire | Ignition sources, flammable materials |
| Mechanical | Cutting, crushing, collision, fall of object |
| Electrical | Electrical shock, electrical arc |
| Working conditions | Temperature, lighting |
| Physiological | Working posture, vibration, manual handling |
| Unexpected events | Loss of control, loss of power |
| Organisation | Constraining schedule, lack of information |
| Psycho-social | Incomplete and monotonous activities |

The equipment, activities, and substances are identified by location (surface site and building or underground areas) and lifecycle phase. The hazards emerging from each equipment, activity, and substance are identified with a specific list of hazards based on Ref. [425], modified and complemented for accelerator facilities, and published in Ref. [426]. The hazard description links to relevant sources of standard practice (see Section 9.3.2) at the time of the Feasibility Study. These sources must be updated regularly during the project life cycle.

In a large particle accelerator, most equipment, activity, and substance and their associated hazards occur repetitively at various locations and in different phases. To avoid tedious and error-prone updates of a written document, the hazard register is kept in a relational database where every piece of information needs to be updated only at a single location and is automatically propagated. Different types of database reports produce documents and reports summarising the hazards for a specific location and phase or the recommended sources of appropriate standard practice (Section 9.3.2). Hazards for which no standard practice exists (or where standard practice from *conventional* industry is not applicable) require a detailed risk assessment and identification of mitigation measures (see Section 9.3.3).

### 9.3.2 Standard practices

Standard practices are based on hazard identification, risk descriptions, and mitigating measures applying to technologies and related activities in industry that are similar to those encountered in an accelerator. Examples are electrical distribution of standard current and voltage and transport.

Suitable sources of standard practice are national and international legislation on hazardous equipment, activities, and substances. These documents are prescriptive, and their application is usually mandatory. A hazard is often considered under control when such mandatory regulations are implemented and applied.

CERN's safety policy [422] is to apply European Directives and associated standards where available. Numerous types of consumer and industrial products are subject to European Directives, which include chapters on Essential Health- and Safety Requirements (EHSR). Their purpose is to guarantee equal safety standards for workers and for consumers (who buy and use products) throughout the member states of the European Union. Only under this condition can goods be freely traded within the European Economic Area (EEA). Suppliers from third countries must also apply the European safety standards to introduce their products into the common market.

Another source of standard practice is the publications by occupational health and safety organisms, e.g., *SUVA*[1], *INRS*[2], *HSE UK*[3], *CTPI-AEAI CH*[4], and of industrial associations e.g., *BG ETEM*[5]. Their recommendations provide the practical elements required to meet the legal prescriptions at the workplace. They are often borne from common sense and may be easy to apply, effectively eliminating the hazard and making further risk assessment superfluous.

Product documentation by manufacturers is another source of standard practice. The manufacturer, having designed the equipment so that it meets legal requirements, gives the necessary information for safe use in the form of user manuals, video tutorials, and other training materials. Over time, CERN has built up a wealth of knowledge about safety, based on the experience from operating the existing infrastructure. The source material defining standard practices is assembled and classified in the hazard register.

### 9.3.3 Performance-based design

For hazards for which standard practices are not available or are not fully applicable to the particularities of an underground accelerator complex, the study follows a performance-based design (PBD) approach.

PBD is a state-of-the-art risk assessment approach, which is used in areas such as fire safety and introduced in national legislation in several countries, including Switzerland [427]. It is a process of defining alternative design options in an iterative way and assessing the impact against a predefined set of safety objectives. For this study, a PBD methodology introduced by the Society of Fire Protection Engineers (SFPE) [428] was adopted. According to SFPE, the PBD process can be divided into 6 steps (see Fig 9.1) after having defined the scope of the assessment:

1. Identify the safety goals (see Section 9.2.2).
2. Define the design objectives and develop the performance criteria (i.e., safety requirements).
3. Develop the accident scenarios based on the system-level risk assessment.
4. Develop the trial design with the proposed safety systems.
5. Evaluate the trial design against the performance criteria.
6. Create new trial designs until they meet the criteria - (iterative process).

[1] https://www.suva.ch/
[2] https://www.inrs.fr/
[3] https://www.hse.gov.uk/
[4] www.bsvonline.ch/fr/publications/det
[5] https://www.bgetem.de/

The process ends when a trial design that meets the safety objectives is retained as the baseline solution.

Different methodologies can be used for the evaluation of trial designs, ranging from simple assessments to highly complex algorithms [429] and numerical simulations. A qualitative analysis was performed during the conceptual study phase (2014 to 2018). For the feasibility study, a more in-depth evaluation was performed using quantitative (i.e., worst credible scenarios) and probabilistic approaches.

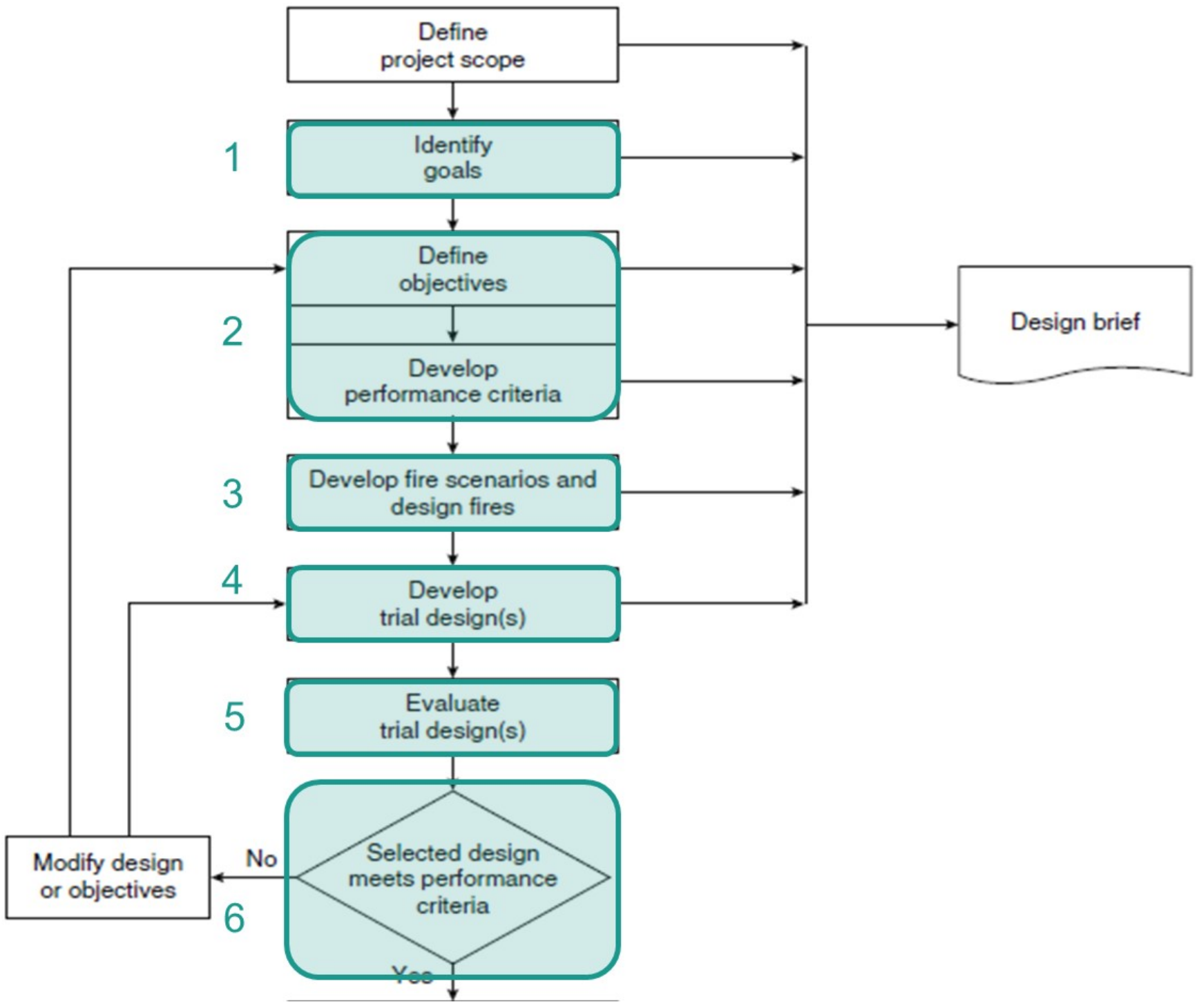

Fig. 9.1: Performance-based design flowchart. Example for fire scenarios from SFPE [428]

### 9.3.4 Safety systems

For the safety concept described in this document, a *safety system* refers to a set of engineering controls aimed primarily at ensuring life safety by incorporating both preventive and protective measures during normal and accident scenarios. A system that is used primarily for technical purposes but also performs safety functions can be considered a safety system, provided that it adheres to the required safety standards.

The following Sections (9.4 and 9.5) provide a brief summary of the safety systems proposed as a result of the study. Additional details are provided as references to reports, notes, etc.

## 9.4 Safety concept for the operation phase

As mentioned in Section 9.2.1, the strategy is to cover conventional hazards with standard practices and integrate accelerator-specific hazards in a performance-based design approach. The outcome is an integrated approach towards safety for the project.

### 9.4.1 Safety systems supporting the concept

The combination of the safety systems described hereafter constitutes the present safety concept. The suppression and/or major modification of any given safety system will have an impact on the global safety concept, which might require partial or total redesign of the concept.

The Safety concept is built on the following main pillars:

- Static confinement.
- Air management & dynamic confinement.
- Personnel transport.
- Emergency response.
- Secure power network.
- Access control.

In addition to these pillar systems, several high-integrity safety systems complement the safety concept. They permanently monitor the safety conditions and trigger appropriate safety actions from other systems or personnel.

- Access Safety System.
- Hazard detection.
- Emergency call.
- Evacuation alarm.

The particle accelerators have the following modes:

- Run mode: the beam is circulating or could be circulating; personnel can only be present in very limited areas of the facility.
- Access mode: beam cannot circulate. Personnel are authorised to access the facility. There are different types of access periods:
  - Short access: for interventions over a period of typically one hour or more and less than a day.
  - Technical stop: for planned interventions over a period of several days up to a few months.
  - Long shutdown: for planned interventions lasting several years for major consolidation and upgrade works.

Compared to other phases, the operation phase constitutes the longest part of the project's lifetime and during which the majority of occupational hazards (see Section 9.3.1) are present simultaneously. Depending on the state of the accelerators, the occupants may be exposed to different hazards. The average occupancy density in underground facilities is low, although occupancy during technical stops and long shutdowns can be comparable to that of the installation phase.

#### *Static confinement (Compartmentalisation)*

The fire safety concept relies on passive fire compartments as the principal strategy to sectorise the fire load and prevent fire and smoke propagation across the facility whilst enabling safe evacuation of occupants and intervention of the emergency teams.

The concept has partitions along the main tunnel and in all areas with a specific fire risk (e.g., electrical alcoves), using a dedicated fire-resistant partition. Moreover, protected areas for evacuation (e.g., safe areas[6]) will also be compartmentalised. Doors, dampers, and hatches installed in the compartment must also be fire-resistant and match the same resistance level as the fixed partition. For services penetration of those partitions (e.g., pipes, cable trays), the openings must be effectively sealed, following safety standards, to match the fire resistance required.

[6] Areas free from smoke and hazardous gases as a buffer for occupants to wait for the lift, *not to be confused with a temporary refuge shelter or equivalent*.

As a fundamental part of the fire safety concept, the fire rating of the compartments must guarantee that the fire and smoke are contained for the time needed to enable safe evacuation and intervention. Moreover, to limit property losses and recovery time, the integrity of the compartments must withstand all foreseeable fire scenarios. For this, the rating will be based on various factors: the fire load, the possibility of reaching flashover conditions in the affected compartments, and the time needed to ensure the safe evacuation and intervention of the emergency teams. The fire rating must also be consistent with the minimum structural fire resistance requirements assigned by standard practices and Host State regulations (see Table 9.6).

The following fire-resistance rating is proposed as a baseline for the feasibility study of the underground infrastructure. The resistance stated below is based on ISO 834 curves [430] for design purposes as defined in EN-13501-2 [431]. Section 9.4.3 provides more information on the difference between standard and natural curve-based fire resistance.

- Accelerator tunnel (arc and straight sections): EI90 partition (every 400 m) with interlocked door normally open.
- RF sector - connection between klystron gallery and accelerator tunnel: EI90 partition at the bottom and top of connection staircases.
- Service caverns - safe area (lift shaft): EI120 partition.
- Connection tunnels - connections between experiment and service caverns: EI120 .
- Alcoves - connection to the accelerator tunnel: EI120 partition.
- Other underground spaces with specific risks (UPS, transformers): EI120.

In the accelerator tunnel, fire compartments also serve as an important element of the evacuation strategy by limiting the distances travelled within the affected compartment, and they allow occupants to quickly move into an adjacent (non-affected) compartment which serves as a safe area (free from smoke, gas and fire consequences) to continue the evacuation until reaching the safe area joined to the pressurised lift shaft and, ultimately, the assembly point at the surface.

Different requirements in terms of evacuation distances exist for tunnels. Table 9.2 provides references for railway and road infrastructures in Europe, as well as for other particle accelerator infrastructures.

Table 9.2: Overview of distances between emergency exits in transport, technical, and accelerator tunnels. EU Regulation 1303/2014/EC and Directive 2004/54/EC apply to EU network of rail and road tunnels, respectively. XFEL compartments are made of solid fire-proof walls with a triple water mist enclosing system instead of a fire door.

| **Source** | **Length** [m] | **Comment** |
|---|---|---|
| EU road tunnels [432] | 500 | Between emergency exits (if required) |
| EU railway tunnels [433] | 500 | Between connections to adjacent tube |
| OLT4 [434] | 500 | Between access shafts in non-frequented technical galleries |
| NFPA 520 [435] | 610 | To reach a safe exit, refuge, or portal |
| XFEL [436] | 600 | Between two compartments (wall + water curtain) |
| ILC [437] | 500 | Passage connecting the two galleries |

These requirements are understood to be the distance to reach an area of relative safety where the occupants are no longer exposed to untenable conditions. Moreover, once the occupants reach the adjacent (unaffected) compartments, the remaining length needed to evacuate using motorised means is not considered in the maximum distance requirement. This is the case of the Gotthard and Lyon-Turin base tunnels [438] where evacuation is guaranteed through the adjacent (non-affected) double-tube concept.

The distance between two compartments must consider the maximum acceptable length (Table 9.2), as well as the layout constraints and practical aspects of the facility. For the FCC, this can be summarised as:

$$\begin{cases} d \leq 500\,\mathrm{m} \\ d = \text{distance between alvoves} \div \text{integer number of compartments between two alcoves} \end{cases}$$

Solving the system with $1600\,\mathrm{m}$ between two alcoves results in 4 compartments, hence one every $400\,\mathrm{m}$. With this layout, each alcove will house safety-related systems for two compartments to the left and two to the right. The final implementation of each individual compartment will depend on the exact location of the beamline elements of both the collider and the booster. For example, one should avoid the installation of a compartment in the middle of a quadrupole girder, or close to a radiation absorber. A compartment assembly is planned to be included in the arc half-cell mockup to test and identify the best location. Figure 9.2 shows a possible integration layout of such a partition. The challenges emerging from integrating such a compartment around different services, including the hatches for overhead robots, need to be considered. A subsequent design phase has to develop detailed plans.

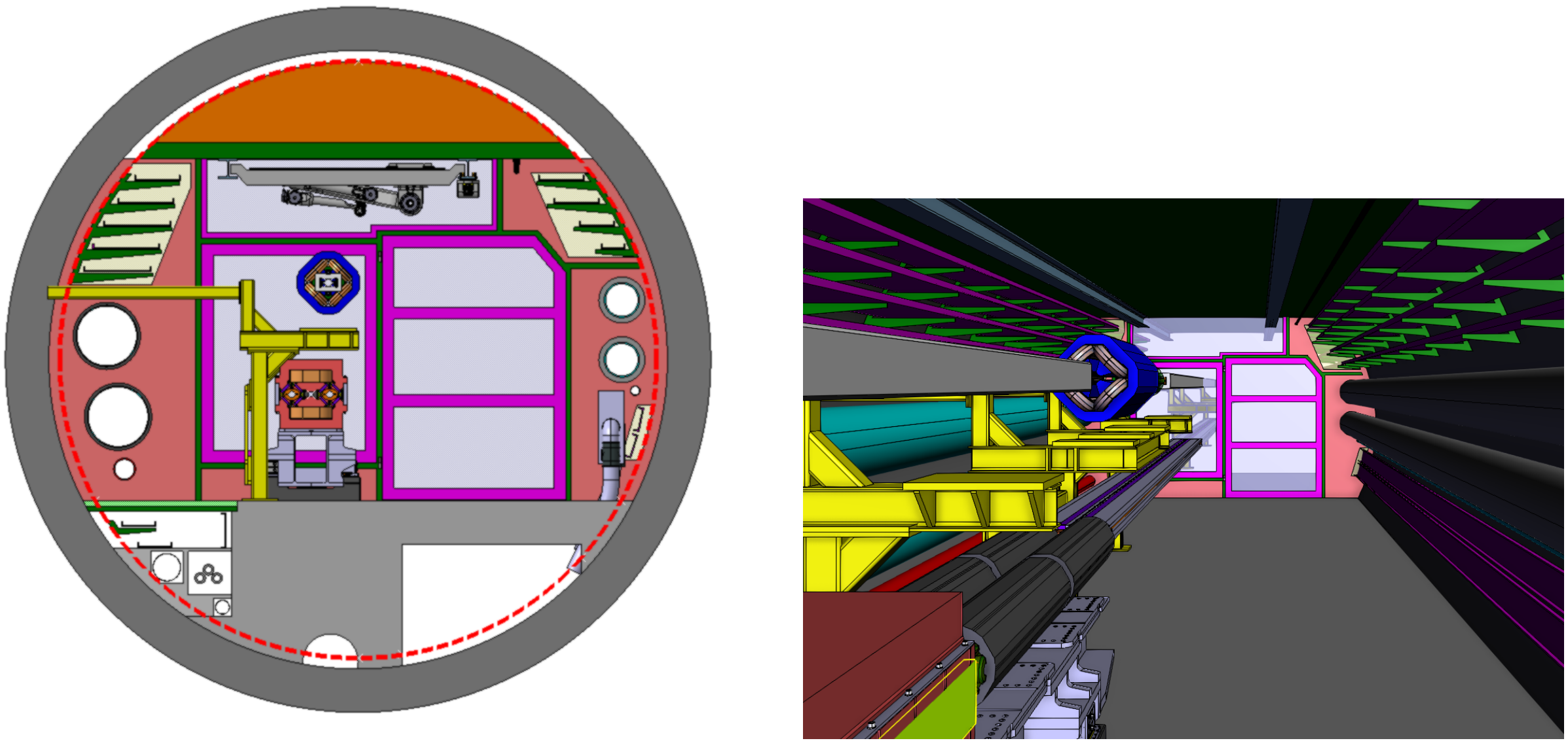

(a) 2D view

(b) 3D view

Fig. 9.2: Illustration of the integration of a fire compartment in the cross-section of the FCC tunnel. Light red: fixed partition walls; Purple: frames with grey fill: movable parts - fire door, robot hatch, installation hatch.

#### *Air management & dynamic confinement (ventilation)*

Due to the nature of underground facilities, adequate indoor air quality is required for both the occupants and the equipment. A robust ventilation system ensures the appropriate temperature and humidity conditions in the tunnel during operation and when people are present. During access mode, it provides a supply of fresh air for occupants with an appropriate air exchange rate.

Five different working modes for the tunnel ventilation are envisaged:

– *Run mode*: the accelerators are in operation and most electrical systems are powered; a heat load is transferred to the air in this mode; radiation protection aspects are taken into account for the air flow. No personnel are present in the main tunnel, the alcoves, and the experiment caverns.
– *Access mode*: the accelerators are not operational. Personnel can access and work in the tunnels. Occupational health & safety requirements must be considered.
– *Flushing mode*: Completely renew the air underground when moving from the Run to Access mode. Personnel are not present in underground structures.
– *Economy mode*: Reduced airflow for energy-saving reasons in situations other than Access mode, reducing the air exchange to a minimum. The presence of personnel is not foreseen in any underground structures.
– *Emergency mode*: activated in the event of a fire or gas release to guarantee dynamic confinement between the affected and adjacent compartments.

*Tunnel arcs*

For tunnel arcs, two technical solutions were studied [439]:

1. *Semi-transverse* ventilation scheme where the air is supplied to each sector from both endpoints via a specific duct throughout the sector below the transport zone, supplied to each compartment transversely by air diffusers, and extracted either through the tunnel itself or by an emergency extraction duct on the ceiling of the tunnel;
2. *Longitudinal ventilation* scheme where the air is supplied at one end of the sector and extracted at the other end.

In addition to air management during nominal situations, an emergency extraction system is proposed for the main underground areas. This system will be activated in case of fire or cryogenic gas leak to:

– Allow safe evacuation from the affected compartment while providing enough safety margin (Section 9.4.3).
– Create a dynamic confinement to prevent smoke from spreading to adjacent compartments.
– Remove heat and smoke in the affected compartment, reducing the risk of structural damage and reinforcing property protection.
– Allow for a safe and efficient intervention from the emergency teams, allowing emergency responders the ability to manipulate and control the system to their advantage.

Since there are still two concepts for the ventilation system being considered, there are also two different concepts for the emergency extraction system under consideration, one for each ventilation scenario (semi-transverse & longitudinal).

In the semi-transverse ventilation concept, the sector between the two shafts is considered to be split in the middle. Air is supplied to the tunnel through four wall-mounted air diffusers in each compartment. During access mode $54\,000\,\mathrm{m^3h^{-1}}$ ($2 \times 27\,000\,\mathrm{m^3h^{-1}}$) of fresh air is supplied to both halves of the sector. In the event of a fire, the smoke and hot air are extracted from both ends of the sector and exhausted to the outside. In such emergency mode, the extraction dampers located under the ceiling of the tunnel are open in all three compartments (affected + two adjacent). The air in the affected compartment will be extracted at a flow rate of $10\,000\,\mathrm{m^3h^{-1}}$ and $3500\,\mathrm{m^3h^{-1}}$ in the adjacent compartments, which yields a total extraction flow rate of $17\,000\,\mathrm{m^3h^{-1}}$. In the event of an emergency (e.g., fire), the fire doors of the affected compartment and the two adjacent ones are closed. Hence, to compensate for the extracted volume, an equal amount of fresh air (*make-up air*) is ensured via the nominal supply diffusers connected to the slab duct (Fig. 9.3). This concept creates a pressure difference between the different compartments, ensuring that smoke and gas are confined, providing enough time for the occupants to evacuate from them (see Section 9.4.3).

In the case of longitudinal ventilation, under nominal conditions $54\,000\,\mathrm{m^3h^{-1}}$ of fresh air is supplied to the tunnel from one end of the sector and extracted at the other. In case of emergency, the same extraction flow rates as per the semi-transverse are used in the affected and adjacent compartments. The fire doors of the affected compartment and the two adjacent ones are also closed, but in this case, the make-up air is supplied from both ends of the sector towards the adjacent compartments ($2 \times 8500\,\mathrm{m^3h^{-1}}$), ensured via bidirectional dampers installed on the walls of the compartment. These dampers allow make-up air to enter the affected compartment, sweeping the smoke towards open extraction dampers located under the vault, providing the necessary compensation. The isolation of the compartments creates a pressure difference, ensuring the confinement of smoke/gas and providing enough time for the occupants to evacuate.

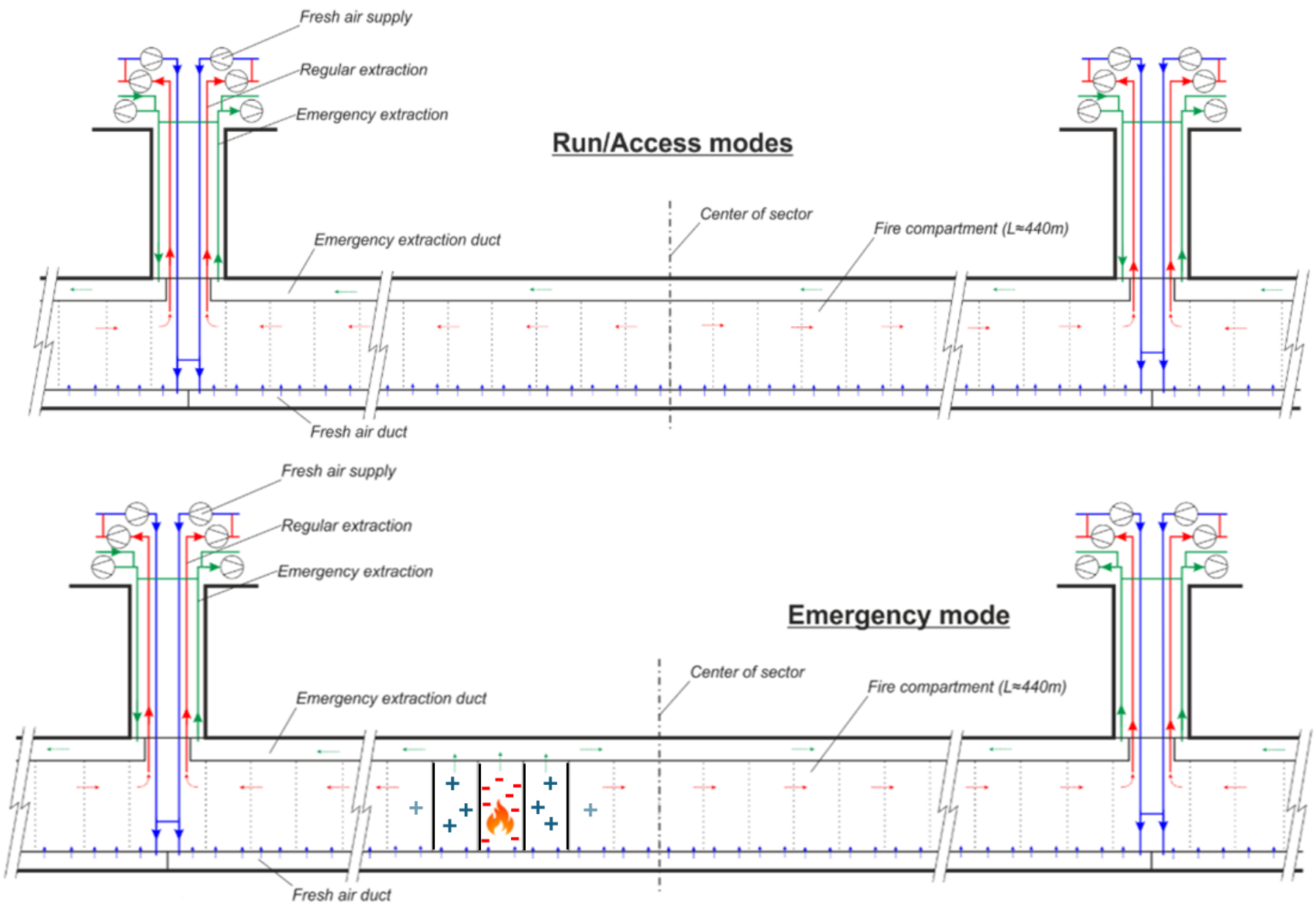

Fig. 9.3: Semi-transversal ventilation mode (top) and emergency mode (bottom). The + and - signs illustrate the pressure cascade in the affected compartment and the adjacent ones.

Regardless of the ventilation concept adopted, it is important to consider the full operability of the fire compartment doors at all times. The risk of overpressure due to nominal, emergency or degraded modes will be carefully studied to ensure that the evacuation concept is not compromised if these doors cannot be opened normally by applying a force less than 70 N. Compensatory measures, such as installing pressure-breaking systems, sliding fire doors, etc., must be considered if such a risk is non-negligible.

A comparison of smoke extraction system performance with respect to life safety of occupants is discussed in Section 9.4.3.

*Alcoves*

The electrical alcoves are ventilated via dedicated ducts using the same system as the main tunnel. Emergency extraction is also available in these areas to support evacuation and intervention. However, due to the reduced evacuation distance inside the alcove, this system is not deemed necessary for the life safety of occupants; hence, it is not expected to be triggered automatically. However, in case of a fire, the smoke dampers must isolate the normal ventilation to prevent any propagation outside the compartment.

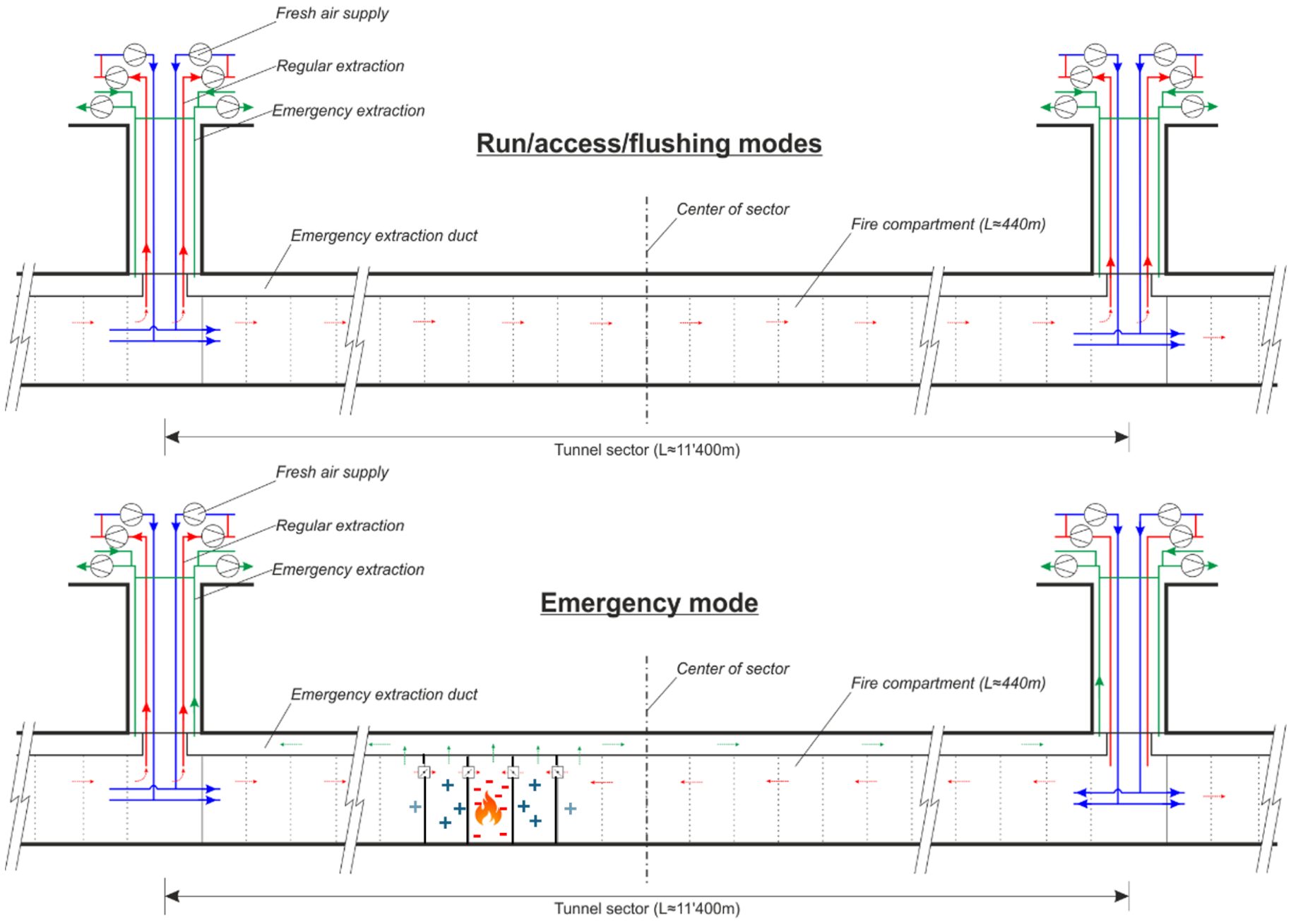

Fig. 9.4: Longitudinal ventilation mode (top) and emergency mode (bottom). The + and - signs illustrate the pressure cascade in the affected compartment and the adjacent ones.

*Experiment caverns*

The experiment caverns are ventilated by a dedicated system that also ensures dynamic confinement between the experiments and the accelerator tunnel, as well as providing the necessary airflow to ensure hygienic conditions and proper indoor air quality.

Several Computational Fluid Dynamics (CFD) analyses of CMS and ATLAS [440, 441], experiment detectors at the present LHC, have proven that the evacuation in experiment caverns can be safely performed without any active smoke extraction system due to their large volumes and relatively short evacuation distances. The FCC experiment caverns will be even larger, hence an active emergency extraction system for smoke is not required for life safety. However, the ventilation system should be able to support the extraction of cold smoke. In the case of particle (sub)detectors that require non-negligible amounts of cryogenic fluids, a risk assessment will be performed to determine if a dedicated gas extraction system is needed to cope with the Oxygen Deficiency Hazard (ODH).

*Service caverns*

Service caverns are ventilated with a dedicated system that ensures dynamic confinement between the experiment cavern and the accelerator tunnel, as well as providing the necessary airflow to ensure hygienic conditions. Like the experiment caverns, a smoke extraction solution is required to support intervention and recovery time after a fire, but it is not deemed necessary for the occupants' life safety.

*Safe areas & protected staircases*

As a basis of the safety concept, certain areas must be kept free of smoke. Consequently, safe areas, protected lift shafts, klystron gallery staircases, and personnel tunnels connecting service and experiment caverns will benefit from independent dynamic confinement. This over-pressurising system will, in general[7], be fed fresh air and designed to guarantee its performance throughout the duration of the worst credible intervention case. Among other requirements, this implies that such a system must be supplied

[7] Alternative solutions might be studied case-by-case, such as using air from adjacent areas.

by a safety power network with fire-rated cables and redundant fans.

*Surface buildings*

The surface buildings are ventilated by dedicated air handling units designed according to state-of-the-art best practices. Smoke extraction systems will be required for rooms with a floor area exceeding $300\,\mathrm{m}^2$ (in France according to the 'Code de Travail' [442]) or $600\,\mathrm{m}^2$ (in Switzerland according to AEAI DPI 21-15 [443]), in application of prescriptive design standards.

***Personnel transport***

Due to the distances between the access shafts ($11.4\,\mathrm{km}$), occupants need to rely on a dedicated transport system to commute and evacuate the tunnel in an emergency. At any given time, the capacity of the personnel transport system available underground must ensure the evacuation of all occupants. The same transport system can be used during normal conditions for safe and efficient transport from the service caverns to workplaces along the arcs, with the possibility of parking the vehicles at the entrance of the alcoves and in the service caverns. In some situations, autonomous vehicles can go directly to the location of the workplace, drop-off the occupants (and material) before driving to the nearest alcove where it will stay parked until the end of the shift/activity. In contrast to the nominal scenario, during an emergency (i.e., evacuation alarm), the occupants are requested to walk towards the nearest safe alcove and take a seat in one of the parked vehicles.

The parking capacity for these vehicles will be ensured by the lay-by zones at the entrance of the alcoves. The baseline layout has parking spaces for up to 7 vehicles per alcove.

The simultaneous circulation of pedestrians and vehicles in the transport zone requires robust detection systems (e.g., stop in case of an obstacle or reduce speed near people or other vehicles) and a centralised communication system to ensure full autonomous driving.

The study included the feasibility assessment of an autonomous vehicle (Fig. 9.5) with a bi-directional drive. It is narrow enough to allow meeting traffic and bypassing of other vehicles within the $2.20\,\mathrm{m}$ wide transport zone of the cross-section [444].

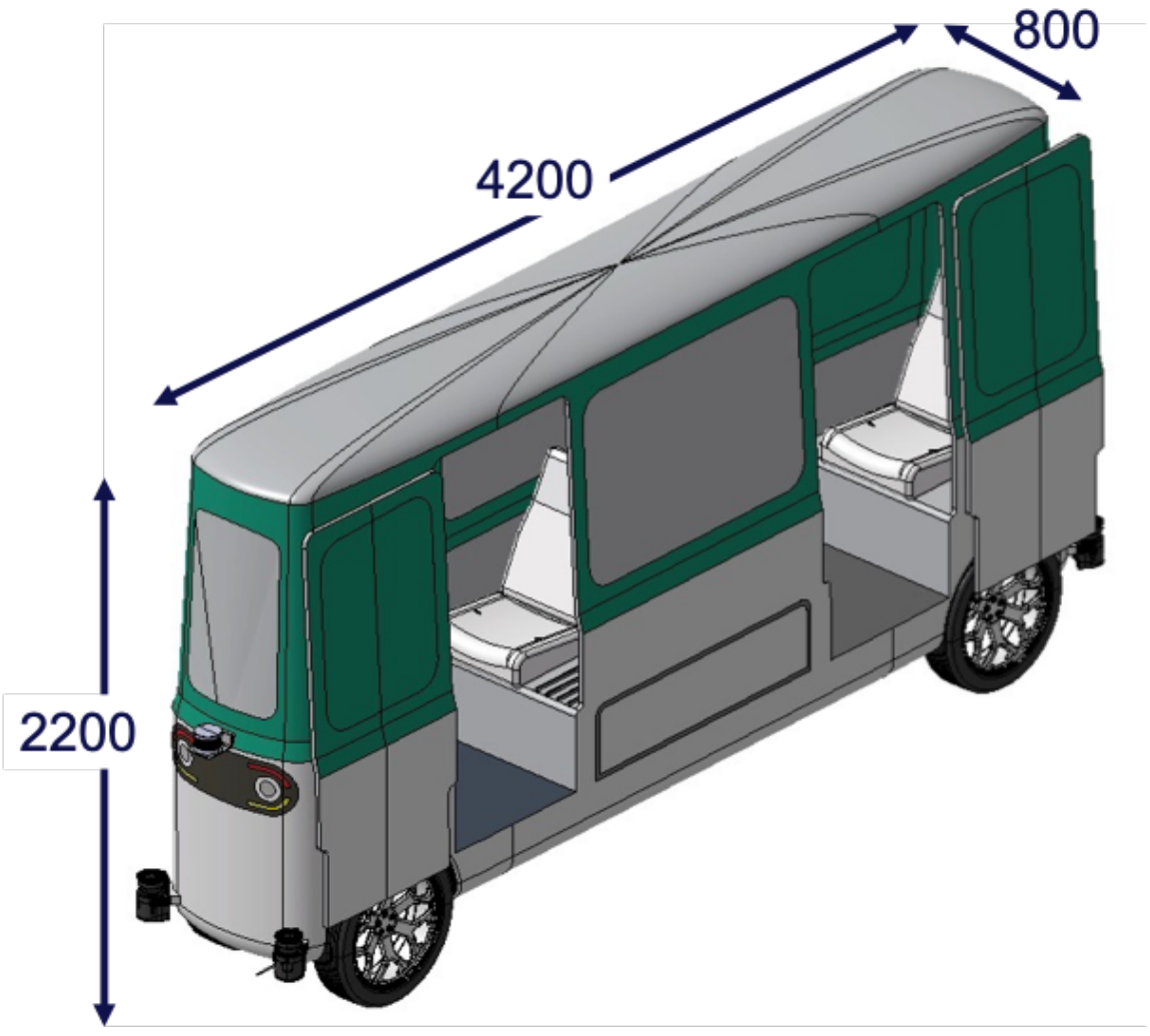

(a) Transport vehicle with 4 seats.

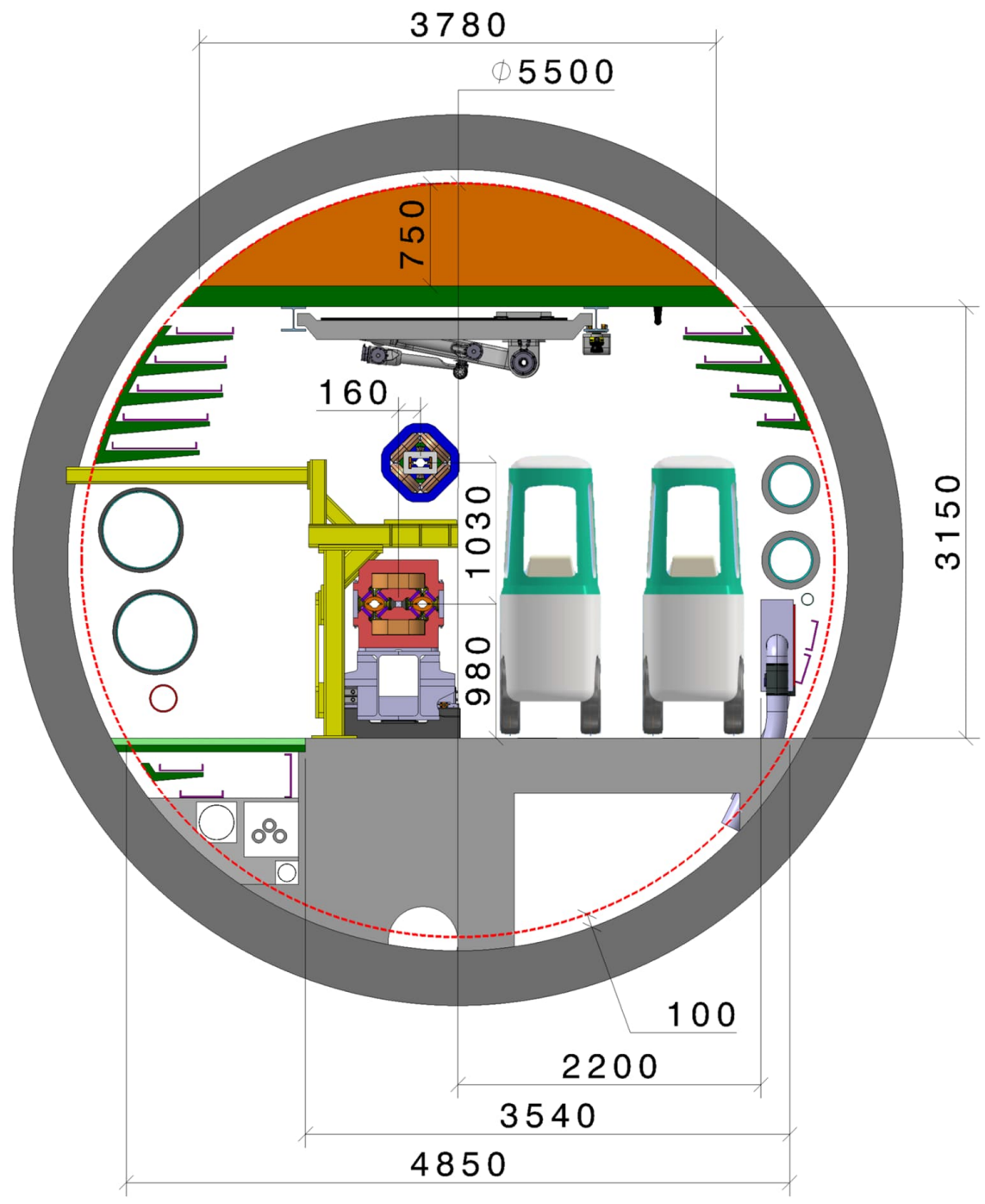

(b) Cross-section layout with the two vehicles passing each other within the 2.20 m transport zone.

Fig. 9.5: Personnel transport system [445].

The space on either side of a vehicle in the transport zone is ≈ 70 cm left and right. However, when two vehicles overtake or pass, the margin reduces to ≈ 20 cm [445].

Personnel transport to and from the surface is ensured by at least two independent lifts situated in an over-pressured elevator and stair compartment of the shafts. In the experiment points, two sets of two lifts are planned to provide the necessary reliability and ensure safe evacuation (see Sections 9.4.1 and 9.4.4). As requested by the European Lift Directive and the associated harmonised standard [446], access hatches connecting a staircase to the lift shaft must be installed every 11 m for rescue purposes. The lifts must also conform with ISO 8100-1 [447].

***Means of emergency response***

A simple scale-up of the current emergency response strategy in place for the LHC is not feasible due to the size of the infrastructure. A dedicated emergency preparedness and intervention concept will need to be developed in a subsequent design phase as well as a further extension of the existing Mutual Assistance Agreements with the Host States [448]. The feasibility study prepared the basis and carried out a first analysis of the concept. It consists of the following approach (see also Section 9.4.5):

1. A central emergency response coordination team.
2. First response from trained personnel.
3. Robotic intervention with firefighting capabilities.
4. On-site support from local emergency services, particularly for surface incidents.
5. Intervention from CERN Fire & Rescue Service (CFRS) and if relevant, other appropriately trained emergency services.

A dedicated approach is proposed for the installation and construction phases (see Sections 9.5.2 and 9.5.3).

### *Secure power network*

Safety systems must maintain their functions at all times. A loss of the general electrical network, e.g., due to a power cut, would lead to a loss of electrically powered safety systems, hence the need for a secured power network to ensure the availability of safety systems at all times.

The secured power network is part of the global electrical network powered with backup sources in case the main power supply is lost, to allow the critical and safety-related loads to be powered at all times. To avoid duplicating medium voltage (MV) lines in the tunnel, it is proposed to optimise the infrastructure using the same MV link in the tunnel for powering general services and the secured network, applying a logic of load shedding in case of a power outage. The aim is to minimise the impact in the tightly integrated area and to avoid distributing the power at Low Voltage (LV), dedicated to safety systems, from the surface to the middle of the tunnel, due to the large cable distances that can induce high power losses and critical voltage drops [449].

Two backup sources are planned for the secured power network:

– A connection to an alternative public electrical utility.
– Emergency power supplies on each surface site.

In case of a failure, maintenance, or any other unavailability of the general service network, the electrical network is supplied using the backup power sources mentioned above. In this case, the couplings between the secured network and the general services network are automatically disconnected when the voltage from the main sources is no longer available. This logic of load shedding is used on the surface and underground, all the way down to the electrical alcoves, providing power only to the safety systems, leaving the other loads de-energised.

All active and passive components involved in the load shedding must be properly designed, maintained, inspected, and tested regularly to ensure a level of reliability which is consistent with that of the safety systems it serves. A description of the network and the single-line diagram is available in Ref. [449].

### *Access control*

Control of access to the installations at CERN today is based on several layers of protection, each one only allowing a more restricted number of people to enter. On-site access is permitted for identified affiliated personnel; entering a surface building housing an accelerator access point requires special authorisations, and finally, entering a beam facility is subject to the most rigorous control.

Before granting access to a beam facility, and once a user is identified, several database verifications are performed, which include:

– Authorisation is granted by the facility responsible to a person recognised by the organisation and with a professional need to access this facility.
– All mandatory and specific Safety training for the facility is valid.
– The activity has been declared and approved; the organisational unit in charge of works' coordination keeps a detailed list of planned and approved activities together with the assigned personnel, as well as a list of on-call personnel likely to intervene at short notice.

Access control, therefore, constitutes the first element in the chain of personnel safety systems, installed to protect the personnel and guarantee that only identified, trained, and authorised personnel enter the underground premises.

Access control is also used to ensure that the number of persons accessing the facility does not exceed the maximum number that has been set for safety or operational reasons. For example, the pressurised safe zones at the bottom of the shafts can only host a limited number of people in case of emergency. Therefore, once this number is reached, further access is blocked until some personnel leave the facility. Similarly, the limited capacity of the underground vehicle fleet limits the number of people allowed to enter, and this limit is also ensured by the access control system.

Access control is formed by two sets of access points; one at the surface and one underground.

**Surface access points**

The access points at the surface, installed just before reaching the pressurised lifts, give access to the underground areas that are accessible in both beam and access modes, i.e., the service caverns and the klystron galleries.

These access points are composed of one or several personnel access devices (PAD) and a material access device (MAD). The devices are linked with commercial access control products, such as access rights and biometry databases, using computer networks and state-of-the-art human-computer interfaces.

A personnel access device (PAD) performs all the checks in one place. It is an inviolable barrier that guarantees that only one identified person can enter at a time. The user's identity is confirmed by a biometric check, relevant access authorisations are verified, and equipment checks (for example, possession of passive or active dosimeters) are performed. All these actions take place while the user is inside the PAD.

The PADs are built using commercial-off-the-shelf components, integrated to form one system, following the same principles that have been applied throughout the CERN accelerator complex for almost 20 years. Future PADs, although conceptually similar to the PADs of existing CERN accelerators, will leverage technological advances that can be seen, for example, in the domain of border control, with similar airlock-like booths installed in recent years at major international airports.

In addition to the PAD(s), each access point is equipped with at least one MAD. This device allows a safe way to bring bulky material in and out of the accelerator premises. The current concept uses a booth of the same size as the lift in the shaft. Its doors can only be operated from outside. After opening the doors on one side, the material is placed inside the device, the doors are closed and locked, and a scanning process ensures that no one is present within the device. Once scanning is completed successfully, the door on the other side of the MAD can be opened and the material removed. The MADs at existing CERN facilities are equipped with a combination of standard volumetric detectors and custom image processing algorithms. The MADs for a future project will certainly also benefit from technological advances, especially in the field of image processing.

**Underground access points**

The service caverns are sufficiently shielded from radiation emitted, while the collider is operated with beams. However, access to the accelerator tunnels and experiment caverns requires passing through additional access points, which are interlocked with the presence of beams and other specific hazards. Constructed with the same components as their surface counterparts, the underground access points leading to areas subject to the presence of beams will thus form part of the Access Safety system.

Given the size of the FCC and the distance between two adjacent sites, the granularity of the access control system needs further enhancements. Having a detailed count of people inside the facility, as well as a good vision of their location within the facility, is important for safety. Individual tracking devices will be worn by all personnel entering the underground premises. Antennas distributed at regular intervals (in each alcove and throughout the access tunnels and service caverns) will allow the emergency response team to locate people within the underground complex.

### *Access Safety System*

The Access Safety System is an interlock mechanism that acquires the status of and acts on elements important for safety (EIS) of two types: EIS-access and EIS-beam. It prevents the beam from being present at the same time as personnel. The EIS-access consists of the personnel and material access devices (access control components), emergency exit doors, movable shielding walls etc. The EIS-beam consists of accelerator components that can stop the circulation and the injection of beams. These components will include the beam dumps of the collider and the booster. Additional EIS-beam devices will allow redundancy for each interlock chain with technological diversity (e.g., a bending magnet or a moving stopper obstructing the beam aperture). A functional safety approach, e.g., following the IEC 61511 [450] standard, will be used throughout the lifecycle of the system.

The main functions of the Access Safety System include the following:

**Search mechanism.** After a prolonged operation with access, the machine volumes must be visually inspected for the absence of human presence. To facilitate this task, each site is divided with doors into smaller, more manageable units called access sectors. Once the patrol is completed, the search memory of each access sector is armed. The division of the FCC into access sectors takes into account the civil engineering constraints, accessibility of equipment, etc. Short accesses thereafter are permitted as long as members of the personnel are in possession of a safety token. Should an intrusion occur or access without a token take place, the search of the corresponding access sector is disarmed, and a new patrol is required. Unlike the preceding accelerators, the patrol will be largely automated, with the long portions of the tunnel swept by a robot equipped with a camera. Algorithms to detect the presence of humans, combined with remote supervision, allow unambiguous confirmation of the absence of personnel.

**Monitoring of *safe for beam* conditions.** Whenever equipment is powered and there is ongoing or imminent injection or circulation of beams, the system monitors the state of all EIS-access. The moment the EIS-access quit their safe state, or there is a detected failure of the EIS-access equipment, the EIS-beam of the corresponding interlock chain is systematically and automatically activated to stop the beam.

**Establishing *safe for access* conditions.** Prior to permitting access to the FCC, the accelerator equipment is stopped using its conventional control system. The role of the access safety system is then to inhibit the possibility of starting the beam by interlocking the EIS-beam. Access can begin only when the interlock actions have been confirmed.

**Monitoring of *safe for access* conditions.** When access to the FCC is allowed, the safety system continuously monitors the state of the EIS-beam. Degradation of the protection barriers results in actions ranging from blocking access to launching evacuation and requesting the stoppage of an upstream accelerator that could potentially inject a beam in a downstream accelerator.

**Establishing *safe for beam* conditions.** In order to permit the operation of the FCC with beam, it must be empty of personnel and its search memories armed. To further secure the transition from access to beam operation, an audible warning signal will sound to announce the imminent injection of beams. The personnel are trained to press emergency power cut buttons if there is an audible beam imminent warning signal. Only once the warning has correctly sounded and no abnormal presence in the facility is detected will the conditions for removing the inhibition from EIS-beam be established.

### *Hazard detection*

The access safety system prevents any human entry into certain zones in case of radiation and electrical hazards by blocking access while the accelerators are operated or are about to be operated with beam. Access is only allowed when the beam is not present and when the pre-conditions for access are met. In addition, the access safety system monitors its own elements and launches evacuation alarms should a hazardous situation occur.

However, even with a complete stop of beam operation, it is important to monitor the situation for other hazards. The principal ones identified for the FCC accelerator are fire, oxygen deficiency, and ionising radiation.

*Fire detection system*

The Fire Detection (FD) system must satisfy the requirements resulting from fire risk studies. These imply generalised early fire detection. Several fire detection technologies meet this requirement, e.g., smoke detection with opacimeters or with laser beams. However, not all are well suited to be deployed in the FCC tunnel (low height, material obstructing lines of sight, electronics sensitive to ionising radiation). The baseline solution consists of equipping each alcove with aspirating smoke detectors (ASD). A variant – long range ASD (L-ASD) - which is suitable for the FCC, has been put in service at CERN in the SPS ring. It is capable of detecting aspirated smoke with a 100 m long sampling segment that can be extended by up to 700 m of non-sampling tube.

With the proposal of having alcoves every $\sim$1600 m, it is possible to install 16 L-ASDs in each alcove and aspirate air samples from the 800 m left and right of the alcove. Additional ASDs will be installed for detection within the alcoves themselves, as well as within the service and experiment caverns. Control and monitoring equipment racks, interconnected with optical fibres, will be installed in each alcove. All components of the fire detection system must meet the requirements of the applicable European standard, EN 54 [451].

It should be noted that this solution requires about 400 km of aluminium aspirating tubes. Should further analysis prove that having thermal detection is sufficient in terms of performance, newer technologies, more suited for long tunnels, will be considered. The thermal detection method most adapted to the accelerator environment consists of using distributed temperature sensing (DTS). It is based on optical fibre line-type temperature variation detection. This solution is particularly well suited for long tunnel sections, where an integrator unit connected to an optical fibre pulled inside the 11.4 km tunnel segment can be installed in each service area. The spatial resolution of the detection is of the order of a few metres and the algorithms can trigger an alarm both in case of a temperature rise (fire) or a temperature drop (e.g., due to cryogenic leak) of a few degrees. However, the suitability of the technology in an environment with high levels of ionising radiation must first be proven.

*Oxygen deficiency hazard*

The two technical sites PH and PL, where the superconducting RF cavities are located, require particular attention due to the presence of cryogenic coolants. Commercial oxygen deficiency hazard (ODH) detectors that are installed today in the LHC complex can be used to monitor the oxygen level in the vicinity of cryogenic installations at the two RF points of the FCC. They are based on electrochemical cell sensors installed inside the tunnel, connected to detector electronics which can be housed up to 900 m away. Signals from several detectors are then fed to the control and monitoring equipment rack.

Alternative detection techniques can be used (aspirating detector, fibre sensors etc.). In addition to oxygen level monitoring in precise locations, larger parts of the FCC could be monitored for a sudden temperature drop using the fibre-based distributed temperature sensing.

*Radiation monitoring system*

A radiation monitoring system will be installed to measure prompt radiation levels at the interfaces to the accelerator tunnel, the experiment caverns, and the injector complex. These detectors are equipped with alarm functions that will signal any increased ambient dose equivalent rates. Radiation detectors will be installed at strategic locations in the tunnel and experiments to continuously measure residual ambient dose equivalent rates after beam stop.

Detectors will be installed to monitor the activation levels in the air of the tunnel. The radioactivity in the water circuits is expected to be negligible and will be checked during regular sampling campaigns. Air and water releases, radioactivity in the environment, and levels of stray radiation on surface sites will be surveyed by an environmental monitoring system with real-time monitoring where required.

### *Emergency Call Points*

*Manual call points*

Manual call points (break-the-glass devices) are installed throughout the facility. They will be on each side of an alcove entry door, as well as on each side of the fire partition doors in the tunnel. In addition, CERN best practices require placing manual call points at regular intervals in the tunnels, grouped with other safety equipment: emergency lights, evacuation loudspeakers, etc., fixed on easily identifiable modular panels.

*Emergency communication system*

In the past, the CERN underground facilities have been equipped with emergency telephone sets (known as red-telephones). Based on analogue technology, the devices contain no active electronic components and could, therefore, be even placed permanently in areas possibly exposed to high levels of ionising radiation. Mobile phones come with more functions, for instance, precise localisation and video transmission, but they rely on a communication infrastructure that comprises electronic equipment, which is not radiation-hard.

Although the need for reliable, safe communication is clear throughout the complex, no decision has been made as to which specific technology is the most appropriate. The installation of fixed red-telephones could be one of the solutions. Having a leaky feeder cable with a fire-resistant jacket and circuit integrity properties ($P_{ca}$90-PH90, [452]), and fed by a secure power network, would allow emergency communication with mobile telephony services throughout the subsurface infrastructure. Today, this seems to be the most viable option. Such an approach could benefit from the developments required to provide secure data communication means to control autonomous vehicles as well as emergency surveillance robots. Moreover, this might also respond to the need for safe communication for interventions. In this case, its performance and reliability must be compatible with the intervention time (see Section 9.4.1), and some adaptations would be necessary (double loop, additional fire rating, etc.).

### *Evacuation alarms*

Whenever a hazardous situation (fire, oxygen deficiency, possible problem with access safety system) is detected, the concerned zones of FCC are evacuated. To this end, the FCC tunnel will be equipped with loudspeakers placed every $\sim$25 m connected to amplifier racks of a voice alarm system located in the alcoves.

The choice of voice alarm system is dictated by the added possibilities for emergency response teams to play pre-recorded messages or directly address the personnel underground with ad hoc instructions. An alternative solution, based on using more sparsely installed sirens, exists. Although less versatile, allowing only an evacuation siren with no accompanying voice messages, this solution is more suited to very large (e.g., experiment caverns) or noisy (e.g., compressor room) locations.

The system control and monitoring equipment racks, installed at regular intervals, acquire evacuation requests from the detection systems. The racks are connected by optical fibre and configured with appropriate evacuation matrices. Thus, in case of a hazardous event, evacuation is launched not only in the same location but also at other sites, as required by the evacuation matrices. This is possible as the control racks communicate between themselves. The components of the evacuation system must be compliant with the appropriate European standard EN 54 [451].

Due to the circular shape of the underground structure, most locations can be evacuated in two directions. However, in emergency situations, evacuating personnel can often use only one of the two possible escape routes. Despite the fact that most personnel movements (including evacuation) will be made using autonomous vehicles, all alcoves and fire compartment doors will be equipped with adaptable direction signage. It must clearly indicate the direction in which to go to escape a detected danger.

#### *Emergency lighting and route guidance*

All of the underground facility is equipped with emergency lighting as defined in the corresponding CERN safety guideline [453]. This includes:

- A general ambient safety lighting, ensuring at least 5 lux ($lm/m^2$) at floor level,
- Marked evacuation routes (exit ways, doors, direction) with illuminated and luminescent signs. This wayfinding (guidance) system will also be dynamic and inform occupants if an evacuation way is no longer safe (for example, smoke present in adjacent compartment).

All these systems need to be connected to the secure power network (see Section 9.4.1) or have an autonomy of at least 120 minutes to guarantee the life safety of occupants. They also need to be designed to ensure an overall system fire-rated integrity of at least 90 minutes (E90).

Moreover, robotic systems can be deployed to guide the occupants in their evacuation, by providing an indication of the safe direction to evacuate (projecting arrows on the floor of the transport zone) [454] and providing audio assistance to the emergency response team. Evacuation signs will also be visible on the compartment doors.

### 9.4.2 Occupational hazards during nominal conditions

While the facility is in operation, the occupants are exposed to the hazards that are present under nominal conditions. The hazards and safety requirements during nominal operation are mentioned here, whereas the accident scenarios are discussed in Section 9.4.3.

#### *Ionising radiation*

For the mitigation of risks associated with ionising radiation, the standard prescriptive methods of the existing CERN radiation protection rules and procedures are used. However, optimisation of the design of civil engineering infrastructure is more assimilated in a performance-based approach, where the proposed design has to be evaluated to see whether it will comply with the radiation protection objectives.

*Radiological hazards*

High-energy particle beams interact with each other in the experiments or with beamline components in the accelerator, creating secondary particles that may pose radiological risks during operation (prompt radiation). This interaction also produces residual radioactivity that will present a potential hazard during maintenance (residual radiation). Workers may be exposed to ionising radiation from activated equipment, gases, or fluids by external exposure. The radioactive contamination hazards and the resulting internal exposure from incorporation are expected to be negligible.

The FCC-ee will operate in four phases, with beam energies rising from 45 GeV in Z mode to 182.5 GeV in $t\bar{t}$ mode, and beam currents dropping from 1.27 A to 4.9 mA. Thus, radiological risks vary with the phase and the most conservative scenarios have been used for their assessments.

At the stage of the feasibility study, detailed integration, such as the routing of ventilation ducts and water pipes, is not yet done. Infrastructure integration will be optimised in later design phases. The same is valid for the material moved from the accelerator to storage areas or workshops for maintenance and repair or for elimination. As in other industrial facilities, radioactive sources and X-ray generators may be used for various purposes. These activities will follow standard prescriptive methods, as already commonly applied at CERN or in industry.

All results of the radiation transport simulations presented here were obtained using the FLUKA code [331] with FLAIR [455].

*Radiation area classification*

Dose constraints for the design of the infrastructure as well as active radiation monitoring will ensure that radiation doses received by personnel working on site remain below the regulatory limits [456] under nominal and accident operating conditions.

The radiation area classification scheme [457] allows radiation risks to be managed and ensures that optimisation processes, training, access control, and collective and personal protection means for personnel intervening in the respective areas are adequate for the expected risks. A summary of the radiological area classification adopted at CERN is provided in Table 9.3 .

Table 9.3: Synopsis of the work area classification at CERN [457].

| **Classification** | **Ambient dose equivalent rate limit** | | **Annual effective dose limit** |
|---|---|---|---|
| | **permanent workplaces** | **low-occupancy areas** | |
| Non-designated | 0.5 μSv/h | 2.5 μSv/h | 1 mSv |
| Supervised | 3 μSv/h | 15 μSv/h | 6 mSv |
| Simple controlled | 10 μSv/h | 50 μSv/h | 20 mSv |
| Limited stay | not permitted | 2 mSv/h | 20 mSv |
| High radiation | not permitted | 100 mSv/h | 20 mSv |
| Prohibited | not permitted | - | 20 mSv |

*Protection objectives*

The risks resulting from ionising radiation are analysed very early in the design phase among others, with the objective of developing mitigation approaches that ensure that accessibility requirements are met for each area.

Areas with unacceptably high radiation levels from prompt radiation (high radiation and prohibited radiation areas) are inaccessible during operation, and essential risk control measures are employed as described above. The access control system prevents entry to such hazardous areas, whereas an interlock system halts beam operation if unauthorised access occurs. An emergency stop system ensures the rapid and controlled termination of beam operation if necessary.

Where feasible, areas accessible during beam operation will generally be designed to allow a classification as a non-designated area, avoiding any radiation protection restrictions. The planned civil engineering infrastructure complies with the radiation protection requirements for the operation of FCC-ee. However, the present infrastructure may not necessarily be ready in all parts for FCC-hh, especially where compliance can only be reached by significant costs at the current stage or where infrastructure is only required for FCC-hh. The required infrastructure modifications or additions will need to be implemented after the operation of FCC-ee. One example is the dedicated beam dump caverns that will be only needed for FCC-hh and would not be used by the FCC-ee.

Residual dose rates in the accelerator tunnel have been estimated for several areas, and these will allow a first evaluation of potential constraints on the operation of the facility and maintainability, as well as the related optimisation of the design of beamline components.

*Injector complex*

The injector complex injects electron and positron beams alternately into the booster. Radiological hazards during nominal operation come mainly from prompt radiation from the positron source, beam dumps, and loss locations in the linear accelerators and the damping ring. It is necessary to determine the minimum soil thickness needed to contain the prompt radiation, aiming to keep the klystron hall above the accelerators as a non-designated area.

An initial set of beam parameters and assumed loss terms has been used to estimate the shielding thickness required above the injector complex. Figure 9.6 shows the ambient dose equivalent rates at 90° through shielding from a beam loss on a thin long target, indicating the thickness required to achieve the dose rate objectives. Approximately 6 m of soil cover is deemed adequate for the specified loss term.

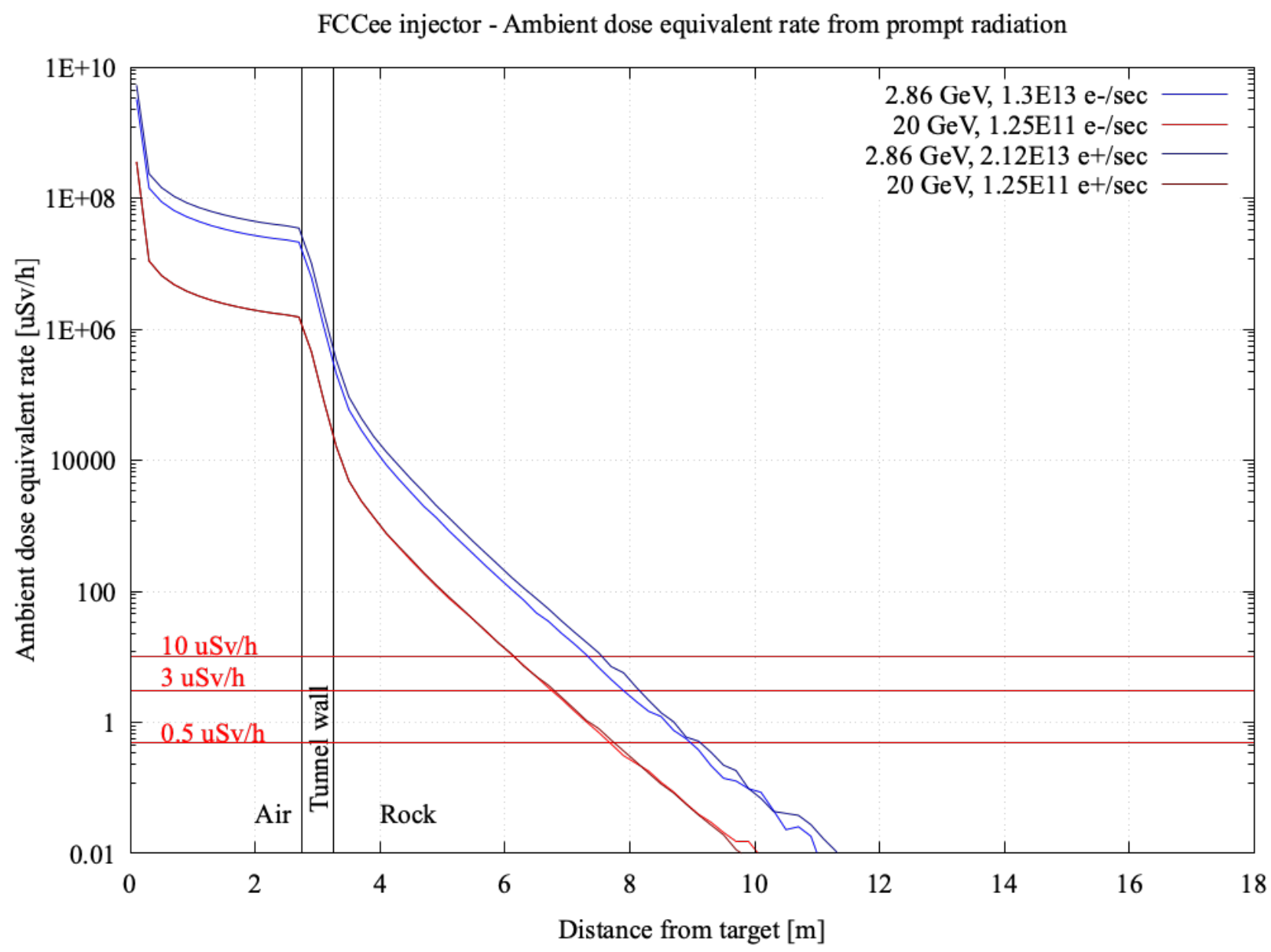

Fig. 9.6: Ambient dose equivalent rate through rock (from 3.25 m distance onwards), resulting from a continuous beam loss in the FCC-ee injector complex based on initial assumptions for potential beam losses.

A conventional fixed target, where 2.86 GeV electrons interact in a tungsten rod, is used to produce the positrons. The target station will have specific shielding. Activation of the target and downstream accelerator elements also generates sources of radiation which have an impact on workers during maintenance work. These residual radiation levels are reduced by local shielding and optimisation of the elements concerned during the technical design phase.

The highest levels of activation in the injector complex will occur at the positron production target, and initial studies show that these levels will impose strict restrictions on access and maintenance, with extended cool-down times. Remote handling and shielding are indicated, aligned with intervention and maintenance protocols.

*Underground infrastructure*

The collider will be housed in the tunnel with a limited number of access points and connection tunnels, minimising the interfaces between accessible and inaccessible areas that need to be studied and adapted in the design to meet radiation protection objectives. The lateral shielding of 10 m rock effectively contains stray radiation during nominal operation and accident events for both FCC-ee and FCC-hh. More penetrating radiation from loss points in the forward direction is effectively shielded by hundreds of metres of rock, ensuring protection for other potential underground infrastructures.

The general design assumptions taken for the underground infrastructure are as follows.

- The accelerator tunnel and experiment caverns are not accessible during beam operation.
- The service caverns and klystron galleries must be accessible during beam operation for both FCC-ee and FCC-hh. The minimum shielding between the accelerator tunnel and accessible areas is about 10 m, as required by FCC-hh and also driven by stability considerations from civil engineering. Connection tunnels should be long ($\sim$ hundreds of metres) or, if short ($\sim$ tens of metres), include chicanes and/or shielded doors. Chicanes are preferred over shielded doors which have a risk of failure and because chicanes have fewer maintenance issues.
- Service caverns should be non-designated areas (see Table 9.3) for unrestricted access.
- Klystron galleries will be classified as non-designated areas or supervised radiation areas.
- Access to the arcs must be possible without requiring passage through high-radiation areas.
- Alcoves, i.e., technical areas attached to the accelerator tunnel, must be shielded to prevent activation of the installed equipment and are inaccessible during beam operation.
- Areas with high radiation and activation levels must be separated from low activation areas in terms of ventilation and cooling systems (compartments) and personnel access (bypass routes, local shielding).
- The FCC-ee booster will be located on top of the collider ring. As it will operate at about only 10% of the collider current and at varying energies, the studies conducted so far were limited to the collider itself.

The items identified in the collider that will likely show higher activation levels and hence higher residual dose rates are:

- The beamstrahlung dumps next to the interaction points; two dumps on each side of each experiment in points PA, PD, PG, and PJ;
- collimators and absorbers installed in a dedicated insertion region in point PF and next to the interaction points;
- the matching regions next to the interaction points impacted by scattered radiation from the beam collisions in the experiments;
- the main beam dumps in point PB;
- the inner detectors in the interaction points;
- the EM separators installed at point PH.

The following sections highlight some results of the radiation protection studies conducted during the feasibility study phase.

***Arc sections***

The arc sections make up the largest part of the accelerator. Activation and average residual dose rate levels can have a potential impact on personnel, depending on the time it takes in the tunnel to reach a particular workplace.

The main relevant source terms are beam interactions with the residual gas molecules in the vacuum chamber and the synchrotron radiation. The latter contributes only significantly to activation in the $t\bar{t}$ mode, where synchrotron radiation is energetic enough to produce relevant photoneutron fluence and, thus, induced activity.

Figure 9.7 illustrates the expected residual dose rates after $t\bar{t}$ operation. The dose rate levels induced by the other modes are two to three orders of magnitude lower. In the $t\bar{t}$ mode, some waiting time may be required before performing interventions in the tunnel to reduce the dose to personnel. However, the decay is relatively fast, and within a few days, it is down to very low levels acceptable for long interventions. These figures show dose rates without any shielding installed around the photon absorbers. It is now planned to install this shielding. It will have an important mitigating effect so that dose rates will be considerably lower in the $t\bar{t}$ phase.

The ventilation system is designed to recycle air, limiting the release of short-lived radioisotopes during operation. Inhalation and immersion doses were calculated for interventions immediately after the beam stop without air renewal (worst case), with insignificant doses in all modes, except for $t\bar{t}$, where a few µSv per hour of stay could be expected. However, flushing is always required for safety before entering the tunnel.

In conclusion, the FCC-ee arcs will exhibit some activation, but the low dose rates should not lead to significant doses to individuals passing or intervening in the arc sections.

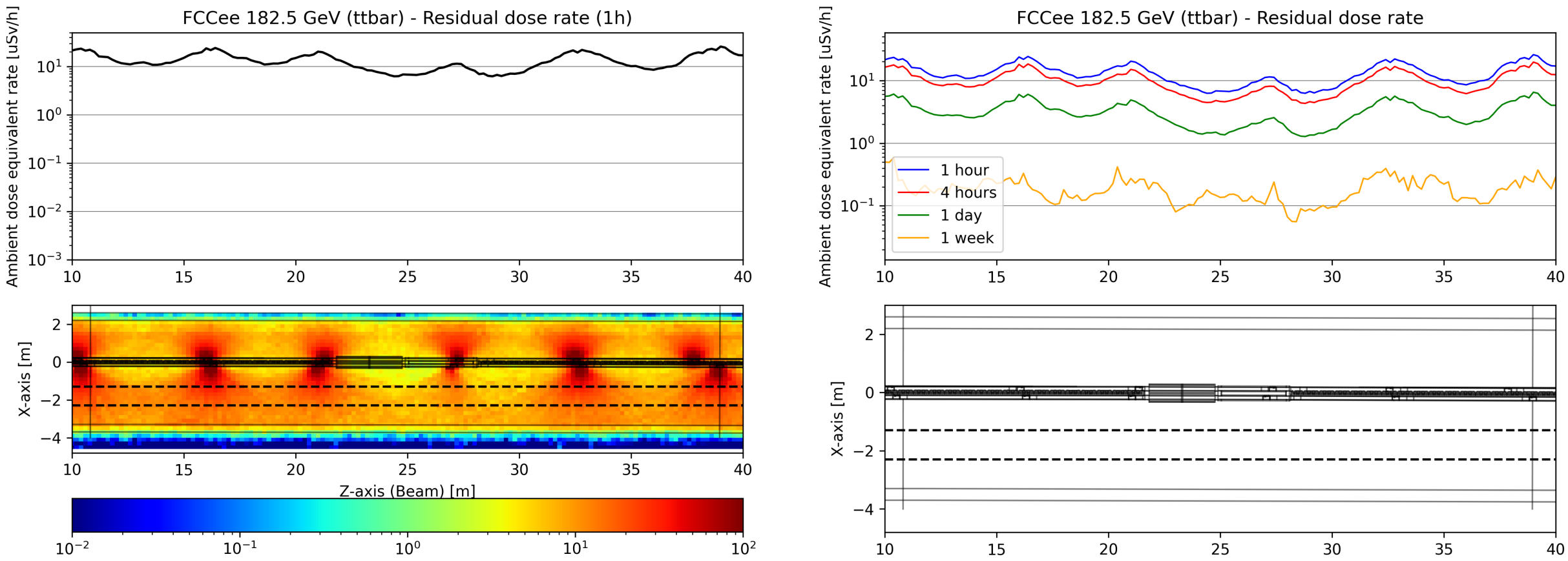

(a) (b)

Fig. 9.7: The graphs represent the distribution of the ambient dose equivalent rate (µSv/h) in the arcs after the $t\bar{t}$ operation phase, averaged over the y-axis from -50 cm to +50 cm, approximately 1 m above the floor. (a): with a decay time of 1 hour, (b): Average values after various decay times: 1 hour, 4 hours, 1 day, and 1 week. The irradiation profile considers all years of $t\bar{t}$ operation phase, taking into account shutdown periods.

### *Connection tunnels between experiment and service caverns*

Colliding beams at interaction points create secondary particles that spread throughout the experiment cavern. Although the detector absorbs much of this radiation, some can reach areas such as the service cavern, which must remain accessible during operation. It has been assessed whether accessing the service cavern is feasible during operation with measures such as shielding and chicanes in the connection tunnels.

This study is for FCC-hh, as proton-proton collisions produce higher dose rates than the electron-positron collisions in FCC-ee. A layout compatible with FCC-hh will, therefore, be compliant with FCC-ee. This approach is justified because the caverns and connection tunnels will not change between the two colliders.

Radiation transport simulations were performed for the entire geometry, including the experiment cavern, the connection tunnels, and the service cavern. The assessment considered the most conservative operational FCC-hh scenario ('ultimate'), with a collision rate of $3.24 \times 10^{10}$ p/s[8].

The spatial distribution of the dose rate through the service tunnel and within the service gallery is shown in Fig. 9.8. The results demonstrate that the proposed design will meet the target ambient dose equivalent rate limit, i.e., $<0.5$ µSv/h inside the service cavern, allowing access during operation with permanent occupancy (non-designated area classification).

The feasibility of designing and building the following mitigation measures was confirmed by integration and civil engineering: (1) mobile shielded doors with a thickness of 1 m at the entrance of the transport tunnel and (2) a two-legged chicane for the access tunnels for personnel. Additionally, the current design includes a three-legged chicane (not included in the simulation studies shown here), for connecting the main accelerator tunnel with the experiment service area, to minimise the dose rate contribution from the accelerator tunnel. The schematic layout proposed by civil engineering and integration includes chicanes for direct access tunnels. This will ensure that the radiation protection objectives for the service cavern are met.

[8]Based on a peak luminosity of $30 \times 10^{34}$ cm$^{-2}$s$^{-1}$

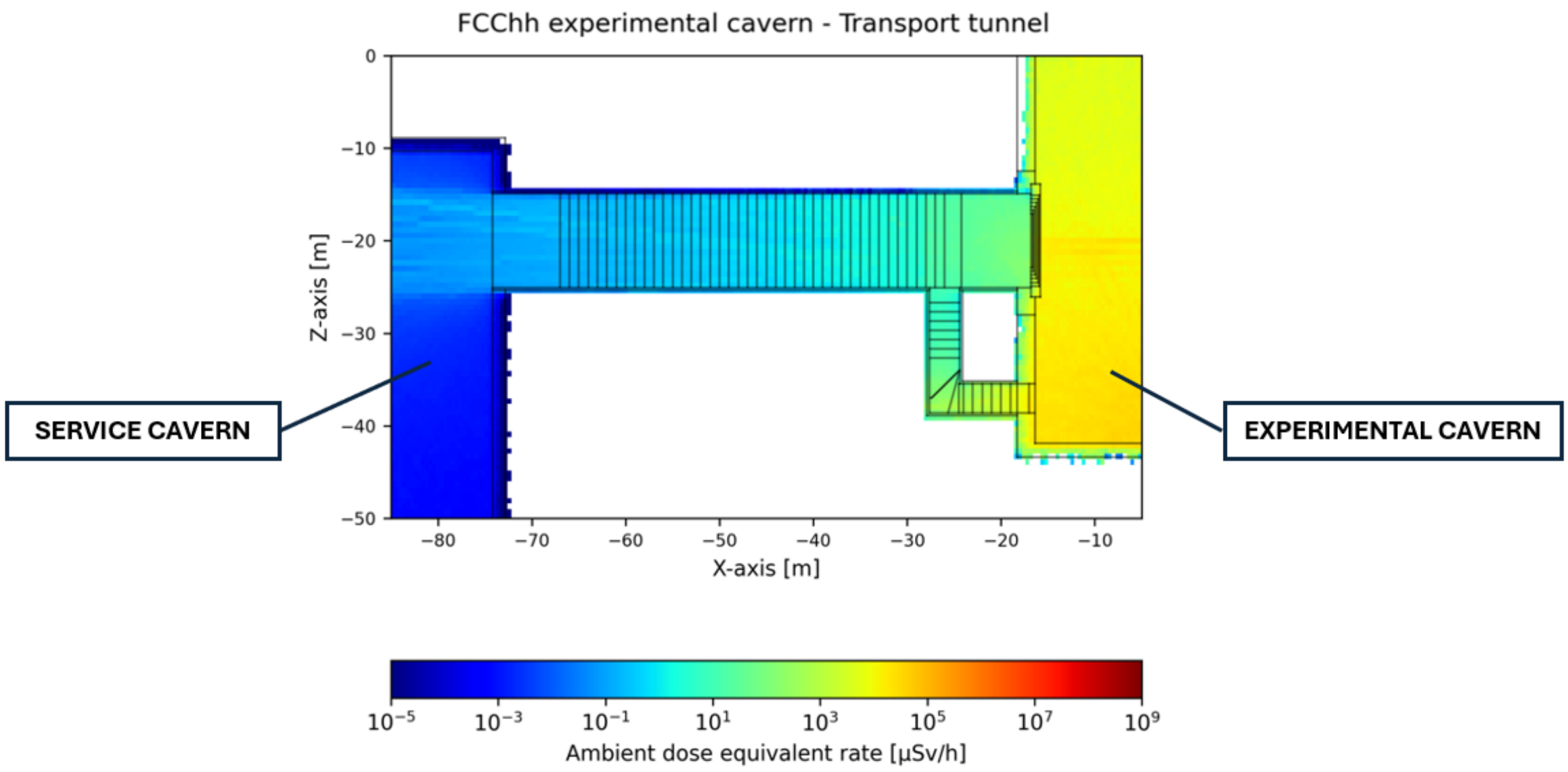

Fig. 9.8: The plot shows the ambient dose equivalent rate averaged over 3 m in height at the FCC experiment points from pp collisions along the transport tunnel between the service and experiment caverns. At the tunnel-experiment cavern interface, a 1 m shielding wall and a two-legged chicane for personnel passage are implemented.

***Bypass tunnels***

Bypass tunnels allow access from the service cavern to the accelerator tunnel, bypassing the experiment caverns. Bypass tunnels can be designed to allow access to areas with low activation levels while avoiding passing through areas of increased residual radiation risk. The bypass tunnels must reduce prompt radiation streaming towards the service cavern down to acceptable levels, ideally without the need for additional mobile shielding walls for easier access.

Two bypass tunnels are planned at each experiment point to connect the service cavern with the accelerator tunnel. FCC-ee and FCC-hh feature different radiation sources near the interaction regions, with radiological hazards that must be considered when designing the layout of bypass tunnels and their junctions with the accelerator tunnel. The constraints must be assessed for both accelerators:

– *FCC-ee*. Beamstrahlung, synchrotron radiation (SR), and radiative Bhabha (RBB) electrons increase the residual radiation levels in the straight sections next to the interaction points. Residual dose rate levels in the personnel and transport passage are in the range between 0.1 and 100 µSv/h after decay times of 1 to 4 days, as determined in some preliminary simulation studies (see Fig. 9.9). The integration of SR and RBB shielding on the outgoing beamline, which is still pending, will further help to reduce residual dose rates. The beamstrahlung dump must be shielded so that the impact on the personnel and transport passage is reduced to sufficiently low levels in the passageway, located on the opposite side of the tunnel.
– *FCC-hh*. Residual radiation levels are increased in the straight sections next to the IPs, from collision debris in the IP. The ambient dose equivalent rate levels in the personnel and transport passage range between 0.1 and 10 mSv/h after decay times of 1 day to 1 week (see Fig. 9.10). The area has to be considered as a high radiation area. Given the enlarged tunnel cross section, it seems feasible to integrate additional shielding along the beam line, effectively reducing residual dose rates to acceptable levels in the personnel and transport passage. Alternatively, the bypass tunnels could be extended before the operation of the FCC-hh.

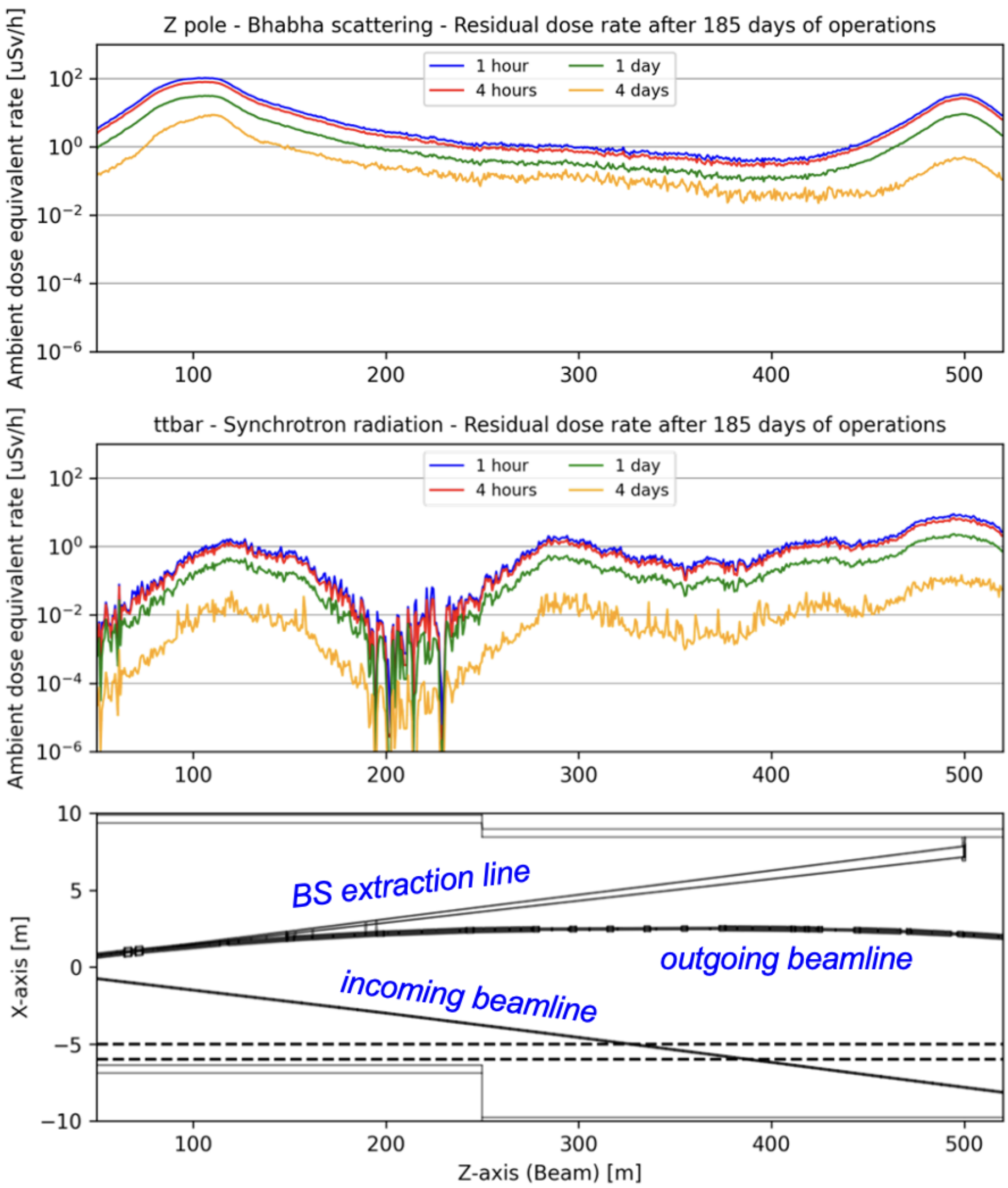

Fig. 9.9: The top and centre plots show residual dose rates averaged within the dashed lines in the straight section geometry next to the FCC-ee IP (bottom plot, distance from the IP). These were computed for two radiation sources: (1) radiative Bhabha during Z pole and (2) synchrotron radiation during $t\bar{t}$, assuming one year of operation (185 days).

### *Beam dump*

The beam dumps will receive a rather small fraction of the accelerated particles, considering the top-op operation mode and approximately only one dump per day, and will therefore become radioactive to a level that will likely not require a compartmentalisation of the area for separate ventilation. The dumps are shielded and placed in a section where the tunnel cross-section is increased. The passage for transport and personnel in the straight section is sufficiently far from the dumps to achieve acceptable dose rate levels at the passage. Dedicated dump caverns and extraction beam lines will need to be added

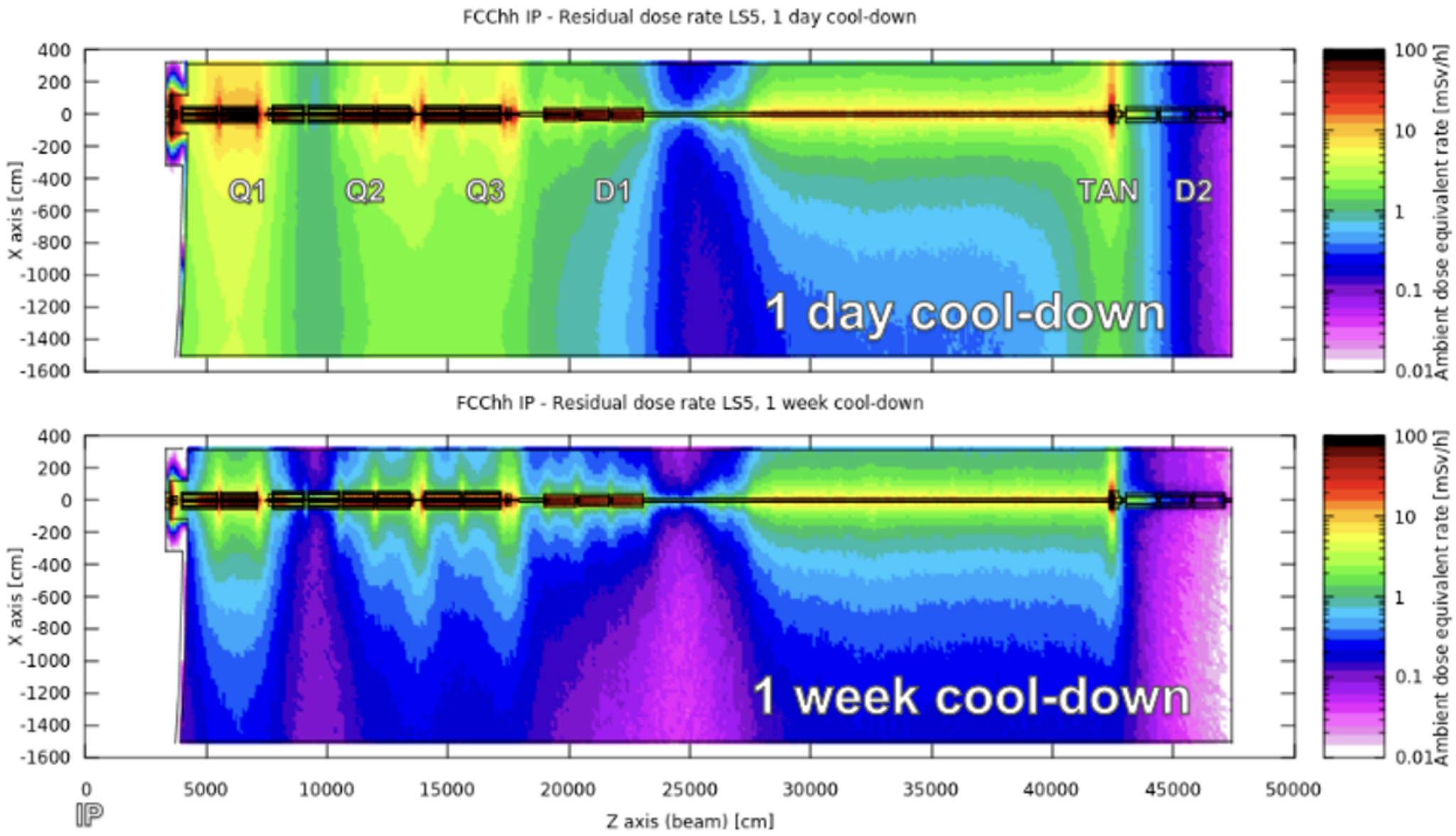

Fig. 9.10: Results showing residual dose rate at FCC-hh IP straight sections after an irradiation of 25 y, with an average collision rate of $5.4{\times}10^9$ p/sec (10 y) + $3.2{\times}10^{10}$ p/sec (15y). Acceptable positions for the connection of the bypass tunnel are at ∼200 m and at ∼450-500 m.

for FCC-hh.

The radiological study of the beam dumps examines the residual radiation levels next to the beam dumps and is based on a preliminary dump proposal that still lacks a detailed technical design. The study was done for the Z pole scenario, which is considered to be the most conservative. Each beam dump is composed of a cylindrical graphite structure of a length of 600 cm. The structure consists of three layers of graphite with varying densities of 1.1 $\mathrm{g/cm^3}$ and 1.8 $\mathrm{g/cm^3}$. An additional layer consisting of a high-density absorber made of CuCrZr is integrated at the end of the structure. The dump is encapsulated with a 1 cm thick layer of titanium alloy. The extraction line is equipped with three successive cylindrical spoilers made of graphite (with densities ranging from 1.7 to 1.8 $\mathrm{g/cm^3}$) and 3 cm thick. A first dump shielding was implemented, consisting of a first layer of iron (20 cm) and a second layer of concrete (40 cm). Although this is still a conceptual shielding design, it illustrates the expected effects of potential shielding.

FLUKA Monte Carlo simulations were performed in order to assess the residual dose rate after one year of operation (185 days), considering one beam dump per day, where all particles stored in the collider are discharged. The number of particles dumped annually is $4.44 \times 10^{17}\mathrm{p/y}$ (Z pole). The results show that the residual dose rate reaches values around 45-50 $\mu\mathrm{Sv/h}$, at a distance of 1-2 m from the shielded dump, after 185 days of operation (one dump per day) and 1 hour of cool-down; after 1 day, it reduces to 15 $\mu\mathrm{Sv/h}$. The dose rate values are relatively low, indicating that even a minimal shielding around the dumps effectively reduces residual radiation to almost acceptable levels.

***Beamstrahlung dump***

The radiological impact of the beamstrahlung dump using liquid lead as an absorber was assessed. FLUKA Monte Carlo simulations were performed for the Z pole operation mode, as it generates the highest beamstrahlung power among all operational modes. Initial simulations at $\mathrm{t\bar{t}}$ mode indicate as well considerable dose rates and more important activation due to the higher energetic beamstrahlung spectrum.

The model considered an inclined beam dump with a layer of liquid lead. An initial shielding design was tested to mitigate the risks associated with residual radiation. The shielding is 1 m thick in all directions and extends 5 m upstream around the vacuum chamber to mitigate backscattering.

Figure 9.11 shows the average dose rate values in the transverse plane across the tunnel, at the level of the beamstrahlung dump, considering an irradiation profile of 1 year at Z pole operation (185 days). Residual dose rate values, after 1 hour of cool-down time, may reach 7 Sv/h inside the shielding. Approximately 3 m away from the shielding, the dose rates decrease to below 100 µSv/h.

At these very high dose rates, optimised shielding is mandatory to contain the residual dose rate at an acceptable level outside the dump. The current shielding must be enlarged and optimised to reduce the dose rate to $1-10$ µSv/h in the passage area after 4 hours of cool-down time. Appropriate shielding will certainly be more massive, but can be integrated. The dose rate from activated lead in the pumping circuit and reservoir will be further assessed and included inside a shielded casing. Waiting times before manual interventions on the dump will be considerable (days/weeks), and remote operation will be used to minimise doses to personnel while handling components and performing maintenance. The activation of the lead absorber will require particular attention and a dedicated study during the technical design phase to assess the radiological risks and implications for handling and elimination.

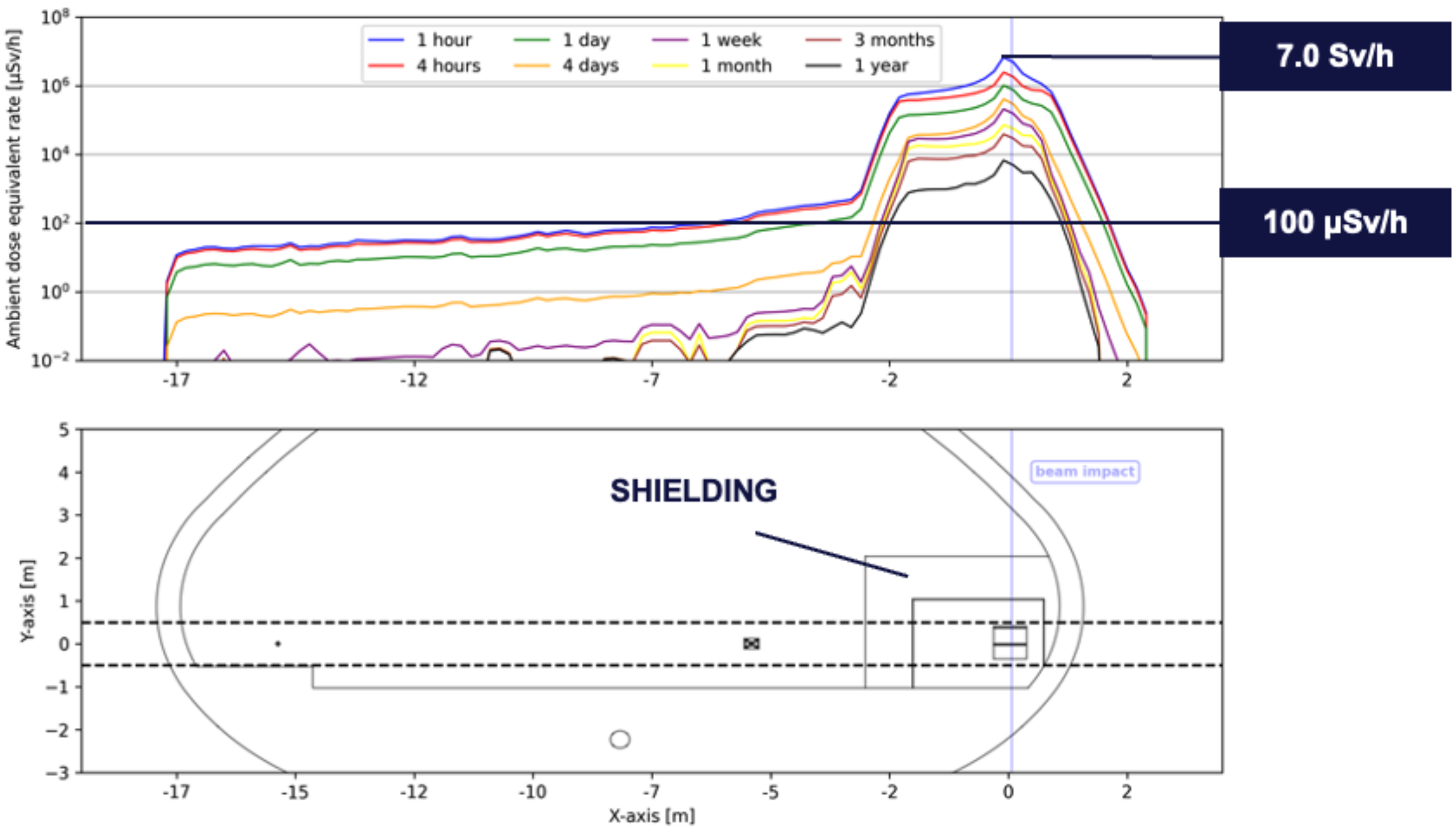

Fig. 9.11: Residual ambient dose equivalent rate profile at the FCC-ee beamstrahlung dump, after 185 days of Z pole operation, for different cool-down times. The profiles are averaged in the volume enclosed by the two dashed lines, which extends 1 m on Y-axis and Z-axis.

### *Collimation*

FLUKA Monte Carlo simulations were conducted to evaluate residual radiation in the collimation section (PF) and the potential impact on the design, integration, civil engineering and cooling/ventilation. The model encompasses the positron betatron collimation section, including 12 quadrupole magnets, 6 collimators (both primary / secondary and horizontal / vertical) and 2 shower absorbers. A standard tunnel cross section is used without detailed integration of additional infrastructure, such as ducts, cables, and piping.

The Z pole is considered the operational mode with the highest loss rate of $8.5\times10^{11}$ p/s, based on a beam lifetime of 30 minutes.

Figure 9.12 presents the residual dose rate profiles along the straight section of the positron be-tatron collimation after 185 days at Z pole operation, evaluated for different cool-down times (ranging from 1 hour to 1 year). The results indicate the highest peak occurring in front of the secondary horizontal collimator. The high dose rates span several tens to hundreds of metres, with maximum residual dose rate values reaching 1 mSv/h after one week of cool-down.

Based on the parameters provided, the collimation section exhibits very high dose rates, making immediate hands-on maintenance in this part of the tunnel impossible. Evaluating and improving collimator shielding could help reduce activation levels in the surrounding infrastructure. With further optimisation of design parameters, the implementation of appropriate bypass tunnels, and a hybrid maintenance approach (combining personnel and robotics), it appears feasible to allow access after one day of cool-down. This time frame is also essential to ensure the decay of short-lived radioactive isotopes in the air, enabling safe ventilation and the controlled release of air into the environment through flushing.

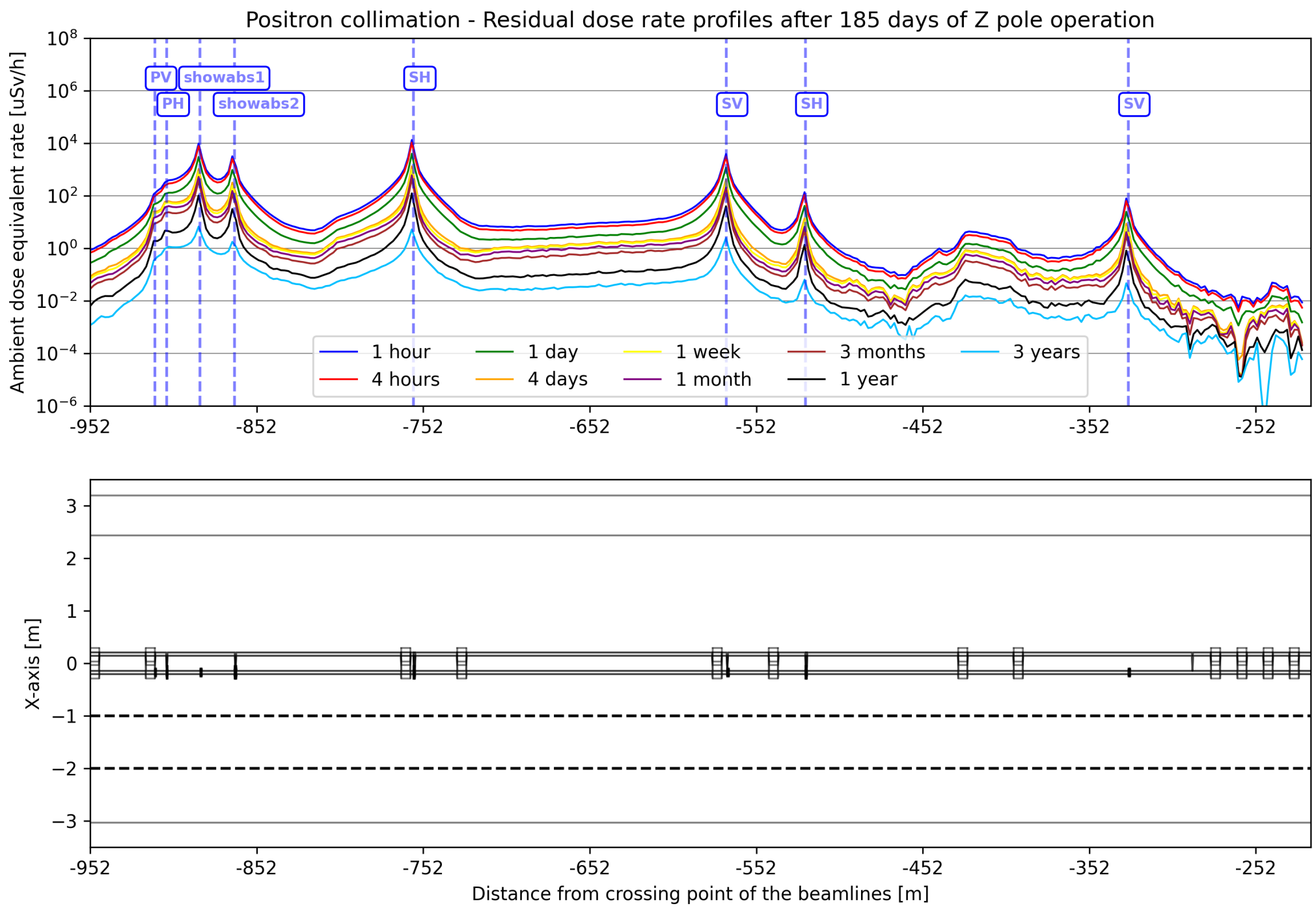

Fig. 9.12: The plot displays the residual dose rate profiles along the positron betatron collimation straight section after 185 days of Z pole operation, evaluated for different cool-down times (ranging from 1 hour to 1 year). The labels correspond to the components on the beamline: PV (primary vertical), PH (primary horizontal), showabs1 (first shower absorber), showabs2 (second shower absorber), SV (secondary vertical), and SH (secondary horizontal).

*Surface Sites*

Radiological hazards at surface sites during nominal beam operation are potentially induced by prompt radiation that can propagate from the experiment cavern up through the shaft.

The same parameters and radiation sources used to calculate the prompt radiation in the connection tunnels, described above, are applied to calculate the propagation of the prompt radiation from the experiment cavern to the surface along the shaft. The study considers FCC-hh, where proton-proton collisions (‘ultimate’ scenario, with a collision rate of $3.24 \times 10^{10}$ p/s) produce a dose rate higher than that of electron-positron collisions in FCC-ee.

The results are shown in Fig. 9.13. The colour map on the right displays the values of the ambient dose equivalent rate along the shaft using a colour gradient (in µSv/h). The left graph depicts the profile of the ambient dose equivalent rate, with values averaged over the volume defined by the two dashed lines in the colour map on the right. The two horizontal red dashed lines represent the minimum and maximum depths of the shafts for the four experimental points. This allows visualisation of the range of ambient dose equivalent rates at the different surface sites.

The plot shows surface dose rates ranging from 10 to 70 µSv/h. Shielding installed at the machine-detector interface, and a 1-metre thick concrete shielding at the top of the shaft are adequate mitigation measures.

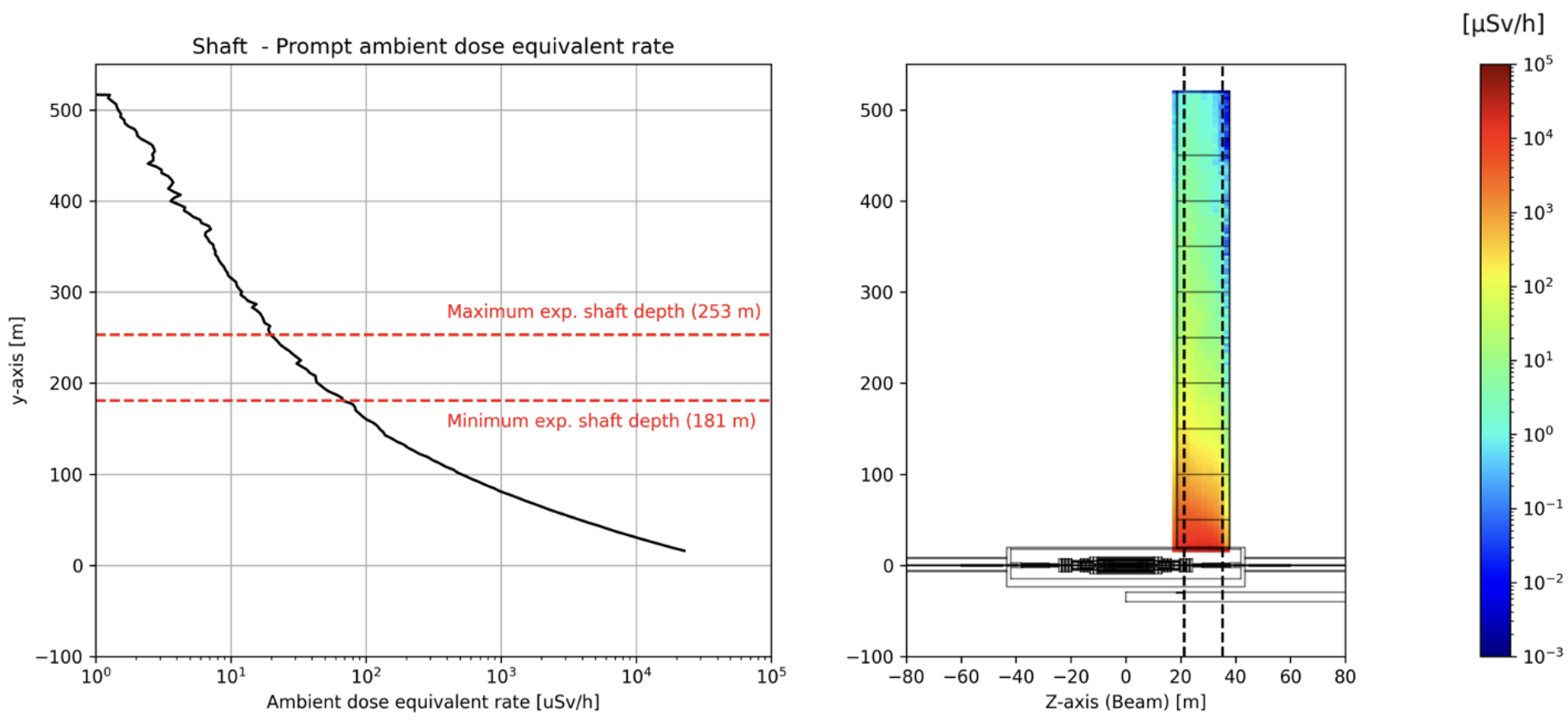

Fig. 9.13: Ambient dose equivalent rate through the shaft resulting from pp collisions in the experiment cavern.

Other radiological hazards at the surface site may arise from the handling of activated material that has been removed from the tunnel. Standard prescriptive methods are used to limit the radiological risk to personnel by having dedicated areas to store and work on radioactive items. Buffer zones are established at the border between areas where activation can occur and non-activation areas to control the flow of material and to guarantee the correct classification of items removed from the accelerator areas.

The release of radioactive air and water from the accelerator tunnel and the experiment cavern poses a negligible risk. This is mitigated by recycling the air during operation and having dedicated water handling installations to avoid any potential contamination or uncontrolled releases.

The use of radioactive sources and X-ray-generating devices in surface facilities may occasionally occur during the construction and operation phases. These common practices are covered by standard prescriptive methods and rigorous regulations that have been widely applied at CERN and in industry.

### *Non-ionising radiation (NIR)*

Risk of non-ionising radiation includes exposure to static magnetic fields, time-varying electromagnetic fields (e.g., RF), lasers and non-coherent light sources (e.g., UV & Infrared, light, LED). These risks are currently addressed by standard practices, namely by European Directives [458, 459]. It is possible, at this stage, to conclude that the design of NIR-generating accelerator and detector components will respect the exposure limit values of the directives.

For static magnetic fields (e.g., from permanent or electro-magnets), the exposure limit values (ELV) and action levels (AL) are given by CERN's General Safety Instruction GSI-NIR-1 [460].

*Stray magnetic field from experiment detectors*

A study was performed simulating typical fringe fields from the experiment cavern housing an FCC-hh detector representing a worst-case scenario and concluded that the residual magnetic field at the surface of experiment points is well below current limits. At a distance of $200\,\mathrm{m}$ above the detector, the stray field is of the order of $0.1\,\mathrm{mT}$, whereas the limit for wearers of active medical devices is $0.5\,\mathrm{mT}$ [461].

*Stray magnetic field from the collider*

A study was performed simulating the extent of stray fields for evaluation against the ELVs and ALs . All stray field envelopes at the ELVs stay within 1 m from the magnet's centre, except for the quadrupole, which emits a stray field envelope at $0.5\,\mathrm{mT}$ to most of the cross-section of the tunnel (Fig. 9.14). This is due to the fact that the quadrupole has substantial field strength ($\approx 0.45\,\mathrm{T}$ at the aperture) and an open iron yoke. In comparison, the dipole yoke is also open but with low field in the aperture ($60\,\mathrm{mT}$), while the sextupole also has a relatively high field ($0.5\,\mathrm{T}$) but confined within a closed iron yoke. Since quadrupoles will be spread throughout the arc, people with active implantable medical devices (AIMD) would not be able to access the tunnel, without a valid medical certificate [460]. Even when the magnets are not powered, there is still a residual risk of remanent fields stemming from magnetised metal in the surroundings. Occasional access in specific areas may be granted, based on a risk assessment and fenced off from any individual $0.5\,\mathrm{mT}$ sources.

Access is possible for the public and workers without any particular risk[9] ($B < 40\,\mathrm{mT}$). Any work near the dipoles and quadrupoles when the magnetic source is on would require a specific workplace risk assessment and training.

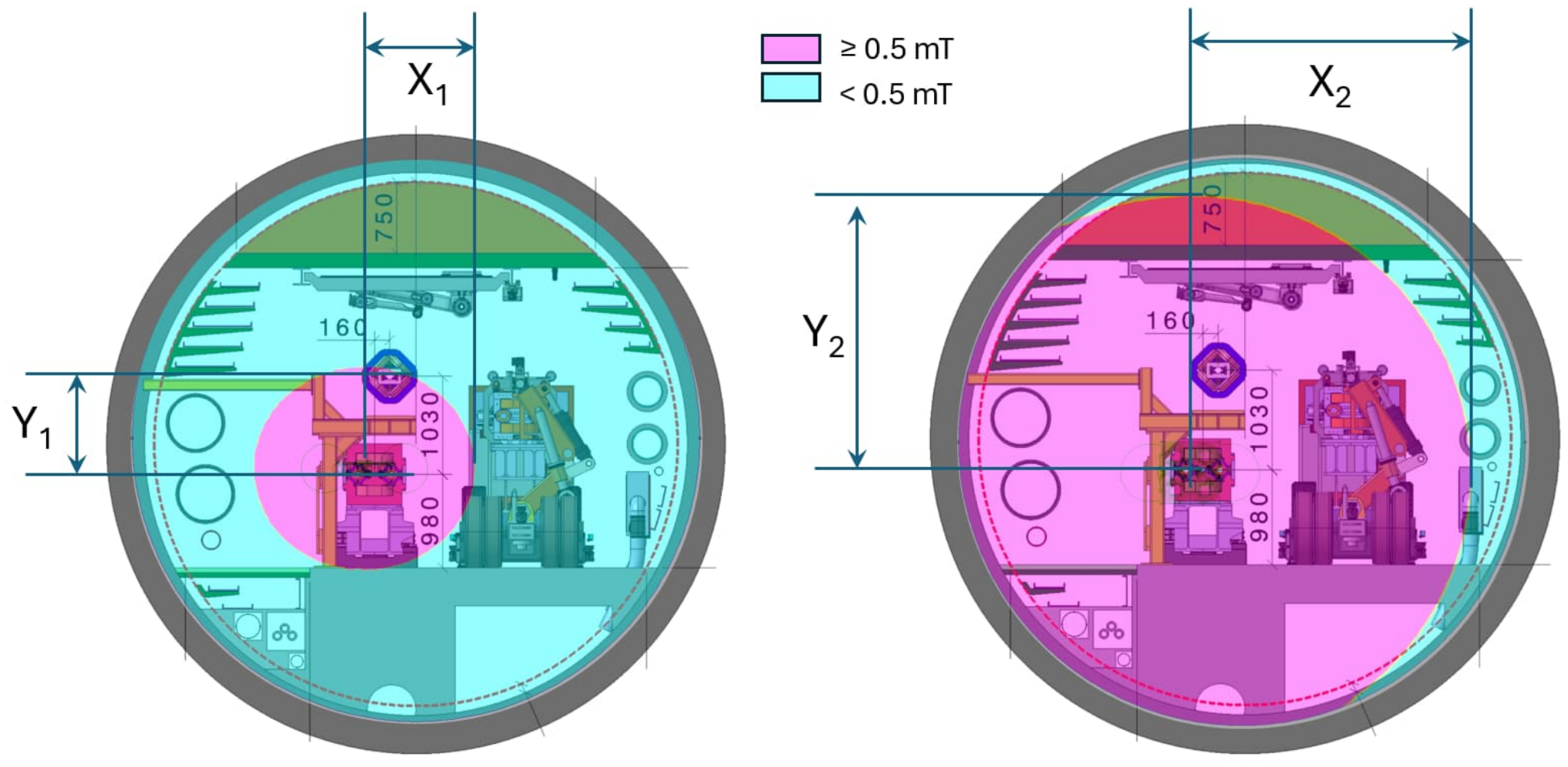

(a) (b)

Fig. 9.14: Simulation of stray field envelope at 0.5 mT for the FCC-ee collider arc. a) Dipole: ($X_1$, $Y_1$) = (1070, 980) mm; b) Quadrupole: ($X_2$, $Y_2$) = (2690, 2720) mm

.

*Lasers for beam polarisation*

Fixed laser installations are planned in some alcoves for beam polarisation and energy calibration. The laser installation will be installed within a dedicated designated laser area (DLA), with interlocks to cut off the source in case of unauthorised access. The classification of the laser systems, as well as the engineering control measures, will follow the applicable International Standards, i.e., IEC 60825 series [462].

[9] Such as wearers of passive metal implants and pregnant workers.

*Optical fibre network*

The infrastructure will deploy an optical fibre network (OFN) for internal communication and data transmission. Depending on the maximum output power of the transmitter ($\mathrm{dB\,m}$), the OFN will be classified within a given hazard level in accordance with IEC 60825 [462]. For any given hazard level, a set of engineering control measures are required.

***Electrical safety***

The electrical infrastructure will cover everything from low voltages (e.g., lighting) to high voltages (e.g., power transmission). Despite this wide range, electrical safety may still be covered by standard practices. Most of the electrical equipment will be housed in surface buildings, in the service caverns, and the alcoves. The FCC accelerator and experiment detectors are expected to be designed according to the appropriate international standards (IEC) where they obey the safety by design principle, i.e., ensuring a proper degree of inherent protection (e.g., IPx rating according to the IEC 60529 standard [463]) preventing live parts being accessible. Any non-standard equipment will be subject to a rigorous risk assessment design process to ensure conformity with the applicable legislation. The electrical network will be designed according to national and international standards for all voltage levels (low, medium, and high). A secure power network is foreseen as mentioned in Section 9.4.1. The location of cable trays in the tunnel arc and the alcoves is chosen to minimise interference with the occupants.

Occupants and emergency teams will be authorised and required to activate an emergency stop in case of an unsafe situation or incident, cutting off the electrical power to the general (non-safety-related) services. In the LHC, this is achieved by a spatial distribution of general emergency stop (AUG) buttons along the tunnels and caverns. The exact technical solution for the FCC will be studied in the next phases, during the detailed design of the electrical network.

For works or activities in which the protection elements are temporally removed, there are a set of technical and administrative procedures to ensure the safe execution of such activities. For example, lock-out, tag-out and electrical work permits to ensure safety during interventions on electrical equipment. A safety-by-design approach is enforced, considerably reducing the technical and administrative burden of specific safety prescriptions.

***Chemical safety***

Only a few chemicals are required for the operation of the FCC infrastructure. Most of them will be used in the closed circuits of the cooling water network:

– Biocides - chlorine dioxide, produced from hydrochloric acid and sodium chloride.
– Corrosion inhibitors - mainly sulphuric acid.

Significant quantities of synthetic lubricants (e.g., BREOX) are expected on the surface sites for the cryogenic compressors. Cooling refrigerants (coolants) are also needed for the fan coils in the tunnels and chillers on the surface. Some traces of oil are expected underground, but these will be in minimal quantities (e.g., vacuum pumps, motors, and other machinery).

At this stage, only standard chemicals are intended to be used, hence standard practices and regulations are applicable when handling these chemicals. For each chemical product, the supplier must deliver a safety data sheet (SDS) or an extended safety data sheet (eSDS) and a technical data sheet (TDS) with information on product usage, on hazards during use and during emergency procedures.

#### *Mechanical safety*

The feasibility study focused on the technical infrastructure at large and its possible impact on the civil engineering layout. The mechanical design of the beamlines and equipment inside the infrastructure will be detailed in the next phases of the project. The necessary safety studies and assessments will be performed in parallel.

The use of robotics (both overhead and floor-driven) will be widely present in the FCC infrastructure. The cohabitation of robots and personnel poses a safety hazard that will be dealt with during the design phase, where the robotic systems are designed and manufactured according to the EU Regulation 2023/1230/EU on machinery [464] and associated harmonised standards.

#### *Indoor air quality*

The indoor air quality in the underground infrastructure is guaranteed by the ventilation system described in Section 9.4.1. During access mode, the air renewal for hygienic purposes is guaranteed by the air exchange in the concerned sector. The values of the flow rates and air velocities in the tunnel are shown in Table 9.4. In the technical points (PB and PF), the supply flow rate for the service caverns and connection tunnels is 20 000 $m^3\,h^{-1}$. For the experiment points (PA, PD, PG and PJ), the supply flow rate for the service caverns and connection tunnels is also 20 000 $m^3\,h^{-1}$, whereas the experiment caverns are supplied with 50 000 to 70 000 $m^3\,h^{-1}$. The RF points (L and H), are supplied with 20 000 $m^3\,h^{-1}$ in the service areas (cavern and connections) and 30 000 $m^3\,h^{-1}$ in the RF sector in the tunnel [439].

The ventilation of the underground infrastructure is a pre-condition for access. Hence the correct functioning of the ventilation system must be ensured before granting access to personnel.

Table 9.4: Metrics of the ventilation system in the accelerator tunnel during access mode [439].

| **Scheme** | **Flow rate** $[m^3\,h^{-1}]$ | **Velocity** $[m\,s^{-1}]$ | **ACH**[a] $[h^{-1}]$ |
|---|---|---|---|
| Semi-transverse | 27 000 (per half arc) *(54 000 per sector)* | 0.25 *from 0 (midpoint) to 0.5* | 0.3 *in a sector* |
| Longitudinal | 54 000 per sector | 1 | 0.3 |

[a] Air changes per hour

#### *Noise*

As far as reasonably possible, the owners of noise-generating equipment installed both in the underground areas or surface sites must apply technical solutions to achieve a daily exposure level below 80 dB(A) [465]. These limits are considered state-of-the-art and would avoid the mandatory use of personal protective equipment (e.g., earplugs). In the case of specific equipment emitting higher sound levels (e.g., cryogenic compressor building at the surface), specific engineering and personal protective measures must be put in place.

#### *Workplace ergonomics*

Despite the fact that there are no permanent workplaces in the underground infrastructure, occupants may be requested to spend considerable amounts of time performing repetitive tasks therein. The integration of the tunnel arc, as well as the experiment and services caverns, must take into account the requirements to ensure proper workplace ergonomics.

In the tunnel arc, the passage from the transport area towards the opposite side of the accelerator(s), i.e., the external side, passes underneath the beamline in between two beam elements (e.g., dipoles, corrector magnets, etc.). Such a passage must take into account the ergonomics of occupants crossing from one side to the other several times a day. Standard practices as used in industrial installations are available in EN 547 [466] to cover this occupational hazard. If passage underneath the beamline remains the baseline approach, the minimum vertical clearance from the floor to the vertical obstacle (i.e., the beam pipe) is 1123 mm [10].

### 9.4.3 Occupational hazards during accident scenarios

During the feasibility study phase, only some selected accident scenarios were analysed, mainly those deemed to have an impact on the layout and civil engineering structure of the underground infrastructure. Additional accident scenarios will be assessed during the next phases of the study.

#### *Ionising radiation*

Accident scenarios that may be relevant in terms of radiological protection will not produce different hazards from those described for nominal operation, i.e., external exposure from prompt or residual radiation or internal exposure from the incorporation of radioactive contamination. The relevant scenarios considered are the following.

- Beam losses: refers to the inadvertent deviation or depletion of particle beams from their intended trajectory within the accelerator, leading to secondary radiation from interactions with accelerator components. A machine protection system must ensure that such events are detected and the beam will be directed either to the beam dumps or absorbed by the collimation system or protection devices.
- Fire: fire can release radioactivity from combustible activated materials, with smoke that carries radioactive isotopes. As it spreads and settles, contamination may occur. The radioactivity content in combustible materials is low and will not lead to a relevant exposure of the intervening personnel. The radiological risk to workers or emergency responders is negligible, given the greater risks that such emergency situations present, such as oxygen deficiency, smoke toxicity, and fire.
- Leakages: leaks from water circuits can release large amounts of water that may contain radioactive isotopes. The radioactive concentration of the water circuits is low and will not lead to any relevant exposure scenarios.

The main accident scenario considered is an accidental beam loss, where the civil engineering infrastructure is critical to contain prompt radiation and protect workers in adjacent areas. Although unlikely and mitigated by the machine protection system, such scenarios are considered in the design. Beam loss scenarios are assessed at critical locations, that are, the injector complex, bypass tunnels, and RF klystron galleries. The dose objective of 1 mSv per incident[11] is respected in the following loss cases studied.

*Injector complex*

Dose equivalents from secondary radiation during full beam loss scenarios of one-hour duration remain largely inferior to those of the continuous loss scenario with a limited loss rate, as described above.

The doses behind 6 m of soil will be less than 100 μSv as shown in Fig. 9.15. Therefore, the constraining scenario is determined by a sustained limited beam loss rate.

[10]Based on the tallest population in CERN's members states, with a wrist to shoulder blade distance of 823 mm ($95^{\text{th}}$ percentile) and an additional 300 mm allowance covering the helmet, and head movement clearances.

[11]Design target values for a non-designated area. It refers to the maximum effective dose per event with a probability <0.01 per year.

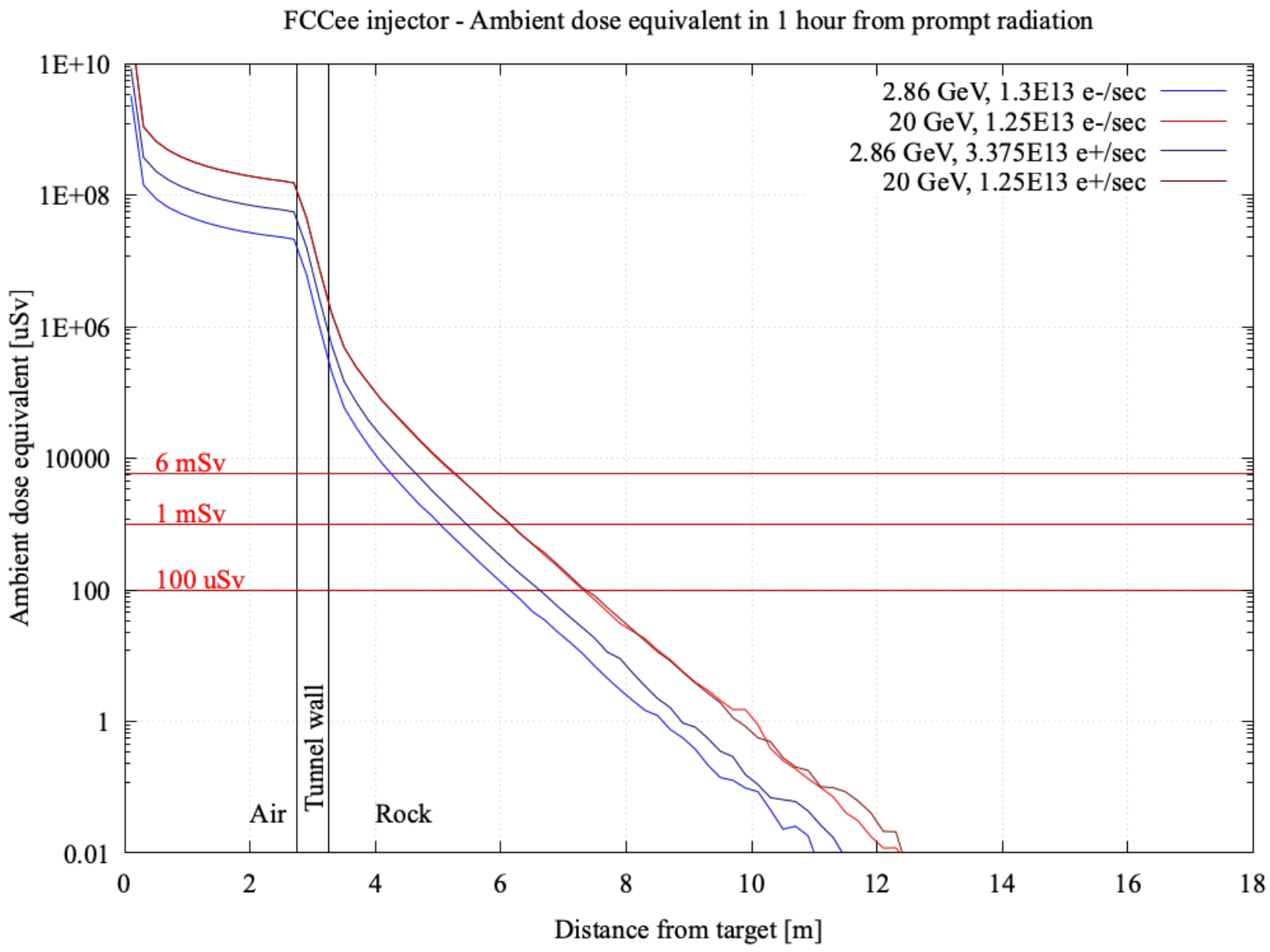

Fig. 9.15: Ambient dose equivalent through the shielding at the FCC-ee injector complex, resulting from a 1-hour beam loss scenario.

*Klystron gallery*

During the operation of the collider and booster, the klystron galleries remain accessible. Given that radiation can propagate through waveguide ducts and emergency exit staircases, prompt radiation propagation was assessed in the event of an accident scenario, i.e., beam loss.

In the absence of well-defined beam loss scenarios, the study considers the most conservative case, which is the complete beam loss, where all the particles stored in the ring (both electrons and positrons) impinge on a single cryomodule located below the waveguide duct. Although this scenario is highly improbable, it can serve as an upper limit for the analysis of a catastrophic event. Radiation transport simulations were performed to determine the ambient dose equivalent in the klystron gallery.

The beam loss scenario is analysed for the Z mode (beam energy 45 GeV) as it has the highest beam intensity of all operational modes, which is not compensated for by the higher energy at $t\bar{t}$ mode. The resulting dose was normalised using the total number of particles stored in the ring, that is, 15 800 bunches of $1.51 \times 10^{11}$ particles per bunch, for 2 beams giving $4.8 \times 10^{15}$ particles in total.

The results, shown in Fig. 9.16 and Fig. 9.17, include two types of plots: a 2D colour map illustrating the dose distribution within a specific tunnel subcell, highlighting the transverse cut used in the simulation model, and a 1D profile averaging the values from the colour map along the waveguide duct profile.

The results show that, during a full beam loss scenario in Z mode, the dose in the klystron gallery is of the order of $10\,\mu\mathrm{Sv}$. This value is well below the annual dose limit of $1\,\mathrm{mSv}$ for a non-designated area, as indicated in Table 9.3.

The access requirement to the klystron gallery during FCC-hh operation still needs to be decided. If the gallery remains accessible, the radiological conditions must be compatible with the radiation source presented by the FCC-hh.

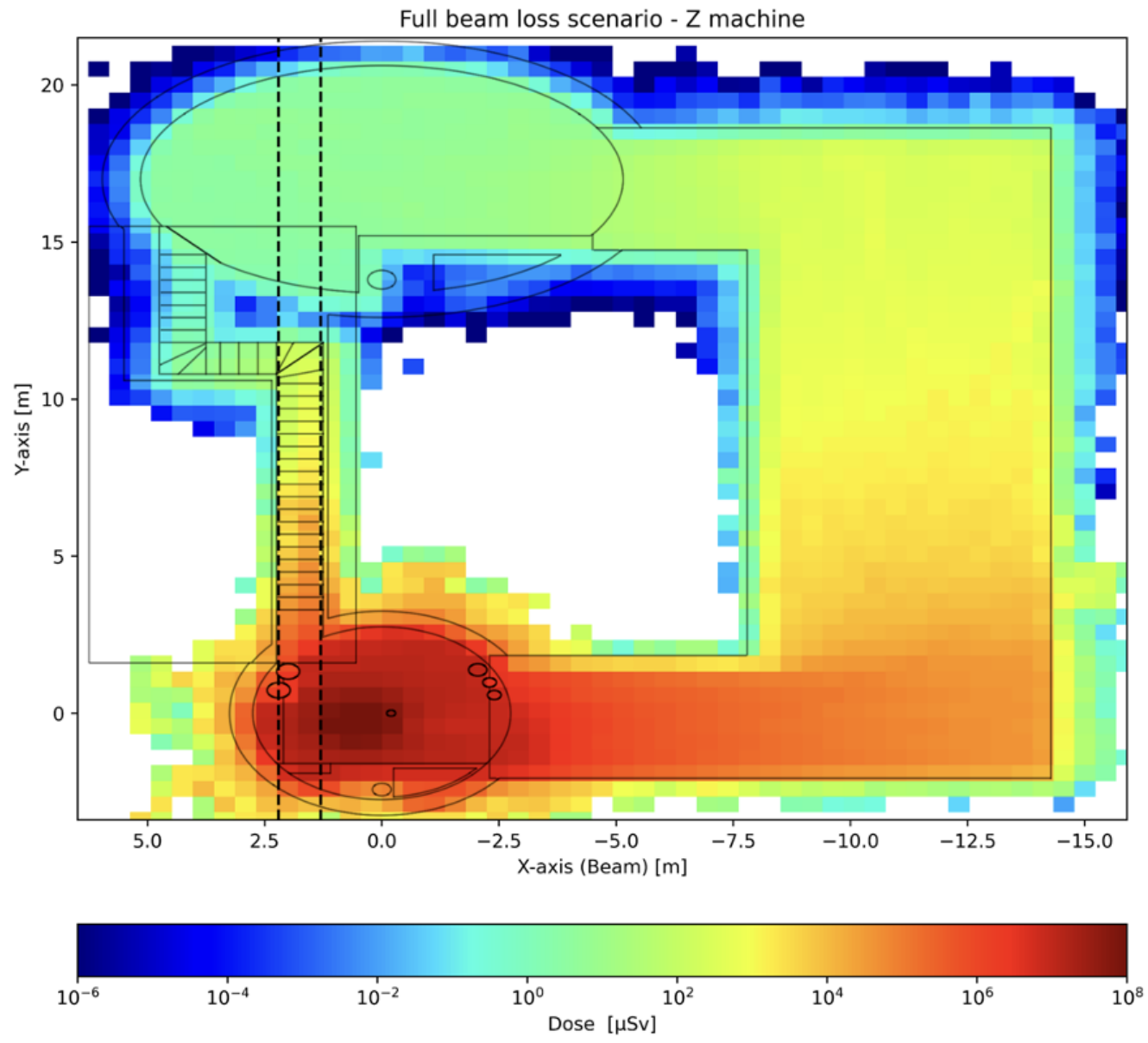

Fig. 9.16: The ambient dose equivalent is calculated for a catastrophic accident scenario, considering that all particles stored in the FCC-ee collider are lost against an RF cryomodule in a single event. The colour map illustrates that the estimated ambient dose equivalent inside the klystron gallery is approximately 10 μSv.

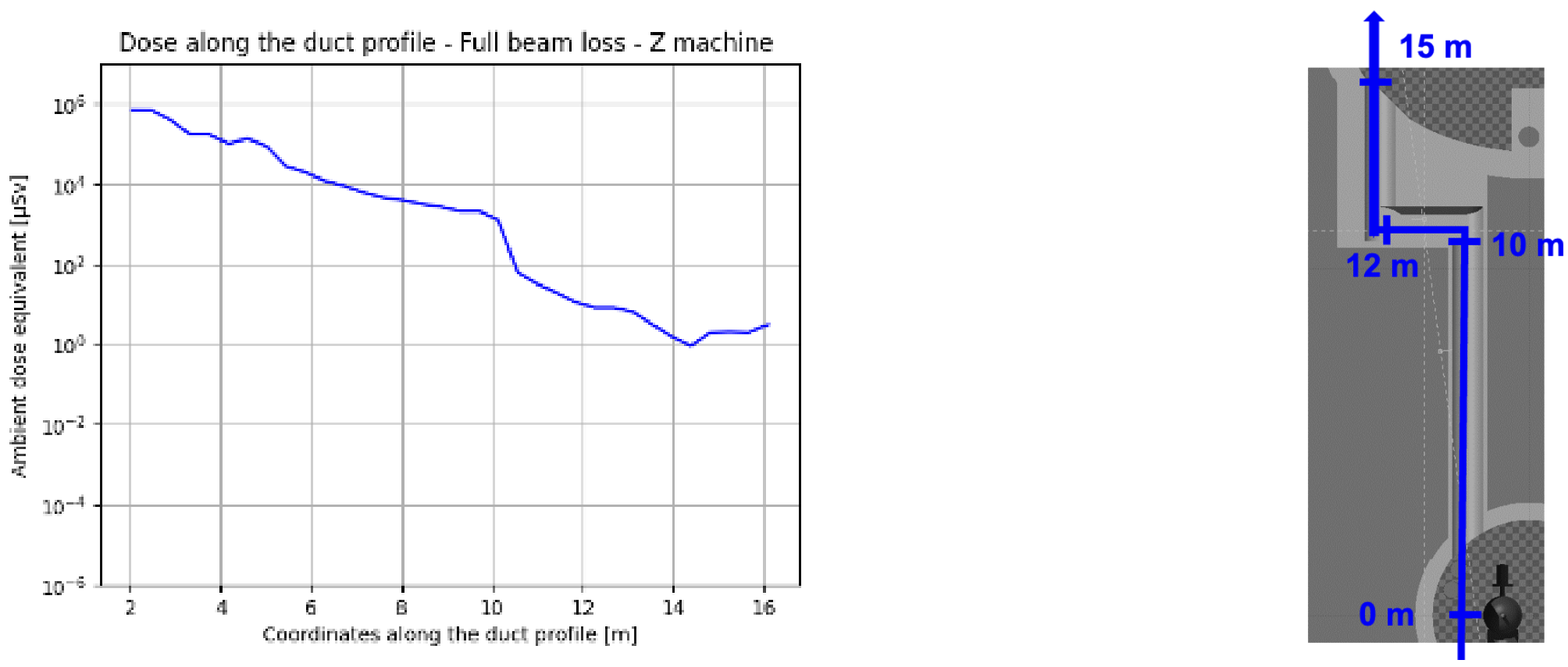

(a) (b)

Fig. 9.17: The plot presents the dose projected along the duct during a complete beam loss scenario at the FCC-ee RF section. Prior to the second leg (at x = 10 m), the ambient dose equivalent reaches 2 mSv. Then, the values decrease down to 1 μSv, due to the presence of the chicane. At the far end, the dose increases again as a result of the contribution of radiation coming from the staircase.

*Bypass tunnels*

The S-shaped bypass tunnels at the experiment points, approximately 100 m in length, were evaluated for their compatibility to ensure low doses in the service cavern during catastrophic beam losses in the accelerator tunnel. In the absence of defined beam loss scenarios, the study assumes the most conservative case: a full beam loss of all stored particles of both beams impinging on an accelerator component in front of the junction with the bypass tunnels. Figure 9.18 shows the results of the radiation transport sim-ulations, where the service cavern is connected to the bottom left, and the accelerator tunnel is running at the top of the plot. Ambient dose equivalents remain below 1 µSv for FCC-ee in Z mode and at about 10 µSv for FCC-hh for a catastrophic full beam loss in the collider. The results demonstrate compliance with the design objectives.

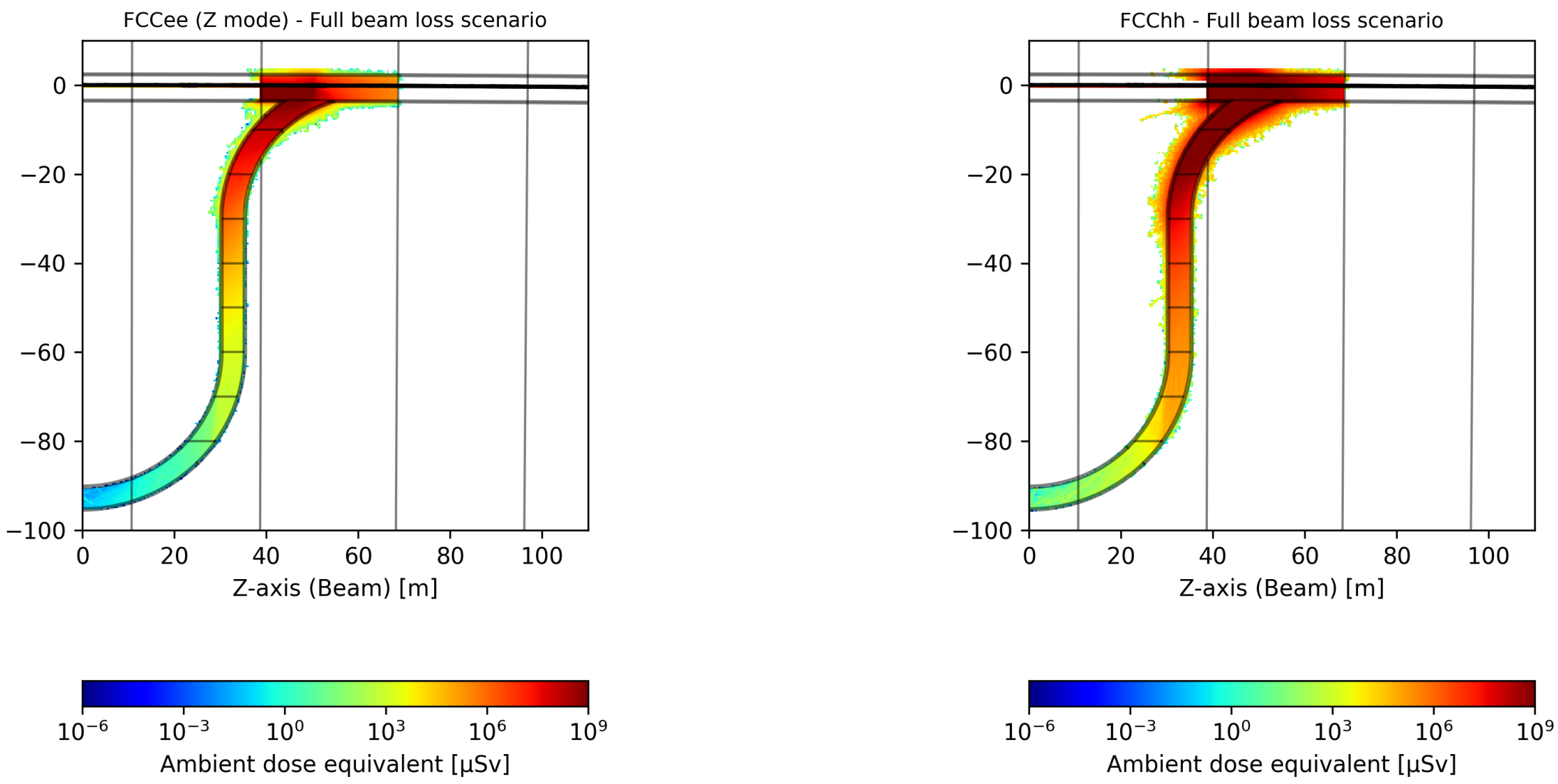

(a) (b)

Fig. 9.18: These plots show the stray radiation scattering along a bypass tunnel from a full beam loss of both beams in FCC-ee, Z-mode (beam energy $45$ GeV) (a) and FCC-hh (b).

***Fire and smoke***

Fire and smoke propagation pose a significant risk to the FCC infrastructure, particularly in underground areas. The likelihood of a fire can vary based on factors such as location, activity, layout, fire load, and occupancy levels. However, fires in confined spaces are especially dangerous due to the accelerated growth caused by thermal radiation feedback and the rapid spread of smoke.

As an outcome of the initial hazard registry analysis, the following list describes possible ignition sources or fire-specific hazards across FCC areas:

1. High and low voltage installation, including cables and power wet/dry transformers (overheat, short-circuit, arc) (For LHC, the frequency of an electrical fire was estimated as $\approx 4 \times 10^{-3}$/year [467])
2. Transport vehicles (accident, battery malfunction)
3. Hot works (welding, grinding)
4. Cryogenic systems (oxygen enrichment of air, condensed air dripping on combustible materials)
5. Pumps and mechanical devices (overheat)
6. Flammable and explosive atmospheres (in surface building only)
7. Glowing cigarette
8. Arsonist (considered as unlikely for underground areas due to access control and active surveillance)

These hazards are mitigated by means of a fire safety concept that describes the coordinated set of civil-engineering, technical and organisational measures to reduce fire risk to the level that meets safety objectives (see Section 9.2.2).

The fire safety concept follows two strategies:

– *prescriptive approaches* based on Host States laws: this *deem-to-satisfy* solution ensures that at least life safety criteria are met, and safety measures are commensurate with standard industrial fire risks. This will be the default strategy for surface buildings.
– *performance-based design*: as explained in Section 9.3.3, whenever the technical prescriptions cannot be implemented, are not appropriate or simply out of scope, the safety objectives need to be ensured by guaranteeing the safety performance of the facility in case of accidental scenarios.

The following subsections summarise the safety studies to characterise fire hazards in FCC and describe the prevention and mitigation measures included in the main areas of the infrastructure. Fire resistance, reaction to fire and organisational measures to ensure fire safety are discussed at the end of the chapter. The proposed fire safety concept will be refined in the subsequent phases of the study and updated to reflect any future changes to the baseline used in the feasibility study.

*Tunnel Arcs*

The accelerator tunnel represents the largest volume and floor area of the entire facility. Thus, specific attention has been devoted to studying how a fire will impact the evacuation and the life safety of the occupants. The entire analysis is available in a technical report [468], with a summary of the main assumptions and outcomes below.

To verify that the proposed compartmentalisation and smoke extraction strategies ensure life safety objectives, computational fluid dynamics (CFD) simulations using the NIST's Fire Dynamics Simulator (FDS) (v6.8.0) [469] have been performed for a complete $400\,\mathrm{m}$ long compartment.

The tenability limits below (based on ISO 13571 [470]) define the performance criteria used throughout the study:

– Visibility of the evacuation path (at $2\,\mathrm{m}$ height) $< 10\,\mathrm{m}$
– Fractional Effective Dose (FED) $> 0.3$
– Temperature $> 60\,^{\circ}\mathrm{C}$

Additional criteria are analysed for further comparison among trial designs such as: FED$> 0.1$, hot smoke layer temperature of $200\,^{\circ}\mathrm{C}$, smoke layer spread velocity and heat flux $> 2.5\,\mathrm{kW\,m^{-2}}$.

The time to reach these untenable conditions constitutes the 'Available Safe Egress Time' (ASET). The 'Required Safe Egress Time' (RSET) is defined as the time required for the occupants to leave the compartment. The comparison of both times provides the safety margin used for the analysis of life safety criteria, as illustrated in Fig. 9.19. The ASET > RSET condition is a safety requirement for all scenarios studied.

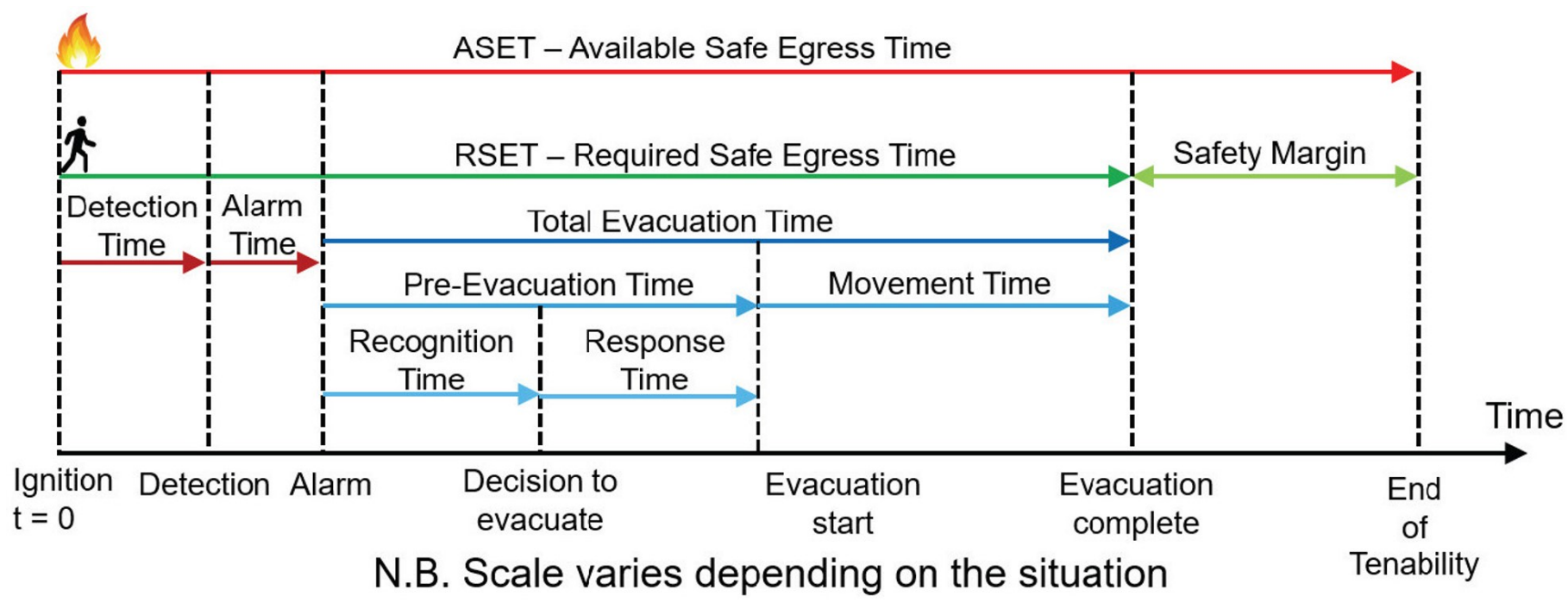

Fig. 9.19: Illustration of the different time stamps and input parameters to evaluate the ASET and RSET, starting at the ignition ($t = 0$).

The baseline pre-movement time and walking speed are taken from BS PD 7974-6 [471] and set as 30 s and $1.2\,\mathrm{m\,s^{-1}}$, respectively. The 90th (pre-movement) and 10th (walking) percentiles of the standard distribution associated to those values (i.e., 120 s and $0.8\,\mathrm{m\,s^{-1}}$, respectively) have also been used to explore slower responses. Table 9.5 summarises the list of input parameters required to perform the evaluation.

Table 9.5: Input values to perform the ASET-RSET analysis. Occupants are considered to be awake and familiar with the premises - Cat. A, and assumed to be trained to a high level of safety management - level M1 (ensured by appropriate training before access to underground facilities is granted (see Section 9.4.5). Whilst the complexity of the building is considerable, having a wayfinding system (see Section 9.4.1) and only two possible routes (left or right) allows the selection of a building level B2. Level A1 is considered since a robust detection system is to be installed across underground premises. The occupants' location is relative to the location of the fire.

| Parameter | Value | Justification |
|---|---|---|
| Detection time | 120 s | Performance requirement for such fire |
| Pre-movement time | 30 s (120 s) | BS PD 7974-6 [471] |
| Ventilation ramp up | 0 - 60 s | Ventilation design [439] |
| Walking speed | $1.2\,\mathrm{m\,s^{-1}}$ ($0.8\,\mathrm{m\,s^{-1}}$) | BS PD 7974-6 [471] |
| Occupant's location | -100 m / 0 m / +150 m | Credible scenarios |

Several design fire scenarios were developed during the conceptual study phase, during brainstorming sessions conducted with accelerator safety experts and using reference data from the literature [472]. Among these, a fire developing from a transport vehicle loaded with pallets was shown to be the worst-credible scenario with respect to life safety in the tunnel. The heat release rate (HRR) curve obtained shows a medium-fast growing phase that reaches a first peak of 4 MW in 8 min. After a short dip, the HRR continues to grow up to a 8 MW peak reached after 17 min, as shown in Fig. 9.20.

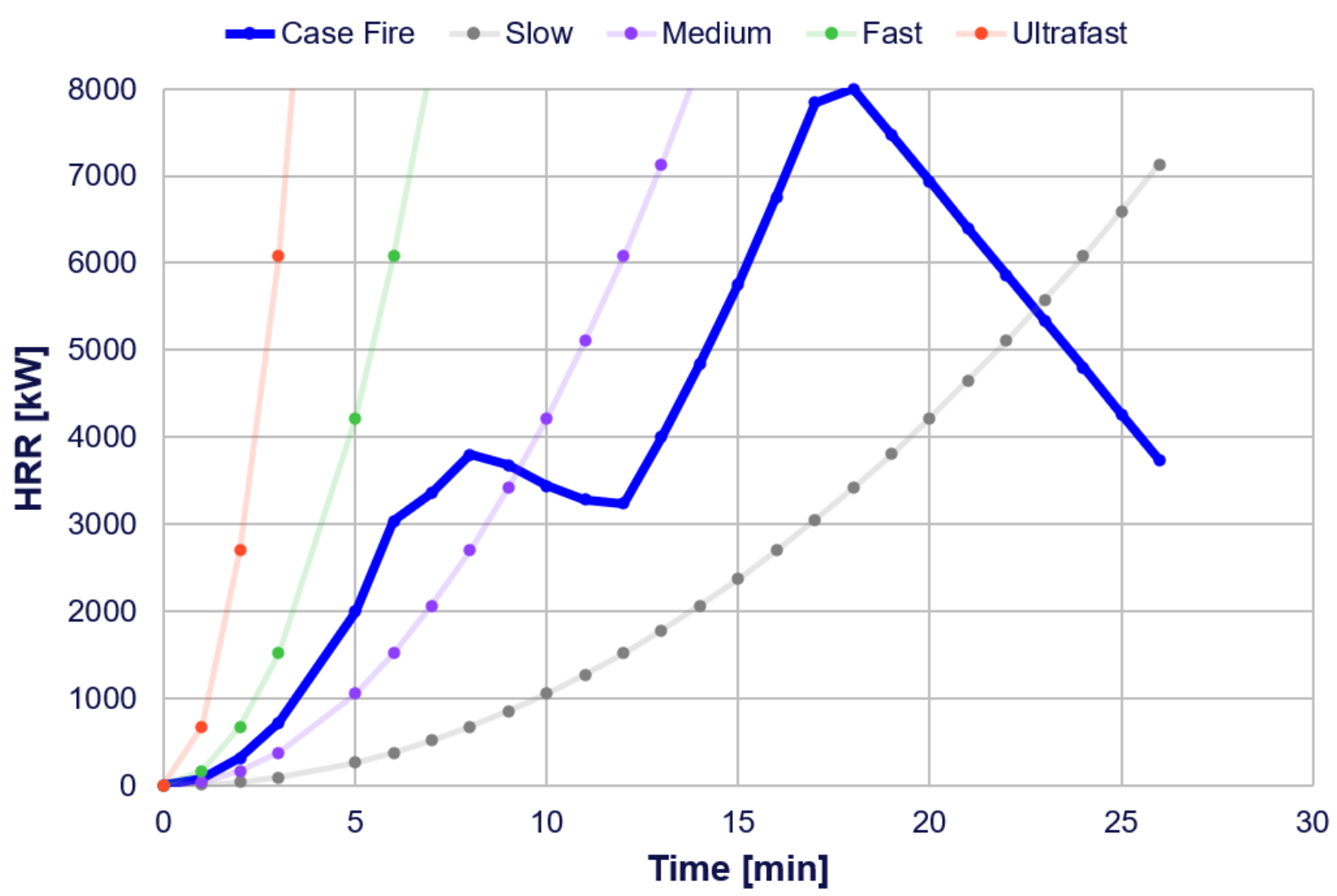

Fig. 9.20: Fire scenario of transport vehicle considered for PBD assessment against the life safety objective in the tunnel arc. Standard $t^2$-curves (ISO 16733-1 [473]) are also depicted for comparison (from slow to ultra-fast growth). The analysis is stopped after 27 min to match the simulated time relevant for life safety.

The study also investigated the two main ventilation strategies proposed above as well as the impact on the final available safety margin time (ASET − RSET) of several different design criteria such as: smoke extraction flow, fresh air supply, door behaviour, delay times, background initial flow. All the cases (nominal and degraded) are described in a specific technical report [468].

The performance criteria are examined using time-position plots for all scenarios studied, as illustrated in Figs. 9.21 and 9.22. The evacuation time-position profiles (extensively used in tunnel fire safety [474, 475]) allow the ASET $>$ RSET condition to be verified as well as quantifying the safety margin when the occupants reach the end of the compartment.

The analysis of all scenarios confirmed that both of the proposed smoke extraction concepts allow enough time to safely evacuate the compartment under nominal conditions. Some degraded modes (i.e., no detection, and therefore no automatic activation of any action) do not meet ASET $>$ RSET condition.

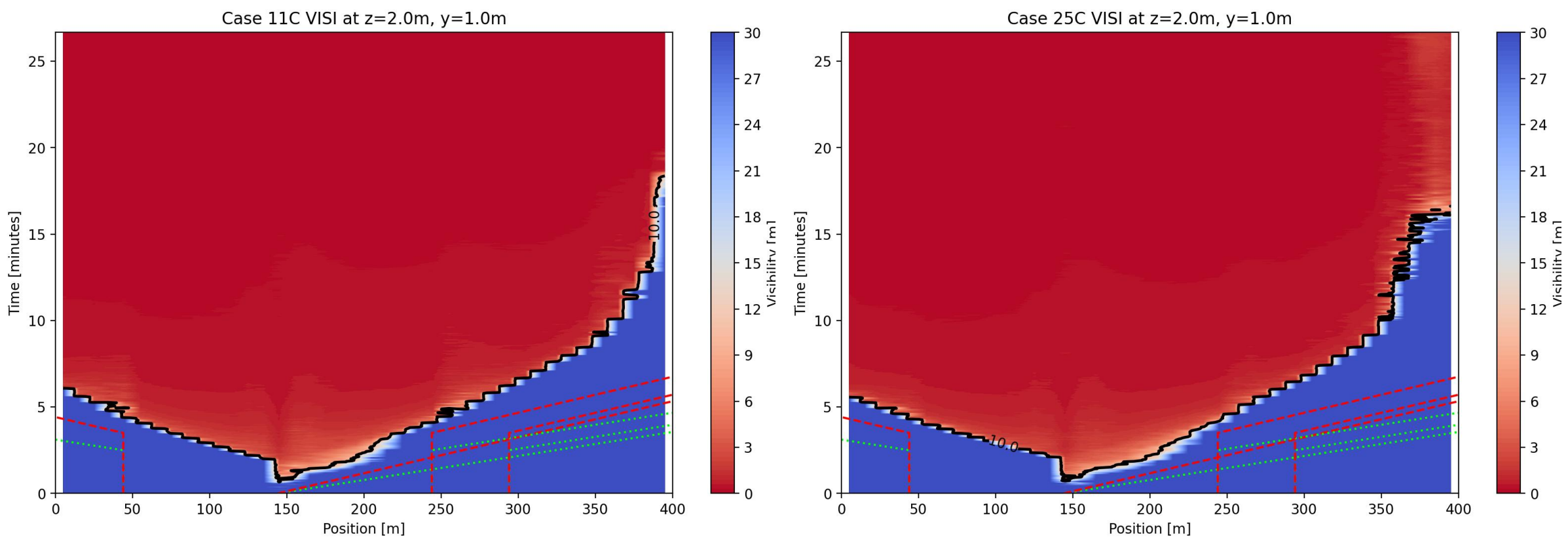

(a) Semi-transverse Case 11C [468]

(b) Longitudinal Case 25C [468]

Fig. 9.21: Visibility time-position plot for semi-transverse (ST) and longitudinal ventilation (LT) strategies. The $10\,\mathrm{m}$ visibility threshold is clearly depicted as well as different possible evacuation pathways (shown as dashed lines) for reduced and nominal human behavioural cases. Ventilation vector: left-to-right. Both strategies ensure that occupants can evacuate before reaching untenable visibility conditions; ST is the most performant option, namely downstream the seat of the fire.

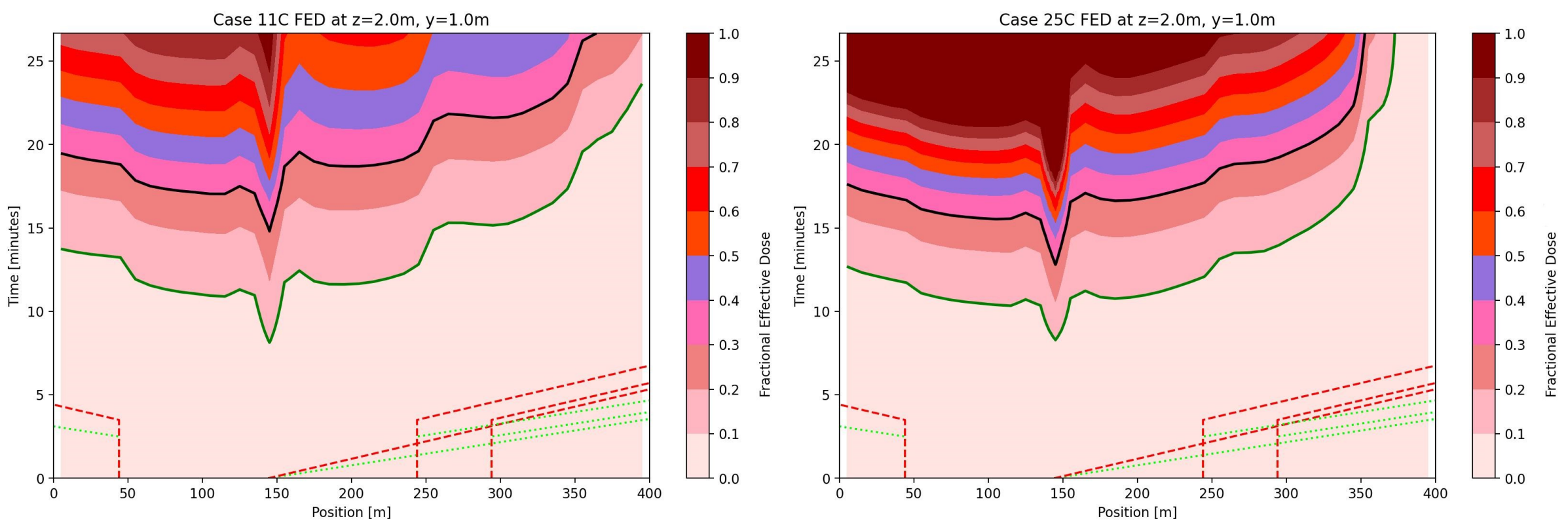

(a) Semi-transverse Case 11C [468]

(b) Longitudinal Case 25C [468]

Fig. 9.22: Fractional effective dose (FED) plot for semi-transverse (ST )and longitudinal (LT) ventilation strategies. Ventilation vector: left-to-right. The FED 0.1 and 0.3 thresholds are clearly depicted, as well as evacuation lines for reduced and nominal human behavioural cases.

Finally, the safety margin for nominal and reduced speed evacuees provided by nominal conditions (no degraded modes) of cross-sectional and longitudinal smoke extraction strategies are compared in

Fig. 9.23. Several extraction capacities from 0 to 25 000 $m^3$/h are also explored. The semi-transverse solution shows better smoke-sweeping capacity, providing a better margin for the same extracted flow. The comparison also demonstrates that the absence of smoke extraction induces a negative safety margin, i.e., the occupants are unable to reach the compartment due to the lack of visibility.

In conclusion, from a safety point of view, the dynamic confinement offered by the semi-transverse scheme is found to be more efficient and robust (including degraded modes) for the safe evacuation of occupants in case of an incident. Thus, for this feasibility study phase, the semi-transverse option is considered as the baseline for the FCC tunnel.

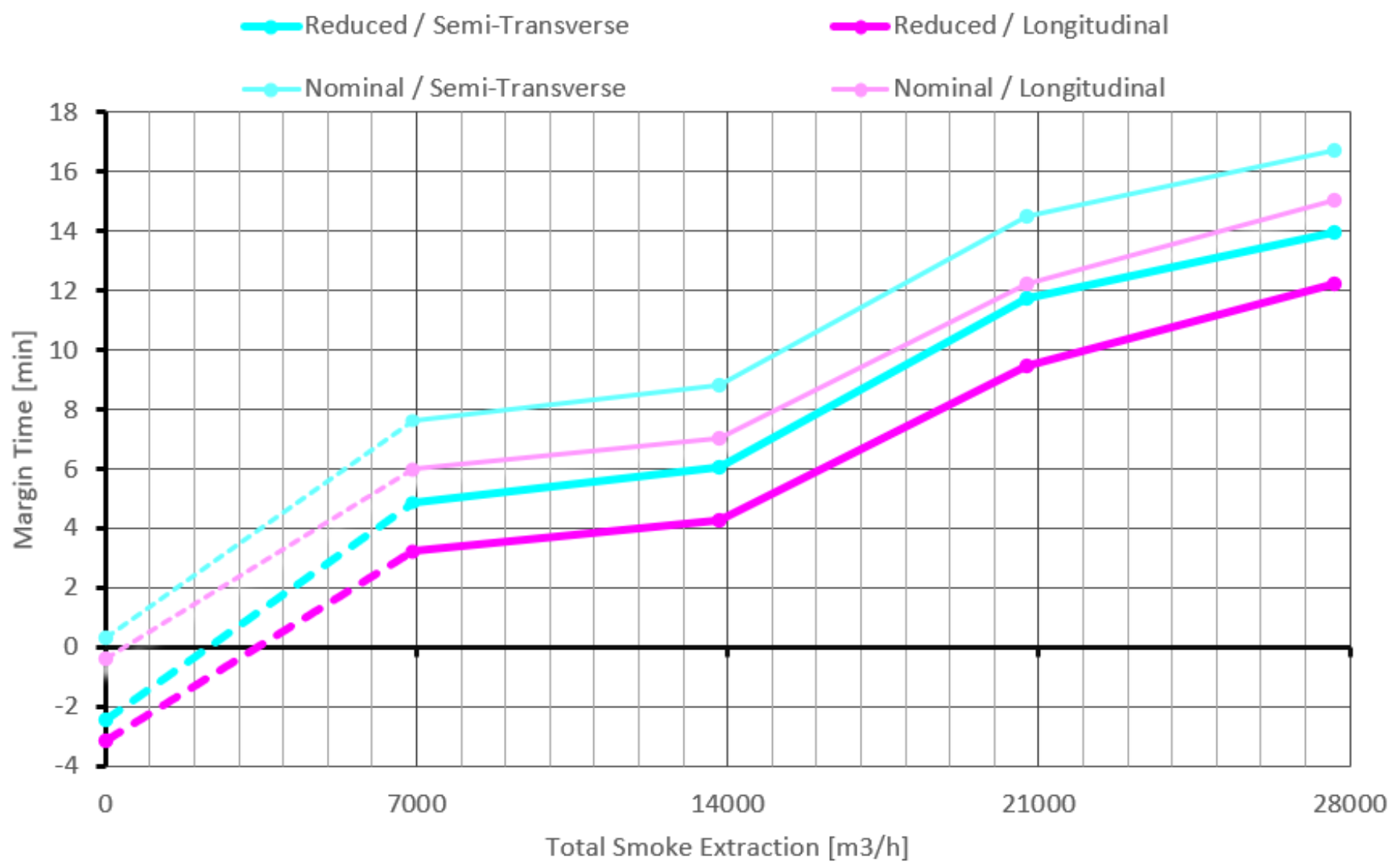

Fig. 9.23: Safety margin (ASET-RSET) available for reduced (120 s pre-movement time and $0.8\,\mathrm{m\,s^{-1}}$) and nominal (30 s pre-movement time and $1.2\,\mathrm{m\,s^{-1}}$) evacuees and different smoke extraction (and supply) flows.

*Shafts and safe areas*

Within the shaft and their waiting areas (see Section 9.4.4), the fire risk will be limited to fire loads stemming from the safety systems themselves. These areas are critical for safe evacuation and must not be compromised; hence, no storage of combustibles is allowed. As mentioned above, the pressurised volume and an EI120 compartment ensure that these areas are free of smoke at all times.

Conversely, the transport zone of the vertical shaft poses a significant fire risk. Due to its chimney-type configuration, a fire at the bottom of the shaft will quickly propagate upwards, making it difficult to reach and extinguish. To mitigate this, additional fire safety measures should be considered, such as cable enclosures, fire protective barriers at regular intervals, etc. Although this scenario does not pose a major threat to life safety (since the transport zone is separated from the lift shaft), it is important to consider how safety systems are protected at the surface. In a later stage of the study, the impact on property protection and continuity of operation of such scenarios will be considered in a cost-benefit analysis to identify the most appropriate measures.

*Experiment caverns*

Fire risk in experiment caverns is primarily associated with the large quantity of data and power cables feeding the detector and the ignition sources and combustibles integrated within its sub-systems.

The reaction to the fire of cables will be controlled to limit the energy, smoke, and acidity production and all the cable trays exiting in the experiment caverns (via the connection galleries) must be sealed and covered with intumescent paint to prevent propagation. In localised areas with vertical dense cable tray arrangements, automatic extinguishing systems (e.g., water mist) will be considered as a compensatory measure for property protection and business continuity.

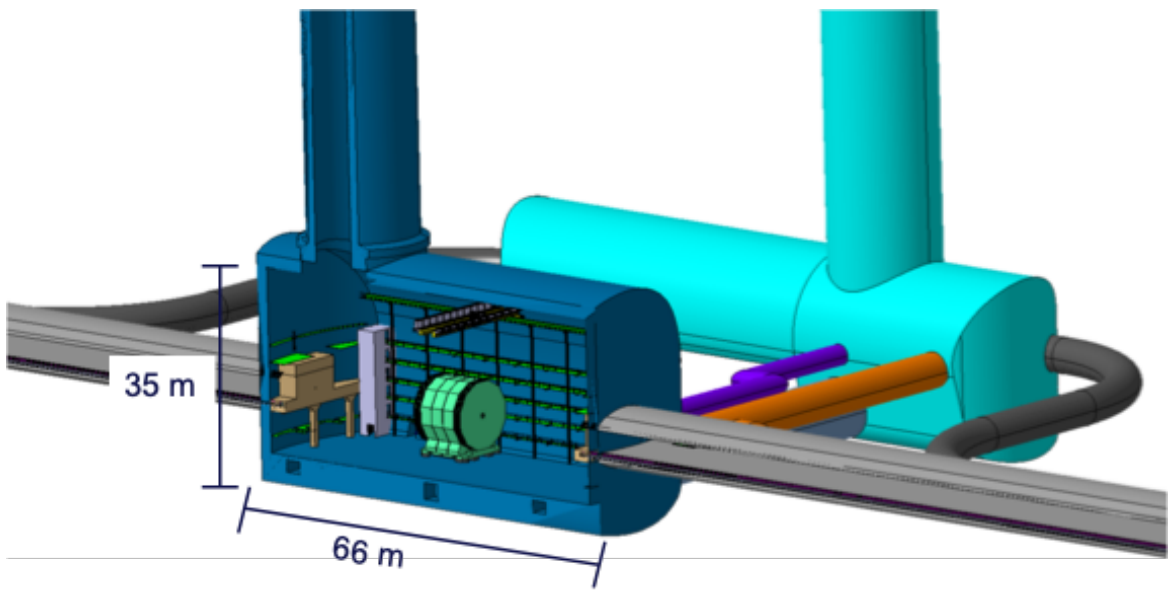

(a) FCC-ee

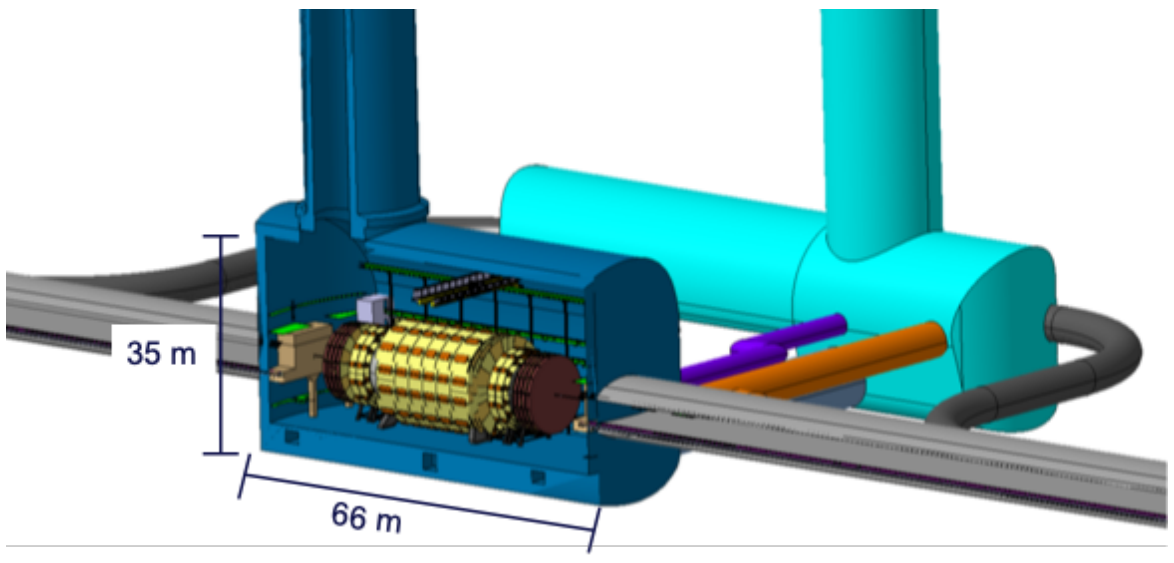

(b) FCC-hh

Fig. 9.24: Illustration of the FCC detectors at the experiment point PG.

Regarding the detector, a dedicated risk assessment will be conducted for each detector subsystem, identifying the most appropriate compensatory measures to maintain the fire risk within acceptable limits. The LHC detectors currently employ several effective safety measures to mitigate the fire risks, including:

– **Inert nitrogen atmosphere**: the inner detector volumes are filled with nitrogen to eliminate ignition sources in combustible areas.
– **Inert gas extinguishing systems**: flooding systems are in place to suppress potential fires in outer layers, with manual activation available from the control room.
– **Dedicated detection systems**: localised detection systems monitor for anomalies and can trigger power cuts to prevent malfunctions. These systems utilise inner temperature monitoring and multi-parameter gas sampling to ensure precise oversight within the detector.
– **Over flooding foam system**: a water-based foam system allows flooding of the cavern in case of extensive fire. This system is not aimed at life safety and might be destructive to the detector electronics. Technical and cost-benefit analyses are required to evaluate the appropriateness for the FCC detectors.

From a life safety standpoint, the lower-level safe-passage connections to protected areas in the service cavern (purple galleries in Fig. 9.24) allow quick evacuation in case of fire. Due to the volume of the FCC-ee detector compared to the large dimensions of the cavern and its shaft (Fig. 9.24), smoke extraction is not required for life safety. This was closely studied for ATLAS and CMS in the LHC [440, 441] and remains valid for the FCC experiment caverns at this stage. A smoke extraction system is still necessary to ensure safe and efficient fire-fighting intervention, as well as to reduce potential property loss in case of fire.

*Service caverns*

The service caverns house the majority of the technical infrastructure to feed the underground facilities. They are areas with significant fire loads and ignition sources, such as power transformers, electrical racks, and computing rooms, which must be enclosed within equipment-specific EI120-rated fire compartments. The proximity to pressurised safe areas and the efficient vertical means of evacuation (shafts) ensure the life safety of the occupants, as verified by the studies performed for the LHC [440, 441].

The evacuation passages from the experiment area and the main tunnel are also completely enclosed within an EI120 fire-rated over-pressurised compartment and will not be impacted by any fire in the service cavern.

A cost-benefit analysis will determine if dedicated extinguishing systems, such as inert gases or water mist, will be needed to protect property in critical areas such as computing rooms or power transformer zones.

*Klystron galleries*

The klystron galleries concentrate high electrical energy and combustible loads comprising power cables and the klystrons themselves. While dry technologies exist, the infrastructure is being designed for the possibility of hosting oil-filled klystrons. These galleries are equipped with a ceiling-mounted emergency extraction system, similar to the main tunnel, and evacuation distances are limited $< 200\,\mathrm{m}$ (i.e., $400\,\mathrm{m}$ in between two exits). Several pressurised connection staircases to the main tunnel, are distributed along the gallery, ensuring the evacuation of occupants in case of fire (see below).

*Alcoves*

Alcoves will contain large amounts of active equipment needed for the operation of the accelerator in a small volume: transformers, control and power racks, electrical components, etc. They are considered as high fire risk areas due to the combination of fire load with multiple electrical ignition sources. Hence, they must be compartmentalised with respect to the accelerator tunnel, with a fire door, fire dampers and sealing of all services which cross.

The maximum combustion time of a $1\,\mathrm{MW}$ fire in the alcoves is $\approx 1\,\mathrm{h}\;30\,\mathrm{min}$. This is considering the relatively small and enclosed volume ($\leq 40\,\mathrm{m}^3$) and the fact that fire cannot be sustained below $13\%$[12] $O_2$ concentration [476]. Hence, the fire rating for the alcove compartments is EI120, considering that flashover conditions cannot be discarded at this stage.

The racks and services feeding the safety systems of the tunnel arc will be installed in the alcoves. This equipment will be contained in separate volumes within the alcoves, protected with a dedicated fire compartment. This limits the risk of compromising those safety systems in case of a fire inside the alcove.

The evacuation distance to reach the fire door at the entrance of the alcove remains limited (below $40\,\mathrm{m}$), so the occupants can quickly reach the adjacent compartment in the main tunnel.

The smoke extraction ducts installed in the alcoves are connected to the tunnel smoke extraction system with fire-rated dampers. As mentioned above, smoke extraction from the alcoves is deemed not necessary for life safety but is important for the emergency intervention teams, property protection and business continuity in case of a fire.

In some specific alcoves, oil-filled transformers might be installed. In these cases, localised extinguishing systems (e.g., $CO_2$, inert gas, automatic powder extinguisher, water mist) will be considered case by case to mitigate the impact on property protection and continuity of operations to an acceptable level. Moreover, the rail-mounted robotic intervention solution that is planned for the tunnel areas might not be suited to the alcove geometry and justifies further study for automatic extinguishing of high-fire risk-specific equipment.

*Surface buildings*

The surface buildings will house most of the underground technical systems. The fire load is considered to be present in the form of electrical cables, racks, power transformers, and other industrial equipment (oil compressors, pumps, fans, etc.). Prescriptive approaches for industrial facilities as per Swiss AEAI [443] and French Code du Travail [442] provide the design requirements to ensure life safety goals. In addition to the minimum host state requirements, areas considered as high-risk (with high fuel load density or ignition sources) will be equipped with automatic fire detection systems, enabling the prompt detection and activation of first responders. In such cases, the evacuation alarms will also be automatically triggered.

[12] Conservative value to account for hydrocarbon fuels

*Reaction to fire*

The underground inner lining will be made of non-combustible materials (A1/A2 as per EN 13501-1 [477]) and will not increase the fire load in the facility. Partitions must also be non-combustible (including sandwich panels, false ceilings/floors, and finishing). This is in line with the requirements for underground transport tunnels [433, 478, 479].

The combustibility of material contained in the premises will be strictly controlled to limit combustibility and the consequences of any ignition. CERN Safety Instruction IS 41 [480] limits the spread of flames, droplets behaviour and smoke production of all materials to a minimum or low contribution to fire (i.e., A,B and C Euroclasses [477]) and prevents the use of halogenated materials. Whenever these provisions cannot be met, a dedicated risk analysis is performed to determine compensatory measures to limit such materials.

Specific attention is to be paid to the reaction to fire of cables. CERN Specific Safety Instruction SSI-FS-2-1 [481] already requires the installation of cables that are certified to, at least, Cca-s1,a2,d1 class or equivalent. Cca-s1,a2,d1 [452] must be retained as the minimum class for cables in the FCC, as it implies:

– Limited self-propagation.
– Low thermal energy released.
– Low smoke production.
– Limited fire growth rate.
– Non-acidity of gases (i.e., halogen-free).
– No sustained flaming droplets (>10 s).

*Fire resistance*

In the event of fire, the integrity of the underground structure must be maintained for the period of time required to:

– Ensure self-rescue and complete evacuation of occupants.
– Allow safe intervention of emergency responders.
– Limit tunnel damage and recovery time.

This fire-resistant requirement must be applied to the civil structures as well as to the technical systems that need to operate in case of fire (e.g., cables, cables trays, ventilation ducts). It will be evaluated using natural curves[13] and engineering methods. Natural curves will be developed with the worst possible scenarios impacting the infrastructure (localised and distributed fires) and will consider all phases (growth, developed fire and decay). It is important to note that these natural curves might not be the same as used in Section 9.4.3 for life safety assessment due to the main safety objective being assessed: a critical scenario for evacuation (a rapid growing fire) is not necessarily the most critical scenario for structural resistance (long-lasting fire). The resistance assessment must consider the serviceability limits for the load bearing and spalling requirements to ensure that the above-mentioned objectives are met, despite the absence of extended collapse of the structure. For convenience, the final resistance can be expressed as per ISO 834 [430] using a time-temperature equivalent method as available in [482] and [483].

[13] Temperature-time curve that describes the behaviour of a fire in an uncontrolled or natural environment, without external intervention or standardised conditions. This curve is derived from real fire scenarios and reflects the actual growth, peak and decay phases of a fire.

Table 9.6 highlights the minimum prescriptive requirements for fire resistance used in the Host States, as well as in other particle accelerator infrastructures. The risk levels N1, N2, and N3 for French road tunnels are determined by the potential impact of structural collapse on both the infrastructure and its surroundings [484]. In particular, risk level N3 is indicated for immersed tunnels and when the risk of local collapse can represent a major risk of flooding. While the corresponding prescriptive fire resistance (R240 + HCM120) appears disproportionate for FCC due to the lack of comparable fire loads, the risk of water intake as a consequence in case of local collapse will be studied in the next phases of the project.

Table 9.6: Overview of minimum fire resistance prescribed in underground tunnels. R refers to fire resistance as per ISO 834 curve and EN 13501-2 [431]. RWS/HCM refer to the normalised hydrocarbon time-temperature curves with faster growth and higher temperatures than the ISO-834 curve as depicted in Fig 9.25a. $q$ stands for fire load density. Clearance is the maximum allowable height of the tunnel.

| **Source** | **Fire Resistance** [m] | **Comment** |
|---|---|---|
| FR road tunnels [478] | R60 | If clearance $< 3.5$ m for all risk levels |
| | R120 | If clearance $> 3.5$ m for risk levels N1 |
| | HCM120 | If clearance $> 3.5$ m for risk levels N2 |
| | R240 + HCM120 | If clearance $> 3.5$ m for risk level N3 |
| FR railway urban tunnels [485] | R120 | for public transportation |
| CH Industrial Building [486] | R60 (R90) | Height $> 11$ m $< 30$ m (if $q > 1200$ MJ/m$^2$) |
| | R90 (R120) | Height $> 30$ m $< 100$ m (if $q > 1200$ MJ/m$^2$) |
| CH road tunnels [487] | R60 | For lightweight vehicles |
| | RWS/HCM 120 | If heavy goods |
| CH railway tunnels [488] | HCM 120 | If submerged or below water tables |
| | R120 (RWS/HCM 120) | For unstable tunnel (if >trains/day) |
| | R120 | For stable ground |
| XFEL [436] | R90 | For shaft and underground infrastructure |
| HL-LHC [489] | R60 | tunnel |
| | R120 | safe areas and protected shaft |

As indicated above, the proposed fire resistance for underground infrastructure expressed as per ISO 834 is:

- R90 for tunnel and experiment caverns.
- R120 for alcoves, service caverns and areas with specific fire load.
- R120 for safe areas and protected shafts.

The different fire scenarios studied in the tunnel areas show that even in degraded mode, temperatures remain below 300 °C for a large part of the tunnel (Fig. 9.25b). Close to the seat of the fire, the maximum recorded air temperatures for most unfavourable scenarios remain below 600 °C. This is substantially below the ISO 834 curve (Fig. 9.25b), demonstrating that the proposed fire resistance ensures the life safety of occupants during the evacuation phase. The next phase of the study will evaluate whether greater resistance is needed to ensure the objective of property protection and business continuity.

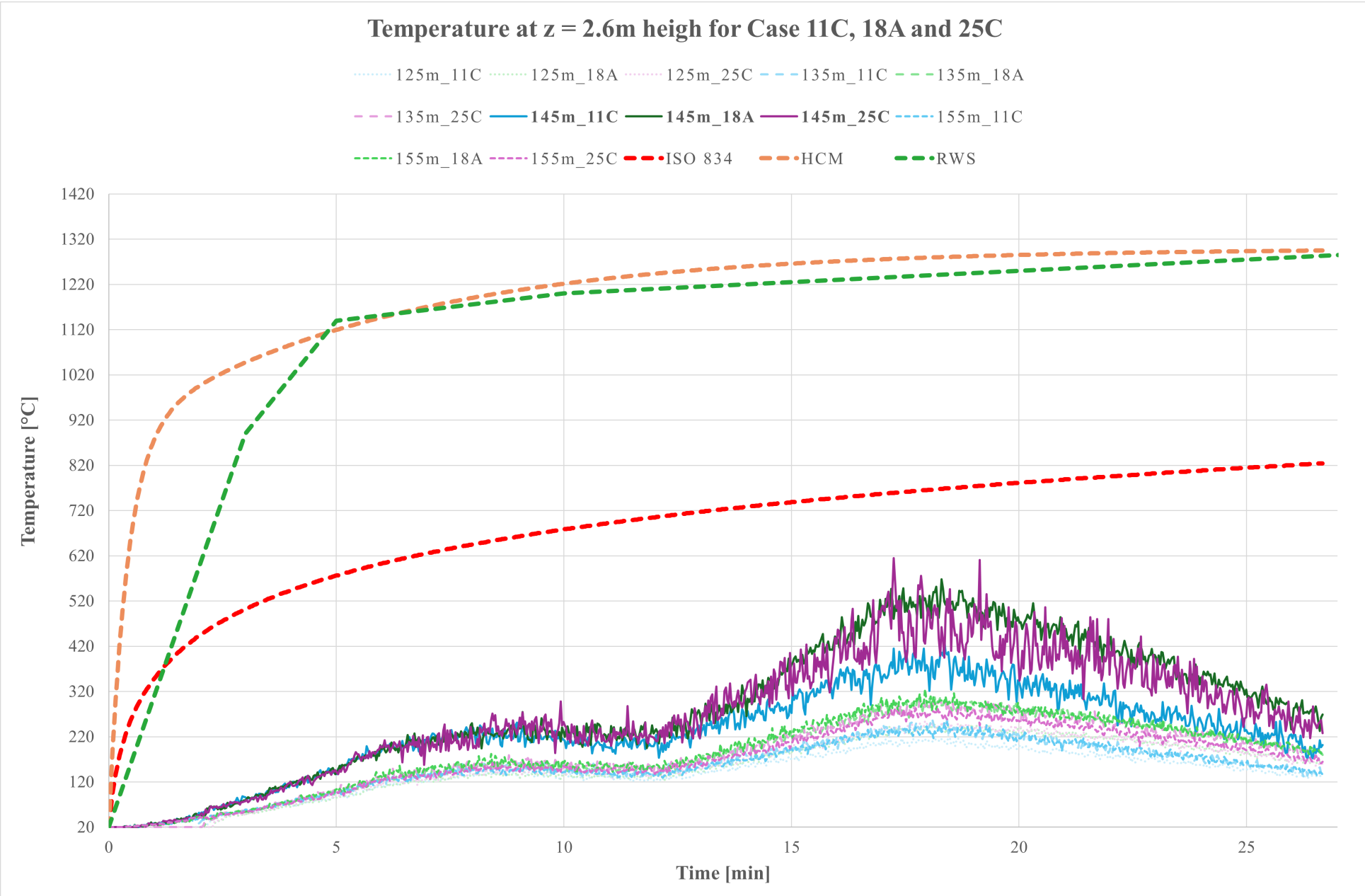

(a) Temperature evolution for transport vehicle fire with three different ventilation conditions [468].

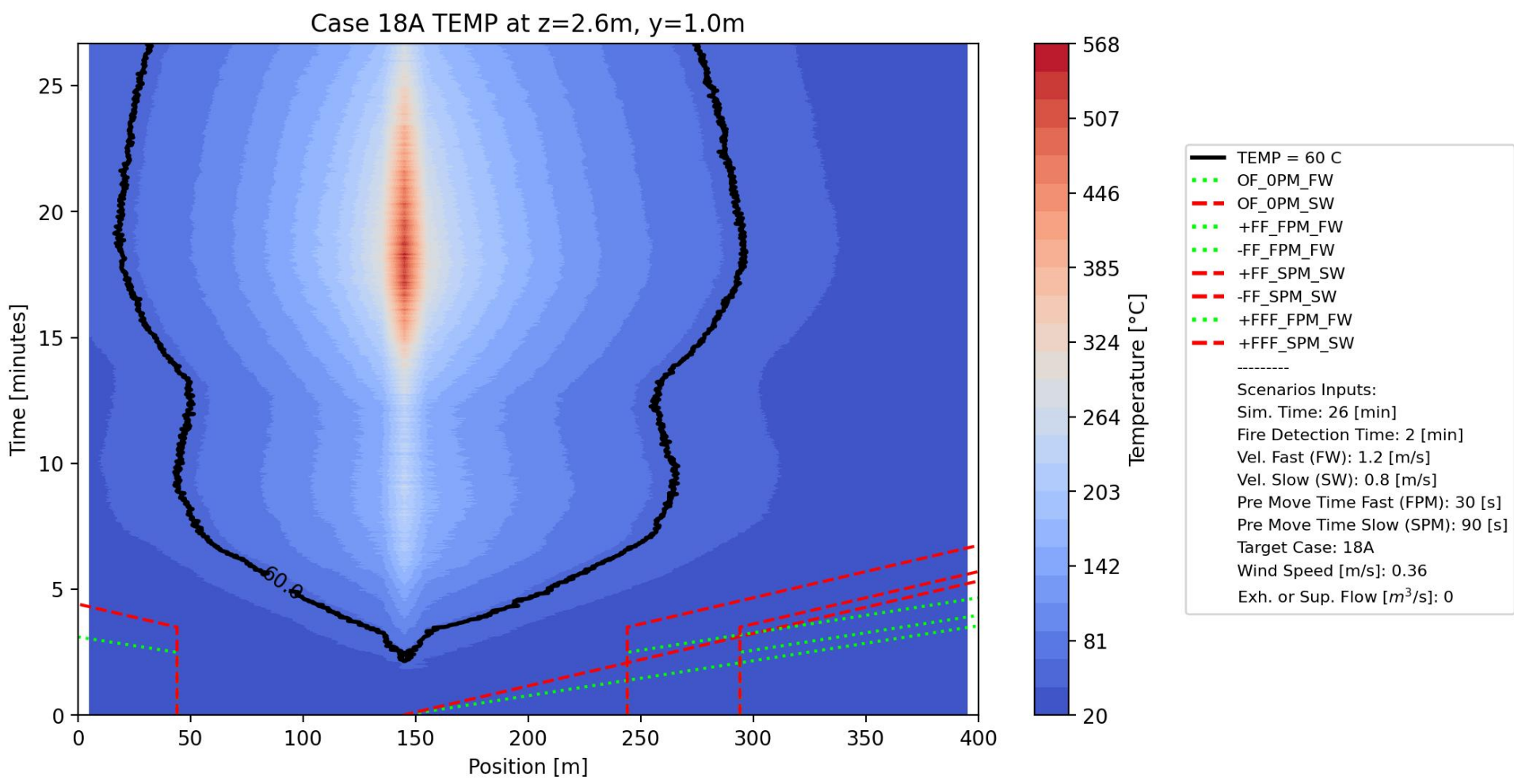

(b) Temperature time-position plot for case 18A [468]. Ventilation vector: left-to-right. Air/smoke temperature is measured at 2.6 m from the ground. This case represents a failure of the smoke extraction system and maximises the thermal impact to the tunnel for the vehicle fire design. The black line corresponds to the 200 °C threshold. The red and green dashed/dotted lines are evacuation trajectories for worst and nominal human behaviour cases.

Fig. 9.25: Temperature profiles in the case of a fire in the tunnel.

*Organisational measures for fire prevention*

The following general requirements to prevent fires and limit their consequences will be implemented:

- All hot works are to be conducted under a fire permit approval stipulating preventive and compensatory measures to be put in place (most of them are already implemented in the current CERN safety policy);
- Smoking will be strictly forbidden in all buildings at CERN, including underground areas.
- Fire safety awareness must be part of the access training ranks. This shall include instructions on prompt manual alarm triggering, behavioural reflex, and first response training.
- Maintenance and control. Patrol (automated) will check for unusual cues (hot spots, improperly placed material).
- Regular tests and fire drills. Fire safety systems must be permanently monitored and checked. Unplanned evacuation drills will regularly verify that the global safety concept is successfully implemented and if there are discrepancies, propose actions and update the concept.

***Oxygen deficiency hazard (ODH)***

The SRF cryomodules of the FCC-ee rely on liquid helium to reach their superconducting state. The 400 MHz cavity cryomodule will be cooled using $115\,\mathrm{kg}$ of helium at $4.5\,\mathrm{K}$ (He-I), whereas the 800 MHz cryomodule will use $55\,\mathrm{kg}$ of superfluid helium (He-II) at $2\,\mathrm{K}$. Following a risk assessment [490], a few accident scenarios were identified as potential sources of helium release in the FCC tunnel. Such a release poses considerable risks for people working underground. The helium gas released into the environment would displace the air (i.e., the oxygen) and lead to a possible asphyxiation of occupants underground; this is referred to as an Oxygen Deficiency Hazard (ODH). Moreover, low-temperature helium flow can also cause severe internal/external cold burns. A performance-based design approach, using numerical simulations in the form of computational fluid dynamics (CFD), provided an analysis of the ODH in the RF section of the FCC-ee accelerators. The outcome of the safety studies is summarised below, based on a series of reports that contain the full details [491, 492].

For simulations, a full fire compartment ($400\,\mathrm{m}$) in the RF sector at point PH was selected as the control volume. The CFD simulations were performed in two dedicated batches, corresponding to the two accident scenarios chosen for the analysis: 1) sustained RF quench and 2) loss of beamline vacuum. The FCC-ee SRF system will be designed to avoid a sustained RF quench, leaving the loss of beamline vacuum to be considered as the most credible incident (MCI) at this stage, with a mass flow rate of $20\,\mathrm{kg\,s^{-1}}$ [490]. The scenarios studied are summarised in Table 9.7. The case and mesh settings for the CFD solver are detailed in [491, 492].

Table 9.7: List of simulation scenarios and the main input parameters.

| Ref. | Scenario | Relief points [per CM] | Burst disc $\phi$ [mm] | Mass flow [$\mathrm{kg\,s^{-1}}$] | Pressure & Temperature [bar(a)] / [K] | Sim time [s] |
|---|---|---|---|---|---|---|
| SC1 | Sustained RF quench | 1 | 100 | 2.9 | 2.0 / 5.32 | 57 |
| SC2 | | 2 | 100 | | | |
| SC3 | | 4 | 50 | | | |
| SRF01 | Loss of Beamline Vacuum | 1 | 100 | 20 | | 26 |
| SRF02 | | 2 | 100 | | | |

The simulations were run on CERN's HPC cluster and post-processed, focussing on analysing the following results:

- Comparison between input scenarios.
- Dynamic behaviour of the oxygen levels in the cross-section of the tunnel.

– Time needed to reach ODH conditions in the evacuation path.
– Propagation speed of the Helium cloud front.
– Propagation speed of the Helium Plug [14].
– Longitudinal size of the Helium Plug.

The findings are summarised in Table 9.8, showing the time at which untenable conditions are reached in the transport/evacuation zone (i.e., $O_2 \leq 18\%$), the propagation speed of the helium cloud front, and the behaviour of the helium plug formed in the tunnel. This includes scenarios where gas is still being relieved from the burst disk ($\dot{m}_{He} \neq 0$) and when the inventory is empty ($\dot{m}_{He} = 0$), along with the size of the corresponding plug at the end of the simulation. Figure 9.26 illustrates the results of the cross-section and longitudinal propagation of the helium cloud.

Table 9.8: Results of the CFD simulations with the scenarios described in Table 9.7.

| Ref. | Time to reach 18% $O_2$ [s] | Propagation speed [m/s] Cloud front | He plug ($\dot{m}_{He} \neq 0$) | He plug ($\dot{m}_{He} = 0$) | He plug size [m] |
|---|---|---|---|---|---|
| SC1 | 7 | 1.52 | 0.66 | 0.58 | 32 |
| SC2 | 10 | 1.33 | 0.57 | 0.7 | 29 |
| SC3 | 15 | 1.66 | 0.63 | 0.81 | 35 |
| SRF01 | 4 | 3.3 | 5.3 | 3.3 | 73 |
| SRF02 | 2 | 2.57 | 3.8 | 2.57 | 53 |

The simulations confirm the turbulent flow effects observed in real-life scenarios [493]. Untenable conditions are reached as soon as 2 seconds after the opening of the pressure relief device, and after an additional 4 seconds the cloud has propagated another 12 m downstream. The turbulent flow leads to the creation of a helium plug, which was observed in all 5 scenarios, right next to the release point and extending up to 70 m in length within 26 seconds (Table 9.8). The plug can obstruct the entirety of the transport zone, posing a heavy constraint for the evacuation of the occupants towards the adjacent compartment where they would retrieve the transport vehicle to evacuate from the underground area.

Given these observations, with the cryomodules at nominal conditions (i.e., maximum inventory), the access to the RF section of the tunnel must be blocked when the risk of such accident events is present. Access to the klystron gallery is allowed, however, the bore holes for the wave guides need to be leak tight. Additional future studies may determine if there is a minimum inventory where access to the RF sector would be possible. In this case, occasional access might be granted with prior approval (i.e., procedures) and by requesting the use of personal protective equipment (PPE), such as self-rescue masks. It is intended to extend these studies, for the further design of the FCC study to include the use of the emergency (smoke) extraction duct and measure the impact on the extent of the helium plug and cloud propagation, in view of iterating on the access conditions to the RF sector.

It is also preferable to have a cryomodule design with multiple release points, although a single release point might still be studied in more detail to improve the outcome compared to SRF01.

The ODH assessment for the experiment caverns will be performed in the next phase of the study, when the detector technologies are chosen and which cryogenic gas will be utilised (if any).

#### *Seismic hazard*

The region is characterised by a moderate seismic level. With 11 earthquakes of magnitude > 5 estimated to have occurred within a radius of 100 km around Geneva in the last 500 years the hazard is considered

[14] Area of the tunnel that is completely filled with helium gas

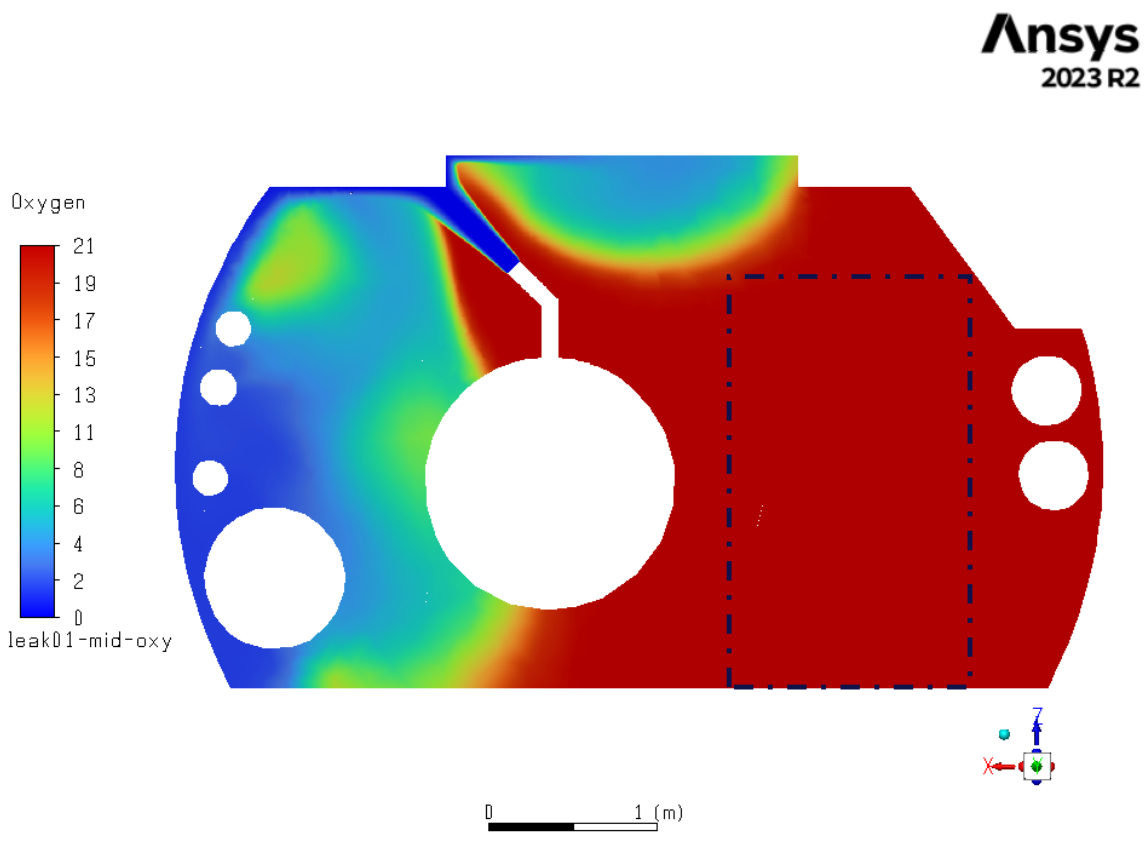

(a) $t = 1$ s

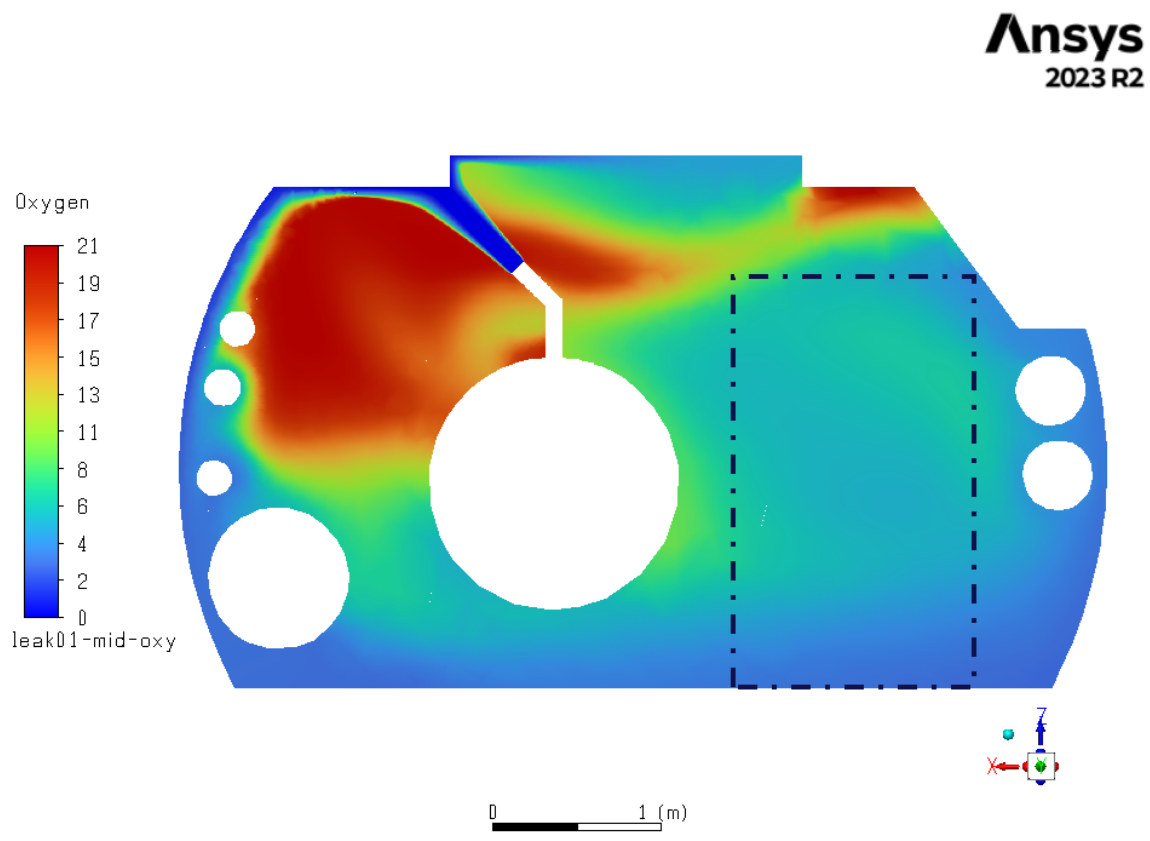

(b) $t = 3$ s

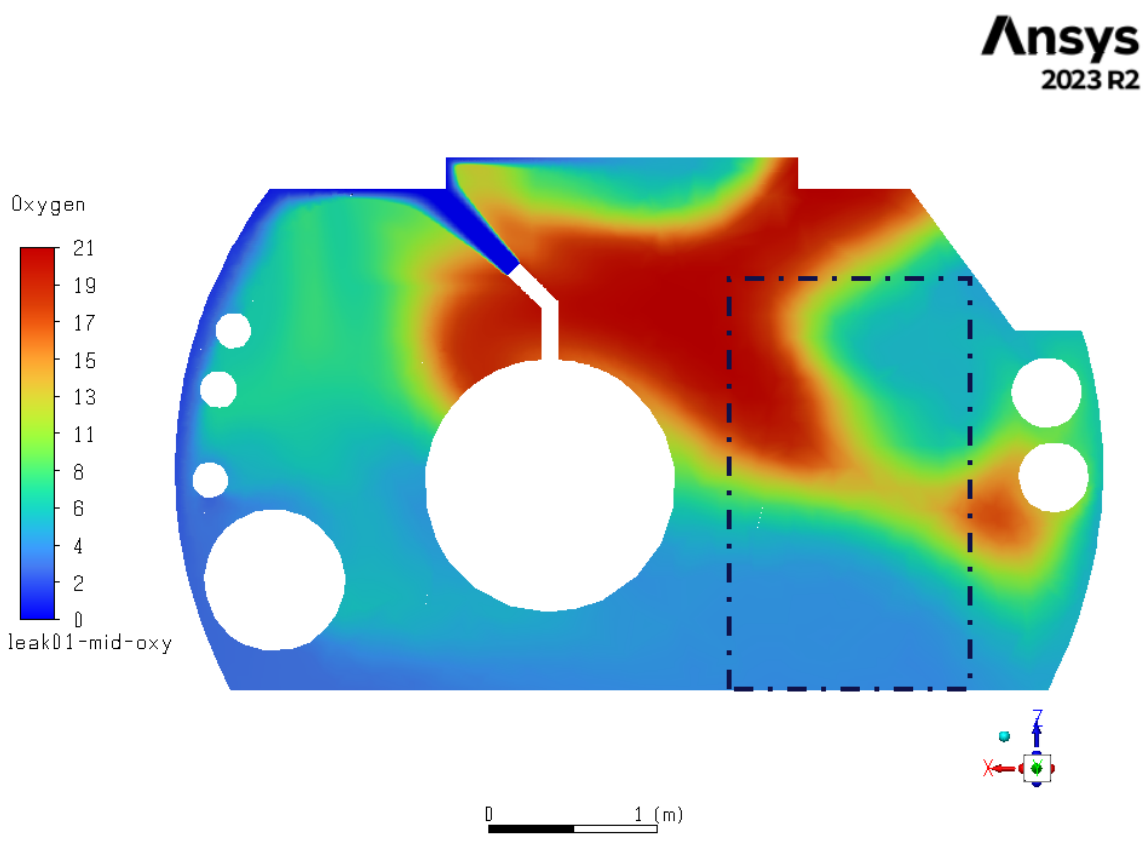

(c) $t = 6$ s

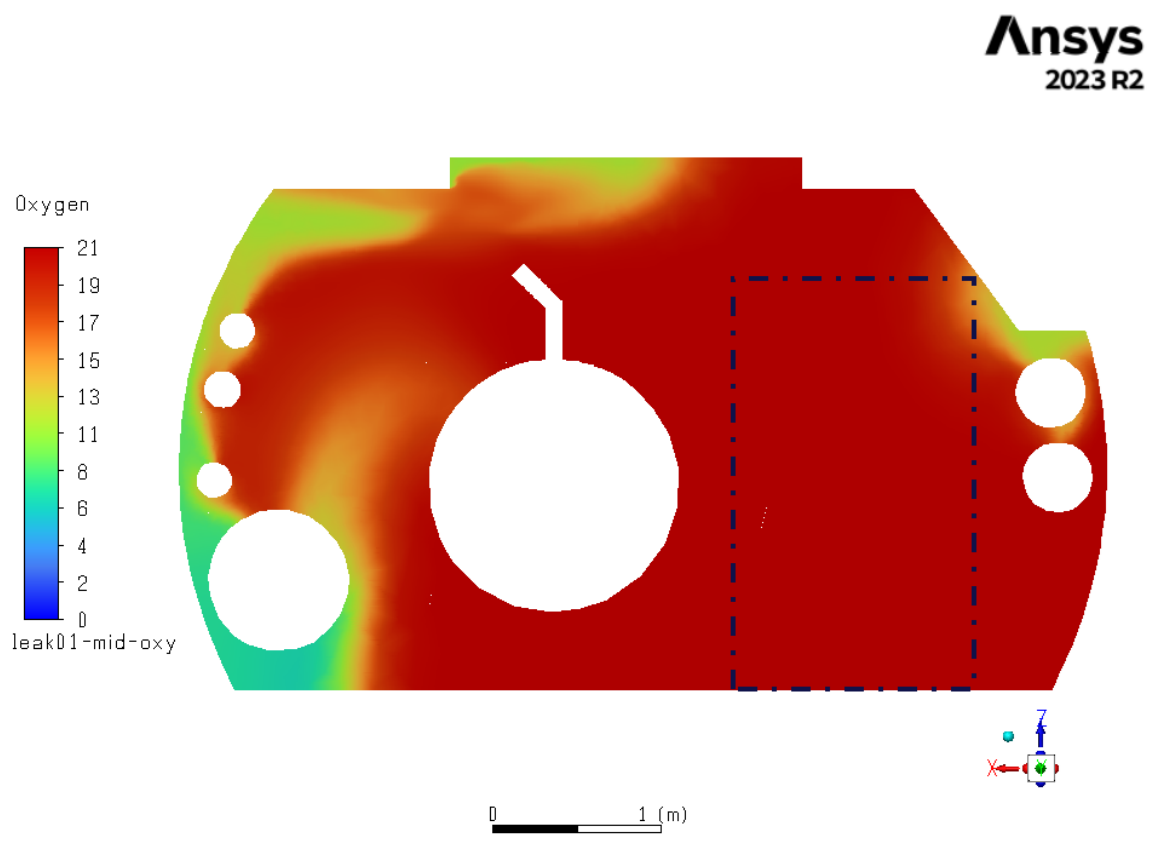

(d) $t = 26$ s

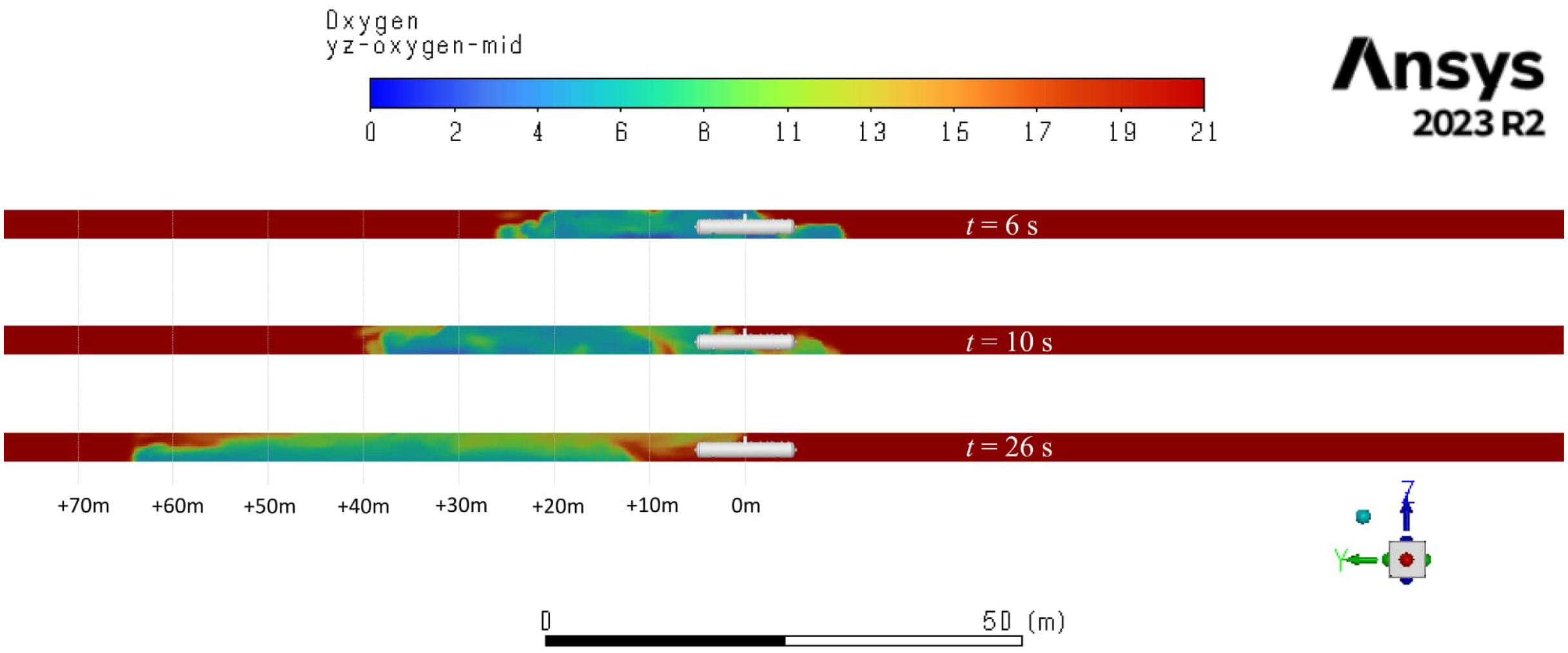

(e) He plug propagation at $t = 6, 10, 26$ s

Fig. 9.26: Results for of the CFD simulations of scenario SRF02 at a time $t$, following the opening of the burst disk. a) - d): Cross-section view just 1 m away from the burst disk. e): longitudinal view of the helium propagation. Colour scale shows the oxygen content in the 2D plane (passage from red to orange: the transition to 18% and below). Ventilation flows right-to-left.

and analysed. At least two seismically active faults stretch about 10 and 30 km southwest of Geneva, within the area of the current reference implementation scenario. In order to provide suitable seismic safety requirements for the design of the civil and mechanical structures, equipment, and installations foreseen in the underground facilities, subsequent preparatory phases need to assess seismic hazards at the depths concerned. To do so, a collaboration with the Swiss Seismological Service (SED), hosted by the Swiss Federal Institute of Technology Zurich (ETHZ), the University of Geneva (UNIGE), and the French Université Grenoble Alpes (UGA) is proposed for further studies. So far, a simplified approach based on the adaptation of the legal seismic safety requirements for ordinary surface buildings on French territory was used to design and assess new mechanical structures and equipment in the Large Hadron Collider (LHC) and Super Proton Synchrotron (SPS) complexes, meaning that the structural elements could be over- or under-sized to resist the required earthquake loads. This proposal will build upon a project to draw a probabilistic-seismic-hazard-oriented and tailor-made assessment for the proposed underground facilities. The main products of the projects are:

- Seismic Hazard Curves: a family of seismic hazard curves that represent the annual frequency of exceeding different levels of ground motion, essential for designing earthquake-resistant structures and assessing the safety of existing facilities.
- Site Response Analyses: uniform hazard spectra (UHS) and scenario earthquakes, including acceleration time-histories, for seismic design and safety evaluations of the civil infrastructure and the equipment structures hosted there.

  The overall study is planned for a 3-year period (2025 - 2028)[15].

### 9.4.4 Evacuation

A series of evacuation studies were undertaken for the underground infrastructure.

#### *General requirements and considerations*

The following list summarises the minimal requirements for all FCC areas, in line with the prescriptive requirements in France's Code du Travail [442] and Switzerland's Norme de Protection Incendie [443] used for standard facilities:

- At any given point, occupants must be able to choose at least two distinct evacuation paths. The maximum dead-end evacuation distance is $\leq 40\,\mathrm{m}$ (e.g., in alcoves).
- The width of any evacuation doors must be $\geq 0.9\,\mathrm{m}$.
- The maximum evacuation distance:
  - for surface buildings: in accordance with [442, 443];
  - for underground installations: in accordance with performance-based design and inline with best practices in other underground tunnel infrastructures.
- Emergency shelters (refuges) without safe escape routes connecting to the surface are not considered in the safety concept for the operation phase. This is in line with the provisions of Directive 2004/54/EC for road tunnels [432].

#### *Tunnel arc: evacuation modelling*

The evacuation from the underground infrastructure will rely on both horizontal and vertical means. Personal transport vehicles will ensure the horizontal evacuation from the nearest alcove to the safe area at the bottom of the access shaft. Vertical evacuation will be ensured via pressurised lift shafts equipped with a secured set of lifts (composed of 2 lifts for redundancy and availability), to bring occupants to the assembly points at the surface.

[15] Note that the French and Swiss seismic hazard model currently in force for surface buildings will be superseded by the new European Seismic Hazard Model (ESHM 2020) [494] that will be incorporated in the second generation of the Eurocode 8 [495].

At the bottom of every lift shaft a fire-, smoke- and gas-proof safe area is needed to ensure that occupants can safely wait for the arrival of the lift. Defining the size (surface area) of the safe area and its impact on tunnel occupancy is crucial to verify the safety concept during evacuation processes, ultimately defining the maximum number of occupants allowed underground at the same time. For a safe evacuation, maximum admissible crowding in safe areas should be under 3 occupants/m$^2$ [496].

Many conditions with different probabilities (e.g., occupancy distribution in the tunnel, walking speed or transportation speed) need to be evaluated in order to define the required surface area and reserve enough space near the shaft in the integration drawings. At this phase of the study these parameters are not exhaustively known; hence a stochastic approach with random variables (in a given set ranges) is more suitable. Therefore, to determine the size of the safe area, an evacuation model was developed using plain Monte Carlo simulations.

Depending on the location of the fire, one or more sectors of the tunnel arc will need to be evacuated. A worst-case scenario is assumed where the fire is just next to the connection to a service cavern, blocking access from one sector only, which means that all occupants of an entire sector must evacuate via the other access shaft (10 – 11 km away). In addition, it is a common practice to also evacuate the adjacent sectors that are not affected by the fire (as in the LHC scenario). To summarise, occupants from one and a half sector (1.5 sectors) will be evacuating through a single shaft (Fig. 9.27).

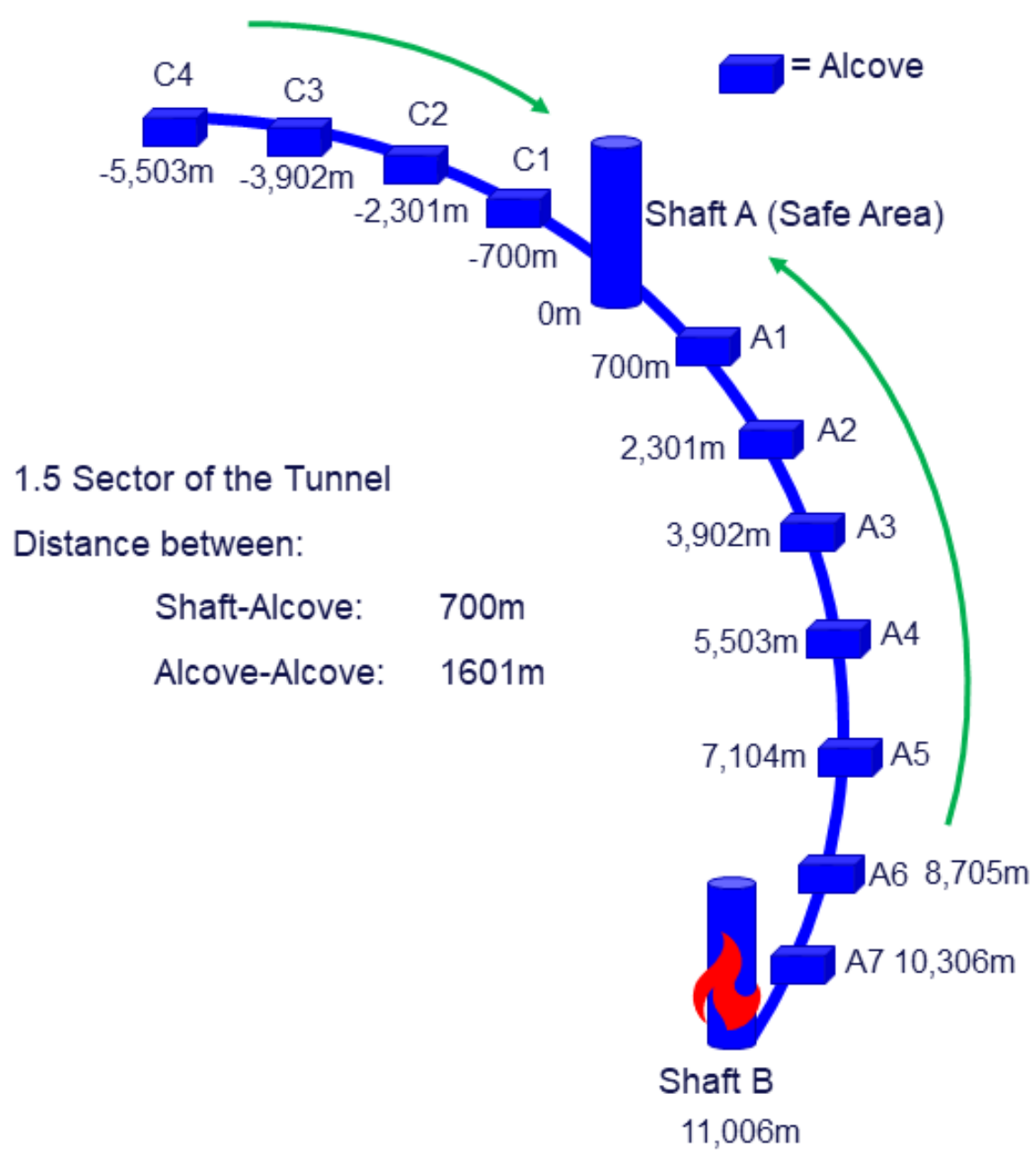

Fig. 9.27: Simplified schematic of the evacuation model.

In this model, the number and location of the alcoves are important, indicating the total parking capacity. For the feasibility study, the baseline number of alcoves is 7 per sector, but alternative studies were carried out assuming 9 alcoves.

It is assumed that all occupants inside the tunnel will have access to a transport vehicle. Therefore, the maximum number of occupants allowed inside the sector is limited depending on the capacity of the transport vehicles and the parking space in the sector. A total of 4 different simulation scenarios were created (see Table 9.9) with 2 different distribution methods for the occupancy distribution in the tunnel. The number of occupants per sector is comparable to a scale-up (factor 3) of the maximum number of occupants seen during Long-Shutdown 2 (LS2) in the LHC [497].

Table 9.9: Evacuation scenario parameters.

| Scenario Number | Vehicle occupant capacity | Occupancy distribution | N. of alcoves (per Sector) | Vehicle Parking Capacity (Alcove/Shaft) | Occupants / 1.5 Sector | Occupants / Sector |
|---|---|---|---|---|---|---|
| 1 | 2 | Binomial | 7 | 10/20 | 260 | **174** |
| 2 | 3 | Binomial | 7 | 10/20 | 390 | **260** |
| 3 | 4 | Uniform | 7 | 4/4 | 192 | **128** |
| 4 | 4 | Uniform | 9 | 4/4 | 240 | **160** |

The distribution of occupants in the tunnel is an important variable in determining the size of the safe area. Since lone working is not allowed by default, it is assumed that occupants are working in groups. The number of occupants per group and their position is treated as random. They are placed either in between two adjacent alcoves or between the service cavern (Shaft A) and the nearest alcove. After the evacuation alarm, both the occupants between Shaft A and Shaft B and the occupants in the half sector on the other side of Shaft A start to evacuate to the safe area in Shaft A (Fig. 9.27). To do this, occupants between the two alcoves must first walk to the alcove where they parked their vehicles and then travel to the safe area of Shaft A with their vehicles. Occupants in the area between Shaft A and the nearest alcove will walk directly. Two different occupancy distributions (Fig. 9.28) were used to indicate the initial positions of groups.

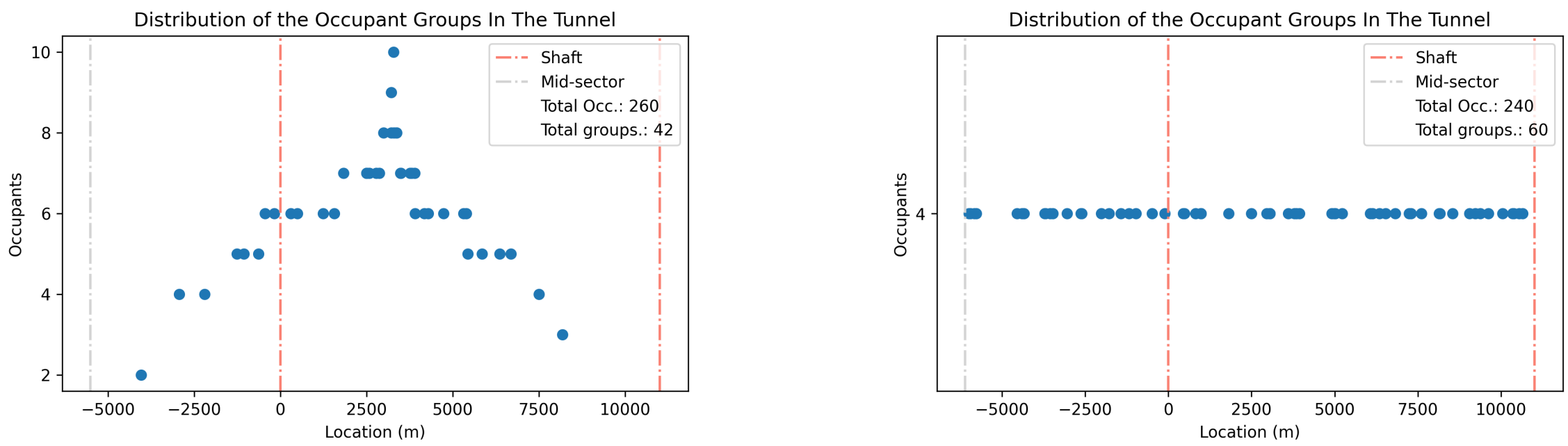

(a) (b)

Fig. 9.28: Example of occupant group distribution along 1.5 sectors. Shaft A location at position 0 m. a) Binomial distribution, restricted to 2 and 10 occupants per group; b) Uniform random distribution, restricted to groups of 4 occupants.

The simulations consist of 50 different random occupancy distribution scenarios, each run with Monte-Carlo (1 000 samples). The results are shown in Fig. 9.29, with an occupant walking velocity as a normal distribution (mean(SD) of $1.2(0.3)\ \mathrm{m\,s^{-1}}$) and a transport velocity as a uniform distribution ($[20-30]\ \mathrm{km\,h^{-1}}$). The maximum number of occupants that reached the safe area at the end of each simulation (50 000 in total) is recorded and is also shown in Fig. 9.29. Scenarios 3 and 4 require the smallest safe area size. This is mainly due to the uniform distribution of the occupant groups in the tunnel, meaning that they will all arrive at the safe area in a distributed fashion, thus crowding is limited by the travel time of the lift. Scenarios 1 and 2 have a binomial distribution with an agglomeration of occupants in the middle of the arc (this could be the case during a major repair or maintenance in a particular arc cell). This would lead to a higher crowding at the safe area once these occupants reach the lift, hence a larger safe area is required to maintain a density below 3 occupants/$m^2$. The results show that having a minimum safe area of $50\,\mathrm{m}^2$ would be suitable safety. It is feasible to implement such an area near the lifts of all eight service caverns. Additional details are available in the technical report [397].

***Technical points: service caverns***

The service caverns at the technical points are connected to the accelerator tunnel via compartmentalised passages that will be used as emergency escape routes. In some points (such as PF), the service shaft is not directly connected to the service cavern. Therefore, an additional horizontal connection tunnel is required. It will also be used as a safe evacuation path.

The service shafts at the technical points are equipped with one set of lifts to ensure the vertical evacuation to the surface. Each set is composed of two lifts for redundancy and availability. Lifts have a dedicated pressurised fire-resistant compartment (lift shaft), separating the air volume from the surrounding environment (i.e., service cavern) to protect the integrity of the evacuation to the surface in case of emergency. The pressurised compartment is extended to the safe area, with its size as determined above.

***Experiment points: experiment and service caverns***

The experiment points will enable independent dedicated flows of occupants up to the surface, separating those evacuating from the experiment and service caverns from those evacuating from the accelerator tunnel.

The experiment cavern is connected to the service cavern through three separate personnel passages that can be used as emergency escape routes: two at the level of the beamline and one at the ground floor. These personnel passages will contain little or no combustible material. Additional bypass chicanes are integrated whenever a personnel passage requires heavy mobile shielding for radiation protection purposes. A second connection tunnel on the ground floor is only for the passage of material and will not be considered a safe evacuation path.

The main tunnel is connected to the service cavern by two bypass tunnels about 100 m from the interaction point and two connection tunnels at the end of the long straight section close to the experiment cavern (Fig. 9.30).

Airlocks are planned in the connections between the tunnel/experiment cavern and the bypass tunnels/ service cavern, separating and isolating the ventilation volumes to prevent the mixing of radioactive air with accessible areas.

Similar to the other points, the occupants rely on lifts for vertical evacuation to the surface site, which will be done by a single shaft located in the service cavern. This is a new concept, different from the experiment points at LEP/LHC, which profited from more than one shaft. However, the separation of the flow from the accelerator tunnel and the experiment cavern is still ensured by having two sets of two lifts (four in total), in the shaft: one set to house the occupants evacuating from the experiment and one set for occupants evacuating from the accelerator (Fig. 9.30). Each set is composed of 2 lifts for redundancy and availability. Each set of lifts has a pressurised fire-resistant compartment, separating the air volume from the surrounding environment (i.e., shaft and service cavern) to protect the integrity of evacuation to the surface. The pressurised compartment is extended to the dedicated experiment and accelerator (machine) safe areas (Fig. 9.30). The safe area is where the occupants wait for the lift to evacuate to the surface. The size of the accelerator safe area, as previously mentioned, is a minimum of $50\,\mathrm{m}^2$, while the exact footprint of the experiment safe area is yet to be integrated and will dictate the maximum occupancy allowed in the experiment cavern (3 occupants / $m^2$).

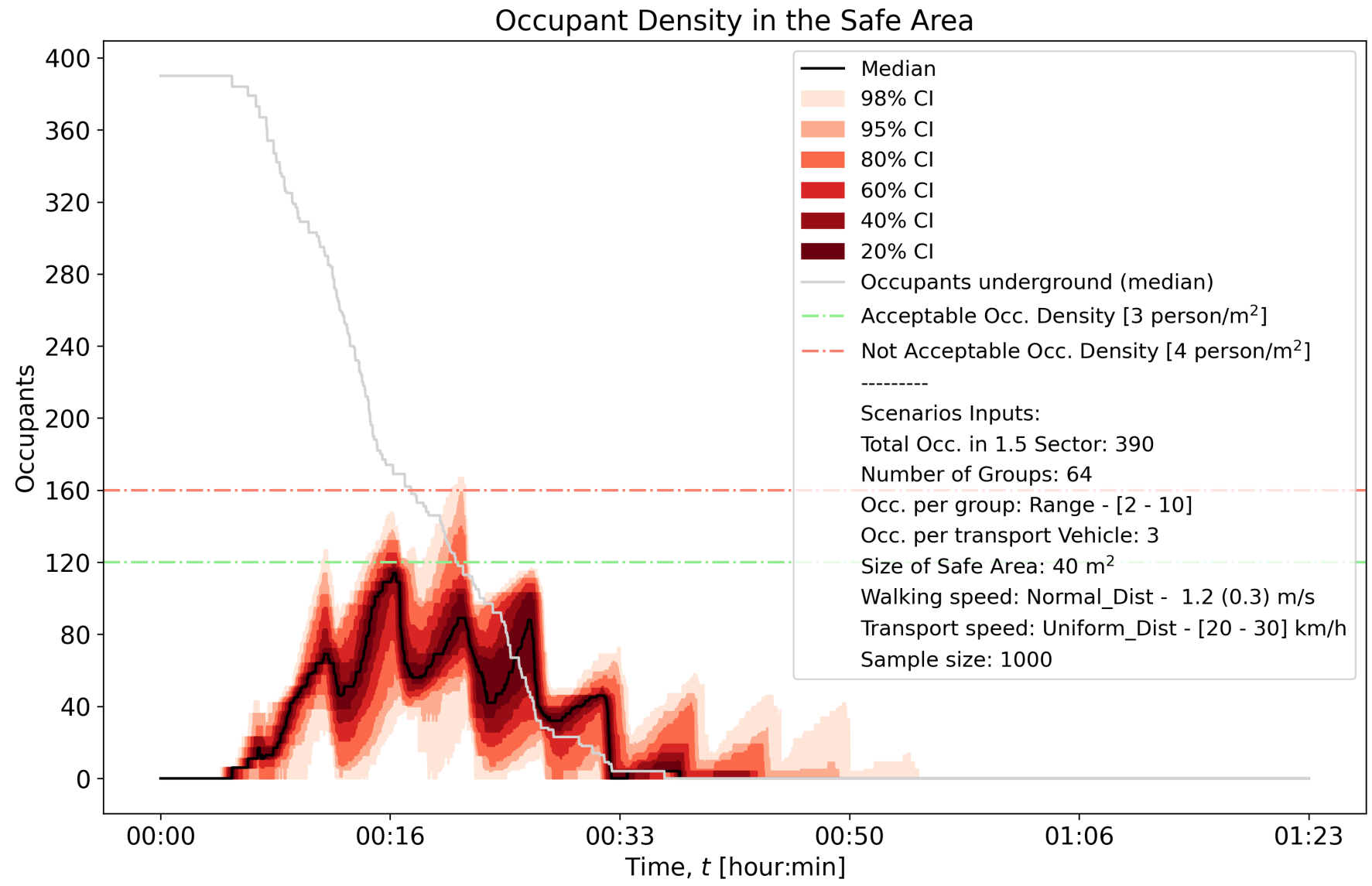

(a) Result of one simulation for scenario 2

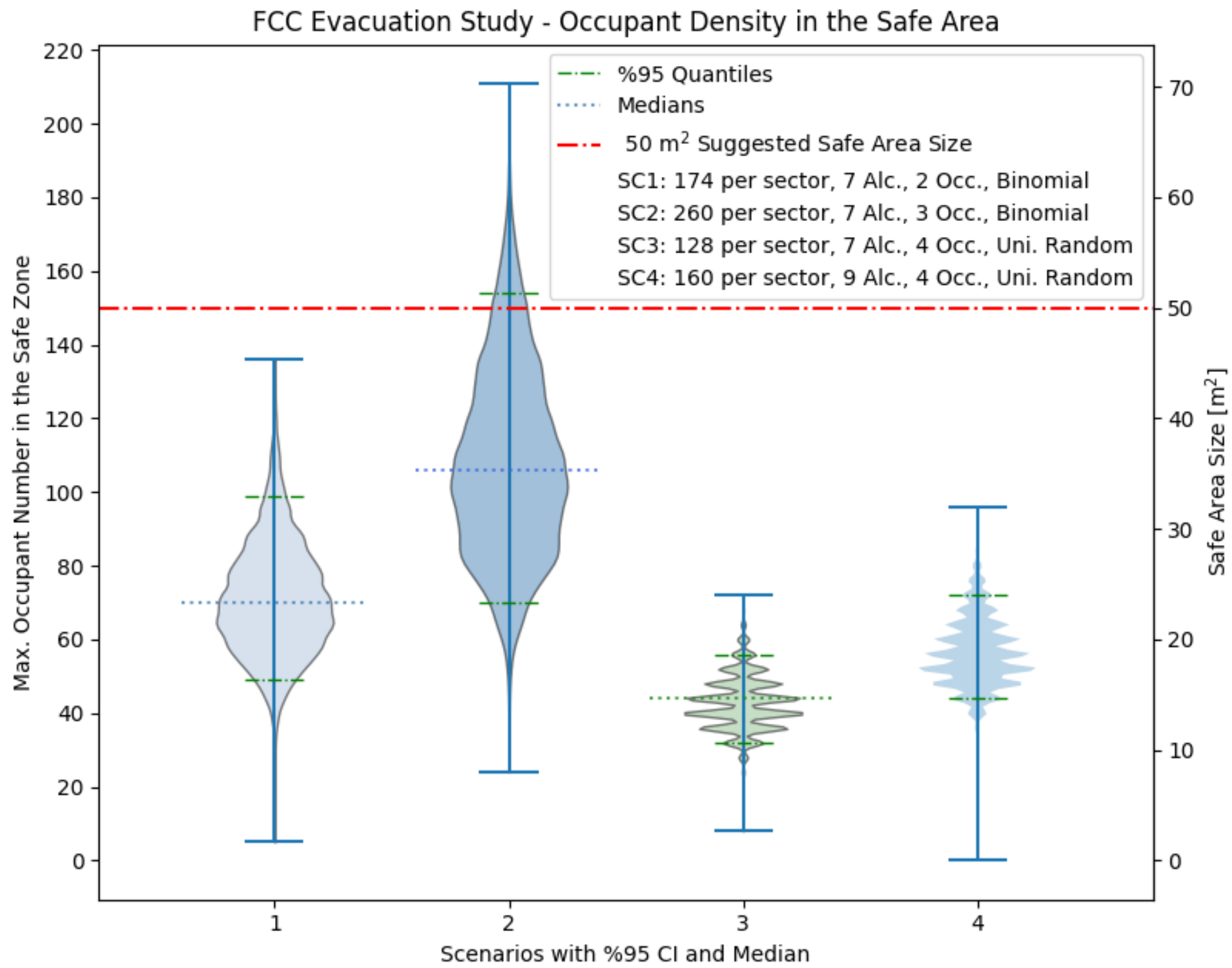

(b) Result of 50 000 simulation runs for each scenario.

Fig. 9.29: Simulation results of the evacuation study. a) results of one simulation run of scenario 2, with the occupancy density (crowding) in the safe area over time; b) Maximum number of occupants (crowding) in the safe area for each scenario.

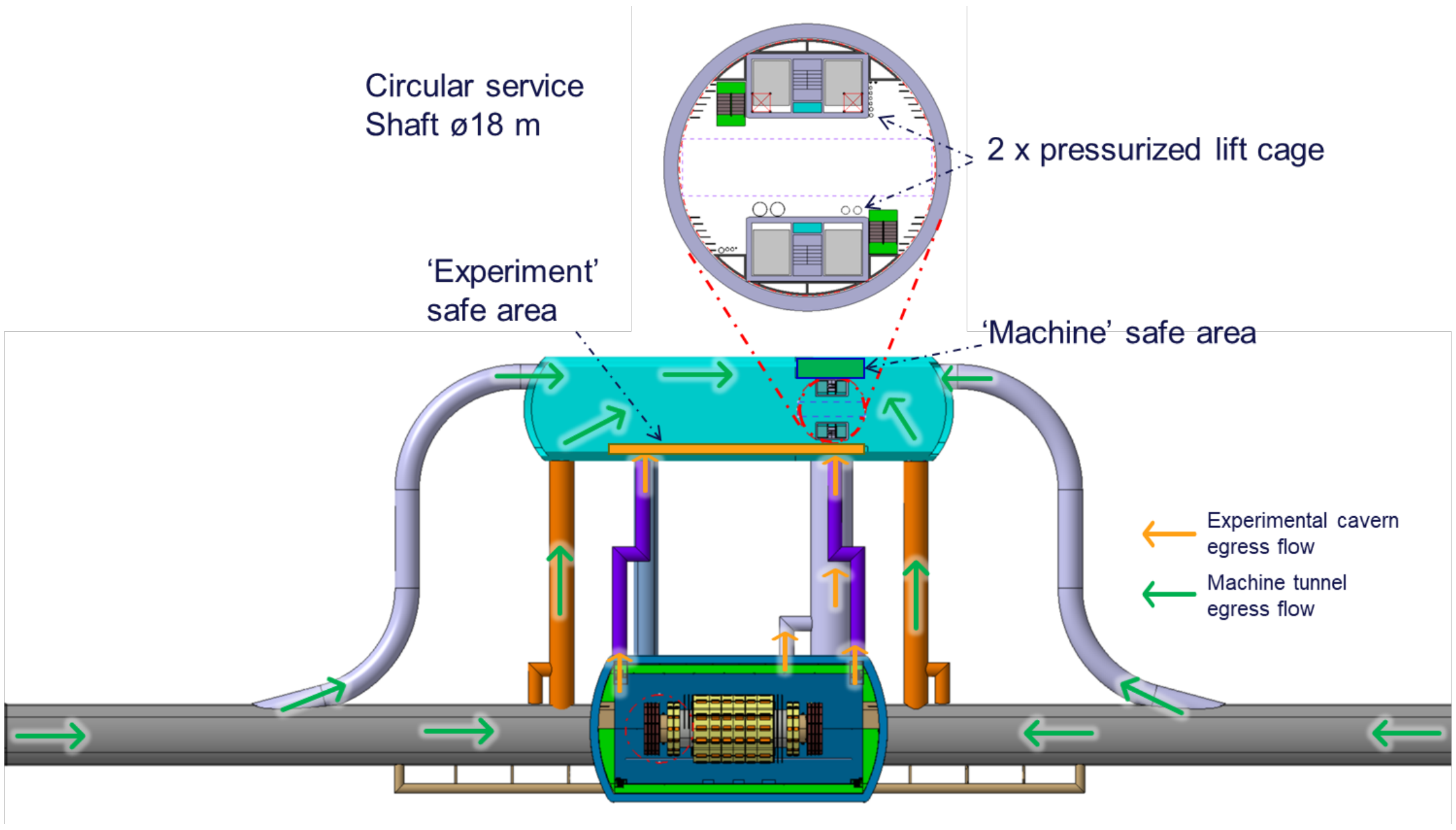

Fig. 9.30: Integration of the evacuation flow of occupants in an experiment point. The illustration shows the example of the experiment cavern housing an FCC-hh detector.

***Klystron galleries***

The superconducting RF (SRF) cavities in the RF sections will be fed by waveguides coming from klystron galleries located above the main tunnel. The length of these klystron galleries is roughly 2010 m in point PH and 1445 m in point PL. A series of connections to the main accelerator tunnel via staircases are available at regular intervals and at both extremities, providing the necessary means of evacuation from the klystron galleries in case of emergency (e.g., fire in the modulators) and thus avoiding dead ends. These connections are distributed as follows:

- At PH, there are a total of 6 tunnel connection staircases, one every 341 m.
- At PL, there are a total of 4 tunnel connections, one every 354 m.

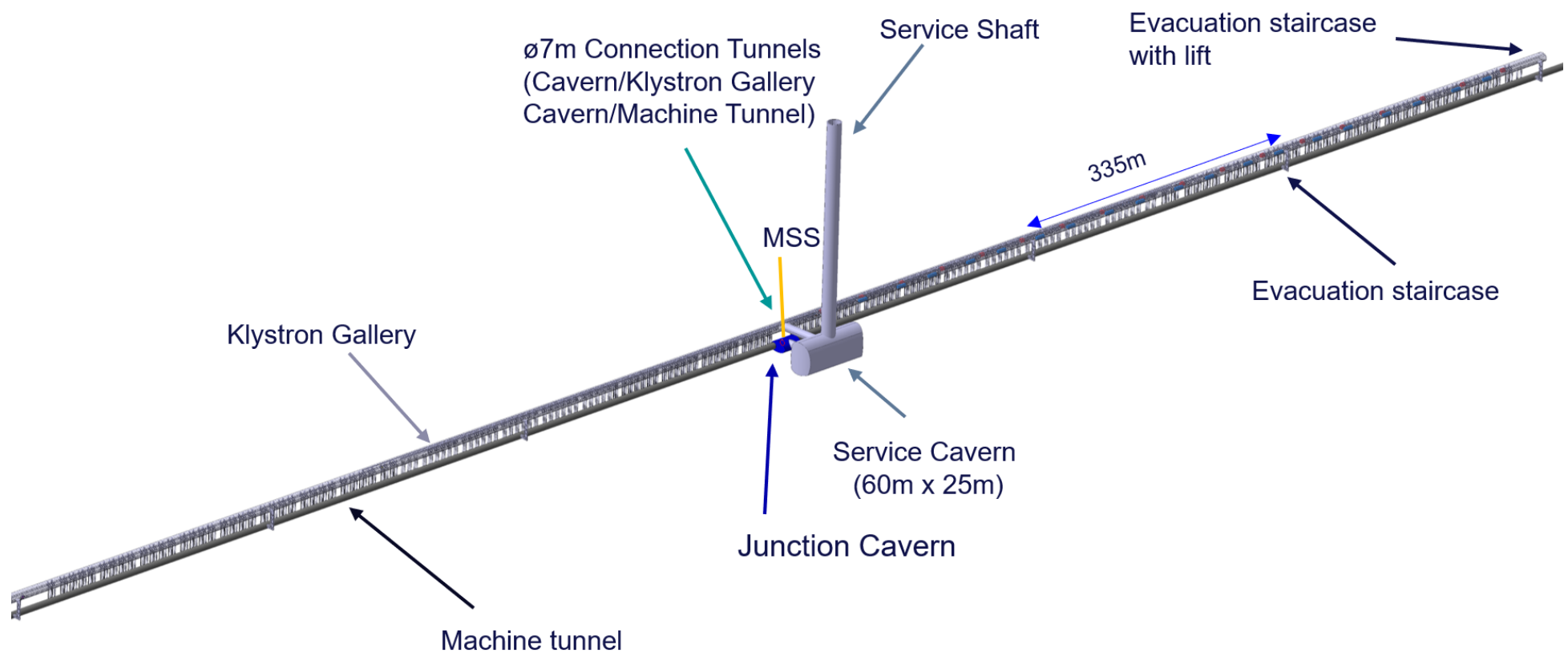

(a) Point PH - $341\,\mathrm{m}$ in between staircases

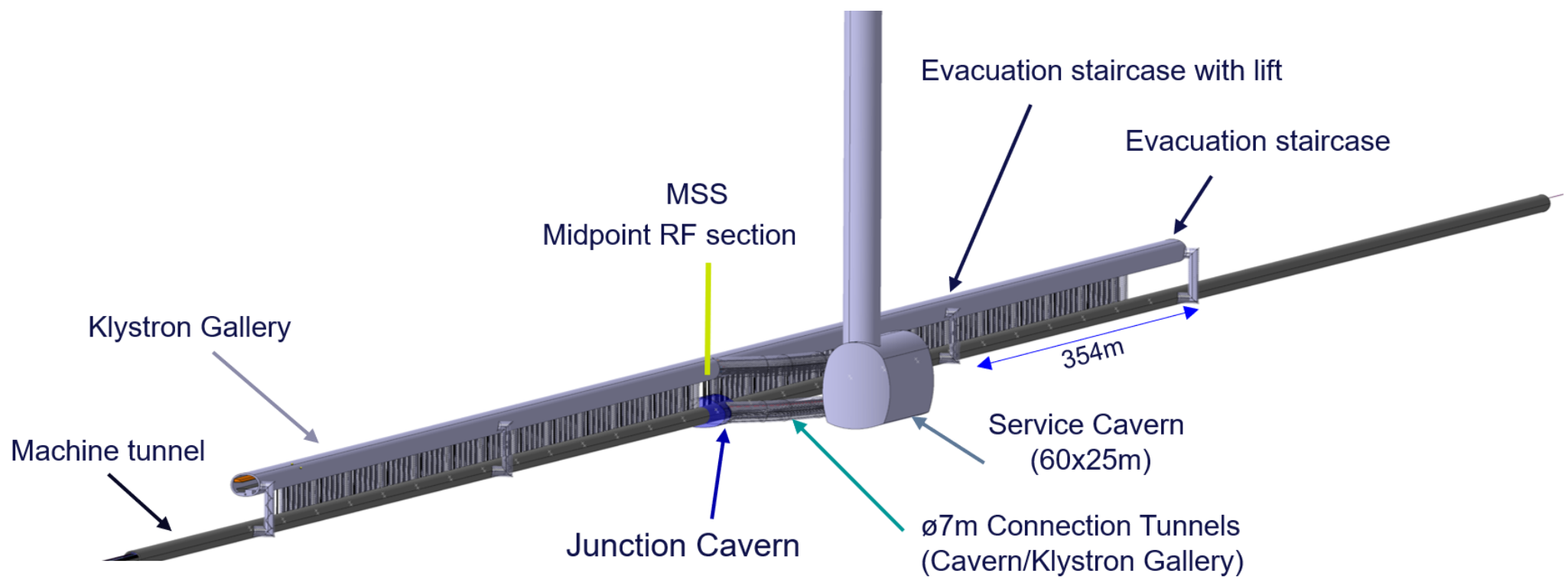

(b) Point PL- $354\,\mathrm{m}$ in between staircases

Fig. 9.31: Klystron galleries and their evacuation connections in point PH (a) and PL (b)

The frequency of these connections helps to minimise the distance that any occupant would need to walk to reach another fire compartment in the main tunnel. Each staircase will be equipped with fire doors at both ends (tunnel and gallery). In addition to the static confinement properties, each connection will also be pressurised, ensuring a dynamic confinement between the main tunnel and the gallery. A more in-depth risk analysis will be conducted for this area during ta subsequent design phase to assess if the distance and number of staircases are sufficient.

In addition to the staircases, at the two ends of each klystron gallery, lifts are available for transporting personnel and material, avoiding having to pass through the RF section to work in the arc. These lifts will not be used in an emergency (e.g., fire). Access to the Klystron gallery is also possible from the service cavern in the middle of the RF sector.

### *Alcoves*

Alcoves will be spread along the accelerator tunnel (minimum 7 alcoves per sector): two big alcoves and five small alcoves. These alcoves house electrical racks and equipment and will form a dedicated EI120 fire-rated compartment, separated from the main tunnel. They will have a single access (in and out), through a fire door connecting to the main tunnel, creating a dead-end in terms of evacuation. The maximum allowable distance from the end of the alcove to the fire door must not exceed 40 m. This is based on the maximum evacuation distance to reach a protected staircase in France [442], ensuring that the adjacent fire compartment (i.e., the main tunnel, relatively free from smoke) can be reached safely in time.

In the baseline layout, all the alcoves comply with this maximum distance. The layout also includes two mezzanines, one on each side of the alcove. Staircase connections from the mezzanine to the ground floor will be installed at both extremities of the alcove to ensure that the evacuation length is not doubled and to avoid being trapped on the mezzanine.

#### *Surface buildings*

The surface buildings will rely on applicable national and international standard practices for the evacuation requirements applicable to industrial infrastructures within the Host States. For PB the Swiss AEAI Directives 15-15 [486] and 16-15 [498] will apply. For all the other sites, the French Labour codes are applicable [499].

### 9.4.5 Emergency preparedness and intervention concept

An underground particle accelerator of 91 km in circumference presents unique challenges in terms of emergency response. The threefold size increase compared to the LHC presents challenges, including much larger distances between access points (11 km) and surface distance between shafts. Hence, a new emergency preparedness and intervention concept is required.

The emergency response concept proposed is divided into four distinct phases, starting from a confirmed accident event (e.g., fire) [500]:

- First Response: trained personnel or contractors at the accident site are responsible to raise an alarm, provide first aid, deliver first response (e.g., fire extinguishers) and initiate evacuation procedures.
- Robotic Intervention: remote-operated robotic vehicles gather situational data and conduct preliminary firefighting actions before human intervention. Ceiling-mounted and floor-based robots can fulfil these tasks, for example.
- Local Fire Services: on-site support from local emergency services, primarily for surface-level interventions.
- CERN Fire & Rescue Service (CFRS): CFRS to coordinate emergency interventions, if relevant with other emergency services, with the use of equipment stored both in surface and underground locations. This includes for instance foam-based firefighting systems and compressed air foam (CAFS) units. Rapid-deployment solutions are required to ensure a quick intervention to all eight surface sites.

#### *Support infrastructure for emergency response*

The concept will also rely on a robust support infrastructure for an efficient emergency response:

- Surface site: landing zone in the vicinity for airborne emergency intervention, a command post, a casualty room, and a logistical hub for storing specialised intervention vehicles, firefighting robots, and extinguishing agents.
- Underground: compartments with designated safe waiting areas, compartmentalisation, smoke extraction (with CFRS override and remote control), priority intervention vehicles and robotic assistance units. A reliable communication system, vertical dry risers connecting the surface and service cavern, and underground water reservoirs will ensure adequate firefighting resupply.

Unlike current baseline practice in the LHC, the use of robots will play a critical role in the emergency response for the FCC by enabling the possibility of quick fire response within the underground infrastructure, where human access may be restricted during accelerator operations.

***Training and implementation strategy***

To fulfill the concept, all underground personnel will be trained in emergency response and first aid. A structured transition plan will be developed to align with this new approach. CERN has already begun testing robotic solutions to validate their potential in real emergency scenarios. Further technical details will be made available in the relevant reports [500].

Given the critical partnership with the local fire services in fulfilling the concept, we emphasize the importance of CFRS being actively involved in the joint work with these key partners. This includes the establishment of joint training programmes, workshops, and drills building on the current practices.

By implementing these measures, the FCC will enhance its ability to manage emergency situations efficiently while prioritising the safety of personnel and infrastructure.

## 9.5 Safety during the construction and installation phases

### 9.5.1 Safety coordination

The baseline concept for safety coordination is based on French legislation, as 7 out of the 8 surface sites are located within French territory. For the surface site PB in Switzerland, Swiss legislation applies. The detailed design will include a proposal to harmonise processes across all worksites to increase efficiency.

In accordance with French law [501], the project owner is responsible for safety coordination throughout all phases of the project [423].

Specialised external service providers (i.e., safety coordinators) will assist the project owner in this safety coordination mission.

Safety coordinators are responsible for preparing coordination plans and the PGCSPS (Plan Général de Coordination en matière de Sécurité et de Protection de la Santé) for the relevant worksite(s). These plans must be provided to contractors during the tendering phase to clearly outline their safety obligations and expectations, as well as to specify the applicable laws and regulations governing the project. All worksite-related safety documents, procedures, and methods must be written in French to ensure compliance with national regulations. Each contractor remains responsible for the safety of their own employees, according to their operating methods. Any accident or incident occurring within the worksites must be reported by the contractors using a form-based system. Today, at CERN, this reporting is governed by CERN's safety rule SR-SIM [502].

### 9.5.2 Construction phase

During the construction phase, civil engineering works on the surface and underground are executed by specialised firms under contract with CERN. It ends with the handover of the buildings and underground structures to CERN from the contractor.

The evacuation plan and emergency procedures in the event of incidents within the worksite(s), will be included in the PGCSPS. Such procedures are developed in consultation with CERN, the contractor and the local emergency services. It indicates, among others, the assembly points for emergency services, access conditions to the various zones means to evacuate victims and the need for specific fire-fighting / response equipment. Joint training, reconnaissance and drills should be organised with the CFRS to ensure a tailored response and strengthen the collaboration with the local services.

### 9.5.3 Installation phase

The installation phase begins with the handover of the surface buildings and underground structures from the civil engineering contractor to CERN.

This will happen at different moments for each access point and the associated tunnel sectors: in the underground area of an access point, the installation phase can start after the total or partial release of the adjacent tunnel sectors leading to the neighbouring access point with lift facilities. In a tunnel sector, installation activities are allowed as long as there are two adjacent access points from which the occupants can reach safely to evacuate. These rules will guarantee a two-way evacuation possibility, with secured horizontal and vertical means of escape at all times. The technical and organisational details of safety during the installation phase will be engineered in the project preparatory phase.

Two sub-phases of installation can be distinguished: the service installation (electricity, lighting, ventilation, cooling, detection and alarm systems, and transport system) and the accelerator installation (beamline elements, magnets, RF cavities, cryogenic systems, power converters, and control systems).

#### ***Service installation***

Service installation begins after the release (handover) of the civil engineering (CE) structures from the contractor to CERN. The temporary lighting, ventilation, lifts and other services installed by the civil engineering contractor could potentially be taken over and used by CERN for the initial phases of the service installation. Obviously, no work can be executed by CERN and its contractors without these essential services. It is therefore advisable to foresee the transfer of these assets to CERN at the time of handover, saving the time and cost of CERN installing essential services in a provisional manner. The transfer should include the assets, as well as the technical documentation and maintenance procedures. Similarly, robust construction site access control devices (e.g., turnstiles) could be transferred from the contractor and connected to the CERN access control system. This would allow verification of CERN imposed training, etc., access control and the enforcement of work coordination.

##### *Safety organisation during service installation*

The services mentioned above are largely installed by contractors mandated by CERN. In order to co-ordinate work and occupational safety coherently, experience from the High Luminosity LHC shows that the work of a pairing (*binome*) consisting of an activity coordinator employed by CERN and an independent safety coordinator (see Section 9.5.1) is very effective. The FCC project should adopt the same approach. Given the geographical span of the facility and the unavailability of most of the safety and automation systems in this early phase of the installation works, each of the access point sites will be permanently manned by guards. Their primary role will be to ensure smooth material entry without compromising access control procedures. In addition, during working hours, a site supervisor will ensure the respect of safety procedures and act as a liaison with service and equipment groups at CERN in case behavioural or technical anomalies are observed.

##### *Installation order during service installation*

The underground ventilation system is essential for the well-being and safety of workers, especially in the 11.4 km long tunnel sectors. The definitive ventilation system will be installed with priority. This includes cooling water, which is required for the air-handling units along the tunnel. Once the smoke extraction duct is equipped, the dampers and extractors are powered and commissioned, local control of smoke development is possible. At the same time, while avoiding co-activities, the definitive electrical supply and lighting will be installed. This includes the installation of detection systems' control and indicating equipment (CIE) racks in the alcoves, high-voltage cables to the alcoves and transformers. Cable trays will be installed to guide the numerous cables in an orderly manner.

The installation of definitive detection systems can proceed with lower priority. If smoke detection is performed by aspiration tubes, then this system can only be made operational after the accelerator installation, because of its sensitivity to dust and welding fumes, which would raise numerous false alarms. Manual call points distribution and a dedicated procedure will ensure prompt alarm activation during this degraded phase.

For the FCC, the transport system has a safety role during evacuation, it is also indispensable for the next phase in which the accelerator is installed. As described in Section 9.4.1, the transport system is based on autonomous vehicles, both for the transport of personnel and for large and heavy loads. All infrastructure to allow autonomous movement of the vehicles must be available at the end of the service installation phase.

*Emergency response during service installation*

During the service installation, a provisional alarm system is required since the automatic fire detection would not yet be available. Manual call points (break-the-glass devices) and provisional evacuation sirens will be installed and cabled to the CIE racks. Provisional communication infrastructure is necessary to transmit the alarms to the central emergency coordination team.

Due to the large distances in the tunnel, and on the surface between the surface sites and a CERN site, where the emergency coordination takes place, the possibility for timely intervention of trained emergency services has to be implemented before the installation phase starts. Consequently, early development of the integrated emergency concept is needed.

*Occupational safety during service installation*

During service installation the safety systems may not be installed or not be fully operational, and strict measures for occupational safety must be implemented. According to preliminary estimates, up to 200 occupants per sector are expected during the installation phase. This is significant but less than in the worst-case scenario during the operation phase that was studied (Section 9.4.4).

The focus is on the self-protection of the workers:

- Access is only granted to trained personnel.
- All workers must be trained in the use of the self-rescue mask (for use in case of fire or smoke), and the use of the transport vehicles for evacuation.
- A sufficient number of certified first-aiders at the workplace [16] must be trained.
- All workers must also be trained in raising an alarm by pressing the alarm button if there is a fire or smoke development and to react correctly to the evacuation sirens.
- The installation schedule and planning must have shifts dedicated to high-risk activities where the human presence is limited to the absolute minimum. For example, using night shifts for heavy transport activities, such as cable drums or other bulky loads, where the access to the sector concerned is restricted to the transport team only.
- The underground worksites must be kept clear of combustible materials at all times. Any such material (packaging, cable drums) must be evacuated to the surface immediately after its use.
- The transport zone of about $2.5\,\mathrm{m}$ width must be kept free of stored material throughout the tunnel to permit evacuation.

***Accelerator Installation***

Once the basic services - ventilation, lighting, electricity, smoke extraction, detection, and alarm systems and transport - are commissioned, the sub-phase of accelerator installation may begin.

*Safety organisation during accelerator installation*

The same concept as the previous phase is assumed, with an update of the existing PGCSPS, if necessary.

[16] A first-aid qualification with a focus on the most frequent types of personal accidents at industrial workplaces. The training lasts two days and is concluded by a proficiency test.

*Installation order during accelerator installation*

Drawing on experience from LHC, the accelerator will be installed in a sequence of

- Transporting the magnets and other beamline elements like collimators, and RF cavities to their designated location.
- Making provisional geometrical alignment.
- Connecting the beamlines with each other.
- Connecting the electrical supply and the controls to the magnets.
- Concluding with a final geometrical alignment.

The partition walls separating the tunnel into 400 m long fire compartments can be installed as soon as the magnets are placed. A phased approach, where only certain partitions are initially installed or where the partitions leave openings for the installation of further cables and services, allows the flexibility needed during installation while improving the safety of degraded modes. Passive protections in the form of partition walls are the most reliable system for this phase. Finally, the automatic detection and alarm system can be commissioned.

Similar to the service installation sub-phase, the planning will include shifts dedicated to high-risk activities, such as magnet transport, and avoid co-activity by restricting access to only the teams concerned.

*Transport during accelerator installation*

Magnets will be lowered underground with overhead cranes through the large shafts of the service caverns at sites PA, PD, PG and PJ. Then, a magnet transport vehicle will move them past the already installed magnets to their final location. During this journey, the magnet transport vehicle effectively blocks the transport zone of 2.20 m width, thus blocking the evacuation exit. Access during transport to a sector will be restricted to the transport team, reducing the number of occupants underground during this period. Once the magnet transport has passed the midpoint of the sector, a returning vehicle can pass an arriving one in the extended lay-by area at the central alcove. The same rule applies to the transport of other large items which are wider than 80 cm and thus cannot be bypassed by a personnel vehicle.

From the moment when all magnets are positioned in a sector, the transport, and evacuation rules applicable during normal operation become valid, because only the transport zone is available. In particular, special, narrow personnel transport vehicles must be used for personnel and light loads.

*Emergency response during accelerator installation*

The provisional alarm system with break-glass push buttons installed during the service installation sub-phase remains active until the automatic detection and alarm system has been commissioned. If the detection technology chosen is aspiration smoke detectors, this can only be done after all dust and fume generating works are concluded. The alarm system must be configured to trigger the evacuation alarm (sirens or voice messages) and alert the central emergency coordination team, which will analyse the situation and engage the most appropriate emergency response with the different emergency services.

*Occupational safety during accelerator installation*

The type of work during accelerator installation is mostly of a mechanical and electro-mechanical nature. This includes hot work (grinding, welding and brazing), which is a potential ignition source. The compartment walls, installed after magnet placement, may have a higher leakage rate for smoke than in their final configuration because some openings for cables and other services cannot be tightened before all installation work is finished. The mitigating measures personnel have to apply to take account of this are identical to those in the service installation sub-phase: minimisation of combustible material, rigorous training of workers performing hot work in extinguishing and alarm procedures, self-rescue mask and evacuation training for all workers.

## 9.6 Conclusion

The safety concept outlined in this report emphasises a comprehensive and integrated approach to safety throughout all phases of the proposed facility. From the initial construction and installation phases to the long-term operation of the collider, safety systems to manage risks associated with both normal and accident scenarios have been studied.

This concept demonstrates the feasibility, in terms of life safety, of the current baseline layout and design for the FCC-ee. However, any modification of the proposed baseline will require an update of the concept and the re-assessment of its safety effectiveness.

Moving forward, a continuous risk assessment and iterative enhancement of the concept is essential for the evolving nature of the project. Additionally, it is crucial to address the remaining open points and research gaps that were identified during the development of the concept.

As the effort progresses, the safety concept will incorporate other safety objectives, maintaining a strong safety-oriented approach in the design of this unprecedented facility.

# Chapter 10

# FCC-hh collider design and performance

## 10.1 Design and performance

The Future Circular Collider (FCC) integrated programme offers the most powerful post-LHC experimental infrastructure proposed to address key open questions in particle physics. It envisions an initial electron-positron collider phase, FCC-ee, which will later be followed by a proton-proton collider, both of which will be installed in the same tunnel of approximately $91\,\mathrm{km}$ in circumference, close to CERN.

The hadron collider, FCC-hh, would operate at a centre-of-mass energy of about $85\,\mathrm{TeV}$ (or above), extending the energy frontier by almost an order of magnitude compared with the LHC, and providing a 5- to 10-times higher integrated luminosity than the upcoming High-Luminosity LHC. The mass reach for direct discovery at FCC-hh will amount to several tens of TeV, and it will allow, for example, the direct production of new particles, whose existence could already be indirectly exposed by precision measurements at FCC-ee. The FCC-hh hadron collider can also accommodate ion and lepton-hadron collision options, allowing complementary physics explorations.

The FCC-hh accelerator configuration described in this chapter, including its layout and injector options, preserves the rich potential for a diverse physics programme, ranging from heavy ion collisions to a dedicated flavour or forward physics programmes, as are familiar from the LHC. Examples of this potential were amply discussed in the CDR and are briefly reviewed in Section 1.2 of Volume 1. Furthermore, it is expected that the 4-fold symmetry of the interaction regions will add flexibility and boost performance to the exploitation of the two lower-luminosity interaction points.

The FCC integrated programme follows the successful example of the past Large Electron Positron Collider (LEP) and Large Hadron Collider (LHC) projects at CERN, which used one and the same infrastructure for successively realising two large collider projects. Inspired by the sequence of LEP and LHC, the comprehensive long-term FCC programme maximises the opportunities of physics. The lepton and hadron colliders, FCC-ee and FCC-hh, would profit from common civil engineering and also from sharing much of the technical infrastructure. The updated baseline layout of the FCC-ee electron-positron collider and its injector design are fully compatible with the demands of the future hadron collider (FCC-hh), and do not compromise the latter's performance. The main ring optics and RF configurations for both colliders were refined to simplify operation across the energy range of FCC-ee and to enable a smooth transition to FCC-hh after the completion of the FCC-ee research programme.

The FCC-hh baseline assumes $14\,\mathrm{T}$ $Nb_3Sn$ magnets, that are compatible with a collision centre-of-mass energy of $85\,\mathrm{TeV}$. An R&D path towards HTS magnets would enable higher collision energies. The present design allows four collision points and experiments. The experience gained from the Phase II upgrades of the LHC detectors for the HL-LHC, developments for further exploitation of the LHC and ongoing detector R&D for future Higgs factories will be important stepping stones for the development of the FCC-hh experiments.

In the next sections, the layout for the collider, the injector options, and the injection lines will be described. Following this, Section 10.4 describes the pathway to developing the key technology required for the FCC-hh: the development of high-field magnets based on both the $14\,\mathrm{T}$ $Nb_3Sn$ technology and higher-field HTS, that could enable higher collision energies and/or cost savings. Teams around the world are pursuing the development of the requisite magnet technology. The primary regions involved in the advancement of magnet technology include the US, Europe, Japan, and China. In each region, there are growing efforts to coordinate and integrate research internally among universities, laboratories, and, to some degree, industry.

An R&D plan is outlined which shall close the gap between the near- to medium-term scope (5–10 years) of the European High Field Magnet (HFM) Programme and the US Magnet Development Programme on the one hand, and the long-term needs of the FCC integrated programme on the other hand. It complements the bottom-up, technology-driven approach of the HFM Programme with a top-down, long-term strategic roadmap derived from the requirements of the FCC-hh. Moreover, this plan will provide a framework for the coordination of R&D efforts on a global scale and over a sustained period of time.

Finally, this chapter closes with a discussion of some other accelerator systems that would be required for the FCC-hh, including cryogenic plants and their distribution.

## 10.2 FCC-hh layout and optics

### 10.2.1 Layout of the FCC-hh ring

Since the publication of the Conceptual Design Report (CDR) [10] several studies have been carried out to provide a revised and improved version of the ring layout. The key concepts that have been used to determine the main properties of the new layout can be summarised as follows: the outcome of the placement studies; the proposal to equip the new layout of the FCC-ee ring with four experiment points; the considerations on the harmonic number of the FCC-hh ring that should be made compatible with that of its injector.

The placement studies provided strong indications that the length of the FCC-hh ring had to be shortened to satisfy the multiple constraints that emerged during the investigations. Most of the reduction compared with the CDR was achieved by reducing the lengths of the arcs, keeping the total length of the straight sections constant. However, a final adjustment also reduced the length of the straight sections.

The FCC-ee design introduced the important constraint that the ring should assume a four-fold symmetry and four-fold superperiodicity, to enable four experiment insertions with optimum luminosity performance. This feature, which must also be fulfilled by – and may similarly benefit – the FCC-hh ring, radically changes the configuration with respect to what was presented in the CDR [10]. The functions of the various technical insertions had to be reviewed and different functions to be combined differently, which resulted in a modified layout and optics design for most of the insertions. Another major change in the layout is the radial displacement of the interaction points (IP) of the FCC-hh ring to align with the IP positions of the FCC-ee ring. This geometric modification has a significant impact, as the dispersion suppressor is employed to control the new ring geometry. This adjustment must be implemented while ensuring the feasibility of the optics, particularly maintaining proper dispersion matching. This change has an important side effect on the optimisation of size and layout of the experiment caverns, with the option of sharing the detector's services, infrastructure, and possibly components between the lepton and the hadron FCC colliders [503]. A further major modification to the ring layout is the reduction of the number of surface sites from twelve to eight, which is beneficial in terms of overall costs and which also facilitated the placement. The last, but certainly not least, change with respect to the original baseline configuration is the proposal to locate the last part of the transfer lines inside the ring tunnel. This choice has the benefit of reducing the length of the tunnel needed for the transfer lines from the injector to the FCC-hh.

The proposed layout of the FCC-hh ring is shown in Fig. 10.1. The circumference is $90.66\,\mathrm{km}$, with $2032\,\mathrm{m}$ long technical insertions and experiment insertions accommodated in a straight tunnel of length of $1400\,\mathrm{m}$, with approximately $965\,\mathrm{m}$ distance between the last dipole magnets on either side of the IP. The experiment insertions are located in Point A (PA), Point D (PD), Point G (PG), and Point J (PJ). The IPs of the hadron and lepton rings are superimposed within a few millimetres.

Figure 10.2 shows a comparison of the three rings on the right side of an experiment insertion. The crossing angle between the two FCC-ee rings is clearly visible, with the FCC-hh straight section in between the two rings. Further away from the IP, the three rings converge and are fully compatible with a single tunnel. Detailed studies of the geometry of the hadron ring optimised the current layout so as to reduce the radial distance between the three rings, with a beneficial impact on the integration. This important result was achieved by carefully displacing the main dipoles in the region of the dispersion suppressors (see the next section). Applying the same recipe, also in the technical insertions the geometries of the hadron and lepton rings were matched as closely as possible, with a radial distance between the rings of approximately $30\,\mathrm{cm}$).

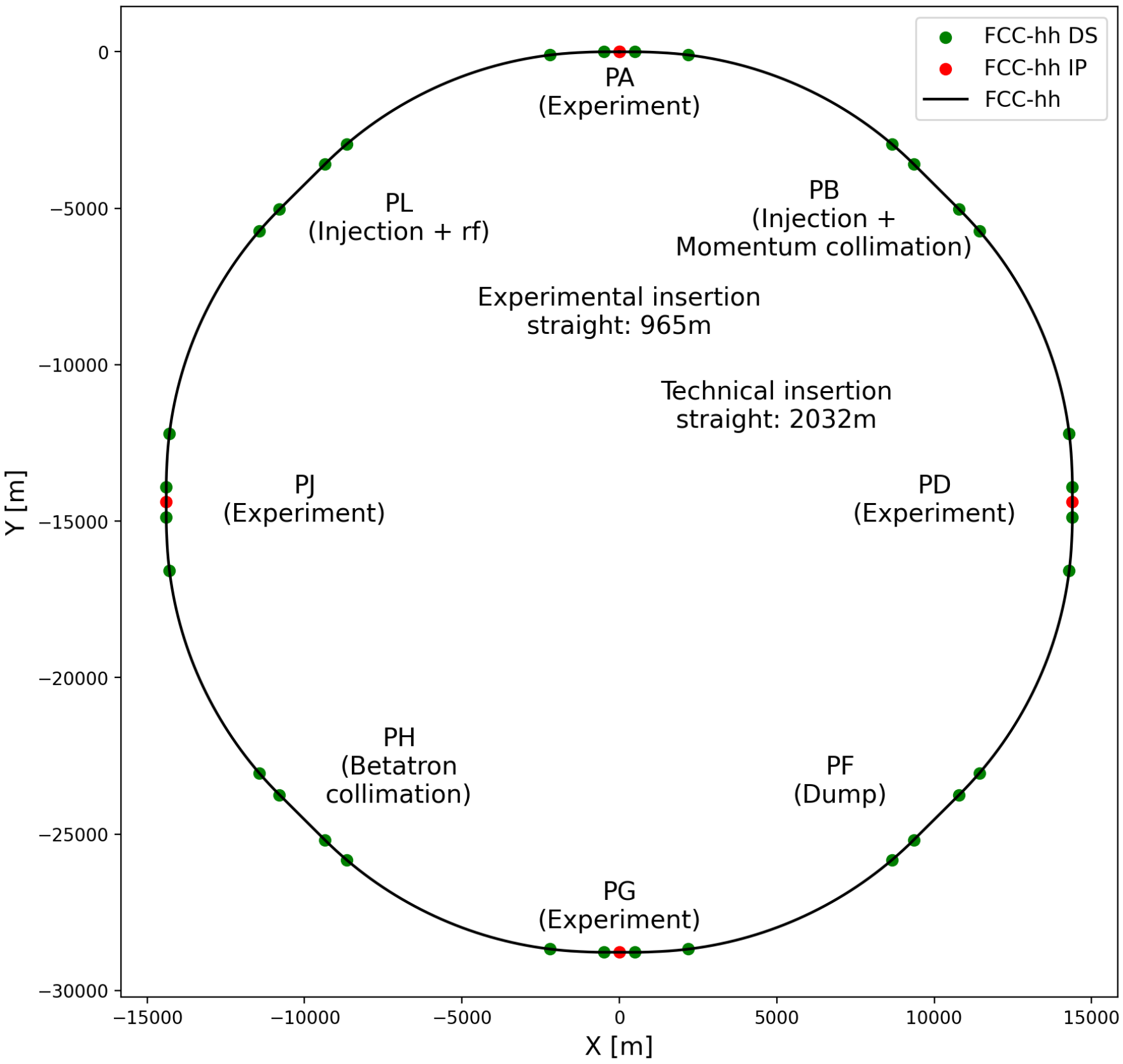

Fig. 10.1: Overall view of the layout of the new FCC-hh ring configuration. The functions of the various insertion regions are mentioned. The experiment straight sections are located with respect to the four-fold symmetry of the ring. The length of the straight sections is also mentioned.

The merging of secondary experiments with injection systems, which resembles the concept implemented in the LHC, was abandoned because it would have generated transfer lines that were too long. The injection of the clockwise beam is performed in PB, and it is combined with the momentum collimation, whereas the injection of the counter-clockwise beam is performed in PL, and it is combined with the RF system. The beam is always injected into the outer beam channel of the FCC-hh accelerator. The two other technical insertions, points PF and PH, each have a single role: to house the beam dump and the betatron collimation systems.

A review of the FCC-hh quadrupole families, including their magnetic properties and the mechanical apertures, has been carried out. The list of magnet families used in the current layout of the FCC-hh ring is presented in Table 10.1. Several of the proposed families in the insertions could still be further optimised as discussed below.

Before entering the details of the discussion of the design of the magnetic lattices of the new layout of the FCC-hh, it is important to mention a crucial change with respect to the CDR. The nominal field of the main dipoles is $14\,\mathrm{T}$, which corresponds to the centre-of-mass energies of $84.6\,\mathrm{TeV}$. The geometry of the proposed layout is fully compatible with the new ranges of dipole field and centre-of-mass energy.

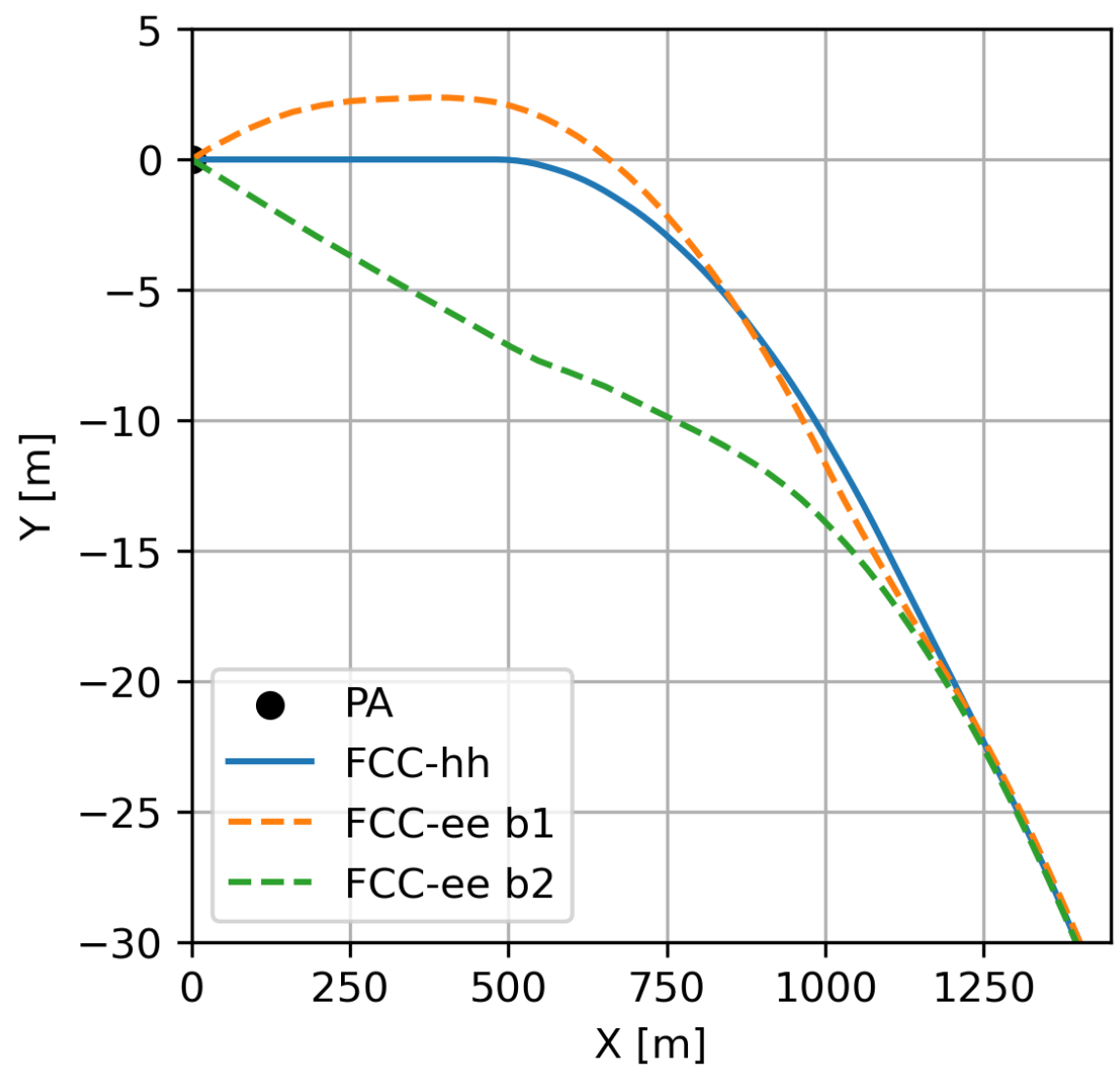

Fig. 10.2: Comparison of the geometry of the FCC-hh ring and for the two FCC-ee rings in the vicinity of PA. The same situation occurs in the other three experiment insertions.

In the rest of the document, the strength of the quadrupoles is expressed for a beam energy of $45\,\mathrm{TeV}$, which amounts to adding a 6% safety margin to the available magnetic strength. When citing minimum magnet strengths, an injection energy of $3.3\,\mathrm{TeV}$ is considered, as had been adopted by the CDR. These value could be scaled linearly to alternative lower injection energies of $1.7$ or $1.3\,\mathrm{TeV}$.

### 10.2.2 Design of regular arcs and dispersion suppressors

The review of the FCC-hh lattice imposed by the placement studies was used to reassess some of the principal assumptions made for the former CDR layout [10]. This is the case for the choice of regular arc cell length. For highest-energy hadron accelerators, designing the lattice with a longer regular cell offers one clear advantage. Namely, the longer cell leads to an increase in the dipole filling factor (for the FCC-hh, this factor was increased from 0.8 as in the CDR to 0.82 at present), with an accompanying increase in the values of beta-functions and dispersion. The larger optical functions enhance the efficiency of the correction systems, such as the chromatic sextupoles and Landau octupoles, but they also render the beam more sensitive to magnetic field errors, in particular at injection, putting additional demands on the magnet design and the correction systems. An obvious upper limit to the cell length is set by the physical dimensions of the vacuum chamber, the injected beam emittance, and the required beam aperture. In Ref. [504], the target aperture values are given as $13.4\,\sigma$ at injection energy and $15.5\,\sigma$ at collision energy, where $\sigma$ denotes the nominal rms beam size computed for a normalised rms emittance of $2.2\,\mu\mathrm{m}$).

In Fig. 10.3, the regular cell with optical parameters and available beam aperture is shown for the current layout. The number of main dipoles per cell has increased from 12 in the CDR [10] to 16. The overall cell layout with the dipole (blue) and quadrupole (red) magnets is indicated at the top. The phase advance per cell was kept at $90^{\circ}$. The cell length is about $275.8\,\mathrm{m}$, to be compared with about $213.0\,\mathrm{m}$ for the CDR. The change in the arc cell results in a larger value of the beta-functions and dispersion, which has several positive and also a few negative side effects. Even for this longer cell, two cryo-jumpers, connected to the quadrupole cryostats, are sufficient to feed all superconducting magnets from the cryogenic line [505]. The beneficial increase in cell length and the induced increase in the values of optical parameters require a small adaptation of the dimensions of the beamscreen. Figure 10.3 shows the cross section of the beam screen from the CDR [10] (dotted line) and a proposed new version with larger horizontal aperture (continuous line). These changes in the beamscreen dimensions will require a more detailed design validation in the next study phase.

Table 10.1: Main parameters of the dipole and quadrupole families used in the entire ring. All fields are referred to 45 TeV to include a safety margin on the required magnetic strength. Note that the orientation of the various aperture types might change according to the polarity of the quadrupole, in order to optimise the beam aperture.

| NAME | Magnetic [1] length [m] | Nominal field [T] | Nominal gradient [T/m] | Aperture cross section | Aperture dimensions [mm] | Coil aperture diameter [mm] | Number of magnets |
|---|---|---|---|---|---|---|---|
| MB | 14.187 | 14 | NA | Beamscreen [2] | – | 50 | 4464 |
| MBX | 10.0 | 14.6 | NA | Octagon [3] | 33.1/33.1/13.5/76.5 | 116 | 8 |
| MBR | 13.0 | 5.6 | NA | Octagon | 33.1/33.1/13.5/76.5 | 80 | 24 |
| MBW | 14.0 | 1.7 | NA | Ellipse [4] | 29.5/22.0 | | 20 |
| MQ | 6.4 | NA | 375.0 | Beamscreen | – | 50 | 536 |
| MQ | 9.6 | NA | 375.0 | Beamscreen | – | 50 | 8 |
| MQ | 10.0 | NA | 375.0 | Beamscreen | – | 50 | 8 |
| MQ | 12.0 | NA | 375.0 | Beamscreen | – | 50 | 24 |
| MQM | 6.0 | NA | 375.0 | Rectellipse [5] | 15.0/13.2, 15.0/15.0 | 50 | 17 |
| MQ1 | 14.3 | NA | 130.0 | Circle | 36.49 | 164 | 16 |
| MQ2 | 12.5 | NA | 105.0 | Circle | 58.24 | 210 | 32 |
| MQ3 | 14.3 | NA | 105.0 | Circle | 58.24 | 210 | 16 |
| MQ4 | 12.0 | NA | 175.0 | Rectellipse | 28.9/24.0, 28.9/28.9 | 70 | 8 |
| MQ5 | 12.0 | NA | 260.0 | Rectellipse | 20.0/18.2, 20.0/20.0 | 60 | 8 |
| MQR | 6.4 | NA | 280.0 | Rectellipse | 20.0/18.2, 20.0/20.0 | 60 | 8 |
| MQY | 6.0 | NA | 200.0 | Rectellipse | 28.9/24.0 - 28.9/28.9 | 60 | 2 |
| MQY | 9.1 | NA | 200.0 | Rectellipse | 28.9/24.0, 28.9/28.9 | 60 | 14 |
| MQYL | 12.8 | NA | 200.0 | Rectellipse | 28.9/24.0, 28.9/28.9 | 60 | 6 |
| MQW | 4.0 | NA | 40.0 | Ellipse | 26.1/15.3 | | 44 |

[1] The specified magnetic length corresponds to an individual magnet. Multiple magnets may be concatenated to form a longer structure if necessary.

[2] This cross-section corresponds to the shape depicted in Fig. 10.3 (right) and is defined by points.

[3] An octagon is defined by four numbers: the half width and half height along main axes, two angles sustaining the cut corner in the first quadrant, given in radians and in order of increasing values.

[4] An ellipse is defined by two numbers specifying the horizontal $a_{\mathrm{h}}$ and vertical $a_{\mathrm{v}}$ semi-axes, given in the form $a_{\mathrm{h}}/a_{\mathrm{v}}$.

[5] A rectellipse is a shape obtained by the intersection of a rectangle and an ellipse [116,211] and is defined by four numbers specifying the half-width $r_{\mathrm{w}}$ and half-height $r_{\mathrm{h}}$ of the rectangle, horizontal $a_{\mathrm{h}}$ and vertical $a_{\mathrm{v}}$ semi-axes of the ellipse, given in the form $r_{\mathrm{w}}/r_{\mathrm{h}}, a_{\mathrm{h}}/a_{\mathrm{v}}$. Note that, in general, the ellipse is replaced by a circle.

The performance of the corrector systems that are present in the arc was also evaluated and reviewed [506], and the new layout of a short straight section of a periodic FODO cell is shown in Fig. 10.4. As a direct consequence of the cell lengthening, each arc consists of fewer cells, reducing the number of available slots for installing the required correctors. However, the larger beta-functions and dispersion enhance the effectiveness of some correctors in their roles. The properties of the correction circuits in the arcs are listed in Table 10.2, where $N_{\mathrm{c}}$ represents the number of circuits per arc, while $N_{\mathrm{m}}$ stands for the number of magnets per circuit. The symbol $n$ in the units represents the order of the magnetic multipole, which is 0 for a dipole.

The analysis of corrector efficiencies has identified several additional optimisations that can be applied to the layout of the Short Straight Sections (SSS). These improvements will enhance the dipole filling factor of the collider and will be incorporated into the next iteration of the FCC-hh ring lattice. In general, a shortening of each SSS of about 4 m can be envisaged. Another possible source of gains in the energy reach of FCC-hh can come from increasing the length of the main dipole. Even for this aspect, the change in the design of the periodic FODO cell will be part of the next version of the magnetic lattice.

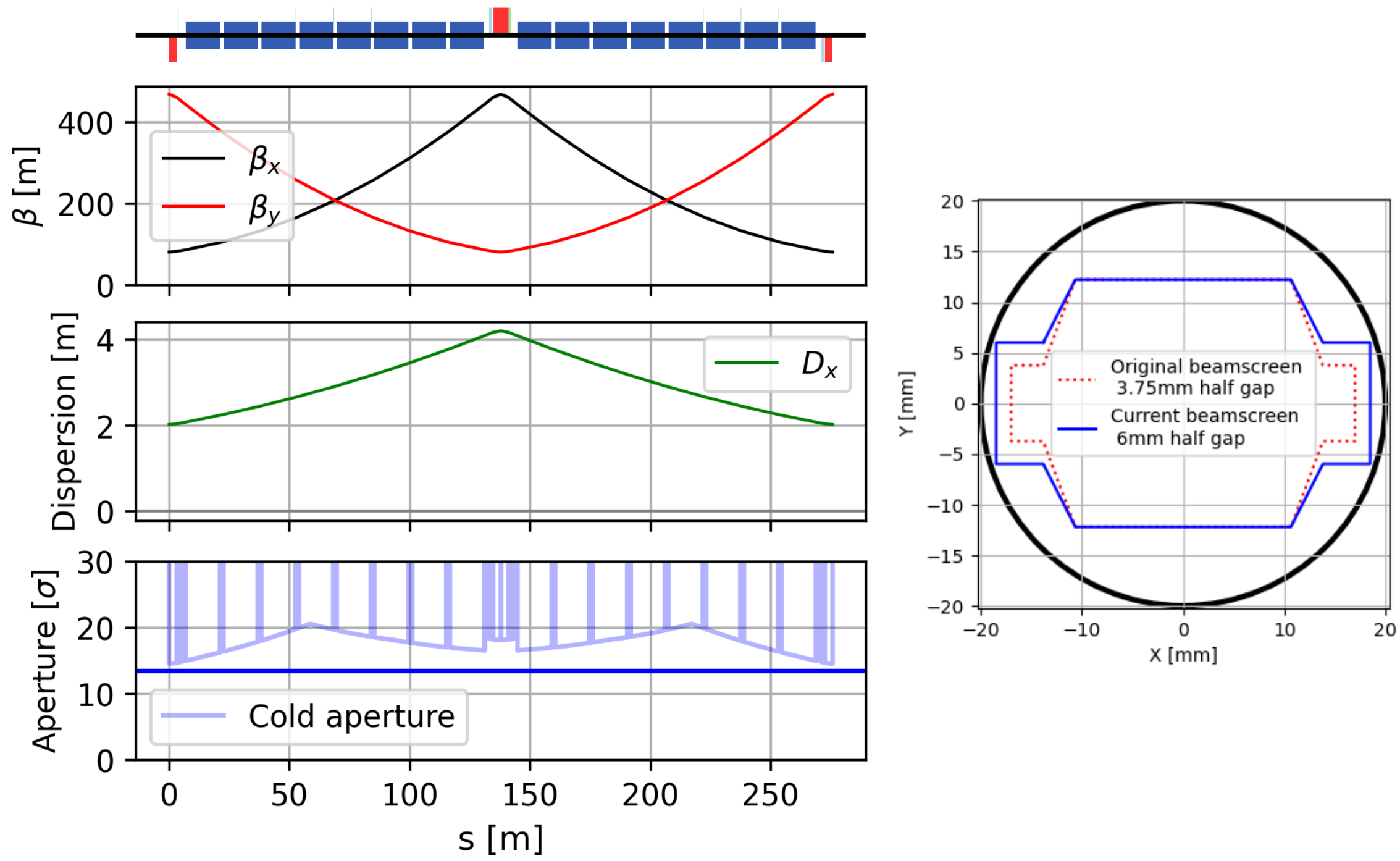

Fig. 10.3: Layout, optical parameters, and dispersion of the 16-dipole periodic FODO cell design (left) and cross section of the beamscreen compatible with the aperture requirements of the arc cell (right). The horizontal line in the aperture plot represents the minimum value acceptable.

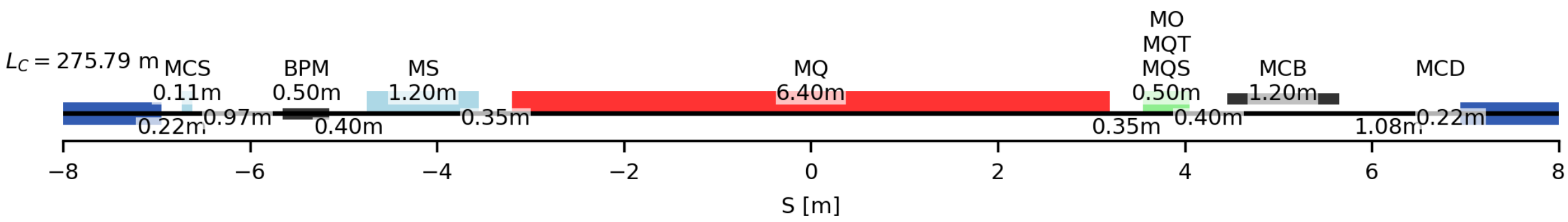

Fig. 10.4: Layout of the periodic FODO cell short straight section (left to right: sextupole spool piece MCS, beam position monitor BPM, chromatic sextupole MS, main quadrupole MQ, Landau octupole MO, trim quadrupole MQT, skew quadrupole MQS, orbit corrector MCB, and decapole spool piece MCD)

The dispersion suppressors (DS) were reviewed starting from the CDR design [10]. The presently proposed layouts are shown in Fig. 10.5, where the DS for the transition between the regular arc and experiment insertion is displayed at the top, and the one connecting the regular arc and a technical insertion at the bottom.

The complexity of the design of the DS for the experiment insertions resides in the need to control the whole geometry and to radially displace the IP position. The control of the geometry is achieved by selectively displacing blocks of the main dipoles, whereas the quadrupoles are kept in regular positions to ensure control of the optics and dispersion. The DS starts just behind Q19 and ends with Q6, and its extent is an essential parameter to ensure optimal geometry control. The quadrupoles in the DS are assumed to have independent powering, which ensures enough optical flexibility. In practice, this can be achieved with main quadrupoles, powered in series with the other main quadrupoles in the regular part of the arc, and trim quadrupoles to ensure independent control of the gradient generated at a given ring location. This scheme has been successfully implemented in the LHC [116] and provides optimised hardware use (quadrupoles and power converters). However, on most occasions, these trim quadrupoles must be set to the opposite polarity of that of the main quadrupole they accompany, trading efficiency for space and simplicity. The details of the quadrupole families and their powering will be studied in more detail at a later stage.

Table 10.2: Correctors circuits properties and capabilities for the FCC-hh Ring. For the correction of the linear coupling, the resonance strength indicated by $C^-$ is defined as $C^- = \frac{1}{2\pi}\int\sqrt{\beta_x\beta_y}k_s e^{i(\mu_x-\mu_y)}\mathrm{d}s$, where $\beta_{x,y}$, $k_s$ and $\mu_{x,y}$ are the beta functions, skew quadrupole gradient, and phase advances, respectively.

| | Length [m] | Strength [$\mathrm{Tm}^{-n}$] | Integral strength [$\mathrm{Tm}^{1-n}$] | $N_\mathrm{c}$ | $N_\mathrm{m}$ | Correctors capabilities |
|---|---|---|---|---|---|---|
| Dipole orbit correctors MCB | 1.2 | 4.5 | 5.4 | 61 | 1 | residual rms orbit $\leq 0.3\,\mathrm{mm}$ |
| Trim quadrupoles MQT | 0.5 | 220 | 110 | 2 | 8 | up to 0.15 tune shift with $\beta$-beating $\leq 1\%$ |
| Skew quadrupoles MQS | 0.5 | 220 | 110 | 2 | 2 | $C^- < 10^{-4}$ |
| Chromatic sextupoles MCS | 1.2 | 7000 | 8400 | 2 | 28 | control around $Q' = 10$ with squeezed optics |
| Landau octupoles MO | 0.5 | $2.2 \times 10^5$ | $1.1 \times 10^5$ | 2 | 18 | Same amplitude detuning as in CDR |

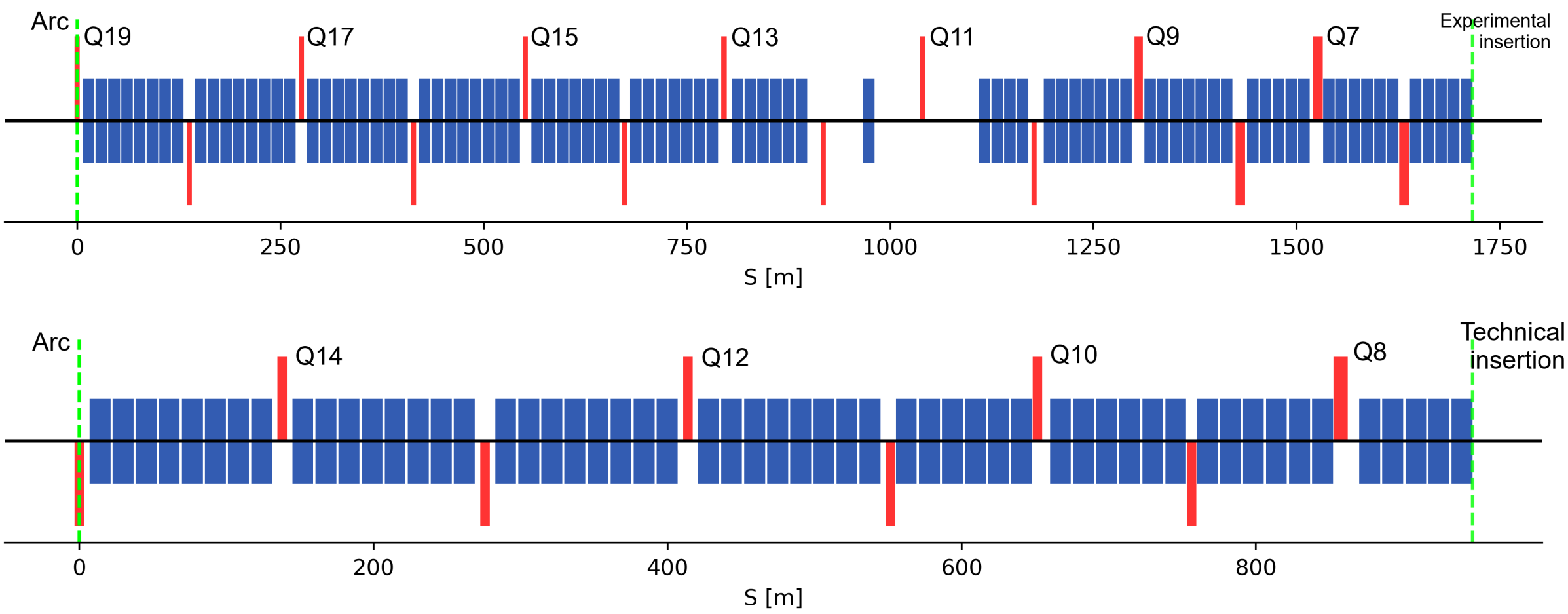

Fig. 10.5: Layout of the dispersion suppressor on the left of an experiment insertion (top) and technical insertion (bottom).

The DS for the technical insertions is shorter, running between, but not including, Q15 and Q7. In this case, the main role of the DS is to control the dispersion since the ring geometry does not need to be changed in the technical insertions. It should also provide enough optical flexibility to enable the matching of the optical parameters of the regular arc with those of the various technical insertions.

Both types of DS have short gaps (of the order of $2\,\mathrm{m}$) to allow the installation of collimators, the so-called TCLD collimators, which should absorb particles that have lost energy in interactions with the primary and secondary collimators.

### 10.2.3 Experiment straight sections

The experiment insertions are the core of the FCC-hh ring design, both for the complex geometry needed to match the radial position of the IPs with those of the FCC-ee and for the requirement to achieve the low $\beta^*$ value needed to reach the target luminosity performance. At injection energy, $\beta^* = 10\,\mathrm{m}$ (like the LHC case [116]), which is then reduced to $\beta^* = 30\,\mathrm{cm}$ at collision energy. The squeeze of $\beta^*$ will take place, at least partially, during the energy ramp, based on considerations of the available aperture that indicate the minimum value $\beta^*$ for a given beam energy.

The layout of the four experimental insertions is identical to maintain consistent optics. However, the polarity of the separation dipoles varies, as both beams change between outer and inner magnet aperture at the different insertions. This causes a slight variation in the horizontal dispersion function, which manifests itself in two distinct forms. Two important differences with respect to the layout presented in Ref. [10] were implemented: the strictly straight beam-line section is about $965\,\mathrm{m}$ instead of $1400\,\mathrm{m}$ as a consequence of the radial displacement of the IP. Furthermore, the separation (MBX type) and recombination (MBR type) dipoles are superconducting to reduce their total length, which is important given the reduced length of the straight section. This set up resembles what is being implemented for the HL-LHC [8,9], where the normal-conducting D1 of the LHC will be replaced by a superconducting one. The D2 recombination dipole is already superconducting in the LHC and will remain so in the HL-LHC, although with an increased coil aperture.

In terms of the magnetic field, the integrated strength of the separation and recombination dipoles is $146\,\mathrm{Tm}$ at $45\,\mathrm{TeV}$, and the nominal fields of the D1 and D2 magnets are $14.6\,\mathrm{T}$ for and $5.6\,\mathrm{T}$, respectively. The radius of the coil aperture has been determined and found to be $58\,\mathrm{mm}$ for D1 and $40\,\mathrm{mm}$ D2. For comparison purposes, the coil radius of the HL-LHC D1 and D2 separation and recombination dipoles is $75\,\mathrm{mm}$ and $52.2\,\mathrm{mm}$, respectively.

The optical parameters, dispersion, and beam aperture for the layout of the PD experiment insertion are shown in Fig. 10.6 for the solution with $\beta^* = 10\,\mathrm{m}$ for the injection energy (left) and $\beta^* = 30\,\mathrm{cm}$ for collision energy. The most striking difference is the maximum value of the beta-functions that reaches about $2000\,\mathrm{m}$ and $70\,000\,\mathrm{m}$ for injection and collision optics, respectively. The target aperture values for the two cases differ and so do the bottleneck locations. For the injection optics, aperture limits appear in the dispersion suppressors surrounding the straight section. Conversely, for the collision optics, including the orbit from the crossing angle, the aperture limit is first reached in the separation dipole (D1), followed by the inner triplet, while the rest of the insertion provides aperture values that largely exceed the target. This suggests that, in the future, the aperture of the D1 separation dipole should be increased. On the positive side, in view the available aperture margin in Q1, a further reduction of $\beta^*$ could be envisaged.

Note that the ring chromaticity varies during the squeeze due to the contribution of the experimental insertions to the overall ring chromaticity. The strength of the chromatic sextupoles needed to obtain $Q' = 10$ when $\beta^* = 10\,\mathrm{m}$ is approximately $19\,\%$ of that required to obtain the same $Q'$ when $\beta^* = 30\,\mathrm{cm}$.

A solution has been found for the squeeze from $\beta^* = 10\,\mathrm{m}$ to $\beta^* = 30\,\mathrm{cm}$ and the corresponding quadrupole strengths are shown in Fig. 10.7. The starting point for this transition is not compatible with the aperture requirements at injection energy. Therefore, the transformation of the nominal injection optics to the starting point of the squeeze sequence is envisioned during the energy ramp[1], as the squeeze sequence satisfies the aperture requirements when the beam energy reaches $10\,\mathrm{TeV}$ or higher.

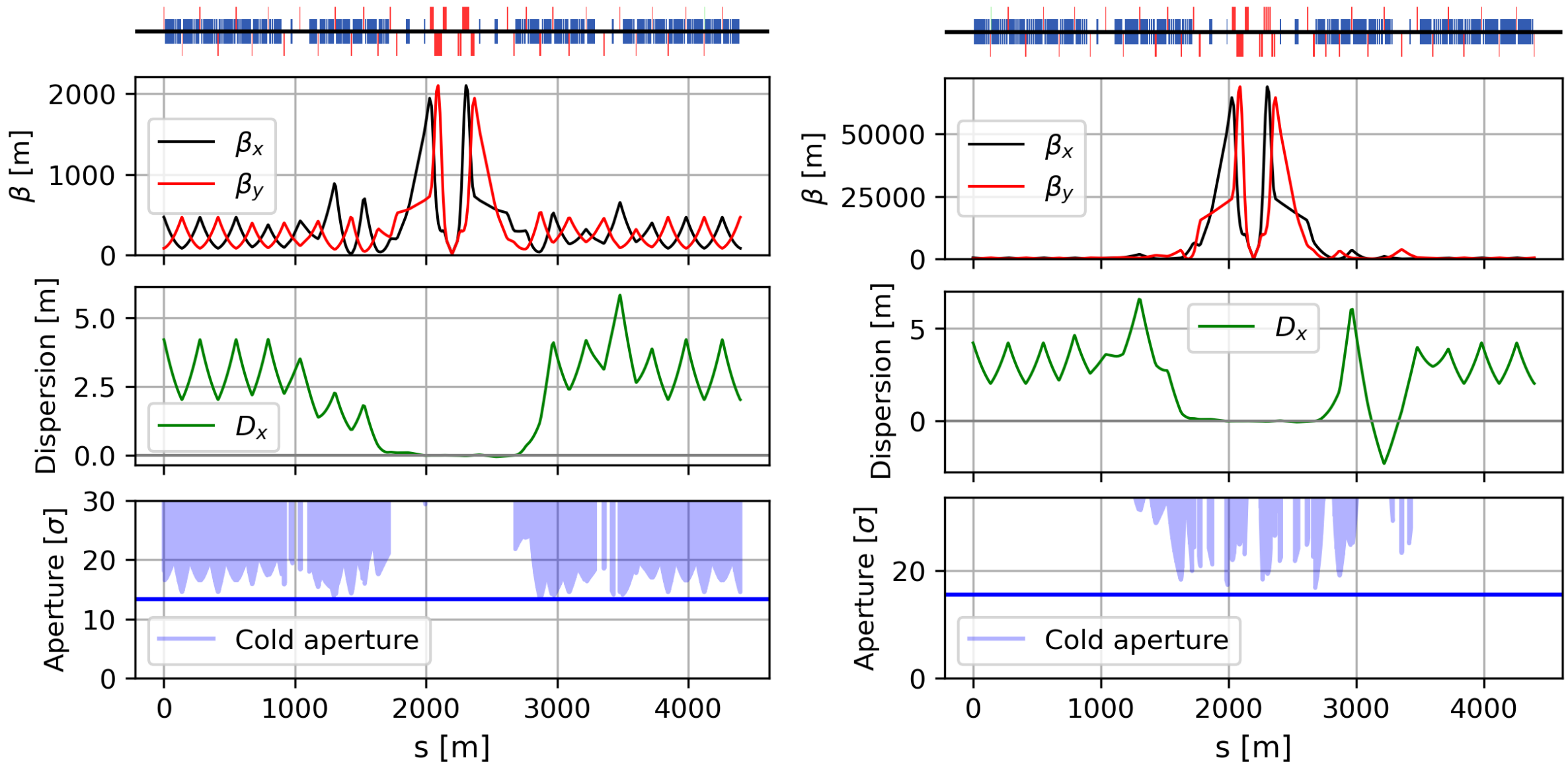

Fig. 10.6: Optical parameters, dispersion, and beam aperture for the injection optics of the experiment insertion at PA (left, $\beta^* = 10\,\mathrm{m}$) and the collision optics (right, $\beta^* = 30\,\mathrm{cm}$) including crossing angle. The horizontal line in the aperture plot represents the minimum value acceptable.

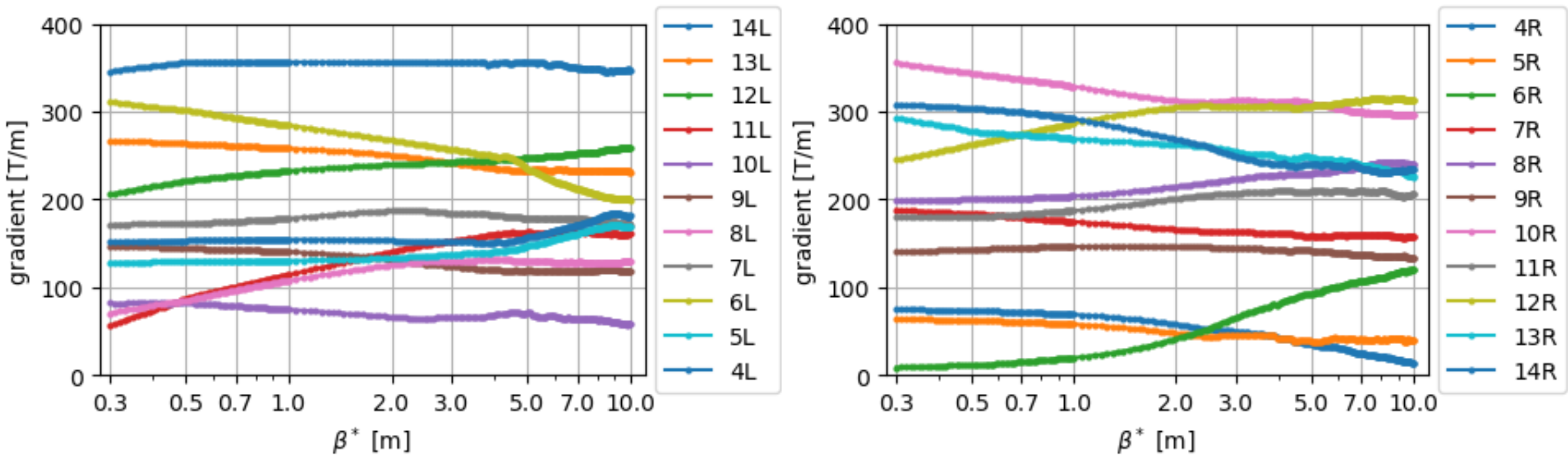

Fig. 10.7: Quadrupole strength during the $\beta^*$ squeeze for the left and right sides of the experiment insertion (left and right plots, respectively).

The strength values of the quadrupoles that generate the two optical configurations of the experiment insertions are summarised in Table 10.3.

A cornerstone feature for the performance of the experiment insertions is the crossing scheme. A set of correctors generates a local orbit bump around the IP to mitigate the effect of long-range beam-beam interactions [10]. The orbit also separates the beams, preventing collisions until the beams are at top energy and the optics is squeezed. Although the bump is fully corrected and, therefore, transparent to the rest of the ring, the horizontal and vertical dispersion generated must be taken into account. The strategy followed is similar to that employed in the LHC [116]. At injection, the quadrupoles in the dispersion suppressors are used to rematch the horizontal dispersion to the neighbouring arcs, while the standing vertical dispersion is small enough that it does not necessitate countermeasures. When transitioning to collision optics, a set of orbit correctors in the arc is used to create a closed-orbit bump that exploits the field of the chromatic sextupoles to cancel the vertical dispersion. This is illustrated in Fig. 10.8, where the crossing and separation bumps are presented at injection energy (top) and at top energy (bottom) for the nominal value of the half-crossing angle and half-separation that are $100\,\mu\mathrm{rad}$ and $1.5\,\mathrm{mm}$, respectively.

[1]The duration of the energy ramp is approximately $20\,\mathrm{min}$.

Table 10.3: Strength of the insertion quadrupoles for the injection optics of the experiment insertion at PA (second column, $\beta^* = 10\,\mathrm{m}$) and the collision optics (third column, $\beta^* = 30\,\mathrm{cm}$). All the gradients correspond to $45\,\mathrm{TeV}$.

| NAME | Gradient [T/m] | Gradient [T/m] | Gradient [T/m] | Gradient [T/m] | Maximum gradient [T/m] |
|---|---|---|---|---|---|
| MQ.14L | −127.9 | −172.3 | −127.9 | −172.3 | 375.0 |
| MQ.13L | 119.9 | 133.1 | 119.9 | 133.1 | 375.0 |
| MQ.12L | −114.1 | −102.9 | −114.1 | −102.9 | 375.0 |
| MQ.11L | 94.9 | 28.2 | 94.9 | 28.2 | 375.0 |
| MQ.10L | −116.6 | −40.9 | −116.6 | −40.9 | 375.0 |
| MQ.9L | 190.9 | 146.5 | 190.9 | 146.5 | 375.0 |
| MQ.8L | −176.4 | −70.1 | −176.4 | −70.1 | 375.0 |
| MQ.7L | 222.7 | 170.6 | 222.7 | 170.6 | 375.0 |
| MQ.6L | −129.1 | −311.2 | −129.1 | −311.2 | 375.0 |
| MQY.5L | 149.3 | 127.8 | 155.1 | 128.0 | 260.0 |
| MQY.4L | −88.0 | −151.7 | −90.9 | −151.6 | 175.0 |
| MQXE.3L | 94.8 | 94.8 | 94.8 | 94.8 | 105.0 |
| MQXD.2L | −92.9 | −92.9 | −92.9 | −92.9 | 105.0 |
| MQXC.1L | 114.1 | 114.1 | 114.1 | 114.1 | 130.0 |
| MQXC.1R | −114.1 | −114.1 | −114.1 | −114.1 | 130.0 |
| MQXD.2R | 92.9 | 92.9 | 92.9 | 92.9 | 105.0 |
| MQXE.3R | −94.8 | −94.8 | −94.8 | −94.8 | 105.0 |
| MQY.4R | 60.0 | 75.8 | 60.0 | 74.6 | 175.0 |
| MQY.5R | −72.7 | −64.8 | −70.0 | −63.5 | 260.0 |
| MQ.6R | 124.4 | 8.6 | 124.4 | 8.6 | 375.0 |
| MQ.7R | −149.2 | −187.9 | −149.2 | −187.9 | 375.0 |
| MQ.8R | 166.7 | 198.9 | 166.7 | 198.9 | 375.0 |
| MQ.9R | −117.2 | −140.0 | −117.2 | −140.0 | 375.0 |
| MQ.10R | 85.4 | 178.1 | 85.4 | 178.1 | 375.0 |
| MQ.11R | −101.2 | −90.3 | −101.2 | −90.3 | 375.0 |
| MQ.12R | 108.2 | 122.7 | 108.2 | 122.7 | 375.0 |
| MQ.13R | −120.8 | −146.0 | −120.8 | −146.0 | 375.0 |
| MQ.14R | 107.2 | 153.9 | 107.2 | 153.9 | 375.0 |

The orbit bump at top energy that created the dispersion bump is clearly visible, and the overall correction of the spurious vertical dispersion is good. The primary constraint is the alternating crossing plane between PA and PG. The crossing and separation schemes in all experimental insertions are designed to accommodate all possible combinations of crossing and separation planes while ensuring compatibility with this constraint.

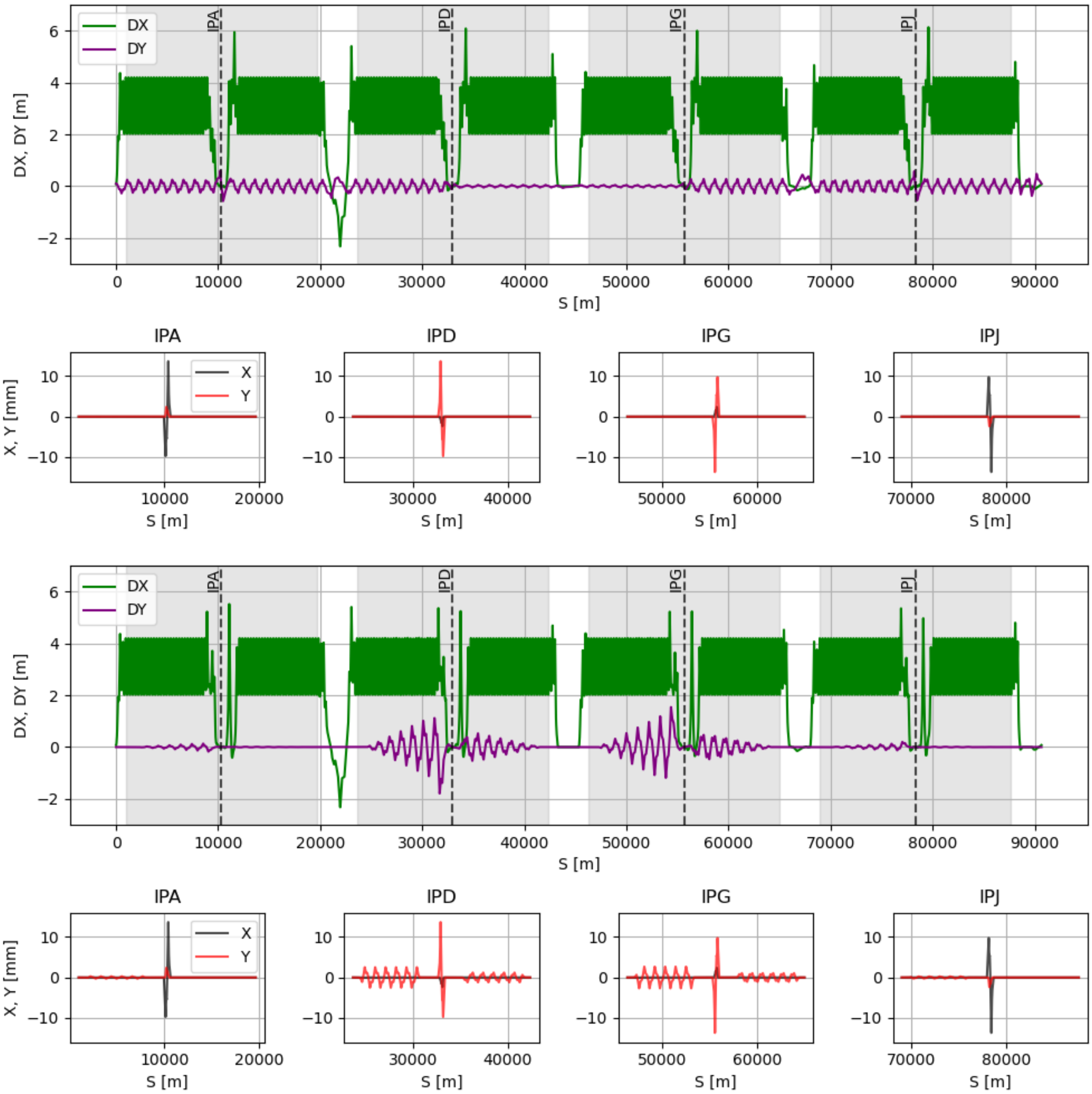

Fig. 10.8: Closed dispersion and orbit with crossing angle and separation at injection (top) and collision (bottom) energies with a closer look at the experimental insertions.

### 10.2.4 Beam dump

As described above, changes in the ring layout with respect to CDR [10] reduced the length of the technical straight sections from 2800 m to 2032 m. The new layout accommodates the dump systems for both beams at PF, as a result of civil engineering considerations. At an intermediate stage [507], the possibility of merging the dump and injection systems for the clockwise beam in PB was considered. However, this option was deemed unfeasible, primarily due to machine protection concerns arising from the complexities of handling combined failure scenarios for the injection and dump systems. The necessary layout modifications could only be resolved by dedicating a technical insertion exclusively to the beam dump. The system is illustrated in Fig. 10.9 (left), where the circulating and extracted beam envelopes are shown for the nominal extraction case, ensuring sufficient clearance.

The externally activated dumping system has the function to extract and dilute the beam, and to dispose the entire extracted beam onto an external absorber located 2.5 km from the extraction point in a dedicated cavern. The design of the complete dump system is driven both conceptually and hardware-wise by machine protection considerations [508].

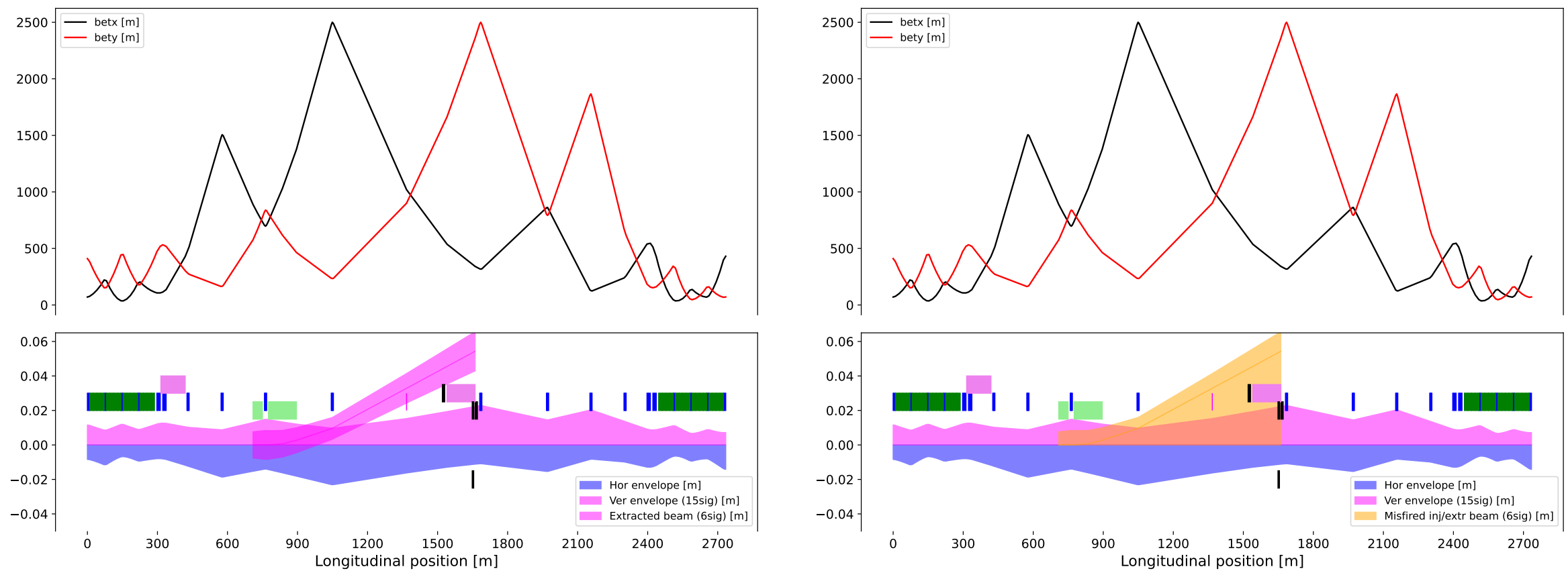

Fig. 10.9: Left: Optics (top) and beam envelopes and active elements (bottom) for the extraction straight section around PF. Right: Dump system failure scenario: asynchronous beam dump. The configurations shown refer to collision energy.

The extraction system design had to be adapted to the reduced length of the straight section, leading to an increased length of the superconducting extraction septa (utilising superconducting shield and truncated cosine-theta technologies), as well as higher switch voltage and an extended system length for the extraction kickers. At this stage, it is challenging to predict the switch technology that will be available on the FCC-hh timescale. However, if significant difficulties arise in achieving the required hardware parameters, an alternative approach could allow the extracted beam to pass through an aperture in the cryostat of the downstream quadrupole, reducing the required septum deflection angle by approximately one-third of its nominal strength.

The design of systems involving fast pulsed devices acting in a single turn on a beam with unprecedented power, such as that of the FCC-hh, is driven by their failure scenarios. A relevant failure scenario concerning the asynchronous beam dump is shown in Fig.10.9 (right), where the misfiring of an extraction or dilution kicker leads to an immediate trigger of the full system that, however, is not synchronised with the beam abort gap. Dedicated protection absorbers are expected to protect the downstream machine from particle spray. The damage limit of the absorbers drives the extraction kicker rise time. This limit is most critical for global machine protection, in particular in the case that a beam abort occurs with the full beam at top energy. For this specific failure scenario, only a single dumped beam is shown. The systems for the other beam are arranged symmetrically around the centre of the straight section.

The general optical parameters and the beam aperture are shown in Fig. 10.10

The strength and aperture values for the quadrupoles that generate the optical configuration of the beam dump are summarised in Table 10.4.

### 10.2.5 RF and injection straight section

The technical insertion at PL is used to accommodate the RF system and the injection of the counter-clockwise beam. The main constraints imposed by the RF system are the increased inter-beam distance, which must reach 420 mm to accommodate the accelerating cavities, and the cancellation of dispersion in the straight section where the cavities are located to minimise synchro-betatron coupling. These constraints are met by introducing two doglegs that both increase the inter-beam distance and ensure that the dispersion and its derivative are zero in the central part of the insertion.

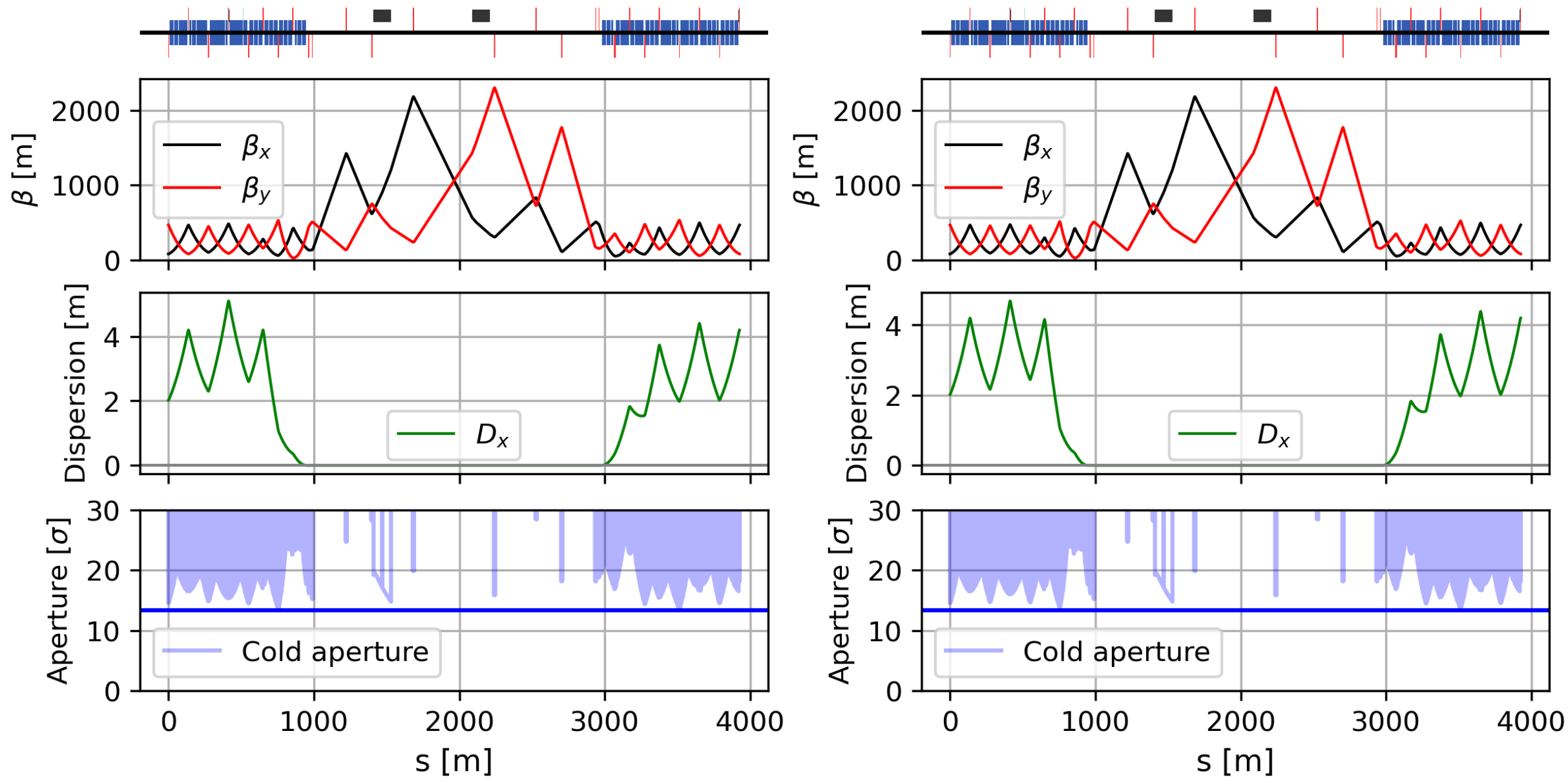

Fig. 10.10: Optical parameters, dispersion, and beam aperture at injection of two configurations of the dump insertion with different settings in the dispersion suppressors to match the ring to injection (left) and collision (right) tunes. MKD and MSD positions are highlighted in black. The horizontal line in the aperture plot represents the minimum value acceptable.

The beam is injected in the vertical direction, as the transfer line will run above the main ring magnets, just on top of the external aperture of the circulating beam. Optimal injection conditions require a phase advance of $90^\circ$ between the kicker and the injection protection dump, and maximising the value of $\sqrt{\beta_x\beta_y}$ at the location of the injection protection dump. An additional constraint in the design of the injection of the counter-clockwise beam is to use the same hardware as installed in PB for the injection of the clockwise beam: this does not pose any serious challenge for the overall design of the insertion.

Figure 10.11 shows the vertical trajectory of the injected beam, assuming a difference in height between the circulating and injected beams of $1250\,\mathrm{mm}$. This first block of bending magnets represents the end of the transfer line, followed by a downward bending dipole, a second dipole that starts to deflect the trajectory to become parallel with the plane of the collider and then the septa, with a field that decreases with the reduction of the thickness of the septum blade (the details are given in Table 10.5). The last block of bending magnets represents the kickers that cancel the residual angle of the injected beam. The second dipole is a recent addition to the magnetic sequence that helps to reduce the longitudinal footprint of the insertion dedicated to the injection elements and is labelled MBWI in Table 10.5. It is installed in a region where the injected and the circulating beams are separated enough that a conventional dipole can be used rather than a septa-type magnet. This allows using a stronger magnetic field of $1.4\,\mathrm{T}$ to accelerate the geometrical transition between the plane of the transfer line to that of the collider.

Note that the quadrupole located in between the two groups of bending devices is used to provide an additional dipole kick exploiting the off-axis traversal of the beam.

Figure 10.12 shows the optical parameters, dispersion, and beam aperture at injection, corresponding to the most critical configuration. The part of the insertion close to the DS on the right-hand side (light-blue box in the figure) is where the injection system is located: it is left empty here, as the injection elements are inactive for the circulating beam, and they do not affect the beam whose optics are shown here. The section envisaged for the RF system (green box in the figure) has a length of approximately $870\,\mathrm{m}$.

Table 10.4: Strength of the insertion quadrupoles for the optics of the technical insertion at PF (beam dump) and its dispersion suppressors. All the gradients correspond to $45\,\mathrm{TeV}$.

| NAME | Gradient $[\mathrm{T/m}]$ | Maximum gradient $[\mathrm{T/m}]$ |
|---|---|---|
| MQ.14LF.B1 | 116.47 | 375.0 |
| MQ.13LF.B1 | −120.36 | 375.0 |
| MQ.12LF.B1 | 116.72 | 375.0 |
| MQ.11LF.B1 | −115.10 | 375.0 |
| MQ.10LF.B1 | 155.11 | 375.0 |
| MQ.9LF.B1 | −160.22 | 375.0 |
| MQ.8LF.B1 | 62.91 | 375.0 |
| MQM.B7LF.B1 | −151.48 | 375.0 |
| MQM.A7LF.B1 | −151.48 | 375.0 |
| MQY.6LF.B1 | 43.49 | 200.0 |
| MQY.5LF.B1 | −47.69 | 200.0 |
| MQY.4LF.B1 | 24.98 | 200.0 |
| MQY.4RF.B1 | −24.98 | 200.0 |
| MQY.5RF.B1 | 47.69 | 200.0 |
| MQY.6RF.B1 | −43.49 | 200.0 |
| MQM.A7RF.B1 | 136.32 | 375.0 |
| MQM.B7RF.B1 | 136.32 | 375.0 |
| MQ.8RF.B1 | −49.48 | 375.0 |
| MQ.9RF.B1 | 169.34 | 375.0 |
| MQ.10RF.B1 | −127.26 | 375.0 |
| MQ.11RF.B1 | 142.57 | 375.0 |
| MQ.12RF.B1 | −116.81 | 375.0 |
| MQ.13RF.B1 | 125.28 | 375.0 |
| MQ.14RF.B1 | −131.00 | 375.0 |

Table 10.6 summarises the strength of the quadrupoles that generate the optical configuration of RF and injection insertion.

### 10.2.6 Betatron collimation

The collimation system is meant to clean the betatron halo and off-momentum particles surrounding the main beam. These two functions are performed by two subsystems located in two insertions to allow for dedicated optical conditions. Cleaning of the betatron halo requires small values of the dispersion function and appropriate phase advances between the various collimation stages. Cleaning off-momentum particles requires large normalised dispersion[2] values and is discussed in further detail in Section 10.2.7.

A side effect of the collimation system is the beam impedance originating from the jaws that are close to the beam edge. Recently, for the HL-LHC, mitigation measures for the beam impedance based on local changes to the beam optics were proposed (see Ref. [509, 510] and references therein), which aim at increasing the beta functions at the collimators. This novel concept has also already been incorporated into the optics design of the FCC-hh betatron collimation insertion.

Two additional novel features were implemented in the layout of the collimation insertions, both related to the geometry of the doglegs that are part of the insertions. The main function of doglegs in the LHC collimation insertions is to filter out the neutral particles generated by the beam-matter interaction in the jaws of the collimators [511]. Thanks to the doglegs, the neutrals are prevented from depositing energy in the coils of the first superconducting dipole on the opposite side of the insertion, i.e., at the beginning of the DS. Hence, the LHC collimation geometry need not be exported to the FCc-hh, by rescaling it to the higher beam energy. Instead, it is sufficient to introduce the minimum deflection that brings the neutrals out of the cross-section of the coils of the first superconducting dipole at the beginning of the downstream DS. This approach simplifies the layout of the doglegs and relaxes the requirements in terms of dipole strength needed to generate the doglegs. Based on these considerations, each of the two doglegs of the PH insertion is made of two dipole blocks (composed of two MBW-type dipoles) with opposite polarities, and each block generates $46.0\,\mathrm{Tm}$ at collision energy.

[2] The normalised dispersion is defined as $D_x/\sqrt{\beta_x}$.

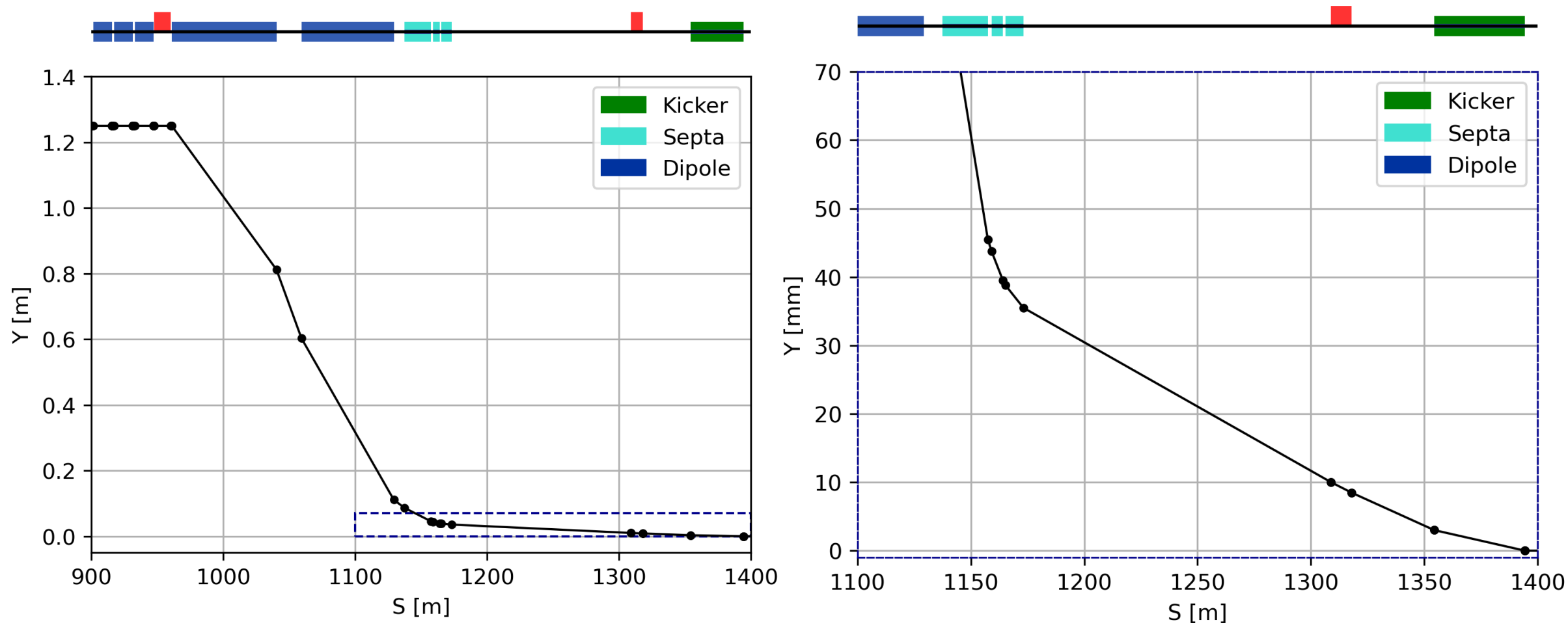

Fig. 10.11: Trajectory in the vertical plane of the injected counter-clockwise beam from the transfer line arc to the injection kicker (left) and a detailed view of the path through the septa (right). Exceptionally, the description of the layout is made using the counter-clockwise beam as a reference.

Table 10.5: Electromagnetic and aperture parameters of the septa, dipole, and kickers assumed for the injection of the counter-clockwise beam.

| NAME | Active length [m] | Blade thickness [mm] | Magnetic field [T] | Aperture cross section | Aperture dimensions [mm] |
|---|---|---|---|---|---|
| MSI1 | 8.0 | 8 | 0.7 | Ellipse | 12.0 / 11.0 |
| MSI2 | 5.0 | 12 | 1.0 | Circle | 16.0 |
| MSI3 | 20.0 | 18 | 1.2 | Circle | 16.0 |
| MBWI | 70.0 | NA | 1.4 | Circle | 16.0 |
| MKI | 40.0 | NA | 0.024 | Circle | 18.3 |

The second change is keeping the inter-beam distance constant along the whole straight insertion. In fact, in the LHC, the arc inter-beam distance of $194\,\mathrm{mm}$ is increased in the collimation insertions to $224\,\mathrm{mm}$ by the geometry of the doglegs of the two counter-rotating beams. In the layout of the FCC-hh collimation insertions, the inter-beam distance remains constant and has the same value as in the arcs, namely $250\,\mathrm{mm}$. This choice has a fundamental consequence, as in this case, not only the optical parameters are the same for the counter-rotating beams, but also the dispersion functions are the same. Hence, this design choice facilitates having exactly identical layouts and optics for the two beams.

The Q6 quadrupoles, previously composed of six MQTLH magnets as in the LHC [116], have been replaced in the new layouts by a single MQY quadrupole. This change optimises the beam aperture, eliminating an unnecessary bottleneck.

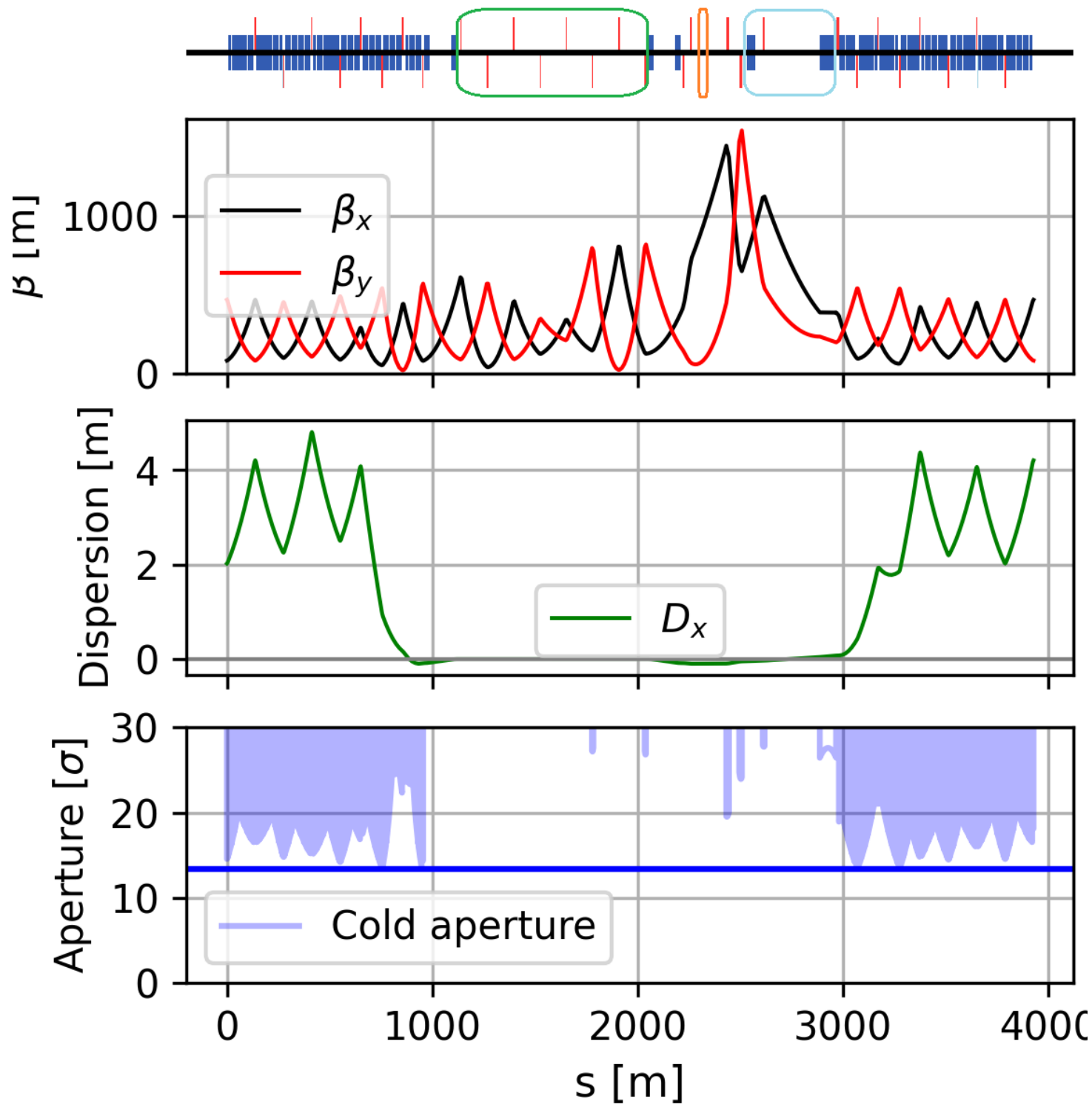

Fig. 10.12: Optical parameters, dispersion and beam aperture at injection, corresponding to the most critical configuration for the optics of the RF and injection insertion for the anti-clockwise beam. Sections dedicated to a specific function are highlighted: the area where the RF cavities may be located (green), injection protection devices (orange) and injection elements (light blue). The horizontal line in the aperture plot represents the minimum value acceptable.

Additionally, to maximise the available beam aperture, the orientation of the elliptical beam pipe in the MQW quadrupoles is adjusted according to the quadrupole polarity.

The betatron collimation subsystem is located in PH, and the corresponding layout and optical functions are shown in Fig. 10.13. As discussed above, a novel optical configuration featuring high-beta values has been designed to mitigate the beam impedance generated by the collimators. However, this optics is not suitable for the injection energy because here it would violate the aperture requirements. Furthermore, cleaning performance is essential at collision energy and much less important at injection. Therefore, two optical configurations were designed: one with mid-range values of the beta-function to be used at injection, and one with large values of the beta-functions that should be used at collision energy. These two configurations are those shown in Fig. 10.10 (left, for the injection configuration and right, for the collision configuration). An optical transition between the two configurations is envisioned during the energy ramp. This same type of transition will first be tested at the LHC during Run 3, with the goal of becoming operational for the HL-LHC. Consequently, the FCC-hh will benefit from existing operational experience with dynamic optical changes in the collimation insertions. It is important to highlight that, in both LHC and FCC-hh, the target aperture differs between the two configurations, and, in all cases, the aperture requirements are satisfied.

Table 10.7 summarises the strength and aperture values for the quadrupoles generating the two optical configurations of the insertion for betatron collimation.

Table 10.6: Strength of the insertion quadrupoles for the RF insertion at PL and its dispersion suppressors. All the gradients correspond to $45\,\mathrm{TeV}$.

| NAME | Gradient $[\mathrm{T/m}]$ | Maximum gradient $[\mathrm{T/m}]$ |
|---|---|---|
| MQ.14LL.B1 | 117.49 | 375.0 |
| MQ.13LL.B1 | −115.18 | 375.0 |
| MQ.12LL.B1 | 115.75 | 375.0 |
| MQ.11LL.B1 | −110.70 | 375.0 |
| MQ.10LL.B1 | 158.84 | 375.0 |
| MQ.9LL.B1 | −170.55 | 375.0 |
| MQ.8LL.B1 | 75.86 | 375.0 |
| MQM.7LL.B1 | −353.09 | 375.0 |
| MQR.6LL.B1 | 261.98 | 280.0 |
| MQR.5LL.B1 | −252.75 | 280.0 |
| MQR.4LL.B1 | 274.79 | 280.0 |
| MQR.3LL.B1 | −180.75 | 280.0 |
| MQR.2LL.B1 | 187.29 | 280.0 |
| MQR.1LL.B1 | −274.64 | 280.0 |
| MQR.01LL.B1 | 233.43 | 280.0 |
| MQR.1RL.B1 | −259.94 | 280.0 |
| MQY.2RL.B1 | −78.80 | 200.0 |
| MQY.3RL.B1 | 71.95 | 200.0 |
| MQYL.4RL.B1 | 87.58 | 200.0 |
| MQYL.5RL.B1 | −132.28 | 200.0 |
| MQY.6RL.B1 | 67.81 | 200.0 |
| MQYL.7RL.B1 | 96.84 | 200.0 |
| MQ.8RL.B1 | −48.36 | 375.0 |
| MQ.9RL.B1 | 150.32 | 375.0 |
| MQ.10RL.B1 | −113.83 | 375.0 |
| MQ.11RL.B1 | 139.02 | 375.0 |
| MQ.12RL.B1 | −100.32 | 375.0 |
| MQ.13RL.B1 | 116.73 | 375.0 |
| MQ.14RL.B1 | −116.30 | 375.0 |

### 10.2.7 Momentum collimation and injection

The technical insertion at PB serves two functions, namely injection of the clockwise circulating beam and momentum collimation. A transition region of approximately $530\,\mathrm{m}$ (from the kicker to the primary collimator) is used to match the optics between these two systems, while also providing shielding between the kickers and momentum collimation debris, particularly in the event of injection losses.

This insertion presents two main challenges from a beam optics perspective. First, the optimal beam properties for maximising the performance of each system individually are conflicting. Ideally, the injection region should have zero dispersion to prevent the need for very tight injection protection absorber settings, which must otherwise account for momentum offsets while maintaining alignment with the collimation hierarchy.

Conversely, a large value of the normalised dispersion is necessary for the momentum collimation systems to function effectively. Additionally, to protect each system from potential losses originating from the other, the longitudinal footprint of each subsystem must be carefully allocated, including a buffer area with shielding between them.

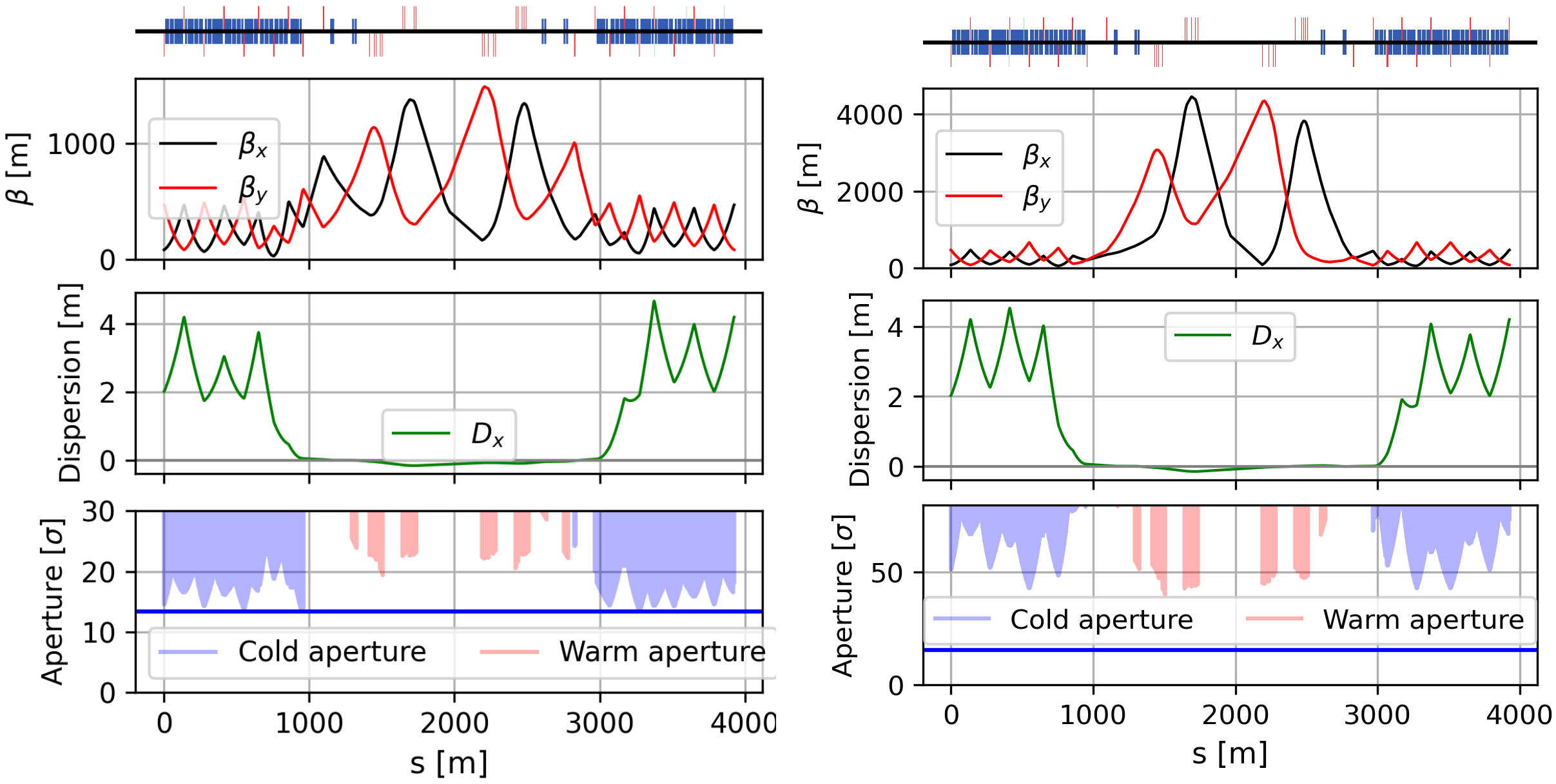

Fig. 10.13: Optical parameters, dispersion and beam aperture for the low-beta optics (for injection) of the betatron collimation insertion PH (left) and the high-beta (for collision) variant (right). The blue and the red aperture regions indicate superconducting and normal-conducting magnets, respectively. The horizontal line in the aperture plot represents the minimum value acceptable.

The first challenge is addressed by fine-tuning the upstream dispersion suppressor to ensure that the residual dispersion from the arc meets these constraints. The second challenge is managed by optimising the injected beam trajectory, as discussed in the previous section.

In this insertion, the optics of the two beams can differ if necessary since the counter-clockwise beam is not subject to injection constraints. This flexibility allows its optics to be configured in a way that minimises the impact of momentum collimation losses on the injection elements.

Doglegs are also needed for the momentum collimation system, and each block of dipoles (composed of three MBW-type dipoles) generates 69.0 Tm at collision energy.

The layout and hardware configuration of the injection system are identical, but mirror reflected, of that shown in Fig. 10.11 and Table 10.5.

Figure 10.14 shows the optical parameters, dispersion, and beam aperture at injection that corresponds to the most critical configuration. The dispersion is matched so that it is small as the beam reaches the injection kicker and then grows as the beam reaches the dogleg where the primary collimator is located.

The strength values for the quadrupoles that generate the optical configuration of the insertion for the off-momentum collimation are summarised in Table 10.8.

### 10.2.8 Impedance considerations

Beam-coupling impedance remains largely unchanged with respect to the CDR [10, Section 2.4.7]). The main implications from the change of layout are related to a shorter ring circumference, which implies a slightly larger revolution frequency and a reduced total length of beam screens and vacuum pipe, along with lower maximum beam energy, a lower magnetic field on the beam screen at top energy, and possibly a higher operating temperature of the beamscreen. The vertical dimension of the beam screen remains unchanged from the CDR, with a half height of 12.2 mm.

Table 10.7: Strength of the insertion quadrupoles for the betatron insertion at PH (second column, low-beta, third column, high-beta) and its dispersion suppressors. All the gradients correspond to 45 TeV.

| NAME | Gradient [T/m] | Gradient [T/m] | Maximum gradient [T/m] |
|---|---|---|---|
| MQ.14LH.B1 | 117.3 | 127.6 | 375.0 |
| MQ.13LH.B1 | −103.8 | −110.4 | 375.0 |
| MQ.12LH.B1 | 113.2 | 116.3 | 375.0 |
| MQ.11LH.B1 | −105.7 | −124.0 | 375.0 |
| MQ.10LH.B1 | 152.8 | 168.7 | 375.0 |
| MQ.9LH.B1 | −115.9 | −108.8 | 375.0 |
| MQ.8LH.B1 | 52.4 | 61.4 | 375.0 |
| MQM.7LH.B1 | −57.5 | −206.0 | 375.0 |
| MQY.6LH.B1 | 28.3 | 79.3 | 200.0 |
| MQW.5LH.B1 | −25.5 | −28.5 | 44.0 |
| MQW.4LH.B1 | 38.6 | 34.8 | 44.0 |
| MQW.4RH.B1 | −33.2 | −31.9 | 44.0 |
| MQW.5RH.B1 | 38.2 | 34.8 | 44.0 |
| MQY.6RH.B1 | −154.3 | −83.9 | 200.0 |
| MQM.7RH.B1 | 215.7 | 207.8 | 375.0 |
| MQ.8RH.B1 | −58.2 | −36.7 | 375.0 |
| MQ.9RH.B1 | 159.5 | 142.6 | 375.0 |
| MQ.10RH.B1 | −114.2 | −114.5 | 375.0 |
| MQ.11RH.B1 | 139.5 | 140.0 | 375.0 |
| MQ.12RH.B1 | −88.9 | −97.3 | 375.0 |
| MQ.13RH.B1 | 114.6 | 113.5 | 375.0 |
| MQ.14RH.B1 | −103.9 | −113.2 | 375.0 |

The reduction in tunnel length results in an 8% increase in the revolution frequency, shifting the first unstable betatron line to higher frequencies. However, this has no significant impact on stability, as the associated rigid-mode instability is effectively suppressed by the transverse feedback damper [10]. Additionally, the reduced tunnel length decreases the resistive-wall impedance of the vacuum structure—particularly of the beam screens—by the same 8%. The reduction of the magnetic field from 16 T to 14 T also leads to a reduction of the resistivity of the copper beam screens, through magnetoresistance, of about 8% (assuming a residual resistivity ratio of 70 as in the LHC beam screens [512], and standard copper properties for magnetoresistance [513] and temperature dependency [514], as implemented in XWAKES [515]), which leads to a reduction of their resistive-wall impedance by 4%. [3] By contrast, the lower flat-top energy (42.3 TeV instead of 50 TeV) has a more pronounced effect, as both tune shifts and growth rates are inversely proportional to $\gamma$, leading to an 18% increase.

Since at top energy, the resistive-wall impedance of the beam screens accounts for only 10% of the total impedance [10], the stability at the flat top is degraded with respect to the CDR by approximately 20%, primarily due to the lower beam energy. At injection energy, where the beam screen contributes 50% of the total impedance, stability improves by 4% if the beam screen temperature remains unchanged. Even in a worst-case scenario, the increase in instability growth rates and tune shifts remains manageable at both injection and flat top [10]. If the injection energy is lowered to 1.3 TeV, the multi-bunch instabil-ity growth times decrease by a factor of almost three to 20–25 turns [10]. Also, here, a safety margin still exists since even faster multi-bunch instabilities, e.g. with rise times about 10 turns, could be damped by an LHC-type feedback system along with Landau octupoles and non-zero chromaticity [516].

[3] If the beam screens were operated at 70 K instead of 50 K, their resistive-wall impedance would increase by 21% with respect to the CDR configuration (50 K, 16 T). At injection energy (corresponding to a dipole field of 1.09 T, almost unchanged with respect to the CDR), the corresponding increase due to temperature would be of 48%, as the temperature effect is more pronounced at low values of the magnetic field.

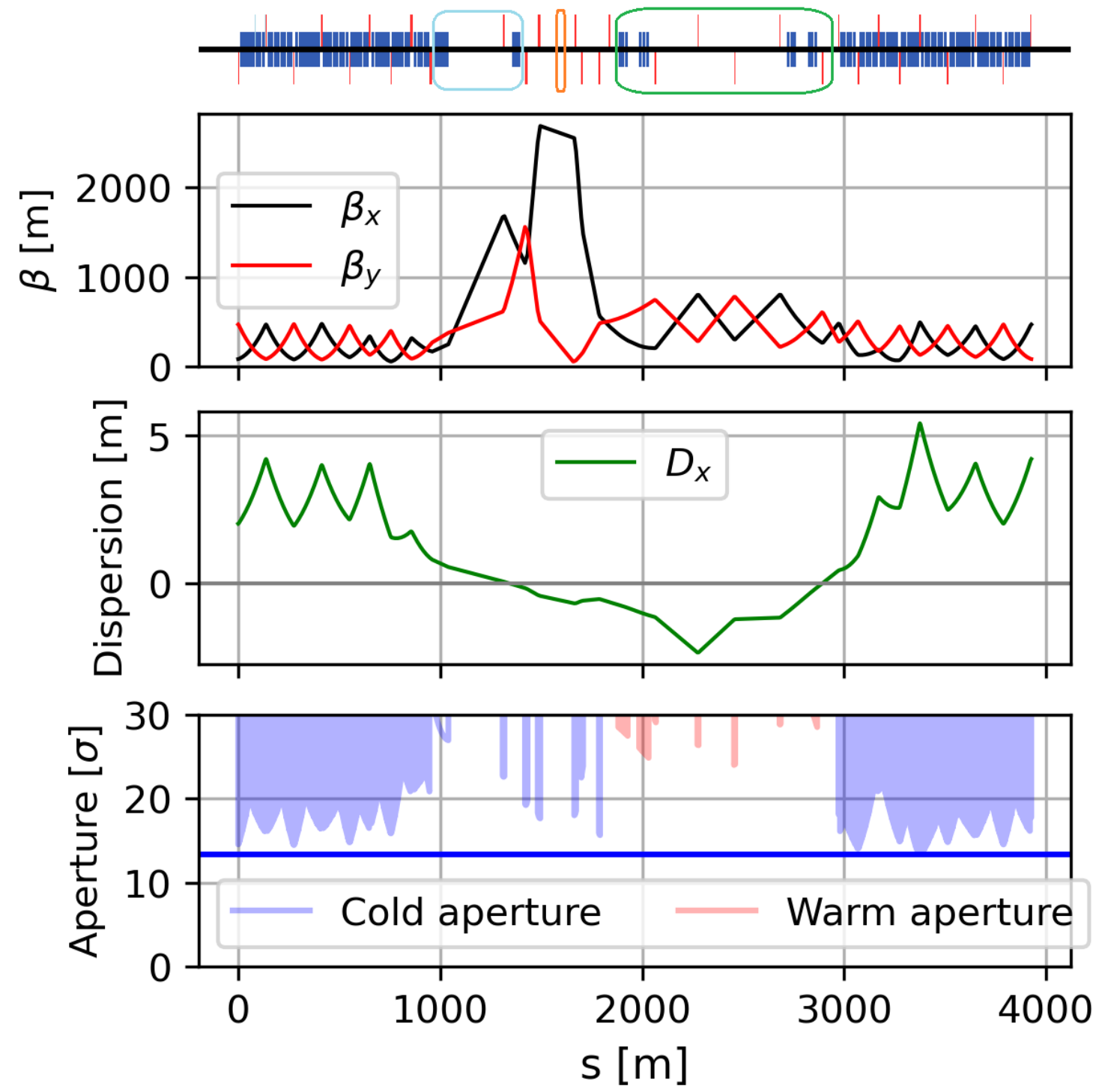

Fig. 10.14: Optical parameters, dispersion and beam aperture at injection for the momentum collimation and injection insertion PB for the clockwise beam. Sections dedicated to a specific function are highlighted: injection elements (light blue), injection protection devices (orange) and momentum collimation elements (green). The horizontal line in the aperture plot represents the minimum value acceptable.

### 10.2.9 Ongoing studies

The cell magnet layout and the number of dipoles per cell are already optimised for maximum beam energy. An even higher dipole filling factor might still be achieved by optimising the length of the main dipoles, which is subject to various constraints, such as the size of the shaft used to lower the dipoles.

A non-standard option for the FCC-hh ring configuration utilises combined-function magnets [517]. Further investigations of this approach (see, e.g., Ref. [518]), including corrector systems and refined length of the combined-function magnets, is essential for determining whether if this setup could serve as a viable alternative to the present baseline lattice.

## 10.3 FCC-hh injection

### 10.3.1 Hadron injectors in the SPS or LHC tunnel

Machines located in three different tunnels were initially considered as potential hadron injectors: a booster within the collider tunnel, a modified LHC or a new superconducting machine in the LHC tunnel, and a superconducting SPS (scSPS) in the SPS tunnel. The option of placing a booster in the collider tunnel, using superferric magnets at 1.1 T with a 70% filling factor, was discarded at an early stage due to the need for long bypass tunnels —approximately 15 km in length—around the experiment caverns.

The remaining options, involving a machine in either the LHC or SPS tunnel, differ in their achiev-able injection energy for the collider. An scSPS limits the injection energy to 1.3 TeV, whereas a machine in the LHC tunnel could exceed the collider's baseline injection energy of 3.3 TeV.

Table 10.8: Strength of the insertion quadrupoles for the momentum collimation insertion at PB and its dispersion suppressors. All the gradients correspond to 45 TeV.

| NAME | Gradient [T/m] | Maximum gradient [T/m] |
|---|---|---|
| MQ.14LB.B1 | 123.6 | 375 |
| MQ.13LB.B1 | −124.0 | 375 |
| MQ.12LB.B1 | 118.1 | 375 |
| MQ.11LB.B1 | −125.3 | 375 |
| MQ.10LB.B1 | 150.5 | 375 |
| MQ.9LB.B1 | −129.6 | 375 |
| MQ.8LB.B1 | 57.9 | 375 |
| MQYL.7LB.B1 | −20.1 | 200 |
| MQY.6LB.B1 | 67.8 | 200 |
| MQYL.5LB.B1 | −122.6 | 200 |
| MQYL.4LB.B1 | 68.8 | 200 |
| MQY.3LB.B1 | 103.2 | 200 |
| MQY.2LB.B1 | −56.8 | 200 |
| MQM.1LB.B1 | −131.8 | 375 |
| MQY.1RB.B1 | 1.6 | 200 |
| MQW.2RB.B1 | −20.2 | 44 |
| MQW.3RB.B1 | 32.2 | 44 |
| MQW.4RB.B1 | −31.3 | 44 |
| MQW.5RB.B1 | 27.8 | 44 |
| MQY.6RB.B1 | −169.8 | 200 |
| MQM.7RB.B1 | 231.0 | 375 |
| MQ.8RB.B1 | −40.5 | 375 |
| MQ.9RB.B1 | 129.9 | 375 |
| MQ.10RB.B1 | −111.3 | 375 |
| MQ.11RB.B1 | 131.8 | 375 |
| MQ.12RB.B1 | −103.3 | 375 |
| MQ.13RB.B1 | 110.2 | 375 |
| MQ.14RB.B1 | −115.6 | 375 |

In addition to the injection energy of 3.3 TeV the main requirements of the hadron injector are the delivery of the beam parameters in intensity, emittance and bunch spacing, as well as a fill duration of around 30 min.

A detailed study was carried out to evaluate the feasibility of modifying the existing LHC into a fast-ramping 3.3 TeV injector. This analysis included assessing layout modifications in the straight sections, optimising the powering segmentation to enable faster ramping, and developing a simplified optics design [519]. The FCC-hh Conceptual Design Report [10] concluded that, at 3.3 TeV, the threshold for FCC-hh injection absorbers to survive beam impact requires a burst-mode transfer of 130 batches from the modified LHC to the FCC-hh, each batch consisting of 80 bunches. This configuration would allow the full LHC beam to be transferred within a few seconds.

Alternatively, the LHC could be replaced by a 4 T superconducting machine based on magnet designs from RHIC, Tevatron, or FAIR (SIS200/300). To avoid excessively long transfer lines, this machine would require either polarity reversal or a twin-aperture magnet design. The RF system is assumed to limit the ramp-up time to no more than 50 s, resulting in a total collider filling time of 39 minutes. Additionally, this option would necessitate a burst-mode injection transfer lasting several seconds.

Another approach to replacing the current LHC within its tunnel involves a superferric, iron-dominated machine operating at 2T, capable of achieving a transfer energy of approximately 1.75 TeV. However, neither of these options has been studied in further detail at this stage.

If the collider can accommodate an injection energy of 1.3 TeV, a superconducting machine in the SPS presents an intriguing option. This possibility has been explored at the conceptual level and is documented in Ref. [520].

This study considered following the present SPS lattice, which naturally is well-matched to the existing tunnel geometry, using missing bend dispersion suppressors. A collimation system needs to be integrated into the straight-section layout. The beam transfer from the scSPS to the collider is limited to 640 bunches per transfer due to absorber limits. In this case, no burst-mode transfer is required. The filling time considered in Ref. [520] needs to be revised in view of the reduced number of collider bunches and a possibly higher ramp rate.

Overall, the feasibility of an scSPS is mainly determined by the design of large aperture, fast ramping dipoles, and quadrupoles. The design study compares 2D designs of main dipoles and quadrupoles for both single- and double-layer coil configurations. Among these, the 4.2 K-compatible double-layer coil design emerges as the most promising. The study also evaluated a challenging 80 mm aperture, primarily motivated by also allowing a slow-extraction based fixed-target programme reaching up to 1.3 TeV. Recent advancements in crystal-assisted slow extraction provide confidence that the current spiral step of 20 mm could be reduced to approximately 5 mm, directly impacting the required magnet aperture.

The initial magnet study also highlights the significant benefits of doubling the injection energy in the SPS to 50GeV, as this would reduce the critical energy/field swing and AC losses. In addition, this approach could eliminate the current need for transition crossing.Given that the PS will be more than 100 years old at the time of FCC-hh, one might consider replacing the PS with a 50 GeV machine - a superconducting machine in the PS tunnel or a normal conducting PS2 (studied as an option to renovate the LHC injector chain).

#### Injector option summary and outlook

The three options for an FCC hadron injector are summarised in Table 10.9. All options can deliver the main beam parameters required, such as intensity, emittance, and bunch spacing. The main difference is the implied FCC-hh injection energy. The modified LHC option stands out for the filling time, in particular, since the filling time of the scSPS was calculated with a rather conservative ramp rate. The LHC would also need a large effort in continuous consolidation over decades before being repurposed as an FCC hadron injector.

If the collider injection energy stays above the reach of an scSPS, an attractive option seems to be a new 4 T machine in the LHC tunnel. Another possibility, with an intermediate beam energy, is a 2-T superferric machine. If 1.3 TeV is acceptable for the collider, an scSPS becomes appealing, also in view of a possible synergy of beam transfer tunnels.

Given the timescale for all options, it seems valid to consider renovating the entire injector chain, allowing the removal of certain bottlenecks in beam production.

For the next study steps, it is planned to establish a parametric model for the main cost drivers of a 3.3 TeV superconducting machine and a 1.75 TeV superferric machine in the LHC tunnel, and a 1.3 TeV superconducting machine in the SPS tunnel.

Table 10.9: Comparison of high-energy booster options, a superconducting machine in the SPS tunnel (scSPS), a modified faster-ramping LHC and a new 4 T machine in the LHC tunnel. A further option (not shown) with an intermediate energy of 1.75 TeV would be a superferric machine in the LHC tunnel.

| | Unit | scSPS | mod. LHC | 4 T LHC |
|---|---|---|---|---|
| Circumference | km | 6.9 | 26.7 | 26.7 |
| Apertures | | 1 | 2 | 1 |
| Injection energy | GeV | 26 | 450 | 450 |
| Extraction energy | TeV | 1.3 | 3.3 | 3.3 |
| Injection field | T | 0.12 | 0.6 | 0.6 |
| Maximum field | T | 6 | 4 | 4 |
| Energy/field swing | | 50 | 7 | 7 |
| Individual dipole length | m | 12 | 14.3 | 14.3 |
| Dipole filling factor | | 0.65 | 0.66 | 0.66 |
| Number of dipoles | | 372 | 1232 | 1232 |
| Number of quadrupoles | | 216 | 480 | 480 |
| Number of bunches | | 640 | 2600 | 2600 |
| Stored energy | MJ | 15 | 167 | 167 |
| Booster filling time | min | 0.5 | 7.5 | 3.8 |
| Ramp rate | T/s | 0.4 | 0.026 | 0.08 |
| Cycle length | min | 1.1 | 12 | 4.9 |
| Booster cycles per FCC fill | | 34 | 4 | 8 |
| Collider filling time | min | 37 | 46 | 39 |

### 10.3.2 Layout of the FCC-hh injection

In the case of using the present LHC or a new 4 T machine in the LHC tunnel as the FCC hadron injector, the beam would be transferred at 3.3 TeV, which, in Ref. [10], was considered the baseline injection energy.

With the new geometry of the lepton transfer lines from the surface to the collider tunnel (see Section 7.7), there is no longer a synergy between the lepton and hadron transfer tunnels to simplify the civil engineering work for the lepton tunnels.

For hadron beams from the LHC tunnel, it is proposed that both beams be extracted from P8, as illustrated in Fig. 10.15. Compared to the current LHC dump system, designed for 7 TeV and operated at 6.8 TeV, an extraction system for FCC beams would require only half the deflection while benefiting from at least 50 years of technological advancements. Additionally, such an extraction system could be designed with a simpler triggering logic compared to a fail-safe dump system.

Transfer lines from LHC P8 to the collider tunnel require superconducting magnets well within the reach of current Nb-Ti technology (see Table 10.10). As soon as the transfer lines join the collider tunnel, the magnetic field required drops to the injection field of the collider magnets, as described in detail in Section 10.3.1.

Table 10.10: Lengths of transfer lines and tunnels from the LHC with the required magnet fields in the connection tunnel and inside the collider tunnel, respectively.

| | TL length [km] | Tunnel length [km] | Dipole fields [T] |
|---|---|---|---|
| LHC P8 to PB | 9.5 | 1.9 | 6.6/1 |
| LHC P8 to PL | 7.5 | 3.8 | 4.7/1 |

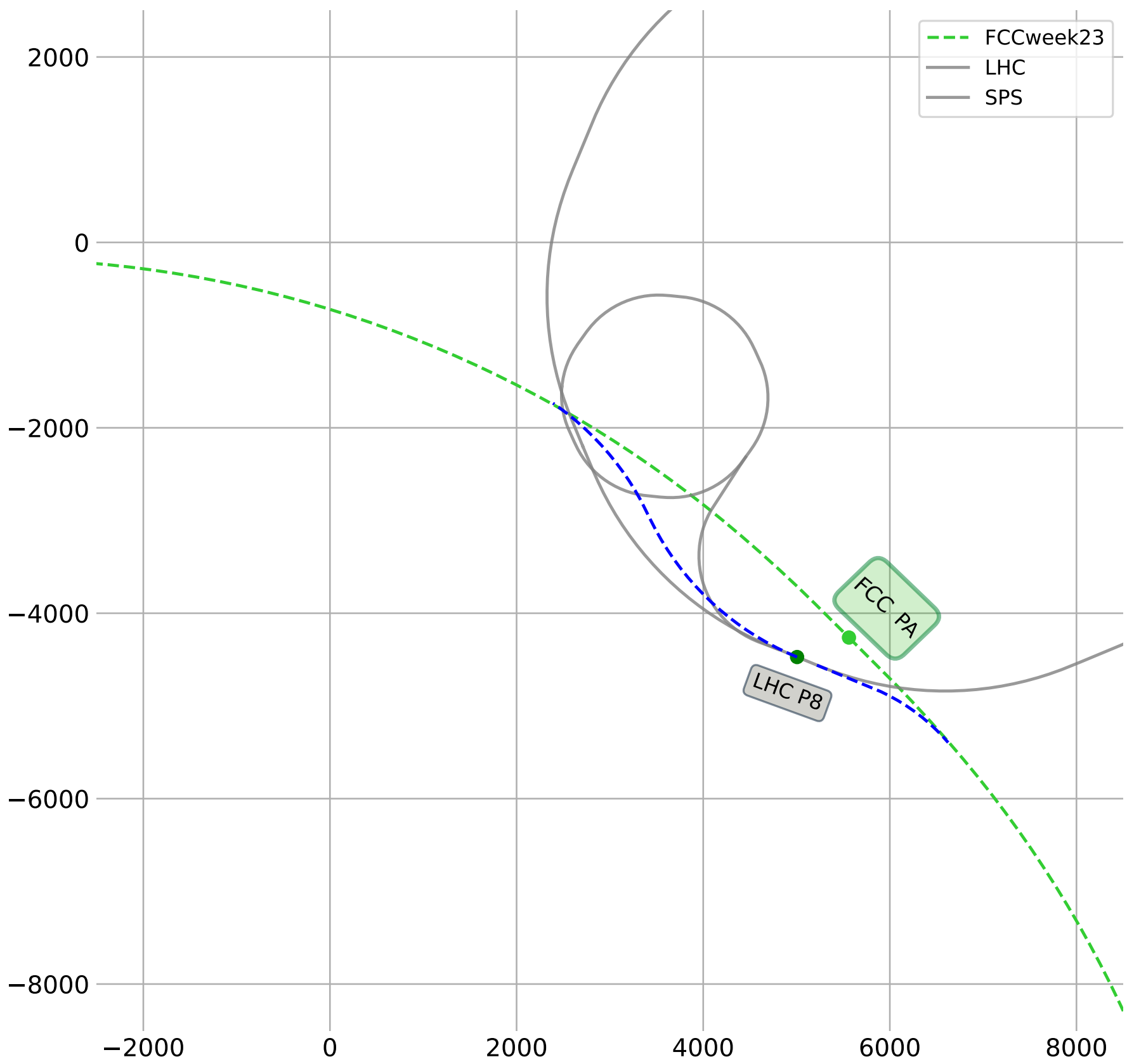

Fig. 10.15: Transfer tunnels (blue dashed) for 3.3 TeV hadron beams from LHC P8 to the collider. The transfer lines continue inside the collider tunnel for several km to reach the injection straights PB and PL.

In case a lower injection energy into the collider is attainable, the option of a super-ferric iron-dominated machine in the LHC tunnel allows a beam transfer at around 1.75 TeV. This option could use the same extraction system layout as described above. The reduced transfer energy would be beneficial for the system design. Also, the transfer line geometry could be the same as described above, with lower fields in the transfer dipoles.

### 10.3.3 Layout of the FCC-hh injection lines from a superconducting SPS

For a superconducting SPS, based on $\sim$ 6 T magnets, the transfer energy is limited to 1.3 TeV. At this energy, the use of the existing TI8 tunnel for hadron transfer is within reach of the current technology, which reduces the new tunnelling for the clockwise beam injected in PB by 3 km (see Fig. 10.16 and Table 10.11). The dipole fields required amount to 5.9, 2.3 and 7.4 T for the different arcs, respectively. As soon as the transfer line is in the collider tunnel, a field of 0.44 T is required.

For the anti-clockwise beam injected in PL, the straight section 6 (LSS6) of the SPS is favoured to have a short connection of less than 1 km to the collider tunnel with dipole fields of 4.2 T.

### 10.3.4 Transfer lines in the ring tunnel

Two sections of the FCC-hh transfer lines are inside the collider tunnel, in the arcs that join PA and PB (for the clockwise beam) and PA and PL (for the counter-clockwise beam). The design of the transfer line will inherit the features of the regular arcs of the ring over this length. It is assumed that periodic cells of the transfer line will be copies of those in the hadron collider, with each containing 16 dipole magnets. Furthermore, the focusing structure will also be applied to the part of the transfer lines outside the ring tunnel, whereas the geometry will impose a different choice of dipole magnets.

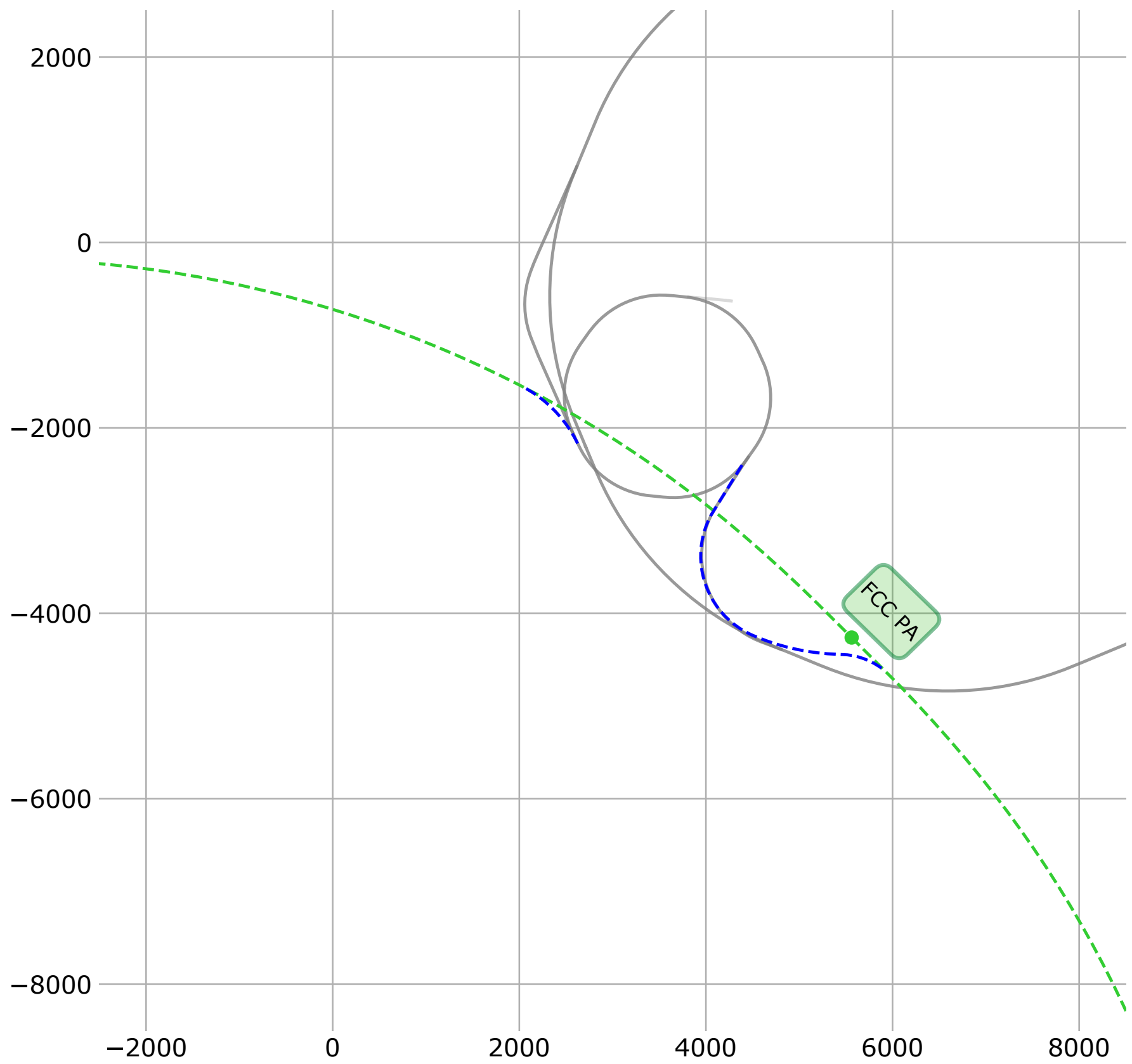

Fig. 10.16: Transfer tunnels (blue dashed) for 1.3 TeV hadrons from SPS LSS4 to the collider in clockwise direction, and from SPS LSS6 to the collider in anticlockwise direction. The transfer line continues inside the collider tunnel for several km to reach the injection straights PB and PL.

Table 10.11: Lengths of transfer lines and tunnels from the scSPS with the required magnet fields in the connection tunnel and inside the collider tunnel, respectively.

| | TL length [km] | Tunnel length [km] | Dipole fields [T] |
|---|---|---|---|
| SPS-LSS4 to PB | 12.9 | 1.6 | 7.4/0.44 |
| SPS-LSS6 to PL | 6.9 | 0.9 | 4.2/0.44 |

Two injection energies have been considered, 3.3 TeV, which is the baseline assuming the LHC as an injector, and an alternative of 1.3 TeV, assuming a superconducting SPS. The higher injection energy is considered for the magnet design parameters in Table 10.12).

In view of the large scale of production for the transfer line magnets, assembly and manufacturing will be a key focus of future development efforts.

The aim of the following preliminary assessment was to consider existing magnets at CERN and elsewhere, firstly to generate multiple design concepts and secondly to make an initial assessment of whether these could satisfy the transfer line requirements. Ultimately, three concepts, highlighted in Fig. 10.17, were selected for further development. These three concepts will be developed until the most suitable design can be identified, also taking into account resource effectiveness. Typically, this

Table 10.12: FCC-hh transfer line magnets parameters for injection energy of 3.3 TeV and for the section located inside the ring tunnel. The total length of the tunnel section of the two transfer lines is 19.6 km. Such an estimate is an upper bound based on the FCC-hh arc length. The FCC-hh ring regular cell is 276 m long.

<table>
<tr><th></th><th></th><th>Electromagnet</th><th>Permanent magnet</th></tr>
<tr><td>Interconnections</td><td>Longitudinal gap (mm)</td><td>550</td><td>50</td></tr>
<tr><td>Aperture and field quality</td><td>Beam pipe inner diameter (mm)</td><td colspan="2">30</td></tr>
<tr><td></td><td>Good field region (GFR) diameter (mm)</td><td colspan="2">20</td></tr>
<tr><td></td><td>Field linearity in GFR (units)</td><td colspan="2">2</td></tr>
<tr><td>Dipoles</td><td>Integrated dipole field per cell (T m)</td><td>230</td><td>230.4</td></tr>
<tr><td></td><td>Dipole magnet field strength (T)</td><td colspan="2">1</td></tr>
<tr><td></td><td>Magnet magnetic length (m)</td><td>5</td><td>1.2</td></tr>
<tr><td></td><td>Number of magnets per cell</td><td>46</td><td>192</td></tr>
<tr><td></td><td>Total dipole length (km)</td><td>8.51</td><td>8.52</td></tr>
<tr><td>Quadrupoles</td><td>Integrated quadrupole gradient per cell (T)</td><td colspan="2">118</td></tr>
<tr><td></td><td>Quadrupole magnet gradient (T/m)</td><td>20</td><td>25</td></tr>
<tr><td></td><td>Magnet magnetic length (m)</td><td>3</td><td>1.2</td></tr>
<tr><td></td><td>Number of magnets per cell</td><td>4</td><td>8</td></tr>
<tr><td></td><td>Total quadrupole length (m)</td><td>444</td><td>355.2</td></tr>
</table>

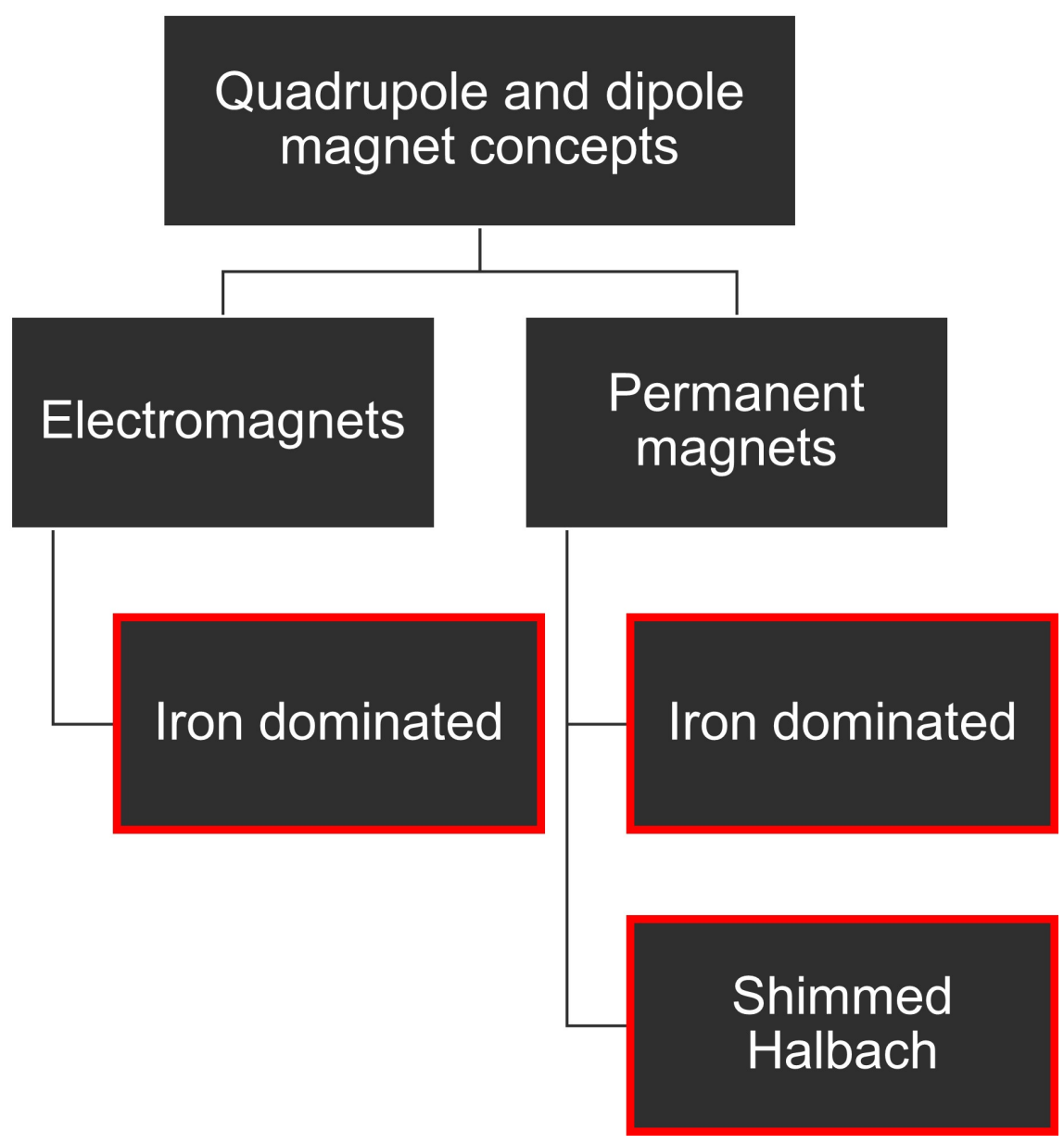

Fig. 10.17: Sketch of the possible conceptual designs for the transfer line magnets. The three magnet design concepts highlighted in red were selected.

assessment will be based on lifetime cost, but it will also consider other factors, such as climate impact and ethical issues regarding where raw materials are sourced.

***Electromagnet design concepts***

Iron-dominated normal-conducting accelerator magnets are a mature and well-understood technology. Furthermore, the field strength required is compatible with the normal-conducting magnet capabilities, making them suitable for this application. The volume of the yoke will be optimised to reduce weight and material cost while retaining sufficient torsional rigidity and magnetic performance. The most critical optimisation involves determining the optimal size of the coil cross-section to minimise the total lifetime cost. This requires balancing capital investment costs, primarily driven by the conductor, with overhead expenses, mainly related to electrical power consumption.

One dipole magnet design provides a $1.2\,\mathrm{T}$ field in the $30\,\mathrm{mm}$ aperture and is largely inspired by the TI2/TI8 transfer line magnets that connect the SPS ring to the LHC (see Fig. 10.18). An H-shaped yoke geometry was selected because it has better torsional rigidity and saturation performance compared to other options. This minimises the dipole cross-section dimensions, particularly the height. The exact length of each magnet unit has not yet been determined, but a $5\,\mathrm{m}$ dipole would weigh $4600\,\mathrm{kg}$.

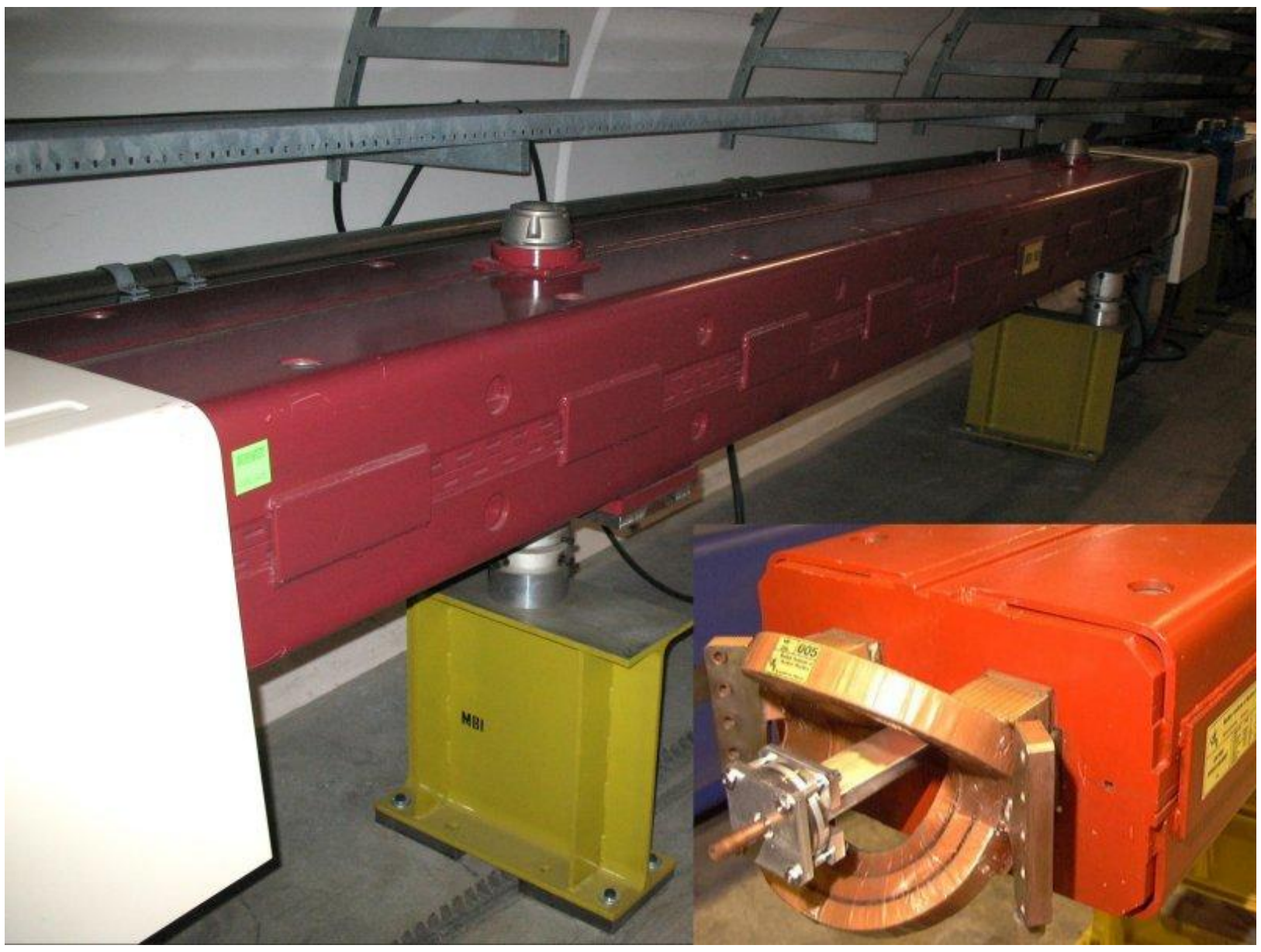

Fig. 10.18: Photograph of a $6.3\,\mathrm{m}$ long dipole magnet (HCMBI_001) used in the LHC transfer line.

The yoke is made of laminated electrical steel. The laminations are punched and then stacked together, which is the most cost-effective assembly method for producing a large series of long magnets. This configuration also minimises eddy current losses when the magnets are ramped, reducing running costs.

The $58\,\mathrm{kAturns}$ required are provided by two bedstead coils. This coil shape reduces the interconnection length between magnets, thus maximising the longitudinal filling factor. The coils are wound with a hollow copper conductor and cooled with demineralised water. Figure 10.19 (left) shows the 2D field distribution in the yoke cross-section, while the 3D magnet geometry is visible in the right plot of the same figure. Note the length of the magnet relative to its cross-section.

The quadrupole concept is also inspired by the TI2/TI8 transfer line magnets. It would be an assembly of laminated steel quadrants with water-cooled coils. Each of the four coils provides $7.4\,\mathrm{kAturns}$, which produces a $25\,\mathrm{T/m}$ gradient in the $30\,\mathrm{mm}$ diameter aperture, with a $1.5\,\mathrm{m}$ long quadrupole weighing $1000\,\mathrm{kg}$. The 2D field distribution in the yoke cross-section is shown in Fig. 10.20.

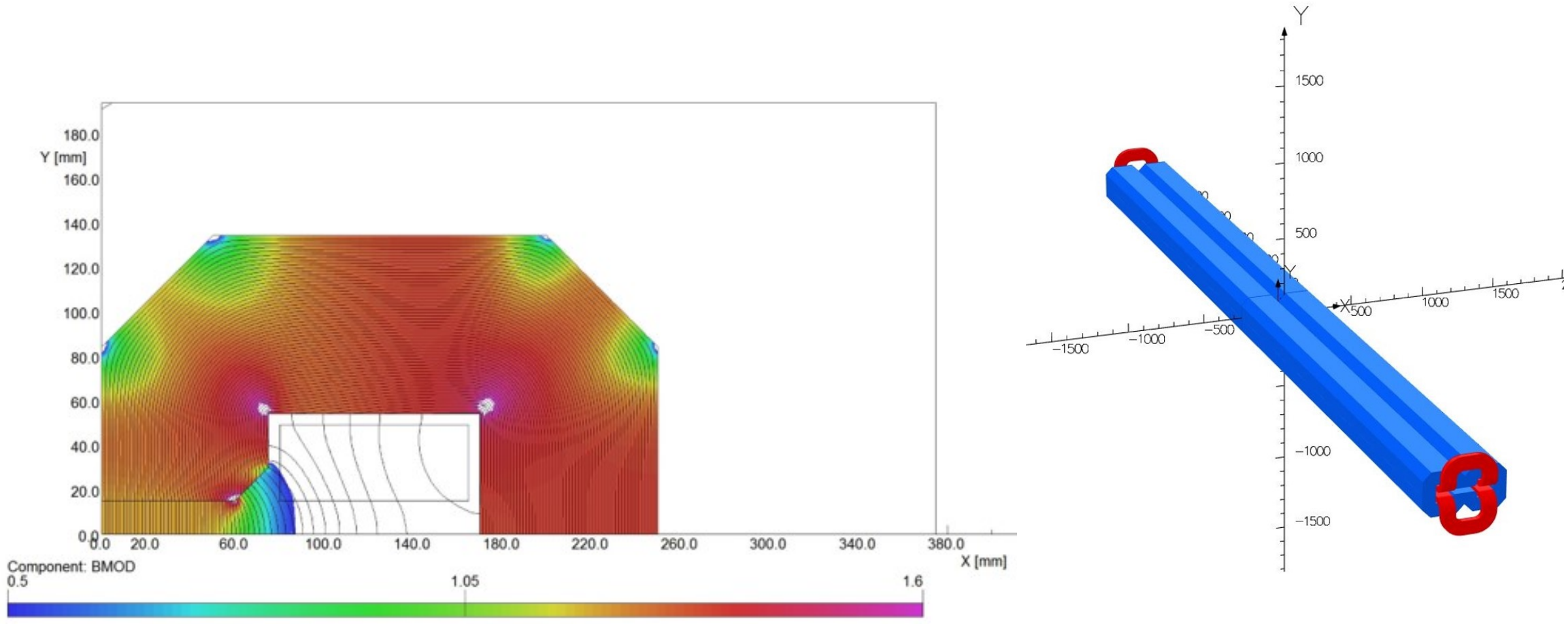

Fig. 10.19: Left: Normal conducting dipole magnet 2D field distribution (four-fold symmetry). Right: Normal conducting dipole magnet 3D model.

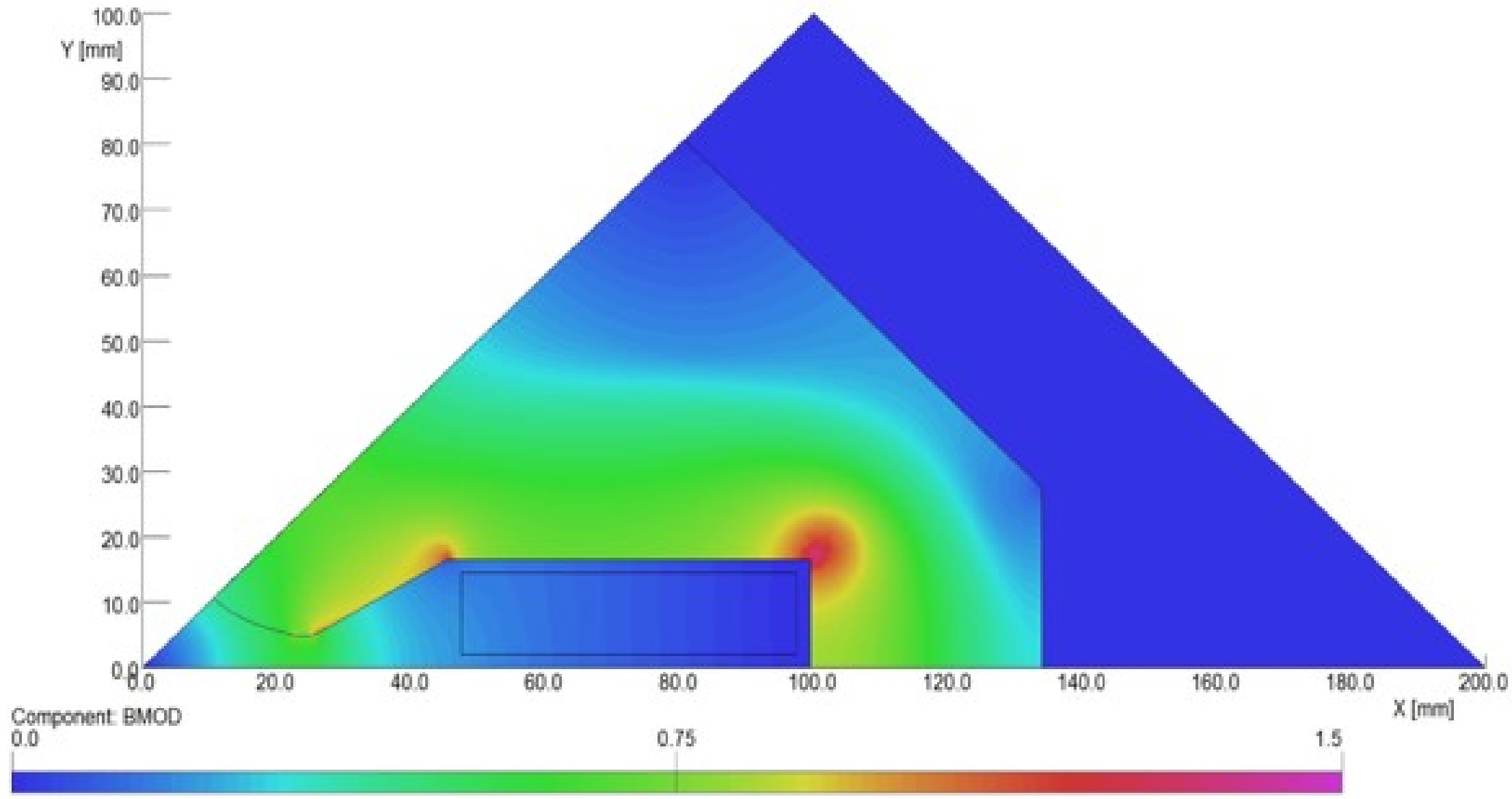

Fig. 10.20: Normal conducting quadrupole magnet 2D field distribution (eight-fold symmetry).

***Permanent Magnet Solutions***

Permanent magnets (PM) for accelerators are especially well suited to the transfer line specifications. As the FCC-hh beam will always be injected at the same energy, the magnet strength does not need to be varied. Additionally, the field quality requirements are less stringent than those in the collider ring because the beam only makes a single pass. It is worth mentioning that PMs are already used in accelerators, albeit on a much smaller scale. For instance, at CERN, there are permanent accelerator magnets in Linac4 [521] and the beamline of the FASER [522] experiment. At Fermilab, the 3.4 km recycler ring is entirely based on PMs [523, 524]. Furthermore, PMs can also be found in insertion device magnets for electron synchrotrons around the world (see Ref. [525] for a review).

Note that PM blocks/wedges are the magnetised elements that are assembled to create an accelerator magnet (see Fig. 10.21). During their manufacture, they should gain homogeneous magnetisation (magnitude and direction) oriented relative to an external mechanical datum surface. However, man-ufacturing tolerances introduce magnetisation errors that must be monitored through quality assurance testing and mitigated by the accelerator magnet designs.

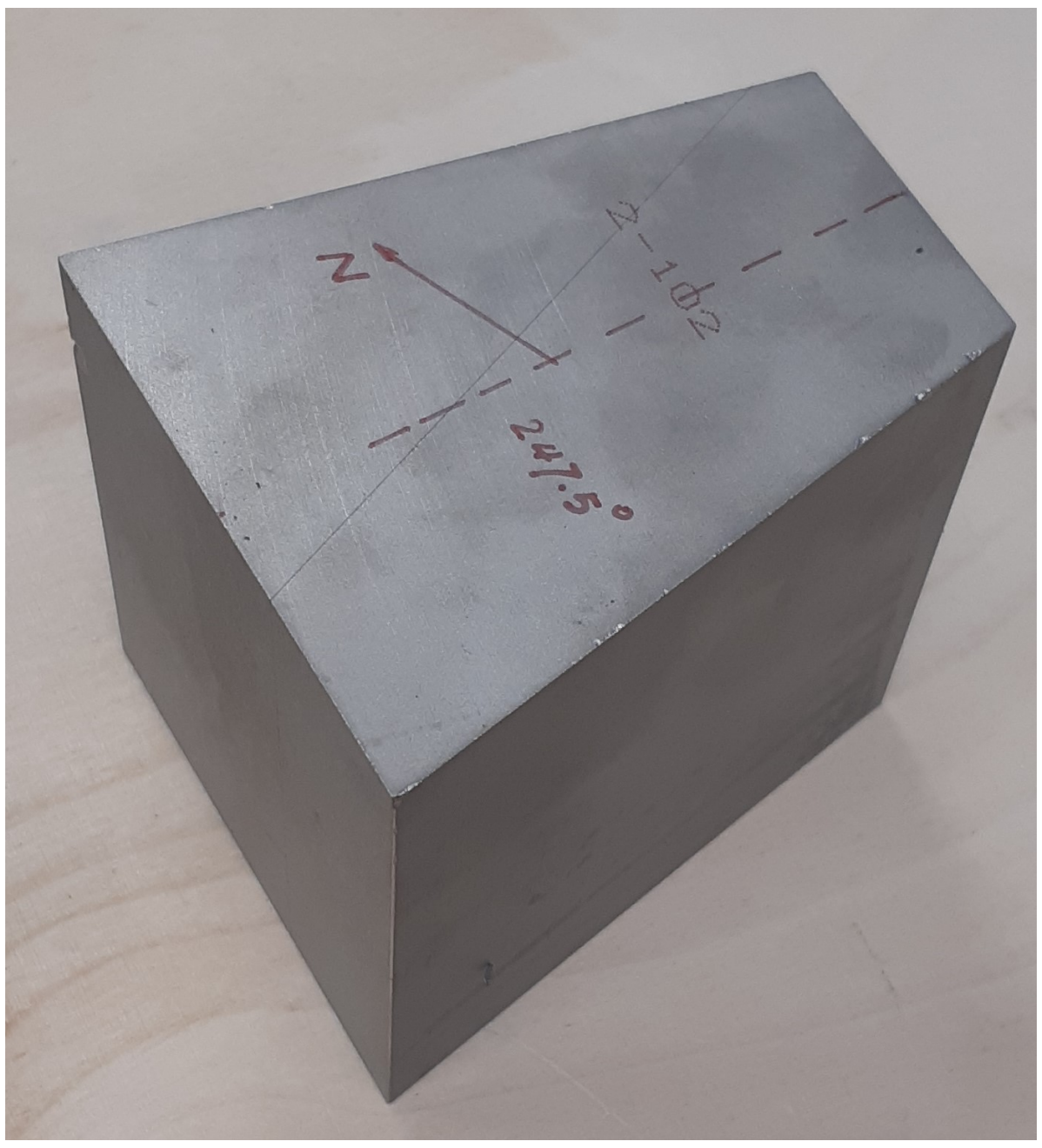

Fig. 10.21: Samarium Cobalt permanent magnet wedge used in the FASER Halbach magnet at CERN.

Magnetic forces between PM elements make it challenging to assemble them in a precise and safe way. However, the design concepts presented here are not novel, and multiple assembly techniques are already available. More research is needed to determine whether these approaches are suitable for the industrial scale required by the FCC-hh or whether new techniques will need to be developed.

The design of the transfer line will also include active correctors, i.e., normal-conducting magnets that are meant to compensate for, e.g., trajectory errors in the transfer lines due to the extraction conditions in the FCC-hh injector or magnetic-field errors from the PMs of the transfer line.

#### ***Risks Associated with Permanent Magnets***

PMs have two key weaknesses: the magnetic field they produce is susceptible to radiation dose and variations in temperature, both of which will affect the transfer lines. Therefore, it is important to examine these vulnerabilities and make an initial assessment of whether realistic operating conditions could prevent the transfer line magnets from functioning adequately.

The first vulnerability is the temperature dependence of PMs, which could cause a drift in magnetic field strength as the ambient temperature varies. Annual temperature data from the LHC tunnel [526], which is assumed to behave similarly to the FCC-hh ring tunnel, and PM material data were used to calculate the maximum possible integrated dipole error per cell (see Table 10.13). The magnitude is small enough to be managed using active corrector magnets.

The second vulnerability is PM demagnetisation due to irradiation. Demagnetisation mechanisms are complex, but damage is primarily a function of dose and energy [527], with different levels of PM having different susceptibilities. For this initial assessment, it is assumed that the PM elements of the transfer line receive a uniform dose indirectly (i.e., from the collider beam). Moreover, the direct dose from the transfer line beam is neglected, and it is assumed that there is no realistic failure scenario where the beam could be lost and dumped into the transfer line. This assumption will be reviewed when the details of the machine protection configuration are discussed. Another aspect that will be reviewed in later studies is the possible impact of localised losses occurring in the collider ring, such as the so-called unidentified falling object (UFO) events that have been observed and studied in detail in the LHC (see, e.g., Ref. [528] and references therein for a review of this topic). Losses from these events are typically on the order of $1 \times 10^8$ protons. While they could potentially trigger a quench in a superconducting magnet, they may also impact the PMs. This possibility will be investigated further in future studies.

Table 10.13: Permanent magnet temperature dependence parameters with annual LHC tunnel temperature variation.

| | Unit | SmCo | NdFeB |
|---|---|---|---|
| Temperature dependence | $\%/°C$ | $0.035$ | $-0.120$ |
| Temperature variation (annual) | $°C$ | $\pm 0.87$ | |
| Integrated dipole field per cell | $\mathrm{Tm}$ | $235$ | |
| Correction field magnitude per cell | $\mathrm{Tm}$ | $0.716$ | $2.435$ |

There are significant uncertainties in calculating an estimate for the accumulated lifetime dose. Three key variables are: residual gas density in the collider vacuum tube; longitudinal position (dose rate is higher between collider magnets); and number of years spent in commissioning vs. in operation (dose rate is lower in the latter phase). Using the FCC-hh particle transport simulation studies [529] and taking conservative assumptions, a lifetime dose of $1.7\,\mathrm{kGy}$ was calculated (see Table 10.14). For SmCo or NdFeB magnets, this dose should not alter the magnetic field on the axis by more than 1 unit (factor of $10^{-4}$) so the transfer line requirements would still be satisfied. In conclusion, whilst radiation damage can be considered a low risk, this assertion will need to be rigorously tested before committing to a PM transfer line design. In addition, a comprehensive literature search will be conducted to understand the resistance of PMs to radiation. Potentially, the FCC will need to conduct its own PM irradiation test campaign.

Table 10.14: Radiation dose rate parameters for the assessment of the FCC-hh transfer line magnet.

| | Unit | Value |
|---|---|---|
| Commissioning dose rate | Gy/year | 200 |
| Commissioning duration | year | 5 |
| Operation dose rate | Gy/year | 20 |
| Operations duration | year | 35 |
| Total lifetime | year | 40 |
| Lifetime dose | kGy | 1.7 |

An additional risk of using a PM design is insufficient supply from manufacturers. For that reason, although they are more sensitive to temperature and less resistant to radiation [530], NdFeB PMs were selected because they are produced at a lower cost and on a larger scale than SmCo PMs [531]. Table 10.15 shows that the forecast global supply of NdFeB should be sufficient to cover the required volume for both PMs concepts. Note that the iron-dominated permanent magnet design has not yet been optimised, so this value represents an upper limit.

#### ***Iron-Dominated permanent magnet design concepts***

Iron-dominated permanent accelerator magnets generate a magnetic field using PM elements and function similarly to conventional iron-dominated electromagnets. The iron yoke serves as a low-reluctance path, enhancing magnetic flux density. The pole helps shape the magnetic field distribution within the aperture and can be shimmed to improve field quality. This is particularly beneficial in PM accelera-tor designs, where magnetisation and assembly errors pose significant challenges. The iron also helps homogenise the magnetic field, mitigating these errors.

Table 10.15: Permanent magnet material need and forecast supply.

| | | Unit | |
|---|---|---|---|
| NdFeB global supply | 2019 | k-tonnes/year | 130 |
| | Forecast for 2030 | k-tonnes/year | 200 |
| Required volume of NdFeB magnet blocks/wedges for concepts | Iron-dominated concept designs | tonnes | 950 |
| | Shimmed Halbach concept designs | tonnes | 625 |

The dipole and quadrupole design concepts both feature parallelepiped-shaped PM blocks. These are easier to produce and have better magnetisation direction tolerances compared to other geometries. First, the low-carbon steel yokes and non-magnetic components are assembled, and then the PM blocks are added one by one into the assembly. The blocks are inserted into cavities where some low-carbon steel material is present on either side. This reduces magnetic forces, making the assembly process simpler and faster.

The dipole design concept has an iron length of $1.2\,\mathrm{m}$ and an overall cross-section dimension of $0.2{\times}0.2\,\mathrm{m}$, it provides a field of $1\,\mathrm{T}$ in the aperture of $30\,\mathrm{mm}$. The field integral can be adjusted by a few percent by inserting magnetic shunts at the magnetic measurement stage, mainly to compensate for irregularities of PM blocks. The dipole magnet cross section is presented in Fig. 10.22 (left), and the 2D magnetic field distribution is shown in the right part of the same figure.

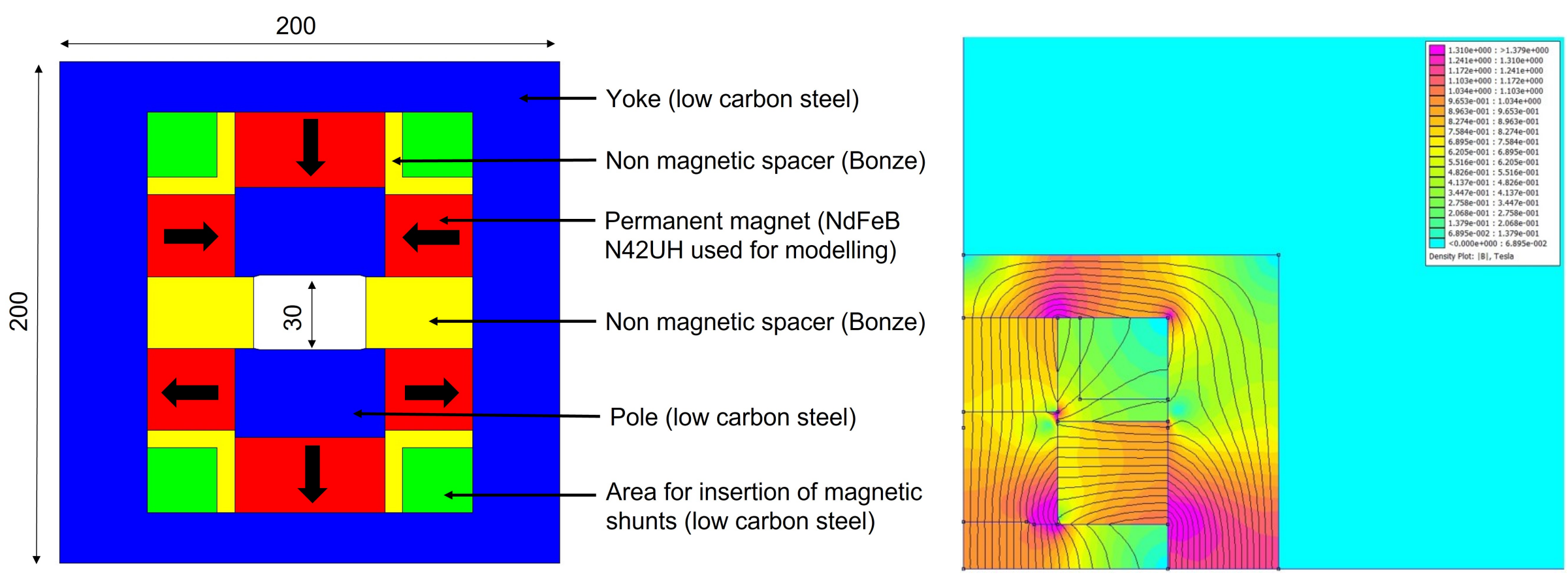

Fig. 10.22: Left: Iron-dominated permanent dipole magnet cross-section. Right: Iron-dominated permanent dipole magnet 2D field distribution (four-fold symmetry).

The quadrupole concept design has an iron length of $1.1\,\mathrm{m}$ and an overall cross-section dimension of $0.2{\times}0.2\,\mathrm{m}$. The gradient in the aperture diameter of $30\,\mathrm{mm}$ can be adjusted from $25-30\,\mathrm{T/m}$ using radially adjustable shims. This operation is performed in the magnetic measurement stage. The cross section is presented in Fig. 10.23 (left) and the 2D magnetic-field distribution is shown in the right part of the same figure.

The price of PM material will be a significant factor in the overall accelerator magnet cost. It is, therefore, crucial to minimise the PM volume in each design, first by shimming/narrowing the poles, and secondly by optimising the working point of the PM elements on their magnetic hysteresis curves.

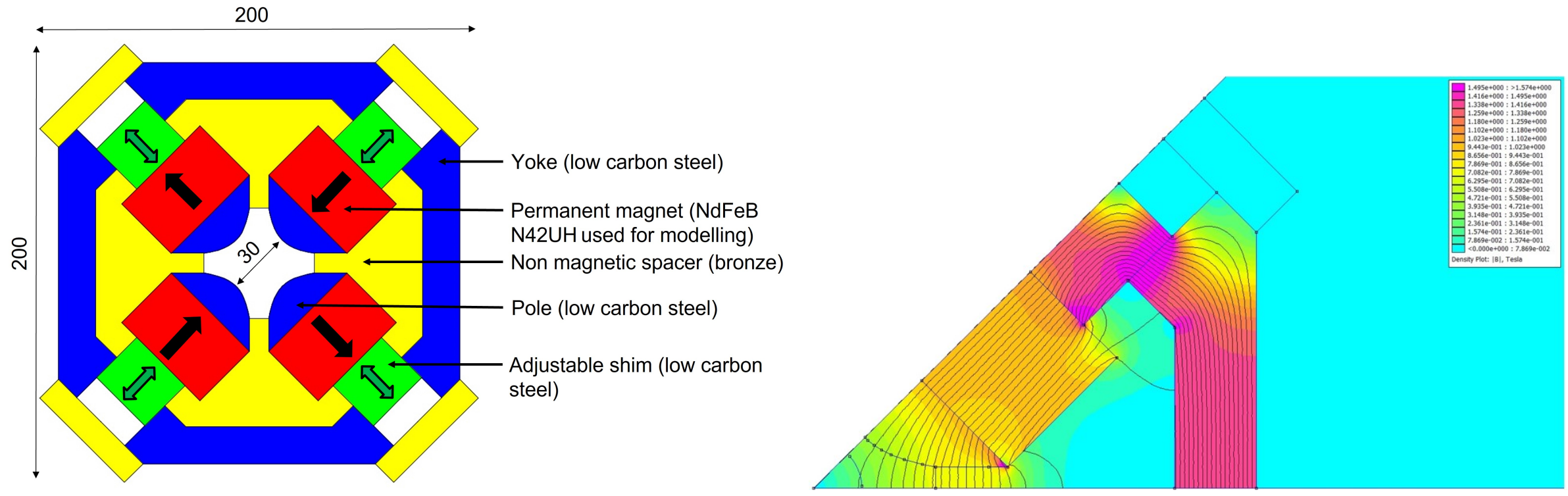

Fig. 10.23: Left: Iron-dominated permanent quadrupole magnet cross-section. Right: Iron-dominated permanent quadrupole magnet cross-section (four-fold symmetry).

***Shimmed Halbach magnet design concepts***

In Halbach arrays, PM elements are arranged to minimise the magnetic field on one side of the array whilst maximising it on the other. Indeed, cylindrical Halbach arrays are an exceedingly compact configuration of permanent multipole accelerator magnets, i.e., they require minimal PM element volume to produce a given field strength over a circular aperture [532]. However, the absence of iron to homogenise the magnetic field means that manufactured Halbach magnets are highly sensitive to PM magnetisation and assembly errors. Consequently, significant resources are typically invested in the manufacture and assembly of Halbach accelerator magnets to ensure satisfactory field qualities. This can include a range of measures: first, it requires extensive quality assurance, measuring the magnetic field of PM wedges (either individually or as a representative sample). Second, it may be necessary to match problematic PM wedges around the circumference of the cylinder to partially cancel field errors [532]. Finally, when assembling PM wedges, a mechanical structure may be needed to guide them in position; or mechanical shims might be used for small adjustments [533]. Both of these options reduce the PM packing factor, which affects magnetic performance. Even with these expensive mitigation measures, the field quality that can realistically be achieved is limited. Based on CERN experience building Halbach accelerator magnets, it is anticipated that a standard design would not satisfy the current FCC-hh transfer line requirements for field uniformity of about $\pm 2$ units.

Fortunately, cylindrical Halbach accelerator magnets can be shimmed to substantially decrease harmonic field errors. This technique was successfully developed and implemented in the CBETA 500 MeV electron energy recovery linac at Cornell University [534]. Initially, conventional Halbach magnets were manufactured with a slightly increased bore size to accommodate the shim assembly. Following that, standard rotating coil measurements are used to evaluate the magnetic field error harmonics (see Fig. 10.24 (left)).

An automated code was subsequently developed to optimise the distribution of magnetic shims around the bore circumference. These shims become magnetised, functioning as small dipole sources. Finally, the optimisation code determines the required length for each iron shimming rod before insertion into the magnetically transparent collar (see Fig. 10.24 (right)).

Working with suppliers to create suitable processes, the CBETA team produced an initial production run of 216 magnets. Initially, the magnets had a relative field error of 18.2 units on average, which decreased to 2.2 units after shimming. This gives confidence that by making the manufacturing tolerances less stringent during the initial build phase, the overall cost of production can be reduced without impacting the quality of the magnetic field [534] (see Fig. 10.25).

Figure 10.26 shows a cross section of the cylindrical FCC-hh transfer line Halbach dipole and quadrupole magnet concepts, respectively. Note that the inner bores of the magnets have been increased beyond what is required for the beam pipe to accommodate the shim assembly. The designs demonstrate that the required field strengths can be achieved using sensible PM wedge geometries and volumes. However, it is more challenging to determine what level of field quality can realistically be achieved because this is a function of manufacturing tolerances that are not yet known. Currently, extrapolating from the initial production run of CBETA indicates that the FCC-hh field quality requirements can be met. More research and development will be needed to confirm this statement. There are two main avenues of investigation that need not be mutually exclusive. A prototyping campaign would have the joint benefits of expanding knowledge of manufacturing processes and tolerances in the real world. Alternatively, stochastic modelling tools could be used to assess the sensitivity of the unshimmed Halbach field quality to different manufacturing tolerances, including PM wedge magnetisation magnitude/directional errors and wedge radial/azimuthal positional errors. For example, the ROXIE 2D magnetic modelling package [535] has a tool to perform randomised error studies. It is primarily used to investigate the effect of varying the azimuthal position of conductors in superconducting magnets but could also be used to study a range of Halbach parameters.

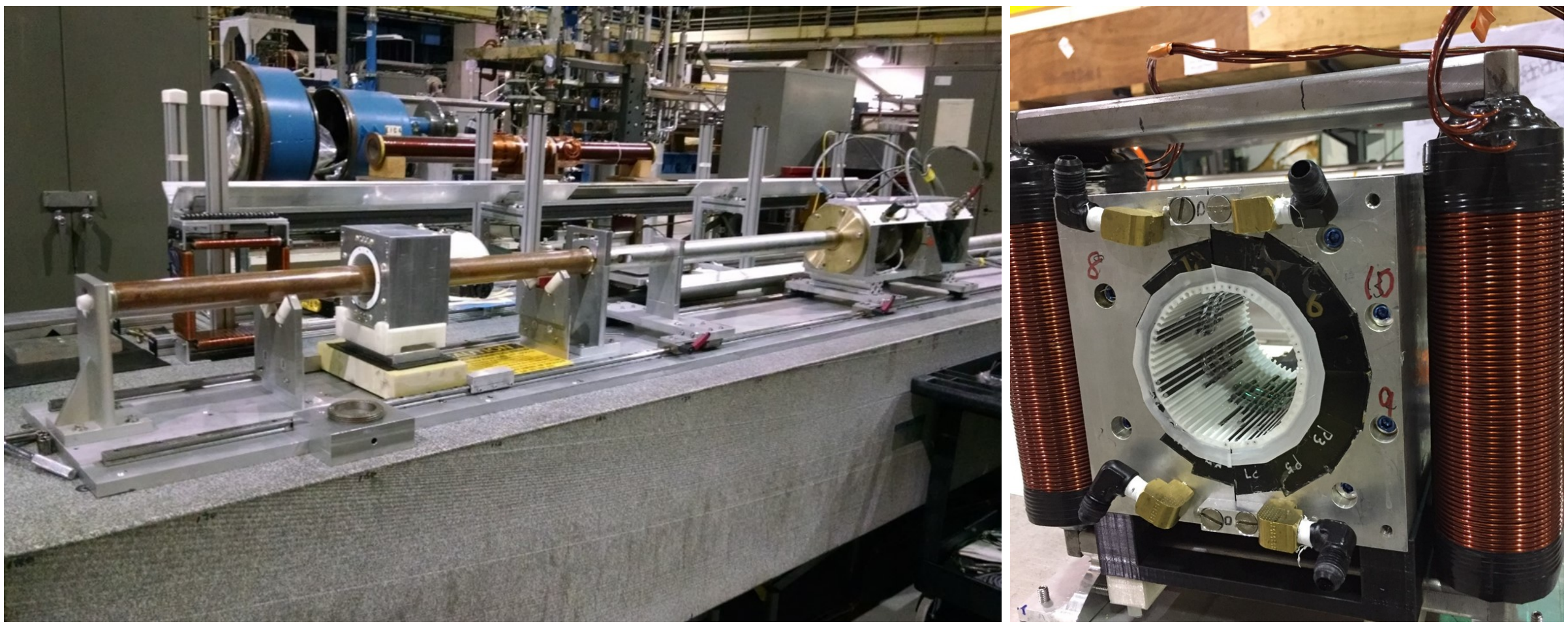

Fig. 10.24: Left: Performing rotating coil magnetic field quality measurements on an unshimmed Halbach magnet. Right: Shimmed Halbach magnet. Note the iron shimming rods of varying lengths (black) that have been assembled and glued to the plastic collar (white).

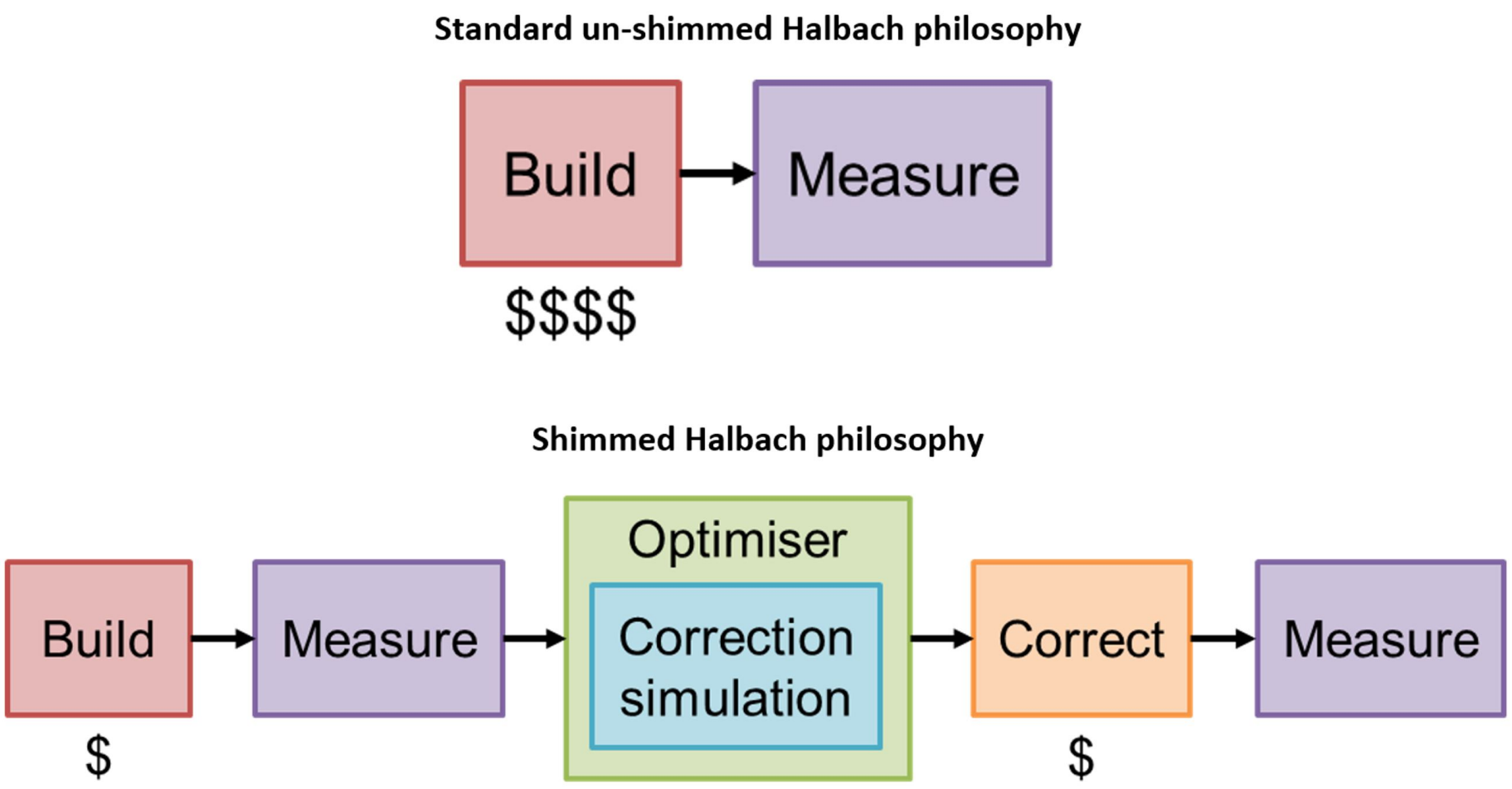

Fig. 10.25: Flow chart describing the production of conventional and shimmed cylindrical Halbach magnets, respectively.

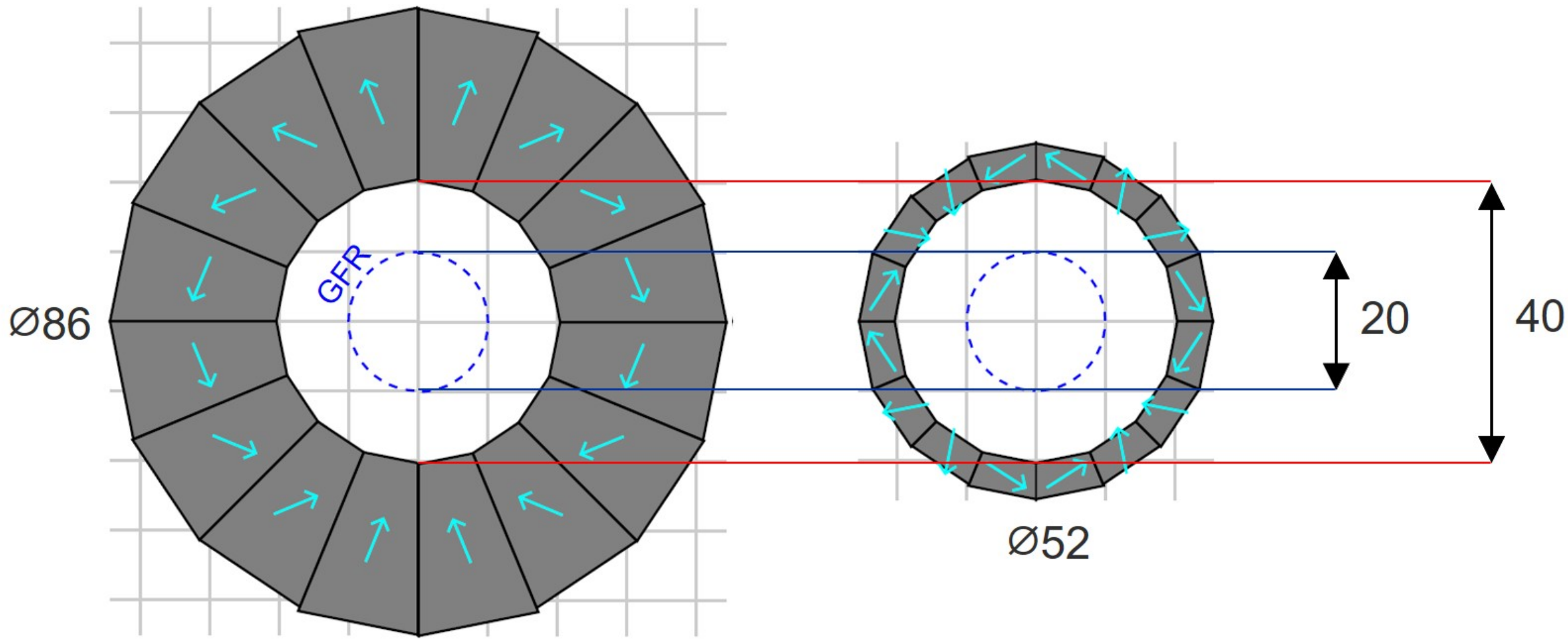

Fig. 10.26: Halbach dipole (left) and quadrupole (right) for the FCC-hh transfer line magnet concepts.

***Conclusions on FCC-hh transfer line magnets and future work***

An initial assessment resulted in three viable candidate magnet concepts. The next phase of investigation will develop these designs to a level where the most suitable can be evaluated. Given the large scale of FCC-hh compared to existing accelerators, a significant focus of this work will be on manufacturing and assembly challenges.

Obtaining accurate lifetime cost estimates will be key to making an informed choice. Currently, there are too many uncertainties to perform reliable calculations. However, it is still useful to begin considering how the contribution of different capital and overhead costs will affect the final total, which is qualitatively shown in Fig. 10.27. Although it should not be used as a tool to evaluate each concept, this information is useful to compare costs within each category.

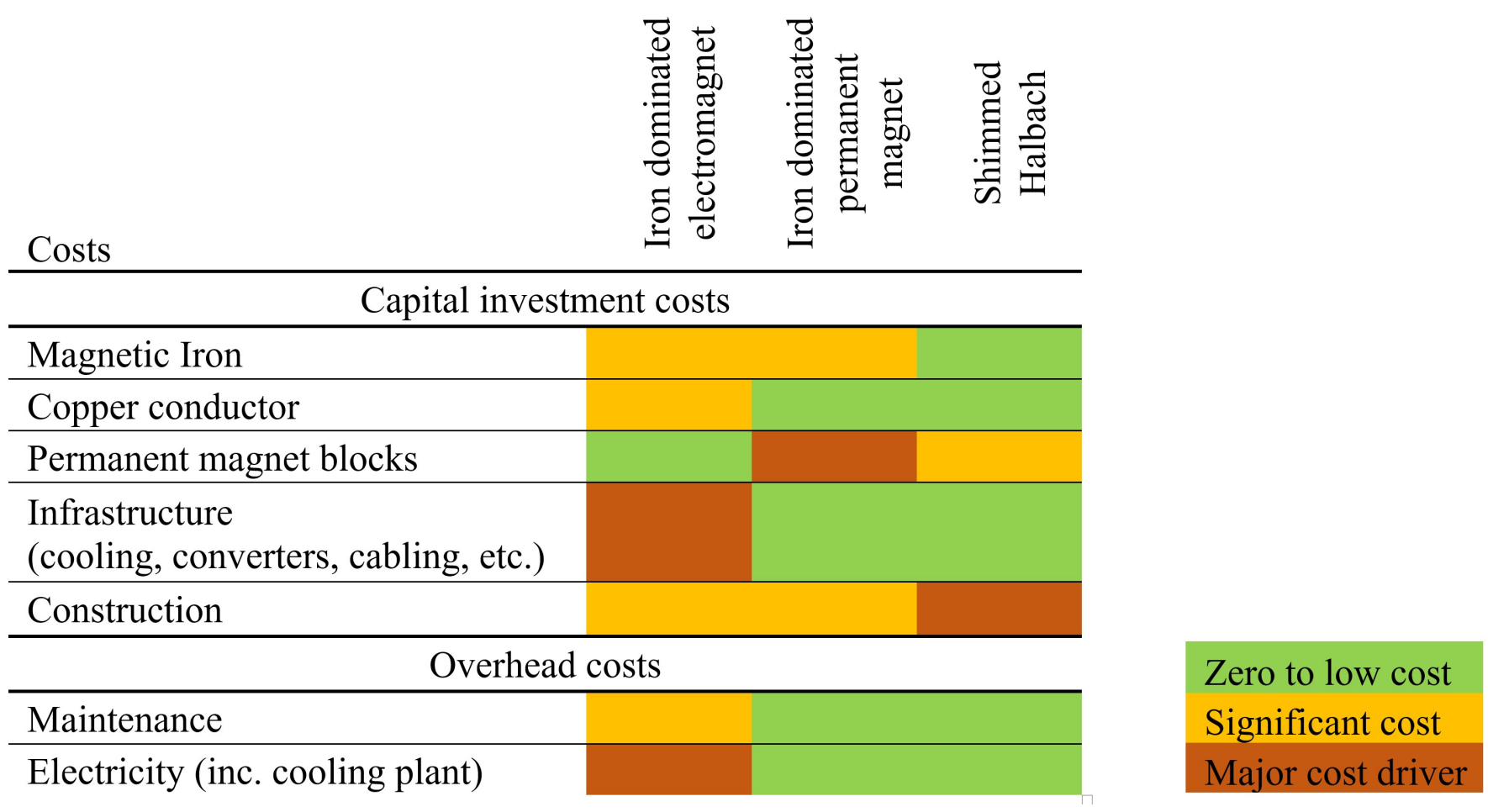

Fig. 10.27: Qualitative considerations on the magnet costs for the various concepts considered.

A permanent magnet research and development programme has the potential for significant benefits for accelerator engineering and beyond. Collaboration agreements can also accelerate progress, with different partners contributing various expertise and capabilities. Collaborators at other research institutes could offer valuable contributions to the electromagnetic design of accelerator magnets. Meanwhile, private-sector manufacturers are well-placed to develop new industrial-scale production processes.

Integration of the transfer lines in the FCC-hh tunnel was studied, considering the various options for the magnet design. The result of these studies is shown in Fig. 10.28, where the normal-conducting (left) and PM (right) solutions are presented. In both cases, the nominal tunnel size does not need to be increased to allow the installation of the transfer lines. However, it is evident how the smaller size of the cross-section of the PMs renders this concept highly interesting, also with regard to integration.

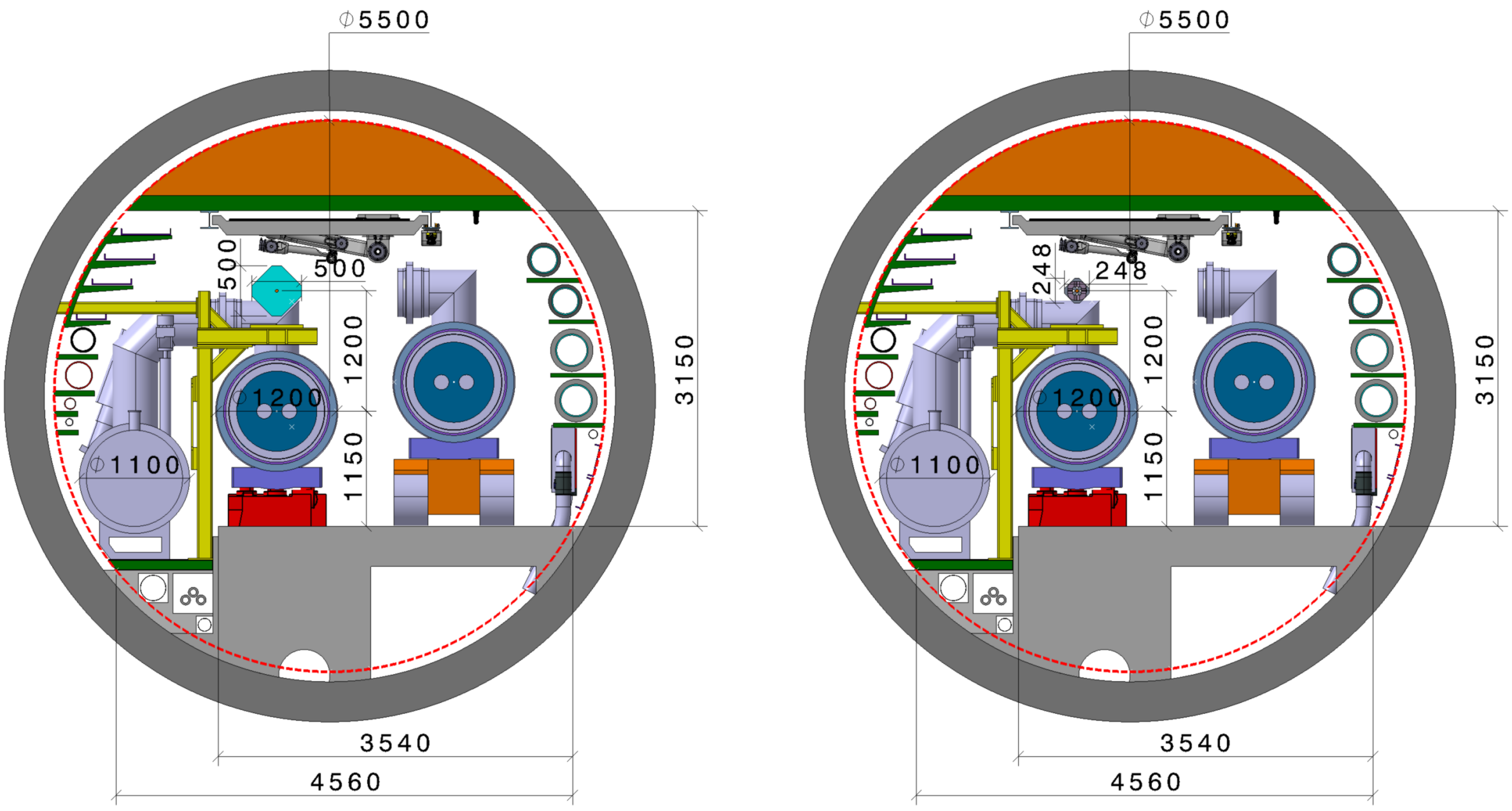

Fig. 10.28: Cross-section of the FCC-hh ring tunnel with a sketch of the integration of the transfer lines for the two options under study, namely with normal conducting magnets (left) and permanent magnets (right). Integration of the transfer lines into the ring tunnel does not require any increase in the tunnel cross-section.

## 10.4 High-field magnets

The following is organised in four sections. Section 10.4.1 provides an update to the baseline magnet parameters for FCC-hh dipoles, and outlines the changes with respect to the 2019 conceptual design report. Section 10.4.2 mentions alternative choices for the magnet system and how they would affect other accelerator subsystems; Section 10.4.3 presents design options that will validate key magnet parameters, places them in the context of past and present projects, and provides key figures for a block-coil design; finally, opportunities and challenges of HTS options for FCC-hh dipole magnets are discussed in Section 10.4.4. Considerations of cost and timeline are not given here, but will be addressed by FCC and the HFM Programme in the input to the update of the European Strategy for Particle Physics.

### 10.4.1 Baseline with $Nb_3Sn$ magnets

***Evolution of main parameters***

The main parameters of the FCC-hh with $Nb_3Sn$ main dipoles and its evolution with respect to the design report [10] is given in Table 10.16. As discussed in Section 10.2.1, from the point of view of the conceptual layout and optics, the main change is the 7% reduction of the accelerator circumference, from 97.76 km to 90.66 km. The relevant parameter for the energy reach, i.e., the length of the arcs, is reduced by 5%. However, an iteration on the optics layout, thanks in particular to the use of longer cells, allows for an increase of the arc filling factor by 4% from 0.80 to 0.83 [518].

Table 10.16: Parameters of the FCC-hh lattice, 2019 and 2025 values.

| | | CDR 2019 | 2025 $Nb_3Sn$ |
|---|---|---|---|
| Dipole field | (T) | 16.0 | 14.0 |
| Dipole aperture | (mm) | 50 | 50 |
| Dipole magnetic length | (m) | 14.3 | 14.3 |
| Operational temperature | (K) | 1.9 | 1.9 |
| Tunnel length | (km) | 97.76 | 90.66 |
| Arc length | (km) | 81.0 | 76.9 |
| Arc filling factor | - | 0.80 | 0.83 |
| C.m. energy | (TeV) | 50+50 | 42.5+42.5 |
| Loadline fraction | - | 0.86 | 0.80 |
| jc at 16 T and 4.2 K | (A/mm2) | 1500 | 1200 |
| Number of dipoles | - | 4587 | 4463 |
| Number of quadrupoles | - | 760 | 520 |

For the magnet baseline parameters, the main change is the reduction of the operational field from 16 T [536] to 14 T [537]. This choice enables the following accompanying measures, which together provide a consistent baseline with high confidence level based on HL-LHC experience:

- An increased margin for magnet operation: the loadline fraction (nominal current divided by maximum theoretical current) is decreased from 86% [536] (the same value as for the LHC Nb-Ti dipoles at 7 TeV [538, 539], but considered risky for the production of more than 4000 $Nb_3Sn$ dipoles) to 80% (2% above the baseline for the HL-LHC $Nb_3Sn$ triplet). A discussion of the margins associated to this loadline fraction is given in Section 10.4.1;
- It is assumed that the conductor critical non-copper current density required to produce 14 T at 80% loadline fraction can be attained with the best $Nb_3Sn$ conductor available today, corresponding to 1200 A/mm2 at 16 T, 4.22 K; see Section 10.4.3 for the fit and expected values at other fields and temperatures. The target defined for an FCC-hh conductor in 2015 of 1500 A/mm$^2$ at 16 T, 4.22 K [540] has been achieved and exceeded in laboratory samples. Development of usable wires for magnets, based on this technology, is under way, but efforts have not yet started to industrialise the technology.
- EuroCirCol studies showed that the design of 16 T magnets is marginal with respect to the target of maximum 200 MPa of stress in the coil [536][4]; for a classical magnet design, the 12.5% reduction in field from 16 T to 14 T corresponds to a 25% reduction in stress, and therefore the 14 T magnet design can satisfy a condition of maximum stress of 150 MPa, which is more appropriate for a large scale production.

Other relevant parameters such as operational temperature, aperture, inter-beam distance, maximum size of the magnet, are unchanged. With the reduction of the magnetic field and of the arc length, and the increase of the arc filling factor, the centre of mass (c.m.) energy is 84.6 TeV.

***Filling factor, cell quadrupoles***

Cell quadrupoles in the LHC lattice have a magnetic length of 3.15 m, and a gradient of 220 T/m, thus providing ∼700 T of integrated gradient [538]. The cell contains six dipoles, whose magnetic length is 14.3 m, and total cell length is 107 m: the ratio between total magnetic length of the dipoles and cell length is the filling factor of the arc: 14.3×6/107=0.80. The remaining 20% is left for interconnections and spool pieces (occupying 1.35 m for each dipole, i.e., about 10%), cell quadrupoles (5.9%) and other magnets.

[4]Note that the peak stress in the magnet coil is not a direct observable of the magnet; here, as in all the previous literature, peak stresses in the finite element model are referred to, with some assumptions on magnet preload, etc.

For the FCC-hh, the seven-fold increase in energy compared with the LHC would require a seven times larger integrated gradient, i.e., ∼5000 T, and, using the same technology and cell layout, one would need 22.5-m-long quadrupoles occupying nearly 50% of the arcs. To avoid this adverse effect on the filling factor, one has to choose longer cells; this makes not only more space for dipoles because quadrupoles are more spaced, but also shortens the quadrupoles since the integrated gradient is inversely proportional to the cell length [537].

In the first version of the FCC-hh optics the number of dipoles per cell has been doubled w.r.t. the LHC from six to twelve, with a total cell length of 210 m [10]. A recent further step brought the number of dipoles to 16, with a total cell length of 276 m [541]: the 1600 T of integrated gradient can be achieved by 4.2-m-long quadrupoles giving 375 T/m [10], occupying 3% of the cell. In this way the filling factor is increased to 0.83, and the resulting energy with 14 T dipoles is 85 TeV COM; see Table 10.16, fourth column. Further optimisations could give filling factors up to 0.87, thus reaching 90 TeV; this can be considered an upper limit, since the assumption for the distance between the magnetic lengths of two consecutive dipoles is 1.5 m (10% larger than in the LHC, requiring more massive end supports for the coil ends of 14 T magnets) and it appears difficult to reduce.

#### ***Injection energy and magnet aperture***

A possible FCC-hh injection energy is 3.3 TeV as in the Conceptual Design Report [10]; this gives an energy increase by a factor 14.2 between injection and collision energy, slightly smaller than in the LHC, where this factor is 15.6. The larger injection energy compensates for the larger beta functions due to the longer cell length, and facilitates fitting the beam within a 50 mm aperture, and leaving space for a beam screen to intercept the synchrotron radiation[5]. Since the beam size is proportional to the square root of cell length divided by the injection energy, compared to the LHC injection, the beam size in the FCC is reduced by 30%. Differently the LHC dipoles, thanks for the much larger bending radius, the FCC-hh dipoles do not require a curved geometry since the aperture gain would be negligible (of the order of 1 mm). Protons at 3.3 TeV for injection into the FCC-hh could be provided either from a modified LHC or from a new 4 T machine in the LHC tunnel.

Alternative FCC-hh injector options with 1.75 TeV and 1.3 TeV injection energy based on a superferric ring in the LHC tunnel or a faster cycling superconducting machine in the SPS tunnel are also being considered [10, Chapter 6.4]. However, these would pose major challenges for the accelerator and magnet systems, with an unprecedented energy increase for a high-energy hadron collider (factors of 25 or 33 from injection to top energy, respectively, to be compared with a record energy swing by a factor of 23 for the HERA proton ring), which would require a renegotiation of requirements on the control of field quality and magnet reproducibility at injection. Moreover, these could also require a larger magnet aperture (thus increasing the mass of the conductor), or a shorter cell length (thus reducing the filling factor and the energy reach). Further studies are needed to converge on the choice of the optimum injector for the FCC-ee.

#### ***Main dipole margins***

The short sample condition is defined as the combination of the peak field in the coil and the current density in the superconductor that corresponds to the measured critical surface of a strand short sample at the operational temperature: this is the maximum theoretical field achievable in the magnet. Superconducting magnets operate at a fraction of this value, called *loadline fraction*. The loadline margin is defined as one minus the loadline fraction. The *current fraction* is the ratio between the maximum theoretical current that can be carried at the operational field and the operational current. The current sharing temperature is the temperature where the pair of nominal field and current density lie on the critical surface. The *temperature margin* is the difference between the current sharing temperature and the operational temperature.

[5]Note that the beam screen presented in the baseline [10] corresponds to 100 TeV c.m. energy, with double the synchrotron radiation at 85 TeV.

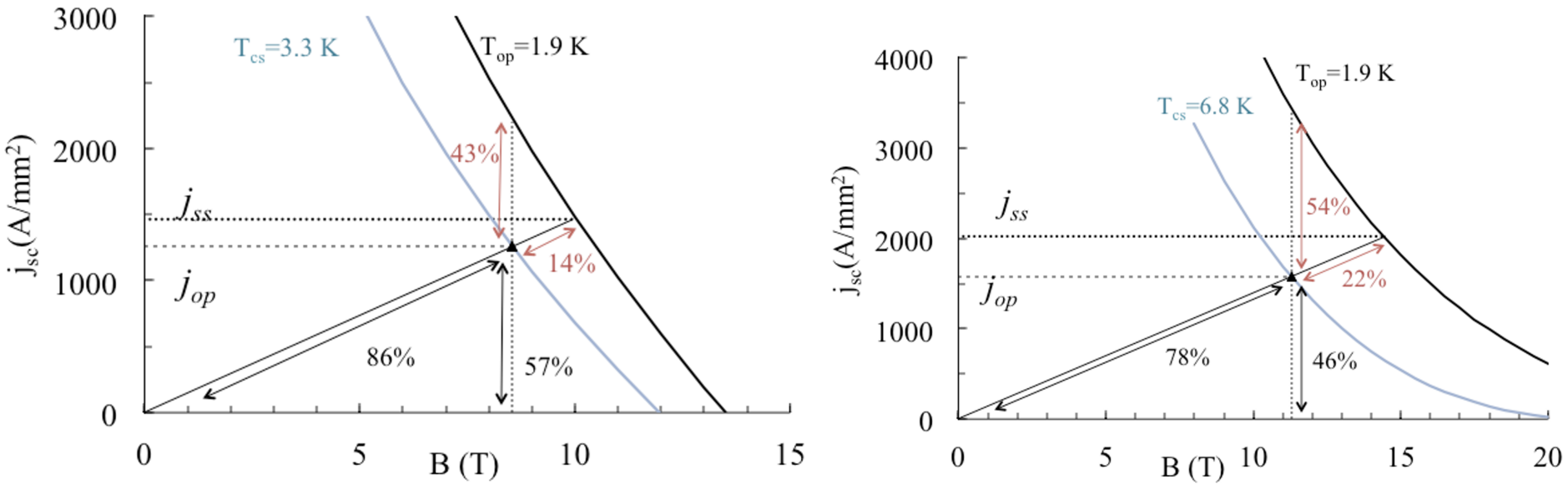

Fig. 10.29: Margins for Nb-Ti LHC dipoles at 7 TeV and 1.9 K (left) and for the HL-LHC quadrupoles at 7 TeV and 1.9 K (right).

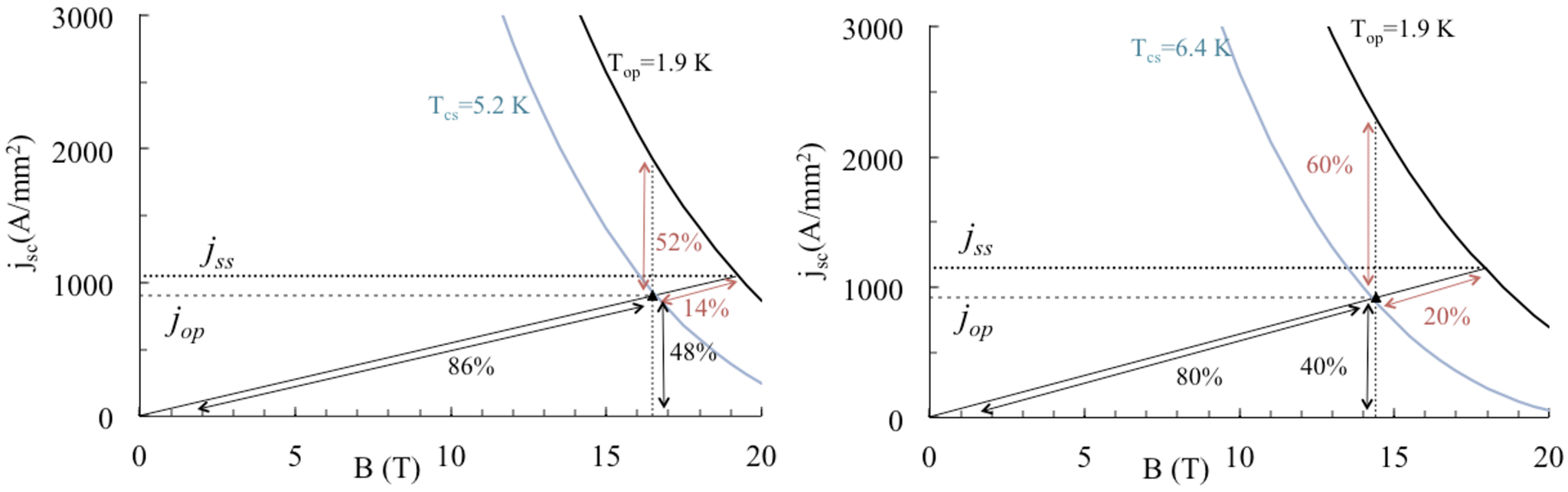

Fig. 10.30: Margins for FCC-hh $Nb_3Sn$ dipoles at 1.9 K: baseline of 2019 (left) and of 2025 (right).

Figure 10.29 presents a visualisation of these margins for (HL-)LHC magnets based on two different types of Low Temperature Superconductor (LTS), namely for the LHC arc dipoles operating at 7 TeV, and for the HL-LHC triplet quadrupoles at 7 TeV. LHC Nb-Ti dipoles have 14% loadline margin, 43% current margin and 1.3 K temperature margin [538,539,542]. The HL-LHC $Nb_3Sn$ quadrupole production is proving that a 22% loadline margin can be achieved systematically, with very limited training and no retraining after a thermal cycle [543,544]; the current margin is 54%, and the temperature margin is 4.9 K, i.e., about three times larger than in the LHC dipoles. All quadrupoles systematically reach operational current at 4.5 K, thus proving the existence of a temperature margin larger than 2.6 K.

Margins for the 2019 and 2025 baseline for the FCC-hh dipoles are shown in Fig. 10.30. The 2025 baseline for FCC-hh guarantees a 4.6 K temperature margin, cf. the previous value of 3.3 K, and a 60% current margin. Margins for the 2019 and 2025 FCC-hh $Nb_3Sn$ dipoles are compared with those for the LHC Nb-Ti arc dipoles and for the HL-LHC $Nb_3Sn$ triplet magnets in Table 10.17.

### *Hypothesis on hysteresis losses*

In the case of the FCC-hh $Nb_3Sn$ dipoles, the main source of the so-called AC losses are the hysteresis losses due to the magnetisation of the superconductor (also called persistent currents); they are peculiar to superconductors, and they are proportional to the filament size. The energy dissipated over a full cycle is independent of the ramp rate, and therefore the power losses scale with the ramp rate $\mathrm{d}B/\mathrm{d}t$. Note that other sources of AC losses as currents induced in loops that are partially resistive through the inter-strand resistance in cables, or the inter-filament resistance are proportional to $\mathrm{d}^2B/\mathrm{d}t^2$. Losses are dissipated in the magnet coil during the ramp, and the cryogenic power needed to deal with them is related to the magnet temperature: higher operational temperatures imply lower power consumption for cryogenics. On the other hand, the magnet temperature has little influence on the power needed to deal with synchrotron radiation, since this is intercepted on the beam screen at or above 50 K; see Table 10.20.

Table 10.17: Margins for LHC dipoles at 7.0 TeV, HL-LHC triplet, FCC-hh dipoles, all operating at 1.9 K.

| | Energy (TeV) | Bore field (T) | Peak field (T) | Loadline margin (%) | Current margin (%) | Temperature margin (K) |
|---|---|---|---|---|---|---|
| LHC dipole | 7.0 | 8.3 | 8.7 | 14% | 43% | 1.3 |
| HL-LHC triplet | 7.0 | $9.9^{\dagger}$ | 11.3 | 22% | 54% | 4.9 |
| FCC-hh dipole 2019 | 50 | 16 | 16.5 | 14% | 52% | 3.3 |
| FCC-hh dipole 2025 | 42.5 | 14 | 14.5 | 20% | 60% | 4.5 |

† The quadrupole gradient times the aperture radius is given here.

For the FCC-hh magnet, a target of 5 kJ/m over the ramp (i.e., 10 kJ/m for the complete cycle) has been set in the CDR [10]. With the present status of technology for $Nb_3Sn$ conductors, having filaments of the order of 50 µm, $Nb_3Sn$ dipoles at 14 T have about twice this, i.e., 20 kJ/m for the full cycle [545]. Several paths are possible to close this gap between targets and technology:

- Reducing the filament size; the drawback in this option is that it requires R&D on the conductor. The need was identified [540, 546] and attempts in this direction were carried out. The resulting technology should clearly not increase cost or decrease critical current, which is not granted;
- Using artificial pinning centre (APC) technology [547], that gives lower critical current density at low field, and therefore lower hysteresis losses, satisfies the 10 kJ/m target. APC also have the potential of giving larger current densities in the 14 T-18 T range, see next section;
- Increasing the cooling power of the cryogenic system, that only has to absorb these losses during the ramp;
- Provide additional temperature margins so the magnet cold mass can buffer the energy deposited during the ramp; superfluid helium in the cold mass plays an important role in the buffering mechanism;
- Doubling the ramping time, with a corresponding reduction of the integrated luminosity.

### 10.4.2 Alternative choices for the LTS magnet system

Several alternative choices and how they would impact other accelerator subsystems are described below. They are: (i) operation at 4.5 K, (ii) a combined function lattice, (iii) longer dipoles and (iv) a lower operational field of 12 T. More design options that meet the above baseline parameters and do not affect other accelerator subsystems are discussed in Section 10.4.3.

***4.5 K operation***

The data of HL-LHC $Nb_3Sn$ quadrupole magnets shows that they can all operate at 4.5 K [543, 544]. Moreover, instabilities limiting performance are more severe at 1.9 K than at 4.5 K. This could open the possibility of operating at 4.5 K. Keeping the same magnets with 20% loadline margin at 1.9 K would imply a reduced loadline margin of 10% at 4.5 K, and a current density margin of 42% and a temperature margin of 1.9 K; see Fig. 10.31. Even though the HL-LHC magnets are routinely reaching this condition in individual tests, this baseline for the FCC-hh appears promising but extremely daring, given that the magnet system must also be robust against beam-induced effects that may be more severe than in the LHC due to the increased beam energy. The 4.5 K option is under study, and would present the following advantages: (i) Reduced complexity of the cryogenic system and its cost; (ii) Reduction of the cryogenic power to remove heat from the cold mass of a factor two to three; since about half of the heat has to be removed from the cold mass (the other half on the beam screen), this would imply a reduction of the cryogenic power by 30%.

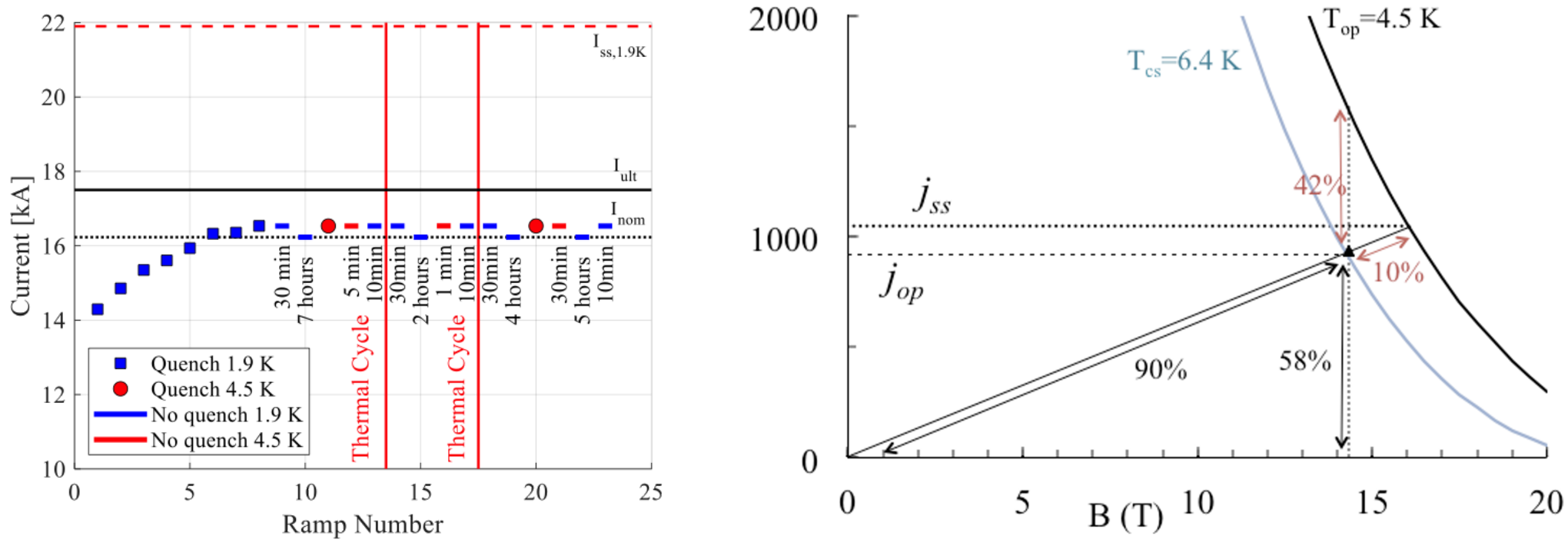

Fig. 10.31: A typical HL-LHC $Nb_3Sn$ quadrupole magnet training, reaching nominal current also at 4.5 K (left) and margins for an option of FCC-hh $Nb_3Sn$ dipoles designed for 1.9 K operating at 4.5 K (right).

The 4.5 K option could rely on dry magnets (cold mass not filled with liquid helium (LHe)); this would reduce the He inventory; however, it would imply challenges to magnet cooling during the ramp, and to magnet insulation (LHe is a much more reliable insulator than vacuum because the Paschen effect in residual gas could lead to a dielectric breakdown during a quench). To prove the viability of this option it is necessary to develop a baseline for the cooling at 4.5 K, and determine whether the theoretical temperature margin of ∼2 K and ∼40% in current density is sufficient to operate the ∼4400 magnets in the FCC, or whether to increase the margins. Note that in case of a 4.5 K baseline it is quite probable that all magnets should be individually tested at 1.9 K before installation, since training is much faster at 1.9 K, because during training at 4.5 K performance plateaus (that can be overcome at 1.9 K) may appear.

#### *Combined function Lattice*

A combined function lattice relies on removing the main quadrupoles and having a systematic quadrupole component ($b_2$) in the dipoles. This layout reduces the flexibility of operation, but it is used in some low-energy accelerators. The main advantage is to remove the need for two families of quadrupole magnets, even if their number (500) is fairly moderate with respect to that of the dipoles. The combined function lattice implies more stringent requirements on dipole alignment, since it contains the quadrupolar component; the option was considered, but eventually discarded for the LHC [548]. If the conductor peak field is kept constant, the combined function option would allow a further increase in beam energy of approximately 0.5%. The reason is that the larger filling factor would be partially compensated by the need to lower the main dipole field because part of the peak field would be created by the quadrupolar component [549].

#### *20 m long magnets*

In the late 2000s, the US LARP programme achieved the scaling of a $Nb_3Sn$ quadrupole from 1.5 m to 3.4 m long magnets [550]; HL-LHC quadrupoles at CERN prove the scaling to 7.5 m [544]. Scaling in length of FCC-hh dipoles is currently planned in two steps, first to 5 m (with the possibility of vertical testing of the magnet) and a second to 14.3 m.

The limit in magnetic length for LHC and FCC of 14.3 m is related to the European Union regulations for standard transport. For the SSC project, 17.5 m-long Nb-Ti dipoles were planned for the lattice and 15 were built and reached requirements [551]. Having 20 m-long magnets would have three positive consequences:

- A reduction of the total number of magnets to manufacture from ∼4400 to ∼3200 units, with, possibly, a 30% shorter production (about two years), or 30% fewer production lines (see Table 10.16);
- A reduction of the magnet cost; assuming 25% of the total cost is due to manufacturing and a reduction of 30% in the magnet number, the total cost of the dipoles would be reduced by 8%;
- An increase of the filling factor of 3.5%, either used to increase the energy to 88 TeV COM, or to reduce the dipole cost (via a 0.5 T field reduction and less conductor).

In case this route is further pursued, detailed studies are required to assess the impact of 20 m long magnets on several other areas, such as transport and integration.

***12 T magnets***

A layout based on a 12 T operational field would provide 73 TeV COM energy, with the same hypothesis on the filling factor as in the baseline shown in Table 10.16. Using the same loadline margin as for the 14 T baseline, 12 T dipoles would have a similar current and temperature margin, but require 30% less conductor. 12 T dipole short models are planned in the HFM programme, manufactured by INFN [552], as well as at CERN and PSI.

### 10.4.3 $Nb_3Sn$ dipole designs

***Requirements***

Based on the experience of the HL-LHC triplet quadrupole magnets, the following targets have been set:

- *Field*: To provide an operational field in the aperture of 14 T, short models must be able to systematically reach 15 T (so-called ultimate field). For the operational field of 12 T, an ultimate field of at least 13 T is required.[6]
- *Loadline and temperature margin*: The dipoles operated at 1.9 K should provide a loadline margin of at least 20%, and a temperature margin larger than 2.6 K at operational field to reach operational field also at 4.5 K.
- *Mechanics*: The structure must be able to withstand forces corresponding to the ultimate field. The coil, in the case of a design including a coil preload, must be not in tension at nominal field.[7] The target maximum stress must not exceed 150 MPa in all phases (assembly, cool-down, and powering to nominal).
- *Protection*: the hotspot temperature for the protection system in the nominal scenario (without failures) must not exceed ∼270 K. The system should guarantee a maximum hotspot temperature of 350 K in case of realistic and conservative failure scenarios.

***Conductor properties***

The parametrisation used for the $Nb_3Sn$ conductor is the following

[6]Note that LHC dipoles, operating at 8.1 T in the LHC at an equivalent energy of 6.8 TeV, had short models systematically reaching 9.5 T.

[7]Note that this is a less stringent requirement than previously used (10 MPa of compression at nominal or at ultimate field); this is justified on the grounds that many experimental data show that partial unloading does not prevent reaching nominal field.

Table 10.18: Assumptions for critical current density of $Nb_3Sn$ superconductor.

| Critical current density in superconductor (A/mm$^2$) | 12 T | 15 T | 18 T |
|---|---|---|---|
| 1.9 K | 3655 | 2170 | 1185 |
| 4.22 K | 2800 | 1515 | 705 |

$$J_{\mathrm{sc}}(T,B) = \frac{C_0}{B}\left(1-\left[\frac{T}{T_{\mathrm{c0}}}\right]^{1.52}\right)^{0.96}\left(1-\left[\frac{T}{T_{\mathrm{c0}}}\right]^{2}\right)^{0.96}\left(\frac{B}{B_{\mathrm{c20}}\left(1-[T/T_{\mathrm{c0}}]^{1.52}\right)}\right)^{2.5} \tag{10.1}$$

with $T_{\mathrm{c0}}$=16 K, $B_{\mathrm{c20}}$=29.38 T, and $C_0$ is the variable parameter to scale the conductor performance: $C_0$= 214 000 AT/mm$^2$ for the improved HL-LHC considered as the baseline. The specification for HL-LHC has been given at 12 T and at 15 T. For an FCC-hh dipole at 14 T, specifications should be given at 15 T and at 18 T (at 0.5 T larger than peak field in operational condition and at short sample field, respectively). The critical current densities at different fields and temperatures are given in Table 10.18.

***Quantity of conductor***

The driving term of the magnet cost is the quantity of conductor: it accounts for about one third of the cost for the LHC dipoles, and half of the cost for the HL-LHC magnets. The quantity of conductor is parametrised via the equivalent coil width $w_{\mathrm{eq}}$, i.e., the width of the 60° sector coil having the same surface area as the insulated coil, see Fig. 10.32, left. This quantity has the advantage of being related to the field and to the current density via the approximated expression

$$B[\mathrm{T}] \approx 0.007\, J[\mathrm{A/mm^2}] w_{\mathrm{eq}}[\mathrm{mm}]. \tag{10.2}$$

The volume of the insulated conductor is given by

$$V \approx \frac{4\pi l_{\mathrm{m}}}{3}\left[(r+w_{\mathrm{eq}})^2 - r^2\right] = \frac{4\pi l_{\mathrm{m}}}{3}\left[2rw_{\mathrm{eq}} + w_{\mathrm{eq}}^2\right] \tag{10.3}$$

where $l_{\mathrm{m}}$ is the magnet length, and $r$ is the magnet aperture radius (see also Fig. 10.32, right). For FCC-hh, a coil width of up to 55 mm is considered, thus giving a quantity of conductor 2.5 times larger than in the LHC dipoles.

***Previous achievements in bore field***

$Nb_3Sn$ accelerator dipole models have been developed since the late 80's. A short summary of the achievements is given below. All cases refer to 1 to 2 m long magnets (usually called short models), except for the HL LHC magnets (11 T dipoles and final triplet quadrupole).

In 1990 the CERN-Elin dipole reached 9.5 T at 4.3 K in a 50 mm aperture, with a two-layer coil based on $\cos(\theta)$ geometry and grading [541]. This option was not retained for the LHC due to the higher cost with respect to the Nb-Ti option at 1.9 K and the complexity of the technology.

In the 1990s the MSUT dipole reached 11.3 T at 4.5 K in a 50 mm aperture, with a two-layer coil based on $\cos(\theta)$ geometry and grading [553]. The magnet was tested again in 2020 at 1.9 K, achieving the power converter limit (11.8 T bore field).

In the 1990s the D20 LBNL dipole reached 13.4 T at 1.9 K in a 50 mm aperture, with a four-layer coil based on $\cos(\theta)$ geometry and strong grading [554, 555].

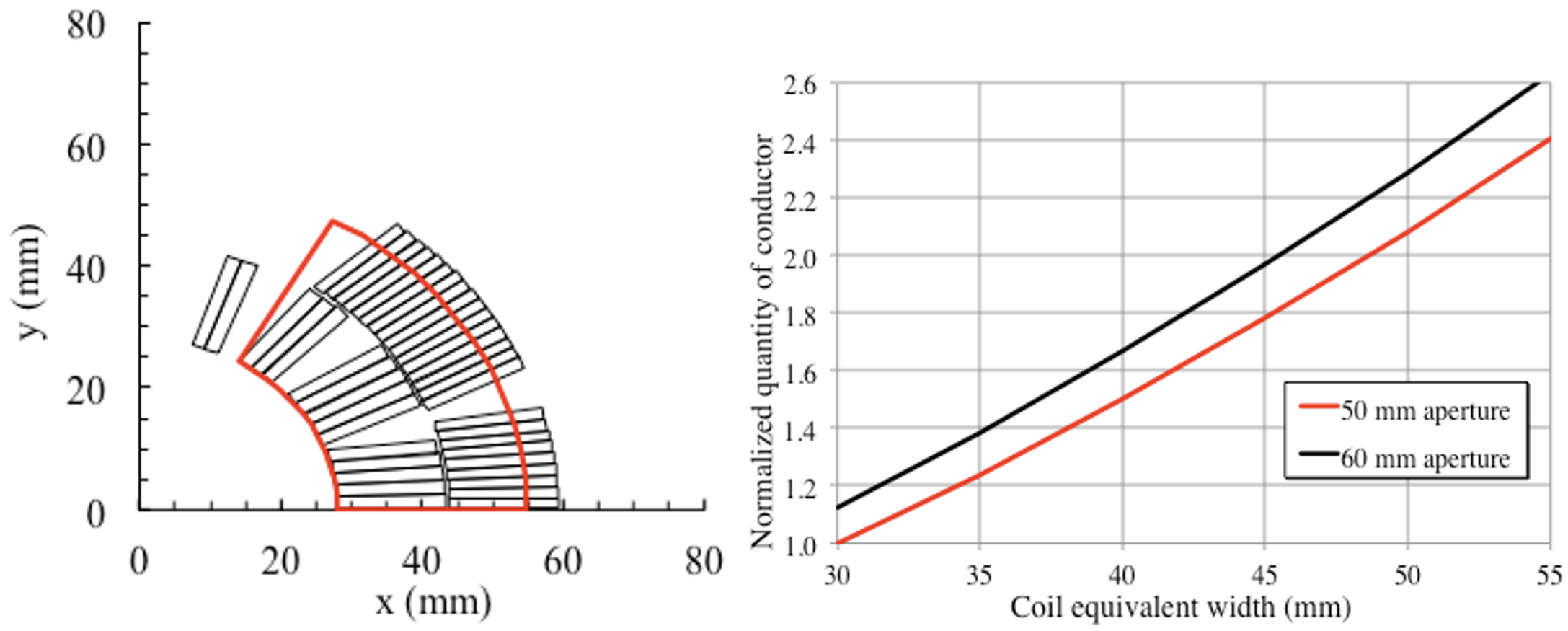

Fig. 10.32: Coil layout for the LHC dipole, and equivalent coil width (width of the red sector, left) and quantity of conductor versus equivalent coil width (right).

In the 2000s the HD2 LBNL dipole reached 13.8 T at 4.5 K, with a novel coil configuration based on a two-layer block design. Unlike to previous models, which all had an aperture of 50 mm, this magnet has a free bore of 36/43 mm diameter [556].

In the second half of the 2010s, the Fresca2 CERN-CEA dipole reached 14.6 T at 1.9 K, with a coil configuration based on a four-layer block design. This magnet, conceived as a cable test station operating at 13 T, has a free bore of 100 mm diameter [557,558]. All of its features conform to accelerator requirements, except for the very large coil width (80 mm, see Fig. 10.33) that makes it unaffordable.

In the second half of the 2010s, some 11 T dipoles short models reached 12.0 T [559, 560] both at CERN and in FNAL in a 60 mm aperture at 1.9 K.[8] Note that the coil has the same width as the LHC dipoles, and therefore the higher field is obtained by higher current density (540 A/mm$^2$ rather than 360/440 A/mm$^2$ at nominal current), see Fig. 10.33. The scaling to 5 m showed degradation of performance after thermal cycles in the long magnets; nevertheless, the programme achieved two significant results: (i) the first double aperture $Nb_3Sn$ dipole magnet, fully compatible with machine operation and (ii) the first scaling of $Nb_3Sn$ coils to 5 m lengths, reaching a bore field above 11 T.

In the second half of the 2010s, MDPCT1 dipole reached 14.5 T at 1.9 K in a 60 mm aperture. This magnet, made at FNAL, is a four-layer $\cos(\theta)$ dipole with grading. This is the first dipole to reach fields above 14 T at 4.5 K [561], and also the first to reach more than 14 T with a coil width of the order of 50 mm (as compared to 80 mm of Fresca2). After reassembly and after a thermal cycle the performance degraded irreversibly by more than 10% [561].

Since 2015, the HL-LHC interaction region triplet quadrupoles [543, 544, 562] showed very good reproducibility of performance (11.3 T peak field in nominal conditions), nominal currents were also achieved at 4.5 K in all models, 13 T was achieved in many short models, scaling to 7.15 m lengths, and there was no degradation of performance after thermal cycle. The project is halfway through production at the moment of writing.

#### *Magnet designs:* $\cos(\theta)$*, block coil, common coil*

As shown during the EuroCirCol studies [536], the target magnetic field can be achieved with different designs, each one presenting opportunities and challenges. The more classical way is to follow the $\cos(\theta)$ configuration; a two-layer design can give 12 T dipoles [552] with a coil width of the order of 35-40 mm, and a third and fourth layer are needed to reach 14 T operational fields as in MDPCT1. The first option is being developed in INFN and at CERN (FalconD) and a proposal for a four layer coil was put forward

[8] Note the possible confusion between the magnet name (11 T) and the maximum achieved field (12.0 T).

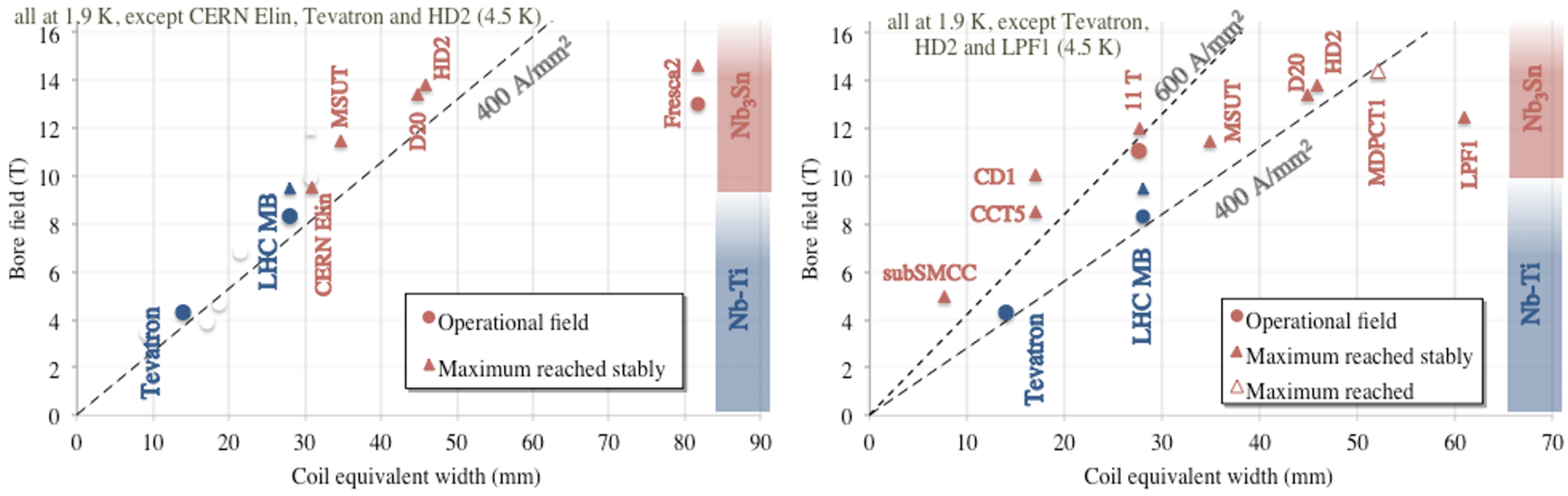

Fig. 10.33: Operational and achieved field versus coil width in LHC and Tevatron colliders, and in $Nb_3Sn$ short models (left), and 11 T, MDPCT1 dipoles, and in common coil and CCT demonstrators (right).

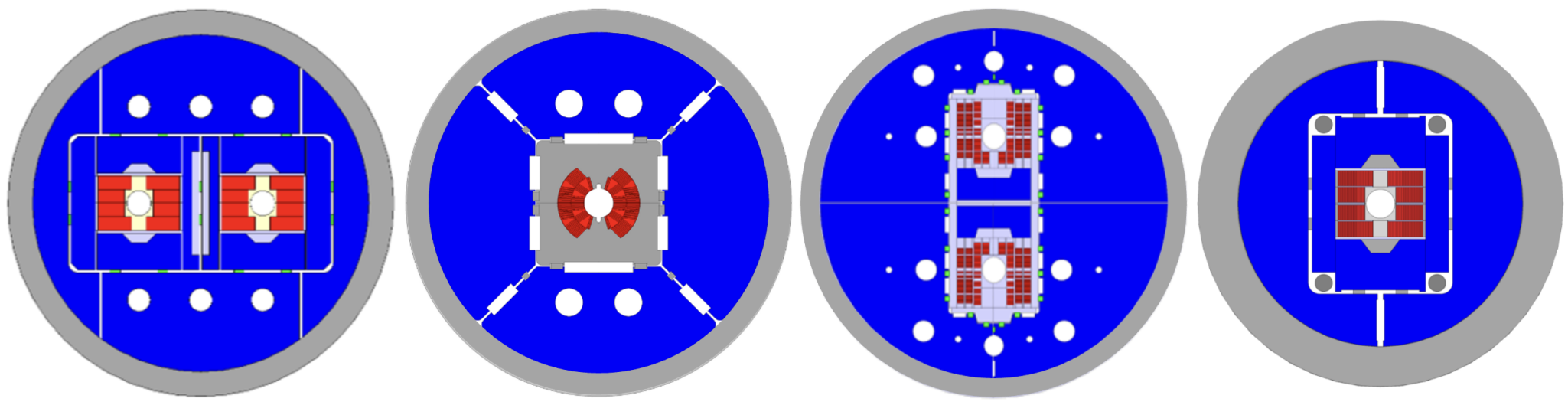

Fig. 10.34: Conceptual design of 14 T and 12 T magnets: BOND (left), FalconD (centre left), SMACC1 (centre right), F2D2 (right).

by INFN for EuroCirCol [563].

The design based on block coils holds the record in field today and therefore is a natural alternative to the $\cos(\theta)$ design. This option is being pursued by CERN, with a two-layer coil and a 25 mm wide cable (BOND [564]), and by CEA-Saclay with a four-layer coil including grading (F2D2 [565]) and an intermediary step with flat racetrack coils (R2D2 [566]). As an example, the cross-section and the main parameters of the BOND magnet are given in Fig. 10.34 and Table 10.19. The coil width, as for other designs, is of the order of 50-55 mm. This is a first list of parameters of a design that could be further optimised to reduce the conductor mass.

The main differences with respect to the designs presented in 2019 [536], are (i) 20% lower stored energy due to the lower field, (ii) a maximum hotspot temperature of 270 K in case of a quench with nominal protection, and (iii) a coil stress below 150 MPa. The conductor mass is 10% lower, and current densities of the 2025 baseline are very similar to the current densities of the 2019 baseline in the inner layer (all 2019 designs use grading).

A third option is the common coil design, which is an intrinsically double aperture magnet, based on racetrack coils plus non-planar correction coils; the idea has been proposed in the 90's [567], and today is the design that has been adopted for the Chinese SppC dipole. In 2022, IHEP built a a hybrid $Nb_3Sn$/Nb-Ti technology demonstrator magnet called LFP1-U that reached 12.5 T in two small 14 mm diameter apertures [568], and 90% of short sample at 4.5 K; see Fig. 10.33. The next step for IHEP is LFP3, aiming to produce 16 T, with 13 T obtained from $Nb_3Sn$ plus 3 T from HTS [569]. The common coil path is being followed by CIEMAT, aiming at a 14 T operational field with $Nb_3Sn$, based on a common coil design in a 50 mm aperture ( [570] and DAISY design [571]. Recently, the PSI team has made a significant improvement in the design, finding an asymmetric configuration of the coil that allows having planar correction coils of the same type as the common coils [572]. This design will be applied to CIEMAT magnets.

Table 10.19: Parameters for possible designs of 12 T and 14 T, scaled to 14.3 m magnetic length.

| | | FalconD | SMACC1 | BOND | F2D2 | DAISY |
|---|---|---|---|---|---|---|
| Coil type | | $\cos(\theta)$ | Stress-Managed Common Coil | Block Coil | Block Coil Graded | Common Coil |
| Field | (T) | 12 | 12 | 14 | 14 | 14 |
| Current | (kA) | 19.9 | 12.7 | 19.3 | 9.2 | 15.4 |
| Peak field | (T) | 12.5 | 12.6 | 14.8 | 14.6 | 14.6 |
| Loadline margin | (%) | 25 | 23 | 18 | 24 | 19 |
| Equivalent coil width | (mm) | 37.4 | 20.1+19.1 | 53.4 | 23.9+27.5 | 31.7+23.0 |
| $J$ overall | (A/mm$^2$) | 418 | 411/603 | 348 | 299/438 | 326/346 |
| $J$ superconductor | (A/mm$^2$) | 1204 | 1205/2129 | 967 | 875/1968 | 933/982 |
| $J$ copper | (A/mm$^2$) | 1337 | 1339/1774 | 1074 | 973/1093 | 1037/909 |
| Stored energy$^\dagger$ | (MJ) | 15.5 | 22.9 | 29.5 | 30.0 | 31.9 |
| Inductance$^\dagger$ | (mH) | 64 | 266 | 146 | 692 | 252 |
| Coil energy density | (J/mm$^3$) | 0.079 | 0.109 | 0.089 | 0.098 | 0.079 |

$^\dagger$ FalconD and F2D2 designs are single aperture; here stored energy and inductance are scaled to two apertures.

***Magnet designs based on stress management***

A stress managed magnet [573] design can be defined as a design where the supporting structure is not external, as in the $\cos(\theta)$ layout, or external and internal, as in the block and common coil designs, but it is spread within the coil as a metallic winding former. The advantage is twofold: (i) the structure can intercept forces and avoid stress accumulation and (ii) the winding former acts as a winding, reaction and impregnation tooling, thus, accelerating manufacturing processes. This can allow the use of higher current densities in the windings, though this advantage is compensated by the winding former, which dilutes the effective current density. A disadvantage could be that after epoxy impregnation, bonded interfaces are loaded in tension and may break, leading to long training. Alternatives to epoxy impregnation are being studied.

Stress-managed structures may be the only way to reliably reach fields above 15 T. The US-MDP research programme is strongly investing in this direction [574–576], proposing two designs both based on stress management. A stress managed $\cos(\theta)$ , where a $\cos(\theta)$ coil is wound on a former that includes the wedges, is being studied at FNAL. A $Nb_3Sn$ coil in mirror configuration has reached 12.7 T at 87% of short sample limit [576].

A fully stress-managed magnet is the CCT (canted cosine theta) configuration, where each turn is supported by a rib integrated to a former. This design, also called tilted solenoid or double helix, dates back to the early 1970s [577] and has been adopted by LBNL since the mid 2000's, aiming to reach fields towards 20 T [578]. A drawback of CCT with respect to the previous option is that the use of conductor is less optimised, since part of it is used to generate a solenoidal field that is cancelled by the other coil. However, the structure is totally mixed with the coil – a kind of endoskeleton – making easier protection and mechanics and possibly allowing higher current densities.

The short model CCT5 built at LBNL reached 8.5 T with a $Nb_3Sn$ winding using a 10 mm cable over a 90 mm aperture [578]. At PSI, using a similar design and the same cable, 10.1 T were reached in a 60 mm aperture short model named CD1 [579]. In both cases the overall current densities (over the insulated coil) are very large (800 and 1000 A/mm$^2$ respectively): a field of 8-10 T is reached with an equivalent coil width of the order of 17 mm; see Fig. 10.33. PSI is currently pursuing a stress-managed asymmetric common coil as a candidate for 14 T (SMACC1) [572].

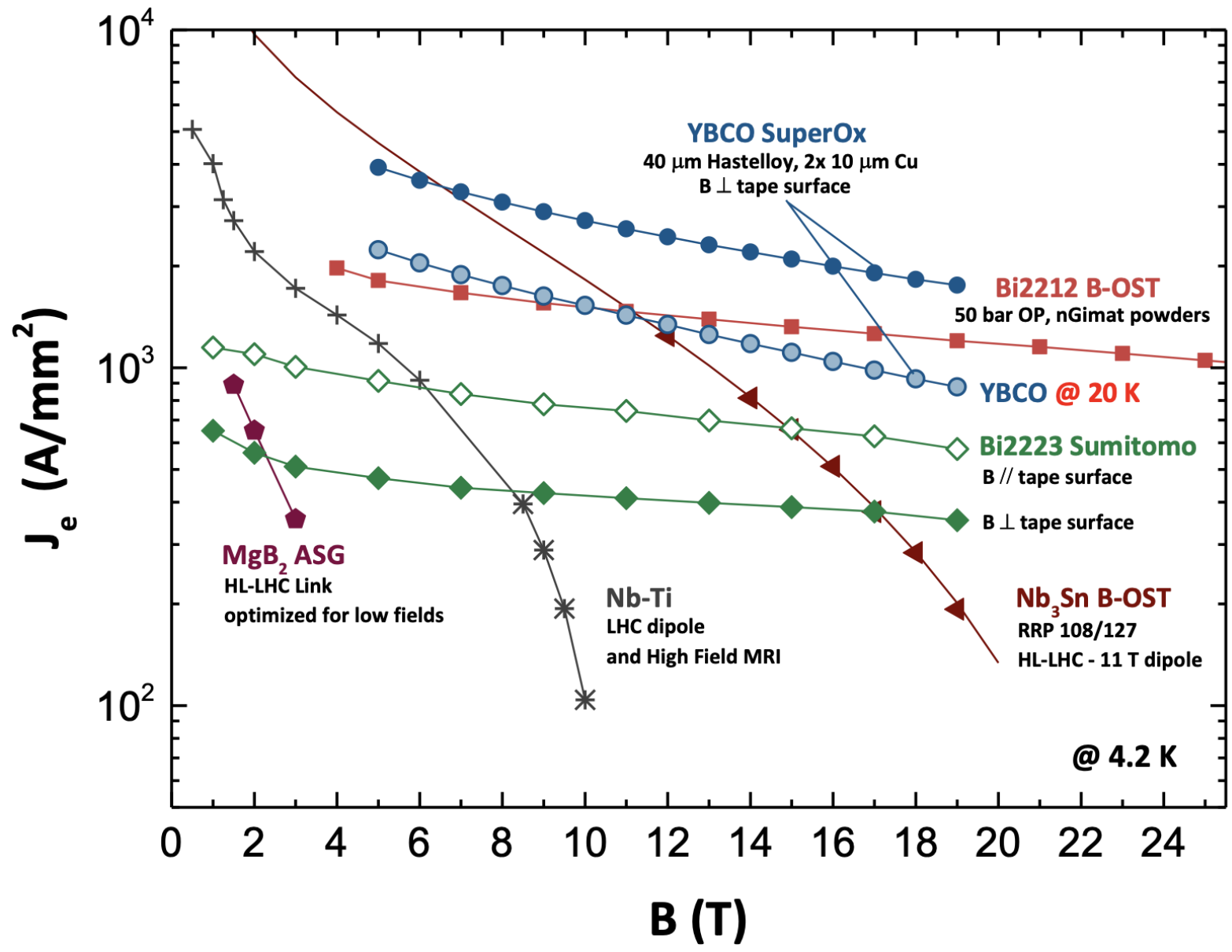

Fig. 10.35: Critical (engineering) current densities of different HTS conductors, $Nb_3Sn$, and Nb-Ti. Reproduced from [581].

#### *Hybrid Nb-Ti/$Nb_3Sn$ magnets*

Material grading is an option that consists of using a cheaper and less performant material in the low field region. It has been used for the D19H magnet [580], and is being used in demonstrators (for instance LPF1 in the IHEP programme [568]). Unlike the current grading, where higher current density is used for low field regions (as in CERN Elin, MSUT, D20, and in the LHC dipole), here similar current densities are used in the inner and outer layer, but the outer layer is made with Nb-Ti. Nb-Ti at 1.9 K with an overall current density of the order of 400 A/mm2 can be used in coil regions not exceeding 8 T; this could lead to a significant reduction of the mass of $Nb_3Sn$ conductor (>30%); however, this prevents operating at 4.5 K, since the temperature margin of Nb-Ti at 80% of the loadline at 1.9 K is only about 2 K.

### 10.4.4 High Temperature Superconductor: perspectives and challenges

#### *High Temperature Superconductor: opportunities*

High Temperature Superconductor (HTS), discovered in the mid 80's in the family of cuprates, has many interesting features that make it a game changer in superconducting technology, see Fig. 10.35 [581]: (i) high values of critical current density (>2000 A/mm$^2$) also above 20 T, opening the way to high fields, (ii) the ability of sustaining high current densities and fields at and above 20 K, opening the path to cheaper cooling systems (even though liquid nitrogen cooling is still considerably out of reach for high field magnets), and (iii) a much higher stability versus thermo-mechanical perturbation due to the increase in enthalpy margin at higher operating temperatures. The HFM programme is exploring the options of dipole magnets in the 14 to 20 T range, at temperatures between 4.5 K and 20 K, thus covering a COM energy in the range of 85 to 120 TeV.

Two types of high-temperature superconductors are commercially available: BSCOO and REBCO. BSCOO [582], produced mainly by one company in the US, has the advantage of being available as round wire with small filaments. Its use is complex since, like $Nb_3Sn$, it needs a reaction after winding, but at 900°C in oxygen rich ∼50 bar atmosphere. A process-compatible electrical insulation system must be selected. Moreover, BSCOO is brittle and, hence, requires careful handling in coil manufacturing and appropriate mechanical structures to respect stress limitations. Several short magnet models have successfully been built in the US; see Section 10.4.4.

REBCO [583] is produced by multiple suppliers worldwide; this conductor is produced in the geometry of a tape, and its industrialisation recently profited from large private investments aiming at ultra-compact magnet systems for fusion (solenoids and toroidal field coils) with ∼20 T coil field. This has reduced the price in the past ten years by a factor three. The conductor does not require a heat treatment after winding. On the other hand, (i) unit lengths are still typically short ( 300 m) and longer lengths today affect the cost, (ii) REBCO is a highly anisotropic conductor (anisotropy factor of ∼5). The challenge is not only to make a fully transposed (Rutherford-like) cable from tape to wind a coil, but also to deal with mechanical properties: while very robust against tension and transverse compressive stress, delamination of the tape under tensile transverse stress causes degradation. Another challenge is to control hysteresis losses and field quality as the equivalent filament size is very large in the plane of the tape (typically 4-12 mm).

Development of iron-based superconductors [584, 585], is being strongly pursued in China. The critical current density achieved is still lagging a factor 2-3 behind the expectations set ten years ago, but based on raw-material prices the material has the potential of a very low cost; this could become a significant advantage for a high-field collider magnet, where about half of the price is the superconductor.

Given their potential, the HFM Programme is investing in REBCO tape technology development in the KIT/CERN Collaboration on Coated Conductor (KC4), as well as in iron-based superconductor technology (CNR-SPIN/CERN), aiming at a round wire with powder-in-tube layout [586].

***Fields achieved in EuCARD2 and US-MDP dipole models***

Between 2015 and 2020, three dipole magnets based on REBCO were built within the framework of FP7-Eucard2 programme [587]. In CEA, a dipole technology demonstrator based on double REBCO ribbons and three racetrack coils, without aperture, reached 5.4 T [583]. In the same period, two magnets were built with Roebel cable: in CEA a 40 mm aperture dipole was manufactured with a $\cos(\theta)$ configuration, 12 mm coil width, reached 1.16 T due to one degraded coil [588]. This magnet will be re-tested with a replacement coil. At CERN a dipole was built with the same cable, with an aligned block dipole configuration in a 40 mm aperture, reaching 3.35 T at 5 K and 4.3 T at 4.5 K (FeatherM2.12 and 2.34 [587, 589]).

More recently, the US programme, US-MDP, has developed several Bi-2212 racetrack coils and a CCT magnet (Bin5) which reached 1.6 T in a 31 mm aperture [590]. A CCT with a CORC® cable based on a REBCO tape reached 2.9 T in a 65 mm aperture [591], with a more recent test exceeding 5 T.

***Plans for 20 T: hybrid or full-HTS?***

The FP-7 Eucard2 programme aimed at an HTS insert for a $Nb_3Sn$ external coil [587]; the same strategy has been adopted by US-MDP and IHEP. US-MDP proposes to achieve 20 T via a magnet using different designs [575]: on the one hand an all-CCT hybrid, and on the other hand a COMB (Conductor on Molded Barrel) HTS part, and stress managed $\cos(\theta)$ for the inner $Nb_3Sn$ layer and a classical $\cos(\theta)$ for the outer layer. The IHEP common coil design is also based on an HTS inner coil and $Nb_3Sn$ outer coils [569], with HTS contributing 5 T to the field.

The hybrid option HTS/$Nb_3Sn$ does not allow the higher operational temperature that would be possible with a full-HTS magnet (e.g., 20 K). For the time being, $Nb_3Sn$ is still much better mastered than HTS, and therefore both US and China are working on the hybrid option, at least to make technology demonstrators. It must be noted, though, that the performance-related cost for HTS at 20 K is today two to three times larger than at 4.5 K, and since the conductor is the driving parameter of the estimated magnet cost, this can be a showstopper for the 20 K option. Improved performance at 20 K is a development objective for REBCO conductor.

***Challenges: field reproducibility, mechanics, hysteresis losses, protection***

A stringent requirement for accelerator main magnets is the *precision and reproducibility of the transfer function* not only at high field, but also at injection and during the ramp ($10^{-4}$ to $10^{-5}$ needed relative to main field). This aspect is strongly related also to the design of insulation and protection: some HTS coils that are being made for fusion operate in DC mode and make use of non-insulated coils, allowing current redistribution that eases reaching higher fields. This option is not suitable for accelerator magnets, which require a dielectric insulation. Metal insulations are also being studied, but these are not yet proven to be suitable for accelerator magnets in terms of field control, ramp losses, and protection.

Another issue is related to transverse (to the tape surface) tensile stress in the winding that causes delamination of the superconducting layer from the supporting substrate, degrading the cable performance. Careful design may control this effect.

The high temperature margins of HTS have the drawback that quench velocities are slow, and the voltage rise that is monitored to detect the quench can take too long. Consequently, quench protection and detection strategies need to significantly different from those of LTS magnets. Moreover, at 20 T the energy density to be absorbed by the coil is well above the enthalpy limit: detection of, and protection from, quenches in HTS coils are still major challenges, and coils that reached 20 T for fusion applications were lost during quenches. Systems that can provide a partial extraction of the stored energy for long magnets are being studied. Another possibility would be to avoid quenches, i.e., having a system that, once a temperature increase is detected, manages to prevent a resistive transition or at lest a thermal runaway.

Finally, stability can be achieved by very large filaments: HTS tapes used for fusion have a width of 4 to 12 mm. This width leads to a large magnetisation, giving rise to hysteresis losses, field quality distortions, and field drifts, that are expected to be larger than for LTS, with long field quality drifts expected in simple tape-stack cables. These effects should be included in the conception of a novel cable that could make REBCO compatible with HEP requirements.

In conclusion, it can be said that the technological readiness level (TRL) of HTS technology for accelerators lags considerably behind that of LTS conductors. Vigorous R&D is therefore needed, also in concert with other fields and applications, over the coming years to demonstrate how HEP can benefit from the potential advantages of HTS technology.

## 10.5 FCC-hh accelerator systems and technical infrastructures

### 10.5.1 Cryogenics requirements with 1.9 K $Nb_3Sn$ magnets

In the latest accelerator baseline, the tunnel length has been reduced to $90.7\,\mathrm{km}$ with eight access points, requiring an update of the layout of the cryogenic system, which is proposed and described in Fig. 10.36. The current cross-sectional view of the cryogenic distribution line (QRL), and the layout of the individual pipes (lines) can be found in Fig. 10.37. Both figures refer to an FCC-hh operation scenario with 14 T dipole magnets [592].

The layout of the cryogenic system seen in the figure consists of two accelerator cryoplants at points PB, PD, PF, PH, PJ, and PL. Each of these cryoplants has a cooling capacity of $53.1\,\mathrm{kW}$ at $4.5\,\mathrm{K_{eq}}$, including $9.9\,\mathrm{kW}$ at $1.9\,\mathrm{K}$.

Two cryoplants are foreseen in the high-luminosity insertion region at points PA and PG, each of them with a capacity of $78.9\,\mathrm{kW}$ at $4.5\,\mathrm{K_{eq}}$, including $17.6\,\mathrm{kW}$ at $1.9\,\mathrm{K}$. Detector cryoplants will be installed in PA, PD, PG and PJ, with a $1.5\,\mathrm{kW}$ at $4.5\,\mathrm{K_{eq}}$, similar to the existing helium cryoplant of the CMS experiment. The cryogenic system also includes twenty-four boil-off re-liquefiers.

Without considering detector cryoplants and the boil-off re-liquefiers, this represents a total cryogenic cooling capacity required of $950\,\mathrm{kW}$ at $4.5\,\mathrm{K_{eq}}$. It includes $100\,\mathrm{kW}$ at $4.5\,\mathrm{K_{eq}}$ for the cryogenics of the 2 high-luminosity insertion regions (inner triplets).

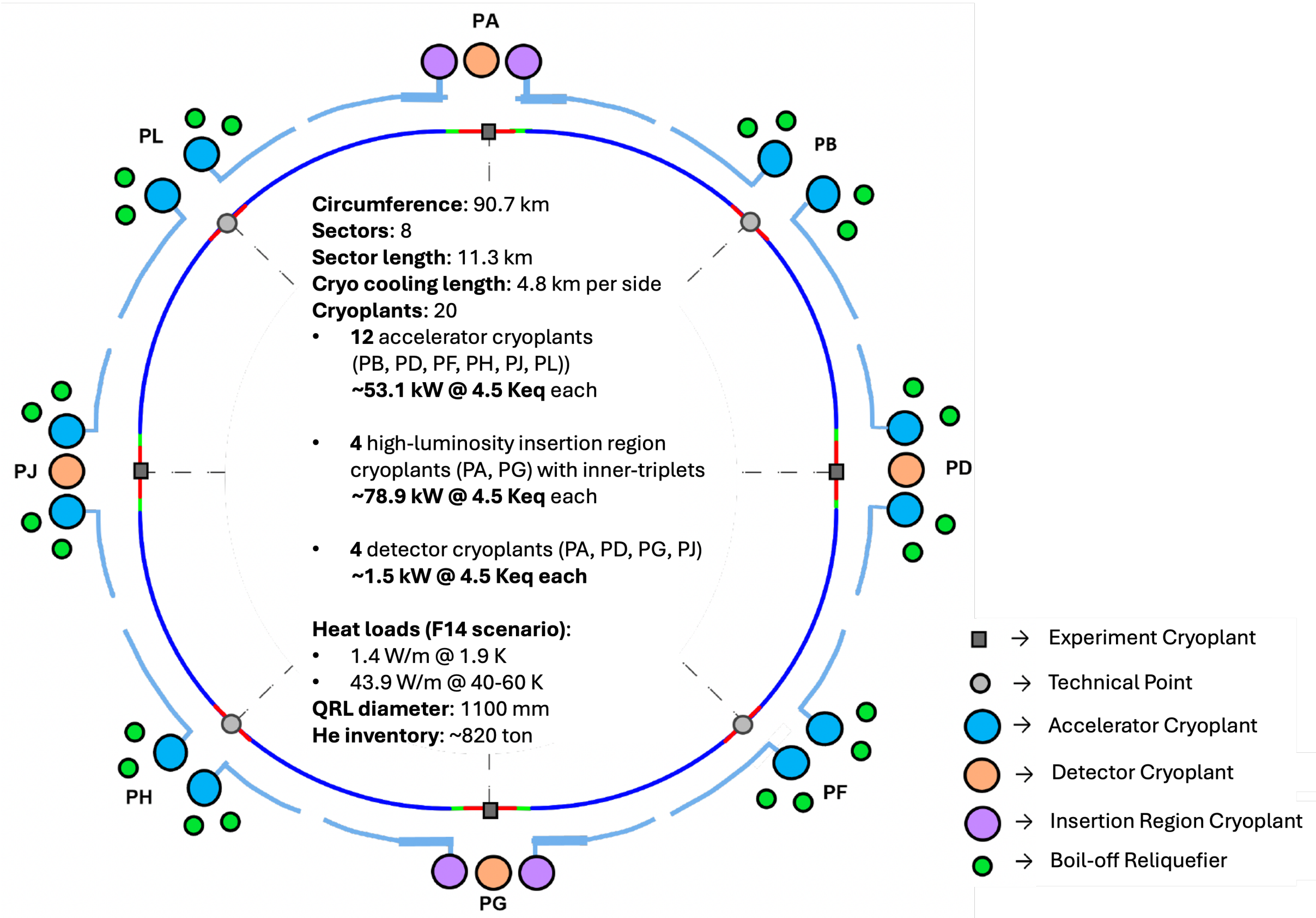

Fig. 10.36: Updated layout of the FCC-hh cryogenic system layout.

This translates into a total electrical power requirement of about 210 MW to operate the cryogenic systems. Neglecting the two high-luminosity insertion regions, the total electrical power required for the FCC-hh cryogenic system decreases to 186 MW.

The advantage of this layout is the reduced cryogenic sector length, decreasing to 4.8 km of cooling length per cryoplant compared to 8.4 km in the CDR. This reduction directly translates into a smaller pumping line. Consequently, the cryogenic distribution line is reduced from DN1350 to DN1100 type. The total helium inventory for the cryogenic system is estimated at 820 t.

The key parameters for the FCC-hh accelerator cryogenic system discussed above are summarised in Table 10.20 and neglect the detector needs and cryogenics associated with the high-luminosity insertion regions in PA and PG.

### 10.5.2 Power converters

A preliminary study has been conducted to assess power converter requirements, focusing on the necessary footprints and volumes in alcoves and surface areas.

The study specifically examines the powering of the superconducting magnets. To stabilise power consumption from the utility grid and minimise the sizing of upstream electrical components (e.g., transformers, cables), energy storage systems integrated with the power converters are required. These systems store energy during the accelerator's ramp-down phase and supply it during beam acceleration. Given their substantial space requirements, these storage elements must be placed in alcoves to minimise power losses from long cable runs.

Preliminary findings, based on FCC-hh magnet specifications, suggest that the space allocated for FCC-ee power converters (in terms of footprint and volume) is comparable to what is needed for the FCC-hh magnet powering including energy storage.

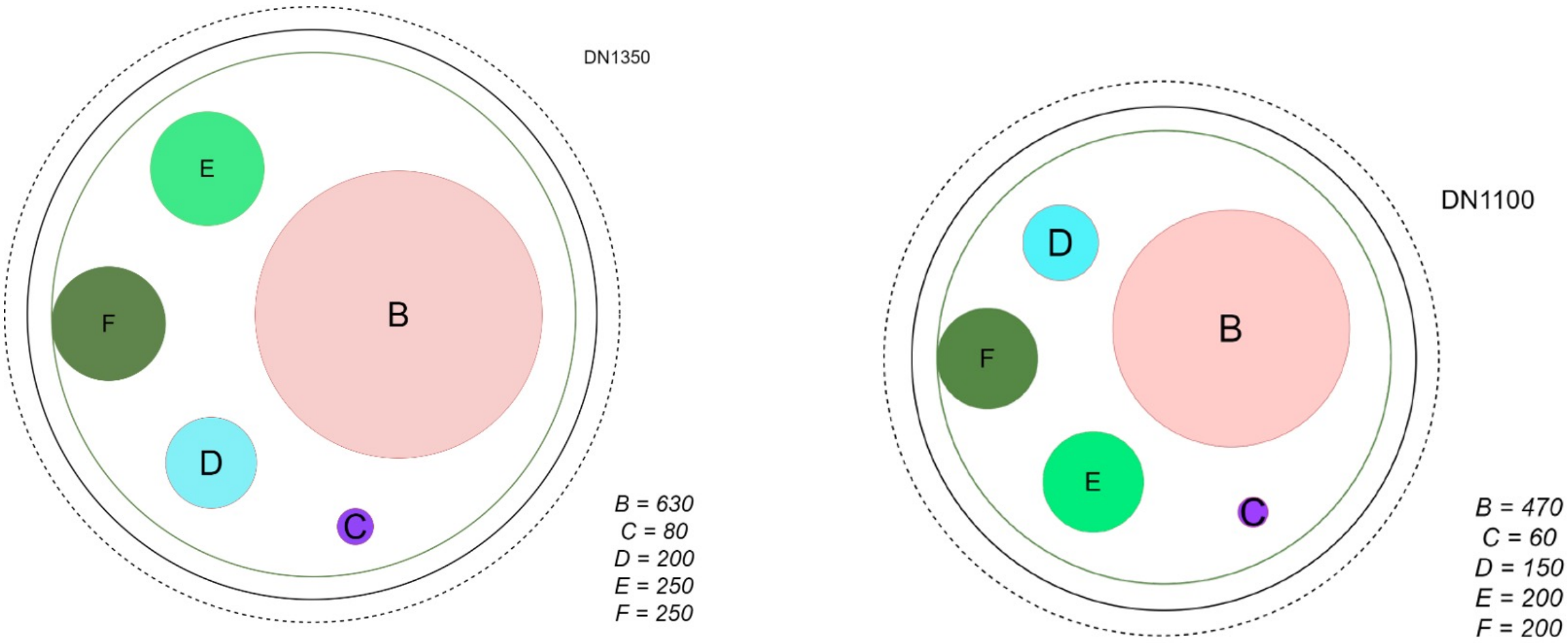

Fig. 10.37: Layout of the cross section of the FCC-hh QRL as of the CDR (left) and in the current configuration (right). The line B represents the very low pressure helium return line; the line C represents the supercritical helium supply line; the line D represents the low pressure helium return line; the line E represents the helium thermal shield supply line; the line F represents the helium thermal shield return line.

Table 10.20: Key parameters for the cryogenic system of the FCC-hh. A safety factor of 1.2 is included in the helium cryogenic capacity. These figures do not include the high-luminosity insertion regions with the triplet quadrupoles, or the requirements of the experiment detectors.

| **FCC-hh Parameters** | **CDR [2018]** | **FSR [2025]** |
|---|---|---|
| Circumference [km] | 97.75 | 90.7 |
| Dipole field [T] | 16 | 14 |
| Centre of mass energy [TeV] | 100 | 85 |
| Synchr. radiation for two beams [MW] | 4.8 | 2.4 |
| Magnet temperature [K] | 1.9 | 1.9 |
| Beam screen temperature [K] | [40 - 60] | [40 - 60] |
| Helium cryogenic capacity at 4.5 K equivalent [kW] | 1214 | 947 |
| Number of cryo islands | 6 | 8 |
| Total number of cryoplants | 10 | 16 |
| Arc cooling length [km] | 84 | 76.8 |
| Electrical consumption [MW] | 266 | 208 |
| Helium inventory [t] | 880 | 820 |

### 10.5.3 Technical systems

The technical infrastructures described in this section, designed to meet the needs of the FCC-ee, also already incorporate the requirements for FCC-hh.

#### *Electrical networks*

The electrical connections to the 400 kV lines of RTE (France's Transmission System Operator) have been designed to accommodate the power demands of both FCC-ee and FCC-hh, with the latter requiring about the same power as the FCC-ee in the $t\bar{t}$ mode. Three connection points are planned, each rated at $220\,\mathrm{MVA}$, and will be interconnected through an internal high-voltage network.

For FCC-ee, the estimated peak power demand for $t\bar{t}$ operation is approximately $360\,\mathrm{MW}$, which is well below the $660\,\mathrm{MVA}$ capacity of the grid connection. This setup provides redundancy in the power sources, ensuring reliable operation. However, the power demand is not evenly distributed between all points. Specifically, PH, which houses the RF systems, has a substantial power requirement of $200\,\mathrm{MW}$. To address this, a dedicated $400\,\mathrm{kV}$ substation will be allocated for the RF power supply. The two remaining connection points, PA and PD, will handle all other loads via the internal high-voltage network and can operate redundantly.

For FCC-hh, the primary power consumers will be the cryogenic systems, with their load more evenly distributed across all three connection points. The eight points of the FCC-hh ring will require a power rating between 40 and $60\,\mathrm{MW}$ each.

Unlike FCC-ee, all three grid connections will be essential for powering FCC-hh, leaving no redundancy in power sources. In the event of a power source failure, accelerator operation would not be possible. However, critical systems could continue running in a degraded mode to maintain minimum cryogenic conditions.

#### *Cooling and ventilation*

The redistribution of thermal loads from FCC-ee to FCC-hh will require the installation of some additional cooling towers at most points, except for PH, where a lower capacity will suffice. However, the overall thermal load in the tunnel will be reduced as the new loads will be located mainly in service caverns and surface buildings.

These adjustments will be limited to designated service areas, ensuring that modifications to cooling networks remain localised. In addition, the tunnel ventilation systems will remain unchanged, with only minor adaptations required for specific service and experiment caverns.

#### *Safety systems*

For FCC-hh, additional safety measures will be necessary. Unlike FCC-ee, where oxygen deficiency is not a concern in the arc tunnel, the presence of cryogenic systems in FCC-hh requires dedicated monitoring and protective measures. Consequently, oxygen deficiency hazard detection systems will be installed in the tunnel, alcoves, and service caverns to maintain a safe working environment.

### 10.5.4 3D integration studies

The 3D integration studies of the FCC-hh have been detailed for all areas of the underground tunnels and caverns, fitting the requirements from the work packages. This section presents the results of the configuration and layout of the accelerators and infrastructure systems that were studied for the feasibility study report. This is an evolution of the previous studies made for the conceptual design report. [13,307] The 3D integration studies take into account both of the configurations of the FCC-ee and FCC-hh to ensure space compatibility.

#### *Integration of point PA and point PG*

Point PA and point PG will serve as large experiment points, hosting detectors in the cavern in both the FCC-ee and the FCC-hh phases of operation. The cavern size and extent meet the needs of the detectors for both FCC-ee and FCC-hh. In addition, for the FCC-ee, the long straight sections (LSS) on either side of point PA and point PG will host the beamstrahlung dumps and the polarimeter system (see Fig. 10.38 and Fig. 10.39). The LSS of point PA will also host the end of the transfer line from the injector linac, into the booster.

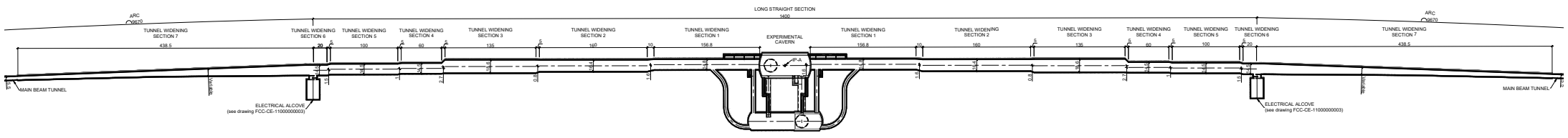

Fig. 10.38: FCC underground - civil engineering in point PA

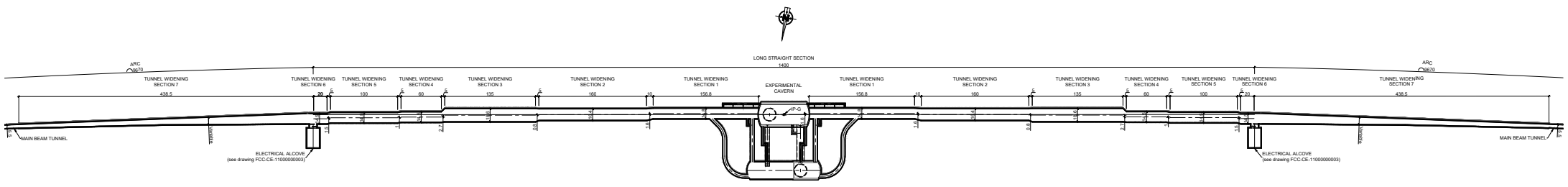

Fig. 10.39: FCC underground - civil engineering in point PG

The figures included in this chapter represent the results of the 3D integration studies for point PA and point PG.

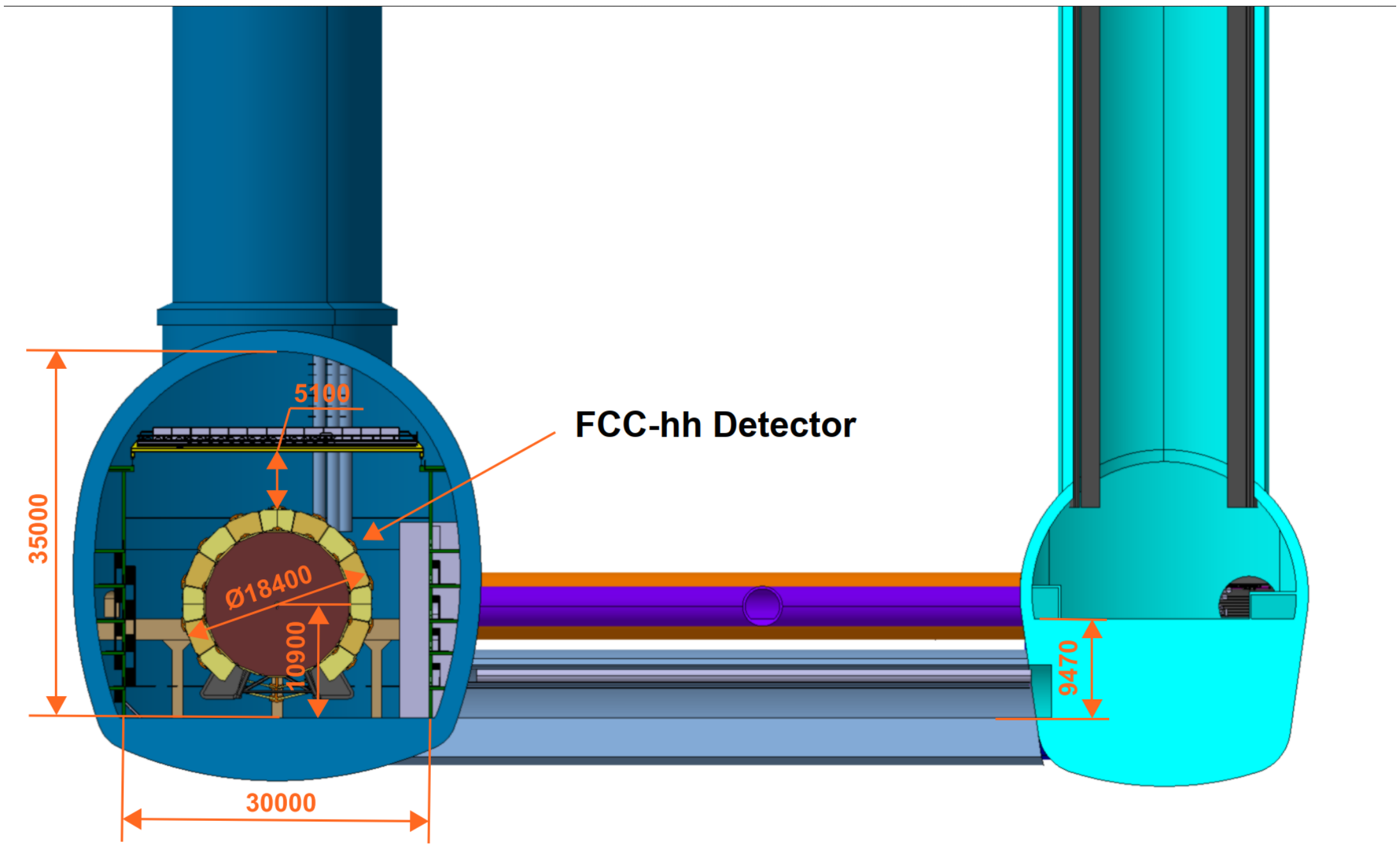

Fig. 10.40: FCC-hh point PA - experimental cavern cross-section

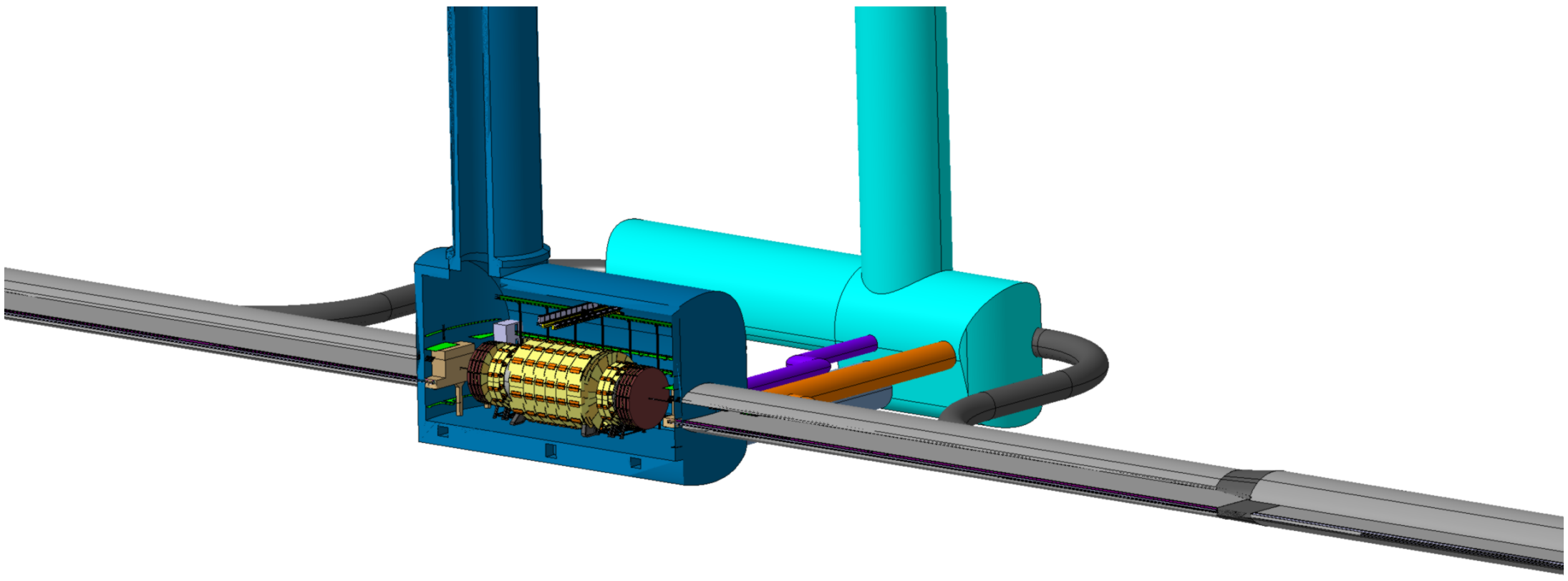

Fig. 10.41: FCC-hh point PA - experimental cavern iso view

***Integration of point PD and point PJ***

Point PD and point PJ will serve as small experiment points, hosting detectors in the cavern in both the FCC-ee and the FCC-hh phases of operation (see Fig. 10.42 and Fig. 10.43). The cavern size and extent meet the needs of the detectors for both FCC-ee and FCC-hh. In addition, for the FCC-ee, the long straight section (LSS) on either side of point PD and point PJ will host the beamstrahlung dumps.

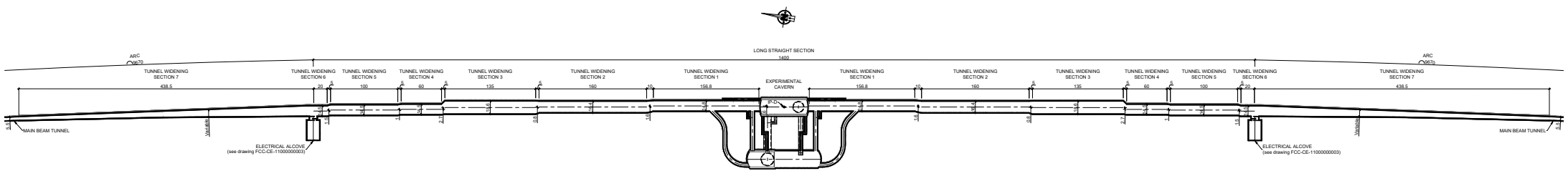

Fig. 10.42: FCC underground - civil engineering in point PD

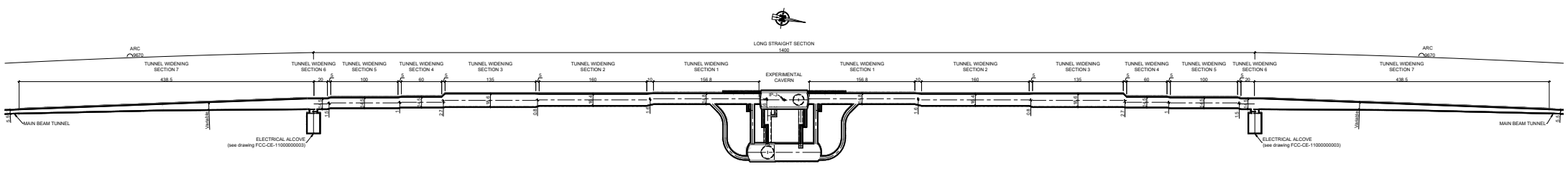

Fig. 10.43: FCC underground - civil engineering in point PJ

The figures included in this chapter represent the results of the 3D integration studies for the point D and the point J.

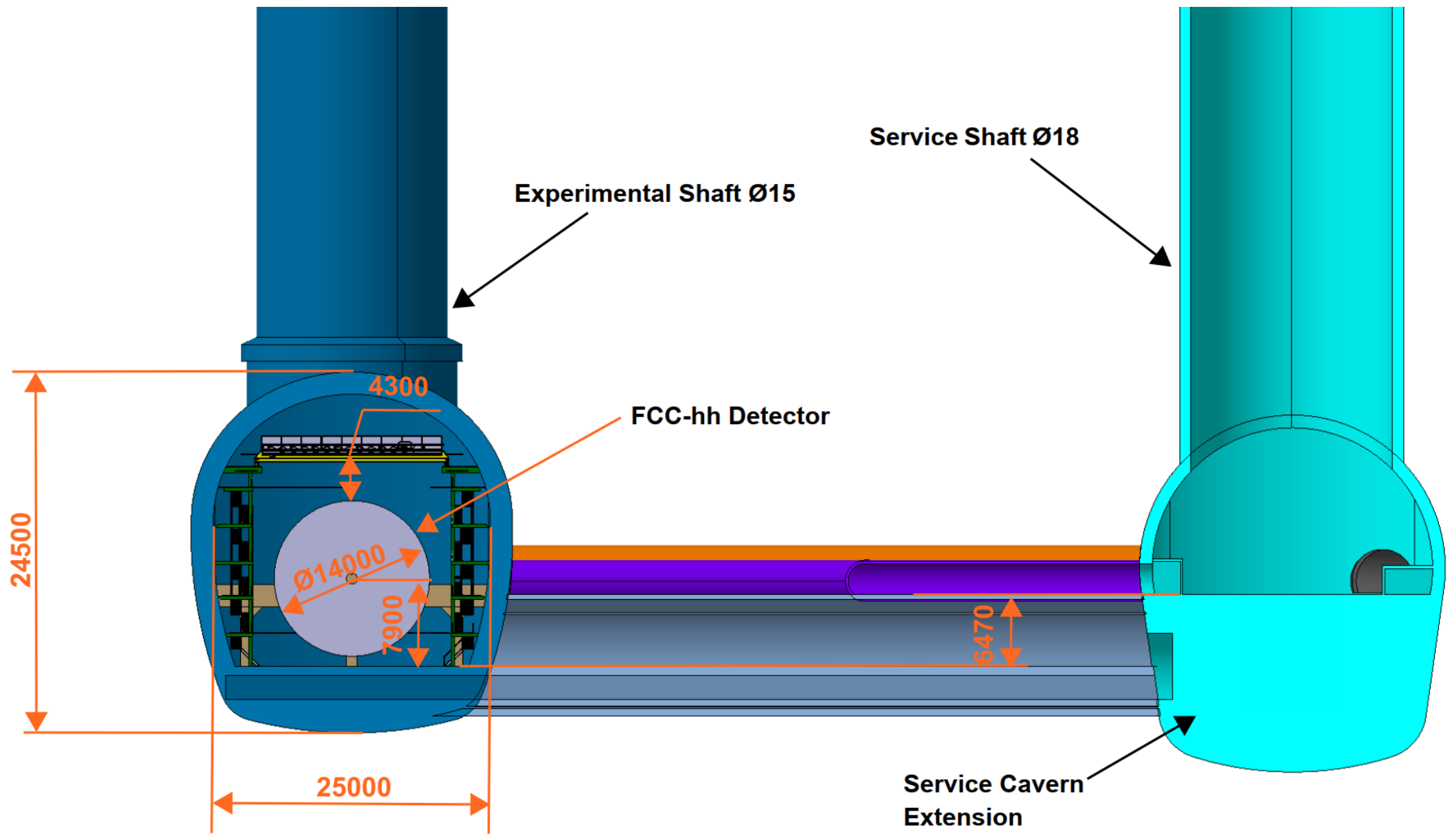

Fig. 10.44: FCC-hh point PD and PJ - experiment cavern cross-section

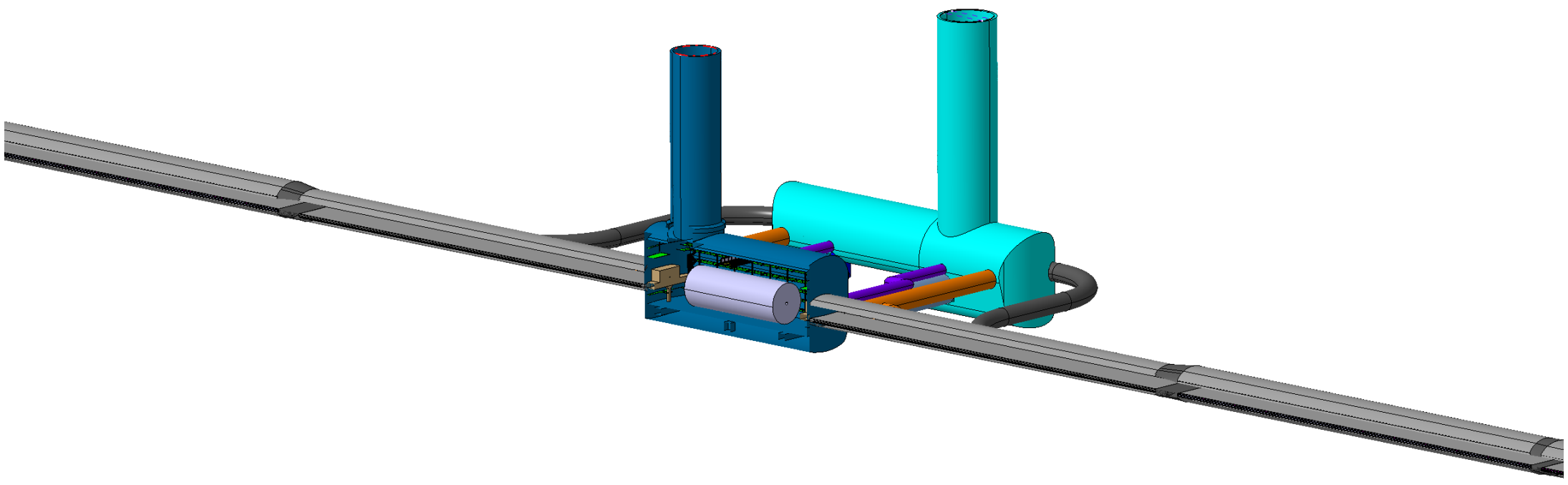

Fig. 10.45: FCC-hh point PD and PJ - experiment cavern iso view

### *Integration of the arcs*

The FCC-hh superconducting magnets will have a maximum diameter of $1.2\,\mathrm{m}$, regardless of the technology used ($Nb_3Sn$, HTS, etc.), resulting in the tunnel cross section shown in Fig. 10.46. For the FCC-ee ring, the transport area is $2.2\,\mathrm{m}$ wide, allowing personal vehicles to pass each other freely throughout the tunnel. In contrast, for FCC-hh, this area may be reduced to $2\,\mathrm{m}$, restricting vehicle crossings to designated alcove lay-bys spaced out every $1.6\,\mathrm{km}$. Consequently, traffic control will require a different approach compared to FCC-ee.
The figures included in this chapter represent the results of the 3D integration studies in the arcs.

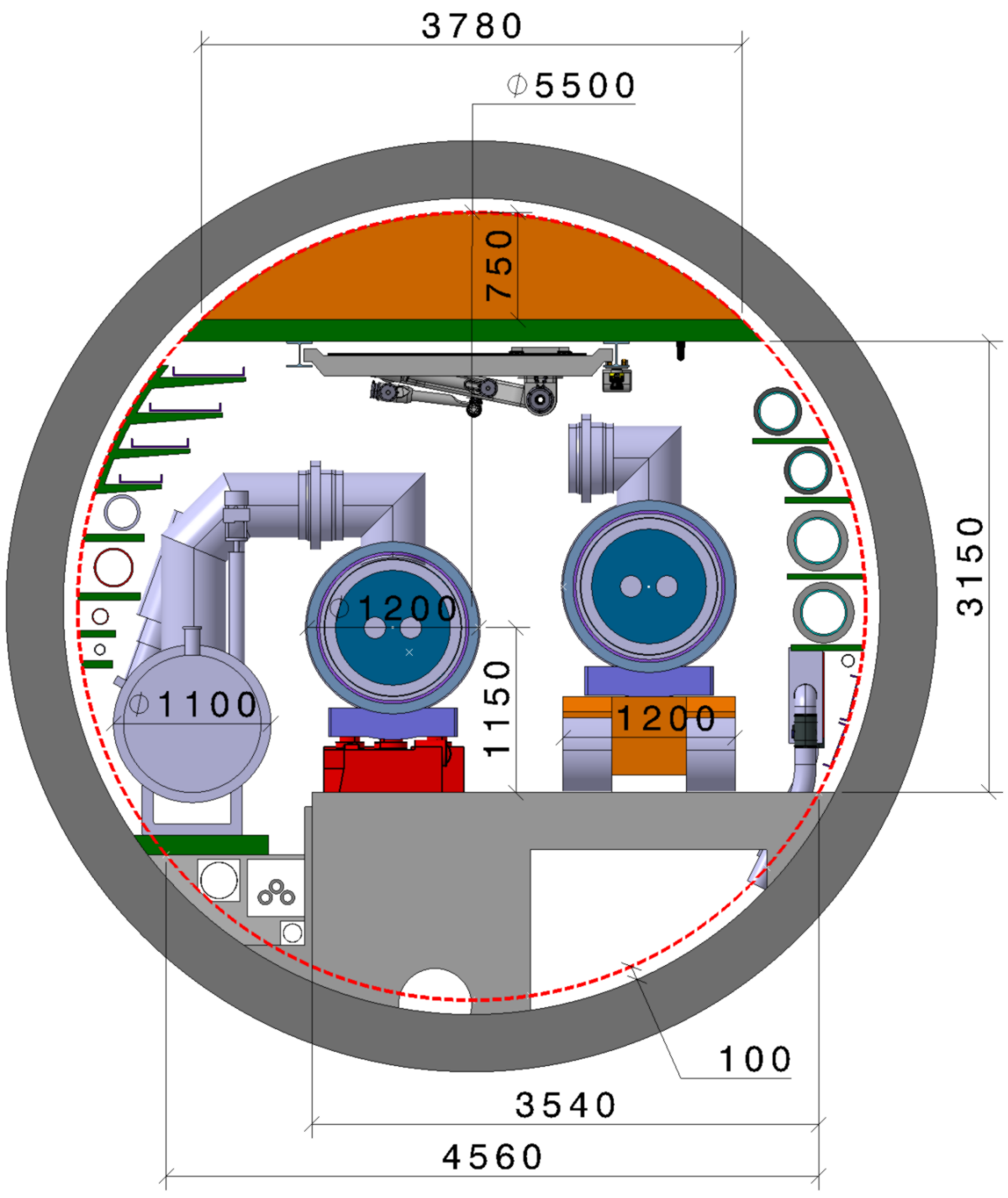

Fig. 10.46: Cross section of the tunnel for the FCC-hh layout in the arcs.

### *Next phase for the integration studies*

The integration studies have been made with pre-design of volumes and will evolve with detailed technical design in the next phase of the FCC studies. Optimisation of the 3D integration studies with respect to the requirements from work packages will continue for all the areas following the identified technical and space needs:

– Optimisation of the large experiment points PA and PG [375] [376] and small experiment points PD and PJ [377], integrating more detailed detectors design in the experiment caverns,
– Optimisation of the arc half-cell [382].

### 10.5.5 FCC-hh power demand

In FCC-ee, the primary power-consuming systems are the RF systems, which compensate for synchrotron radiation losses. These systems must supply approximately 100 MW to the beams across different operating modes. In Z mode, RF power accounts for 65% of the total power demand, while in $t\bar{t}$ mode, it represents 37.5%. These values reflect the energy requirements necessary to sustain beam stability and luminosity performance.

For FCC-hh, the dominant power consumer will be the cryogenic systems required for magnet cooling. The cryogenic capacity is determined primarily by two factors: the static heat load of the magnets and the synchrotron radiation deposited on the beamscreen. The cryogenic power demand varies depending on the operating temperature of the magnets and the beamscreen. Table 10.21 presents the power demand for the baseline scenario where 14 T $Nb_3Sn$ magnets are cooled to 1.9 K, with a total synchrotron radiation power of 2.4 MW and a beamscreen temperature ranging in the interval 40 – 60 K.

The cryogenic systems for the arc magnets alone require 185 MW of electrical power, with an additional 22.5 MW needed for the high-luminosity insertions inner-triplet magnets (two sets for two experiments). The RF power requirement is estimated at 23 MW during the energy ramp and decreases to 12 MW at flat top, during beam collisions.

Table 10.21: FCC-hh power demand by technical systems at 85 TeV beam centre-of-mass energy.

| **System** | **85 TeV** |
|---|---|
| Radio frequency [MW] | 17 |
| Cryogenics [MW] | 207.5 |
| Cooling & ventilation [MW] | 40 |
| Magnet powering [MW] | 33 |
| Experiments [MW] | 24 |
| Data centre [MW] | 8 |
| General services [MW] | 26 |
| **Total power [MW]** | **355** |

Magnet powering will be managed from alcoves with interconnection boxes, and power losses will be minimised through short cable sections, reducing demand to 33 MW. The estimated power needs for experiments are based on the LHC experiments (ATLAS and CMS), while general services power requirements have been extrapolated from LHC operations. Although some figures require further refinement, these estimates align with expected scaling from the LHC.

For a configuration using $Nb_3Sn$ magnets cooled to 1.9 K and generating 2.4 MW of synchrotron radiation, the total power demand is approximately 360 MW, comparable to that of FCC-ee. The FCC-hh power demand might further be reduced to below 300 MW, if the magnet temperature can be raised to 4.5 K.

#### *Operational model*

The accelerator operational model determines the energy consumption. The power demand varies depending on the time of year and the operational state of the machine. The machine schedule defines six operational periods: (1) Shutdown; (2) Commissioning; (3) Physics operation; (4) Short downtime (without machine access); (5) Technical stops (longer downtimes); and (6) Machine development. The power demand for each period is shown in Table 10.22.

Table 10.22: FCC-hh power demand during different operational periods.

| **Period** | **85 TeV** |
|---|---|
| Shutdown [MW] | 122 |
| Technical stop [MW] | 122 |
| Downtime [MW] | 122 |
| Commissioning [MW] | 324 |
| Machine development [MW] | 324 |
| Beam operation [MW] | 355 |

The introduction of a *eco-mode* for the cryogenic systems during shutdown, reducing power consumption to 83 MW, significantly decreases the power demand during non-operational periods. The annual schedule consists of:

- 120 days of shutdown,
- 30 days of commissioning,
- 20 days of machine development,
- 10 days of technical stops, and

– 185 days of physics (beam operation).

The FCC-hh machine availability is expected to be similar to that of LHC, with an operational efficiency of approximately 80%, corresponding to an expected period of beam collisions of 1700 hours.

***Energy consumption***

Based on the operational schedule and the expected power demand throughout the FCC-hh programme, the estimated annual energy consumption at 85 TeV is shown in Table 10.23.

Table 10.23: FCC-hh annual energy consumption at 85 TeV collision energy.

| **Beam Energy** | **42.3 TeV** |
|---|---|
| Annual energy consumption [TWh/y] | 2.34 |

For FCC-hh, total annual energy consumption is slightly higher compared to the energy consumption of FCC-ee because of the continuous operation of cryogenic systems. Furthermore, power demand remains relatively high even during shutdown periods. However, introducing an *eco-mode* has significantly reduced cryogenic power consumption during non-operational phases.

### 10.5.6 Other systems

Other accelerator systems for the FCC-hh, e.g., beam-transfer systems, the cryogenic beam vacuum system, beam diagnostics, etc., are described in the Conceptual Design Report (CDR) [593].

### 10.5.7 On-going studies

A key feature highlighted during the CDR phase is the potential application of an HTS coating to the beam screen. This approach was explored as a mitigation measure to address the increased resistivity of the beam screen due to its higher operating temperature (around $50\,\mathrm{K}$), which was chosen to reduce the demands of cooling power.

Following the publication of the FCC CDR, extensive research has focused on conducting comprehensive analyses of the coating's performance, examining its electromagnetic properties and other related factors.

A prototype beam screen incorporating the suggested HTS coating is currently under construction. This will facilitate a comprehensive examination of its magnetic characteristics, such as the overall field quality of both the beam screen and the HTS coating. The tests will include dipole and quadrupole external magnetic fields. Impedance measurements will also be attempted, if feasible, to fully characterise the HTS coating.

The application of an amorphous carbon coating to the HTS presents an opportunity to elevate the beam screen's operating temperature, as it inherently resolves vacuum instability issues. The optimal operating temperature will be determined by balancing multiple factors, including HTS performance, beamscreen resistivity, effects on beam dynamics, and benefits for the cryogenic system. These aspects will be further studied in the next project phase to identify the optimum operating temperature.

Finally, FCC-hh will require the installation of the cryogenic distribution line (QRL) that has a diameter below $1.1\,\mathrm{m}$ regardless of the operational temperature of the magnets ($1.9\,\mathrm{K}$, or $4.5\,\mathrm{K}$).

# Appendix A

# Costs

## A.1 FCC-ee construction

The capital cost for construction of the FCC-ee is summarised in Table A.1. This cost includes construction of the entire new infrastructure and all equipment for operation at the Z, WW and ZH working points. Operation of the FCC collider at the $t\bar{t}$ working point will require later installation of additional RF cavities and associated cryogenic cooling infrastructure with a corresponding total cost of 1,260 MCHF.

Table A.1: Estimated investment costs for different construction domains in 2024 Swiss Francs.

| **Domain** | **Cost [MCHF]** |
|---|---|
| Civil engineering | 6160 |
| Technical infrastructures | 2840 |
| Injectors and transfer lines | 590 |
| Booster and collider | 4140 |
| CERN contribution to four experiments | 290 |
| **FCC-ee total** | **14 020** |
| + Four experiments (non-CERN part) | 1300 |
| **FCC-ee total, including four experiments** | **15 320** |

The costs indicated include materials and personnel costs of contractors and suppliers. They do not include the cost of scientific, engineering, technical and administrative personnel at the host organisation or at the collaborating institutes.

The total construction cost for CERN amounts to 14 020 MCHF (covering Z, WW and ZH working points), dominated by 45% or 6160 MCHF by civil engineering, which also comprises all site-related investments. The capital cost for the technical infrastructures, including site-external connections, is 2840 MCHF, corresponding to 20% of the total construction cost. Another 30%, or 4,140 MCHF, corresponds to the collider and full-energy booster construction. The cost estimate for the injector complex and transfer lines is 590 MCHF. In addition, the CERN contribution to the four experiments is estimated at 290 MCHF, while the non-CERN part of the total cost of these four experiments would amount to approximately 1300 MCHF, as shown in Table A.1. The investments are distributed over a time frame of about 15 years.

The indicated figures include materials and personnel costs of contractors and suppliers. They do not include the cost of scientific, engineering, technical and administrative personnel at the host organisation or the collaborating institutes. The engagement of this personnel directly leads to the generation of noteworthy socio-economic benefits via the pathways of education, training, direct, indirect and induced economic value added.

## A.2 FCC-ee operation costs

### A.2.1 Annual cost of operations of each stage

A material cost of 200 MCHF has been estimated by individually categorising each item of equipment according to its nature: fixed installation (1% OPEX), technical infrastructure (5% OPEX), and con-sumables (3% OPEX). The 200 MCHF figure represents the weighted sum of these categories across all subsystems. Adding an annual electricity cost of 100 MCHF (with some variation depending on the collision energy point) and approximately 300 MCHF for personnel (1,400 FTEs), the total average annual operating cost of FCC-ee across all stages amounts to around 600 MCHF.

## A.3 FCC-hh Construction and Operational Costs

### A.3.1 FCC-hh construction

Table A.2 splits the construction costs into the creation of the infrastructures (subsurface, surface and territorial developments), the enabling technical infrastructures, and the accelerators.

The total construction cost amounts to 18 880 MCHF. It is dominated by the cost for the collider magnets, amounting to about 10 000 MCHF. This cost, however, will need to be reviewed in light of ongoing R&D efforts. In particular, further research into novel types of superconducting materials is essential—not only to potentially reduce costs, but also to unlock broader technological benefits and cross-sectoral applications.

The cost of the four FCC-hh experiments can be estimated only at a later stage, once more detailed designs of the detectors exist. The construction investments are distributed over a time frame of about 15 years.

Table A.2: Estimated investment costs for different construction domains in 2024 Swiss Francs.

| **Domain** | **Cost (MCHF)** |
|---|---:|
| Civil engineering | 520 |
| Technical infrastructures | 3960 |
| Injectors and transfer lines | 1000 |
| Collider | 13 400 |
| **FCC-hh total** | **18 880** |

The costs indicated include materials and personnel costs of contractors and suppliers. They do not include the cost of scientific, engineering, technical and administrative personnel at the host organisation or at the collaborating institutes.

### A.3.2 FCC-hh operation

Operational costs for the FCC-hh phase foreseen as a second step in 2070 have not yet been determined.

# Appendix B

# Installation

## B.1 Installation planning

### B.1.1 The technically-limited timeline for FCC-ee construction

Preparatory placement studies for the FCC are well advanced. The further timeline is mainly determined by the preparation of the civil engineering (subsurface investigations, civil engineering design and tendering, and the project authorisation processes with the host states, planned to advance in parallel), and, afterwards, by the civil construction and the subsequent installation of technical infrastructure and accelerator components. The timeline of FCC-ee in Table B.1, including work already accomplished. More details on the construction and installation aspects are discussed below.

Table B.1: Timeline for FCC-ee design, construction, and operation.

| Milestone / phase | years |
|---|---|
| Conceptual Design Study | 2014 – 2018 |
| Territorial implementation studies | 2016 – 2025 |
| Definition of the placement scenario | 2022 |
| Feasibility Report ready | 2025 |
| Earliest Project Approval by CERN Council | 2027/28 |
| Environmental evaluation & project authorisation processes | 2026 – 2031 |
| Main technologies R&D completion | 2031 |
| Technical Design Report ready | 2032 |
| Civil engineering - Collider | 2033 – 2041 |
| TI Installation – Collider | 2039 – 2043 |
| Accelerator Installation – Collider | 2041 – 2045 |
| HW Commissioning – Collider | 2042 – mid 2046 |
| Start Beam Operation – Collider | mid 2046 – 2047 |
| Nominal Beam Operation – Collider | 2048 – 2062 |

### B.1.2 FCC-ee technical infrastructure and machine

This section outlines a high-level schedule for installing the FCC-ee machine and its supporting infrastructure. Based on the feasibility study, it adopts a 90.7 km circumference layout with eight surface sites. The timeline extends from the start of civil engineering excavation scheduled for January 2033 to the anticipated readiness for beam commissioning in July 2046. The plan is divided into two main phases:

- Civil construction phase 2033-2041
- Installation phase 2038-2046

***Civil engineering strategy integration***

*Synergies found between installation planning and civil engineering phase and updated civil engineering strategy*

The baseline presented in the midterm review outlined eight handovers of individual shaft sectors. Initially, all installation work within an arc could only commence after the release of a single designated point (see Fig. B.1).

To enhance coordination, synergies between civil engineering and installation planning were explored. As a result, the civil engineering team restructured the handover process, breaking down the singular shaft-sector handover into multiple phased handovers. The revised civil engineering strategy (see Fig. B.2) involves removing the TBMs through the experiment shaft (PX), which allows an earlier release of the machine shaft (PM) at a technical point.

Additionally, a gradual handover approach for arc installation was introduced. The tunnel floor and smoke extraction duct will be the final construction activity completed by civil engineering along the tunnel arc. To coordinate with the machine installation, the tunnel arc is split into sectors corresponding to the alcove locations. Therefore, once civil engineering has completed construction activities up to an alcove, the sector can be released for the installation of machine infrastructure.

For each shaft sector, the following elements now have distinct release dates:

1. PM of the experiment point
2. PM of the technical point
3. Staged release of the arc done alcove by alcove. 5 to 7 staged handovers are foreseen for one arc sector going from the technical point to the experiment point. A separating door will be moved at every handover to separate the two worksites
4. PX of the experiment point

The civil engineering construction durations across each of the sites vary primarily due to the different structural layouts, depths and construction methods.

*Impact on installation organisation and planning*

Breaking down the handover process of civil engineering enables an earlier start for installation activities. The early handover of the PM technical points allows for the installation of essential shaft components, including lifts, cranes, and technical infrastructure. Each staged release of the arc facilitates the progressive installation of technical infrastructure in the tunnel, shifting part of the work into the shadow of the critical path.

The staged release of a sector between two access points means that, on the technical point side, technical infrastructure installation will be underway, while civil engineering work will still be ongoing on the opposite side of the separating doors. Safety regulations require two evacuation routes in case of an emergency, which is ensured by the separating door between the two worksites. This door is exclusively for emergency evacuation. In such an event, personnel from the civil engineering side will be responsible for assisting in the evacuation of those from the installation side.

Since the overall arc and final ventilation equipment will not yet be in place during the staged release, the installation of technical infrastructure will be carried out using a temporary ventilation system.

#### ***Handover dates***

The handover dates were provided by the civil engineering team based on the TBMs advancing at a rate of 16 m/day, which is 10% less than the previous civil engineering baseline.

#### ***Installation phase***

*Installation phase for a sector*

The installation planning was developed through a bottom-up approach, beginning with a detailed analysis of the sequence and duration required for each sector. The installation process for each of the eight sectors will be carried out in four distinct phases. First, the installation of technical infrastructure and lifting equipment inside the shaft. Second, the installation of the technical infrastructures in the tunnel. Third, the installation of the accelerator itself. Fourth, the commissioning period of the technical infrastructures and the accelerators (see Fig. B.3).

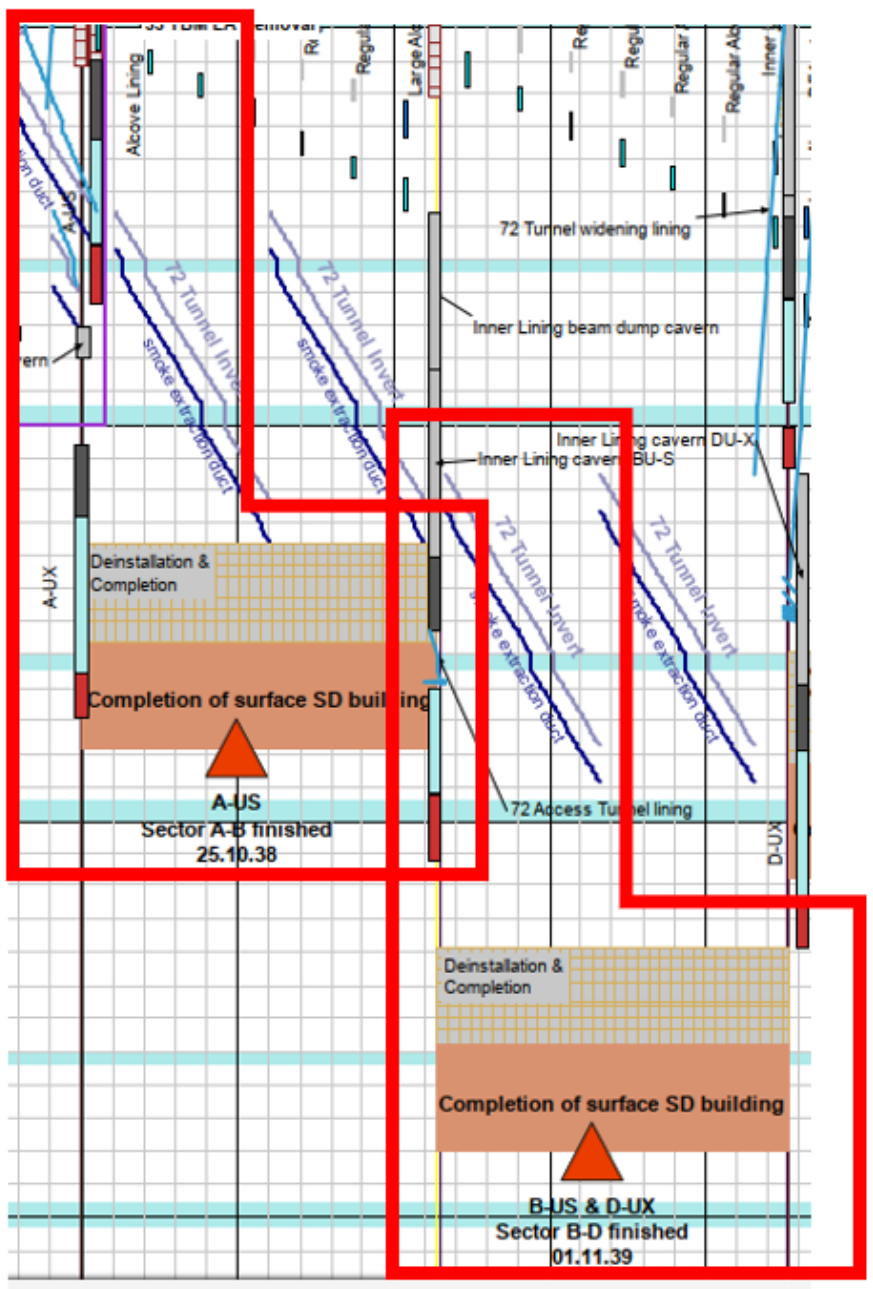

Fig. B.1: Civil engineering previous baseline: one shaft-sector release.

The technical infrastructure must be installed before the accelerator to ensure unobstructed access to the walls and floors and to prevent potential damage from heavy construction activities. Furthermore, most of this infrastructure will remain in place throughout the entire lifespan of the FCC, from the electron collider to the hadron machine. While the installation of technical infrastructure does not require immediate ease of access, all maintenance constraints must be carefully considered.

The installation starts with the shaft which has to be equipped with all services (lifts, crane, ventilation, electricity, pipes . . . ). This phase is quite slow due to the limited number of workers (expected to be around 30) who can work inside the shaft, because they all depend on the shaft crane. The shaft installation is expected to last for close to 12 months for a 250 m deep shaft. The duration of shaft installation works depends on the shaft depth.

During the installation phase, certain technical infrastructure elements, such as lighting and power supply, will be commissioned progressively. However, a dedicated six-month period will be required at the end of the installation phase to conduct comprehensive testing and final commissioning of all systems.

Once the technical infrastructure of an arc is fully completed, a thorough cleaning will be carried out before the installation of accelerator components begins. The magnet installation process is organized into three shifts, with the shaft crane serving as the primary bottleneck, as each magnet requires 90 minutes to be lowered. To complete magnet installation within a year, at least two trailers will be necessary.

All cabling will be pre-installed, allowing for system connections to take place after the installation of magnets and vacuum systems. Fire safety doors will be installed at the final stage once all magnets and cavities at points PH and PL are in place. As a result, the final safety systems will only become operational after the full installation of the accelerator. Until then, only temporary safety systems will be in place during the installation phase.

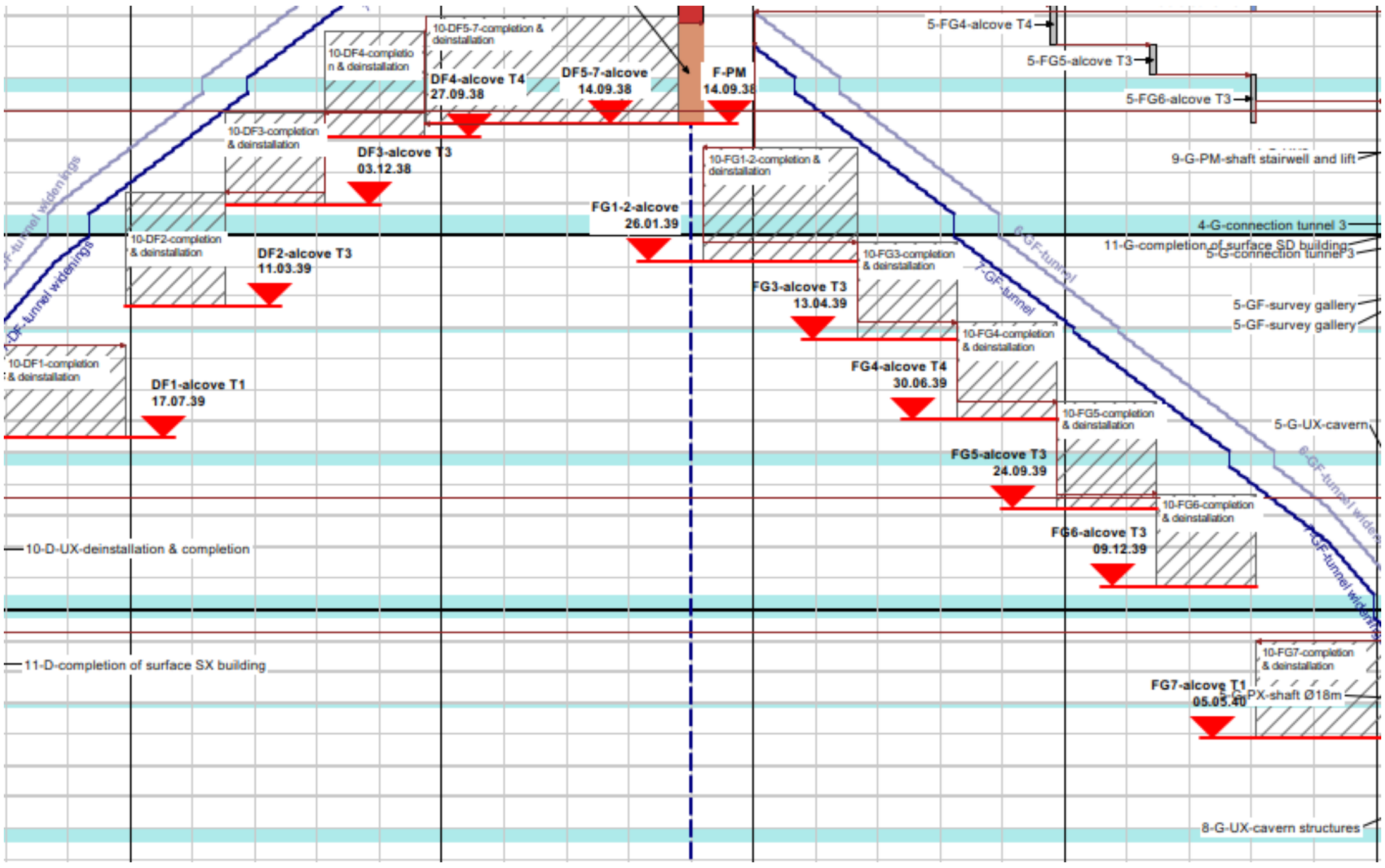

Fig. B.2: Civil engineering current baseline: the staged handover of the arc.

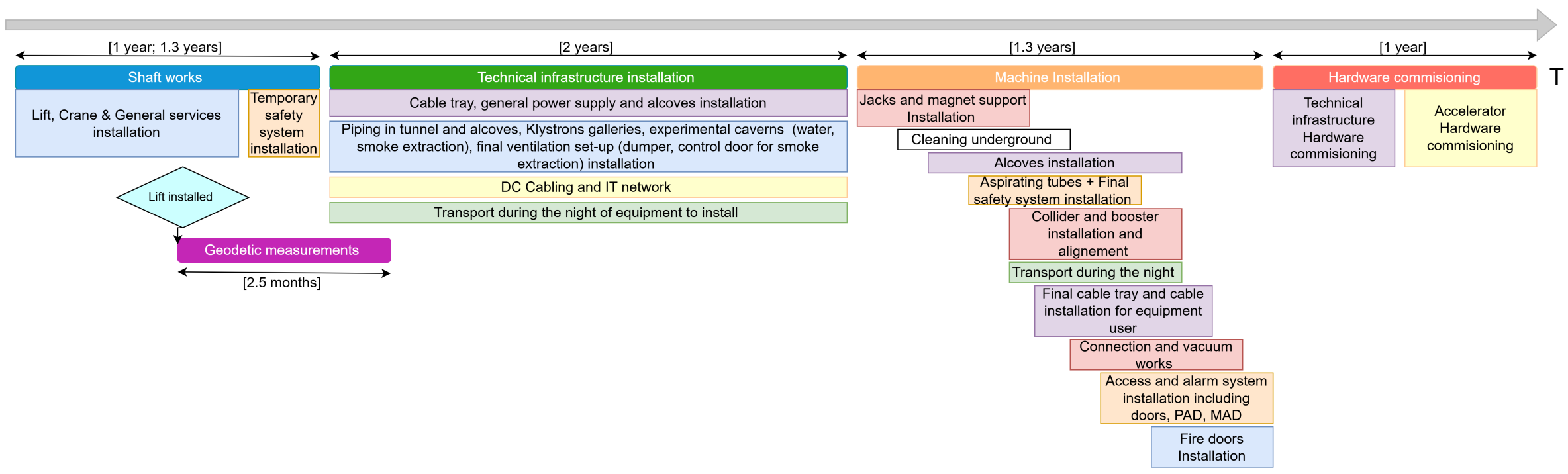

Fig. B.3: Sector sequence for installation.

Once all accelerator systems are installed, the commissioning phase will begin, starting with the technical infrastructure. The safety systems will be the first to undergo commissioning, with this phase expected to take approximately six months. Following this, the accelerator commissioning will commence, which is also estimated to last around six months, after which the system will be ready for the first beam tests.

*Experiment point installation*

Experiment shaft installation mainly comprises the installation of the ventilation duct. The installation is expected to be completed within six months. The installation of the experiments is scheduled to take place after the overall accelerator hardware commissioning. A minimum of two years has been allocated for installing each experiment within the cavern. The total available time for the installation of each experiment ranges between four and five years. Table B.3 provides an overview of the start dates and duration of the experiment installations.

Table B.2: Handover dates from civil engineering for the machine shaft (PM), experiment shaft (PX) and first alcove.

| **Location** | **Handover to CERN (2025 update)** |
|---|---|
| PA PX | Mar - 41 |
| PA PM | Feb - 39 |
| A - B Alcove | Jan - 39 |
| PB PM | Nov - 38 |
| B - D Alcove | Nov - 38 |
| PD PX | May - 40 |
| PD PM | April - 38 |
| D - F Alcove | Sep - 38 |
| PF PM | Sep - 38 |
| F - G Alcove | Jan - 39 |
| PG PX | Apr - 41 |
| PG PM | Feb - 39 |
| G - H Alcove | May - 40 |
| PH PM | Dec - 39 |
| H - J Alcove | Sep - 39 |
| PJ PX | Sep - 40 |
| PJ PM | Jun - 38 |
| J - L Alcove | Jun - 39 |
| PL PM | Jun - 39 |
| L - A Alcove | Jun - 39 |

Table B.3: Experiment installation start date and duration.

| **Point** | **Start date** | **Duration in years** |
|---|---|---|
| PA | 11/09/2041 | 4.4 |
| PD | 11/11/2040 | 5.2 |
| PG | 12/10/2041 | 4.2 |
| PJ | 14/03/2041 | 4.8 |

*Installation at points with radiofrequency systems*

Radiofrequency equipment will be installed in point PH and in point PL. The installation planning focuses on two main areas: the klystron galleries and the long straight section (LSS). Cavities and klystrons will be installed to support all stages of FCC-ee operation, ensuring readiness for the full accelerator lifecycle.

In the klystron gallery, the technical infrastructure will be installed first to ensure the availability of general services in this area. The rate of installation is assumed to be the same as in the tunnel. Following this, the cores will be equipped with waveguides and cable trays using the space from the centre of the galleries to the edges. Two months after the start of core installation, klystron and the low-level RF and control installation will commence.

In the LSS, once the technical infrastructure has been installed, the cryogenic distribution line (QRL) installation can start. Since the cavities will be delivered from point PJ and that the sector H-J is being installed from PH to PJ with magnets, the cavity installation will therefore only start when the transport of magnets has been completed for space reasons. Once the cavities have been placed, the connection under clean room conditions between the cavities will take place and two weeks later, other general connection work for the waveguides and controls will be done. Then, cryogenic hardware commissioning will be carried out. After the sector hardware commissioning of the technical infrastructure is completed, a one-year period is planned for cavity conditioning. The sequence is shown in Fig. B.4. The duration depends on the number of systems to be installed at each point.

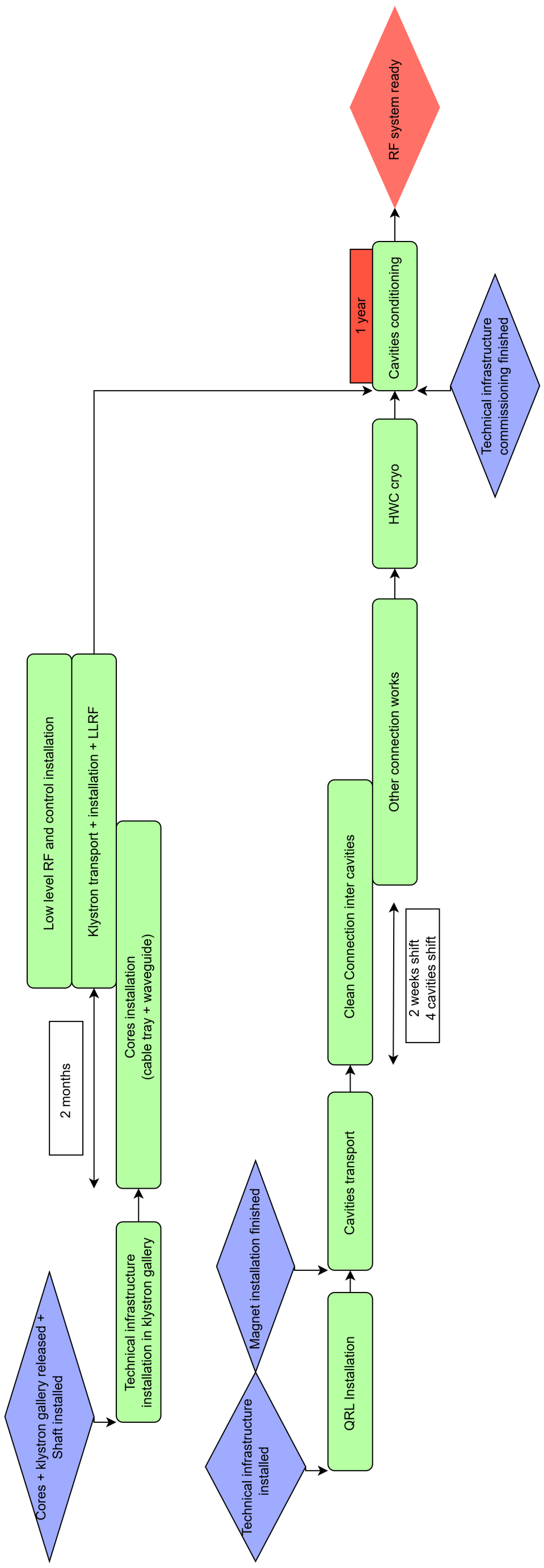

Fig. B.4: Radiofrequency equipment installation sequence.

*Strategy for global installation schedule phase*

The installation strategy involves deploying a maximum of four teams working in parallel on the same type of task, such as shaft works, technical infrastructure, and machine installation. As a result, the start of certain activities may be delayed until the necessary resources become available.

In the previous baseline, the resource usage was organised on a first ready, first installed basis. Including the radiofrequency sequence into the overall schedule has an impact on the critical path and to optimise the latter, resources were allocated in priority to points PH, PJ and PL which will host radiofrequency installation or its transport.

For most sectors benefiting from the staged release for the installation of technical infrastructure, the installation will proceed in the same direction as the civil engineering handover, moving from the technical point to the experiment point. The machine installation follows a sequential 'train' approach to streamline transport logistics, adhering to the same direction—from the technical point to the experiment point.

*Overall planning installation*

Figure B.5 shows overall planning for FCC-ee with the location in the machine indicated across the top and the timeline on each side. Main blocks of activities shown are:

- Civil engineering works.
- Shaft and sector civil engineering release date.
- Shaft works, technical infrastructure installation, machine installation and hardware commissioning depicted with a direction line.

The critical path is driven by two sequences:

- The resource levelling from sectors J-L and L-A to sectors D-F and F-G
- The handover date of PH PM and the RF installation at point PH

The current schedule gives the overall machine readiness for beam in **July 2046**. The durations of all activities are provided as calendar days.

### B.1.3 Dismantling FCC-ee

***Introduction***

This section presents a high-level schedule for the period covering the dismantling of FCC-ee to the installation of the FCC-hh machine and its related infrastructure. The time period covered is from the stop of the beam at the end of 2062 to the readiness for beam commissioning in 2073. The key phases are the following:

- Stop of FCC-ee beam at the end of 2062
- Dismantling activities end 2063 – end 2065
- Installation phase 2066 - 2073

***Dismantling FCC-ee***

Since the FCC-ee and FCC-hh machines are different in terms of equipment inside the machine, the whole machine in the arc, the LSS and the experiment will be dismantled. A specific study [324] addressed the dismantling activities foreseen to transition from FCC-ee to FCC-hh. Figure B.6 from this study indicates that it is possible to complete the dismantling of FCC-ee in 2-years. The study also found that it should be possible to dismantle the experiments within this time frame as well.

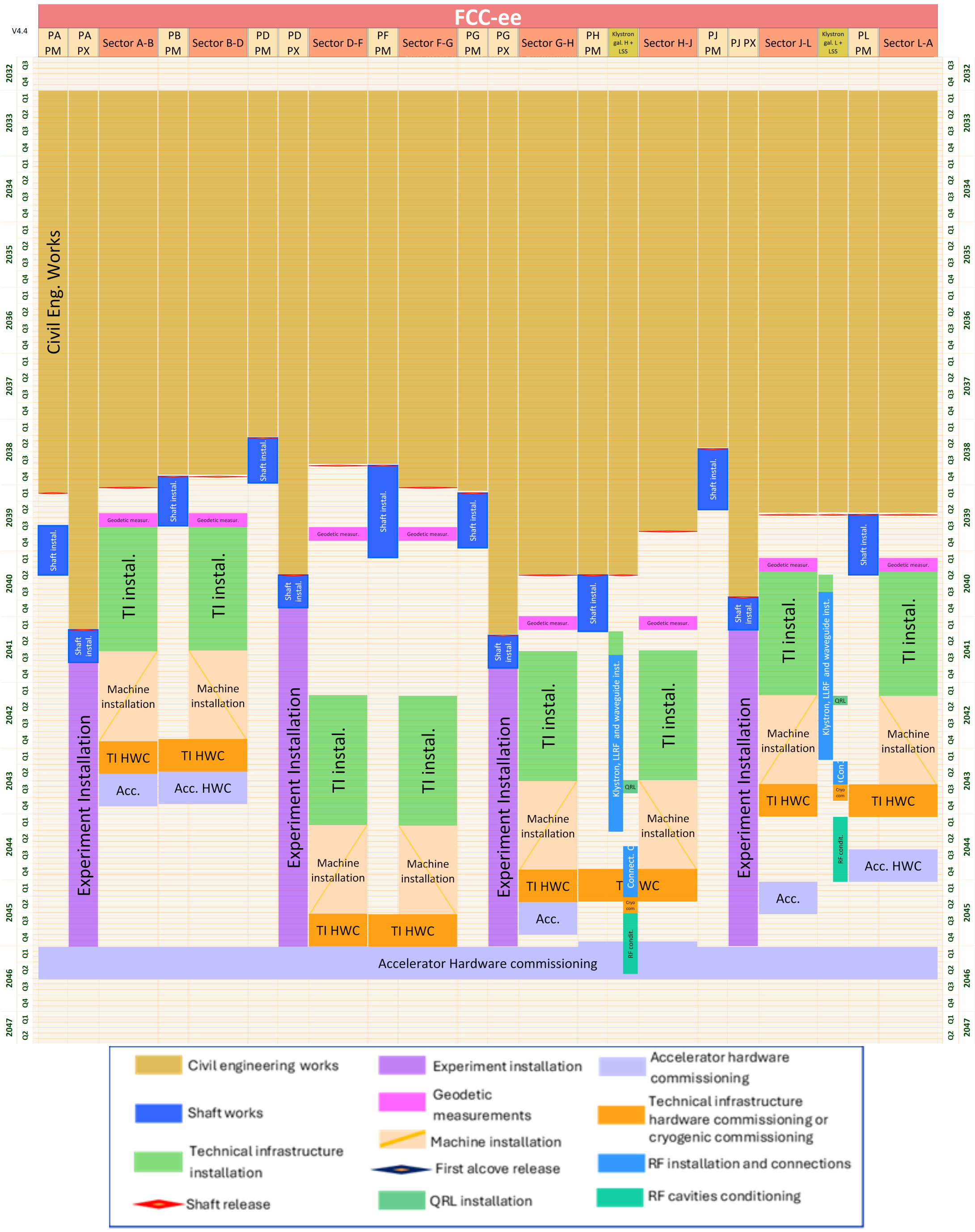

Fig. B.5: Overall linear planning for FCC-ee. The durations of all activities are counted as calendar days. PM is the machine shaft and PX, the experiment shaft.

For the overall planning, a 2-year duration for dismantling was considered independently of the study. In the arc, most of the equipment will be removed from the tunnel: instrumentation, collider and booster magnets, vacuum equipment, cabling for magnets, magnet supports and alignment structures, power converters and related cabling located in the alcoves and the cooling circuit for magnets. In the FCC-ee RF points, point PH and PL, the cavities and QRL will be dismantled.

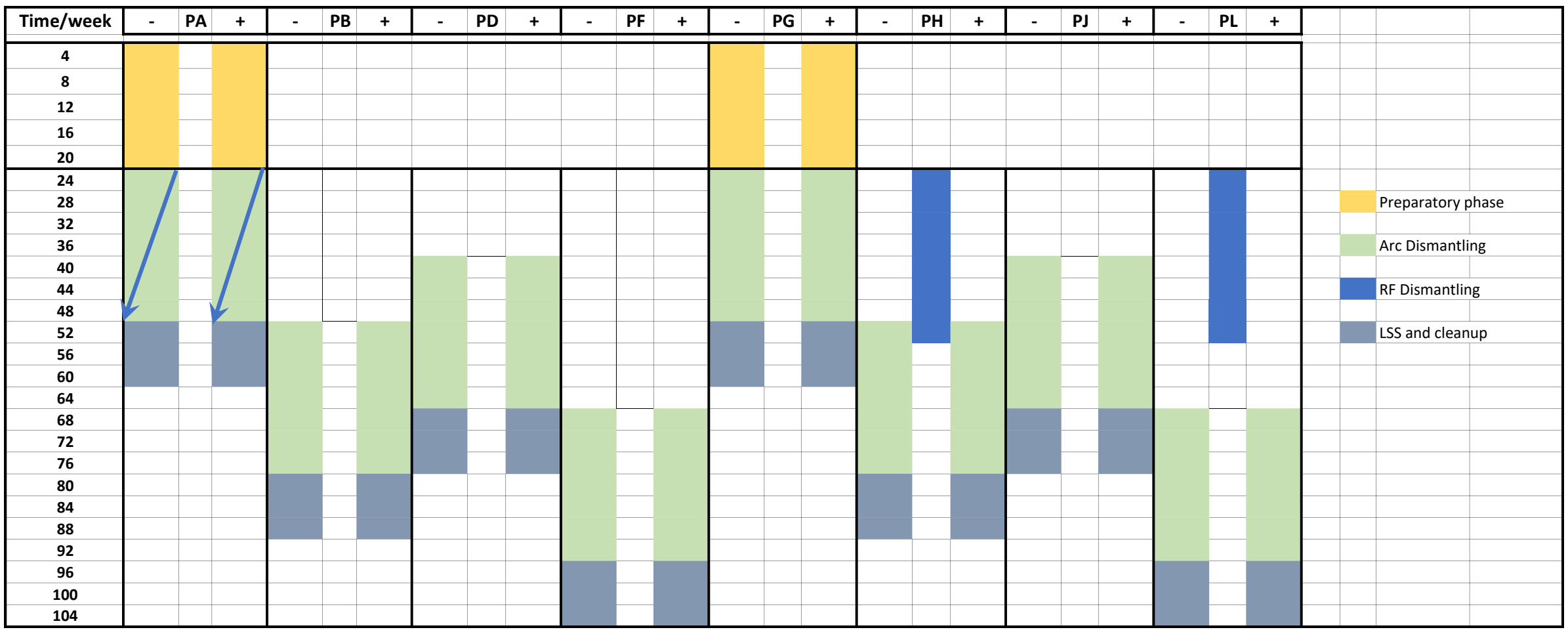

Fig. B.6: Planning for dismantling FCC-ee.

#### *Civil engineering construction*

To support the operation of FCC-hh, additional civil engineering constructions will be undertaken.

- Transfer lines
- Beam dump at point PF

Civil engineering has not yet conducted a detailed study of the durations, but the beam dump at point PF has been identified as the most time-critical. As long as the beam dump at point PF is completed within 3.5 years—starting concurrently with dismantling at the end of 2063—it will not impact the schedule.

### B.1.4 Installation of FCC-hh

#### *Equipment kept and updated from FCC-ee*

The technical infrastructures (ventilation, power, network) from FCC-ee will remain for FCC-hh.

#### *Installation of a sector*

Given the long operational lifetime of FCC-ee, an upgrade of the safety systems will be required. This will be followed by installation of specific cabling and piping for FCC-hh. Next the Quench Recovery Line (QRL) will be installed, to provide cryogenics for the superconducting magnets of the machine. The magnet supports will be installed next, followed by the magnets and their interconnection.

#### *Installation at an RF point*

In point PL, which will be the point where the radiofrequency equipment will be installed, there will also be an upgrade of safety systems, cabling and piping and then QRL installation followed by installation of the cavities.

There will be technical infrastructure hardware commissioning, cryogenic cool down in each sector and in point PL cavity conditioning.

To finish installation activities, each sector will have accelerator hardware commissioning and finally there will be global hardware commissioning for the whole machine.

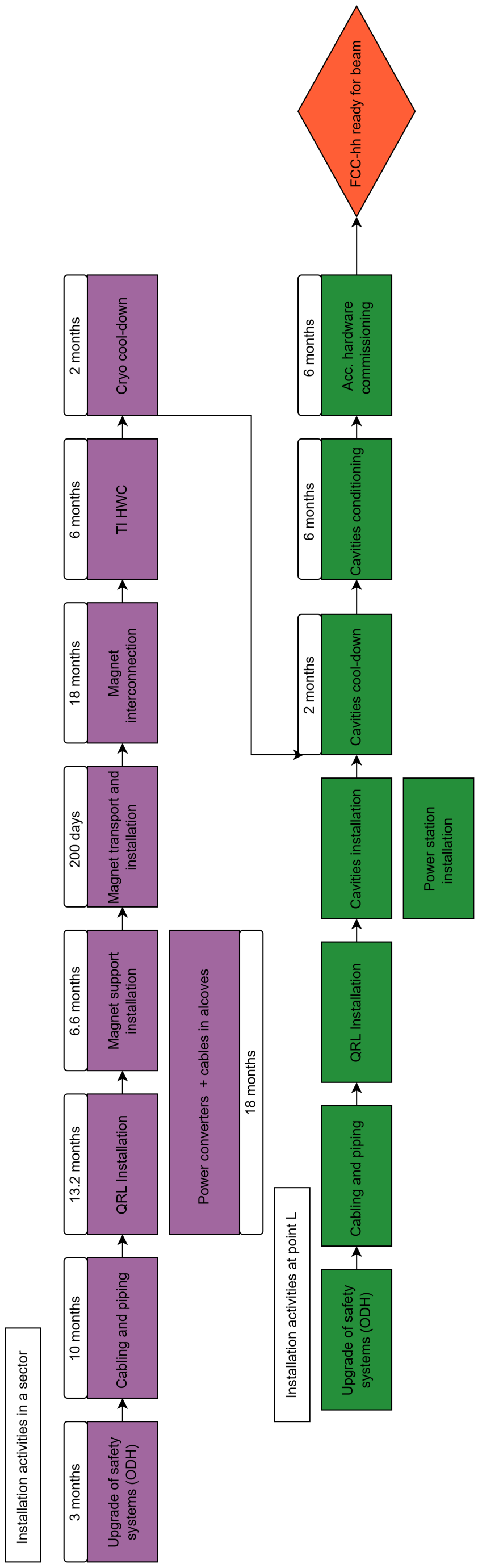

Fig. B.7: Installation sequence for FCC-hh.

***General strategy for installation (priorities concerning civil engineering works)***

Similar to the FCC-ee machine, a maximum of four teams can work on the same type of activity simultaneously. The planning is, therefore, driven by the resource levelling and the shift pattern it implies.

Priorities for resource usage have been determined considering the radiofrequency equipment to be installed in point PL and the beam dump (the most critical construction) in point PF. The first sectors to start installation activities are A-B, B-D, J-L and L-A, allowing point PL to be installed as soon as possible. Then sectors D-F, F-G, G-H and H-J can be installed once the resources are available.

This plan creates some contingency for civil engineering to build the beam dump at point PF and limits its potential impact on the readiness for beam commission.

There are 3.5 years starting from the dismantling activities at the end of 2063 which are idle time as a result of the resource levelling in point PF. These 3.5 years can be used to build the beam dump at point PF without having an impact on the beam commissioning date.

The overall planning is divided into two phases:

- Dismantling activities for 2 years
- Installation activities for 8 years

The transition from FCC-ee to FCC-hh is planned to be completed within 10 years, ensuring a timely and efficient upgrade process.

The FCC-hh would be ready for beam commissioning in **2073**.